%% file: tmdbook.tex
\documentclass[12pt]{article}
\pdfoutput=1
\usepackage[textwidth=6.8in,textheight=9.in]{geometry}
\usepackage{graphicx}
\usepackage{epsfig}
\usepackage{longtable}
\usepackage{imakeidx}
\makeindex[columns=2,  
            title=Index, 
          options= -s index-style.ist,
            intoc]
\usepackage{amsmath,amssymb}
\usepackage{hyperref}
\usepackage{wrapfig}
\usepackage{dsfont} 
\usepackage{multirow}   
\usepackage{color}
\usepackage{verbatim}
\usepackage{rotating}
\usepackage{subfig}
\usepackage{colortbl}
\usepackage[dvipsnames]{xcolor}
\usepackage{url}
\usepackage{bm}
\usepackage{slashed}
\usepackage{placeins}
\usepackage{makecell}
\usepackage{slashed}
\usepackage[normalem]{ulem}
\usepackage{hhline}
\usepackage{tikz}
\usepackage[mathscr]{eucal}  
\usepackage[compat=1.0.0]{tikz-feynman}

\usepackage[font={small}]{caption}

\usepackage{bigstrut}
\usepackage{slashbox}
\usepackage{tabularx}

\usepackage{ wasysym }

\numberwithin{equation}{section}
\numberwithin{figure}{section}
\numberwithin{table}{section}

\input{./macros/macros}

\usepackage{newpxtext,newpxmath}

\usepackage{titlesec,fancyhdr,enumitem}

\titleformat{\section}[hang]{\Huge\bfseries}{\color{NavyBlue}\thesection{\color{NavyBlue}\,-\,}}{0pt}{\color{NavyBlue}\Huge\bfseries}

\titlespacing{\subsection}{0pt}{1ex}{0.5ex}
\titleformat*{\subsection}{\color{NavyBlue}\large\bfseries}
\titlespacing{\subsubsection}{0pt}{1ex}{0.5ex}
\titleformat*{\subsubsection}{\color{NavyBlue}\bfseries}

\usepackage{times,multicol,graphicx}
\usepackage{xcoffins}
\usepackage{times,color}

\begin{document}


\vspace{-.3cm}
\rightline{\small
\begin{minipage}[r]{4.2in}
\noindent{Preprints:} JLAB-THY-23-3780, LA-UR-21-20798, MIT-CTP/5386 \\
\end{minipage}
}

\vspace{-1.2cm}
\hspace{-0.4cm}
\includegraphics[width=1.3in]{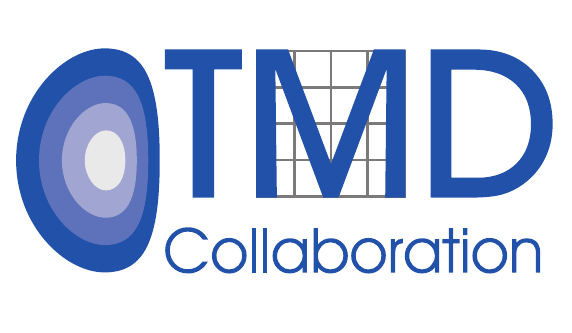} 

\vspace{-.4cm}
\begin{center}
{\LARGE \bf TMD Handbook}
\end{center}
\vspace{-0.1cm}

\input{./authors/authors}

\vspace{-.2cm}
\input{./abstract/abstract}
\thispagestyle{empty}

\newpage

   \NewCoffin \result
  \NewCoffin \name
  \NewCoffin \affiliations
  \NewCoffin \logo
  \NewCoffin \subname
  \NewCoffin \authors
  \NewCoffin \rulei

\SetHorizontalCoffin \result {}
\SetHorizontalCoffin \name {\fontsize{52}{50}\sffamily\bfseries TMD Handbook}

\SetHorizontalCoffin \affiliations {\fontsize{10}{10}\sffamily 
%
April 6, 2023}
                     
\SetVerticalCoffin \subname {440pt} {\raggedleft\fontsize{20}{22}\sffamily 
\color{NavyBlue}
A modern introduction 
to the physics of \\
Transverse Momentum Dependent 
distributions}

\SetVerticalCoffin \logo {400pt} {
\includegraphics[width=8cm]{./tmd-logo/tmd_collaboration.pdf} 
}

\SetVerticalCoffin \authors {140pt}
{\raggedright \fontsize{12}{14}\sffamily
Renaud~Boussarie\\
Matthias~Burkardt \\
Martha~Constantinou \\
William~Detmold\\
Markus~Ebert\\
Michael~Engelhardt\\
Sean~Fleming\\
Leonard~Gamberg\\
Xiangdong~Ji\\
Zhong-Bo~Kang\\
Christopher~Lee\\
Keh-Fei~Liu\\
Simonetta~Liuti\\
Thomas~Mehen $^*$\\
Andreas~Metz\\
John~Negele\\
Daniel~Pitonyak\\
Alexei~Prokudin\\
Jian-Wei~Qiu \\
Abha~Rajan \\
Marc~Schlegel \\
Phiala~Shanahan\\
Peter~Schweitzer\\
Iain~W.~Stewart $^*$\\
Andrey~Tarasov\\
Raju~Venugopalan\\
Ivan~Vitev\\
Feng~Yuan\\
Yong~Zhao \\[8pt]
$*$ - Editors
}

\RotateCoffin \affiliations {270}

\SetHorizontalCoffin \rulei  {\color{NavyBlue}\rule{6.25in}{1pc}}

\JoinCoffins \result                \name 
\JoinCoffins \result[\name-b,\name-l] \subname     [t,l](-24mm,-14mm)
\JoinCoffins \result[\name-t,\name-r] \rulei   [b,r](0mm,4mm)
\JoinCoffins \result[\name-b,\name-r] \rulei     [B,r](0mm,-8mm)
\JoinCoffins \result[\name-B,\name-r] \affiliations     [B,l](12mm,20mm)
\JoinCoffins \result[\name-B,\name-r] \authors     [t,r](10mm,-52mm)
\JoinCoffins \result[\name-b,\name-l] \logo   [t,l](-10mm,-100mm)
\TypesetCoffin \result
\newpage

\clearpage
\tableofcontents
\clearpage

\pagestyle{fancy}
\lhead{TMD Handbook}
\rhead{\thepage}
\cfoot{}

%
\index{factorization|see {TMD factorization}}
\index{rapidity anomalous dimension|see {Collins-Soper evolution kernel}}
\index{twist-3 TMDs|see {subleading power TMDs}}
\index{soft drop|see {groomed jet observable}}
\index{model!parton model|see {parton model}}
\index{soft factor|see {TMD soft function}}
\index{Efremov-Teryaev-Qiu-Sterman matrix element|see {Qiu-Sterman (QS) function}}

\newpage
\input{./sec-introduction/sec-introduction}

\newpage
\input{./sec-definition/sec-definition}

\newpage
\input{./sec-factorization/sec-factorization}

\newpage
\input{./sec-evolution/sec-evolution}

\newpage
\input{./sec-phenomenology/sec-phenomenology}

\newpage  
\input{./sec-latticeQCD/sec-latticeQCD}

\newpage
\input{./sec-models/sec-models}

\newpage
\input{./sec-smallx/sec-smallx}

\newpage
\input{./sec-jets/sec-jets}

\newpage
\input{./sec-twist3/sec-twist3}


\newpage
\input{./sec-gtmds/sec-gtmds}

\newpage
\input{./sec-outlook/sec-outlook}

\newpage
\input{./sec-acknowledgement/sec-acknowledgement}

\newpage
\appendix
\input{./sec-appendix/app-glossary}
\input{./sec-appendix/app-Feynman-rules}

\input{./sec-appendix/app-Fourier}
\input{./sec-appendix/app-definitions}
\input{./sec-appendix/app-RGkernels}

\input{./sec-appendix/acronyms}

\input{./sec-appendix/chaptercontacts}

\newpage
\printindex

\newpage  
\bibliography{tmd} 
\end{document}

%% file: macros/macros.tex
\usepackage{mathtools}
\usepackage{hhline}
\DeclareRobustCommand{\eq}[1]{Eq.~\eqref{eq:#1}}
\DeclareRobustCommand{\eqs}[2]{Eqs.~\eqref{eq:#1} and \eqref{eq:#2}}
\DeclareRobustCommand{\fig}[1]{Fig.~\ref{fig:#1}}

\DeclareRobustCommand{\app}[1]{appendix~\ref{app:#1}}
\DeclareRobustCommand{\sec}[1]{Sec.~\ref{sec:#1}}
\DeclareRobustCommand{\secs}[2]{Secs.~\ref{sec:#1} and \ref{sec:#2}}
\DeclareRobustCommand{\chap}[1]{Chapter~\ref{sec:#1}}
\DeclareRobustCommand{\chaps}[2]{Chapters~\ref{sec:#1} and \ref{sec:#2}}

\newcommand{\nslash}{\kern 0.2 em n\kern -0.50em /}
\newcommand{\kslash}{\kern 0.2 em k\kern -0.45em /}
\newcommand{\pslash}{\kern 0.2 em p\kern -0.50em /}
\newcommand{\Sslash}{\kern 0.2 em S\kern -0.50em /}
\newcommand{\Pslash}{\kern 0.2 em P\kern -0.50em /}
\newcommand{\Dslash}{\kern 0.2 em D\kern -0.65em /\kern 0.15em}

\newcommand{\nn}{\nonumber}
\newcommand{\eps}{\epsilon}

\newcommand{\nocontentsline}[3]{}
\newcommand{\tocless}[2]{\bgroup\let\addcontentsline=\nocontentsline#1{#2}\egroup}

\newcommand{\MSbar}{\overline{\mathrm{MS}}}
\newcommand{\LQCD}{\Lambda_\mathrm{QCD}}
\newcommand{\lqcd}{\Lambda_\mathrm{QCD}}
\newcommand{\df}{\mathrm{d}}
\newcommand{\cO}{\mathcal{O}}
\newcommand{\cT}{\mathcal{T}}
\newcommand{\bn}{{\bar{n}}}
\newcommand{\GammaC}{\Gamma_{\rm cusp}}
\newcommand{\as}{\alpha_s}
\newcommand{\unsub}{{\rm (u)}}
\newcommand{\subt}{{\rm subt}}
\newcommand{\img}{i}
\newcommand{\diff}[1]{\mathrm{d}#1}

\newcommand\TMD{\mathrm{TMD}}
\newcommand{\bt}{\mathbf{b}_T}
\newcommand{\bts}{\mathbf{b}_{*}}
\newcommand{\kt}{\mathbf{k}_T}
\newcommand{\lt}{\mathbf{l}_T}
\newcommand{\pt}{\mathbf{p}_T}

\newcommand{\qt}{\mathbf{q}_T}
\newcommand{\rt}{\mathbf{r}_T}
\newcommand{\deltat}{\mathbf{\Delta}_T}

\newcommand{\ays}[2]{{\rm AY}^{#2}(\Tsc{q},Q_{#1})}

\newcommand{\bef}{\begin{figure}[htb]\centering}
\newcommand{\eef}{\end{figure}}
\newcommand{\vc}[1]{\boldsymbol{#1}}
\newcommand{\mcdot}{\!\cdot\!}

\newcommand{\Ma}{M}
\newcommand{\Mb}{M_h}
\newcommand{\cP}{\mathcal{P}}
\newcommand{\cB}{\mathcal{B}}
\newcommand{\cG}{\mathcal{G}}
\newcommand{\cF}{\mathcal{F}}
\newcommand{\cH}{\mathcal{H}}
\newcommand{\Pa}{P_N}            
\newcommand{\Pb}{P_h}            
\newcommand{\gtilde}[1]{\tilde{\widetilde{#1}}}
\newcommand{\w}{\omega}
\DeclareMathOperator*{\SumInt}{%
\mathchoice%
  {\ooalign{$\displaystyle\sum$\cr\hidewidth$\displaystyle\int$\hidewidth\cr}}
  {\ooalign{\raisebox{.14\height}{\scalebox{.7}{$\textstyle\sum$}}\cr\hidewidth$\textstyle\int$\hidewidth\cr}}
  {\ooalign{\raisebox{.2\height}{\scalebox{.6}{$\scriptstyle\sum$}}\cr$\scriptstyle\int$\cr}}
  {\ooalign{\raisebox{.2\height}{\scalebox{.6}{$\scriptstyle\sum$}}\cr$\scriptstyle\int$\cr}}
}
\newcommand{\na}{n_a}
\newcommand{\nb}{n_b}

\newcommand{\T}[1]{\boldsymbol{#1}_{\text{T}}}
\newcommand{\Tsc}[1]{#1_{\text{T}}}

\definecolor{dpmagenta}{rgb}{0.8, 0.0, 0.8}

\newcommand{\no}{\nonumber \\}
\newcommand{\bmax}{b_{\rm max}}

\newcommand{\bi}{\begin{itemize}}
\newcommand{\ei}{\end{itemize}}

\newcommand{\tf}{\tilde{f}}

\newcommand{\beq}{\begin{equation}}
\newcommand{\eeq}{\end{equation}}

\newcommand{\bsc}{b_T}

\newcommand\bstarsc{b_*}

\newcommand\bone{b_c}
\newcommand\mubstar{\mu_{\bstarsc}}
\newcommand\muQ{\mu_Q}

\newcommand{\appor}[1]{{\rm T}_{\rm #1}}

\newcommand{\bT}{{{b}_T}}
\newcommand{\cs}[2]{ \Gamma_{#2}(\Tsc{q},Q_{#1}) }

\newcommand{\fixo}[2]{{\rm FO}^{#2}(\Tsc{q},Q_{#1})}
\renewcommand{\T}[1]{\boldsymbol{#1}_{\text{T}}}
\newcommand{\TT}[2]{W^{#2}(\Tsc{q},Q_{#1})}
\newcommand{\TTnew}[2]{W_{\rm New}^{#2}(\Tsc{q},Q_{#1};\eta,C_5)}
\newcommand{\TTa}[2]{W_{\rm a}^{#2}(\Tsc{q},Q_{#1};\eta,C_5)}

\newcommand{\YY}[2]{Y^{#2}(\Tsc{q},Q_{#1})}

\newcommand{\arrowcom}[1]{\textcolor{red}{\textbf{$\Longrightarrow$ #1}} \\}

\def\??#1{\mbox{}\\\arrowcom{#1}}

\newcommand{\Del}{\bm{\nabla}}
\newcommand{\plus}{\!+\!}
\newcommand{\minus}{\!-\!}
\DeclareMathOperator{\Imag}{Im}

\def\T{_{_T}}

\newcommand{\ot}{\leftarrow}

\newcommand{\nnu}{\nonumber\\}
\newcommand{\bea}{\begin{eqnarray}}
\newcommand{\eea}{\end{eqnarray}}
\newcommand{\be}{\begin{equation}}
\newcommand{\ee}{\end{equation}}
\newcommand{\ba}{\begin{eqnarray}}
\newcommand{\ea}{\end{eqnarray}}
\newcommand{\la}{\langle}
\newcommand{\ra}{\rangle}

\newcommand{\vmod}{ {v} }

\newcommand{\uncomment}[1]{{}}

\newcommand{\vect}[1]{\ensuremath{{\bm{#1}}}}

\def\bfkperp{{\bm k}_T}
\def\bfpperp{{\bm p}_T}
\def\bfhp{\hat{\bm h}}
\def\bfPhperp{{\bm P}_{hT}}
\def\Phperp{P_{hT}}

\def\kperp{k_T}
\def\pperp{p_T}
\def\avkperp{\la \kperp^2 \ra}

\newcommand{\mh}{M_h }

\newcommand{\hzeta}{{\tilde\zeta}}

\def\vkTpi{\kTpi}
\def\vkTN{\kTN}

\def\kTpi{{\bm p}_{T\pi}}
\def\kTN{{\bm p}_{Tp}}
\def\qT{\bm q_T^{ } } 

\def\hhat{\hat{\bm h}}

\newcommand{\xbj}{\ensuremath{x}}
\newcommand{\zh}{\ensuremath{z_{h}}}

\def\bea#1\eea{\begin{align}#1\end{align}} 

\def\avk{\langle k_\perp ^2\rangle}
\def\avp{\langle p_\perp ^2\rangle}
\def\avPT{\langle P_T^2\rangle}

\newcommand{\MAe}[3]{\Bigl\langle#1\Bigr\rvert#2\Bigr\rvert#3\Bigr\rangle}

\newcommand\newhalfstaple[2][1]
{
 \begin{tikzpicture}[scale=#1]
  \draw [line width=0.25mm] (0,0) -- (0.2,0) -- (0.2,#2*0.2);
 \end{tikzpicture}
}

\newcommand\halfstaple[1][1] { \newhalfstaple[#1]{-1} }     
\newcommand\halfstapleu[1][1]{ \newhalfstaple[#1]{+1} }     

\newcommand\newsoftstaple[2][1]
{
 \begin{tikzpicture}[scale=#1]
  \coordinate (dx) at (#2*0.15, 0);
  \coordinate (dy) at (0   , 0.1);
  \coordinate (bT) at (0.  , 0.05);
  \draw [line width=0.15mm] (0,0) -- ($(dx) + (dy)$) -- ($2*(dy)$) -- ($2*(dy)+(bT)$) -- ($(dx) + (dy) + (bT)$) -- ($(bT)$) -- (0,0);
  \draw [line width=0.12mm] (0,0) -- (0, -0.0055);
 \end{tikzpicture}
}

\newcommand\leftsoftstaple[1][1] { \newsoftstaple[#1]{-1} }     
\newcommand\rightsoftstaple[1][1]{ \newsoftstaple[#1]{+1} }     
\newcommand\softstaple[1][1]     { \rightsoftstaple[#1]   }     

\newcommand{\cS}{\mathscr{S}}
\newcommand{\cW}{\mathscr{W}}

\newcommand{\sub}[1]{ \begin{subequations} #1 \end{subequations} }

%% file: authors/authors.tex

\begin{center}

{\small 
Renaud~Boussarie$^{1}$,
Matthias~Burkardt$^{2}$,
Martha~Constantinou$^{3}$, 
William~Detmold$^{4}$,
Markus~Ebert$^{4,5}$,
Michael~Engelhardt$^{2}$,
Sean~Fleming$^{6}$,
Leonard~Gamberg$^{7}$,
Xiangdong~Ji$^{8}$,
Zhong-Bo~Kang$^{9}$,
Christopher~Lee$^{10}$,
Keh-Fei~Liu$^{11}$,
Simonetta~Liuti$^{12}$,
Thomas~Mehen$^{13}$,
Andreas~Metz$^{3}$,
John~Negele$^{4}$,
Daniel~Pitonyak$^{14}$,
Alexei~Prokudin$^{7,16}$,
Jian-Wei~Qiu$^{16,17}$,
Abha Rajan$^{12,18}$,
Marc~Schlegel$^{2,19}$, 
Phiala~Shanahan$^{4}$,
Peter~Schweitzer$^{20}$,
Iain~W.~Stewart$^{4}$,
Andrey~Tarasov$^{21,22}$,
Raju~Venugopalan$^{18}$,
Ivan~Vitev$^{10}$,
Feng~Yuan$^{23}$,
Yong~Zhao$^{24,4,18}$
}

\vspace{0.2in}
{\it \footnotesize
~$^{1}$CPHT, CNRS, Ecole Polytechnique, Institut Polytechnique de Paris, 91128 Palaiseau, France\\
~$^{2}$Department of Physics, New Mexico State University, Las Cruces, NM~88003 \\
~$^{3}$Department of Physics,  Temple University,  Philadelphia,  PA 19122 - 1801,  USA\\
~$^{4}$Center for Theoretical Physics, Massachusetts Institute of Technology, Cambridge, MA~02139\\
~$^{5}$Max Planck Institut f\"ur Physik, F\"ohringer Ring 6, 80805 Munich, Germany\\
~$^{6}$Department of Physics, University of Arizona, Tucson, AZ 85721\\
~$^{7}$Division of Science, Penn State University Berks,
Reading, PA~19610\\
~$^{8}$Maryland Center for Fundamental Physics, University of Maryland, College Park, 20742, USA\\
~$^{9}$Department of Physics and Astronomy, University of California, Los Angeles, CA 90095\\
~$^{10}$Theoretical Division, Los Alamos National Laboratory, Los Alamos, NM~87545 \\
~$^{11}$Department of Physics and Astronomy, University of Kentucky, Lexington, KY 40506\\
~$^{12}$Department of Physics, University of Virginia, Charlottesville, VA 22904\\
~$^{13}$Department of Physics, Duke University, Durham, NC 27708\\
~$^{14}$Department of Physics, Lebanon Valley College, Annville, Pennsylvania 17003 \\
~$^{15}$Department of Physics, Old Dominion University, Norfolk, VA 23529 \\
~$^{16}$Theory Center, Jefferson Lab,
Newport News, Virginia 23606\\
~$^{17}$Department of Physics, William \& Mary, Williamsburg, Virginia 23187\\

~$^{18}$Physics Department, Brookhaven National Laboratory, Upton, NY 11973\\

~$^{19}$Institute for Theoretical Physics, T\"{u}bingen University, 72076 T\"{u}bingen, Germany\\
~$^{20}$Department of Physics, University of Connecticut, 
  Storrs, CT 06269 \\
~$^{21}$Department of Physics, North Carolina State University, Raleigh, NC 27607 \\
~$^{22}$Joint BNL-SBU Center for Frontiers in Nuclear Science at Stony Brook University, Stony Brook, NY 11794 \\

~$^{23}$Nuclear Science Division, Lawrence Berkeley National Laboratory, Berkeley, CA 94720\\
~$^{24}$Physics Division, Argonne National Laboratory, Lemont, IL 60439\\
}

\end{center}

%% file: abstract/abstract.tex
\begin{center}
    Abstract
\end{center}

This handbook provides a comprehensive review of transverse-momentum-dependent parton distribution functions and fragmentation functions, commonly referred to as transverse momentum distributions (TMDs).  TMDs describe the distribution of partons inside the proton and other hadrons with respect to both their longitudinal and transverse momenta. They provide unique insight into the internal momentum and spin structure of hadrons, and are a key ingredient in the description of many collider physics cross sections. Understanding TMDs requires a combination of theoretical techniques from quantum field theory, nonperturbative calculations using lattice QCD, and phenomenological analysis of experimental data. The handbook covers a wide range of topics, from theoretical foundations to experimental analyses, as well as recent developments and future directions. It is intended to provide an essential reference for researchers and graduate students interested in understanding the structure of hadrons and the dynamics of partons in high energy collisions.

%% file: sec-introduction/sec-introduction.tex
\section{Introduction}
\label{sec:Intro}

Nucleons are the fundamental building blocks of all atomic nuclei and make up essentially all the visible matter in the universe, including the stars, the planets, and us. However, the nucleon is not static but has a complex and dynamic internal structure,
the details of which are only beginning to be revealed in modern experiments. A deeper understanding of this building block of matter therefore requires that we understand the nucleon's internal structure in terms of its constituents. 

In Quantum Chromodynamics (QCD), the theory of the strong interactions, the nucleon emerges as a strongly interacting, relativistic bound state of quarks and gluons (referred to collectively as partons). Fifty years of experimental investigations into the nucleon's internal structure have provided remarkable insight into the dynamics of quarks and gluons~\cite{Brambilla:2014jmp}. However, many outstanding questions remain. This is largely because of the {\it color confinement} -- a fundamental property of QCD. Even the most advanced detector can not see quarks and gluons in isolation as they are forever bound inside nucleons, or in general, in all hadrons.  It is an unprecedented intellectual challenge to probe the quark-gluon dynamics, quantify the quark-gluon structure of hadrons, and study the emergence of hadrons from quarks and gluons, given that we can not see quarks and gluons directly by any modern tools.  

On the other hand, QCD has another equally important and fundamental property, {\it asymptotic freedom} -- the strong force is weakly coupled if it is probed at a sufficiently short-distance~\cite{Gross:1973id,Politzer:1973fx}.  It is asymptotic freedom that makes it possible for us to develop the powerful theoretical formalism, known as QCD factorization~\cite{Collins:1989gx}, that links the quarks and gluons at sub-femtometer scales to the hadrons measured by modern detectors in high energy experiments with a set of well-defined and fundamental distribution functions that encode rich information on nucleon's internal structure.  It is the universality of these distribution functions and the precision that we can achieve in determining them from known experimental data that ensures the predictive power of QCD and allows us to explore and study the dynamics of quarks and gluons and the structure of 
the nucleon by  
performing experiments in which the nucleon receives
large momentum transfers, even though we never ``see" quarks and gluons directly.  QCD along with its factorization formalism has been extremely successful in interpreting all available data from high energy scatterings when probing distances less than 0.1~fermi (fm) (or equivalently, with larger than 2~GeV momentum transfer in the collision), which has provided us the confidence and the tools to discover the Higgs particle at the LHC and to explore the new physics beyond QCD and the Standard Model in high energy hadronic collisions.  

Interestingly, many typical hard probes for distance scales
 less than 0.1~fm 
in high energy scattering are not very sensitive to the confined spatial landscape and motion of quarks and gluons inside the bound nucleon with a radius about one fermi. 
For this reason, the results of generations of experiments have only provided one-dimensional snapshots of the longitudinal momentum distributions of quarks and gluons inside a colliding nucleon, utilizing the well-developed collinear QCD factorization formalism to encode them in universal parton distribution functions (PDFs).
In recent years we have begun to capture more detailed information about nucleon structure
due to our ability to precisely measure new types of observables in high energy scattering with two distinctive momentum scales:  one hard scale with a large momentum transfer to pin down the particle nature of quarks and gluons along with one soft scale to be sensitive to the confined motion and the spatial landscape of the quarks and gluons inside the nucleon. Most importantly, theoretical advances over the past decade have resulted in the development of a powerful  transverse momentum dependent (TMD) QCD factorization formalism that  
enables
us to extract the 3-dimensional (3D) motion of quarks and gluons inside a colliding nucleon. Information from these new and precise data enables the determination of universal transverse momentum dependent parton distributions (or simply, TMDs).  With additional data from experiments around the world and the future Electron-Ion Collider (EIC)~\cite{Accardi:2012qut}, a much sharper and detailed picture of the nucleon's internal landscape will  become available to shed light on the 
dynamics of confined quarks and gluons that form the nucleon
- the building block of our visible world.

\begin{wrapfigure}{r}{0.35\textwidth}
\vspace{-0.35in}
\begin{center}
\includegraphics[width=0.33\textwidth]{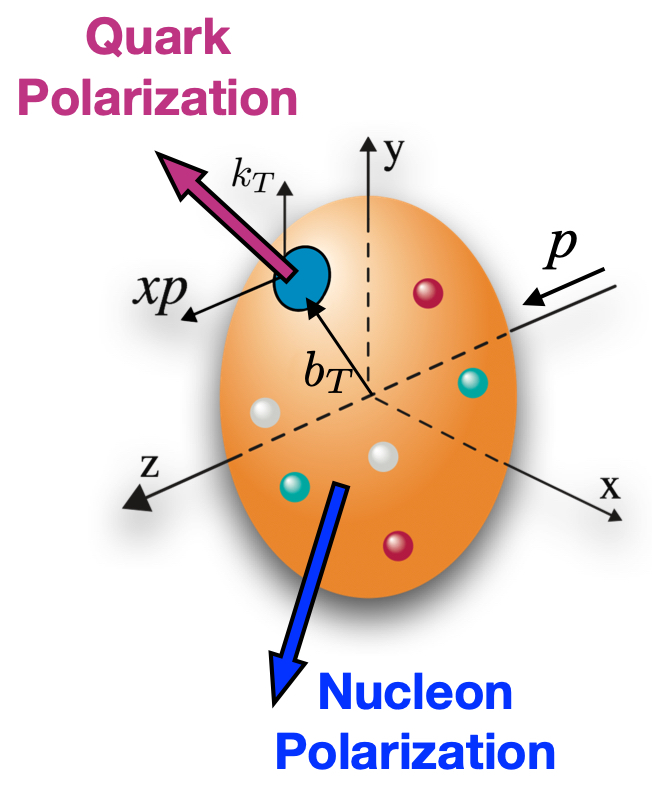}
\end{center}
\vspace{-0.2in}
 \caption{Illustration of the momentum and spin variables probed by TMD parton distributions.}
\vspace{-0.3in}
 \label{fig:TMDPDFcartoon}
\end{wrapfigure}

In this TMD Handbook, we provide a modern introduction to the physics of transverse momentum dependent distributions -- the TMDs, which encode the quantum correlations between hadron polarization and the motion and polarization of quarks and gluons inside it, as sketched in Fig.~\ref{fig:TMDPDFcartoon}.  We cover the precise definition of these fundamental and universal TMDs and their properties, the TMD factorization formalisms to match these quantum distributions to physical observables measured in high energy scattering experiments, and phenomenological approaches for extracting them from precise experimental data.  We introduce new advances in ab initio lattice QCD (LQCD) calculations, as well as various model calculations of the TMDs. We discuss what 
we can
learn from the TMDs to understand better how the dynamics of QCD determines the properties of the nucleon and its internal landscape.  This TMD handbook is a project of the TMD Collaboration -- a Topical Collaboration in Nuclear Theory for the Coordinated Theoretical Approach to Transverse Momentum Dependent Hadron Structure in QCD [\href{https://sites.google.com/a/lbl.gov/tmdwiki/}{\textcolor{blue} {https://sites.google.com/a/lbl.gov/tmdwiki/}}], supported by the Office of Nuclear Physics of the U.S. Department of Energy.

In the rest of this 
Chapter,
we provide an intuitive introduction to the TMDs and their role in describing the hadron's internal structure, and the role of lattice QCD for calculating these intrinsically nonperturbative but fundamental distribution functions.  
An outline for the material in the remaining chapters of this handbook can be found in \sec{outline}.

\subsection{Hadrons, Partons and QCD}

\begin{figure}[t]
\begin{center}
\vspace{-0.08in}
  \includegraphics[width=0.27\textwidth]{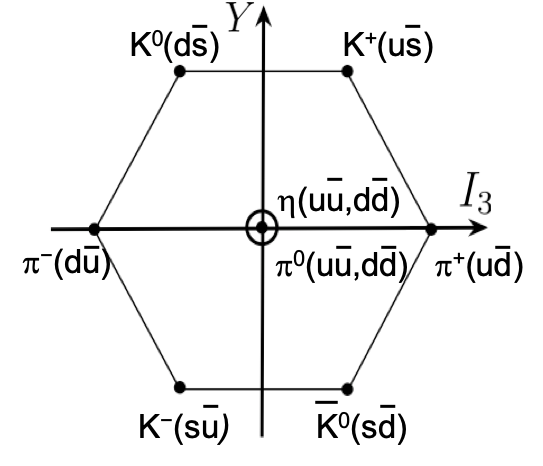}
  \ \
  \includegraphics[width=0.29\textwidth]{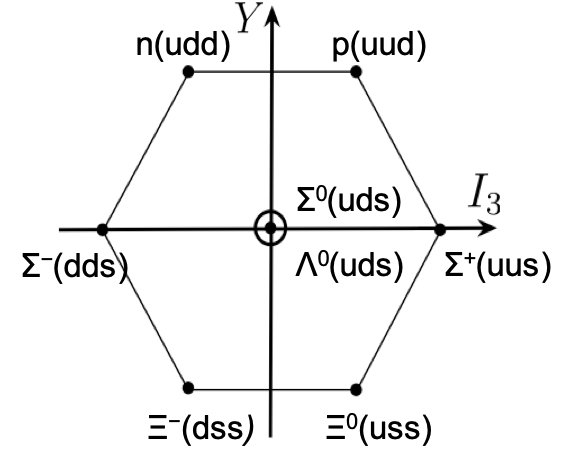}
  \ \
  \includegraphics[width=0.32\textwidth]{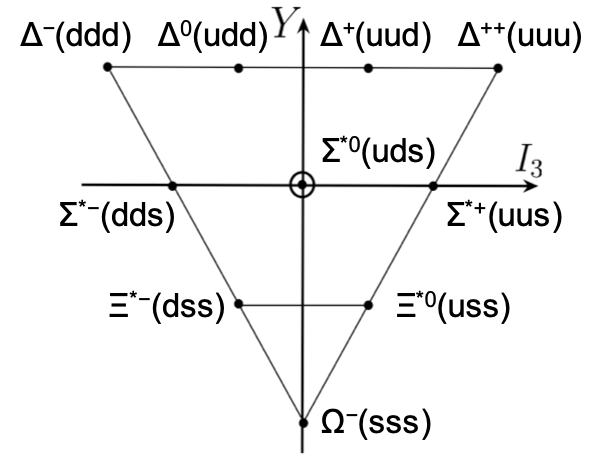} 
  \\
{\hskip -0.038\textwidth} (a) {\hskip 0.255\textwidth} (b)  {\hskip 0.29\textwidth} (c) 
\end{center}
\vspace{-0.1in}
 \caption{The Quark Model of hadrons in the eight-fold way. (a) mesons (flavor 8 representation); (b) baryons (flavor 8 representation); and (c) baryons (flavor 10 representation).}
 \label{fig:QM}
\end{figure}

The discovery of the neutron by Chadwick in 1932 heralded the strong interactions, as a force much stronger than the electromagnetic Coulomb repulsion between the protons in a nucleus was needed to keep atomic nuclei together. In the twenty years following this discovery tremendous progress was made in understanding the interactions between two nucleons, however particle physics was still rather simple with the only additions being the pions ($\pi$) as expected from Yukawa theory and the muons ($\mu$) (``Who ordered that?"~\cite{muon-Times}). In the next decade this simple state of affairs was blown apart by the proliferate  discovery of different particles, which led to the development of the eight-fold way by Gell-Mann and Ne`eman that put light hadrons and mesons into multiplets of flavor $SU(3)$, as shown in Fig.~\ref{fig:QM}. The eight-fold way was given deeper meaning by Gell-Mann and Zweig with the introduction of the quark model (QM) of hadrons: mesons were bound states of quark-antiquark pairs and baryons bound states of three quarks. 

Quarks with spin-$\frac{1}{2}$ and fractional charges were not taken seriously as fundamental particles at that  
time, but rather were
regarded as a convenient bookkeeping device.  Taking the quarks as real degrees of freedom requires 
understanding
how the $\Delta^{++}(1232)$ could be a low-lying state. In the 
original QM, this particle is a bound state of three ``up'' quarks, {\tt uuu} as shown in Fig.~\ref{fig:QM}(c). If it were presumed to be the product of three S-wave quarks to make it the lowest lying state, its wave function would be 
symmetric under the interchange of any two quarks, which is not allowed for fermions. The puzzle was resolved after the color of quarks was introduced. In 1964 O.\ W.\ Greenberg proposed adding to the quarks a new quantum number
called color that could take on  $N_c$ values, where  $N_c$ is the number of colors. Choosing  $N_c$ = 3 made
the $\Delta^{++}(1232)$ wave function antisymmetric under interchange of two quarks with different colors as required by fermi statistics. Thus was born the seed of QCD.

\begin{figure}[t]
\begin{center}
\includegraphics[width=0.16\textwidth]{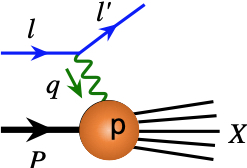}  
  {\hskip 0.03\textwidth} 
\includegraphics[width=0.16\textwidth]{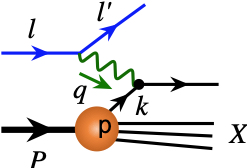} 
  {\hskip 0.05\textwidth} 
\includegraphics[width=0.56\textwidth]{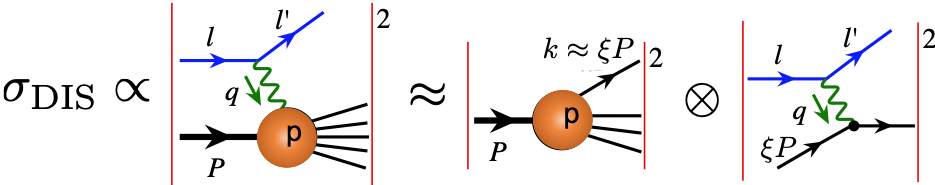}
\\  {\hskip -0.13\textwidth}
  {(a)} {\hskip 0.15\textwidth} {(b)} {\hskip 0.405\textwidth} {(c)} 
  \vspace{-0.05in}
  \caption{
  Inelastic electron-proton DIS with one-photon exchange (a), via a single point-like particle (b), and the DIS cross section in the parton model (c). 
  }
  \vspace{-0.2in}
\label{fig:dis-pm}
\end{center}
\end{figure}

The discovery that the nucleon is composed of spin-$\frac{1}{2}$ point-like particles from the experimental measurements of inclusive electron-proton deeply inelastic scattering (DIS) cross section, $e(l)+p(P) \to e(l')+X$, performed at SLAC over 50 years ago, confirmed the existence of quarks~\cite{Bloom:1969kc}.  By measuring the scattered lepton momentum $l'$ to define the momentum transfer of the collision, $q = l-l'$, as sketched in Fig.~\ref{fig:dis-pm}(a), and keeping $Q \equiv \sqrt{-q^2} \gg 1/R$, where $R$ is the proton radius, DIS \index{deeply inelastic scattering (DIS)} experiments provided a short-distance electromagnetic probe for the charged point-like particles inside the proton. 
With the momentum transfer, 
$Q\gg 1/R$, 
and the effective size of the hard collision 
$\sim {\mathcal O} (1/Q) \ll R$, 
pulling two or more point-like particles 
out from
the same hard collision would be penalized by  powers of 
$1/(QR)$,
and therefore, the DIS cross section is dominated by the scattering off a single point-like particle, as indicated in Fig.~\ref{fig:dis-pm}(b).  Furthermore, the time of the hard collision 
$\sim 1/Q$ 
is much shorter than the characteristic time scale $R$ for the dynamics inside the proton, implying that interactions inside the proton are effectively frozen during the hard collision.  Although the inclusive DIS cross section 
$E'\df\sigma_{ep\to e' X}/\df^3l'$
is 
invariant under boosts along the collision axis,
the scattering is
best pictured in the infinite momentum frame in which the proton moves very fast and all point-like particles inside it move in parallel. The momentum of the active point-like particle is $k\approx \xi P \sim Q$ with a momentum fraction $\xi$, while its  typical transverse momentum 
$k_T \sim 1/R \ll Q$.
Let $f_{i/p}(\xi)$ be the probability distribution density to find a type ``$i$'' point-like particle inside the fast moving proton carrying momentum fraction $\xi$, then the DIS cross section on the proton can be approximated as sketched in Fig.~\ref{fig:dis-pm}(c), and expressed as
\begin{align}
E'\frac{\df\sigma_{ep\to e'X}}{\df^3l'} 
& \approx
\sum_i \int d\xi\, f_{i/p}(\xi)\, E'\frac{d\hat{\sigma}_{ei\to e'X}}{d^3l'} 
=  \sum_i e_i^2 \left\{
\frac{2\alpha^2}{Q^2 s}\left[\frac{1+(1-y)^2}{y^2}\right] \right\}
 f_{i/p}(x)
\nonumber\\
&\equiv \sum_i f_{i/p} \otimes  \hat{\sigma}_i \,.
\label{e:dis-pm}
\end{align}
Here $\alpha=e^2/4\pi$
is the electromagnetic fine structure constant, $\sum_i$ sums over all possible types of point-like particles weighted by the fractional charge squared $e_i^2$, $s=(l+P)^2\simeq 2P\cdot l$ is the 
center of mass energy squared,
$y=P\cdot q/P\cdot l$ 
is the fractional energy loss of the electron in the proton rest frame
, and 
$x=Q^2/2P\cdot q$ 
is the Bjorken variable. The $\hat{\sigma}_i$ in the abstract notation of Eq.~(\ref{e:dis-pm}) represents the partonic cross section, $E'd\hat{\sigma}_{ei\to e'X}/d^3l'$ and $\otimes$ refers to the convolution over momentum fraction $\xi$.  
Eq.~(\ref{e:dis-pm}) is Feynman's parton model 
formula \cite{Feynman:1969ej} for the inclusive DIS cross section, where  
$f_{i/p}(\xi=x)$ 
are the parton distribution functions (PDFs), and the partons are spin-$\frac{1}{2}$ point-like particles, which are now identified as quarks. The \index{parton model} 
parton model
formula in Eq.~(\ref{e:dis-pm}) shows that other than the lepton-parton scattering, $E'd\hat{\sigma}_{ei\to e'X}/d^3l'$, the rest of DIS cross section on the proton is inpendent of $Q^2$ and depends only on 
$x$. This phenomenon is 
known as Bjorken scaling, and was verified by the early SLAC experiments. 

\begin{figure}[t]
\begin{center}
\includegraphics[width=0.9\textwidth]{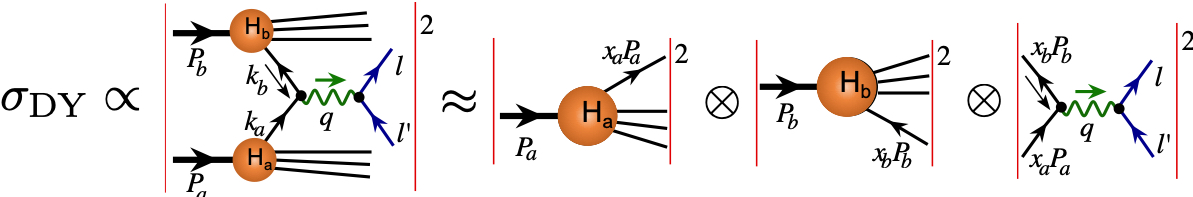} 
  \vspace{-0.05in}
  \caption{
The Drell-Yan cross section in the parton model. 
}
  \vspace{-0.2in}
\label{fig:dy-pm}
\end{center}
\end{figure}

The success of the 
parton model
formula in Eq.~(\ref{e:dis-pm}) to interpret the DIS data verifies the existence of spin-$\frac{1}{2}$ point-like particles inside the proton and provides a way to extract the PDFs.  However, it does not provide an independent test of this parton picture.  The inclusive production of massive lepton pairs in hadron-hadron collisions, $H_a(P_a)+H_b(P_b) \to \gamma^{*}(q)[\to l\overline{l}(q)]+X$, known as the Drell-Yan process \cite{Drell:1970wh}, provided the much needed test.  With the invariant mass of the lepton pair $Q^2\equiv q^2 \gg 1/R^2$, the inclusive cross section should be dominated by the probability to find a quark in one hadron and an antiquark in another hadron convolved with the annihilation of the quark-antiquark pair into the observed lepton pair, as sketched in Fig.~\ref{fig:dy-pm}, 
\begin{align}
\frac{\df \sigma_{H_a+H_b\to l\bar{l}+X}}{\df Q^2 \df Y}
&\approx \sum_{i,j} f_{i/H_a} \otimes f_{j/H_b} \otimes \hat{\sigma}_{ij}
= \frac{4\pi\alpha^2}{3N_cQ^2 s} 
\sum_{i}\, e_i^2\,
f_{i/H_a}(x_a)\, f_{\bar{i}/H_b}(x_b)
\,.
\label{e:dy-pm}
\end{align}
Here 
$Y=\frac{1}{2}\ln(x_a/x_b)$ 
is the rapidity of the lepton pair, 
$\sum_i$ sums over all possible types of quark and antiquark,
the parton momentum fractions are given by 
$x_a = Q\, e^{Y}/\sqrt{s}$ and 
$x_b = Q\, e^{-Y}/\sqrt{s}$ 
with total center of mass collision energy squared ${s} = (P_a+P_b)^2$, and 
$1/N_c=1/3$ is the color factor, which was missing in the original Drell-Yan formula 
since it predated QCD.  With PDFs extracted from the DIS measurements, 
the Drell-Yan formula in Eq.~(\ref{e:dy-pm}) has effectively no free parameter 
for predicting the massive lepton pair production in hadronic collisions including 
its dependence on $Q^2$, $Y$ and collision energy $\sqrt{s}$.
However, a somewhat large 
$K_{\rm factor}=\sigma_{\rm Exp}/\sigma_{\rm Thy}\sim 2$ was found~\cite{Berger:1982vs},
which indicates that the normalization of the predicted cross section is off 
by roughly a factor of 2.  

A better microscopic picture of the strong interactions emerged in the form of QCD, and the quarks are held together by the strong color force via gluons.  As a quantum field theory, the QCD Lagrange density is constructed from two types of particle fields: spin-$\frac{1}{2}$ Dirac fields (quarks), $\psi^i_f$ with color $i=1,2,3=N_c$ and flavor $f=u,d,s,c,b,t$ 
for up, down, strange, charm, bottom, and top quarks, respectively,
and massless spin-1 vector fields (gluons), $A_{\mu}^a$ with color $a=1,2,\dots,8=N_c^2-1$, with SU(3) local color gauge symmetry,
\begin{equation}
{\cal L}_{QCD}(\psi_f, A_{\mu}) 
= -\frac{1}{4} G_{\mu\nu,a}^2[A] +
\sum_f \overline{\psi}_f \left(iD_{\mu}[A]\gamma^{\mu} - m_f \right) \psi_f
\label{e:qcd}
\,.
\end{equation}
Here the gluon field strength 
$G_{\mu\nu,a}[A] = \partial_{\mu}A_{\nu,a} -
  \partial_{\nu}A_{\mu,a}-g\,f_{abc}\, A_{\mu,b}\, A_{\nu,c} $,
the covariant derivative 
$D_{\mu}[A] = \partial_{\mu}+ig\, A_{\mu,a}\, t_a $,
the generator $t_a$ and structure constant $f_{abc}$ define the SU(3) color 
\begin{wrapfigure}{r}{0.5\textwidth}
\begin{center}
\vspace{-0.26in}
\includegraphics[width=0.14\textwidth]{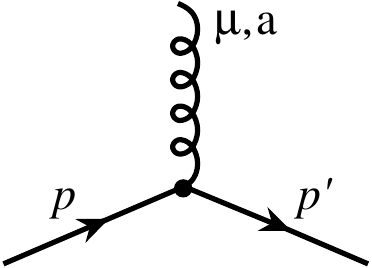} 
  {\hskip 0.1in}
\includegraphics[width=0.14\textwidth]{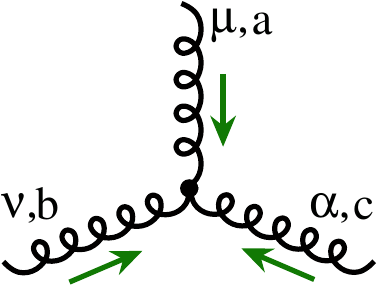} 
 {\hskip 0.1in}
\includegraphics[width=0.14\textwidth]{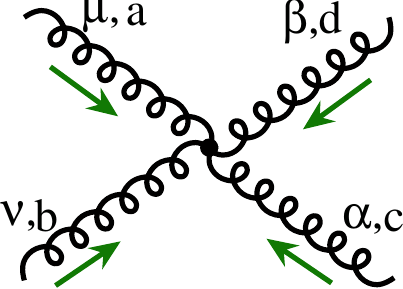}  
 \\
 {\footnotesize {\hskip 0.005\textwidth} (a) {\hskip 0.13\textwidth} (b) {\hskip 0.13\textwidth} (c)}
  \vspace{-0.05in}
  \caption{
  QCD interactions: quark-gluon (a), three-gluon (b) and four-gluon (c).
  }
\label{fig:feynman-rule}
\end{center}
\vspace{-0.3in}
\end{wrapfigure} 
algebra, $\left[t_a,\, t_b\right] = i f_{abc}\, t_c $, and 
$g$ is the strong coupling constant.  Symbolically, the elementary interaction between quark (solid-line) and gluon (curly-line) can be represented by the Feynman diagram in Fig.~\ref{fig:feynman-rule}(a), and three-gluon and four-gluon interactions in Fig.~\ref{fig:feynman-rule}(b) and (c), respectively, with corresponding Feynman rules given in the Appendix~\cite{Collins:2011zzd}. 
QCD is a renormalizable quantum field theory and its effective interaction strength is characterized by a running coupling $\alpha_s(\mu) = g^2(\mu)/(4\pi)$, 
depending on the renormalization scale $\mu$ which corresponds with the scale at which the interaction was probed.  
At the lowest order, $\alpha_s(\mu) = 
4\pi/\beta_0\ln(\mu^2/\Lambda_{\rm QCD}^2)$ 
with $\beta_0=\frac{11}{3}C_A - \frac{4}{3}T_F\, n_f$, where $C_A=N_c$, $T_F=\frac{1}{2}$, $n_f$ is number of active quark flavors, and the fundamental strong interaction scale $\Lambda_{\rm QCD} \sim 1/R$ depends on $n_f$.  Unlike the electromagnetic force, $\alpha_s(\mu)$ decreases as the $\mu$ increases, that is, the strong interaction becomes weaker at a shorter distance or with a larger momentum transfer. This property is called  
asymptotic freedom, and makes it possible to perform controlled perturbative evaluation
of strong interaction dynamics at short distances using QCD perturbation theory.

However, the nucleon (in general, all hadrons) has internal strong dynamics taking place at the characteristic scale of $1/R\sim 200$~MeV~$\sim\Lambda_{\rm QCD}$. At these scales the strong coupling  is so large that QCD perturbation theory is not applicable.  That is, any high energy scattering cross section with identified hadron(s), even with a large momentum transfer $Q$, can not be fully calculated in QCD perturbation theory. In the 
parton model,
as shown in Eqs.~(\ref{e:dis-pm}) and (\ref{e:dy-pm}), the leading nonperturbative information of the hadron is embedded into the universal PDFs.  
Using QCD, we can identify the parton constituents $i$ that are probed by $f_{i/p}$ as the quarks, antiquarks and gluons, 
and define PDFs in terms of matrix elements of gauge invariant quark or gluon operators (see Eqs.~(\ref{eq:barepdf}) and (\ref{eq:renpdf}) below).
It is QCD factorization that allows us to consistently separate physics taking place at different momentum scales, to organize the leading process-{\it independent} nonperturbative contributions to high energy scattering cross sections with identified hadron(s) into universal distribution functions, such as PDFs, combined with perturbatively calculable short-distance partonic cross sections $\hat{\sigma}(Q)$, and to justify that all process-{\it dependent} nonperturbative information can be neglected as power suppressed corrections.  On the other hand, physically observed cross sections should not depend on 
the precise scale at which short and long distance contributions are factorized from each other, 
which leads naturally to renormalization group improved QCD factorization and the scale-dependence of the factorized universal distribution functions, including PDFs and TMDs (as discussed in Sec.~\ref{sec:evolution} of this handbook). That is, like the strength of strong interaction, $\alpha_s$, the PDFs, and in general, the factorized universal distribution functions depend on the scale at which they are probed. 
The renormalization group improved QCD factorization formalism allows us to calculate the partonic cross section $\hat{\sigma}(Q)$ order-by-order in perturbation theory to improve its precision, to extract the nonperturbative, but, universal PDFs from experimental data, to predict the scale dependence of the PDFs, and to test QCD dynamics and factorization by verifying the universality of these PDFs.  
As discussed in \chaps{TMDdefn}{Factorization} of this handbook, the parton
model formulas in Eqs.~(\ref{e:dis-pm}) and (\ref{e:dy-pm}) can be derived from QCD as the leading power contribution in both $\alpha_s$ and $1/Q$ to the corresponding QCD processes. Furthermore, corrections from higher orders in $\alpha_s$ can be systematically included.
The nonperturbative PDFs can also be studied and extracted from lattice QCD calculations of hadron matrix elements of quark and gluon correlators, as discussed in the Sec.~\ref{subsec.intro-lattice}, 
complementary to what can be extracted from experimental data.

\subsection{Structure of the Nucleon}

As a relativistic bound state of quarks and gluons, the nucleon's internal structure can not be described by any kind of ``still picture'' that is often used to describe atomic or molecular structure.  As shown in Eqs.~(\ref{e:dis-pm}) and (\ref{e:dy-pm}), the leading structure information of the nucleon, which we can extract from experimental data of high energy scattering with a large momentum transfer $Q$, is embedded in the PDFs as the probability densities 
to find a collinear quark, or antiquark, or gluon carrying a momentum fraction $\xi$ inside a colliding nucleon. This is pictured in Fig.~\ref{fig:dis-pm} or Fig.~\ref{fig:dy-pm}.  
In QCD, the nucleon's rich internal structure can be described by its matrix elements of gauge-invariant partonic operators composed of quark and gluon fields with various spin projections, similar to the definition of quark distribution in Eqs.~(\ref{eq:barepdf}) and (\ref{eq:renpdf}). 
Although none of these matrix elements are direct physical observables, owing to the fact that no quark and gluon can be seen in isolation, the QCD factorization formalism does link these matrix elements to physically measured cross sections with power suppressed and controllable approximations.

\begin{figure}[t!]
\begin{center}
  \includegraphics[width=0.42\textwidth]{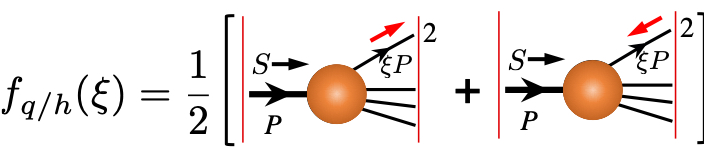}
  {\hskip 0.08\textwidth}
  \includegraphics[width=0.45\textwidth]{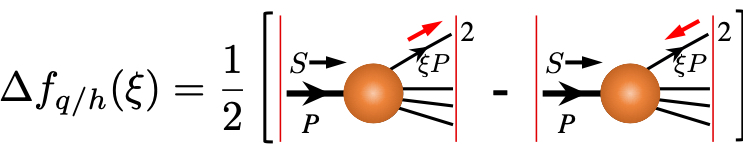} 
  \\
  \includegraphics[width=0.45\textwidth]{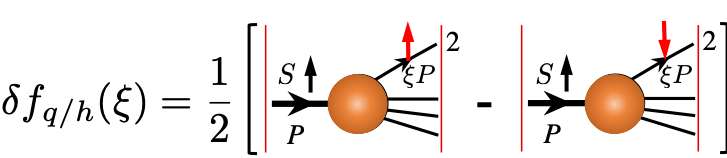}
\end{center}
\vspace{-0.2in}
\caption{Graphic representation of leading order collinear quark distributions of hadron $h$: 
$f_{q/h}(\xi)$ (unpolarized),
$\Delta f_{q/h}(\xi)$ (longitudinally polarized) and
$\delta f_{q/h}(\xi)$ (transversely polarized).  The red and black arrow represent spin direction of the quark and hadron, respectively.  See the text for more details.
 }
 \label{fig:quark-pdfs}
\end{figure}

\index{transversity!introduction|(}
When a spin-$1/2$ quark of longitudinal momentum fraction $\xi$ is probed in a scattering experiment with polarized nucleons, we can access 
four possible quark polarization states, which are often referred as unpolarized, longitudinally polarized and two transversely polarized states.  With a spin-$\frac{1}{2}$ nucleon, we can define the unpolarized collinear quark distribution of an unpolarized nucleon $f_{q/h}(\xi)$, the collinear quark helicity distribution of a longitudinally polarized nucleon $\Delta f_{q/h}(\xi)$, and the collinear transversity distribution of a transversely polarized nucleon with the direction of quark transverse polarization parallel to the direction of nucleon's polarization 
$\delta f_{q/h}(\xi)$ (equivalent to the notations $h_1(\xi)$ or $\delta q (\xi)$ 
that are also used in the literature)
as sketched in Fig.~\ref{fig:quark-pdfs}.  Similarly, we have three collinear antiquark distributions for the nucleon.  However, the nucleon has only two collinear gluon distributions: the unpolarized collinear gluon distribution of a unpolarized nucleon $f_{g/h}(\xi)$ and the collinear gluon helicity distribution of a longitudinally polarized nucleon $\Delta f_{g/h}(\xi)$.  Transverse spin-flip of a spin-$\frac{1}{2}$ nucleon can not generate a two-unit spin-flip needed to define the collinear gluon transversity of the nucleon, and therefore the nucleon does not have a gluon transversity distribution.  Unpolarized collinear PDFs have been well-determined from experimental data on QCD factorizable high energy scattering observables with large momentum transfers \cite{Constantinou:2020hdm}.  Helicity distributions and transversity distributions are expected to be better determined from future precision data from Jefferson Lab and the EIC.
\index{transversity!introduction|)}

With a large momentum transfer $Q\gg 1/R$, 
the scattering takes place at such a short time $\sim 1/Q$ that the hard probe
is not very sensitive to the physics at the scale of $\Lambda_{QCD}\sim 1/R$, 
including the active parton's confined transverse motion ($k_T$) in momentum space and its spatial distribution ($b_T$) in position space, as shown in Fig.~\ref{fig:TMDPDFcartoon}. 
On the other hand, both the confined motion and the spatial distribution of quarks and gluons inside a bound nucleon 
are important part of the nucleon's 3D internal structure, which is 
an immediate consequence of QCD dynamics.  
To probe such 3D internal structure, 
we need a new type of {\it two-scale} ``hard-probes'', which are physical observables with a large momentum transfer $Q_1\gg \Lambda_{\rm QCD} \sim 1/R$ to localize the probe to see the particle nature of quarks and gluons, while they also have an additional well-measured soft momentum scale, $Q_2 \ll Q_1$, so that they are much more sensitive to the details of hadron's internal structure. 
For example, as described in \chap{TMDdefn}, the Drell-Yan cross section is an ideal two-scale observable if we measure the differential cross section as a function of the pair's transverse momentum $q_T = |{\bf q}_T|$ in addition to measuring the invariant mass of the lepton pair, $Q$, 
in particular because
the production rates are dominated by the region where $q_T \ll Q$. 
In terms of the parton model picture in Fig.~\ref{fig:dy-pm}, the pair's transverse momentum is a sum of the transverse momentum of the active quark and antiquark, 
$\qt = {\bf k}_{aT}+{\bf k}_{bT}$.  
The parton model formula in Eq.~(\ref{e:dy-pm}) is then modified as,
\begin{align}
\frac{\df \sigma_{H_a+H_b\to l\bar{l}+X}}{\df Q^2 \df Y \df^2\qt}
&= \frac{4\pi\alpha^2}{3N_cQ^2 s} 
\sum_{i}\, e_i^2\,
\int \df^2{\bf k}_{aT}\, \df^2{\bf k}_{bT}\, \delta^{(2)}(\qt - {\bf k}_{aT} - {\bf k}_{bT})
\nonumber \\
& {\hskip 1.4in} \times
f_{1({i/H_a})}(x_a,{\bf k}_{aT})\,  f_{1({\bar{i}/H_b})}(x_b,{\bf k}_{bT})
\nonumber\\
&=  \hat{\sigma}_{q\bar{q}\to l\bar{l}}\;
\otimes\,
f_1\, 
\widetilde{\otimes}\,
f_1
\,.
\label{e:dyqt-pm}
\end{align}
where $f_{1}(x_a,{\bf k}_{aT})$ is the TMD version of the collinear quark distribution $f(\xi=x_a)$, and 
$\widetilde{\otimes}$ represents the convolution of both longitudinal momentum fraction and transverse momenta of the active quark and antiquark, different from $\otimes$ that represents only the convolution of longitudinal momentum fraction as in Eq.~(\ref{e:dy-pm}).
The 
transverse momenta, which are expected to be much smaller than the hard scale $Q$, are neglected in evaluating the hard part $\hat{\sigma}_{q\bar{q}\to l\bar{l}}$.

That is, the $q_T$-distribution of Drell-Yan cross section is directly sensitive to the transverse momentum of the active partons, and a good probe for the TMD PDFs (or simply, TMDs).  

\begin{figure}[pt]
 \centering
  \includegraphics[width=0.85\textwidth]{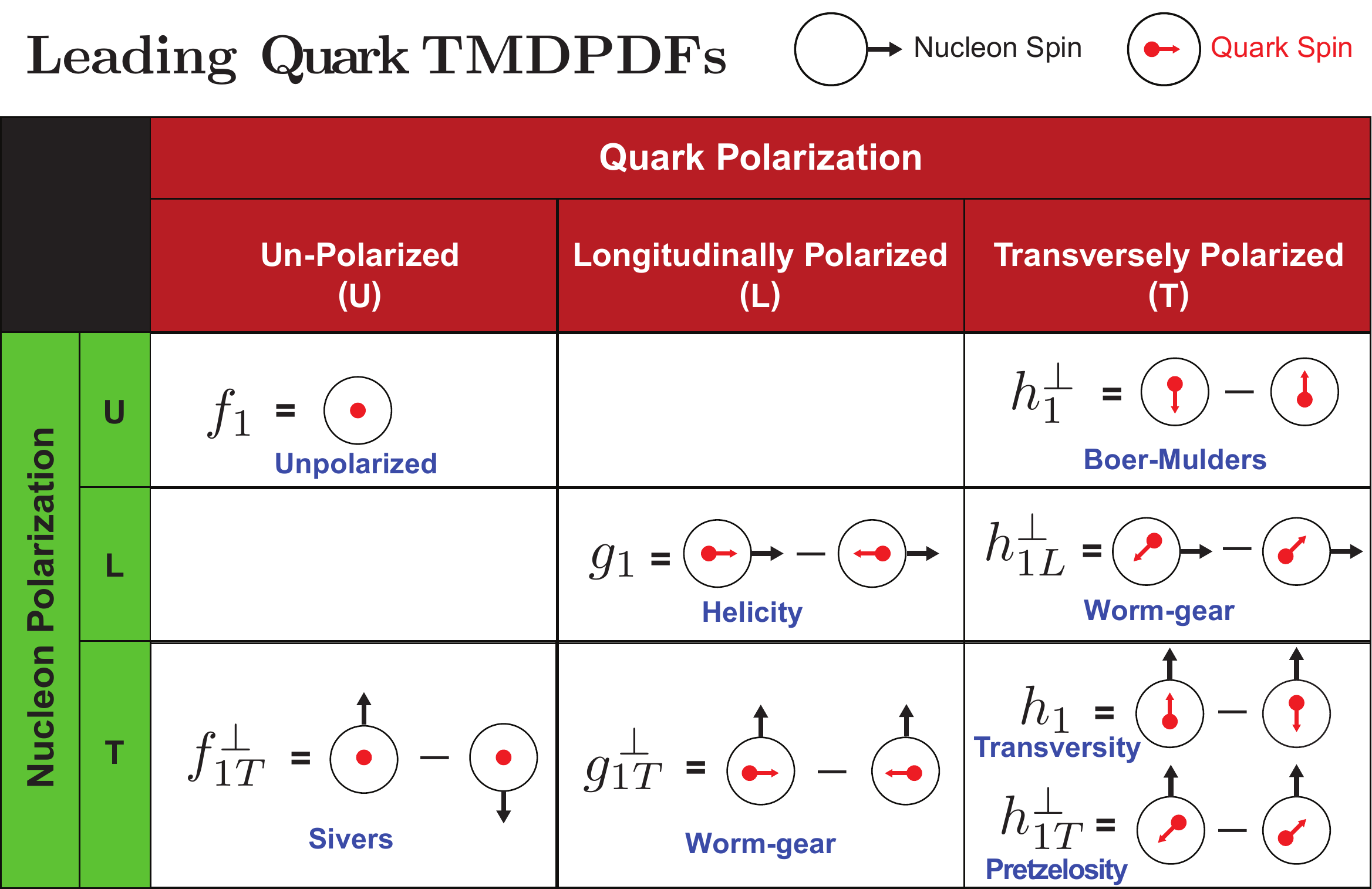}
 \caption{Leading power spin dependent quark TMDPDFs.
 The red dot and black circle represent the quark and nucleon, while the red and black arrow represent their spin direction, respectively. 
}
 \label{fig:qTMDPDFsLP}
\end{figure}

Like the collinear PDFs, the TMDs are distribution densities to find a quark or a gluon carrying a longitudinal momentum fraction $\xi$ and transverse momentum ${\bf k}_T$ inside a colliding nucleon.  The detailed definitions of TMDs in QCD will be given in \chap{TMDdefn}.  
With the dependence on the active parton's transverse momentum, the TMDs carry much more information on hadron structure than what longitudinal PDFs can provide.   The TMDs provide the leading information on quantum correlation between nucleon's spin and active parton's polarization as well as its motion. Instead of three independent quark PDFs in Fig.~\ref{fig:quark-pdfs}, we will have eight independent and non-vanishing quark TMDs because of the quark's transverse motion, as summarized in Fig.~\ref{fig:qTMDPDFsLP}. Here the TMDs are organized in terms of the correlation between polarization states of quark and nucleon: unpolarized ($U$), longitudinally polarized (L), and transversely polarized (T).  Similarly, we have TMDs for gluons and antiquarks, introduced in \chap{TMDdefn}.

Like the unpolarized quark PDF, $f_{q/h}(\xi)$ in Fig.~\ref{fig:quark-pdfs}, its TMD version, $f_1(\xi,{\bf k}_T)$ in Fig.~\ref{fig:qTMDPDFsLP} represents the probability density to find an unpolarized quark carrying collinear momentum fraction $\xi$ and transverse momentum ${\bf k}_T$ inside an unpolarized nucleon.
On the other hand, some TMDs have no correspondence to collinear PDFs.  For example, the Sivers function $f_{1T}^{\perp}$ represents a quantum correlation between the transverse spin direction of the nucleon and the strength and direction of transverse motion of a unpolarized active quark, as well as its flavor dependence.  Another interesting TMD, with no collinear correspondence, is the pretzelosity $h_{1T}^{\perp}$ that
\index{pretzelosity $h_{1T}^{\perp}$}
represents how the correlation of nucleon spin and quark spin can influence the quark's transverse motion, and approximately, its moment 
in a model dependent way
can be interpreted as quark orbital angular momentum contribution to the proton's spin~\cite{She:2009jq,Avakian:2010br,Efremov:2010cy}, see 
Sec.~\ref{subsec:pretzel-and-orbital-angular-momentum}.

In summary, TMDs are fundamental distributions which carry novel information about the nucleon's internal momentum and spin structure, beyond what is known from high precision determinations of the classic PDFs.

\subsection{Matching Cross Section to the Structure}

The TMDs, and in general any parton distributions or correlation functions, can not be directly measured in physical experiments since we can not directly detect quarks and gluons in isolation.  We need QCD factorization formulas to relate TMDs to physical observables, such as cross sections or spin asymmetries defined in terms of ratios of polarized and unpolarized cross sections.  
Like the parton model formula for inclusive Drell-Yan cross section in Eq.~(\ref{e:dy-pm}), sketched in Fig.~\ref{fig:dy-pm},
we have an extended parton model factorization formula 
in Eq.~(\ref{e:dyqt-pm})
to express the differential Drell-Yan cross section, $\df\sigma/\df^4q$, in terms of TMDs when $q_T\ll Q$.  
A similar and more rigorous QCD factorization formula for the differential Drell-Yan cross section will be introduced in \chap{TMDdefn}.

\begin{figure}[thb]
\begin{center}
  \includegraphics[width=0.9\textwidth]{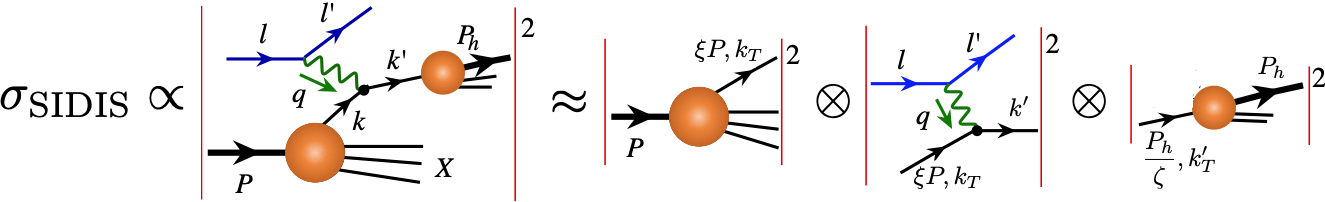} 
  \vspace{-0.05in}
  \caption{The semi-inclusive DIS cross section in the parton model. 
  }
  \vspace{-0.2in}
\label{fig:sidis-pm}
\end{center}
\end{figure}

However, with the Drell-Yan process alone in Eq.~(\ref{e:dyqt-pm}), it is impossible to extract and disentangle various quark TMDs listed in Fig.~\ref{fig:qTMDPDFsLP}, not to mention the antiquark and gluon TMDs.  We need more well-defined and factorizable two-scale observables 
to be able to probe all TMDs.  By detecting a hadron (or jet) of momentum $P_h$ in the final state of electron-proton DIS in addition to the scattered electron,
as sketched in Fig.~\ref{fig:sidis-pm}, this semi-inclusive DIS (SIDIS) process provides more well-defined two-scale observables, where the hard scale $Q\gg \Lambda_{\rm QCD}$ and the soft scale is the transverse momentum of the observed final-state hadron $P_{hT}$ in the photon-hadron frame where the exchanged virtual photon and the colliding hadron define the $z$-axis.
In this virtual photon-hadron frame, the produced leading hadron in the most events of SIDIS is very likely to go in the direction opposite to the colliding hadron and to have a very small $P_{hT}$.  So that, the $P_{hT}$-distribution of lepton-hadron SIDIS is another natural {\it two-scale} observable.  
In particular, it forms an important part of the physics program at a future electron-ion collider~\cite{Accardi:2012qut}, where it will be fully explored.

In the parton model picture, the lepton-proton SIDIS cross section can be factorized, as sketched in Fig.~\ref{fig:sidis-pm}, when 
$Q\gg P_{hT}/z_h \gtrsim \Lambda_{\rm QCD}$ with $\zh = P\cdot P_h / P \cdot q$,  
\begin{align}
E'E_h\frac{\df\sigma_{ep\to e' h X}}{\df^3l' \df^3P_h}
&\approx 
\hat{\sigma}_{eq\to e' q'}
\otimes f_1\, 
\widetilde{\otimes}\, D_{h/q'} ,
\label{e:sidis-pm}
\end{align}
where $f_1(\xi,k_T)$ is defined in Fig.~\ref{fig:qTMDPDFsLP} and $D_{h/q}(\zeta,{\bf k}'_T)$ is a new type of TMDs, which are quark TMD fragmentation functions (FFs) for a quark of flavor $q$ to hadronize into an observed hadron $h$ carrying the momentum fraction $\zeta$ of the fragmenting quark momentum, as sketched in Fig.~\ref{fig:sidis-pm}. 
Like the quark TMD PDFs in Fig.~\ref{fig:qTMDPDFsLP}, the quark TMD FFs are also organized in terms of the correlation between the polarization and transverse momentum of the fragmenting quark and the properties of the hadron observed in the final-state, as summarized in Fig.~\ref{fig:TMDFFsLP}.  With well-defined antiquark and gluon TMD FFs, these distributions provide the fundamental information on how hadrons with measured transverse momentum emerge from energetic quarks and gluons.  
In Eq.~(\ref{e:sidis-pm}), $\widetilde{\otimes}$ represents the convolution of both momentum fraction $\xi$ and $k_T$ or $\zeta$ and $k'_T$ in case of FFs. Like the Drell-Yan $q_T$ distribution, the lepton-proton SIDIS cross section can also be factorized in QCD in terms of TMDs \cite{Ji:2004wu}. More detailed formulae for the SIDIS factorization theorem are given in \chap{TMDdefn}, and more detailed arguments for this factorization are given in Chapter~\ref{sec:Factorization}.

\begin{figure}[t!] 
\begin{center}
\includegraphics[width=1.0\textwidth]{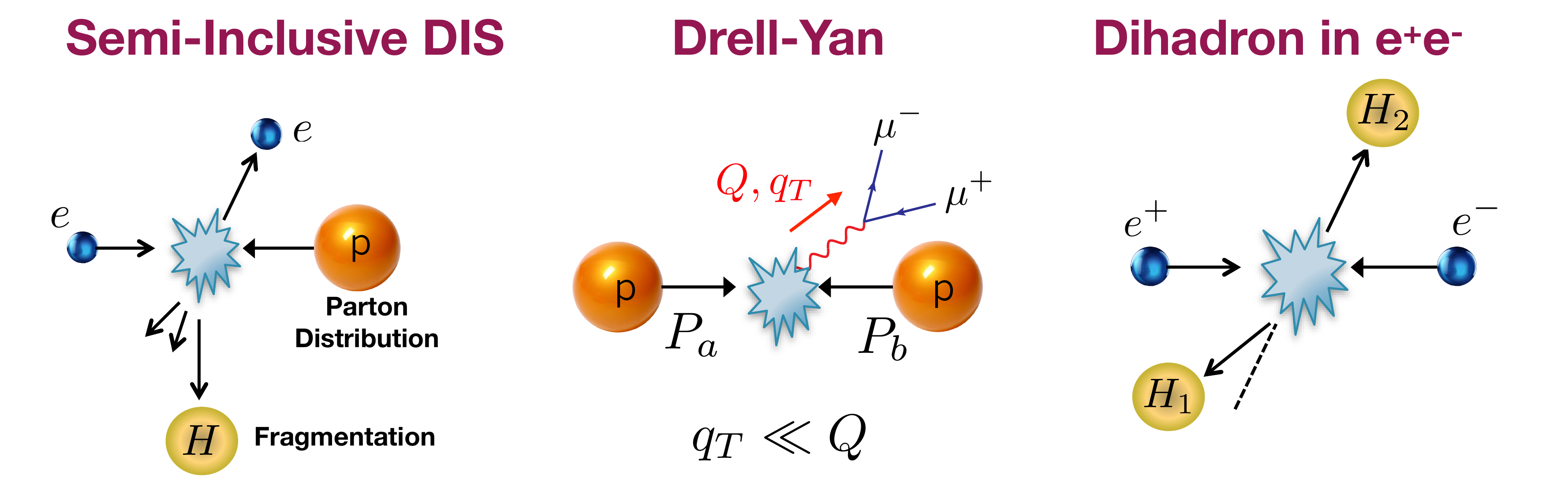} 
\end{center}
\vspace{-0.1in}
 \caption{Schematic illustration of TMD parton distributions and fragmentation functions which appear in key cross sections.
}
 \label{fig:TMDcrosssect}
\end{figure}

With SIDIS, we obtain a new type of two-scale observables that are sensitive to the TMD PDFs in addition to the Drell-Yan process.  However, SIDIS also introduces the capability to probe a new type of TMD physics encoded in the universal TMD FFs.  In order to extract these TMDs from experimental data, clearly, we need to identify more factorizable two-scale observables that are sensitive to the same TMDs, including both TMD PDFs and TMD FFs.  Another natual two-scale observable is the di-hadron production in $e^+e^- $ collisions: $e^+ + e^- \to H_1(P_1) + H_2(P_2) +X$.  In the region where the two produced high momentum hadrons are almost back-to-back, the momentum imbalance, $\bar{p}\equiv |\vec{P}_1 + \vec{P}_2| \ll |\vec{P}_1 - \vec{P}_2|/2 \equiv \overline{P} $, defines a soft momentum scale $\bar p$ together with a hard scale $\overline{P}$. A TMD factorization theorem can then be derived to express the di-hadron production in this region in terms of two TMD FFs together with a perturbatively calculable hard part~\cite{Collins:1981va}
\begin{align}
E_1E_2\frac{\df\sigma_{e^+e^-\to H_1 H_2 X}}{\df^3P_1 \df^3P_2}
&\approx \sum_{i,j}
\hat{\sigma}_{e^+e^-\to ij } \otimes 
D_{H_1/i}\, \widetilde{\otimes}\, D_{H_2/j}.
\label{e:ee-pm}
\end{align}
In Fig.~\ref{fig:TMDcrosssect}, we summarize how TMD PDFs and TMD FFs are connected to the three classical two-scale factorizable observables with parton model factorization formalisms in Eqs.~(\ref{e:dyqt-pm}), (\ref{e:sidis-pm}) and (\ref{e:ee-pm}), respectively.  Corresponding QCD factorization formalisms are presented in the later Chapters in this Handbook.

Much of the predictive power of perturbative QCD factorization approach to experimentally measured cross sections relies on the universality of these TMDs and our ability to calculate the short-distance hard parts $\hat{\sigma}$'s.  However, extracting the 3D hadronic structure information, encoded in these TMDs, from experimentally measured data of these cross sections requires to solve an inverse problem to deconvolute TMDs from the factorization formalisms, such as those in Eqs.~(\ref{e:dyqt-pm}), (\ref{e:sidis-pm}) and (\ref{e:ee-pm}).  
As discussed in Chapter~\ref{sec:phenoTMDs}, with the experimental data of these two-scale observables we can extract these TMDs simultaneously by QCD global analysis of all available data \cite{Cammarota:2020qcw}.  A typical procedure for global analysis involves following necessary steps:
\begin{enumerate}
\item 
Identify good ``two-scale'' observables, such as cross sections or spin asymmetries that are defined as ratios of polarized and unpolarized cross sections, that can be factorized into convolution of TMDs along with perturbatively calculable short-distance hard parts, like those three classical examples in Fig.~\ref{fig:TMDcrosssect}; 
\item 
Make a choice of experimental data sets for these good observables, such that the data set can give the best constraints on a close set of TMDs; 
\item 
Calculate and/or verify the perturbative short-distance hard parts for these good observables;
\item
Develop a program for solving the scale dependence of the TMDs, which depend on the hard scale at which they are probed, just like PDFs;
\item 
Choose an algorithm to minimize the difference between the data and theoretical calculations based on the factorization formalisms to extract the set of universal TMDs that can best describe the data within the experimental uncertainties.  
\end{enumerate} 
However, with the limited data on the ``two-scale'' observables, our knowledge on the TMDs is still limited~\cite{Cammarota:2020qcw}.  More detailed methods and procedures for extracting TMDs from QCD global analysis are given in Chapter~\ref{sec:phenoTMDs}.

\begin{figure}
\begin{center}
\vspace{-0.3in} 
\includegraphics[width=0.75\textwidth]{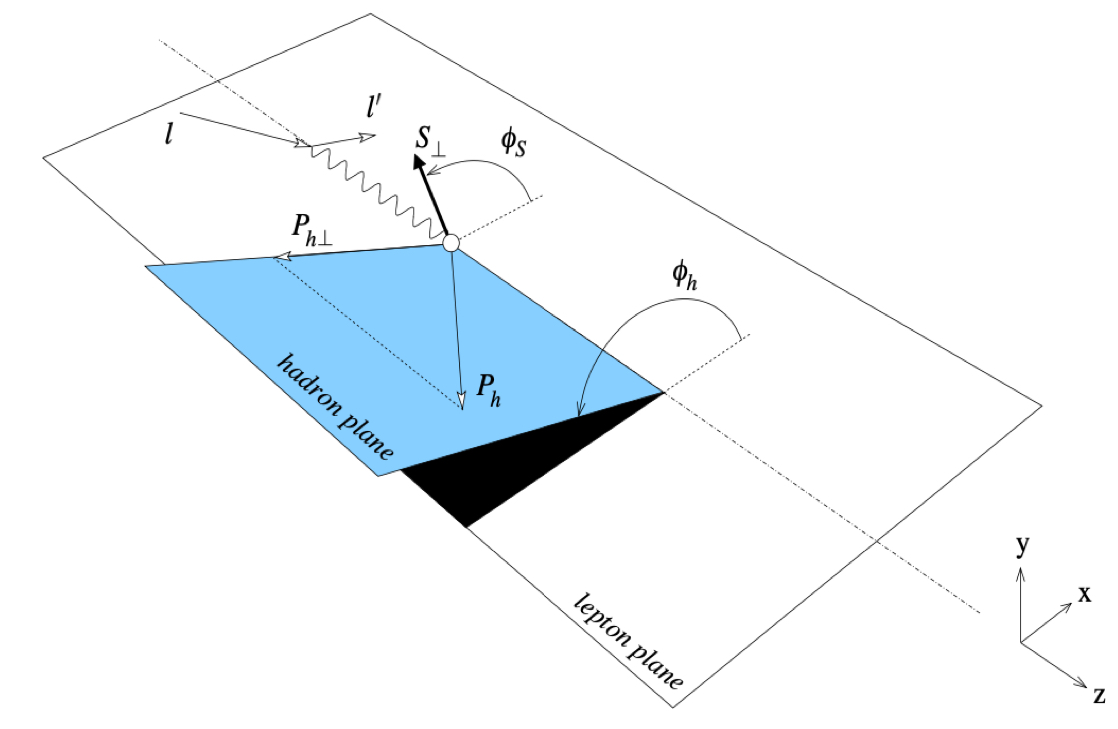} 
\vspace{-0.1in}
  \caption{\footnotesize The Trento convention for the ``photon-hadron" frame of SIDIS~\cite{Bacchetta:2004jz}.
  }
\label{fig:sidis-trento}
\end{center}
\vspace{-0.25in}
\end{figure}

With the large number of independent TMD PDFs and TMD FFs, and the rich information of TMD physics, it is a challenge to find more two-scale observables and new approaches to isolate different TMDs.  Since all TMDs are classified in terms of the correlation of hadron polarization and the polarization of quarks and gluons, spin asymmetries of cross sections with polarized beams can provide many more independent observables from the three classical two-scale observables alone.  In addition, with the well-defined leptonic plane, defined by the colliding and scattered lepton, and the hadronic plane, in terms of the colliding hadron and the observed final-state hadron, as shown in Fig.~\ref{fig:sidis-trento}, 
we can measure the lepton-hadron SIDIS cross section as a function of the angle $\phi_h$ between these two planes, and the angle $\phi_S$ between the direction of hadron spin and the leptonic plane.
Measuring various angular modulations of these two planes in SIDIS provides a unique way to isolate the information on various combinations of TMDs, 
as demonstrated in \chap{phenoTMDs}.
For example, for SIDIS with an unpolarized electron beam scattered off a transversely polarized proton (${\bf S}_\perp$), we can define the single transverse spin asymmetry as 
$
A_{UT}=[ \sigma_{UT}({\bf S}_\perp) - \sigma_{UT}(-{\bf S}_\perp)]
/ [\sigma_{UT}({\bf S}_\perp) + \sigma_{UT}(-{\bf S}_\perp)]
$
which is a function of the angle between the leptonic and hadronic plane 
and the direction of hadron spin.
By measuring different angular modulations, 
\begin{align}
A_{UT}^{Collins} &\propto \langle \sin(\phi_h+\phi_S)\rangle_{UT} \propto h_1 \, 
\widetilde{\otimes}\,
H_1^\perp\, ,
\nonumber\\
A_{UT}^{Sivers} &\propto \langle \sin(\phi_h-\phi_S)\rangle_{UT} \propto f_{1T}^\perp \, 
\widetilde{\otimes}\,
D_1\, ,
\nonumber\\
A_{UT}^{Pretzelosity} &\propto \langle \sin(3\phi_h-\phi_S)\rangle_{UT} \propto h_{1T}^\perp \, 
\widetilde{\otimes}\, 
H_1^\perp\, ,
\label{e:sidis-modulation}
\end{align}
where TMD PDFs and TMD FFs are defined in Figs.~\ref{fig:qTMDPDFsLP} and \ref{fig:TMDFFsLP}, respectively, we can obtain enhanced sensitivities to different TMDs.  
With experiments carried out at HERMES and COMPASS, the
on-going experiments at Jefferson Lab and more at the future EIC, lepton-hadron SIDIS experiments with various beam polarizations will provide rich information on the TMDs
and the 3D hadron structure in momentum space. 

However, above angular modulations rely on our ability to determine the ``photon-hadron ($\gamma^* P$)'' frame where the TMD factorization and the lepton and hadron planes are defined.  The large momentum transfer between the colliding lepton and hadron can trigger QED photon radiation from the colliding and scattering leptons to make it difficult, if not impossible, to fully determine the momentum of the exchanged virtual photon and the photon-hadron frame, which immediately impact the meaning of above angular modulations as well as the measured value of $P_{hT}$ \cite{Liu:2020rvc,Liu:2021jfp}. More discussion on the role of QED radiation in extracting TMDs from SIDIS will be given in \sec{QED} of Chapter~\ref{sec:phenoTMDs}.

\subsection{Calculation of Hadron Structure in Lattice QCD}
\label{subsec.intro-lattice}

Hadron structure, encoded in the universal PDFs, TMDs and other quark-gluon correlation functions of the hadron, can not be calculated in QCD perturbation theory.
The numerical technique of lattice QCD in principle provides a way in which to calculate nonperturbative QCD information about the properties of hadrons by directly evaluating the QCD path integrals that define such quantities. For many years, this technique has been used to access static and quasi-static quantities such as magnetic moments and electroweak form factors of the proton. Calculations of some of these quantities are now reaching a level of sophistication where the small effects of QED and isospin-breaking must be included in the LQCD calculations and a community consensus of such results is  maintained by the Flavour Lattice Averaging Group \cite{Aoki:2021kgd}.

PDFs and TMDs 
and related objects are defined precisely in QCD  by operators that involve correlations of quark and gluon fields with lightlike separations in spacetime.  It is therefore very natural to ask whether given sufficient computing power we could calculate the PDFs and TMDs, and in general, the leading quark-gluon correlations inside a bound nucleon {\it directly} in LQCD. If it were possible,   the quantum correlations between a hadron's mass and  spin and the motion of quarks and gluons inside it could be determined,  shedding  light on how quarks and gluons are confined inside the hadrons. 
However for these partonic  quantities, an impediment to LQCD calculations is raised by the light-cone nature of their definition. 
Since LQCD is most practically formulated in Euclidean space, direct determinations of such lightlike separated correlations are not possible. For that reason, most QCD studies of partonic physics have concentrated on the $x^n$ weighted Mellin moments of PDFs. However for technical reasons, these calculations have been restricted  to the lowest few moments, $n\in\{1,2,3\}$.

Despite the challenges, various approaches to go beyond calculations of moments and extract quark-gluon correlation functions from lattice QCD have been  proposed and investigated over the years~\cite{Liu:1993cv,Liu:1999ak,Liu:2016djw,Aglietti:1998ur,Abada:2001if,Detmold:2005gg,Braun:2007wv,Musch:2010ka}.  Stimulated by the quasi-PDFs approach introduced in Ref.~\cite{Ji:2013dva} (the approach was later formulated in a large-momentum effective field theory \cite{Ji:2014gla,Ji:2020ect}), many new ideas and approaches have been proposed for the extraction of PDFs, TMDs and other quark-gluon correlation functions from LQCD calculations. These approaches include the pseudo-PDFs~\cite{Radyushkin:2017cyf}, current-current correlators in momentum space  \cite{Chambers:2017dov} and current-current correlators in position space~\cite{Ma:2017pxb}.  The central idea of all these new approaches is to identify quantities that can  be reliably calculated in LQCD as well as being objects from which the PDFs, TMDs or other quark-gluon correlation functions can be extracted with controllable approximations~\cite{Ma:2014jla}. 
While calculations of these quantities are still being refined, tremendous progress has been made and these developments hold the exciting potential for accurate, model-independent determinations of (TMD) PDFs directly from QCD. 

In Chapter~\ref{sec:lattice}, the recent developments in LQCD calculations of the PDFs and TMDs will be covered.  Although extracting the fundamental PDFs, TMDs and other quark-gluon correlation functions from LQCD calculations is analogous to
extracting such distributions from  observables that are measured precisely in experiments and factorizable in QCD, LQCD calculations may provide additional complementary information on hadrons that is difficult to extract from experiment. For example, calculations can cover parameter values and kinematics
that are difficult for experiments to reach. Moreover in LQCD, we have the freedom to choose the combinations of operators that are calculated in order to determine aspects of hadron structure that might not be readily accessible in experiments. Despite the so-far insurmountable challenges for {\it direct} LQCD calculations of PDFs, TMDs and other leading quark-gluon correlation functions, the various LQCD approaches that will be discussed below  definitively enhance our ability to explore the rich, nonperturbative structure of hadrons and the dynamics of quarks and gluons at the QCD scale.


\subsection{Guide to Reading the Handbook}
\label{sec:outline}

As this handbook is quite comprehensive, a guide to the reader that goes beyond the section titles in the table of contents is likely to be of benefit to all readers. With that in mind we will introduce various subjects in the chapters, making editorial comments regarding each section with the aim of guiding a reader in the desired direction. 

If there is one chapter that lies at the heart of this document it is \chap{TMDdefn} ``Definitions of TMDs'', which should not be skipped, though some sections may be omitted depending on the depth of knowledge that is being sought. Keep in mind that studying the definition of TMDs is like peeling off the layers of an onion with each new layer exposing further subtle facts. It would not be inappropriate to take this metaphor literally as well and there is no shame in keeping some tissues handy. The layers start with an overview based on the parton model in \sec{naivedef} and become successively more detailed. The complexity of the later sections is softened by explicit examples that are worked out in the text.  
We recommend that everyone read Secs.~\ref{sec:naivedef}--\ref{sec:tmdpdfs_new}, \ref{sec:TMDFFs}, \ref{sec:universality}, \ref{sec:leadingTMDPDF}, \ref{sec:latt_def_connection}, and \ref{sec:TMDfactSIDISee}. Those interested in getting to the core of the onion should also read Secs.~\ref{sec:tmd_defs}, \ref{sec:other_tmd_defs}, the remaining parts of \sec{qgspinTMDFF}, and Secs.~\ref{sec:largeqT} and \ref{sec:integratedTMDs}.
In \sec{TMDforDY} the factorization theorem for Drell-Yan is covered, which is important because it introduces many of the concepts and much of the notation that is used later. Section~\ref{sec:tmdpdfs_new} introduces the high-level definitions of the TMD functions, including the two broad categories of definitions widely used in the literature.  
Section~\ref{sec:tmd_defs}  covers the concept of rapidity regulators in TMDs. The need to regulate rapidity divergences is an important feature of TMD functions and they can not be fully understood without an understanding of this topic. This section also contains an explicit one-loop example that illustrates rapidity regulation in action. 
Section~\ref{sec:other_tmd_defs} connects the main definition of the TMD PDF used in this handbook to some others in the literature. 
Section~\ref{sec:TMDFFs} introduces the TMD fragmentation functions.  
Section~\ref{sec:qgspinTMDFF} discusses the universality of TMDs for different processes and introduces the full complement of spin dependent TMDs. 
The connection between TMDs and PDFs at perturbative transverse momentum is discussed in \sec{largeqT}, and the extent to which integrating TMDs over their transverse momentum results in longitudinal PDFs is taken up in \sec{integratedTMDs}. 
The connection of continuum TMDs and lattice friendly definitions for TMD correlators is introduced in \sec{latt_def_connection}, including the Lorentz invariance and large momentum effective theory (LaMET) approaches.
Finally, \sec{TMDfactSIDISee} dives into TMD factorization for DY, SIDIS and $e^+ e^-$ cross sections including the definitions of kinematic variables that are used throughout the handbook.
 
\chap{Factorization} provides a broad view of the ideas behind proofs of factorization for TMD processes, without diving too deep into the details. Readers interested in more technical details are referred to the referenced literature, while those seeking to get an intuitive understanding of the key concepts will find what they are looking for.  A novel aspect of this section is the simultaneous treatment of both the traditional QCD factorization approach of Collins-Soper-Sterman and the more recent SCET approach, with parallels also drawn. 

\chap{evolution} covers evolution and resummation of large logarithms in TMD physics. Again this chapter peels like an onion with the broadest view being the outer layer, and successive layers zooming in on the details. An effort is made to cover and connect approaches from QCD factorization and SCET. 
Section~\ref{sec:evolutionintro} gives a historical overview of the development of TMD evolution starting with QCD factorization and then the SCET approach. 
In \sec{EvolResum} one finds an overview of what the goal of resummation is and what can be achieved, while in \sec{TMDEvol} gets into the beautiful nitty-gritty of resummation, and also provides a short one-loop example. 
Section~\ref{sec:CoordEvol} covers solutions of the evolution equations used in the CSS formalism, while the path followed to solve evolution equations in SCET is taken up in \sec{evolution_SCET}. 
\sec{2dRRGE}
on ``Two-dimensional evolution'' reviews novel insights that can be gained from treating simultaneously the evolution in invariant mass and rapidity and is a recommended read. 
Section~\ref{sec:nonsingular} is a small but important section on how the resummed results can be smoothly matched onto fixed order results, so that final cross sections are accurate in regions where resummation is or is not important. 
Finally, \sec{momentumspace} revisits evolution but this time with all or parts of the evolution in momentum space, rather than only using position space as in the previous sections.

\chap{phenoTMDs} is a thorough tour of the phenomenology involved in the extraction of TMDs. 
Section~\ref{sec:phenoTMDs_intro} gives a historical overview of the rich phenomenology of extracting TMD functions. Not only is this section interesting reading it also will really help readers make sense of the current state of affairs, and hence is recommended. 
Sections~\ref{sec:unpol_observables}--\ref{sec:phenomelology-other}
give an overview of phenomenology for TMD functions ranging from the unpolarized distributions, 
\index{Sivers function $f_{1T}^{\perp}$!introduction}
\index{Collins function $H_1^{\perp}$!introduction}
to the important Sivers and Collins functions, to the interesting Boer-Mulders and worm-gear distributions. These sections all involve processes that are dominated by the quark substructure of hadrons, where the wealth of available data means that we are currently very capable of probing the strong interactions. 
Gluon TMDs are not nearly as well probed by experimental measurements, as described in 
\sec{gluonTMD_obs}.  Experimental analysis of nuclear TMDs are also in their infancy, with plenty of open opportunities, as described in \sec{phenoTMDs_nuclei}.
In \sec{QED} we discuss the importance of accounting for QED radiation when extracting TMD functions, and \sec{pheno_outlook} gives a glimpse into future phenomenological directions for TMD physics.

\chap{lattice} focuses on how nonperturbative knowledge about TMDs can be obtained from first principles with lattice QCD. We begin in \sec{lqcd} with a brief review of lattice QCD techniques, to paint the stage for those unfamiliar with lattice calculations, albeit with broad brush strokes. The goal of lattice QCD calculations in the context of this handbook is to determine various aspects of TMDs. To build up to this challenging endeavor we discuss several topics that serve as important stepping stones. 
Section~\ref{sec:lattice:structure} broadly covers the structure of the proton on the lattice, including the current status of classic results like the analysis of moments of PDFs, and the decomposition of the proton's spin.  This is followed by \sec{lattice:xdep} which gives an extensive overview of the currently very active program of determining longitudinal PDFs and  structure functions on the lattice.  This section sets the stage for lattice extractions of TMD functions, for example by reviewing work on the LaMET approach involving quasi-PDFs.
Section~\ref{sec:lattice_tmd_calcs} then dives into lattice and TMD functions, including the Lorentz Invariants approach, calculation of TMD soft functions, and Lattice QCD information for TMD evolution.  This section will be of interests to those planning to do research on TMD functions on the lattice, as well as non-experts looking for a good overview of what information can currently be determined by lattice QCD calculations, as well as prospects for the future.

\chap{models} covers models of hadronic physics applied to TMDs. This is well worth reading as models have played an important role in the development of this field. In particular a significant result was a model calculation by Brodsky, Hwang and Schmidt~\cite{Brodsky:2002cx} of the single spin asymmetry (SSA) that demonstrated a nonzero transverse SSA in SIDIS, as discussed in Sec.~\ref{Subsec-models:review-Brodsky-Hwang-Schmidt}.  Sections~\ref{Sec:models-limits-in-QCD}--\ref{Sec:models-T-odd-quark-TMDs} cover frameworks for both T-even and T-odd TMDs, including various types of parton and quark models, the bag model, Nambu--Jona-Lasinio models, AdS/QCD models, and soliton models, and we also cover models for gluons TMDs (Sec.~\ref{Sec:models-gluon-TMDs}) and for fragmentation functions (Sec.~\ref{Sec:models-fragmentation}). We also discuss in Sec.~\ref{Sec:formal-constraints} more general results for TMDs that so far lack QCD derivations, namely positivity constraints and sum rules. Finally, we constrast relations derived from models with results obtained from QCD in Sec.~\ref{Sec:relations-in-models}.
 
\chap{smallx} focuses on the small-$x$ kinematic region of TMDs with important implications for our understanding of QCD.  An overview of why this region is of such interest, and the connection between TMDs and saturation phenomena for hadronic systems is given in \sec{saturation}. It also gives a foundation to the sections that follow and is recommended reading for anyone interested in the small-$x$ regime. Section~\ref{sec:WW-DGD} takes a closer look at the gluon distribution functions at small $x$, including both the Weiz\"acker-Williams and dipole distributions. In \sec{TMDsmallxevol} we discuss the evolution of TMD distributions when simultaneously accounting for small-$x$ resummation. Section~\ref{sec:smallxSDtmds} concentrates on the more advanced topic of spin-dependent TMDs at small-$x$. Finally, \sec{satMS} leads us to the frontier of the field in terms of the physics of saturation and multiple scattering effects.  A road map of future research that needs to be done is given in \sec{smallxOutlook}. 
 
\chap{JetFrag} takes a different approach to the extraction of TMD functions by considering measurements of jet observables; specifically jet fragmentation. The chapter begins with an overview of jets to a level that is needed to understand subsequent sections. \sec{jet-TMDPDFs} considers jets as probes of TMD PDFs, which while more complicated than the processes already considered can provide a wealth of data from hadron colliders. The following sections are refinements on this idea: \sec{jetsubjetfrag} considers jet substructure, \sec{jetquarkonia} studies jets with heavy quarkonium and \sec{TEEC} introduces transverse energy-energy correlators. \sec{mediumjets} takes jets into the realm of in-medium effects as applied to either cold QCD matter or the QGP. This is a rich field so that this section only provides a broad view. 
 
\chap{twist3} is recommended for anyone interested in azimuthal asymmetry observables  whose structure functions enable us to probe novel subleading-power TMDs, such as quark-gluon-quark correlators. These observables depend on sixteen new TMD PDFs and four new TMD FFs, in addition to those that already appeared at leading power, making the subject somewhat daunting. 
After an introduction in \sec{subTMDintro}, we describe in \sec{subTMDobs} to discussing observables, focussing on terms in the SIDIS cross section that are sensitive to subleading TMDs, and giving their general decomposition in terms of hadronic structure functions.  Historically SIDIS provided our first view of these asymmetries.
In \sec{subTMDdistns} we define the subleading power TMD distributions as operator matrix elements, and then in \sec{subTMDfact} we present the factorization formula that relates the structure functions to leading and subleading TMDs.  In \sec{subTMDexpt} we give a review of experimental measurements of subleading power TMD observables. 
Section~\ref{sec:subTMDcalc} discusses both lattice and model based methods for estimating the contribution of subleading TMDs to different processes. 
Many things remain to be worked out for the subject of subleading power TMDs, and \sec{subTMDoutlook} gives a summary and outlook. 

In \chap{gtmd} we zoom out to consider a more general class of multidimensional functions which probe quark substructure, of which the TMDs are just an important case. Section~\ref{sec:Wigner} introduces the Wigner distribution and how it reduces to more familiar TMD PDFs, impact parameter-dependent parton distributions, PDFs, and form factors. Section~\ref{sec:GTMD} introduces generalized TMDs (GTMDs) through a Fourier transform in transverse position space of the Wigner distribution. Section~\ref{sec:GTMD_observables} discusses observables which can be used to measure the GTMDs. Section~\ref{sec:GTMD_OAM} connects the GTMDs and associated GPDs (generalized parton distributions) to the orbital angular momentum of partons. Section~\ref{sec:GTMD_OAM_lattice} discusses the measurement of GTMDs on the lattice, while \sec{GTMD_models} considers the evaluation of the GTMDs and Wigner distributions in specific models.

%% file: sec-definition/sec-definition.tex
\section{Definition of TMDs}
\label{sec:TMDdefn}

In this chapter we introduce theoretical background as well as complete definitions of transverse momentum distribution functions. For simplicity, we  focus on the Drell-Yan process in proton-proton collisions, $pp \to \gamma^*/Z X \to \ell^+\ell^- X$ with unpolarized protons, as a physical example that connects TMD PDFs to experiment. Here $X$ is the hadronic debris from the deeply inelastic collision. We also consider SIDIS, $e^- p \to e^- H X$, where $H$ is the fragmentation hadron. We start with basic ideas from the parton model in \sec{naivedef}, followed by an overview of results obtained from TMD factorization in \sec{TMDforDY}. In \sec{tmdpdfs_new} we discuss the basic ingredients necessary for the most popular definition of TMD PDFs, which can be constructed using various different rapidity regulators, as reviewed in \sec{tmd_defs}. In \sec{other_tmd_defs} we discuss alternate definitions where both the TMD PDFs and the short distance part of the Drell-Yan factorization theorem depend on an additional rapidity variable.  

Next, we generalize the discussion to include polarized protons which gives us access to the full range of spin dependent leading power TMD functions, displayed in \fig{qTMDPDFsLP}, and the analogous TMD fragmentation functions. The complete field theory definitions for leading power spin dependent TMD PDFs and TMD FFs for both quarks and gluons are given in \sec{qgspinTMDFF}.  We then consider two theoretical methods to obtain insight into these distributions.
For perturbative $k_T\gg\Lambda_{\rm QCD}$, connections between these TMD PDFs and longitudinal PDFs are discussed in \sec{largeqT}.  
In \sec{integratedTMDs} we discuss the relationship between collinear PDFs and TMD PDFs.
Then in \sec{latt_def_connection} we discuss the use of matrix elements employed in Lattice QCD computations that can be connected to TMD PDFs, providing an introduction to the more detailed discussion in \sec{lattice}.

Finally, in \sec{TMDfactSIDISee} we give full leading power results for TMD cross sections with polarized protons, discussing the Drell-Yan process, semi-inclusive deep inelastic scattering (SIDIS) $e^- p\to e^- H X$ which involves both a TMD PDF for the proton $p$ and a TMD FF for the hadron $H$, and $e^+e^-$ collisions with the back-to-back production of two identified hadrons $H_1$ and $H_2$, $e^+e^-\to H_1 H_2 X$, which involves two TMD FFs.  

Some fundamental aspects of our notation are also introduced in this chapter. A summary of our notation including relations to alternates used in the literature can also be found in \app{glossary}.

\subsection{Basic Ideas from the Parton Model}\label{sec:naivedef}

\index{parton model|(}

Before turning to the modern definitions of TMD PDFs, we start with a historical review with the goal of building intuition about the physics encoded in TMDs. 
The intuitive concepts of both parton distributions and TMD parton distributions significantly predate QCD, by a number of years~\cite{Gardiner:1970wy}. 
They arise naturally whenever the kinematics of a process, viewed from a parton model perspective, imply sensitivity to the longitudinal momentum fraction that the colliding partons have relative to the bound state that contains them, as well as their small intrinsic transverse motion in the bound state. 
In this section we base our discussion on the generalized parton model, which allows for the presence of gluon radiation.

\tocless \subsubsection{Drell-Yan in the Parton Model}\label{sec:partonDY}

\index{Drell-Yan}
To facilitate discussion of this parton model (and its generalization to include transverse
momentum dependence), we first consider the unpolarized Drell-Yan process, where two protons collide to produce a lepton pair, 
\begin{align} \label{eq:DrellYan}
  p(P_A)+p(P_B) \to \ell^+ + \ell^- +  X \,.
\end{align}
Here $X$ denotes all the other final-state particles, including the proton remnants and those produced along with the leptons.
In this process, the measurement of the leptonic momentum probes the kinematics of the colliding quark and anti-quark partons in the protons through the hard process $q\bar q\to \gamma^*(q)/Z(q)\to \ell^+\ell^-$.  Here $P_A^\mu$ and $P_B^\mu$ are the proton momenta, and $q^\mu$ is the momentum of the virtual photon or $Z$-boson.  The nature of this probe is made precise through factorization formulas which describe the cross section for the hadronic collisions. 

Let us start by reviewing the intuition embodied in the 
most basic collinear version for inclusive Drell-Yan integrated over all transverse momenta:
\begin{equation}
\label{e.collexample}
\frac{\diff{\sigma}{}}{\diff{Q^2} \, \diff{Y}
 } 
= \sum_{i,j} \int_{x_a}^1 \diff{\xi_a}  \int_{x_b}^1 \diff{\xi_b}  \, f_{i/H_a}(\xi_a)
\, f_{j/H_b}(\xi_b)
\frac{\diff{\hat{\sigma}_{ij}(\xi_a,\xi_b)}{}}{\diff{Q^2}  \, \diff{Y}
} \biggl[ 1 + \cO\biggl(\frac{\lqcd^2}{Q^2}\biggr)\biggr]
\, .
\end{equation}
Here $Q^2 = q^2$ is the invariant mass of the $\ell^+\ell^-$ pair, $Y$ is their rapidity, a variable that is related to their polar angle from the collision axis (precise definitions can be found in \sec{TMDforDY}), and 
\begin{align}
  x_{a}= Q e^{+Y}/\sqrt{s}\,, \qquad
  x_{b}= Q e^{-Y}/\sqrt{s}\,,
\end{align}
where $s=(P_A+P_B)^2$ is the invariant mass for the incoming pair of hadrons.
Equation \eqref{e.collexample} expresses the natural classical intuition for scattering of composite particles with 
point-like constituents for $Q^2\gg  \Lambda_{\rm QCD}^2$\,, and has corrections suppressed by ${\cal O}(\Lambda_{\rm QCD}^2/Q^2)$ as indicated.
The total cross section $\diff\sigma$ contains the cross section $\diff\hat \sigma_{ij}$ for the partonic process
\begin{align}
 i(p_a) + j(p_b) \to \ell^+ + \ell^- +  X
\,.\end{align}
Here, we scatter
partons of type $i$ and $j$ with momenta $p_a^\mu$ and $p_b^\mu$, and these momenta have longitudinal momentum fractions $\xi_a$ and $\xi_b$ relative to the longitudinal components of $P_A$ and $P_B$ respectively.
In Eq.~\eqref{e.collexample}, this partonic cross section is multiplied by a probability density $f_{i/H_a}(\xi_a)$
for finding a parton $i$ in hadron $H_a$ with momentum fraction $\xi_a$, 
times a probability density $f_{j/H_b}(\xi_b)$ for finding a parton $j$ in hadron $H_b$ with momentum 
fraction $\xi_b$. These are combined with an integral over all possible momentum fractions and a sum over all parton types, which includes both quarks and antiquarks of various flavors, and gluons. 
In an observable like \eqref{e.collexample}, that has been averaged over the 
large allowed physical range for the transverse momentum of the dilepton pair, it is reasonable that the exact 
transverse momentum dependence of the partons in the convolution integral is not numerically 
important. Thus it is sensible that what appears is the average of the small 
transverse momentum within the target structures, so that the densities 
$f_{i/H_a}(\xi_a)$ and $f_{j/H_b}(\xi_b)$ are functions of only the longitudinal momentum components.

The situation changes if one considers a more detailed cross section, differential in the transverse momentum $\qt$ of the dilepton pair.
Here a measurement of $Q$, $Y$, and $\qt$ is equivalent to a measurement of the full dilepton four-momentum $q$.   If the transverse momentum is large (e.g.,~$q_T \sim Q$), the simplest generalization
of \eqref{e.collexample} is adequate, with the partonic cross section made differential in the transverse 
momentum,
\begin{align} \label{eq:collexample2}
\frac{\diff{\sigma}{}}{\diff{^4 q}
} 
= \sum_{i,j} \int_{x_a}^1\!\! \diff{\xi_a}
  \int_{x_b}^1\!\! \diff{\xi_b}  \, f_{i/H_a}(\xi_a)
\, f_{j/H_b}(\xi_B) \frac{\diff{\hat{\sigma}_{ij}(\xi_a,\xi_b)}{}}{\diff{^4 q} 
}
\biggl[ 1 + \cO\biggl(\frac{\lqcd^2}{q_T^2},\frac{\lqcd^2}{Q^2}\biggr)\biggr]
 & \, . \\
& \quad \text{(Large $q_T$)} 
 \nn
\end{align}
Since the transverse momentum is large, the comparatively
small transverse momentum 
generated by bound state effects inside the targets can still be ignored in calculations of the 
differential partonic cross section.  Thus, the distribution functions still depend only on longitudinal components of momentum, $\xi_A$ and $\xi_B$.

As smaller transverse momenta are considered, this becomes less reasonable, and in the 
vicinity of $q_T \sim \Lambda_{\rm QCD}$, it becomes clear from momentum conservation alone 
that the differential cross section is very sensitive to the small transverse momentum inside the 
colliding bound states. 
For the regime where $\Lambda_{\rm QCD} \lesssim q_T \ll Q$ a different partonic picture is needed, wherein the probability densities 
describing the incoming colliding bound states include dependence on the small transverse momenta. 
This TMD version of the parton model is
\begin{align}
\label{eq:tmdexample}
\frac{\diff{\sigma}{}}{\diff{^4 q}}
&= \frac{1}{s} \sum_{i\in{\rm flavors}}
 \!\!\! \hat{\sigma}_{i\bar i}^{\rm TMD}(Q) \!
\int\!\! \diff{^2 \kt} \, f_{i/H_a}(x_a,\kt)
 \, f_{\bar i/H_b}(x_b,\qt-\kt)
 \biggl[ 1 + \cO\biggl(\frac{q_T^2}{Q^2},\frac{\lqcd^2}{Q^2}\biggr)\biggr]
 \,. \\
&\phantom{x} \hspace{5.5in}
 \text{(Small $q_T$)} \nn 
\end{align}
This equation again represents a natural model rooted in classical intuition. A partonic cross section 
(represented by $\hat{\sigma}_{i\bar i}^{\rm TMD}$)  multiplies a product of probability densities $f_{i/H_a}$ and $f_{\bar i/H_b}$ for finding partons $i$ and $\bar i$.
Now, however, these densities depend on both longitudinal momentum fractions ($x_a$ and $x_b$) and transverse
($\kt$ and $\qt-\kt$) components of the incoming parton momenta. 
Since we impose that $q_T\ll Q$, all other particles (denoted as $X$ in \eq{DrellYan}) must themselves have small transverse momentum.
These restrictions imply that the longitudinal momentum fractions are fixed to $x_a$ and $x_b$ and there is no longer an integral over the fractions $\xi_{a,b}$, but there is now an integral over transverse momenta which are constrained to add up to $\qt$.
Due to the restriction to the leading terms in the small $q_T$ limit, the parton types $i$ and $\bar i$ in the sum in \eq{tmdexample} are restricted to quarks and anti-quarks of the same flavor ($\bar i$ being the charge conjugate of $i$).
This region of small transverse momentum is evidently more sensitive to details of the target structure
than either the large transverse momentum region described by \eqref{eq:collexample2} or
the transverse integrated cross section in \eqref{e.collexample}.

In practice, one often works with the Fourier-transformed TMD PDF, which is defined as
\begin{align} \label{eq:tmdpdf_bspace}
 \tilde f_{i/H}(x, \bt) &= \int\df^2\kt \, e^{-\img \bt \cdot \kt} f_{i/H}(x, \kt)
\,.\end{align}
Here, $\bt$ is Fourier-conjugate to the transverse momentum $\kt$, and $\tilde f_{i/H}(x, \bt)$ is referred to as the transverse position space or coordinate space distribution.
We provide some additional details on the Fourier transform in \app{Fourier_transform}.
Inserting \eq{tmdpdf_bspace} into \eq{tmdexample}, one obtains the equivalent result
\begin{align} \label{eq:tmdexample_bspace}
 \frac{\diff{\sigma}{}}{\diff{^4 q}} &
 = \frac{1}{s} \sum_{i\in{\rm flavors}} \!\! 
   \hat{\sigma}_{i\bar i}^{\rm TMD}(Q) \!
   \int\! \frac{\df^2\bt}{(2\pi)^2} \, e^{i \bt \cdot \qt} \,
   \tilde f_{i/H_a}(x_a, \bt) \tilde f_{\bar i/H_b}(x_b, \bt)
   \biggl[ 1 + \cO\biggl(\frac{q_T^2}{Q^2},\frac{\lqcd^2}{Q^2}\biggr)\biggr]
\,.\end{align}
Compared to \eq{tmdexample}, we have traded the convolution integral in $\pt$ for the inverse Fourier transform in $\bt$,
which in practice is more convenient to work with and thus the preferred choice for most phenomenological applications.
Note however that despite working in $\bt$ space, \eq{tmdexample_bspace} receives the same corrections in $q_T/Q$ and $\lqcd/Q$ as \eq{tmdexample},
and thus is valid only at small $q_T \ll Q$.

Eqs.~(\ref{e.collexample}-\ref{eq:tmdexample}) alone are already useful as phenomenological models,
even without the introduction of field theoretic concepts. We will see that when we treat the problem field-theoretically, there is more than one type of leading-order TMD PDF, even in the situation discussed here with unpolarized hadrons $H_{A,B}$, cf. \fig{qTMDPDFsLP}.  It is worth pausing to 
remark on several points of interpretation. First, the parton densities are understood here to be 
intrinsic to the structure of the colliding hadrons, insensitive to the type of process,  
and thus universal. This will be important, as the universality is a major part of the predictive power 
of the parton model.
Second, the differential cross section $\hat \sigma_{i\bar i}$ for partonic scattering is of course very sensitive to 
the specific type of overall cross section of which it is a part. However, it also involves a large 
$Q$, and this ultimately ensures that it is sensitive only to the dynamics of small time and distance 
scales of order $\sim Q^{-1}$. This turns out to make it ideal for calculations in perturbation theory, made possible by the asymptotic freedom of QCD.

The study of TMDs is the study of the small $q_T$ behavior in \eq{tmdexample}, motivated 
largely by the expectation that the small transverse momentum dependence in the TMD PDFs,
$f_{i/H_a}(x_a,\kt)$ and $f_{j/H_b}(x_b,\qt-\kt)$, carries more information
about nucleon structure than the more standard collinear PDFs  $f_{i/H_A}(\xi_a)$ and $f_{j/H_b}(\xi_b)$.

\tocless
\subsubsection{SIDIS in the Parton Model}
\label{sec:partonSIDIS}

\index{SIDIS}
In addition to the parton distributions $f_{i/h}$, another important set of distribution functions for probing hadronic structure are the fragmentation functions $D_{h/j}$, which describe the process whereby a parton $j$ is converted to a final state hadron $h$.  To introduce them we consider the semi-inclusive DIS (SIDIS) process where an electron and proton collide inelastically, with a measured final state hadron $h$,
\begin{align} \label{eq:SIDIS_intro}
 e^-(l) + p(P) \to e^-(l') + h(P_h) + X  \,.
\end{align} 
Here, $l^\mu$ and $l^{\prime\mu}$ are the initial and final electron momenta, $P^\mu$ is the proton momentum, and $P_h^\mu$ is the final hadron's momentum.  Once again $X$ denotes hadronic debris from this deep inelastic collision. 
This process probes the short distance scattering of the electron and a quark of flavor $i$ in the proton, $e^- i \to e^- i$, through exchange of a virtual photon or $Z$-boson with spacelike momentum $q^\mu$, so that
\begin{align}
 q^\mu &=l^\mu-l^{\prime\mu} \,, \qquad
  q^2 = -Q^2 <0 \,.
\end{align}
The final state quark $i$ then fragments to the hadron $h$. Key variables for describing the SIDIS cross section include
\begin{align} \label{eq:xyz}
   x = \frac{Q^2}{2P\cdot q} \,, \qquad
   y = \frac{P\cdot q}{P\cdot l} \,,\qquad
   z_h = \frac{P\cdot P_h}{P\cdot q} \,.
\end{align} 
Here, $x$ is the standard DIS Bjorken scaling variable. In the proton rest frame, $y$ is the fractional energy loss of the electron, and $z_h$ is the ratio of the energy of the hadron to that of the $\gamma^*/Z^*$ in the proton rest frame.

Again we start with the basic collinear version of the fragmentation process, with an unpolarized proton and without a measurement of final state transverse momentum:
\begin{align} \label{eq:sidiscollin}
  \frac{\diff\sigma}{\diff x\diff y\diff z_h}
  &=  \sum_{i,j} 
   \int_x^1\!\! d\xi \int_{z_h}^1\!\! d\zeta\, f_{i/p}(\xi)\: D_{h/j}(\zeta) \:
 \frac{\diff\hat\sigma_{ij}(\xi,\zeta)}{\diff x \diff y\diff z_h}
 \: \biggl[ 1 + \cO\biggl(\frac{\lqcd^2}{Q^2}\biggr)\biggr] 
\,.
\end{align}
Here, $\diff\hat\sigma_{ij}$ is the cross section for scattering a parton of type $i$ into a parton of type $j$,
i.e., it corresponds to the partonic process
\begin{align}
 e^-(l) + i(k) \to e^-(l') + j(p) + X
\,.\end{align}
For the incoming parton $i$ with momentum $k^\mu$, the momentum fraction $\xi$ is defined as the ratio of the longitudinal momentum component of $k^\mu$ relative to the proton momentum $P^\mu$. For the outgoing parton $j$ with momentum $p^\mu$, the momentum fraction $\zeta$ is defined as the ratio of the longitudinal momentum component of $P_h^\mu$ relative to $p^{\mu}$. 
In \eq{sidiscollin} the partonic cross section $\diff\hat\sigma_{ij}$ is combined with a probability density $f_{i/p}(\xi)$ for finding the parton $i$ in the proton with momentum fraction $\xi$. In addition it is combined with $D_{h/j}(\zeta)$ for the fragmentation process, which is the probability density for the parton $j$ to fragment to a hadron $h$, where $h$ has a fraction $\zeta$ of the parton's momentum. Equation~(\ref{eq:sidiscollin}) is the direct analog of the Drell-Yan cross section in Eq.~(\ref{e.collexample}), except with one parton distribution function and one fragmentation function, rather than two parton distribution functions.

To make the process more differential, we consider measuring in addition the transverse momentum ${\bf P}_{hT}$ of the hadron $h$.
We choose to define ${\bf P}_{hT}$ in the $\gamma^* p$ center of mass frame, with 3-momenta ${\bf q}$ and ${\bf P}$ aligned along the $z$-axis, such that it satisfies ${\bf q}\cdot {\bf P}_{hT} = {\bf P}\cdot {\bf P}_{hT}=0$. (For further details about this frame, see the extended discussion in \sec{TMDfactSIDISee}.) 
For simplicity, we continue to consider unpolarized protons and measure only the magnitude of the transverse momentum, $P_{hT}=|{\bf P}_{hT}|$.  

For large $P_{hT}\sim Q$ the transverse momentum of the hadron is inherited from the transverse momentum of the parton $j$ at leading order in $\Lambda_{\rm QCD}^2/P_{hT}^2\ll 1$. This yields a parton model cross section that is similar in form to \eq{sidiscollin}, but with the cross section $\diff\hat\sigma_{ij}(\xi,\zeta)/\diff x\diff y\diff z_h \diff P_{hT}^2$ in the integrand.  This is the exact analog of the Drell-Yan generalization, in going from Eq.~(\ref{e.collexample}) to Eq.~(\ref{eq:collexample2}).

On the other hand, for small $P_{hT}$ we begin to probe transverse momentum in the fragmentation process, while at the same time becoming sensitive to the  transverse momentum of the initial state parton inside the proton. In particular for $\Lambda_{\rm QCD} \lesssim P_{hT} \ll Q$ the TMD version of the parton model cross section is 
\begin{align}
  \label{eq:sidisUnpolTMD}
  \frac{\diff\sigma}{\diff x\diff y\diff z_h \diff^2 {\bf P}_{hT}}
  &= \sum_{i} 
    \hat\sigma^{\rm TMD}_{ii}(Q,x,y)
    \int\!\! \diff^2\pt  \, \diff^2\kt \,
    \delta^{(2)}\bigl({\bf P}_{hT} - z_h \kt - \pt \bigr) \,
    f_{i/p}(x,\kt) \: D_{h/i}(z_h, \pt) \:
 \nn\\
 &\qquad\times \biggl[ 1 + \cO\biggl(\frac{P_{hT}^2}{Q^2},\frac{\lqcd^2}{Q^2}\biggr)\biggr] 
 \qquad\qquad\quad \text{(Small $P_{hT}$)}  \,.
\end{align}
Once again with the parton model description, we have a factor $\hat\sigma^{\rm TMD}_{ii}$ determined by the partonic cross section multiplying probability densities $f_{i/p}$ and $D_{h/i}$, which now depend on both longitudinal momentum fractions ($x$ and $z_h$) and transverse momenta. The restrictions on final state radiation fix the longitudinal momentum fractions appearing in the TMDs, and imply that it is the same parton flavor $i$ appearing in both TMD functions in \eq{sidisUnpolTMD}.
Here the TMD fragmentation function $D_{h/i}(z_h,\pt)$ gives the probability of parton $i$ fragmenting to hadron $h$ with longitudinal momentum fraction $z_h$, where the hadron $h$ has a transverse momentum $\pt$ relative to the direction of motion of the parton $i$ .
In the frame where it is the proton and outgoing hadron that are aligned along the $z$-direction, the transverse momentum conservation is given by the partonic formula $\kt + \qt = -\pt/z_h$. In the $\gamma^*p$ rest frame used here, we replace $\qt \to -{\bf P}_{hT}/z_h$, which yields the $\delta$-function in \eq{sidisUnpolTMD}.

Once again it is useful to work with the Fourier-transformed TMD FF,
\begin{align}\label{eq:FF}
  \widetilde D_{h/i}(z,{\bf b}_T) 
 &= \frac{1}{z^2} \int\!\! \diff^2 \pt\, e^{-i{\bf b}_T\cdot \pt/z}\,
D_{h/i}(z,\pt)  \\
 &=  \int\!\! \diff^2 \pt'\, e^{+i{\bf b}_T\cdot \pt'}\,
   D_{h/i}(z,-z\pt') 
   \,. \nn
\end{align}
Here we see that for the fragmentation function $\widetilde D_{h/i}(z,{\bf b}_T)$ the transverse position $\bt$ is defined as the Fourier conjugate variable to $\pt'$, the momentum of the incoming quark in a frame where the transverse momentum of the hadron $h$ vanishes.

Together with \eq{tmdpdf_bspace}, this enables us to write \eq{sidisUnpolTMD} in an equivalent fashion as
\begin{align} \label{eq:FFfact0}
\frac{\diff\sigma}{\diff x\diff y\diff z_h \diff P_{hT}^2}
  &= \sum_{i\in{\rm flavors}} \!
   \hat\sigma^{\rm TMD}_{ii}(Q,x,y)
   \int_0^{2\pi}\!\!\!\!\! \diff \phi_h \!
   \int\!\! \diff^2{\bf b}_T \, e^{+i{\bf b}_T\cdot {\bf P}_{hT}/z_h }\:
    \tilde f_{i/p}(x,{\bf b}_T)\: \widetilde D_{h/i}(z_h,{\bf b}_T) \:
 \nn\\
 &\qquad\times \biggl[ 1 + \cO\biggl(\frac{P_{hT}^2}{Q^2},\frac{\lqcd^2}{Q^2}\biggr)\biggr] 
  \,. 
\end{align}
Here $\phi_h$ is the transverse angle of the vector ${\bf P}_{hT}$. 
We will take up an extended version of this formula, which is derived from QCD and applies for polarized protons and with additional angular measurements, in \sec{TMDfactSIDISee}.

\vspace{0.2cm}
\tocless \subsubsection{Beyond the Parton Model}\label{sec:beyondparton}

With this introduction of the concepts, several key points need to be addressed. Parton model descriptions like 
Eqs.~(\ref{e.collexample}-\ref{eq:tmdexample}) need to be justified in QCD.
This is the topic of factorization theorems, to be discussed in \sec{Factorization}. A related question that must be addressed is exactly how to define the PDFs and TMD PDFs (and similar objects) in quantum field
theory.  While many 
aspects of the parton picture remain valid, there are a number of important caveats that arise, such as the dependence on parameters associated with the renormalization scheme, like the renormalization scale $\mu$. 
A central goal of the next few sections will be to flesh out in detail the quantum field theory definition of TMD PDFs and TMD FFs that arise from the proof of factorization theorems for TMD sensitive cross sections.  

\index{collinear PDF}
As a prelude to some of the extra ingredients that appear, we will briefly review the field theory definition of the unpolarized collinear PDF for a parton of flavor $i$ in a hadron $H$. The definition for these PDFs is much simpler than the corresponding definition for TMD PDFs.
In quantum field theory the starting point is the definition of a 
bare parton distribution 
\begin{equation}\label{eq:barepdf}
f^0_{i/H}(\xi) = \int \frac{d w^-}{2 \pi} \, e^{-i \xi P^+ w^-}
\; \bigl\langle H(P) \bigr| \, \bar{\psi}_{i}^0(0,w^-,{\bf 0}_T ) {\frac{\gamma^+}{2}}  W_{n_a}(w^- ,0)
\psi_{i}^0(0,0,{\bf 0}_T) \, \bigl| H(P) \bigr\rangle \, .
\end{equation}
This formula involves a Wilson line 
operator $W_{n_a}(w^- ,0)$ connecting  the points $0$ and $w^-$ along the light-cone, which ensures gauge invariance (see \eq{Wilson_lines}, explicit definitions of the notation used here are left to \secs{TMDforDY}{tmdpdfs_new}).
The $0$ superscripts in \eq{barepdf} denote bare quantities. 
The bare fields  obey canonical commutation relations and thus give a true number density interpretation. 
(In a free field theory we can set $W_{n_a}=1$ and \eq{barepdf} becomes a literal number density.)  Implicit in this definition is the presence of an ultraviolet regulator, like dimensional regularization. Of course, for a renormalizable interacting theory like QCD, this bare definition needs to be replaced by something 
involving renormalized quantities for it to be useful. In the most commonly used $\MSbar$ scheme, this process is carried out by (minimal) removal of ultraviolet divergences with a renormalization factor $Z^{\rm PDF}_{ij}$, and introduces dependence on the renormalization scale $\mu$, yielding
\begin{equation}\label{eq:renpdf}
 f_{i/H}(\xi,\mu)
= \sum_{j} \int_\xi^1\! \frac{dz}{z} 
\, Z^{\rm PDF}_{ij}(z,\mu) \, f^0_{j/H}(\xi/z) 
 \, . 
\end{equation}
Since the renormalization involves a mixing of parton types, it contains a sum over $j$. 
We see that the renormalized PDF $ f_{i/H}(\xi,\mu)$ is obtained by a type of generalized multiplicative renormalization
of the bare PDF. Here the parameter $\mu$ plays the role of a momentum cutoff on the fluctuations from quantum fields that are retained in the PDF. It is effectively speaking akin to a cutoff on invariant mass, $|p^2|\lesssim \mu^2$, but where the cutoff $\mu$ has been introduced in a gauge invariant manner. 

The renormalization procedure also introduces a dependence on $\mu$ into the short distance partonic cross section, such as $\diff{\hat{\sigma}_{ij}(\xi_a,\xi_b,\mu)}/\diff{^4 q}$ in \eq{collexample2}.
Ultimately, the choice of $\mu$ is dictated by the requirement that the 
partonic scattering cross sections are well-behaved perturbatively (with no large logarithms in $\diff{\hat\sigma_{ij}}$). Therefore the PDFs are not literally
process independent, since different processes will require different choices of $\mu$. 
However, the dependence on $\mu$ can be systematically calculated with perturbative evolution equations, which for the PDF are known as the DGLAP equations~\cite{Gribov:1972ri,Lipatov:1974qm,Altarelli:1977zs,Dokshitzer:1977sg}. Once this is accounted for, the PDFs can be understood to be effectively universal. 

For the TMD PDFs (and fragmentation functions) extra subtleties enter beyond the need for ultraviolet renormalization, both formally and in their interpretation, and these issues are among the main topics of sections~\ref{sec:TMDforDY}--\ref{sec:other_tmd_defs} in this chapter. 
Here we give a brief review of the historical landmarks that characterized the development of the current rigorous understanding. 
It was realized by the late 1990s or early 2000s that existing definitions were not adequate for some applications, especially those associated 
with hadron structure.  
(Readers reviewing the relevant literature from the 1970s-1990s should be aware that terminology has evolved significantly since that time and, for example, the ``TMD'' label only became pervasive comparatively recently. In some parts of this earlier literature, terms like ``unintegrated'' PDF are used interchangeably with ``TMD PDF.'')
A useful review of the status of TMD PDF definitions and associated open problems as they were understood around 2003 is found 
in \cite{Collins:2003fm}, and it provides a useful context for the last two decades of development. An issue highlighted there that will be relevant to the discussions below is the appearance of so-called rapidity (or light-cone) divergences in the most natural 
candidate definitions for TMD PDFs. Rapidity divergences correspond to configurations of partons moving with infinite rapidity in a direction \emph{opposite} the direction of motion of the parent hadron.  They are regulated by neither the nonperturbative infrared physics nor by the ultraviolet regulators, and so they signal a significant challenge to any proposed definition. Ways of dealing with them will be discussed in much more detail in coming sections. The basic problem of light-cone divergences and the need to regulate them was recognized very early on. For example, Refs.~\cite{Soper:1979fq} and~\cite{Ralston:1980pp} pointed out that an extra parameter they called $\zeta = (2 P \cdot n)^2 / (-n^2)$ appears in some QCD calculations, where the ``$\zeta$'' notation is meant to be reminiscent of the Mandelstam $s$ and thereby evoke a kind of evolution with collision energy. In their definition of $\zeta$,  $P$ is a target hadron four-momentum and 
$n$ is a non-lightlike gauge fixing vector with $n^2 \neq 0$. The $\zeta$ acts effectively as a rapidity regulator and the need to fix it ultimately becomes associated with a new type of evolution.  
Collins derived the corresponding evolution for the Sudakov form factor in Ref.~\cite{Collins:1980ih}, and in the Collins-Soper-Sterman (CSS) formalism the analogous 
behavior appears as the Collins-Soper (CS) equation. Fundamentally, the rapidity divergences are artifacts of approximations at the level of the factorization derivations that place the Wilson lines appearing in the gauge invariant form of TMD 
definitions exactly on the light cone. 
Regulating them while maintaining explicit gauge invariance in definitions can be accomplished by shifting the Wilson lines slightly off the light cone, 
see Refs.~\cite{Collins:1981tt,Collins:1982wa} for early discussions of Wilson lines in TMD PDFs. The importance of including a transverse gauge link at light-cone infinity to obtain fully gauge invariant results was pointed out in Ref.~\cite{Belitsky:2002sm}.

The role of Wilson lines was also important in early discussions concerning the use of TMD correlation functions for describing non-trivial polarization dependence, and it was one of the motivating factors that led to later refinements to TMD definitions. 
A now famous TMD mechanism called the Sivers effect~\cite{Sivers:1990cc} 
\index{Sivers effect}
was proposed in 1990 to explain the larger than expected 
transverse single spin asymmetries in experiments like~\cite{Klem:1976ui,Dragoset:1978gg,Antille:1980th,Apokin:1990ik,Saroff:1989gn,Adams:1991rw,Adams:1991cs}.\footnote{See also Ref.~\cite{Aidala:2012mv,Chen:2012taa}}
An argument presented in~\cite{Collins:1992kk}, however, appeared to show using the time-reversal and parity ($TP$) invariance of QCD that the Sivers TMD must be zero. 
A later model calculation in 2002 by Brodsky, Hwang, and Schmidt (BHS)~\cite{Brodsky:2002cx} showed that a Sivers-like asymmetry \emph{does} arise at leading power in processes 
like SIDIS, and they interpreted this as indicating a conflict with factorization. A more detailed description of this influential calculation can be found in Sec.~\ref{Subsec-models:review-Brodsky-Hwang-Schmidt}. 
Work in~\cite{Collins:2002kn} addressing the BHS result demonstrated that TMD factorization actually does hold, and is 
not in contradiction with the definition of a Sivers TMD, despite the earlier proof appearing to show it must vanish by $TP$ invariance. The loophole is that the $TP$-based argument neglected a non-trivial role for the Wilson lines in TMD  correlation functions.  Once they are taken into account in the factorization derivation, $TP$ invariance shows not that the Sivers function vanishes, but that it acquires a process-dependent overall sign~\cite{Collins:2002kn}. In other words, the apparent proof that the Sivers function vanishes by $TP$ invariance is the consequence of an overly literal interpretation of the TMD PDF as a simple number density. 

While the Wilson lines in collinear correlation functions can sometimes appear to be largely formalistic, 
the examples above, of the light-cone divergences and 
the process-dependent sign on some TMD PDFs, highlight the central role Wilson line structures play in TMD factorization (see chapter~\ref{sec:Factorization}).
Another driving motivation to revisit the issue of TMD PDF definitions in the early 2000s was that their domain of practical application 
began to broaden. The focus of early applications was to a handful of specific processes at collider energies, where the role played by intrinsic 
nonperturbative transverse structure was of less direct interest, and could possibly even be viewed as a nuisance in some applications.
However, the TMD concept was being used increasingly in 
hadron structure studies. (An extensive classification of the various polarization structures allowable in a TMD approach was developed by 
Boer, Mulders, and Tangerman in 
the mid 1990s in, for example, Refs.~\cite{Boer:1997nt,Tangerman:1994eh,Mulders:1996dh}.)
Some of the work needed to orient CSS-based treatments more toward hadron structure was simply  organizational.  
For example, in the practical cross section formulas like Eq.~(1.1)
of Ref.~\cite{Collins:1984kg}, tracing the various factors back to 
the separate operator definitions for specific TMD correlation functions is non-obvious. (Indeed, nonperturbative transverse momentum contributions are 
only explicitly introduced later in Eq.~(5.6).) It was pointed out in Refs.~\cite{deFlorian:2000pr,deFlorian:2001zd} that the original CSS-like organization placed process 
dependent perturbative contributions not in an overall explicitly factorized hard part, but in exponential factors that resemble evolution contributions for separate TMD functions.
In the context of resummation approaches to transverse momentum distributions, later work reorganized these non-universal perturbative contributions into explicitly 
separate hard factors~\cite{Catani:2000vq,Catani:2009sm,Catani:2012qa,Bozzi:2010xn,Catani:2013tia}.

Other complications appeared to be more fundamental, such as the non-trivial dependence on the structure of Wilson lines structures discussed above, and the 
realization that CSS-type factorization might break in cases where it is reasonable to conjecture that it might hold (see Sec.~\ref{sec:factviolation} for more on this).

Proposals for refining the TMD definitions during this period can be found in, for example, Refs.~\cite{Collins:1999dz,Belitsky:2002sm,Collins:2003fm,Boer:2003cm,Bomhof:2004aw,Ji:2004wu,Ji:2004xq,Bomhof:2006dp,Cherednikov:2008uk,Cherednikov:2007tw,Cherednikov:2008ua,Hautmann:2007uw,Collins:2008ht,Hautmann:2009zzb}. The treatment that has since been settled upon was provided in the textbook of Collins in 2011~\cite{Collins:2011zzd}, and an application to hadron structure phenomenology was presented in Ref.~\cite{Aybat:2011zv}.

\index{parton model|)}

\subsection{TMD Factorization Theorem for Drell-Yan} 
\label{sec:TMDforDY}

\index{Drell-Yan}
In this section we give a basic introduction to the TMD factorization theorems that describe the Drell-Yan process, $pp\to \gamma^*/Z\to \ell^+\ell^-$ with unpolarized protons, which serves to set up basic notation and concepts for TMD factorization.

For the analysis of hard scattering processes it is useful to use light-cone coordinates since the hadronic dynamics are always preferentially probed along the collision axis and involve partons whose dynamics are described by fluctuations near the light-cone. We choose the $\hat z$-axis for the incoming protons in the $pp$ collision and in terms of $(t,x,y,z)$ components define the lightlike basis vectors
\begin{align} \label{eq:nab}
  n_a^\mu = \frac{1}{\sqrt2} (1,0,0,1)   \,,
    \qquad\qquad 
  n_b^\mu = \frac{1}{\sqrt2} (1,0,0,-1)  \,,
\end{align}
with $n_a^2=n_b^2=0$ and $n_a\cdot n_b=1$.
Any four vector can then be decomposed in terms of these basis vectors as
\begin{align} \label{eq:lc_coord}
 p^\mu
 = (n_b \cdot p) n_a^\mu + (n_a \cdot p) n_b^\mu +  p_T^\mu
 \equiv (p^+, p^-, \pt)
\,.\end{align}
In the second equation, we introduced a common short-hand notation for the light-cone decomposition.
Its components are defined as
\begin{align}
 p^+ \equiv n_b \cdot p = \frac{1}{\sqrt2} \bigl(p^0 + p^z \bigr)
\,,\qquad
 p^- \equiv n_a \cdot p = \frac{1}{\sqrt2} \bigl(p^0 - p^z \bigr)
\,,\qquad
 \pt = (p_x, p_y)
\,.\end{align}
Note that $\pt$ is treated as a standard two-dimensional vector in Euclidean space,
as opposed to the corresponding Minkowski vector $p_T^\mu = (0, p_x, p_y, 0)$.
This leaves an ambiguity in defining its magnitude, which we define as
\begin{align}
 p_T \equiv |\pt| = \sqrt{- p_T \cdot p_T} 
\,,\end{align}
where only the latter expression involves a scalar product in Minkowskian signature.

In light-cone coordinates, Lorentz-invariant scalar products take the simple form
\begin{align}
 p \cdot b = p^+ b^- + p^- b^+ - \pt \cdot \bt
\,,\qquad
 p^2 = 2 p^+ p^- - \pt^2 = 2 p^+ p^- - p_T^2
\,.\end{align}
These coordinates are particularly convenient to discuss energetic hadrons.
For example, the momenta $P_A$ and $P_B$ of the incoming protons in the Drell-Yan process are given by
\begin{align} \label{eq:PA_PB}
 P_A^\mu 
         = P_A^+ \left(1, e^{-2 Y_A} , \mathbf{0}_T \right)
\,,\qquad
 P_B^\mu 
         = P_B^- \left(e^{+2 Y_B} , 1, \mathbf{0}_T \right)
\,,\end{align}
where the components are $p^\mu =(p^+,p^-,\bf{p_T})$ and the proton rapidities are defined as
\begin{align}
 Y_A = \frac12 \ln\frac{P_A^+}{P_A^-} = \frac12 \ln\frac{2 (P_A^+)^2}{m_p^2}
\,,\qquad
 Y_B = \frac12 \ln\frac{P_B^+}{P_B^-} = \frac12 \ln\frac{m_p^2}{2 (P_B^-)^2}
\,.\end{align}
\eq{PA_PB} makes it clear that the momenta $P_{A,B}$ are aligned along the $n_{a,b}$ directions.
In the limit of taking the protons massless, $m_p \to 0$, we have $Y_{A,B} \to \pm \infty$
and the protons are exactly aligned along $n_{a,b}$.

Consider the production of a Drell-Yan pair $\ell^+\ell^-$ with total momentum $q^\mu$, and invariant mass $Q^2=q^2$.
Decomposing $q^\mu$ in light cone coordinates
we can then define the lepton pair's rapidity $Y$ and transverse momentum $q_T$ by 
\begin{align}
 Y=\frac12 \ln\Bigl(\frac{n_b\cdot q}{n_a\cdot q}\Bigr) 
  =\frac12 \ln\Bigl(\frac{q^+}{q^-}\Bigr)  \,,
 \qquad
 q_T^\mu = q^\mu - n_a^\mu n_b\cdot q - n_b^\mu n_a\cdot q
   = (0,q_T^x,q_T^y,0)
 \,.
\end{align}
We denote the Euclidean transverse momentum as $\qt$, and also use a plain $q_T$ to denote the magnitude of the Euclidean vector $q_T = |\qt|$, but write the magnitude squared as $\qt^2 = (q_T^x)^2+(q_T^y)^2$. (This avoids the notational issue of using $q_T^2$, which can be mistaken as a four-vector squared.)
We assume $Q^2\gg \Lambda_{\rm QCD}^2$, but for the transverse momentum we allow both $q_T\sim \Lambda_{\rm QCD}$ and $q_T\gg \Lambda_{\rm QCD}$. 
We decompose the cross section as
\begin{align} \label{eq:sigma}
 \frac{\df\sigma}{\df Q \df Y \df^2\qt} = \biggl( \frac{\df\sigma^{\rm W}}{\df Q \df Y \df^2\qt} + \frac{\df\sigma^{\rm Y}}{\df Q \df Y \df^2\qt} \biggr) \biggl[ 1 + \cO\biggl(\frac{\lqcd^2}{Q^2}\biggr)\biggr]
\,.\end{align}
Here, $\df\sigma^{\rm W}$ denotes the most singular part of the cross section, which dominates at small $q_T$. It is defined such that at any order in a strict $\alpha_s$ expansion it includes all terms that exhibit $1/q_T^2$ behavior as $q_T \to 0$. In practice this singular behavior is tamed by the resummation of large logarithms, see \chap{evolution}.
We use the superscript $W$ since these contributions are often referred to as the $W$ term. 
In contrast, the $Y$ term, denoted $\df\sigma^{\rm Y}$, are nonsingular terms that are suppressed by $\cO(q_T^2/Q^2)$ relative to $\df\sigma^{\rm W}$.%
\footnote{More generally, fiducial experimental cuts on the phase space of the final state leptons will induce linear $\cO(q_T/Q)$ corrections~\cite{Ebert:2019zkb}, but these can be computed with the same leading-power factorization technology~\cite{Ebert:2020dfc}.}
These nonsingular components of the cross section are necessary to reproduce the full results for the partonic cross sections $\diff\hat\sigma_{ij}/\diff Q\diff Y \df^2\qt$ in a fixed order $\alpha_s$ expansion, and are often referred to as the $Y$ term \cite{Collins:1981uk,Collins:1981va,Collins:1984kg}.
Methods for carrying out the resummation of large logarithms in $\df\sigma^{\rm W}$ sometime incorporate nonsingular terms, with a compensating modification to $\df\sigma^{\rm Y}$.
As indicated both contributions receive corrections in $\lqcd^2/Q^2$, analogous to the collinear factorization result in Eq.~\eqref{e.collexample}.
In this chapter, we only discuss $\df\sigma^{\rm W}$, whose factorization into a piece describing physics at the hard scale $Q$ and universal TMD PDFs describing physics at the low scale $q_T$ is well understood, and neglect corrections from $\df\sigma^{\rm Y}$. A dedicated discussion of the $\df\sigma^{\rm Y}$ contributions can be found in \sec{nonsingular}.

\index{TMD factorization!introduction}
TMD factorization was originally derived by Collins, Soper and Sterman (CSS)
in \cite{Collins:1981uk,Collins:1981va,Collins:1984kg}.
Refs.~\cite{Collins:1985ue,Collins:1988ig,Collins:1989gx,Collins:2011zzd,Diehl:2015bca} showed the cancellation of potentially factorization-violating Glauber modes,
and the factorization was further elaborated on and extended
in \cite{Catani:2000vq,deFlorian:2001zd,Catani:2010pd,Collins:2011zzd}.
It has also been considered in the framework of Soft-Collinear Effective Theory (SCET)
\cite{Bauer:2000ew, Bauer:2000yr, Bauer:2001ct, Bauer:2001yt}
by various authors \cite{Becher:2010tm, Becher:2011xn, Becher:2012yn, GarciaEchevarria:2011rb, Echevarria:2012js, Echevarria:2014rua, Chiu:2012ir}.
For a relation of the different approaches to each other, see e.g.~\cite{Collins:2017oxh, Ebert:2019okf}.
In the original formulation by Collins and Soper \cite{Collins:1981uk,Collins:1981va,Collins:1984kg}
and its modification by Ji, Ma and Yuan \cite{Ji:2004wu}, $\sigma^{\rm W}$ is written as
\begin{align} \label{eq:sigma_old}
 \frac{\df\sigma^{\rm W}}{\df Q \df Y \df^2\qt}
 &= \sum_{{\rm flavors}~i}\!\! H_{i\bar i}(Q^2,\mu;\rho) \int\! \df^2\bt \, e^{i \bt \cdot \qt} \,
   \tilde f_{i/p}(x_a, \bt, \mu, \tilde\zeta_a; \rho)\,
   \tilde f_{\bar i/p}(x_b, \bt, \mu, \tilde\zeta_b; \rho)
\,,\end{align}
where the $\tilde\zeta$ and $\rho$ variables are discussed below. 
In the modern definition by Collins \cite{Collins:2011zzd}, which yields a factorization theorem that is equivalent to many SCET based definitions \cite{Becher:2010tm, Becher:2011xn, Becher:2012yn, GarciaEchevarria:2011rb, Echevarria:2012js, Echevarria:2014rua, Chiu:2012ir,Li:2016axz}, the singular cross section can be written as
\begin{subequations} \label{eq:sigma_new}
\begin{align}
 \frac{\df\sigma^{\rm W}}{\df Q \df Y \df^2\qt} &
 = \sum_{{\rm flavors}~i}\!\! H_{i\bar i}(Q^2,\mu) \int\! \df^2\bt \, e^{i \bt \cdot \qt} \,
   \tilde f_{i/p}(x_a, \bt, \mu, \zeta_a) \,
   \tilde f_{\bar i/p}(x_b, \bt, \mu, \zeta_b)
\label{eq:sigma_new_a}
\\&
 = \sum_{{\rm flavors}~i}\!\! H_{i\bar i}(Q^2,\mu) \int\! \df^2\bt \, e^{i \bt \cdot \qt} \,
   \tilde B_{i/p}(x_a, \bt, \mu, \zeta_a/\nu^2)\,
    \tilde B_{\bar i/p}(x_b, \bt, \mu, \zeta_b/\nu^2)
   \nn\\&\hspace{5.5cm}\times
   \tilde S_{n_a n_b}(b_T,\mu,\nu)
\label{eq:sigma_new_b}
\,.\end{align}
\end{subequations}
In \eqs{sigma_old}{sigma_new} we use the notation $f=f_1$ and $\tilde f=\tilde f_1$ for the unpolarized TMD PDF, and this should be understood as the case in what follows, unless otherwise indicated.
We will start by describing the most important ingredients common to Eqs.~(\ref{eq:sigma_old},\ref{eq:sigma_new_a},\ref{eq:sigma_new_b}), and then return to comparisons between these three equivalent expressions for the cross section.
In both \eqs{sigma_old}{sigma_new}, the factorization is written in Fourier space, with $\bt$ being Fourier-conjugate to the measured transverse momentum $\qt$, and in both cases the hard function $H_{i\bar i}$ encodes virtual corrections to the underlying hard process $q_i q_{\bar i} \to \gamma^*/Z \to l^+ l^-$, with the quark flavors $i,\bar i$ being summed over. Here $i$ is a quark flavor and $\bar i$ is the charge conjugate of $i$, since other flavor combinations and cases involving gluons occur only in $\diff\sigma^Y$.
Note that, whenever possible, we will neglect target mass corrections from $m_P^2\ll Q^2$, together with other $\LQCD^2/Q^2$ power corrections.

Compared to our parton model discussion in \eq{tmdexample}, the TMD PDFs in \eq{sigma_new} have dependence on two additional variables, the renormalization scale $\mu$ and Collins-Soper scales $\zeta_{a,b}$~\cite{Collins:1981uk,Collins:1981va}. These dependences arise from defining the renormalized TMD PDFs in quantum field theory, while being careful about the treatment of rapidity dependence. A more detailed discussion of the relation between bare and renormalized TMD PDFs is given below in \sec{tmdpdfs_new}, while methods of handling rapidity divergences that appear in intermediate steps of the TMD PDF definitions, and which are related to the appearance of $\zeta_{a,b}$, are treated in \sec{tmd_defs}.
The dependences of the TMD PDFs on 
both $\mu$ and $\zeta$ are governed by evolution equations, which are discussed in \sec{evolution}.  In particular this enables a TMD PDF $f_{i/H}(x,\bt,\mu_0,\zeta_0)$ to be evolved from initial scales $\mu_0$ and $\zeta_0$ to final scales $\mu$ and $\zeta$, yielding $f_{i/H}(x,\bt,\mu,\zeta)$. In this context the scales appearing in \eq{sigma_new} can be interpreted as the final scales after this evolution. Taking a $\mu\sim Q$ then minimizes large logarithms in $H_{i\bar i}(Q,\mu)$. 
Likewise, the final Collins-Soper scales $\zeta_{a,b}$ are given by
\begin{align} \label{eq:zeta}
 \zeta_a &= 2(x_a P_A^+)^2 e^{-2y_n} = x_a^2 m_p^2 e^{2(Y_A-y_n)}
 \,, \qquad
 \zeta_b = 2(x_b P_B^-)^2 e^{2y_n}= x_b^2 m_p^2 e^{-2(Y_B-y_n)}
\,,\end{align}
such that their product yields the invariant mass of the hard process,
\begin{align} \label{eq:zetaazetab}
 \zeta_a \zeta_b &= (2 x_a x_b P_A^+ P_B^-)^2 = Q^4
\,.\end{align}
Here $2 P_A^+ P_B^- \approx (P_A + P_B)^2 = s$ is the center-of-mass energy of the proton-proton collision,
while $Y_A$ and $Y_B$ are the rapidities of the two protons (which are equal in the center-of-momentum frame, $Y_A=Y_B=y_P$).
The rapidity variable $y_n$ in \eq{zeta} controls an additional scheme dependence which cancels between the two TMD PDFs.
While this allows one to derive evolution equations with respect to $\zeta_{a,b}$, there does not appear to be a great benefit from exploiting the $y_n$ dependence otherwise, and often the simplest choice $y_n=0$ is adopted.

In \eqs{sigma_old}{sigma_new_a}, the result is written in terms of renormalized TMD PDFs $f_{i/p}$ and $f_{\bar i/p}$ which give rise to the transverse momentum $\qt$. In contrast, in \eq{sigma_new_b} it arises from three renormalized functions, namely two beam functions $B_{i/p}$ and $B_{\bar i/p}$~\cite{Stewart:2009yx} which describe collinear radiation close to the proton, and the soft function $S_{n_a n_b}$ encoding soft exchange between the colliding partons $i$ and $\bar i$, but which is independent of the quark flavors $i$ and $\bar i$.
The two cases can be trivially related through
\index{TMD parton distribution function!quark definition}
\begin{align} \label{eq:fisBS}
 \tilde f_{i/p}(x,\bt,\mu,\zeta) = \tilde B_{i/p}(x,\bt,\mu,\zeta/\nu^2) \sqrt{\tilde S_{n_a n_b}(b_T,\mu,\nu)}
\,,\end{align}
where the so-called rapidity renormalization scale $\nu$ cancels between $B_{i/p}$ and $S_{n_a n_b}$.
The relation in \eq{fisBS} is common in the SCET based constructions in Refs.~\cite{Becher:2010tm, Becher:2011xn, Becher:2012yn, GarciaEchevarria:2011rb, Echevarria:2012js, Echevarria:2014rua, Chiu:2012ir,Li:2016axz}, since renormalized beam and soft functions are constructed before combining them into renormalized TMD PDFs. In contrast, in the modern Collins construction~\cite{Collins:2011zzd} only the bare analogs of $B_{i/p}$ and $S_{n_a n_b}$ appear, which are directly used to define the renormalized TMD PDFs. Further discussion of these various constructions can be found in \sec{tmd_defs}.
In the discussion in \sec{tmdpdfs_new} we will treat \eqs{sigma_new_a}{sigma_new_b} on the same footing.

Finally, we return to discussing the differences between \eqs{sigma_old}{sigma_new}.
Here the crucial difference is the scheme employed for the hard function $H_{i\bar i}$, which from the scheme independence of $\df\sigma^{\rm W}$ automatically defines the scheme for the product of the two $f_{i/p}$s, or the product of the two $B_{i/p}$s and $S_{n_a n_b}$. Thus two categories of definitions of TMD PDFs can be identified according to the definition of $H_{i\bar i}$ appearing in their associated factorization theorems.
In \eq{sigma_new}, $H_{i\bar i}$ is defined purely in the $\MSbar$ scheme, and thus only depends on the hard scale $Q$ and the renormalization scale $\mu$. In particular, this $H_{i\bar i}(Q,\mu)$ can be computed by a partonic form factor calculation in dimensional regularization with $d=4-2\eps$, by simply using $\MSbar$ subtractions for $1/\eps$ poles. We will refer to approaches that fit within this framework as the $\MSbar$ class of schemes. In contrast, in \eq{sigma_old} $H_{i\bar i}(Q,\mu;\rho)$ also depends on an additional rapidity scale $\rho$, and the TMDs use different definitions for the Collins-Soper scales, which are therefore denoted by $\tilde\zeta_{a,b}$.

In the following we will focus on the most popular TMD PDF schemes, corresponding to the category involving the $\MSbar$ hard function $H_{i\bar i}(Q,\mu)$. This includes the discussion in Secs.~\ref{sec:tmdpdfs_new} and \ref{sec:tmd_defs}. A complete description of the alternate schemes involving $H_{i\bar i}(Q,\mu;\rho)$, including definitions of the variables $\rho$, $\tilde\zeta_{a,b}$, will then be taken up in \sec{other_tmd_defs}.

\subsection{Basic Definition of TMD PDFs}
\label{sec:tmdpdfs_new}

The goal of this section is to provide basic rigorous field theory definitions of the TMD PDFs with an emphasis on aspects that are universal across all approaches to handling issues associated with regulating so-called rapidity singularities, leaving differences to the discussion in \sec{tmd_defs}. We focus here entirely on constructions which yield the factorization theorem in \eq{sigma_new}. 
  
The construction of complete TMD definitions is driven by the following 
constraints~\cite{Collins:2011zzd}, which all of the definitions discussed here will satisfy (up to any exceptions which we will note explicitly):
\begin{enumerate}
\item The definition should follow from, and be constrained by, the steps needed to factorize a class 
of physical processes. In our case, this includes at least those processes for which 
TMD factorization theorems are most easily derivable: 
\begin{itemize}
\item $e^+ e^-$-annihilation into a pair of nearly back-to-back hadrons.
\item hadron-hadron production of lepton pairs (Drell-Yan scattering) or weak bosons.
\item Semi-inclusive deep inelastic scattering.
\end{itemize}
\item It should work at both perturbative and nonperturbative levels. Namely, it should be possible to 
use it with nonperturbative models of partons, and provide rigorous connections of the definition to calculations done with fundamental nonperturbative methods like lattice QCD.
\item Gauge invariance should be preserved, ideally before regulators (UV or rapidity) are removed.
\item Unphysical contributions not present in the 
unfactorized physical processes should cancel naturally in the definition. 
These include, for example, 
Wilson line self energies or interactions with the Wilson line at $\infty$~\cite{Collins:2008ht}.
\item Renormalization is multiplicative and evolution equations are exactly homogeneous in the power expansion which yields \eqs{sigma_old}{sigma_new}. 
\item A final practical consideration is that definitions should simplify multi-loop fixed order partonic calculations, in order to make it easier to build in the transition between the nonperturbative and perturbative $k_T$ regimes.
\end{enumerate}      
In some cases these conditions may clash, such as 2. and 6., which makes apparent the importance of having available multiple constructions that can be demonstrated to be equivalent for the final TMD PDFs.      

\index{TMD parton distribution function!quark definition}
\index{TMD soft function}
The small transverse momentum described by a TMD PDF $f_{i/p}$ arises from two physical sources.
Firstly, it arises from energetic radiation close to each proton, which is described by a proton matrix element, which is equivalently referred to as either an unsubtracted TMD PDF $f_{i/p}^\unsub$ or as an unsubtracted beam function.
Secondly, one has to consider soft exchange between the two partons $i$ and $j$ involved in the hard collision,
which is encoded in a soft vacuum matrix element $S_{n_a n_b}$.
Unlike in the inclusive factorization theorem leading to Eq.~(\ref{e.collexample}), these soft radiation effects do not cancel out, and encode important eikonal soft dynamics between the two directions defined by identified hadrons. 
In practice, there can also be a double counting between the two matrix elements, which is removed by dividing by a soft subtraction factor $S_{n_a n_b}^{\subt}$.
As indicated by the notation, this factor is closely related to the soft function itself.\footnote{In the approach of CSS~\cite{Collins:1982wa} and Collins~\cite{Collins:2011zzd} these subtractions ensure there is no double counting of momentum regions, and also the proper cancellation of singularities. In SCET these subtractions are known as zero-bin subtractions~\cite{Manohar:2006nz} and arise from ensuring fluctuations encoded by collinear fields do not have singular overlap with those of the soft fields, thus also ensuring there is no double counting of infrared regions.}
The generic definition for a TMD PDF can thus be written as 
\begin{align} \label{eq:tmdpdf_1}
 \tilde f_{i/p}(x, \bt, \mu , \zeta) &
 =  \lim_{\substack{\eps\to 0 \\ \tau\to 0}} Z_{\rm uv}^i(\mu,\zeta,\eps) \,
    \frac{\tilde f_{i/p}^{0\,\unsub}\bigl(x, \bt, \eps, \tau, x P^+ \bigr)}{\tilde S^{0\,\subt}_{n_a n_b}(b_T,\eps,\tau)}
    \sqrt{\tilde S_{n_a n_b}^{0}(b_T,\eps,\tau)}
\,.\end{align}
Here, the superscript $^0$ denotes that the functions on the right hand side of \eq{tmdpdf_1} are bare quantities. They suffer from both ultraviolet (UV) divergences, which can be regulated using dimensional regularization with $d=4-2\eps$ dimensions, and so-called rapidity divergences which require a dedicated regulator~ \cite{Soper:1979fq,Collins:1981uk,Collins:1992tv,Collins:2008ht,Becher:2010tm,GarciaEchevarria:2011rb,Chiu:2011qc,Chiu:2012ir}, which in \eq{tmdpdf_1} is generically denoted as $\tau $. The rapidity divergences cancel between the various factors in the right hand side of \eq{tmdpdf_1}, such that the renormalization counterterm $Z_{\rm uv}^i$ in \eq{tmdpdf_1} only subtracts divergences in $\eps$.
As usual, the UV divergences give rise to the renormalization scale $\mu$, which is defined in the $\overline{\rm MS}$ scheme. Likewise, the rapidity divergences give rise to sensitivity to the Collins-Soper scale $\zeta$~\cite{Collins:1981va,Collins:1981uk}, whose precise definition depends on the employed regulator $\tau$.
Note that in the definition in \eq{tmdpdf_1}, one effectively absorbs half of the soft function into the TMD PDF $f_{i/p}$, while the other half is absorbed into the TMD PDF $f_{j/p}$ for the other proton.
The ratio $f_{i/p}^{0\,\unsub} / S^{0\,\subt}_{n_a n_b}$ is constructed such that it describes collinear radiation, as discussed above.

\index{TMD parton distribution function!beam function}
\index{TMD soft function}
Before giving explicit definitions of the functions in \eq{tmdpdf_1}, we briefly connect to TMD factorization as given in \eq{sigma_new_b}. In this approach, one separately constructs renormalized beam and soft functions, which can either be used to directly give the cross section (or combined as in \eq{fisBS} to give the TMD PDF). In this case, one has to renormalize both UV and rapidity divergences, which is achieved through
\begin{align} \label{eq:B_renorm}
 \tilde B_{i/p}(x, \bt, \mu, \zeta/\nu^2) 
 &= \lim_{\substack{\eps\to 0 \\ \tau\to 0}}
   \tilde Z_B^i(b_T, \mu, \nu, \eps, \tau, x P^+)\, 
   \tilde B_{i/p}^{0}\bigl(x, \bt, \eps, \tau, x P^+ \bigr)
 \nn\\&
 = \lim_{\substack{\eps\to 0 \\ \tau\to 0}} 
   \tilde Z_B^i(b_T, \mu, \nu, \eps, \tau, x P^+) \,
   \frac{\tilde f_{i/p}^{0\,\unsub}\bigl(x, \bt, \eps, \tau, x P^+ \bigr)}{\tilde S^{0\,\subt}_{n_a n_b}(b_T,\eps,\tau)}
\,, \\ \label{eq:S_renorm}
 \tilde S_{n_a n_b}(b_T,\mu,\nu) 
  &= \lim_{\substack{\eps\to 0 \\ \tau\to 0}}
   \tilde Z_S(b_T, \mu, \nu, \eps, \tau)\,
  \tilde S_{n_a n_b}^{0}(b_T,\eps,\tau)
\,.\end{align}
Here, $\nu$ is the rapidity renormalization scale arising from subtracting poles in $\tau$.
The TMD PDF obtained by combining collinear and soft matrix elements can be equivalently defined from \eqs{B_renorm}{S_renorm} as
\begin{equation} \label{eq:fBS_relation}
 \tilde f_{i/p}(x,\bt,\mu,\zeta) = \tilde B_{i/p}(x,\bt,\mu,\zeta/\nu^2) \sqrt{\tilde S_{n_a n_b}(b_T,\mu,\nu)}
\,.\end{equation}
Here, the $\nu$ dependence cancels between both functions, leaving only the Collins-Soper scale $\zeta$,
whose precise definition again depends on the definition of the regulator $\tau$ and thus is scheme dependent (cf. the $y_n$ dependence in \eq{zeta}).

\begin{figure}[pt]
 \centering
\includegraphics[width=0.45\textwidth]{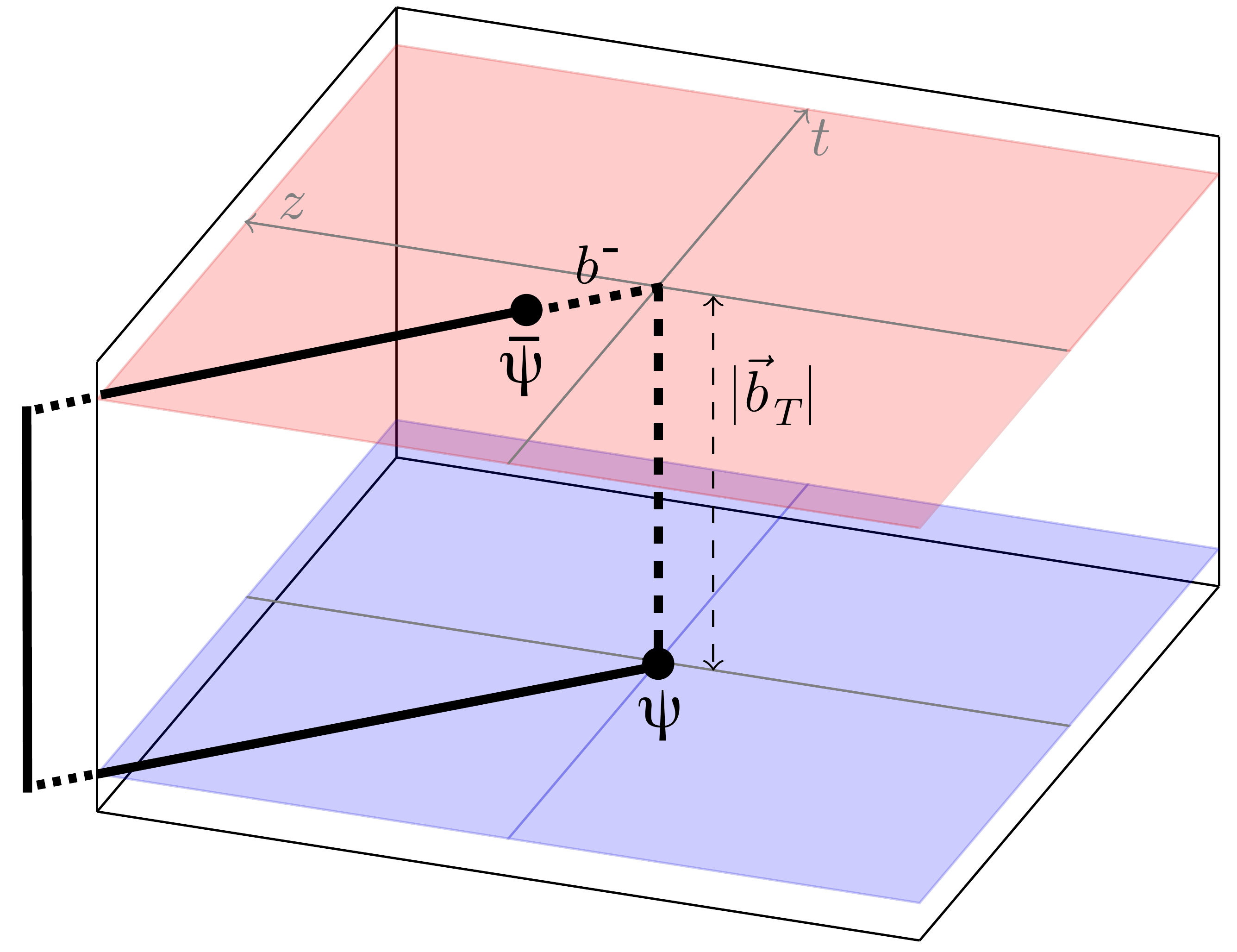} 
  \hfill
\includegraphics[width=0.45\textwidth]{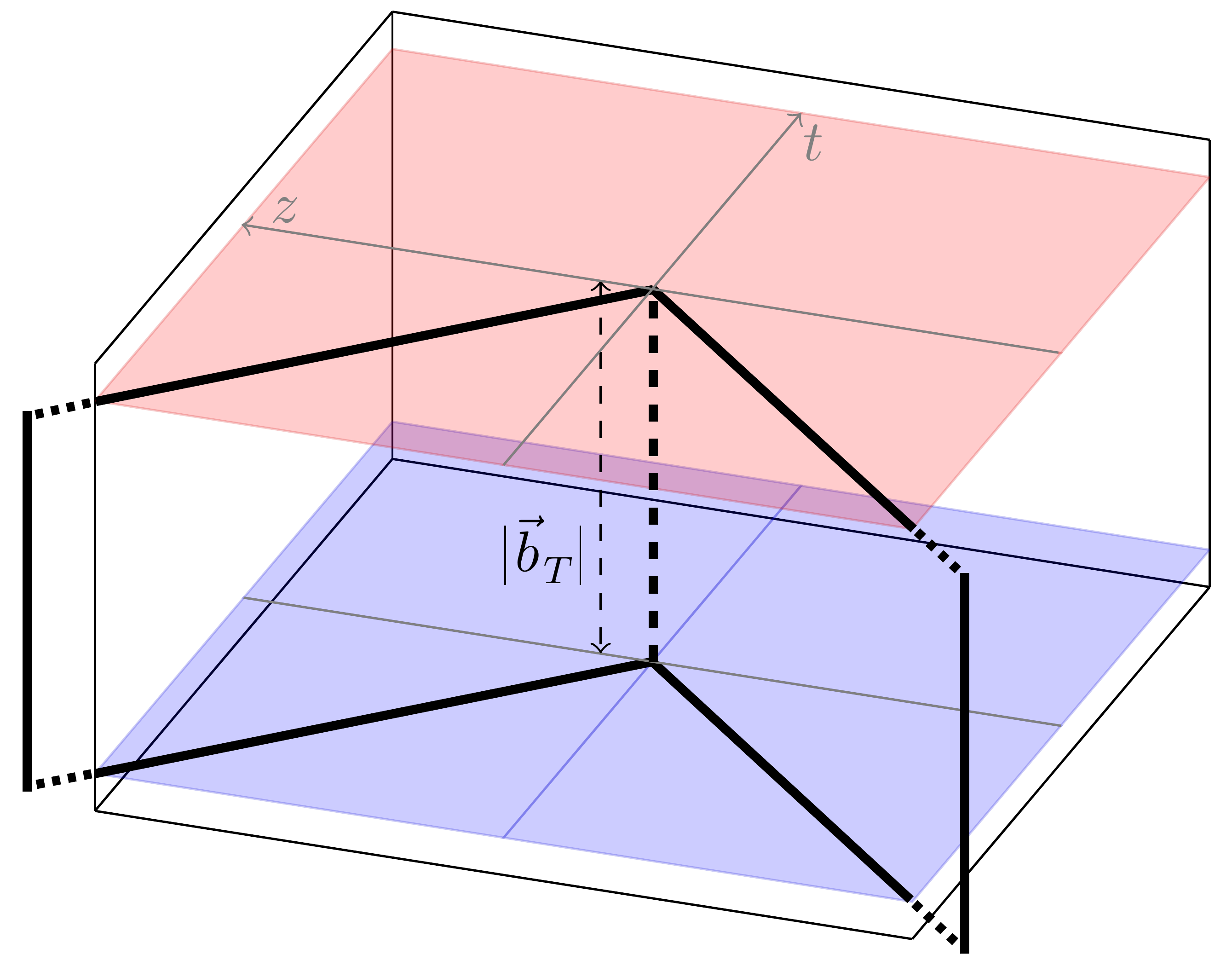}
 \caption{Graphs of the Wilson line structure $W_{\sqsubset}(b^\mu,0)$ of the unsubtracted TMD PDF $f_{i/p}^{0\,\unsub}$ (left)
 and of $W_{\softstaple}(b_T)$ for the soft function $S^{0}_{n_a n_b}$ (right), defined in \eqs{beamfunc}{softfunc}.
 The Wilson lines (solid) extend to infinity in the directions indicated.
 Adapted from \cite{Li:2016axz}. }
 \label{fig:wilsonlines}
\end{figure}

We now give explicit definitions of the proton and soft matrix elements relevant for quark TMD PDFs.
We consider a proton $p$ moving close to the $n_a^\mu = (1,0,{\bf 0}_T)$ direction with momentum $P^\mu = P^+ (1, e^{-2y},{\bf 0}_T)$.
The corresponding definitions for a proton moving along the $n_b^\mu = (0,1,{\bf 0}_T)$ direction are obtained by exchanging $n_a \leftrightarrow n_b$. 
The bare unsubtracted TMD PDF (or equivalently the bare beam function) and the bare soft function are defined as
\begin{align} \label{eq:beamfunc}
 \tilde f_{i/p}^{0\,\unsub}(x,\bt,\eps,\tau,x P^+) 
 &= \int\frac{\df b^-}{2\pi} e^{-\img b^- (x P^+)}
 \Bigl< p(P) \Bigr|  \Bigl[ \bar \psi^{0}_i(b^\mu)
 W_{\sqsubset}(b^\mu,0)  \frac{\gamma^+}{2}
  \psi^{0}_i(0) \Bigr]_\tau \Bigl| p(P) \Bigr>
,\\ \label{eq:softfunc}
 \tilde S^{0}_{n_a n_b}(b_T,\eps,\tau) &= \frac{1}{N_c} \bigl< 0 \bigr| {\rm Tr} \bigl[ W_{\softstaple}(b_T)\bigr]_\tau \bigl|0 \bigr>
\,.\end{align}
Here the brackets $[\cdots]_\tau$ denote that the operators inside are considered with an additional rapidity regulator $\tau$, where the details on methods for how this is done are left to \sec{tmd_defs} below.
Note that by Poincar\'e invariance, the proton matrix element in \eq{beamfunc} only depends on the difference $b^\mu - 0 = b^\mu$
of the positions of the quark fields. In parts of the literature, the correlator is defined as
$\bar \psi^{0}_i(0)  W_{\sqsubset}(0, b^\mu)  \frac{\gamma^+}{2} \psi^{0}_i(b^\mu)$,
which thus is related to our convention by $b^\mu \to -b^\mu$. In particular, this also reverses the sign in the Fourier transform.

\index{Wilson lines}
In \eqs{beamfunc}{softfunc} we have $b^\mu = (0, b^-, \bt)$, and the staple shaped Wilson lines $W_{\sqsubset}(b^\mu,0)$ and $W_{\softstaple}(b_T)$ are defined by products of straight line segments,
\begin{align} 
 \label{eq:Staple_Wilson_line}
 W_{\sqsubset}(b^\mu,0) 
 &= W[0 \to -\infty n_b \to  -\infty n_b + \bt \to b]
 \nn \\ 
 &= W_{n_b}(b^\mu;-\infty,0)
  W_{\hat b_T}\!\bigl(-\infty n_b; 0, b_T\bigr)
  W_{n_b}(0^\mu;0,-\infty)
\,,  \\[5pt]
 W_{\softstaple}(b_T)
   &= W[0 \to -\infty n_b \to -\infty n_b + \bt \to \bt \to -\infty n_a + \bt \to -\infty n_a
      \to 0 ]
    \nn \\ 
   &= W_{n_a}(b_T;0,-\infty) W_{n_b}(b_T;-\infty,0) 
   W_{\hat b_T}(-\infty n_b; 0,b_T)
   \nn\\&\quad\times
   W_{n_b}(0;0,-\infty) W_{n_a}(0;-\infty,0)
   W_{\hat b_T}(-\infty n_a; b_T,0)
\label{eq:Soft_Wilson_Loop}
\,,\end{align}
with $\hat b_T^\mu = b_T^\mu/b_T$. 
For later use we also define a generalized version of the first product of Wilson lines, where we take $x^\mu=(0,x^-,{\bf x}_T)$ and $y^\mu=(0,y^-,{\bf y}_T)$ as the two endpoints,
\begin{align} \label{eq:Staple_Wilson_line_gen}
 W_{\sqsubset}(x^\mu,y^\mu) 
 &= W[x\to -\infty n_b+x \to -\infty n_b + y \to y ]
 \nn\\
 &= 
  W_{n_b}(x^\mu;-\infty,0)
  W_{\hat\Delta}\!\bigl(-\infty n_b^\mu+y_T^\mu; 0, |{\bf x}_T-{\bf y}_T|\bigr)
  W_{n_b}(y^\mu;0,-\infty)
 \,,
\end{align}
and here $\hat\Delta^\mu = (x_T-y_T)^\mu/|{\bf x}_T-{\bf y}_T|$.
Here the Wilson line along a generic path $\gamma$ is defined by the path-ordered exponential
\begin{align}
	W[\gamma]
	= P \exp\biggl[ -\img g_0 \int_{\gamma} \df x^\mu A^{c0}_\mu(x)\, t^c\biggr]
	\,,
\end{align}
where $t^c$ are the generators of $\mathrm{SU}(3)$ in the fundamental representation.
The individual Wilson lines $W_n(x;a,b)$ are defined as path-ordered exponentials connecting the point $x^\mu+ a n^\mu$ to $x^\mu+b n^\mu$ along the direction $n$,
\begin{align}  \label{eq:Wilson_lines}
 W_n(x^\mu; a,b) &= P \exp\biggl[ -\img g_0 \int_a^b \df s \, n \cdot A^{c0}(x^\mu + s n^\mu) t^c\biggr]
\,.
\end{align}
Note that for $W_n$ the subscript $n$ is always a four vector. 
Also note that here $W[\gamma]$ and $W_n$ are defined using the bare strong coupling $g_0$ and bare gluon fields $A^{c0}$. 
The $P$ in $W_n(x^\mu;a,b)$ denotes path ordering for the expanded exponential, where the matrices $t^c$ are ordered by their corresponding values along the path from $s=a$ to $s=b$, starting from right and going to the left.
For reference we note that $W_n(x^\mu; c,b) W_n(x^\mu; a,c) = W_n(x^\mu; a,b)$, and since $W_n^\dagger(x^\mu; a,b) = W_n(x^\mu; b,a)$, we have $W_n^\dagger(x^\mu;a,b) W_n(x^\mu;a,b) = 1$. 
The Wilson line structures appearing in the unsubtracted TMD PDF and soft function are illustrated in \fig{wilsonlines}.  
Note that the transverse Wilson lines $W_{b_T}$ follow a straight line path in the transverse plane at light-cone $-\infty$ as indicated. These segments are needed for the full gauge invariance of the operators. (In perturbative calculations the transverse links can be neglected in nonsingular gauges such as Feynman gauge, where the gluon field strength vanishes at infinity, but are known to be important in certain singular gauges, see~\cite{Ji:2002aa,Belitsky:2002sm,Idilbi:2010im,GarciaEchevarria:2011md}.)

On first encountering the operator definition of the unsubtracted TMD PDF, a reader may be struck by the fact that the simplest straight Wilson line between two points is not what appears in $f_{i/p}^{0\,\unsub}$. Instead, the proof of factorization that leads to these objects dictates that it is the staple shaped Wilson line in \eq{beamfunc} which connects the quark fields at the two different spacetime points and ensures gauge invariance of the composite operator. To understand physically why it is this staple shape that appears, we note that having a quark field whose color is transported off to spacetime infinity by a Wilson line along the light-cone is the closest approximation to a parton by operators in QCD.  This concept is made very explicit in the construction of the soft collinear effective theory~\cite{Bauer:2000yr,Bauer:2001ct}, where the composite operator obtained by attaching a quark to this type of Wilson line plays a fundamental role. Thus the staple shaped Wilson line is the natural QCD consequence of taking partons at the spacetime points $0$ and $b^\mu$ and connecting them by a transverse Wilson line to obtain a gauge invariant operator.

\subsection{Definitions with Rapidity Regulators}
\label{sec:tmd_defs}

In the previous section, the basic definition of TMD PDFs has been given.
As indicated there, definitions of TMD PDFs not only require the specification of a UV renormalization scheme,
which we take to be the standard $\MSbar$ scheme, but also to define an additional rapidity regularization for individual ingredients.
A large variety of such rapidity regulators have been suggested in the literature,
giving rise to various different constructions of TMD factorization.
Here, we will briefly summarize these definitions for TMD PDFs in the $\MSbar$ class of schemes,
while TMD PDF defined in additional schemes will be discussed in \sec{other_tmd_defs}.
For more details on results with the different rapidity regulators, we refer to appendix \ref{app:TMDdefn}.

The origin of rapidity divergences is intimately connected to the derivation of TMD factorization,
which will be discussed in more detail in \sec{Factorization}.
Roughly speaking, TMD factorization is based on organizing the cross section into hard, collinear and soft regions.
At the perturbative level, this corresponds to expanding Feynman diagrams in these regions.
For example, one would expand
\begin{align} \label{eq:rapidity_example}
 \underbrace{\int_{q_T}^Q \frac{\df k}{k}}_{\mathrm{full}}
 = \lim_{\tau\to0} \biggl[ \underbrace{\int_{0}^Q \frac{\df k}{k} R_c(k,\tau)}_{\mathrm{collinear}}
   + \underbrace{\int_{q_T}^\infty \frac{\df k}{k} R_s(k,\tau)}_{\mathrm{soft}} \biggr]
 = \ln\frac{Q}{q_T}
\,.\end{align}
Here, the full theory contains the integral on the left hand side of \eq{rapidity_example}.
In the collinear region, one expands away the transverse momentum $q_T$, which is considered small compared to $Q$, $q_T \ll Q$, while in the soft region the large momentum $Q \to \infty$ is expanded away. These expansions are unavoidable, since they are necessary to derive the factorization theorem.
This renders the collinear and soft integrals logarithmically divergent, and their separate evaluation requires the introduction of a regulating function $R(k,\tau)$, for which $R_c(k,\tau)$ and $R_s(k,\tau)$ are the versions appropriate for the collinear and soft calculations. 
Upon combining the two contributions, one can remove the regulator, $\tau\to0$, and obtain the correct final result.

In the following, we first give an overview of the rapidity regulators employed in the literature,
before explicitly illustrating the application of such a regulator at one loop.

\subsubsection{Overview of rapidity regulators}
\label{sec:tmd_defs_overview}

\index{rapidity regulator}
Here, we collect key properties of the different rapidity regulators encountered in the literature.
The different notations for the rapidity-regularized unsubtracted TMD PDF and soft functions in each case are summarized in table \ref{tbl:overview_tmdpdfs}.
We also collect explicit expressions for the corresponding one-loop results of the quark TMD PDF in \app{TMDdefn},
which explicitly illustrates their equivalence.
\begin{itemize}
 \item \textbf{Space-like Wilson-lines}:
       The modern definition by Collins~\cite{Collins:2011zzd} plays a key role in the all order proof of TMD factorization discussed in \sec{Factorization}. Here the lightlike directions $n_a$ and $n_b$, for the paths of the Wilson lines in the definitions in \eqs{beamfunc}{softfunc}, are replaced by spacelike reference vectors
	\begin{align} \label{eq:Collins_rap}
		& n_a^\mu \to n_A^\mu(y_A) \equiv n_a^\mu - e^{-2 {y_A}} n_b^\mu = (1,-e^{-2 y_A},{\bf 0}_T)
		\,,\nn\\
		& n_b^\mu \to n_B^\mu(y_B) \equiv n_b^\mu - e^{+2 {y_B}} n_a^\mu = (-e^{+2 y_B},1,{\bf 0}_T)
   	  \,,
    \end{align}
    which ensures maximum universality for the TMD PDF definitions~\cite{Collins:2011zzd,Aybat:2011zv}.  With this rapidity regulator the limit $\tau\to 0$ corresponds to $y_A\to \infty$ and $y_B\to -\infty$. The ratio of bare soft function and soft subtractions appearing in \eq{tmdpdf_1} is given by~\cite{Collins:2011zzd}
	\begin{align}
	  \frac{\sqrt{\tilde S^0_{\rm JC}}}{\tilde S^{0\subt}_{\rm JC}} &=
	    \sqrt{\frac{\tilde S^{0}_{n_A(y_A) n_B(y_n)}(b_T,\eps, y_A-y_n)}{\tilde S^{0}_{n_A(y_A) 	n_B(y_B)}(b_T,\eps,y_A-y_B) \tilde S^{0}_{n_A(y_n) n_B(y_B)}(b_T,\eps, y_n-y_B)}}
 	   \,.
	\end{align}
	Here, the additional rapidity $y_n$ governs how the split of the soft function is made when combining it with each of the two TMDs appearing in the factorized cross section. 
	This combination can be simplified into a single Wilson line using~\cite{Buffing:2017mqm}
	\begin{align}\label{eq:ColinsSoft}
        \lim_{y_A\to\infty} \frac{\sqrt{\tilde S^0_{\rm JC}}}{\tilde S^{0\subt}_{\rm JC}}
      &= \frac{1}{\sqrt{\tilde S^{0}_{n_A(2y_n) n_B(2y_B)}(b_T,\eps, 2y_n-2y_B)}}
      \,.
	\end{align}
    In addition, the rapidity regulator is implemented in the bare unsubtracted TMD PDF by replacing $n_b\to n_B(y_B)$ in the staple shaped Wilson line, so that $[\cdots]_\tau$ in \eq{beamfunc} is enforced by using
	\begin{align} \label{eq:Staple_Wilson_line_JC}
	  W_{\sqsubset}^{n_B}(b^\mu,0) &= 
	  W_{n_B}(b^\mu;-\infty,0)
	  W_{b_T}\!\bigl(-\infty n_B; 0, 1\bigr)
	  W_{n_B}(0;0,-\infty)
	  \,,
	\end{align}
    and this then gives $\tilde f_{i/p}^{0\,\unsub}(x,\bt,\eps,y_B,x P^+)$.
    The final renormalized TMD PDF is then obtained as
       \begin{align} \label{eq:CSf}
        \tilde f_{i/p}(x, \bt, \mu, \zeta) &
        = \lim_{\eps\to0} Z_{\rm uv}(\mu,\zeta,\eps) \lim_{y_B \to -\infty}
          \frac{\tilde f_{i/p}^{0\,\unsub}(x,\bt,\eps,y_B,x P^+)}{\sqrt{\tilde S^{0}_{n_A(2y_n) n_B(2y_B)}(b_T,\eps, 2y_n-2y_B)}}
       \,,\end{align}
     and here the Collins-Soper scale arises as $\zeta =2 (x P^+ e^{-y_n})^2$.
 \item The \textbf{\boldmath $\delta$ regulator}
       was introduced in \cite{GarciaEchevarria:2011rb,Echevarria:2012js} by Echevarria, Idilbi and Scimemi (EIS) and later modified in \cite{Echevarria:2015usa,Echevarria:2015byo,Echevarria:2016scs}.
       The regulator modifies the Feynman rules of the Wilson lines $W_{n_a}$ and $W_{n_b}$. At one loop this simply shifts the eikonal propagators $(n_a\cdot k + i0)$ and $(n_b\cdot k+i0)$ by an infinitesimal amount $i \delta^+$ and $\img \delta^-$, respectively.
       In this scheme the bare regulated soft function is split into two parts to associate with the two TMD PDFs as
       \begin{align}
        \tilde S^0_{\rm EIS}(b_T, \eps, \sqrt{\delta^+ \delta^-})
        = \sqrt{ \tilde S^0_{\rm EIS}(b_T,\eps, \delta^+ e^{-y_n})} \sqrt{ \tilde S^0_{\rm EIS}(b_T,\eps, \delta^- e^{+y_n})} \,,
       \end{align}
       where $y_n$ regulates the amount of the soft function combined with each of the unsubtracted TMD PDFs.
       Here the subtraction factor is equal to the corresponding soft function component, $\tilde S^{0\subt}_{\rm EIS}(b_T,\eps,\delta^+e^{-y_n})= \tilde S^{0}_{\rm EIS}(b_T,\eps,\delta^+e^{-y_n})$ so the physical TMD PDF is obtained as
       \begin{align}
         \tilde f_{i/p}(x, \bt, \mu, \zeta) &
         =  \lim_{\substack{\eps\to 0 \\ \delta^+\to 0}} Z_{\rm uv}^i(\mu,\zeta,\eps) \,
            \frac{\tilde f_{i/p}^{0\,\unsub}\bigl(x, \bt, \eps, \delta^+/(x P^+) \bigr)}
                  {\sqrt{ \tilde S_{{\rm EIS}}^{0}(b_T,\eps, \delta^+ e^{-y_n})}}
       \,.\end{align}
        The Collins-Soper scale in this approach is given by $\zeta =2 (x P^+ e^{-y_n})^2$, which is the same as in the modern Collins construction.
 \item The \textbf{\boldmath $\eta$ regulator} due to Chiu, Jain, Neill and Rothstein (CJNR) in \cite{Chiu:2011qc,Chiu:2012ir} separately modifies the Feynman rules of Wilson lines appearing in the unsubtracted TMD PDF and soft function. 
 It introduces regulating factors of $|k^+/\nu|^{-\eta}$ in the Wilson lines appearing in $W_{\sqsubset}(b^\mu,0)$, and regulating factors $|k^z/\nu|^{-\eta/2}$ in the Wilson lines appearing in $S_{n_a n_b}$.
 Amplitudes are expanded in the limit $\eta\to 0$, and
       Rapidity divergences become manifest as poles in $\eta$, similar to UV divergences that arise as poles in $\eps$.
       This regulator is commonly applied by separately renormalizing the unsubtracted TMD PDF and soft functions, giving rise to renormalized beam functions and renormalized soft functions. Here, poles in $\eta$ are cancelled by a rapidity renormalization factor, giving rise to a (dimension-1) rapidity scale $\nu$ (which is analogous to $\mu$ in the $\MSbar$ scheme).
       In this construction the TMD PDF is obtained in either one of two ways, from the bare or renormalized quantities:
       \begin{align}
         \tilde f_{i/p}(x, \bt, \mu, \zeta) &
            = \lim_{\substack{\eps\to0 \\ \eta\to0}} Z_{\rm uv}^i(\mu,\zeta,\eps) \,
              \tilde f_{i/p}^{0\,\unsub}(x, \bt, \eps, \eta, x P^+) \sqrt{\tilde S^0_{\rm CJNR}(b_T,\eps,\eta)}
         \nn\\&
            = \tilde B_{i/p}^{\rm CJNR}\bigl(x, \bt, \mu, \sqrt{\zeta}/\nu\bigr) \sqrt{\tilde S^{\rm CJNR}(b_T,\mu,\nu)}
      \,.\end{align}
      In this construction $\zeta = 2(x P^+)^2$, corresponding to taking $y_n=0$ in the modern Collins result.
      Note that in this regulator, there is no zero-bin subtraction factor, so $S_{\rm CJNR}^{0\,\subt}=1$.
 \item The \textbf{exponential regulator} due to Li, Neill and Zhu (LNZ) \cite{Li:2016axz} inserts an $\exp[-\tau e^{-\gamma_E} k^0]$ factor into the phase space of each real emission, where $k^0$ is the energy of the emission. 
Since the energies of the emissions are additive, this can be viewed as utilizing the fully differential distribution function in $(x,k^0,\kt)$ and then performing a Laplace transform over $k^0$ with an infinitesimal $\tau$. Thus this regulator is defined at the operator level without modifying the Wilson lines and is clearly gauge invariant at intermediate stages.
       Similar to the $\eta$ regulator, it can be used to define separately renormalized beam and soft functions, so the final renormalized TMD PDF can either be obtained from bare or renormalized quantities
       \begin{align}
         \tilde f_{i/p}(x, \bt, \mu, \zeta) &
            = \lim_{\substack{\eps\to0 \\ \tau\to0}} Z_{\rm uv}^i(\mu,\zeta,\eps) \,
              \frac{\tilde f_{i/p}^{0\,\unsub}(x, \bt, \eps, \tau, x P^+)}{\sqrt{\tilde S^0_{{\rm LNZ}}(b_T,\eps,\tau)}}
         \nn\\&
            = \tilde B_{i/p}^{\rm LNZ}\bigl(x, \bt, \mu, \sqrt{\zeta}/\nu \bigr) \sqrt{\tilde S^{\rm LNZ}(b_T,\mu,\nu)}
      \,,\end{align}
      where here again $\zeta = 2(xP^+)^2$.
      Since for this construction the soft function is equal to the  subtraction $\tilde S_{\rm LNZ}^{0\,\subt}=\tilde S_{\rm LNZ}^0$, it enters on the first line in the denominator, while in the second line a factor of $1/\tilde S_{\rm LNZ}^{0\,\subt}$ is already contained inside the rapidity-renormalized beam function $\tilde B_{i/p}^{\rm LNZ}$. 
 \item The \textbf{analytic regulator} was first introduced by Becher and Neubert (BN) \cite{Becher:2010tm}.  So far this regulator has been primarily considered for Drell-Yan, so we focus on that case here.
       In the modified version due to Becher and Bell \cite{Becher:2011dz}, one inserts a factor $(\nu/k^+)^\alpha$ for each real emission.
       In this regulator, the soft function is absent, $\tilde S_{\rm BN}^0 = 1$, and subtractions are also absent $\tilde S^{0\,\subt}_{\rm BN}=1$. Due to the asymmetry of the regulator under $n_a\leftrightarrow n_b$, one has different constructions for the $n_a$-collinear and $n_b$-collinear unsubtracted TMD PDFs, which can be combined to obtain the product of the renormalized TMD PDFs,
       \begin{align}
       \label{eq:BNdef}
        & \lim_{\substack{\eps\to0\\\alpha\to0}} \Bigl[ \tilde f_{i/p}^{0 \unsub, \rm BN}(x_1,\bt,\eps,\alpha,x_a P_A^+) \tilde f_{j/p}^{0 \unsub, \rm BN}(x_2,\bt,\eps,\alpha,x_b P_B^-) \Bigr]
         \nn\\ &
        = \biggl(\frac{b_T^2 Q^2}{b_0^2}\biggr)^{-\gamma_\zeta^q(\mu,b_T)} \Big[ \tilde f_{i/p}^{\rm BN}\bigl(x_1, \bt, \mu,\zeta=b_0^2/b_T^2\bigr)\, \tilde f_{j/p}^{\rm BN}\bigl(x_2, \bt, \mu,\zeta=b_0^2/b_T^2\bigr) \Big]
       \,,\end{align}
       where 
       $b_0 = 2 e^{-\gamma_E} = 1.12292$. In this definition one exponentiates the dependence on $\zeta_1 \zeta_2 = Q^4$ through the CS kernel $\gamma_\zeta$ (see \sec{evolution}) and correspondingly fixes $\zeta = b_0^2/b_T^2$ in the remaining TMD PDFs. For this reason there is often no  $\zeta$ variable written in the renormalized TMD PDFs.  Note that this definition does not suffice to separately define $\tilde f_{i/p}^{\rm BN}$ and $\tilde f_{j/p}^{\rm BN}$ uniquely, and hence is on a different footing compared to other constructions.
 \item The \textbf{pure rapidity regulator} was introduced by Ebert, Moult, Stewart, Tackmann, Vita, and Zhu (EMSTVZ)~\cite{Ebert:2018gsn}. It is defined similarly to the analytic regulator by inserting a factor $|k^+/k^-|^{-\eta/2}$ for each real emission, but is analogous to the eta and exponential regulators in that renormalized $n_a$ and $n_b$-collinear TMD PDFs can be defined separately.
 It shares the feature that the soft function and subtraction function are absent, $\tilde S^0_{\rm EMSTVZ} = \tilde S^{0\,\subt}_{\rm EMSTVZ} = 1$. Here the $n_a$ and $n_b$-collinear TMD PDFs are simply related by $\eta \leftrightarrow -\eta$, so although they may be defined separately, only the product of renormalized TMD PDFs is the same as this product in the eta and exponential regulator constructions. This pure rapidity regulator has been used to study power corrections to TMD factorization at one loop, since the regulator itself does not induce power suppressed contributions~\cite{Ebert:2018gsn}.
\end{itemize}

{
\begin{table*}[b!]
 \centering
 \begin{tabular}{| l | c | c | c | c |}
 \hline
 Regulator & \!\thead{Unsubtracted \\TMD PDF}\! & Soft function
 & \! \thead{Subtracted \\TMD PDF $f_{i/p}$}\!  & CS scale
 \\ \hline
 \makecell{\small Space-like \\ 
    \small Wilson lines \cite{Collins:2011zzd}}
   & $\tilde f_{i/p}^{0\,\unsub}(y_B)$
   & \!$\tilde S^{0}_{n_A n_B}(2y_n-2y_B)$\!
   & $\lim\limits_{y_B\to-\infty} \frac{\tilde f_{i/p}^{0\,\unsub}(y_B)}{\sqrt{\tilde S^{0}_{n_A n_B}(2y_n-2y_B)}}$
   & $\zeta = 2(x P^+ e^{-y_n})^2$
  \\ \hline
 \makecell{\small $\delta$ regulator \cite{Echevarria:2012js}}
   & $\tilde f_{i/p}^{0\,\unsub}(\delta^+)$
   & $\tilde S^{0}_{n_a n_b}(\sqrt{\delta^+ \delta^-})$
   & $\lim\limits_{\delta^+\to 0} \frac{\tilde f_{i/p}^{0\,\unsub}(\delta^+)}{\sqrt{\tilde S^0(\delta^+ e^{-y_n})}}$
   & $\zeta = 2(x P^+ e^{-y_n})^2$
 \\ \hline
 \makecell{\small $\eta$ regulator$^*$ \cite{Chiu:2012ir} }
   & $\tilde f_{i/p}^{0\,\unsub}(\eta)$
   & $\tilde S^{0}_{n_a n_b}(\eta)$
   & $\lim\limits_{\eta\to0} \tilde f_{i/p}^{0\,\unsub}(\eta) \sqrt{\tilde S^0(\eta)}$
   & $\zeta = 2(x P^+)^2$
 \\ \hline
 \makecell{\small Exponential \\ \small regulator$^*$ \cite{Li:2016axz}}
   & $\tilde f_{i/p}^{0\,\unsub}(\tau)$
   & $\tilde S^{0}_{n_a n_b}(\tau)$
   & $\lim\limits_{\tau\to0} \frac{\tilde f_{i/p}^{0\,\unsub}(\tau)}{\sqrt{\tilde S^0(\tau)}}$
   & $\zeta = 2(x P^+)^2$
 \\ \hline
 \end{tabular}
 \caption{
   Summary of different schemes for defining a TMD PDF $\tilde f_{i/p}(x, \bt, \mu , \zeta)$ as in \eq{tmdpdf_1}.
   In all functions, we drop all arguments except for the regulator.
   In the subtracted TMD PDF, we only show the limit of taking the rapidity regulator to zero,
   but not the UV subtraction.
   Schemes denoted with an asterisk ($^*$) are also defined with renormalized beam and soft functions.
 }
 \label{tbl:overview_tmdpdfs}
\end{table*}
}

\subsubsection{Illustration at one loop}
\label{sec:tmd_defs_nlo}

In this section, we study the quark TMD PDF perturbatively at one loop,
which will show concretely how rapidity divergences arise and how they are regulated in practice.
As shown in \eq{tmdpdf_1}, we need to consider both the unsubtracted quark TMD PDF and the soft function, which we can then combine into the TMD PDF.
Since the unsubtracted TMD PDF is defined with an external hadronic state and thus is genuinely nonperturbative,
we need to replace the external hadron with a parton, which allows us to use standard Feynman rules to perturbatively study this matrix element.
In \sec{largeqT}, we will see that this calculation is of practical relevance, as it allows us to perturbatively relate the TMD PDF to the standard PDF whenever $q_T \gg \lqcd$ is a perturbative scale.

\paragraph{Unsubtracted TMD PDF.}
We begin our calculation with the unsubtracted TMD PDF at NLO, and as discussed calculate its matrix element
with the hadron state replaced by an external quark of lightlike momentum $p^\mu = (p^+, 0, \mathbf{0})$.
Here we consider
\begin{align} \label{eq:beam_nlo_0}
 \tilde f_{q/q'}^{0\,\unsub}(x,\bt,\eps,\tau) &
 = \int\frac{\df b^-}{2\pi} e^{-\img b^- (x p^+)} \Bigl< q'(p) \Bigr|  \Bigl[ \bar \psi^{0}_q(b^\mu)
   W_{\sqsubset}(b^\mu,0)  \frac{\gamma^+}{2} \psi^{0}_q(0) \Bigr]_\tau \Bigl| q'(p) \Bigr>
\,.\end{align}
Here, the subscript on $f_{q/q'}$ indicates that we analyze the contribution from an external parton of flavor $q'$ to the quark TMD PDF of flavor $q$.
At one loop, $q=q'$ of identical flavor, while starting at two loops, one can also have different flavors $q \ne q'$. 
There can also be contributions from quarks mixing with gluons at one loop, which we will not consider in this section. To ensure that gluon contributions drop out,  we can consider $q$ to be a non-singlet combination of quark flavors.

\begin{figure*}
 \centering
\subfloat[]{
\includegraphics[trim={1cm 0 1cm 0}, clip, width=3.75cm]{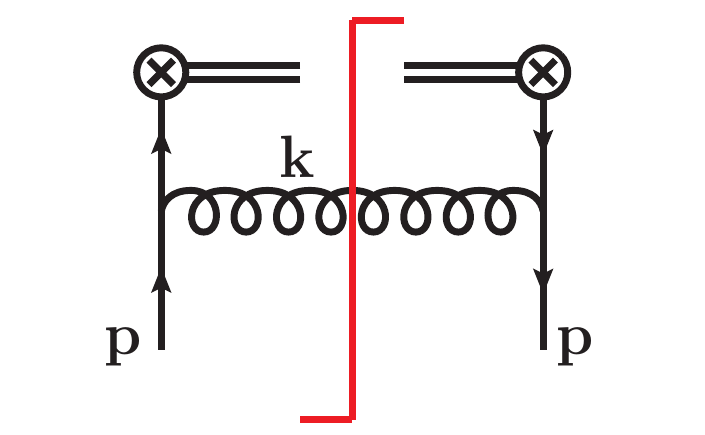}
}
 \hfill
\subfloat[]{
\includegraphics[trim={1cm 0 1cm 0}, clip, width=3.75cm]{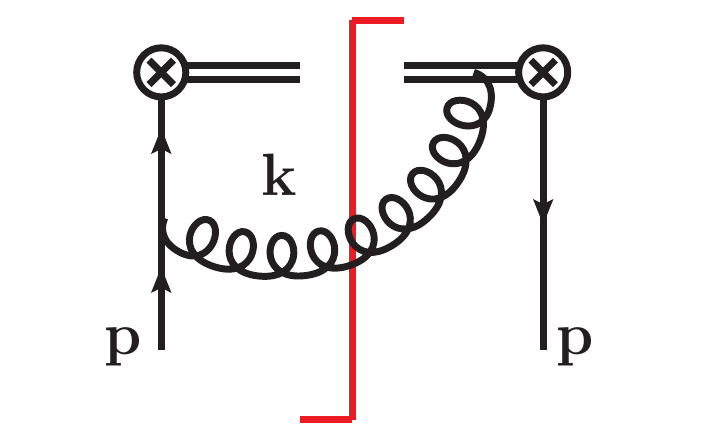}
}
 \hfill
\subfloat[]{
\includegraphics[trim={1cm 0 1cm 0}, clip, width=3.75cm]{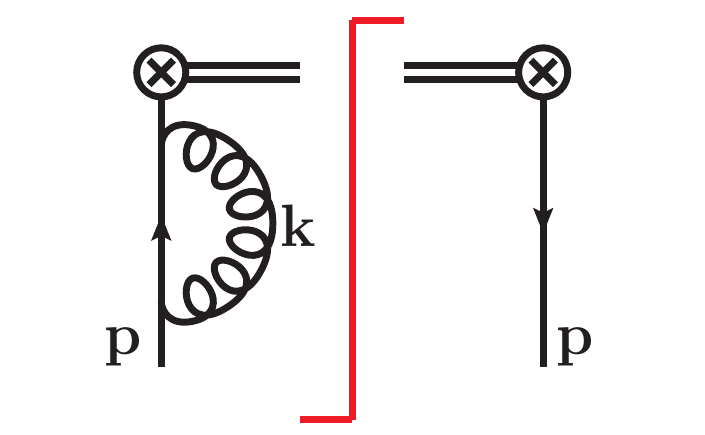}
}
 \hfill
\subfloat[]{
\includegraphics[trim={1cm 0 1cm 0}, clip, width=3.75cm]{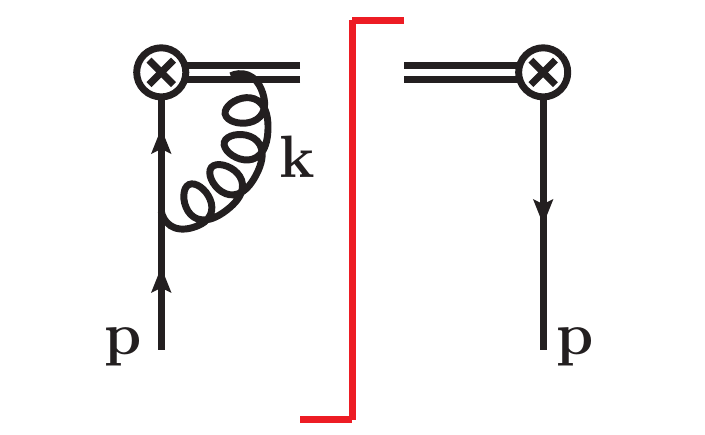}
}
 \caption{One-loop contribution to the unsubtracted quark-TMD PDF.
 The $\otimes$ denote the two quark fields, the double line the staple shaped Wilson lines connecting the quark fields, and the red line the on-shell cut.
 The diagrams (b)--(d) have mirror diagrams that are not explicitly shown.
 In pure dimensional regularization, the virtual diagrams (c) and (d) are scaleless and vanish. }
 \label{fig:beam_nlo}
\end{figure*}

At one loop, there are only four types of diagrams and their mirror diagrams, which are shown in \fig{beam_nlo}.
In these diagrams, the $\otimes$ denotes the quark fields, with the left field positioned at the origin
and the right field located at $b^\mu$. The two double lines represent the Wilson line segments along $n_b$.
Note that we do not consider the transverse gauge links, as they do not contribute in a physical gauge
such as Feynman gauge (in general gauges they do matter,  see e.g.~\cite{Ji:2002aa,Belitsky:2002sm,Idilbi:2010im,GarciaEchevarria:2011md}).
Physically, this reflects that the Wilson line at infinite distance does not impact the physics at the finite distance $b^\mu$.

The different diagrams in \fig{beam_nlo} arise from different ways to exchange a gluon between the quark fields
and Wilson lines in \eq{beam_nlo_0}. To evaluate these diagrams, we need to know the Feynman rules for the connection
of a gluon to a Wilson line. They are given by
\begin{align}
W_{n_b}(b^\mu;-\infty,0) \,\text{:}\qquad
\begin{tikzpicture}[baseline=(current bounding box.center)]
\begin{feynman}
   \vertex (x) {};
   \vertex[right=1.5cm of x] (y) {$b^\mu$};
   \vertex[right=0.74cm of x] (g1);
   \vertex[below=1cm of g1] (g2) {$k,\mu$};
   \diagram* { (g1) -- [gluon] (g2) };
   \draw[double, double distance = 1.2pt, thick] (x) -- (y);
 \end{feynman}
\end{tikzpicture}
\quad=&\quad -g_0 n_b^\mu t^a \frac{e^{-\img k \cdot b}}{n_b \cdot k + \img0}
\,,\nn\\
 W_{n_b}^{\dagger}(0;-\infty,0) \,\text{:}\qquad
\begin{tikzpicture}[baseline=(current bounding box.center)]
\begin{feynman}
   \vertex (x)  {$0^\mu$};
   \vertex[right=1.5cm of x] (y);
   \vertex[right=0.9cm of x] (g1);
   \vertex[below=1cm of g1] (g2) {$k,\mu$};
   \diagram* { (g1) -- [gluon] (g2) };
   \draw[double, double distance = 1.2pt, thick] (y) -- (x);
 \end{feynman}
\end{tikzpicture}
\quad=&\quad +g_0 n_b^\mu t^a \frac{1}{n_b \cdot k - \img0}
\,.\end{align}
As indicated, the first line gives the Feynman rule for the Wilson line stretching from $b^\mu$ to light-cone infinite,
while the second line shows the Feynman rule for the Wilson line stretching back from light-cone infinite to the origin.
In both cases, the gluon momentum $k$ is incoming, and the gluon has color index $a$ and polarization vector $\eps^\mu$.
Note that a gluon exchange between these two segments is proportional to $n_b^2 = 0$, and thus vanishes.
For this reason, \fig{beam_nlo} does not contain a diagram where the gluon is exchanged between the Wilson line segments.

In our calculation, we will regulate both infrared (IR) and UV divergences by extending spacetime to $d=4-2\eps$ dimensions.
The quark momentum is chosen as $p^\mu = (p^+, 0, 0)$, which simplifies the calculation due to $p^2 = 0$.
We are now ready to write down the explicit expressions for the diagrams in \fig{beam_nlo}:
\begin{align} \label{eq:beam_nlo_1}
 {\cal M}_a &
 = -\img g_0^2 C_F  \int\!\frac{\df^d k}{(2\pi)^d} \int \frac{\df b^-}{2\pi} e^{-\img b^- (x p^+)} e^{\img (p-k) \cdot b}
   \frac{\bar u(p)  \gamma^\mu (\slashed{p}-\slashed{k}) \gamma^+(\slashed{p}-\slashed{k}) \gamma_\mu u(p)}
        {2 [(p-k)^2 + \img0]^2 (k^2 + \img0)}
\,,\\
 {\cal M}_b &
 = -2 \img g_0^2 C_F  \int\!\frac{\df^d k}{(2\pi)^d} \int \frac{\df b^-}{2\pi} e^{-\img b^- (x p^+)} e^{\img (p-k) \cdot b}
   \frac{\bar u(p) \gamma^+ (\slashed{p}-\slashed{k})  \gamma^+ u(p) }{2 (k^+ + \img0) [(p-k)^2 + \img0] (k^2 + \img0)}
\,.\end{align}
Note that the quark field sitting at $b^\mu$ induces a phase $e^{\img q \cdot b}$,
where $q^\mu$ is the momentum flowing out of the right vertex $\otimes$.
We have not given diagrams (c) and (d), as they vanish in dimensional regularization,
i.e.~they involve scaleless integrals of the type  $\int\df^d k\, f(k^2) = 0$ in dimensional regularization.
The overall factor of $2$ in ${\cal M}_b$ arises from the mirror diagram.

To proceed, we evaluate the $b^-$ integral as
\begin{align} \label{eq:bm_integral}
 \int \frac{\df b^-}{2\pi} e^{-\img b^- (x p^-)} e^{\img (p-k) \cdot b}
 = \int \frac{\df b^-}{2\pi} e^{-\img b^- [(1-x) p^+ - k^+] + \img \bt \cdot \kt}
 = \delta[(1-x) p^+ - k^+] e^{\img \bt \cdot \kt}
\,,\end{align}
where we used that $b^+ = p^- = 0$, such that the remaining phase arises purely from the transverse momentum.
This result has a simple interpretation: the emitted gluon carries away the longitudinal momentum $k^+ = (1-x) p^+$,
such that the leftover momentum $x p^+$ is absorbed by the quark field.
In other words, the parton participating in the hard interaction will carry the momentum fraction $x$ of the external parent hadron.

Using light-cone coordinates, the integration measure becomes $\df^d k = \df k^+ \df k^- \df^{d-2} \kt$,
whose $k^+$ integral is already fixed by \eq{bm_integral}.
Performing the standard Dirac algebra in the numerators in \eq{beam_nlo_1}, we obtain
\begin{align} \label{eq:beam_nlo_2}
 {\cal M}_a &
 = \img g_0^2 C_F  \int\frac{\df^{d-2} \kt}{(2\pi)^d} \, e^{\img \bt \cdot \kt}
   \int\! \df k^- \frac{(2-d) (1-x) p^+}{[(p-k)^2 + \img0] (k^2 + \img0)}
\,,\\
 {\cal M}_b &
 =\img g_0^2 C_F  \int\frac{\df^{d-2} \kt}{(2\pi)^d} \, e^{\img \bt \cdot \kt} \int\! \df k^-
   \frac{-4 x / (1-x)p^+}{[(p-k)^2 + \img0] (k^2 + \img0)}
\nn
\,.\end{align}
The remaining integral over $k^-$ can be evaluated using the residue theorem,
\begin{align} \label{eq:residue_integral}
 \int\df k^-  \frac{1}{[(p-k)^2 + \img0] (k^2 + \img0)} &
 = \int\df k^- \frac{1}{[-2 x p^+ k^- - \kt^2 + \img0][2 (1-x) p^+ k^- - \kt^2 + \img0]}
 \nn\\&
 = \frac{\img \pi}{p^+ \kt^2} \theta(x) \theta(1-x)
\,.\end{align}
From the first line, we see that if $x$ and $1-x$ have different signs,
then the residues of $k^-$ will lie on the same complex half plane,
and one can deform the $k^-$ contour into the other half plane such that the integral vanishes.
(Here, the signs of the Feynman $\img0$ prescription are crucial.)
Hence, the only physical contribution arises if $0 < x < 1$, the expected physical range of the momentum fraction.
We then choose the residue at $k^- = \kt^2 / (2 k^+) > 0$, which is equivalent to choosing $k^2 = 0$.
Thus, we can interpret this choice as setting the gluon in \fig{beam_nlo} on shell.
In fact, we could have started with this choice right away by using the Cutkosky rule~\cite{tHooft:1973wag}
\begin{align} \label{eq:cutkosky}
 \frac{1}{k^2 + \img0} 
  \to 2\, {\rm Im}\Big(  \frac{1}{k^2 + \img0}  \Big)
   =  - 2\pi\img \theta(k^0) \delta(k^2)
   \equiv -2\pi\img \delta_+(k^2) 
\,.\end{align}
Our more exhaustive derivation shows how this constraint naturally arises from the definition of the unsubtracted TMD PDF.

Combining the two matrix elements in \eq{beam_nlo_2} with \eq{residue_integral},
we obtain the one-loop contribution to the bare unsubtracted quark TMD PDF as
\begin{align} \label{eq:beam_nlo_3}
 {\cal M}_a + {\cal M}_b &
 = \frac{g_0^2 C_F}{2\pi}  \left[ \frac{1+x^2}{1-x} - \eps (1-x)\right]
   \int\frac{\df^{d-2} \kt}{(2\pi)^{d-2}} \, \frac{e^{\img \bt \cdot \kt}}{\kt^2}
\,.\end{align}
To evaluate the remaining $\kt$ integral, we have to fix how we want to treat $\kt$ and $\bt$ in $2-2\eps$ dimensions.
There is no unique choice, but ultimately every choice leads to equivalent TMD PDFs.
Following \cite{Li:2016axz} we extend $\bt = (b_T, 0, \vec 0_{-2\eps})$ and $\kt = k_T(\cos\theta, \sin\theta, \vec 0_{-2\eps})$,
such that the phase only picks out the purely two-dimensional piece.
This yields
\begin{align} \label{eq:angle_integral}
 \int\frac{\df^{d-2} \kt}{(2\pi)^{d-2}} \, \frac{e^{\img \bt \cdot \kt}}{\kt^2}
 &= \frac{\Omega_{-2\eps}}{(2\pi)^{2-2\eps}} \int_0^\infty \df k_T \, k_T^{1-2\eps} \int_0^\pi \df\theta \, \sin^{-2\eps}\theta \frac{e^{\img b_T k_T \cos\theta}}{k_T^2}
  = \frac{\Gamma(-\eps)}{4 \pi} (\pi \bt^2)^\eps
\,,\end{align}
where $\Omega_n = 2 \pi^{(n+1)/2} / \Gamma[(n+1)/2]$ is the area of a unit $n$-sphere.
Thus, we finally arrive at
\begin{align} \label{eq:beam_nlo_4}
 {\cal M}_a + {\cal M}_b &
 = \frac{\as(\mu) C_F}{2\pi} \left[ \frac{1+x^2}{1-x} - \eps (1-x)\right]
   \Gamma(-\eps) \left(\frac{\bt^2 \mu^2 }{4e^{-\gamma_E}} \right)^\eps
\,,\end{align}
where we also replaced the bare by the renormalized coupling in the $\overline{\mathrm{MS}}$ scheme,
\begin{align}  \label{eq:gbareinMSbar}
  g_0= Z_g \mu^\eps g(\mu) \, \Bigl(\frac{e^{\gamma_E}}{4\pi}\Bigr)^{\eps/2}
\,,\qquad
 \as(\mu) = \frac{g(\mu)^2}{4 \pi}
\,.\end{align}
Here $Z_g=1 + {\cal O}(g^2)$ is the strong coupling counterterm which can be set to one for this one-loop calculation. The inclusion of the factor of $(e^{\gamma_E}/4\pi)^{\eps/2}$ implements the use of the $\MSbar$ scheme rather than MS scheme.%
\footnote{Note that another, slightly less popular, definition of $\MSbar$ replaces $e^{\eps \gamma_E}\to 1/\Gamma(1-\eps)$ in \eq{gbareinMSbar}. One must be careful about which convention is being used when examining perturbative results in the literature.}

\eq{beam_nlo_4} seems satisfactory, as we apparently only need to expand in $\eps\to0$
to obtain the desired bare result. This will yield poles in $1/\eps$ that arise from regulating the $k_T \to 0$ region in \eq{angle_integral}.
However, there is still one problem: the result in \eq{beam_nlo_4} diverges as $x\to1$,
i.e.~in the limit when the struck quark carries all the energy of the parent hadron,
or equivalently where the energy of the emitted gluon vanishes, $k^+ \to 0$.
This is precisely the manifestation of the rapidity divergence at one loop in the unsubtracted TMD PDF,
which will only cancel when combining \eq{beam_nlo_4} with the soft function,
which has a similar divergence as $k^\mu \to 0$. In order to correctly combine the two results,
we need to regulate this divergence. Then, after combination we can remove the regulator and obtain the desired finite result.

\index{rapidity regulator}
To illustrate this in practice, in the following we employ the $\eta$ regulator~\cite{Chiu:2011qc,Chiu:2012ir}, which modifies the formula for Wilson lines, and which can be implemented directly at the level of \eq{beam_nlo_4}. The regulator results in adding the following factor to the integral%
\footnote{In \cite{Chiu:2011qc,Chiu:2012ir}, the regulator is denoted as $\eta$.
For continuity of the presentation, here we denote it as $\tau$.
The factor of $\sqrt2$ compensates for a different light-cone convention in \cite{Chiu:2011qc,Chiu:2012ir}.}
\begin{align} \label{eq:R_eta_beam}
 R_c(k,\tau) = \Bigl|\frac{\sqrt2 k^+}{\nu}\Bigr|^{-\tau} = \left(\frac{(1-x) p^+}{\nu/\sqrt2}\right)^{-\tau}
\,.\end{align}
It allows us to regulate the divergent term in \eq{beam_nlo_4} through the identity
\begin{align} \label{eq:plusexpn}
 \frac{1+x^2}{1-x} (1-x)^{-\tau} &
 = -\left(\frac{2}{\tau} + \frac32 \right) \delta(1-x) + \left[\frac{1+x^2}{1-x}\right]_+ + \cO(\tau)
\,.\end{align}
Here, the plus prescription is defined such that
\begin{align} \label{eq:plusdef}
 \bigl[ f(x) \bigr]_+ = f(x) ~~\mathrm{for}~~x \ne 1
\,,\qquad
 \int_0^1 \df x \, \bigl[ f(x) \bigr]_+ = 0
\,,\end{align}
such that it only modifies the limit $x\to1$ in a way that yields a well-defined integral up to $x = 1$.
Applying \eq{R_eta_beam} to \eq{beam_nlo_4} and using \eq{plusexpn}, we finally obtain the bare unsubtracted TMD PDF
\begin{align} \label{eq:beam_nlo_5}
 \tilde f_{q/q}^{0\unsub\,(1)}(x, \bt, \eps, \tau) &
 = \frac{\as(\mu) C_F}{2\pi} \biggl\{
   -\Bigl(\frac{1}{\eps} + L_b\Bigr) [P_{qq}(x)]_+  + (1-x)
   \nn\\&\hspace{1.8cm}
   + \delta(1-x) \left(\frac{1}{\eps} + L_b\right) \left(\frac32 + \frac{2}{\tau} - 2 \ln\frac{x p^+}{\nu/\sqrt2}\right)
   + {\cal O}(\tau) + {\cal O}(\eps) \biggr\}
\,.\end{align}
Here, we introduced the shorthand notation
\begin{align} \label{eq:Lb}
 L_b = \ln\frac{\bt^2 \mu^2}{b_0^2}  \,, 
  \qquad \text{with}\quad b_0 = 2 e^{-\gamma_E} \,,
\end{align}
for the canonical logarithm encoding the $\bt$-dependence, and introduced notation for the quark-quark one-loop splitting function which reads
\begin{align} \label{eq:Pqq}
 P_{qq}(x) = \frac{1+x^2}{1-x}
\,.\end{align}
\eq{beam_nlo_5} is our desired final result: the divergence as $x\to1$ is regulated through the plus distribution,
with the divergence now manifest as a pole in $1/\tau$.
In addition, it contains a $1/\eps$ pole from the $k_T\to0$ region of the integral in \eq{angle_integral}.
Note that the divergence in the first line in \eq{beam_nlo_5} is proportional to the quark-to-quark splitting function $P_{qq}$.
In fact, one encounters the identical divergence for the collinear PDF itself, illustrating the universality from the collinear limit of QCD.

The bare result in \eq{beam_nlo_5} depends somewhat on the employed rapidity regulator, and is not universal.
So that results with other regulators can be easiliy compared, we collect explicit bare results for all regulators discussed above in \sec{tmd_defs_overview} in \app{TMDdefn}.

\paragraph{Soft function.}

\begin{figure*}
 \centering
\subfloat[]{
\includegraphics[width=4.25cm]{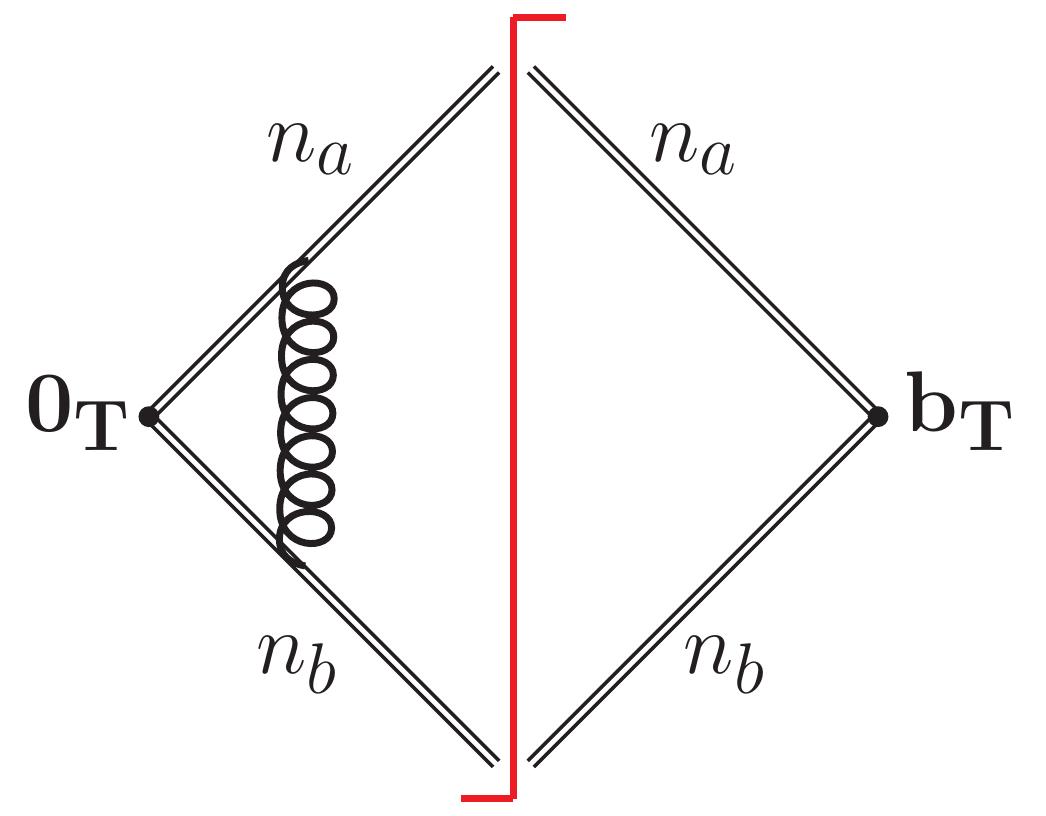}
}
 \qquad
\subfloat[]{
\includegraphics[width=4.25cm]{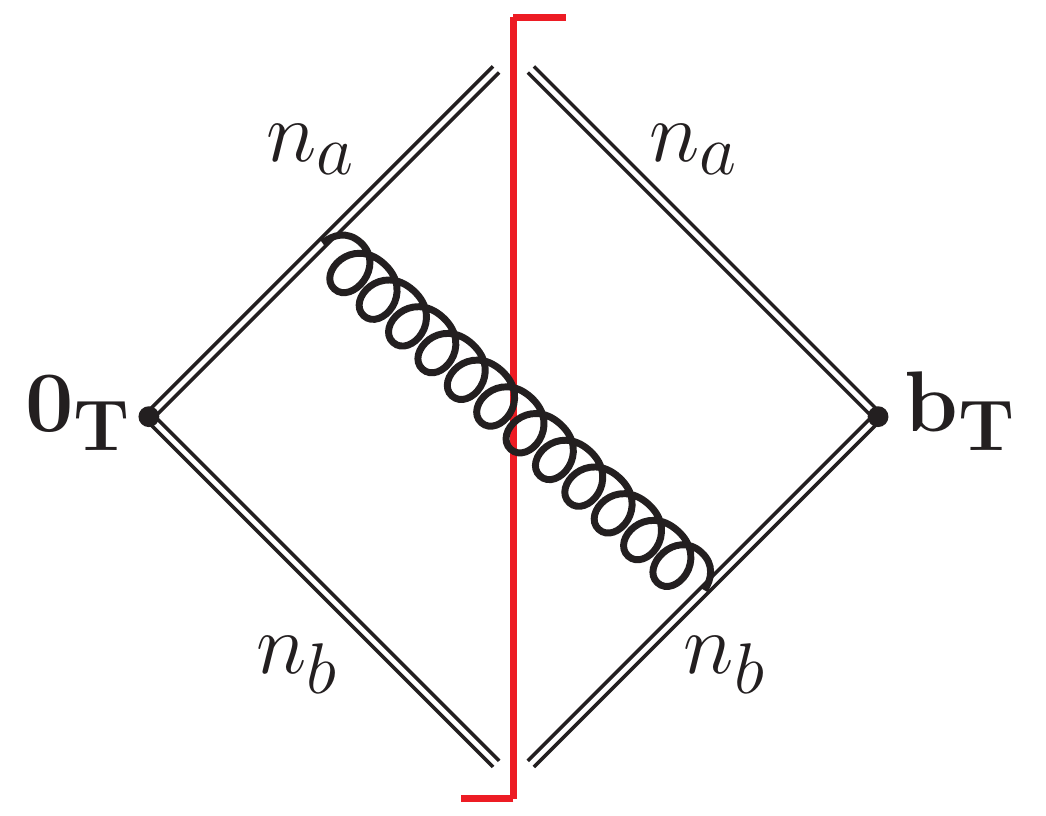}
}
 \caption{One-loop contributions to the soft function, with mirror diagrams obtained by a left-right swap of the exchanged gluon not shown.
 The double lines denote the Wilson lines from the transverse positions $\mathbf{0_T}$ and $\bt$ stretching to light-cone infinity as indicated.
 The red line denotes the on-shell cut. Diagram (a) is scaleless and vanishes in pure dimensional regularization.}
 \label{fig:soft_nlo}
\end{figure*}

Let us now study the corresponding one-loop calculation of the soft function.
The relevant diagrams are shown in \fig{soft_nlo}, up to mirror diagrams, and can be evaluated in the same fashion as shown explicitly for the unsubtracted TMD PDF.
As before, we first give the generic bare result without any rapidity regulator,
\begin{align} \label{eq:soft_nlo_1}
 {\cal M}_S 
 &= 2 g_0^2 C_F  \int\frac{\df^d k}{(2\pi)^d} \, e^{i \bt \cdot \kt}
   \frac{-i}{(2k^+k^- -\kt^2+i0)} \frac{1}{(k^+-i0)(-k^-+i0)}
 \nn\\
 &= 2 g_0^2 C_F  \int\frac{\df^d k}{(2\pi)^d} \, e^{i \bt \cdot \kt}
   (2\pi) \delta_+(k^2) \frac{1}{k^+ k^-}
 \nn\\&
 = \frac{g_0^2 C_F}{\pi} 
   \int \frac{\df^{2-2\eps} \kt}{(2\pi)^{d-2}} \, \frac{e^{i \bt \cdot \kt}}{k_T^2}
   \int_0^\infty \frac{\df k^-}{k^-}
\,.\end{align}
Since the result from \fig{soft_nlo}(a) is scaleless, here we show only the contribution from \fig{soft_nlo}(b) and its mirror image. With the expression in the first line we can do the $k^+$ integral by contours, fixing $k^+ = \kt^2/(2k^-)-i0$ with $k^->0$. This gives an equivalent result to the expression in the second line, which uses \eq{cutkosky} to express the integral with the on-shell constraint for the cut graph.
Clearly, \eq{soft_nlo_1} is divergent as either $k^- \to 0$ or $k^- \to \infty$.
Since the rapidity of the emission $k$ is given by $y_k = \frac12 \ln(k^+/k^-)$,
these limits correspond to $y_k \to \pm\infty$, which explains the terminology ``rapidity divergence''.
To regulate it in a manner consistent with the above calculation of the unsubtracted TMD PDF,
we again use the $\eta$ regulator~\cite{Chiu:2011qc,Chiu:2012ir}, which for the soft function inserts the factor
\begin{align} \label{eq:R_eta_soft}
 R_s(k,\tau) = \biggl|\frac{k^+ - k^-}{\nu / \sqrt2} \biggr|^{-\tau}
    w^2(\tau,\nu)
\,.\end{align}
(The absolute value is important.) Here $w(\tau,\nu)$ is a bookkeeping parameter for the rapidity divergence, related to a bare parameter by $w^0=w(\tau,\nu)\nu^{\tau/2}$.  It satisfies $w(0,\nu)=1$ and $\nu \partial/\partial \nu w(\tau,\nu) = -(\tau/2) w(\tau,\nu)$. Thus, the regulated integral becomes
\begin{align}
 \int_0^\infty \frac{\df k^-}{k^-} &
 \to w^2 \Bigl(\frac{\nu}{\sqrt2}\Bigr)^\tau \int_0^\infty \frac{\df k^-}{k^-} \biggl| \frac{\kt^2}{2k^-} - k^- \biggr|^{-\tau}
 = \frac{\nu^\tau k_T^{-\tau}}{2^\tau \sqrt{\pi}} \Gamma\Bigl(\frac12 - \frac\tau2\Bigr) \Gamma\Bigl(\frac\tau2\Bigr)
\,.\end{align}
Inserting this into \eq{soft_nlo_1}, we obtain the bare rapidity-regulated soft function as
\begin{align} \label{eq:soft_nlo_2}
 \tilde S_q^{0\,(1)}(b_T,\eps,\tau) &
 = \frac{g_0^2 C_F}{\pi} \frac{\nu^\tau}{2^\tau \sqrt{\pi}} \Gamma\Bigl(\frac12 - \frac\tau2\Bigr) \Gamma\Bigl(\frac\tau2\Bigr)
   \int \frac{\df^{2-2\eps} \kt}{(2\pi)^{d-2}} \, \frac{e^{i \bt \cdot \kt}}{k_T^{2+\tau}}
\nn\\&
 = \frac{g_0^2 C_F}{\pi} \frac{\nu^\tau}{2^\tau \sqrt{\pi}} \Gamma\Bigl(\frac12 - \frac\tau2\Bigr) \Gamma\Bigl(\frac\tau2\Bigr)
   \frac{\pi^{\eps} \Gamma(-\eps-\tau/2)}{4 \pi 2^\tau \Gamma(1+\tau/2)} b_T^{2\eps+\tau}
\,,\end{align}
where the integral over $k_T$ is easily obtained similar to \eq{angle_integral}. 
Expanding in $\tau\to 0$ and $\eps\to 0$ and using \eq{gbareinMSbar}, we obtain
\index{TMD soft function}
\begin{align} \label{eq:soft_nlo_3}
 \tilde S_q^{0\,(1)}(b_T,\eps,\tau) &
 = \frac{\as(\mu) C_F}{2\pi} \left[ \frac{2}{\eps^2} + 4 \left(\frac{1}{\eps} + L_b\right) \left(-\frac{1}{\tau} + \ln\frac{\mu}{\nu}\right) - L_b^2 - \frac{\pi^2}{6} \right] + {\cal O}(\tau)+ {\cal O}(\eps)
\,.\end{align}

\paragraph{TMD PDF.}
Having calculated the unsubtracted TMD PDF and the soft function at one loop, we can now combine them into the TMD PDF following \eq{tmdpdf_1}.
To do so, we first note that in the $\eta$ regulator we have chosen for illustration, the soft subtraction factor is equal to unity,
$\tilde S_{n_a n_b}^{0\subt}  = 1$~\cite{Chiu:2012ir}, so from \eq{tmdpdf_1} the physical TMD PDF is constructed as
\begin{align} \label{eq:tmdpdf_eta_1}
 \tilde f_{i/H}(x, \bt, \mu , \zeta) &
 =  \lim_{\substack{\eps\to 0 \\ \tau\to 0}} Z_{\rm uv}^i(\mu,\zeta,\eps) \,
    \tilde f_{i/H}^{0\,\unsub}\bigl(x, \bt, \eps, \tau, x P^+ \bigr) \sqrt{\tilde S_{n_a n_b}^{0}(b_T,\eps,\tau)}
\,.\end{align}
Comparing the one-loop results \eqs{beam_nlo_5}{soft_nlo_2}, we see that all poles in $\tau$ precisely cancel in this combination.
Taking the product $f_{q/q}^{0\unsub} \sqrt{S^0_{n_a n_b}}$ the final bare result for the physical TMD PDF at one-loop is 
\begin{align} \label{eq:tmdpdf_nlo}
 \tilde f_{q/q}^{0\,(1)}(x, \bt, \eps, \zeta) &
 = \frac{\as(\mu) C_F}{2\pi} \biggl[ - \biggl(\frac{1}{\eps} + L_b\biggr) \bigl[P_{qq}(x)\bigr]_+ +  (1-x) \biggr]
 \\\nn&
  + \frac{\as(\mu) C_F}{2\pi} \delta(1-x) \biggl[ \frac{1}{\eps^2} - \frac{L_b^2}{2}
   + \biggl(\frac{1}{\eps} + L_b\biggr) \biggl(\frac{3}{2} + \ln\frac{\mu^2}{\zeta} \biggr)
   - \frac{\pi^2}{12} \biggr] + {\cal O}(\eps)
\,.\end{align}
The first line in \eq{tmdpdf_nlo} contains an infrared $1/\eps$ pole, while the second line has ultraviolet $1/\eps$ poles that will be removed by renormalization.
Here, $\zeta \propto (\bn \cdot p)^2 = 2 (x P^+)^2$ corresponds to the light-cone momentum carried by the struck quark,
and the proportionality factor may be rapidity scheme dependent.

To obtain the renormalized TMD PDF $f_{q/q}$, we cancel all $1/\eps$ poles in \eq{tmdpdf_nlo} that are of ultraviolet origin with the counterterm that appears in \eq{tmdpdf_1}, which in $\MSbar$ yields
\begin{align}
\label{eq:Zq_factor}
 Z^q_{\rm uv}(\mu,\zeta,\eps) = 1 - \frac{\as(\mu) C_F}{2\pi} \biggl[ \frac{1}{\eps^2}
   + \frac{1}{\eps} \biggl(\frac{3}{2} + \ln\frac{\mu^2}{\zeta} \biggr) \biggr] + \cO(\as^2)
\,.\end{align}
Combining the ingredients in \eq{tmdpdf_1} this yields the quark-to-quark contribution to the renormalized TMD PDF at one-loop as 
\index{TMD parton distribution function!one-loop calculation}
\begin{align} \label{eq:tmdpdf_nlo_renom}
 \tilde f_{q/q}^{(1)}(x, \bt, \mu, \zeta) &
 = \frac{\as(\mu) C_F}{2\pi} \biggl[ - \biggl(\frac{1}{\eps} + L_b\biggr) [P_{qq}(x)]_+ +  (1-x)
  - \frac{L_b^2}{2}  + L_b \biggl(\frac{3}{2} + \ln\frac{\mu^2}{\zeta} \biggr) - \frac{\pi^2}{12} \biggr]
\,.\end{align}
Note that the remaining $1/\eps$ pole here is of infrared origin and thus must not be absorbed in the UV counterterm.
It is the same collinear divergence that is present for the PDF, which enables the TMD PDF to be matched on to the PDF for perturbative $b_T$, see \sec{largeqT}.
The results in \eqs{tmdpdf_nlo}{tmdpdf_nlo_renom} are independent of the chosen rapidity regulator, and different regulators only differ by the explicit intermediate expressions for the bare unsubtracted TMD PDF and soft function.%
\footnote{When comparing results from the literature, care has to be taken concerning the employed definition of the $\MSbar$ scheme, see \eq{gbareinMSbar} and the following discussion.}

In the SCET literature, one often separately renormalizes the unsubtracted TMD PDF, in this case referred to as beam function, and the soft function, see \eqs{B_renorm}{S_renorm}.
In this case, one reproduces the same TMD PDF when combining the renormalized beam and soft functions as given in \eq{fBS_relation}.
Here, we explicitly illustrate this at one loop.
In the $\eta$ regulator scheme used above, $S^{0\subt}=1$, so the renormalized beam function is given by
\begin{align} \label{eq:B_renorm_eta}
 \tilde B_{i/p}(x, \bt, \mu, \zeta/\nu^2) &
 = \lim_{\substack{\eps\to 0 \\ \tau\to 0}} \tilde Z_B^q(b_T, \mu, \nu, \eps, \tau, x P^+) \,\tilde f_{i/p}^{0\,\unsub}\bigl(x, \bt, \eps, \tau, x P^+ \bigr)
\,.\end{align}
The counterterm can be easily read off from the result in \eq{beam_nlo_3} for the quark-to-quark channel. To do this one
expands first in $\tau\to0$, adding a term to $\tilde Z_B^q$ to cancel the $1/\tau$ divergence to all orders in $\eps$, and then expands in $\eps\to0$, adding additional terms to $\tilde Z_B^q$ to cancel $1/\eps$ divergences.
This is necessary to ensure that the coefficient of the $1/\tau$ terms is $\mu$-independent, which is important when deriving the corresponding renormalization group evolution equations in $\mu$ and $\nu$~\cite{Chiu:2012ir}.
Recall that the $1/\eps$ pole in the first line of \eq{beam_nlo_3} is of infrared origin and hence must not subtracted by the counterterm. This yields,
\begin{align} \label{eq:beam_Z}
 \tilde Z_B(b_T, \mu, \nu, \eps, \tau, x P^+) 
 = 1 &- \frac{\as(\mu) C_F}{2\pi} \biggl[ 
  \frac{2w^2(\tau,\nu)}{\tau} \left( \frac{1}{\eps} + L_b + \ldots \right) + 
\frac{1}{\eps} \left(\frac32  - 2 \ln\frac{x p^-}{\nu} \right) \biggr]
  \nn\\
 & + {\cal O}(\alpha_s^2)
\,.\end{align}
Here the ellipsis $+\ldots$ denotes the fact that all orders in $\eps$ are kept in the function multiplying the $1/\tau$ divergence in this counterterm (and we have displayed only the expanded form for brevity).
Combining \eqs{beam_nlo_3}{beam_Z} as in \eq{B_renorm_eta}, we obtain the quark-to-quark contribution to the renormalized beam function in the $\eta$ regulator scheme,
\begin{align} \label{eq:B_renorm_eta_2}
 \tilde B_{q/q}(x, \bt, \mu, \zeta/\nu^2) &= \delta(1-x) 
 +\frac{\as(\mu) C_F}{2\pi} \biggl[
   -\Bigl(\frac{1}{\eps} + L_b\Bigr) [P_{qq}(x)]_+  + (1-x)
 \nn \\
 &\ \  + \delta(1-x) L_b \left(\frac32 - 2 \ln\frac{x p^-}{\nu}\right) \biggr]
 + {\cal O}(\alpha_s^2)
\,.
\end{align}
The renormalized soft function is similarly constructed following \eqs{S_renorm}{soft_nlo_2},
from which we can read off the soft function counterterm and the renormalized soft function as
\begin{align} \label{eq:soft_renorm_eta}
 \tilde Z_S(b_T, \mu, \nu, \eps, \tau) &
 = 1-\frac{\as(\mu) C_F}{2\pi} \left[ 
  -\frac{4w^2(\tau,\nu)}{\tau} \left(  \frac{1}{\eps} + L_b + \ldots \right)
  + \frac{2}{\eps^2} + \frac{4}{\eps}  \ln\frac{\mu}{\nu}  \right]
  + {\cal O}(\alpha_s^2)
\,,\nn\\
 \tilde S_{n_a n_b}(b_T,\mu,\nu) &
 = 1+ \frac{\as(\mu) C_F}{2\pi} \left[ - L_b^2  + 4 L_b \ln\frac{\mu}{\nu} - \frac{\pi^2}{6} \right]
  + {\cal O}(\alpha_s^2)
\,.\end{align}
Again the ellipsis $+\ldots$ denotes all higher-order terms in $\eps$ which are kept in the $1/\tau$ coefficient in the $\tilde Z_S^{(1)}$ counterterm. These terms are identical for the counterterms in \eqs{beam_Z}{soft_renorm_eta}. 
Combining the renormalized results for $\tilde B_{q/q}^{(1)}$ and $\tilde S_{n_a n_b}^{(1)}$ from \eqs{B_renorm_eta_2}{soft_renorm_eta} following \eq{fBS_relation}, one reproduces the renormalized TMD PDF in \eq{tmdpdf_nlo_renom}. This illustrates the equality of the two approaches.

\FloatBarrier

\subsection{Additional TMD PDF definitions}
\label{sec:other_tmd_defs}

Here we discuss the original CS definition and the JMY definition, which have an extra variable in the renormalized TMD PDF, $f_{i/p}(x_a, \bt, \mu, \hat\zeta_a; \rho)$.  We also discuss the connection between these definitions and the earlier ones described in \secs{tmdpdfs_new}{tmd_defs}.

\index{TMD parton distribution function!JMY scheme}
For these definitions, the unsubtracted unpolarized bare quark distribution is again defined as
\begin{align} \label{eq:tmdCorig}
 \tilde f_{i/p}^{0(u)}(x,\bt,\eps,v,xP^+)
  &= \int   \frac{db^-}{2\pi}e^{-ib^-(xP^+)}
  \Bigl\langle  p(P)\Big|\bar\psi^{0}_i(b^\mu) W_{\sqsubset}^v(b^\mu,0)
   \frac{\gamma^+}{2}  \psi^{0}_i(0)\Big| p(P) \Bigr\rangle\ ,
    \nn\\
 W_{\sqsubset}^v(b^\mu,0)
 &= W[0\to -\infty v \to -\infty v+\bt \to b ] \nn\\
 &=W_{v}(b^\mu;-\infty,0) 
      W_{\hat b_T}\!\bigl(-\infty v; 0,b_T\bigr)
       W_{v}(0;0,-\infty)
  \,,
\end{align}
with the Wilson lines defined as in \eq{Wilson_lines}. Once again we use here past pointing Wilson lines, as is suitable for Drell-Yan. The key difference compared to the definitions discussed in \secs{tmdpdfs_new}{tmd_defs} is the choice of the Wilson line direction $v$. 

Physically, a parton interpretation is most natural if 
the gauge link vector $v$ is chosen along the conjugate light-cone
direction to $P^\mu$, i.e., $v^\mu = n_b^\mu$. However, as described above,
the light-cone gauge link introduces rapidity (or light-cone)
singularities for the TMD distribution,
where the radiated gluon (virtual or real) has vanishing 
minus momentum $\ell^-$, or large rapidity $\ln (\ell^+/\ell^-)$.  In the original definition of TMD PDFs by Collins and Soper~\cite{Collins:1981uk,Collins:1981uw} this problem was circumvented by the use of a physical gauge $n\cdot A=0$ for the gauge fields, where $n\approx n_b$, but one keeps $n^2\ne 0$ to regulate the singularities. This procedure induces dependence of the TMD PDF on the Collins-Soper scale
\begin{align}
 \hzeta_a^2 = \frac{(2P_A\cdot n)^2}{|n^2|}
  \,.
\end{align}
Note that $\hzeta_a$ has mass dimension-$1$, and hence differs from the dimension-$2$ parameter $\zeta_a$ used in \secs{tmdpdfs_new}{tmd_defs}.
In Refs.~\cite{Collins:1981uk,Collins:1981uw} a square-root of the hard function, $\sqrt{H_{i\bar i}(Q,\mu;\rho)}$ displayed in \eq{sigma_old} was absorbed into the definition of the TMD PDF, in which case the dependence on $\rho$ cancels out (for an explicit definition of $\rho$, see \eq{defnrho} below). In the modern use of schemes that build off of the original Collins-Soper definitions, this process dependent hard function is instead factored out, and thus enters cross sections as a term that multiplies the TMDs.
Ref.~\cite{Collins:2017oxh} derives explicit relations between the original Collins-Soper TMD PDF definition discussed here, and the modern Collins definition~\cite{Collins:2011zzd} with space-like Wilson lines, discussed in \sec{tmd_defs_overview}. 

The Ji-Ma-Yuan (JMY) scheme~\cite{Ji:2004wu} builds on the original CS TMD PDF definition by relaxing the restriction to a particular gauge, and instead regulating the rapidity divergences by choosing directions $v$ in the Wilson lines to be slightly off-light-cone, with
\begin{align}  \label{eq:vJMY}
v &=(v^+,v^-,0_T) \,,\ \text{ with }\ v^-\gg v^+ >0 \,.
\end{align} 
The use of this $v$ in \eq{tmdCorig} implements the analog of the regulator $\tau$ from \sec{tmdpdfs_new}. 
Unlike the modern Collins definition, here $v$ is a time-like vector.
Physically, the virtual gluons with
rapidity smaller than $\ln v^+/v^-$ are excluded from the parton
distribution. With this definition a dimensionful Collins-Soper scale $\hzeta_a$ emerges as a parameter for the distribution, with\footnote{Note that we reserve the notation $\hat\zeta_a$ for the dimensionless Collins-Soper scale introduced in \sec{latt_def_lorentz}, and hence use $\hzeta_a$ here. }
\begin{align}
 \hzeta_a^2 = \frac{(2P_A\cdot v)^2}{v^2}
  \simeq \frac{2 v^- (P_A^+) ^2}{v^+} \,.
\end{align}
The approximation indicated by the $\simeq$ here is exact only when taking the limit in \eq{vJMY}.
The limit of entirely lifting the cut-off, $v^-/v^+\rightarrow \infty$, corresponds to $\hzeta_a\rightarrow \infty$. 
Similarly, for the TMD distribution for the opposite proton, a Wilson line path parameter $\bar v$
is introduced, and the distribution gains a dependence on an additional parameter $\hzeta_b$, where
\begin{align}  \label{eq:vbarJMY}
  \bar v &=(\bar v^+,\bar v^-,0_T) \,,\ 
   \text{ with }\ \bar v^+\gg \bar v^- >0 \,, \\
   \hzeta_b^2 &= \frac{(2\bar{v}\cdot P_B)^2}{\bar{v}^2 }
    \simeq \frac{2 \bar v^+ (P_B^-)^2}{\bar v^-} 
 \nn \,.
\end{align}
Again the limit of entirely lifting the cut-off, $\bar v^+/\bar v^-\to \infty$, corresponds to $\hzeta_b\to \infty$. 

The corresponding soft function for this scheme is~\cite{Ji:2004wu}
\begin{align} \label{eq:JMYsoft}
 \tilde S^0_{v\bar v}(b_T,\eps,\rho)
 &= \frac{1}{N_c} \bigl< 0 \bigr| {\rm Tr} 
     \bigl[ W^{v\bar v}_{\softstaple}(b_T) \bigr]
     \bigl|0 \bigr>
  \,,\\
 W^{v\bar v}_{\softstaple}(b_T) 
  &= W[0\to -\infty \bar v \to -\infty \bar v+ b_T \to b_T \to -\infty v + b_T \to -\infty v \to 0 ] \nn\\
  &= W_{\bar v}(b_T;0,-\infty) W_{v}(b_T;-\infty,0) 
   W_{\hat b_T}(-\infty v; 0,b_T)
   \nn\\&\qquad\times
   W_{v}(0;0,-\infty) W_{\bar v}(0;-\infty,0) 
   W_{\hat b_T}(-\infty \bar v; b_T,0)
   \,.\nn
\end{align}
Because of the definitions of $v$ and $\bar v$ there is an additional invariant
\begin{align} \label{eq:defnrho}
  \rho^2= \frac{(2v\cdot \bar v)^2}{v^2\,\bar{v}^2}
    \simeq  \frac{v^-\bar v^+}{v^+ \bar v^-} \gg 1 \,,
\end{align}
which appears as a variable in the soft function in \eq{JMYsoft}.
The approximation indicated by the $\simeq$ in \eq{defnrho} is exact only when taking the leading term from the limits in \eqs{vJMY}{vbarJMY}.
Furthermore, in this construction the soft overlap subtraction is equal to the soft function, $S^{0\subt}_{v\bar v}(b_T,\eps,\rho) =S^0_{v\bar v}(b_T,\eps,\rho) $.
Using the direct analog of \eq{tmdpdf_1} this then gives the renormalized TMD PDF for this construction as 
\begin{align} \label{eq:tmdpdf_0}
 \tilde f_{i/p}(x_a, \bt, \mu , x_a\hzeta_a;\rho) &
 =  \lim_{\substack{\eps\to 0}} 
    Z_{\rm uv}^i(\mu,\rho,\eps) \,
    \frac{\tilde f_{i/p}^{0\,\unsub}\bigl(x_a, \bt, \eps, v, x P^+ \bigr)}{\sqrt{\tilde S^{0}_{v\bar v}(b_T,\eps,\rho)}}
    + {\cal O}(v^+,\bar v^-) \,.
\end{align}
In this relation the ${\cal O}(v^+,\bar v^-)$ indicates that the result is expanded in the limits given in \eqs{vJMY}{vbarJMY}. In a similar manner, the other TMD PDF appearing in the Drell-Yan cross section will be $\tilde f_{\bar i/p}(x_b,{\bf b}_T,\mu, x_b\hzeta_b; \rho)$. Here the analog of the relation $\zeta_a\zeta_b=Q^4$ from \eq{zetaazetab} is given by
\begin{align}  \label{eq:QforJMY}
 \frac{x_a \hzeta_a}{\sqrt{\rho}}\, \frac{x_b \hzeta_b}{\sqrt{\rho}} 
   = Q^2  = \sqrt{\zeta_a \zeta_b} \,.
\end{align}
In the JMY scheme the three variables $x_a \hzeta_a$, $x_b\hzeta_b$, and $\rho$ are all large, but the ratio in \eq{QforJMY} is fixed to be $Q^2$.

Note that $\rho$ is an extra variable that is present in the TMD PDF with this JMY definition. In the Drell-Yan cross section the dependence on $\rho$ in this TMD PDF cancels with the $\rho$ dependence of the other TMD PDF and the hard function $H_{i\bar i}(\mu,Q;\rho)$ to yield a $\rho$ independent result, see \eq{sigma_old}. 
To relate individual TMD PDFs in the JMY and $\MSbar$ class of  definitions we must simplify the dependence on the extra variable that appears in the JMY case. This can be accomplished by relating the limits in \eqs{vJMY}{vbarJMY} by choosing
$x_a \tilde\zeta_a = x_b\tilde \zeta_b$ and $y_n=0$ which implies
\begin{align} \label{eq:JMYtoMSbar}
  \Bigl( \frac{v^+ \bar v^+}{v^-\bar v^-} \Big)^{1/4} =  e^{Y_P+Y} = e^{y_n} = 1
  \,,\qquad
  \frac{x_a\hzeta_a}{\sqrt{\rho}} = \sqrt{\zeta_a} 
  =
  \frac{x_b\hzeta_b}{\sqrt{\rho}} = \sqrt{\zeta_b} = Q
  \,.
\end{align} 
The first equation here implies the last two equations. 
Note that the first equation is a ratio of two large numbers that is  fixed to $1$, where $y_n=0$ is the rapidity scheme parameter that appeared in \eq{zeta} for the $\MSbar$ class of schemes.  
In this case we still take the limit of $\hzeta_a,\hzeta_b,\rho\to \infty$, but holding the ratios in \eq{JMYtoMSbar} fixed. This constraint on the limits reduces the number of variables in the JMY definition by one. Comparing \eqs{sigma_old}{sigma_new} we see that with the constraint in \eq{JMYtoMSbar}  the relation between the two definitions is given by
\begin{align} \label{eq:JMYtoCS}
   \tilde f_{i/p}(x_a, \bt, \mu , x_a\hzeta_a=\sqrt{\rho}Q,\rho) 
  &= \sqrt{\frac{H_{i\bar i}(Q^2,\mu)}{H_{i\bar i}(Q^2,\mu;\rho)} } \ \ 
   \tilde f_{i/p}(x_a, \bt, \mu , \zeta_a=Q^2) \,.
\end{align}
Here the $\sqrt{H_{i\bar i}(Q^2,\mu)/H_{i\bar i}(Q^2,\mu;\rho)}$ prefactor acts as a scheme conversion factor, and can be written out as a perturbative series in $\alpha_s$ that is dominated by the scale $\mu\sim Q$. Since this factor has non-trivial $\mu$ dependence, we will see in \sec{evolution} that the TMD evolution equations differ between the $\MSbar$ class of schemes discussed in \secs{tmdpdfs_new}{tmd_defs}, and the schemes discussed here that have $\rho$ dependence. The evolution equations can also be used to relate these TMDs at different values of $\tilde \zeta_a$ and $\zeta_a$ than used in \eq{JMYtoMSbar}.

\subsection{TMD Fragmentation Functions}
\label{sec:TMDFFs}

So far, we have only discussed TMD PDFs in detail, which describe the extraction of a quark from an incoming hadron,
where the quark carries a longitudinal momentum fraction $x$
and a transverse momentum $\kt$ relative to the parent hadron.
The corresponding final-state process is described by a quark
that is produced in a hard interaction and then nonperturbatively
fragments into a detected hadron, for example a pion.
In this case, the hadron carries a longitudinal momentum fraction $z$
and a transverse momentum relative to the fragmenting quark.
This nonperturbative process is encapsulated in a TMD fragmentation function (TMD FF),
the final-state analog of the TMD PDF.
In this section, we provide a brief introduction to unpolarized TMD FFs,
similar to our general discussion of unpolarized TMD PDFs in \sec{tmdpdfs_new}.
In the following \sec{qgspinTMDFF}, we generalize both distributions
to polarized processes, allowing for polarizations of both the quark and the hadron state.

To contrast TMD PDFs and FFs, recall the factorization of the Drell-Yan process in \eq{DrellYan},
\begin{align}
 p(P_A)+p(P_B) \to \ell^+(l) + \ell^-(l') +  X
\,.\end{align}
When measuring the momentum $q = l + l'$ of the $\ell^+ \ell^-$ final state at small $q_T$,
the factorization theorem in \eq{sigma_new} is appropriate,
\begin{align} \label{eq:sigma_DY_rep}
 \frac{\df\sigma^{\rm W}}{\df Q \df Y \df^2\qt} &
 \sim \int\! \df^2\bt \, e^{i \bt \cdot \qt} \,
      \tilde f_{i/p}(x_a, \bt, \mu, \zeta_a) \,
      \tilde f_{\bar i/p}(x_b, \bt, \mu, \zeta_b)
\,,\end{align}
where for brevity we focus only on the Fourier integral of \eq{sigma_new}.
Here, $f_{i/p}$ are the TMD PDFs that describe the extraction of
a parton of flavor $i$ at small transverse momentum from the hadron $p$.
The SIDIS process discussed in \sec{partonSIDIS},
\begin{align}
 e^-(l) + p(P) \to e^-(l') + h(P_h) + X
\,,\end{align}
is closely related to the Drell-Yan process, as one merely exchanges
the roles of the incoming hadron and outgoing lepton,
and the momentum transfer is given by $q = l - l'$.
When the transverse momentum $P_{Th}$ of the detected hadron is small,
the cross section obeys factorization similar to \eq{sigma_DY_rep},
\begin{align} \label{eq:sigma_SIDIS}
 \frac{\df\sigma^W}{\df x \df y \df z_h \df^2 {\bf P}_{hT}} &
 \sim \int\! \df^2\bt \, e^{i \bt \cdot  {\bf P}_{hT} / z} \,
  \tilde f_{i/p}(x, \bt, \mu, \zeta_a) \,
  \tilde D_{h/i}(z_h, \bt, \mu, \zeta_b)
\,,\end{align}
where $x, y, z_h$ are the standard SIDIS observables defined in \eq{xyz}.
Compared to \eq{sigma_DY_rep}, here we have a different hard function $H^{\rm SIDIS}_{i\bar i}$,
which only differs from $H_{i \bar i}$ because $q^2 > 0$ in SIDIS,
while $q^2 < 0$ for Drell-Yan. The TMD PDF $\tilde f_{i/p}$ in \eq{sigma_SIDIS}
is identical to that in Drell-Yan, as it describes the same physics
of extracting a parton at small transverse momentum from the hadron.%
\footnote{More precisely, the definition of the TMD PDF in Drell-Yan
involves Wilson lines extending to $-\infty$, while in SIDIS they
extend to $+\infty$, but this does not impact the unpolarized TMD PDF.
This will be discussed in more detail in \sec{qgspinTMDFF}.}
In contrast, the second TMD PDF in \eq{sigma_DY_rep} was replaced by the
TMD fragmentation function $\tilde D_{h/i}(z_h, \bt, \mu, \zeta_b)$,
which precisely encodes the fragmentation of the final-state parton of
flavor $i$ into the hadron $h$, where $h$ carries the longitudinal
momentum fraction $z_h$, and $\bt$ is Fourier conjugate to its transverse momentum.

\begin{figure*}
 \centering
\subfloat[photon-hadron frame]{ 
\includegraphics[width=0.45\textwidth]{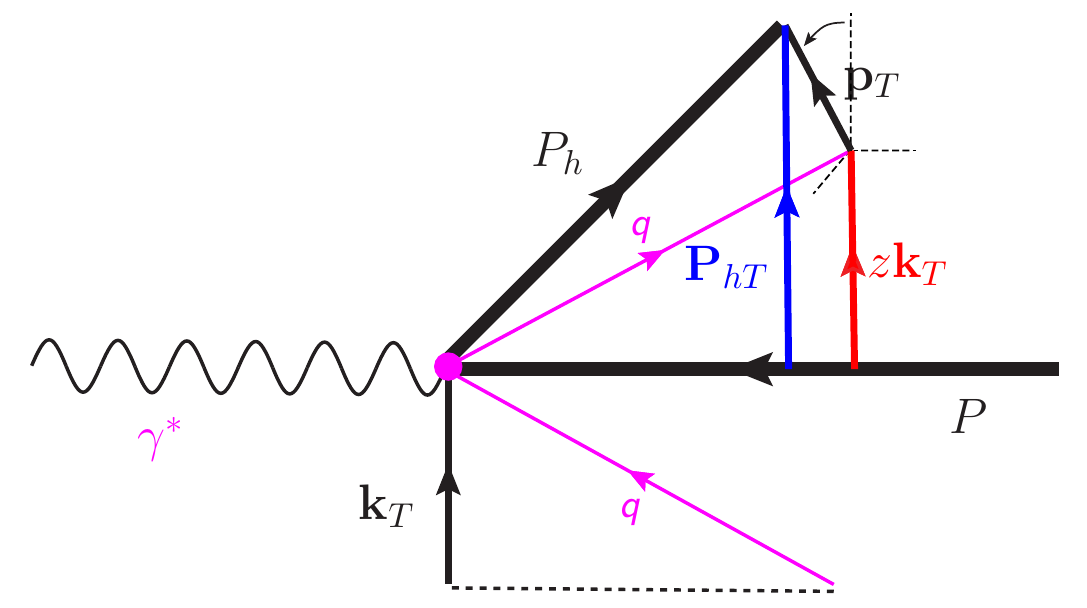} 
}
 \hspace{1cm}
\subfloat[hadron-hadron frame]{ 
\includegraphics[width=0.45\textwidth]{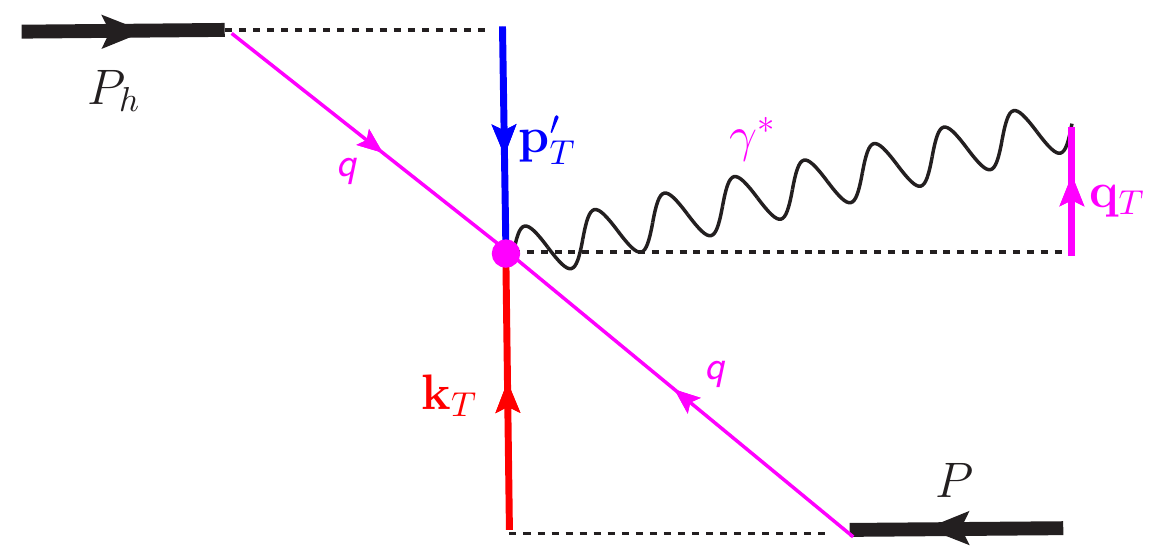} 
}
 \caption{Illustration of the different frames used to describe the kinematics of the SIDIS process, as discussed in the text.}
 \label{fig:SIDIS_frames}
\end{figure*}

Before proceeding, let us discuss one important subtlety in the precise
definition of transverse momentum. In general, the notion of ``transverse''
is defined with respect to two reference directions.
For Drell-Yan, it is natural to use the proton directions as the reference directions,
and consequently it is natural to discuss the transverse momentum of the quark
relative to its parent hadron. In contrast, in SIDIS there are two useful choices:
\begin{enumerate}
 \item[\textcircled{\raisebox{-.3pt}{\footnotesize 1}}] 
 Hadron-Hadron frame: Similar to Drell-Yan, one can define the transverse momentum relative to
       the directions of the incoming proton $p$ and outgoing hadron $h$, 
       both of which thus have vanishing transverse momentum, ${\bf P}_T = {\bf P}_{hT} = 0$.
       For the TMD FF, this implies that $\qt \ne 0$, and the fragmenting parton
       has a non-vanishing transverse momentum $\pt'$ relative to the detected hadron.
       Whenever we use $\qt$ or $\pt'$, we refer to this frame. In summary:
       \begin{align} {\rm hadron{-}hadron~frame:~specified~by}\quad
        \qt \ne 0 \,,\quad
        \pt' \ne 0
       \,.\end{align}
       If we denote the transverse momentum of the parton from the TMD PDF by
       ${\bf k}_{T}$, then the measured momentum transfer is given by
       \begin{align} \label{eq:qT_hadron_frame}
        \qt^{\scriptsize\textcircled{\raisebox{-.4pt}{\tiny 1}}} 
          = -{\bf k}_{T}^{\scriptsize\textcircled{\raisebox{-.4pt}{\tiny 1}}}
            + \pt^{\prime\scriptsize\textcircled{\raisebox{-.4pt}{\tiny 1}}}
       \,.\end{align}
       Here we have included a superscript $\raisebox{.6pt}{\scriptsize\textcircled{\raisebox{-.3pt}{\tiny 1}}}$ to indicate the frame being used.
 \item[\textcircled{\raisebox{-.3pt}{\footnotesize 2}}] 
  Photon-Hadron frame: In the experiment, it is common to define transverse momenta
       relative to the momenta of the incoming and outgoing leptons, 
       as these can be measured very well. Since this definition does not
       uniquely specify the angles, the measurements are commonly performed
       in the so-called Trento frame~\cite{Bacchetta:2004jz}, where $\bf q$
       is aligned along the $z$ axes. In this case, the incoming hadron
       still has vanishing transverse momentum ${\bf P}_T = 0$, 
       and the same interpretation of the TMD PDF applies.
       However, the outgoing hadron $h$ now has transverse momentum $\pt \ne 0$
       relative to the fragmenting parton.
       Whenever we use $\pt$, we refer to this frame. In summary:
       \begin{align} {\rm photon{-}hadron~frame:~specified~by}\quad
         \pt \ne 0
       \,.\end{align}
       If we denote the transverse momentum of the parton from the TMD PDF by
       ${\bf k}_{T}$, then the measured hadron momentum is given by
       \begin{align} \label{eq:PTh_parton_frame}
        {\bf P}_{hT}^{\scriptsize\textcircled{\raisebox{-.3pt}{\tiny 2}}}
         = z_h {\bf k}_{T}^{\scriptsize\textcircled{\raisebox{-.3pt}{\tiny 2}}}
          + \pt^{\scriptsize\textcircled{\raisebox{-.3pt}{\tiny 2}}}
       \,.\end{align}
       Again, the superscript $\raisebox{.6pt}{\scriptsize\textcircled{\raisebox{-.3pt}{\tiny 2}}}$  indicates the frame being used.
\end{enumerate}
These two frames are illustrated in \fig{SIDIS_frames}.
Importantly, these two frames are related by
\begin{align} \label{eq:relation_frames}
 \pt^{\scriptsize\textcircled{\raisebox{-.3pt}{\tiny 2}}} 
  = - z_h {\bf p}_T^{\prime\scriptsize\textcircled{\raisebox{-.3pt}{\tiny 1}}}
\,,\qquad
 {\bf P}_{hT}^{\scriptsize\textcircled{\raisebox{-.3pt}{\tiny 2}}} 
 = - z_h \qt^{\scriptsize\textcircled{\raisebox{-.3pt}{\tiny 1}}}
\,,\end{align}
where the second relation follows from the first relation
together with \eqs{qT_hadron_frame}{PTh_parton_frame}.
\eq{relation_frames} is to be understood in terms of the components
in the two different frames. Namely, $\pt$ and ${\bf P}_{hT}$
are specified in the parton frame, while $\kt$ and $\qt$ are specified in
the hadron frame. \eq{relation_frames} then allows one to easily
transform between these two frames~\cite{Boglione:2019nwk,Collins:2011zzd}.

Having clarified this important subtlety in the definition of
the transverse direction, we are now in a position to define the
TMD FF in analogy to the TMD PDF. We first recall the definition
of the unsubtracted TMD PDF in \eq{beamfunc},
\begin{align} \label{eq:beamfunc_rep}
 \tilde f_{i/p}^{0\,\unsub}(x,\bt,\eps,\tau,x P^+)
 &= \int\frac{\df b^-}{2\pi} e^{-\img b^- (x P^+)}
 \Bigl< p(P) \Bigr|  \Bigl[ \bar \psi^{0}_i(b^\mu)
 W_{\sqsubset}(b^\mu,0)  \frac{\gamma^+}{2}
  \psi^{0}_i(0) \Bigr]_\tau \Bigl| p(P) \Bigr>
\,,\end{align}
where the incoming proton is aligned along the $n_a$ direction, and $b^\mu = (0, b^-, \bt)$.
In the SIDIS process, one still encounters a $n_a$-collinear proton,
while the outgoing hadron is aligned along the orthogonal direction $n_b$.
For ease of comparison to \eq{beamfunc_rep}, here we provide the definition
of the TMD FF for a hadron along the $n_a$ direction with large $P^+$ momentum.
The corresponding definition for an outgoing hadron aligned along the $n_b$-direction
is easily obtained by replacing $n_a \leftrightarrow n_b$, and consequently $b^- \to b^+$ and $P^+ \to P^-$.

\index{TMD fragmentation function!quark definition}
\index{TMD soft function}
The bare unsubtracted TMD FF for a parton of flavor $i$ inside a hadron $h$ is defined as
\begin{align} \label{eq:TMDFF_unsub}
 \tilde \Delta_{h/i}^{0\unsub}(z, \bt, \eps, \tau, P^+/z) &
 = \frac{1}{4 N_c z} {\rm Tr} \!\int\!\frac{\df b^-}{2\pi} \sum_X e^{\img b^- (P^+/z)} \gamma^+_{\alpha\alpha'}
  \nn\\&\times
   \MAe{0}{ \big[ (W_{\halfstapleu} \psi_i^{0\alpha})(b) \big]_\tau }{h(P),X}
   \MAe{h(P),X}{\,\big[ (\bar\psi_i^{0\alpha'} W_{\halfstaple})(0)\big]_\tau }{0}
\,.\end{align}
As for the TMD PDF, here $P$ is the hadron momentum, $b^\mu = (0, b^-, \bt)$,
the superscript $0$ denotes bare fields, and $\tau$ is a generic rapidity regulator.
The TMD FF is normalized by $1/z$ and $1/(4 N_c)$,
the latter corresponding to the number of colors and spin states,
and the trace is over color and spin indices $\alpha, \alpha'$.
The key difference between \eqs{beamfunc_rep}{TMDFF_unsub} is that the hadron state appears
as an out-state in the matrix element, with $\sum_X$ denoting the sum
over all additional hadronic final states $X$.
(In the TMD PDF, the proton appears in the initial state, and the complete sum over $X$
in the final state can be eliminated by unitarity.)

Recall that \eq{TMDFF_unsub} is defined in a frame where
$P$ has no transverse momentum, while the quark field $\psi$
acquires a transverse momentum $\pt'$ conjugate to $\bt$.
Thus, the corresponding momentum-space TMD FF is given with respect to $\pt'$,
\begin{align} \label{eq:TMDFF_FT} &
 \tilde \Delta_{h/i}^{0\unsub}(z, \bt, \eps, \tau, P^+/z)
 = \int \df^2 \pt' \, e^{+\img \pt' \cdot \bt}
   \Delta_{h/i}^{0\unsub}(z, -z\pt', \eps, \tau, P^+/z)
\,.\end{align}
The sign of the Fourier phase is fixed by the corresponding sign
in \eq{TMDFF_unsub}. Note that it differs by a sign from the convention
for the TMD PDF, compare \eq{tmdpdf_bspace}, reflecting that
\eqs{beamfunc_rep}{TMDFF_unsub} differ in the sign of their Fourier phase.

The Wilson lines $W_{\halfstaple}$ and $W_{\halfstapleu}$  correspond to ``half''
of the staple-shaped Wilson line defined in \eq{Staple_Wilson_line},
as indicated by their symbols, and again are important for the gauge invariance
of the fragmentation function.  In the fragmentation functions the transverse link
that appears in the TMD PDFs is split in two parts, and
as will be discussed in \sec{TMDFFs}, the Wilson lines extend to $+\infty$
as opposed to the $-\infty$ for TMD PDFs in Drell-Yan, so we can write 
$W_{\hat b_T}(+\infty n_b; 0,b_T) = W_{\hat b_T}(+\infty n_b;+\infty,b_T)W_{\hat b_T}(+\infty n_b;0,+\infty)$.
Explicitly, when taken on the light-cone the Wilson lines appearing in the
unsubtracted TMD FFs in \eq{unsubTMDFFspin} are defined as~\cite{Collins:1981uk,Collins:2011zzd}
\begin{align} \label{eq:half_staple_Wilson_line}
 W_{\halfstapleu}(b) &
 = W_{n_b}(b^\mu;+\infty,0)
   W_{\hat b_T}\!\bigl(+\infty n_b; +\infty, b_T \bigr)
\,,\nn\\
 W_{\halfstaple}(0) &
 = W_{\hat b_T}\!\bigl(+\infty n_b; 0, +\infty \bigr)
   W_{n_b}(0;0,+\infty)
\,,\end{align}
where the individual Wilson-line segments are defined in \eq{Wilson_lines}.
As indicated by the brackets $[\cdots]_\tau$ in \eq{unsubTMDFFspin}, the same rapidity regulators discussed in \sec{tmd_defs} must also be implemented for the unsubtracted fragmentation function. As we have seen, most often this modifies the precise definition of the Wilson lines.

Finally, it remains to combine the unsubtracted TMD FF
with a soft factor and UV renormalization factor as in \eq{tmdpdf_1},
\begin{align} \label{eq:TMDFF_ren}
 \tilde \Delta_{h/i}(z,\bt,\mu,\zeta) &
 =  \lim_{\substack{\eps\to 0 \\ \tau\to 0}} Z_{\rm uv}^i(\mu,\zeta,\eps) \,
    \frac{\tilde \Delta_{h/i}^{0\unsub}(z,\bt,\eps,\tau,P^+/z)}{\tilde S^{0\,\subt}_{n_a n_b}(b_T,\eps,\tau)}
    \sqrt{\tilde S_{n_a n_b}^{0}(b_T,\eps,\tau)}
\,.\end{align}
Here $\zeta=2 (p_h^+)^2 e^{-2y_n}/z^2$ for the fragmentation case, where $y_n$ is
the rapidity cutoff parameter defined (for example) in the Collins approach with
 space-like Wilson lines for the rapidity regulator in \sec{tmd_defs}.
By inverting \eq{TMDFF_FT} and employing \eq{relation_frames},
we thus obtain the desired TMD FF in momentum space in either frame as
\begin{align} \label{eq:unsubTMDFFspin_pT}
 \Delta_{h/i}(z, \pt = -z\pt', \mu, \zeta) &
 = \int\frac{\df^2\bt}{(2\pi)^2} e^{-\img \pt' \cdot \bt}
   \tilde \Delta_{h/i}(z, \bt, \mu, \zeta)
\,.\end{align}

\subsection{Quark and Gluon Spin Dependent TMDs and FFs}
\label{sec:qgspinTMDFF}

In this section we provide a number of generalizations of the field theory definition of the unpolarized TMD PDF discussed above.
This includes both measuring the quark spin structure and allowing for the hadron to be polarized,
and the corresponding TMD PDFs were summarized above in \fig{qTMDPDFsLP}.
In addition, we also consider spin-dependent gluon TMD PDFs, as well as spin-dependent TMD fragmentation functions (TMD FFs) that are needed in SIDIS.
Note that we limit ourselves to TMD PDFs and TMD FFs at leading power in the transverse momentum (often referred to as leading twist).
Their extension at subleading twist will be discussed in \chap{twist3}.

We will limit our discussion to spin-$1/2$ hadrons such as the proton. For a review which includes results for hadrons of other spin, see Ref.~\cite{Manohar:1992tz}. The spinor $u(P,S)$ for a spin-$1/2$ hadron with polarization vector $S^\mu$ satisfies
\begin{align}  \label{eq:Hspinorproduct}
  u(P,S) \bar u(P,S) & = (\slashed{P}+M) \frac12 \bigl(1+\gamma_5 \slashed{S}\bigr)
  \,,
\end{align}
where $M$ is the hadron mass. The spin vector can be decomposed in a covariant fashion as
\begin{align} \label{eq:spin_vector}
 S^\mu = S_L \frac{ P^+ n_a^\mu - P^- n_b^\mu}{M} + S_T^\mu
\,,\qquad
 S^2 = -(S_L^2 + S_T^2)
\,,\end{align}
where $S_L$ and $S_T^\mu$ denote the longitudinal and transverse spin components.
To see the connection between the spin vector $S^\mu$ and notions that are familiar from the treatment of spin in quantum mechanics, we can work in the hadron rest frame where  $S^\mu=(0,{\bf S})=(0,{\bf S}_T,S_L)$. In this frame  \eq{Hspinorproduct} can be written in terms of the standard spin density matrix $\rho$ for a spin-$1/2$ particle,
\begin{align}
 u(P,S) \bar u(P,S) 
   & = 2M \left( \begin{array}{cc} \rho & 0 \\ 0 & 0 \end{array} \right) 
   \,,
  \qquad \rho = \frac12 \bigl(1 + \bm{\sigma}\cdot {\bf S}\bigr)
  \,,
\end{align}
where $\bm{\sigma}$ is the usual vector of Pauli spin matrices. For a pure spin state we have ${\bf S}^2 = -S^2 = 1$, whereas for a mixed polarization state one has ${\bf S}^2<1$.  In the following sections we will make extensive use of $S_L$ and $S_T^\mu$ when discussing the TMDs that are probed by longitudinal and transversely polarized hadronic targets.

For polarized hadrons, there are two distinguished transverse directions, namely $\pt$ and ${\bf S}_T$.
To describe all possible transverse structures that can be built out of these quantities,
it will be useful to introduce a transverse metric $g_T^{\alpha\beta}$ and the transverse fully-antisymmetric tensor $\eps_T^{\alpha\beta}$.
Following \cite{Bacchetta:2006tn}, we define these tensors as
\begin{align} \label{eq:def_gT_epsT}
 g_T^{\alpha\beta} = g^{\alpha\beta} - \bigl( n_a^\alpha n_b^\beta + n_a^\beta n_b^\alpha\bigr)
\,,\qquad
 \eps_T^{\alpha\beta} = \eps^{\alpha\beta\rho\sigma} n_{a\,\rho} n_{b\,\sigma} = \eps^{\alpha\beta-+}
\,.\end{align}
With our choice of $n_{a,b}^\mu$ as given in \eq{nab}, the only non-vanishing components of these tensors are given by
\begin{align} \label{eq:transverse_metric}
 g_T^{11} = g_T^{22} = -1 \,,\qquad \eps_T^{12} = - \eps_T^{21} = 1
\,.\end{align}

Throughout this section, we will always consider the case of a $n_a$-collinear hadron, both for the quark and gluon TMD PDFs
and the corresponding TMD fragmentation functions introduced in the following.
The corresponding expressions for $n_b$-collinear hadrons can easily obtained by replacing $n_{a} \leftrightarrow n_{b}$,
which also exchanges $x^\pm \to x^\mp$ for each four vector $x^\mu$.
We caution the reader that in principle, this would also affect the definitions of the longitudinal proton spin $S_L$ in \eq{spin_vector}
and the transverse tensor $\eps_T^{\alpha\beta}$ in \eq{def_gT_epsT}.
In practice, it is of course more convenient to always use the same definition of $S_L$ and $\eps_T^{\alpha\beta}$
for both $n_a$-collinear and $n_b$-collinear hadrons, which can be compensated by a sign flip of all  $S_L$ and $\eps_T^{\alpha\beta}$
appearing in the $n_b$-collinear case. This subtle yet important sign is often not stated explicitly in the literature,
and has to be kept in mind when comparing explicit expressions.

\subsubsection{Universality of TMD PDFs and TMD FFs}
\label{sec:universality}

\index{universality}

In the following sections we present suitable generalizations of the definition of TMD PDFs to cases with spin polarization and to processes other than Drell-Yan.  We will also give the analogous definitions for TMD FFs.  Here we address a key ingredient needed for this generalization, namely the proper treatment of the process dependent incoming or outgoing directions for the Wilson line operators. 
As discussed above in \sec{beyondparton},  this dependence is important for obtaining a nonzero Sivers function, since this distribution would otherwise vanish by the $TP$ invariance of QCD~\cite{Collins:2002kn}.
\index{Sivers function $f_{1T}^{\perp}$!process dependence}

To generalize our notation to incorporate Wilson line paths that come in from $-\infty$ or go out to $+\infty$, we consider unsubtracted TMD PDFs with the staple shaped Wilson lines
\begin{align}  \label{eq:stapleinout}
 W_{\sqsubset}(b^\mu,0)
   &= W_{n_b}(b^\mu;-\infty,0) 
      W_{\hat b_T}\!\bigl(-\infty n_b; 0,b_T\bigr)
       W_{n_b}(0;0,-\infty) \,,
       \nn\\
 W_{\sqsupset}(b^\mu,0)
 &= W_{n_b}(b^\mu;+\infty,0) 
 W_{\hat b_T}\!\bigl(+\infty n_b; 0,b_T\bigr)
 W_{n_b}(0;0,+\infty) \,,    
\end{align}
where the incoming or outgoing staple is indicated by the subscripts on the LHS.
Similarly, soft functions involve Wilson loops with paths that extend to $\pm$ infinity, given by matrix elements of 
\begin{align}  \label{eq:softinout}
W_{\rightsoftstaple}(b_T) 
 &= 
  W_{n_a}(b_T;0,-\infty) W_{n_b}(b_T;-\infty,0)
  W_{\hat b_T}(-\infty n_b; 0,b_T)
  \nn\\&\qquad\times
  W_{n_b}(0;0,-\infty) W_{n_a}(0^\mu;-\infty,0) 
  W_{\hat b_T}(-\infty n_a; b_T,0)
\,,
 \nn\\
 W_{\leftsoftstaple}(b_T) 
 &= 
 W_{n_a}(b_T;0,+\infty) W_{n_b}(b_T;+\infty,0)
 W_{\hat b_T}(+\infty n_b; 0,b_T)
 \nn\\&\qquad\times
 W_{n_b}(0;0,+\infty) W_{n_a}(0^\mu;+\infty,0) 
 W_{\hat b_T}(+\infty n_a; b_T,0)
   \,.
\end{align}
This generalizes the results quoted above in \eq{Staple_Wilson_line} and drawn in \fig{wilsonlines}, where only the cases $W_{\sqsubset}$ and $W_{\rightsoftstaple}$ were considered. In general the Wilson lines in \eqs{stapleinout}{softinout} also involve additional rapidity regulators, as discussed in detail in \sec{tmd_defs}. The soft function involves a vacuum matrix element of $W_{\rightsoftstaple}$ or $W_{\leftsoftstaple}$ and hence is not sensitive to the hadron state, or the quark or gluon operator polarization. It is therefore universal up to the direction for the Wilson lines and their color representation. On the other hand, the unsubtracted TMD PDF does depend on the choice of hadron state, the quark or gluon operator polarization, and in principal on the Wilson line directions.

To be definite we consider the three TMD process of Drell-Yan, SIDIS, and back-to-back hadron production in $e^+e^-$ annihilation, as illustrated in \fig{TMDcrosssect}. 
For the analysis of these processes we follow Refs.~\cite{Collins:2004nx,Collins:2011zzd}, where the structure of the Wilson line operators that are consistent with the derivation of factorization have been analyzed using the space-like Wilson line regulator $n_a\to n_A$ and $n_b\to n_B$ given above in \eq{Collins_rap}. A key ingredient in this analysis is considering which directions are consistent with momentum contour deformations out of the so-called Glauber region (see \chap{Factorization}).
A summary of the key results is given in Table~\ref{tbl:wilson_directions}. 
For each TMD appearing in a given process, the $\pm\infty$ directions are correlated between the soft function and the hadronic matrix elements giving various unsubtracted TMD PDFs $\tilde f_{i/p_S}^{[\Gamma]0\unsub}$ or
unsubtracted TMD FFs $\tilde D_{i/p_S}^{[\Gamma]0\unsub}$, see Ref.~\cite{Collins:1999dz}. 
In particular, for the TMD PDFs the hadronic and vacuum matrix elements are taken to involve different Wilson line paths as follows
\begin{align}
  & \tilde F_{i/p_S}^{-\infty}:  
  &&  \tilde f_{i/p_S}^{[\Gamma]0\unsub} \text{ defined using } W_{\sqsubset} \,,
  && \tilde S^{0}_{n_a n_b} \text{ defined using } W_{\rightsoftstaple}
  \\
  & \tilde F_{i/p_S}^{+\infty}:  
  &&  \tilde f_{i/p_S}^{[\Gamma]0\unsub} \text{ defined using } W_{\sqsupset} \,,
  && \tilde S^{0}_{n_a n_b} \text{ defined using } W_{\leftsoftstaple}
  \,. \nn
\end{align}
Here the first line gives the same definition as in \eqs{beamfunc}{softfunc}, while the second line modifies the Wilson lines used in these definitions.
(For simplicity we do not discuss the freedom in the direction dependence of soft subtractions, and instead refer the interested reader to Ref.~\cite{Collins:2004nx}.) For the Drell-Yan process one obtains Wilson lines from $-\infty$ in the TMD PDFs, while for SIDIS one finds lines extending out to $+\infty$. The results for these two cases can be related by a combination of time-reversal and parity ($TP$) symmetry~\cite{Collins:2002kn}. This gives equality for the T-even distributions and an extra minus sign for the T-odd distributions, as shown in the last two rows of Table~\ref{tbl:wilson_directions}.  A more detailed presentation of the physical argument for this sign flip is given below in \sec{leadingTMDPDF}. 

The process dependence of the Wilson lines appearing in the  fragmentation functions in SIDIS and $e^+e^-$ annihilation have also been analyzed in Ref.~\cite{Collins:2004nx}. Here is it found that all lines can be consistently taken to point out to $+\infty$, and hence that all leading power fragmentation functions are universal without sign flips.  For this reason we postpone giving definitions of the Wilson lines appearing in the unsubtracted TMD FFs, which are denoted by 
$W_{\halfstaple}$ and
$W_{\halfstapleu}$, in \sec{LQTMDFF} below. 

{
	\renewcommand{\arraystretch}{1.4}
	\begin{table*}[t!]
		\centering
		\begin{tabular}{| p{6cm} | c | c | c | c |}
			\hline
			\multirow{2}{*}{Function} & Drell-Yan & SIDIS  & SIDIS & $e^+e^-$ \\
			& TMD PDF    & TMD PDF & TMD FF & TMD FF
			\\ \hline
			Wilson lines in $\tilde S_{n_a n_b}^{0\pm\infty}$
			& $W^{-\infty}_{\rightsoftstaple}$
			& $W^{+\infty}_{\leftsoftstaple}$ 
			& $W^{+\infty}_{\leftsoftstaple}$
			& $W^{+\infty}_{\leftsoftstaple}$
			\\[1ex] \hline
			Wilson lines in 
			$\tilde f_{i/p_S}^{[\Gamma]0\unsub}$ or
			$\tilde D_{i/p_S}^{[\Gamma]0\unsub}$
			& $W_{\sqsubset}$
			& $W_{\sqsupset}$
			& $W_{\halfstaple}$,
			$W_{\halfstapleu}$
			& $W_{\halfstaple}$,
			$W_{\halfstapleu}$
			\\[0.5ex] \hline
			TMDs
			& $\tilde F_{i/p_s}^{\mbox{\tiny $-\infty$}}$
			& $\tilde F_{i/p_s}^{\mbox{\tiny $+\infty$}}$
			& $\tilde D_{i/p_s}$
			& $\tilde D_{i/p_s}$
			\\[1ex] \hhline{|=|=|=|=|=|}
			T-even\ $\tilde F=\tilde f_1, \tilde g_{1}, \tilde g_{1T}^\perp,$
			& \multicolumn{2}{c|}{ \multirow{2}{*}{$\tilde F^{-\infty}_{i/p_S}(x,b_T) = \tilde F^{+\infty}_{i/p_S}(x,b_T)$} }
			& \multicolumn{2}{c|}{ \multirow{3}{*}{universal} }
			\\
			\hspace{2.1cm}$\tilde h_{1L}^\perp, \tilde h_1, \tilde h_{1T}^\perp$
			& \multicolumn{2}{c|}{  }
			& \multicolumn{2}{c|}{  }
			\\ \hhline{|-|-|-|}
			T-odd\ $\tilde F=\tilde f_{1T}^\perp, \tilde h_1^\perp$
			& \multicolumn{2}{c|}{$\tilde F^{-\infty}_{i/p_S}(x,b_T) = -\tilde F^{+\infty}_{i/p_S}(x,b_T)$ }
			& \multicolumn{2}{c|}{}
			\\[1ex] \hline
		\end{tabular}
		\caption{Summary of Wilson line directions for TMD PDFs and TMD FFs in different processes~\cite{Collins:2004nx,Collins:2011zzd}.  The TMD FFs are seen to be universal between SIDIS and $e^+e^-\to H_1 H_2 X$. For the TMD PDFs the distributions in Drell-Yan and SIDIS can be related by time-reversal symmetry as indicated.}
		\label{tbl:wilson_directions}
	\end{table*}
}

Since all leading power TMD PDFs can be expressed in terms of the same eight functions, the underlying hadronic distributions apply equally well to both Drell-Yan and SIDIS once the extra sign for the T-odd distributions is taken into account, and hence are universal. 
To keep track of the extra process dependent minus sign in subsequent sections, we define 
\begin{align} \label{eq:kappaDYSIDIS}
 \kappa = \bigg\{  \begin{array}{cl} +1 & \ \ (\text{Drell-Yan}) \\ -1 & \ \ ({\rm SIDIS}) 
 	  \end{array} \,,
\end{align}
and will include factors of $\kappa$ in front of the TMD PDFs $h_1^\perp$ and $f_{1T}^\perp$ in suitable places. This implies that we will always use the relations shown in Table~\ref{tbl:wilson_directions} to express the TMD PDFs in terms of those defined with Wilson lines that extend from $-\infty$ as in Drell-Yan. With this choice there is no longer any ambiguity in expressions which require specifying a Wilson line direction. Therefore in subsequent sections we will drop the $\pm\infty$ superscripts on the Wilson lines that were used for the discussion in this section.

\subsubsection{Leading Quark TMD PDFs}
\label{sec:leadingTMDPDF}

\index{TMD parton distribution function!quark definitions with spin}

To generalize the definition in \eq{tmdpdf_1} of the unpolarized leading power quark TMD PDF to include other quark spin structures and to include polarized protons, we modify the definition of the unsubtracted TMD PDF $f_{i/p}^{0\unsub}$ in \eq{beamfunc} to include a general spin matrix $\Gamma$:
\begin{align} \label{eq:unsubTMDPDFspin}
 \tilde f_{i/p_S}^{[\Gamma]0\unsub}(x, \bt, \eps, \tau, x P^+) &
 = \int\!\frac{\df b^-}{2\pi} e^{-\img b^- (x P^+)}
   \Bigl< p(P,S) \Big| \Bigl[
    \bar \psi^i(b^\mu) W_\sqsubset(b^\mu,0) \frac{\Gamma}{2} \psi^i(0) \Bigr]_\tau
   \Big| p(P,S) \Bigr>
  \,.
\end{align}
Here $S$ in the states indicates the spin of the proton.
The full set of TMD PDFs at leading order in small transverse momentum, so-called leading twist, is obtained by considering the Dirac structures 
\begin{align} \label{eq:Gamma_LP}
 \Gamma \in \{\, \gamma^+ \,,\, \gamma^+ \gamma_5 \,,\, \img \sigma^{\alpha+} \gamma_5 \,\}
\,,\end{align}
where $\sigma^{\mu\nu} = \frac{\img}{2} [\gamma^\mu, \gamma^\nu]$. To see why there are only these three structures at leading power, we note that the leading power operators are built out of the ``good-components'' of the fermion field~\cite{Kogut:1969xa}, which obey $\frac12 \gamma^-\gamma^+\psi^i = \psi^i$. In SCET these projection relations are obeyed by the collinear fermion fields that are used to construct leading power operators, see~\cite{Bauer:2002nz}. It is straightforward to check that any other choices for $\Gamma$ will either give zero or can be reduced to one of those in \eq{Gamma_LP} when sandwiched between these projectors in $\bar \psi^i (\frac12 \gamma^+\gamma^-) \Gamma (\frac12 \gamma^-\gamma^+)\psi^i $.

Inserting \eq{unsubTMDPDFspin} into \eq{tmdpdf_1} with the corresponding soft function, subtraction function, and UV renormalization factors yields the renormalized spin-dependent TMD PDF $\tilde f_{i/p_S}^{[\Gamma]}(x, \bt, \mu, \zeta)$.  
These TMD PDFs can be affected by whether their Wilson lines point to $\pm\infty$, as discussed in \sec{universality}.  We work in a convention where symmetry relations are always used to convert TMD PDFs to the versions obtained with Wilson lines from $-\infty$, and encode the extra process dependent sign that appears for the time reversal odd TMD PDFs in a coefficient $\kappa =\pm 1$, see \eq{kappaDYSIDIS}.

\begin{figure*}[t]
 \centering
  \includegraphics[width=0.85\textwidth]{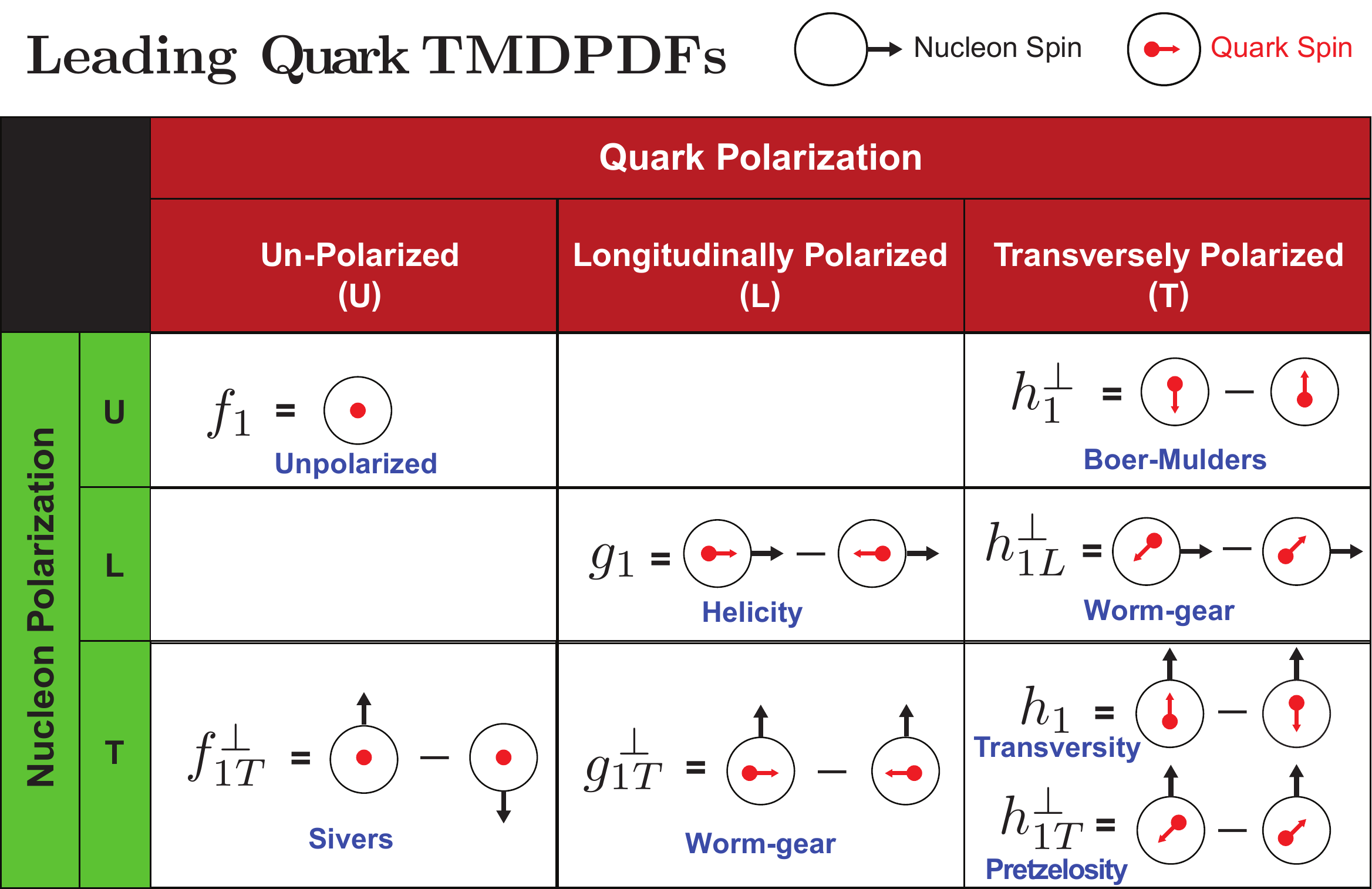}
 \caption{Leading power quark parton distribution functions for the proton or a spin-$1/2$ hadron.}
 \label{fig:qTMDPDFsLP2}
\end{figure*}

For a spin-$1/2$ particle such as the proton, using the spin vector from \eq{spin_vector}, the spin-dependent TMD PDFs can be further decomposed into eight independent structures~\footnote{There exist different notations for the corresponding TMDs, for instance those used by Torino-Cagliari group. The notations and relations to our notations can be found  in Refs.~\cite{Bacchetta:2004jz, Anselmino:2005sh, DAlesio:2007bjf, Anselmino:2011ch}.} as follows~\cite{Ralston:1979ys,Tangerman:1994eh,Boer:1997nt,Mulders:1995dh}\footnote{Notice that in the original naming convention for TMDs, symbol $\perp$ was used for TMDs with at least one uncontracted index $k_T^\alpha$  in the correlator. In our convention, we use $\perp$  to denote all leading-twist TMDs proportional to $k_T$ in the correlator. An example of the consequence of such a convention is $g_{1T}^\perp(x,k_T)$ which was originally denoted as $g_{1T}(x,k_T)$ in Ref.~\cite{Mulders:1995dh}.}:
\begin{align} \label{eq:tmd_decomposition}
 f_{i/p_S}^{[\gamma^+]}(x,\kt,\mu,\zeta) &
 = f_1(x,k_T) - \frac{\eps_T^{\rho\sigma} k_{T\rho} S_{T\sigma}}{M} \kappa  \, f_{1T}^\perp(x,k_T)
\,,\\
 f_{i/p_S}^{[\gamma^+\gamma_5]}(x,\kt,\mu,\zeta) &
 = S_L \, g_{1}(x,k_T)
 - \frac{k_T \cdot S_T}{M} g_{1T}^\perp(x,k_T)
\,,\nn\\
 f_{i/p_S}^{[\img \sigma^{\alpha+}\gamma_5]}(x,\kt,\mu,\zeta) &
 = S_T^\alpha h_1(x,k_T)
 + \frac{S_L k_T^\alpha}{M} h_{1L}^\perp(x,k_T)
 \nn\\&\quad
 - \frac{\kt^2}{M^2} \biggl(\frac12 g_T^{\alpha\rho} + \frac{k_T^\alpha k_T^\rho}{\kt^2}\biggr) S_{T\,\rho} h_{1T}^\perp(x,k_T)
 - \frac{\eps_T^{\alpha\rho} k_{T\rho}}{M} \kappa\,  h_1^\perp(x,k_T)
\,.\nn
\end{align}
Here for brevity on the right hand side we have dropped the arguments $\mu$ and $\zeta$, as well as the flavor and hadron subscripts $i/p_S$.
Note that all functions on the right-hand side only depend on the magnitude $k_T = |\kt|$, but that all $k_T^\mu$ with an explicit Lorentz index or scalar products are evaluated in Minkowskian metric.
In \eq{tmd_decomposition}, $M$ denotes the nucleon mass, which is inserted such that all distributions on the right-hand side have the same mass dimension.
The displayed greek indices are transverse, and the transverse tensors $g_T^{\alpha\beta}$ and $\eps_T^{\rho\sigma}$ are defined in \eq{def_gT_epsT}.
The spin vector of the proton $p_S$ is decomposed as given in \eq{spin_vector}.

\eq{tmd_decomposition} gives the eight leading spin-dependent TMDs for a spin-$\frac12$ hadron~\cite{Ralston:1979ys,Tangerman:1994eh,Boer:1997nt,Mulders:1995dh} following the conventions of~\cite{Bacchetta:2006tn}.  
This general decomposition is obtained by considering the most general decomposition of the correlator in \eq{unsubTMDPDFspin} with open spinor indices on the fermion fields that satisfy the good component projection relations $\frac12 \gamma^-\gamma^+ \psi^i=\psi^i$, and which is linear in the hadron spin-polarization vector. The contractions with the Dirac structures in \eq{Gamma_LP} then suffice to project out all the leading power TMDs, as given in \eq{tmd_decomposition}.
The complete decomposition at subleading-twist, which also contains the Dirac structures not included in \eq{Gamma_LP}, can be found in~\cite{Bacchetta:2004zf,Goeke:2005hb,Bacchetta:2006tn}, and is discussed further in \chap{twist3} below.
As summarized in \fig{qTMDPDFsLP2}, the different structures correspond to specific polarizations for the quark operator and hadron, three of which have collinear counterparts (i.e.~integrated over $\kt$):
\begin{itemize}
 \item $f_1(x,k_T)$ describes an unpolarized quark inside an unpolarized hadron, similar to the unpolarized collinear distribution $f_1(x)$.
 \item $g_{1}(x,k_T)$  is the helicity distribution which describes a longitudinally polarized quark inside a longitudinally polarized hadron,
       similar to the collinear helicity distribution $g_{1}(x)$.
 \item $h_1(x,k_T)$ is the transversity distribution which describes a transversely polarized quark inside a transversely polarized hadron,
       similar to the collinear transversity distribution $h_{1}(x)$.
\end{itemize}
The remaining distributions only arise when measuring transverse momenta and have no collinear counterpart:
\begin{itemize}
 \item $f_{1T}^\perp(x,k_T)$  is the Sivers function~\cite{Sivers:1989cc} which describes an unpolarized quark inside a transversely polarized hadron. 
 \index{Sivers function $f_{1T}^{\perp}$!introduction}
       Since it is $T$-odd, it was originally believed to vanish due to symmetry arguments~\cite{Collins:1992kk}.
       It was later clarified that it is non-vanishing when correctly taking the Wilson lines in the definition of the unsubtracted TMD PDF and soft function into account~\cite{Brodsky:2002cx, Collins:2002kn, Brodsky:2002rv}.
 \item The function $g_{1T}^\perp(x,k_T)$ describes longitudinally polarized quarks in a transversely polarized hadron,
       and vice versa $h_{1L}^\perp(x,k_T)$ describes transversely polarized quarks in a longitudinally polarized hadron~\cite{Kotzinian:1997wt}.
       They are referred in the literature as  ``worm-gear'' T and L functions or Kotzinian-Mulders~\cite{Tangerman:1994eh,Kotzinian:1995cz} functions.
 \item  $h_1^\perp(x,k_T)$ is the Boer-Mulders function~\cite{Boer:1997nt} which describes a transversely polarized quark in an unpolarized hadron.
       Like the Sivers function $f_{1T}^\perp$, it is time-reversal odd.
 \item $h_{1T}^\perp(x,k_T)$ is the pretzelosity
 \index{pretzelosity $h_{1T}^{\perp}$}
 function, which contributes to the distribution of a transversely polarized quark in a transversely polarized hadron~\cite{Mulders:1995dh}, in addition to the transversity $h_1(x,k_T)$. Curiously, the name of this function
 stems from its expected shape~\cite{Miller:2003sa} published by G.~Miller, which was also highlighted in the New York Times \cite{nytimes03}, exhibiting the unusual shape of the proton due to the presence of this function.
\end{itemize}
Following the discussion in Ref.~\cite{Kang:2009bp} we review in detail the argument for the sign change of the Sivers function.
\index{Sivers function $f_{1T}^{\perp}$!process dependence|(}
Let $|\alpha\rangle = |p(P,S)\rangle$
and $\langle \beta|$ be equal to the rest of the matrix element
in~\eq{unsubTMDPDFspin}.  The definition in~\eq{unsubTMDPDFspin} is suitable for  the Drell-Yan process with past pointing Wilson lines while for SIDIS one defines a similar matrix element with future pointing Wilson lines $W_{\sqsupset}(b^\mu,0)$ as in~\eq{stapleinout}. 
From the parity and time-reversal 
invariance of QCD,
$\langle \alpha_P|\beta_P \rangle=
 \langle \alpha|\beta\rangle$ and 
 $\langle \beta_T|\alpha_T \rangle=
 \langle \alpha|\beta\rangle$, where
 $|\alpha_P\rangle$ and $|\beta_P\rangle$, and
 $|\alpha_T\rangle$ and $|\beta_T\rangle$ are the parity and 
 time-reversal transformed states from the states $|\alpha\rangle$
 and $|\beta\rangle$, 
 respectively. Thus one derives~\cite{Kang:2009bp} that the only difference between  $f_{i/p_S}^{[\gamma^+]}(x,\kt,\mu,\zeta)$ for SIDIS and Drell-Yan is $S\to -S$. Therefore one immediately concludes that 
the spin-averaged TMD quark distributions are process independent
\begin{align}
f_1^{\rm SIDIS}(x,k_T) = f_1^{\rm DY}(x,k_T) 
\end{align} 
while Sivers function changes the sign
\begin{align}
f_{1T}^{\perp {\rm SIDIS}}(x,k_T) = - f_{1T}^{\perp {\rm DY}}(x,k_T) 
\end{align} 
The sign change of the Sivers function is a property of the 
gauge invariant TMD parton distributions. 
Similar arguments can also be made for the other TMD PDFs listed in Table~\ref{tbl:wilson_directions}. 
In order to make sure that a single definition (that of Drell-Yan) can be used for both SIDIS and Drell-Yan we previously introduced the coefficient $\kappa =\pm 1$ in \eq{kappaDYSIDIS} to explicitly account for this sign change in SIDIS processes.  In this notation, tests of the sign-flip prediction between SIDIS and DY become tests of $\kappa^{\rm DY}=-\kappa^{\rm SIDIS}$.  \index{Sivers function $f_{1T}^{\perp}$!process dependence|)}

One can obtain the position-space version of \eq{tmd_decomposition} by a Fourier transform with respect to $\kt$.
In contrast to \eq{tmd_decomposition}, which is commonly adopted in the literature,
there are different conventions for the spin-decomposition in position space.
Historically, it was common to simply Fourier-transform \eq{tmd_decomposition} as is, and this was used for example in~\cite{Boer:2011xd}
and in the lattice studies in~\cite{Musch:2011er,Engelhardt:2015xja,Yoon:2017qzo} (see \sec{lattice}).
In this case, one decomposes $\tilde f$ as~\cite{Boer:2011xd}%
\footnote{Note that we have accounted for a relative minus sign in $b^\mu$ when relating the definition in \cite{Boer:2011xd} to our convention.
See also \cite{Scimemi:2018mmi}, and the comment below \eq{beamfunc}.}
\begin{align} \label{eq:tmd_decomposition_2}
 \tilde f_{i/p_S}^{[\gamma^+]}(x,\bt,\mu,\zeta) &
 = \tilde f_1(x,b_T)
 + \img \eps_{\rho\sigma} b_T^\rho S_T^\sigma M \tilde f_{1T}^{\perp}(x,b_T)
\,,\nn\\
 \tilde f_{i/p_S}^{[\gamma^+\gamma_5]}(x,\bt,\mu,\zeta) &
 = S_L \tilde g_{1}(x,b_T)
 + \img b_T \cdot S_T M \tilde g_{1T}^{\perp}(x,b_T)
\,,\nn\\
 \tilde f_{i/p_S}^{[\img  \sigma^{\alpha+}\gamma_5]}(x,\bt,\mu,\zeta) &
 = S_T^\alpha \tilde h_1(x,b_T)
 - \img S_L b_T^\alpha M \tilde h_{1L}^{\perp}(x,b_T)
 + \img \eps^{\alpha\rho} b_{\perp\rho} M \tilde h_1^{\perp}(x,b_T)
 \nn\\&\quad
 + \frac12 \bt^2 M^2 \biggl(\frac12 g_T^{\alpha\rho} + \frac{b_T^\alpha b_T^\rho}{\bt^2}\biggr) S_{\perp\,\rho} \tilde h_{1T}^{\perp}(x,b_T)
\,.\end{align}
Here, the explicit factors of $\img$ ensure that all functions on the right-hand side are manifestly real.
Due to the $\kt$-dependent prefactors in \eq{tmd_decomposition}, the $\tilde f, \tilde g$ and $\tilde h$ in \eq{tmd_decomposition_2}
are now $\bt$-dependent derivatives of Fourier transformations of the corresponding $f, g$ and $h$ in \eq{tmd_decomposition}, namely
\begin{alignat}{3} \label{eq:relations_f_tildef}
 &\tilde f_1(x,b_T) \equiv \tilde f_1^{(0)}(x,b_T)
\,,\qquad
 && \tilde f_{1T}^\perp(x,b_T) \equiv \tilde f_{1T}^{\perp\,(1)}(x,b_T)
\,,\qquad
 && \tilde h_{1T}^\perp(x,b_T) \equiv \tilde h_{1T}^{\perp\,(2)}(x,b_T)
\,,\nn\\
 &\tilde g_{1L}(x,b_T) \equiv \tilde g_{1L}^{(0)}(x,b_T)
\,,\qquad
 && \tilde h_1^\perp(x,b_T) \equiv \tilde h_1^{\perp\,(1)}(x,b_T)
\,,\nn\\
 & \tilde h_1(x,b_T) \equiv \tilde h_1^{(0)}(x,b_T)
 && \tilde g_{1T}(x,b_T) \equiv \tilde g_{1T}^{(1)}(x,b_T)
\,,\nn\\
 & && \tilde h_{1L}^\perp(x,b_T) \equiv \tilde h_{1L}^{\perp\,(1)}(x,b_T)
\,.\end{alignat}
The derivatives are defined as
\begin{align} \label{eq:TMD_bt_derivative}
 \tilde f^{(n)}(x, b_T, \mu, \zeta) &
 \equiv n! \left(\frac{-1}{M^2 b_T} \partial_{b_T} \right)^n \tilde f(x, b_T, \mu, \zeta)
\nn\\&
 = \frac{2\pi\, n!}{(b_T M)^n} \int_0^\infty \df k_T \, k_T \left(\frac{k_T}{M}\right)^n J_n(b_T k_T) \, f(x, k_T, \mu, \zeta)
\,.\end{align}
Note that for real $f$ this Hankel transform is real as well, and that $\tilde f^{(n)}$ have the same mass dimension for all $n$.
This convention with the explicit factor of $n!$ was introduced in \cite{Boer:2011xd} such that taking $b_T \to 0$
recovers the moments obtained by integrating over $\kt$ as defined in~\cite{Mulders:1995dh,Boer:1997nt}.
The inverse transformation is given by
\begin{align} \label{eq:TMD_bt_derivative_inv}
 f^{(n)}(x, k_T, \mu, \zeta) &
 = \frac{M^{2n}}{2 \pi n!} \int_0^\infty \df b_T \, b_T \left(\frac{b_T}{k_T}\right)^n J_n(b_T k_T) \, \tilde f^{(n)}(x, k_T, \mu, \zeta)
\,.\end{align}
For more details on deriving these Fourier relations, see \app{Fourier_transform}.

In~\cite{Gutierrez-Reyes:2017glx}, it was proposed to absorb the hadron masses used in the normalization in \eq{tmd_decomposition} into the Fourier-transformed distributions. Compared to the above decomposition, in this convention one replaces the distributions $\tilde F(x,b_T)$ by the alternate definitions $\tilde F^\prime(x,b_T)$ where the relationship is
\begin{align}
 b_T^\alpha M \tilde F^{(1)} = \frac{b_T^\alpha}{b_T} \tilde F^{\prime (1)}
\,,\quad
 b_T^\alpha b_T^\rho M^2 \tilde F^{(2)} = \frac{b_T^\alpha b_T^\rho}{b_T^2} \tilde F^{\prime (2)}
\,,\quad
 \tilde F^{(n)} = (b_T M^2)^{-n} \tilde F^{\prime(n)}
\,.\end{align}
Here the last equality is applied when there are no $b_T$ dependent factors in front of the distribution $\tilde F^{(n)}$.

It is natural to ask whether the TMD PDFs are positive definite functions. This has been explored recently in Ref.~\cite{Collins:2021vke}.
\index{positivity of TMDs}
For further discussion in the context of phenomenological analyses, see \chap{phenoTMDs}, and for a discussion in the context of various models for TMDs, see Sec.~\ref{Sec:positivity-constraints}

\subsubsection{Leading Quark TMD FFs}
\label{sec:LQTMDFF}

\index{TMD fragmentation function!spin dependent}

We now discuss the generalization of the TMD FFs, introduced in \sec{TMDFFs} for unpolarized processes,
to also allow for polarizations of the fragmenting quark and detected hadron.
The generalization of \eq{TMDFF_unsub} is given by
\begin{align} \label{eq:unsubTMDFFspin}
 \tilde \Delta_{h/i}^{[\Gamma]0\unsub}(z, \bt, \eps, \tau, P^+/z) &
 = \frac{1}{4 N_c z} {\rm Tr} \!\int\!\frac{\df b^-}{2\pi} \sum_X e^{\img b^- (P^+/z)} \Gamma^+_{\alpha\alpha'}
  \nn\\&\times
   \MAe{0}{ \big[ (W_{\halfstapleu} \psi_i^{0\alpha})(b) \big]_\tau }{h(P,S),X}
   \MAe{h(P,S),X}{\,\big[ (\bar\psi_i^{0\alpha'} W_{\halfstaple})(0)\big]_\tau }{0}
\,.\end{align}
Compared to \eq{TMDFF_unsub}, the hadron state now carries the spin $S$,
and the Dirac structure $\Gamma$ is chosen analogous to \eq{Gamma_LP} as
\begin{align} \label{eq:Gamma_LP_FF}
 \Gamma \in \{\, \gamma^+ \,,\, \gamma^+ \gamma_5 \,,\, \img \sigma^{\alpha+} \gamma_5 \,\}
\,.\end{align}
\eq{unsubTMDFFspin} is combined with the soft function as in \eq{TMDFF_ren}
to obtain the UV-renormalized and soft-subtracted TMD FF.
After taking the Fourier transform with respect to $\bt$, we then obtain the renormalized TMD FF
in momentum space,

\begin{align}
 \Delta_{h/i}(z, \pt = -z\pt', \mu, \zeta) &
 = \int\frac{\df^2\bt}{(2\pi)^2} e^{-\img \pt' \cdot \bt}
   \tilde \Delta_{h/i}(z, \bt, \mu, \zeta)
\,.\end{align}
As discussed at length in \sec{TMDFFs}, here $\pt$ is the hadron transverse momentum
relative to the fragmenting quark in the photon-hadron frame, while $\pt'$ is the parton
momentum relative to the detected hadron in the hadron-hadron frame.

\begin{figure*}[t]
 \centering
  \includegraphics[width=0.85\textwidth]{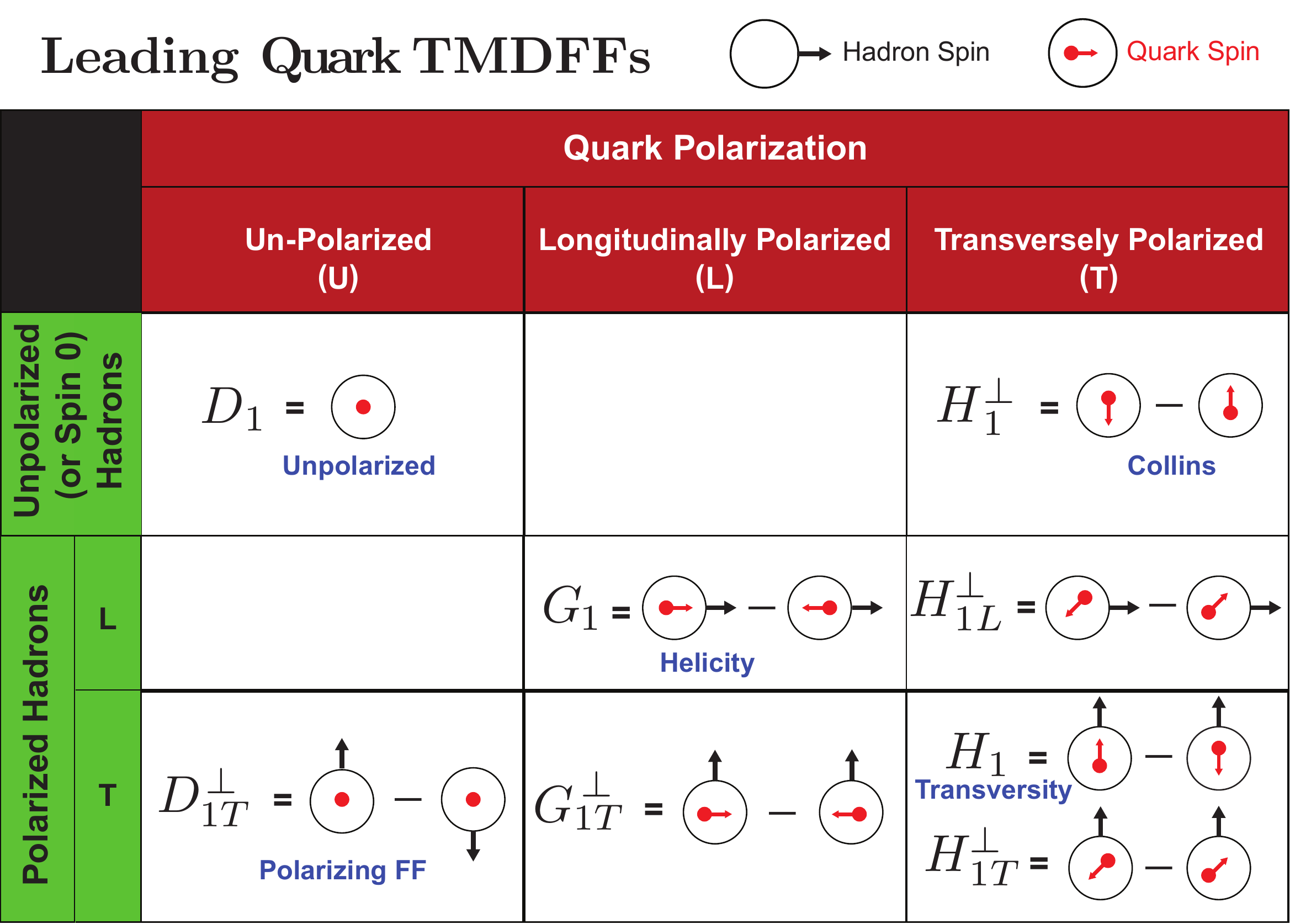}
 \caption{Leading power quark TMD fragmentation functions for a spin-$1/2$ (or for an unpolarized or spin 0) hadron.}
 \label{fig:TMDFFsLP}
\end{figure*}

The spin-dependent TMD FF can be decomposed into eight independent structures in the same fashion as the spin-dependent TMD PDF,
\begin{align} \label{eq:tmdff_decomposition}
	\Delta_{h/i}^{[\gamma^+]}(z, -z\pt',\mu,\zeta) &
	= D_1\bigl( z, z p_T' \bigr) - \frac{\eps_T^{\rho\sigma} p'_{T\rho} S_{T\sigma}}{M_h} D_{1T}^\perp\bigl( z, z p_T' \bigr)
	\,,\\
	\Delta_{h/i}^{[\gamma^+\gamma_5]}(z, -z\pt',\mu,\zeta) &
	= S_L \, G_{1}\bigl( z, z p_T' \bigr)
	- \frac{p'_T \cdot S_T}{M_h} G_{1T}^\perp\bigl( z, z p_T' \bigr)
	\,,\nn\\
	\Delta_{h/i}^{[\img \sigma^{\alpha+}\gamma_5]}(z, -z\pt',\mu,\zeta) &
	= S_T^\alpha \, H_1\bigl( z, z p_T' \bigr)
	+ \frac{S_L p_T^{\prime\alpha}}{M_h} H_{1L}^\perp\bigl( z, z p_T' \bigr)
	\nn\\&\quad
	- \frac{\pt^{\prime\,2}}{M_h^2} \biggl(\frac12 g_T^{\alpha\rho} + \frac{p_T^{\prime\alpha} p_T^{\prime\rho}}{\pt^{\prime\,2}}\biggr) S_{T\,\rho} H_{1T}^\perp\bigl( z, z p_T' \bigr)
	- \frac{\eps_T^{\alpha\rho} p'_{T\rho}}{M_h} H_1^\perp\bigl( z, z p_T' \bigr)
	\nn\,.
\end{align}
Once again on the right-hand side we suppress the arguments $\mu$ and $\zeta$ as well as the subscripts $h/i$.
This decomposition is analogous to \eq{tmd_decomposition}, where the TMD FFs are distinguished from the TMD PDFs by using capital symbols,
but have similar interpretations in terms of the quark and  hadron polarization, as summarized in \fig{TMDFFsLP}.
Note that the functions appearing in \eq{tmdff_decomposition} are written as a function of $z p'_T$,
while the prefactors only involve $\pt'$, as is common in the literature, see e.g.~\cite{Bacchetta:2006tn}.
We again encounter two $T$-odd functions, namely $D_{1T}^\perp$ and the Collins function $H_1^\perp$.

As before, we can equivalently consider the decomposition of the spin-dependent TMD FF in position space.
Following our presentation of the TMD PDF, we write the decomposition analogous to \eq{tmd_decomposition_2} as~\cite{Boer:2011xd}%
\footnote{When comparing \eq{tmdff_decomposition_2} to the corresponding expression in \cite{Boer:2011xd},
one has to account for a sign change $b^\mu \to -b^\mu$ due to a different definition of the TMD correlator
as well as $\eps_T^{\alpha\beta} \to -\eps_T^{\alpha\beta}$ because we consider a $n_a$-collinear hadron.}
\begin{align} \label{eq:tmdff_decomposition_2}
 \tilde \Delta_{h/i}^{[\gamma^+]}(z,\bt,\mu,\zeta) &
 = \tilde D_1(z,b_T)
 - {\img} \eps_{T\, \rho\sigma} b_T^\rho S_T^\sigma M_h \tilde D_{1T}^{\perp}(z,b_T)
\,,\nn\\
 \tilde \Delta_{h/i}^{[\gamma^+\gamma_5]}(z,\bt,\mu,\zeta) &
 = S_L \tilde G_{1}(z,b_T)
 + {\img} b_T \cdot S_T M_h \tilde G_{1T}^{\perp}(z,b_T)
\,,\nn\\
 \tilde \Delta_{h/i}^{[\img  \sigma^{\alpha+}\gamma_5]}(z,\bt,\mu,\zeta) &
 = S_T^\alpha \tilde H_1(z,b_T)
 + {\img} S_L b_T^\alpha M_h \tilde H_{1L}^{\perp}(z,b_T)
 - {\img} \eps_T^{\alpha\rho} b_{\perp\rho} M_h \tilde H_1^{\perp}(z,b_T)
 \nn\\&\quad
 + \frac12 \bt^2  M_h^2 \biggl(\frac12 g_T^{\alpha\rho} + \frac{b_T^\alpha b_T^\rho}{\bt^2}\biggr) S_{\perp\,\rho} \tilde H_{1T}^{\perp}(z,b_T)
\,.\end{align}

Due to the $\pt'$-dependent prefactors in \eq{tmdff_decomposition}, the $\tilde D, \tilde G$ and $\tilde H$ in \eq{tmdff_decomposition_2}
are now $\bt$-dependent derivatives of Fourier transformations of the corresponding $D, G$ and $H$ in \eq{tmdff_decomposition}.
They are given by,
\begin{alignat}{3}
 &\tilde D_1(z,b_T) \equiv \tilde D_1^{(0)}(z,b_T)
\,,\qquad
 && \tilde D_{1T}^\perp(z,b_T) \equiv \tilde D_{1T}^{\perp\,(1)}(z,b_T)
\,,\qquad
 && \tilde H_{1T}^\perp(z,b_T) \equiv \tilde H_{1T}^{\perp\,(2)}(z,b_T)
\,,\nn\\
 &\tilde G_{1}(z,b_T) \equiv \tilde G_{1}^{(0)}(z,b_T)
\,,\qquad
 && \tilde D_1^\perp(z,b_T) \equiv \tilde D_1^{\perp\,(1)}(z,b_T)
\,,\nn\\
 & \tilde H_1(z,b_T) \equiv \tilde H_1^{(0)}(z,b_T)
 && \tilde G_{1T}(z,b_T) \equiv \tilde G_{1T}^{(1)}(z,b_T)
\,,\nn\\
 & && \tilde H_{1L}^\perp(z,b_T) \equiv \tilde H_{1L}^{\perp\,(1)}(z,b_T)
\,,\end{alignat}
where the derivatives are defined as
\begin{align} \label{eq:TMDFF_bt_derivative}
 \tilde D^{(n)}(z, b_T, \mu, \zeta) &
 \equiv n! \left(\frac{-1}{M_h^2 b_T} \partial_{b_T} \right)^n \tilde D(z, b_T, \mu, \zeta)
\nn\\&
 = \frac{2\pi\, n!}{(M_h^2)^n} \int_0^\infty \df p_T' \, p_T' \left(\frac{p_T'}{b_T}\right)^n J_n(b_T p_T') \, D(z, z p_T', \mu, \zeta)
\,.\end{align}

\subsubsection{Leading Gluon TMD PDFs}\label{sec:gluonTMD_def}

\index{TMD parton distribution function!gluon definitions}

So far, we have only considered quark TMD PDFs and TMD FFs.
One can similarly define the corresponding gluon distributions. The unsubtracted gluon TMD PDF is defined via
\begin{align} \label{eq:fg}
 \tilde f_{g/p_S}^{\alpha\beta\, 0\unsub}(x, \bt, \eps, \tau, x P^+) &
 = \frac{1}{x P^+} \int\!\frac{\df b^-}{2\pi} e^{-\img b^- (x P^+)}
   \bigl< p(P,S) \big|
    G^{+\alpha}(b^\mu) {\cW}_\sqsubset(b^\mu,0) G^{+\beta}(0)
   \big|  p(P,S) \bigr>
\,.\end{align}
Compared to the TMD PDF in \eq{beamfunc}, the quark fields have been replaced by the gluon field strength tensor $G^{\alpha\beta} = \partial^\alpha A^\beta - \partial^\beta A^\alpha  - i g [A^\alpha, A^\beta]$.
In contrast to the quark fields, which transform in the fundamental representation of the QCD gauge group, $G^{\alpha\beta}$ transforms in the adjoint representation.
To compensate for this, the Wilson line $W_\sqsubset$ in the fundamental representation has been replaced by the Wilson line ${\cW}_\sqsubset$ in the adjoint representation. It is defined as in \eqs{Staple_Wilson_line}{Wilson_lines}, up to taking the color matrix in \eq{Wilson_lines} in the adjoint representation.
The different mass dimension of $G^{\alpha\beta}$ compared to $\psi^i$ is compensated
by the overall normalization factor $1/x P^+$.

Due to the tensor structure of the gluon field strength $G^{\alpha\beta}$, the gluon TMD PDF in \eq{fg} carries a tensor structure as well.
In principle, due to the presence of two gluon field strengths, it is a rank-four tensor.
However, two of the indices are fixed to be $+$, see \eq{fg}, as all other choices are power suppressed, leaving only the rank-two TMD PDF $f_{g/p_S}^{\alpha\beta}$.
In addition, $\alpha$ and $\beta$ have to be transverse indices, which is kept implicit in \eq{fg}.

The TMD PDF is then obtained similar to \eq{tmdpdf_1} by combining the unsubtracted gluon TMD PDF with the corresponding gluon soft function,
\begin{align} \label{eq:tmdpdf_g}
 \tilde f^{\alpha\beta}_{g/p_S}(x, \bt, \mu , \zeta) &
 =  \lim_{\substack{\eps\to 0 \\ \tau\to 0}} Z_{\rm uv}^i(\mu,\zeta,\eps) \,
    \frac{\tilde f_{g/p}^{\alpha\beta\,0\,\unsub}\bigl(x, \bt, \eps, \tau, x P^+ \bigr)}{\tilde \cS^{0\,\subt}_{n_a n_b}(b_T,\eps,\tau)}
    \sqrt{\tilde \cS_{n_a n_b}^{0}(b_T,\eps,\tau)}
\,.\end{align}
In analogy to \eq{Soft_Wilson_Loop} but using adjoint Wilson lines, the gluon soft function is defined as
\begin{align} \label{eq:Soft_Fn_adj}
 {\tilde \cS}^{0}_{n_a n_b}(b_T,\eps,\tau)
 &= \frac{1}{N_c^2 - 1} \bigl< 0 \bigr| {\rm Tr} \bigl[
  \cW_{\softstaple}(b_T) 
 \bigr]_\tau \bigl|0 \bigr>
\,.\end{align}
Compared to the quark soft function in \eq{softfunc}, we again take all Wilson lines in the adjoint instead of the fundamental representation,
and have adjusted the overall normalization factor to $N_c^2 - 1 = 8$, the total number of independent generators of the adjoint representation.
Note that here we have chosen incoming Wilson lines from $-\infty$.  The process dependence of gluon TMD PDFs is more complicated than the quark case and has been explored in Ref.~\cite{Buffing:2013kca,Boer:2015vso}.

\begin{figure}[pt]
 \centering
  \includegraphics[width=0.8\textwidth]{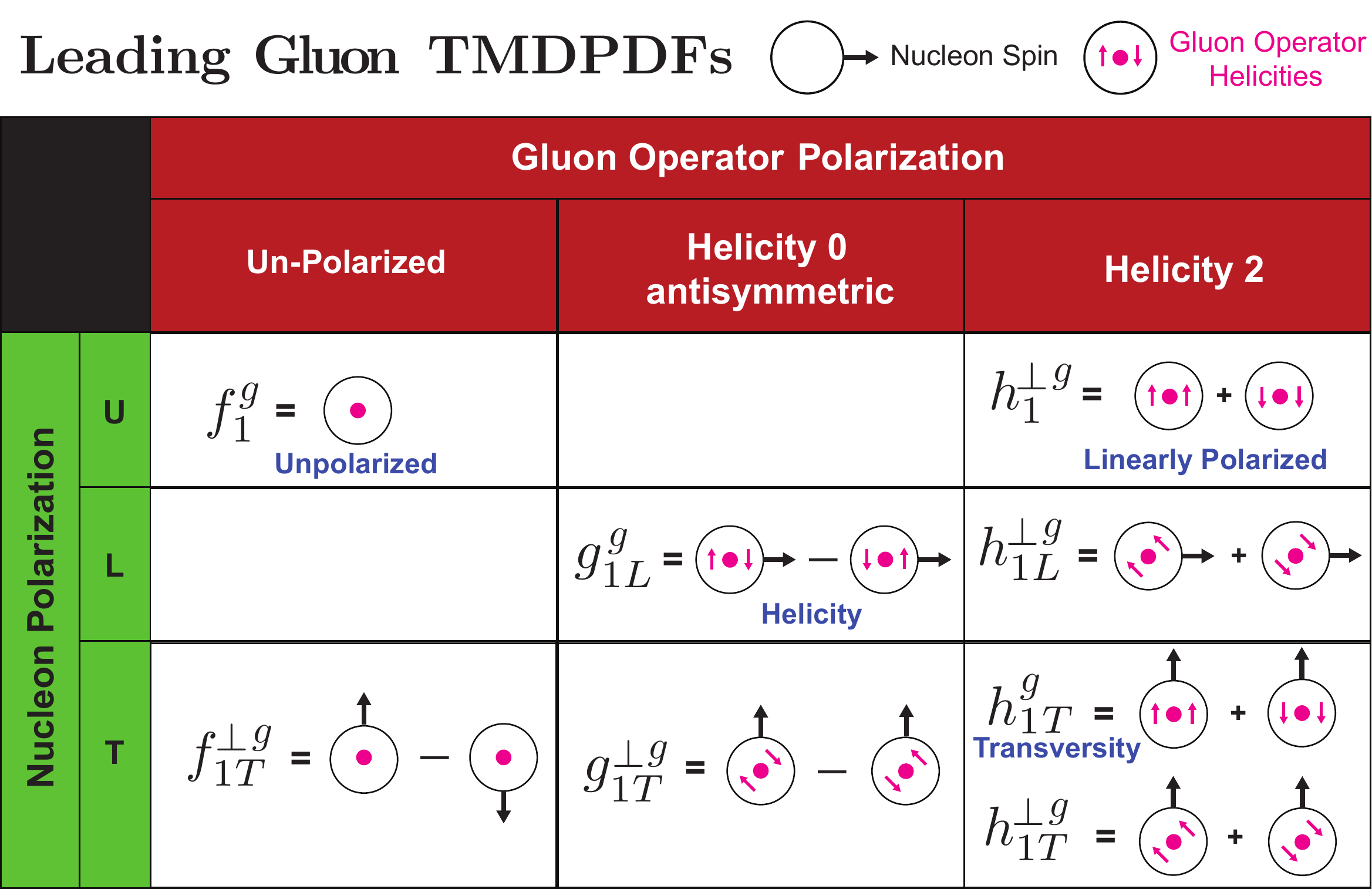}
 \caption{Leading power gluon TMD parton distribution functions for a spin-$1/2$ hadron (or unpolarized hadron).}
 \label{fig:gTMDPDFsLP}
\end{figure}

As for the quark TMD PDF, we can decompose the spin-dependent gluon TMD PDF $f_{g/p}^{\alpha\beta}$ into independent structures.
Due to the spin-$1$ nature of the gluon, this has a different structure than for the spin-$\frac12$ quark.
The decomposition in momentum space was first given in \cite{Mulders:2000sh}.
Here we follow the conventions of \cite{Meissner:2007rx}, with lower case $f^g_X$, $g^g_X$ and $h^g_X$ functions, which parallels the notation used for the quark TMD PDFs, and enables us to reserve capital letters for TMD FFs.
Thus we write the spin decomposition as
\begin{align} \label{eq:tmd_decomposition_g_1}
 f_{g/p}^{\alpha\beta}(x, \kt, \mu, \zeta) &
 = \frac12 \biggl[ - g_T^{\alpha\beta} f_1^g(x, k_T) + \frac{\kt^2}{M^2} \left(\frac{g_T^{\alpha\beta}}{2} + \frac{k_T^\alpha k_T^\beta}{\kt^2} \right) h_1^{\perp\,g}(x, k_T) \biggr]
 \\&
 + \frac{S_L}{2} \biggl[ -\img \eps_T^{\alpha\beta} g_{1L}^g(x, k_T) 
  - \frac{k_{\rho}\eps_T^{\rho \{\alpha} k_T^{\beta\}}}{2 M^2} h_{1L}^{\perp\,g}(x, k_T) \biggr]
 \nn\\&
 + \frac12 \biggl\{
     g_T^{\alpha\beta} \frac{k_{T\rho}S_{T\sigma}\eps_T^{\rho \sigma}}{M} f_{1T}^{\perp g}(x, k_T)
    - \img \eps_T^{\alpha\beta} \frac{k_T \cdot S_T}{M} g_{1T}^{g}(x, k_T)
    \nn\\&\qquad
   + \frac{k_{T\rho}\eps_T^{\rho \{\alpha} k_T^{\beta\}}}{2 M^2} \frac{k_T \cdot S_T}{M} h_{1T}^{\perp g}(x, k_T)
   - \frac{k_{T\rho}\eps_T^{\rho\{\alpha} S_T^{\beta\}} + S_{T\rho}\eps_T^{\rho\{\alpha}k_T^{\beta\}}}{4 M} h_{1T}^{g}(x, k_T)
   \biggr\}
\,.\nn
\end{align}
Here, the notation $a^{\{\alpha} b^{\beta\}} = a^\alpha b^\beta + a^\beta b^\alpha $ indicates symmetrization in the indices.
As we have done previously, for brevity we suppress the $\mu$ and $\zeta$ scales and the subscript $g/p$ on the right hand side of \eq{tmd_decomposition_g_1}. The function $f_{g/p}^{\alpha\beta}$ has two transverse indices, $\alpha,\beta=1,2$. Denoting basis vectors by $\hat e_1$ and $\hat e_2$, they can be decomposed into the $\pm 1$ gluon helicities, denoted by $\uparrow=(\hat e_1+i\hat e_2)/\sqrt{2}$ and $\downarrow=(\hat e_1-i\hat e_2)/\sqrt{2}$.
The irreducible representations are simple products of helicities, for which it is convenient to form symmetric and antisymmetric combinations (see also Ref.~\cite{Lyubovitskij:2021qza}). This is summarized in \fig{gTMDPDFsLP}. The symmetric helicity $0$ combination $\uparrow\downarrow+\downarrow\uparrow$ gives the unpolarized configurations that appear for unpolarized and transversely polarized hadrons, with distributions $f_1^g$ and $f_{1T}^{\perp g}$. The antisymmetric helicity $0$ combination $\uparrow\downarrow-\downarrow\uparrow$ yields the helicity distributions $g_{1L}^g$ and $g_{1T}^{\perp g}$ for longitudinally and transversely polarized hadrons, respectively. Finally, the helicity $2$ combinations $\uparrow\uparrow+\downarrow\downarrow$ and $\uparrow\uparrow-\downarrow\downarrow$ are given in \eq{tmd_decomposition_g_1} by the $h_1^{\perp g}, h_{1L}^{\perp g}, h_{1T}^g$ and $h_{1T}^{\perp g}$ terms for the spin-$1/2$ hadron polarizations indicated in \fig{gTMDPDFsLP}. They are symmetric and traceless combinations of $\alpha,\beta=1,2$.
In \fig{gTMDPDFsLP} the orientation of gluon helicity arrows indicates the degree to which they are correlated with the direction of the momentum $p_T$ or hadron spin. 

The corresponding expression in position space was first given in \cite{Echevarria:2015uaa}.%
\footnote{Compared to \eq{tmd_decomposition_g_1}, the expression in \cite{Echevarria:2015uaa} is missing a factor $1/2$ in all terms involving $S_T$, which we have restored. They also have a typo in their $\tilde h_{1T}^{\perp g\,(2)}(x, b_T)$, which should read $\tilde h_{1T}^{\perp g\,(3)}(x, b_T)$.}
Here, we deviate from their notation to use the same conventions as in the quark case
presented in \sec{leadingTMDPDF}. Namely, we define the Fourier transform and its inverse
by \eqs{TMD_bt_derivative}{TMD_bt_derivative_inv}, allowing us to use the explicit expressions
in \app{Fourier_transform}. This yields
\begin{align} \label{eq:tmd_decomposition_g_2}
 \tilde f_{g/p}^{\alpha\beta}(x, \bt, \mu, \zeta) &
 = \frac12 \biggl[
   - g_T^{\alpha\beta} \tilde f_1^g(x, b_T)
   - \frac{1}{2} b_T^2 M^2 \left( \frac{g_T^{\alpha\beta}}{2}  +  \frac{b_T^\alpha b_T^\beta}{\bt^2} \right) \tilde h_1^{\perp\,g}(x, b_T) \biggr]
 \\&
 + \frac{S_L}{2} \biggl[
   -\img \eps_T^{\alpha\beta} \tilde g_{1L}^g(x, b_T)
   + \frac{M^2}{4}  b_{T\rho}\eps_T^{\rho \{\alpha} b_T^{\beta\}} \tilde h_{1L}^{\perp\,g}(x, b_T) \biggr]
 \nn\\&
 + \frac12 \biggl[
    -\img M  g_T^{\alpha\beta} b_{T\alpha}S_{T\beta}\eps_T^{\alpha \beta} \tilde f_{1T}^{\perp g}(x, b_T)
    - M \eps_T^{\alpha\beta} b_T \cdot S_T \tilde g_{1T}^{g}(x, b_T)
    \nn\\&\qquad
   + \frac{\img M^3}{12} b_{T\rho}\eps_T^{\rho \{\alpha} b_T^{\beta\}} b_T \cdot S_T \tilde h_{1T}^{\perp g}(x, b_T)
  \nn\\&\qquad
   + \frac{\img}{4} M \Bigl(b_{T\rho} \eps_T^{\rho\{\alpha} S_T^{\beta\}} + S_{T\rho}\eps_T^{\rho\{\alpha}b_T^{\beta\}}\Bigr) \tilde h_{1T}^{g}(x, b_T)
   \biggr]
\,.\nn
\end{align}
The relation between the functions in momentum and position space is given by
\begin{alignat}{3}
 &\tilde f_1^g(x,b_T) \equiv \tilde f_1^{g (0)}(x,b_T)
\,,\qquad
 && \tilde f_{1T}^{\perp g}(x, b_T) \equiv \tilde f_{1T}^{\perp g\,(1)}(x, b_T)
\,,\qquad
 && \tilde h_{1L}^{\perp\,g}(x, b_T) \equiv \tilde h_{1L}^{\perp\,g\,(2)}(x, b_T)
\,,\nn\\
 & \tilde g_{1L}^g(x, b_T) \equiv \tilde g_{1L}^{g\,(0)}(x, b_T)
\,,\qquad
 && \tilde g_{1T}^{g}(x, b_T) \equiv \tilde g_{1T}^{g\,(1)}(x, b_T)
\,,\qquad
 && \tilde h_{1}^{\perp\,g}(x, b_T) \equiv \tilde h_{1}^{\perp\,g\,(2)}(x, b_T)
\,,\nn\\
 &
 && \tilde h_{1T}^{g}(x, b_T) \equiv \tilde h_{1T}^{g\,(1)}(x, b_T)
 && \tilde h_{1T}^{\perp g}(x, b_T) \equiv \tilde h_{1T}^{\perp g\,(3)}(x, b_T)
\,,\end{alignat}
where the $\tilde f^{(n)}$ are defined in \eq{TMD_bt_derivative}.

\begin{figure}[t]
 \centering
  \includegraphics[width=0.8\textwidth]{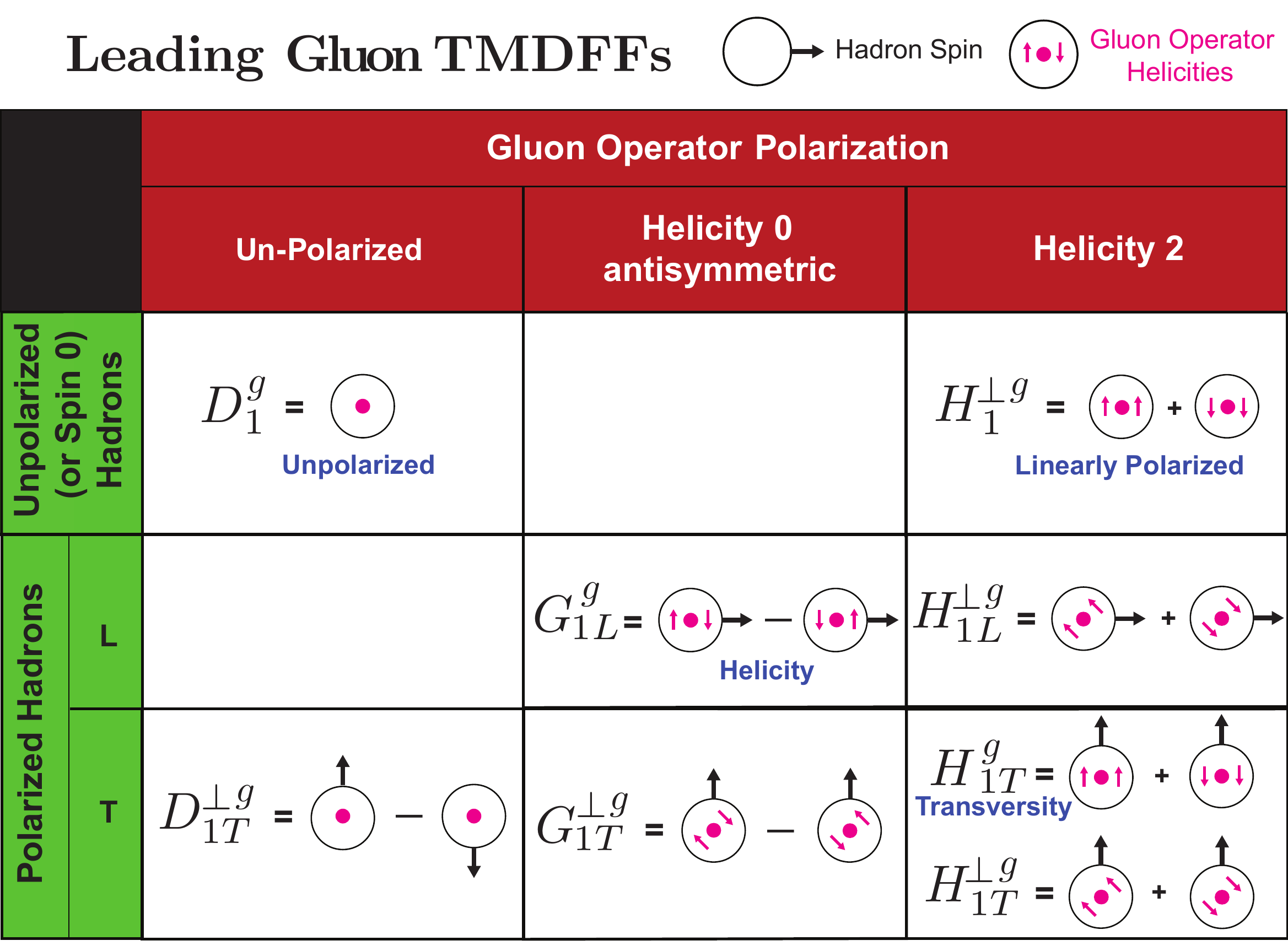}
 \caption{Leading power gluon TMD fragmentation functions for an unpolarized or spin-$0$ hadron or a polarized spin-$1/2$ hadron. }
 \label{fig:gTMDFFsLP}
\end{figure}

\subsubsection{Leading Gluon TMD FFs}

\index{TMD fragmentation function!gluon}

We now discuss the definition of polarized gluon TMD FFs.
Similar to quark TMD FFs, the gluon TMD FF $\Delta_{h/g}^{\mu\nu}(z, \pt)$ encodes the fragmentation of a gluon produced in the underlying hard scattering into a hadron of type $h$, carrying the longitudinal momentum $z$ and transverse momentum $\pt$ relative to the fragmenting gluon.

To contrast TMD FFs and TMD PDFs, we first recall the definition of the spin-dependent unsubtracted TMD PDF in \eq{fg},
\begin{align} \label{eq:fg_2}
 \tilde f_{g/h}^{\alpha\beta\, 0\unsub}(x, \bt, \eps, \tau, x P^+) &
 = \frac{1}{x P^+} \int\!\frac{\df b^-}{2\pi} e^{-\img b^- (x P^+)}
   \bigl< h(P,S) \big|
    G^{+\alpha}(b^\mu) {\cW}_\sqsubset(b^\mu,0) G^{+\beta}(0)
   \big|  h(P,S) \bigr>
\,.\end{align}
Here, the incoming proton is aligned along the $n_a$ direction, $b^\mu = (0, b^-, \bt)$,
and $\alpha$ and $\beta$ are transverse indices.
The corresponding unsubtracted TMD FF is defined as
\begin{align} \label{eq:TMDFF_g}
 \tilde \Delta_{h/g}^{\alpha\beta\, 0\unsub}(z, \bt, \eps, \tau, z P^+)
 = \frac{1}{2z} \frac{1}{z P^+} \int\!\frac{\df b^-}{2\pi} \sum_X e^{\img b^- (z P^+)}
   &\bigl< 0 \big| G^{+\alpha}(b) \cW_{\halfstapleu}(b) \big| h(P,S), X\bigr>
   \nn\\\times&
   \bigl< h(P,S), X \big| G^{+\beta}(0) \cW_{\halfstaple}(0) \big|  0 \bigr>
\,.\end{align}
As for the quark TMD FF, the matrix element is normalized by an additional factor of $1/z$ and $1/2$,
the latter accounting for the gluon spin states.
The hadron $h$ appears as an out-state in the matrix element, with $\sum_X$ denoting the sum over all additional hadronic final states $X$,
which in contrast to the TMD PDF can not be eliminated by unitarity.
In \eq{TMDFF_g}, the Wilson lines $\cW_{\halfstapleu}$ and $\cW_{\halfstaple}$ are defined as in \eq{half_staple_Wilson_line},
up to replacing the fundamental Wilson lines $W_{n}$ by Wilson lines $\cW_n$ in the adjoint representation.

The gluon TMD FF is obtained by combining \eq{TMDFF_g} with the soft function and UV renormalization as in \eq{tmdpdf_g}.
In momentum space, the TMD FF can be decomposed analogously to \eq{tmd_decomposition_g_1} as
\begin{align} \label{eq:tmdff_decomposition_g_1}
 \Delta_{h/g}^{\alpha\beta}(z, -z \pt', \mu, \zeta) &
 = \frac12 \biggl[
  - g_T^{\alpha\beta} D_1^g(z, z p'_T)
  + \frac{\pt^{\prime\,2}}{M^2} \left( \frac{g_T^{\alpha\beta}}{2}  + \frac{p_T^{\prime\alpha} p_T^{\prime\beta}}{\pt^{\prime\,2}} \right) H_1^{\perp\,g}(z, z p'_T) \biggr]
 \\&\hspace{-0.2cm}
 + \frac{S_L}{2} \biggl[ 
  -\img \eps_T^{\alpha\beta} G_{1L}^g(z, z p'_T) 
  - \frac{p_{T\rho}' \eps_T^{\rho \{\alpha} p_T^{\prime\beta\}}}{2 M^2} H_{1L}^{\perp\,g}(z, z p'_T) \biggr]
 \nn\\&\hspace{-0.2cm}
 + \frac12 \biggl\{
  g_T^{\alpha\beta} \frac{p_{T\rho}'S_{T\sigma} \eps_T^{\rho\sigma}}{M} D_{1T}^{\perp g}(z, z p'_T)
  - \img \eps_T^{\alpha\beta} \frac{p'_T \cdot S_T}{M} G_{1T}^{g}(z, z p'_T)
   \nn\\&\hspace{-0.2cm} \qquad \!\!\! 
  + \frac{p'_{T\rho} \eps_T^{\rho \{\alpha} p_T^{\prime\beta\}}}{2 M^2} \frac{p'_T \!\cdot\! S_T}{M} H_{1T}^{\perp g}(z, z p'_T)
  - \frac{p'_{T\rho}\eps_T^{\rho \{\alpha} S_T^{\beta\}} \!+\! S_{T\rho}\eps_T^{\rho\{\alpha}p_T^{\prime\beta\}}}{4 M} H_{1T}^{g}(z, z p'_T)
   \biggr\}
 .\nn
\end{align}
As in \eq{tmd_decomposition_g_1}, the notation $a^{\{\alpha} b^{\beta\}} = a^\alpha b^\beta + a^\beta b^\alpha $ indicates symmetrization in the indices,
and on the right hand side we suppress explicit $\mu$ and $\zeta$ scales and the subscript $g/h$.
The relation of the different structures to the gluon and hadron spin is summarized in \fig{gTMDFFsLP}.

Taking the Fourier transform of \eq{tmdff_decomposition_g_1}, we obtain the decomposition in position space,
\begin{align} \label{eq:tmdff_decomposition_g_2}
 \tilde \Delta_{h/g}^{\alpha\beta}(z, \bt, \mu, \zeta) &
 = \frac12 \biggl[
   - g_T^{\alpha\beta} \tilde D_1^g(z, b_T)
   - \frac{1}{2} b_T^2 M^2 \left( \frac{g_T^{\alpha\beta}}{2}  +  \frac{b_T^\alpha b_T^\beta}{\bt^2} \right) \tilde H_1^{\perp\,g}(z, b_T) \biggr]
 \\&
 + \frac{S_L}{2} \biggl[
  -\img \eps_T^{\alpha\beta} \tilde G_{1L}^g(z, b_T)
  + \frac{M^2}{4}  b_{T\rho}\eps_T^{\rho \{\alpha} b_T^{\beta\}} \tilde H_{1L}^{\perp\,g}(z, b_T) \biggr]
 \nn\\&
 + \frac12 \biggl[
  - \img M  g_T^{\alpha\beta} b_{T\alpha}S_{T\beta}\eps_T^{\alpha \beta} \tilde F_{1T}^{\perp g}(z, b_T)
  - M \eps_T^{\alpha\beta} b_T \cdot S_T \tilde G_{1T}^{g}(z, b_T)
    \nn\\&\qquad
  + \frac{\img M^3}{12} b_{T\rho}\eps_T^{\rho \{\alpha} b_T^{\beta\}} b_T \cdot S_T \tilde H_{1T}^{\perp g}(z, b_T)
  \nn\\&\qquad
  + \frac{\img}{4} M \Bigl(b_{T\rho} \eps_T^{\rho\{\alpha} S_T^{\beta\}} + S_{T\rho}\eps_T^{\rho\{\alpha}b_T^{\beta\}}\Bigr) \tilde H_{1T}^{g}(z, b_T)
   \biggr]
\,.\nn
\end{align}
The relation between the functions in momentum and position space is given by
\begin{alignat}{3}
 &\tilde D_1^g(z,b_T) \equiv \tilde D_1^{g (0)}(z,b_T)
\,,\qquad
 && \tilde D_{1T}^{\perp g}(z, b_T) \equiv \tilde D_{1T}^{\perp g\,(1)}(z, b_T)
\,,\qquad
 && \tilde H_{1L}^{\perp\,g}(z, b_T) \equiv \tilde H_{1L}^{\perp\,g\,(2)}(z, b_T)
\,,\nn\\
 & \tilde G_{1L}^g(z, b_T) \equiv \tilde G_{1L}^{g\,(0)}(z, b_T)
\,,\qquad
 && \tilde G_{1T}^{g}(z, b_T) \equiv \tilde G_{1T}^{g\,(1)}(z, b_T)
\,,\qquad
 && \tilde H_{1}^{\perp\,g}(z, b_T) \equiv \tilde H_{1}^{\perp\,g\,(2)}(z, b_T)
\,,\nn\\
 &
 && \tilde H_{1T}^{g}(z, b_T) \equiv \tilde H_{1T}^{g\,(1)}(z, b_T)
 && \tilde H_{1T}^{\perp g}(z, b_T) \equiv \tilde H_{1T}^{\perp g\,(3)}(z, b_T)
\,,\end{alignat}
where the $f^{(n)}$ are defined in \eq{TMDFF_bt_derivative}.

\FloatBarrier
\subsection[TMD PDFs and TMD FFs at small \texorpdfstring{$b_T$}{bT}]{\boldmath TMD PDFs and TMD FFs at small $b_T$}
\label{sec:largeqT}

\index{small $b_T$ expansion}

Much of the focus of our discussion is on the case where $q_T$, or equivalently $b_T$, are nonperturbative scales, i.e.~$q_T \sim b_T^{-1} \sim \LQCD$, where TMD PDFs and TMD FFs are genuinely nonperturbative objects encoding nonperturbative transverse-momentum distributions of quarks and gluons inside hadrons.
However, it turns out that contributions from perturbative scales 
$q_T \sim b_T^{-1} \gg \LQCD$ also often play an important role in predictions for colliders.  Hence, in this section we investigate this regime where the dependence on transverse variables can be computed perturbatively in QCD. In general, we will need to smoothly connect the nonperturbative and perturbative regimes to make phenomenological predictions.  

For the perturbative regime, $q_T \sim b_T^{-1} \gg \LQCD$, one can relate the unpolarized TMD PDFs to collinear PDFs through a type of operator expansion~\cite{Collins:1981uw,Collins:1984kg},
\begin{align} \label{eq:tmdpdf_matching}
 \tilde f_{1\,i}(x, b_T, \mu, \zeta)
 = \sum_j \int_x^1 \frac{\df y}{y} \tilde C_{ij}\Bigl(\frac{x}{y}, b_T, \mu, \zeta\Bigr) f_{j}(y, \mu)
   + \cO(b_T^2 \LQCD^2)
\,.\end{align}
In \eq{tmdpdf_matching}, the sum runs over all parton flavors $j$, and the $\tilde C_{ij}$ are perturbative matching kernels, such that the only nonperturbative input on the right hand side of \eq{tmdpdf_matching} is given by the collinear PDF $f_j(y,\mu)$.
As indicated, this equation holds up to corrections in $b_T^2 \LQCD^2$.
Taking the Fourier transform yields an equivalent result in transverse momentum space,
\begin{align} \label{eq:tmdpdf_matching_qT}
 f_{1\,i}(x, k_T, \mu, \zeta)
 = \sum_j \int_x^1 \frac{\df y}{y} C_{ij}\Bigl(\frac{x}{y}, k_T, \mu, \zeta\Bigr) f_{j}(y, \mu)
   \times \Big[ 1 + \cO(\LQCD^2/k_T^2) \Big]
\,.\end{align}

Relations similar to Eqs.~(\ref{eq:tmdpdf_matching}) and (\ref{eq:tmdpdf_matching_qT})
also hold for the other polarized TMD PDFs and for the TMD FFs, which can be related to polarized collinear PDFs and FFs, as discussed below.

\eq{tmdpdf_matching_qT} can be understood by noting that for perturbative $q_T$, both \eqs{collexample2}{tmdexample} are valid descriptions of the Drell-Yan process, where \eq{collexample2} should be used when $q_T \sim Q$ and \eq{tmdexample} should be used when $q_T \ll Q$.
In a strict perturbative expansion of the cross section in terms of the strong coupling constant $\as$,
\eqs{collexample2}{tmdexample} yield identical results when expanded for small $q_T \ll Q$.
This can be used to calculate the perturbative matching kernels $C_{ij}$, and Refs.~\cite{Catani:2011kr,Catani:2012qa} used this approach to obtain these kernels at NNLO.

Alternatively, one can calculate the matching kernels $\tilde C_{ij}$ directly from the matrix element definitions in \eqs{beamfunc}{softfunc}.
In this approach, one replaces the proton state in \eq{beamfunc} with a quark state $\psi_j$ of flavor $j$ or a gluon state $j=g$. 
Since the soft function is defined as a vacuum matrix element without hadronic states, it can be straightforwardly calculated perturbatively from \eq{softfunc}.
The construction of the TMD PDFs for parton $i$ is then either achieved by combining the bare unsubtracted TMD PDF and bare soft function, as in \eq{tmdpdf_1}, or by combining renormalized beam and soft functions as in 
\eq{fBS_relation}.
As explained in \sec{tmd_defs}, the bare perturbative results require the use of a dedicated rapidity regulator $\tau$,
which cancels upon combining the bare unsubtracted TMD PDF and bare soft function in \eq{tmdpdf_1}. Likewise, the individual renormalized beam and soft functions depend on a rapidity renormalization scale $\nu$, which cancels out when these functions are combined in \eq{fBS_relation} to obtain the renormalized TMD PDF.
Finally we subtract the perturbative results for the longitudinal PDF $f_{j/P}(y,\mu)$ itself, thus obtaining the desired kernels $\tilde C_{ij}$ and $\tilde C_{ig}$.
This calculation was explicitly illustrated at one loop for the quark TMD PDF using the $\eta$ regulator in \sec{tmd_defs_nlo}.

This strategy was employed in \cite{Aybat:2011zv, Collins:2011zzd} using Wilson lines off the light-cone at NLO.
NNLO results were first obtained in in \cite{Gehrmann:2014yya} using the analytic regulator,
and later also in \cite{Echevarria:2015byo,Echevarria:2016scs} and \cite{Luebbert:2016itl} using the $\delta$ and $\eta$ regulator, respectively.
Recently, N$^3$LO results were independently obtained by two groups in \cite{Li:2016ctv,Luo:2019szz, Ebert:2020yqt, Luo:2020epw} employing the exponential regulator.
The calculations in \cite{Ebert:2020yqt} were based on a hybrid approach, where the full QCD cross section was expanded in the collinear limit,
which was then combined with the exponential regulator to obtain the TMD PDFs.
We remark that calculations based on the exponential or $\eta$ regulator are often used to separately calculate renormalized beam and soft functions.

For illustration, we show how to obtain the perturbative matching of the quark TMD PDF onto the quark PDF, omitting contributions from matching onto the gluon TMD PDF. The required perturbative result for the TMD PDF was calculated in \sec{tmd_defs_nlo}, with the final result given in \eq{tmdpdf_nlo_renom}. The corresponding perturbative result for the collinear quark PDF is given by
\begin{align} \label{eq:PDF_nlo}
 f_{q/q}(x, \mu) = \delta(1-x) - \frac{\as(\mu) C_F}{2\pi} \frac{1}{\eps}  [P_{qq}(x)]_+ + \cO(\as^2)
\,,\end{align}
where the displayed $1/\eps$ pole is of infrared (collinear) origin.
Comparing \eq{PDF_nlo} to \eq{tmdpdf_nlo_renom}, one sees that this term exactly cancels between the perturbative results for the TMD PDF and PDF.
The remaining terms then yield the matching kernel
\begin{align} \label{eq:TMDPDF_matching_nlo}
 \tilde C_{qq}(x,b_T,\mu,\zeta)
 &= \delta(1-x) + \frac{\as(\mu) C_F}{2\pi} \biggl\{ - L_b \bigl[P_{qq}(x)\bigr]_+ + (1-x)
 \\\nn&\hspace{3.8cm}
   + \delta(1-x) \biggl[ - \frac{1}{2} L_b^2
   + L_b \biggl(\frac{3}{2} + \ln\frac{\mu^2}{\zeta} \biggr)
   - \frac{\pi^2}{12} \biggr] \biggr\}
 + \cO(\as^2)
\,,\end{align}
where $L_b$ and $P_{qq}(x)$ are given in \eqs{Lb}{Pqq}, respectively.
As discussed in \sec{tmdpdfs_new} and \app{TMDdefn}, the final results for 
definitions of the TMD PDF agree across many choices of the rapidity regulators that are used at intermediate steps, and hence \eq{TMDPDF_matching_nlo} holds for all rapidity regulators discussed in \sec{tmd_defs_overview}.  
A different result for $\tilde C_{qq}$ is obtained if one uses one of the TMD PDF definitions discussed in \sec{other_tmd_defs}, since these TMD PDFs and the corresponding $\tilde C_{qq}$ depend on the different parameter $\tilde\zeta$ and an extra parameter $\rho$.

\paragraph{Matching of spin-dependent TMD PDFs and TMD FFs.}
In a similar fashion, one can match all spin-dependent quark and gluon TMDs onto their collinear counterparts for perturbative $q_T \sim b_T^{-1} \gg \LQCD$.
For example, the quark helicity distribution $g_{1}(x,b_T,\mu,\zeta)$ 
can be matched onto the collinear helicity distribution $g_1(x, \mu)$, instead of the collinear PDF $f_1(x,\mu)$.
The collinear distributions that a TMD PDF matches onto are classified by so-called twist, related to the twist of operators defining the collinear distribution. 
In particular, the non-local operator defining the collinear distributions can be expanded in terms of a tower of local operators, for example for a pair of quarks of flavor $i$
\begin{align} \label{eq:twistOps}
  P_{S,N}^{\mu_0\mu_1\cdots\mu_N} \: \bar\psi_i^0\, \Gamma_{\mu_0} iD_{\mu_1} \cdots 
     iD_{\mu_N} \, \psi_i^0 \,,
\end{align}
where we have an infinite set of local operators indexed by $N$, and $P_{S,N}$ is a dimensionless projector onto the spin-$S$ combination of the Lorentz indices $\mu_0\cdots \mu_N$. The operator in \eq{twistOps} has dimension $D=3+N$, and twist $t=D-S$.
Typically, the contributions from longitudinal distributions are suppressed with increasing twist, such that one only needs to consider the lowest values of twist.
For example, the longitudinal PDF $f_1(x,\mu)$ is twist $2$, which is thus also referred to as leading twist.
Other twist $2$ longitudinal PDFs include the helicity $g_1(x)$ and transversity $h_1(x)$. 
For small $b_T$, some TMD distributions do not match onto twist $2$ distributions, but obtain their first contribution at twist $3$, where the so-called Qiu-Sterman function~\cite{Qiu:1991pp} $T(x_1, x_2, x_3)$ plays the role of the unpolarized twist-2 PDF, and also has a chiral-odd counterpart $\delta T_\eps(x_1, x_2, x_3)$. See for example~\cite{Braun:2009mi} for a discussion of twist-$3$ PDFs.

In tables~\ref{tbl:TMDPDF_matching_q} and \ref{tbl:TMDPDF_matching_g}, we summarize the knowledge, at the time of this writing, of these matching relations
for quark TMD PDFs and gluon TMD PDFs.
For each TMD distribution function, the table shows which collinear distributions that they match onto (up to twist $3$), the perturbative order to which the matching is known, and a list of references for these matching calculations.
Care has to be taken when using these results due to a variety of different conventions used in the literature, including whether results are expressed in momentum or position space, how the different distributions are normalized (cf.~\eq{tmd_decomposition_2} for the convention used here), and the precise definition of the $\MSbar$ scheme.

The unpolarized TMD PDFs have the simplest structure, and their matching is already known at N$^3$LO~\cite{Luo:2019szz, Luo:2020epw, Ebert:2020yqt, Ebert:2020qef}.
Much less is known for the other TMDs, in particular those matching onto subleading-twist PDFs.
For the quark TMD PDF, recently a comprehensive calculation of the twist-3 matching at LO was presented in \cite{Moos:2020wvd}. 

For TMD FFs the situation is more complicated, as a similar relation between TMD FFs and collinear FFs
by means of an operator product expansion (OPE) can not be established in the same way, see e.g.~\cite{Balitsky:1990ck,Moos:2020wvd} for a detailed discussion.
However, at a perturbative level, one can relate the two objects by demanding that the cross sections
calculated within TMD and collinear factorization are consistent. This has been used to obtain
a matching of the unpolarized TMD FF onto the collinear FF up to N$^3$LO~\cite{Echevarria:2015usa, Echevarria:2016scs, Luo:2019hmp, Luo:2020epw, Ebert:2020qef}.
There is also an interesting relation between TMD PDFs and TMD FFs through analytic continuation~\cite{Chen:2020uvt},
which so far has only been employed for the unpolarized TMDs~\cite{Luo:2020epw}.

\begin{table}
 \centering
 \begin{tabular}{|l|c|c|c|r|c|}
 \hline
 \multirow{2}{*}{Name} & \multirow{2}{*}{Function} & Twist-2  & Twist-3  & \multicolumn{1}{c|}{Known}   & \multirow{2}{*}{Refs.} \\
 & & matching & matching & \multicolumn{1}{c|}{order} &
 \\ \hhline{|=|=|=|=|=|=|}
 unpolarized   & $\tilde f_1(x,b_T)$          & $f_1(x)$ & --       & N$^3$LO $(\as^3)$ &  \cite{Aybat:2011zv, Catani:2012qa, Collins:2011zzd, Gehrmann:2014yya,Echevarria:2016scs}
 \\ & & & & &
\cite{Luebbert:2016itl, Luo:2019hmp, Luo:2019szz, Luo:2020epw, Ebert:2020yqt}
 \\ \hline
 helicity      & $\tilde g_{1}(x,b_T)$       & $g_1(x)$ & $\cT_g(x)$ & NLO $(\as^1)$     & \cite{Bacchetta:2013pqa, Gutierrez-Reyes:2017glx}
 \\ \hline
 worm-gear $T$ & $\tilde g_{1T}^\perp(x,b_T)$ & $g_1(x)$ & $\cT_g(x)$ & LO $(\as^0)$      & \cite{Kanazawa:2015ajw, Scimemi:2018mmi}
 \\ \hline
 Sivers        & $\tilde f_{1T}^\perp(x,b_T)$ & --       & $T(-x,0,x)$ & NLO $(\as^1)$     & \cite{Boer:2003cm,Ji:2006ub, Ji:2006vf, Koike:2007dg}
\\ & & & & &
 \cite{Kang:2011mr, Sun:2013hua, Dai:2014ala, Scimemi:2019gge}
 \\ \hhline{|=|=|=|=|=|=|}
 transversity  & $\tilde h_1(x,b_T)$          & $h_1(x)$ & $\cT_h(x)$ & NNLO $(\as^2)$ & \cite{Bacchetta:2013pqa, Gutierrez-Reyes:2017glx, Gutierrez-Reyes:2018iod}
 \\ \hline
 worm-gear $L$ & $\tilde h_{1L}^\perp(x,b_T)$ & $h_1(x)$ & $\cT_h(x)$ & LO $(\as^0)$   & \cite{Kanazawa:2015ajw, Scimemi:2018mmi}
 \\ \hline
 Boer-Mulders  & $\tilde h_1^\perp(x,b_T)$    & --       & $\delta T_\eps(-x,0,x)$ & NLO $(\as^0)$ & \cite{Scimemi:2018mmi}
 \\ \hline
 pretzelosity  & $\tilde h_{1T}^\perp(x,b_T)$ & --       & $\cT_h(x)$ & LO $(\as^0)$   & \cite{Moos:2020wvd}
 \\ \hline
 \end{tabular}
 \caption{Collinear matching of the quark TMD PDFs up to collinear twist $3$ at perturbative $b_T^{-1} \gg \LQCD$. 
 For brevity, the renormalization scale $\mu$ and the Collins-Soper scale $\zeta$ are suppressed.
 The upper four rows of the table show chiral-even TMDs, while the bottom four rows show chiral-odd TMDs.
 $\cT_g(x)$ and $\cT_h(x)$ are abbreviations for specific combinations of the twist-$3$ distributions. Table adapted from Ref.~\cite{Moos:2020wvd}, to which we refer for further details.}
 \label{tbl:TMDPDF_matching_q}
\end{table}

\begin{table}
 \centering
 \begin{tabular}{|l|c|c|c|r|c|}
 \hline
 \multirow{2}{*}{Name} & \multirow{2}{*}{Function} & Twist-2  & Twist-3  & \multicolumn{1}{c|}{Known}   & \multirow{2}{*}{Refs.} \\
 & & matching & matching & \multicolumn{1}{c|}{order} &
 \\ \hhline{|=|=|=|=|=|=|}
 unpolarized         & $\tilde f_1(x,b_T)$    & $f_1(x)$ & -- & N$^3$LO $(\as^3)$ &  \cite{Catani:2011kr, Gehrmann:2014yya, Echevarria:2016scs, Luebbert:2016itl}
 \\ & & & & &
 \cite{Luo:2019bmw, Luo:2020epw, Ebert:2020yqt}
 \\ \hline
 linearly polarized  & $\tilde h_1^{\perp g}(x, b_T)$  & $f_1(x)$ & -- & NNLO $(\as^2)$ &  \cite{Catani:2011kr, Chiu:2012ir, Becher:2012yn}
 \\ & & & & &
\cite{Echevarria:2015uaa, Gutierrez-Reyes:2017glx, Gutierrez-Reyes:2019rug, Luo:2019bmw}
 \\ \hline
  helicity & $\tilde g_{1L}^g(x, b_T)$ & $g_1(x)$ & & NLO ($\as^1)$ & \cite{Echevarria:2015uaa}
 \\ \hline
  & $\tilde g_{1T}^g(x, b_T)$  & & & &
 \\ \hhline{|=|=|=|=|=|=|}
 Sivers & $\tilde f_{1T}^{\perp g}(x, b_T)$  & -- & & &
 \\ \hline
  & $\tilde h_{1T}^g(x, b_T)$  & & & &
 \\ \hline
  & $\tilde h_{1L}^{\perp g}(x, b_T)$  & & & &
 \\ \hline
  & $\tilde h_{1T}^{\perp g}(x, b_T)$  & & & &
 \\ \hline
 \end{tabular}
 \caption{Collinear matching of the gluon TMD PDFs up to collinear twist $3$ at perturbative $b_T^{-1} \gg \LQCD$. 
 For brevity, the renormalization scale $\mu$ and the Collins-Soper scale $\zeta$ are suppressed.
 The upper four rows of the table show chiral-even TMDs, while the bottom four rows show chiral-odd TMDs.
 Empty entries indicate that for most gluon TMDs the matching has not yet been considered in the literature, and these rows presumably obtain their first contributions at twist $3$.
 }
 \label{tbl:TMDPDF_matching_g}
\end{table}

\FloatBarrier
\subsection{Relating Integrated TMDs to Collinear Functions}
\label{sec:integratedTMDs}

An interesting question to ask about TMD PDFs is how they relate to longitudinal PDFs.
Naively they might be thought of as simply more differential TMDs that yield the longitudinal PDFs upon integration.
One might thus expect that
\begin{align} \label{eq:relation_TMDpdf_pdf_naive}
 \int\!\df^2\kt \, f_{i/H}^{0\,(u)}(x, \kt) &
 \stackrel{?}{=}
 f^{0}_{i/H}(x)
\,.\end{align}
Indeed, it is easy to see that this expectation is fulfilled
at the \emph{bare} level,
\begin{align} \label{eq:relation_TMDpdf_pdf}
   \int\!\df^2\kt \, f_{i/H}^{0\,(u)}(x, \kt,\eps,\tau,x P^+) &
 = \int\!\df^2\kt \, \int\frac{\df^2\bt}{(2\pi)^2} \, e^{\img \bt \cdot \kt} \tilde f_{i/H}(x, \bt)
 \nn\\&
 = \tilde f_{i/H}^{0\,(u)}(x, \bt = {\bf 0},\eps,\tau, x P^+)
 \nn\\&
 = f^{0}_{i/H}(x,\eps)
\,.\end{align}
In the first step, we inserted the definition of the Fourier transform in \eq{tmdpdf_bspace},
and then used the standard identity $\int\!\df^2\kt e^{\img \bt \cdot \kt} = (2\pi)^2 \delta(\bt)$
to evaluate the bare TMD PDF at $\bt = 0$. With this choice, the Wilson lines in \eq{beamfunc}
can be simplified using $W_n(a,b) W_n^\dagger(b,a) = 1$, such that the staple-shaped Wilson line
$W_\sqsubset$ in \eq{Staple_Wilson_line} reduces to the straight Wilson line encountered in the definition
of the PDF in \eq{barepdf}.
Similarly, one can show that the bare TMD soft function defined in \eq{softfunc} reduces to unity
when integrated over all $\kt$,
\begin{align}
 \int\!\df^2\kt \, S^{0}_{n_a n_b}(k_T,\eps,\tau) = \tilde S^{0}_{n_a n_b}(\bt=0,\eps,\tau)  = 1
\,.\end{align}
Thus, at the bare level both unsubtracted and subtracted TMD PDF recover the longitudinal PDF
when integrated over all $\kt$.%
\footnote{Note that there are many more spin-dependent TMD PDFs than longitudinal PDFs.
However, most of these are proportional to $\bt$, see e.g.~\eq{tmd_decomposition_2}
and thus vanish exactly at the bare level for $\bt = 0$. Thus, only the
unpolarized ($f_1$), helicity ($g_{1}$) and transversity ($h_1$) distributions remain,
for which there are corresponding longitudinal PDFs.}

However, after renormalization the naive expectation of \eq{relation_TMDpdf_pdf_naive} is broken, i.e.
\begin{align} \label{eq:breaking_TMD_PDF_relation}
 \int\df^2\kt \, f_{i/p}(x, \kt, \mu , \zeta)
 \ne f_i(x, \mu)
\,.\end{align}
This is easy to see from the scale dependence on the two sides of \eq{breaking_TMD_PDF_relation}. Evolution equations for the TMDs are discussed in detail in \sec{evolution}. We have
\begin{align} \label{eq:breaking_TMD_PDF_relation2}
 \mu \frac{\df}{\df\mu} f_{i/p}(x, \kt, \mu , \zeta) &
 = \gamma_\mu^i(\mu, \zeta) f_{i/p}(x, \kt, \mu , \zeta)
\,,\nn\\
 \mu \frac{\df}{\df\mu} f_i(x, \mu) &
 = \sum_j \int_x^1 \frac{\df x'}{x'} P_{ij}(x', \mu)  f_j \Bigl(\frac{x}{x'}, \mu\Bigr)
\,.\end{align}
Importantly, the $\mu$ evolution of the TMD PDF is diagonal both in flavor $i$
and momentum fraction $x$, while for the PDF it sums over all flavors $j$ and
involves a convolution in $x$. Even more strikingly, only the TMD PDF
depends on the CS scale $\zeta$, while the PDF is independent of it.
Clearly, these observations forbid a simple relation between TMD PDF and PDF  at the renormalized level,
and thus break \eq{relation_TMDpdf_pdf_naive}.

It is then natural to ask why \eq{relation_TMDpdf_pdf_naive} holds at the bare level,
but not at the renormalized level, see e.g.~Refs.~\cite{Ji:2004wu,Collins:2011zzd,GarciaEchevarria:2011rb}.
The fundamental reason is that the limit $b_T \to 0$
that is the source of the simple relation in \eq{relation_TMDpdf_pdf}
corresponds to large momenta $k_T \sim 1/b_T \to \infty$, and thus an ultraviolet region.
When defining the renormalized TMDs, this ultraviolet region has been regulated
by a UV regulator, and UV divergences are removed yielding results in the $\MSbar$ scheme. The act of renormalization, taking
$\eps \to 0$ and absorbing UV divergences into counterterms, does not commute
with taking the limit $b_T \to 0$.

This fact was nicely illustrated in Ref.~\cite{GarciaEchevarria:2011rb},
where the integral in \eq{relation_TMDpdf_pdf_naive} was carried out
using perturbative one-loop results, similar to those presented in \sec{tmd_defs_nlo}.
In this case, one actually has to perform the $\kt$ integral in $d-2 = 2 - 2\eps$ dimensions,
cf.~\eq{angle_integral}. Only then does one find that all perturbative corrections
to the TMD PDF vanish up to terms identical to the longitudinal distribution,
thereby confirming \eq{relation_TMDpdf_pdf} at the one-loop order.

Since the above argument is based on integrating the TMD PDF into the UV region $k_T \to \infty$,
one can also ask what happens if one limits the integral in \eq{relation_TMDpdf_pdf_naive}
to some large but finite value, $|\kt| \le k_T^{\rm cut} \gg \lqcd$. Intuitively this should avoid some of the issues with the UV region, and
one can hope to at least find a good \emph{approximation} of the form
\begin{align} \label{eq:relation_TMDpdf_pdf_cut}
 \int_{k_T \le k_T^{\rm cut}} \df^2\kt \, f_{i/p}(x, \kt, \mu , \zeta)
 \stackrel{?}{\approx} f_i(x, \mu)
\,.\end{align}
Technically, the integral in \eq{relation_TMDpdf_pdf_cut} is sensitive to both
nonperturbative $k_T \sim \lqcd$ and to perturbative $k_T \sim k_T^{\rm cut}$.
The nonperturbative region is expected to have little impact for large $k_T^{\rm cut}$,
which can be tested numerically by suitable models.
In contrast, the perturbative region is calculable in terms of longitudinal PDFs,
as explained in \sec{largeqT}. This makes \eq{relation_TMDpdf_pdf_cut}
a reasonable guess, and studies of this relation have been performed for example in~Refs.~\cite{Qiu:2000hf, Berger:2002ut, Bacchetta:2013pqa}.

The formalism for carrying out such an analysis was taken a step further in Ref.~\cite{Ebert:2022cku} by developing a  method to carry out the $k_T\le k_T^{\rm cut}$ integration in \eq{relation_TMDpdf_pdf_cut} in a model-independent way by using a position space cutoff $b_T\le b_T^{\rm cut}$. This provides control over the required approximations since the long-distance contributions from $b_T \sim 1/\lqcd$ are organized into a systematic expansion in $1/(b_T^{\rm cut} k_T^{\rm cut}) \ll 1 $ by inducing higher-order surface terms at the cutoff. This approach yields a method for evaluating the integral of renormalized TMD functions where all sources of uncertainty  can be determined. These sources include perturbative uncertainty from missing higher order terms in the $\alpha_s$ expansion at small $b_T$, nonperturbative TMD effects that appear as coefficients in a $(b_T^{\rm cut}\Lambda_{\rm QCD})^2$ expansion, and the overall impact of the precise choice for the cutoffs $k_T^{\rm cut}$ and $b_T^{\rm cut}$.
To test this setup, one can examine unpolarized TMDs and use the state-of-the-art OPE result with \eq{tmdpdf_matching} evaluated at three-loop logarithmic accuracy~\cite{Ebert:2022cku}. (These orders of renormalization group improved perturbation theory are explained in \chap{evolution}.)
\begin{figure}[t!]
 \centering
 \includegraphics[width=0.65\textwidth]{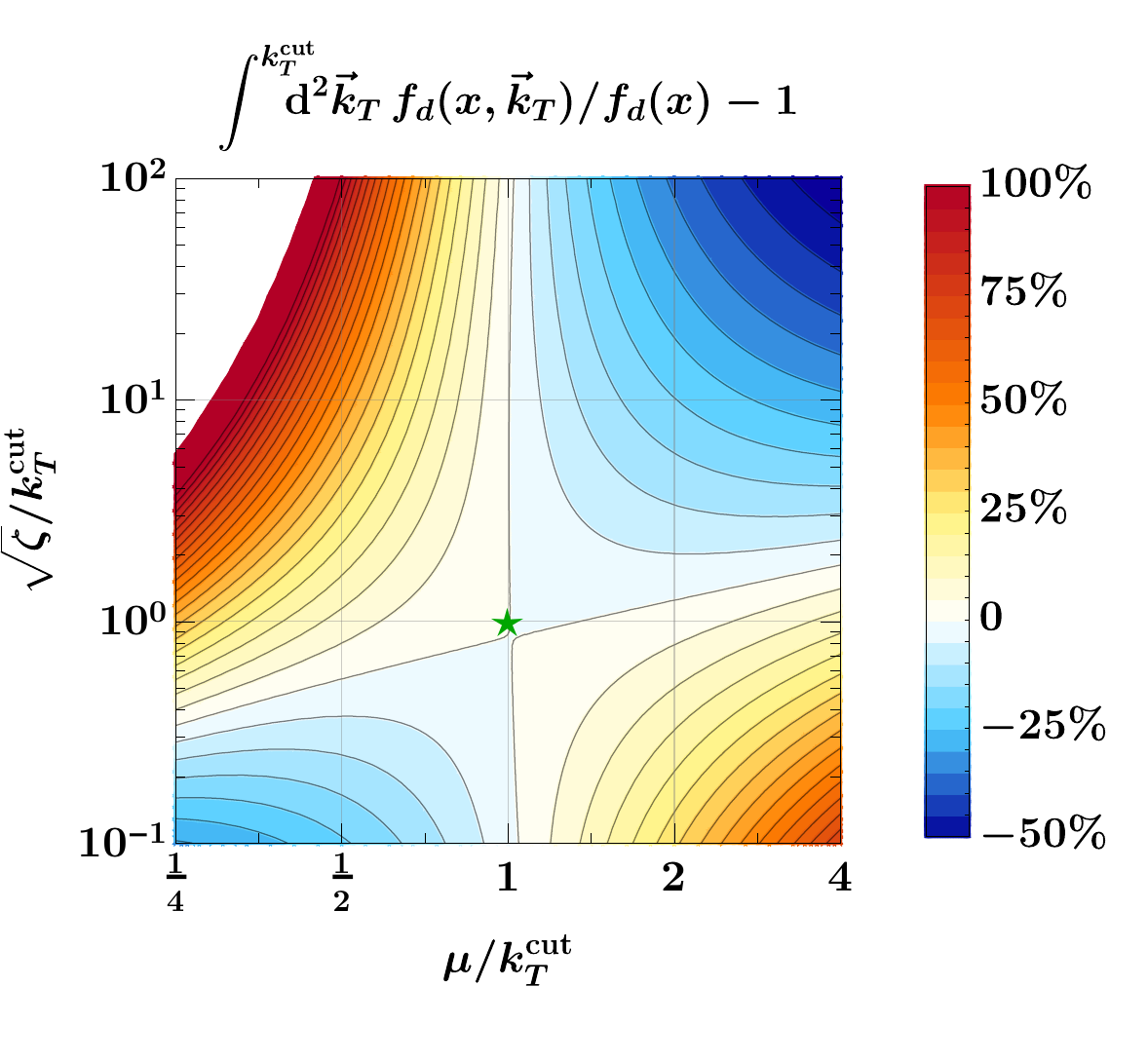}
\vspace{-0.6cm}
 \caption{Comparison of the cumulative integral of the TMD PDF over $k_T \le k_T^{\rm cut}$ to the longitudinal PDF
 for the $d$-quark with $x=0.01$ and $k_T^{\rm cut} = 10~\mathrm{GeV}$ as a function of both $\mu$ and $\zeta$.
 The star denote the special point $\mu = \sqrt\zeta = k_T^{\rm cut}$.
 The contours increase in steps of $5\%$, such that the innermost shaded regions indicate deviations of $\pm 5\%$.
 Taken from Ref.\cite{Ebert:2022cku}.}
 \label{fig:tmd_heatmap}
\end{figure}
In \fig{tmd_heatmap}, \eq{relation_TMDpdf_pdf_cut} is tested
by plotting the relative deviation between the cumulative integral over $k_T\le k_T^{\rm cut}$ and the longitudinal PDF.
The plot uses a $d$-quark with fixed $x=0.01$ and $k_T^{\rm cut} = 10~\mathrm{GeV}$, and then varies the renormalization scale $\mu$ and Collins-Soper scale $\zeta$.

Astonishingly, one finds that even in the presence of perturbative corrections
and renormalization group running the TMD PDF and PDF agree extremely well; for the natural choice for the two renormalization scale parameters $\mu = \sqrt\zeta = k_T^{\rm cut}$ the agreement is at the percent level.
Thus we can conclude that
\begin{align} \label{eq:relation_TMDpdf_pdf_cut_scales}
 \int_{k_T \le k_T^{\rm cut}} \df^2\kt \, f_{i/p} \Big(x, \kt, \mu=k_T^{\rm cut} , \sqrt{\zeta} = k_T^{\rm cut}\,\Big)
 \simeq f_i\big(x, \mu=k_T^{\rm cut}\big)
\,.\end{align}
This gives justification to the original physical picture underlying \eq{relation_TMDpdf_pdf_cut}.
The contour bands in \fig{tmd_heatmap} also illustrate that the dependence on variations
of either $\zeta$ or $\mu$ around $k_T^{\rm cut}$ is quite moderate, while there is a rather large effect
of varying both scales simultaneously. As explained in \chap{evolution}, a simultaneous variation, such as along the diagonal directions,  induces
large double logarithms predicted by the hard evolution, which can not be compensated by evolution of the collinear PDF.
\begin{figure}[t!]
 \centering
 \includegraphics[width=0.48\textwidth]{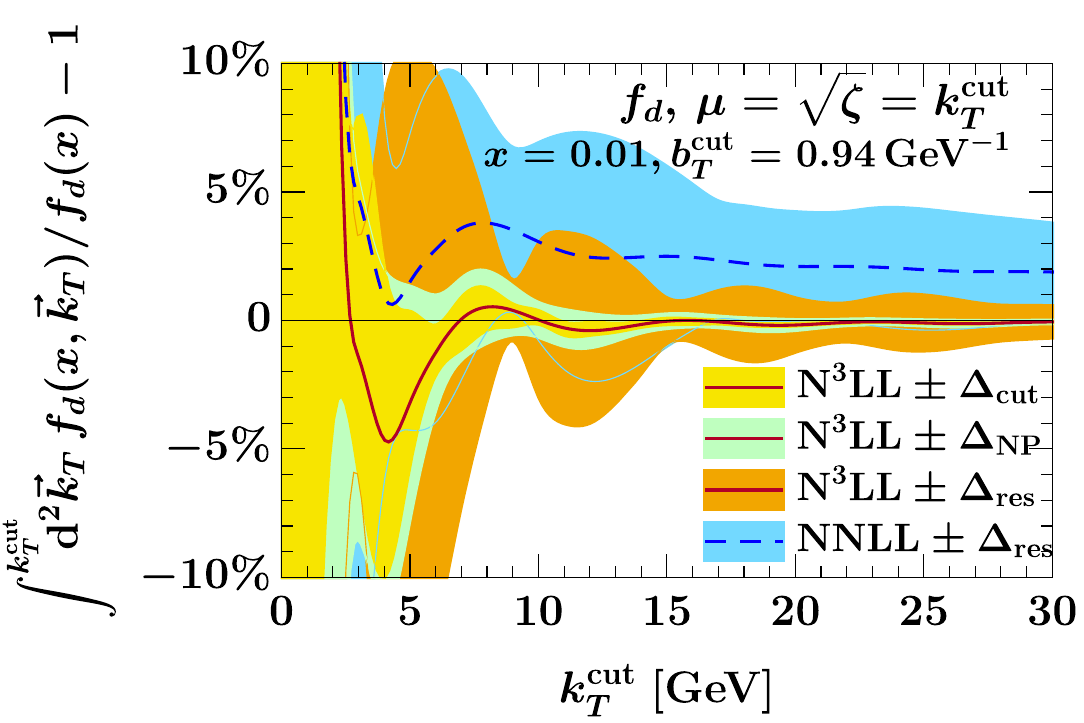}
 \quad
 \includegraphics[width=0.48\textwidth]{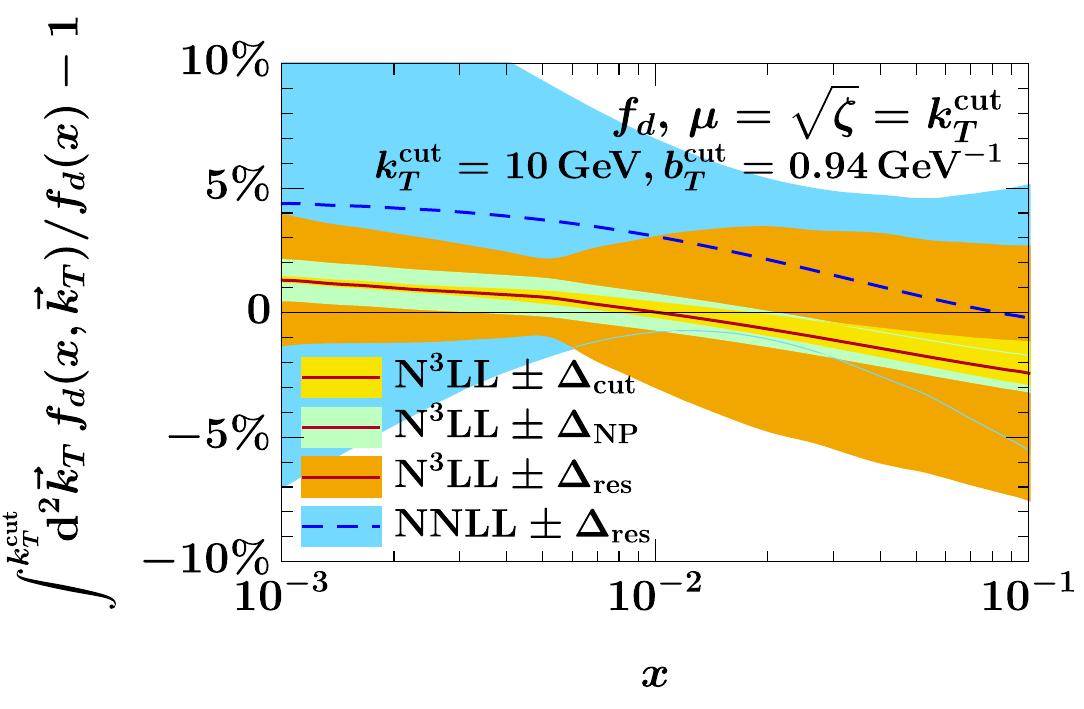}
 \caption{Comparison of the cumulative integral of the TMD PDF over $k_T \le k_T^{\rm cut}$ to the longitudinal PDF
 for the $d$-quark as a function of $k_T^{\rm cut}$ (left) and $x$ (right).
 Taken from Ref.\cite{Ebert:2022cku}.}
 \label{fig:tmd_normalization}
\end{figure} 

To verify that the above observation is not an accidental feature
of the values $x=0.01$ and $k_T^{\rm cut}=10\,{\rm GeV}$ used so far, in \fig{tmd_normalization}
results for the comparison as a function of $k_T^{\rm cut}$ (left figure) and $x$ (right figure) are given. Here the various sources of uncertainty are also assessed, as indicated by the different colored bands.  The yellow band shows very small uncertainties from terms beyond third order in the $1/(b_T^{\rm cut}k_T^{\rm cut})$ expansion, which are assessed by varying the choice of $b_T^{\rm cut}$ used in the analysis. The green band shows the quite small uncertainties from treating the leading
${\cal O}(b_T^{\rm cut\,2} \Lambda_{\rm QCD}^2)$ nonperturbative corrections as unknowns, and varying them in the analysis. Finally, the largest uncertainties are from the truncation of perturbation theory, which can be observed to decrease fairly significantly when going from second order (blue band at NNLL) to third order (orange band at N$^3$LL). 
Overall, for all $x$ values and for $k_T^{\rm cut}\ge 10\,{\rm GeV}$ one sees   that \eq{relation_TMDpdf_pdf_cut_scales} is satisfied as an equality within the uncertainties. Furthermore, these uncertainties are only $\cO(5\%)$ for a large range of $x$ and $k_T^{\rm cut}$ values.
For the small $k_T^{\rm cut}$ region it is not surprising that
the relation in \eq{relation_TMDpdf_pdf_cut_scales} breaks down since the result becomes sensitive to the nonperturbative nature of the $k_T$ distribution.
To conclude, we observe that \eq{relation_TMDpdf_pdf_cut_scales} holds to an excellent approximation
over a large range of $x$ and $k_T^{\rm cut}$, and thus can be understood as a  practical
version of the naive expectation in \eq{relation_TMDpdf_pdf_naive}.

While this method was discussed here only in the context of the unpolarized TMD,
the strategy is completely general and thus also applies to spin-dependent TMDs.
For example, Ref.~\cite{Bacchetta:2013pqa} already studied the $k_T^{\rm cut}$
dependence of the helicity and transversity distributions at one loop.
For those TMDs that vanish at bare level, i.e.~at $b_T = 0$,
the method provides a way to test model-independently how fast they vanish as a function of $k_T^{\rm cut}$.  This approach therefore holds potential for determining the extent to which bare relations, like the TMD positivity constraints discussed in Sec.~\ref{Sec:positivity-constraints}, 
\index{positivity of TMDs} can be extended to formulas for renormalized distributions in QCD. It should also enable more rigorous treatment of the Burkardt and Sch\"afer-Teryaev sum rules discussed in Secs.~\ref{Sec:Burkardt-sum-rule} and \ref{Sec:Schafer-Teryaev-sum-rule}, including assessing the precise conditions under which they are valid and the size of power corrections to these results.

\subsection{Connection to Lattice QCD}
\label{sec:latt_def_connection}

\index{lattice QCD calculations!TMDs}
Lattice QCD is the only currently available method to obtain nonperturbative hadron structure information from the underlying field theory, QCD, without uncontrolled model assumptions.
It is therefore important to develop methods for calculating TMD observables within Lattice QCD.

Since the Euclidean spacetime signature employed in Lattice QCD does not straightforwardly accomodate real-time separations, the TMDs defined using lightlike Wilson-line operators are not directly calculable on the Euclidean lattice.
One method to circumvent this problem is to use space-like Wilson lines in the definition of the TMD correlators in \eq{tmdCorig} and exploit Lorentz covariance to relate their matrix elements to equal-time ones on the lattice~\cite{Hagler:2009mb,Musch:2010ka,Musch:2011er,Engelhardt:2015xja,Yoon:2016dyh,Yoon:2017qzo}.
Since this method does not apply to the soft factor, most often ratios of matrix elements are considered where the soft factors drop out.
Another method that has led to much progress in the lattice calculation of collinear PDFs is the large-momentum effective theory (LaMET)~\cite{Ji:2013dva,Ji:2014gla,Ji:2020ect}, extensions of which to TMDs have been constructed over the past few years~\cite{Ji:2014hxa,Ji:2018hvs,Ebert:2018gzl,Ebert:2019okf,Ebert:2019tvc,Ji:2019sxk,Ji:2019ewn,Vladimirov:2020ofp,Ebert:2020gxr,Ji:2020jeb}.

\subsubsection{Lorentz-invariant approach}
\label{sec:latt_def_lorentz}

\begin{figure*}
 \centering
 \includegraphics[width=0.45\textwidth]{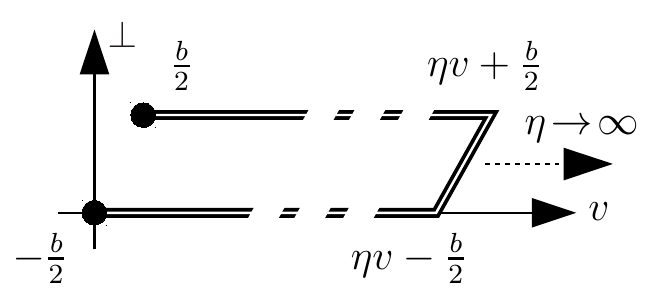}
 \caption{Staple-shaped path for the gauge connection $W_{\sqsupset \eta}^v$ in \eq{latt_corr_def}.}
 \label{fig:stapleLattice}
\end{figure*}

For use in lattice calculations, the correlator from which the unsubtracted TMD PDFs are obtained after Fourier transformation is generalized in several ways,
\begin{equation} \label{eq:latt_corr_def}
\widetilde{\Phi}^{[\Gamma ]}_i (b,P^{\prime },P,S,v,\eta,a ) 
  = \frac{1}{2} \Bigl\langle p(P^{\prime},S) \Big| 
    \bar\psi^{0}_i (b^{\mu }/2) \Gamma W_{{\sqsupset}\eta}^{v} 
    (b^{\mu }/2,-b^{\mu }/2) \psi^{0}_i (-b^{\mu }/2) 
     \Big| p(P,S) \Bigr\rangle \,.
\end{equation}
On the lattice, the UV regulator (previously denoted by $\eps$) is realized by the lattice spacing $a$.
Since in a concrete lattice calculation the Wilson lines attached to the quark operators $\bar{\psi }, \psi $ can not extend to infinity, a staple-shaped gauge connection of finite extent $\eta$ is used, 
\begin{align}  \label{eq:stapleLattice}
 W^{v}_{\sqsupset\eta}(b^\mu/2,-b^\mu/2)
   &= W_{v}^{\dagger} (b^\mu/2;0,\eta) 
      W_{\hat{b}}\!\bigl(\eta v-b/2; 0, |b|\bigr)
       W_{v} (-b^\mu/2;0,\eta) 
   \,. 
\end{align}
This gauge connection is shown in \fig{stapleLattice}, which corresponds to a generalization of the illustration given in \fig{wilsonlines} (left) above in order to include finite length and to introduce more flexible variables for the endpoints.
Apart from the quark operator separation $b^{\mu } $, the staple link is described by the direction of the staple $v^{\mu}$ and the length of the staple $\eta $.
In a concrete lattice calculation, an extrapolation $\eta \rightarrow -\infty$ (Drell-Yan) or $\eta\rightarrow +\infty$ (SIDIS) must be performed from data obtained at finite $\eta $.  
(See \sec{universality} for discussion on the two cases $\eta=\pm\infty$.)  
The staple direction $v$ is taken off the light cone into the space-like region.
This specification is crucial in order to make the definition amenable to lattice computation; the reason is that standard Lattice QCD methods to calculate matrix elements of the form in \eq{latt_corr_def} are restricted to operators that are defined at a single time. As already indicated further above, the temporal lattice direction is Euclidean, and therefore no operators with a real, Minkowski time extent can be accommodated.
Consequently, it is imperative that all separations in the operator, i.e., $b$ and $\eta v$, be space-like. For this reason we take
\begin{align}
  \frac{v^+}{v^-} = - e^{2 y_v} < 0 \,.
\end{align}
Only then is there no obstacle to boosting the problem to a Lorentz frame in which the operator in \eq{latt_corr_def} exists at a single time, with $y_v=0$.
The lattice calculation is then carried out in that particular frame.

The correlator (\ref{eq:latt_corr_def}) furthermore depends on the momenta $P,P^{\prime } $ of the in- and outgoing states as well as their spin $S$.
TMDs are obtained in the forward limit, $P=P^{\prime } $, which we will assume for the remainder of the discussion in this section.
(The generalization to nonzero momentum transfer yields the Generalized Transverse Momentum-Dependent parton distributions (GTMDs)
\index{GTMD} discussed in Chapter~\ref{sec:gtmd}.)
A useful parameter to characterize the rapidity of the staple link direction $v$ relative to the hadron is the dimensionless Collins-Soper type evolution parameter
\begin{align} \label{eq:hatzeta}
 \hat{\zeta } = \frac{v\cdot P}{\sqrt{|v^2| P^2}} 
 \,. 
\end{align}
This parameter characterizes the staple link connecting the quark operators. It therefore differs from the variable $\zeta_a$ defined in \eq{zeta}, which involves a combination of variables inherited from the proton matrix element ($m_p$ and $y_A$) and the TMD soft factor ($y_n$).

Using Lorentz covariance, the matrix element in \eq{latt_corr_def} can be decomposed into independent tensors constructed from $P^\mu$, $b^\mu$ and $v^\mu$, with the coefficients (or amplitudes) uniquely determined by the Lorentz scalars $P\cdot b$, $b^2$, $\hat{\zeta }$, $v\cdot b/\sqrt{-v^2}$, and $\eta^2 v^2$~\cite{Musch:2011er}. (Following standard conventions, we do not treat the dependence on $m_p^2=P^2$ as a variable.)
Such decompositions will be presented in Sec.~\ref{sec:TMDratios}. In Table~\ref{tab:lia1} we list these Lorentz scalars, together with their values in two reference frames for comparison. TMD PDFs are originally defined in a frame where $b^{+} =0$ and $v_T =P_T =0$. This constrains one of the five Lorentz scalars, since it implies the relation, expressed in Lorentz-invariant form,
\begin{align}
  \frac{v\cdot b}{v\cdot P}
 & = \frac{P\cdot b}{m_N^2} \Bigl[ 1- \sqrt{1+\hat\zeta^{-2}} \Big]
\,.
\end{align}
In Table~\ref{tab:lia1}
the column labeled  Modern CS $(y_B)$ corresponds to the frame choice used in the modern Collins-Soper definition with space-like Wilson lines of infinite extent, \eq{Staple_Wilson_line_JC} inserted into \eq{beamfunc}, with finite but large $|y_B|$.  The column labeled Euclidean Lattice gives the values in the frame where $v^\mu$ has no time component ($y_v=0$), in which the lattice calculation is performed.  Since all the Lorentz scalars can be determined in this Euclidean frame, one can obtain full information about the unsubtracted TMD PDF.
In order to make full contact with the modern Collins definition of the unsubtracted TMD PDF, which is considered in the limit $\eta\to \infty$ and eventually with large $y_B\to -\infty$, the lattice results obtained at finite values  must ultimately be extrapolated towards a large rapidity difference $\hat{\zeta }\to \infty$ and large $\eta\to \infty$.

{
\renewcommand{\arraystretch}{1.4}
\begin{table}
   \centering
   \begin{tabular}{|c|c|c|}
      \hline
      Lorentz Invariant &   Modern CS ($y_B$)&  Euclidean Lattice  \\
      \hline
      $P\cdot b$ & $P^+ b^-$ & $-P_z b_z $
      \\ \hline
      $b^2$ & $- {\bf b}_T^2$ & $-b_z^2 - {\bf b}_T^2$
      \\ \hline
      $\hat\zeta = \dfrac{v\cdot P}{m_p \sqrt{-v^2}}$ &
      $\sinh(y_P-y_B)$ & 
      $\hat\zeta = \dfrac{-v_z P_z }{m_p \sqrt{v_z^2 + v_T^2 } } $ 
      \\ \hline
      $\dfrac{v\cdot b}{\sqrt{-v^2}}$ & $\dfrac{- e^{y_B} b^-}{\sqrt{2}}$ &
      $\dfrac{-v_z b_z
      -{\bf v}_{T} \cdot {\bf b}_{T} }{\sqrt{v_z^2+v_T^2}}$
      \\ \hline
      $\eta^2 v^2 $ & $-\infty$ & $ -\eta^2 (v_z^{2}+v_T^2)$
      \\ \hline
   \end{tabular}
   \caption{Comparison in position space of the Lorentz invariant variables between the Euclidean lattice approach and the modern CS definition prior to taking the $y_B\to -\infty$ limit. In modern CS we have $b^\mu=(0, b^-, b_T)$ in light-cone coordinates where $v=n_B(y_B)$ from \eq{Collins_rap}. The Euclidean lattice construction takes $b^\mu=(0,b_T^x,b_T^y,b^z)$ in Cartesian coordinates.  }
   \label{tab:lia1}
\end{table}
}

An important corollary of this discussion is that the soft factor (\ref{eq:softfunc}), cf.~\fig{wilsonlines} (right), cannot be straightforwardly calculated in Lattice QCD in a completely analogous fashion.
Since it contains two staple directions with two different rapidities, there exists no Lorentz transformation that simultaneously renders both directions purely spatial. In the modern Collins definition the soft function is combined with the unsubtracted TMD PDF as in \eq{CSf}, which is necessary for the $y_B\to -\infty$ limit that yields the full TMD PDF to exist. One way to deal with this obstacle is to circumvent it by constructing observables in the form of ratios in which soft factors cancel. These may be, e.g., ratios between matrix elements with different Dirac structures $\Gamma $, allowing one to access spin physics, or ratios between matrix elements with different external momenta, allowing one to access nonperturbative TMD evolution.
Examples of suitable spin physics observables are given in Sec.~\ref{sec:TMDratios}.
There, the Lorentz-invariant calculational scheme is laid out in further detail, and an overview is given of the numerical results obtained in the computational program based on this approach. The systematic dependence of lattice TMD observables with respect to various parameters, e.g., the staple length $\eta $ and the evolution parameter $\hat{\zeta } $
is exhibited using selected twist-2 TMD observables. Initial results pertaining to twist-3 TMDs are presented in Sec.~\ref{sec:latt_tmd_e}, and results for GTMDs, yielding, in particular, quark orbital angular momentum in the proton, are discussed in Sec.~\ref{sec:GTMD_OAM_lattice}.

\subsubsection{Large-momentum effective theory}
\label{sec:latt_def_qtmd}

\begin{figure*}
 \centering
\subfloat[\label{fig:quasibeam}]{
\includegraphics[width=6.25cm]{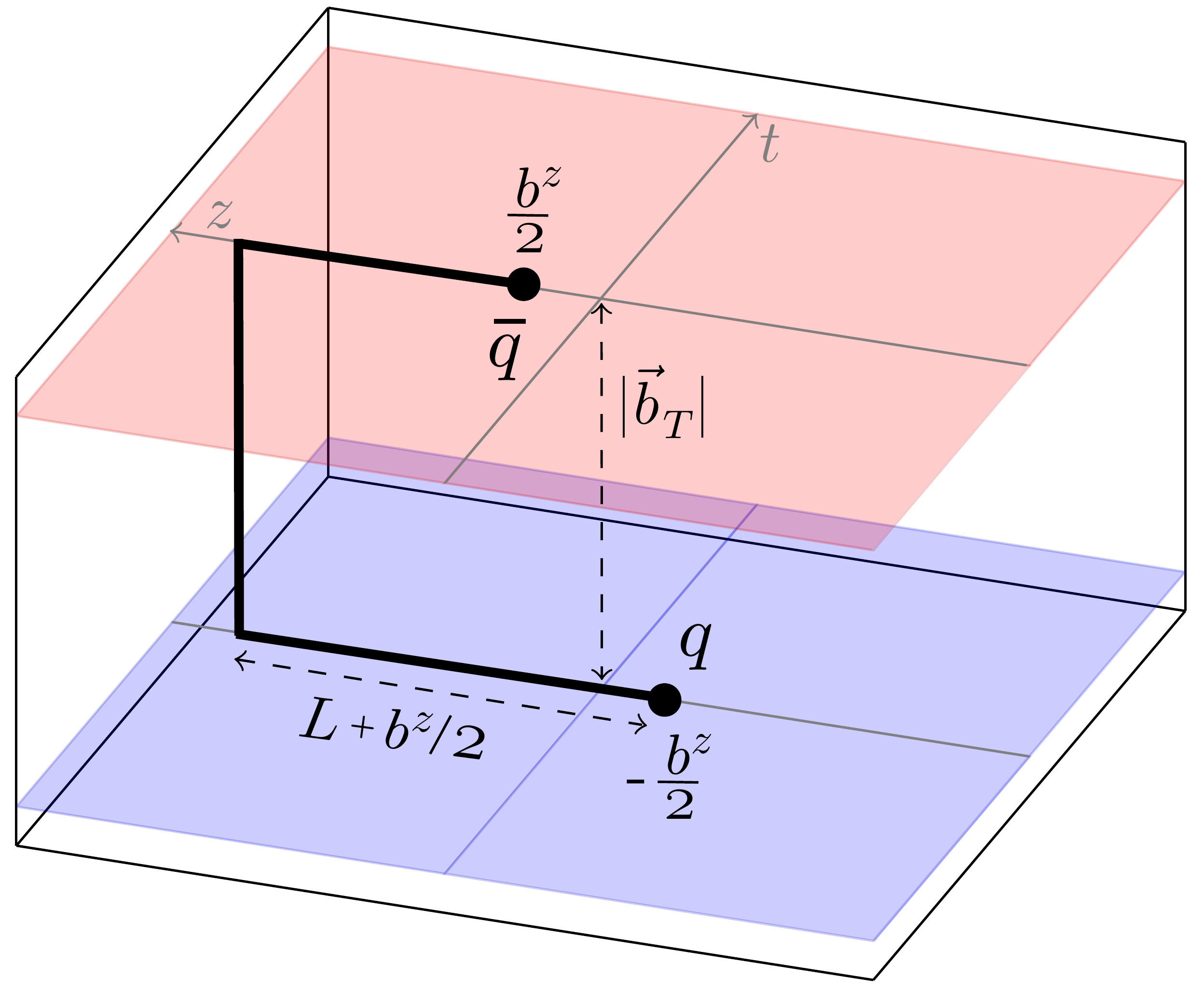}
}
 \hspace{3cm}
\subfloat[\label{fig:boost}]{\phantom{x}\vspace{0.3cm}
\includegraphics[width=6.25cm]{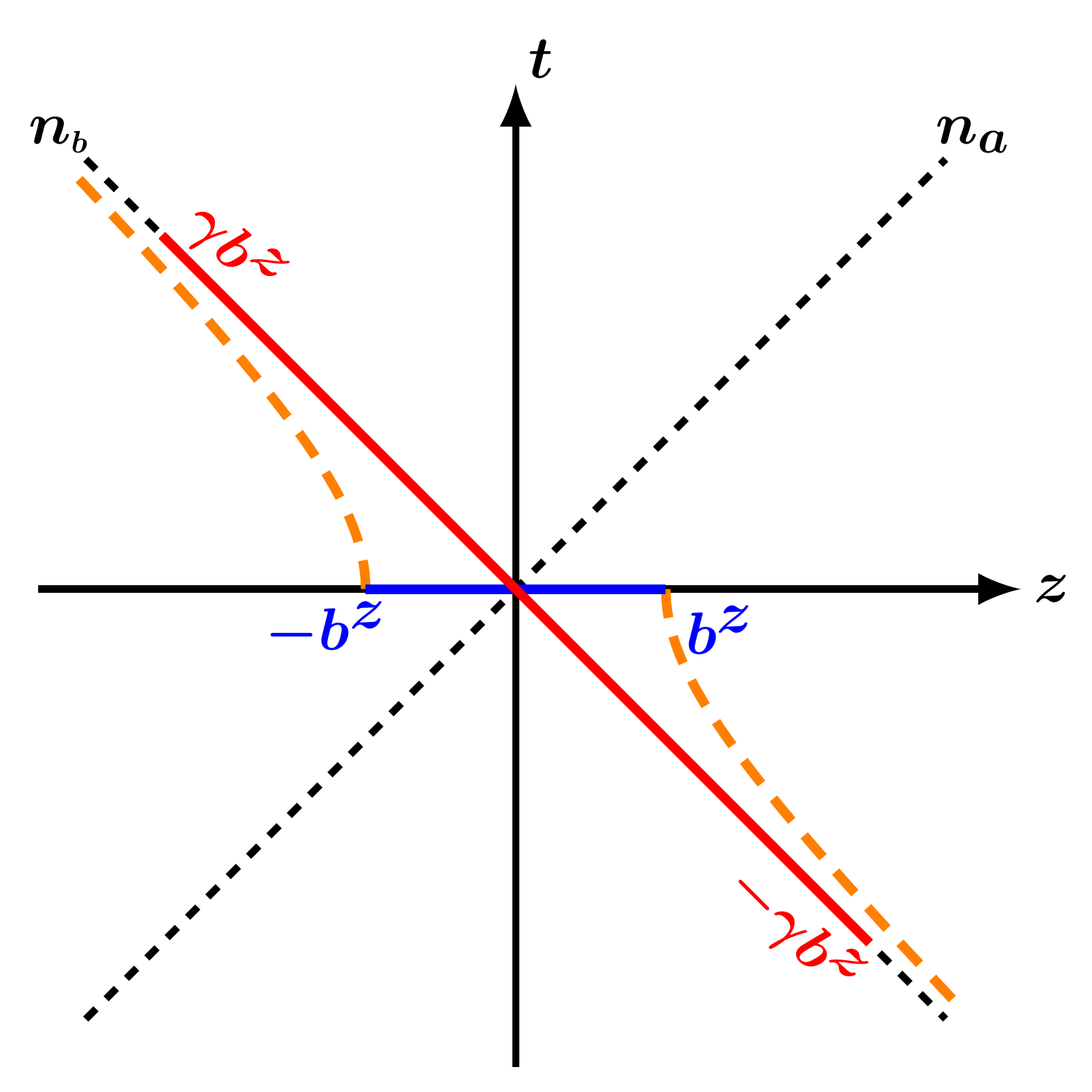}
}
 \caption{(a) Staple-shaped path for the gauge connection used in the LaMET calculation of TMD PDFs. (b) The behavior of the longitudinal separation under a Lorentz boost along the $z$ direction. Here $\gamma b^z$ is the length of the projection of the boosted Wilson line along the $n_b$ light-cone direction.}
\end{figure*}

\index{quasi TMD}
The idea of \index{large-momentum effective theory (LaMET)}large-momentum effective theory (LaMET) is to approximate light-cone correlations for parton physics by the equal-time correlations in a boosted hadron state. In the TMD case, one starts from the matrix element in (\ref{eq:latt_corr_def}) where both the hadron momentum $P^\mu$ and Wilson line direction $v^\mu$ are along the $z$ direction~\cite{Ji:2014hxa,Ji:2018hvs,Ebert:2019okf}, as shown in \fig{quasibeam}, and studies
\begin{align} \label{eq:qbeam_def1}
	\hat B_i(b^z,\bt,a,P^z,\eta) &= \frac{1}{2} \Bigl\langle p(P,S) \Big| 
    \bar\psi^{0}_i (b^{\mu }/2) \Gamma W_{{\sqsupset}\eta}^{\hat{z}} 
    (b^{\mu }/2,-b^{\mu }/2) \psi^{0}_i (-b^{\mu }/2) 
     \Big| p(P,S) \Bigr\rangle \,.
\end{align}
The Wilson line $W_{{\sqsupset}\eta}^{\hat{z}} (b^{\mu }/2,-b^{\mu }/2)$ has finite length, and 
closes in the transverse direction at the end of the staple, along $b_T^\mu$, which is different from that in the Lorentz-invariant approach, as the latter requires the Wilson line be parallel to the full $b^\mu$. (The use of $b^\mu$ provides more symmetries that are used to reduce the number of independent amplitudes.) Here $\hat B_i(b^z,\bt,a,P^z,\eta)$ is referred to as the unsubtracted quasi TMD~\cite{Ji:2014hxa,Ji:2018hvs} or quasi beam function~\cite{Ebert:2019okf}. The renormalized quasi TMD in $x$ space is defined with the inclusion of a quasi soft function $\hat{S}^i(b_T, a, \eta)$,
 \begin{align} \label{eq:qtmd0}
 \hat f_i(x, \bt,\mu,P^z) =  \int \frac{\df b^z}{2\pi} \, e^{\img b^z (x P^z)}\,
 &\hat {Z}'_i(b^z,\mu,\tilde \mu) \hat {Z}_{\rm uv}^i(b^z,\tilde \mu, a)
 \nn\\&\times
 \hat B_i(b^z,\bt,a,P^z,\eta) \Big/\sqrt{\hat{S}^i(b_T, a, \eta)}
\,,\end{align}
where $\hat{S}^i$ cancels the large $\eta$ dependence in the quasi beam function.
$\hat{Z}_{\rm uv}$ is the lattice renormalization factor, and $\hat{Z}'$ converts from the lattice renormalization scheme with scheme parameter $\tilde\mu$, to the $\MSbar$ scheme with $\MSbar$ renormalization scale $\mu$.

Instead of using Lorentz covariance, in LaMET one boosts the equal-time correlator to the light-cone direction by calculating in a large-momentum hadron state. A picture of the Lorentz boost is shown in \fig{boost}. When $P^z\gg m_p$, one can perform a large-momentum expansion of the lattice construction to extract the parton physics, where the leading power contribution includes short distance matching and scale running~\cite{Ji:2020ect}.
In Fourier space, the distribution defined by boosting the matrix element of the equal-time correlator is also called the unsubtracted quasi TMD~\cite{Ji:2014hxa} or the quasi beam function~\cite{Ebert:2019okf}. After inclusion of the quasi soft function and renormalization~\cite{Ebert:2019tvc,Shanahan:2019zcq}, one can take the $\eta\to\infty$ limit and obtain the quasi TMD.

The corresponding quasi soft function has the same issue as encountered in the Lorentz-invariant approach, namely, no single Lorentz boost can transform it into the soft function used in TMD factorization. Therefore, the difference between the quasi and physical TMD PDFs includes a perturbative matching coefficient and a nonperturbative contribution from the mismatch of the soft functions~\cite{Ebert:2019okf}. Nevertheless, one can still form ratios of the quasi TMDs in different hadron states or for different spin structures to obtain information on TMD evolution or ratios of TMD PDFs in $x$-space~\cite{Ebert:2019okf,Ebert:2018gzl,Ebert:2020gxr}. Recently, methods have been proposed to calculate this nonperturbative contribution~\cite{Ji:2019sxk}, which is called the reduced soft function, from lattice QCD. With this development,
a full determination of the TMD PDF as well as the Drell-Yan cross section appears within reach in lattice
QCD~\cite{Ji:2019ewn}.

A more detailed discussion of developments in the LaMET approach is given in Sec.~\ref{sec:latt_soft_func}. 

\begin{figure*}
 \centering
 \includegraphics[width=0.8\textwidth]{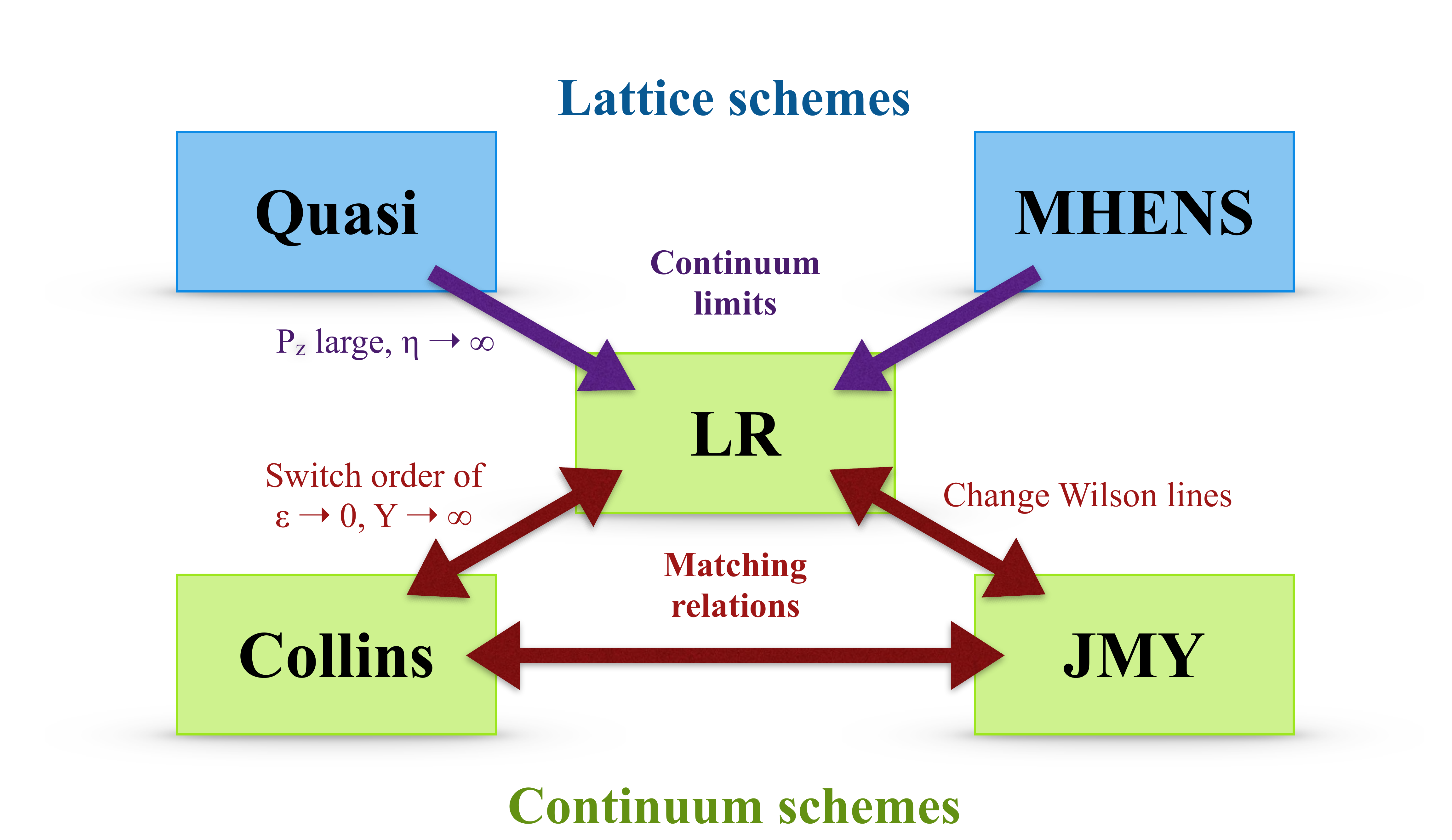}
 \caption{An overview of lattice and continuum TMD schemes and their relationships. Figure from Ref.~\cite{Ebert:2022fmh}.}
 \label{fig:schemes}
\end{figure*} 

\subsubsection{Relations between lattice and continuum TMDs}
\label{sec:latt_schemes}

Since both the Lorentz-invariant approach, or the Musch-H\"agler-Engelhardt-Negele-Sch\"afer (MHENS) scheme, and the quasi TMD use off-the-light-cone Wilson lines, they are closely related to the Collins scheme. In fact, using Lorentz invariance, one can show that both the MHENS~\cite{Musch:2011er} and quasi~\cite{Ebert:2022fmh} beam functions with infinite Wilson lines are equivalent to that in the Collins scheme.
Therefore, the Collins soft function can be used to subtract the rapidity divergences in the MHENS and quasi beam functions to define the relevant TMDs. In lattice QCD calculations, the light-cone limit in the MHENS and quasi TMDs are achieved by boosting the hadron momentum $P^z$ to infinity. Since the lattice theory has a natural UV momentum cutoff, the $P^z\to\infty$ limit has to be taken after the UV regularization, which, however, is opposite to the Collins scheme where the UV regularization is done before the light-cone limit (see \eq{CSf}). Therefore, the lattice TMDs correspond to a new scheme, which is called the large-rapidity (LR) scheme~\cite{Ebert:2022fmh}, and differs from the Collins scheme by the order of $\epsilon\to0$ and $y_B\to-\infty$ limits. 

Due to the asymptotic freedom of QCD, the large rapidity or momentum limit only affects the UV region, so changing the order of $\epsilon\to0$ and $y_B\to-\infty$ limits leave infrared physics intact.
Using the LaMET formalism~\cite{Ji:2013dva,Ji:2014gla,Ji:2020ect}, behind which is the general principle for effective field theories, one can relate the two different orders of limits with a factorization formula or perturbative matching~\cite{Ebert:2022fmh}, which has been proposed in Refs.~\cite{Ebert:2018gzl,Ebert:2019okf,Ji:2019sxk,Ji:2019ewn}. Moreover, the matching is diagonal in the parton flavor space and independent of the spin structure, and there is no mixing between the gluon and quark channels~\cite{Ebert:2022fmh}, which makes their individual lattice calculations easier.

Besides, the JMY scheme is related to the LR scheme by replacing the spacelike Wilson lines with timelike ones, so one can derive the matching coefficients for the JMY and LR schemes to the Collins scheme from one another through such an analytical continuation~\cite{Ebert:2022fmh}. The relations of both lattice and continuum off-the-light-cone schemes are shown in \fig{schemes}.

\subsection[Complete TMD Factorization for DY, SIDIS, and \texorpdfstring{$e^+e^-$}{e+ e-}]
           {Complete TMD Factorization for DY, SIDIS, and \boldmath $e^+e^-$}
\label{sec:TMDfactSIDISee}

In this section, we extend our previous discussion of TMD factorization for the unpolarized Drell-Yan process
to polarized Drell-Yan  in \sec{TMDDrellYan}, to Higgs production at hadron colliders in \sec{TMDHiggs}, to
polarized Semi-Inclusive Deep-Inelastic Scattering (SIDIS)  in \sec{TMDSIDIS},
and to dihadron production at $e^+ e^-$ colliders  in \sec{TMDee}.

\subsubsection{Polarized Drell-Yan cross section}
\label{sec:TMDDrellYan}

\index{Drell-Yan}
We now consider the polarized Drell-Yan process,
\begin{align}
 H_1(P_1, S_1) + H_2(P_2, S_2) \to \gamma^*/Z \to \ell^+ \ell^-
\,,\end{align}
where the hadrons $H_{1,2}$ have momenta $P_{1,2}$ and spin $S_{1,2}$. 
By measuring the angular distributions of the detected lepton pair, one can study the polarization of the struck quarks, which in turn allows one to study correlations of the quark and hadron polarizations.
After some preliminaries, below we will specialize to the special case of Drell-Yan for a pion-proton collision, $H_1=\pi$ and $H_2=p$.

\begin{figure*}
 \centering
 \includegraphics[height=5cm]{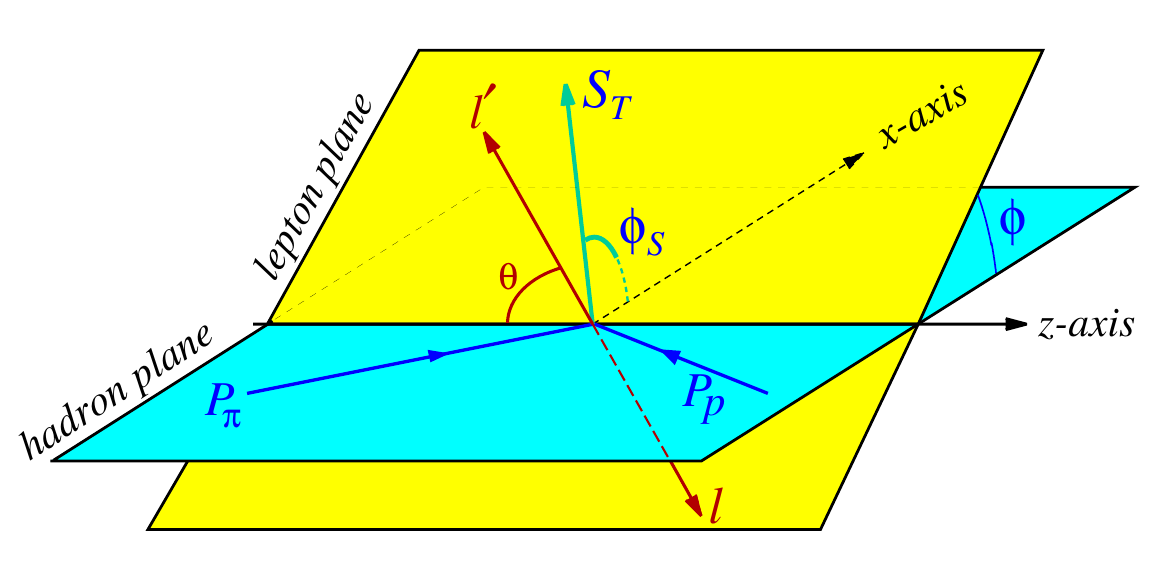}
 \caption{The DY process in the Collins-Soper frame where the pion and the proton come in with different momenta $P_\pi$, $P_p$, but each carries the same transverse momentum $\frac12\,\qT$, and the produced lepton pair is at rest. The angle $\phi$ describes the inclination of the leptonic frame with respect to the hadronic plane, and $\phi_S$ is the azimuthal angle of the transverse-spin vector of the proton. 
 Here $\theta$ is the angle between the final state lepton and $z$-axis, where the $z$-axis is defined by the intersection of the hadron and lepton planes.
 The figure is from Ref.~\cite{Bastami:2020asv}.}
 \label{fig:CS_frame}
\end{figure*}

One needs to define a reference frame in which to measure the leptonic angles, which is commonly achieved in the Collins-Soper frame~\cite{Collins:1977iv}.
It can be obtained from the lab frame, where the incoming pion is aligned along the $z$ axis, by first boosting along $z$ such that the virtual boson has vanishing longitudinal momentum. Subsequently, one performs a transverse boost such that the virtual boson has vanishing transverse momentum, i.e.~is produced at rest.
In such a rest frame, the two leptons are produced back-to-back with transverse momenta $\pm \qt/2$.
In the Collins-Soper frame, the lepton momenta span the $(x,z)$-plane, and one defines the angle $\phi$ as the inclination of the hadron to the lepton plane.
Likewise, one defines the angle $\phi_S$ as the azimuthal angle of the proton spin vector with respect to the lepton plane.
These kinematics are illustrated in \fig{CS_frame} for pion-proton scattering.

Following~\cite{Arnold:2008kf}, we write the Drell-Yan cross section in the one-photon approximation as
\begin{align}
 \sigma = \int\frac{\df^3 \vec l}{l^0}\frac{\df^3 \vec l'}{{l'}^0} \frac{\alpha_{\rm em}^2}{\mathscr{F}\, Q^4} L_{\mu\nu} W^{\mu\nu}
\,,\end{align}
where $l$ and $l'$ are the lepton momenta,
\begin{align} 
  \mathscr{F} = 4 \sqrt{ (P_1 \cdot P_2)^2 - M_1^2 M_2^2}\,,
\end{align}
is the flux factor of the incoming hadrons, and $Q^2 = q^2$ with $q = l + l'$ is the photon virtuality.
The kinematics of the lepton pair are described by the spin-averaged leptonic tensor
\begin{align}
 L^{\mu\nu}
 = \sum_{\lambda,\lambda'} \bigl[\bar u(l,\lambda) \gamma^\mu v(l', \lambda')\bigr] \bigl[v(l',\lambda') \gamma^\nu u(l, \lambda)\bigr]
 = 4 \Bigl( l^\mu {l'}^\nu + l^\nu {l'}^\mu - \frac{Q^2}{2} g^{\mu\nu} \Bigr)
\,.\end{align}
The hadronic physics is encoded in the hadronic tensor
\begin{align}
 W^{\mu\nu} = \int\frac{\df^4 x}{(2\pi)^4} e^{\img q \cdot x} \langle P_1, S_1; P_2, S_2 | J_{\rm em}^\mu (0) J_{\rm em}^\nu (0) | P_1, S_1; P_2, S_2 \rangle
\,,\end{align}
where $J_{\rm em}^\mu$ is the electromagnetic current.
By decomposing the Lorentz tensors $L^{\mu\nu}$ and $W^{\mu\nu}$ into all independent angular and spin structures,
one can derive the most general decomposition of the Drell-Yan cross section.
In the most general case with two arbitrarily polarized hadrons, there are a total of 48 independent structures~\cite{Arnold:2008kf}, out of which 24 are suppressed at small $q_T$.

For brevity of our discussion, we only focus on the case of pions
scattering off polarized protons, $\pi p \to \gamma^* \to \ell^+ \ell^-$, as measured by the COMPASS Collaboration~\cite{Aghasyan:2017jop}, and refer to \cite{Arnold:2008kf} for the fully generic result.
We also neglect contributions from $Z$ exchange, which are suppressed at low energies.
At small $q_T$, this process is described by only six independent structures, and can be written as~\cite{Arnold:2008kf}
\begin{align} \label{eq:sigma_polarized_DY}
 \frac{\df\sigma}{\df^4q \df\Omega}
 = \frac{\alpha_{\rm em}^2}{\mathscr{F} \, Q^2} \Bigl\{ &
    \Bigl[ (1 + \cos^2\theta) F_{UU}^1 + \sin^2\theta \cos(2\phi) F_{UU}^{\cos2\phi} \Bigr]
    \nn\\&
    + S_L \sin^2\theta \sin(2\phi) F_{UL}^{\sin 2\phi}
    \nn\\&
    + S_T (1-\cos^2\theta) \sin\phi_S F_{UT}^{\sin\phi_S}
    \nn\\&
    + S_T \sin^2\theta \Bigl[ \sin(2\phi+\phi_S) F_{UT}^{\sin(2\phi+\phi_S)} + \sin(2\phi-\phi_S) F_{UT}^{\sin(2\phi-\phi_S)} \Bigr]
   \Bigr\}
\,,\end{align}
where $\Omega$ is the solid angle of the dilepton system in the Collins-Soper frame, with the angles $\phi$, $\theta$ and $\phi_S$ defined accordingly, see \fig{CS_frame}.
The first subscript on the structure functions $F$ indicates that the pion is unpolarized ($U$),
while the second subscript corresponds to the proton polarization, which can be unpolarized ($U$), longitudinally ($L$) or transversely ($T$) polarized.
It is also common to measure the individual structure functions normalized to the unpolarized case, i.e.,
\begin{align} \label{eq:Asym.def}
 A_{XY}^{\rm weight}(x_\pi, x_p , q_T, Q^2) = \frac{F_{XY}^{\rm weight}(x_\pi, x_p , q_T, Q^2)}{F_{UU}^1(x_\pi, x_p , q_T, Q^2)}
\,.\end{align}
As made explicit here, all structure functions only depend on the longitudinal momentum fractions $x_\pi$ and $x_p$,
as well as the transverse momentum $q_T^2$ and photon virtuality $Q^2$.

\index{TMD factorization!Drell-Yan}
The structure functions in \eq{sigma_polarized_DY} can be expressed in terms of the spin-dependent TMDs introduced in \sec{qgspinTMDFF} as follows~\cite{Arnold:2008kf}:
\begin{align} \label{eq:SFs}
 F_{UU}^1 &= \phantom{-} {\cal C}  \bigl[f_{1,\pi} \; f_{1,p} \bigr]
\,,\\\nn
 F_{UU}^{\cos 2\phi} &= \phantom{-} {\cal C} \bigg[ \frac{2(\hhat \cdot \vkTpi)(\hhat \cdot \vkTN) - \vkTpi \cdot \vkTN}{M_\pi \; M_p} \; h_{1,\pi}^{\perp} \; h_{1,p}^{\perp} \bigg]
 \,,\\\nn
 F_{UL}^{\sin 2\phi} &= -  {\cal C} \bigg[ \frac{2(\hhat \cdot \vkTpi)(\hhat \cdot \vkTN) - \vkTpi \cdot \vkTN}{M_\pi \; M_p} \; h_{1,\pi}^{\perp}\; h_{1L, p}^{\perp} \bigg]
\,,\\\nn
 F_{UT}^{\sin \phi_S} &= \phantom{-} {\cal C} \bigg[ \frac{\hhat \cdot \vkTN}{M_p} \; f_{1,\pi} \; f_{1T, p}^{\perp} \bigg]
\,,\\\nn
 F_{UT}^{\sin (2\phi - \phi_S)} &= -  {\cal C}  \bigg[ \frac{\hhat \cdot \vkTpi}{M_\pi} \; h_{1,\pi}^{\perp} \;h_{1, p} \bigg]
\,,\\\nn
 F_{UT}^{\sin (2\phi + \phi_S)} &= -  {\cal C}  \bigg[ \frac{2(\hhat \cdot \vkTN) \big[2(\hhat \cdot \vkTpi)(\hhat \cdot \vkTN)
       - \vkTpi \cdot \vkTN \big] - \vkTN^2(\hhat \cdot \vkTpi)}{2 \; M_\pi \; M_p^2} \, h_{1, \pi}^{\perp}\, h_{1T, p}^{\perp} \bigg]
\,.\end{align}
Here, $\hhat=\qT/q_T$ points along the $x$-axis in the CS frame, $M_{\pi,p}$ are the pion and proton masses, and the convolution integrals are defined as
\begin{align} \label{eq:convolution-integral}
 {\cal C}\bigl[\omega \, f_{\pi} \, f_p \bigr] &
 =  \sum_i H_{i\bar i}(Q^2,\mu) \int d^2\kTpi \, d^2\kTN \, \delta^{(2)}(\qT- \kTpi-\kTN)
\\\nn&\qquad \times
    \omega(\qT,\kTpi,\kTN,\ldots) \; f_{i/p}(x_a,p_{TN},\mu,\zeta_a)\, f_{\bar i/\pi}(x_b,p_{T\pi},\mu,\zeta_b)
\,,\end{align}
where the sum runs over all quark flavors $i = u, \bar u, d, \bar d, \dots$, 
the hard function $H_{i\bar i}$ encodes physics at the hard scale $Q$, and $\omega$ is a kinematic weight function given by the prefactors in \eq{SFs}. For a virtual photon up to two-loop order, its only flavor dependence is given by the quark charges $e_i$ which are each proportional to the electromagnetic coupling $|e|$, 
so 
\begin{align}
H_{i\bar i}(Q^2,\mu) = e_i^2\, H(Q^2,\mu) + {\cal O}(\alpha_s^3)
 \,.
\end{align}
The convolution variables $\kTpi$ and $\kTN$ correspond to the momenta of struck quarks in the pion and proton, respectively.
We have also restored all arguments of the TMD functions, where as usual the Collins-Soper scales obey $\zeta_\pi \zeta_N = Q^4$.

Similar to our treatment of unpolarized Drell-Yan in \sec{TMDforDY}, one can also express \eq{SFs} more compactly in $b_T$ space as~\cite{Boer:2011xd, Bastami:2020asv}
\begin{align} \label{eq:SFs_bt}
 {F}_{UU}^1 &= \phantom{-}   {\cal B} \bigl[\tilde f_{1, \pi}^{(0)}\; \tilde f_{1,p}^{(0)} \bigr]
\,,\\\nn
 {F}_{UU}^{\cos 2\phi} &= \phantom{-}   M_\pi M_p \; {\cal B} \bigl[\tilde h_{1,\pi}^{\perp(1)}\; \tilde h_{1,p}^{\perp(1)} \bigr]
\,,\\\nn
 {F}_{UL}^{\sin 2\phi} &= - M_\pi  M_p \; {\cal B} \bigl[\tilde h_{1,\pi}^{\perp(1)}\; \tilde  h_{1L,p}^{\perp(1)} \bigr]
\,,\\\nn
 {F}_{UT}^{\sin \phi_S} &= \phantom{-} M_p \;{\cal B} \bigl[\tilde f_{1,\pi}^{(0)}\; \tilde  f_{1T,p}^{\perp(1)} \bigr]
\,,\\\nn
 {F}_{UT}^{\sin (2\phi -\phi_S)} &= -M_\pi \;{\cal B} \bigl[\tilde h_{1,\pi}^{\perp(1)}\; \tilde  h_{1,p}^{(0)} \bigr]
\,,\\\nn
 {F}_{UT}^{\sin (2\phi + \phi_S)} &= -\frac{M_\pi M_p^2}{4} \; {\cal B} \bigl[\tilde h_{1,\pi}^{\perp(1)}\; \tilde  h_{1T,p}^{\perp(2)} \bigr]
\,.\end{align}
Here, the analog of the convolution integral in \eq{convolution-integral} is the Bessel transform
\begin{align} \label{eq:convBess}
     {\cal B}[\tilde f_{\pi}^{(m)}\; \tilde f_{p}^{(n)}] & \equiv 
     \sum_{i} H_{i\bar i}(Q^2,\mu)\! \int_0^\infty\!
     \frac{d b_T}{2\pi} \, b_T \, b_T^{m+n} \, J_{m+n}(q_T b_T)
     \, \tilde f_{i/p}^{(m)}(x_a,{b_{T}},\mu,\zeta_a)
     \, \tilde f_{\bar i/\pi}^{(n)}(x_b,{b_{T}},\mu, \zeta_b)
,\end{align}
where $J_{m+n}(x)$ is the Bessel function of the first kind of order $m+n$,
and the $\tilde f^{(n)}$ are derivatives of the Fourier transform of $f$, as defined in \eq{TMD_bt_derivative}.

\subsubsection{Higgs production in gluon fusion}
\label{sec:TMDHiggs}

So far, we have only discussed the (polarized) Drell-Yan process, which is a key benchmark for both low-energy experiments such as JLab or COMPASS as well as for high-energy colliders such as Tevatron or the LHC.
A key property of the Drell-Yan process is that it is initiated by quark annihilation, but does not directly probe the gluonic structure of the proton.
Gluon-induced scatterings become much more important at high-energy colliders, in particular for the production of Higgs bosons.

\begin{figure*}
 \centering
 \includegraphics[width=5cm]{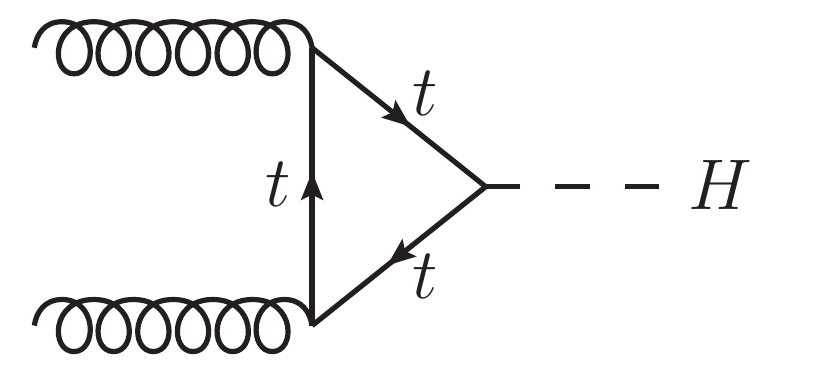}
 \caption{Higgs-boson production in gluon fusion mediated by a top-quark loop. Contributions from lighter quarks are suppressed by their quark masses.}
 \label{fig:ggH}
\end{figure*}

While the Higgs boson does not directly couple to gluons, it can be produced in gluon fusion through a closed quark loop, as depicted in \fig{ggH}.
Since the Yukawa coupling of the Higgs boson to a quark is proportional to the quark mass, this process is dominated by a virtual top loop,
while contributions from lighter quarks such as the $b$ quark are suppressed.

\index{TMD factorization!Higgs from gluon fusion}
The transverse momentum distribution of the Higgs boson is a key observable for probing its production mechanism,
and thus has been extensively measured at the LHC~\cite{Aad:2014lwa,Aad:2014tca,Aad:2016lvc,Aaboud:2017oem,Aaboud:2018xdt,ATLAS:2020wny,Khachatryan:2015rxa,Khachatryan:2015yvw,Khachatryan:2016vnn,Sirunyan:2018kta,Sirunyan:2018sgc}.
At small transverse momentum $q_T \ll m_H$ of the Higgs boson, its theoretical description requires the use of TMD factorization.
The analog of \eq{sigma_new} for a gluon-induced process in the collision of unpolarized protons reads
\begin{subequations} \label{eq:sigma_gg}
\begin{align} \label{eq:sigma_gg_a}
 \frac{\df\sigma^{\rm W}}{\df Q \df Y \df^2\qt} &
 = 2 H_{\rho\sigma\rho'\sigma'}(Q,\mu) \int\! \df^2\bt \, e^{i \bt \cdot \qt} \,
   \tilde f_{g/p}^{\rho\sigma}(x_a, \bt, \mu, \zeta_a) \tilde f_{g/p}^{\rho'\sigma'}(x_b, \bt, \mu, \zeta_b)
\\ \label{eq:sigma_gg_b}
 &= 2 H_{\rho\sigma\rho'\sigma'}(Q,\mu) \int\! \df^2\bt \, e^{i \bt \cdot \qt} \,
   \tilde B_{g/p}^{\rho\sigma}(x_a, \bt, \mu, \zeta_a/\nu^2) \tilde B_{g/p}^{\rho'\sigma'}(x_b, \bt, \mu, \zeta_b/\nu^2)
   \nn\\&\hspace{5.5cm}\times
   \tilde {\cS}_{n_a n_b}(b_T,\mu,\nu)
\,.\end{align}
\end{subequations}
As in \eq{sigma_new}, we present this result both using the approach of renormalized TMD PDFs $f_{g/p}^{\rho\sigma}$ and by separately considering renormalized beam functions $B_{g/p}^{\rho\sigma}$ and the soft function $\cS_{n_a n_b}$.
Here the soft function has Wilson lines in the adjoint representation, and hence differs from the soft function $S_{n_a n_b}$ in quark-initiated Drell-Yan. The definition of the corresponding bare gluon soft function $\cS_{n_a n_b}^0(b_T,\eps,\tau)$ is given above in \eq{Soft_Fn_adj}. 
As before, these functions depend on the transverse separation $\bt$, which is Fourier-conjugate to $\qt$,
the longitudinal momentum fractions $x_{a,b}$, the renormalization scale $\mu$ and the Collins-Soper scale $\zeta_{a,b}$.
The latter obeys $\zeta_a \zeta_b = Q^4$, where for on-shell Higgs production $Q^2 = m_H^2$.
The beam and soft functions also depend on the rapidity renormalization scale $\nu$, which cancels exactly between them.

The key difference between \eq{sigma_new}, the TMD factorization for unpolarized Drell-Yan, and \eq{sigma_gg} is the Lorentz structure of the TMD PDFs (or beam functions).
This arises because even in an unpolarized proton, the spin-$1$ nature of the gluon induces a non-trivial Lorentz structure, as was pointed out~in \cite{Mantry:2009qz}, see also~\cite{Catani:2010pd}.
From \eq{tmd_decomposition_g_1}, the gluon TMD PDF in an unpolarized hadron reads
\begin{align} \label{eq:tmd_g_unpol}
 \tilde f_{g/p}^{\rho\sigma}(x,\bt,\mu,\zeta) &
 = -\frac{g_T^{\rho\sigma}}{2} \tilde f_1^g(x, b_T,\mu,\zeta)
   + \Bigl( \frac{g_T^{\rho\sigma}}{2} + \frac{b_T^\rho b_T^\sigma}{\bt^2} \Bigr) \tilde h_1^{\perp g}(x, b_T,\mu,\zeta)
\,,\end{align}
where $g_T^{\rho\sigma}$ is the transverse metric and $b_T$ on the right-hand side is a Minkowskian four-vector.

\eq{sigma_gg} holds for generic gluon-induced processes, where the process-dependence is carried by the hard function $H_{\rho\sigma\rho'\sigma'}$.
For the case of Higgs production discussed here, the spin-$0$ nature of the Higgs boson forbids any non-trivial spin correlations, such that the hard function simplifies to
\begin{align}
 H^{\rho\sigma\rho'\sigma'}(Q,\mu) = H_{ggH}(Q,\mu) g_T^{\rho\sigma} g_T^{\rho'\sigma'}
\,,\end{align}
such that the cross section only depends on the combination
\begin{align}
 \tilde f_{g/p}^{\rho\sigma}(x_a,\bt) \tilde f_{g/p\,\rho\sigma}(x_b,\bt) &
 = \frac12 \Bigl[ \tilde f_1^g(x_a,\bt) \tilde f_1^g(x_b,\bt) + \tilde h_1^{\perp g}(x_a,\bt) \tilde h_1^{\perp g}(x_b,\bt) \Bigr]
\,,\end{align}
where we suppressed the scales for brevity.
Inserting this into \eq{sigma_gg_a}, one obtains the simple result
\begin{align}
 \frac{\df\sigma^{\rm W}}{\df Q \df Y \df^2\qt}
 = H_{ggH}(Q,\mu) \int\! \df^2\bt \, e^{i \bt \cdot \qt} \,
   \bigl[& \tilde f_1^g(x_a, \bt, \mu, \zeta_a) \tilde f_1^g(x_b, \bt, \mu, \zeta_b)
   \nn\\&
        + \tilde h_1^{\perp g}(x_a, \bt, \mu, \zeta_a) \tilde h_1^{\perp g}(x_b, \bt, \mu, \zeta_b) \Bigr]
\,,\end{align}
and similarly for the form in \eq{sigma_gg_b}.

Finally, we remark that Higgs production at the LHC is dominated by perturbative $\LQCD \ll q_T \ll m_H$,
in which case one can relate the gluon TMD PDFs to collinear PDFs as discussed in \sec{largeqT},
supplemented by resummation of large logarithms as outlined in \chap{evolution}.

\subsubsection{Polarized SIDIS cross section}
\label{sec:TMDSIDIS}

\index{SIDIS}
We now consider Semi-Inclusive Deep-Inelastic Scattering (SIDIS),
\begin{align} \label{eq:SIDIS}
 \ell(l) + p(P) \to \ell(l') + h(P_h) + X
\,,\end{align}
where the incoming lepton (an electron, positron or muon) with momentum $l$ scatters off a proton with momentum $P$, both of which can be polarized.
One measures both the outgoing lepton with momentum $l'$ and a hadron of type $h$ (such as a pion or kaon) and momentum $P_h$, but is inclusive over any additional hadronic radiation $X$.

\begin{figure*}
 \centering
 \includegraphics[width=8cm]{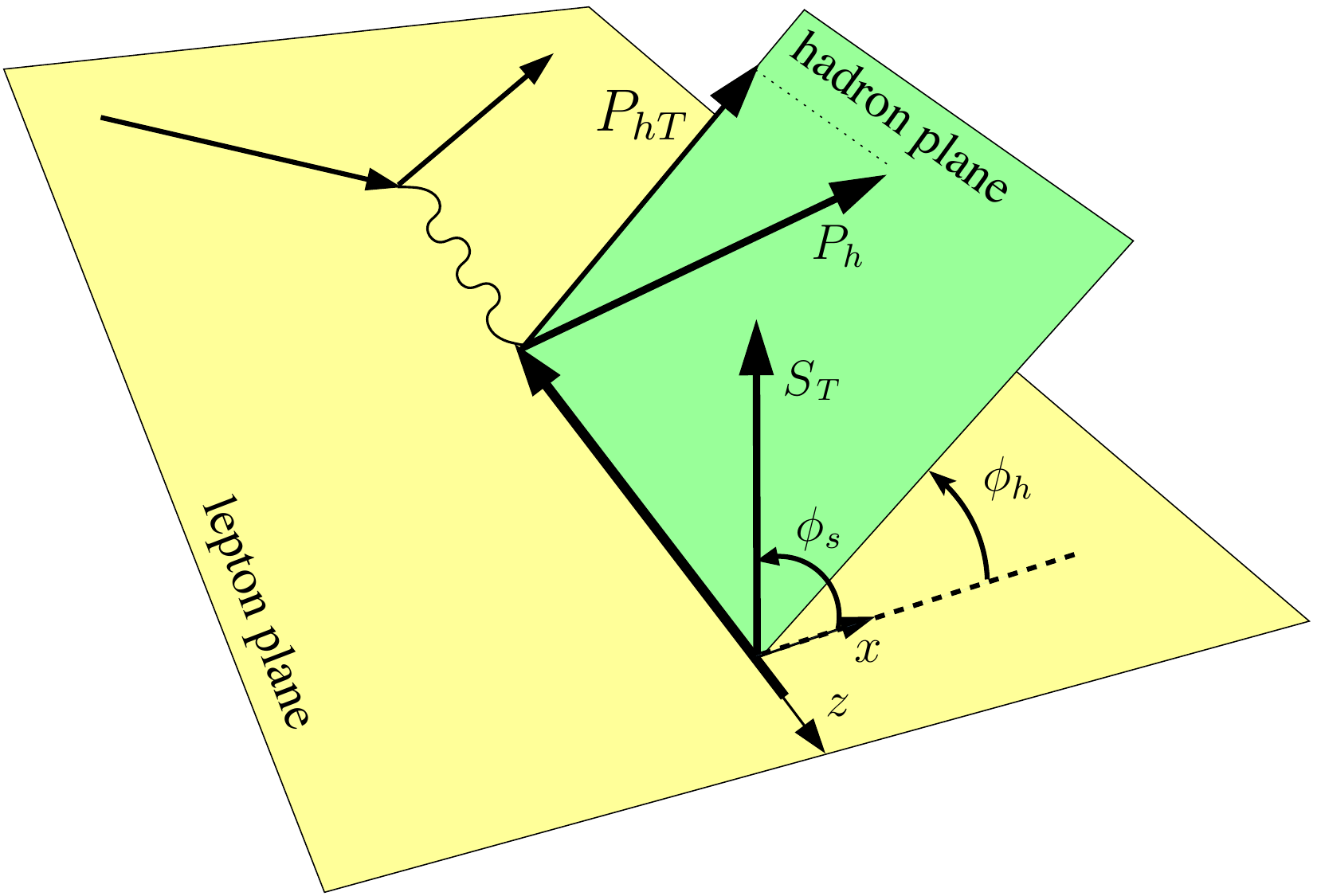}
 \caption{Semi Inclusive Deep Inelastic Scattering process (SIDIS) in $\gamma^* p$
 center of mass frame. The plot is from Ref.~\cite{Kang:2015msa}, adapted to the notation used here.}
 \label{fig:sidis}
\end{figure*}

As in the case of polarized Drell-Yan discussed in \sec{TMDDrellYan}, we are interested in measuring angular correlations in order to extract correlations between the polarization of the struck quark and the spin of the proton.
This requires defining a reference frame in which to specify angular measurements, which is commonly chosen according to the Trento conventions~\cite{Bacchetta:2004jz}.
In this frame, the spacelike momentum $q$ defines the $z$ axis, which together with the lepton momenta defines the $(x,z)$-plane, with respect to which all angles are defined. This is illustrated in \fig{sidis}.

We are interested in measuring the momentum component $\Phperp$ and azimuthal angle $\phi_h$ of the detected hadron in this frame.
In addition, there is an azimuthal angle $\psi_l$  characterizing the overall orientation of the lepton scattering plane around the incoming lepton direction.
The angle is calculated with respect to an arbitrary reference axis, which in the case of transversely polarized targets is chosen to be the direction of the polarization vector $S_T$.
In the DIS limit $\psi_l \approx \phi_S$, where the latter is the azimuthal angle of the spin-vector of the struck hadron.
These observables are also illustrated in \fig{sidis}.

In the limit that $Q \ll m_{W,Z}$, the SIDIS process can be described in the single-photon exchange approximation, and is characterized by 18 independent structure functions~\cite{Bacchetta:2006tn}.
At leading order in a $1/Q$ expansion, only a subset of 8 structure functions contributes, and the SIDIS cross section can be written as~\cite{Bacchetta:2006tn, Bastami:2018xqd}
\begin{align} \label{eq:SIDIS-leading}
 \frac{\df^6\sigma}{\df\xbj \, \df y \, \df \zh \, \df \phi_S \, \df\phi_h \, \df\Phperp^2} &
 = \frac{\alpha_{\rm em}^2}{\xbj\,y\,Q^2} \biggl(1-y+\frac12y^2\biggr) \, \biggl[
   F_{UU,T} +  \cos(2\phi_h)\,   p_1\,F_{UU}^{\cos(2\phi_h)}
\nn\\&\quad
   + S_L \sin(2\phi_h)\,p_1\,F_{UL}^{\sin(2\phi_h)} + S_L \, \lambda p_2\,F_{LL}
\nn\\&\quad
   + S_T\sin( \phi_h-\phi_S)\, F_{UT,T}^{\sin( \phi_h-\phi_S)}  
\nn\\&\quad
   + S_T\sin( \phi_h+\phi_S)\,p_1\,F_{UT}^{\sin( \phi_h+\phi_S)}
   + \lambda\,S_T\cos(\phi_h-\phi_S)\,p_2\,F_{LT}^{\cos( \phi_h-\phi_S)}
\nn\\&\quad
   + S_T\sin(3\phi_h-\phi_S)\,p_1\,F_{UT}^{\sin(3\phi_h-\phi_S)} \biggr]
\,,\end{align}
Up to corrections suppressed as $1/Q^2$, the kinematic prefactors $p_i$ in \eq{SIDIS-leading} are given by~\cite{Bastami:2018xqd}
\begin{align} \label{eq:y-prefactors}
 p_1 &= \frac{1-y}{1-y+\frac12\,y^2}
\,,\quad
 p_2 = \frac{y(1-\frac12\,y)}{1-y+\frac12\,y^2}
\,,\quad
 p_3 = \frac{(2-y)\sqrt{1-y}}{1-y+\frac12\,y^2}
\,,\quad
 p_4 = \frac{y\sqrt{1-y}}{1-y+\frac12\,y^2}
\,.\end{align}
The factors $p_3$ and $p_4$ are listed here for completeness, but only appear in the subleading power cross sections, given for SIDIS in Eq.~(\ref{e:SIDIS_subleading}).

The structure functions $F_{XY}^{\rm weight}$ in \eq{SIDIS-leading} implicitly depend on $\xbj$, $\zh$, $\Phperp^2$ and $Q^2\simeq \xbj y S$.
Their superscripts indicate the azimuthal dependence, while the subscripts encode the beam  and target polarizations.
The first subscript $U$ ($L$) denotes the unpolarized beam (longitudinally polarized beam with twice helicity $\lambda$ ).
The second subscript $U(L\text{ or }T)$ refers to the target, which can be unpolarized (longitudinally ($S_L$) or transversely ($S_T$) polarized with respect to virtual photon).
$F_{UU,T}$ is the structure function due to transverse polarization of the virtual photon (indicated by the third sub-index $T$). The subleading terms in the SIDIS cross section can be found in Ch.~\ref{sec:twist3} in Eq.~\eqref{e:SIDIS_subleading}.

\index{TMD factorization!SIDIS at leading power}
The structure functions in \eq{SIDIS-leading} are described in terms of convolutions of TMDs and FFs,
similar to the case of polarized Drell-Yan, see \eqs{sigma_polarized_DY}{SFs}.
They are given at leading power by~\cite{Bacchetta:2006tn}
{\allowdisplaybreaks
\begin{align} \label{eq:structure-functions-twist-2}
 F_{UU,T}	&= {\cal C} \bigl[ f_1 D_1 \bigr]
\,,\nn\\
 F_{UU}^{\cos 2\phi_h} &= {\cal C}\left[ \frac{2\, \bigl(\bfhp\cdot\bfpperp\bigr) \bigl(\bfhp\cdot\bfkperp\bigr) - \bfpperp\cdot\bfkperp}{\zh M_N\mh}  \,h_{1}^{\perp }\,H_{1}^{\perp } \right]
\,,\nn\\
 F_{UL}^{\sin 2\phi_h} &= {\cal C}\left[ \frac{2\, \bigl(\bfhp\cdot\bfpperp\bigr) \bigl(\bfhp\cdot\bfkperp\bigr) - \bfpperp\cdot\bfkperp}{\zh M_N\mh}  \,h_{1L}^{\perp } H_{1}^{\perp } \right]
\,,\nn\\
 F_{LL}	&={\cal C} \bigl[ g_1 D_1 \bigr]
\,,\nn\\
 F_{LT}^{\cos(\phi_h -\phi_S)} &= {\cal C}\left[\,\frac{\bfhp\cdot\bfkperp}{M_N} \,g_{1T}^\perp D_1\right]
\,,\nn\\
 F_{UT}^{\sin\left(\phi_h +\phi_S\right)} &= {\cal C}\left[\frac{\bfhp\cdot\bfpperp}{\zh \mh} \,h_{1} H_1^{\perp} \right]
\,,\nn\\
 F_{UT}^{\sin\left(\phi_h -\phi_S\right)} &= {\cal C}\left[-\,\frac{\bfhp\cdot\bfkperp}{M_N} \,f_{1T}^{\perp } D_1\,\right]
\,,\nn\\
 F_{UT}^{\sin\left(3\phi_h -\phi_S\right)} &= {\cal C}\left[\frac{4\,(\bfhp\cdot\bfpperp)\,(\bfhp\cdot\bfkperp)^2 -2\,\bigl(\bfhp\cdot\bfkperp\bigr)\, \bigl(\bfkperp\cdot\bfpperp\bigr) -\bigl(\bfhp\cdot\bfpperp\bigr)\,\bfkperp^2\ }{2 \zh M_N^2 \mh} \, h_{1T}^{\perp } H_1^{\perp } \right]
\,,\end{align}
}where we always abbreviate $F_{XY}^{\rm weight}\equiv F_{XY}^{\rm weight}(x,\zh,\Phperp,Q^2)$,
and $\bfhp   = \bfPhperp/\Phperp$ is the unit vector along the $x$-axis.
The convolution is defined analogously to \eq{convolution-integral} as~\cite{Bacchetta:2006tn}
\begin{align} \label{eq:def-convolution-integral}
 {\cal C}\left[\omega\;f\;D\right] &
 = x \sum_i H_{ii}(Q^2,\mu)\int d^2\bfkperp \, d^2\bfpperp
 	\; \delta^{(2)}(\zh \bfkperp+ \bfpperp-\bfPhperp)
 \nn\\&\quad \times
   \omega \,f_{i/p_S}(x,k_T,\mu,\zeta_1)\ D_{h/i}(\zh,p_T,\mu,\zeta_2)
\,,\end{align}
where $\omega$ is a weight function, which in general depends on $\bfkperp$ and $\bfpperp$, and the sum runs over all quark and anti-quark flavors $i=u,\bar u, d, \bar d$, etc. 
Here, the hard function for the SIDIS process is denoted by $H_{ii}(Q^2,\mu)$, and is related to that for the Drell-Yan process by
\begin{align}
 H_{ii}(Q^2,\mu) = H_{i\bar i}(-Q^2,\mu)
\,.\end{align}
One can also express the convolutions in \eq{structure-functions-twist-2} through Fourier transforms of products of TMDs in $b_T$ space~\cite{Boer:2011xd},
\begin{align}
\label{eq:structure-functions-sidis-bspace}
 {F}_{UU}(x,\zh,\Phperp,Q^2) &= \phantom{-}   {\cal B}\left[\tilde f_{1}^{(0)}\, \tilde D_{1}^{(0)}\right]
\,,\nn\\
 F_{UU}^{\cos 2\phi_h} (x,\zh,\Phperp,Q^2) &= \phantom{-} M_N\, M_h\; {\cal B}\left[\tilde h_{1}^{\perp (1)}\,\tilde H_{1}^{\perp (1)}\right]
\,,\nn\\
 F_{UL}^{\sin 2\phi_h} (x,\zh,\Phperp,Q^2) &= \phantom{-}   M_N\, M_h\; {\cal B}\left[\tilde h_{1L}^{\perp (1) }\,\tilde H_{1}^{\perp (1)}\right]
\,,\nn\\
 F_{LL}(x,\zh,\Phperp,Q^2) &= \phantom{-}   {\cal B}\left[\tilde g_{1}^{(0)}\, \tilde D_{1}^{(0)}\right]
\,,\nn\\
 F_{LT}^{\cos(\phi_h -\phi_S)}( x,\zh,\Phperp,Q^2) &= \phantom{-}  M_N\; {\cal B}\left[\tilde g_{1T}^{\perp (1)}\, \tilde D_1^{(0)}\right]
\,,\nn\\
 F_{UT}^{\sin\left(\phi_h +\phi_S\right)}( x,\zh,\Phperp,Q^2) &= \phantom{-}  M_h\; {\cal B}\left[\tilde h_{1}^{(0)}\, \tilde H_1^{\perp (1)}\right]
\,,\nn\\
 F_{UT}^{\sin\left(\phi_h -\phi_S\right)}( x,\zh,\Phperp,Q^2) &= -  M_N\; {\cal B}\left[\tilde f_{1T}^{\perp (1)}\, \tilde D_1^{(0)}\right]
\,,\nn\\
 F_{UT}^{\sin\left(3\phi_h -\phi_S\right)}( x,\zh,\Phperp,Q^2) &= \phantom{-}  \frac{M_N^2\, M_h}{4} \; {\cal B}\left[\tilde  h_{1T}^{\perp (2)}\, \tilde H_1^{\perp (1)} \right]
\,,\end{align}
where the Fourier transform corresponding to \eq{def-convolution-integral} is given analogously to \eq{convBess} as
\begin{align} \label{eq:convBessSIDIS1}
 {\cal B}[\tilde f_{}^{(m)}\; \tilde D_{}^{(n)}] 
 \equiv  & x \sum_i H_{ii}(Q^2,\mu) \int_0^\infty \frac{\df b_T}{2\pi}\; b_T\, b_T^{m+n} \, J_{m+n}(q_T b_T) \nonumber \\
& \times \tilde f_{i/N}^{(m)}(x,{b_{T}},\mu,\zeta_1) \; 
 \tilde D_{h/i}^{(n)}(\zh,{b_{T}},\mu,\zeta_2)
\,.\end{align}
The Fourier-transformed TMD PDFs $\tilde f$ and their derivatives $\tilde f^{(n)}$ are defined in \eqs{tmd_decomposition_2}{TMD_bt_derivative},
and the corresponding definitions for the TMD FFs $\tilde D$ and their derivatives $\tilde D^{(n)}$ are given in \eqs{tmdff_decomposition_2}{TMDFF_bt_derivative}.

\subsubsection[Back-to-back hadron production in \texorpdfstring{$e^+ e^-$}{e+ e-}]
              {Back-to-back hadron production in \boldmath $e^+ e^-$}
\label{sec:TMDee}

\index{ee to back-to-back hadrons@$e^+e^-$ to back-to-back hadrons}
The first process where TMD factorization was proven is back-to-back hadron production in $e^+ e^-$ annihilation~\cite{Collins:1981uk},
\begin{align}
 e^+(P_{e^+}) + e^-(P_{e^-}) \to h_1(P_{h_1}) + h_2(P_{h2}) + X
\,,\end{align}
where $h_{1,2}$ are the identified hadrons with momenta $P_{h 1,2}$, and one is inclusive over additional hadronic final states $X$.
Similar to the detected outgoing hadron in SIDIS, see \sec{TMDSIDIS}, these hadrons arise from the fragmentation of quarks in the underlying partonic process, and are described by fragmentation functions characterized by the longitudinal momentum fractions
\begin{align}
 z_{h1} = \frac{2 |\mathbf{P}_{h1}|}{Q} \,,\quad z_{h2} = \frac{2 |\mathbf{P}_{h2}|}{Q}
\,,\end{align}
where the center-of-mass energy $Q^2=(P_{e^+}+P_{e^-})^2$
defines the hard scale of the process.

At leading order, the hadrons are produced exactly back to back, which is spoiled at higher orders due to the additional radiation $X$, which thus gives rise to an imbalance between the hadron momenta.
The near back-to-back region is characterized by a small transverse momentum of the dihadron system compared to $Q$, which is the realm of TMD factorization.

\begin{figure*}
 \centering 
 \includegraphics[width=8.5cm]{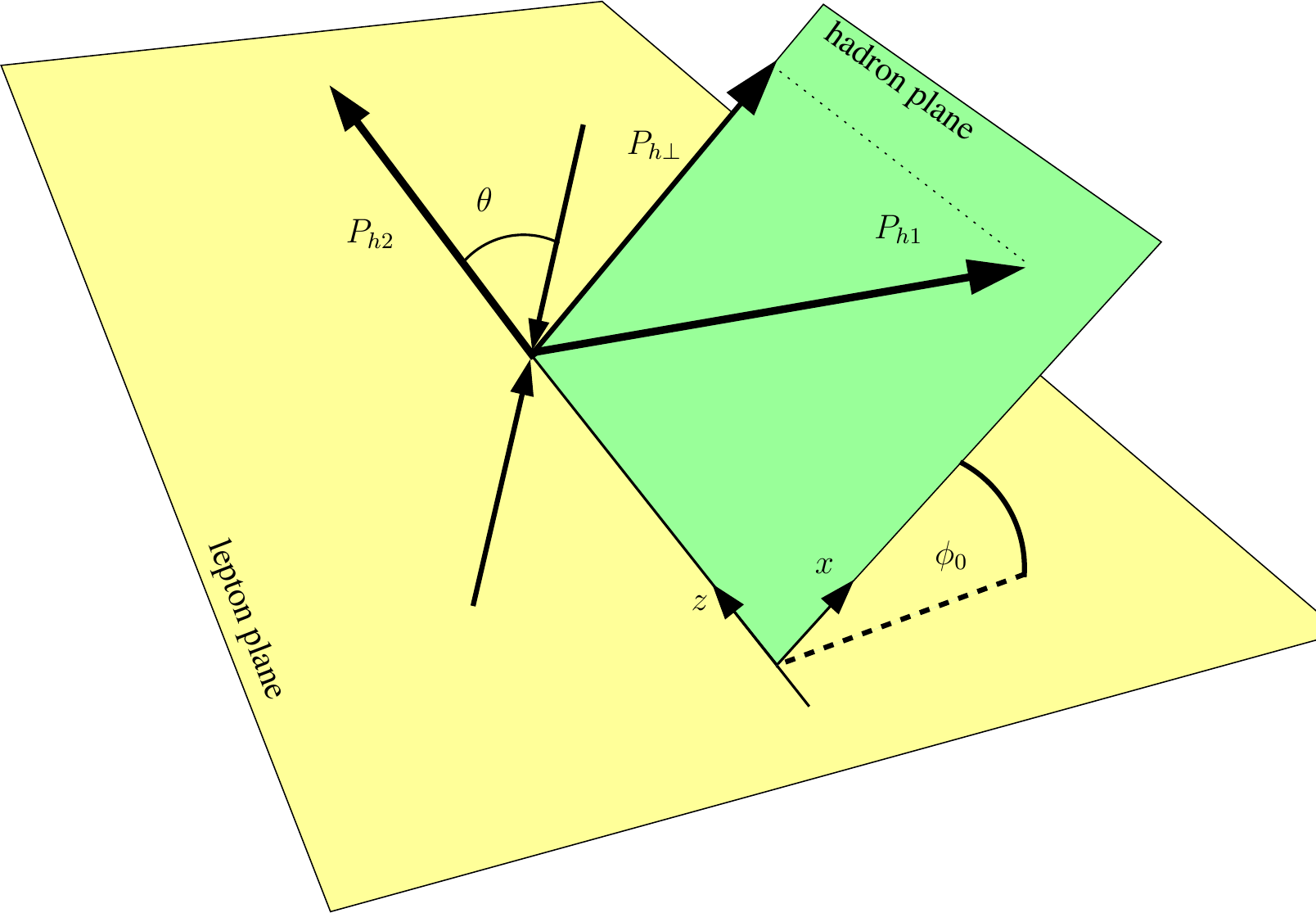}
 \caption{$e^++e^-\to h_1+h_2+X$ process in the frame of method (2). The figure is from~\cite{Kang:2015msa}.}
 \label{fig:epem}
\end{figure*}

As before, measuring angular distributions of the final-state hadrons can give access to spin correlations in the fragmenting hadrons,
most famously in the form of the Collins effect that gives rise to an azimuthal asymmetry of the form $\cos(2\phi)$~\cite{Collins:1992kk}.
To define the azimuthal angle $\phi$, two different reference frames have been proposed in the literature~\cite{Boer:1997mf}:
\begin{enumerate}
 \item One defines the thrust axis of the $e^+e^-$ annihilation and measures the relative azimuthal angular correlation between the two hadrons in
the two back-to-back jets. In this case, one has to measure two azimuthal angles $\phi_1$ and $\phi_2$, and the Collins effect manifests itself as a $\cos(\phi_1+\phi_2)$ asymmetry, and is referred to as the $A_{12}$ asymmetry.
 \item One aligns the $z$ axis along one of the detected hadrons, and measures the azimuthal angle $\phi_0$ of the other hadron with respect to this axis and the lepton plane, as illustrated in \fig{epem}. The Collins effect then appears as a $\cos(2\phi_0)$ asymmetry, and is referred to as the $A_0$ asymmetry.
\end{enumerate}
The $A_{12}$ asymmetry can not be directly described within TMD factorization, as one needs to define the jet directions, which goes beyond standard TMD factorization.
Hence, we will only consider the second approach of the $A_0$ asymmetry, which is described within TMD factorization in terms of the Collins function.

In the limit of small transverse momentum $P_{h\perp}$, the cross section as predicted by TMD factorization reads~\cite{Boer:2008fr,Pitonyak:2013dsu}
\begin{align} \label{eq:xs_e+e-}
 \frac{\df^5\sigma^{e^+e^-\to h_1 h_2 + X}}{\df z_{h1} \df z_{h2} \df^2\mathbf{q}_{T} \df \cos \theta}
 = \frac{N_c \pi \alpha_{\rm em}^2}{2 Q^2}z_{h1}^2z_{h2}^2\left[\left(1+\cos^2\theta\right)F_{UU}
   +\sin^2\theta \cos(2 \phi_0) F_{UU}^{\cos 2 \phi_0}\right] \ 
\,.\end{align}
As illustrated in \fig{epem}, the transverse momentum $\mathbf{P_{h\perp}}$ is defined as the component of $P_{h1}$ transverse to $P_{h2}$,
its azimuthal angle $\phi_0$ is measured relative to the lepton plane, and $\theta$ is the polar angle between the hadron $h_2$ and the $e^+ e^-$ beam.

\index{TMD factorization!ee to back-to-back hadrons@$e^+e^-$ to back-to-back hadrons}
Here the structure function $F_{UU}$ is a convolution of unpolarized TMD fragmentation functions for a quark and an anti-quark, and the polarized structure function $F_{UU}^{\cos 2\phi}$ is a convolution of Collins fragmentation functions for a quark and an anti-quark
\begin{align}
   F_{UU} &= \phantom{-}
   {\cal C}_{ee}[D_1 \bar D_1], \nonumber \\
   F_{UU}^{\cos 2\phi} &=  - {\cal C}_{ee}\left[\frac{2(\hat{h}\cdot {\bf k}_{T1})(\hat{h}\cdot{\bf k}_{T2})-{\bf k}_{T1}\cdot{\bf k}_{T2}} {z_{h1} z_{h2} M_{h1} M_{h2}} H_1^\perp \bar H_1^\perp\right] \; \label{Eq:SFepems}
\end{align}
where the operation ${\cal C}_{ee}$ is defined by   
\begin{align} \label{eq:def-convolution-integral-ee}
	&\hspace{-0.25cm}{\cal C}_{ee}\left[w(k_{T1},  k_{T2}) D_1\;\bar D_2\right] 
	= \sum_i  H_{i\bar i}(Q^2,\mu)\int \frac{d^2{\bf k}_{T1}}{z_{h1}^2} \, \frac{d^2{\bf k}_{T2}}{z_{h2}^2}
	\; \delta^{(2)}\left( -\frac{{\bf k}_{T1}}{z_{h1}}-\frac{{\bf k}_{T2}}{z_{h2}}  -\qT \right)w(k_{T1},  k_{T2})
	\nonumber \\
&\hspace{0.5cm}\times\,\left[D_1^{h_1/i}(z_{h1},k_{T1},\mu,\zeta_1)D_2^{h_2/\bar i}(z_{h2},k_{T2},\mu,\zeta_2)+ \,D_1^{h_1/\bar{i}}(z_{h1},k_{T1},\mu,\zeta_1) D_2^{h_2/i}(z_{h2},k_{T2},\mu,\zeta_2)\right]
 .
\end{align}
In $b_T$ space this becomes
\begin{align}\label{e:sfepem}   
{F}_{UU}(z_{h1},z_{h2},q_T,Q^2) &=
   \phantom{-}   {\cal B}[\tilde D_{1}^{(0)}\; \tilde {\bar D}_{1}^{(0)}] \; ,  \\
{F}_{UU}^{\cos 2\phi_0}(z_{h1},z_{h2},q_T,Q^2) &= 
  -   M_{h1}\, M_{h2} \; {\cal B}[\tilde H_{1}^{\perp(1)}\; \tilde {\bar H}_{1}^{\perp(1)}]\; ,
  \nn
\end{align}
where
\begin{align}
    {\cal B}[\tilde D_{1}^{(n)}\; \tilde {\bar D}_{2}^{(m)}] & 
   \equiv  \sum_i H_{i\bar i}(Q^2,\mu) \int_0^\infty \frac{d b_T\, b_T}{2\pi}\; b_T^{n+m} \, J_{n+m}(q_T b_T)
   \nonumber \\ &\hskip -2cm \times \left( \tilde D_{1}^{(n) \bar i}(z_{h1},b_{T},\mu,\zeta_1) \; \tilde D_{2}^{(m) i}(z_{h2},b_{T},\mu,\zeta_2)  +  \tilde D_{1}^{(n) i}(z_{h1},b_{T},\mu,\zeta_1) \; \tilde D_{2}^{(m)\bar i}(z_{h2},b_{T},\mu,\zeta_2)\right).
   \label{Eq:convBess1epem}
\end{align}

\subsubsection{TMD cross sections for other processes}
\label{sec:TMDextensions}

In this chapter we have discussed TMD PDFs which describe the distribution of light partons (up, down, and strange, which can be treated as massless) within an initial state hadron and TMD FFs which describe the hadronization of a light parton to a final state hadron. The simplest and most frequently considered cross sections that are sensitive to these TMD distributions have been described in Secs.~\ref{sec:TMDDrellYan} to \ref{sec:TMDee}. These distributions will be the focus of the next several chapters of the handbook.

In particular, note that we will discuss phenomenology of TMD PDFs and FFs in \chap{phenoTMDs}. We discuss in detail unpolarised observables both for SIDIS in \sec{SIDISmult} and DY in \sec{DYpheno}. In \sec{PolarizedObservables}, \sec{phenomelology-Boer-Mulders-and-beyond}, and \sec{phenomelology-other} we will discuss the progress in understanding polarised TMDs from SIDIS, DY and $e^+e^-$ data. We will discuss observables in proton-proton scattering in twist-3 formalism in \sec{AN}. Observables and corresponding cross-sections for gluon TMDs will be considered in \sec{gluonTMD_obs}.

Beginning in \chap{JetFrag} we will discuss important generalizations involving heavy quarks (typically charm or bottom quarks) and final states involving jets. Jets are collimated showers of energetic  hadrons that are frequently observed in high energy collisions. The concept of jets and some of the algorithms that are used to define them are discussed in \chap{JetFrag}. \sec{jet-TMDPDFs} shows how cross sections with final state jets can be used to extract TMD PDFs. Jet fragmentation and substructure involving the measurement of an identified hadron and its momentum transverse to the jet axis is an important new application of the TMD formalism and is discussed in \sec{jetsubjetfrag}. Applications to the theory of the  production of heavy quarkonium (bound states of heavy quarks and antiquarks) are the subject of \sec{jetquarkonia}. Transverse energy-energy correlations, which provide a new method to study TMD dynamics, are described in \sec{TEEC}. \sec{mediumjets} discusses the medium modification of jets, which requires the TMD formalism, and can be used to probe both cold nuclear matter as well as the quark gluon plasma. 

%% file: sec-factorization/sec-factorization.tex
\section{Factorization Theorems}
\label{sec:Factorization}

\index{TMD factorization!basic ingredients|(}

\subsection{Factorization Basics}
\label{sec:factbasics}

Ultimately, intuitive partonic pictures (like many of those discussed in the Introduction)
need to be justified or derived in real QCD. 
The challenge is that 
borders separating effects that are genuinely intrinsic to bound-state particles (the hadrons) 
from effects specific to the 
interactions between them is ambiguous in relativistic quantum field theories like QCD. 
In QCD, the notion of a parton as a nearly freely 
propagating point-like quantum of the quark or gluon field is most meaningful in contexts where 
asymptotic freedom applies. To put partonic intuition on a firm theoretical footing, therefore, 
it is important to be able to 
isolate a short-distance part of an interaction and calculate it with small-coupling techniques, 
independently of nonperturbative details of physics at large distance scales. Complications can arise because actual  
physical processes generally involve a complicated interplay between large-distance, nonperturbative dynamics 
and short-distance, process-specific dynamics. For maintaining predictive power it is necessary, if possible, to 
factor these different categories of QCD physics into pieces that can be calculated and interpreted independently, and then to
reassemble them into calculations of physical observables. This is what QCD factorization theorems 
aim to do. 

Factorization theorems have many important practical consequences. For instance, they constrain the possible definitions of 
parton densities and similar nonperturbative objects, and they ultimately lead to the evolution equations 
that relate objects at different scales. Below we will summarize some of the main issues that must be 
confronted in a factorization derivation generally, with a focus on aspects specific to TMD factorization.

For the majority of this chapter, including Secs.~\ref{sec:factelements}-\ref{sec:factviolation}, we follow the CSS methods for deriving factorization, as in Refs.~\cite{Collins:1981uk,Collins:1984kg,Collins:2011zzd,Collins:2011zzd}, since this provides foundational and complete results for factorization proofs.  In \sec{factSCET} we discuss factorization and factorization violation from the point of view of SCET following Ref.~\cite{Rothstein:2016bsq}, and also make direct correspondences between ingredients in the CSS and SCET formalisms. 

It should be understood that the discussion of factorization in this section is only an outline, and many important subtleties can not be 
discussed in the limited space of a handbook. There remain many open questions related to understanding the applications, limits, and interpretation of factorization theorems, especially for processes with sensitivity to the details of hadron structure or nonperturbative effects.

\begin{figure}[pt]
 \centering
 \begin{tabular}{c@{\hspace*{5mm}}c}
  \includegraphics[width=0.45\textwidth]{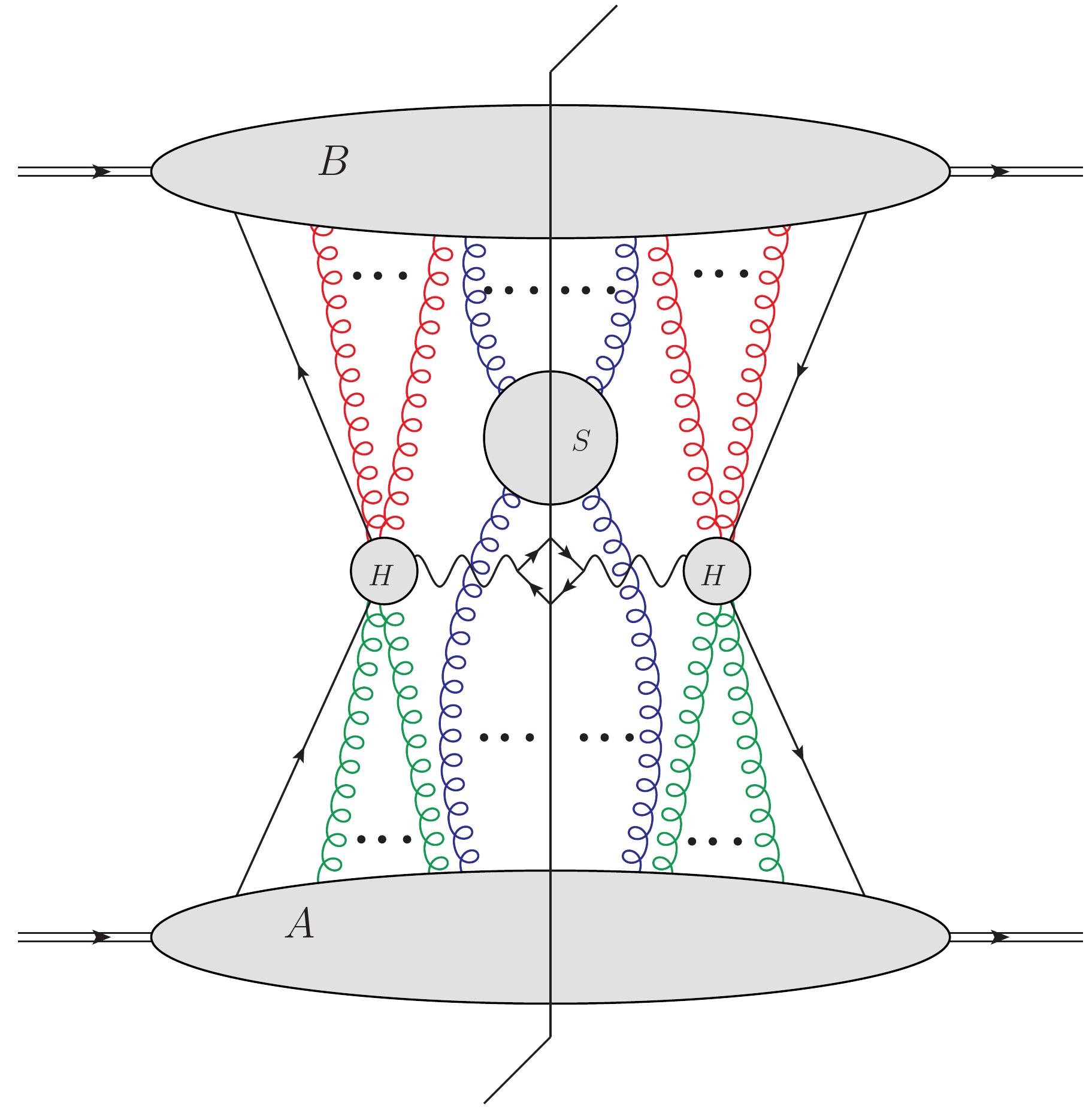}
  \hspace{0.45cm}
  &
  \hspace{0.45cm}
  \includegraphics[width=0.45\textwidth]{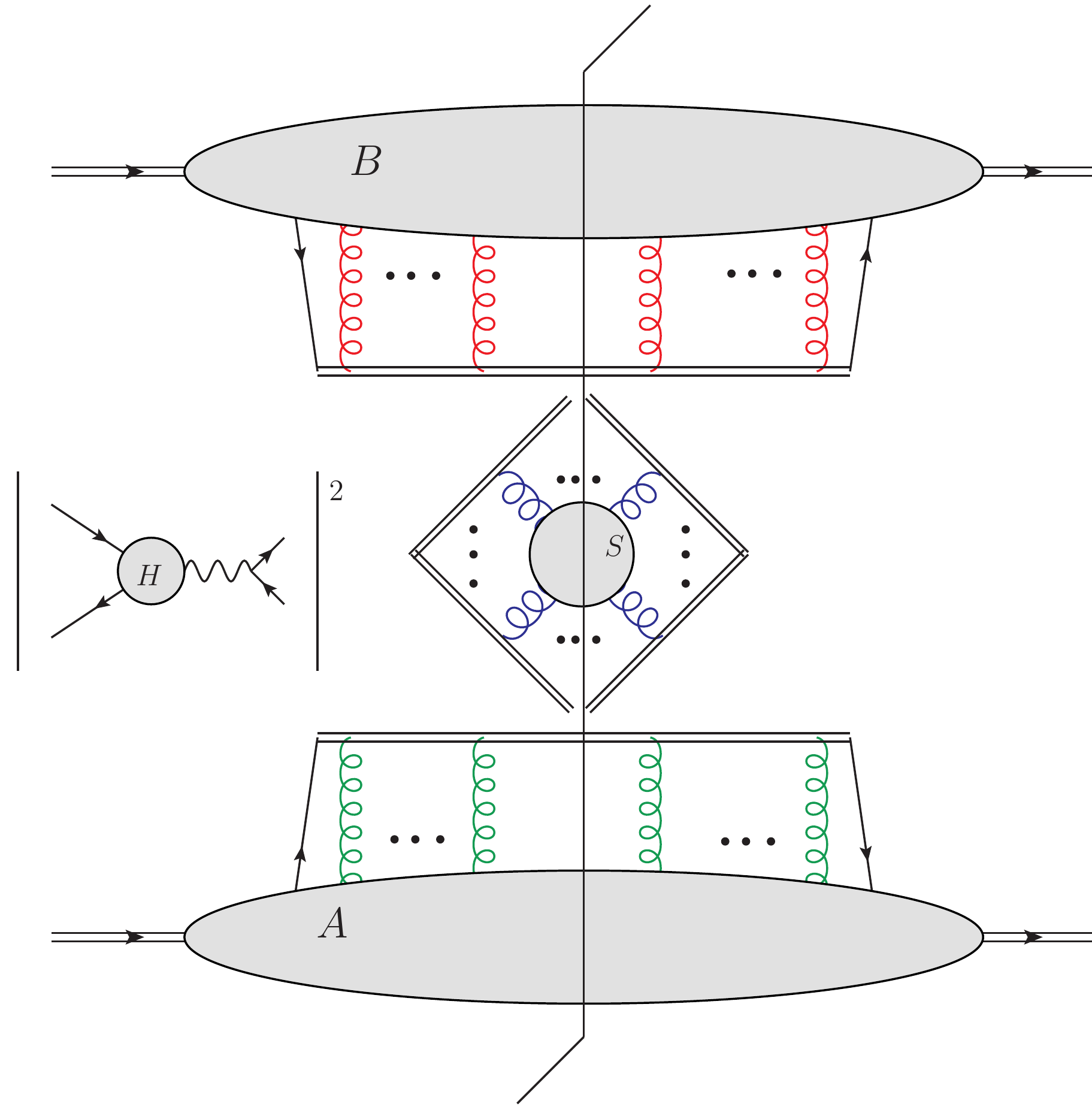} 
 \\
  (a) & (b) 
\end{tabular}
 \caption{(a) Graphical structure corresponding to leading regions in Drell-Yan scattering, before 
 factorization. Green gluons are collinear to lines in the $A$-blob, red gluons are collinear to lines in the $B$-blob, and 
 blue gluons have nearly zero momentum (soft).
 (b) Separation into hard, soft, and collinear parts after approximations 
 and Ward identities---see Sec.~\ref{subsubsec:finalsteps}.}
 \label{fig:leadingregions}
\end{figure}
 
\subsection{Elements of Factorization} 
\label{sec:factelements}

The starting point for deriving factorization is a study of the asymptotic behavior in general graphical structures 
at arbitrary order in perturbation 
theory as some particular hard scale (we will always call it $Q$) approaches infinity. 
In principle, this needs to be done 
separately for different processes, and the details of a specific process can be important, as will be discussed in more detail below. The factorization derivation for Drell-Yan scattering (already analyzed in detail in Chapter~\ref{sec:TMDdefn}) is a prototypical example, and we will continue to refer to it for illustrative purposes. 
For Drell-Yan scattering, the hard scattering scale $Q$ is the invariant mass of the produced lepton-antilepton pair. 
For definiteness, let us assume that the cross section is differential in the transverse momentum $q_T$ of the produced dilepton pair, 
and we will further assume that $q_T \sim \Lambda_{QCD}$ so that the relevant factorization is TMD factorization with 
sensitivity to nonperturbative hadron structure. 
The basic steps for deriving the factorization formula in the large $Q$ limit, both for the Drell-Yan example and for those other processes for which factorization theorems exist, can be summarized according to the following steps:

\subsubsection{Region analysis}  \label{sec:factregions}

For an arbitrary Feynman graph contributing to a specific process, certain configurations of internal 
momentum for internal parton lines dominate in the asymptotically large $Q$ limit. 
The first step, then, is to identify and catalogue all these ``leading regions.'' A systematic approach to region 
analysis was developed by Libby and Sterman~\cite{Libby:1978bx}, (also see \cite[Chapter 5]{Collins:2011zzd}), 
and its key ideas are that:
i.) there is a correspondence between mass divergences in Feynman graphs and their $Q \to \infty$ 
asymptotes and ii.) the mass divergences correspond to surfaces in the higher dimensional space of the
momentum of all lines in a general graph that are trapped between propagator poles. These ``pinched 
singular surfaces'' (PSSs) can not be deformed away from the poles that trap them. In the Libby-Sterman approach, the 
identification and characterization of PSSs becomes a largely geometric problem, and they are often 
summarized in graphical form as in Fig.~\ref{fig:leadingregions}(a) for Drell-Yan scattering at small 
transverse momentum for the produced lepton pair. The $H$ blobs 
contain lines that are off shell by at least order $Q$, while the $A$ and $B$ blobs contain parton 
lines that are collinear to one or the other incoming hadron momentum. The $S$ blob represents lines with nearly 
zero momenta in the center-of-mass system. The gluon lines shown attaching $A$ and $B$ to $H$ 
represent gluons collinear to $A$ and $B$ respectively and attaching to the interior lines of $H$. 
The gluon lines attaching $S$ to $A$ and $B$ are soft, having nearly zero momentum in the center-of-mass system.  To summarize, an arbitrary Feynman graph contributing at leading power in $\Lambda_{QCD}/Q$ to 
Drell-Yan scattering at low $q_T$ matches the structure of Fig.~\ref{fig:leadingregions}(a) if it contributes to a PSS. 
The leading regions for TMD factorization of classic processes are summarized in Table~\ref{tbl:leading_regions}.

It needs to be emphasized that, while Fig.~\ref{fig:leadingregions}(a) corresponds to a mass divergence, 
the lines in actual Feynman graphs are integrated over all momenta. Therefore, different regions can overlap 
in non-trivial ways, and this creates additional work in the factorization derivation. Ultimately, however, 
the $H$ subgraphs will correspond roughly to hard factors, and the $A$, $B$, and 
$S$ factors will be factored away. The extra gluon lines shown entangling the $H$, $A$, 
$B$, and $S$ blobs represent additional collinear-to-$A$ (green) and collinear-to-$B$ (red) gluon lines that can attach inside 
$H$, as well as soft lines (blue) that can attach inside both collinear $A$ and $B$ blobs.  These ``extra'' lines also 
indicate that more work is needed before factorization is achieved.

{
\renewcommand{\arraystretch}{1.4}
\begin{table*}[t!]
\centering
\begin{tabular}{| l | c | c | c |}
	\hline
	Leading regions & Momentum scaling & CSS QFT blobs & SCET objects         
	\\ \hline
    Hard & $p^2 \gg Q^2 \lambda^2$ & $H$ & off-shell, $C$ 
    \\ \hline
    Collinear-a & $p^\mu\sim Q(\lambda^2,\lambda^0,\lambda)$ & $A$ & $\xi_{n_a}, A_{n_a}^\mu$ 
	\\ \hline
    Collinear-b & $p^\mu\sim Q(\lambda^0,\lambda^2,\lambda)$ & $B$ & $\xi_{n_b}, A_{n_b}^\mu$ 
	\\ \hline
	Soft & $p^\mu \sim Q(\lambda,\lambda,\lambda)$ & \multirow{2}{*}{$S$} & $\psi_s, A_s^\mu$
	\\ \hhline{--~-} 
	Glauber & $p^+p^- \ll {\bf p}_\perp^2\sim Q^2\lambda^2$ & & off-shell, ${\cal L}_G^{(0)}$    
	\\[1ex] \hline
	\end{tabular}
	\caption{Summary of the leading momentum regions for the classic TMD observables (Drell-Yan, SIDIS, and back-to-back hadron production in $e^+e^-$) and their corresponding QFT blobs in the CSS formalism and objects in SCET.  Matrix elements of the SCET fields in the last column yield functions equivalent to the evaluation of the final $A,B,S$ blobs in CSS. In the momentum scaling column the parentheses refer to $(p^+,p^-,p_\perp)$ components, and $\lambda\ll 1$ is a small expansion parameter.}
	\label{tbl:leading_regions}
\end{table*}
}

\subsubsection{Approximations}
\label{subsub:apps}

Identifying the graphical structures that contribute to leading regions does not immediately produce
factorization, but it does suggest the necessary approximations. Within each leading region, a specific 
power-law expansion in $1/Q$ applies, giving region-specific approximations. These approximations 
allow the internal kinematics of different parts of a graph to be disentangled. (Note that, without 
approximations, all components of a parton's four-momentum can flow through both $A$ and $B$ in 
\ref{fig:leadingregions}(a)). The exact details of the physical observable under consideration generally play 
a role in determining which power-law approximations are applicable. In the Drell-Yan example with 
$\Tsc{q}{} \sim \Lambda_{QCD}$, for instance, the power-law expansion includes an expansion in powers of 
$q_T/Q$. If instead $q_T \sim Q$, a different expansion applies. 
\begin{figure}[pt]
 \centering
 \begin{tabular}{c@{\hspace*{5mm}}c}
  \includegraphics[width=0.45\textwidth]{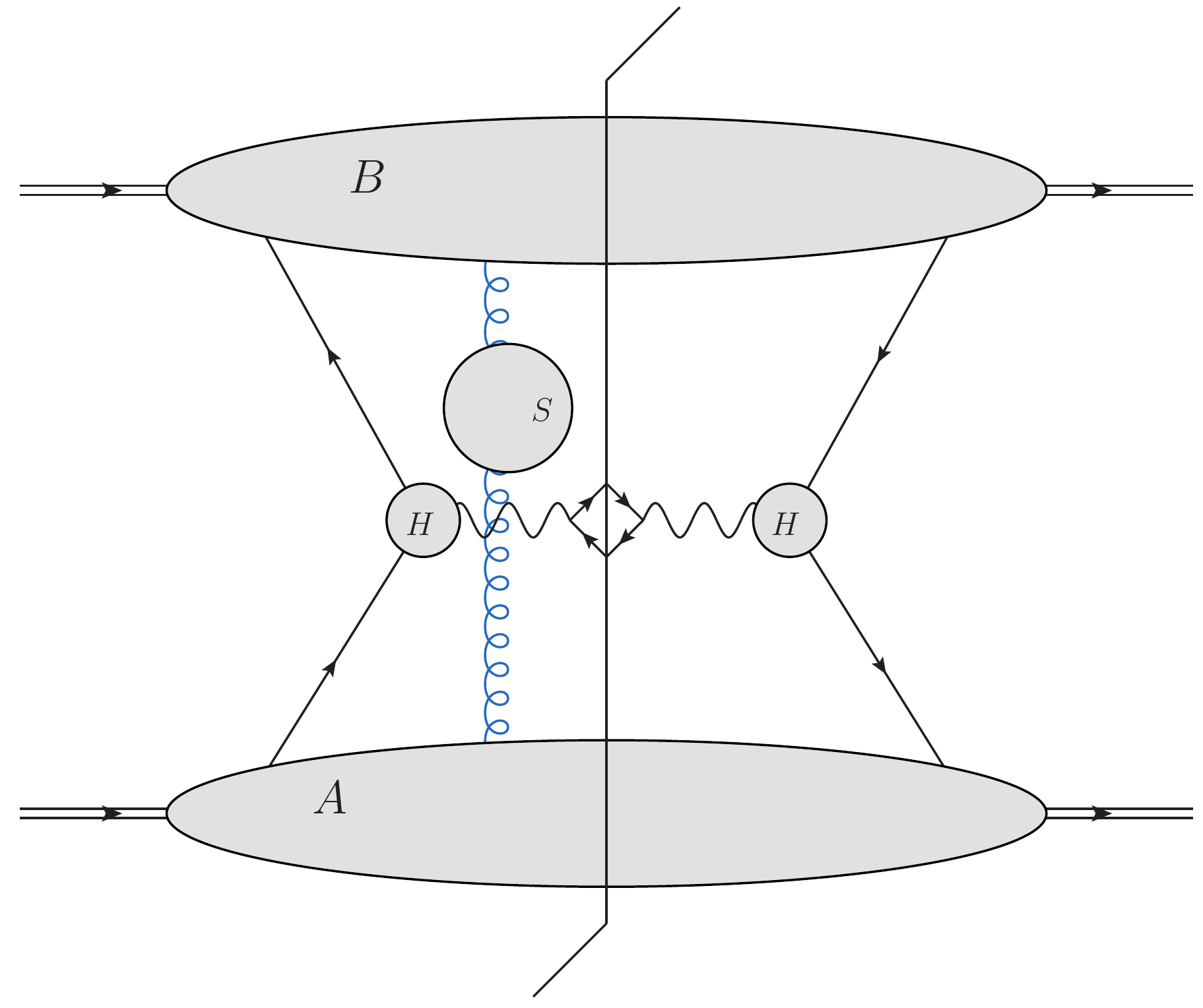}
  \hspace{0.45cm}
  &
  \hspace{0.45cm}
  \includegraphics[width=0.45\textwidth]{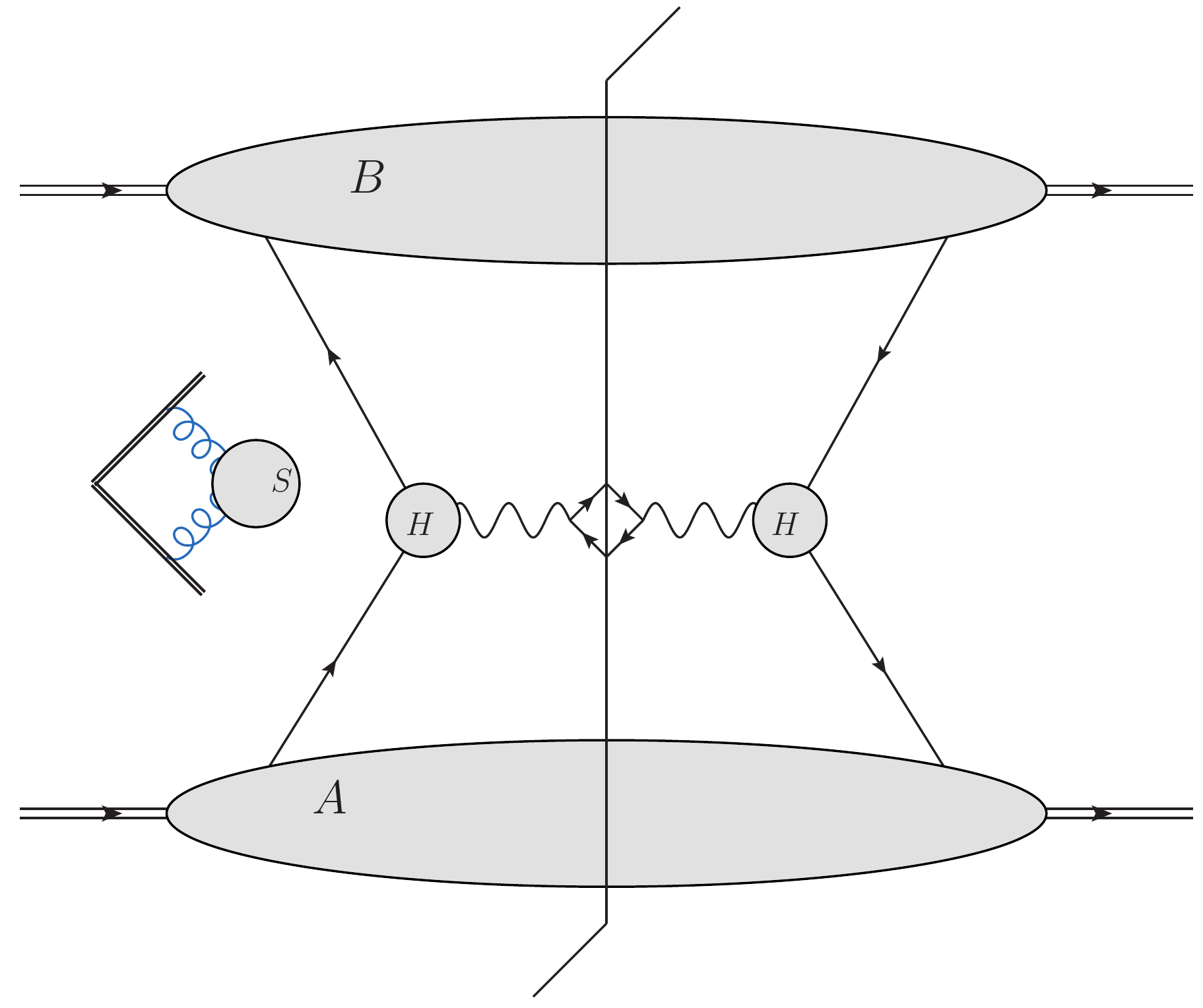} \\
  (a) & (b) 
\end{tabular}
 \caption{Example of a soft gluon being factorized. See Eq.~\eqref{gyapprox}. }
 \label{fig:fact_example}
\end{figure}

These approximations allow the ``extra'' gluon attachments to 
$H$ in \ref{fig:leadingregions}(a)), which appear at first to spoil factorizability, be identified with eikonal attachments, after application of Ward identities. Ultimately, the eikonal 
attachments are to be identified with Wilson line operators in soft and collinear factors. 

As an example, consider the single gluon connecting the $A^\mu$ and $B^\nu$ blobs via the 
$S^{\mu \nu}$ blob in Fig.~\ref{fig:fact_example}(a), where the $\mu$ and $\nu$ are the Lorentz indices associated 
with the gluon coupling. Assume the momentum carried by this gluon is soft In the center of mass 
system, $l \sim (0,0,{\bf 0}_T)$. Also, in the center of mass frame $A^+$ and $B^-$ are the largest components of the 
collinear blobs. So the contraction of factors in the integrand has the leading behavior:
\begin{equation}
A^\mu B^\nu S_{\mu \nu} \approx A^+ B^- S^{+ -} \, , 
\end{equation}
 with errors being power suppressed. There are further simplifications if we multiply by 1 by 
 including a factor of $(l^+ l^-)/(l^+ l^-)$. Then, 
 \begin{equation} 
A^\mu B^\nu S_{\mu \nu} \approx A^+ l^- B^- l^+ \frac{S^{+ -}}{(l^+ + i \epsilon) (l^- - i \epsilon)} \approx 
A^\mu l_\mu B^\nu l_\nu \frac{n_{b,\mu} n_{a,\nu} S^{\mu  \nu}}{(l^+ + i \epsilon) (l^- - i \epsilon)}  \, . \label{gyapprox}
\end{equation}
Here $n_a$ and $n_b$ are the auxiliary vectors defined in Chapter~\ref{sec:TMDdefn}, Eqs.~\ref{eq:nab}.  
Note that after the second $\approx$, we have had to assume that all components of $l^\mu$ are of comparable size, and we 
have inserted the $\pm i \epsilon$ in the denominator without comment. These steps will be discussed more below. 
Now that $l$ four-momenta are contracted exactly with $A^\mu$ and $B^\nu$, Ward identities reduce $l_\mu A^\mu$ and 
$l_\nu B^\nu$ to simple blobs independent of any ``extra'' external gluon momentum. The only memory of the soft 
gluon is in the last factor involving the ``eikonal''  propagators $1/(l^+ + i \epsilon) $ and $1/(l^- - i \epsilon) $, and all of this 
has been factored away from the rest of the graph. These last factors are denoted by the $S$ blob with the double lines 
in Fig.~\ref{fig:fact_example}(b).

(Note carefully that none of the momentum integrals, including the integrals over $l$ components, 
have been made explicit in \eqref{gyapprox}.)
 
\subsubsection{The Glauber region}  \label{sec:Glauber}

\index{Glauber region}
One relies on the sorts of approximations discussed in the last subsection to convert extra gluon attachments into eikonal 
lines when the gluons are collinear or soft. In cases where they are soft, there is also a requirement that the longitudinal 
components are not small relative to the transverse components. If a soft gluon momentum $l$ 
is pinched in a region where $|l^+ l^-| \ll |{\bf l}_{\perp}|^2$, then it is said to be trapped in 
the ``Glauber region.'' (Note that multiple Glauber gluon interactions between spectator remnants are reminiscent of the 
multiple nucleon interactions that give rise to shadowing in the classic Glauber model \cite{Glauber:1955qq} of high 
energy nuclear scattering.) 
Glauber gluons create complications for factorization derivations 
because, when a gluon is pinched in the Glauber region, the Ward identities that would 
normally disentangle it from
$A$ or $B$, as in the example of \ref{gyapprox} do not apply. 
If $l^+$ or $l^-$ is small relative to $l_T$, the approximation in the 
last $\approx$ of Eq.~\eqref{gyapprox} fails.  

In processes like semi-inclusive deep inelastic scattering, with at most one hadron in the initial state, 
the Glauber region can be avoided by an 
appropriate choice of integration contours. This is related to the choice of $\pm i\epsilon$ in Eq.~(\ref{gyapprox}). In hadron-hadron collisions, the situation is more complicated, and 
the importance of Glauber gluons depends on the details of the process. In Drell-Yan scattering, there are in fact 
Glauber pinches graph by graph. The solution to the Glauber gluon problem for more complicated processes like Drell-Yan scattering 
is discussed below. 

\subsubsection{Inclusivity of processes}

The kinds of factorization theorems that emerge (or fail to emerge) 
from a derivation are sensitive to the level of inclusivity of the 
process under consideration. The above Drell-Yan example includes a sum over all final states, 
excluding the momentum of the lepton pair. This ultimately leads to the cancellation of the Glauber pinches 
discussed above, and so is critical to the derivation.  (Useful reviews of this cancellation can be 
found in~\cite[Chs.14.3-14.5]{Collins:2011zzd}, and see also~\cite{Collins:1998ps} and the introduction 
to \cite{Diehl:2015bca}.)

\subsubsection{Last steps}
\label{subsubsec:finalsteps}

After the cancellation of Glauber poles, the approximations discussed in Sec.~\ref{subsub:apps} can finally be applied, 
and a cross section separates into factors. This step is often represented graphically as in 
Fig.~\ref{fig:leadingregions}(b). The double lines with gluon attachments represent Wilson line operators in hadronic matrix 
elements for $A$ and $B$ and for a vacuum matrix element for $S$. To finalize the factorization 
derivation, the overlap of momentum in integrals from one region to another need to be accounted for. 
The important aspects of the final form of factorization are that the hard part be calculable to fixed order 
in perturbative QCD, while the nonperturbative factors should be identifiable with interpretable matrix elements 
like PDFs. 
In Fig.~\ref{fig:leadingregions}(b), the separate soft factor connecting $A$ and $B$ is awkward for a TMD factorization 
formula, and the formulas for Fig.~\ref{fig:leadingregions}(b) generally include an array of arbitrary 
cutoffs. For this and other reasons, there is generally still room for refinement and optimization, which has led to a lot of work on this topic.  These topics go beyond the scope of this section, and are largely the topic of Chapter~\ref{sec:TMDdefn}. 

\subsection{Process Dependence}
\label{sec:procdep}

Even when there are no Glauber pinches in a process, the necessity to avoid the Glauber region places constraints on the 
types of contour deformations that can be used to derive factorization, and this translates into constraints on the Wilson lines 
that can be used to define TMD PDFs. As a consequence, there can be interesting instances of non-trivial process dependence. 
The most well-known case of 
this is the Sivers function in Drell-Yan scattering (at small $q_T$) and SIDIS. 
\index{Sivers function $f_{1T}^{\perp}$!process dependence} 
The gluon attachments 1
that ultimately correspond to 
Wilson lines require contour deformations in opposite directions in the complex plane to avoid the Glauber region. 
The end result is a future-pointing Wilson line in the TMD PDF for SIDIS and a past-pointing Wilson line for TMD PDFs in Drell-Yan scattering. The different directions for the Wilson lines amounts to an overall minus sign change for the Sivers function when comparing Drell-Yan scattering and SIDIS. See \secs{universality}{leadingTMDPDF} for further details.

\subsection{Factorization Violations}
\label{sec:factviolation}

\index{TMD factorization!factorization violation}

Factorization can be violated for a number of reasons. Since the derivations are based on power-law expansions in 
ratios of invariant scales (e.g., $\Lambda_\text{QCD}/Q$ or $q_T/Q$), then factorization can fail if hard scales like $Q$ 
become too small and power corrections are important or the power expansion fails outright. Understanding quantitatively 
which combinations of scales allow for the safe application of particular factorization theorems is an important practical task, especially 
given that many of the experiments designed to probe hadronic structure correspond to rather small $Q$. Much research remains 
to be done in this area. 

Particularly interesting cases of factorization breaking occur when final states are made too exclusive, so that Glauber gluons fail 
to cancel. This can lead to very large phenomenological consequences, as in diffractive hard 
scattering~\cite{Khoze:2000dj,Kaidalov:2001iz,Klasen:2008ah,Kaidalov:2009fp}. In hadron-hadron collisions with measured 
transverse momentum for hadrons in 
the final state, the contour deformations analogous to those discussed above in \sec{procdep} that avoid the Glauber region do not lead to separate  
Wilson lines for separate TMD PDFs or fragmentation functions~\cite{Collins:2007nk,Rogers:2010dm}. Instead of a simple sign change 
for a single TMD PDF, the process dependence involves the details of the whole process. 

\begin{figure}[t!]
 \centering
 \begin{tabular}{c@{\hspace*{5mm}}c}
  \includegraphics[width=0.4\textwidth]{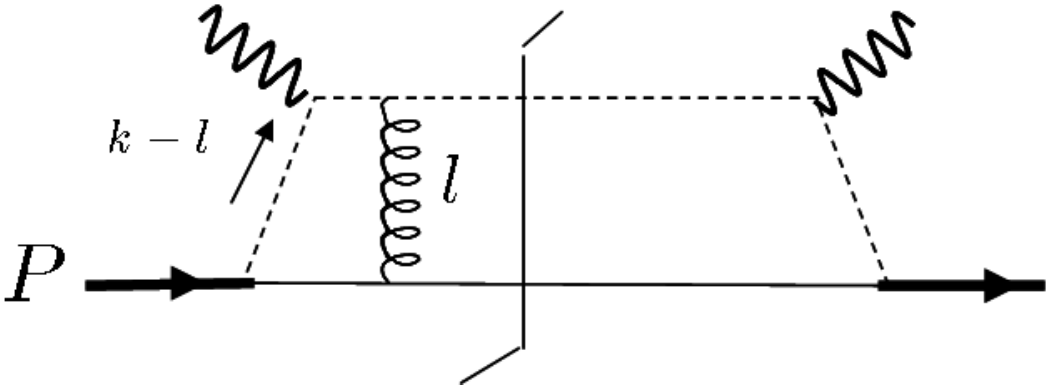}
  \hspace{0.4cm}
  &
  \hspace{0.4cm}
  \includegraphics[width=0.4\textwidth]{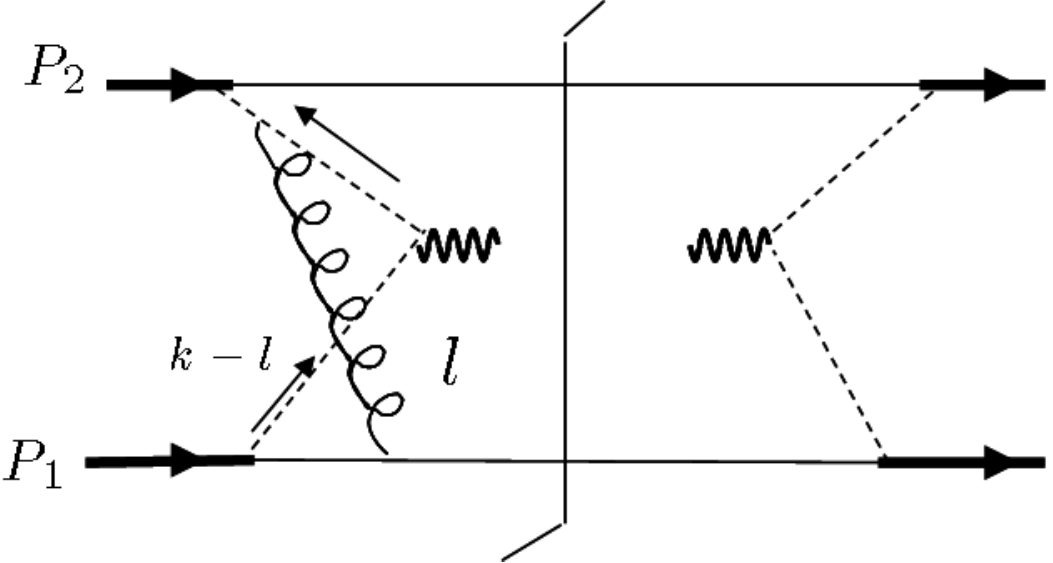} \\
  (a) & (b) 
\end{tabular}
 \caption{A single ``extra'' gluon attachment as it appears in (a) SIDIS and (b) the Drell-Yan process. (Figures taken from~\cite{Mulders:2011zt}.)}
 \label{fig:DYvSIDIS}
\end{figure}

%
\begin{figure}[t!]
\centering
\includegraphics[width=.8\textwidth]{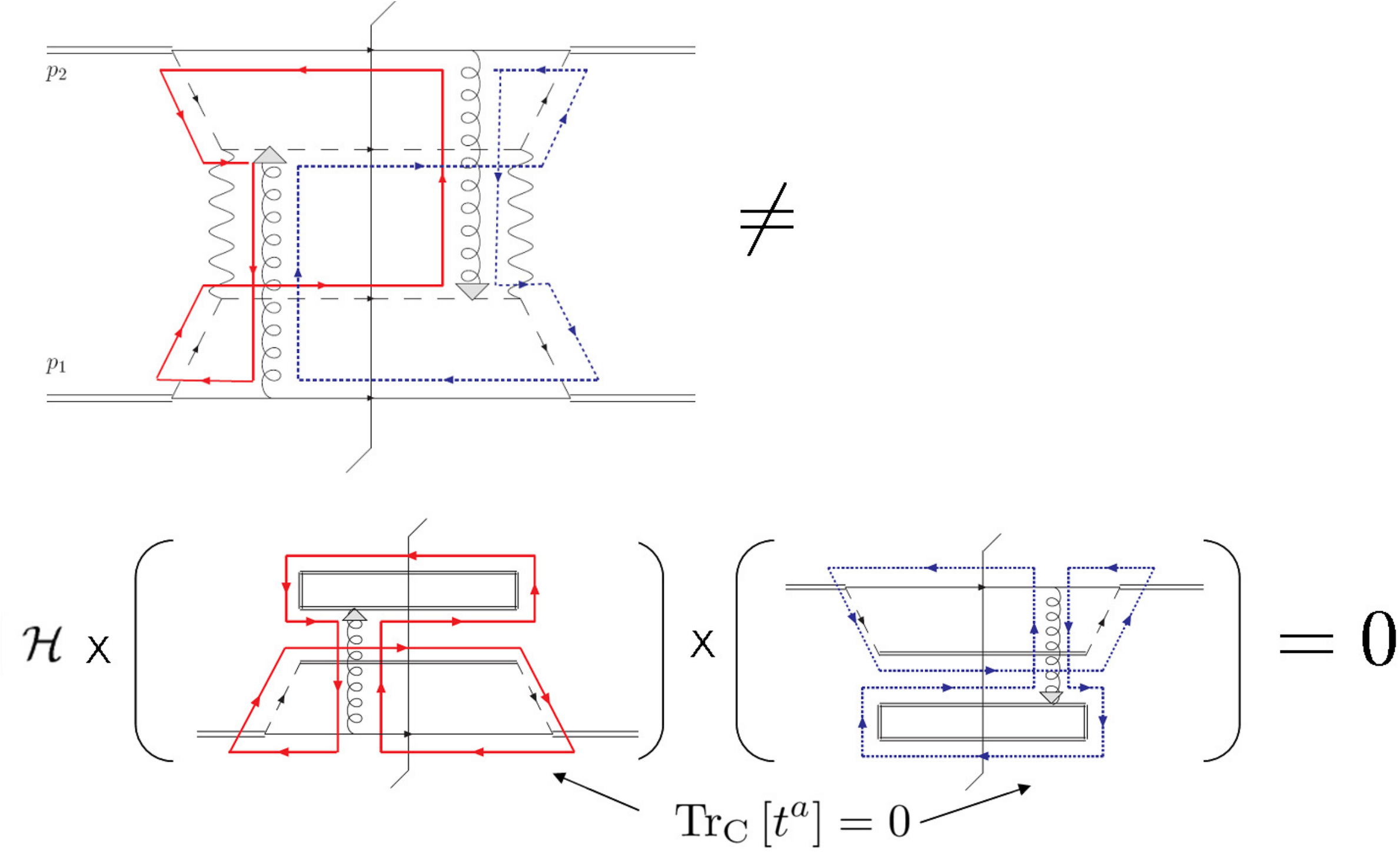}
\caption{A visualization of the failure of color flow to factorize into independent Wilson line structures for separate hadrons in a process involving color in both the initial and final states of the hard part. The Wilson loop structures in the second line vanish due to the traceless single color matrices. (Figure taken from Ref.~\cite{Mulders:2011zt}.)}
\label{fig:factviol}
\end{figure}

This type of factorization breaking and/or process dependence arises from complications in the Ward identity arguments needed to separate long-distance interactions into gauge invariant correlation functions with appropriate Wilson line structures. The sign dependence of the Sivers function, for example, can be understood at the level of Feynman graphs by noting that the extra 
collinear gluon attachments that result in a Wilson line attached to a quark coming in from the distant past in the case of Drell-Yan scattering while they attach to an outgoing quark in the SIDIS case. (See Fig.~\ref{fig:DYvSIDIS}). The fact that attachments are to an \emph{incoming} line in the Drell-Yan process is critical in determining the shape of the Wilson line in \eqref{eq:Staple_Wilson_line}. In the case of SIDIS, the same Wilson line is used, but pointing in the ``$+$''-infinity rather than the ``$-$''-infinity directions. In much of the original work on process dependence, the Wilson line in a TMD PDF definition was therefore notated with a ``$[+]$'' or ``$[-]$'' superscript to indicate which direction was relevant to a particular process~\cite{Bomhof:2004aw}. The quantum-mechanical phase of the quark wavefunction is shifted in an opposite way depending on whether the quark propagates in from the distant past or out to the distance future---see \cite{Collins:2008ht} for an optical analogy.

One notes that it is the direction of flow of the color charge (incoming or outgoing) that determines the Wilson line direction, and this suggests that factorization theorems for more complicated TMD processes can be constructed, with increasingly complex Wilson line structures for the TMD correlation functions~\cite{Bomhof:2006dp}.
However, this tends to fail for somewhat interesting reasons~\cite{arXiv:1001.2977} that can be visualized in the Feynman diagram shown in Fig.~\ref{fig:factviol}.
There, two hadrons collide and a colorless particle is exchanged in the hard part.
The graph before factorization is nonzero, but after factorization, the only possible Wilson line structure for each PDF is a Wilson loop. With a single gluon attachment, however, each Wilson loop gives a factor of zero. The right side of the graph thus fails to reproduce the non-vanishing unfactorized graph. In other words, the quantum-mechanical phase shift associated with extra collinear gluon attachments is a consequence of the presence of \emph{both} hadrons simultaneously, and not a simple sum of Wilson line phase shifts associated with each hadron. (This captures the essence of the problem with color flow arguments, though more details are needed to show that it represents an unavoidable problem for factorization---see, for example, Ref.~\cite{arXiv:1001.2977}.)

\subsection{Factorization in SCET}
\label{sec:factSCET}

\index{Soft Collinear Effective Theory (SCET)}
\index{TMD factorization!SCET}

In the SCET~\cite{Bauer:2000ew,Bauer:2000yr,Bauer:2001ct,Bauer:2001yt,Bauer:2002nz} approach to factorization, an effective field theory is set up with fields that describe the infrared momentum regions of QCD, which typically have either collinear or soft scaling. 
The effective Lagrangian encodes self interactions of these fields, as well as their interaction with each other and with the hard region of momentum space encoded in Wilson coefficients. 

For the classic TMD processes described in \sec{TMDfactSIDISee} the relevant modes are soft, $n_a$-collinear, and $n_b$-collinear, in one-to-one correspondence with the regions $S$, $A$, and $B$ discussed in \sec{factregions}. Indeed, although the formal setup is different, there is a close parallel between many items appearing in the CSS formalism and SCET formalism. In our brief review of factorization in SCET we will highlight these parallels.  The relevant SCET Lagrangian for TMD observables involving at most two energetic jets or hadrons is
\begin{align}
  {\cal L} = 
    {\cal L}_{n_a}^{(0)} + {\cal L}_{n_b}^{(0)} + {\cal L}_S^{(0)}
   + {\cal L}_{\rm hard}^{(0)} + {\cal L}_{G}^{(0)} 
   +{\cal O}(\lambda) \,,
\end{align}
where $\lambda\sim q_T/Q \ll 1$ is the TMD power counting parameter, and only leading power terms are kept for the discussion here as denoted by the superscripts $(0)$. 
The term ${\cal L}_{n_a}^{(0)}$ describes interactions between $n_a$-collinear quark and gluon fields describing momenta with the scaling $(n_a\cdot p,n_b\cdot p, p_\perp)\sim Q(\lambda^2,1,\lambda)$, ${\cal L}_{n_b}^{(0)}$ does the same for $n_b$-collinear fields where $(n_b\cdot p,n_a\cdot p, p_\perp)\sim Q(\lambda^2,1,\lambda)$, and ${\cal L}_S^{(0)}$ describes interactions between soft quark and gluon fields with momenta scaling as $p^\mu \sim Q \lambda$.
Each of the Lagrangians ${\cal L}_{n_a}^{(0)}$, ${\cal L}_{n_b}^{(0)}$,  and ${\cal L}_S^{(0)}$ is equivalent to a copy of the QCD Lagrangian for its fields, up to the fact that the fields are setup so that they induce subtractions that enable them to correctly capture their momentum region while avoiding double counting of other infrared regions~\cite{Manohar:2006nz}. In SCET these induced subtractions are referred to as zero-bin contributions, while in the CSS approach they are referred to as soft subtractions. In SCET for TMDs the leading power interactions between modes in different momentum regions are entirely contained in the Lagrangians ${\cal L}_{\rm hard}^{(0)}$  and ${\cal L}_{G}^{(0)}$, which describe the off-shell short-distance hard scattering process and off-shell long-distance Glauber interactions respectively.  A summary of the way that the leading momentum regions for classic TMD observables are described by objects in SCET is given in Table~\ref{tbl:leading_regions}.

For TMD cross sections in Drell-Yan, SIDIS, or back-to-back hadron production in $e^+e^-$ collisions, the QCD current $\bar{\psi} \Gamma \psi$ is matched onto SCET to obtain a leading power hard interaction which involves a quark current with Wilson lines
\begin{align} \label{eq:Lhard}
    {\cal L}_{\rm hard}^{(0)} &= \int \!\! \df\omega_a\df\omega_b\: C^{(0)}(\omega_a,\omega_b) \: \bar\chi_{n_a,\omega_a} \Gamma (S_{n_a}^\dagger S_{n_b}) \chi_{n_b,\omega_b} \,.
\end{align}
Here the $n_a$-collinear field $\chi_{n_a,\omega_a}= \delta(\omega_a-in_b\cdot \partial) (W_{n_a}^\dagger[n_b\cdot A_{n_a}] \xi_{n_a})$ involves a quark field $\xi_{n_a}$ attached to a Wilson line built from collinear gluon fields $A_{n_a}$ that extends to infinity in the direction $n_b$, and this product of fields has minus-momentum $\omega_a$. The description is then directly analogous for $\chi_{n_b,\omega_b}$.  For intuition these $\chi_{n_a,\omega_a}$ fields are the closest possible analog of fields for partons in the parton model. \index{parton model} From the SCET point of view, the presence of the Wilson lines is necessary in order to satisfy the full structure of gauge transformations allowed in this effective theory. The $S_{n_a}[n_a\cdot A_s]$ and $S_{n_b}[n_b\cdot A_s]$ in \eq{Lhard} are Wilson lines involving soft gluon fields $A_s$. They describe the fact that soft interactions with an energetic color source in direction $n$ and a given overall color representation are described by a Wilson line in this representation along $n$. These soft Wilson lines encode the eikonal soft interactions, as discussed for CSS in Sec.~\ref{subsub:apps}.  Finally the Wilson coefficient $C^{(0)}$ encodes contributions from the hard momentum region.  
The SCET hard scattering Lagrangian in \eq{Lhard} is derived by integrating out off-shell momentum regions with $p^2\gg \lambda^2$, which is done to all orders in perturbation theory. Integrating out hard fluctuations with $p^2\sim Q^2$ and hard-collinear fluctuations with $p^2\sim Q^2\lambda$ leads to the $W_{n_a}$, $W_{n_b}$, $S_{n_a}$, $S_{n_b}$ Wilson lines and the hard Wilson coefficient $C^{(0)}$. The structure of the resulting terms is also constrained by gauge symmetries of the effective theory. Since the hard interaction Lagrangian encodes the coupling to the leptonic currents, it is always included perturbatively, namely once for each hard interaction.

\index{Glauber region}
In SCET all interactions that can potentially spoil factorization are encoded in the Glauber Lagrangian ${\cal L}_{G}^{(0)}$, whose detailed form can be found in Ref.~\cite{Rothstein:2016bsq}.  It contains leading power long-distance interactions between both $n_a$-$n_b$-soft, $n_a$-soft, and $n_b$-soft modes, all of which are forward scattering in nature with $1/p_\perp^2$ type potentials.  Since insertions of this Lagrangian are not suppressed it can be inserted an arbitrary number of times at leading power, and these interactions have the potential to spoil factorization since they recouple collinear and soft regions in a non-trivial manner. Thus the influence of ${\cal L}_{G}^{(0)}$ must be shown to be either of a form which can be absorbed into a soft or collinear matrix element, or to fully cancel out. 
In SCET both the soft and collinear Wilson lines along a direction $n_i$ can be chosen to extend either from $-\infty n_i^\mu + x^\mu$ to $x^\mu$, or from $x^\mu$ to $x^\mu+\infty n_i^\mu$. This affects the signs $\pm i\epsilon$ of eikonal propagators like those shown in Eq.~(\ref{gyapprox}). In SCET the subtractions from the Glauber region guarantee that results are independent of this choice. However, sometimes the only non-trivial impact of the Glauber region is to influence the direction of collinear and soft Wilson lines, and in that case their effects can be absorbed into collinear and soft matrix elements with precisely specified Wilson line directions.%
~\footnote{Technically this amounts to simultaneously not considering certain  Glauber interactions, nor corresponding soft and collinear Glauber-region subtractions. These are in 1-to-1 correspondence, such that the sum of the direct Glauber interactions and Glauber-region subtractions give zero.}
For TMD factorization both occur, the cancellation of certain ${\cal L}_G^{(0)}$ contributions and the absorption of other ${\cal L}_G^{(0)}$ effects. In particular, Glauber interactions between so-called active partons can be absorbed into the direction of soft Wilson lines, forcing them to extend out to either $+\infty$ or $-\infty$, while analogously the interactions between active and spectator partons can be absorbed into the direction of the collinear Wilson lines~\cite{Rothstein:2016bsq}. 
In the CSS formalism the same result is obtained but in a different manner, since the Glauber and soft regions are not separated from the start. Instead certain propagators are left in a form that can handle simultaneously the Glauber and soft contributions. The results for collinear and soft regions are determined by contour deformations that are done to put these contributions in the collinear and soft regions (that end up in their matrix elements), and expansions are carried out at this point. 
These contour deformations are done in order to avoid the Glauber region when possible.  Finally, there are spectator-spectator Glauber interactions which cancel out due to unitarity and the inclusive sum over hadronic states in TMD observables, which has been worked out in detail in CSS~\cite{Collins:1988ig,Collins:2011zzd} and also occurs in SCET~\cite{Rothstein:unpublished}. See Sec.~\ref{sec:Glauber} for further discussion in context of CSS.  With these considerations in hand, ${\cal L}_{G}^{(0)}$ can be dropped for the remaining analysis of TMD factorization in SCET. 

To derive the form of the TMD factorization, for example for Drell-Yan $pp\to X\ell^+\ell^-$, one considers the hadronic matrix elements $\langle p| {\cal L}_{\rm hard}^{(0)\dagger}| X\rangle\langle X | {\cal L}_{\rm hard}^{(0)} | p \rangle$. Since  the decoupled Lagrangians ${\cal L}_{n_a}^{(0)}$, ${\cal L}_{n_b}^{(0)}$,  and ${\cal L}_S^{(0)}$ enter as a direct sum, the state $|X\rangle$ can be factorized into soft and collinear components and the dynamics of the soft and collinear matrix elements factorizes. Finally \eq{Lhard} involves a simple product of fields from the different sectors.  This enables the cross section to be factorized without relying on perturbation theory, leading to the TMD factorization theorems discussed in \sec{TMDfactSIDISee}.   The $n_a$-collinear quark fields and Wilson lines lead to the operators with the staple-shaped path giving the bare beam function $\tilde B_{i/p}^{0}$, with another $\tilde B_{\bar i/p}^0$ from the $n_b$-collinear sector, while the soft Wilson lines give the soft function $\tilde S^0_{n_a n_b}$, which is the vacuum matrix element of the closed loop given in \eq{softfunc} and \fig{wilsonlines} (right panel). The beam functions can be further written as unsubtracted TMDPDFs and soft subtractions, $\tilde B_{i/p}^{0} = \tilde f_{i/p}^{0}/\tilde S^{0\subt}_{n_a n_b}$, where $\tilde f_{i/p}^{0}$ is given in \eq{beamfunc} and \fig{wilsonlines} (left panel), and $\tilde S^{0\subt}_{n_a n_b}$ encodes the zero-bin subtractions which stop the unsubtracted TMDPDF matrix element from double counting the soft regime.  

While this discussion is only at the broad outline level, and hence leaves out many of the details and subtleties associated with actually carrying out the derivation of the TMD factorization theorems (such as the rapidity regularization), it provides the basic picture of how the factorization comes about in SCET.

\index{TMD factorization!basic ingredients|)}

%% file: sec-evolution/sec-evolution.tex
\newpage
\section{Evolution and Resummation}
\label{sec:evolution}

\subsection{Introduction}
\label{sec:evolutionintro}

In this Handbook,
in ~\chaps{TMDdefn}{Factorization},  we have introduced 
the theoretical tools to investigate 
hadrons as a dynamical system of quarks and gluons (partons) from QCD field theory in the context of TMD observables.
In this chapter we continue this field theoretic treatment and cover
the subject of TMD evolution. \index{evolution!TMD} We will  emphasize how 
the QCD defintions of TMDs, coupled with TMD factorization theorems   
 yield  the TMD evolution equations. 
We will review two general classes of approaches to TMD evolution. One class of approaches formulated more directly in traditional QCD, and another in the language of Effective Field Theory, i.e. Soft Collinear Effective Theory (SCET). In both, evolution of TMDs follows from the methods to regulate and define them as reviewed in \chap{TMDdefn}.

Having established the   QCD field theory definitions 
of the TMD PDFs that arise from the modern proofs of factorization~\cite{Collins:2003fm,Collins:2008ht,Ji:2004wu,Ji:2004xq,Collins:2011zzd,GarciaEchevarria:2011rb,Echevarria:2012js,Collins:2017oxh} 
we find that the TMDs depend on two auxiliary variables. One is the renormalization scale $\mu$,
arising from renormalizing the UV divergence,
 which separates high and low energy or mass scales from one another.
The second is the rapidity evolution scale,  $\nu$ or $\zeta$, associated with regulating rapidity divergences, separating soft and collinear momentum regions from one another.  Thus, what is unique about TMD evolution is that it takes place in \emph{two} dimensions, as opposed to one for usual collinear DGLAP evolution. 

As we saw in~\chap{TMDdefn} 
the interplay of these scales makes it possible to use  QCD factorization to 
express TMD observables  as a convolution of a hard scattering cross section and renormalized TMDs at leading power in the hard scale. We refer to this  as the $W$ term, \eq{sigma_new}.
The requirement of independence of the $W$ term 
on these regulator scales leads to renormalization group (RG) and \emph{rapidity renormalization group} (RRG) evolution equations---the Collins-Soper (CS) equations---relating TMD PDFs and other ingredients at different scales.~\footnote{We note that the RRG encompasses a more general class of evolution that includes not only TMD evolution but also other types of evolution such as BFKL evolution~\cite{Fleming:2014rea,Rothstein:2016bsq}.\index{BFKL evolution equation}}

The solutions of the TMD evolution equations will generically lead to solutions for transverse position space 
(Fourier transform) TMD PDFs of the schematic form:
 \begin{equation}\label{eq:tmd_sol_schematic}
     \tilde f_{i/P}(x,\bt,\mu,\zeta) = \tilde f_{i/P}(x,\bt,\mu_0,\zeta_0) U_\text{RG}(\mu_0,\mu;\zeta_0)V_\text{RRG}(\zeta_0,\zeta;b_T,\mu)\,,
 \end{equation}
 where $U_\text{RG}$ evolves the TMD PDF from an energy/mass scale $\mu_0$ to another scale $\mu$, and $V_\text{RRG}$ evolves it from a rapidity scale $\zeta_0$ to another scale $\zeta$.  
 (See \fig{regions}.) Explicit forms for these evolution kernels, and methods to obtain them, are the topic of the rest of this Chapter. Such evolution is essential to relate the TMD PDFs at a hadronic or low scale,  to cross sections measured at large collision energies $Q$.  In Chapter~\ref{sec:phenoTMDs} the  status of  predictions and tests of the TMD formalism from  phenomenological studies is covered. Central to the phenomenology of extracting TMDs from 
SIDIS, Drell-Yan, weak gauge boson production, $e^+e^-$ annihilation into hadron pairs, including corresponding azimuthal and spin modulations of cross sections, is the implementation of TMD evolution. This technology provides much of the predictive power of TMD factorization.

Another powerful consequence of evolution is the possibility to resum large logs of ratios of scales such as $q_T/Q$,  or $Qb_T$ that appear in the perturbative  expansion of  the $W$ term.
  We review this connection here in this Chapter, in particular in \sec{EvolResum}. Before doing so,  we present a short historical overview.

 \subsubsection{Historical overview}

Much of the literature  on TMD factorization\index{TMD factorization!CSS} and evolution was pioneered by Collins, Soper, and Sterman (CSS)~\cite{Collins:1981uk,Collins:1981uw, Collins:1984kg} \index{evolution!CSS},  was expanded upon in recent years~\cite{Ji:2004wu,Ji:2005nu,Collins:2011zzd,Aybat:2011zv},  and further elaborated upon and extended
in~\cite{Catani:2000vq,deFlorian:2001zd,Catani:2010pd,Collins:2011zzd}. It has also been cast in the framework of Soft Collinear Effective Theory (SCET)
\cite{Bauer:2000ew, Bauer:2000yr, Bauer:2001ct, Bauer:2001yt} 
by numerous authors \cite{Becher:2010tm, Becher:2011xn, Becher:2012yn, GarciaEchevarria:2011rb, Echevarria:2012js, Echevarria:2014rua, Chiu:2012ir}.\index{TMD factorization!SCET}
For a relation of the different approaches to each other, see e.g.~\cite{Collins:2017oxh}, and for a historical review on TMD PDFs we refer the reader to~\cite{Collins:2011zzd}.

The CSS based construction of TMD PDFs satisfy the property of maximum universality~\cite{Collins:2004nx,Collins:2011zzd,Aybat:2011ta}, meaning that the same correlation functions appear in a large number of processes.  This universality 
provides  the predictive power of the TMD formalism. 
Modern treatments in SCET~\cite{Becher:2010tm, Becher:2011xn, Becher:2012yn, GarciaEchevarria:2011rb, Echevarria:2012js, Echevarria:2014rua, Chiu:2012ir,Ebert:2019tvc} cast factorization and evolution in the framework of effective field theory and the matching and running of EFT operators and matrix elements, and have proven to be useful for obtaining higher order perturbative results for anomalous dimensions and resummed cross sections.
The equivalence between various constructions which leads to a factorized cross section like \eq{sigma_new} has been reviewed  in Chapters~\ref{sec:TMDdefn} and \ref{sec:Factorization}.

We remark that it is somewhat common to refer loosely to the  CSS formalism~\cite{Collins:2003fm,Belitsky:2002sm,Collins:2008ht,Collins:2011zzd} and its modern implementations~\cite{Ji:2004wu,Ji:2004xq,Collins:2011zzd,GarciaEchevarria:2011rb,Echevarria:2012js}
as a $q_T$ resummation method. Resummation methods however, generally do not take into account the nonperturbative physics that becomes important in regimes where logarithms are so large that perturbative expansions break down and nonperturbative physics becomes relevant (see \sec{EvolResum}).  However, TMD factorization formalisms exploit the renormalization group and the CS equation to
calculate the cross section such that point by point in $b_T$,   $\forall\,  b_T$ (and $q_T$ via the Fourier transform) asymptotic freedom is exploited to maintain a small $\alpha_s$ 
and a valid perturbative expansion in the hard scattering cross section.
These methods are more powerful than resummation methods since they constitute a true TMD factorization formalism using a pQCD perturbation expansion for all $b_T$, even well into the nonperturbative large $b_T$ region~\cite{Collins:1984kg} where details  
 of hadron structure become important. We will review this approach and its correspondence to the SCET formulation in \sec{TMDEvol}.

Below, we will review the derivation of the TMD evolution equations both from the ``direct'' QCD approaches like CSS, and from the EFT framework of SCET. Both routes to the evolution equations are ultimately equivalent, but highlight complementary aspects of TMD evolution and inspire various methods to obtain their explicit solutions.

Before delving into the details of TMD evolution equations and their solutions, we proceed next to an introductory discussion of resummation of large logarithms in perturbative expansions of TMD cross sections. 
We will concentrate on the Drell-Yan process \eq{sigma} as the physical example connecting TMD PDFs to experiments, as the methods  illustrated here  easily carry over to other TMD processes.

\subsection{TMD Evolution and Resummation}
\label{sec:EvolResum}

We emphasize, once again, evolution serves two primary purposes. The first is to relate the TMD PDFs themselves at nonperturbative scales to cross sections measured at large energies.  The second equally important, is to provide a route to the summation of large perturbative logs appearing in the fixed-order expansions of TMD cross sections in QCD.\index{resummation} The tools and approaches we review in this Chapter aim at achieving both purposes. The actual extraction, computation, or modeling of nonperturbative TMD PDFs are the focus of later Chapters, and we will dive into the technology of TMD evolution itself in the next Section. Before doing so, we take the opportunity in this Section to present a bit more general introduction to the connection between evolution and perturbative resummation, whose history of development and application in QCD is long and impressive.

  The relation of Sudakov resummation to factorization was emphasized in the early works of \cite{Mueller:1979ih,Collins:1980ih,Collins:1981uk,Sen:1981sd}. It  was further developed and given a unified treatment in the work of \cite{Contopanagos:1996nh}, which derived the resummation of Sudakov logarithms from factorization properties of QCD cross sections and renormalization group evolution of the factorized contributions. This approach ties very naturally to the framework of effective field theories like SCET, which soon emerged, with their built-in tools of matching and running between scales.  

Early work on resumming large logs 
 in the context of evolution and transverse momentum factorization was carried out by Collins, Soper, and Sterman (the CSS approach)~\cite{Collins:1981uk,Collins:1981va,Collins:1984kg}.\index{evolution!CSS}\index{factorization!CSS}
 For early work on summation of large perturbative logs in the context the of the transverse momentum distribution of the Drell-Yan cross section  at moderate to high transverse momentum, see Refs.~\cite{Altarelli:1984pt,Davies:1984hs}, and for  a review of the topic, see  Refs.~\cite{Arnold:1990yk}.

 \subsubsection{The goal of resummation}
\label{sec:TMDResum}
\index{resummation!accuracy}

Let us introduce now the relation between evolution of TMD PDFs, or more generally of elements of a factorized cross section in QCD, and the resummation of large logs that arise in perturbative QCD predictions of cross sections that depend on more than one physical scale, separated by a hierarchy.

The large logs that appear in the perturbative expansion of the Drell-Yan cross section are in the ``$W$-term'' of \eq{sigma_new}, and are easiest to count in $\bsc$ space.\index{$W$ term} Expressing the cross section,
\begin{equation} 
\label{eq:DY_bspace}
\frac{d\sigma^W}{dQ dY d^2\mathbf{q}_T} = \int d^2 \mathbf{b}_T \, e^{i\mathbf{b}_T\cdot\mathbf{q}_T} \widetilde \sigma^W (\mathbf{b}_T)\,,
\end{equation}
for perturbative values of $q_T \sim b_T^{-1} \gg \lqcd$, we can schematically express the perturbative expansion of $\widetilde\sigma$ as
\begin{align}
\label{eq:sigma_b_exp}
    \widetilde\sigma^W(\vect{b}_T) = f_i(x_1)f_j(x_2)\biggl\{ 1 &+  \phantom{\Bigl(}\frac{\as}{4\pi}\phantom{\Bigr)}^{\phantom{1}}  \Bigl( c_{12} L_b^2 + c_{11} L_b + c_{10}\Bigr) \\
     &+ \Bigl(\frac{\as}{4\pi}\Bigr)^2\Bigl( c_{24} L_b^4 + c_{23} L_b^3 + c_{22} L_b^2 + c_{21} L_b +  c_{20} \Bigr) \nn \\
     &+ \Bigl(\frac{\as}{4\pi}\Bigr)^3\Bigl( c_{36} L_b^6 + c_{35} L_b^5 + c_{34} L_b^4 + c_{33} L_b^3 + c_{32} L_b^2 + c_{31} L_b +  c_{30}  \Bigr) \biggr\} + \cdots \,, \nn
\end{align}
where the ellipses  indicate terms of higher order in $\as$, and 
 where we suppress for the moment the scale dependence of PDFs and $\as$ for our heuristic illustration here (but which are accounted for in the methods described below). At every order, powers of $\as^n$ are accompanied by logs of order up to $L_b^{2n}$, where $L_b = \ln(Q b_T/b_0)$, where $b_0 = 2e^{-\gamma_E}$. These logs become prohibitively large for $b_T\gg 1/Q$ (equivalently, $q_T\ll Q$). In this case  the standard perturbative expansion in small $\as \ll 1$ breaks down, thus requiring their resummation to all orders in $\as$, which requires predicting the coefficients $c_{nm}$ systematically. It turns out to be much more straightforward and systematic to predict logs in the \emph{logarithm} of the perturbative cross section \eq{sigma_b_exp}:
\begin{align}
\label{eq:log_orders}
    \widetilde \sigma^W(\vect{b}_T) = f_q(x_1)f_{\bar q}(x_2) C[\as] \exp\biggl\{ \phantom{\Bigl(}\frac{\as}{4\pi}\phantom{\Bigr)}^{\phantom{1}}  \Bigl( d_{12} L_b^2 &+ d_{11} L_b\Bigr)\\
    + \Bigl(\frac{\as}{4\pi}\Bigr)^2\Bigl( d_{23} L_b^3 &+ d_{22} L_b^2 + d_{21} L_b\Bigr) \nn \\
     + \Bigl(\frac{\as}{4\pi}\Bigr)^3\Bigl( d_{34} L_b^4 &+ d_{33} L_b^3 + d_{32} L_b^2 + d_{31} L_b\Bigr)
    \biggr\}+ \dots\,, \nonumber \\
    \small \text{LL}\ & \small\qquad \text{NLL} \quad \text{NNLL} \quad \text{N$^3$LL} \nonumber
\end{align}
where $C[\as]$ collects the constant coefficients:
\begin{equation}
    C[\as] = 1 + \sum_{n=1}^\infty \Bigl(\frac{\as}{4\pi}\Bigr)^n C_n\,,
\end{equation}
and the logs in \eq{log_orders} now organize themselves into exponentiated towers beginning with the leading log (LL) tower of terms $\as^n L_b^{n+1}$, then next-to-leading log (NLL) $\as^n L_b^n$, then NNLL $\as^n L_b^{n-1}$, etc. Heuristically, if one counts a large log as $L_b \sim 1/\alpha_s$, these towers correspond to terms of order $1/\alpha_s$ (LL), order 1 (NLL), order $\alpha_s$ (NNLL), etc. The constant coefficients $C_n$ may be included according the same heuristic counting, or included to one higher order of accuracy, which is sometimes called ``primed'' counting~\cite{Abbate:2010xh,Almeida:2014uva}.
It is the coefficients $d_{nm}$ that turn out to be most simply related to coefficients in the perturbative expansions of anomalous dimensions of objects in factorization theorems written in Chap.~\ref{sec:TMDdefn}. Achieving resummation of each tower of logs requires knowing these anomalous dimensions in TMD evolution to appropriate orders, shown later in Table~\ref{tbl:resum_orders}. 

The logs in \eq{log_orders} arise from ratios of widely separated energy or rapidity scales that contribute to the Drell-Yan cross section. The power of factorization as reviewed in \chaps{TMDdefn}{Factorization} is to separate the logs $L_b$ into separate, single-scale contributions,  e.g.
\begin{equation}
\label{eq:separation_scales}
    \ln^2 (Q b_T) = \ln^2\frac{\mu}{Q} + \Bigl[\ln^2(\mu b_T) +  2\ln(\mu b_T)\ln\frac{\nu}{\mu}\Bigr] - 2\ln(\mu b_T)\ln\frac{\nu}{Q}\,,
\end{equation}
where for illustration we have organized the logs into contributions to the double log that come from hard, soft, and beam functions in the factorized form of the cross section in \eq{sigma_new_b}. The factorized logs are now of ratios of the arbitrary scale or rapidity boundaries $\mu,\nu$ and physical scales $Q$ or $b_T$. Logs of $\mu/Q$ or $\mu b_T$ are associated with regulation of UV divergences in hard functions and the TMD PDFs, at high and low scales $Q$ and $1/b_T$. Logs of $\nu/\mu$ and $\nu/Q$ are associated with regulation of rapidity divegences and separation of collinear ($\nu\sim Q$) and soft ($\nu\sim \mu\sim 1/b_T$) degrees of freedom contributing to TMD PDFs. The evolution equations of TMD PDFs in these scales that we review in this Chapter admit solutions that take exponentiated forms that achieve the resummation of logs in \eq{log_orders}.

\subsubsection{A first glance at resummation from evolution}

Before going into the full details of TMD evolution and resummation of all the logs in a TMD cross section in \sec{TMDEvol}
 we begin with a simplified discussion of scale evolution, focusing only on the single-scale hard function describing physics above the hard scale $Q$, to illustrate the basic idea of resummation from evolution.
First, recall \eq{sigma_new},
\begin{align} \label{eq:sigma_W_reminder}
 \frac{\df\sigma^{\rm W}}{\df Q \df Y \df^2\qt} &
 = \sum_{{\rm flavors}~i}\!\! H_{i\bar i}(Q^2,\mu) \int\! \df^2\bt \, e^{i \bt \cdot \qt} \,
   \tilde f_{i/p}(x_a, \bt, \mu, \zeta_a) \,
   \tilde f_{\bar i/p}(x_b, \bt, \mu, \zeta_b)
\,.\end{align}
In the following, we will focus on the role of the UV renormalization scale $\mu$,
while the evolution in the Collins-Soper scale $\zeta_{a,b}$ is discussed below.
A priori, $\mu$ is completely arbitrary, and formally cancels exactly between the
ingredients on the right-hand side.
In practice, both the hard function and the TMD PDFs
are known only at a certain (perturbative) order, and the choice of $\mu$ becomes important.
To understand how one choses it in practice, let us inspect the first-order  perturbative
result of the hard function, which is given by
\begin{align} \label{eq:hard_nlo}
 H_{i\bar{i}}(Q^2, \mu)&=\delta_{i\, \bar i}\, \sigma_0\,  H(Q^2,\mu)\\ 
H(Q^2,\mu)& =  \left[ 1 + \frac{\as(\mu)}{4\pi} \left( - 2 \Gamma_0 \ln^2\frac{Q}{\mu} - \gamma_0 \ln\frac{Q}{\mu} + H_1 \right) \right] + \cO(\as^2) \nonumber
\,.\end{align}
Here, $\sigma_0$ is the Born cross section, $\Gamma_0$ and $\gamma_0$ are coefficients of the so-called cusp and noncusp anomalous
dimensions, while $H_1$ is a process-dependent constant. Here, a precise definition of these quantities
is not needed; we only need to know that they are fixed numeric constants.
In order to truncate \eq{hard_nlo} at $\cO(\as)$, or more generally at some finite perturbative order
$\cO(\as^n)$, we \emph{must} ensure that the coefficient in square brackets is small,
such that an expansion in $\as \ll 1$ is justified. Clearly, this can not be fulfilled
for  any arbitrary choice of $\mu$. When  $Q/\mu \sim 1$, the logarithmic terms in \eq{hard_nlo}
are indeed small, and the expansion in $\as \ll 1$ is applicable. This suggests that
in order for perturbation theory to be reliable, we must choose $\mu \sim Q$.

We can repeat the same strategy for the TMD PDFs in \eq{sigma_W_reminder}.
Since the expressions are rather cumbersome, we will not do so explicitly.
However, it is quite intuitive that since the TMD PDFs are sensitive to the scale $b_T$,
one needs to set $\mu \sim 1/b_T$ . Recalling that intuitively $q_T \sim 1/b_T$,
this implies $\mu \sim q_T$. However, this choice leads to a problem for the hard function,
since the expansion in \eq{hard_nlo} breaks down for $\mu \sim q_T$ in the region
we are interested in, namely $q_T \ll Q$ for which $\ln(Q/q_T) \gg 1$.

Fortunately, we can use evolution equations to solve this apparent conundrum.
Anticipating later results in this chapter, 
we note that the hard function
obeys the simple renormalization group equation (RGE)
\begin{align} \label{eq:hard_RGE}
 \frac{\df}{\df \ln\mu} \ln H(Q, \mu)
 = \gamma_\mu^H(Q, \mu)
 = 4  \GammaC[\as(\mu)] \ln\frac{Q}{\mu} + \gamma_\mu^H[\as(\mu)]
\,,\end{align}
which follows from studying the factorization of the cross section \eq{sigma_W_reminder} as a function of the arbitrary separation scale $\mu$ between high- and low-scale physics. Here, $\GammaC$ and $\gamma_\mu^H$ are the cusp and noncusp anomalous dimensions, whose one-loop coefficients
we already noted in \eq{hard_nlo}. \index{cusp anomalous dimension}
$\GammaC$ appears universally in QCD as the coefficient of the log-enhanced piece of the anomalous dimension of operators built out of Wilson lines meeting at an angle, or, ``cusp'', as in the matrix elements defining TMD PDFs or soft functions, to which the hard function is related through invariance of the factorized cross section as a function of $\mu$.

Now,
\eq{hard_RGE} can be easily solved as
\begin{align} \label{eq:hard_RGE_sol}
 H(Q, \mu) = H(Q, \mu_0) \exp\left[ \int_{\mu_0}^\mu \frac{\df\mu'}{\mu'} \gamma_\mu^H(Q, \mu') \right]
\,,\end{align}
where $\mu_0$ is an \emph{arbitrary} reference scale.
\eq{hard_RGE_sol} is the solution to our above problem: we can simply set
$\mu_0 \sim Q$, such that $H(Q, \mu_0)$ is reliably calculable as discussed above,
while choosing $\mu \sim q_T$ as required for the calculation of the TMD PDFs.
To see how the apparent problem of large logarithms has disappeared,
we can evaluate the integral in \eq{hard_RGE_sol}. For simplicity, we set $\mu_0 = Q$ and neglect
the $\mu$ dependence of the running coupling $\as(\mu)$, which yields
\begin{align} \label{eq:hard_RGE_sol_2}
 H(Q, \mu) &
 = H(Q, Q) \exp\left[ \int_{Q}^{\mu} \frac{\df\mu'}{\mu'} \gamma_\mu^H(Q, \mu') \right]
 \nn\\&
 \approx H(Q, Q) \exp\left[- 2 \GammaC(\as) \ln^2 \frac{Q}{\mu} - \gamma_\mu^H(\as) \ln\frac{Q}{\mu} \right]
\,.\end{align}
If we were to re-expand this exponential, we would recover the potentially large double logarithms
$\as \ln^2(Q/\mu)$ in \eq{hard_nlo} we were worried about spoiling perturbation theory.
However, as they now appear only in the exponential, they do not deteriorate our perturbative results,
and we are free to set $\mu \sim q_T$ as desired.
(Note the overall minus in the exponential, which guarantees exponential suppression as $q_T / Q \to 0$.)
This is referred to as \emph{resummation}, as a whole class of logarithms $\as^n L^{2n}$
have been summed to all orders in perturbation theory as anticipated in \eq{log_orders}. 

In the following sections we will complete the discussion of evolution of the TMD PDF factors in \eq{sigma_W_reminder} and the summation of all logs of $q_T/Q$ or $Qb_T$ appearing in the perturbative expansions of the TMD cross section \eq{DY_bspace}.

\subsection{TMD Evolution}
\label{sec:TMDEvol}
\index{evolution!TMD}
In this section we will review the derivation of the TMD evolution equations both from the ``direct'' QCD approaches like CSS, and from the EFT framework of SCET. Both routes to the evolution equations are ultimately equivalent, but highlight complementary aspects of TMD evolution and inspire various methods to obtain their explicit solutions.

\begin{figure}[t]
\centering
\vspace{-12pt}
\includegraphics[width=.5\textwidth]{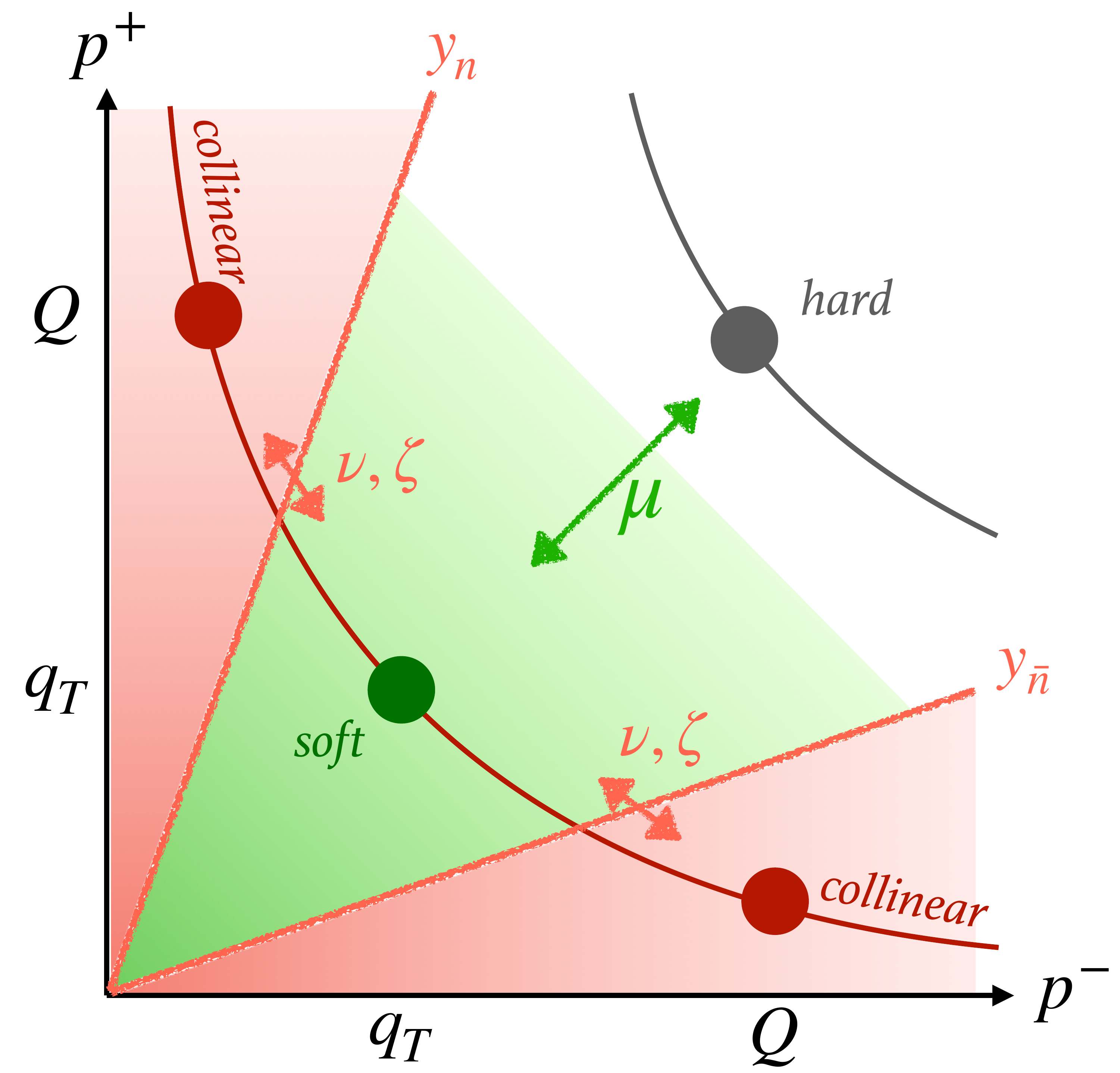}
\vspace{-6pt}
\caption{Momentum regions for TMD factorization and evolution. UV regulators like $\MSbar$ separate hard momentum regions from IR collinear/soft regions. The variation of this arbitrary boundary leads to the $\mu$-RGEs. Rapidity regulators such as those reviewed in Chapter \ref{sec:TMDdefn} define schemes for separating collinear and soft regions from one another (e.g. $\nu,\zeta,y_{n,\bar n}$), and variation of these boundaries leads to rapidity RGEs, i.e., Collins-Soper equation. The rapidity evolution kernel has its own $\mu$-RGE, capturing variation of the scale $\mu$ where the rapidity factorization/evolution occurs. Solutions of the $\mu$ and rapidity RGEs sum large logs of ratios of mass and rapidity scales of these separated regions that appear in perturbative expansions of TMD cross sections.}\vspace{-1em}
\label{fig:regions}
\end{figure}

In \chap{TMDdefn} the quantum field theory
definitions of TMD PDFs as composite operators were established 
with the unique role played by the soft factors which are  essential for the  consistency  of  TMD definitions and their validity in a factorization formula like \eq{sigma_new}.   Essential to these definitions are the subtraction of UV and rapidity divergences  resulting in the renormalized TMD PDFs. As a consequence (as stated earlier), the TMD PDFs depend on two auxiliary parameters, the rapidity and renormalization  scales, $\zeta$ and $\mu$ respectively.
For removal of rapidity divergences, various schemes were summarized in \sec{tmd_defs} with  corresponding scheme dependent rapidity scales; mainly depending on the implementation of rapidity subtraction  through the soft factor.
For a summary of the various rapidity regulator schemes, see Table \ref{tbl:overview_tmdpdfs} in \sec{tmd_defs_overview}, and  \app{TMDdefn}.
The role of these regulators in separating UV/IR and soft/collinear momentum regions is illustrated schematically in \fig{regions}. The  invariance of factorized cross section ~$d\sigma^W$  with respect to these scales
results in a system of differential equations that determines the scale dependence of the TMDs. These are the TMD evolution equations.

\begin{table}[t]
\small
 \centering
 \bgroup
\def\arraystretch{1.5}
\begin{tabular}{|c|c|c|c|}
 \hline
  \emph{Anomalous dimensions:}  &  TMD PDF $\mu$ RG \eqref{eq:RG.TMD.pdf} & CS kernel/RRG \eqref{eq:CSS.evol} & Cusp \eqref{eq:RG.K}
 \\ \hhline{|=|=|=|=|}
 CSS       & $\gamma_q[\as(\mu);\zeta/\mu^2]$  & $\tilde K(b_T;\mu)$   & $-\gamma_{K}[\as(\mu)]$ \\ \hline
 SCET RRG \cite{Chiu:2011qc,Chiu:2012ir} &  $\gamma_\mu^q(\mu,\zeta)$ & $\gamma_\zeta^q(\mu,b_T) = -\gamma_\nu^q(b_T,\mu)$ &   $-2\GammaC^{q}[\as(\mu)]$ \\ \hline
 EIS \cite{GarciaEchevarria:2011rb,Scimemi:2018xaf} &  $\gamma_F^q(\mu,\zeta)$ & $-2\mathcal{D}^q(\mu,b_T)$ &   $-2\GammaC^{q}(\mu)$ \\ \hline
 BN \cite{Becher:2010tm} & $-\GammaC^q[\as]L_\perp +2\gamma^q[\as]$ & $-F_{q\bar q}(L_\perp,\as)$ &   $-2\GammaC^q[\as]$ \\ \hline
 \hline
\emph{Alternate organization:}  &  \eq{JMY.muRGE} & \eq{zetarg} & \eq{reno} \\ \hline
 JMY \cite{Ji:2004wu}      & $2\gamma_F[\as(\mu)]-\gamma_S(\mu,\rho)$  & $2[K(\mu,b_T) + G(\mu,\zeta)]$   & $-\gamma_{K}[\as(\mu)]$ \\ \hline\hline
 \end{tabular}
 \egroup
 \caption{Common notations for anomalous dimensions of quark TMD PDFs in TMD evolution,  \eq{TMD.evol}. The first column of anomalous dimensions are for $\mu$-RG from UV renormalization of the TMD PDFs in \eq{RG.TMD.pdf}. The second column is the Collins-Soper kernel or rapidity anomalous dimension of the TMD PDFs in \eq{CSS.evol}. The final column gives different names for the ``cusp'' anomalous dimension, which appears in \eq{RG.K} as the $\mu$-RG anomalous dimension of the CS kernel itself, i.e., the mixed $\mu,\sqrt{\zeta}$ derivative of the TMD PDF. For the BN row, $L_\perp = \ln(\bt^2 \mu^2/b_0^2)$,
 and the $\zeta$ dependence in the evolution is not actually explicit, see comments below \eq{BNdef}. 
  For the JMY row, note that the pieces of the evolution are organized slightly differently, as expressed in 
 \eqs{JMY.muRGE}{zetarg}, and the results are given for the subtracted TMD PDF in \eq{tmdpdf_0}, which also has dependence on an extra scheme parameter $\rho$. The universal anomalous dimension $\gamma_K$ of the CS kernel is the same in \eqs{RG.K}{reno}. (A similar table can made for gluon TMDs, with $q\to g$ where appropriate and CS kernel for gluons.)
 }
 \label{tbl:anom_dim_notation}
\end{table}

Both the
CSS~\cite{Collins:1981uk,Collins:1981uw,Collins:2011zzd,Aybat:2011zv}
and SCET~\cite{Becher:2010tm, Becher:2011xn, Becher:2012yn, GarciaEchevarria:2011rb, Echevarria:2012js, Echevarria:2014rua, Chiu:2012ir,Ebert:2019tvc} 
formalisms lead to a common set of evolution equations for the generic TMD PDF defined in \eq{tmdpdf_1}:
\index{evolution!CSS}
\index{evolution!SCET}
\index{evolution!TMD}
\begin{subequations}
\label{eq:TMD.evol}
\begin{align}
\label{eq:RG.TMD.pdf}
   \frac{ d \ln \tf_{i/p}(x,{\bt};\mu,\zeta) }{ d \ln \mu } &\quad\overset{\text{CSS}}{=} \quad \gamma_{q}[\alpha_s(\mu);\zeta/\mu^2] & \overset{\text{SCET}}{=} & \quad  \gamma_\mu^q(\mu,\zeta)\,, \hspace{2em} \\
\label{eq:CSS.evol}
  \frac{ \partial \ln \tf_{i/p}(x,{\bt}; \mu, \zeta) }
       { \partial \ln \sqrt{\zeta} }
  &\quad\overset{\text{\phantom{CSS}}}{=} \quad
  \tilde{K}({b_T};\mu) & \overset{\text{\phantom{SCET}}}{=} & \quad \gamma_\zeta^q(\mu,b_T) \, , \hspace{2em}\\
\label{eq:RG.K}
   \frac{ d \tilde{K}({b_T};\mu) }{ d \ln \mu  }
  &\quad\overset{\text{\phantom{CSS}}}{=} \quad -\gamma_{K}[\alpha_s(\mu)] & \overset{\text{\phantom{SCET}}}{=} & \quad -2 \GammaC^{q}[\as(\mu)]\, ,\hspace{2em}
   \end{align}
  \end{subequations}
where we have shown typical names given to each anomalous dimension in much of the CSS- and SCET-based literature. These and some other common notations are summarized in Table~\ref{tbl:anom_dim_notation}. These equations are for quark TMD PDFs of flavor $q$; analogous equations, with appropriate anomalous dimensions, hold for gluon TMD PDFs.  

The first equation \eqref{eq:RG.TMD.pdf} expresses the usual RG evolution in $\mu$ from UV renormalization. The second equation \eq{CSS.evol},  is the Collins-Soper equation, which  expresses the evolution in the Collins-Soper scale $\zeta$ resulting from regulating rapidity divergences.  $\tilde{K}$, or $\gamma_\zeta^q$, the rapidity anomalous dimension is the Collins-Soper kernel.  It is independent of $x$ and $\zeta$ and the flavor of the parton and the hadron in the PDF, however it does depend on the color representation for the parton; there is one for quarks and another for the gluon~\cite{Collins:2012ss}. Noting that the only dependence of $\tf_{i/p}(x,{\bt};\mu,\zeta)$ on $\zeta$   (or $y_n$) (see Eq.~\eqref{eq:zeta})  
 is through the soft factor, from the
 definition of $\tf_{i/p}(x,{\bt};\mu,\zeta)$ one obtains  by direct computation~\cite{Collins:2011zzd,Aybat:2011zv},
 \begin{align}\label{eq:CSkern}
 \tilde K(b_T;\mu) & =\lim_{\substack{y_A\to+\infty\\y_B\to-\infty}}\frac{1}{2}\frac{\partial}{\partial y_n}
\ln\left(\frac{\tilde S^{0}_{n_A n_B}(b_T,\epsilon, y_n- y_B)}{\tilde S^{0}_{n_A n_B}(b_T,\epsilon, y_A-y_n)}\right) + {\rm UV~counterterm}\  .
 \end{align}
 In SCET these equations will arise similarly from RG and rapidity RG evolution of beam and soft functions. 
The third equation expresses the $\mu$-RG evolution of the Collins-Soper kernel $\tilde K$ or rapidity anomalous dimension $\gamma_\zeta$.  Taking the derivative of  \eqref{eq:RG.TMD.pdf} with respect to $\ln\zeta$ 
and   using  that the mixed $\mu,\zeta$ second derivatives of the TMD PDF are equal, we immediately find,
\begin{align}
\label{eq:mu_zeta_consistency}
\frac{\partial \gamma_q[\as(\mu);\zeta/\mu^2]}{\partial \ln\sqrt{\zeta}}&=-\gamma_{K}[\as(\mu)].
\end{align}
This   imposes a consistency condition on the anomalous dimensions in \eqs{RG.TMD.pdf}{CSS.evol}. They imply a relation between $\gamma_q,\gamma_{K}$ (i.e., $\gamma_\mu^q,\gamma_\zeta^q$). Now one can easily integrate this equation with respect to $\zeta$, where we choose $\zeta_0\sim \mu^2$. As a result the anomalous dimension of the TMD PDF $\gamma_q$ has linear dependence on  $\ln\left(\zeta/\mu^2\right)$; expressing this in terms of the CSS and SCET  notation from Table \ref{tbl:anom_dim_notation} one obtains, 
\begin{subequations}
\label{eq:mu_anom_dim}
\begin{align}
    \gamma_{q}[\alpha_s(\mu);\zeta/\mu^2] & =-\frac{1}{2}\gamma_{K}[\alpha_s(\mu)]\ln\frac{\zeta}{\mu^2}
    +
    \gamma_{q}[\alpha_s(\mu);1] \, , \label{eq:SCET.mu.anomdima}\\
    &\hspace{-6cm} \text{or}\nn \\  
    \gamma_\mu^q(\mu,\zeta) &=-\Gamma_\text{cusp}^q[\as(\mu)] \ln \frac{\zeta}{\mu^2} + \gamma_\mu^q[\as(\mu)]\,, 
    \label{eq:SCET.mu.anomdim}
\end{align}
\end{subequations}
that is, the coefficient of the log in the UV anomalous dimension of the TMD PDF is the
anomalous dimension $\gamma_{K}$ or $\GammaC^q$ of the Collins-Soper kernel in \eq{RG.K}. 
\index{cusp anomalous dimension}

Further, from  \eqref{eq:RG.K} we can straightforwardly integrate with respect to $\mu$ and thus, the Collins-Soper kernel or rapidity anomalous dimension itself  take the form,
\index{Collins-Soper evolution kernel}
\begin{subequations}
\label{eq:CSkernel_form}
\begin{align}
\tilde K(b_T;\mu)&= -\int_{1/\bar b_T}^\mu \frac{d\mu'}{\mu'} \gamma_{K} [\as(\mu')] + \tilde K(b_T,1/\bar b_T) \,, \\
&\hspace{-4.5cm} \text{or}\nn \\  
 \gamma_\zeta^q(\mu,b_T) &= -2\int_{1/\bar b_T}^\mu \frac{d\mu'}{\mu'} \Gamma_\text{cusp}^q [\as(\mu')] + \gamma_\zeta^q[\as(1/\bar b_T)]\,.
 \label{eq:gamma_zeta_SCET_form}
\end{align}
\end{subequations}
where  $\bar b_T = b_T/b_0$, recalling $b_0= 2e^{-\gamma_E}$. 
The first terms on the right-hand side of each expression in \eq{CSkernel_form} are predicted by the RGE \eq{RG.K} of the Collins-Soper kernel, while the latter term is a boundary condition, which is not predicted by the RGE itself. It can in principle be specified at any scale $\mu_0$ but it is conventional to choose it at $\mu_0=1/\bar b_T$ as in \eq{CSkernel_form}. With this choice, explicit logs of $\mu_0 \bar b_T$ in the perturbative expansion of the non-cusp anomalous dimension are eliminated, hence its sole dependence on $\as(1/\bar b_T)$ in the form \eq{gamma_zeta_SCET_form}.\footnote{There is an unfortunate historical convention in much of the SCET literature that the full anomalous dimension, as on the LHS of \eq{gamma_zeta_SCET_form}, and its non-cusp piece, as on the RHS of the same equation, are given the same symbolic name, e.g. $\gamma_\zeta^q$, but distinguished by the form of their arguments, i.e. the non-cusp piece being specified $\gamma_\zeta^q[\alpha_s]$ signifying that it is given by an expansion in $\as$ with pure numerical coefficients, \eq{cusp-non}.}
For perturbative values of $1/\bar b_T$, it can be predicted in fixed-order perturbation theory. For nonperturbative $1/\bar b_T$, it should be obtained via a nonperturbative model 
via ab initio calculations such as by lattice QCD, nonperturbative models, and global analyses of related experimental data.
The latter program is the topic of Chap.~\ref{sec:phenoTMDs}.
Identical relations to the above hold for gluon TMD PDF anomalous dimensions.

The cusp anomalous dimension $\Gamma_\text{cusp}$ 
\index{cusp anomalous dimension}
in \eq{CSkernel_form}, to which $\gamma_K$ is equivalent, is a universal object appearing ubiquitously in QCD (see, e.g., \cite{Polyakov:1980ca,Korchemsky:1985xj,Korchemsky:1987wg,Collins:1989bt}). It appears due to divergences in matrix elements of operators built out of Wilson lines in different directions meeting at an angle, forming a ``cusp'', such as between two jets or hadronic beams. The angle may be $\pi$ for back-to-back configurations. The hard function, for example, in \eq{hard_RGE} is associated with Wilson lines in the fundamental representation of SU(3) in two lightlike directions, and has an anomalous dimension whose log-enhanced piece is known to have the coefficient $\Gamma_\text{cusp}$. It is known to two \cite{Korchemsky:1987wg}, three \cite{Moch:2004pa}, and recently even four-loop \cite{Henn:2019swt} order in QCD. The consistency relations between anomalous dimensions of hard and TMD PDF pieces of the cross section \eq{sigma_W_reminder}, as well as between UV and rapidity evolution of the TMD PDFs expressed in \eq{sigma_W_reminder}, guarantee the further universality between the cusp anomalous dimension and the rapidity anomalous dimension.

Finally, we  
obtain the generic solution for the evolved TMD PDF in \eq{tmd_sol_schematic}  by 
performing  integration  on the rapidity parameter $\zeta$ in~\eq{CSS.evol},  
and integration on the renormalization scale $\mu$ in \eq{RG.TMD.pdf},
where we have evolved the TMD PDF from the pair of initial to final scales
$\{\mu_0,\zeta_0\}\rightarrow \{\mu,\zeta\}$, 
and illustrated by \fig{commute},
\begin{align}\label{eq:solution_tmd_intro}
\tf_{i/P}(x, \bt, \mu, \zeta)
    &=\tf_{i/P}(x,\bt,\mu_0,\zeta_0)\  \exp\left\{\int_{\mu_0}^\mu\frac{d\mu'}{\mu'}\gamma_q\bigl[\alpha_s(\mu');\zeta_0/\mu'^2\bigr]\right\}
    \exp\left\{\tilde{K}(b_T;\mu)\ln\sqrt{\frac{\zeta}{\zeta_0}}\right\}\, ,
\end{align}
which gives definitions to the RG and RRG evolution kernels $U$ and $V$ in \eq{tmd_sol_schematic}. The RG evolution between scales $\mu_0,\mu$ is governed by the anomalous dimension $\gamma_q$, and the rapidity evolution between the rapidity scales $\zeta_0,\zeta$ by the Collins-Soper kernel $\tilde K$.
  
Below we will review how the evolution equations \eq{TMD.evol}, and solutions \eqref{eq:solution_tmd_intro} are applied  to exploit  the universality properties of the nonperturbative content which emphasize the intrinsic properties associated with hadronic structure in \eqref{eq:solution_tmd_intro}, 
as well as the 
perturbative content which are optimized to have no large logarithms in their expansion in powers of $\alpha_s$.
First we consider  the treatment in the CSS formalism, in which the Wilson lines in the definition of the TMD PDF are tilted away from the light-cone. Then we cover the treatment in the SCET framework, in which the TMD PDFs are factored into beam and soft functions, each with their own RG and rapidity RG evolution equations, which combine to give the same TMD evolution equations \eq{TMD.evol}. We will also review how explicit forms for the solutions of these equations can be written, first in $\bt$ space and then transformed to momentum space, and also directly in momentum space. The difference in various prescriptions or approaches to doing this amounts to alternative choices (implicit or explicit) for the low scales from which the TMD PDFs are evolved.

\begin{table}[t!]
\small
 \centering
 \bgroup
\def\arraystretch{1.5}
\begin{tabular}{|c|c|c|c|c|c|c|}
 \hline
$\gamma_K\left(\as(\mu)\right) $ & $\beta[\as(\mu)]$  &  $\gamma_q\left(\as(\mu);1\right)$ & $\tilde K(\bar b_T; 1/\bar b_T)$  & $\tilde C_{j/j'}$ & accuracy & accuracy (SCET)
 \\ \hhline{|=|=|=|=|=|=|=|}
 ---      & ---  &  ---  & --- & 0 &  QPM  &      \\ \hline
 1      & 1 & --- &  --- & 0&   LO-LL & LL  \\ \hline
 2 &  2 & 1 &   1 &  0&   LO-NLL & NLL    \\ \hline
 3 & 3& 2&   2 & 0 &   LO-NNLL &   \\ \hline
 2 &  2& 1 & 1 & 1 &  NLO-NLL & NLL$'$   \\ \hline
 3    & 3  & 2  & 2 & 1    &  NLO-NNLL & NNLL  \\ \hline
 3    & 3  & 2  & 2 & 2 & NNLO-NNLL & NNLL$'$ \\ \hline
4   & 4  & 3  & 3 & 2 & NNLO-N$^3$LL & N$^3$LL \\ \hline
4   & 4  & 3  & 3 & 3 & N$^3$LO-N$^3$LL & N$^3$LL' \\ \hline
 \end{tabular}
 \egroup
 \caption{Orders of accuracy needed for evolution of TMD PDFs and other ingredients entering  the transverse momentum dependent  $W$ term needed to achieve given orders of logarithmic accuracy (LL, NLL, NNLL, etc.). The numbers refer to the
 loop order $k$ 
  to which the quantity needs to be computed, e.g. $k=n+1$ in terms of the coefficients in the expansions \eqs{cusp-non}{beta_expansion} of the anomalous dimensions or beta function.
 (A dash ``---'' indicates the quantity does not exist at $\as^0$.) The names of anomalous dimensions are those in the CSS row of Table~\ref{tbl:anom_dim_notation}, same counting applies to other rows. We also include the needed accuracies for coefficient functions $\tilde C$ that will appear in Eq.~\eqref{e:OPE_f1_old} (which corresponds to perturbative expansions of beam and soft functions in SCET in \eq{DYbspace}.) 
``QPM'' refers to the \emph{quark parton model}, \index{parton model} i.e., Born-level.
This table describes the accuracy of the resummed $W$ term; a full prediction for a TMD cross section will include matching to a fixed-order ``$Y$'' term whose accuracy is specified separately, see \sec{nonsingular}.
 }
 \label{tbl:resum_orders}
\end{table}

First, however in Table~\ref{tbl:resum_orders} we summarize the orders of accuracy to which the  anomalous dimensions and other relevant quantities (the beta function for running of $\as$ and fixed-order coefficient functions in, e.g., Eq.~\eqref{e:OPE_f1_old}), need to be known, in powers $k$ of $\as^k$, to achieve the orders of accuracy in resummed logs illustrated in \eq{log_orders}. In the next section we illustrate the calculations to leading order in $\alpha_s$.

\subsubsection{One-loop examples}
From the calculations of quark TMD PDFs in perturbation theory in \sec{tmd_defs_nlo}, we can illustrate how to obtain their UV and rapidity anomalous dimensions to one-loop order. For higher-order results, see \app{kernelexpansions}.

We recall that the UV anomalous dimension is associated with the behavior of the TMD PDF as the arbitrary boundary between hard scales $Q$ and low scales $q_T$ or $1/b_T$ is varied, while the rapidity anomalous dimension is associated with its behavior as the (arbitrary) boundaries between forward/backward and central rapidities is varied. 
At one-loop order, the variation with respect to these boundaries can be obtained from the soft and collinear divergences in the one-loop graphs shown in \fig{beam_nlo}. From the one-loop result for the bare TMD PDF in \eq{beam_nlo_5} using $\overline{\text{MS}}$ to regulate the UV divergences and the $\eta$ regulator for rapidity divergences, one obtains the UV renormalization factor \eq{Zq_factor}, from which the anomalous dimension of the renormalized TMD PDF in \eq{tmdpdf_nlo_renom} can be obtained by the condition of $\mu$-independence of the bare TMD PDF,
\begin{equation}
\label{eq:gammamuq_from_Z}
\gamma_\mu^q(\mu,\zeta) = -(Z_\text{uv}^q)^{-1} \mu\frac{d}{d\mu}Z_\text{uv}^q\,,
\end{equation}
which to one loop gives
\begin{equation}
\gamma_\mu^q(\mu,\zeta) = \frac{\as(\mu)C_F}{\pi} \biggl(\frac{3}{2} + \ln\frac{\mu^2}{\zeta}\biggr)\,,
\end{equation}
where in evaluating the $\mu$ derivative in \eq{gammamuq_from_Z} it is important to remember the relation \eq{gbareinMSbar} between the bare and renormalized coupling constants. Equivalently one can just take the $d/d\ln\mu$ derivative of the renormalized TMD PDF itself, \eq{tmdpdf_nlo_renom}. 

Meanwhile the Collins-Soper kernel, or equivalently the rapidity anomalous dimension, is obtained from the $\ln\sqrt{\zeta}$ derivative of the renormalized TMD PDF in \eq{tmdpdf_nlo_renom}, which gives to one loop
\begin{equation}
\gamma_\zeta^q(\mu,b_T) = -\frac{2\as(\mu)C_F}{\pi} \ln\frac{\mu b_T}{b_0}\,,
\end{equation}
so that at one loop in perturbation theory the non-cusp part of the Collins-Soper kernel is zero. (It begins at two loops.)  Alternatively, the anomalous dimension $\gamma_\zeta$ can be reconstructed from those of the beam and soft functions in \eq{B_renorm_eta_2} and \eq{soft_renorm_eta}. For these pieces, the anomalous dimensions can be obtained from the rapidity renormalization in the RRG, differentiating the $\tilde Z_{B,S}$ renormalization factors in \eq{beam_Z} and \eq{soft_renorm_eta} with respect to $\ln\nu$, keeping in mind the $\nu$ dependence of the bookkeeping parameter $w(\tau,\nu)$ appearing in \eq{R_eta_soft}. We leave the explicit formulae for the anomalous dimensions $\gamma_\nu^{B,S}$ for \sec{RGandRRG}.

We see also at one loop the consistency relation \eq{RG.K} between mixed derivatives of the TMD PDFs, giving the RG evolution of the Collins-Soper kernel itself, is explicitly satisfied,
\begin{equation}
\sqrt{\zeta}\frac{d}{d\sqrt{\zeta}}\gamma_\mu^q = \mu\frac{d}{d\mu}\gamma_\zeta^q = - \frac{2\as(\mu)C_F}{\pi} = -2\Gamma_\text{cusp}[\as]\,.
\end{equation}
Results for anomalous dimensions to two loops and beyond are given in \app{kernelexpansions}.

In the next subsections we review how to solve the TMD evolution equations in general, to arbitrary orders in the perturbative expansions of anomalous dimensions, in both CSS and SCET formalisms.

\subsection{CSS Formalism}
\label{sec:CoordEvol}
\index{evolution!CSS}
In this section we will consider rapidity regularization based on TMD factorization of Collins~\cite{Collins:2011zzd,Aybat:2011zv,Collins:2014jpa,Collins:2017oxh} and also the earlier scheme of Ji, Ma, and Yuan~\cite{Ji:2004wu,Ji:2004xq}.
We take as a starting point the Drell-Yan cross section in terms of the TMD $d\sigma^{\rm W}$ term.   
In Collins'  formulation~\cite{Collins:2011zzd}, TMD factorization is carried out in $\bt$ configuration space\footnote{See \sec{momentumspace} for some methods to perform resummation in momentum space.}, 
where the cross section is expressed in terms of  Fourier transforms of the TMD PDFs, ~\eqref{eq:sigma_W_reminder}.

\subsubsection{The CSS Solution}

Here we present the solution optimized for perturbative calculations~\cite{Collins:2011zzd,Aybat:2011zv,Collins:2014jpa}.
  We begin with \eq{sigma_W_reminder}. 
  Since the solution to the evolution equation is 
   independent of the path traced out in the $\{\mu,\zeta\}$ phase space 
  of rapidity and energy, other choices are also possible~\cite{Chiu:2012ir,Scimemi:2018xaf}; for a discussion, 
  see section~\ref{sec:2dRRGE}.  
  
First, we point out that if we insert the solution \eqref{eq:solution_tmd_intro} into the $W$ term, Eq.~\eqref{eq:sigma_W_reminder} and perform the Fourier integral where $\bsc$
extends from 
$\bsc=[0,\infty)$,  one can not avoid using the parton densities and TMD evolution kernel in the nonperturbative large $\bsc$ region. Furthermore, a  fixed order perturbative expansion of $\tilde K(\bsc ,\mu)$ will    encounter large logarithmic contributions from higher order terms and thus, a perturbative treatment of $\bsc$ is not reliable. 
  For these reasons, Collins,~et.~al.~\cite{Collins:1984kg,Collins:2011zzd,Collins:2014jpa} provided a prescription  that aims to maximize the use of perturbation theory for small $\bT$ and at the same time, combine nonperturbative information; that is to match  perturbative and nonperturbative properties of the TMD factorization formulation.
A widely used scheme to  separate perturbative and nonperturbative contributions
partitions the  large and small $\Tsc{b}$  via a function $\bstarsc$ that 
freezes above some $\bmax$ and equals $\Tsc{b}$ for small $\Tsc{b}$:
\begin{equation}
\bstarsc(\Tsc{b}) \longrightarrow
\begin{dcases}
\Tsc{b} & \Tsc{b} \ll \bmax \\
\bmax & \Tsc{b} \gg \bmax \, . \label{eq:bdef}
\end{dcases}
\end{equation}
 Here we adopt the 
$\bstarsc$ prescription~\cite{Collins:2014jpa} by replacing $\bt$ in the solution to the evolution equations  by the function
\begin{align}
\label{eq:bstardef}
    \bts \equiv \bts(\bt) = \frac{\bt} {\sqrt{1+b_T^2/b_{max}^2}}.
\end{align}
 Note that $\bstarsc$ freezes at $b_{max}$ when $b_T$ is large so that $\bstarsc$ is always small (i.e., in the perturbative region).
Other choices can be found in~\cite{Bacchetta:2015ora,Bacchetta:2017gcc}. This definition is constructed so that it equals $\bt$ for small values and smoothly approaches the upper cutoff $\bmax$ when $\bt$ becomes large. Typical values of $\bmax\sim 1 \, {\rm GeV^{-1}}$ and can be thought of as characterizing a boundary the perturbative and nonperturbative $\bt$-dependence~\cite{Collins:2011zzd,Aybat:2011zv}.
We can now use this to  match the perturbative and nonperturbative pieces of the TMD PDF.
 To do this, we partition the left-hand side of \eq{solution_tmd_intro}  into the part of $\tf$ at $\bts$
 through the identity,
\begin{align}
\label{eq:np}
\tf_{i/p}(x,\bt,\mu,\zeta)=\tf_{i/p}(x,\bts,\mu,\zeta)\frac{\tf_{i/p}(x,\bt,\mu,\zeta)}{\tf_{i/p}(x,\bts,\mu,\zeta)}
\end{align}
which we calculate perturbatively and a part that we deem "intrinsically nonperturbative".

At large $\bt$ the expression for the evolved TMD PDFs are defined by the deviation of $f_{j/p}(x,\bt,\mu,\zeta)$ and 
$\tilde K(\bsc;\mu)$ between $\bt$ and $\bts$ in terms of the nonperturbative universal and scale independent functions  $g_{j/p}(x,\bsc,\bmax)$ and $g_k(\bsc;\bmax)$. They are defined through the ratio in Eq.~\eqref{eq:np},
\begin{align}
\label{eq:rat}
\frac{\tf_{i/p}(x,\bt,\mu,\zeta)}{\tf_{i/p}(x,\bts,\mu,\zeta)}&
=\frac{\tf_{i/p}(x,\bt,\mu_0,\zeta_0^\prime )}{\tf_{i/p}(x,\bts,\mu_0,\zeta_0^\prime)}
\exp \left[\ln\sqrt{\frac{\zeta}{\zeta_0^\prime}}\left(\tilde K(\bsc,\mu)-\tilde K(\bstarsc,\mu)\right)\right]\nonumber\\
&=\exp\left[-g_{i/p}(x,\bsc)\right]
\exp \left[-\ln\sqrt{\frac{\zeta}{\zeta_0^\prime}} g_k(\bsc;\bmax)\right]\, ,
\end{align}
where 
RRG and RG transformations are performed,  $\zeta\rightarrow \zeta_0^\prime$ and $\mu\rightarrow \mu_0$~\cite{Collins:2011zzd} 
respectively ({\em n.b.} the effects of anomalous dimension, $\gamma_q$,  cancel), and  where  the nonperturbative part of the Collins Soper kernel $g_k(\bsc;\bmax)$ is, 
\begin{align}
\label{eq:gk}
    g_k(\bsc;\bmax)&=\tilde K(\bstarsc, \mu_0)- \tilde K(\bsc, \mu_0)\, ,
\end{align}
and the intrinsic transverse momentum distribution (in Fourier space) is given  by the exponent of $g_{i/p}(x,\bsc)$. 
 The arbitrary reference scale, $\zeta_0^\prime$  
determines how much of the of the TMD density is in $g_{i/p}(x,\bsc)$ 
and how much is put into the exponential of $g_k$ times the log in Eq.~\eqref{eq:np}~\cite{Collins:2014jpa}.
From Eq.~\eqref{eq:rat} both $g_{i/p}$ and $g_k$ vanish as $\bsc\rightarrow 0$~\cite{Collins:1984kg,Collins:2014jpa}.  Also, both functions are independent of $\zeta$ and $\mu$ because there is an exact cancellation
in terms obtained by applying $\mu$-RG and $\zeta$-RRG transformations  to Eqs.~\eqref{eq:gk} and~\eqref{eq:rat}. 
Now substituting Eqs.~\eqref{eq:rat} and \eqref{eq:gk} and  using \eqref{eq:SCET.mu.anomdima} in \eqref{eq:np}, \eq{solution_tmd_intro} can be expressed as, 
\begin{align}\label{eq:sol1}
    \tf_{i/p}(x,\bt,\mu,\zeta)&=\tf_{i/p}(x,\bts,\mu_0,\zeta_0)\nonumber \\
    &\times \exp \left[\ln\sqrt{\frac{\zeta}{\zeta_0}} \tilde K(\bstarsc,\mu_0)+\int_{\mu_{0}}^\mu\frac{d\mu'}{\mu'}\left(\gamma_{q}[\alpha_s(\mu');1] 
    -\ln\frac{\sqrt{\zeta}}{\mu'} \gamma_{K}[\alpha_s(\mu')]\right)\right]\nonumber \\
   & \times \exp\left[-g_{i/p}(x,\bsc)-\ln\left(\sqrt{\frac{\zeta}{\zeta_0^\prime}}\right) g_k(\bsc;\bmax)\right].
\end{align}
Finally to optimize the solution for perturbative calculations, 
 RG and RRG  transformations are
performed,  $\mu_0\rightarrow 1/\bstarsc $ and $\zeta_0\rightarrow 1/\bstarsc^2$ respectively, 
permitting perturbative calculations of  $\tilde K$ and $\tilde f$~\cite{Collins:2011zzd},
where now \eqref{eq:sol1} becomes
\begin{align}\label{eq:sol2}
    \tf_{i/p}(x,\bt,\mu,\zeta)&=
    \tf_{i/p}(x,\bts,\mubstar,\mubstar^2)\nonumber \\
    &\times \exp \left[\ln\frac{\sqrt{\zeta}}{\mubstar}\tilde K(\bstarsc,\mubstar)+
    \int_{\mubstar}^\mu\frac{d\mu'}{\mu'}\left(\gamma_{q}[\alpha_s(\mu');1] 
    -\ln\frac{\sqrt{\zeta}}{\mu'} \gamma_{K}[\alpha_s(\mu')]\right)\right]\nonumber \\
   & \times \exp\left[-g_{i/p}(x,\bsc)-\ln\left(\sqrt{\frac{\zeta}{\zeta_0^\prime}}\right) g_k(\bsc;\bmax)\right]\, ,
\end{align}
with 
\begin{equation}
\mu_{b_*} \equiv \frac{C_1} {\bstarsc}\,,
\end{equation}
where $C_1/\bstarsc$ is the hard scale. It is chosen to allow perturbative calculations of $\bstarsc$-dependent quantities and where $C_1 $ is a constant 
of order unity chosen to allow for perturbative calculations
without large logarithms~\cite{Collins:2011zzd,Collins:2014jpa}.

Thus, 
we can express the TMD parton densities at small $\bt$ in terms of the integrated PDFs using an operator product expansion as expressed in \sec{largeqT}, where now, Eq.~\eqref{eq:tmdpdf_matching} takes the form,
\begin{align}
\tf_{i/p}(x,\bts,\mubstar,\mubstar^2) &= \sum_{j}\int_x^1\!\frac{d\hat{x}} {\hat{x}}\,
 \tilde{C}_{i/j}(x/\hat{x},\bsc;\mubstar,\mubstar^2,\alpha_s(\mubstar))\,f_{j/p}(\hat{x};\mubstar)
+ {O((m\,b_*(b_T))^p)}\, .\label{e:OPE_f1_old}
\end{align}
Thus,  the first line of \eqref{eq:sol2}
is expressed in terms of the collinear pdfs using an OPE in terms of collinear PDFs~\cite{Collins:2011zzd,Aybat:2011zv,Collins:2014jpa},
\begin{align}
    \tf_{i/p}(x,\bt,\mu,\zeta)&= \sum_{j}\int_x^1\!\frac{d\hat{x}} {\hat{x}}\,\tilde{C}_{i/j}(x/\hat{x},\bsc;\mubstar,\mubstar^2,\alpha_s(\mubstar))\,f_{j/p}(\hat{x};\mubstar) \nonumber \\ \nonumber
    &\times \exp \left[\ln\frac{\sqrt{\zeta}}{\mubstar}\tilde K(\bstarsc,\mubstar)+
    \int_{\mubstar}^\mu\frac{d\mu'}{\mu'}\left(\gamma_{q}[\alpha_s(\mu');1] 
    -\ln\frac{\sqrt{\zeta}}{\mu'} \gamma_{K}[\alpha_s(\mu')]\right)\right]\nonumber \\
   & \times \exp\left[-g_{i/p}(x,\bsc)-\ln\left(\sqrt{\frac{\zeta}{\zeta_0^\prime}}\right) g_k(\bsc;\bmax)\right]\, .
\end{align}
The sum is over all flavors $j$ of partons: quarks, anti-quarks and gluons
and $f_{j}(\hat{x};\mu_{b_*})$ is understood to be renormalized at the scale $\mubstar$.

Finally, we mention that there are some alternatives in the literature to the $\bstarsc$-prescription. 
Here, we sketch out the approach  propsed by Qiu and Zhang. In Refs.~\cite{Qiu:2000ga,Qiu:2000hf} the authors separate the perturbative and nonperturbative contribution through the parameter $b_{max}$ such that,  
\begin{equation}
    d\tilde{W}(b_T,Q)=d\tilde{W}(b_T,Q), \; {\rm for}\; b_T \leq b_{max}\, ,
\end{equation}
and 
\begin{equation}
d\tilde{W}(b_T,Q)=d\tilde{W}(b_{max},Q)\,d\tilde{W}^{\rm NP}_{QZ}(b_T,Q;b_{max}), \qquad{\rm for}\;b_T > b_{max}\, ,
\end{equation}
where $d\tilde{W}^{\rm NP}_{QZ}(b_T,Q;b_{max})$ includes power corrections to improve the matching between the perturbative and nonperturbative regions of $d\tilde{W}(b_T,Q)$.
This approach  attempts to minimize
the influence of the nonperturbative piece of $d\tilde{W}(b_T,Q)$, which contains several parameters and does not have a fixed functional form, at small $b_T$ where one should be driven by perturbatively calculable effects.  In the context of the ``resummation approach''~\cite{Laenen:2000de,Kulesza:2002rh}, 
one avoids the  Landau pole encountered
in  performing Fourier transforms ($b$-space integrations)
by extending $\bsc$ to the complex plane
and exploiting the analytic structure of the running coupling.  Phenomenological parameters then appear only as nonperturbative power corrections.

\subsubsection{Ji-Ma-Yuan scheme}\label{sec:JMYevol}

\index{evolution!JMY }
The Ji-Ma-Yuan (JMY) scheme~\cite{Ji:2004wu} is similar to that proposed by Collins-Soper~\cite{Collins:1981uk}, i.e., off-light-front gauge link is applied to regulate the rapidity divergence. Instead of a space-like gauge link used in Collins-Soper~\cite{Collins:1981uk}, a time-like off-light-front gauge is adopted in JMY scheme.
The rapidity regulator is defined as $\tilde\zeta^2=(v\cdot P)/v^2$, where $v$ is the direction of the Wilson line with $v^2$ a small parameter but positive. See \sec{other_tmd_defs} for a detailed review of the definition of TMD PDFs in this scheme.

There is a UV evolution equation for TMD PDFs with respect to
the factorization scale $\mu_F$. For example, for the un-subtracted momentum-space
TMD quark distribution, the
renormalization group equation becomes very simple,
\begin{equation}
\label{eq:JMY.muRGE}
       \mu \frac{df_q^{(unsub.)}(x,\kt, \mu,\zeta)}{d\mu}
        = 2\gamma_F[\as(\mu)] f_q^{(unsub.)}(x,\kt,\mu, \zeta) \ ,
\end{equation}
where $\gamma_F$ is the anomalous dimension of the quarks in the
axial gauge and at one-loop order $\gamma_F = (3\alpha_s/4\pi)C_F$.
The subtracted TMD distribution in \eq{tmdpdf_0}, then, satisfies
\begin{equation}
\label{eq:JMYsub.muRGE}
       \mu \frac{df_q^{(sub.)}(x,\kt, \mu,\zeta,\rho)}{d\mu}
        = \big\{2\gamma_F[\as(\mu)]-\gamma_S(\mu,\rho) \big\} 
        f_q^{(sub.)}(x,\kt,\mu, \zeta,\rho) \ ,
\end{equation}
where $\gamma_S(\mu,\rho)$ is the anomalous dimension of the soft factor,
\begin{equation}
\label{eq:JMY.muRGE3}
    \mu\frac{\partial S(b_T,\mu,\rho)}{\partial\mu} = \gamma_S(\mu,\rho) S(b_T,\mu,\rho)\,,
\end{equation}
where this $S$ is the renormalized version of \eq{JMYsoft}, see \cite{Ji:2004wu} for details. The $\rho$ dependence of $\gamma_S$ in \eq{JMY.muRGE3} cancels against $\rho$ dependence that appears in the hard function in the factorized hard-scattering cross sections in this formalism, see e.g.  \eqs{sigma_old}{JMYtoCS}.

The evolution of the TMD PDF, now in $\bt$ space, with respect to $\zeta$ 
takes the form,
\begin{equation}
    \tilde\zeta\frac{\partial}{\partial \tilde\zeta}\tilde{f}_q^{(sub.)}(x,\bt,\mu,x\tilde\zeta,\rho)
      = \left(K(\mu, b_T)+G(\mu, x\tilde \zeta)\right)
       \tilde{f}_q^{(sub.)}(x,\bt,\mu,x\tilde\zeta,\rho)
\label{eq:zetarg}
\end{equation}
where $K$ depends on UV renormalization scale $\mu$ and infrared
impact parameter $b_T$, and is nonperturbative when $b_T$ is large;
$G$ is perturbative for large $\mu$ and $\zeta$; and both are free
of gluon and quark mass singularity. The sum $K+G$, however, is
independent of explicit dependence on the UV scale $\mu$ and hence,
\begin{equation}
         \mu\frac{d}{d\mu} K  = - \gamma_K[\as(\mu)] = -\mu\frac{d}{d\mu}G
\label{eq:reno}
\end{equation}
where $\gamma_K$ is the cusp anomalous dimension
and is a perturbation series in $\alpha_s(\mu)$ free of infrared
singularities. 
The one-loop anomalous dimension $\gamma_K$ is
given by $ \gamma_K = \frac{\alpha_s}{\pi}2C_F$.
Using the above renormalization group
equation \eq{reno}, one can sum over large logarithms
$\ln{\zeta^2b_T^2}$ in $K+G$ when $b$ is small (otherwise $K$ is
nonperturbative). Substituting the result into \eq{zetarg}, one finds an expression that resums double-leading logarithms in
$\tilde\zeta b_T$.

Solving the evolution equations also resums the large logarithms
in the TMDs. The procedure follows Collins-Soper 81~\cite{Collins:1981uk,Collins:1981uw}, and later
Collins-Soper-Sterman 85~\cite{Collins:1984kg}.
First of all, there are large logarithms in $K+G$ (which is
independent of the renormalization scale). To sum it, we solve the
renormalization group equation to get
\begin{equation}
     K(b_T,\mu) + G(x\zeta, \mu)
      = K(b_T,\mu_L) + G(x\zeta, \mu_H) - \int^{\mu_H}_{\mu_L}
      \frac{d \mu}{ \mu} \gamma_K(\alpha(\mu)) \
      .
\end{equation}
To isolate the large logarithms, one has to choose $\mu_L$ to be
on the order of $\Lambda_{\rm QCD}$ and $\mu_H$ to be on the order
of $\zeta$. Therefore, we let
\begin{equation}
    \mu_L= C_1 M_p; ~~~\mu_H = C_2x\zeta =C_2 Q\sqrt{\rho}\ .
\end{equation}

Substituting the above into the Collins-Soper equation for $\tilde f_q^{(sub.)}$, the large logarithms in $\zeta$ can be
factorized,
\begin{eqnarray}
     \tilde{f}_q^{(sub.)}(x,\bt,\mu,x\zeta,\rho) &=& \exp\left\{-\int^{C_2x\zeta}_{\mu_L} \frac{d\mu}{\mu}
         \left[\ln\left(\frac{C_2x\zeta}{\mu}\right)\gamma_K(\alpha(\mu))
          - K(b_T,\mu_L) - G(\mu/C_2, \mu)\right]\right\}
     \nonumber \\
    && \times \tilde{f}_q^{(sub.)}(x,
          \bt,\mu,x\zeta_0=\mu_L/C_2, \rho) \ ,
\end{eqnarray}
where the exponential factor contains the entire dependence on
$\zeta$, in particular, the large Sudakov double logarithms.
However, the above expression contains much more than just the
leading double logarithms; it contains all the subleading logs as
well.

\subsection{Evolution in SCET}
\label{sec:evolution_SCET}
\index{evolution!SCET}
In this section we review the formulation and derivation of TMD PDF evolution equations, and their solutions, in the framework of Soft Collinear Effective Theory, using the tools of RG evolution and rapidity RG evolution of beam and soft functions describing the dynamics of collinear and soft modes in the EFT.

\subsubsection{RG and RRG}
\label{sec:RGandRRG}

Let us take as a starting point the Drell-Yan cross section given in \eq{sigma_new_b}, in terms of separate beam and soft functions, with a clear separation of the rapidity evolution of the two pieces in $\nu$. 
We work for now with the expression in $\mathbf{b}_T$ space, and for simplicity keep one flavor channel in this section, that is,
\begin{equation} 
\label{eq:DYbspace}
\begin{split}
\frac{d\sigma^W}{dQ dY d^2\mathbf{q}_T} &= \int d^2 \mathbf{b}_T \, e^{i\mathbf{b}_T\cdot\mathbf{q}_T} \widetilde \sigma (\mathbf{b}_T) \\
\widetilde \sigma(\mathbf{b}_T) &= H(Q,\mu)\tilde B(x_a,\mathbf{b}_T,\mu,\zeta_a/\nu^2) \tilde B(x_b,\mathbf{b}_T,\mu,\zeta_b/\nu^2)\tilde S(b_T,\mu,\nu)\,,
\end{split}
\end{equation}
which is related to the form \eq{sigma_new_a} 
through the relation \eq{fBS_relation} between beam/soft functions and the TMD PDFs. In this section we are suppressing all flavor indices. 
The UV and rapidity divergences in these functions are renormalized according to \eqs{B_renorm}{S_renorm}. Anomalous dimensions for $\mu$ and $\nu$ evolution of the beam and soft functions are defined by
\begin{align}
\gamma_\mu^{B}(\mu,\zeta/\nu^2) &= -(\tilde Z_B)^{-1} \mu\frac{d}{d\mu} \tilde Z_B (\mathbf{b}_T,\mu,\nu,xP) &\qquad \gamma_\mu^S (\mu,\mu/\nu)&= -(\tilde Z_S)^{-1} \mu\frac{d}{d\mu}\tilde Z_S(b_T,\mu,\nu) \nn \\
\gamma_\nu^{B}(b_T,\mu) &= -(\tilde Z_B)^{-1}\nu\frac{d}{d\nu} \tilde Z_B(\mathbf{b}_T,\mu,\nu,xP) &\qquad \gamma_\nu^S(b_T,\mu) &= -(\tilde Z_S)^{-1} \nu\frac{d}{d\nu}\tilde Z_S(b_T,\mu,\nu)\,.
\end{align}
where $x=x_a$ or $x_b$ and $P=P_a^+$ or $P_b^-$ as appropriate for each beam function. The dependences on regulators $\epsilon,\tau$ and the limits $\epsilon,\tau\to 0$ that appear in \eqs{B_renorm}{S_renorm} are also implicit here. The exact relation of the Collins-Soper scale $\zeta$ to $xP$ depends on the regulator used, as explained in \sec{tmd_defs}.
The hard function has only a $\mu$ anomalous dimension,
\begin{align}
\gamma_\mu^H = - (Z_H)^{-1} \mu\frac{d}{d\mu} Z^H(Q,\mu)\,,
\end{align}
where $Z_H$ is the UV renormalization counterterm for the hard function. We can define it here in terms of $Z_{B,S}$:
\begin{equation}
    Z_H(Q,\mu) = \bigl[\tilde Z_B(b_T,\mu,\nu,x_a P_a^+) \tilde Z_B(b_T,\mu,\nu,x_b P_b^-) \tilde  Z_S(b_T,\mu,\nu)\bigr]^{-1}\,,
\end{equation}
where the $b_T,\nu$ dependences on the right-hand side will cancel, and $Z_H$ will also depend only on the combination $2x_a P_a^+ x_b P_b^- = Q^2$ in \eq{zetaazetab}.

The renormalized functions all then satisfy the RG and rapidity RG (RRG) equations,
\begin{subequations}
\begin{align}
\label{eq:beamRGE}
\mu\frac{d}{d\mu}\tilde B(x,\mathbf{b}_T,\mu,\zeta/\nu^2) &= \gamma_\mu^{B}(\mu,\zeta/\nu^2) \tilde B(x,\mathbf{b}_T,\mu,\zeta/\nu^2) \\
\label{eq:beamRRGE}
\nu\frac{d}{d\nu}\tilde B(x,\mathbf{b}_T,\mu,\zeta/\nu^2) &= \gamma_\nu^{B}(b_T,\mu) \tilde B(x,\mathbf{b}_T,\mu,\zeta/\nu^2)
\end{align}
\end{subequations}
for the beam functions, and
\begin{subequations}
\begin{align}
\label{eq:softRGE}
\mu\frac{d}{d\mu}\tilde S(b_T,\mu,\nu) &= \gamma_\mu^S(\mu,\mu/\nu)\tilde S(b_T,\mu,\nu) \\
\label{eq:softRRGE}
\nu\frac{d}{d\nu}\tilde S(b_T,\mu,\nu) &= \gamma_\nu^{S}(b_T,\mu)\tilde S(b_T,\mu,\nu)
\end{align}
\end{subequations}
for the soft function. The hard function just satisfies the RG equation
\begin{equation}
\mu\frac{d}{d\mu} H(Q,\mu) = \gamma_\mu^H(Q,\mu) H(Q,\mu)\,.
\end{equation}

\begin{figure}
\centering
\includegraphics[width=.45\textwidth]{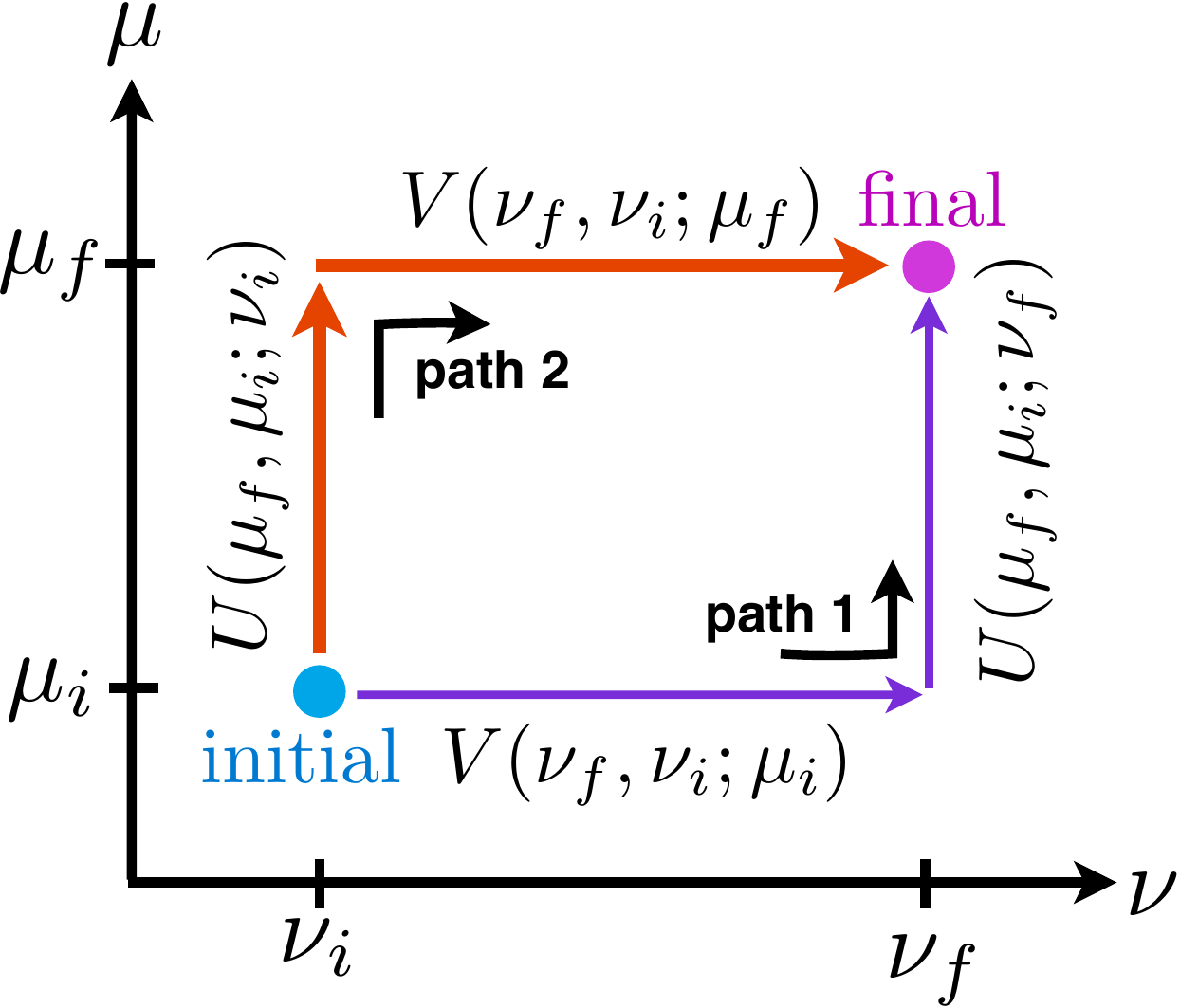}
\caption{Two equivalent paths of RG and RRG evolution. The beam and soft functions in \eqs{Bevolved}{Sevolved} can be along either path, with \eq{secondderivatives} guaranteeing path independence of the combined evolution. Figure taken from \cite{Chiu:2012ir}.}
\label{fig:commute}
\end{figure}

The independence of the physical cross section \eq{sigma_new_b} on $\mu,\nu$ imposes constraints on the beam, soft, and hard anomalous dimensions:
\begin{subequations}
\begin{align}
\label{eq:muconsistency}
0 &= \gamma_\mu^H(Q,\mu)  + \gamma_\mu^S(\mu,\mu/\nu) + \gamma_\mu^{B}(\mu,\zeta_a/\nu^2) + \gamma_\mu^{B}(\mu,\zeta_b/\nu^2) \\
\label{eq:nuconsistency}
0 &= \gamma_\nu^{S}(b_T,\mu) + 2\gamma_\nu^{B}(b_T,\mu) \,.
\end{align}
\end{subequations}
Since \eq{nuconsistency} requires $\gamma_\nu^B = -\gamma_\nu^S/2$, we will speak of a single ``rapidity anomalous dimension'' $\gamma_\nu = \gamma_\nu^B$.
The commutativity of $\mu$ and $\nu$ derivatives also imposes a very powerful constraint on the beam and soft anomalous dimensions, namely,
\begin{align}
\label{eq:secondderivatives}
\mu\frac{d}{d\mu} \gamma_\nu &= \nu\frac{d}{d\nu} \gamma_\mu^{B} = 2\GammaC 
 \,, 
\end{align}
Here $\GammaC=\GammaC^{q,g}$ depending on the flavor channel. These equations \eq{secondderivatives} guarantee that RG evolution along the two paths shown in \fig{commute} is equivalent. In \sec{2dRRGE} we will review an interpretation of \eq{secondderivatives} as the evolution of a conservative vector field.

The appearance of $\Gamma_\text{cusp}$ on the RHS of \eq{secondderivatives} follows from the consistency condition \eq{muconsistency} with the hard anomalous dimension, which takes the form
\begin{equation}
\gamma_\mu^H(\mu) = -4 \GammaC[\as(\mu)] \ln\frac{\mu}{Q} + \gamma_\mu^H[\as(\mu)]\,,
\end{equation}
with the coefficient of the log being proportional to $\Gamma_\text{cusp}$ \cite{Chiu:2012ir}. 
The $\mu$-anomalous dimensions of the beam and soft functions take similar forms:
\begin{subequations}
\begin{align}
\label{eq:gammamuB}
\gamma_\mu^{B}(\mu,\zeta/\nu^2) &=  \GammaC[\as(\mu)]\ln\frac{\nu^2}{\zeta} + \gamma_\mu^{B}[\as(\mu)] \\
\label{eq:gammamuS}
\gamma_\mu^S(\mu,\mu/\nu) &= 4 \GammaC[\as(\mu)]\ln\frac{\mu}{\nu} + \gamma_\mu^S[\as(\mu)]\,,
\end{align}
\end{subequations}
which imply the form for $\gamma_\mu^i$ for the TMD PDF in \eq{SCET.mu.anomdim}.

As for the rapidity anomalous dimensions, obtaining their all-orders form requires some care. At one loop, the (quark) beam and soft functions take the forms \eqs{B_renorm_eta_2}{soft_renorm_eta}, implying the one-loop values for the anomalous dimensions,
\begin{align}
\gamma_\nu(b_T,\mu) =  \frac{\as(\mu)}{4\pi}2\Gamma_0\ln \frac{\mu  b_T}{b_0}\, ,
\end{align}
recalling $b_0 = 2e^{-\gamma_E}$.
The ``non-cusp'' parts are zero at one loop. Evaluated at a low scale $\mu\sim b_T^{-1}$, it is sufficient to evaluate the rapidity anomalous dimension at a fixed order. However, at a larger scale $\mu$, the $\gamma_\nu$'s themselves contain large logs that need to be resummed. The resummed form can be obtained by integrating the consistency conditions \eq{secondderivatives}:
\begin{equation}
\label{eq:gammanuBS}
\gamma_\nu(b_T,\mu) =  2 \int_{b_0/b_T}^\mu \frac{d\mu'}{\mu'} \GammaC[\as(\mu')] + \gamma_\nu[\as(b_0/b_T)]\,,
\end{equation}
where $\gamma_\nu[\as(\mu)]$ is the ``non-cusp'' piece, and is the constant of integration at $\mu=b_0/b_T$, the scale where the log in the anomalous dimension vanishes, i.e.
\begin{equation}
    \gamma_\nu[\as(b_0/b_T)] \equiv \gamma_\nu(b_T,\mu=b_0/b_T)\,.
\end{equation}
Perturbatively this non-cusp piece is nonzero starting at two loops. 

In the next subsection we proceed to derive evolution of the TMD PDF $\tilde f$ from the above evolution equations for beam and soft functions in SCET.

\subsubsection{Combined TMD PDF evolution} \label{combinedBandS}

In SCET it is natural to keep track of the evolution of the separate beam and soft functions each associated with a separate mode in the effective Lagrangian. However their evolution equations, \eqs{beamRGE}{beamRRGE} for the beam function and \eqs{softRGE}{softRRGE} for the soft function  can also be recombined to give the evolution equations \eq{CSS.evol} for the combined TMD PDF \eq{fBS_relation},
which we repeat here: 
\begin{equation}
 \tilde f_{j}(x,\bt,\mu,\zeta) = \tilde B_{j}(x,\bt,\mu,\zeta/\nu^2) \sqrt{\tilde S(b_T,\mu,\nu)}
\,.\end{equation}
It now obeys the evolution equations:
\begin{subequations}
\label{eq:TMDRGEs}
\begin{align}
\mu\frac{d}{d\mu}\tilde f_j(x,\mathbf{b}_T,\mu,\zeta) &= \gamma_\mu^j(\mu,\zeta) \tilde f_j(x,\mathbf{b}_T,\mu,\zeta) \\
\zeta\frac{d}{d\zeta}\tilde f_j(x,\mathbf{b}_T,\mu,\zeta) &= \frac{1}{2}\gamma_\zeta^j(\mu,b_T) \tilde f_j(x,\mathbf{b}_T,\mu,\zeta)\,,
\end{align}
\end{subequations}
where $j=q,g$ as appropriate. The TMD PDF anomalous dimensions are given in terms of the beam and soft anomalous dimensions \eqs{gammamuB}{gammamuS}, first for the $\mu$-anomalous dimension:
\begin{equation}
\begin{split}
\gamma_\mu^j(\mu,\zeta) &= \gamma_\mu^{B_j}(\mu,\zeta/\nu^2) + \frac{1}{2}\gamma_\mu^S(\mu,\mu/\nu) \\
&= \Gamma_\text{cusp}^i[\as(\mu)] \ln \frac{\mu^2}{\zeta} + \gamma_\mu^j[\as(\mu)]\,,
\end{split}
\end{equation}
where the non-cusp part of the anomalous dimension is given by
\begin{equation}
\gamma_\mu^j[\as(\mu)] = \gamma_\mu^{B_j}[\as(\mu)] + \frac{1}{2}\gamma_\mu^S[\as(\mu)] = -\frac{1}{2}\gamma_\mu^H[\as(\mu)]\,.
\end{equation}
In the rapidity evolution, the $\nu$ evolution cancels between $B$ and $S$, but the $\zeta$ evolution of $f$ is inherited from the beam function in \eq{beamRRGE}, giving
\begin{equation}
\label{eq:gammazeta}
\begin{split}
\gamma_\zeta^j(\mu,b_T) &= -\gamma_\nu(b_T,\mu) \\
&= -2\int_{1/\bar b_T}^\mu \frac{d\mu'}{\mu'} \Gamma_\text{cusp}^j [\as(\mu')] + \gamma_\zeta^j[\as(1/\bar b_T)]\,,
\end{split}
\end{equation}
using the resummed form given in \eq{gammanuBS}, and where the non-cusp piece here is given by 
\begin{equation}
\gamma_\zeta^j[\as]
 = \gamma_\nu^{B_j}[\as] + \frac12 \gamma_\nu^{S_j}[\as]
 = -\gamma_\nu[\as]\,.
\end{equation}
The evolution equations \eq{TMDRGEs} coincide with the universal forms given at the beginning of the Chapter, \eq{TMD.evol}.

\subsubsection{RG and RRG solutions}

Now we turn to solving the above evolution equations.
The solutions to the RG and RRG evolution equations \eqs{beamRGE}{beamRRGE} for the beam function and \eqs{softRGE}{softRRGE} for the soft function can  be obtained in straightforward manner. Evolution along the two equivalent paths illustrated in \fig{commute} allows us to write:
\begin{subequations}
\label{eq:Bevolved}
\begin{align}
\label{eq:Bevolved1}
\tilde B(x,\mathbf{b}_T,\mu,\zeta/\nu^2) &= \tilde B(x,\mathbf{b}_T,\mu_L,\zeta/\nu_H^2) U_B(\mu_L,\mu;\nu)V_B(\nu_H,\nu;\mu_L) \\
\label{eq:Bevolved2}
&= \tilde B(x,\mathbf{b}_T,\mu_L,\zeta/\nu_H^2) V_B(\nu_H,\nu;\mu) U_B(\mu_L,\mu;\nu_H) \,,
\end{align}
\end{subequations}
and
\begin{subequations}
\label{eq:Sevolved}
\begin{align}
\label{eq:Sevolved1}
\tilde S(b_T,\mu,\nu) &= \tilde S(b_T,\mu_L,\nu_L) U_S(\mu_L,\mu;\nu)V_S(\nu_L,\nu;\mu_L) \\
\label{eq:Sevolved2}
&= \tilde S(b_T,\mu_L,\nu_L) V_S(\nu_L,\nu;\mu) U_S(\mu_L,\mu;\nu_L) \,,
\end{align}
\end{subequations}
evolving both $B,S$ from their ``natural'' scales where fixed-order logs in their expansions are small. The RG evolution kernels are given by:
\begin{align}
U_B(\mu_L,\mu;\nu) = \exp\biggl[ \int_{\mu_L}^\mu \frac{d\mu'}{\mu'}\gamma_\mu^{B}(\mu',\zeta/\nu^2)\biggr] \,,\quad U_S(\mu_L,\mu;\nu) = \exp\biggl[ \int_{\mu_L}^\mu \frac{d\mu'}{\mu'}\gamma_\mu^{S}(\mu',\mu'/\nu)\biggr]
\label{eq:UBS}
.
\end{align}
In \app{kernelexpansions} we give formulae for evaluating the integral over the anomalous dimensions in an explicit form at given orders of perturbative accuracy, in particular accounting for integrating the functions $\as(\mu')$ that will appear in the expansions of the integrands.
Meanwhile the RRG evolution kernels in \eqs{Bevolved}{Sevolved} are given by:
\begin{align}
\label{eq:VBVS}
V_B(\nu_H,\nu;\mu) &= \exp\biggl[\int_{\nu_H}^\nu \frac{d\nu'}{\nu'} \gamma_\nu^B(b_T,\mu)\biggr] =  \exp\biggl\{ \Bigl[ 2 \eta_\Gamma(b_0/b_T,\mu) + \gamma_\nu^B[\as(b_0/b_T)] \Bigr] \ln\frac{\nu}{\nu_H}\biggr\} \\
V_S(\nu_L,\nu;\mu) &= \exp\biggl[\int_{\nu_L}^\nu\frac{d\nu'}{\nu'} \gamma_\nu^S(b_T,\mu)\biggr] = \exp\biggl\{ \Bigl[ -4\eta_\Gamma(b_0/b_T,\mu) + \gamma_\nu^S[\as(b_0/b_T)] \Bigr] \ln\frac{\nu}{\nu_L}\biggr\} \,. \nn
\end{align}
The $\nu'$ integrals are actually trivial to evaluate since the anomalous dimensions $\gamma_\nu$ in \eq{gammanuBS} have no explicit $\nu$ dependence. We have expressed the integral over the cusp piece in a resummed form $\eta_\Gamma(b_0/\bar b_T,\mu)$, where
\begin{equation}
    \eta_\Gamma(\mu_L,\mu) = \int_{\mu_L}^\mu \frac{d\mu'}{\mu'} \GammaC[\as(\mu')]\,,
\end{equation}
whose explicit forms at specific orders of perturbative accuracy are also given in \app{kernelexpansions}. It is imperative to use its resummed form when $\mu b_T/b_0\gg 1$, e.g. when $\mu$ is taken near the hard scale. Near $\mu\sim b_0/b_T$ one could use a fixed-order truncation of $\eta_\Gamma$.

The hard function, meanwhile, just undergoes $\mu$ evolution, with the solution
\begin{equation}
H(Q,\mu) = H(Q,\mu_H) U_H(\mu_H,\mu)\,,
\end{equation}
where
\begin{equation}
\label{eq:UH}
U_H(\mu_H,\mu) = \exp\biggl\{\int_{\mu_H}^\mu \frac{d\mu'}{\mu'} \gamma_H[\as(\mu')]\biggr\}\,,
\end{equation}
again whose explicit form at given orders of perturbative accuracy is given in \app{kernelexpansions}.

\begin{figure}
\centering
\includegraphics[width=.47\textwidth]{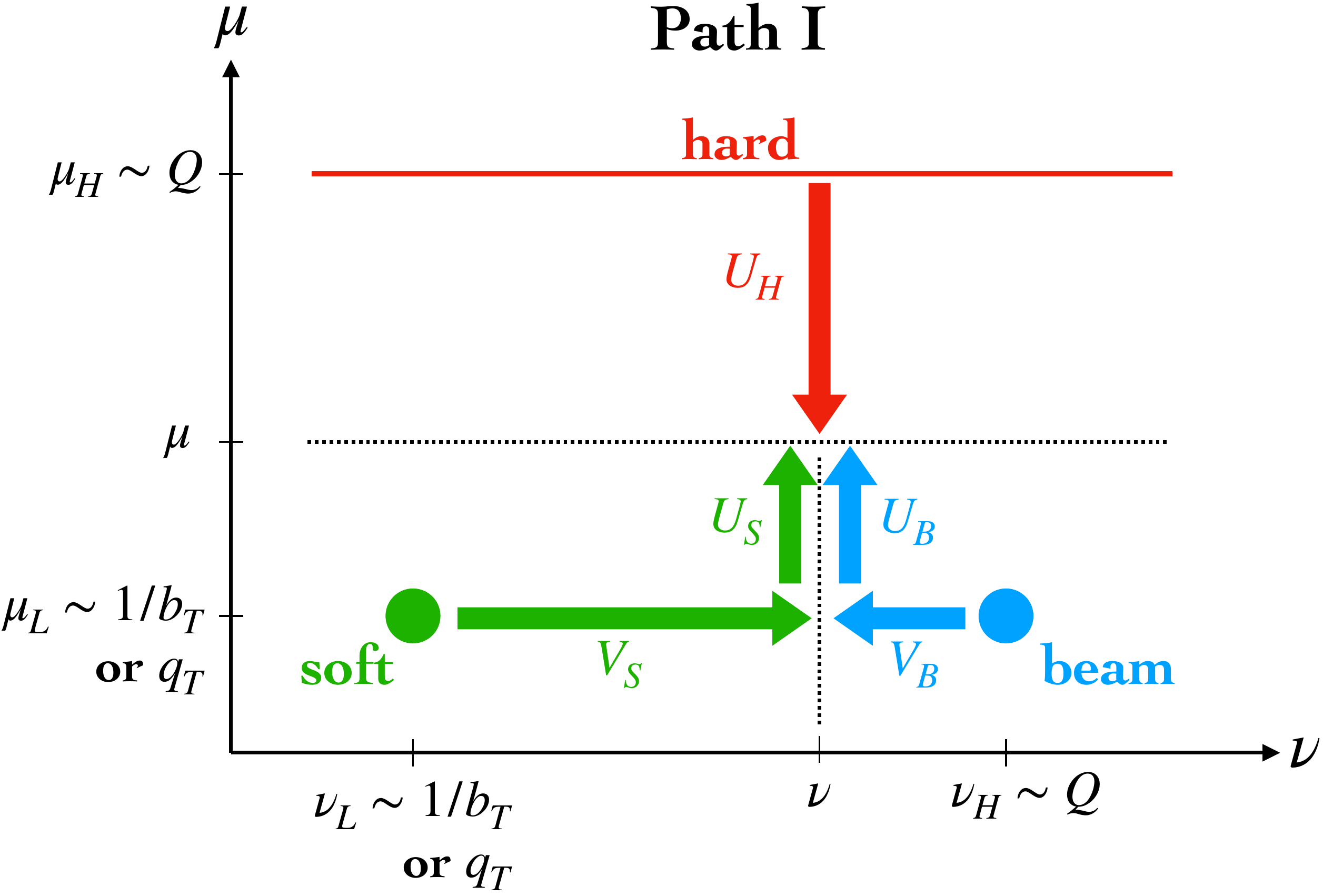}
\qquad
\includegraphics[width=.47\textwidth]{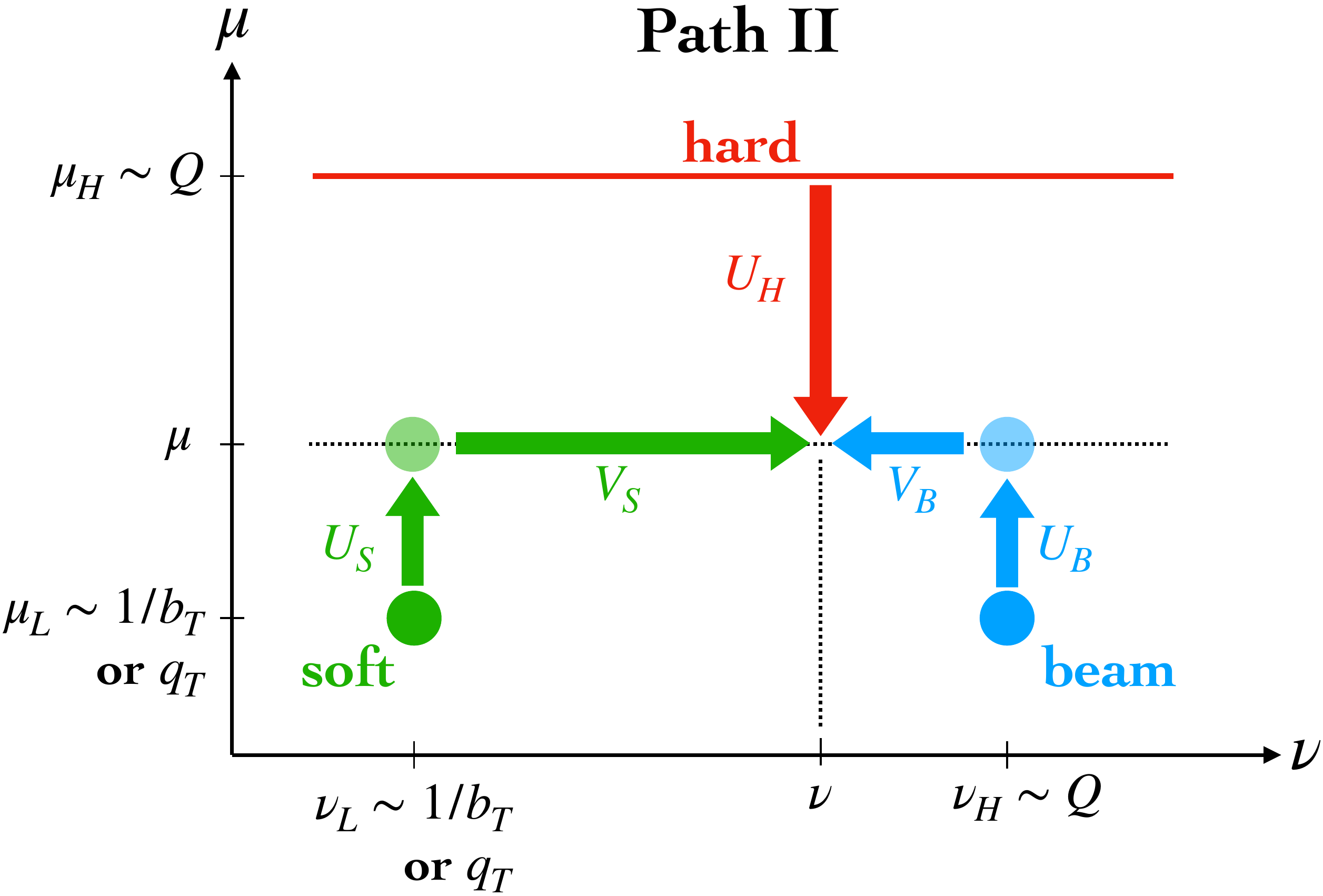}
\caption{Two possible paths for RG and RRG evolution of ingredients of DY cross section for $b_T$ or $q_T$ resummation, from their ``natural'' scales to arbitrary scales $\mu,\nu$. The simplest path arises from choosing $\mu\sim\mu_L$ and $\nu\sim\nu_H$, the natural scales for the beam function. Then we only need to RG evolve the hard function to $\mu_L$ and RRG evolve the soft function to $\nu_H$.}
\label{fig:paths}
\end{figure}
In total then, the resummed Drell-Yan cross section in $b_T$ space \eq{DYbspace} can be written
\begin{align}
\label{eq:resummedcs}
\widetilde\sigma(\mathbf{b}_T,x_a,x_b,Q;\mu_{L,H},\nu_{L,H}) &= U_\text{tot}(\mu_{L,H},\nu_{L,H};\mu,\nu) H(Q^2,\mu_H)\tilde S(b_T,\mu_L,\mu_L/\nu_L) \\
&\quad\times \tilde B(x_a,\mathbf{b}_T,\mu_L,\zeta_a/\nu_H^2) \tilde B(x_b,\mathbf{b}_T,\mu_L,\zeta_b/\nu_H^2) \,. \nn
\end{align}
Note on the LHS of \eq{resummedcs} we have indicated the cross section should be independent of $\mu,\nu$ that appear in the combination of all evolution kernels $U_\text{tot}$. In principle it is also independent of $\mu_{L,H},\nu_{L,H}$, but in practice this is only true if summed to all orders in $\as$; at a truncated resummed order, it retains subleading numerical dependence on these scale choices, which represents the freedom in choosing how to deal with the subleading terms at a given order of resummed accuracy. This variation is also a standard measure of the perturbative uncertainty at a given order of resummed accuracy, as varying the scales $\mu_{L,H},\nu_{L,H}$ probes the size of the missing subleading terms.

We can express $U_\text{tot}$ in two ways, corresponding to the two equivalent paths in \fig{paths}:
\begin{subequations}
\label{eq:Utot}
\begin{align}
U_\text{tot}(\mu_{L,H},\nu_{L,H};\mu,\nu) &\overset{\text{I}}{=} U_H(\mu_H,\mu) U_S(\mu_L,\mu;\nu) V_S(\nu_L,\nu;\mu_L) U_B^2(\mu_L,\mu;\nu)V_B^2(\nu_H,\nu;\mu_L) \\
&\overset{\text{II}}{=} U_H(\mu_H,\mu) V_S(\nu_L,\nu;\mu)U_S(\mu_L,\mu;\nu_L) V_B^2(\nu_H,\nu;\mu)U_B^2(\mu_L,\mu;\nu_H)
\end{align}
\end{subequations}
Using our freedom to choose $\mu,\nu$, the simplest choices arise from choosing $\mu=\mu_L$ and $\nu=\nu_H$, requiring only the hard function to be RG evolved down to $\mu_L$ and the soft function RRG evolved to $\nu_H$:
\begin{equation}
U_\text{tot}(\mu_{L,H},\nu_{L,H}) = U_H(\mu_H,\mu_L) V_S(\nu_L,\nu_H;\mu_L) \,.
\end{equation}
Of course the total evolution should be independent of the path chosen in \eq{Utot}.

At this point exactly where $\mu_{H,L}$ and $\nu_{H,L}$ are chosen is unspecified. The natural choices are that $\mu_H,\nu_H \sim Q$ and $\mu_L\sim\nu_L \sim b_0/b_T$ in $b_T$ space. Making exactly these choices and performing the inverse transform in \eq{sigma_new} to momentum space leads to correspondence with the more traditional picture presented in \sec{TMDEvol}. With these scale choices and the exponentiated forms of $U_H,V_S$ in \eqs{UH}{VBVS}, we recognize that the solution in \eq{resummedcs} achieves the resummation of the towers of fixed-order logs illustrated in \eq{log_orders}, tower by tower, determined by the order of anomalous dimensions included according to Table~\ref{tbl:resum_orders}. By using the forms of evolution kernels $K_{\Gamma,\gamma},\eta_\Gamma$ given in \app{kernelexpansions}, each tower is captured in a simple, closed form in terms of ratios of the running coupling at different scales.

With such scale choices in $b_T$ space, the inverse transform over $b_T$ in \eq{sigma_new} requires a prescription (such as $b_*$ in \eq{bstardef}) to regulate the integral over large $b_T$. In the SCET picture, we are led to consider the scales of the hard, collinear, and soft modes to be freely variable to start with, and in particular $\mu_L,\nu_L$ do not need to be chosen as functions of $b_T$ prior to doing the integral. This leads to alternate methods to perform the resummation directly in momentum space (e.g. \cite{Monni:2016ktx,Ebert:2016gcn}) or to a hybrid scheme to choose the scales partially in $b_T$ and partially in $q_T$ space \cite{Kang:2017cjk}, see \sec{momentumspace}.

\subsection{Two-Dimensional Evolution}
\label{sec:2dRRGE}

\begin{figure}
\centering
\vspace{-12pt}
\includegraphics[width=.47\textwidth]{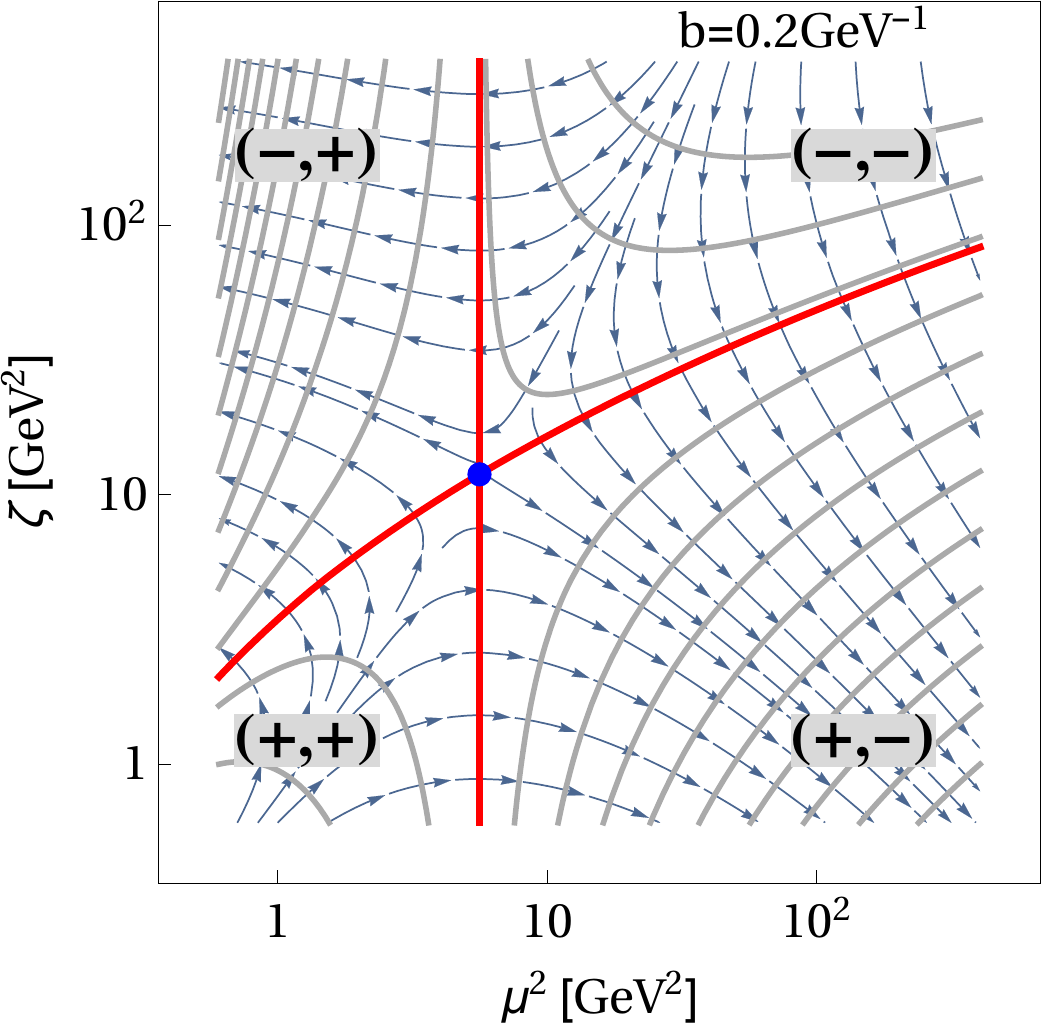}
\qquad\raisebox{1em}{
\includegraphics[width=.45\textwidth]{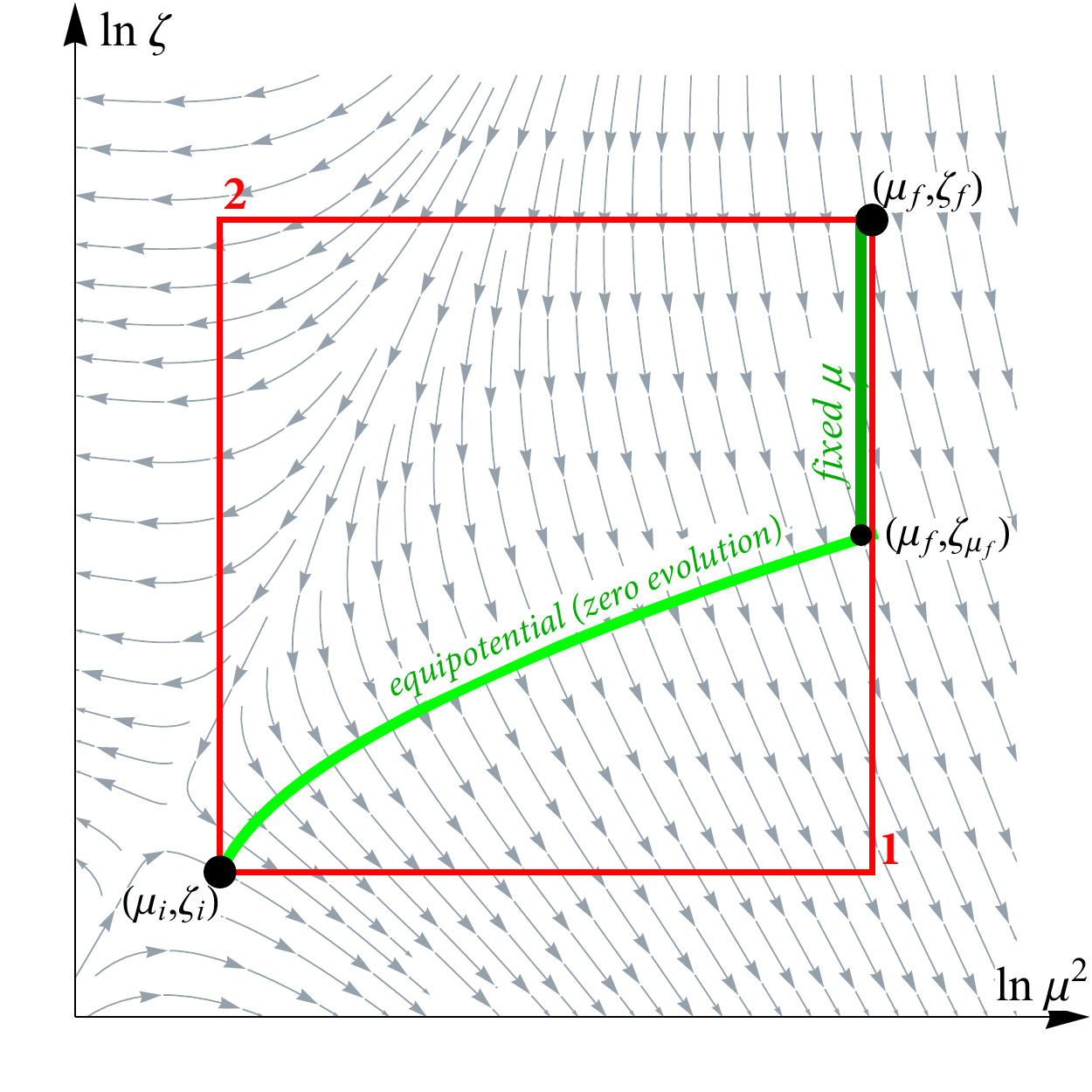}
}
\vspace{-6pt}
\caption{2-D evolution field for TMD evolution. \emph{Left:} Field lines for 2-D anomalous dimensions \eq{Efield} and equipotential lines in grey. The two red lines pass through the saddle point. \emph{Right:} TMD evolution along the straight line paths similar to \fig{paths} in red, and along a path incorporating an equipotential line in green. Figures taken/modified from \cite{Scimemi:2018xaf}.}
\label{fig:equipotential}
\end{figure}

The two-dimensional nature of the TMD evolution equations \eq{TMDRGEs} allows for a nice intepretation with analogues to electromagnetism and other fields of physics, as first presented by \cite{Scimemi:2018xaf}, in which \eq{TMDRGEs} is expressed as a vector differential equation. First, we define the vectors of evolution variables and anomalous dimensions:
\begin{equation}
\bm{\nu} = \Bigl( \ln\mu,\ln\zeta\Bigr)\,,\qquad
\label{eq:Efield}
\mathbf{E}(\bm{\nu},b_T) = \Bigl( \gamma_\mu(\bm{\nu}), \frac{1}{2}\gamma_\zeta(\bm{\nu},b_T)\Bigr)\,.
\end{equation}
The RG and RRG equations in \eq{TMDRGEs} are then expressed as the single vector equation
\begin{equation}
\label{eq:vectorRGE}
\Del f(x,\mathbf{b}_T,\bm{\nu}) = \mathbf{E}(\bm{\nu},b_T) f(x,\mathbf{b}_T,\bm{\nu}) \,,
\end{equation}
where $\Del= d/d\bm{\nu} = ({d}/{d\ln\mu} , {d}/{d\ln\zeta})$.
The consistency relations in \eq{secondderivatives} are then equivalent to the property that $\mathbf{E}$ is a conservative vector field, which is the gradient of a potential:
\begin{equation}
\Del\times\mathbf{E} = 0 \quad\Rightarrow\quad
\mathbf{E}(\bm{\nu},b_T) = \Del U(\bm{\nu},b_T)\,.
\end{equation}
In terms of $U$, the solution of the vector RGE \eq{vectorRGE} for the TMD $f$ can be expressed:
\begin{equation}
\label{eq:potentialdifference}
\ln \frac{f(x,\mathbf{b}_T,\bm{\nu}_f)}{f(x,\mathbf{b}_T,\bm{\nu}_i)} = \int_{\bm{\nu}_i}^{\bm{\nu}_f} d\bm{\nu}\cdot \mathbf{E}(\bm{\nu},b_T) = \int_{\bm{\nu}_i}^{\bm{\nu}_f} d\bm{\nu}\cdot \Del U (\bm{\nu},b_T) = U(\bm{\nu}_f) - U(\bm{\nu}_i)\,.
\end{equation}
An explicit solution for $U$ takes the form:
\begin{equation}
U(\bm{\nu},b_T) = \frac{\ln\zeta}{2}\gamma_\zeta(\mu,b_T) + \int^\mu d\ln\mu' \Bigl\{ \GammaC[\as(\mu')]\ln{\mu'}^2 + \gamma_\mu[\as(\mu')]\Bigr\} + F(b_T)\,,
\end{equation}
where $F$ is a function of only $b_T$. Using the resummed form \eq{gammazeta} for $\gamma_\zeta$, and taking the difference of potentials in \eq{potentialdifference}, we obtain for the total TMD evolution between $\bm{\nu}_i$ and $\bm{\nu}_f$:
\begin{align}
\label{eq:UfUi}
U(\bm{\nu}_f) - U(\bm{\nu}_i) &= \int_{\mu_i}^{\mu_f} d\ln\mu \, \Bigl\{\GammaC[\as(\mu)]\ln\frac{\mu^2}{\zeta_f}  + \gamma_\mu[\as(\mu)]\Bigr\} \\
&\quad + \biggl\{ -\int_{1/\bar b_T}^{\mu_i} d\ln\mu \, \GammaC[\as(\mu)] + \frac{1}{2}\gamma_\zeta[\as(1/\bar b_T)]\biggr\}\ln\frac{\zeta_f}{\zeta_i} \nn \\
&= \int_{\mu_i}^{\mu_f} d\ln\mu \, \Bigl\{\GammaC[\as(\mu)]\ln\frac{\mu^2}{\zeta_i}  + \gamma_\mu[\as(\mu)]\Bigr\} \nn \\
&\quad + \biggl\{ -\int_{1/\bar b_T}^{\mu_f} d\ln\mu \, \GammaC[\as(\mu)] + \frac{1}{2}\gamma_\zeta[\as(1/\bar b_T)]\biggr\}\ln\frac{\zeta_f}{\zeta_i} \,, \nn 
\end{align}
in which we recognize the two forms that come from the one-dimensional evolution along the two equivalent straight-line paths in \fig{equipotential} or \eq{Utot}.

The 2-D picture affords another cute way to illustrate this evolution. We can consider the \emph{equipotential} lines, which are always orthogonal to the evolution field $\mathbf{E}$, along which there is \emph{zero} evolution. Parameterizing such a line in $\bm{\nu}$ space by
\begin{equation}
\bm{\omega}(t;\bm{\nu}_0) = \bigl(t,\ln\zeta_\text{equi}(t)\bigr)\rvert_{\bm{\nu}_0} \,,
\end{equation}
with the equipotential line passing through a specified point $\bm{\nu}_0 = (\ln\mu_0,\ln\zeta_0)$ and $t=\ln\mu$, we solve the equation for an equipotential line,
\begin{equation}
\frac{d\bm{\omega}}{d\ln\mu} \cdot \Del U(\bm{\nu},b_T) = 0 \quad \Rightarrow \quad \frac{\partial U}{\partial{\ln\mu}} + \frac{d\ln\zeta_\text{equi}}{d\ln\mu} \frac{\partial{U}}{d\ln\zeta} = 0\,.
\end{equation}
whose solution is,
\begin{equation}
\label{eq:equipotential}
\ln \zeta_\text{equi}(\mu;\mu_0,\zeta_0) = \frac{-2\int_{\mu_0}^\mu d\ln\mu' \Bigl\{ \GammaC[\as(\mu')] \ln {\mu'}^2 + \gamma_\mu[\as(\mu')] \Bigr\} + \gamma_\zeta(\mu_0) \ln \zeta_0}{\gamma_\zeta(\mu,b_T)} \,.
 \end{equation}
Along the line $\bigl(\mu,\zeta_\text{equi}(\mu;\mu_0,\zeta_0)\bigr)$, the total evolution from $(\mu_0,\zeta_0)$ is \emph{zero}, and $f(x,\mathbf{b}_T,\bm{\nu})$ is the same everywhere on this line. This lets us envision evolution along the green path shown in \fig{equipotential}, so the nonzero evolution is only along the vertical segment from $\zeta_f$ to $\zeta_\text{equi}$ at $\mu_f$:
\begin{equation}
U(\bm{\nu}_f) - U(\bm{\nu}_i) = \frac{1}{2} \gamma_\zeta(\mu_f,b_T) \ln\frac{\zeta_f}{\zeta_\text{equi}(\mu_f;\mu_0,\zeta_0)}\,,
\end{equation}
which is a nice, compact expression. Plugging in \eq{equipotential}, however, we see it is perfectly equivalent to the expressions \eq{UfUi}, encoding the evolution in the horizontal $\mu$ direction instead in the exact location of $\zeta_\text{equi}$ as a function of the initial $\mu_0,\zeta_0$. At any finite order of resummed accuracy, these scales (or at least the final location of $\zeta_\text{equi}(\mu_f;\bm{\nu}_0)$) should still be varied to probe the residual theoretical uncertainty.

\subsection{Connecting Resummation to Fixed Order} \label{sec:nonsingular}

\index{resummation!matching}

A primary goal of transverse-momentum-dependent (TMD) factorization theorems~\cite{Collins:1981uk,Collins:1981uw, Collins:1984kg,Idilbi:2004vb,Collins:2004nx,Ji:2004wu,Ji:2004xq,Collins:2011zzd,GarciaEchevarria:2011rb,Echevarria:2015usa},
is to describe the cross section as a function of the
transverse momentum $\qt$ 
point-by-point, from small $q_T\sim m$ (where $m$ is a typical hadronic mass scale), to large $q_T \sim Q$, where $Q$ is a large momentum or mass in the reaction and sets the hard scale~\cite{Arnold:1990yk,Bozzi:2005wk,Qiu:2000hf,Qiu:2000ga}. 
In order to achieve this, Collins, Soper and Sterman  organized the cross section in the additive form $W+Y$~\cite{Collins:1981uk,Collins:1981uw,Collins:2011zzd,Aybat:2011zv} (see \eq{sigma}).
 As detailed in Section~\ref{sec:TMDforDY} the $W$-term is valid for $q_T\ll Q$ and involves the TMD PDFs. 
They depend on intrinsic transverse momentum as well as the usual momentum fraction variables and since TMD factorization is necessary to describe processes that are sensitive to transverse momentum, that is  small compared to the hard scale. They are the natural quantities that describe the partonic content of target and produced hadrons in deep inelastic (DIS) processes. 
The $Y$-term, which involves collinear PDFs  serves as a correction for larger $q_T$ values and is the difference between the collinear cross section for $q_T\sim Q$ beginning at a fixed order  in the strong coupling,  
and its small transverse momentum asymptotic limit for $m\ll q_T\ll Q$~\cite{Collins:1984kg,Arnold:1990yk,Nadolsky:1999kb,Koike:2006fn,Collins:1984kg,Collins:2011zzd,Collins:2016hqq}.  The latter is called the asymptotic term (AY)~\cite{Collins:1984kg,Collins:2011zzd,Collins:2016hqq}.

Various methods are in use to ensure the two terms match or interpolate smoothly~\cite{Collins:1984kg,Arnold:1990yk,Qiu:2000ga,Qiu:2000hf,Berger:2004cc,Bacchetta:2008xw,Boglione:2014oea,Collins:2016hqq,Collins:2017ybb,Echevarria:2018qyi,Grewal:2020hoc}.  For large $\vect{q}_T$, the resummed $W$ term is re-expanded in fixed-order, yielding the singular log terms at a given order $\alpha_s^n$, the AY term, and the $Y$ term matches it onto the  correct full QCD result at this fixed order. The latter is the fixed order term (FO). 
To do this properly, two things need to be done, for both of which there are multiple valid approaches. First, the resummation in the $W$ term should be ``turned off'' for large $q_T$ so it gets truncated at a fixed order in $\as$; in this region, logs and non-logs are of similar size and the former no longer should be resummed to all orders in $\as$.  Second, since the fixed-order expansion of $W$ will be missing the non-logs at the desired fixed order in $\as$, the missing terms must be added back in the $Y$ term. The transition between resummation and fixed-order regions and the matching onto the full fixed-order QCD result should be achieved in a smooth, well-defined manner. In the first two subsections below we review methods to implement this transition smoothly within the CSS formalism, and in the third, we briefly highlight how this transition is straightforward to implement in the SCET formalism, using the technique of ``profile scales'' \cite{Ligeti:2008ac,Abbate:2010xh,Berger:2010xi,Neill:2015roa,Kang:2017cjk,Lustermans:2019plv,Ebert:2020dfc} to automatically turn off resummation as $\vect{q}_T$ grows from small to large and transition into the correct, truncated fixed-order expansion.

\tocless\subsubsection{Matching in the CSS formalism}
\label{s:CSSrev}

We begin with a synopsis of the $W+Y$ construction 
$q_{T}$-differential cross section, \eq{sigma} which we abbreviate as, 
\begin{align}\label{eq:DY_WY}
\Gamma (\qt,Q,S)
\equiv
 \frac{\df\sigma}{\df Q \df Y \df^2\qt} &=
W(\qt,Q,S)+Y(
\qt,Q,S)+ O((m/Q)^{c}) \,,
\end{align}
where $\qt$ and $Q^{\!2}$ are the transverse momentum
and virtuality, respectively, of the virtual photon.
In \eq{DY_WY}, the
$W$-term factorizes into TMD PDFs (and FFs in SIDIS and SIA)  and is valid for $q_{T}\ll Q$,
while the $Y$-term serves as a correction for larger $q_{T}$ values and uses
collinear PDFs and FFs.

The construction of the cross section in~\eq{DY_WY}
as the
sum of $W(\qt,Q,S)$ and $Y(\qt,Q,S)$ results from applying so-called ``approximators'' in the context of factorization,  to
$\Gamma (\qt,Q,S)$~\cite{Collins:1989gx,Collins:2011zzd,Collins:2016hqq,Collins:2017ybb}
that are designed to be valid for the corresponding momentum  regions of $q_{T}$.

The $W$ term is defined from the TMD approximator, $\mathrm{T}_{\mathrm{TMD}}$
\begin{equation}
\label{eq:wterm}
\TT{}{} \equiv \appor{TMD} \cs{}{}  \, ,
\end{equation}
where for the purposes of this discussion, we   consider the $W$ term for the unpolarized case. 
The $\appor{TMD}$ ``approximator'' is an instruction to replace the object
to its right by an approximation that is designed to be good in the
$\Tsc{q} \ll Q$ momentum region where the approximation has
 fractional errors of the order 
$(\Tsc{q}/Q)^a$ or $(m/Q)^{a'}$.   That is, it replaces the exact $\cs{}{}$ by the
approximate $\TT{}{}$:
\begin{align}
\appor{TMD} \cs{}{}=  \cs{}{} + 
          O \left( \frac{\Tsc{q}}{Q} \right)^a \cs{}{}
        + O \left( \frac{m}{Q} \right)^{a'} \cs{}{}
\, ,
\label{eq:TMDapdef}
\end{align}
with 
where $a, a' >0$.

Another approximator, $\appor{coll}$, handles the
large $\Tsc{q} \sim Q$ region. It replaces $\cs{}{}$ with an
approximation that is good when $\Tsc{q} \sim Q$ with a fractional error of
$(m/\Tsc{q})^b$. That is, 
\begin{align}
 \appor{coll} \cs{}{} = \cs{}{} 
      & + O \left( \frac{m}{\Tsc{q}} \right)^b \cs{}{}
 \, , \label{eq:collapdef}
\end{align} 
where $b>0$.
Since $\appor{coll}$ is to be applied to the
$\Tsc{q} \sim Q$ region, one only needs collinear factorization
at a fixed order~\cite{Collins:1984kg,Altarelli:1984pt} and with a hard scale $\mu \sim Q$.

If $ m \lesssim \Tsc{q} $ and $\Tsc{q} \sim Q$ were the only regions of
interest, then the $\appor{TMD}$ and $\appor{coll}$ approximators would be sufficient. One could
simply calculate using fixed order collinear factorization for the
large $\Tsc{q}$-dependence and TMD factorization for small $\Tsc{q}$-dependence.  
A reasonable description of the full transverse momentum
dependence would be obtained by simply interpolating between the
two descriptions~\cite{Chay:1991jc,Anselmino:2006rv}.

However, the region between large and small $\Tsc{q}$ needs special
treatment if errors are to be  power suppressed point-by-point
in $\Tsc{q}$.  The standard method is to construct a sequence of
nested subtractions~\cite{Collins:2011zzd}. The smallest-size region is a neighborhood of
$\Tsc{q} = 0$, where $\appor{TMD}$ gives a very good approximation.
So, one starts by adding and subtracting the $\appor{TMD}$
approximation:
\begin{align}  \cs{}{} \, =  \, & \appor{TMD} \cs{}{}   +   \Bigg[ \cs{}{} - \appor{TMD} \cs{}{} \Bigg] \, .
\label{eq:nextapp}
\end{align}
From Eq.~\eqref{eq:TMDapdef}, the error term in the square brackets is order $( \Tsc{q}/Q )^a$ and is
only unsuppressed at $\Tsc{q} \gg m$.
Thus, one can apply $\appor{coll}$ and then use a fixed-order
  perturbative expansion in collinear factorization:
\begin{align}
\Gamma(  m \lesssim \Tsc{q} \lesssim Q,Q) 
={}&   \appor{TMD} \cs{}{}
 + \appor{coll} \left[ \cs{}{} - \appor{TMD} \cs{}{} \right]
\nonumber\\
&
 + O\left( \left( \frac{m}{\Tsc{q}} \right)^b \left( \frac{\Tsc{q}}{Q} \right)^a  \right) \cs{}{}
\no
&
 + O\left( \left( \frac{m}{\Tsc{q}} \right)^b \left( \frac{m}{Q} \right)^{a'}  \right) \cs{}{} 
\no
={}& \TT{}{} +  \appor{coll}\cs{}{} -  \appor{coll}\appor{TMD} \cs{}{}
\no 
& + O\left(  \frac{m}{Q}\right)^{\rm c} \cs{}{}
\, ,
\label{eq:powercounting}
\end{align}
where $c = \min(a,a',b)$.  Thus, the cross section is determined
point-by-point in the mid-$\Tsc{q}$ region, up to powers of $m/Q$, by a combination of TMD and
collinear correlation functions.

This construction of $W+Y$ defines $W$, Eq.~\eqref{eq:wterm}, the first term, and  $Y$, to be the second  and third  terms after the second equality in Eq.~\eqref{eq:powercounting}: that is,
\begin{align}\label{eq:yterm}
& \YY{}{}\equiv
\appor{coll} \cs{}{} - \appor{coll} \appor{TMD} \cs{}{}. 
\end{align}
The specific definitions of
  $\appor{coll}$ and $\appor{TMD}$ allowed Eq.\
  (\ref{eq:powercounting}) to work only in the $m
  \lesssim \Tsc{q} \lesssim Q$ region, which we emphasize by the
  argument on the left side of Eq.~\eqref{eq:powercounting}.

In common terminology, the first term in braces in ~\eq{yterm} is
called the ``fixed order'' (FO) contribution, while the second term is 
the ``asymptotic'' (AY) contribution. We will
use the notation
\begin{align}
\fixo{}{} & \equiv \appor{coll} \cs{}{} \label{eq:fodef}  \\
\ays{}{} &\equiv \appor{coll} \appor{TMD} \cs{}{}  \label{eq:asydef} \, .
\end{align}

This corresponds to the terminology in, for example, Ref.~\cite{Nadolsky:1999kb}. The term ``fixed order'' is meant to imply 
that the calculation of $\Gamma$ is done entirely with collinear factorization with hard parts calculated to low order in perturbation theory using $\mu = Q$ and 
with collinear PDFs (and in the case of SIDIS or SIA, FFs) calculated using $\mu = Q$. That is, the hard part and the parton correlation functions are evaluated at the same scale.

The
resulting cross section is accurate up to an error that is of order
$(m/Q)^{c}$, where $c$ is a positive power, and $m$ is a typical
hadronic mass scale. We note that the actual value
for $c$ in the error term $O((m/Q)^{c})$ depends on which structure
function we look at in $\Gamma (\qt,Q,S)$.

In the next subsection we provide some details on the implemenation of the approximators and the 
and we will examine some of the complications involved when combining (matching) TMD factorization with collinear factorization to allow accurate predictions over the whole range of measured transverse momentum in a process like Drell-Yan~\cite{Collins:1984kg,Arnold:1990yk,Berger:2001wr,Bozzi:2003jy}.  We review some improved methods for combining the two types of factorization~\cite{Collins:2016hqq,Gamberg:2017jha}.

\tocless\subsubsection{Improved  Matching TMD and Collinear Factorization}
 
 The error estimates in Eq.~\eqref{eq:powercounting} are inapplicable
  outside this range, i.e., they must not be applied when $\Tsc{q} \gg
  Q$ or $\Tsc{q} \ll m$.  This is because they were extracted from the
  leading power of expansions in relatively small kinematic variables
  $\Tsc{q}/Q$ and $m/\Tsc{q}$
  to give Eqs.~(\ref{eq:TMDapdef}) and~(\ref{eq:collapdef}).
  The issues are illustrated by Eq.\ (\ref{eq:collapdef}).  The
  $(m/\Tsc{q})^b$ estimate is obtained from an expansion in powers of
  mass with respect to the smallest scale in the collinear
  hard-scattering; it is of the order of the first omitted term in the
  expansion.  But once $\Tsc{q}$ gets much smaller, the error can be
  arbitrarily larger.

The above observations do not represent a fundamental breakdown of the
formalism.  They merely indicate that some extra care is needed to
  construct a formalism valid also for 
  $\Tsc{q} \lesssim m$ and $\Tsc{q} \gtrsim Q$.

Let's consider first, $\Tsc{q} \lesssim m$: Clearly collinear factorization is
certainly not applicable for the differential cross section.  But
  this region is actually where the $W$-term in
  Eq.~\eqref{eq:TMDapdef} has its highest validity.  So one simply
  must ensure that the would-be $Y$-term
\begin{equation}
\appor{coll} \cs{}{} - \appor{coll} \appor{TMD} \cs{}{}
\end{equation}
  is sufficiently suppressed in Eq.~\eqref{eq:powercounting} for
  $\Tsc{q} \lesssim m$.  Therefore, we will modify the usual
  definition of $Y$ by inserting a suppression factor at low
  $\Tsc{q}$:
\begin{align}
\label{eq:yterm2}
\YY{}{} 
&{}\equiv \left\{ \appor{coll} \left[ \cs{}{} - \TT{}{} \right] \right\} X(\Tsc{q}/\lambda) \nonumber \\
   &{}= \left\{ \appor{coll} \cs{}{} - \appor{coll} \appor{TMD} \cs{}{} \right\} X(\Tsc{q}/\lambda) \, . 
\end{align}
The smooth cutoff
function $X(\Tsc{q}/\lambda)$  approaches zero for $\Tsc{q}
\lesssim \lambda$ and unity for $\Tsc{q} \gtrsim \lambda$.  It ensures
that the $Y$-term is a correction for $\Tsc{q} \gtrsim m$ only.  As
long as $\lambda = O(m)$, any $\lambda$-dependence must be weak.
This is analogous to the introduction of a $Q_T^{\rm min}$ in Ref.~\cite[Eq.~(2.8)]{Collins:1984kg}.

The exact functional form of $X(\Tsc{q}/\lambda)$ is arbitrary, but is most useful in calculations if it sharply 
suppresses $\Tsc{q} \ll m$ contributions while not affecting $\Tsc{q} \gtrsim m$.  While a step function is acceptable,  
we suggest using a slightly smoother function since one expects the transition from perturbative to nonperturbative physics to 
be relatively smooth. One possible choice is
\begin{equation}
X(\Tsc{q}/\lambda) = 1 - \exp \left\{ -(\Tsc{q} / \lambda)^{a_X} \right\} \ . \label{eq:Xparam}
\end{equation}
This is what is  used in sample calculations in~\cite{Collins:2016hqq,Collins:2017ybb}. A large value for the 
power $a_X$ makes the switching function more like a step function. 

So, now, 
\begin{equation}
\YY{}{} \equiv \left\{ \fixo{}{} - \ays{}{}  \right\} X(\Tsc{q}/\lambda) \, .
\label{eq:Y_}
\end{equation}
The term ``fixed order'' here is meant to imply 
that the calculation of $\Gamma$ is done entirely with collinear factorization with hard parts calculated to low order in perturbation theory using $\mu = Q$ and 
with collinear pdfs (and ffs in the case of SIDIS and SIA) calculated using $\mu = Q$. That is, the hard part and the parton correlation functions are evaluated at the same scale.

Now one  can extend the power
suppression error estimate in~\eq{powercounting} down to
$\Tsc{q} = 0$ to recover~\eq{DY_WY}.  
Equation~\eq{powercounting} becomes
\begin{align}
\label{eq:basic2}
\Gamma(\Tsc{q} \lesssim Q,Q)  = &\TT{}{} +  \YY{}{}
 + O\left( \frac{m}{Q}\right)^{\rm c} \cs{}{},
\end{align} 
which is Eq.~\eqref{eq:DY_WY}, but restricted to $\Tsc{q} \lesssim Q$.

So far, aside from introducing an explicit $X(\Tsc{q}/\lambda)$, we have only 
reviewed the standard $W+Y$ construction. The $\Tsc{q} \lesssim Q$ restriction on 
the left of Eq.~\eqref{eq:basic2} should be emphasized. Since we rely 
on strict power counting in $\Tsc{q}/Q$ and $m/\Tsc{q}$, the region of $\Tsc{q} \gtrsim Q$ is 
not guaranteed to be well-described by the above $W+Y$ construction.

Also, these  modifications to the transition to the $\Tsc{q} / Q \gtrsim 1$ region will leave the  
standard treatment of TMD factorization~\cite{Collins:2011zzd} in the $\Tsc{q} / Q \ll 1$ region
only slightly modified.  In particular, 
the operator definitions for transverse-coordinate-space TMD
functions, along with their evolution properties, are exactly the same as in the usual formalism. 
This is an important aspect of these modifications~\cite{Collins:2016hqq}.

Next, a  modification of  the definition of $W$ was carried out.  This is to provide a
  convenient solution to the problem that with the definition of the $W$ term,  the integral over all $\qt$ of $W(\Tsc{q})$ is zero, because
  $\tilde{W}(\Tsc{b})$ is zero at $\Tsc{b}=0$~\cite{Collins:2016hqq}.

  It would be preferable for the integral to have a normal collinear
  expansion in terms of pdfs and ffs at scale $\muQ$; the lowest order
  term then reproduces the lowest order collinear factorization result
  for the integrated cross section.  At the same time, we wish to
  preserve the results for the Fourier transform of 
  $\tilde{W}(\Tsc{b})$, since these embody the derived factorization
  and evolution properties.   Most importantly, the modified $W$ term must still approximate the
  cross section at low $\Tsc{q}$ to the same accuracy as in 
  \eq{TMDapdef}.  One can achieve the modified $W$ in two stages.

  The first is to modify the Fourier transform of the $W$ term 
  where by $\tilde W(\bsc,Q)$ we refer to the integrand of \eq{sigma_W_reminder}
  to read
\begin{equation}
\label{eq:FTdef1}
  \TTa{}{}
  =
   \int \frac{d^2\bt}{(2 \pi)^2}
   e^{i\qt\cdot\bt} \, \tilde{W}(\bone(\Tsc{b}),Q) \, .
\end{equation}
where
\begin{equation}
  \bone(\Tsc{b}) = \sqrt{ \Tsc{b}^2 + b_0^2/(C_5Q)^2 } \, .
\label{eq:bcut}
\end{equation}
That is, $\tilde{W}(\Tsc{b},Q)$ is replaced by
$\tilde{W}(\bone(\Tsc{b}),Q)$.
 The function $\bone(\Tsc{b})$ is arranged to agree with $\Tsc{b}$
  when $\Tsc{b} \gg 1/Q$, but to be of order $1/Q$ when $\Tsc{b}=0$, ,
  thereby providing a cutoff at small $\Tsc{b}$.
Then, when \eq{FTdef1} is integrated over $\qt$, we
get $\tilde{W}(b_0/(C_5Q),Q)$, instead of the previous value
$\tilde{W}(0,Q)=0$.  We have included an explicit numerical factor of $b_0 \equiv 2
\exp(-\gamma_E)$. 
We have chosen the value of $\bone(0)$ to be
proportional to $1/Q$, so that, 
$\tilde{W}(b_0/(C_5Q),Q)$ has a normal collinear factorization
  property. The numerical constant $C_5$ fixes 
the exact proportionality between $\bone(0)$ and $1/Q$.

Note that the integrand in (\ref{eq:FTdef1}) is nonsingular at
$\Tsc{b}=0$, unlike the unmodifed $\tilde W(\bsc,Q)$.  Thus the large $\Tsc{q}$
behavior is exponentially damped.  Even so, the function still extends
to arbitrarily large $\Tsc{q}$.  

The second and final stage of modification for $W$ is to make an
explicit cutoff at large $\Tsc{q}$, to give:
\begin{align}
  \label{eq:Wnew}
  \TTnew{}{}
  \equiv
  \Xi\left(\frac{\Tsc{q}}{Q},\eta\right)
  \int \frac{d^2 \bt}{(2 \pi)^2}
    e^{i\bt\cdot \qt } \tilde{W}(\bone(\Tsc{b}),Q) \, .
\end{align}
Here $\Xi\left(\Tsc{q}/ (Q\eta)\right)$ is a cutoff function that we
introduce to ensure that $\TTnew{}{}$ vanishes for $\Tsc{q} \gtrsim
Q$, and $\eta$ is a parameter to control exactly where the suppression
of large $\Tsc{q}$ begins. $\Xi\left(\Tsc{q}/Q,\eta\right)$ should
approach unity when $\Tsc{q}\ll Q$ and should vanish for $\Tsc{q}
\gtrsim Q$.  This preserves the required approximation property of
$\TTnew{}{}$ at small $\Tsc{q}$.  At the same time, since the changes
are dominantly at large $\Tsc{q}$, the integral over all $\T{q}$ still
has a normal collinear expansion, as we will make more explicit below.

A simple $\Theta(Q - \Tsc{q})$ step function is acceptable for
$\Xi$.
When we combine $\TTnew{}{}$ with a $Y$ 
one  introduce methods to
minimize sensitivity to the exact form of
$\Xi\left(\Tsc{q}/Q,\eta\right)$.  However, a smoother function is
preferred since the domain of validity of the $W$-term approximation
does not end at a sharp point in $\Tsc{q}$, and thus a smooth function
characterizes general physical expectations.  A reasonable choice is
\begin{equation}
\Xi\left(\frac{\Tsc{q}}{Q},\eta\right) = \exp \left[ -\left( \frac{q_T}{ \eta Q} \right)^{a_\Xi}   \right] \, , \label{eq:Xiparam}
\end{equation}
with $a_\Xi > 2$.  

The only differences between the old and new $W$-term are: i) the use
of $\bone(\Tsc{b})$ rather than $\Tsc{b}$ in $\tilde{W}$, and ii) the
multiplication by $\Xi\left(\Tsc{q}/Q,\eta\right)$. (The second
modification was proposed by Collins~\cite{Collins:2011zzd}. There $\Xi$ is called
$F(\Tsc{q}/Q)$.) Equation~\eqref{eq:Wnew} matches the standard
definition in the limit that $C_5$ and $\eta$ approach infinity~\cite{Collins:2016hqq,Collins:2017ybb}.

\tocless\subsubsection{SCET: profile scales}
\label{sec:profiles}

In SCET the $W$ in \eq{DY_WY} may be viewed as the part of the cross section predicted using the leading-power effective theory Lagrangian, resummed using all the technology reviewed in \sec{evolution_SCET}. The $T$ approximator in \eq{wterm} can be viewed as the set of instructions that describes how to match QCD onto leading-power SCET and compute with it, with power corrections being contained in the corresponding terms of \eq{TMDapdef}. The $Y$ term would be described in SCET as the difference between the prediction of fixed-order perturbative calculation in full QCD with the truncated fixed-order expansion of the resummed leading-power SCET prediction, called the ``nonsingular function'' or ``remainder function''. A smooth interpolation that combines the resummed prediction with this fixed-order remainder is then needed.

In SCET the matching of the resummed result ($W$) onto the fixed-order result for large $q_T$ ($Y$) 
  is naturally achieved by the use of so-called ``profile scales'', see, e.g., \cite{Ligeti:2008ac,Abbate:2010xh,Berger:2010xi}. The $W$-term cross section \eq{resummedcs} sums logs of $\mu_L/\mu_H$ or $\nu_L/\nu_H$, where $\mu_L,\nu_L \sim 1/b_T$ or $q_T$, and $\mu_H,\nu_H\sim Q$. These logs are large and must be resummed when the low and high scales are actually well separated. However, for $1/b_T$ or $q_T\sim Q$, the logs are no longer large, and equally important as the nonsingular terms in the $Y$-term part of the cross section. The resummation of logs in the $W$ term can be smoothly turned off, and properly matched with the nonsingular $Y$ term, by choosing $\mu_L,\nu_L$ to be functions of $1/b_T$ or $q_T$ such that for some value $1/b_T,q_T \lesssim Q$, the low and high scales merge, that is,
\begin{equation}
    \mu_L = \mu_\text{run}(q_T) \,,\quad \nu_L = \nu_\text{run}(q_T)\,,
\end{equation}
(or, alternatively, as functions of $1/b_T$)\,, where $\mu_\text{run},\nu_\text{run}$ are \emph{profile functions}, which must have the behavior
\begin{equation}
\label{eq:profilerun}
    \mu_\text{run}(q_T),\nu_\text{run}(q_T) \sim 
    \begin{cases}
    \mu_0,\nu_0 & \Lambda_\text{QCD} \lesssim q_T\ll Q \\
    q_T & \Lambda_\text{QCD} \ll q_T\ll Q \\
    \mu_H & 1/b_T,q_T \lesssim Q\,,
    \end{cases}
\end{equation}
which freezes the scales at $\mu_0,\nu_0\gtrsim \Lambda_\text{QCD}$ for very low $q_T$, to allow for matching onto a nonperturbative model (a key difference to the $b^*$ prescription is that this only freezes out the scales, not $b_T$ itself); 
achieves perturbative resummation of logs of $q_T/Q$ in the region where these logs are large; and turns off the resummation in the $W$ term by setting all scales equal $\mu_L,\nu_L = \mu_H,\nu_H$ for large $q_T $, automatically turning it into a fixed-order expansion of the log terms. Then in the large $q_T$ region, the $W$ and $Y$ terms automatically combine to give the correct full QCD prediction at a fixed order in $\alpha_s$, where $W$ and $Y$ satisfy the relation
\begin{equation}
\sigma_{\text{FO}(n)}^\text{QCD} = \sigma^W\bigr\rvert_{\text{FO}(n)} + \sigma^Y_{(n)}\,,
\end{equation}
with the subscript $_{(n)}$ indicating the $\alpha_s^n$ term in the fixed-order expansion of the full QCD prediction, the expanded $W$ term, or the nonsingular $Y$ term.

\begin{figure}
\centering
\vspace{-18pt}
\includegraphics[width=.48\textwidth]{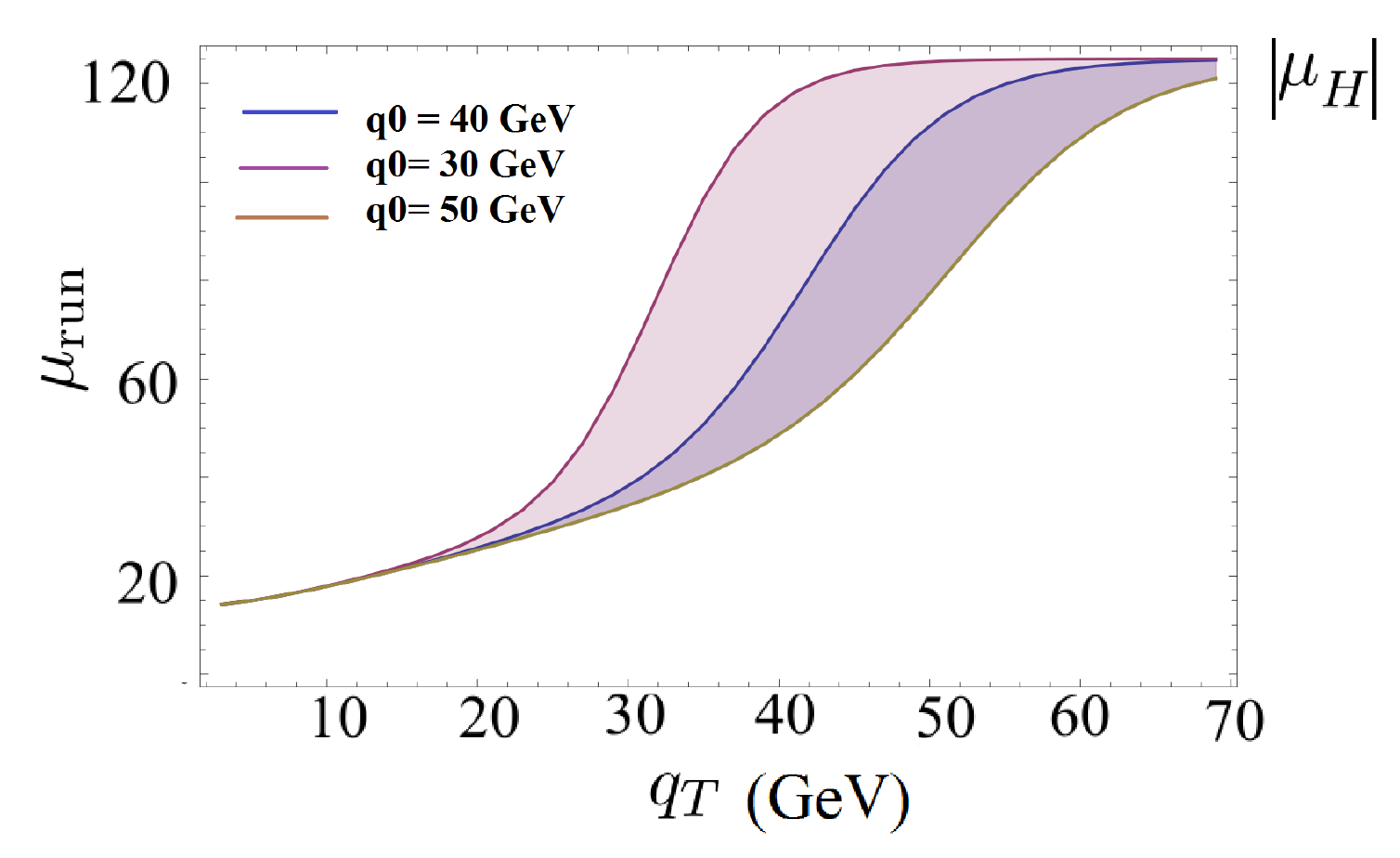}
\qquad\raisebox{-1ex}{
\includegraphics[width=.45\textwidth]{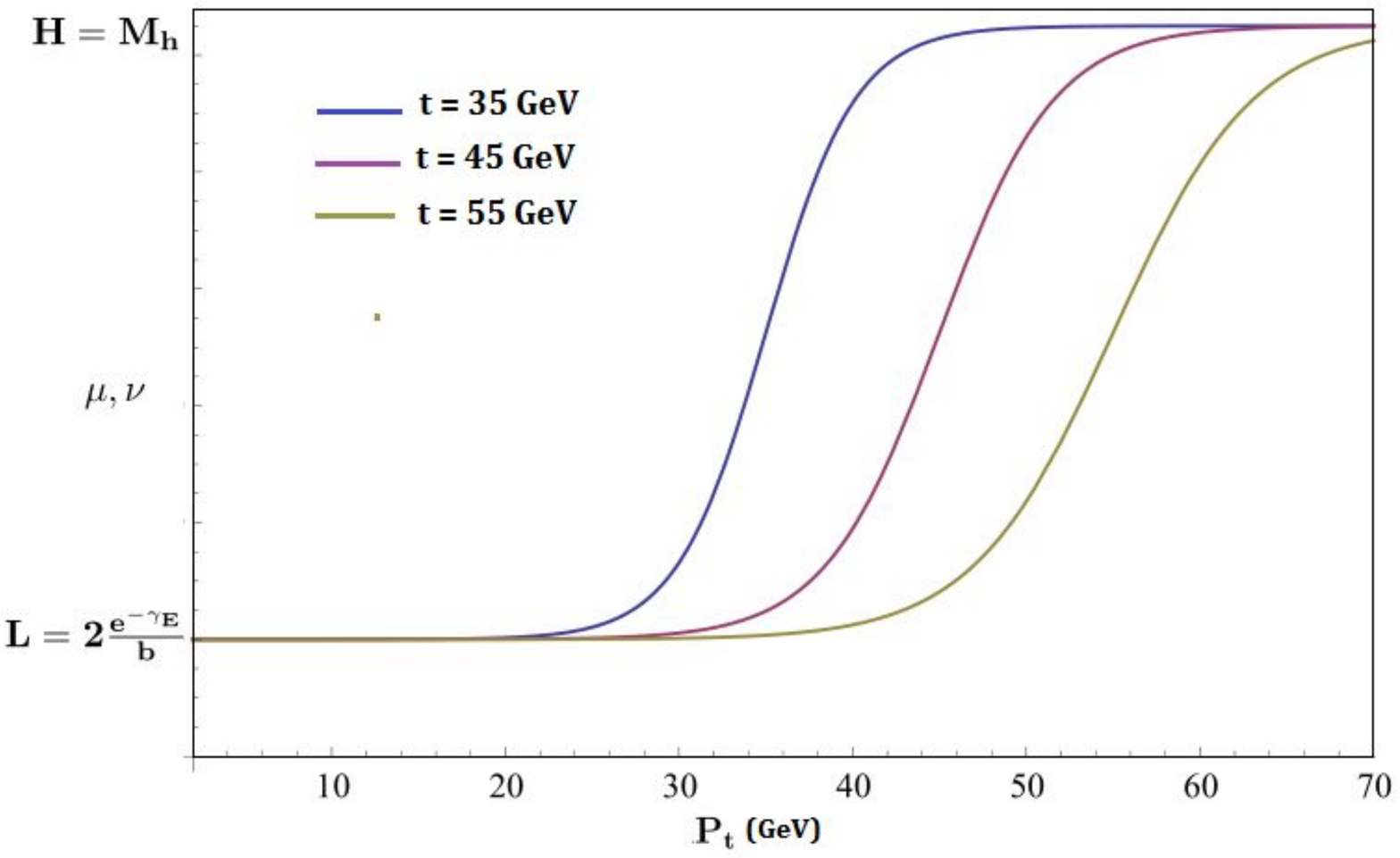}
}
\vspace{-16pt}
\caption{Examples of profile functions for scales $\mu_L,\nu_L$ interpolating between resummation region for small $q_T$ or $1/b_T$ and high $q_T\sim Q$ \cite{Kang:2017cjk,Neill:2015roa}.}
\label{fig:profiles}
\end{figure}
Exactly how the profile scales interpolate between the regions indicated in \eq{profilerun} is a matter of choice. A simple example (by no means unique) from \cite{Kang:2017cjk} interpolating between the resummation and fixed-order region is:
\begin{align}
\label{eq:profile2017}
    \mu_\text{run}(q_T),\nu_\text{run}(q_T) &= q_T^{1-\zeta(q_T)} \mu_H^{\zeta(q_T)} \,,\quad 
    \zeta(q_T) = \frac{1}{2}\biggl\{1 + \tanh\biggl\{\rho\Bigl(\frac{q_T}{q_0} - 1\Bigr)\biggr]\biggr\} \,,
\end{align}
where $\rho$ controls the rate of the rise of the profile from low to high values, and $q_0$ controls the transition point. This point should be chosen to occur close the region where nonsingular terms become comparable in size to the logs in the singular part, see \fig{profiles} (left).
Another example from \cite{Neill:2015roa} is:
\begin{align}
    \label{eq:profile2015}
    \mu_\text{run},\nu_\text{run} = \frac{L}{2}\biggl\{ 1 - \tanh\biggl[s \biggl(\frac{4 q_T}{t}-4\biggr)\biggr]\biggr\} + \frac{H}{2}\biggl\{ 1 + \tanh\biggl[s \biggl(\frac{4 q_T}{t}-4\biggr)\biggr]\biggr\} \,,
\end{align}
with $L,H$ the initial (low) and final (high) values of the scales and $s,t$ determining the rate and location of the transition, respectively, see \fig{profiles} (right).

Many other choices for functional forms of profile scales are possible. Variations in profile parameters can form part of the estimate of theoretical uncertainty. All of this is to illustrate the flexibility afforded by the variable and various scales in the effective field theory framework allowing smooth connections of resummed, fixed-order, and nonperturbative regimes, and robust estimations of theoretical uncertainties.

In the notation of Table~\ref{tbl:resum_orders}, when a resummed $W$ term at a given accuracy is matched onto a $Y$ term at a given fixed-order accuracy, the literature will normally refer to the accuracy of such a matched cross section as N$^k$LL$(')$+N$^n$LO or N$^k$LL$(')+\mathcal{O}(\as^m)$, as the case may be. Recall that the primed $'$ accuracies of the resummed part indicate that the finite coefficient functions $\tilde C$ in Table~\ref{tbl:resum_orders} are kept to one higher power of accuracy in $\as$ than the unprimed. This is often beneficial when matching an N$^k$LL$'$ resummed calculation with a fixed-order calculation to the same accuracy to which $\tilde C$ is known.

\subsection{Resummation in Momentum Space}
\label{sec:momentumspace}
 
 The renormalization group and rapidity evolution equations are simplest when expressed in $b_T$ space. In this form they are ordinary differential equations and can be solved using standard techniques. After the evolution equations are solved, a Fourier integral of a product of the TMDs must be computed in order to obtain the $p_T$ dependence of the cross section. This integration must be done numerically and can be technically challenging. The integral is over all values of $b_T$, including large $b_T$ where perturbation theory is no longer valid. Therefore the resummed expressions involve the running coupling $\alpha_s(\mu)$ evaluated at a scale $\mu \sim 1/b_T$, in which case one will run into the Landau pole where $\alpha_s$ diverges. Earlier we described the most commonly used prescription for dealing with these issues in Eq.~\ref{eq:bstardef}. In this section, we review some recent attempts to avoid these difficulties by formulating the evolution equations directly in momentum space so that one can evolve only between perturbative scales.

\tocless \subsubsection{Distributional momentum-space scheme}

When solving the RGE and RRGE in $b_T$-space one necessarily has to perform scale setting in $b_T$-space. For example, in the last section we evolved the scales $\mu,\nu$ from the scale $Q$ to the scale $b_0/b_T$.
This resums logs of $Q b_T$ in the Fourier transform of the cross section. The goal of Ref.~\cite{Ebert:2016gcn} is to resum logs of the form $[{\rm ln}^2(k_T^2/Q^2)/k_T^2]_+$ directly in momentum space.
As we will see below,  in momentum space the perturbative evaluation of the elements of the factorization theorem leads to distribution functions like $\delta(k_T)$ and $[\theta(k_T)/k_T]_+$. For ordinary functions, RGE is run between the high and low scales to resum logarithms and then in the perturbative part of the calculation $\mu$ is set to a scale that minimizes the logarithms in the perturbative part. Something similar can be done directly in momentum space, but this requires a procedure to set scales in the presence of distributions, as shown in Ref.~\cite{Ebert:2016gcn}.

The work of Ref.~\cite{Ebert:2016gcn} is based on the SCET formalism as presented in \sec{evolution_SCET}.
In momentum space, the evolution in $\mu$ takes the same simple form as in \eqs{beamRGE}{softRGE},
as the corresponding anomalous dimensions are independent of $b_T$,
\begin{subequations}
\begin{align}
\label{eq:beamRGE_kT}
\mu\frac{\df}{\df\mu} B(x,\kt,\mu,\zeta/\nu^2) &= \gamma_\mu^{B}(\mu,\zeta/\nu^2) B(x,\kt,\mu,\zeta/\nu^2) \\
\label{eq:softRGE_kT}
\mu\frac{\df}{\df\mu} S(k_T,\mu,\nu) &= \gamma_\mu^S(\mu,\mu/\nu) S(k_T,\mu,\nu)
\,.\end{align}
\end{subequations}
In contrast, the anomalous dimensions in the $\nu$ evolution in \eqs{beamRRGE}{softRRGE}
depend on $b_T$, and upon Fourier transform the simple product in $b_T$ space
turns into a convolution in momentum space,
\begin{subequations}
\begin{align}
\label{eq:beamRRGE_kT}
\nu\frac{\df}{\df\nu} B(x,\kt,\mu,\zeta/\nu^2) &= \int\df^2\kt' \gamma_\nu^{B}(\kt',\mu) B(x,\kt-\kt',\mu,\zeta/\nu^2) \\
\label{eq:softRRGE_kT}
\nu\frac{\df}{\df\nu} S(k_T,\mu,\nu) &= \int\df^2\kt' \gamma_\nu^{S}(\kt',\mu) S(|\kt - \kt'|,\mu,\nu)
\,.\end{align}
\end{subequations}
The $\nu$-independence of cross section requires $\gamma_{\nu}(\kt,\mu)\equiv \gamma_\nu^S(\kt,\mu) = -2 \gamma_\nu^B(\kt,\mu)$, and commmutativity of $\mu$ and $\nu$ derivatives requires

\begin{align} \label{eq:consistency_kt}
\mu \frac{d}{d\mu} \gamma_\nu^S(\kt,\mu)
 = \nu \frac{d}{d\nu} \gamma_\mu^S(\mu,\nu)\delta(\kt)
 = - 4 \Gamma_{\rm cusp}[\alpha_s(\mu)]\delta(\kt)
\,.\end{align}
\eq{consistency_kt} clearly illustrates that anomalous dimensions,
and consequently the beam and soft functions themselves,
are distributions in momentum space. This complicates the solution
of their (R)RGEs compared to Fourier space. For example,
the formal solution of \eq{consistency_kt} is easily obtained as
\begin{align} \label{eq:sol_consistency_kt}
 \gamma_\nu^S(\kt, \mu)
 = - 4 \delta(\kt) \int_{\mu_0}^\mu \frac{\df \mu'}{\mu'} \Gamma_{\rm cusp}[\alpha_s(\mu')]
 +  \gamma_\nu^S(\kt, \mu_0)
\,.\end{align}
Here, we expect the boundary term to contain logarithms $\ln(k_T / \mu_0)$,
suggesting to choose $\mu_0 = k_T$ to minimize these.
Clearly, this choice clashes with \eq{sol_consistency_kt},
as the $\delta(\kt)$ makes it mathematically ill-defined to
set the lower scale of the integral to $\mu_0 = k_T$.
This is in sharp contrast to Fourier space, where the $\delta(\kt)$ becomes unity,
and one can straightforwardly set $\mu_0 = b_0 / b_T$ to minimize all logarithms
resulting in the simple solution in \eq{gammanuBS}.

To circumvent the above problem, Ref.~\cite{Ebert:2016gcn} developed
a method to solve differential equations such as \eq{consistency_kt}
directly in distribution space. For the simpler case of a one-dimensional
distribution, their prescription is given by
\begin{align}
 D(k,\mu=k|_+)=\frac{d}{dk} \int^k dk' D(k',\mu=k)
\,,\end{align}
where the $\mu = k|_+$ denotes the \emph{distributional scale setting}.
In essence, the prescription is to first take the cumulant of the distribution,
which turns it into a regular function, then set the scale as usual,
and finally take the derivative to go back to distribution space.
In the two-dimensional case relevant for TMD factorization, this becomes
\begin{align}
 D(\pt,\mu=p_T) \equiv
\frac{1}{2\pi p_T}\frac{d}{dp_T}
\left[\int_{|\kt| \leq p_T} d^2\kt D(\kt,\mu=p_T)\right]
\,,\end{align}
where the cumulant and derivative are taken in two dimensions.

Applying this method to \eq{consistency_kt}, we obtain
\begin{align} \label{eq:sol2_consistency_kt}
 \gamma_\nu^S(\pt, \mu) &
 = \left\{- 4 \delta(\pt) \int_{\mu_0}^\mu \frac{\df \mu'}{\mu'} \Gamma_{\rm cusp}[\alpha_s(\mu')]
 +  \gamma_\nu^S(\pt, \mu_0) \right\}_{\mu_0 = p_T|_+}
\nn\\&
 = \frac{1}{2\pi p_T} \frac{\df }{\df p_T}
   \left\{ - 4 \theta(p_T) \int_{p_T}^\mu \frac{\df \mu'}{\mu'} \Gamma_{\rm cusp}[\alpha_s(\mu')]
           + \theta(p_T) \gamma_\nu^S[\as(p_T)] \right\}
\nn\\&
 = \left[ \frac{4 \GammaC[\as(p_T)]}{2 \pi p_T^2} \right]_+^\mu
 + \left[ \frac{1}{2\pi p_T^2} \frac{\df \gamma_\nu[\as(p_T)]}{\df \ln p_T} \right]_+^\mu
 + \delta(p_T) \gamma_\nu^S[\as(\mu)]
\,.\end{align}
In the second step, $\theta(p_T)$ is the Heaviside function
which is crucial for the correct distributional behavior after taking the derivative,
and the boundary term is denoted as $\gamma_\nu^S[\as(p_T)]$, as it only depends on the scale $p_T$.
In the last step, we have taken the derivative; the resulting two-dimensional
distributions are defined as
\begin{align}
 \left[f(\pt)\right]_+^\mu = f(\pt) \quad {\rm for}~p_T > 0
\,,\qquad
 \int_{|\pt| \le \mu} \df^2\pt \left[f(\pt)\right]_+^\mu = 0
\,.\end{align}
One can check that the $\mu$ dependence of the first term in the last line of \eq{sol2_consistency_kt}
precisely obeys \eq{consistency_kt}, while the $\mu$ dependence cancels between
the last two terms. Nevertheless, these two terms have a non-trivial structure,
showing that the boundary term in momentum space is not only located at $p_T = 0$,
but spread throughout $p_T$ space.

\eq{sol2_consistency_kt} illustrates a few key features of the solution
in momentum space, and how it relates to the solution Fourier space.
First, we see that the strong coupling in \eq{sol2_consistency_kt}
is evaluated at $\as(p_T)$ and thus exhibits a Landau pole at $p_T \lesssim \lqcd$,
similar to the result in Fourier space becoming nonperturbative for
$b_0 / b_T \lesssim \lqcd$.
Due to this divergence, one can not Fourier transform \eq{sol2_consistency_kt}.
However, upon expanding \eq{sol2_consistency_kt} in $\as(\mu)$, one can Fourier transform
$\gamma_\nu^S$ order by order in $\as(\mu)$. From this, Ref.~\cite{Ebert:2016gcn}
observed that the fixed-order expansions of $\gamma_\nu^S(\pt, \mu)$ and
$\tilde \gamma_\nu^S(\bt, \mu)$ agree order by order, up to different boundary
terms in their solutions. This implies that to a given resummation accuracy,
both approaches are formally equivalent, despite differing by formally higher-order
terms of nonperturbative origin.

Let us now turn to the solution of the more complicated evolution in $\nu$.
Focusing on the soft function, the formal solution of \eq{softRRGE_kT} is
\bea
S(\pt,\mu,\nu_B) = \int d^2\kt  V(\pt-\kt,\mu,\nu_B,\nu_S)  S(\kt,\mu,\nu_S)\,.
\label{eq:convolution}
\eea
where the rapidity evolution kernel is given by
\bea \label{eq:sol_RRGE_MS}
V(\pt,\mu,\nu_B,\nu_S) = \delta(\pt) +\ln\frac{\nu_B}{\nu_S}\gamma_{\nu}(\pt,\mu)
+\frac{1}{2}\ln^2\frac{\nu_B}{\nu_S}\int d^2\kt  \gamma_{\nu}(\kt,\mu)\gamma_{\nu}(\pt-\kt,\mu)  + ...  .
\eea
where higher order terms involve multiple convolutions.
Similar to the case of $\gamma_\nu^S$, the correct solution
in momentum space is obtained by distributionally setting $\nu_S = p_T|_+$.
This can not be achieved in closed form, and one has to resort to numerical methods,
which so far have not been developed.

  The authors of Ref.~\cite{Monni:2016ktx} developed a numerical resummation of TMD distributions in momentum space based on the general methods of \cite{Banfi:2004yd,Banfi:2014sua} which do not begin from factorization and RGEs per se, but can be applied even to observables that do not manifestly factorize. 
A full review of these methods is outside the scope of this Handbook,
and we refer to \cite{Bizon:2017rah} for more details on their method.
In the following, we briefly discuss their method at NLL accuracy, following the presentation of Ref.~\cite{Monni:2016ktx}.
Their starting point is the cumulant of the $q_T$ distribution,
which at NLL accuracy can be written as follows:
\begin{align} \label{eq:MRT_1}
 \Sigma(q_T) &
 = \int_0^{q_T}\! \df k_T\, \frac{\df \sigma(k_T)}{\df k_T}
 \\ \nn
 &= \sigma_0 \int_0^\infty\! \langle\df k_1\rangle \, R'(k_1)\, e^{-R(\eps k_1)}
   \sum_{n=0}^\infty \frac{1}{n!} \prod_{i=2}^{n+1}
   \int_{\eps k_1}^{k_1}\! \langle{\df k_i}\rangle\, R'(k_i)\,
   \theta\biggl(|\qt| - \biggl\lvert\sum_j \mathbf{k}_j \biggr\rvert\biggr)
\,.\end{align}
Here, the $k_i$ are the real emissions recoiling against the color-singlet final state,
and the largest emission $k_1$ has been singled out. Each emission comes with a measure
$\langle{k_T}\rangle = \frac{\df k_T}{k_T} \frac{\df \phi}{2\pi}$.
The parameter $\eps \ll 1$ indicates that emissions with momenta below $\eps k_1$
are unresolved, i.e.~do not contribute to the observable $q_T$, up to small corrections in $\eps$.
Thus, they can be neglected in the calculation of $q_T$, and have already been integrated over,
with their effect being encoded by the so-called radiator~\cite{Banfi:2012yh},
\begin{equation} \label{eq:MRT_2}
 R(\eps k_T)
 = \int_{\eps k_T}^Q \frac{\df\mu'}{\mu'} \gamma_\mu^H(Q,\mu')
\,.\end{equation}
Its derivative $R'(k_i)$ approximates the full matrix element at NLL,
and is corrected at higher orders.
In essence, \eq{MRT_1} thus constitutes the calculation of the cumulant in $q_T$
by explicitly summing over all possible real emissions weighted by an approximate
matrix element. By working in cumulant space, this can be implemented numerically
via a parton-shower approach. The $q_T$ spectrum is then obtained by a (numerical)
derivative with respect to $q_T$.

Naively, in order to resum logarithms of $\ln(Q/q_T)$, where $Q$ is the hard scale,
one would like to expand $R(\eps k_1)$ and $R'(k_i)$ around $k_i \sim q_T$.
However, this leads to a well known singularity~\cite{Frixione:1998dw}, see \eq{NLL_div} below.
To circumvent this problem, \cite{Monni:2016ktx} proposes to expand around the hardest
emission $k_1$ instead, such that
\begin{align}
 R(\eps k_1) = R(k_1) + R'(k_1) \ln\frac{1}{\eps} + \cdots
\,,\qquad
 R'(k_i) = R'(k_1) + \cdots
\,.\end{align}
Effectively, this approach thus resums logarithms $\ln(Q/k_1)$ rather than
$\ln(Q/q_T)$ in momentum space, and leads to a stable prediction.
On the other hand, it is argued that small $q_T$ can be dominated
by large $k_i$ due to kinematic cancellations in $\qt = \sum_i \mathbf{k}_i$,
and thus $k_1$ is a more appropriate resummation scale than $q_T$ itself.

The structure of \eq{MRT_1} is similar to the convolution structure in \eq{sol_RRGE_MS},
but with the critical difference of scales being set to $k_1$ rather than distributionally.
In Ref.~\cite{Bizon:2017rah}, their approach was also compared to conventional
resummation in Fourier space. Similar to the discussion above, it was found
that formally both approaches are equivalent, up to different terms entering
the boundary conditions of the resummation. Finally, we remark that \eq{MRT_2}
also suffers from the Landau pole when the scales entering the radiator $R(k_1)$
become nonperturbative. This region is simply excluded in the direct-space approach,
and it has not been established yet how to supplement their numerical approach with a nonperturbative model.

\tocless \subsubsection{Hybrid schemes}

Another set of approaches take what we call a \emph{hybrid} approach. Namely, in the expression \eq{DYbspace} for the momentum-space $q_T$ spectrum, one does still choose the low rapidity $\nu$ scale in the soft function to be a function of $b_T$ but chooses the low $\mu$ scale in the beam and soft functions to be purely in momentum space. In \cite{Becher:2011xn,Becher:2012yn}, an early version of this was introduced, with an implicit choice of $\nu_L\sim 1/b_T$ already made but without an actual rapidity scale that can be varied (to properly probe uncertainties). The $\mu_L$ scale was left to be chosen in momentum space, thus avoiding an explicit Landau pole in transforming from $b_T$ to $q_T$ space. In \cite{Neill:2015roa}, the full power of SCET and the RRG was implemented, though with all low scales chosen in $b_T$ space as $\sim 1/b_T$. However, the variable RG and rapidity scales were also made functions of $q_T$ (i.e., profile scales) in such as way as to achieve a smooth matching of the momentum-space cross section onto the fixed-order result for large $q_T$. This paved a path to a more fully ``hybrid'' scheme using SCET and the RRG in \cite{Kang:2017cjk}, where the $\mu_L$ scale was chosen fully in momentum space while $\nu_L$ was left in $b_T$ space. The choice of $\mu_L$ paralleled the choices of \cite{Becher:2011xn,Becher:2012yn} while maintaining full variable dependence on the rapidity scale $\nu_L$ as in \cite{Neill:2015roa}. Without making any attempt to compare the advantages of any of these approaches, we will review the  hybrid approach of \cite{Kang:2017cjk} in some detail here, for its pedagogical value.

In \cite{Kang:2017cjk}, the natural, central choices for the scales are slightly modified from the na\i ve choices $\nu_L\sim 1/b_T$ and $\mu_L\sim q_T$. The Fourier transform integral in \eq{DYbspace} can then either be done   numerically, but much faster than typical in a purely $b_T$-space scale-setting scheme, or be done after a very good approximation to the $b_T$ integrand in \eq{DYbspace} that makes it \emph{analytically} integrable.

After evolution from their natural scales, the $b_T$ integrand of \eq{DYbspace} takes the form \eq{resummedcs}. With the form of the evolution factor in \eq{Utot}, the integral that must be done to bring the cross section back to momentum space is:
\begin{align}
\label{eq:bTintegrand}
\frac{d\sigma^W}{dq_T^2} &\sim \int db_T\, b_T\,J_0(bq_T)S(b_T;\mu_L,\mu_L/\nu_L) B(x_a,\vect{b}_T,\mu_L,\zeta_a/\nu_H^2)B(x_B,\vect{b}_T,\mu_L,\zeta_b/\nu_H^2) \nn \\
&\qquad \times \exp\biggl[ -\Gamma_0\frac{\as(\mu_L)}{\pi}\ln\frac{\nu_H}{\nu_L}\ln\frac{\mu_L b_T}{b_0}\biggr]\,,
\end{align}
where we have truncated the rapidity evolution kernel for now to NLL accuracy (and displayed only terms that have $b_T$ dependence). The high rapidity scale $\nu_H$ here can be considered to be the standard choice $\sim Q$, but the low scale $\nu_L,\mu_L$ are not yet specified. 
We can try to evaluate the integral in \eq{bTintegrand} explicitly at NLL accuracy (setting $S$ and $B$ to tree level):
\begin{equation}
\label{eq:NLL_div}
\frac{d\sigma^W}{dq_T^2} \sim e^{-2\omega_S\gamma_E} \frac{\Gamma(1-\omega_S)}{\Gamma(\omega_S)} \frac{1}{\mu_L^2} \biggl(\frac{\mu_L^2}{q_T^2}\biggr)^{1-\omega_S} f(x_a,\mu_L)f(x_B,\mu_L)\,, \quad \text{where } \omega_S = \Gamma_0\frac{\alpha_s(\mu_L)}{2\pi}\ln\frac{\nu_H}{\nu_L}\,,
\end{equation}
which is a nice simple analytic result, but has a divergence at $\omega_S = 1$. This is a problem since $\omega_S>0$, and typically we do have $\omega_S\sim 1$. This divergence comes from the $\ln \mu_L b_T$ in \eq{bTintegrand}, or from small values of $b_T$, not large, as we would have otherwise expected. This problem was first noted and studied in \cite{Frixione:1998dw}.
Normally choosing $\mu_L\sim b_0/b_T$ solves this issue (while large $b_T$ still requires a regulator/cutoff like those described earlier), but we would like here to leave $\mu_L$ unspecified for now and be free to choose it in momentum space. What we need, then, is a way to regulate the integral in \eq{bTintegrand} for \emph{both} small and large $b_T$.

In \cite{Kang:2017cjk} it was observed that in the low-scale soft function $S(b_T;\mu_L,\mu_L/\nu_L)$, there are terms which if exponentiated would naturally provide a regulator for both the low- and high-$b_T$ regions of the integral. Namely, we would like to include the terms in $S$ that organize themselves into the form:
\begin{equation}
S_\text{exp}(b_T) = \exp\biggl[ -\frac{\as(\mu_L)}{2\pi}\Gamma_0\ln^2(\mu_L b_T/b_0)  - \frac{\as^2(\mu_L)}{4\pi^2}\Gamma_0\beta_0\ln^2(\mu_L b_T/b_0) \ln \frac{\nu_H}{\nu_L}\biggr]\,,
\end{equation} 
which do in fact exist as part of its all-orders expansion. We can shift these terms from the fixed-order expansion of $S$ in \eq{bTintegrand} into the exponent of the rapidity evolution kernel by making a shift in the natural choice of scale $\nu_L$, namely, instead of starting the evolution of $S$ at $\nu_L\sim \mu_L$, we choose to evaluate it instead at the scale:
\begin{equation}
\label{eq:nuLstar}
\nu_L \to \nu_L^* = \nu_L(\mu_L b_T/b_0)^{-1+p}\,,\quad p = \frac{1}{2}\Bigl[ 1- \frac{\as(\mu_L)\beta_0}{2\pi}\ln\frac{\nu_H}{\nu_L}\Bigr]\,,
\end{equation}
This factor now naturally regulates the large (and small) $b_T$ regions of the integral in \eq{bTintegrand}. With $\nu_L$ replaced by $\nu_L^*$ in \eq{bTintegrand}, one can choose $\mu_L$ as a function of a momentum, not of $b_T$, and evaluate the $b_T$ integral numerically fairly quickly, without encountering a Landau pole in $\as(\mu_L)$. There is an optimal choice of $\mu_L$ which is not exactly $q_T$ as one might na\i vely expect but is slightly higher. See \cite{Kang:2017cjk} for details (cf. also \cite{Becher:2011xn,Becher:2012yn}).

One can go further in this approach and obtain an analytic result for the integral in \eq{bTintegrand} that is a very close approximation to the exact numerical result. The approach tackles integrals of the form
\begin{equation}
I_b^0 = \int_0^\infty db\,b J_0(b q_T) e^{-A \ln^2\Omega b}\,,
\end{equation}
in which \eq{bTintegrand} can be put with the choice $\nu_L\to\nu_L^*$ in \eq{nuLstar}. $A,\Omega$ are functions of the various scales \cite{Kang:2017cjk}. By using the Mellin-Barnes representation of the Bessel function,
\begin{equation}
J_0(z) = \frac{1}{2\pi i}\int_{c-i\infty}^{c+i\infty}dt\frac{\Gamma[-t]}{\Gamma[1+t]}\Bigl(\frac{z}{2}\Bigr)^{2t}\,,
\end{equation}
where $c$ lies to the left of all poles of the Gamma function, we obtain a form of $I_b^0$ that is amenable to a series expansion of part of the integrand and an analytic integration:
\begin{equation}
\label{eq:Gaussianintegrand}
I_b^0 = \frac{2}{\pi q_T^2 \sqrt{\pi A}} \Imag\biggl\{e^{-A(L-i\pi/2)^2}\int_{-\infty}^\infty dx\Gamma[-c-ix]^2 e^{-\frac{1}{A}[x+A\pi/2-i(c-t_0)]^2}\biggr\}\,,
\end{equation}
where $L = \ln(2\Omega/q_T)$ and $t_0 = -1+AL$. $c=-1$ turns out to be a stable choice for the integration contour, and the Gamma function admits a useful series expansion in Hermite polynomials:
\begin{equation}
\label{eq:Hermiteexpansion}
\Gamma(1-ix)^2 = e^{-A_0 x^2}\sum_{n=0}^\infty c_{2n}H_{2n}(\alpha x) + \frac{i\gamma_E}{\beta} e^{-B_0 x^2}\sum_{n=0}^\infty c_{2n+1} H_{2n+1}(\beta x)\,,
\end{equation}
$A_0,B_0$ are numerical coefficients chosen so the right-hand side most closely approximates the Gaussian nature of $\Gamma(1-ix)^2$, and $\alpha,\beta$ are also coefficients that are chosen to speed up convergence of the series expansion, while maintaining accuracy of the expansion over a wide enough range in $x$ to capture the range of the Gaussian factor in \eq{Gaussianintegrand}. There is not a unique best choice, but in \cite{Kang:2017cjk} some suggested choices are given. In terms of these choices, the series coefficients $c_{2n,2n+1}$ in \eq{Hermiteexpansion} are uniquely determined. Typically only the first three or so of the even and the odd coefficients are needed for sufficient accuracy. In terms of these coefficients, the result of the integral $I_b^0$ (and thus, the $q_T$ spectrum in momentum space) can be expressed in the explicit form:
\begin{equation}
I_b^0 = \frac{2}{\pi q_T^2} \sum_{n=0}^\infty \Imag \Bigl[ c_{2n} \mathcal{H}_{2n}(\alpha,A_0) + \frac{i\gamma_E}{\beta}c_{2n+1}\mathcal{H}_{2n+1}(\beta,B_0)\Bigr] \,,
\end{equation}
\begin{figure}[t!]
\centering
\vspace{-12pt}
\includegraphics[width=\textwidth]{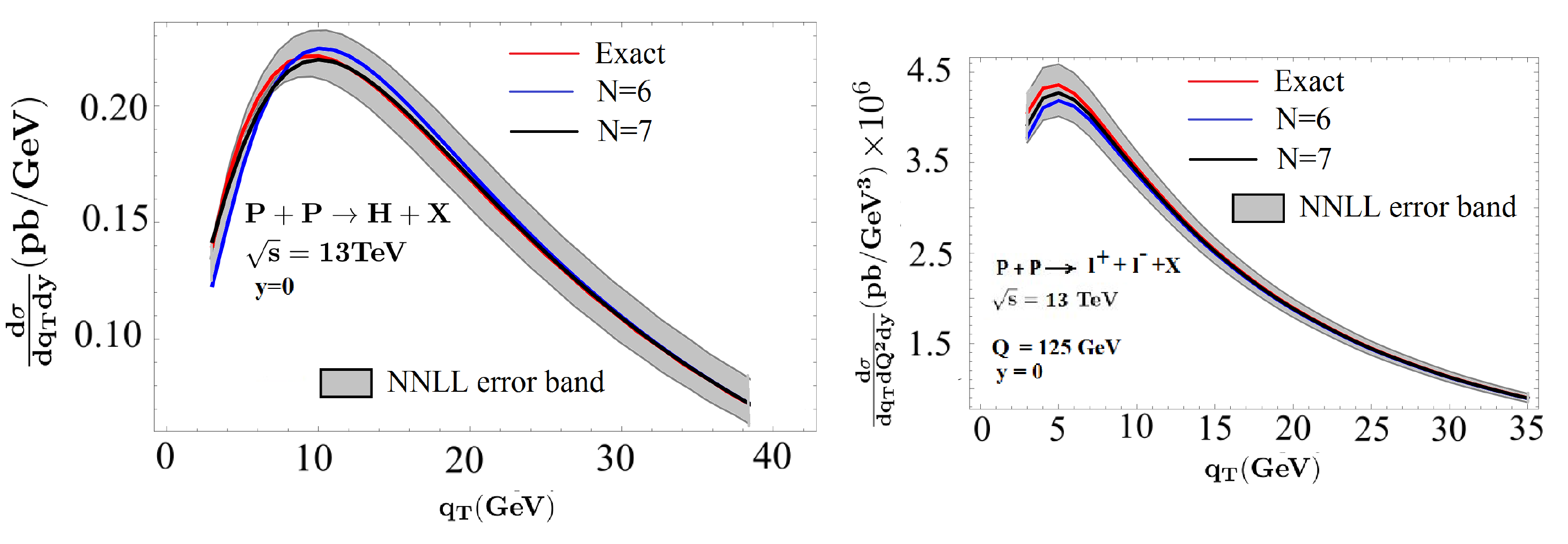}
\vspace{-24pt}
\caption{Systematic improvement in the accuracy of Higgs (left) and DY (right) $q_T$ cross sections with increasing number of terms, with increasing nmber of terms in the Hermite polynomial expansion used in \eq{Hermiteexpansion}. ``Exact'' (red) gives resummed cross section without Hermite expansion (i.e. numerical b integration). $N = 6$ (blue) is the result with six terms in this expansion, three each for real and imaginary terms. $N = 7$ (black) is the result with one more real term. Here we plot only the purely resummed result, i.e. with no matching to the fixed order cross section. From \cite{Kang:2017cjk}.}
\label{fig:Hcomparison}
\end{figure}
where
\begin{equation}
\mathcal{H}_n(\alpha,A_0) = \mathcal{H}_0(\alpha,A_0)\frac{(-1)^n n!}{(1+A_0 A)^n} \sum_{m=0}^{\lfloor{n/2}\rfloor}\frac{1}{m!}\frac{1}{(n-2m)!} \Bigl\{ [ A(\alpha^2-A_0) - 1] (1+A_0 A)\Bigl\}^m(2\alpha z_0)^{n-2m}\,,
\end{equation}
where $z_0 = A(\pi/2+iL)$, the prefactor is
\begin{equation}
 \mathcal{H}_0(\alpha,A_0) = e^{\frac{-A(L-i\pi/2)^2}{1+A_0A}} \frac{1}{\sqrt{1+A_0 A}}\,,
\end{equation}
and the same formulas hold for odd $n$ with $\alpha\to \beta,A_0\to B_0$. Though these expressions are a bit involved, they represent forms of the momentum-space result for the inverse Fourier transform of the $b_T$-space integrand of the form appearing in the cross section \eq{bTintegrand}, retaining the full analytic dependence on all resummation and momentum scales. They are useful for a fast and efficient evaluation of the resummed momentum-space spectrum.

\begin{figure}[t!]
\centering
\vspace{-12pt}
\includegraphics[width=\textwidth]{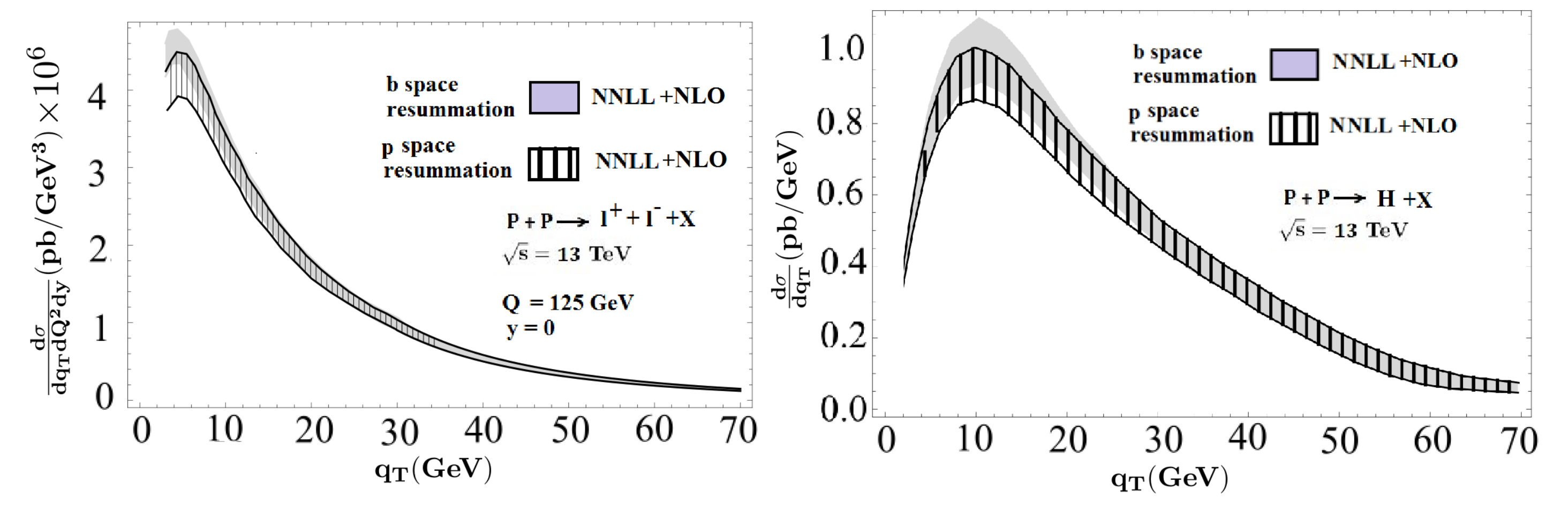}
\vspace{-24pt}
\caption{Comparison of $q_T$ distributions at NNLL accuracy matched to fixed-order $\cO(\as)$, in a $b$-space scheme numerically integrated with a cutoff in $b_T$ to Fourier transform to momentum space (gray solid band) vs. in the hybrid scheme of \cite{Kang:2017cjk}. The overlap shows the consistency of the methods with one another. The fixed-order matching is done with the method of profile scales similar to that described in \sec{profiles} and \cite{Neill:2015roa}.}\vspace{-1em}
\label{fig:bvsp_comparison}
\end{figure}
In \fig{Hcomparison} we show results from \cite{Kang:2017cjk} for the $q_T$ spectrum in Higgs production or Drell-Yan at the LHC, comparing the exact numerical result from evaluating \eq{bTintegrand} numerically, versus the method of Hermite polynomials outlined above, with a total of six or seven basis terms used in the expansion. The discrepancies are smaller than the perturbative uncertainty in the NNLL result. In \fig{bvsp_comparison} we then compare the result of computing the NNLL resummed cross section matched to a fixed-order $\cO(\as)$ prediction, in this hybrid scheme vs. a $b$-space resummation scheme as implemented in \cite{Neill:2015roa}.

\subsection{Summary and Outlook}

In this chapter, we reviewed evolution and resummation for TMDs  which arise as consequences of the renormalization and definitions of TMD PDFs discussed in depth in \chap{TMDdefn} and factorization of TMD cross sections whose demonstration was reviewed in \chap{Factorization}. \sec{evolutionintro} gave the primary motivations to study evolution for TMDs: to compare TMDs extracted at different scales and to resum large logarithms to all orders in perturbation theory.  We also presented a brief historical review of work on deriving the evolution equations. In \sec{EvolResum} we schematically displayed the all orders form of the fixed-order expansion of the perturbative cross section, as well as the reorganized exponentiated form it takes in the resummed perturbative expansion. This was crucial to obtain a well-behaved perturbative expansion for the TMD cross section for small $q_T$. This  schematic description also allows the reader to understand 
which terms are captured by LL, NLL, NNLL, etc.~accuracy  and the order to which individual ingredients in the factorized cross section need to be computed to achieve these accuracies. This section also showed how to solve evolution equations and obtain resummed expressions for the hard function, as an illustrative example. In \sec{TMDEvol} the full TMD evolution was studied. The full form of the CSS equations is given, as well the form of the evolution equations in other approaches including SCET. The anomalous dimensions are extracted at one-loop order from the one-loop calculation of the TMD PDF in \sec{tmd_defs_nlo}. The CSS formalism was further discussed in \sec{CoordEvol}, where the need for a nonperturbative treatment of the large $b_T$ region is explained along with the commonly used $\bstarsc$ prescription for freezing $b_T$ to perturbative scales in the cross section. The perturbative and nonperturbative contributions to the TMD PDF are defined. 

Beginning in \sec{evolution_SCET} the SCET approach to evolution is described in detail. The SCET factorization theorems  emphasize beam and soft functions,  which correspond to matrix elements of collinear and soft operators in the effective theory Lagrangian. Using the relations derived in \chap{TMDdefn} between these beam and soft functions and the TMD PDFs, we were able to derive RG and RRG evolution equations for TMD PDFs that are equivalent to those found in the CSS formalism. Solutions to RG and RRG evolution equations were given and path independence of the solutions was emphasized.  This property motivates the full two-dimensional picture of TMD evolution reviewed in \sec{2dRRGE},  leading to an intuitive geometrical interpretation and  analogies to
 equipotential lines, along which evolution vanishes.
 In addition to the resummed expression for the $W$ term it is important to be able to interpolate between regions in which resummation is needed and regions where fixed order calculations are appropriate. This is the subject of matching the $W$ term onto  large $\qt$
 collinear factorized cross section. This was the subject
 of \sec{nonsingular}, where both CSS and SCET approaches are again reviewed in parallel.
 In the former approach,  $W+Y$ construction is reviewed, where matching of the $W$ term onto the fixed order term is implemented thru the asymptotic term supplemented with cutoff functions designed to respect the errors of the approximators that designate the TMD and collinear momentum regions.  
In the latter approach,  profile functions, which allow one to turn on and turn off resummation as needed, are introduced. \sec{momentumspace}
describes recent proposals to formulate the evolution equations and their corresponding solutions directly in momentum space, or in a hybrid of $b_T$ and momentum spaces, in an attempt to get around some of the   issues encountered in resumming in $b_T$ space, e.g., Landau poles. This is an area in which we expect more work to be done in coming years. 

Evolution and resummation will play a crucial role in subsequent chapters. The following \chap{phenoTMDs}  gives a broad and thorough overview of the phenomenology of TMDs,  and detailed comparison of the predictions TMD formalism with Drell-Yan, SIDIS,  and di-hadron production will be discussed. As we will see, early fits used naive gaussians multiplying collinear PDFs and did not incorporate evolution. As more detailed information about TMDs has become available, the TMD evolution discussed in this  chapter has become essential for properly interpreting and extracting the TMDs.

%% file: sec-phenomenology/sec-phenomenology.tex
\section{Phenomenology and Extraction of TMDs}
\label{sec:phenoTMDs}

\subsection{Introduction: Historical Perspective}
\label{sec:phenoTMDs_intro}

TMD phenomenology plays an important role in testing theoretical ideas about the properties of TMDs, the applicability of QCD factorization theorems, the interplay between perturbative and nonperturbative regimes; and, ultimately, in the analysis of the existing experimental data and predictions for future measurements.

The importance of the transverse motion of partons confined  inside the nucleon  was  pointed out in the 1970s by Feynman, Field, and
Fox~\cite{Feynman:1977yr,Feynman:1978dt}, who realized that the origin of transverse momentum of the produced lepton pair in 
Drell-Yan processes could be either due to the nonzero  
{\em intrinsic} transverse momentum of partons confined in the nucleon (nonperturbative effect) or due to the 
 gluon radiation off active partons (perturbative effect). These studies were precursors of the naive TMD picture such as the Generalized Parton Model, Sec.~\ref{Sec:parton-model-general}, 
 \index{parton model|(}
 developed by the Torino-Cagliari group in pioneering studies of asymmetries in hadron-hadron scattering~\cite{Anselmino:1994tv,Anselmino:1999pw,Boglione:1999dq,Anselmino:2004nk}, 
 and to the rigorous QCD factorization proofs by Collins-Soper-Sterman (CSS)~\cite{Collins:1981uk,Collins:1984kg, Collins:2011zzd}. The modern TMD factorization theorems with  well-defined TMDs are discussed in Chapters~\ref{sec:TMDdefn}- \ref{sec:evolution}.
 
 Azimuthal asymmetries in unpolarized reactions, Drell-Yan, and SIDIS, can be used to test the perturbative and 
nonperturbative aspects of strong interactions, as recognized in early work by 
Georgi and Politzer~\cite{Georgi:1977tv}, Mendez~\cite{Mendez:1978zx}, and Kane, Pumplin, and 
Repko~\cite{Kane:1978nd}.
It was Robert Cahn~\cite{Cahn:1978se,Cahn:1989yf} who first pointed out that intrinsic quark motion can generate an azimuthal $\cos\phi_h$ asymmetry in unpolarized SIDIS, where $\phi_h$ is the azimuthal angle of the hadron plane with respect to the lepton scattering plane. 
This  $\cos \phi_h$ or   ``Cahn effect''  in SIDIS
is a subleading TMD effect and presented in \chap{twist3}.
\index{Cahn effect} 

The systematic description of  SIDIS cross sections in terms of TMD functions began in 1995,
 when Kotzinian \cite{Kotzinian:1994dv} and 
 later 
 Mulders and Tan\-ger\-man \cite{Tangerman:1995hw}, Boer and Mulders \cite{Boer:1997nt} expressed the
unpolarized and polarized SIDIS cross sections in terms of structure functions that are, at tree level, described by convolutions of TMDs. See Ref.~\cite{Bacchetta:2006tn} and \sec{TMDSIDIS} for the modern description of SIDIS in terms of TMDs. The polarized Drell-Yan process was parametrized in terms of TMDs by Mulders and 
Tan\-ger\-man in Ref. \cite{Tangerman:1994eh} and recently the description was completed by Arnold, Metz,  and Schlegel in Ref.~\cite{Arnold:2008kf}, see also \sec{TMDDrellYan}. 
Boer,  Jakob, and Mulders investigated asymmetries in polarized hadron production in $e^+ e^-$ annihilation, see  \sec{TMDee}, Ref.~\cite{Boer:1997mf}, and Pitonyak, Schlegel, and Metz extended upon this work in Ref.~\cite{Pitonyak:2013dsu}.

Simultaneously, the description of asymmetries in terms of multi-parton quantum mechanical correlations, or twist-3 functions, a well-known example being the Efremov-Teryaev-Qiu-Ster\-man 
matrix element~\cite{Efremov:1981sh,Efremov:1983eb,Qiu:1991pp,Qiu:1998ia}, was formulated for processes with only one large hard scale. \index{Qiu-Sterman (QS) function}
These correlations are suppressed relative to the leading power contribution to the unpolarized 
cross-sections, but can be dominant in spin asymmetries. They are key ingredients in collinear approach to factorization~\cite{Qiu:1991pp,Qiu:1991wg}. 
It was later realized that TMD and twist-3 approaches are intimately related~\cite{Ji:2006ub}. The first phenomenological demonstration of the common origin of the transverse  single spin asymmetries in various processes was performed in Ref.~\cite{Cammarota:2020qcw}.

In the 1990s two very important TMD functions encoding correlations of transverse motion and spin were proposed by 
Sivers~\cite{Sivers:1989cc,Sivers:1990fh} and Collins~\cite{Collins:1992kk}. 
In order to describe the large (left-right) single spin asymmetries (SSAs) of pion 
production in hadron-hadron scattering, Sivers suggested that they could originate, at leading 
power, from the intrinsic motion of quarks in the colliding hadrons generating an inner asymmetry  
of unpolarised quarks in a transversely polarized hadron, the so-called ``Sivers effect''. \index{Sivers effect}
He proposed a new TMD  function, now commonly called 
the Sivers function ($f_{1T}^\perp$), which represents the number density of unpolarized partons inside a 
transversely polarized nucleon. \index{Sivers function $f_{1T}^{\perp}$!introduction} This mechanism was criticized at first~\cite{Collins:1992kk} as it seemed to violate 
time-reversal invariance of QCD; however Brodsky, Hwang and Schmidt proved 
by an explicit calculation that initial-state interactions in Drell-Yan  
processes~\cite{Brodsky:2002rv,Brodsky:2013oya} and final-state interactions in SIDIS \cite{Brodsky:2002cx}, 
arising from gluon exchange between the struck quark and the nucleon remnants, can generate a leading (not power suppressed) transverse spin asymmetry.
This model calculation is reviewed in Sec.~\ref{Subsec-models:review-Brodsky-Hwang-Schmidt}.
The situation was further clarified by Collins \cite{Collins:2002kn} who pointed out that this transverse spin asymmetry is due to the Sivers function which,
taking correctly into account the gauge links in the TMDs, is not forbidden by time-reversal but rather enters the descriptions of SIDIS and Drell-Yan processes with opposite signs.

Collins proposed a mechanism based on a spin asymmetry
in the fragmentation of transversely polarized quarks into a spinless hadron~\cite{Collins:1992kk},
\index{Collins function $H_1^{\perp}$!introduction}
\index{transversity!introduction}
which involved a TMD fragmentation function, called the Collins 
function ($H_{1}^\perp$), which generates a typical azimuthal correlation, later denoted as the ``Collins effect''. \index{Collins effect}
This work was preceded by other proposed
methods to measure the polarization state of a parton
coming out of a hard scattering process. Nachtmann
suggested a parity-odd three-particle correlation 
within a jet to determine the longitudinal polarization
of a parton \cite{Nachtmann:1977ek}, and
Dalitz, Goldstein and Marshall discussed how to 
probe the helicity of heavy quarks in $e^+e^-$
annihilation \cite{Dalitz:1988ab,Dalitz:1988aq}.
Efremov, Mankiewicz and T\"ornqvist showed how to
probe {\it transverse} polarization of partons 
using the concept of ``jet handedness'' and
showed how it can be used to measure transversity
\cite{Efremov:1992pe}. This concept was later 
independently elaborated by Collins, Heppelmann
and Ladinsky in Ref.~\cite{Collins:1993kq}. 
(For completeness, it should be mentioned that
transversity can also be accessed in a collinear
factorization approach in terms of the so-called
interference fragmentation functions. 
We refer the interested reader to \cite{Jaffe:1997hf,Bianconi:1999cd,Radici:2001na,Bacchetta:2002ux}.)

The definition of TMDs is gauge invariant and follows from QCD factorization theorems, see the discussion in \chaps{TMDdefn}{Factorization}.  A generic unpolarized TMD PDF $f$ in momentum ${k_T}$-space is related to TMD $\tilde f$ in configuration $b_T$-space  via the inverse Fourier transform. 
Note that theoretically TMDs are usually studied in $b_T$-space, for instance, in studies of the operator definition and evolution of $\tilde f (x, {b_T})$, see \chaps{TMDdefn}{Factorization}, while experimental measurements are carried out in momentum space. Experimentally measured observables, such as cross-sections, are related to the structure functions that encode the dynamics of confined partons and can be expressed in the TMD approximation as convolutions. The convolution in momentum space implies an integration over the unobserved parton momenta, while in configuration space the convolution becomes a simple product~\cite{Boer:2011xd} of TMDs in $b_T$-space. Thus, experimentally measured cross-sections are not a direct measure of TMDs and global QCD fits have to deal with the model dependence and shape bias of TMD parametrizations. 
In addition, the extraction of the hadron structure relies on the  precise reconstruction of the $\gamma^* P$ frame of SIDIS events. The QED radiation distorts the $\gamma^* P$ frame and therefore impacts precise extraction of the underlying hadron structure. The formalism that incorporates both QED and TMD physics will be discussed in \sec{QED}.

In this Chapter we will review phenomenological predictions and extractions of TMDs from SIDIS, Drell-Yan, weak gauge boson production, $e^+e^-$ annihilation into hadron pairs, proton-proton scattering, including corresponding azimuthal and spin modulations of cross sections at leading power. The subleading effects will be discussed in \chap{twist3}. We will also discuss open questions in modern phenomenological studies and outline future directions.

We refer the reader to \chap{JetFrag} for discussions of jets in QCD and the corresponding phenomenology.  Ref.~\cite{Metz:2016swz} is a review on parton fragmentation functions. The 3-D structure of the nucleon is discussed in a topical issue of the European Physical
Journal A~\cite{EPJA}. The path to obtaining
a multi-dimensional ``picture'' of the proton is discussed in Ref.~\cite{Bacchetta:2016ccz},   an overview on the current experimental and phenomenological status of transverse single-spin asymmetries (SSAs) in proton-proton collisions is presented  in Ref.~\cite{Aschenauer:2015ndk}, 
phenomenology of transverse spin is in Ref.~\cite{Boglione:2015zyc}, and experimental results on TMDs are discussed in Ref.~\cite{Avakian:2016rst}.

\subsection{Unpolarized Observables}
\label{sec:unpol_observables}

\subsubsection{SIDIS multiplicities}
\label{sec:SIDISmult}

\index{multiplicities in SIDIS|(}Perhaps one of the most fundamental measurements in SIDIS related to TMD physics is the study of the unpolarized $\Phperp$ differential cross section  obtained by integrating  Eq.~\eqref{eq:SIDIS-leading} over the angle $\phi_h$:
\begin{equation}
    \frac{d^4\sigma_{\rm SIDIS}}{d\xbj\,dy\,d\zh\,d \Phperp}
   =	 4\pi \Phperp \frac{\alpha_{em}^2}{x\,y\,Q^2}\biggl(1-y+\frac12y^2\biggr)\;F_{UU,T} 
	\,. \hspace{12mm} \label{Eq:SIDIS-averaged}
\end{equation}
This cross section contains dependence on  $\Phperp$ and is thus sensitive to transverse momentum dependence of TMDs~\footnote{Notice that if one keeps $1/Q^2$ suppressed terms, then an additional contribution $p_1 F_{UU,L}$ is present in Eq.~\eqref{Eq:SIDIS-averaged}, see Eq.~\eqref{e:SIDIS_subleading}, and 
Eq.~(\ref{Eq:DIS}) contains an additional structure function $F_L(\xbj,Q^2)$
\cite{Bacchetta:2006tn}.  \label{footnote-subsubleading-terms-in-multiplicities}}. The study of $\Phperp$ differential cross section \eqref{Eq:SIDIS-averaged} in a wide range of $\Phperp$ will be crucial for modern and future phenomenology of TMD physics.
 
Two experimental collaborations, HERMES and COMPASS, reported the measurements of the $\Phperp$ differential cross section~\cite{Airapetian:2012ki,Aghasyan:2017ctw}. Both collaborations presented their results in terms of multiplicities.
The HERMES experiment measured 
 pion or kaon production  in the scattering of 27.6 GeV 
positrons from the HERA polarized positron storage ring at DESY 
off proton and deuteron targets in the SIDIS kinematics 
$Q^2 > 1 \,{\rm GeV}^2$, $W^{2}\equiv (P+q)^2 > 10\,{\rm GeV}^{2}$, 
$0.023 < \xbj < 0.4$, $y < 0.85$, $0.2<\zh<0.7$. 
The  
measured multiplicity~\cite{Airapetian:2012ki}  is defined as
\be\label{Eq:multiplicity-HERMES}
	M_n^h  \equiv 
	\frac{d^4\sigma_{\rm SIDIS}/d\xbj\,d Q^2\,d\zh\,d\Phperp}
	{d^2\sigma_{\rm DIS}/d\xbj\,dQ^2} \; ,
\ee
where the DIS cross section in the denominator is
\begin{align}
\label{Eq:DIS}
	\frac{d^2\sigma_{\rm DIS}}{d\xbj\,dQ^2} = \frac{4 \alpha_{em}^2}{\xbj\,Q^4} \left[ \biggl(1-y+\frac{1}{2}y^2\biggr) F_2(\xbj,Q^2)\right]\, .
\end{align}

\begin{figure*}[t!]
  \centering
  \includegraphics[width=0.45\textwidth]{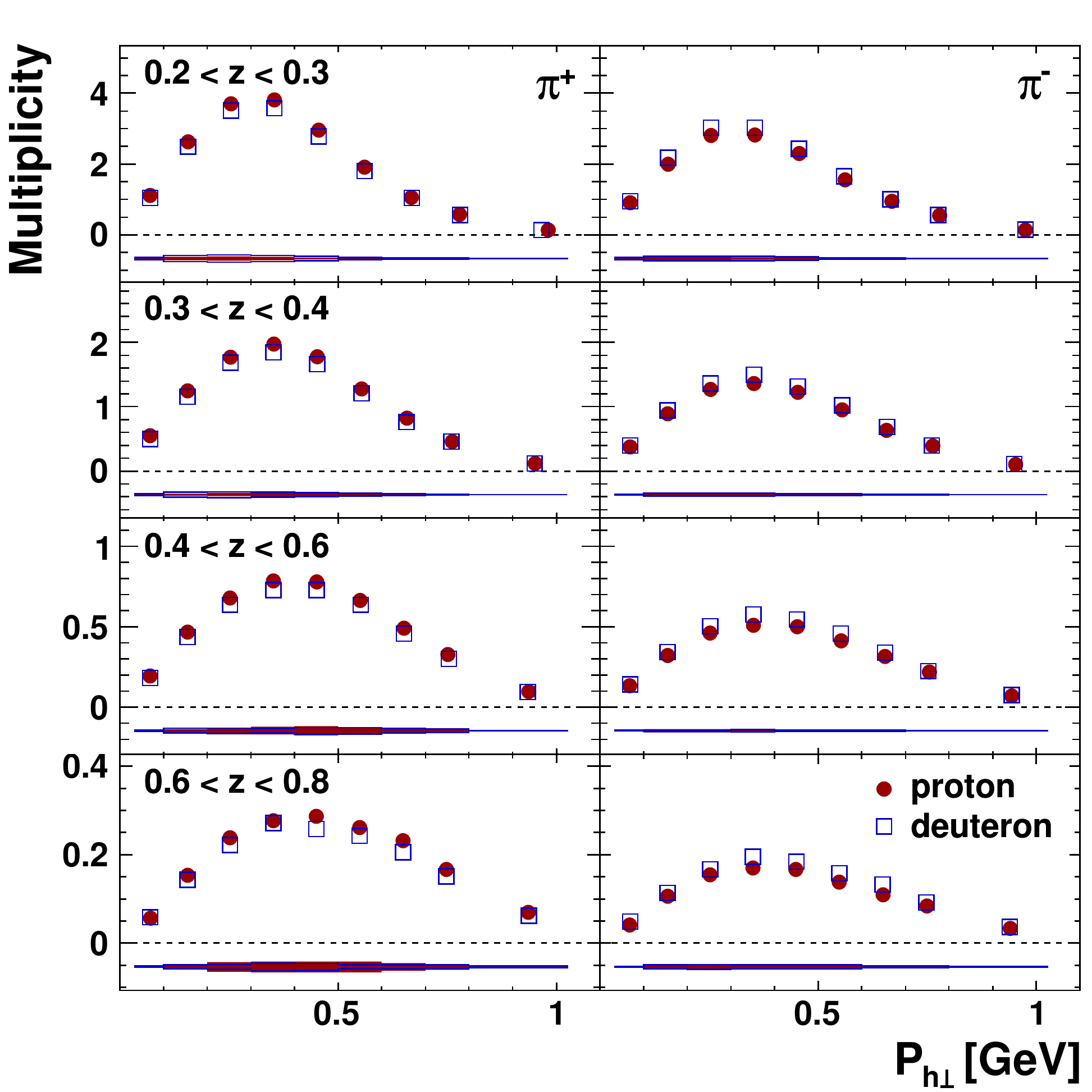}
(a)
  \includegraphics[width=0.45\textwidth]{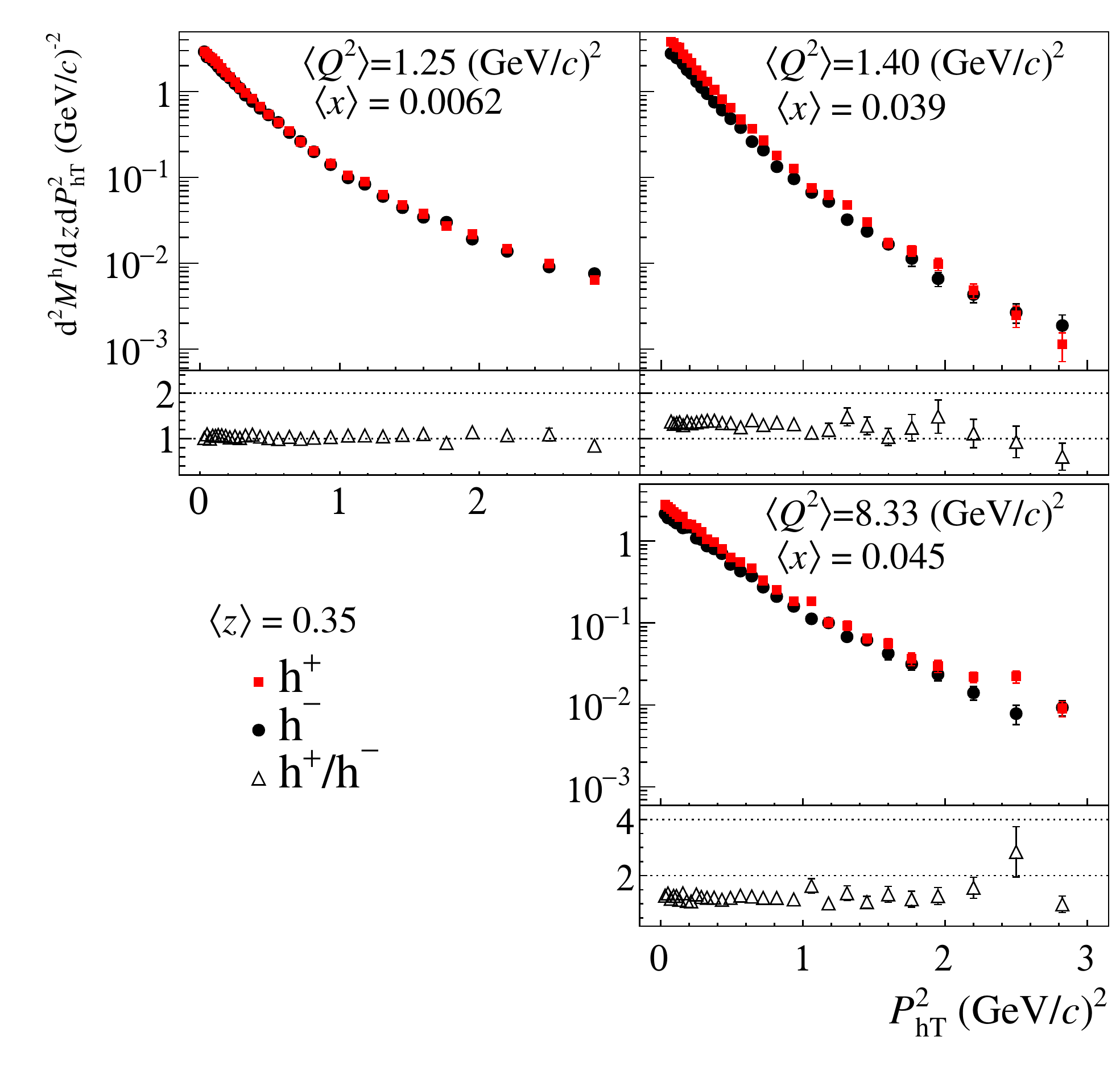}
(b)
  \caption{{\bf(a)} HERMES multiplicities \cite{Airapetian:2012ki} of $\pi^\pm$ pions 
        for the proton and the deuteron as a function of {$\Phperp$}  in four $\zh$ bins. Positive charge is on the left
        and negative charge is on the right of each panel. The figure is from Ref.~\cite{Airapetian:2012ki}. \\
        {\bf(b)} COMPASS results~\cite{Aghasyan:2017ctw}. Top row, upper panels: Multiplicities of positively (full squares) and negatively (full circles) charged hadrons as a function of $\Phperp^2$ at fixed $Q^2$, i.e. $\langle Q^2\rangle \simeq 1.3$~(GeV/$c$)$^2$, for lower (left) and higher (right) $\xbj$ bins. Top row, lower panels: Ratio of multiplicities of positively and negatively charged hadron. Right column: Same at fixed $\xbj$, i.e. $\langle \xbj \rangle \simeq 0.04$, for lower (top) and higher (bottom) $Q^2$ bins. All measured at $\langle \zh \rangle = 0.35$. Only statistical uncertainties are shown. Figures from Ref.~\cite{Aghasyan:2017ctw}
	\label{fig:mult_zpt_pions}}	
\end{figure*} 
Results of HERMES measurements for charged pions are presented in Fig.~\ref{fig:mult_zpt_pions}(a). Notice the kinematical zero at $\Phperp=0$ and the typical shape of the multiplicity which could in principle be described by the TMD approach, provided that TMD approximations hold, for instance $q_T/Q \ll 1$, where $q_T \simeq \Phperp/\zh$.

The COMPASS collaboration 
performed measurements of charged pions, kaons, or charged hadrons produced in collisions of 160 GeV 
longitudinally polarized muons scattered off proton and deuteron 
targets in the typical SIDIS kinematics 
$Q^2 > 1 \,{\rm GeV}^2$, $W > 5\,{\rm GeV}$, 
$0.003 < \xbj < 0.7$, $0.1<y < 0.9$, $0.2<\zh<1$. 
The  COMPASS multiplicity 
\cite{Aghasyan:2017ctw} 
is defined as 
\be\label{Eq:multiplicity-COMPASS}
	M^h  \equiv 
	\frac{d^4\sigma_{\rm SIDIS}/d\xbj\,dQ^2\,d\zh\,d\Phperp^2}
	{d^2\sigma_{\rm DIS}/d\xbj\,dQ^2}\; 
\ee
and shown in Fig.~\ref{fig:mult_zpt_pions}(b).
One can see that HERMES and COMPASS definitions of multiplicity are related by: 
$M_n^h(\xbj,\zh,Q^2,\Phperp) = 2  \Phperp \; M^h(\xbj,\zh,Q^2,\Phperp^2)$.

\subsubsection*{Parton model approximation} The early attempts to describe the unpolarized multiplicities in SIDIS were made in parton model-like approximations to TMDs  in Refs.~\cite{Anselmino:2005nn,Schweitzer:2010tt,Anselmino:2013lza,Signori:2013mda}. 
Here we will discuss the analyses performed in the parton model approximation for TMDs, also known as Generalized Parton Model (GPM).
Refs.~\cite{Anselmino:2005nn,Schweitzer:2010tt,Anselmino:2013lza} assumed factorization of  $\xbj (\zh)$ and $k_T (p_T)$ dependencies,
 and  
 the $k_T$ and $p_T$ dependencies 
 were 
 assumed to be Gaussian, as historically was done for instance in Ref.~\cite{Gardiner:1970wy}, with one free parameter 
which fixes the Gaussian width,
\bea
f_{1\, q/N} (\xbj,k_T)&= f_{1\, q/N} (\xbj)\,\frac{e^{-k_T^2/\avk}}{\pi\avk}
\label{unp-dist}\\
D_{1\, h/q}(\zh,p_T)&=D_{1\, h/q}(\zh)\,\frac{e^{-p_T^2/\avp}}{\pi\avp}\,\cdot
\label{unp-frag}
\eea
The collinear PDFs, $f_{1\, q/N}(\xbj)$ and $D_{1\, h/q}(\zh)$, were taken from the 
available fits of the world data. The widths of the Gaussians could 
depend on $x$ or $z_h$ and might be different for different flavors, and Ref.~\cite{Signori:2013mda} explored flavor dependence of TMDs. 
Ref.~\cite{Anselmino:2013lza} assumes flavor independence and one obtains
\be
F_{UU, T}  =  \xbj \sum_{q} \, e_q^2 \,f_{1\, q/N}(\xbj)\,D_{1\, h/q}(\zh)
\frac{e^{-P_{hT}^2/\avPT}}{\pi\avPT} \label{G-FUU}
\ee
where
\be
\avPT = \avp + \zh^2 \, \avk \>. \label{avPT}
\ee

The Gaussian parameterization of TMDs, used in the GPM, is supported by a number of 
experimental observations~\cite{Schweitzer:2010tt} as well as by dedicated lattice 
simulations~\cite{Musch:2007ya,Orginos:2017kos}. It has the advantage that the intrinsic transverse 
momentum dependence of the cross section can be integrated out analytically. The GPM is a very simple model, which successfully describes a vast body of data and is useful for estimating   the outcome of experimental measurements; however, it is not   the right model for description of  TMD physics. 
Factorization of collinear and transverse momentum dependence as in Eq.~(\ref{unp-frag}) is certainly violated in full TMD evolution beyond the lowest $C^{(0)}$ coefficient function. Also, recent TMD analyses found that more complicated functional forms of intrinsic TMDs are needed to describe the experimental data, see the next subsection.

\begin{figure}[t!]
\begin{center}
\includegraphics[width=0.8\textwidth]{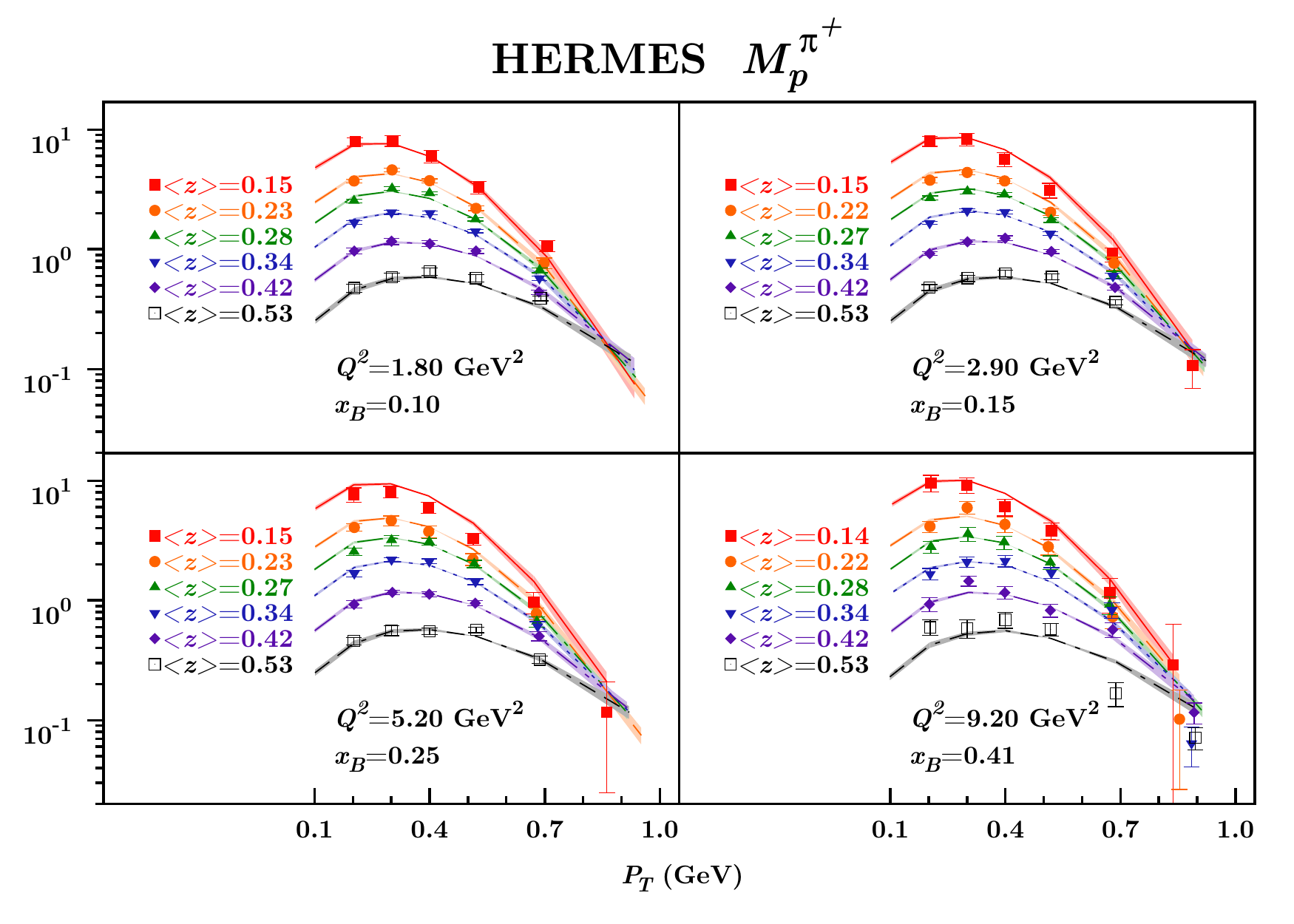}
\caption{\label{fig:hermes-p-pi+}
The multiplicities $M_p^{\pi^+}$ obtained from Eq.~(\ref{Eq:multiplicity-HERMES}), with the parameters of Eq.~(\ref{avPT}), are compared with 
HERMES measurements for $\pi^+$ SIDIS production off a proton 
target~\cite{Airapetian:2012ki}. The shaded uncertainty bands correspond
to a $5$\% variation of the total $\chi^2$. Plot from Ref.~\cite{Anselmino:2013lza}.}
\end{center}
\end{figure}

Nevertheless, this simple TMD Gaussian parameterization, with constant and flavor independent
widths, delivers a satisfactory description~\cite{Schweitzer:2010tt,Anselmino:2013lza,Cammarota:2020qcw} of the
HERMES and COMPASS data points over large ranges of $\xbj$, $\zh$, $P_{hT}$ and $Q^2$. 
These analyses used the following data selection cuts:
 \begin{align}
  0.2< z_h< 0.6,\; Q^2>1.63\,{\rm GeV^2}, \;{\rm and}\;0.2 <P_{hT}<0.9\; \rm (GeV).
\label{eq:cutsJam}
\end{align}
Notice that from the point of view of power counting the conditions $q_T \ll Q$, where $\qt$ is the transverse momentum of the virtual photon in a frame  in which both the target particle and the final-state hadron have no transverse momentum, and $P_{hT} \ll Q$, where $P_{hT}$ is the transverse momentum of the produced hadron in $\gamma^* P$ frame, are equivalent since $q_T \simeq P_{hT}/z_h$. However,  depending on the numerical value for $z_h$, data which satisfy $P_{hT} \ll Q$ may not satisfy $q_T \ll Q$ and therefore be difficult to describe in a TMD approach.
Examples of description of HERMES multiplicities from Ref.~\cite{Anselmino:2013lza} are shown in Fig.~\ref{fig:hermes-p-pi+}.

\subsubsection*{Analyses with TMD evolution} 

In order to go beyond a simple parton model and implement QCD evolution  to connect the different $Q$ scales, one needs to solve the evolution equations, introduce perturbative and nonperturbative Sudakov form factors, and a nonperturbative shape of intrinsic  TMDs, see \chap{evolution}. 

Fits of experimental data from high-energy experiments have been well 
developed in the literature, in particular, in the publications
of the Brock-Laundry-Nadolsky-Yan (BLNY)-type of parameterizations~\cite{Landry:2002ix,Su:2014wpa} utilizing the $b_*$ prescription, see \chap{evolution}.  
Other choices, other than $b_*$ to avoid the Landau pole, have been
made in the literature, see, for example, 
Refs.~\cite{Qiu:2000hf,Qiu:2000ga,Kulesza:2002rh,Kulesza:2003wn,Catani:2000vq, Catani:2003zt,Bozzi:2003jy,Bertone:2019nxa,Scimemi:2019cmh}. 
The HERMES and COMPASS SIDIS data were used for an NLL TMD extraction (in conjunction with Drell-Yan, and Z-boson production data) in Ref.~\cite{Bacchetta:2017gcc}  and also partly used in NLL analysis of Ref.~\cite{Su:2014wpa}.  An N$^3$LO 
description of the SIDIS data was achieved in Ref.~\cite{Scimemi:2019cmh}. 

One of the important nonperturbative functions, $g_K$, encodes the information on large $b_T$ behavior of the evolution kernel $\tilde K$. \index{Collins-Soper evolution kernel}
This function does not depend on the particular process, it does not depend on the scale and has no dependence on the momentum fractions $\xbj$, $\zh$. 
The large-$b_T$ behavior of the CS evolution kernel, $\tilde K$,can be related~\cite{Vladimirov:2020umg} to properties of QCD vacuum and therefore is an important object of study in its own right.
This function should be parametrized phenomenologically and an often-used ~\cite{Nadolsky:1999kb,Landry:2002ix,Konychev:2005iy} parametrization is
\begin{equation}\label{Eq:BLNY-type-parametrization}
g_K(b_T; b_{max}) = g_2 b_T^2 \, ,
\end{equation}
which proved to be very reliable to describe Drell-Yan data and $W^\pm,Z$ boson production. It is often referred to as   the BLNY-type of parameterization~\cite{Nadolsky:1999kb,Landry:2002ix,Konychev:2005iy}.  This Gaussian-type parametrization, $\exp(-g_K(b_T; b_{max}) \log(Q^2/Q_0^2))$, suggests that large $b_T$ region is strongly suppressed \cite{Aidala:2014hva} and in principle can be unreliable to describe data at lower energies which are more sensitive to moderate-to-high values of $b_T$. Other parametrizations were proposed in Refs.~\cite{Aidala:2014hva} and \cite{Su:2014wpa}
and have the form:
\begin{equation} g_K(b_T; b_{max}) =  g_2 \ln\left(\frac{b_T}{b_*}\right) \, ,
\end{equation}
and allows one to describe simultaneously 
unpolarized multiplicities from SIDIS measurements by HERMES, low energy Drell-Yan as well as $Z$ boson production
up to LHC energies~\cite{Su:2014wpa}. 
It was suggested in Ref.~\cite{Collins:2014jpa} that $g_K(b_T; b_{max})$ becomes a constant at large values of $b_T$.

In Ref.~\cite{Scimemi:2019cmh} a global analysis of a large set of DY and SIDIS data, including precision LHC measurements, was performed with N$^3$LO TMD evolution and NNLO matching to the collinear distributions.
   The unpolarized TMD PDFs for the pion were extracted in the same framework in Ref.~\cite{Vladimirov:2019bfa}. In these extractions the Collins-Soper kernel is parameterized as
\begin{eqnarray}\label{def:RAD}
\mathcal{D}(b_T,\mu)=\mathcal{D}_{\text{resum}}(b_*,\mu)+c_0 b_T b_*,
\end{eqnarray}
where $\mathcal{D} \equiv -\tilde K/2$, see Table~\ref{tbl:anom_dim_notation}, and $\mathcal{D}_{\text{resum}}$ is the resummed N$^3$LO expression for the perturbative part of the Collins-Soper kernel, see Ref.~\cite{Vladimirov:2016dll}, and $c_0$ is a free parameter, so that
\begin{eqnarray}\label{def:RAD1}
g_K(b_T; b_{max}) = -2c_0 b_T b_*,
\end{eqnarray}
in our notations. The linear behavior at large-$b_T$ of Eq.~(\ref{def:RAD}) is in agreement with the predicted nonperturbative behavior~\cite{Collins:2014jpa,Vladimirov:2020umg} and  coefficient $c_0$ can be related to the gluon condensate and therefore is exclusively sensitive~\cite{Vladimirov:2020umg} to the structure of QCD vacuum.

\begin{figure}[t!]
\centering
 \includegraphics[width=0.45\textwidth]{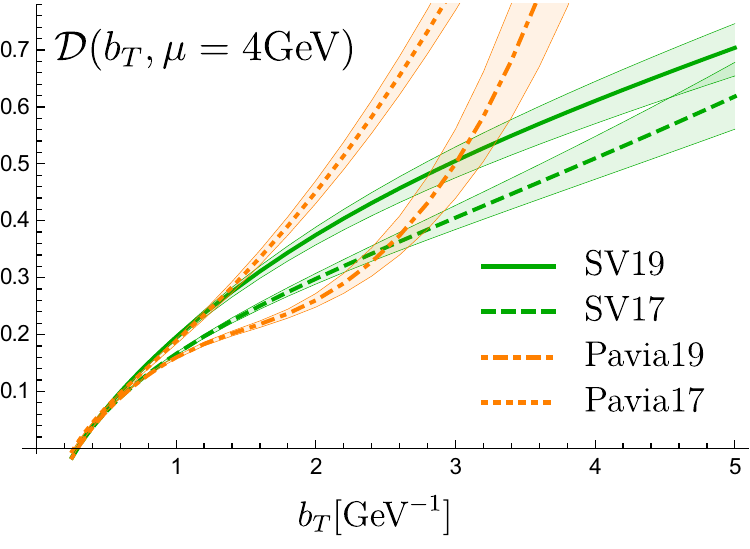}
\caption{\label{fig:phenomenology} Comparison of extracted values of $\cal D$. The lines labeled as SV19, SV17, Pavia19 and Pavia17 correspond to  Refs.\cite{Scimemi:2019cmh},\cite{Scimemi:2017etj},\cite{Bacchetta:2019sam}, and \cite{Bacchetta:2017gcc}. Plot from Ref.~\cite{Vladimirov:2020umg}.}
\end{figure}

The comparison of extractions of the Collins-Soper kernel from the data are shown in Fig.~\ref{fig:phenomenology}. Notice that results differ at large values of $b_T$ because the contribution from this region is additionally suppressed by the intrinsic TMD shape, see Eq.~\eqref{eq:sol1},
therefore more experimental data is needed to explore the large-$b_T$ behavior of the Collins-Soper kernel.
Studies of the Collins-Soper kernel will become increasingly important in future 
for the understanding~\cite{Vladimirov:2020umg} of the universal properties of TMDs and the QCD vacuum.

The first extraction of unpolarized TMDs from a simultaneous fit of available data measured in SIDIS, Drell-Yan and Z boson production was reported in Ref.~\cite{Bacchetta:2017gcc}.
To connect data at different scales, the authors used TMD evolution at NLL accuracy.
The authors of Ref.~\cite{Bacchetta:2017gcc} extracted unpolarized TMDs using 8059 data points with 11 free parameters. 
Ref.~\cite{Bacchetta:2017gcc} used the following data selection criteria:
\begin{align}
Q^2 > 1.4 \,{\rm GeV}^2\, , \;\; 0.2<z_h<0.74\, , \nonumber \\
P_{hT}<{\rm min} [ 0.2 \, Q, 0.7 \, z_h Q] + 0.5 \; \rm {GeV}\, .
\label{eq:cutsPavia}
\end{align}  
The average $\chi^2$/d.o.f. is $1.55 \pm 0.05$ and can be improved up to 1.02 by
restricting the kinematic cuts, without changing the parameters.

\begin{figure}[t!]
\begin{center}
\includegraphics[width=0.65\textwidth]{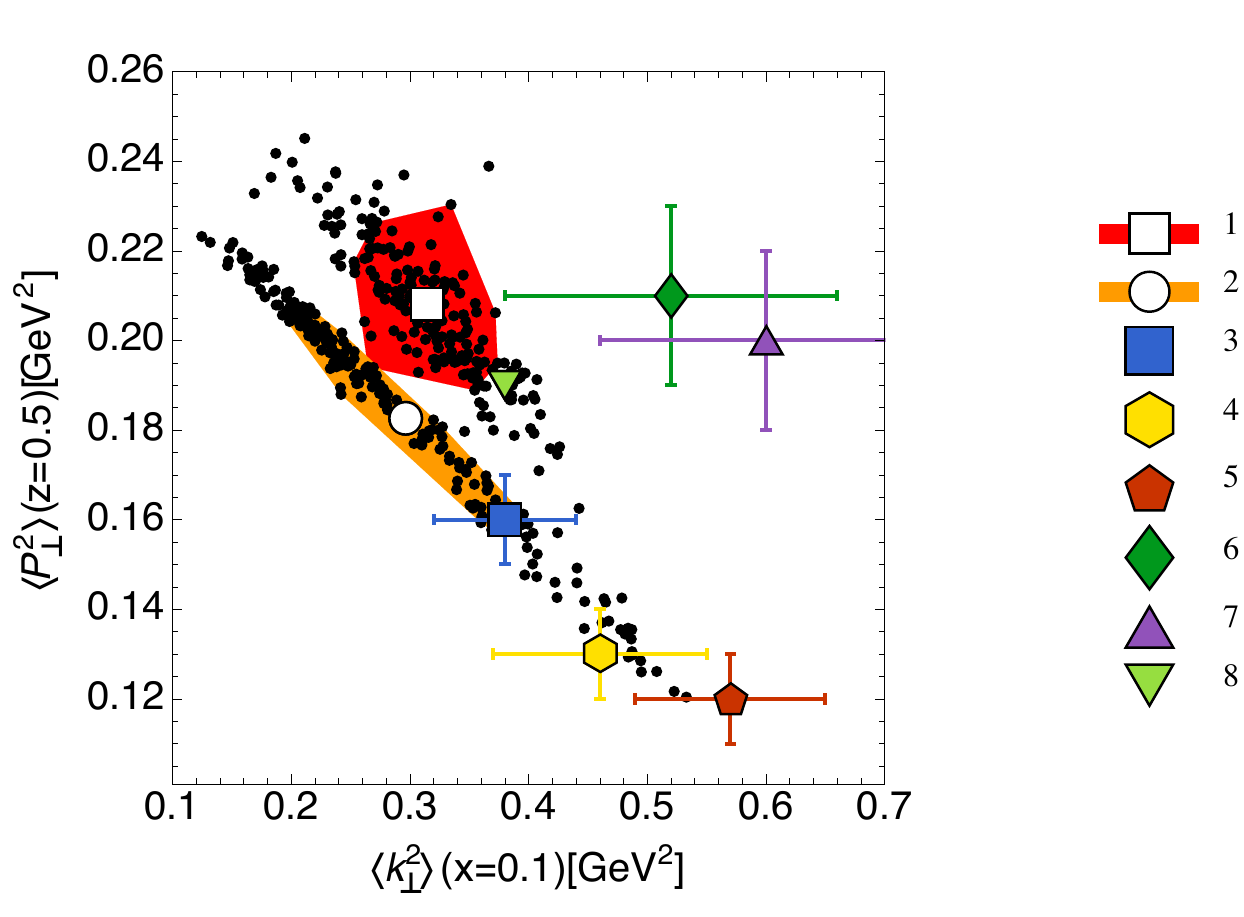}
\end{center}
\vspace{-0.3cm}
\caption{Correlation between transverse momenta in TMD FFs, $\langle P_\perp^2
  \rangle(z=0.5)$, and in TMD PDFs, $\langle k_\perp^2 \rangle(x=0.1)$, in
  different phenomenological extractions. 
 (1): average values (white square) obtained in  Ref.~\cite{Bacchetta:2017gcc}, values
 obtained from each replica (black dots) and
 $68\%$ C.L. area (red); (2) results from Ref.~\cite{Signori:2013mda},
 (3) results from Ref.~\cite{Schweitzer:2010tt}, (4) results from Ref.~\cite{Anselmino:2013lza} for
 HERMES data, 
 (5) results from Ref.~\cite{Anselmino:2013lza} for HERMES data at high $z$, (6) results from Ref.~\cite{Anselmino:2013lza} for normalized COMPASS data, (7) results from Ref.~\cite{Anselmino:2013lza} for normalized COMPASS data at high $z$, (8) results from Ref.~\cite{Echevarria:2014xaa}. Plot from Ref.~\cite{Bacchetta:2017gcc}  } 
\label{f:kT2_vs_PT2}
\end{figure}

The authors used a more complicated shape of intrinsic TMDs compared to the simple Gaussian parameterizations used in Eq.~\eqref{unp-frag}
\begin{align} 
f_{1 {\rm NP}}^a (x, \bm{k}_{\perp}^2) &= \frac{1}{\pi} \  
                        \frac{\big( 1 +\lambda \bm{k}_{\perp}^2\big)}
                         { g_{1a} +\lambda \   g_{1a}^2}
                        \  e^{- \frac{\bm{k}_{\perp}^2}{g_{1a}}} \  ,
\label{e:f1NPk}   \\
D_{1 {\rm NP}}^{a\to h} (z, \bm{p}_{\perp}^2) &=  \frac{1}{\pi} \   
                  \frac{1}{g_{3 a\to h} +
                    \big(\lambda_F/z^2\big) g_{4 a \to h}^{2}}
           \   \bigg( e^{- \frac{\bm{p}_{\perp}^2}{g_{3 a \to h}}}
                            + \lambda_F \frac{\bm{p}_{\perp}^2}{z^2} \  
           e^{- \frac{\bm{p}_{\perp}^2}{g_{4 a \to h}}} \bigg) \  .
\label{e:D1NPk}
\end{align} 
Resulting widths of TMDs are shown in Fig.~\ref{f:kT2_vs_PT2}.  The horizontal axis shows 
the value of the average transverse
momentum squared for the incoming parton, 
$\big \langle \bm{k}_{\perp}^2 \big \rangle$ at $\la x\ra =0.1$. 
The vertical axis shows the value of $\big \langle \bm{p}_{\perp}^2 \big \rangle$ at $\la z\ra=0.5$,
the average transverse momentum squared acquired during the fragmentation process. 
The white square (label 1) indicates the average values of the two quantities
obtained in the analysis of Ref.~\cite{Bacchetta:2017gcc} at $Q^2=1$ GeV$^2$.
Each black dot around the white square is an outcome of one replica. The replica approach consists in creating several replicas of the data points. In each replica each data point in the data set is shifted by a Gaussian noise with the same variance as the measurement and, therefore, represents a possible outcome of an independent experimental measurement. These replicas are then used in the data analysis.

The red region around the white square contains the $68\%$ of the replicas that are closest to the average value.
The same applies to the white circle and the orange region around it (label 2),
related to the flavor-independent version of the analysis in
Ref.~\cite{Signori:2013mda}, obtained by fitting only HERMES SIDIS
data at an average $\langle Q^2 \rangle= 2.4$ GeV$^2$ and neglecting QCD evolution. 
A strong anticorrelation between the transverse momenta is evident in this
older analysis. 
In Ref.~\cite{Bacchetta:2017gcc}, the inclusion of Drell--Yan and $Z$ production data adds physical information
about TMD PDFs, free from the influence of TMD FFs. This reduces significantly the 
correlation between $\big \langle \bm{k}_{\perp}^2 \big \rangle$ at $\la x\ra =0.1$ and $\big \langle \bm{p}_{\perp}^2 \big \rangle$ at $\la z\ra=0.5$.  
The $68\%$ confidence region is smaller than in the older analysis in Ref.~\cite{Signori:2013mda}. 
The average values of $\big \langle \bm{k}_{\perp}^2 \big \rangle$ at $\la x\ra =0.1$  are similar and compatible within error bands. 
The values of $\big \langle \bm{p}_{\perp}^2 \big \rangle$ at $\la z\ra=0.5$ in Ref.~\cite{Bacchetta:2017gcc} analysis
turn out to be larger than in the older Ref.~\cite{Signori:2013mda} analysis by the same group, an effect that is due mainly to COMPASS data.

\begin{figure}[t!]
\begin{center}
\includegraphics[width=0.65\textwidth]{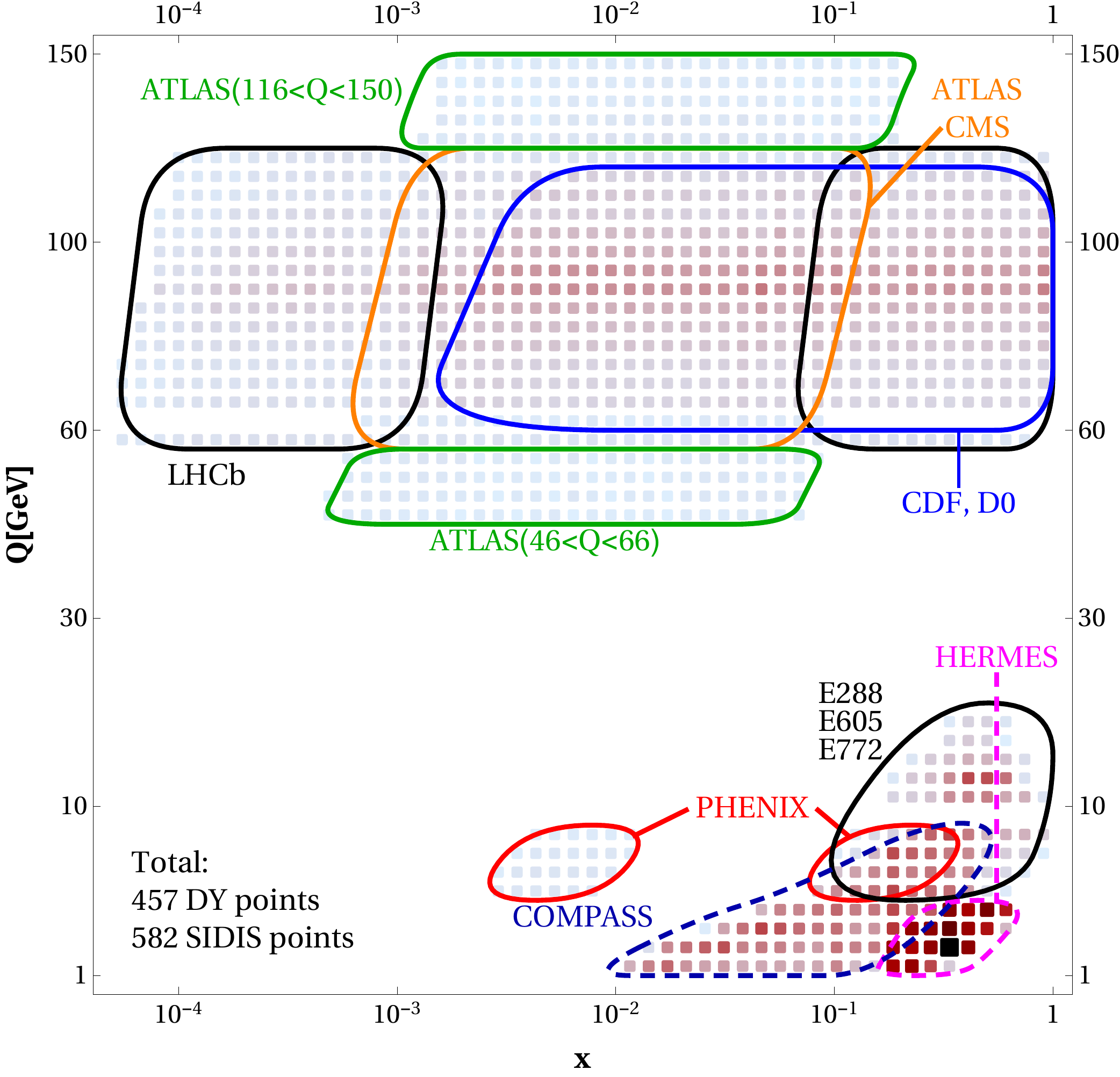}
\caption{\label{fig:dataPoints} Data used in analysis of Ref.~\cite{Scimemi:2019cmh} (a darker color corresponds to a higher density). Plot from Ref.~\cite{Scimemi:2019cmh}.}
\end{center}
\end{figure}

The first NNLO and N$^3$LO analysis of a large body of SIDIS and DY data, see Fig.~\ref{fig:dataPoints}, was presented in Ref.~\cite{Scimemi:2019cmh}. In Ref.~\cite{Scimemi:2019cmh} the hard coefficient function is taken at $\alpha_s^3$-order, the anomalous dimensions are at $\alpha_s^3$-order, and the double-logarithm part  ($\Gamma_{\text{cusp}}$) is at  $\alpha_s^4$-order. It gives N$^3$LO perturbative precision. In the resummation nomenclature, \chap{evolution}, the perturbative input of Ref.~\cite{Scimemi:2019cmh} can be mapped as NNLO-N$^3$LL, see Table~\ref{tbl:resum_orders}, indicating that the order of small-$b_T$ matching for the unpolarized distributions is $\alpha_s^2$.
Altogether the authors of Ref.~\cite{Scimemi:2019cmh} obtain the  global value of $\chi^2/N_{pt}=0.95$ and $1.06$ for NNLO and N$^3$LO respectively of the simultaneous fits of Drell-Yan and SIDIS data, see~\fig{dataPoints}, with the following cuts to select the data: 
\begin{align}
Q > 2 \;{\rm (GeV)},\; q_T/Q < 0.25\, \label{eq:cutsVladimirov}
\end{align}
where $q_T = P_{hT}/z_h$ in SIDIS or $q_T$ in Drell-Yan.

The TMD distribution $F(x,b_T;\mu,\zeta_\mu)$ with $\zeta_\mu$ 
is expressed in the $\zeta$-prescription~\cite{Scimemi:2019cmh}, see \sec{2dRRGE}, as a function of $\mu$ and Collins-Soper kernel $\mathcal{D}$.  
The resulting expression for the evolved TMD distributions reads 
\begin{eqnarray}\label{def:TMD-evolved}
F(x,b_T;Q,Q^2)=\left(\frac{Q^2}{\zeta_Q(b_T)}\right)^{-\mathcal{D}(b_T,Q)}F(x,b_T)\; ,
\end{eqnarray}
where the function $F(x,b_T)$ is the so-called ``optimal" TMD distribution.  The prescription consists in defining the TMD distribution on a null-evolution line $\zeta_Q(b_T)$, see appendix C2 of Ref.~\cite{Scimemi:2019cmh}, that makes evolution factor for TMD distributions to be equal one for all values of the impact parameter $b_T$  such that   $F(x,b_T;Q,\zeta_Q(b_T)) = F(x,b_T)$ becomes independent of any perturbative parameter. This function is completely nonperturbative and one can freely parameterize it without any reference to perturbative order. 
Another important feature of the $\zeta$-prescription used in Eq.~(\ref{def:TMD-evolved}) is that the nonperturbative Soper-Collins kernel $\mathcal{D}$ with its arbitrary functional form at large $b_T$ is the argument of $\zeta_Q(b_T)$. Therefore the evolution can be defined unambiguously in both perturbative and most importantly, the nonperturbative regions. Lastly, no integration is involved in the computation of the evolution exponent and it speeds up numerical computations.

\begin{figure}[t!]
\begin{center}
\includegraphics[width=0.48\textwidth]{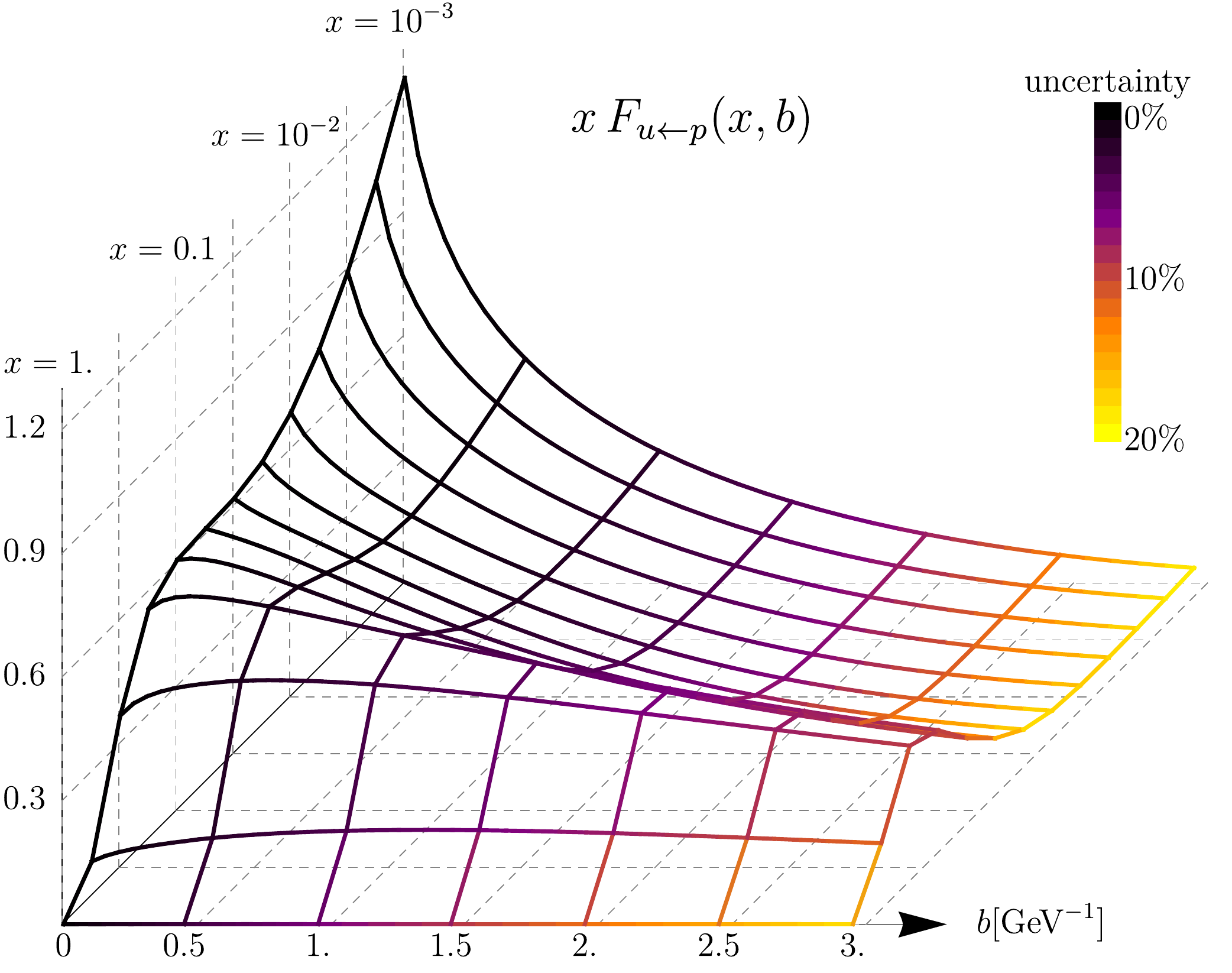}
\includegraphics[width=0.48\textwidth]{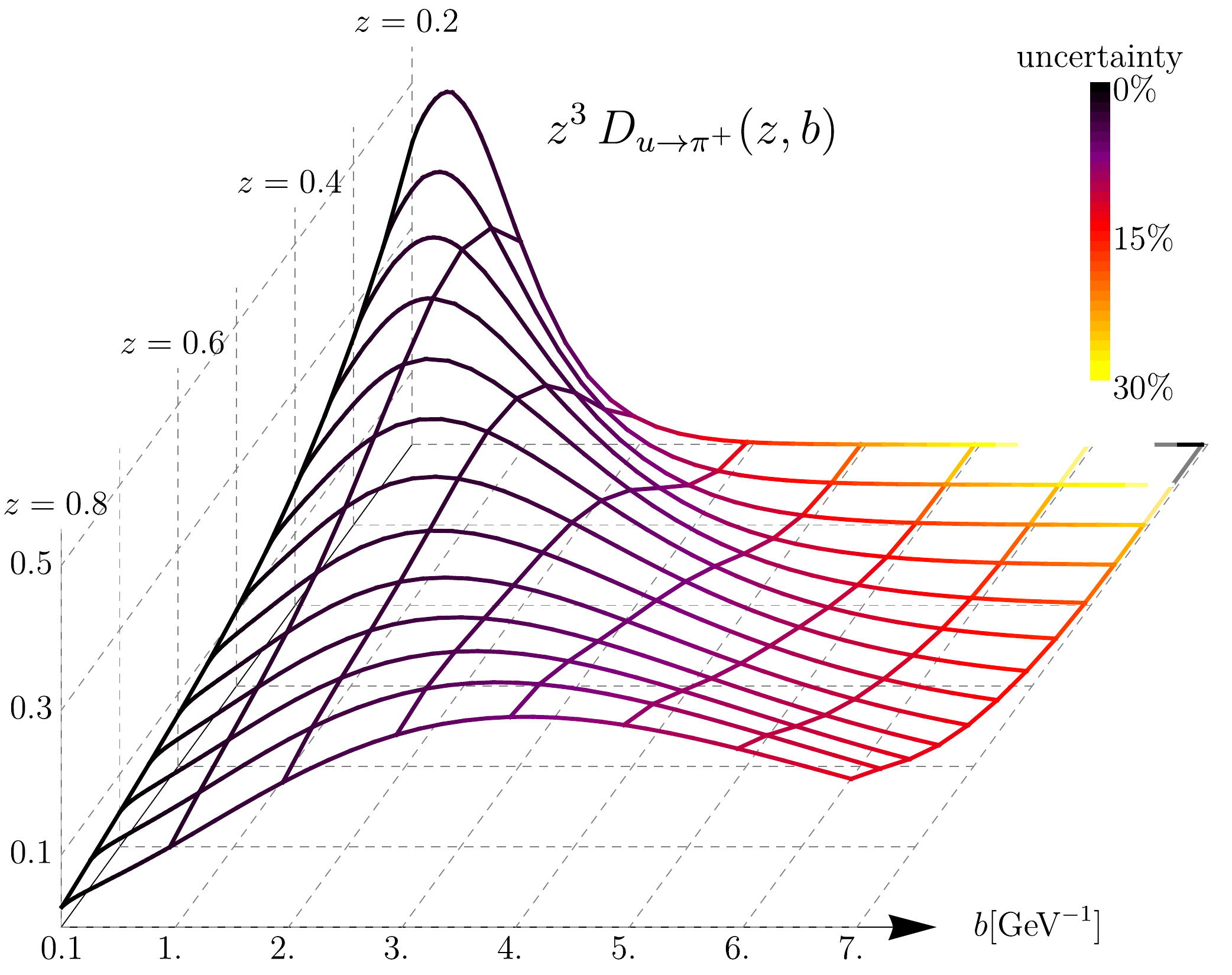}
\caption{\label{fig:TMDs} Example of extracted (optimal) unpolarized TMD distributions. The color indicates the relative size of the uncertainty band. Plot from Ref.~\cite{Scimemi:2019cmh}.}
\end{center}
\end{figure}

The TMD distributions show a non-trivial intrinsic structure. The authors of Ref.~\cite{Scimemi:2019cmh} use the following parameterizations for intrinsic shapes of TMDs
\begin{eqnarray}\label{def:fNP}
f_{NP}(x,b_T)&=&\exp\left(-\frac{\lambda_1(1-x)+\lambda_2 x+x(1-x)\lambda_5}{\sqrt{1+\lambda_3 x^{\lambda_4} b_T^2}}b_T^2\right),
\\\label{def:DNP}
D_{NP}(z,b_T)&=&\exp\left(-\frac{\eta_1 z+\eta_2 (1-z)}{\sqrt{1+\eta_3(b_T/z)^2}}\frac{ b_T^2}{z^2}\right)\left(1+\eta_4 \frac{ b_T^2}{z^2}\right),
\end{eqnarray}
and extract $\lambda_i$ and $\eta_i$. This functional form of $f_{NP}$ was also  used in \cite{Bertone:2019nxa}. It has five free parameters which grant a sufficient flexibility in $x$-space as needed for the description of the precise LHC data. 
An example of distributions in $(x,b_T)$-plane is presented in Fig.~\ref{fig:TMDs}. Depending on the value of $x$, the $b_T$-behavior apparently changes. The authors of Ref.~\cite{Scimemi:2019cmh} observe (the same observation was made in Ref. \cite{Bacchetta:2017gcc}) that the unpolarized TMD FF gains a large $b_T^2$-term in the nonperturbative part. It could indicate non-trivial consequences of hadronization physics, or a tension between collinear and TMD distributions.\index{multiplicities in SIDIS|)}   


\subsubsection{Drell-Yan and weak gauge boson production}
\label{sec:DYpheno}
\index{Drell-Yan|(}

Drell-Yan lepton pair production via either virtual photon or $Z$ boson served in prior chapters of this handbook to set up the basic notation and concepts for TMD factorization.
Factorized in terms of a convolution of two TMD PDFs from each incoming proton at the small transverse momentum $q_T$ as shown in Eq.~\eqref{eq:sigma_new_a}, Drell-Yan production in unpolarized proton-proton collisions is one of the most important processes for extracting unpolarized quark TMD PDFs. 

\begin{figure}[t!]
\begin{center}
\includegraphics[width=0.95\textwidth]{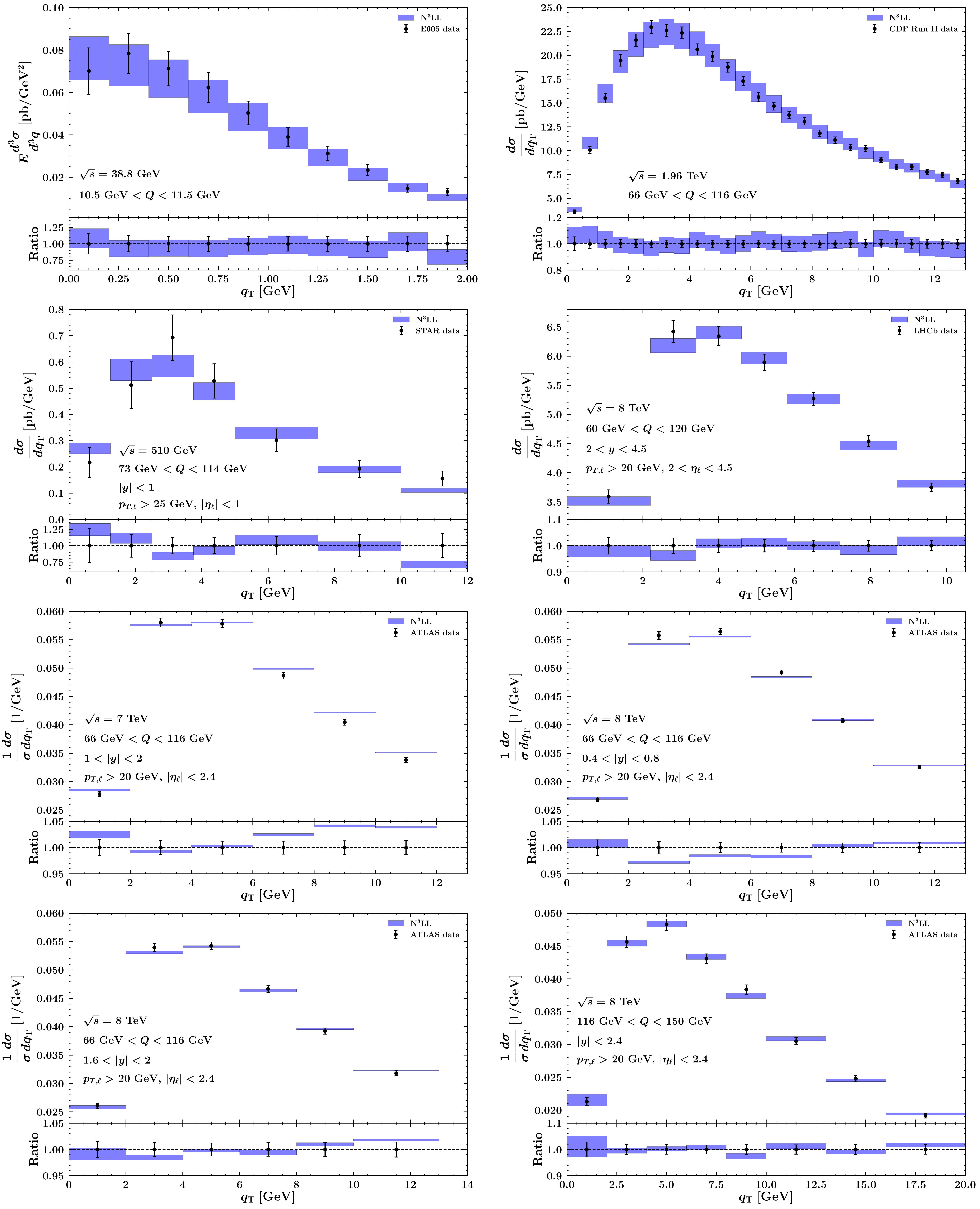}
\caption{\label{fig:DYcomparison} Comparison between experimental data and theoretical predictions from the TMD formalism at ${\rm N^3LL}$ accuracy. Plot from Ref.~\cite{Bacchetta:2019sam}.}
\end{center}
\end{figure}

There is a  tremendous amount of experimental data for Drell-Yan production, ranging from lower energy Fermilab experimments to the highest energy data at the LHC. The lower-energy fixed-target Fermilab data include E605~\cite{Moreno:1990sf} and E288~\cite{Ito:1980ev}, while the higher-energy Fermilab data from collider Tevatron include CDF Run~I~\cite{Affolder:1999jh} and Run~II~\cite{Aaltonen:2012fi}, and D0 Run~I~\cite{Abbott:1999wk} and Run~II~\cite{Abazov:2007ac,Abazov:2010kn}. LHC data include forward $Z$-production data from the LHCb experiment at 7~\cite{Aaij:2015gna}, 8~\cite{Aaij:2015zlq}, and 13~\cite{Aaij:2016mgv} TeV, $Z$-production data from the CMS experiment at 7~\cite{Chatrchyan:2011wt} and 8~\cite{Khachatryan:2016nbe} TeV, $Z$-production data differential in rapidity from the ATLAS experiment at 7~\cite{Chatrchyan:2011wt} and 8~\cite{Aad:2015auj} TeV, and off-peak (low- and high-mass) Drell-Yan data from the ATLAS experiment at 8 TeV~\cite{Aad:2015auj}. Finally, there is also preliminary $Z$ production data from the STAR experiment at 510 GeV. 
  
Earlier description of the small-$q_T$ Drell-Yan data from both fixed-target and collider Fermilab data within the Collins Soper Sterman (CSS) framework has been performed by several groups in, e.g.~\cite{Landry:2002ix,Qiu:2000hf,Qiu:2000ga,Su:2014wpa}, where different ways of implementing nonperturbative contributions have been explored. In recent years, the perturbative precision for the resummation of logarithms of the transverse momentum of the vector boson have been further increased up to the ${\rm N^3LL}$ order. At the same time, a more stringent cut of $q_T/Q\lesssim 0.2$ has been implemented to ensure the data is in the domain of the TMD factorization region. The most recent extractions of the TMD PDFs have been performed by two groups in~\cite{Scimemi:2019cmh} and~\cite{Bacchetta:2019sam} independently. Ref.~\cite{Scimemi:2019cmh} includes both SIDIS and Drell-Yan data, and thus allows simultaneous extraction of both TMD PDFs and TMD FFs. On the other hand, Ref.~\cite{Bacchetta:2019sam} excludes SIDIS data but extends the Drell-Yan data sets, and improves the logarithmic accuracy to the ${\rm N^3LL}$ order. 

It is also important to keep in mind that these two groups use slightly different TMD evolution schemes and also different nonperturbative contributions. Nevertheless, both groups have achieved very good description of the available data. The global analysis of Drell-Yan experimental data generally leads to very good $\chi^2/N_{\rm data}\sim 1$, indicating very good quality of the fit.  Although not available at the moment, it would be highly desirable to compare the extracted unpolarized TMD PDFs from these two groups. Instead here we show in Fig.~\ref{fig:DYcomparison} a comparison between experimental data and theoretical predictions from the TMD formalism at ${\rm N^3LL}$ accuracy. \index{Drell-Yan|)}

\subsection{Polarized Observables}
\label{sec:PolarizedObservables}

\subsubsection{Sivers effect in SIDIS and DY \label{sec:Sivers_SIDIS}}
\index{Sivers function $f_{1T}^{\perp}$!phenomenology|(}
\index{Sivers effect|(}

The  Sivers function $f_{1T}^\perp$~\cite{Sivers:1989cc} encodes the correlation
between the partonic intrinsic motion and the transverse spin of the
nucleon, and it generates a dipole deformation in momentum space and could
not exist without the contribution of orbital angular momentum of
partons to the spin of the nucleon. It arises from interaction of the initial or final state quark with the remnant of the nucleon and thus, many of its features reveal the gauge link structure that reflects the kinematics of the underlining process \cite{Belitsky:2002sm}. 
Above all, the difference between initial and final state gauge link  contours leads to  the opposite signs for Sivers functions in SIDIS and DY kinematics \cite{Brodsky:2002rv,Brodsky:2013oya,Brodsky:2002cx,Collins:2002kn}, see Eq.~\eqref{eq:kappaDYSIDIS}
\begin{eqnarray}\label{eq:sign}
f_{1T}^\perp(x,k_T)_{\text{[SIDIS]}}=-f_{1T}^\perp(x,k_T)_{\text{[DY]}}.
\end{eqnarray}
In the limit of the large transverse momentum the Sivers function is related~\cite{Ji:2006ub} to the key ingredient of collinear factorization of SSAs, the  Qiu-Sterman (QS) function~\cite{Efremov:1981sh,Efremov:1983eb,Qiu:1991pp,Qiu:1998ia}, 
\index{Qiu-Sterman (QS) function|(}
which describes the correlation of quarks with the null-momentum gluon field. Therefore, the measurement of Sivers function and the exploration of its properties is a crucial test of our understanding of the strong force, and one of the goals of polarized SIDIS and DY experimental programs of  future and existing experimental facilities such as the Electron Ion Collider~\cite{Boer:2011fh,Accardi:2012qut}, Jefferson Lab 12~GeV Upgrade~\cite{Dudek:2012vr}, RHIC~\cite{Aschenauer:2015eha} at BNL, and COMPASS~\cite{Gautheron:2010wva,Bradamante:2018ick} at CERN.
It has so far
received the widest attention, from both phenomenological and experimental
points of view.
The Sivers function has been extracted from SIDIS data
by several groups, with consistent results
\cite{Anselmino:2010bs,Anselmino:2005ea,Anselmino:2005an,Collins:2005ie,Vogelsang:2005cs,Anselmino:2008sga,Bacchetta:2011gx,Echevarria:2014xaa,Bacchetta:2020gko,Echevarria:2020hpy,Bury:2020vhj}.

\begin{figure}[t!]
\centering
(a)
\includegraphics[width=0.75\textwidth,,trim=0 0 0 0,clip=true]{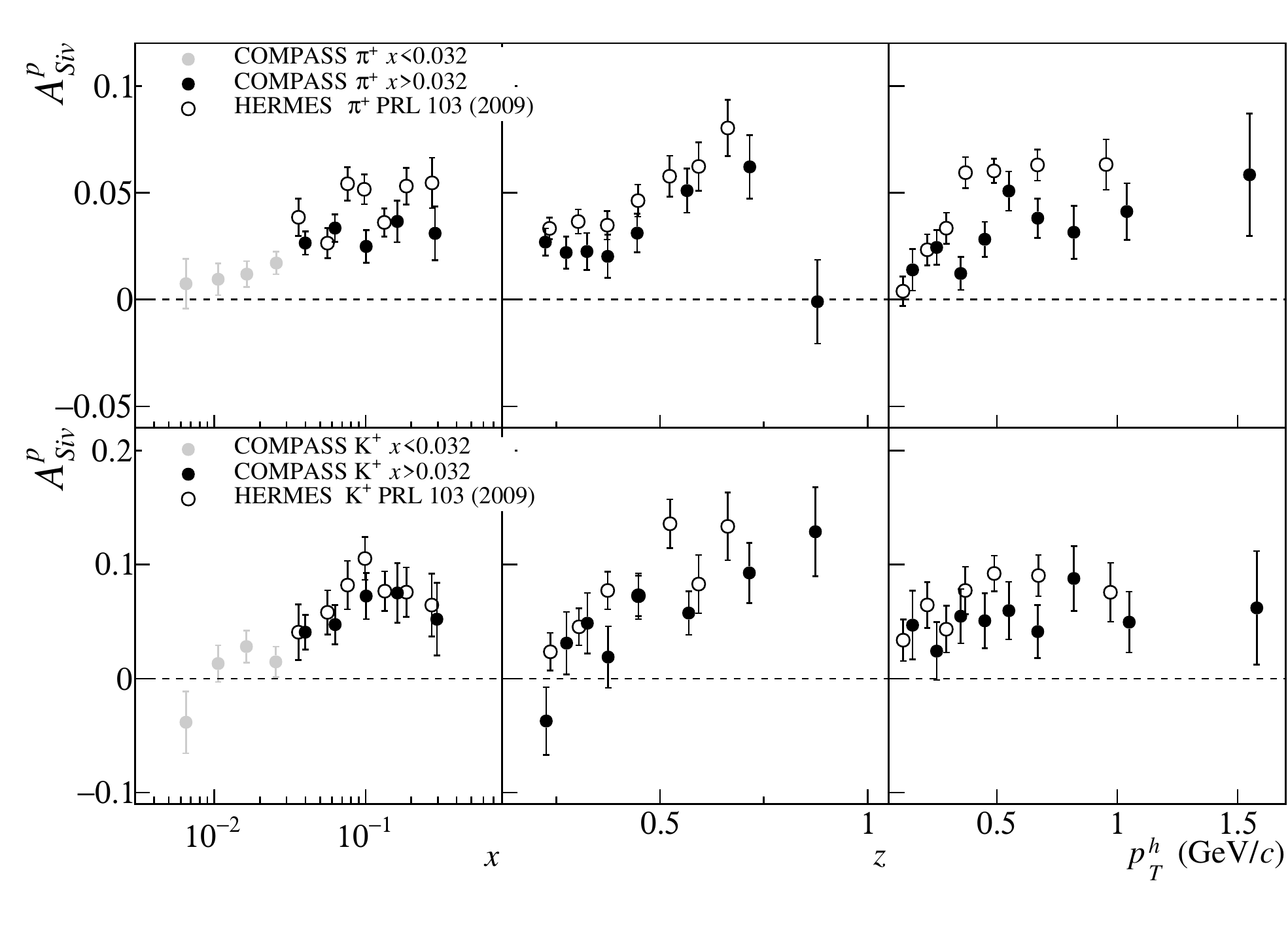}
\\
(b)
\includegraphics[width=0.55\textwidth]{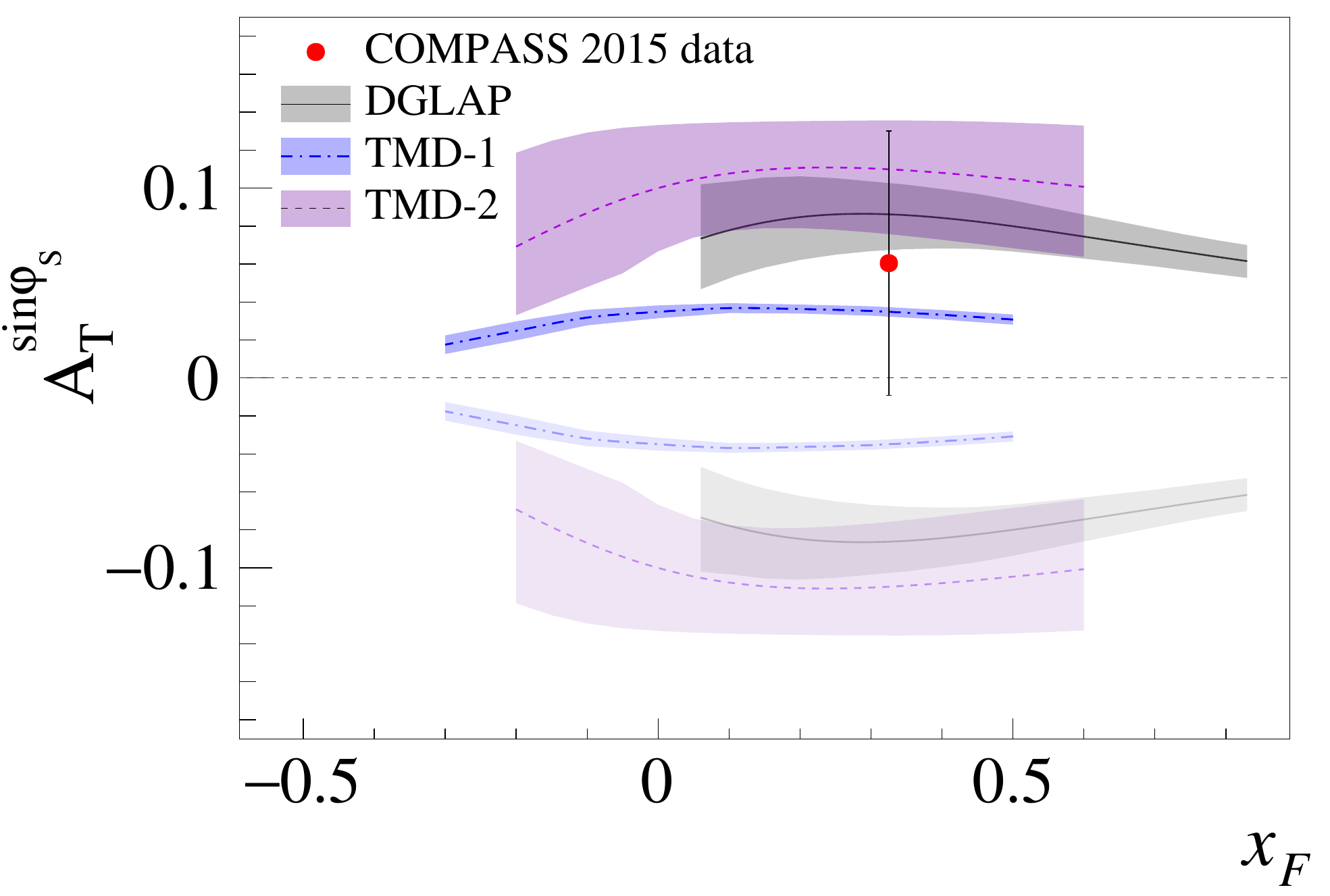}
\caption{{\bf(a)} COMPASS results of Ref.~\cite{Adolph:2014zba} on the Sivers asymmetries for positive pions (top) and kaons (bottom) on
 proton as a function of $\xbj$, $\zh$ and $\Phperp$, requiring $\xbj> 0.032$. The
 asymmetries are compared to HERMES results~\cite{Airapetian:2009ae}. The plot is from Ref.~\cite{Adolph:2014zba}.\\
{\bf(b)} COMPASS experimental result~\cite{Aghasyan:2017jop} for the Sivers asymmetry in Drell-Yan and the
theoretical  predictions for different $Q^2$ evolution schemes from Refs.~\cite{Anselmino:2016uie} (DGLAP),~\cite{Echevarria:2014xaa} (TMD1)
and~\cite{Sun:2013hua} (TMD2).
The dark-shaded (light-shaded) predictions are evaluated with (without) the sign-change prediction. The error bar represents the total experimental uncertainty. The plot is from Ref.~\cite{Aghasyan:2017jop}.
 }
\label{fig-sx32}  
\end{figure}

\begin{figure*}[t!]
  \centering
  \includegraphics[width=0.9\textwidth]{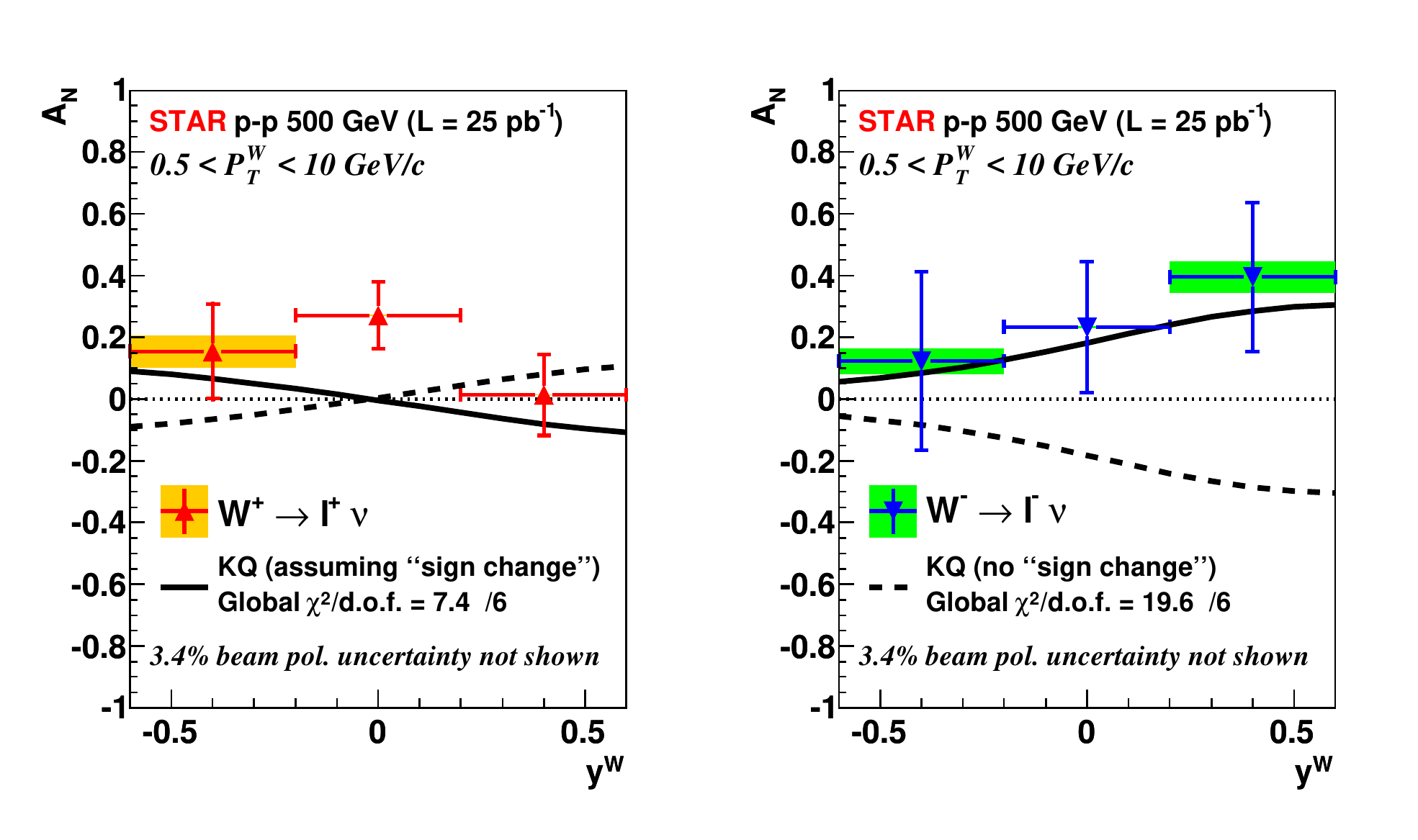}
  \vspace{-0.2cm}
  \caption{STAR results~\cite{Adamczyk:2015gyk} for transverse single-spin asymmetry 
  for $W^{+}$ ({\it left plot}) and $W^{-}$ ({\it right plot}) versus $y^{W}$ compared with the non TMD-evolved KQ~\cite{Kang:2009bp} model, assuming ({\it solid line}) or excluding ({\it dashed line}) a sign change in the Sivers function. The plot from Ref.~\cite{Adamczyk:2015gyk}.}
  \label{Fig:W-An-chi2}
\end{figure*}

The Sivers asymmetry in SIDIS,  $A_{UT,T}^{\sin( \phi_h-\phi_S)}$, is
\begin{eqnarray}\label{eq:sidisaut}
A_{UT,T}^{\sin(\phi_h-\phi_S)}&\equiv&\frac{F_{UT ,T}^{\sin(\phi_h -\phi_S)}}{F_{UU ,T}} = -M_N\frac{\mathcal{B}\left[\tilde f_{1T}^{\perp (1)} \tilde D_{1}^{(0)}\right]}{\mathcal{B}\left[\tilde f_{1}^{(0)} \tilde D_{1}^{(0)}\right]}\; ,
\end{eqnarray}
where $M_N$ is the mass of the nucleon.
In the Drell-Yan process $h_1(P_1,S)+h_2(P_2)\to l^+(l)+l^-(l')+X$  
the experimentally measured transverse spin asymmetry is
\begin{eqnarray}\label{th:AUT}
A_{TU}^{\sin(\phi_h -\phi_S)}\equiv\frac{F_{TU}^{\sin(\phi_h -\phi_S)}}{F_{UU}^1}=
-M\frac{\mathcal{B}[\tilde f_{1T}^{\perp (1)} \, \tilde f_1^{(0)}]}{\mathcal{B}[\tilde f_1^{(0)}\, \tilde f_1^{(0)}]}\, ,
\end{eqnarray}
where $M$ is the mass of the polarised hadron $h_1$.
The Sivers asymmetry has been measured in SIDIS and DY~\cite{Airapetian:2009ae,Airapetian:2020zzo,Alekseev:2008aa,Adolph:2014zba,Qian:2011py,Aghasyan:2017jop,Adamczyk:2015gyk}, Figs.~\ref{fig-sx32},~\ref{Fig:W-An-chi2}.  In particular, these are SIDIS measurements collected in $\pi^\pm$ and $K^\pm$ production off polarized proton target at  HERMES~\cite{Airapetian:2020zzo}, off a deuterium target from COMPASS~\cite{Alekseev:2008aa}, Fig.~\ref{fig-sx32}(a), and $^3$He target from JLab~\cite{Qian:2011py,Zhao:2014qvx}, $h^\pm$ data on the proton target  from COMPASS~\cite{Adolph:2016dvl}. In Drell-Yan the data exist from DY measurements of $W^\pm/Z$ production from STAR~\cite{Adamczyk:2015gyk}, Fig.~\ref{Fig:W-An-chi2}, and pion-induced DY from COMPASS~\cite{Aghasyan:2017jop}, Fig.~\ref{fig-sx32}(b).

\subsubsection*{Parton model approximation} 
Extractions~\cite{Anselmino:2010bs,Anselmino:2005ea,Anselmino:2005an,Collins:2005ie,Vogelsang:2005cs,Anselmino:2008sga,Bacchetta:2011gx,Boglione:2018dqd} of the Sivers functions that utilize parton model approximation, including the Generalized Parton Model, 
generically use the Gaussian model for the $k_T$-dependence 
and generically parametrize the Sivers function as
\begin{align}
    f_{1T}^{\perp a}(x,\kperp^2) =  f_{1T}^{\perp (1) a}(x)   \;
	\frac{2 M^2}{\pi \avkperp_{f_{1T}^\perp}^2} \;
	e^{-\kperp^2/{\avkperp_{f_{1T}^\perp}}}
	\label{Eq:Gauss-f1Tperp}\, ,
\end{align}
where the first moment of the Sivers function $f_{1T}^{\perp (1) a}(x)$,
is, in $k_T$-space and within the Gaussian model approximation,
simply defined  according to
\be\label{eq:define-(1)-mom-in-kT-space}
    f_{1T}^{\perp (n) a}(x)   = \int d^2k_T\;
    f_{1T}^{\perp (n) a}(x,k_T)\, , \quad \quad
    f_{1T}^{\perp (n) a}(x,k_T) = 
    \biggl(\frac{k_T^2}{2M^2}\biggr)^n
	f_{1T}^{\perp a}(x,k_T)\,.
\ee
The exact QCD definition in terms of renormalized functions 
in $b_T$-space is given in Chapter~\ref{sec:TMDdefn}
in Eq.~(\ref{eq:TMD_bt_derivative_inv}), see also 
\app{Fourier_transform}. Since here the meaning of the scales 
$\mu$ and $\zeta$ is undefined, typically no scale dependence 
is indicated in parton model expressions. It is implicitly
understood that parameters, like Gaussian width, refer to the
typical $Q^2$ at which the investigated data was taken. Some of 
these early Sivers function extractions also explored the 
connection to the QS)
function (defined below in Eq.~\ref{e:QS_Siv}, 
see also footnote~\ref{footnote-on-QS-in-pheno-section}).

\subsubsection*{Extractions with TMD evolution} 

Many analyses~\cite{Echevarria:2014xaa,Bacchetta:2020gko,Echevarria:2020hpy} that utilize TMD evolution employ the small-$b_T$ operator product expansion  of the Sivers function via the QS function and parametrize the QS function.\footnote{\label{footnote-on-QS-in-pheno-section} The definitions of the QS function vary in different analyses. 
The following relations can be found for the QS functions used: $-\pi T_q(-x,0,x;\mu)|_{\text{\cite{Bury:2020vhj, Bury:2021sue}}}= f_{1T}^{\perp(1)}(x,\mu)|_{\text{\cite{Bacchetta:2020gko}}}=-\frac{T_F(x,x;\mu)}{2 M}|_{\text{\cite{Echevarria:2020hpy}}}=\pi F_{FT}(x,x;\mu)|_{\text{\cite{Cammarota:2020qcw}}}$.
} This approach avoids possible complications of the  integral relations~\cite{Qiu:2020oqr} however the QS function is not an autonomous function, but mixes with other twist-3 distributions \cite{Braun:2009mi}, and therefore taking into account the correct twist-3 evolution becomes troublesome in these analyses. Ref.~\cite{Echevarria:2020hpy}  uses NNLL resummation while Ref.~\cite{Bacchetta:2020gko} uses NLL resummation.
 
\begin{figure}[t!]
\centering
\includegraphics[width=0.3\textwidth]{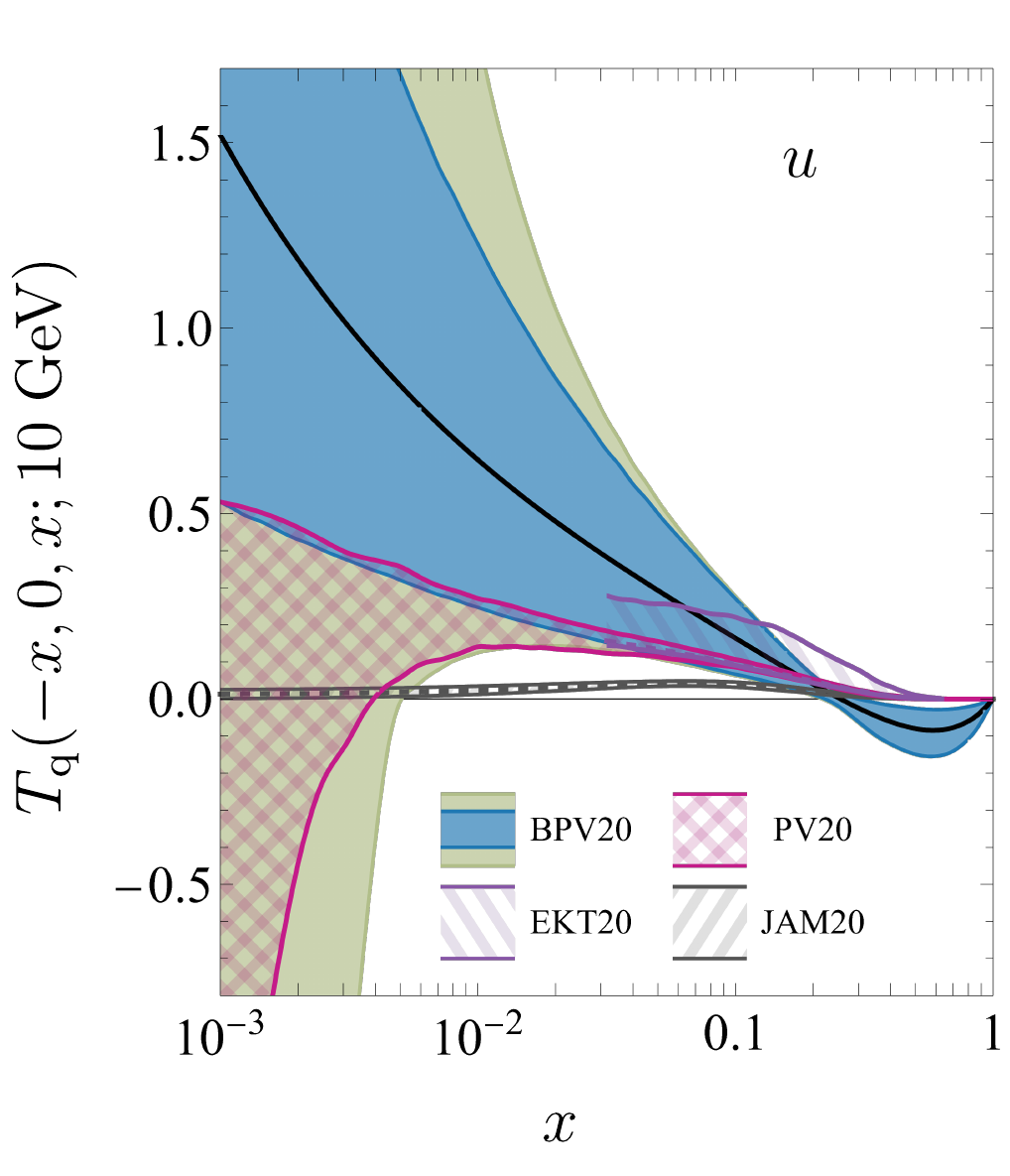}
\includegraphics[width=0.3\textwidth]{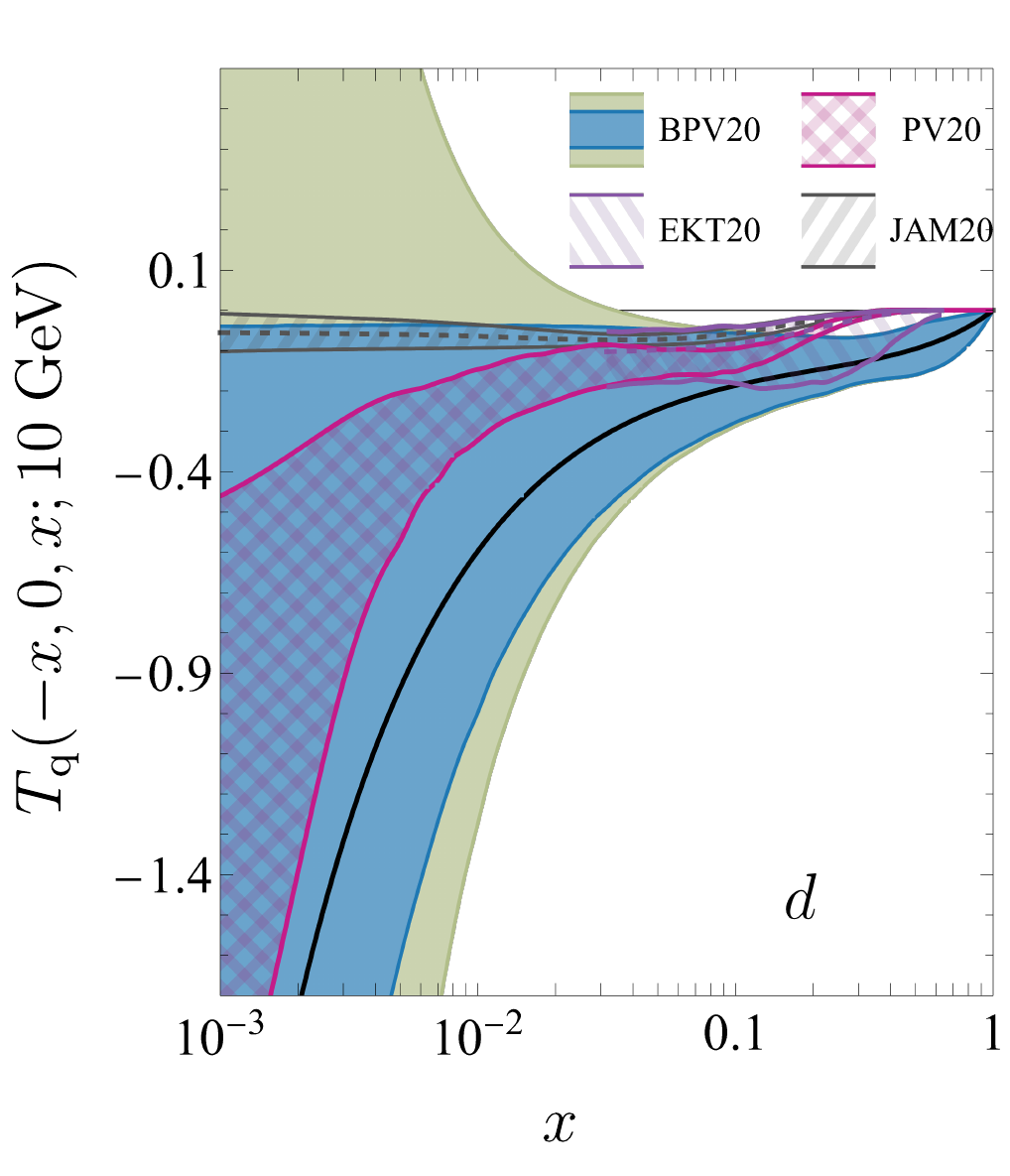}
\\
\centering
\includegraphics[width=0.3\textwidth]{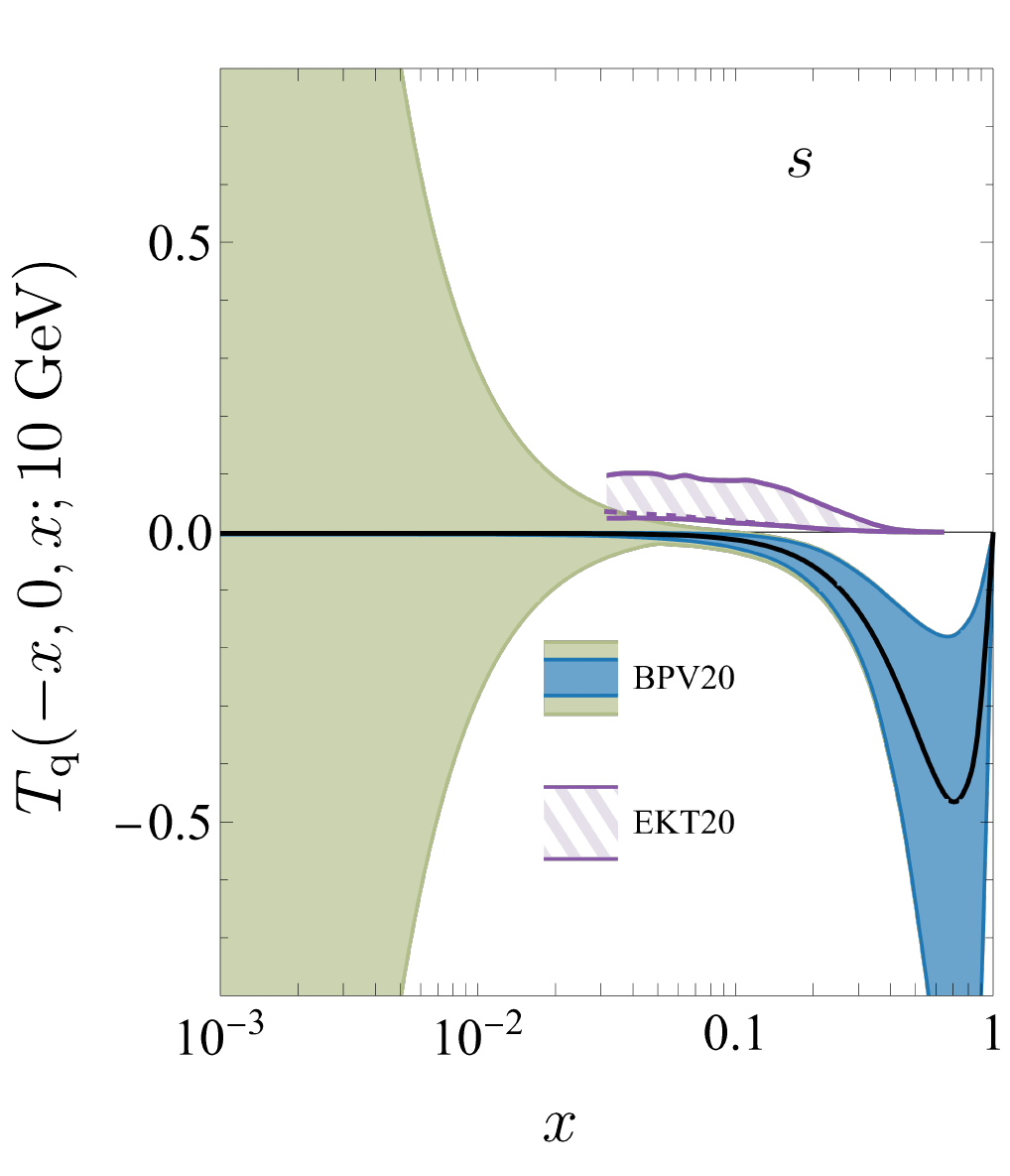}
\includegraphics[width=0.3\textwidth]{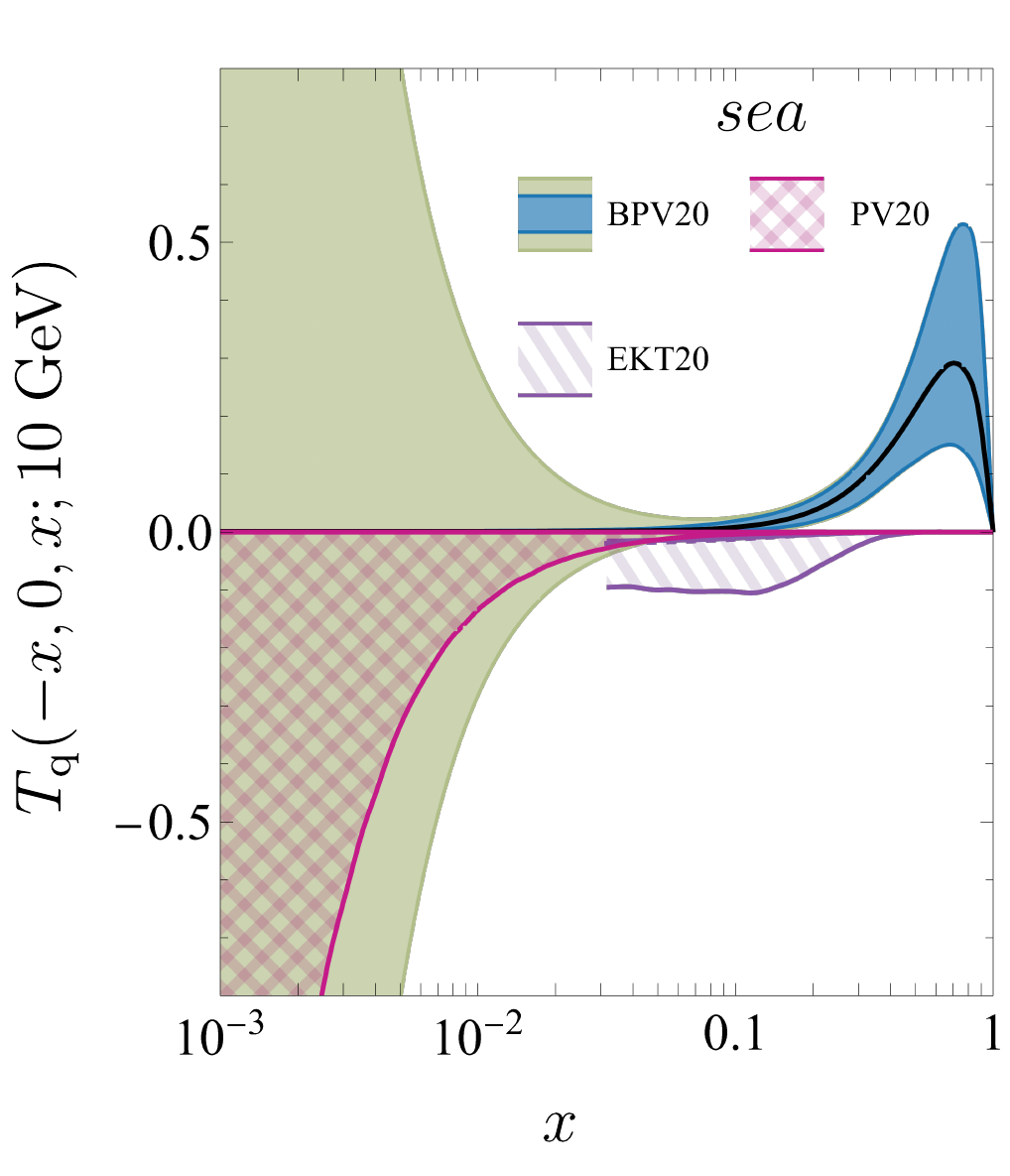}
\caption{\label{fig:QS} Qiu-Sterman function  from Ref.~\cite{Bury:2021sue} at $\mu=10$ GeV for different quark flavors, derived from the Sivers function (\ref{th:QS=f}). Ref.~\cite{Bury:2021sue} results are labeled as BPV20. The black line shows the central value. Blue band shows 68\%CI without gluon contribution added. The green band shows the band obtained by adding the gluon contribution estimated to be $G^{(+)}=\pm (|T_d|+|T_u|)$.  The results are compared to JAM20~\cite{Cammarota:2020qcw} (gray dashed line with the error corridor hatched), PV20~\cite{Bacchetta:2020gko} (magenta hatched region), EKT20~\cite{Echevarria:2020hpy} (violet hatched region, dashed line). Figure from Ref.~\cite{Bury:2021sue}.
}
\end{figure}

The N$^3$LO global analysis of SIDIS and DY data including $W^\pm/Z$ boson production data and extraction of the Sivers function~\cite{Bury:2020vhj,Bury:2021sue}
uses a novel method of inverting the OPE relation and reconstructs QS function from the Sivers functions in a model independent way circumventing the problem of implementation of twist-3 evolution:
\begin{eqnarray}\label{th:QS=f}
T_q(-x,0,x;\mu_b)&=&-\frac{1}{\pi}\left(1+C_F \frac{\alpha_s(\mu_b)}{4\pi}\frac{\pi^2}{6}\right)f_{1T;q\leftarrow h}^\perp(x,b_T)
\\\nonumber &&\!\!\!\!\!\! \!\!\!\!\!\!\!\!\! \!\!\!   -\frac{\alpha_s(\mu_b)}{4\pi^2} \int\limits_{x}^{1} \frac{dy}{y} 
\Big[
\frac{\bar y}{N_c}f_{1T;q\leftarrow h}^\perp\left(\frac{x}{y},b_T\right)+
 \frac{3y^2\bar{y}}{2x}G^{(+)}\left(-\frac{x}{y},0,\frac{x}{y};\mu_b\right)\Big]
+\mathcal{O}(a_s^2,b_T^2)\; ,
\end{eqnarray}
where $\bar{y} = 1-y$,  $\alpha_s$ is the strong coupling constant, $T_q$ and $G^{(+)}$ are QS quark and gluon functions. 
This expression is valid only for small (nonzero) values of $b_T$.
Ref.~\cite{Bury:2020vhj} uses $b_T\simeq 0.11$ GeV$^{-1}$ such that $\mu_b = 10$ GeV. The resulting QS-functions are shown in Fig.~\ref{fig:QS}. \index{Qiu-Sterman (QS) function|)}To estimate the uncertainty due to the gluon contribution,  the gluon contribution is varied as $G^{(+)}=\pm (|T_u| + |T_d|)$.
The resulting 68\% confidence interval uncertainty band and comparison to Refs.~\cite{Cammarota:2020qcw,Bacchetta:2020gko,Echevarria:2020hpy} are also shown in Fig.~\ref{fig:QS}. 

The only global QCD analysis to date that uses SIDIS, DY, $W^\pm/Z$ production data, and $pp\to \pi X$  data on $A_N$ asymmetries is presented in Ref.~\cite{Cammarota:2020qcw} and uses the parton model approximation. This analysis shows universality of the mechanism for spin asymmetries in various processes and extracts the Sivers functions, transversity, and the Collins fragmentation functions from the available experimental data. 

\begin{figure}[t!]
\begin{center}
\includegraphics[width=0.31\textwidth]{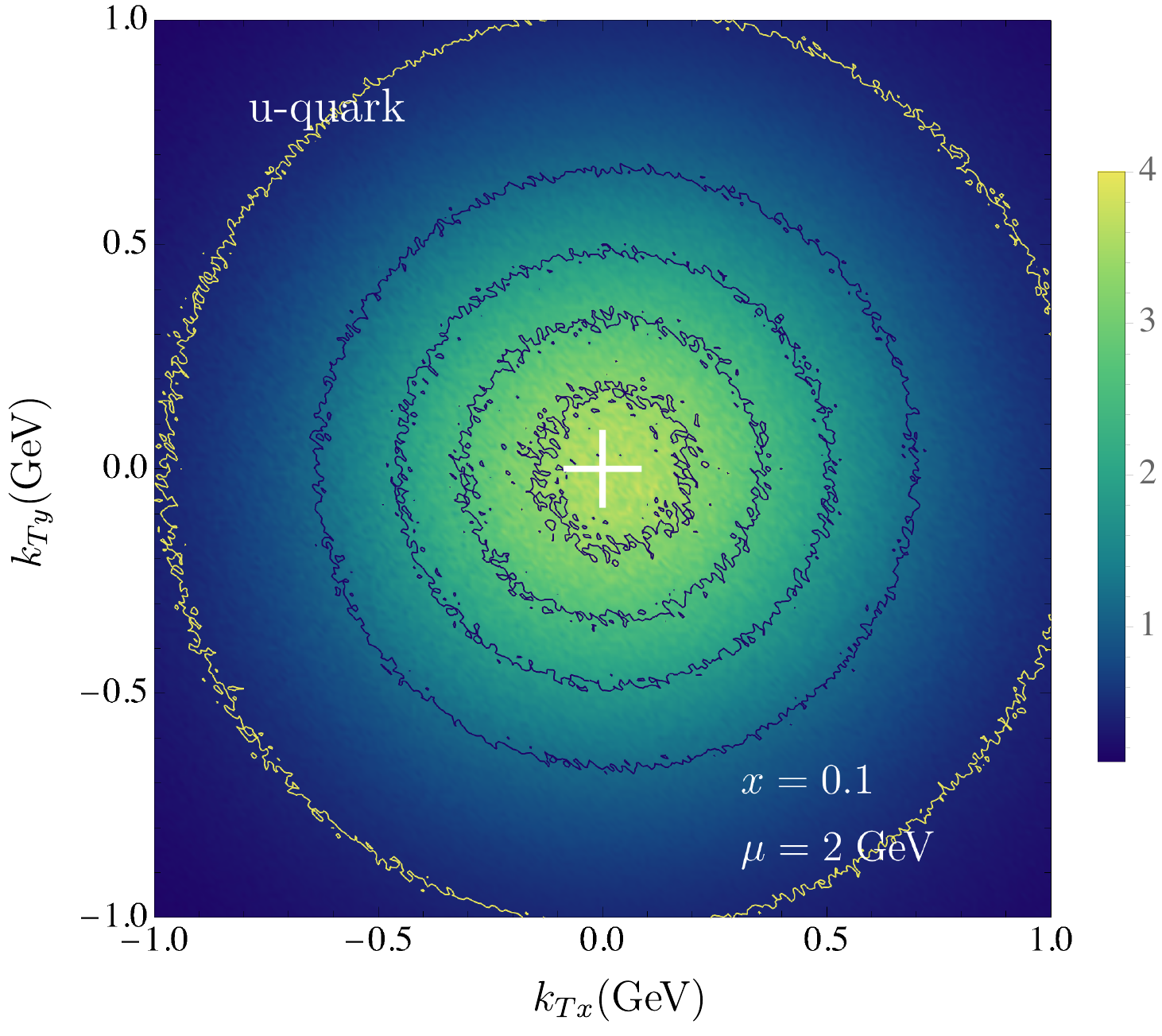}
\includegraphics[width=0.31\textwidth]{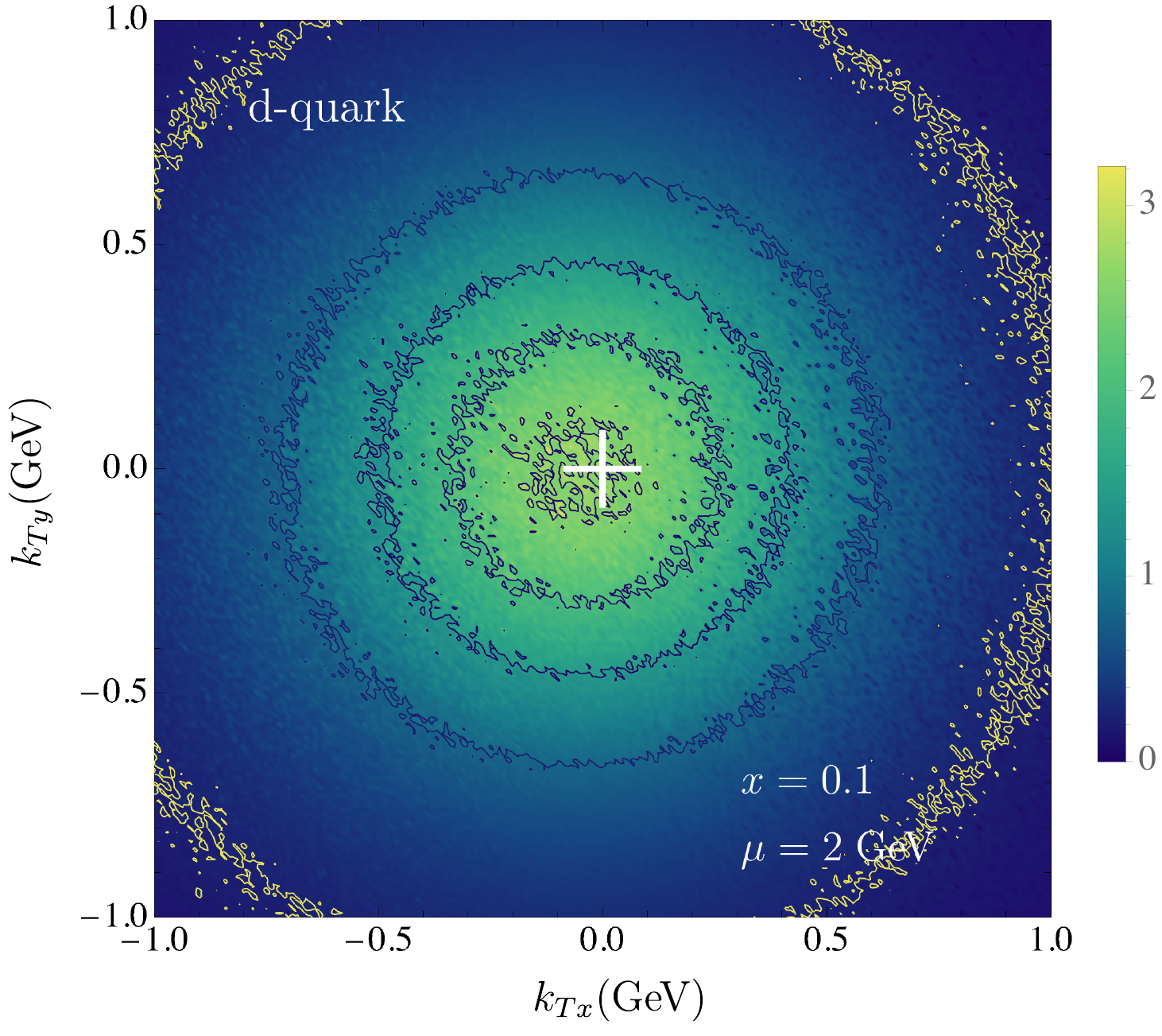}
\end{center}
\caption{\label{fig:tomography} Tomographic scan of the nucleon via the momentum space quark density function $\rho_{1;q\ot h^\uparrow}(x,\vec k_T,\vec S_T,\mu)$ defined in Eq.~(\ref{eq:tomography}) at $x=0.1$ and $\mu=2$ GeV.   Panels are for $u$ and $d$ quarks. The variation of color in the plot is due to variation of replicas and  illustrates the uncertainty of the extraction. The nucleon polarization vector is along $\hat y$-direction.  The figures are from Ref.~\cite{Bury:2021sue}.}
\end{figure}

\begin{figure}[t!]
\begin{center}
\includegraphics[width=0.375\textwidth]{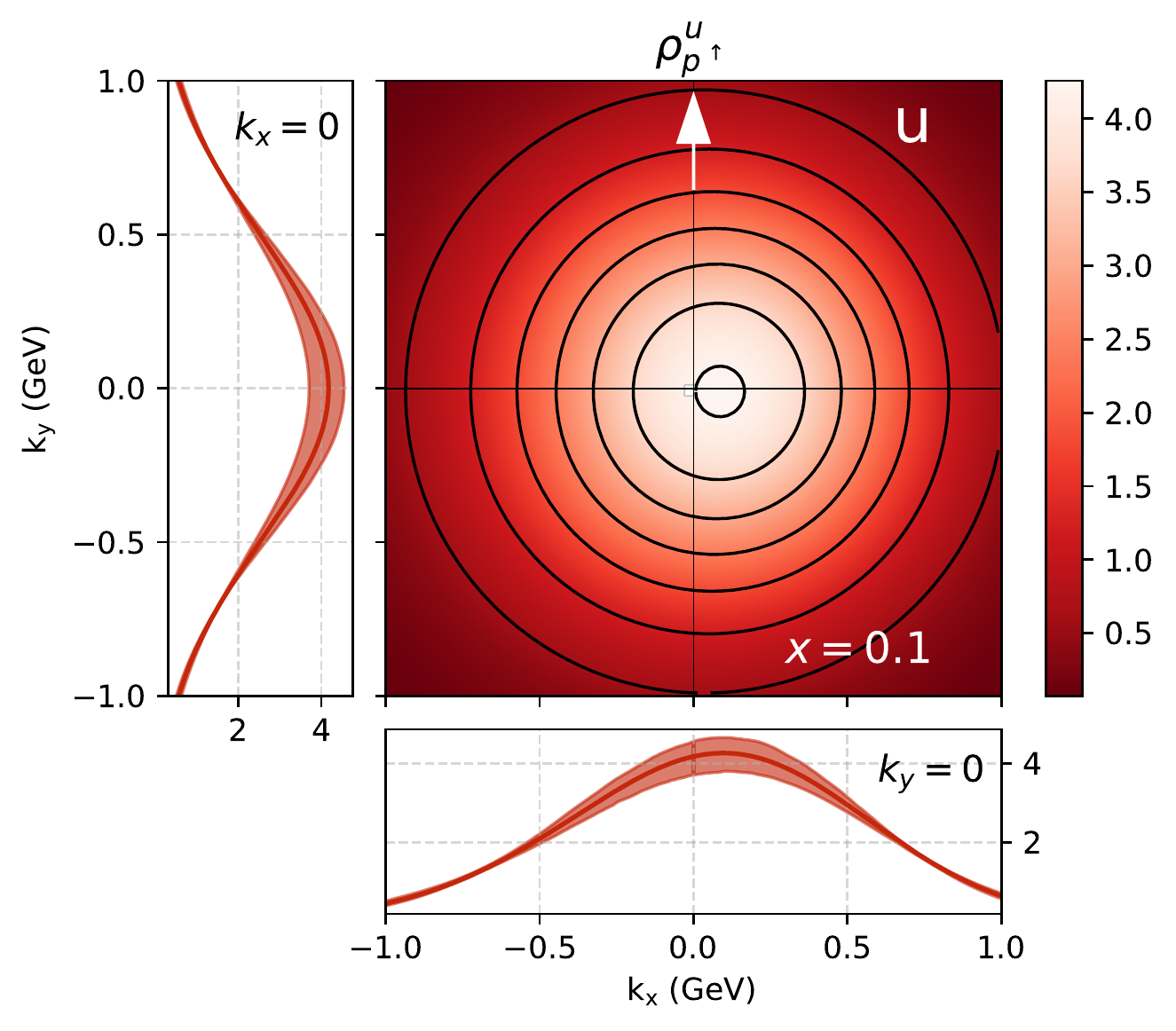}
\includegraphics[width=0.375\textwidth]{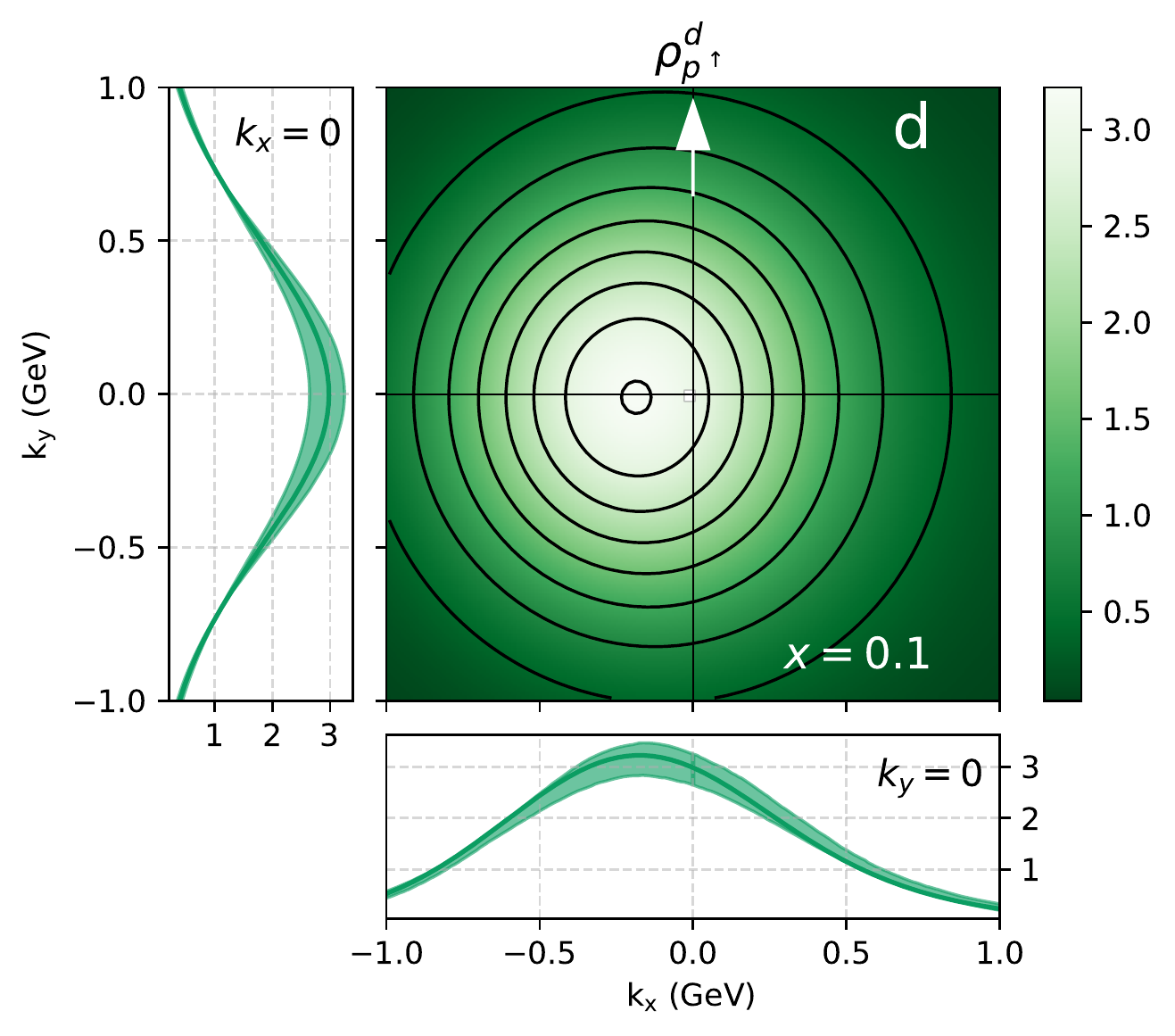}
\end{center}
\caption{\label{fig:tomography1}
  The density distribution $\rho_{p^\uparrow}^a$ of an unpolarized quark
  with flavor $a$ in a proton
  polarized along the $+y$ direction and moving towards the
  reader, as a function of $(k_x, k_y)$ at $Q^2 = 4$ GeV$^2$. The figures are from Ref.~\cite{Bacchetta:2020gko}.}
\end{figure}

\begin{figure}[t!]
\centering
\vskip -2.cm
\includegraphics[width=0.565\textwidth]{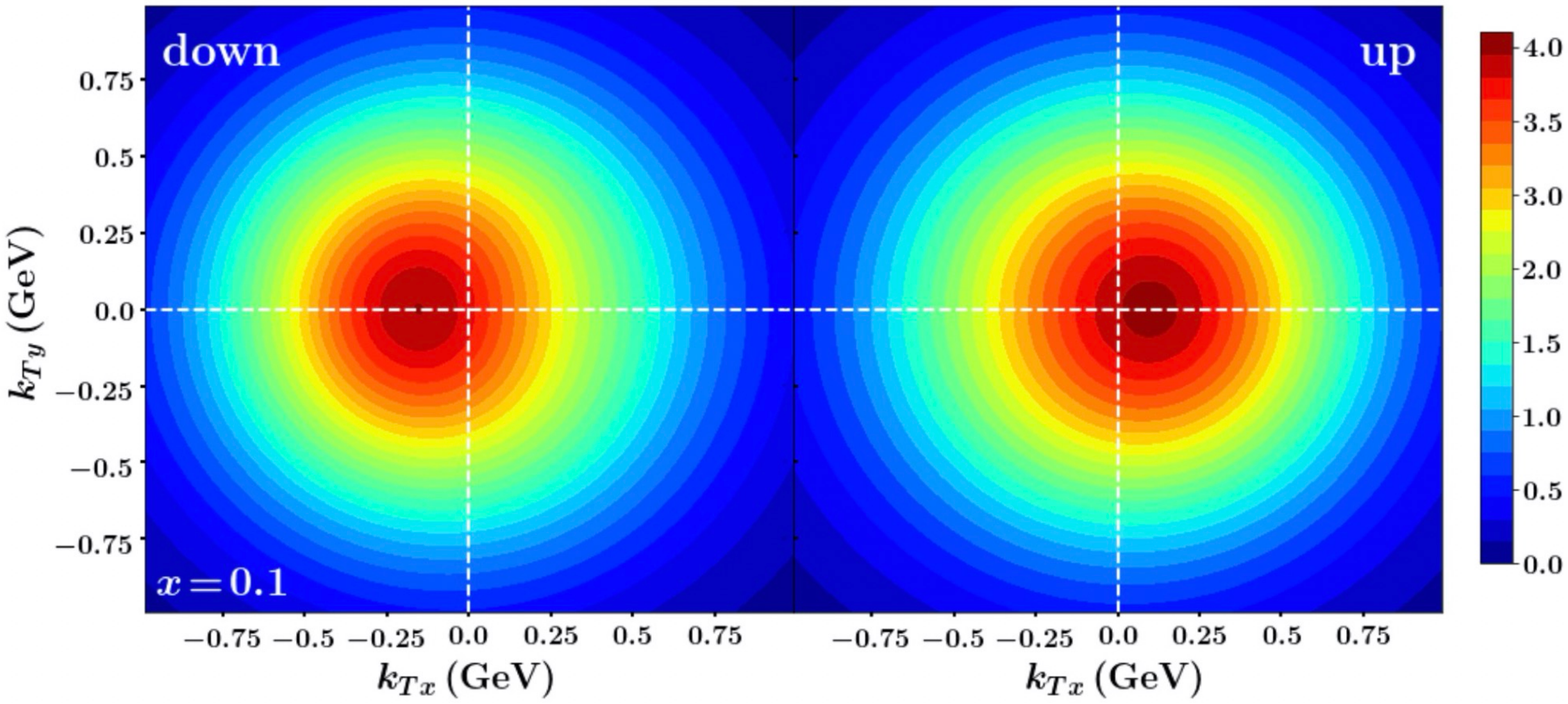}
\vskip -1.5cm
\caption{\label{fig:tomography2} 
The density distribution of an unpolarized up and down quarks
using Sivers functions from Ref.~\cite{Cammarota:2020qcw}.}
\end{figure}
The magnitude of the Sivers function extracted in these fits is generally much smaller than the unpolarized TMD PDF. To present the distortion effect on the unpolarized quarks driven by the hadron polarization, we introduce the momentum space quark density function
\begin{eqnarray}
\rho_{1;q\ot h^\uparrow}(x,\vec k_T,\vec S_T, \mu)=f_{1;q\ot h}(x,k_T;\mu,\mu^2)-\frac{k_{Tx}}{M} f_{1T;q\ot h}^\perp(x,k_T;\mu,\mu^2),
\label{eq:tomography}
\end{eqnarray}
where $\vec k_T$ is a two-dimensional vector $(k_{Tx},k_{Ty})$. This function reflects the TMD density of unpolarized quark $q$ in the spin-$1/2$ hadron totally polarized in $\hat y$-direction, 
{$\vec S_T = (0, 1)$.}
In Figs.~\ref{fig:tomography},~\ref{fig:tomography1},~\ref{fig:tomography2} we plot $\rho$ from Refs.~\cite{Bury:2021sue,Bacchetta:2020gko,Cammarota:2020qcw}  at $x=0.1$ and $\mu=2$ GeV. In Fig.~\ref{fig:tomography}, to present the uncertainty in unpolarized and Sivers function, the authors randomly select one replica for each point of a figure. Thus, the color fluctuation roughly reflects the uncertainty band of their extraction. The presented pictures have a shift of the maximum in $k_{Tx}$, which is the influence of Sivers function that introduces a dipole modulation of the momentum space quark densities. This shift corresponds to the correlation of the orbital angular momentum (OAM) of quarks and the nucleon's spin. One can see from Figs.~\ref{fig:tomography},~\ref{fig:tomography1},~\ref{fig:tomography2}  that $u$ quark has a negative correlation and $d$ quark has a positive correlation. Without OAM of quarks, such a correlation and the Sivers function are zero, and thus we can observe in Figs.~\ref{fig:tomography},~\ref{fig:tomography1},~\ref{fig:tomography2} the evidence for OAM of $u$ and $d$ quarks in the wave function of the nucleon.\index{Sivers effect|)} 

\index{Sivers function $f_{1T}^{\perp}$!phenomenology|)}


\FloatBarrier
\subsubsection[Collins effect in SIDIS and \texorpdfstring{$e^+e^-$}{ep} annihilation]{\boldmath Collins effect in SIDIS and $e^+e^-$ annihilation}
\FloatBarrier
 \index{Collins function $H_1^{\perp}$!phenomenology|(}
\label{subsubsec:Collins_pheno}

\index{Collins effect|(}\index{transversity!phenomenology|(}

Transversity, $h_1$, measures the probability to find a quark in an eigenstate of the transversely projected Pauli-Lubanski operator $s\cdot \gamma_\perp \gamma_5 $ in a transversely polarized nucleon~\cite{Jaffe:1991kp}. The ``transversity'' basis was introduced in hadron-hadron scattering by Goldstein and Moravcsik in Ref~\cite{Goldstein:1974ym}.  Transversity or $h_1(x)$ as a PDF was introduced for the first time by Ralston and Soper in Ref.~\cite{Ralston:1979ys} in their systematic study of the polarized Drell-Yan process. The transversity PDF together with unpolarized and helicity PDFs, describes the structure of a spin-$1/2$ hadron in the leading-power collinear description. The possibility of accessing transversity in double polarized Drell-Yan process and a careful study of its properties and sum rules was explored by Jaffe and Ji in Ref.~\cite{Jaffe:1991kp}. The $Q^2$ evolution of transversity was investigated by Artru and Mekhi in Ref.~\cite{Artru:1989zv} at leading order (LO) in QCD.   Soffer derived a positivity bound for transversity in Ref.~\cite{Soffer:1994ww}, and it was shown by Barone that Soffer inequality is preserved by QCD evolution at LO in Ref.~\cite{Barone:1997fh}. Vogelsang studied NLO evolution of transversity in Ref.~\cite{Vogelsang:1997ak} and showed that Soffer inequality is preserved at NLO QCD.
\index{positivity of TMDs} \index{Soffer bound}

Being chiral-odd, $h_1(x)$ can not be directly accessed in DIS, as another chiral-odd function is needed to form a chiral even observable. Such a function can be a chiral-odd fragmentation function of transversely polarized quark into an unpolarized nucleon, the so-called Collins fragmentation function~\cite{Collins:1992kk}, and transversity can be accessed in  SIDIS. The Collins FF $H_1^\perp$ decodes the
fundamental correlation between the transverse spin of a fragmenting quark
and the transverse momentum of the produced final hadron~\cite{Collins:1992kk}. The measurements that access Collins FF were discussed in Ref.~\cite{Collins:1993kq}. The description of SIDIS in terms of TMD functions was performed by Kotzinian in Ref.~\cite{Kotzinian:1994dv} and by Mulders and Tangerman in Ref.~\cite{Mulders:1995dh}. See Ref.~\cite{Bacchetta:2006tn} for the modern description of SIDIS in terms of TMDs.

The Collins asymmetry in SIDIS is $A_{UT}^{\sin( \phi_h+\phi_S)}$ 
and given by the expression
\begin{equation}
    A_{UT}^{\sin( \phi_h+\phi_S)} \equiv \frac{F_{UT}^{\sin(\phi_h+\phi_S)}}{F_{UU,T}} = \mh \frac{\mathcal{B}[\tilde h_{1} ^{(0)}\, \tilde H_1^{\perp (1)}]}{\mathcal{B}[\tilde f_1^{(0)}\, \tilde D_1^{(0)}]}\; .
\end{equation}
The $F_{UT}^{\sin(\phi_h+\phi_S)}$ structure function of the SIDIS cross section is
given by the convolution of the transversity distribution $h_1$ and the Collins
FF $H_1^\perp$, \eqref{eq:structure-functions-twist-2},\eqref{eq:structure-functions-sidis-bspace},
\begin{align}
F_{UT}^{\sin(\phi_h+\phi_S)}={\cal C}\left[\;\frac{\bfhp\cdot\bfpperp^{ }}{z \mh} \,h_{1} H_1^{\perp}\;\right] = \mh \mathcal{B}[\tilde h_{1}^{(0)} \, \tilde H_1^{\perp (1)}]\; .
\end{align}
The Collins function generates the $\cos 2 \phi_0$ modulation in the $e^+e^-$ cross-section, see Eq.~\eqref{eq:xs_e+e-}, 
 and therefore by combining the data from $e^+e^-$ and SIDIS processes in a global analysis one is able to constrain both transversity and Collins TMD FF.

\begin{figure*}[t!]
\centering 
\includegraphics[width=0.7\textwidth]{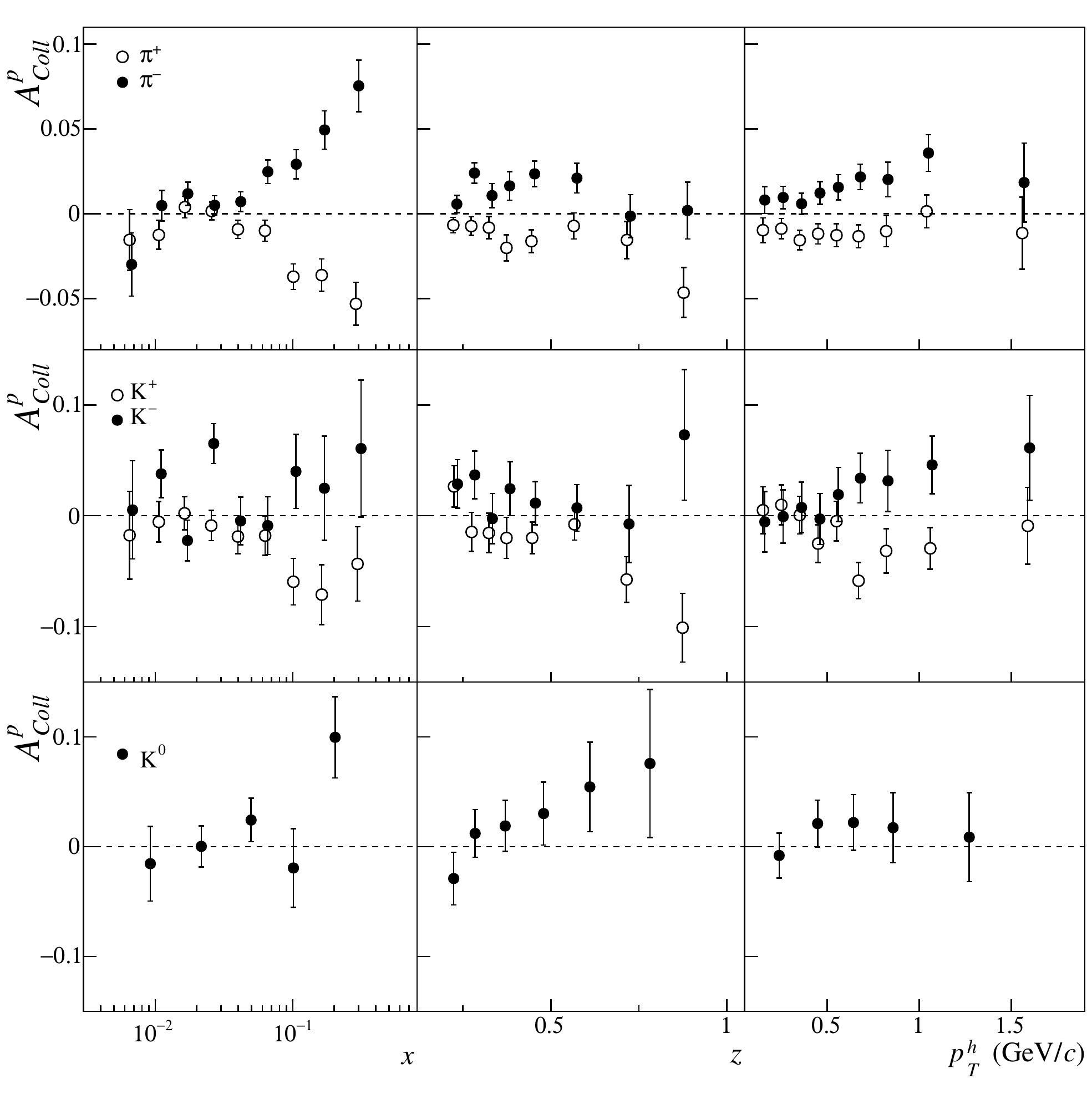}
\caption{ COMPASS results of Ref.~\cite{Adolph:2014zba} on the Collins asymmetries for charged pions (top), charged kaons (middle)
 and neutral kaons (bottom) on proton as a function of $\xbj$, $\zh$ and $\Phperp$. The plot from Ref.~\cite{Adolph:2014zba}.}
\label{fig-call}    
\end{figure*}

The HERMES Collaboration measured Collins asymmetries
in electron proton scattering at the laboratory electron beam energy 27.5 GeV in production of $\pi^+$, $\pi^-$, and $\pi^0$~\cite{Airapetian:2010ds,Airapetian:2020zzo}. The data are presented in bins of $\xbj$, $\zh$, and $\Phperp$ respectively. Clear nonzero asymmetries were found for both $\pi^+$ and $\pi^-$. Large negative asymmetry for $\pi^-$ suggest that unfavored Collins fragmentation function is large and not suppressed with respect to the favored one. Recall that isospin and charge conjugation symmetries suggest that
\begin{align}
H_{1 \,\pi^+/u}^{\perp} = H_{1 \,\pi^+/\bar d}^{\perp} = H_{1 \,\pi^-/\bar u}^{\perp} = H_{1 \,\pi^-/d}^{\perp} \equiv H_{1 \,fav }^{\perp} \nonumber \\
H_{1 \,\pi^+/\bar u}^{\perp} = H_{1 \,\pi^+/d}^{\perp} = H_{1 \,\pi^-/u}^{\perp} = H_{1 \,\pi^-/\bar d}^{\perp} \equiv H_{1 \,unf  }^{\perp}
\end{align}
3D binned data are presented by HERMES in Ref.~\cite{Airapetian:2020zzo}. The favored Collins functions describe valence quarks fragmenting to the pion while unfavored correspond to nonvalence quarks.

HERMES~\cite{Airapetian:2009ae,Airapetian:2020zzo} and JLab Hall A~\cite{Qian:2011py} include the kinematic factor $p_1$ from Eq.~\eqref{eq:y-prefactors} in the measured asymmetry,
\begin{equation}
  A_{UT}^{\sin( \phi_h+\phi_S)}|_{HERMES} \equiv \langle \sin( \phi_h+\phi_S) \rangle = p_1 A_{UT}^{\sin( \phi_h+\phi_S)} \; .
\end{equation}

\begin{figure}[t!]
\includegraphics[width=0.455\textwidth]{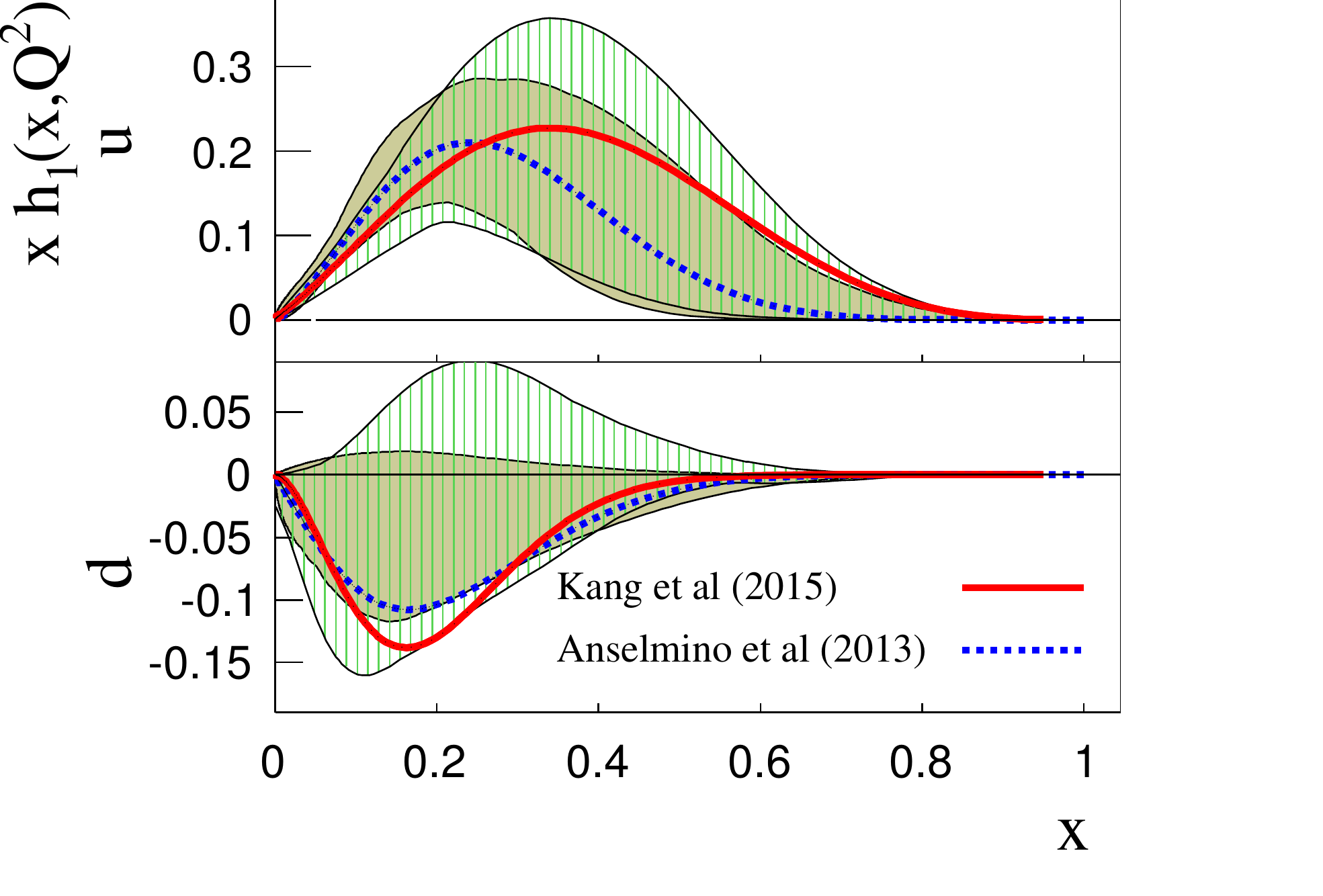}
\includegraphics[width=0.535\textwidth]{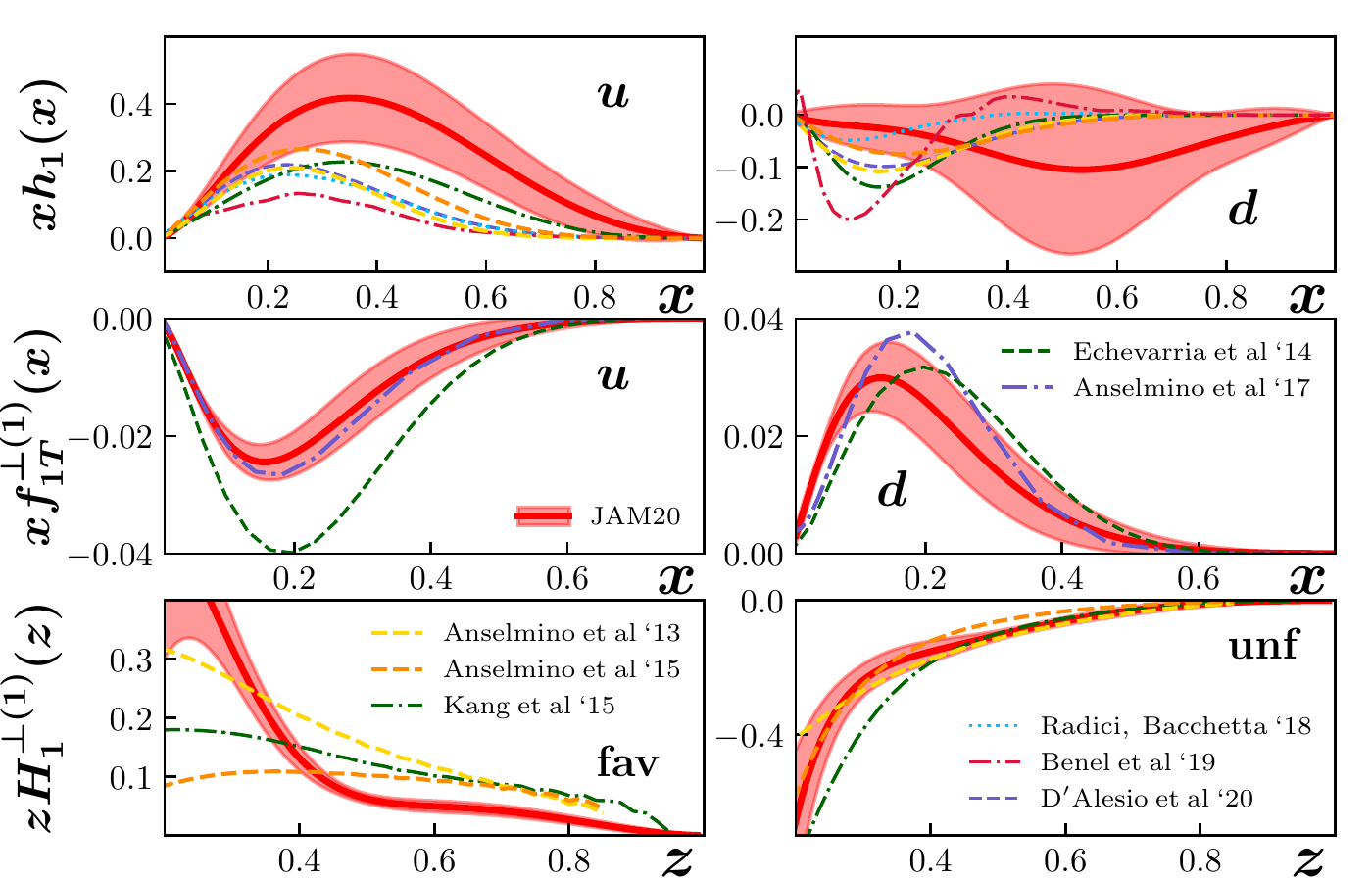}
\caption{Left panel: Comparison of extracted transversity from Refs.~\cite{Kang:2014zza,Kang:2015msa} (solid lines  and  vertical-line hashed region) at $Q^2 = 2.4$ GeV$^2$ with  Torino-Cagliari-JLab 2013 extraction \cite{Anselmino:2013vqa} (dashed lines and shaded region).
Right panel:  The extracted functions $h_1(x)$, $f_{1T}^{\perp(1)}(x)$, and
$H_1^{\perp(1)}(z)$ at $Q^2=4$ GeV$^2$ from JAM20 global
analysis~\cite{Cammarota:2020qcw} (red solid curves with 1-$\sigma$ CL error bands).  The
functions from other groups
~\cite{Anselmino:2013vqa,
       Echevarria:2014xaa,
       Kang:2015msa,
       Anselmino:2015sxa,
       Anselmino:2016uie,
       Radici:2018iag,
       Benel:2019mcq,
       DAlesio:2020vtw} 
are also shown. Plot from Ref.~\cite{Cammarota:2020qcw}} 
\label{f:qcf}
\end{figure}

The COMPASS Collaboration uses muon beam of energy 160 GeV and have measured Collins asymmetries on both NH$_3$
(proton) \cite{Adolph:2014zba}, see Fig.~\ref{fig-call}, and LiD (deuterium) \cite{Alekseev:2008aa} targets. The data are presented as function of $x_B$, $z_h$, and $P_{h\perp}$. Results on the proton target are compatible with HERMES findings and asymmetries are found to be compatible with zero on the deuterium target. The beam energy of COMPASS is higher than the energy of HERMES and thus COMPASS reaches lower values of $\xbj\sim 10^{-3}$. For each point in $\xbj$ the scale $Q^2$ is higher at COMPASS as one has $Q^2 \simeq s \xbj y$. Both experiments consider $Q^2 >1 $ GeV$^2$ in order to be in DIS region and center-of-mass energy of the $\gamma^* p$ system, $W^2 > 10$ GeV$^2$ for HERMES and $W^2 > 25$ GeV$^2$ for COMPASS in order to be outside of the resonance region. 

The COMPASS Collaboration considers $z_h >0.2$ region and the HERMES Collaboration uses $0.2 < z_h < 0.7$ in order to minimize both target fragmentation effects and exclusive reaction contributions. All other experimental cuts are described in Refs.~\cite{Airapetian:2009ae,Alekseev:2008aa,Adolph:2014zba}.
 The definition of azimuthal angle $\phi_S$ of COMPASS experiment is such that
\begin{equation}
  A_{UT}^{\rm Collins}|_{COMPASS} = - A_{UT}^{\sin( \phi_h+\phi_S)}\; . 
\end{equation}
due to a different convention the notation of the Collins angle $\phi_C$ used by the COMPASS Collaboration, see Ref.~\cite{COMPASS:2005csq}.
Jefferson Lab's Hall A published data
on $\pi^\pm$ pion production in 5.9 GeV electron scattering on $^3$He (effective neutron) target \cite{Qian:2011py}.
Jefferson Lab operates at relatively low energy and reaches higher values of $\xbj \sim 0.35$.

Information on Collins fragmentation functions is contained in data from $e^+e^-$ collisions at the energy $\sqrt{s} \simeq 10.6$ GeV  of the BELLE \cite{Seidl:2008xc} and the {\em BABAR} \cite{TheBABAR:2013yha} Collaborations. Both BELLE and {\em BABAR} Collaborations
require the momentum of the virtual photon $P_{h\perp}/z_{h1} < 3.5$ GeV in order to remove contributions
from hadrons assigned to the wrong hemisphere, and it also helps to remove contributions from gluon radiation.  
The analysis of BELLE is performed in ($z_{h1}$,$z_{h2}$) bins with boundaries
at $z_{hi}=$ $0.2$, $0.3$, $0.5$, $0.7$ and $1.0$. The {\em BABAR} Collaboration chooses 6 $z_{hi}$-bins: $[0.15-0.2]$, $[0.2-0.3]$,
$[0.3-0.4]$, $[0.4-0.5]$, $[0.5-0.7]$, $[0.7-0.9]$. A characteristic feature of the asymmetry is growth with $z_{hi}$ which is compatible with the kinematical zero in the limit  $z_{hi}\to 0$.

A recent study presented in Ref.~\cite{Wan:2020lps} considered the $\cos 2\phi_h$ azimuthal asymmetry in $e^+e^-$ due to Collins functions and the acoplanarity to the azimuthal asymmetry due to the digluon radiation. The authors found  that in the region $q_T\ll Q$ region, the asymmetry is dominated by the Collins effect, while the acoplanarity effect dominates in the large $q_T$ region ($q_T/Q>0.5$) and is negligible in the small $q_T$ region. In the intermediate region the two contributions are comparable.

There are many extractions of $h_1$ and $H_1^\perp$ from 
combined fits of SIDIS and $e^+e^-$ data, for instance those of
Refs.~\cite{Anselmino:2013vqa,Kang:2014zza,Kang:2015msa,Anselmino:2015sxa,Cammarota:2020qcw,Vogelsang:2005cs,Efremov:2006qm}. In the extractions in Refs.~\cite{Anselmino:2013vqa,Kang:2014zza,Anselmino:2015sxa,Cammarota:2020qcw,Vogelsang:2005cs,Efremov:2006qm} the parton model approximation is used and TMDs are parametrized as
\begin{align}
h_1^{q}(x,\kperp)
&= 
    h_{1}^{q}(x)\
    \frac{1}{\pi\langle k_T^2\rangle_{h_1^{}}}\;
{\exp\left[{-\frac{k_T^2}{\langle k_T^2\rangle_{h_1^{}}}}\right]}\,,\label{e:h1} \\
H_1^{\perp h/q}(z,\pperp)
&= \frac{2 z^2 \mh^2}{\langle \pperp^2\rangle_{H_1^\perp}}\,
    H_{1\, h/q}^{\perp (1)}(z)\
    \frac{1}{\pi\langle \pperp^2\rangle_{H_1^{\perp}}}\;
{\exp\left[{-\frac{\pperp^2}{\langle \pperp^2\rangle_{H_1^{\perp}}}}\right]}\,,
\label{e:collins1}
\end{align}
while in Refs.~\cite{Kang:2014zza,Kang:2015msa} the OPE is used for TMDs and the extraction is performed at NLL accuracy. 
For completeness we remark that the definition of the transverse moment of the Collins function in (\ref{e:collins1}) in terms of renormalized functions in $b_T$-space, see Chap.~\ref{sec:TMDdefn} and \app{Fourier_transform}, simplifies within the Gaussian model similarly to (\ref{eq:define-(1)-mom-in-kT-space}) as follows
\be\label{eq:define-(1)-mom-in-kT-space-FF}
    H_1^{\perp (n) a}(z)   = \int d^2p_T\;
    H_1^{\perp (n) a}(z,p_T)\, , \quad \quad
    H_1^{\perp (n) a}(z,p_T) = 
    \biggl(\frac{p_T^2}{2M_h^2}\biggr)^n
	H_1^{\perp a}(z,p_T)\,,
\ee
where $\pt = -z \pt'$.\index{Collins effect|)}

\begin{figure}[t!]
\centering
\includegraphics[width=0.75\textwidth]{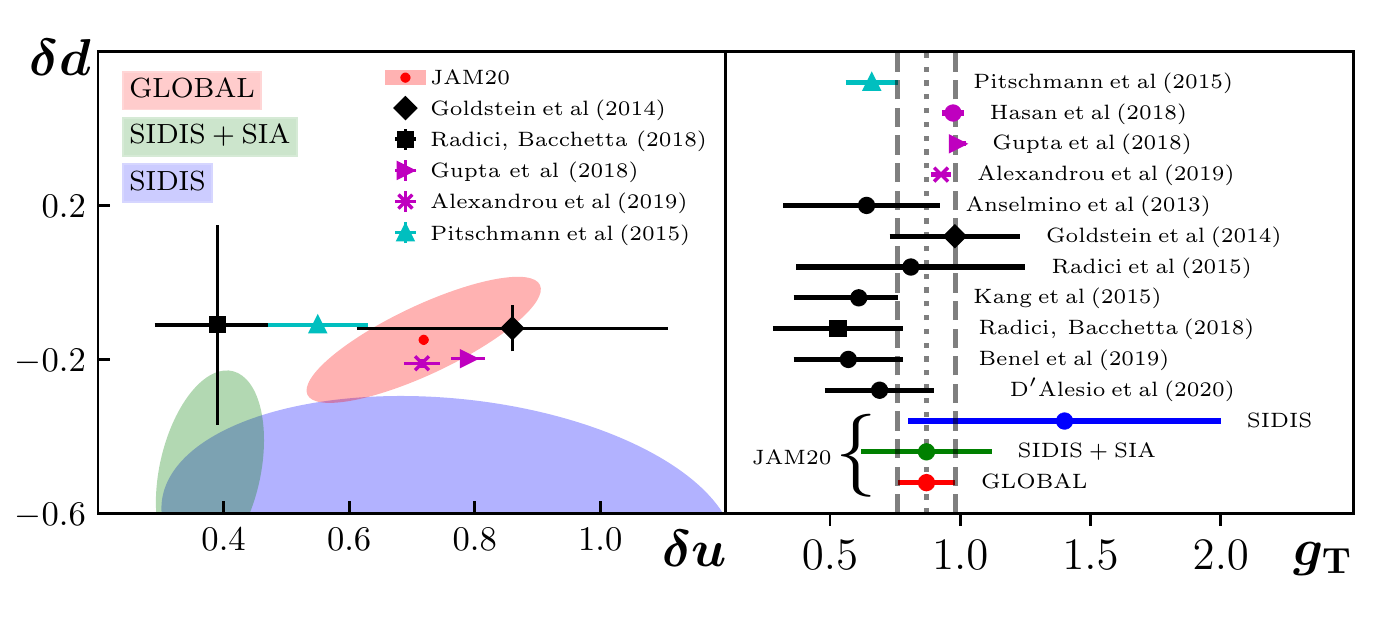}
\caption{
Figure from Ref.~\cite{Cammarota:2020qcw} of the tensor charges $g_T^u$, $g_T^d$, and $g_T^{u-d}$. Note that Ref.~\cite{Cammarota:2020qcw} uses $\delta u$, $\delta d$, and $g_T$, respectively, to denote these quantities.  The JAM20 result refers to that obtained in Ref.~\cite{Cammarota:2020qcw} at
$Q^2=4$~GeV$^2$.  Also shown are other results from phenomenology (black)~\cite{Anselmino:2013vqa,
       Goldstein:2014aja,
       Radici:2015mwa,
       Kang:2015msa,
       Radici:2018iag,
       Benel:2019mcq,
       DAlesio:2020vtw}, lattice QCD (purple)~\cite{Gupta:2018qil,Hasan:2019noy,Alexandrou:2019brg}, and
Dyson-Schwinger (cyan)~\cite{Pitschmann:2014jxa} at the same scale.
}
\label{f:gT}
\end{figure}
\index{tensor charge|(}A quantity of interest that can be calculated from extractions of the transversity function are the presently still not well-known tensor charges $g_T^q$~\cite{Ralston:1979ys,Jaffe:1991kp,Jaffe:1991ra,Cortes:1991ja,Gamberg:2001qc}. The tensor charges play an important role for the understanding of the nucleon spin structure, and are important for the nucleon tomography in momentum space. Unlike the axial charge of the nucleon, which is related by the Bjorken sum rule to the axial coupling constant through which the nucleon couples to weak interactions, the tensor charge is not a conserved charge and has no practical application within the Standard Model. However,  certain hypothetical beyond the Standard Model  (BSM) particles could couple  to the nucleon through the tensor charges. For this reason, $g_T^q$ plays a role for BSM physics, see, e.g., Refs.~\cite{Courtoy:2015haa,Yamanaka:2017mef,Gao:2017ade,Gonzalez-Alonso:2018omy}. 

The tensor charges are computed from the following integrals of $h^q_1 (x)$~\cite{Ralston:1979ys,Jaffe:1991kp,Jaffe:1991ra,Cortes:1991ja,Gamberg:2001qc} over the parton momentum fraction $x$:
\begin{equation}
    g_T^u = \int_0^1 \! dx\,(h_1^u(x)-h_1^{\bar{u}}(x))\,, 
\qquad 
    g_T^d = \int_0^1 \! dx\,(h_1^d(x)-h_1^{\bar{d}}(x))\,, \label{e:gT}
\end{equation}
where $u$ and $d$ represent up and down quarks, respectively.  The isovector combination $g_T^{u-d} \equiv g_T^u - g_T^d$ is also of particular focus. The quantities $g_T^u$, $g_T^d$, and $g_T^{u-d}$ have all been computed in lattice QCD~\cite{Gupta:2018qil,Yamanaka:2018uud,Hasan:2019noy,Alexandrou:2019brg}. Some results for the tensor charges are collected in Fig.~\ref{f:gT}.  The JAM20 results from Ref.~\cite{Cammarota:2020qcw} show what happens to the extracted values for for $g_T^u$, $g_T^d$, and $g_T^{u-d}$ if one includes only SIDIS data (blue), SIDIS and $e^+e^-$ semi-inclusive annihilation (SIA) data (green), and then a global analysis of SIDIS, $e^+e^-$, and $A_N$ data (red).  (See Sec.~\ref{sec:AN}  for more details about the $A_N$ observable in proton-proton collisions.)  Notice how only after a global analysis do the phenomenological values for $g_T^u$, $g_T^d$, and $g_T^{u-d}$ agree with lattice QCD. \index{tensor charge|)}
\index{Collins function $H_1^{\perp}$!phenomenology|)}
\index{transversity!phenomenology|)}

\FloatBarrier
\subsubsection[ \texorpdfstring{$A_N$}{AN} in proton-proton collision]{\boldmath $A_N$ in proton-proton collision}
\FloatBarrier
\label{sec:AN}

\subsubsection*{Cross Section Formulas} \index{$A_N$ in $pp$ collisions|(}Let us consider the production of a single hadron from the collision of two protons $A$ and $B$:
\begin{equation}
    p_A(P) + p_B(P') \rightarrow h(P_h) + X\,. \label{e:pp_reac}
\end{equation}
The differential cross section $d\sigma$ for this reaction can be written in a twist expansion,
\begin{equation}
    d\sigma = d\sigma_{t2} + d\sigma_{t3} + \dots\,,
\end{equation}
where $d\sigma_{t2}$ ($d\sigma_{t3}$) represents the twist-2 (twist-3) term.  This reaction can be analyzed within collinear factorization so long as the hard scale, set by the transverse momentum of the produced hadron $P_{hT}$, satisfies $P_{hT}\gg \Lambda_{QCD}$.

For the case that the initial state protons are unpolarized, the leading-power term in the cross section  $d\sigma_{t2}$ is given at leading-order (in the strong coupling $\alpha_s$) by
\begin{align}
E_h\frac{d\sigma} {d^{3}\vec{P}_{h}} = \frac{\alpha_s^2} {S} \sum_i \sum_{a,b,c}
\int_0^1\!\frac{dz} {z^2}\int_0^1 \!\dfrac{dx'} {x'}\int_0^1 \!\dfrac{dx} {x}\,\,\delta(\hat{s}+\hat{t}+\hat{u})
\,f_1^a(x)\,f_1^b(x')\,D_1^{h/c}(z)\,\mathcal{H}_U^i\,,
\label{e:spin-avg}
\end{align}
where $\sum_i$ is a sum over all partonic interaction channels, $a$ can be a quark, anti-quark, or gluon (and likewise for $b,c$), and $f_1$ ($D_1$) is the usual unpolarized PDF (FF).  The well-known hard factors  for the unpolarized cross section are denoted
by $\mathcal{H}_U^i$~\cite{Owens:1986mp,Kouvaris:2006zy,Kang:2013ufa} 
and can be found in, e.g., appendix A of Ref.~\cite{Kouvaris:2006zy}. They are functions of the partonic Mandelstam variables $\hat{s} = xx' S,\,\hat{t} = xT/z,\,{\rm and}\;\hat{u} = x'U/z$, where $S = (P+P')^2$, $T = (P-P_h)^2$, and $U = (P'-P_h)^2$.

If any of the particles in (\ref{e:pp_reac}) carry a transverse polarization $S_T$, one can then define the SSA $A_N$, 
\begin{equation}
A_N \equiv \frac{d\Delta\sigma(S_T)}{d\sigma}, \label{e:AN}
\end{equation}
where
$d\Delta\sigma(S_T)\equiv \frac{1}{2}\left[d\sigma(S_T)-d\sigma(-S_T)\right]$ and
$d\sigma \equiv\frac{1}{2}\left[d\sigma(S_T)+d\sigma(-S_T)\right]$.
The leading-power contribution to $A_N$ is twist 3,   
and the relevant nonperturbative functions are now twist-3 multi-parton correlators (e.g., quark-gluon-quark or tri-gluon)~\cite{Efremov:1981sh,Efremov:1984ip,Qiu:1991pp,Qiu:1991wg,Qiu:1998ia,Eguchi:2006qz,Kouvaris:2006zy,Eguchi:2006mc,Koike:2009ge, Metz:2012ct,Beppu:2013uda}.  From an experimental standpoint, recent focus has been  on pion production where one of the initial-state protons is transversely polarized.  Schematically, we can write $d\Delta\sigma(S_T)$ as
\begin{align} 
d\Delta\sigma(S_{T}) &= \,\mathcal{H}_A\otimes f_{a(3)}\otimes f_{b(2)}\otimes D_{h/c(2)}\nonumber\\  
&\;\;\;+ \mathcal{H}_B\otimes f_{a(2)}\otimes f_{b(3)}\otimes D_{h/c(2)} \nonumber\\
&\;\;\;+ \mathcal{H}_h\otimes f_{a(2)}\otimes f_{b(2)}\otimes D_{h/c(3)}\,,
\label{e:collfac}
\end{align} 
where $f_{a(t)}$ is the twist-$t$ PDF associated with parton $a$ in proton $A$ (similarly for $f_{b(t)}$), while $D_{h/c(t)}$ is the twist-$t$ FF associated with the hadron $h$ in parton~$c$.  The hard parts are different for each term, depending on which nonperturbative function is kept at twist 3, and are denoted by $\mathcal{H}_A$, $\mathcal{H}_B$, and $\mathcal{H}_h$. In the case of $p^\uparrow p\to \pi\, X$, all three terms in Eq.~(\ref{e:collfac}) enter into the analysis.  

Specifically, one receives twist-3 contributions from (a) the transversely polarized proton, (b) the unpolarized proton, and (c) the (unpolarized) final-state pion.  
For (a), there are two types of terms that arise, a so-called soft-gluon pole (SGP) term and a soft-fermion pole (SFP) term.  These are so named because, since SSAs are a na\"{i}ve time-reversal odd (T-odd) effect, one must pick up a pole in the hard scattering. This pole causes the momentum fraction of either a gluon or quark in the multi-parton correlator to vanish, which leads, respectively, to a SGP or SFP.  The SGP term was calculated in Refs.~\cite{Qiu:1998ia,Kouvaris:2006zy} for $qgq$ correlators and Ref.~\cite{Beppu:2013uda} for tri-gluon ($ggg$) ones, while the SFP term was computed in Ref.~\cite{Koike:2009ge}.  

The contribution from the $qgq$ SGP function $F_{FT}(x,x)$, called the Qiu-Sterman (QS) \index{Qiu-Sterman (QS) function|(} function,\footnote{There are several notations use in the literature for the QS function, e.g., $T_F(x,x)$ and $G_F(x,x)$.} to the spin-dependent cross section reads~\cite{Qiu:1998ia,Kouvaris:2006zy}
\begin{align}
\label{e:QSfinalcr}
\frac{E_hd\sigma_{(T)}^{SGP_{qgq}}(S_T)} {d^{3}\vec{P}_{h}}
&= -\frac{4\alpha_s^2 M} {S}\,\epsilon_{\mu\nu\rho\sigma}P^{'\!\mu}P^\nu P_h^\rho S_T^\sigma\,\sum_i\sum_{a,b,c}\int_0^1\!\frac{dz} {z^3}\int_0^1 \!dx'\int_0^1 \!dx\,\delta(\hat{s}+\hat{t}+\hat{u})\nonumber\\
&\times\, \frac{\pi} {\hat{s}\hat{u}} \,f_1^b(x')\,D_1^{\pi/c}(z)\left[F_{FT}^a(x,x)-x\frac{dF_{FT}^a(x,x)} {dx}\right]\mathcal{H}^i_{F_{FT}}\,,
\end{align}
where the Levi-Civita tensor is defined with $\epsilon^{0123} = +1$. 
The hard factors are denoted by $\mathcal{H}_{F_{FT}}^i$ and can be found in Ref.~\cite{Kouvaris:2006zy}.  
The notation used for the cross section indicates that this is the $qgq$ SGP term for the transversely polarized proton.
\begin{figure}[t!]
\centering
  \includegraphics[width=0.6\textwidth]{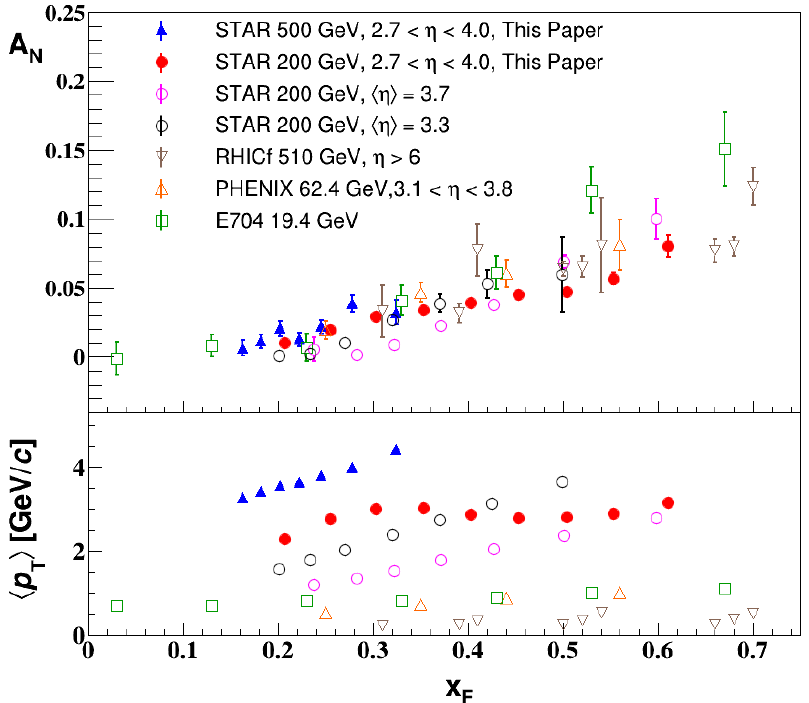}
\caption{Summary of $A_N$ data for $\pi^0$ production from Ref.~\cite{Adam:2020edg}.}
\label{fig:AN_pp}
\end{figure}
The QS function has an important, model-independent relation to the TMD Sivers function~\cite{Sivers:1989cc} $f_{1T}^{\perp}(x,k_T)$ that enters SSAs in processes like SIDIS and Drell-Yan (DY), see ~\sec{Sivers_SIDIS}. 
The identity reads~\cite{Boer:2003cm}
\begin{equation}
 \pi F_{FT}(x,x) =  f_{1T}^{\perp(1)}(x)\big|_{SIDIS} = -f_{1T}^{\perp(1)}(x)\big|_{DY}\,.
 \label{e:QS_Siv}
\end{equation}
The Sivers function is also connected to the QS function through the OPE~\cite{Aybat:2011ge}, see Eq.~\eqref{th:QS=f}.\index{Qiu-Sterman (QS) function|)}

The case of twist-3 effects in the unpolarized proton was analyzed many years ago in Ref.~\cite{Kanazawa:2000hz} and they were found to be negligible.   The twist-3 effects due to the final-state pion were computed in Ref.~\cite{Metz:2012ct} and re-written in Ref.~\cite{Gamberg:2017gle} using Lorentz invariance and equation of motion relations~\cite{Kanazawa:2015ajw}.  The result reads
\begin{align}
\frac{E_hd\sigma^{Frag}(S_T)} {d^{3}\vec{P}_{h}} &=-\frac{4\alpha_{s}^{2}m_{h}} {S}\, \epsilon_{\mu\nu\rho\sigma}P^{'\!\mu}P^\nu P_h^\rho S_T^\sigma\sum_{i}\sum_{a,b,c}\int_{0}^{1}\frac{dz} {z^{3}} \int_{0}^{1}\!dx' \int_0^1 \!dx\,\,\delta(\hat{s}+\hat{t}+\hat{u}) \frac{1} {\hat{s}}\,\nonumber\\ 
&\hspace{-2cm}\times\,h_{1}^{a}(x)\,f_{1}^{b}(x')\left\{\left[H_{1,\pi/c}^{\perp(1)}(z)-z\frac{dH_{1,\pi/c}^{\perp(1)}(z)} {dz}\right]\mathcal{H}_1^i
 +\left[-2H_{1,\pi/c}^{\perp(1)}(z) + \frac{1} {z}\tilde{H}^{\pi/c}(z)\right]\mathcal{H}_2^i\right\}, 
 \label{e:sigmaFrag}
\end{align}
where $m_h$ is the pion mass, $h_1$ is the standard twist-2 transversity function, and the hard factors for each term are given by $\mathcal{H}_1^i$ and $\mathcal{H}_2^i$, which can be found in Ref.~\cite{Gamberg:2017gle}.  In Ref.~\cite{Gamberg:2017gle}, the notation $\tilde{S}^i_{H_1^\perp}$ and $\tilde{S}^i_{H}$ is used for the hard factors, and one has $\mathcal{H}_1^i = \tilde{S}^i_{H_1^\perp}$ and $\mathcal{H}_2^i = \tilde{S}^i_{H}$. 
The notation for the cross section indicates that this is the entire fragmentation term.  The functions $H_1^{\perp(1)}$ and $\tilde{H}$ are unpolarized twist-3 FFs connected to $qgq$ matrix elements~\cite{Kanazawa:2015ajw}.
The function $H_1^{\perp(1)}$ is the first moment of the TMD Collins FF $H_{1}^{\perp}(z,\pperp)$ that enters SSAs in SIDIS and electron-positron annihilation $e^+e^-\!\to h_1\,h_2 \,X$. 
The Collins TMD FF $H_{1}^{\perp}(z,\pperp)$ can also be written in terms of $H_1^{\perp(1)}(z)$ using the OPE~\cite{Kang:2015msa}.

\subsubsection*{Phenomenology}

\begin{figure}[t!]
\centering
 \includegraphics[width=0.8\textwidth]{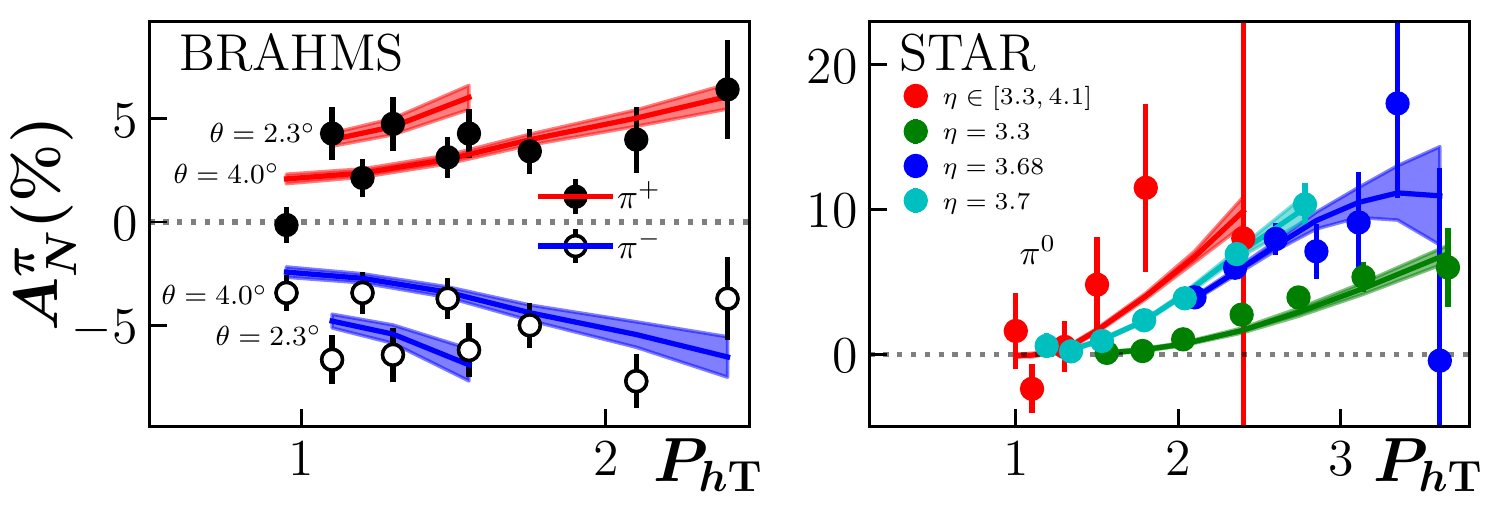}
\caption{Results for the fit in Ref.~\cite{Cammarota:2020qcw} to the $A_N$ data in \cite{Lee:2007zzh,Adams:2003fx,Abelev:2008af,Adamczyk:2012xd}.}
\label{fig:AN_pp_JAM20}
\end{figure}
The experimental measurements of $A_N$ span several decades~\cite{Bunce:1976yb,Klem:1976ui,Adams:1991rw,Krueger:1998hz,Allgower:2002qi,Adams:2003fx,Adler:2005in,Lee:2007zzh,Abelev:2008af,Arsene:2008aa,Adamczyk:2012qj,Adamczyk:2012xd,Bland:2013pkt,Adare:2013ekj,Adare:2014qzo,Acharya:2020opv,Adam:2020edg} and show a characteristic rise at large $x_F$, as one sees for the $\pi^0$ production data in Fig.~\ref{fig:AN_pp}. 
Recent phenomenology  found that the fragmentation piece is dominant~\cite{Kanazawa:2014dca,Gamberg:2017gle,Cammarota:2020qcw}.  The results from the most recent work in Ref.~\cite{Cammarota:2020qcw} are presented in Fig.~\ref{fig:AN_pp_JAM20}, which included only $A_N$ data where $P_{hT}>1\,{\rm GeV}$.  The analysis exploited the relationships between moments of TMDs and twist-3 functions along with a Gaussian ansatz for the TMD transversity, Sivers, and Collins functions, see Ref.~\cite{Cammarota:2020qcw} for details. 
A Monte Carlo framework was used to reliably sample the Bayesian posterior distribution for the parameters.
Such an approach allows the relevant regions in parameter space to be determined, and gives state-of-the-art uncertainty quantification, for the hadronic functions that best describe the data.\index{$A_N$ in $pp$ collisions|(}

\subsection{Boer-Mulders Effect} 
\label{sec:phenomelology-Boer-Mulders-and-beyond}

\index{Boer-Mulders function $h_1^\perp$} \index{Boer-Mulders effect|(}The Boer-Mulders function $h_1^\perp$~\cite{Boer:1997nt}, describing the distribution of transversely polarized quarks in an unpolarized target, 
can be considered the counterpart of the Sivers function $f_{1T}^\perp$
discussed in \sec{Sivers_SIDIS}.
Both functions are T-odd and therefore vanish if the gauge-link is not taken into account in their field-theoretic definition.
In other words, their existence requires initial and/or final state interactions between the active partons of a process and the target remnants.
Both TMDs change sign when going from SIDIS to the Drell-Yan process~\cite{Collins:2002kn}, see~\sec{Sivers_SIDIS}.
Since the Boer-Mulders function is chiral-odd, it is generally harder to measure than the Sivers function, even though no target polarization is required.

Let us first discuss the case of the Boer-Mulders function
in the Drell-Yan process.
It was argued that $h_1^\perp$ could be essential for a full understanding of the data for the angular distribution of the unpolarized Drell-Yan process~\cite{Boer:1999mm}. 
To be more specific, the structure of the Drell-Yan cross section is given by (see~\cite{Boer:2006eq} and references therein)
\begin{equation} \label{e:dy_cs}
\frac{1}{\sigma_{DY}} \frac{d \sigma_{DY}}{d \Omega} = \frac{3}{4
\pi} \frac{1}{\lambda + 3} \Big( 1 + \lambda \cos^2 \theta + \mu
\sin 2 \theta \cos \phi + \frac{\nu}{2} \sin^2 \theta \cos 2\phi
\Big) \,,
\end{equation}
 the angles $\theta$ and $\phi$ characterizing the lepton-pair orientation in a dilepton rest frame like the Collins-Soper frame~\cite{Collins:1977iv}. 
In comparison to Eq.~(\ref{eq:sigma_polarized_DY}), in Eq.~(\ref{e:dy_cs}) all variables but the Collins-Soper angles have been integrated over, while power corrections (for small transverse momenta of the di-lepton pair) have been kept.
Of particular interest in this context is the so-called Lam-Tung relation between the coefficients $\lambda$ and $\nu$~\cite{Lam:1978pu, Lam:1980uc, Collins:1978yt},
\begin{equation} \label{e:lamtung}
\lambda + 2 \nu = 1 \,,
\end{equation}
which is exact at ${\cal O}(\alpha_s)$ in the standard collinear pQCD framework. 
Even at ${\cal O}(\alpha_s^2)$ the numerical violation of~(\ref{e:lamtung}) was originally reported to be small~\cite{Mirkes:1994dp}. 
This was also studied at NNLO (i.e.~$\cO(\as^3)$) in \cite{Gauld:2017tww}. Measurements of the angular coefficients are now available over a wide kinematical range, from fixed-target energies~\cite{Falciano:1986wk, Guanziroli:1987rp, Conway:1989fs, Zhu:2006gx, Zhu:2008sj} to collider Tevatron~\cite{Aaltonen:2011nr} and LHC~\cite{Khachatryan:2015paa} kinematics.
Various beams and targets have been used in the fixed-target regime --- data exist for pion beams scattering off tungsten targets~\cite{Falciano:1986wk, Guanziroli:1987rp, Conway:1989fs}, as well as $pp$ and $pd$ collisions~\cite{Zhu:2006gx, Zhu:2008sj}.
Some of the mentioned data sets show a clear violation of the Lam-Tung relation where, in particular, a large $\cos2\phi$ term was observed. 
Different explanations of this experimental result were then put forward where the most popular one is based on intrinsic transverse motion of partons leading to the Boer-Mulders effect~\cite{Boer:1999mm}. 
The product of two Boer-Mulders functions --- one for each initial-state hadron --- generates a $\cos 2\phi$ term at leading order in $1/Q$~\cite{Boer:1999mm}. 
First extractions of the Boer-Mulders function, based on fixed-target data from Fermilab~\cite{Zhu:2006gx, Zhu:2008sj} were then reported in Refs.~\cite{Zhang:2008nu, Lu:2009ip}.
In Ref.~\cite{Lambertsen:2016wgj} the collinear pQCD calculation of the angular dependence was revisited, and it was argued that the data could actually be largely explained in this framework, including the results from the fixed-target experiments.
Extractions of the Boer-Mulders function should take those results into account.
A geometric picture that has been invoked to explain the angular dependence of the unpolarized Drell-Yan process~\cite{Peng:2015spa} is in qualitative agreement with the finding in Ref.~\cite{Lambertsen:2016wgj}.
More work on this geometric approach and higher-order pQCD calculations, extended to the production of weak gauge bosons, where the cross section has a richer angular dependence, can be found in Refs.~\cite{Chang:2017kuv, Peng:2018tty, Chang:2018pvk, Lyu:2020nul}.

The Boer-Mulders function can also be studied in SIDIS where it appears in combination with the chiral-odd Collins function, giving rise to a $\cos 2 \phi_h$-modulation of the unpolarized cross section~\cite{Boer:1997nt}. 
The relevant structure function takes the generic form \eqref{eq:structure-functions-twist-2}
\begin{equation} \label{e:sf_cos2phi}
F_{UU}^{\cos 2\phi_h} =  {\cal C}\left[\;\frac{2\, \bigl(\bfhp\cdot\bfpperp^{ }\bigr)\,
			\bigl(\bfhp\cdot\bfkperp^{ }\bigr)-\bfpperp^{ }\cdot\bfkperp^{ }}{zM_N\mh} h_1^{\perp q} H_1^{\perp q}\right] +
       \frac{4M_N^2}{Q^2} \; {\cal C}\left[\;\frac{2\,(\bfhp\cdot\bfkperp^{ })^2-\bfkperp^2}{2M_N^2}\,f_{1}\,D_{1}\;
	\right]
 + \ldots  \,
\end{equation}
The second term on the r.h.s~of Eq.~(\ref{e:sf_cos2phi}) is the so-called Cahn effect~\cite{Cahn:1978se, Cahn:1989yf}, 
\index{Cahn effect} which is also caused by intrinsic transverse parton motion, but related to unpolarized TMDs. 
The Cahn effect is a kinematic twist-4 contribution and as such suppressed by a factor $1/Q^2$ relative to the first term. 
On the other hand, since $f_1^q\otimes D_1^q$ is clearly larger than $|h_1^{\perp q}\otimes H_1^{\perp q}|$, 
suppressing the Cahn effect requires very large $Q^2$. 
It should also be noted that the Cahn effect does not represent the entire twist-4 term of a QCD analysis, even though numerically it is most likely the dominant contribution.
The SIDIS structure function $F_{UU}^{\cos 2\phi_h}$ has already been measured in Hall B and Hall A at Jefferson Lab~\cite{Osipenko:2008aa, Yan:2016ods}, at DESY by the HERMES Collaboration~\cite{Airapetian:2012yg}, and at CERN by the COMPASS Collaboration~\cite{Adolph:2014pwc, Moretti:2020uee}.
A first extraction of the Boer-Mulders function based on SIDIS data was reported in Ref.~\cite{Barone:2009hw}.
The results were also used to obtain information about the Boer-Mulders function for antiquarks from the early fixed-target measurements of the angular distribution of the Drell-Yan cross section~\cite{Barone:2010gk}.
More recent attempts to pin down the Boer-Mulders function for the nucleon from SIDIS were presented in Refs.~\cite{Barone:2015ksa, Christova:2017zxa, Christova:2020ahe}.
\begin{figure}[t!]
\centering
\includegraphics[width=0.62\textwidth]{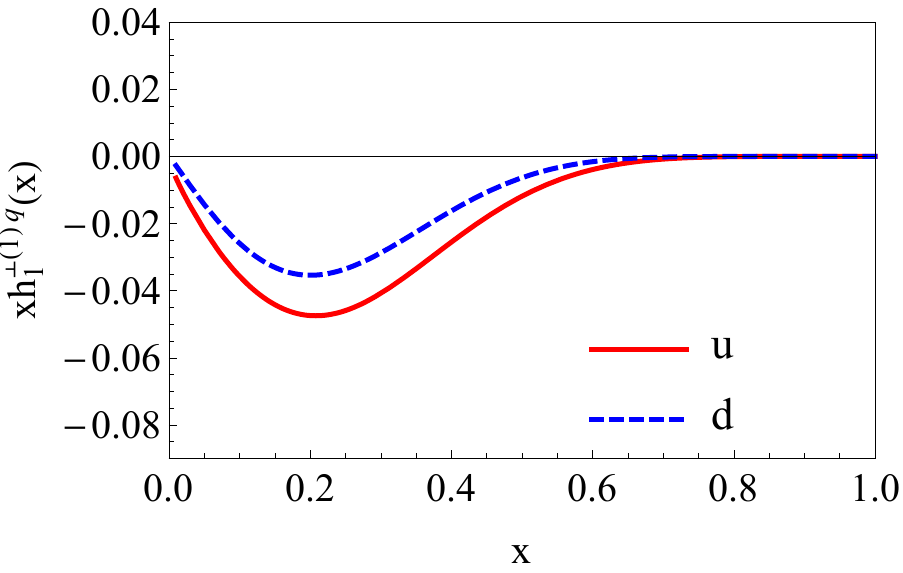}
\caption{\label{basis}
	The first moment of Boer-Mulders functions $
	h_1^{\perp a}$ Ref.~\cite{Barone:2009hw}}
\end{figure}
For the currently available SIDIS data, the Cahn effect is indeed quite large and can even dominate the $\cos 2\phi_h$ term of the cross section.
Similar to the above discussion for the Drell-Yan process, a thorough higher-order collinear pQCD calculation for SIDIS is mandatory in order to be able to draw definite quantitative conclusions about the Boer Mulders function from this process.
Also, new data would be very helpful, where the EIC with its large kinematical coverage could play a crucial role.
The pion and nucleon Boer-Mulders functions were studied intensively in lattice QCD and models, see Chap.~\ref{sec:lattice} and \ref{sec:models}.
Fig.~\ref{basis} shows the first moments of Boer-Mulders functions for up and down quarks extracted from the experimental data in Ref.~\cite{Barone:2009hw}. \index{Boer-Mulders effect|)}

\FloatBarrier
\subsection[  
Worm-gear  \texorpdfstring{$g_{1T}^\perp$}{g1tperp} and \texorpdfstring{$h_{1L}^\perp$}{h1Lperp} and Pretzelosity  \texorpdfstring{$h_{1T}^{\perp}$}{h1Tpeerp} TMD PDFs]{\boldmath Worm-gear  $g_{1T}^\perp$ and $h_{1L}^\perp$ and Pretzelosity  $h_{1T}^{\perp}$ TMD PDFs}
\FloatBarrier
\label{sec:phenomelology-other}
\index{pretzelosity $h_{1T}^{\perp}$}

In this section, we review several other TMDs for which presently the information from experiment is still sparse, yet they are as important as the TMDs discussed above in this chapter for obtaining a complete understanding of the transverse-momentum-dependent nucleon structure and fragmentation process.
Specifically, we consider the worm-gear functions $g_{1T}^\perp$ and $h_{1L}^\perp$, and the pretzelosity TMD $h_{1T}^\perp$.

\subsubsection*{Worm-gear TMD PDFs $g_{1T}^\perp$ and $h_{1L}^\perp$}

\begin{figure}[t!]
  \centering
\includegraphics[width=0.9\columnwidth]{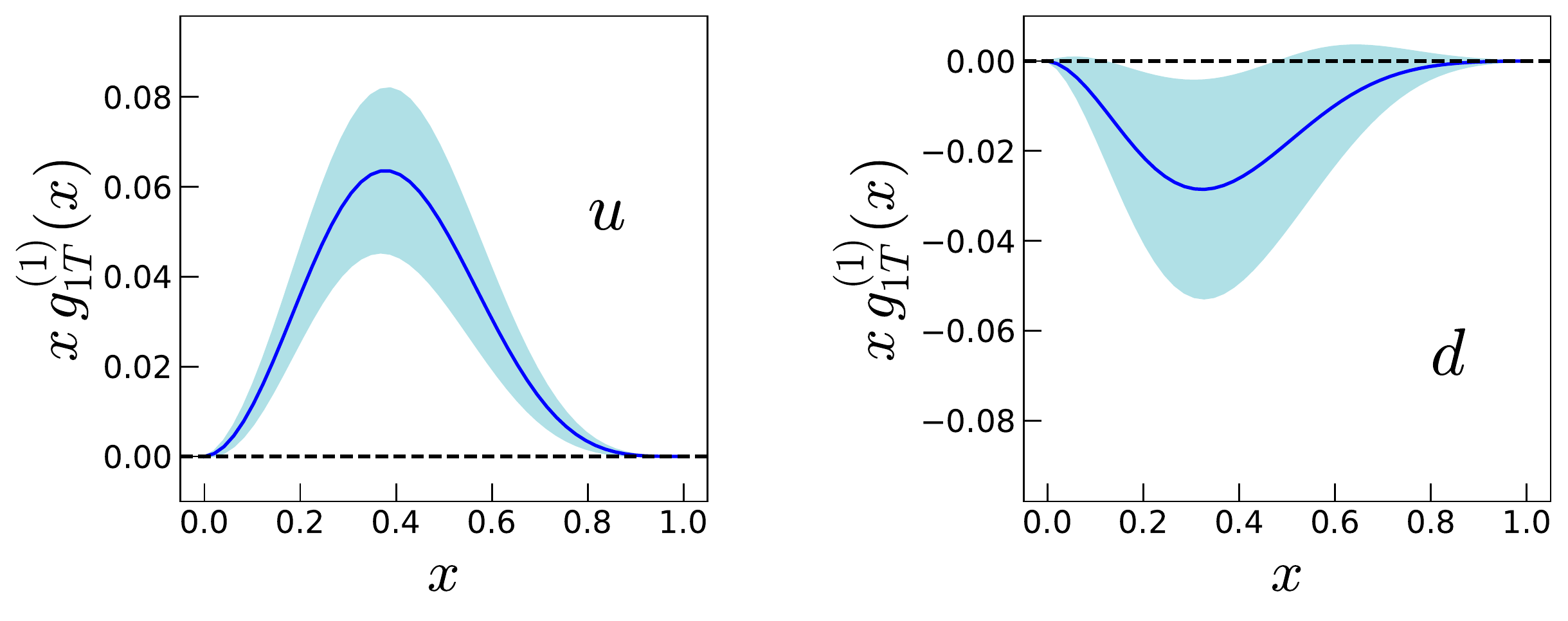}
  \caption{Results of global fit for $xg_{1T}^{\perp(1)}(x)$ at $Q^{2}$ = 4 ${\rm GeV}^2$ for up quarks (left) and down quarks (right).
  The figures are from Ref.~\cite{Bhattacharya:2021twu}.}
  \label{fig:g1Tperp_fit}
\end{figure}

\index{worm-gear functions!phenomenology|(} Let us begin with the worm-gear function $g_{1T}^\perp$, which describes the distribution of longitudinally polarized quarks in a transversely polarized nucleon. 
Very recently, this function was extracted for the first time from a global analysis of SIDIS data using Monte Carlo techniques~\cite{Bhattacharya:2021twu}.
To this end, the asymmetry
(see Eq.~\eqref{eq:structure-functions-twist-2}, \eqref{eq:structure-functions-sidis-bspace}) 
\begin{equation}
A_{LT}^{\cos (\phi_h - \phi_S)} \equiv \frac{F_{LT}^{\cos (\phi_h - \phi_S)}}{F_{UU,T}} =
M \; \frac{{\cal B}\left[\tilde g_{1T}^{\perp (1) }\,\tilde D_{1}^{(0)}\right]}{\mathcal{B}\left[\tilde f_{1}^{(0)} \tilde D_{1}^{(0)}\right]}\; .
\label{eq:g1Tperp}
\end{equation} 
was considered for which data from COMPASS~\cite{COMPASS:2016led, Parsamyan:2018evv}, HERMES~\cite{HERMES:2020ifk}, and Jefferson Lab~\cite{JeffersonLabHallA:2011vwy} exist. 
A Gaussian ansatz was made for $g_{1T}^\perp(x,k_T)$, and the $k_T$-moment
\begin{equation}
g^{\perp(1)}_{1T}(x) = \int d^{2} k_T  \, \dfrac{k^{2}_T}{2M^{2}} \, g_{1T}^\perp(x, k_T)
\label{e:g1Tperp_moment}
\end{equation}
was fitted for the up quark and down quark, with the results of the fit shown in Fig.~\ref{fig:g1Tperp_fit}.

\index{Wandzura-Wilczek (type) approximation|(}
Estimates for the worm-gear TMDPDFs can be obtained using the so-called Wandzura-Wilczek-type (WW-type)
approximation which is based on the QCD equations of motion and consistently neglecting quark-gluon and current quark mass terms \cite{Tangerman:1994bb,Kotzinian:1995cz,Mulders:1995dh,Kotzinian:1997wt,Kotzinian:2006dw,Avakian:2007mv,Metz:2008ib,Teckentrup:2009tk}.
This allows one to approximate 
$g_{1T}^{\perp(1)}(x)$ (and $h_{1L}^{\perp(1)}(x)$) in terms of integral relations involving twist-2 PDFs.
In the case of $g_{1T}^\perp$ one has
\begin{equation}
g^{\perp(1)}_{1T}(x) = \stackrel{\text{WW-type}}{\approx} x \int^{1}_{x} \dfrac{dy}{y} g_{1}(y) \,, \label{e:WWtype}
\end{equation}
where $g_1(x)$ is the helicity PDF.
The WW-type relations become exact in the parton model and
are discussed in Sec.~\ref{Sec:parton-model}. 
The fit results of Ref.~\cite{Bhattacharya:2021twu} are compatible with the WW-type approximation in Eq.~\eqref{e:WWtype}, which is in line with the general finding in Ref.~\cite{Bastami:2018xqd} that the 
WW-type approximation for the worm-gear functions is compatible with available data on the pertinent asymmetries regarding sign and magnitude.
The extracted results for $g_{1T}^{\perp (1)}$ in Ref.~\cite{Bhattacharya:2021twu} also agree within errors with information from lattice QCD~\cite{Yoon:2017qzo}.
Furthermore, it was shown that at present the data for the asymmetry $A_{LT}^{\cos (\phi_h - \phi_S)}$ can not rule out the strict large-$N_c$ approximation, according to which $g_{1T}^{\perp u} = - g_{1T}^{\perp d}$~\cite{Pobylitsa:2003ty}; see also Sec.~\ref{subsubsec:largeNc}.
More precise data are needed in order to move the phenomenology of $g_{1T}^\perp$ to the next level.

The Kotzinian-Mulders or worm-gear distribution $h_{1L}^{\perp}(x,k_T)$~\cite{Kotzinian:1994dv} describes the probability of finding a transversely polarized quark but inside a longitudinally polarized nucleon. 
Since $h_{1L}^{\perp}$ is chiral odd, it has to be coupled to another chiral-odd function to manifest its effects in semi-inclusive processes. In SIDIS, this can be achieved via a $\sin2\phi_{h}$ azimuthal asymmetry~\cite{Kotzinian:1994dv,Kotzinian:1997wt} when $h_{1L}^{\perp}$ is combined with the chiral-odd Collins function $H_1^\perp$~\cite{Collins:1992kk}, see Eqs.~\eqref{eq:structure-functions-twist-2}, \eqref{eq:structure-functions-sidis-bspace}:
\begin{equation}
A_{UL}^{\sin 2\phi_h} \equiv \frac{F_{UL}^{\sin 2\phi_h}}{F_{UU,T}} =
M_N\, M_h \; \frac{{\cal B}\left[\tilde h_{1L}^{\perp (1) }\,\tilde H_{1}^{\perp (1)}\right]}{\mathcal{B}\left[\tilde f_{1}^{(0)} \tilde D_{1}^{(0)}\right]}\; .
\label{eq:wormgearAsym}
\end{equation}  
The early work on the $\sin2\phi_{h}$ asymmetry in the longitudinally polarized SIDIS process have been performed in Refs.~\cite{HERMES:1999ryv,HERMES:2002buj,Avakian:2010ae,Lu:2011pt,
Zhu:2011zza,Boffi:2009sh,Ma:2000ip,Ma:2001ie,Bastami:2018xqd}, showing that the asymmetry is around several percent.

There are no extractions of $h_{1L}^{\perp}$ from the experimental data, however a few analyses used the WW-approximation in which one can write $h_{1L}^{\perp (1)}(x)$  in terms of an integral relation involving the transversity distribution $h_1(x)$ as follows
\begin{align}
h_{1L}^{\perp (1)}(x)\overset{WW-type}{\approx} -x^2\int_{x}^{1}\frac{dy}{y^2}h_1^a(y).
\label{WW-type}
\end{align}
In particular, Ref.~\cite{Li:2021mmi} used NLL TMD factorization formalism to study the $\sin2\phi_h$ asymmetry in SIDIS process at the kinematical configuration of HERMES, CLAS and CLAS12. Good agreement with the existing data was found.
\begin{figure}[t!]
  \centering
  \includegraphics[width=0.45\columnwidth]{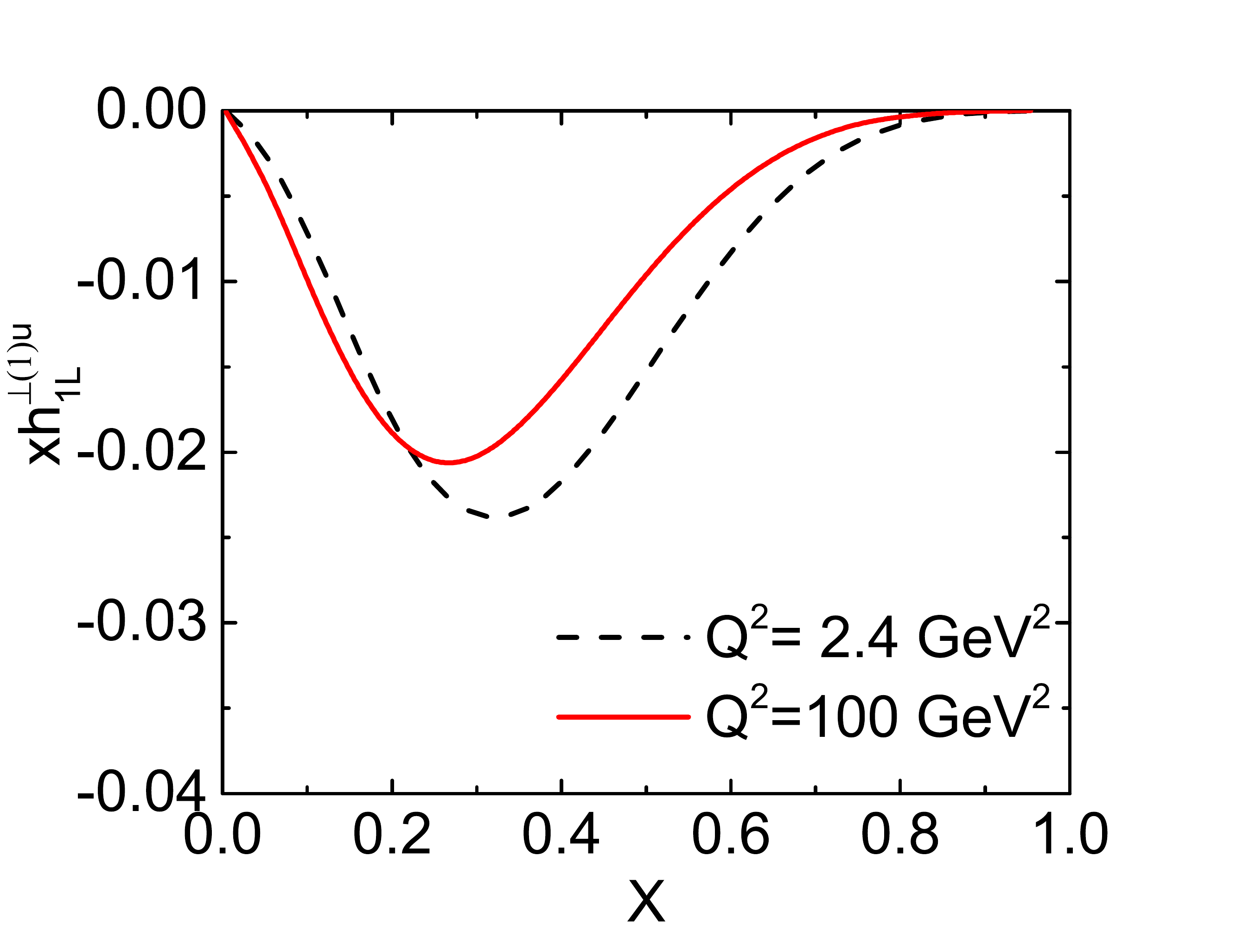}
  \includegraphics[width=0.45\columnwidth]{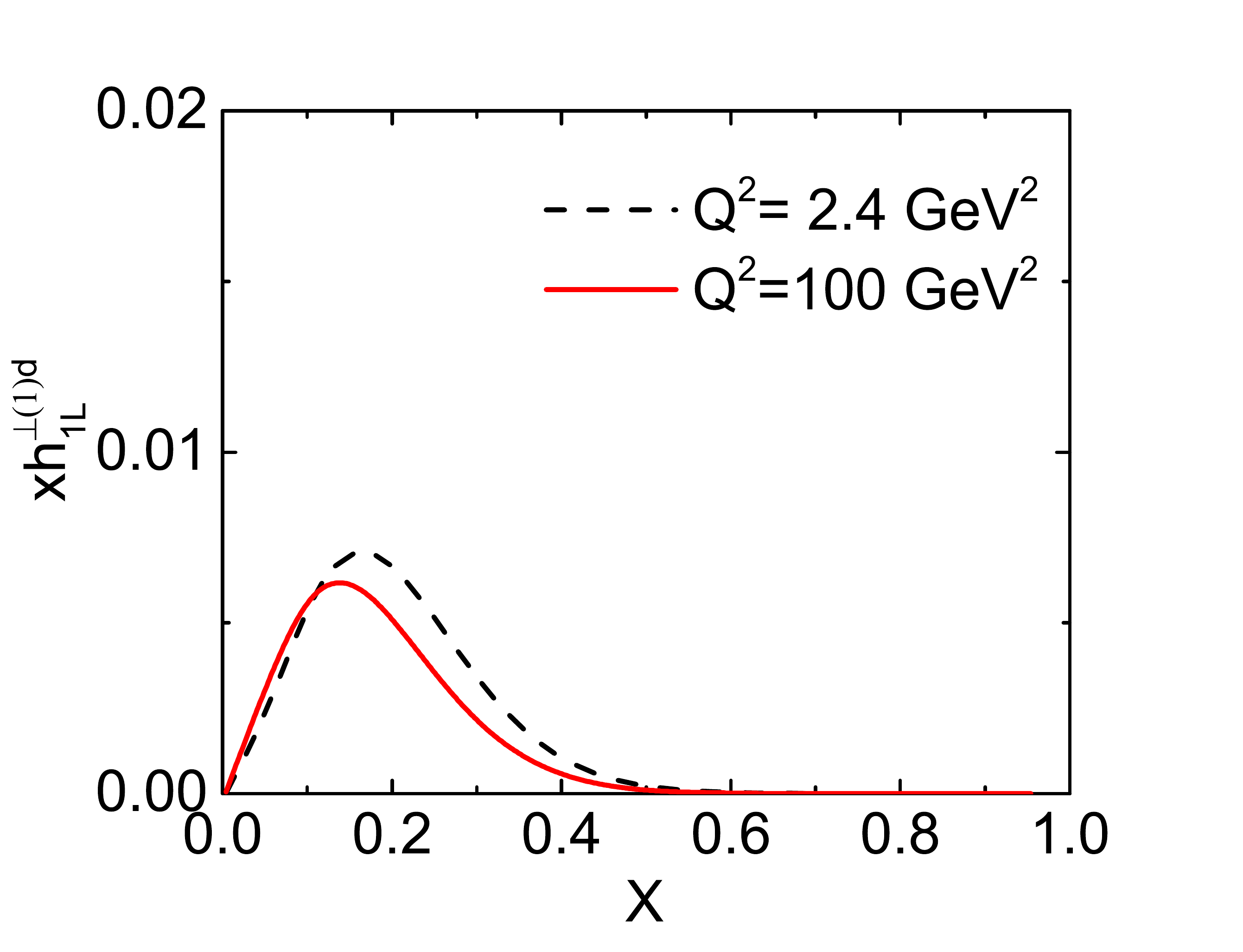}
  \caption{Left panel: $xh_{1L}^{\perp (1)}(x,Q^2)$ of the proton 
  for up quark at $Q^2=2.4\ \textrm{GeV}^2$ and $Q^2=100\ \textrm{GeV}^2$. Right panel: similar to the left panel, but for the down quark. Plots from Ref.~\cite{Li:2021mmi}.
  }
  \label{fig1:xh1l}
\end{figure}
In Fig.~\ref{fig1:xh1l}, we plot the $x h_{1L}^{\perp (1)}(x,Q^2)$  from Ref.~\cite{Li:2021mmi}  for up and down quarks at the initial scale $Q^2$=2.4 GeV$^2$ as well as the evolved scale $Q^2 =100$ GeV$^2$.
The plots show that the $h_{1L}^{\perp(1)}(x,Q^2)$ for the up quark is larger than the one for the down quark in size, and with the opposite sign.\index{worm-gear functions!phenomenology|)}
\index{Wandzura-Wilczek (type) approximation|)}

\subsubsection*{Pretzelosity TMD PDF, $h_{1T}^{\perp}$}
\begin{figure*}[t!]
 \begin{tabular}{c@{\hspace*{-1cm}}c}
 \includegraphics[width=6cm,angle=-90]{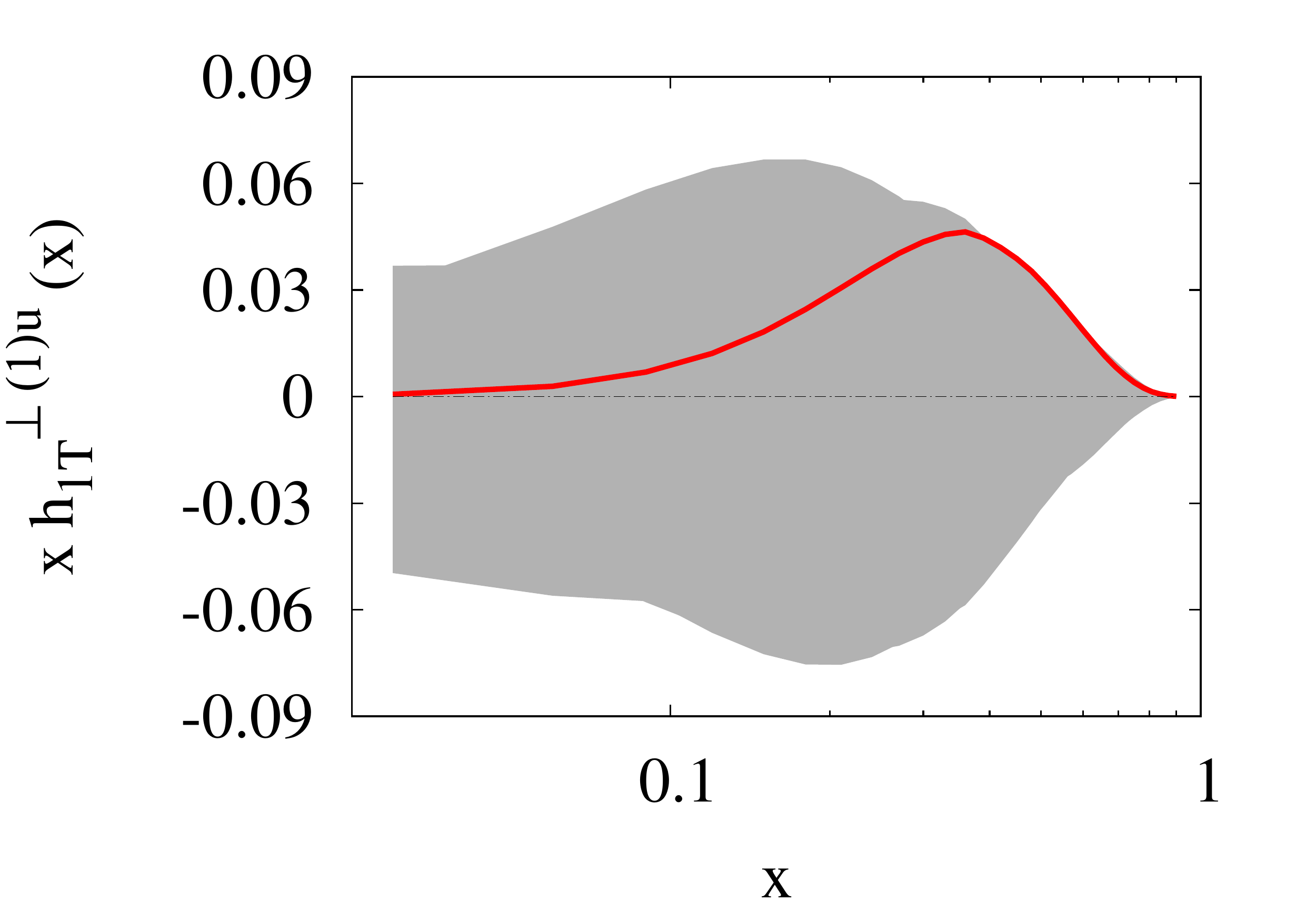}
  &
  \hskip 1cm 
 \includegraphics[width=6cm,angle=-90]{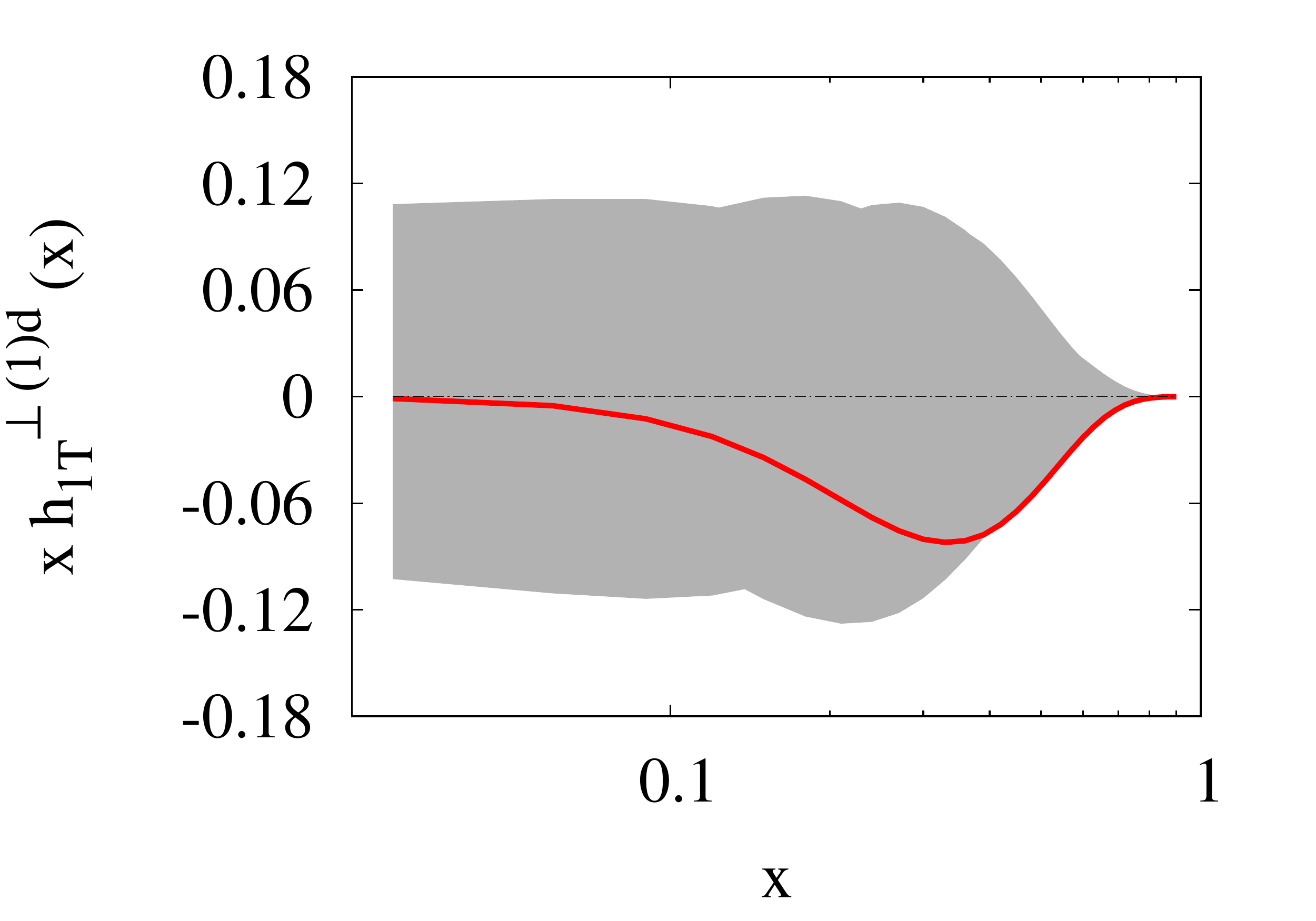}
  \\
  (a) & (b)
  \end{tabular}     
\caption{First moment of the pretzelosity distribution for up (a) and down (b) quarks at $Q^2=2.4$ GeV$^2$. The solid line corresponds to
the best fit and the shadowed region corresponds to the error corridor.}
\label{fig:pretzelosity}
\end{figure*}

The pretzelosity distribution function   
$h_{1T}^{\perp}$~\cite{Lefky:2014eia} describes transversely polarized quarks 
inside a transversely polarized nucleon. The measured asymmetry in SIDIS contains the convolution of pretzelosity $h_{1T}^\perp$ and 
the Collins FF $H_1^\perp$, Eqs.~\eqref{eq:structure-functions-twist-2}, \eqref{eq:structure-functions-sidis-bspace}:
\begin{equation}
A_{UT}^{\sin(3\phi_h -\phi_S)} \equiv \frac{F_{UT}^{\sin(3\phi_h -\phi_S)}}{F_{UU,T}} =
\frac{M_N^2\, m_h}{4} \; \frac{{\cal B}\left[\tilde  h_{1T}^{\perp (2)}\, \tilde H_1^{\perp (1)}\right]}{\mathcal{B}\left[\tilde f_{1}^{(0)} \tilde D_{1}^{(0)}\right]}\; .
\label{eq:pretzelosity}
\end{equation}  
Notice that the knowledge of the Collins FF is needed for the extraction of pretzelosity. 
$h_{1T}^{\perp}$ was extracted in parton model approximation in Ref.~\cite{Lefky:2014eia} and the results are shown in Fig.~\ref{fig:pretzelosity}.
\index{parton model|)}
Weighted SIDIS asymmetries and predictions for the EIC can be found in Ref~\cite{xue:2021svd}.
One interesting aspect of pretzelosity is that it is the only 
leading TMD PDF which is related to orbital angular momentum of quarks, 
even though this relation only holds in quark models,
see Sec.~\ref{Sec:relations-in-models}.

\subsection{Observables for Gluon TMDs} \label{s:gluonTMDs}
\label{sec:gluonTMD_obs}

This section is devoted to presenting an overview on the work that has been done on TMD analyses of experimental observables that may provide information on gluon TMDs. We review the present status on where we stand with respect to gluon TMD phenomenology (see also a fairly recent review \cite{Angeles-Martinez:2015sea}). At the same time we want to emphasize that the theoretical and experimental research on gluon TMDs is an ongoing endeavor. At present, the TMD theory for gluon distributions is less developed compared to their quark counterparts. While precise all-order definitions of gluon TMDs were discussed in the literature (see, e.g., \cite{Echevarria:2015uaa}) rigorous TMD factorization of physical observables involving gluon TMDs has been suggested only for a small number of very simple final states in proton collisions, such as Higgs boson production \cite{Echevarria:2015uaa} or $\eta_{c,b}$-production \cite{Echevarria:2019ynx}. Nonetheless, several spin-independent and spin-dependent observables sensitive to gluon TMDs have been investigated at tree-level (LO) or to one-loop accuracy (NLO) {\it under the working assumption} that rigorous TMD factorization for those observables would hold and be proved in the future. Possible observables sensitive to gluon TMDs have been identified both for lepton-nucleon and proton-proton collisions. Experimentally, a future EIC will be suitable to perform measurements on gluon TMD observables in lepton-nucleon collisions. A limited amount of data for gluon TMD observables in proton collisions has been generated at the LHC. Thus, in principle, one may be able to learn about (unpolarized) gluon TMDs from existing LHC data $\--$ even if this information is rather limited at the moment due to a small number of LHC data points. In future, the knowledge of gluon TMDs can be improved through precise EIC measurements.

Several (color-singlet) final states have been identified in proton collisions that may provide insight into gluon TMDs: photon pair production \cite{Balazs:2006cc,Nadolsky:2007ba,Qiu:2011ai}, Higgs boson production \cite{Boer:2011kf,Sun:2011iw,Boer:2013fca,Echevarria:2015uaa}, single quarkonium production \cite{Boer:2012bt,Ma:2012hh,Ma:2014oha}, associated quarkonium-photon pair production \cite{Dunnen:2014eta}, associated quarkonium-dilepton pair production \cite{Lansberg:2017tlc} and quarkonium pair production \cite{Lansberg:2017dzg,Scarpa:2019fol}. For proton collisions it is important to keep in mind that TMD factorization breaking effects $\--$ the so-called {\it color entanglement} $\--$ may occur for colored final states like jets of hadrons produced within a fragmentation process (see Refs. \cite{Collins:2007nk,Rogers:2010dm} and the following section). Therefore, if TMD factorization is considered for final states in proton collisions that involve quarkonia, those quarkonia need to be produced as color singlet states rather than color octet states \cite{Yuan:2008vn}. We again emphasize that TMD factorization has not been rigorously proven for the associated quarkonium and quarkonium pair final states listed above, even for color singlet quarkonia. On the other hand TMD factorization has not been shown to fail either for color singlet quarkonium final states (except for single $\chi$ production \cite{Ma:2014oha}), and one may consider it as a {\it working hypothesis}.
The correct definition of TMD parton distributions has been discussed in detail in Secs.~\ref{sec:tmdpdfs_new},~\ref{sec:tmd_defs}. Here, we focus on the various gluon TMDs that are relevant for TMD observables. This concerns in particular the TMD distribution of unpolarized gluons in the unpolarized nucleon, $f_1^g(x,{\bm k}_T^2)$, the TMD distribution of linearly polarized gluons in the unpolarized nucleon, $h_1^{\perp g}(x,{\bm k}_T^2)$, and the TMD distribution of unpolarized gluons in a transversely polarized nucleon, $f_{1T}^{\perp g}(x,{\bm k}_T^2)$ $\--$ the gluonic counterpart to the quark Sivers function. Those functions were introduced in Sec.~\ref{sec:gluonTMD_def}.

\subsubsection{Gluon TMDs from proton-proton collisions}

\begin{figure}[t!]
\centering
\includegraphics[width=0.35\textwidth]{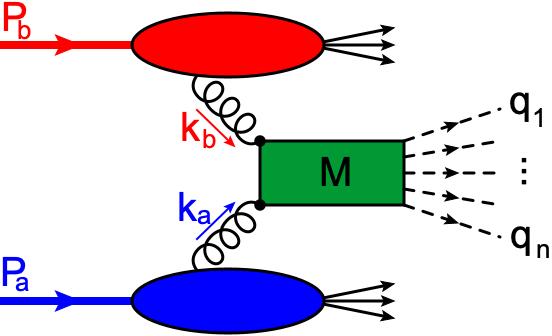}
\caption{General setup of a hard process in proton collisions that is sensitive to gluon TMDs.
}
\label{fig:GeneralGluon}
\end{figure}
The following subsection is devoted to observables in proton-proton collisions that are sensitive to gluon TMDs. The general formalism of how these observables emerge theoretically has been described in Ref. \cite{Lansberg:2017tlc}. Let us consider a general final state of $n$ particles carrying momenta $q_1$, ... , $q_n$ that has been produced in a collision of two protons of momenta $P_a$ and $P_b$, see Fig. \ref{fig:GeneralGluon}. The overall 4-momentum of the final state is labelled as $q^\mu=\sum_{i=1}^n q_i^\mu$. We assume that the production mechanism of the final state is via gluon fusion in a hard scattering amplitude $\mathcal{M}$. The application of TMD factorization to this general process demands (at least) two necessary conditions:
\begin{itemize}
    \item The transverse momentum of the final state, $|{\bm q}_T|$ is much smaller than its invariant mass $Q=\sqrt{q^2}$, i.e., $|{\bm q}_T|\ll Q$. Ideally, $|{\bm q}_T|$ is of the order of some hadronic scale in order to maximize the effect of the intrinsic transverse gluonic motion. At the same time, this condition forbids the production of unobserved partons in the hard scattering amplitude $\mathcal{M}$. In other words, all particles $1$,...,$n$ in the final state must be detected. Otherwise, the final state may recoil against an unobserved parton, such as a radiated gluon. This situation would be described within collinear factorization.
    \item The particles in the final state need to be color singlets, such as leptons, photons, Z-bosons, Higgs bosons, but also quarkonium states like $\eta$, $J/\psi$ or $\Upsilon$ in a color singlet mode. Final state events that are sensitive to color, such as jets, fragmenting hadrons, quarkonia in color octet modes, etc., should be treated with great care as the aforementioned color entanglement may potentially spoil TMD factorization.
\end{itemize}
If these conditions are met one may analyze a general process as in Fig. \ref{fig:GeneralGluon}. According to Ref. \cite{Lansberg:2017tlc} the fully differential cross section for a general process then acquires the following form,
\begin{eqnarray}
    \frac{\mathrm{d}\sigma (|{\bm q}_T|\ll Q)}{\mathrm{dPS}_n} & = & \frac{(2\pi)^4}{4x_ax_bs^2}\,\Big(\hat{F}_1\,\mathcal{C}[f_1^g\,f_1^g]+\hat{F_2}\,\mathcal{C}[w_2\,h_1^{\perp g}\,h_1^{\perp g}]\nonumber\\
    &&\hspace{1.5cm}+\hat{F}_{3a}\,\mathcal{C}[w_{3a}\,h_1^{\perp g}\,f_1^g]+\hat{F}_{3b}\,\mathcal{C}[w_{3b}\,f_1^g\,h_1^{\perp g}]\nonumber\\
    &&\hspace{1.5cm}+\hat{F}_4\,\mathcal{C}[w_4\,h_1^{\perp g}\,h_1^{\perp g}]\Big)+\mathcal{O}(M/Q)\,.\label{eq:GluonTMDMaster}
\end{eqnarray}
In this formula, $s=(P_a+P_b)^2$ denotes the center-of-mass energy. In addition, there are two kinematical scaling variables $x_{a/b}= q\cdot P_{b/a}/P_a\cdot P_b$. 
The $3n$-dimensional phase space element for an $n$-particle final state is denoted by $\mathrm{dPS}_n$. Notice that we consider unpolarized collisions. The structures like, 
e.g., $h_1^{\perp g}\,f_1^g$ are possible because both TMDs 
are chiral-even, since the gluon Boer-Mulders function is not
chiral-odd as in quark case. 
The TMD formula (\ref{eq:GluonTMDMaster}) contains four different convolution integrals of the gluon TMDPDFs $f_1^g$ and $h_1^{\perp g}$, with the general form of the TMD convolution,
\begin{equation}
    \mathcal{C}[w\,f\,g]=\int d^2{\bm k}_{aT}\int d^2{\bm k}_{bT}\,\delta^{(2)}({\bm k}_{aT}+{\bm k}_{bT}-{\bm q}_T)\,w({\bm k}_{aT},{\bm k}_{bT},{\bm q}_T)\,f(x_a,{\bm k}_{aT}^2)\,g(x_b,{\bm k}_{bT}^2)\,.\label{eq:Convg}
\end{equation}
The TMD weights $w_i$ in (\ref{eq:GluonTMDMaster}) read (with $M_N$ being the nucleon mass),
\begin{eqnarray}
w_2 & = & \frac{2({\bm k}_{aT}\cdot {\bm k}_{bT})^2-{\bm k}_{aT}^2{\bm k}_{bT}^2}{4M_N^4}\,,\nonumber\\
w_{3a/b} & = & \frac{{\bm k}_{a/bT}^2{\bm q}_T^2-2({\bm q}_T\cdot{\bm k}_{a/bT})^2}{2M_N^2{\bm q}_T^2}\,,\nonumber\\
w_4 & = & 2\left[\frac{{\bm k}_{aT}\cdot {\bm k}_{bT}}{2M_N^2} - \frac{({\bm k}_{aT}\cdot {\bm q}_T)({\bm k}_{bT}\cdot{\bm q}_T)}{M_N^2{\bm q}_T^2} \right]^2-\frac{{\bm k}_{aT}^2{\bm k}_{bT}^2}{4M_N^4}\,.\label{eq:weightsG}
\end{eqnarray}
The beauty of the "master formula" (\ref{eq:GluonTMDMaster}) is that the TMD convolution integrals that we are ultimately after are completely independent of the final state, and universal  up to the caveats discussed earlier. This, in principle, allows for a combined analysis of gluon induced processes with various final states, each of which have their own peculiarities, advantages and disadvantages.

The factors $\hat{F}_i$ in (\ref{eq:GluonTMDMaster}) are partonic functions that can be calculated in pQCD. In fact, they can be written in terms of helicity amplitudes $\mathcal{M}_{\lambda_a \lambda_b;I}(k_{a/b}=x_{a/b}P_{a/b})$ where $k_{a/b},\lambda_{a/b}$ are the momenta and helicities of the two fusing gluons, respectively (cf. Fig. \ref{fig:GeneralGluon}). Those helicity amplitudes have been well studied in collinear factorization and are often known to higher orders in perturbation theory for a specific final state. According to Ref. \cite{Lansberg:2017tlc} the factors $\hat{F}_i$ acquire the following form,
\begin{eqnarray}
\hat{F}_1 & = & \frac{1}{(N_c^2-1)^2}\sum_{\lambda_a,\lambda_b=\pm 1}\sum_I\mathcal{M}_{\lambda_a,\lambda_b;I}^{ab}(k_{a/b}=x_{a/b}P_{a/b})\,\left(\mathcal{M}_{\lambda_a,\lambda_b;I}^{ab}\right)^\ast (k_{a/b}=x_{a/b}P_{a/b})\,,\nonumber\\
\hat{F}_2 & = & \frac{1}{(N_c^2-1)^2}\sum_{\lambda=\pm 1}\sum_I\mathcal{M}_{\lambda,\lambda;I}^{ab}(k_{a/b}=x_{a/b}P_{a/b})\,\left(\mathcal{M}_{-\lambda,-\lambda;I}^{ab}\right)^\ast (k_{a/b}=x_{a/b}P_{a/b})\,,\nonumber\\
\hat{F}_{3a} & = & \frac{1}{(N_c^2-1)^2}\sum_{\lambda_a,\lambda_b=\pm 1}\sum_I\mathcal{M}_{\lambda_a,\lambda_b;I}^{ab}(k_{a/b}=x_{a/b}P_{a/b})\,\left(\mathcal{M}_{-\lambda_a,\lambda_b;I}^{ab}\right)^\ast (k_{a/b}=x_{a/b}P_{a/b})\,,\nonumber\\
\hat{F}_{3b} & = & \frac{1}{(N_c^2-1)^2}\sum_{\lambda_a,\lambda_b=\pm 1}\sum_I\mathcal{M}_{\lambda_a,\lambda_b;I}^{ab}(k_{a/b}=x_{a/b}P_{a/b})\,\left(\mathcal{M}_{\lambda_a,-\lambda_b;I}^{ab}\right)^\ast (k_{a/b}=x_{a/b}P_{a/b})\,,\nonumber\\
\hat{F}_4 & = & \frac{1}{(N_c^2-1)^2}\sum_{\lambda=\pm 1}\sum_I\mathcal{M}_{\lambda,-\lambda;I}^{ab}(k_{a/b}=x_{a/b}P_{a/b})\,\left(\mathcal{M}_{-\lambda,\lambda;I}^{ab}\right)^\ast (k_{a/b}=x_{a/b}P_{a/b})\,.\label{eq:Fs}
\end{eqnarray}
In these equations, $a,b$ are the color indices of the two fusing gluons, and $N_c=3$ the number of colors. The index $I$ combines all of the quantum numbers of the final state that are summed over.

Equation (\ref{eq:GluonTMDMaster}) becomes particularly simple for a one-particle final state. In this case the partonic factors $\hat{F}_{3a,b}$ and $\hat{F}_4$ vanish. The prefactor $\hat{F}_1$, together with the convolution $\mathcal{C}[f_1^g\,f_1^g]$, leads to the typical $q_T$-transverse momentum spectrum of the final state particle that one would also expect from collinear factorization (using CSS resummation techniques \cite{Collins:1984kg}). This $q_T$-spectrum may be modified by the factor $\hat{F}_2$, together with the convolution integral $\mathcal{C}[w_2\,h_1^{\perp g}\,h_1^{\perp g}]$ of linearly polarized gluon distributions. This modification has been discussed for the production of a Higgs boson \cite{Boer:2011kf} and for the production of color-singlet quarkonia $\eta_{c,b}$ and $\chi_{c,b;0,2}$ \cite{Boer:2012bt}. The scale evolution of the modification of the $q_T$-spectrum from linearly polarized gluons has been studied in Ref. \cite{Boer:2014tka}. It was estimated in Ref. \cite{Boer:2014tka} that the effect of linearly polarized gluons becomes negligible for large scales while it may play a non-negligible role for lower scales in quarkonium production.

\begin{figure}[t!]
\centering
\includegraphics[width=0.50\textwidth]{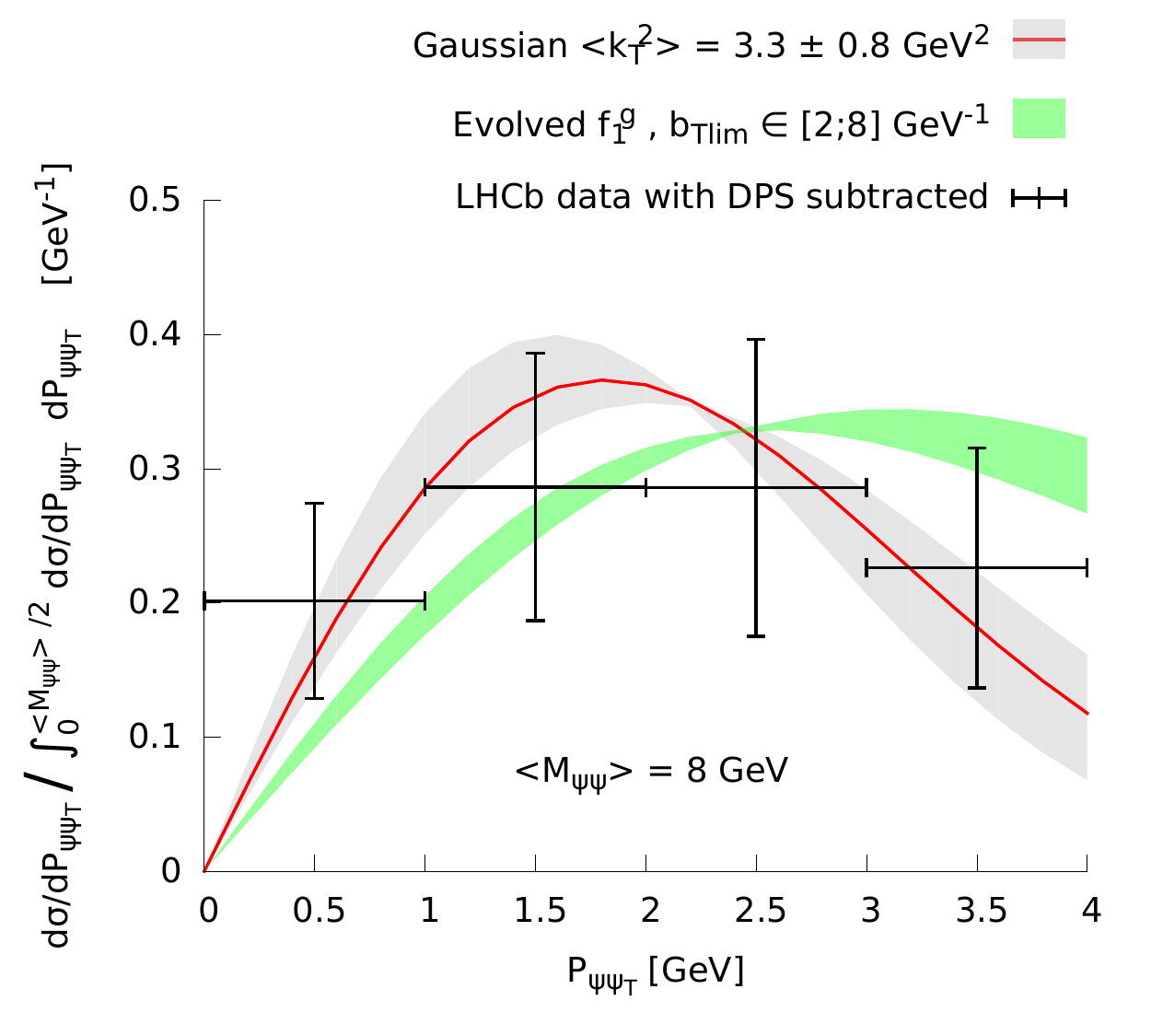}
\caption{Fit of the azimuthally independent term of the $J/\psi$-pair production cross section to LHCb data \cite{Aaij:2016bqq}. Plot is from Ref. \cite{Scarpa:2019fol}.
}
\label{fig:FitunpolG}
\end{figure}

One disadvantage of a one-particle final state, from the point of view of extracting the TMD convolution integrals from experimental data, is that its invariant mass is simply the particle's mass, $Q=M$, so that it can not be tuned. This is however possible for particle pair production. For a two-particle final state one may conveniently work out the partonic prefactors $\hat{F}_i$ in (\ref{eq:Fs}) in the Collins-Soper frame \cite{Collins:1981uk}, i.e., a specific realization of a pair center-of-mass frame where the spatial orientation of the two particles is described by the Collins-Soper angles $\theta$ and $\phi$. In particular the angle $\phi$ is of interest as it describes azimuthal modulations of the differential cross section. It turns out that the partonic factors $\hat{F}_1$ and $\hat{F}_2$ are azimuthally independent while $\hat{F}_{3a/b}$ and $\hat{F}_4$ show a $\cos(2\phi)$ and $\cos(4\phi)$ dependence \cite{Lansberg:2017tlc}, respectively\footnote{Strictly speaking, this statement is only true if the interactions entering the partonic amplitude $\mathcal{M}$ are CP-conserving. If CP symmetry is not conserved one may also find $\sin(2\phi)$ and $\sin(4\phi)$ modulations \cite{Boer:2013fca}.}. The physical explanation for this behaviour is that $\hat{F}_{3a/b}(\phi)\equiv F_3\,\cos(2\phi)$ is generated by a single gluon helicity flip whereas $\hat{F}_4(\phi)\equiv F_4\,\cos(4\phi)$ is related to a double gluon helicity flip, see Eq.~(\ref{eq:Fs}).

Several 2-particle final states have been investigated with the aim of gaining information on gluon TMDs: "background" photon pair production \cite{Qiu:2011ai}, photon pairs as a decay channel of the Higgs boson \cite{Boer:2011kf,Boer:2013fca}, quarkonium-photon pairs \cite{Dunnen:2014eta}, quarkonium-dilepton pairs \cite{Lansberg:2017tlc} and quarkonium pairs \cite{Lansberg:2017dzg,Scarpa:2019fol}. In particular the last final state, i.e., $J/\psi$ pairs, has some advantages over the others. First of all, there exist LHC data on the $q_T$-spectrum of $J/\psi$-pairs, from LHCb \cite{Aaij:2011yc,Aaij:2016bqq}, CMS \cite{Khachatryan:2014iia} and ATLAS \cite{Aaboud:2016fzt}, as well as from D0 at  FermiLab \cite{Abazov:2014qba}. In fact, in Refs. \cite{Lansberg:2017dzg,Scarpa:2019fol} LHCb data was used to fit the convolution integral $\mathcal{C}[f_1^g\,f_1^g]$ in order to extract the TMD distribution of unpolarized gluons for the first time (see Fig. \ref{fig:FitunpolG}).

Even though experimental data is presently not available on the azimuthal dependencies of the differential $J/\psi$-pair cross section, one may theoretically estimate the maximum possible size of the $\cos(2\phi)$ and $\cos(4\phi)$ modulations by assuming the saturation of positivity bounds for gluon TMDs, in particular for the distribution of linearly polarized gluons \cite{Bacchetta:1999kz}. It again turns out that the final state of $J/\psi$ or $\Upsilon$ pairs is exceptionally suitable (compared to other 2-particle final states) for the experimental exploration of the $\cos(2\phi)$ and $\cos(4\phi)$ dependence as the partonic factors $F_{3a/b}$, $F_4$ can become $\--$ in certain kinematical regions $\--$ as large as the one accompanying the unpolarized gluon TMDs, $F_1$ (cf. Eq. (\ref{eq:GluonTMDMaster})). For details we refer the reader to Refs. \cite{Lansberg:2017dzg,Scarpa:2019fol}.

In summary, various final states have been identified theoretically in proton collisions that may be utilized to learn about gluon TMDs, in particular about the gluon TMDs $f_1^g$ and $h_1^{\perp g}$. At present, the lack of experimental data on the $\cos(2\phi)$ and $\cos(4\phi)$ modulations slows down our progress on the exploration of those functions. This will change in the future after the high-luminosity upgrade of the LHC when more precise LHC data can be expected.

The study of polarized gluon TMDs like the gluon Sivers function $f_{1T}^{\perp g}$ (and others) through spin asymmetries in proton collisions requires either a polarized nuclear target or a polarized proton beam.  The latter option has been realized at RHIC, while future polarized targets are being discussed at the LHC, in particular the implementation of such a target at AFTER@LHC \cite{Hadjidakis:2018ifr,Kikola:2017hnp} or at LHCb \cite{Aidala:2019pit}. For more details about the gluon Sivers function we refer the reader to a recent review on that function \cite{Boer:2015vso}. \index{Sivers function $f_{1T}^{\perp}$!gluon}

\subsubsection{Gluon TMDs in lepton-nucleon collisions}

\begin{figure}[t!]
\centering
\includegraphics[width=0.3\textwidth]{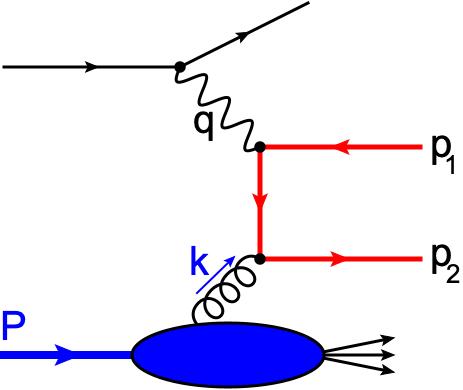}
\caption{Tree-level diagram for heavy back-to-back jet pair production in lepton-nucleon collisions.
}
\label{fig:Gluon2dijet}
\end{figure}

In this subsection we will briefly discuss observables in lepton-nucleon collisions that are sensitive to gluon TMDs. The exploration of the gluonic structure of the nucleon is an important part of the experimental program at the future Electron-Ion Collider, and we will see that this experiment is well suited for the study of gluon TMDs.

The theoretically cleanest process at an EIC that may be used to study gluon TMDs is the production of two back-to-back jets. The two jets need to be produced by two heavy quarks. This process, $ep\to e+\mathrm{jet}+\mathrm{jet}+X$, has been first analyzed in the TMD framework at tree-level in Ref. \cite{Boer:2010zf}. For more formal aspects about TMD factorization of heavy dijet production beyond tree-level we refer the reader to Refs.~\cite{Kang:2020xgk,delCastillo:2020omr,Zhang:2017uiz}.

At tree-level, the only diagram (+ crossed) that contributes in the TMD framework for heavy dijet production in lepton-nucleon collisions, i.e., $\gamma^\ast(q)+n(P)\to \mathrm{jet}(p_1)+\mathrm{jet}(p_2)+X$ is shown in Fig. \ref{fig:Gluon2dijet}. Its calculation is straightforward. If the dijet momentum is labeled as $p\equiv p_1+p_2$, then TMD factorization can be applied if the dijet transverse momentum $p_T$ is much smaller than the hard scale of the process, typically the virtuality of the exchanged photon, $Q$. A small $|{\bm p}_T|$ may be viewed as a small transverse momentum imbalance of the two jets. In other words, the jets are almost back-to-back in transverse space. In the kinematical region $p_T\ll Q$, the differential cross section for dijet production then reads according to Ref.~\cite{Boer:2010zf},
\begin{eqnarray}
\frac{\mathrm{d}\sigma(|{\bm p}_T|\ll Q)}{\mathrm{dPS}_3} & = & A\,f_1^g(x,{\bm p}_T^2)+B\,h_1^{\perp g}(x,{\bm p}_T^2)\,\cos(2\phi)\,.\label{eq:CSdijet}
\end{eqnarray}
In this schematic formula, $A$ and $B$ are partonic factors given in Ref. \cite{Boer:2010zf}, while $\phi$ is the azimuthal angle between lepton plane and the dijet transverse momentum. It is quite remarkable that in Eq.~(\ref{eq:CSdijet}) the gluon TMDs $f_1^g$ and $h_1^{\perp g}$ do not show up in a convolution integral combined with other gluon TMDs as they do in proton collisions (see previous section). Instead, formula in Eq.~(\ref{eq:CSdijet}) suggests that one can probe the support of those functions, in principle point-by-point. In particular the intrinsic transverse gluon momentum ${\bm k}_T$ is replaced by the heavy dijet transverse momentum imbalance ${\bm p}_T$. This is truly a unique feature of this process, and makes it an ideal candidate to be investigated at a future EIC. Since the EIC will also allow for polarized proton beams, this process also offers the opportunity to study polarized gluon TMDs like the gluon Sivers function \cite{Zheng:2018ssm,Kang:2020xgk}.\index{Sivers function $f_{1T}^{\perp}$!gluon}

Certainly, heavy dijet production is not the only final state that can be used to probe gluon TMDs in lepton-nucleon production. As for proton collisions, the production of quarkonium states can also shed information on the gluonic transverse motion in the nucleon. This has been studied recently in a series of papers, see Refs. \cite{Mukherjee:2016qxa,Kishore:2018ugo,Kishore:2019fzb,DAlesio:2019qpk}.

\subsection{TMDs in Nuclei}
\label{sec:phenoTMDs_nuclei}

\begin{figure}[t!]
\centering
\includegraphics[width=0.99\textwidth]{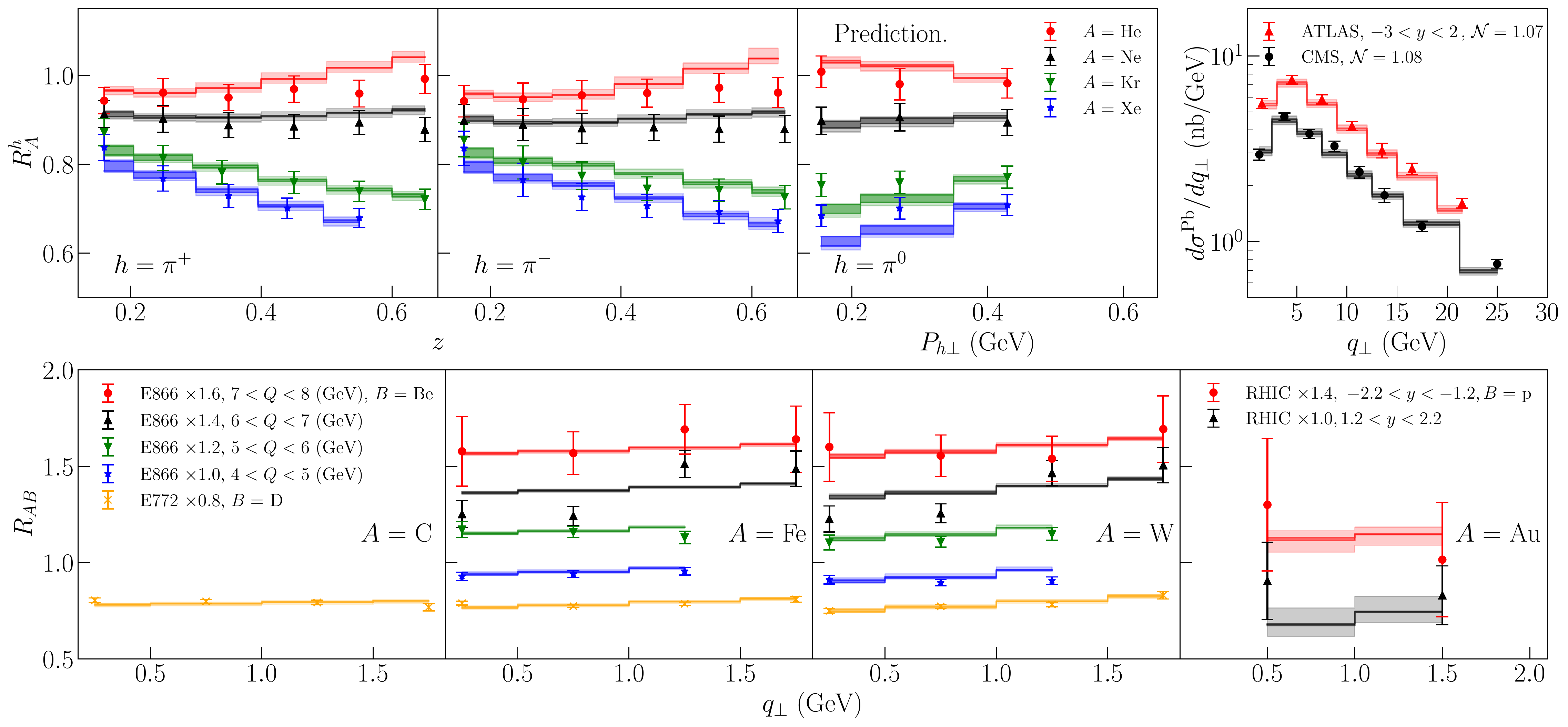}
\caption{Theoretical description of selected experimental data for SIDIS process in lepton-nucleus collisions and DY process in proton-nucleus scatterings. Plot from Ref.~\cite{Alrashed:2021csd}.}
\label{fig:data-inc}
\end{figure}
Significant progress has been made in extracting TMDs for free nucleons from experimental data. On the other hand, the corresponding TMDs in a heavy nucleus is still at a primitive stage. Identifying the partonic structure of quarks and gluons in nuclei has remained as one of the most important challenges confronting the nuclear physics community since the pioneering EMC measurements in 1980s~\cite{EuropeanMuon:1983wih}, and has been regarded as one of the major goals at the future EIC~\cite{Accardi:2012qut,AbdulKhalek:2021gbh}.

The determination of nuclear TMD PDFs and nuclear TMD FFs (collectively called nTMDs) relies on the corresponding TMD QCD factorization. At the moment, experimental data are available from the SIDIS process on a nuclear target, and the Drell-Yan process in proton-nucleus (pA) and pion-nucleus collisions. For a compilation of Drell-Yan data prior to 1993 we refer to the review \cite{Stirling:1993gc}. More recent nTMD measurements were reported by HERMES~\cite{HERMES:2007plz}, JLab~\cite{Dudek:2012vr,Burkert:2008rj,CLAS:2021jhm}, Fermilab~\cite{NuSea:1999egr,Alde:1990im}, RHIC~\cite{Leung:2018tql} and the LHC~\cite{CMS:2015zlj,ATLAS:2015mwq}, and will be further measured by the future EIC with unprecedented precision.

The first simultaneous global QCD extraction of the TMD parton distribution functions and the TMD fragmentation functions in nuclei was performed in Ref.~\cite{Alrashed:2021csd}. The world set of data from semi-inclusive electron-nucleus deep inelastic scattering and Drell-Yan di-lepton production was considered. In total, this data set consists of 126 data points from HERMES, Fermilab, RHIC and LHC. Working at next-to-leading order and next-to-next-to-leading logarithmic accuracy, a very good $\chi^2$/d.o.f. $= 1.045$ was achieved. In this analysis, the broadening of TMDs in nuclei was quantified for the first time .  

In Fig.~\ref{fig:data-inc}, we plot the result of the fit and  the experimental data. In the top row of this figure, we plot the comparison against: the multiplicity ratio measurement at HERMES 
\cite{HERMES:2007plz}
as a function of $z$ (left two columns) and $P_{h\perp}$ (third column from the left), and the DY $q_\perp$ distribution from the LHC (right column). We note that the $P_{h\perp}$ dependent data in the third column is a prediction for those data points. Furthermore, for the LHC data \cite{CMS:2015zlj,ATLAS:2015mwq}, we have provided the $\mathcal{N}_i$ for each of the data sets. In the left three columns of the second row, we plot the comparison against the $R_{AB}$ ratio for the E866~\cite{NuSea:1999egr} and E772~\cite{Alde:1990im} experiments. Finally, in the right column of this row, we plot the $R_{AB}$ at RHIC~\cite{Leung:2018tql}. In each subplot, we have provided the uncertainty from our fit as a dark band, and the uncertainty from the collinear distributions as a light band.

\begin{figure}[t!]
    \centering\offinterlineskip
    \includegraphics[height = 0.22\textheight]{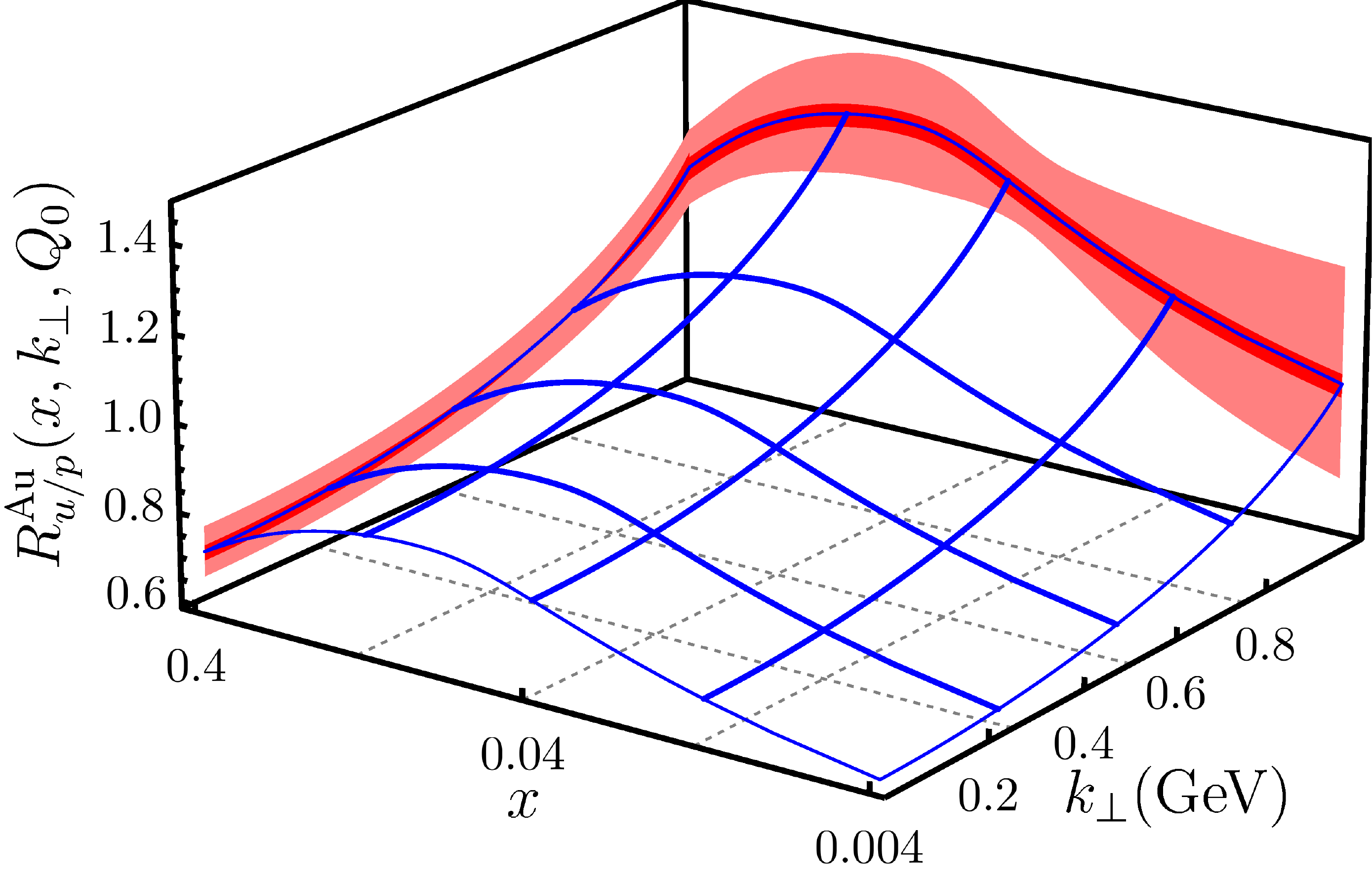}
    \hskip 0.05in
    \includegraphics[height = 0.22\textheight]{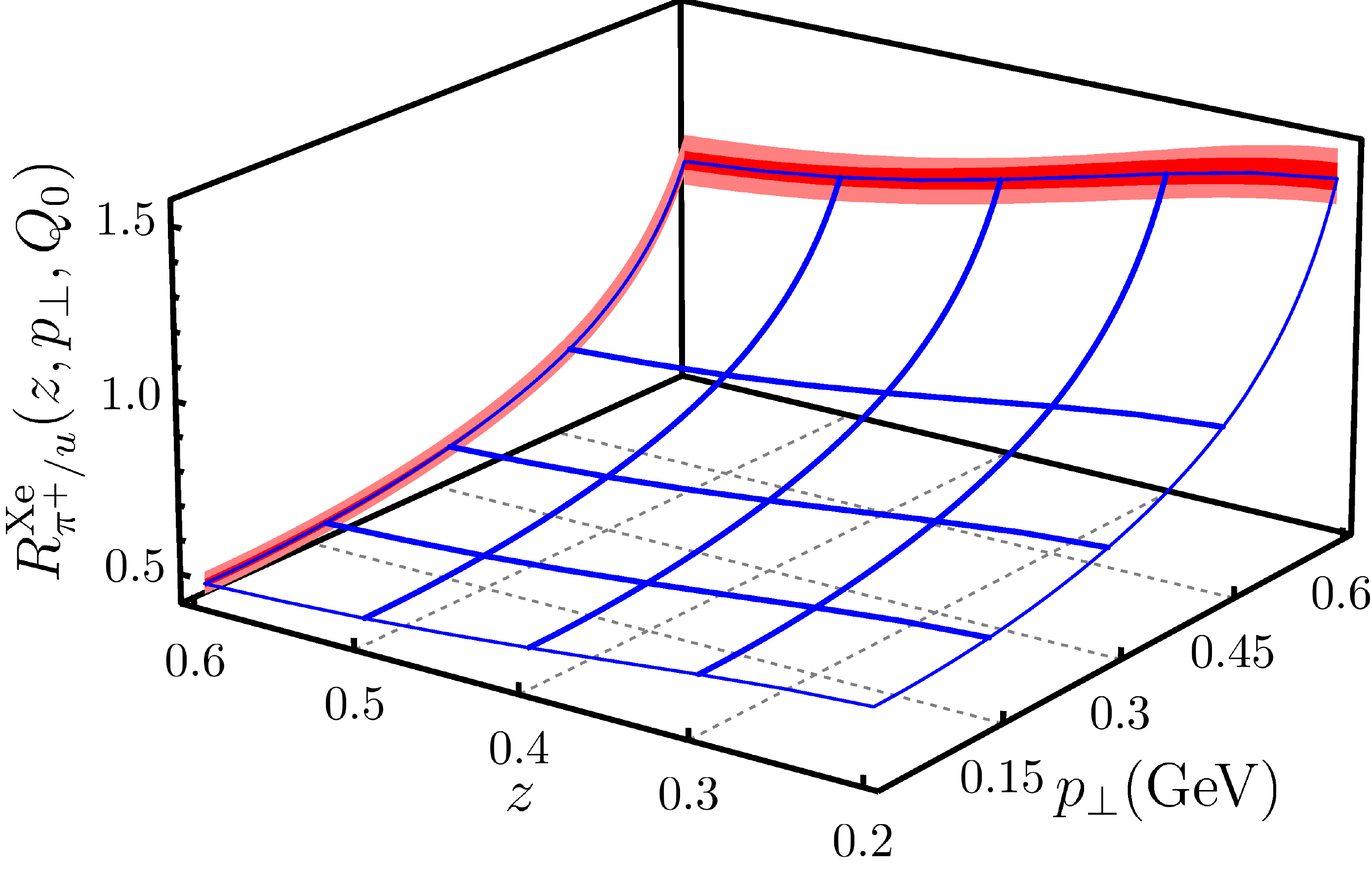}
  \caption{The extracted nuclear ratio for the TMDPDF (left) and the TMDFF (right).}
    \label{fig:nTMDs}
\end{figure}

In the right panel of Fig.~\ref{fig:nTMDs}, we plot the ratio of the $u$-quark TMDPDF of a bound proton in a gold nucleus and that in a free proton as a function of $x$ and $k_\perp$. The curve along the plane for $k_\perp = 1$ GeV demonstrates the shadowing, anti-shadowing, and the EMC effect which originate from the collinear distribution. The curves which lie in planes of constant $x$ increase with increasing $k_\perp$, which indicates the transverse momentum broadening effect, with a suppression at low $k_{\perp}$ and an enhancement at high $k_{\perp}$. In the second panel of this figure, we plot the ratio of the nTMDFF for $u\rightarrow \pi^+$ in a $\rm Xe$ nucleus and that in vacuum as a function of $z$ and $p_\perp$. Analogous to the nTMDPDFs, we see that as $p_\perp$ grows, this ratio becomes larger, indicating that hadrons originating from fragmentation in the presence of a nuclear medium will tend to have a broader distribution of transverse momentum relative to vacuum TMDFFs. 

\subsection{Importance of QED Corrections to SIDIS for Extracting TMDs}
\label{sec:QED}

By evaluating various angular modulations of the angles $\phi_h$ and $\phi_S$, as defined in Fig.~\ref{fig:sidis}, the lepton-hadron SIDIS has a unique advantage in separating contributions from different TMDs, as reviewed in this chapter.  However, lepton scattering at large momentum transfer can be a source of considerable photon radiation, which can significantly distort the inferred hadron structure if it is not properly accounted for. The collision induced radiation can not only affect the momentum transfer $q$ from the colliding lepton to the hadron, preventing a well-defined $\gamma^*(q)P$ frame in Fig.~\ref{fig:sidis} where the TMD factorization is defined, it can also alter the angular modulation between the lepton and hadron planes, making it problematic to define the transverse momentum of the produced hadron, $P_{hT}$, in the ``true'' $\gamma^*P$ frame.  This in turn can induce angular modulations which can mimic those arising from the true hadron structure effects encoded by the TMDs.

With a large momentum transfer, the collision-induced QED radiation is an integral part of the experimentally measured cross section for deep-inelastic lepton-hadron scattering.  Historically, tremendous efforts have been devoted to isolate and remove collision-induced QED contributions from measured cross sections, which would enable one to focus purely on QCD effects in lepton-hadron scattering, by adding a radiative correction factor to the ``Born'' cross section that does not include any collision-induced QED radiation effect.  However, such QED radiative corrections factors that aim to ``correct'' for this QED contamination are necessarily sensitive to the very hadronic physics that we aim to explore in SIDIS reactions~\cite{Liu:2020rvc,Liu:2021jfp}.  This is because the colliding hadron has to experience a range of momentum transfer after we sum over the collision-induced QED radiation. Instead of the single value of momentum transfer $q$ for the Born cross section, QED radiation changes the kinematic variables $(x=Q^2/2P\cdot q,\, Q^2=-q^2)$ for the Born cross section to $(x\in [Q^2/2P\cdot q,1],\, Q^2\in [Q^2_{\rm min}, Q^2_{\rm max}])$ for the SIDIS cross section with the approximation of one-photon exchange.  
With $Q^2_{\rm min}=Q^2(1-y)/(1-x\, y)$ and $Q^2_{\rm max}=Q^2/(1-(1-x)y)$, the radiative corrections factors could be sensitive to nonperturbative hadron physics if $Q^2_{\rm min}$ is sufficiently small even if $Q^2$ is a hard scale.   

The lepton-hadron SIDIS cross section is effectively an inclusive cross section to observe one lepton and one hadron in the final state. It is a well-defined two-scale observable when $Q^2$ is much larger than the momentum imbalance between the observed final-state lepton and hadron, and the imbalance is sensitive to the collision-induced QED and QCD radiation and transverse momentum of the active partons and leptons.  Instead of dealing with a full TMD factorization for all four observed particles (the two leptons and two hadrons) in both QED and QCD, which is likely to be violated~\cite{Collins:2007nk}, a hybrid factorization approach was proposed for treating lepton-hadron SIDIS with collinear factorization for the two leptons and TMD factorization for two hadrons when the SIDIS cross section is in the two-scale regime~\cite{Liu:2020rvc,Liu:2021jfp}.  Such a hybrid factorization approach to SIDIS is possible because the amount of transverse momentum broadening generated by the collision-induced QED and QCD radiation from a ``point-like'' lepton is much smaller than the typical transverse momentum of a colliding parton (which could be further enhanced by QCD radiation from its intrinsic value) for all foreseeable energies of lepton-hadron scattering experiments~\cite{Liu:2021jfp}. The momentum imbalance between the observed lepton and hadron in the final state is therefore dominated by the transverse momentum dependence of the hadron TMDs, which ensures SIDIS to be a useful process for extraction of TMDs.  The key impact of QED radiation on the SIDIS cross section is from the change of the momentum transfer to the colliding hadron, in both its direction and invariant mass, caused mainly by the logarithmic-enhanced collinear QED radiation.

With this hybrid factorization approach, the SIDIS cross section for a colliding lepton of momentum $l$ and helicity $\lambda_l$ and a nucleon of momentum $P$ and spin $S$ can be factorized as~\cite{Qiu:1990xy,Liu:2021jfp}
\begin{eqnarray}
\label{e.sidis-fac-qed} 
    E_{l'} E_{P_h} 
    \frac{d\sigma_{l(\lambda_l) P(S)\to l' P_hX}}
         {d^3 l'\, d^3P_h}
    &\approx&
     \sum_{i j \lambda_k}
    \int_{\zeta_{\rm min}}^1 \frac{d\zeta_l}{\zeta_l^2} \, D_{l'/j}(\zeta_l)
    \int_{\xi_{\rm min}}^1 d\xi_l\, f_{i(\lambda_k)/l(\lambda_l)}(\xi_l)
    \nonumber\\
    &\quad&
    \times
    \left[
    E_{k'} E_{P_h} 
    \frac{d{\hat{\sigma}}_{k(\lambda_k) P(S)\to k' P_hX}}
         {d^3k'\, d^3P_h}
    \right]_{k=\xi_l l, k'=l'/\zeta_l},
\end{eqnarray}
where $D_{l'/j}(\zeta_l)$ and $f_{i(\lambda_k)/l(\lambda_l)}(\xi_l)$ are lepton FFs and lepton PDFs, respectively, the $\zeta_l$ and $\xi_l$ are corresponding lepton momentum fractions, and the integration limits $\zeta_{\rm min} = 1- (1-x)y$ and $\xi_{\rm min}=(1-y)/(\zeta_{\rm min} - x y)$~\cite{Liu:2021jfp}.  In Eq.~(\ref{e.sidis-fac-qed}), the $\hat{\sigma}_{k(\lambda_k) P(S)\to k' P_hX}$ is infrared-safe as lepton mass $m_e\to 0$ with all infrared sensitive collinear radiative contributions along the direction of colliding and observed lepton resummed into the lepton PDFs and lepton FFs, respectively.  In the Born approximation in QED for the $\hat{\sigma}_{k(\lambda_k) P(S)\to k' P_hX}$, which is the LO contribution in $\alpha$, the differential SIDIS cross section in the presence of QED effects can be written in the TMD regime as~\cite{Liu:2021jfp}
\begin{eqnarray}
    \frac{d\sigma_{l(\lambda_l) P(S) \to l' P_h X}}
         {dx\, dy\, d\psi\, dz_h\, d\phi_h dP_{hT}^2}
    &=& \sum_{i j \lambda_k}
       \int_{\zeta_{\rm min}}^1 \frac{d\zeta_l}{\zeta_l^2}
       \int_{\xi_{\rm min}}^1 \frac{d\xi_l}{\xi_l}\, f_{i(\lambda_k)/l(\lambda_l)}(\xi_l)\, D_{l'/j}(\zeta_l)
    \label{e.sidisRC}\\
    & &
    \times
    \frac{\hat{x}}{x \xi_l \zeta_l}
    \bigg[ \frac{\alpha^2}{\hat{x}\, \hat{y}\, \widehat{Q}^2} 
        \frac{\hat{y}^2}{2(1-\hat{\varepsilon})}
    \left(1+\frac{\hat{\gamma}^2}{2\hat{x}}\right)
    \sum_n \hat{w}_n F_n^h(\hat{x},\widehat{Q}^2,\hat{z}_h,\widehat{P}_{hT}^2)
    \bigg],
    \nonumber
\end{eqnarray}
where the kinematic variables with carets in the factorized expression are defined in the same way as those in~\cite{Bacchetta:2006tn} without carets, except the momentum $q = l-l'$ of the exchanged virtual photon (or vector boson in general) is replaced by $\hat{q}(\xi_l,\zeta_l) \equiv \xi_l l - l'/\zeta_l$, and the produced hadron transverse momentum $\widehat{P}_{hT}$ is defined in the virtual $\gamma^*(\hat{q})P$ frame.  In Eq.~(\ref{e.sidisRC}), 
the expression in the square brackets is the one without the presence of collision-induced QED effects with 
the usual 18 SIDIS structure functions, $F_n^h$ with $n=1,\ldots,18$, weighted by factors $\hat{w}_n$ that are functions of the kinematic variables~\cite{Bacchetta:2006tn}, and the factor in front of the square brackets is a Jacobian between the $\gamma^*(q)P$ and $\gamma^*(\hat{q})P$ frame for corresponding variables~\cite{Liu:2021jfp}.

\subsection{Outlook and Future Work}
\label{sec:pheno_outlook}

Phenomenology of TMDs have made a huge leap from partonic model approximations to high precision fits up to N$^3$LO in the last decade. Universal QCD fits including data from SIDIS, DY, $W^\pm/Z$ and $pp$ scattering have been successfully performed.

Phenomenology of TMDs will evolve into global QCD analyses of data at high order perturbative precision and will utilize data from various processes and facilities. Machine learning and AI techniques have been already employed in these types of fits and will continue to be developed and utilized. The 3D structure of the nucleon encoded in TMDs is very intimately related to 1D structure encoded in collinear distributions including twist-3 distributions and in the coming years we will see merging of 1D and 3D phenomenology. We will see increasing impact from ab-initio calculations using lattice QCD
on phenomenological studies of the nucleon structure, see \chap{lattice}.
New precise experimental data will allow studies of other TMDs including subleading TMD functions, see \chap{twist3}. Other observables and measurements will become available that will allow the phenomenology to go beyond the 3D picture and utilize Wigner 5D distributions, see \chap{gtmd}. Gluon and sea-quark structure of the nucleon as well as nuclear modifications to TMDs will be explored with the advent of new experimental measurements, see \chap{smallx}.

Experimental programs of  future and existing experimental facilities such as the Electron Ion Collider~\cite{Boer:2011fh,Accardi:2012qut}, Jefferson Lab 12~GeV Upgrade~\cite{Dudek:2012vr}, RHIC~\cite{Aschenauer:2015eha} at BNL, COMPASS~\cite{Gautheron:2010wva,Bradamante:2018ick} at CERN, BELLE II at KEK, BES III in Beijing, and the LHC at CERN will contribute immensely to our understanding of the hadron structure and progress  of phenomenology in general and 3D structure in particular.

%% file: sec-latticeQCD/sec-latticeQCD.tex
\section{Lattice QCD calculations of TMDs and related aspects of hadron structure}
\label{sec:lattice}

\subsection{Lattice QCD}
\label{sec:lqcd}
\index{lattice QCD|(}
As discussed in Chapter \ref{sec:Intro}, many aspects of hadron structure can not be addressed using perturbative QCD. In Chapter \ref{sec:phenoTMDs} experimental data is used to determine the nonperturbative contributions to the TMD PDFs and FFs. However, aspects of hadron structure can also be calculated from the underlying theory of QCD using lattice field theory techniques, referred to as {lattice QCD} (LQCD).   There are many excellent introductions to LQCD, see for example Refs.~\cite{Rothe:1992nt,Gattringer:2010zz}, to which the reader is referred for full details. In this brief overview, we present relevant aspects of LQCD as they impact the discussion of TMD and hadron structure studies.

LQCD was introduce by Wilson in Ref.~\cite{Wilson:1974sk}, and in this approach,  physical information is extracted from QCD correlation functions that are evaluated from their functional integral representation. At an intermediate stage, a Euclidean spacetime lattice is used to regulate the theory, rendering the functional integral finite-dimensional. The theory is formulated on a discrete, spacetime geometry which in almost all cases is taken to be  a regular four-dimensional hypercubic lattice, $\Lambda_4=\{n_\mu=(n_1,n_2,n_3,n_4) | n_i \in a [0,1,\ldots L_i-1]\}$ where $a$ is the lattice spacing and $L_i$ is the lattice extent in the $i$ direction. Periodic spatial boundary conditions are typically imposed on all fields, and  periodic temporal boundary conditions on the gauge fields and anti-periodic temporal boundary conditions for fermions are used. In some cases an anisotropy is introduced, providing a finer discretization in the temporal direction; furthermore, other geometries have been considered in the past~\cite{Celmaster:1982ht} but are not in current use.  Noting that the exponential of the discretized QCD action is a Boltzmann probability distribution, importance sampling Monte-Carlo methods are then used to stochastically evaluate the requisite integrals. Physics is recovered in the limit that the lattice regulator is removed (the {\it continuum limit})  and the spacetime volume is taken to infinity (the {\it infinite volume limit}).

The QCD action must be implemented approximately on the lattice geometry, with derivatives replaced by finite differences.  For the gauge fields, the simplest action is the Wilson action 
\begin{equation}
    S_{\rm gauge} = \frac{2}{g^2} \sum_{x\in\Lambda_4} \sum_{\mu<\nu} (1-{\rm Re\ Tr}[P_{\mu\nu}(x)]),
\end{equation}
where $P_{\mu\nu}(x)$ defines the elementary plaquette which corresponds to products of gauge link variables $U_\mu(x)$ around a $1\times1$ elementary cell, 
$$
P_{\mu\nu}(x)=U_\mu(x)U_\nu(x+\hat\mu)U_\mu^\dagger(x+\hat \nu) U_\nu^\dagger(x).
$$ 
Here, the link variables $U_\mu(x)=\exp[i a A_\mu(x)]$ are associated with the site $x$ and extend one lattice spacing to the site $x+\hat\mu$, where $\hat\mu$ is a unit vector in the $\mu$ direction. Expanding the LQCD action around the limit $a\to0$ reproduces the continuum QCD action up to ${\cal O}(a^2)$ effects. Variants of the action introduce additional terms that can cancel higher powers of $a$, providing a closer-to-continuum action \cite{Gattringer:2010zz}. 

Naive implementations of lattice fermions have multiple zero-modes for massless fermions, corresponding to ``doubling'' of the light degrees of freedom. These are avoided with the Wilson~\cite{Wilson:1974sk}, the Kogut-Susskind~\cite{Kogut:1974ag} and twisted-mass~\cite{Frezzotti:2003ni} quark actions, but at the expense of  explicitly breaking the chiral symmetry of the massless QCD action. Chirally-symmetric fermion formulations such as the domain-wall fermion (DWF) action~\cite{Kaplan:1992bt}, which avoid this issue by introducing an additional spacetime dimension, and the overlap action~\cite{Narayanan:1994gw}, maintain a lattice chiral symmetry. As with gauge actions, the fermion actions can also be improved to reduce lattice artifacts; in this case, there is a unique dimension-five operator to add to the action~\cite{Sheikholeslami:1985ij}, $i\overline\psi \sigma_{\mu\nu}G^{\mu\nu}\psi$, known as the clover term. Since the (lattice) QCD action is bilinear in the fermion fields, $S_{\rm fermion} \sim\int dx \overline\psi {\cal M} \psi$ where ${\cal M}$ depends on the choice of action and on the gauge field. For the given action encoded in ${\cal M}$, this results in an effective action $S_{\rm fermion}= {\rm Tr\, Ln}[{\cal M}]$.

Given a particular lattice action, LQCD calculations proceed by evaluating the QCD path integrals defining appropriate correlation functions using importance sampling Monte Carlo based on that action.  For a generic multi-local operator ${\cal O}(x_1,x_2,\ldots)$ built from quark and gluon fields,
\begin{equation}
    \langle {\cal O}(x_1,x_2,\ldots)\rangle = \frac{1}{\cal Z}\int {\cal D}U \tilde{\cal O}(x_1,x_2,\ldots) \det[{\cal M}[U]] e^{-S_{\rm gauge}} ,
\end{equation}
where the partition function ${\cal Z} = \int {\cal D}U  \det[{\cal M}[U]] e^{-S_{\rm gauge}}$.  The field operator $\tilde{\cal O}(x_1,x_2,\ldots)$ corresponds to the original operator ${\cal O}$ after fermions are integrated out; this integration results in the ``contraction'' of fermion--anti-fermion pairs in all possible ways, replacing them with quark propagators $S[U]={\cal M}[U]^{-1}$. By sampling the gauge fields according to the probability distribution  ${\cal P}[U] = {\cal Z}^{-1}\det[{\cal M}[U]] e^{-S_{\rm gauge}}$, this can be approximated as

\begin{equation}
    \langle {\cal O}(x_1,x_2,\ldots)\rangle = \frac{1}{N} \sum_{c=1}^N \tilde{\cal O}(x_1,x_2,\ldots)[U_i] 
    + {\cal O}\left(1/\sqrt{N}\right) , 
\end{equation}
where $\{U_1,\ldots U_N\}$ correspond to a properly sampled set (ensemble) of gauge fields. These requisite configurations are produced with the correct distribution through a  Markov chain Monte Carlo process, with the standard algorithm being hybrid Monte Carlo~\cite{Duane:1987de}. Before the year 2000, many studies were performed in the quenched version of QCD in which the quark determinant was neglected for computational expediency. This approximation is not typical in modern calculations, although the freedom 
of using a different quark mass in the quark determinant (referred to as sea quarks in the LQCD context) and the quark propagators (valence quarks in the LQCD context) is sometimes used and referred to as partial quenching.

To undertake a LQCD calculation, the quark masses and the gauge couplings to be used must be specified (along with the values of the coefficients of irrelevant operators used to improve the action) in some way, typically by matching computations of  simple quantities such as the pion and kaon masses to their experimental values. Once this is done, other quantities that are computed are predictions of the theory.
Having performed a set of simulations at different values of the bare parameters, the continuum and infinite volume limits must be taken before physical results are obtained. In addition to the statistical uncertainties of the simulations, the uncertainties implicit in taking these limits must be carefully investigated and accounted for. As discussed above, typical LQCD actions differ from the continuum QCD action by terms of ${\cal O}(a)$ or in some cases ${\cal O}(a^2)$, while for many properties of individual hadrons volume effects are controlled by terms ${\cal O}(e^{-m_\pi L})$ where $L$ is the smallest dimension of the lattice geometry and $m_\pi$ is the mass of the lightest hadron. With few exceptions, LQCD calculations are performed ignoring the up and down quark mass splitting and do not include the effects of electromagnetism, as these contributions are small effects in most cases. Precision calculations must account for these additional systematic effects and do so in relevant contexts~\cite{Aoki:2019cca}.

An important application of LQCD that is centrally relevant to this review is computing the matrix elements of  currents in hadronic states such as the proton. In the continuum, the  currents one might consider are operators such as the axial current $\overline\psi\gamma_\mu\gamma_5\psi$, twist-two operators $i^{n-1}\overline \psi \gamma_{\{\mu_1}D_{\mu_2}\ldots D_{\mu_n\}}\psi$ (where the braces denote symmetrization and trace-subtraction of the enclosed indices), or four-quark operators $\overline\psi\Gamma\psi \overline\psi\widetilde\Gamma\psi$ (where $\Gamma$ and $\widetilde\Gamma$ are Dirac and flavor structures) that typically arise from integrating out physics far above the hadronic scale. As will be discussed below, matrix elements of more complicated nonlocal operators are now also commonly studied. In a discretized lattice theory, these operators must be implemented using the lattice degrees of freedom and differ from the continuum operators by terms ${\cal O}(a)$ (as with the action, improved lattice operators can be constructed that eliminate lattice artifacts to a particular order). 
Since the operators used in lattice calculations are necessarily formulated in terms of the discretized variables, an additional step that must be undertaken to connect to physics in the continuum limit is renormalization of the operators. Even for operators that are scale-independent, such as the isovector axial charge, a finite renormalization is required. This renormalization can be implemented using lattice perturbation theory (see Ref.~\cite{Capitani:2002mp} for an overview) for the appropriate lattice action. Alternatively, nonperturbative renormalization based on momentum subtraction schemes~\cite{Martinelli:1994ty} can also be used and are subject to smaller uncertainties. In such schemes, renormalization constants are fixed by demanding quark and/or gluon two and three point correlation functions take their tree-level values at a particular kinematic point. Ultimately, continuum perturbation theory is then used to convert to standard perturbative schemes such as $\MSbar$. This nonperturbative approach is now standard for local operators and the effects of mixing between operators with the same quantum numbers, such as isoscalar quark operators and corresponding gluon operators, can be incorporated (the more intricate problem of renormalization of nonlocal operators is discussed below).
\index{lattice QCD calculations!nonperturbative renormalization}

To define the notation used below and further introduce LQCD methods, it is useful to overview the calculation of the proton mass. In particular, the proton mass can be determined from the calculations of the two-point correlation function, which can be expressed (assuming an infinite temporal extent of the lattice geometry for simplicity and making use of translational invariance) as
\begin{eqnarray}
C_{\alpha\beta}(t,{\bf p})= a^3 \sum_{\bf x} e^{-i{\bf p \cdot x}} C_{\alpha\beta}(t,{\bf x})= a^3\sum_{\bf x} e^{-i{\bf p \cdot x}}\langle 0 |\chi_\alpha({\bf x},t) \overline\chi_\beta({\bf 0},0) |0\rangle  ,
\label{eq:c2pt}
\end{eqnarray}
where 
\begin{eqnarray}
\chi_\alpha({\bf x},t) = \epsilon^{ijk} u^i_\alpha({\bf x},t) u^j_\gamma({\bf x},t) [C^{-1}\gamma_5]_{\gamma\delta} d^k_\delta({\bf x},t) 
\end{eqnarray}
is an interpolating operator with the quantum numbers of the proton and $C=\gamma_0\gamma_2$ is the charge conjugation matrix ($C\gamma_0^T C^{-1}=-\gamma_0$). In the above expressions, greek indices refer to the Dirac structure while roman indices label color components.

After integrating out the quark fields in the path integral formulation, this correlator is expressed in terms of products of the inverse of the Dirac operator 
\begin{eqnarray}
 C_{\alpha\beta}(t,{\bf p})    &=&  - a^3  \sum_{\bf x} e^{-i{\bf p \cdot x}}
 \epsilon^{ijk} \epsilon^{i'j'k'} [C^{-1}\gamma_5]_{\alpha'\alpha''} [\gamma_5 C]_{\beta'\beta''}
 \\ \nonumber && \hspace*{5mm}\times
\left\langle \left[{\cal M}_d^{-1}\right]_{\alpha''\beta'}^{ki'} \left\{ \left[{\cal M}_u^{-1}\right]_{\alpha'\beta''}^{jj'}\left[{\cal M}^{-1}_u\right]_{\alpha\beta}^{ik'} - \left[{\cal M}_u^{-1}\right]_{\alpha\beta''}^{ij'}\left[{\cal M}^{-1}_u\right]_{\alpha'\beta}^{jk'} \right\}  \right\rangle ,
\end{eqnarray}
where the quark propagator ${\cal M}_f^{-1}={\cal M}^{-1}_f(x,0)$ is the inverse of the Dirac operator for flavor $f$. This correlation function can be evaluated as an average over representative gluon field configurations as discussed above.

Inserting a complete set of states\footnote{Continuum infinite volume states are normalized as ${}_c\langle n, {\bf p}, \sigma| n, {\bf p}, \sigma \rangle_c=2 E({\bf p}) \delta^3(p-p')$, with the lattice states defined from these as $|n, {\bf p}, \sigma\rangle = \sqrt{2V_3 E_n({\bf p})}|n, {\bf p}, \sigma\rangle_c$ where $V_3$ is the spatial volume and $n$ labels excitations.} between the interpolating operators in Eq.~\eqref{eq:c2pt}, it is  straightforward to see that the two-point correlator has time dependence governed by the energies of states with the quantum numbers of the proton and with three-momentum ${\bf p}$:
\begin{eqnarray}
C_{\alpha\beta}(t,{\bf p})&=& a^3 \sum_{n,\sigma} \frac{e^{-E_n({\bf p}) t}}{2 E_n({\bf p})} \langle 0 | \chi_\alpha | n; {\bf p}, \sigma \rangle_c\ {}_c\langle n; {\bf p},\sigma  | \overline\chi_\beta |0\rangle 
\\
&=& a^3 Z({\bf p}) \sum_\sigma u_\alpha(n=0, {\bf p},\sigma) \overline{u}_\beta(n=0, {\bf p},\sigma) \frac{e^{-E_n({\bf p}) t}}{2 E_n({\bf p})} + \ldots
\end{eqnarray}
where $Z({\bf p})$ is an overlap factor\footnote{Defining spinors such that $\overline u(n,{\bf p},\sigma) u(n,{\bf p},\sigma')=2M_n\delta_{\sigma\sigma'}$, the overlap matrix elements are given by $\langle 0 | \chi_\alpha | n; {\bf p}, \sigma \rangle_c = \sqrt{Z({\bf p})} u_\alpha(n,{\bf p},\sigma) e^{i{\bf p\cdot x}}$.} and higher excited states are ignored. Tracing this correlator against a given Dirac structure, often chosen to be $\Gamma^+=\frac{1}{2}(1+\gamma_4)$, leads to
\footnote{Here,  Euclidean $\gamma$-matrices  $\gamma_i^E= -i  \gamma_i^M$ for $i\in\{1,2,3\}$ and $\gamma_4= \gamma_0^M$, written in terms of Minkowski space matrices, are used and satisfy $\left\{\gamma_{\mu}^E, \gamma_{\nu}^E\right\}=2 \delta_{\mu \nu} I$.}
\begin{eqnarray}
\label{eq:CGamma}
C_{\Gamma^+}(t,{\bf p})=\Gamma^+_{\beta\alpha} C_{\alpha\beta}(t,{\bf p}) \stackrel{t\to\infty}{\longrightarrow} 
C e^{-E_n({\bf p}) t},
\end{eqnarray}
where $C$ is a time-independent constant.
Given the discrete time series $C_{\Gamma}(t,{\bf p})$ from the Monte Carlo sampling, the proton mass dispersion relation can be extracted from fits to the time dependence of Eq.~\eqref{eq:CGamma}, either in the asymptotic region where the lowest energy state dominates, or from more general time ranges where excited states must also be accounted for.
\index{lattice QCD|)}

\subsection{Lattice Hadron Structure}
\label{sec:lattice:structure}

\subsubsection{Static structure of hadrons}

While the focus of this handbook is on transverse momentum dependent hadron structure, we begin by discussing lattice calculations of hadron structure in a more general context. 
Since its early days, LQCD has been used as a tool with which to investigate the structure of the proton and other hadrons. Early works focused on static properties such as the magnetic moment~\cite{Martinelli:1982cb} and axial charge~\cite{Fucito:1982ff}, but methods with which to study more complex quantities such as the electromagnetic form factors were initially developed in the 1980s~\cite{Martinelli:1987zd,Martinelli:1988rr,Martinelli:1988xs}. By now, form factors of the vector and axial currents have been studied quite  precisely, and  state-of-the-art calculations make use of the physical values of the quark masses and include multiple lattice spacings and volumes in the calculations~\cite{Aoki:2019cca}. Recent calculations~\cite{Kronfeld:2019nfb} are also providing important insights into understanding weak current interactions of nucleons that are relevant in long-baseline neutrino experiments. 

The collinear PDFs are another key pillar of hadron structure. As discussed above, these are extracted from global fits to experimental data and are known with remarkable precision for the unpolarized and helicity  PDFs, but are less well constrained in the case of the transversity distributions. As an independent theory approach, LQCD has the potential to provide complimentary information to experimental measurements and even has the potential to improve constraints on global fits~\cite{Lin:2017snn}. For partonic structure, the Euclidean metric of LQCD imposes a challenge for direct evaluation of lightlike separated operators. To evade this issue, LQCD studies have i) made use of the operator product expansion (OPE), or more recently ii) considered quantities defined off the light-cone that can be connected to light-cone physics perturbatively as will be discussed  in detail in Sec.~\ref{sec:lattice:xdep}.

The traditional LQCD approach based on the OPE expands nonlocal operators (such as the current-current correlator relevant in DIS) in terms of an infinite sum of local operators
\begin{equation}
J_{\mu}(x) J_{\nu}(0) = \sum_{i,n}  C_{i,n} (x, \mu) x_{\rho_1}...x_{\rho_n} {\cal O}_i^{\mu\nu\rho_1...\rho_n}(\mu),
\end{equation}
where $C_{i,n}(x,\mu)$ are Wilson coefficients and ${\cal O}_i (\mu)$ are  local operators involving quark and gluon fields and covariant derivatives. Hadronic matrix elements of the local operators on the right-hand side are calculable in Euclidean space, and their analytic continuation back to Minkowski space is straightforward. While the Wilson coefficients and the operator matrix elements separately depend on the renormalization scale $\mu$, their product does not. Based on their relevance in high-energy processes such as DIS, the tower of operators is usually expressed in terms of operators of fixed twist (mass dimension minus spin), with the leading twist (twist-two) operators dominating. In particular, the Mellin moments of the leading-twist collinear quark PDFs are given by matrix elements of local gauge-covariant twist-two operators,
\begin{align}
\langle x^n \rangle_{i/H} \equiv \int_{-1}^1 dx\ x^n f_{i/H}(x) \quad \Longleftrightarrow \quad \langle H | {\cal O}_{i,n}(0)|H\rangle = \langle x^n \rangle_{i/H} (p_H\cdot n)^{n+1}\,,
\label{eq:pdfmoments}
\end{align}
where $H$ labels a hadron state of momentum $p_H$, and the leading twist operators
\begin{align}
    {\cal O}_{i,n}(x) = n^{\mu_0}n^{\mu_1}\cdots n^{\mu_n} \bar{\psi}_i \gamma_{\{\mu_0} i\overleftrightarrow{D}_{\mu_1}\cdots i\overleftrightarrow{D}_{\mu_n\}}\psi_i
\end{align}
as in Eq.~\eqref{eq:twistOps}. Here,  $n_{\mu}$ is a lightlike vector such that $n^2=0$, $\overleftrightarrow{D}_\mu= (\overrightarrow{\partial}_\mu - \overleftarrow{\partial}_\mu)/2 + igA_\mu$, and $\{\cdots\}$ indicates symmetrization and trace subtraction of the included Lorentz indices (note that the contraction with the $n$ vectors implies this). Similarly, local twist-2 gluonic operators define moments of gluon PDFs and the off-forward matrix elements of the same sets of operators define the generalized form factors that describe the Mellin moments of generalized parton distributions (GPDs)  \cite{Ji:1996ek,Ji:1996nm,Muller:1994ses,Radyushkin:1997ki,Ji:1998pc,Diehl:2003ny,Belitsky:2005qn,Boffi:2007yc,Kumericki:2016ehc}.

Following the pioneering approach of Refs.~\cite{Martinelli:1987zd,Martinelli:1998hz}, matrix elements of these operators can be determined from three-point correlation functions (suppressing operator indices) 
\begin{eqnarray}
C_{\alpha\beta}^{{\cal O}}(t,\tau,{\bf p},{\bf q})
= a^6\sum_{\bf x} \sum_{\bf y} e^{-i{\bf p \cdot x}}e^{-i{\bf q \cdot y}}\langle 0 |\chi_\alpha({\bf x},t) {\cal O}({\bf y},\tau) \overline\chi_\beta({\bf 0},0) |0\rangle.
\label{eq:c3pt}
\end{eqnarray}
By inserting complete sets of energy eigenstates, this quantity can be expressed in terms of the desired hadronic matrix elements, eigenenergies and  overlap factors that can be determined from corresponding two-point functions (Eq.~\eqref{eq:c2pt}). Fits to the time dependencies of Eqs.~\eqref{eq:c2pt} and \eqref{eq:c3pt} allow the matrix elements to be extracted.

In the context of Eq.~\eqref{eq:pdfmoments}, the LQCD approach is limited to operators of low Lorentz spin because the lattice geometry is invariant under only hypercubic symmetry transformations, elements of $H(4)$, rather than under those of ${\cal O}(4)$ symmetry~\cite{Martinelli:1987zd,Martinelli:1998hz,Gockeler:2000ja,Detmold:2005gg}. Consequently, the operators $O_{i,n}(\mu)$ for $n>3$  mix with lower dimensional operators in the same $H(4)$ irreducible representation with coefficients that diverge with inverse powers of  the lattice spacing as the continuum limit is approached. In order to relate the matrix elements $ \langle p|O_{i,n}(\mu)|p\rangle$ to the moments of PDFs,  the power divergences must be removed which is a significant challenge for $n>3$ operators~\cite{Martinelli:1996pk}. A method to improve this approach through an approximate restoration of the full symmetry has been proposed in Ref.~\cite{Davoudi:2012ya}. Additionally, matrix elements of twist-two operators are statistically more difficult to determine as the number of Lorentz indices, and therefore derivatives, increases.  Recent calculations are reviewed in Refs.~\cite{Lin:2017snn,Aoki:2019cca}.

\subsubsection{Decomposition of the proton momentum and spin} 
\label{decom_spin}

The low moments of PDFs that are accessible in LQCD calculations are already useful to understand various aspects of hadron structure. In particular the $n=1$ moments correspond to matrix elements of the energy momentum tensor (EMT) \index{energy-momentum tensor}and provide insight into the decomposition of the momentum and spin of the proton into its constituent contributions. Understanding these decompositions is a central question in nuclear physics and a major goal for the EIC~\cite{AbdulKhalek:2021gbh}.  As shown in Ref.~\cite{Ji:1996ek}, the matrix elements of the quark and gluon contributions to the EMT in the Belinfante form
\begin{eqnarray}  \label{eq:EMT}
    {\mathcal T}^{\mu\nu}_{q}&=&  \frac{i}{4} \sum_f \bar{\psi}_f \gamma^{\{\mu}\!\stackrel{\leftrightarrow}{D}\!{}^{\nu\}}\psi_f = \frac{i}{4} \sum_f \bar{\psi}_f (\stackrel{\rightarrow}{D}\!{}^{\mu}\gamma^{\nu} + \stackrel{\rightarrow}{D}\!{}^{\nu}\gamma^{\mu} - \stackrel{\leftarrow}{D}\!{}^{\mu}\gamma^{\nu} - \stackrel{\leftarrow}{D}\!{}^{\nu}\gamma^{\mu})\psi_f \\
    {\mathcal T}^{\mu\nu}_{g} &=& - G^{\mu\alpha} G^{\nu}_{\,\,\alpha} + \frac{1}{4} g^{\mu\nu} G^{\alpha\beta} G_{\alpha\beta},
\end{eqnarray}
 between nucleon states can be 
written in terms of gravitational form factors as 
\begin{eqnarray}
\langle p(P',S') | {\mathcal T}^{\mu\nu}_{q,g} | p(P,S),\rangle
  &=&\bar{u}(P',S') \big[T_{1_{q,g}}(q^2)\,\gamma^{\{\mu}\bar{P}^{\nu\}} 
   +\, \frac{i}{2m}T_{2_{q,g}}(q^2)\,\bar{P}^{\{\mu} \sigma^{\nu\}\alpha} q_{\alpha} \nonumber \\
  &+& D_{q,g}(q^2)\, \frac{q^{\mu}q^{\nu}- g^{\mu\nu} q^2}{m} + \bar{C}_{q,g}(q^2) \,g^{\mu\nu}\, m \big] u(P,S) , 
\label{eq:mat_element_1}
\end{eqnarray}
where $P$ and $P'$ are the initial and final momenta of the nucleon,\ respectively,\ and  $\bar{P} = \displaystyle\frac{1}{2}\, (P' + P)$.\ The sum of $\bar{C}_q$ and $\bar{C}_g$ is zero, but not so individually.
The momentum  transfer to the nucleon is $q = P' - P$.\footnote{The nucleon spinor, $u(P,S)$, satisfies the  following normalization conditions $\bar{u}(P,S)\, u(P,S)\, =\, 2m\, , \, \displaystyle\sum_S  u(P,S)\, \bar{u}(P,S)\, = \, P\!\!\!/ + m$.}
In the  $q^2 \rightarrow 0$ limit,\ one obtains
\begin{eqnarray}    \label{eq:momentum_fraction}
   \langle x\rangle_{q,g} &=& T_{1_{q,g}}(0), \\
   \label{eq:angmom}
   J_{q,g} &=& \frac{1}{2} \left[T_{1_{q,g}}(0) + T_{2_{q,g}}(0)\right].
\end{eqnarray}
where $\langle x\rangle_{q,g}$, the second moment of unpolarized PDF, is the momentum fraction carried by the quarks or gluons inside a nucleon.\ The other form factor, $T_{2_{q,g}}(0)$, can be interpreted as anomalous gravitomagnetic moment for quarks and gluons in an analogy to the anomalous magnetic moment~\cite{Teryaev:1999su}. The combination in Eq.~\eqref{eq:angmom} is the total angular momentum $J_{q,g}$ carried by the quarks or gluons.

There are two widely-used formulations of the decomposition of the total angular momentum of the proton.
\begin{itemize}
    \item The Jaffe and Manohar (JM) decomposition~\cite{Jaffe:1989jz}\index{spin decomposition!Jaffe-Manohar} is
 \begin{equation}  \label{eq:JM}
J = J_q + J_g = \frac{1}{2} \Delta \Sigma + L_q^{JM} + \Delta{G} + L_g,
\end{equation}
where $ \frac{1}{2} \Delta \Sigma$ and $\Delta{G}$ are the quark and gluon spin contributions, and $L_q^{JM}$ and $L_g$ are the quark and gluon orbital angular momentum (OAM) contributions. In this case, the energy momentum tensor is defined in the canonical form, not the Belinfante form in Eq.~\eqref{eq:EMT}.
This form is derived in the infinite momentum frame in light-cone gauge where $A^+ =0$. Furthermore, while $\Delta{G}$ can be extracted from high energy experiments, it can not be obtained from a matrix element based on a local operator, rather only from nonlocal correlation functions that are separated along the light-cone. The OAM contributions can be determined from GTMDs~\cite{Lorce:2011kd,Zhao:2015kca,Engelhardt:2017miy} and a corresponding lattice calculation is discussed in Sec.~\ref{sec:GTMD_OAM_lattice}, cf.~also Refs.~\cite{Engelhardt:2017miy,Engelhardt:2020qtg}.

\item The Ji decomposition~\cite{Ji:1996ek}\index{spin decomposition!Ji} is
\begin{equation}  \label{eq:Ji}
J = \frac{1}{2} \Delta \Sigma + L_q^{Ji} + J_g,
\end{equation}
where $\frac{1}{2}\Delta \Sigma$ is the same quark spin as in Eq.~\eqref{eq:JM}, $L_q^{Ji}$ is the quark OAM which can
be obtained from subtracting the quark spin from the quark angular momentum $J_q$ in Eq.~\eqref{eq:angmom}, and $J_g$ is the total gluon angular momentum contribution. Both $J_q$ and $J_g$ can be obtained from the gravitational form factors in Eq.~\eqref{eq:angmom} with the Belinfante energy-momentum tensor in Eq.~\eqref{eq:EMT}. 
Each term in Eq.~\eqref{eq:Ji} is gauge invariant and  defined covariantly by a local operator. The quark OAM can also be calculated as a GTMD observable, which is discussed in Secs.~\ref{sec:GTMD_OAM} and \ref{sec:GTMD_OAM_lattice}, cf.~also Refs.~\cite{Engelhardt:2017miy,Engelhardt:2020qtg}; the GTMD approach allows for a continuous, gauge-invariant interpolation between the Jaffe-Manohar and Ji definitions of OAM.

\end{itemize}

Except for the quark spin, these two decompositions have  different operators for the quark and gluon angular momenta.
This has lead the community to explore their physical meanings, possible relations among them, and respective realizations in experiments for many years.  See Refs.~\cite{Leader:2013jra,Liu:2015xha} for an overview and Ref.~\cite{Ji:2020ena} for a recent review.

The so-called ``proton spin crisis''\index{proton spin crisis} arose from the experimental observation in DIS experiments~\cite{Ashman:1987hv} that the quark spin $\Delta \Sigma$, defined through 
\begin{equation} 
\langle p(P,S)|A_i^0 |p(P,S) \rangle = S_{i} \Delta \Sigma,
\label{eq:singletaxial}
\end{equation}
where $A_i^0$ is the flavor-singlet axial-vector current $ A_i^0 =\sum_{f = u,d,s}\overline{\psi}_f \gamma_i \gamma_5 \psi_f$
contributes only $\sim 30-40\%$ to the total proton spin~\cite{deFlorian:2009vb}. This result is at odds with expectation of the quark model where the proton spin is saturated by the sum of the quark spins. 

Several LQCD calculations  of the forward flavor-singlet axial-vector current matrix element in Eq.~\eqref{eq:singletaxial} have been carried out. Because of the flavor-singlet nature, these calculations necessarily involve the disconnected insertion of the current in nucleon three-point functions which is numerically challenging and is usually undertaken using a stochastic noise estimator such as in Ref.~\cite{Dong:1993pk}. For a recent compilation and evaluation of lattice calculations of $\Delta \Sigma$ and the individual flavor contributions $\Delta u, \Delta d$ and $\Delta s$, see Refs.~\cite{Lin:2017snn,Aoki:2019cca}. It is found that the disconnected insertion contributions are negative which makes the total $g_A^0$ to be $\sim 0.3 - 0.4$.

Analyses~\cite{deFlorian:2014yva,Nocera:2014gqa} of the high-statistics 2009 STAR~\cite{Djawotho:2013pga} and PHENIX~\cite{Adare:2014hsq} {experiments at RHIC} showed evidence of nonzero gluon helicity distribution, $\Delta g(x,Q^2)$, in the proton. For $Q^2=10$ GeV$^2$, the gluon helicity distribution was found to be positive over the region $0.05\leq x \leq0.2$. However, outside this region the results have very large uncertainties that preclude definitive conclusions on $\Delta G=\int_0^1 dx \Delta g(x,Q^2)$. 

The gap between the light-front formulation of  $\Delta G$ and the Euclidean metric of lattice calculations has prevented direct calculation. However, it has been shown in Ref.~\cite{Ji:2013fga} that the matrix elements of appropriate equal-time local operator, when boosted to the infinite momentum frame, are the same as those of the gauge-invariant but nonlocal operator on the light-cone that defines $\Delta G$. However, it is found~\cite{Hatta:2013gta}  that the infinite boost ($P_z \rightarrow \infty$) and infinite
loop momentum ($k_{\mu} \rightarrow \infty$) limit in the renormalization of the operator do not commute. Since lattice calculation can only be carried out at finite nucleon momentum $P_z$, the large momentum effective field theory (LaMET)~\cite{Ji:2013fga,Ji:2014gla,Ji:2014lra} is formulated to match the finite $P_z$ matrix elements to those at the infinite momentum frame perturbatively.   In particular, a lattice calculation of large momentum matrix elements of the local operator $\vec{S}_G = \int d^3 x\, {\rm Tr} (\vec{E}\times \vec{A}_{\rm phys})$ is needed, where $\vec{A}_{\rm phys}$ satisfies the non-Abelian transverse condition $\mathcal{D}^i A_{\rm phys}^i = 0$~\cite{Chen:2008ag}. Noting that $A_{\rm phys}^i$ is related to $A_{c}^i$ in the Coulomb gauge via a gauge transformation~\cite{Zhao:2015kca}, matrix elements of the gluon spin operator
\begin{equation}
\vec{S}_G = \int d^3 x\, {\rm Tr} (\vec{E}\times \vec{A}_{\rm phys}) = \int d^3 x\, {\rm Tr} (\vec{E}_c\times \vec{A}_c)
\end{equation}
can be calculated with both $\vec{E}$ and $\vec{A}$ in the Coulomb gauge. A first lattice calculation in Ref.~\cite{Yang:2016plb}, when extrapolated to the infinite momentum limit, determined $\Delta G = 0.251(47)(16)$, which suggests that the gluon spin contributes about half of the proton spin. However in this calculation, the finite piece in the one-loop LaMET matching coefficient is very large, which indicates a  convergence problem for the perturbative series even after resummation of large logarithms and further investigation is required.
\index{lattice QCD calculations!matching}

An alternative way to determine $\Delta G$ is to calculate the polarized gluon distribution function $\Delta g(x,Q^2)$  through the quasi-PDF approach that is discussed below and then integrate to obtain $\Delta G$. 

Besides the quark and glue spins, there are quark and gluon orbital angular momenta (OAM) as part of the proton spin. The OAM can be extracted experimentally from GPDs and 
GTMDs~\cite{Liu:2015xha}. Ji's quark OAM $L_q^{Ji}$ can be obtained from the form factors of the
energy-momentum tensor (EMT) by subtracting the spin
from $J_q^{Ji}$ in Eq.~\eqref{eq:angmom}~\cite{Mathur:1999uf,Hagler:2003jd,Bratt:2010jn}, and the calculation of Jaffe-Manohar's
$L_q^{JM}$ and $L_g$ on the lattice from GTMDs has been formulated in Refs.~\cite{Zhao:2015kca,Engelhardt:2017miy}.
More details on OAM and lattice calculations of TMD and GTMD observables are discussed in Secs.~\ref{sec:lattice_tmd_calcs} and \ref{sec:GTMD_OAM_lattice}.

When the normalized and renormalized quark angular momentum $Z_{T,q} J_q^{\overline{MS}}$ is calculated, the quark OAM can be obtained by subtracting the quark spin from it. The nonperturbative renormalization of the EMT operator has been carried out in the context of proton mass decomposition~\cite{Yang:2018nqn} where the quark and gluon momentum fractions $\langle x\rangle_q$ and $\langle x\rangle_g$ are calculated. It would be essential to have lattice calculations of the momentum and angular momentum fractions  of the quarks and gluons in the nucleon with both renormalization and normalization taken into account.

\begin{figure}[t!]   
\centering
{
\includegraphics[width=0.4\hsize]{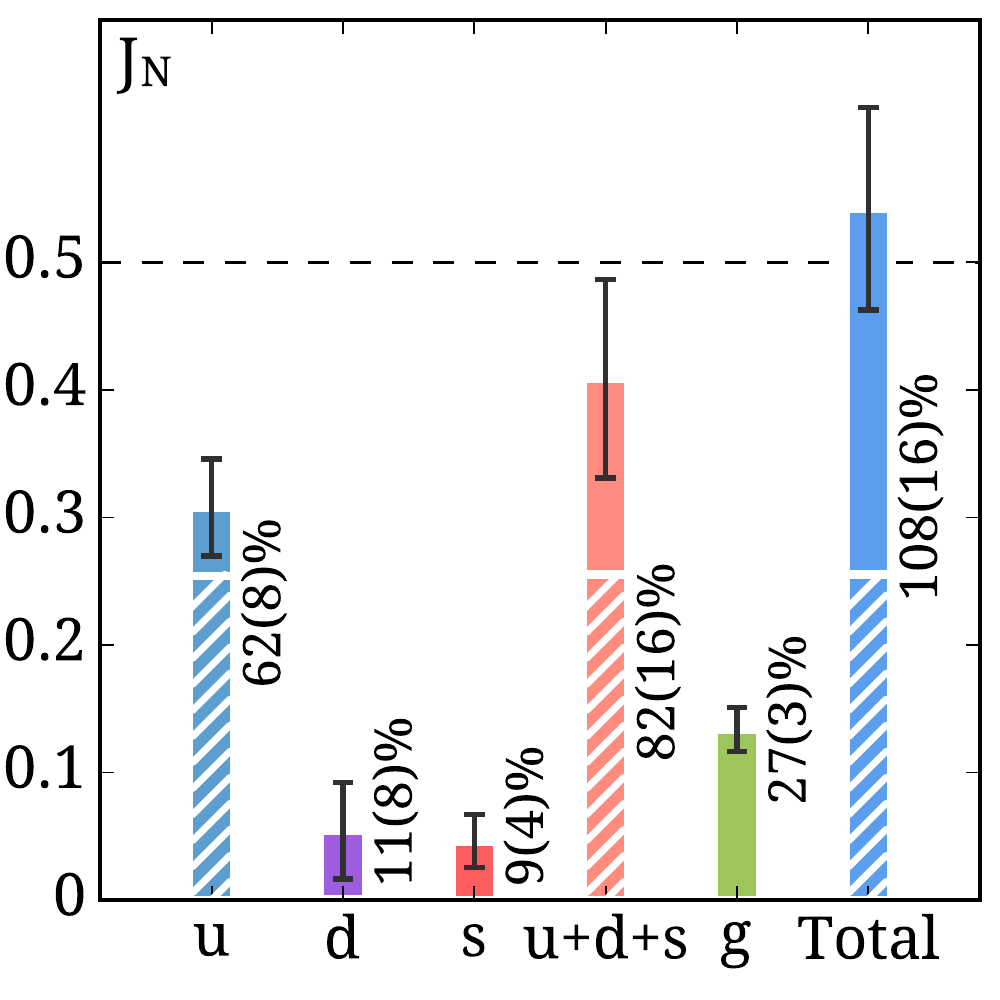}
  \label{nf2}}
{
\includegraphics[width=0.5\hsize]{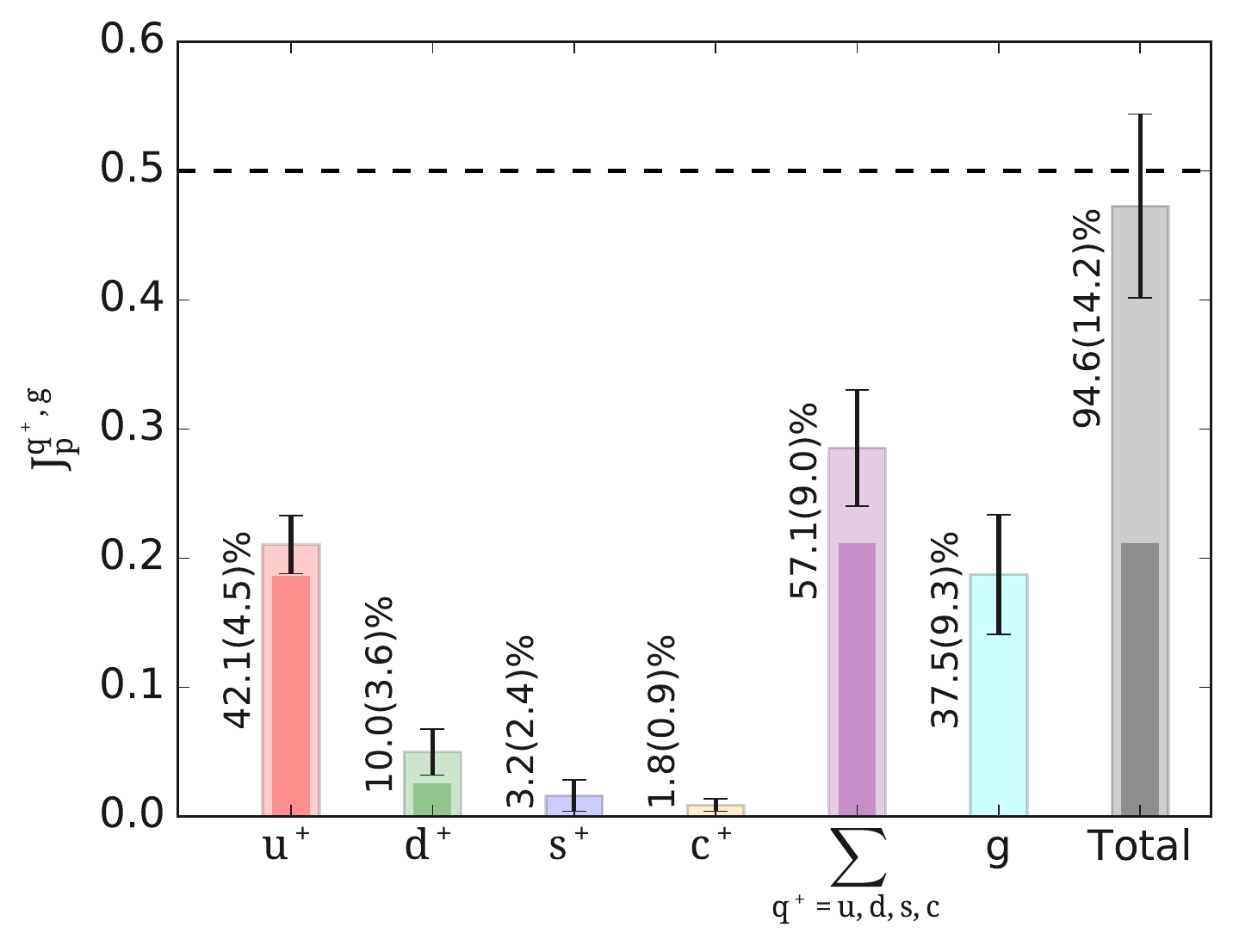}
  \label{nf2+1+1}}
\caption{Proton spin decomposition in terms of the angular momentum $J_q$ for the $u,d$ and $s$ quarks and 
the gluon angular momentum $J_g$ in Ji's decomposition. Left panel is for $n_f =2$ calculation~\cite{Alexandrou:2017oeh}
and right panel is for $n_f = 2+1+1$ calculation. Plot taken from Ref.~\cite{Alexandrou:2020sml}. \label{fig.nf2-4}}
\end{figure}
There have been lattice calculations to tackle Ji's
proton spin decomposition\index{spin decomposition!Ji} in Eq.~\eqref{eq:Ji}. The quark angular momenta $J_q^{Ji}$ for the $u,d$ and $s$ quarks and the gluon angular momentum $J_g$ are plotted in the left panel of Fig.~\ref{fig.nf2-4} for the $n_f =2$ case~\cite{Alexandrou:2017oeh}
and right panel for the $n_f = 2+1+1$ case~\cite{Alexandrou:2020sml}.

\begin{figure}[t!]     
\centering
{
\includegraphics[width=0.4\hsize]{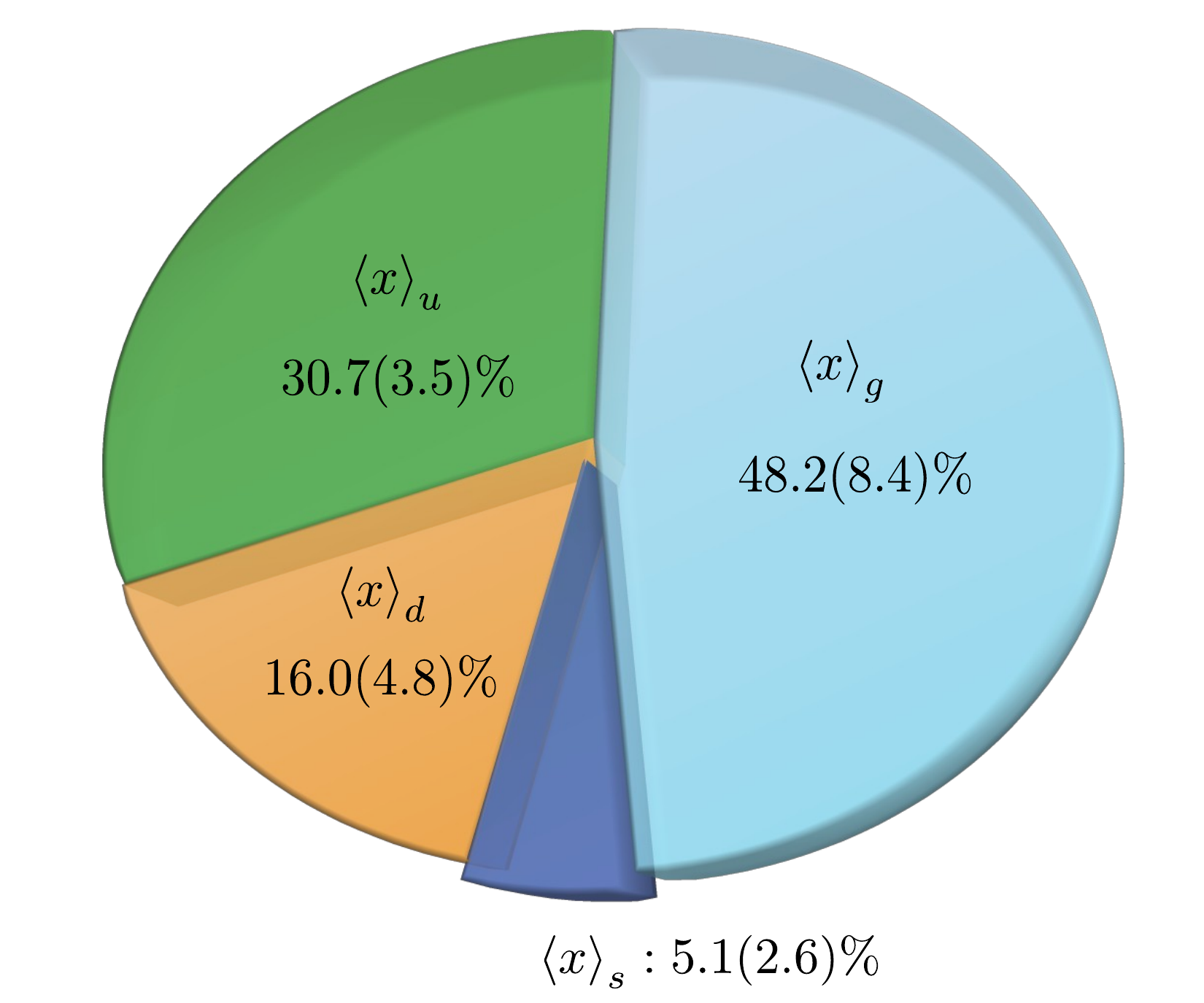}
  \label{x}}
{
\includegraphics[width=0.45\hsize]{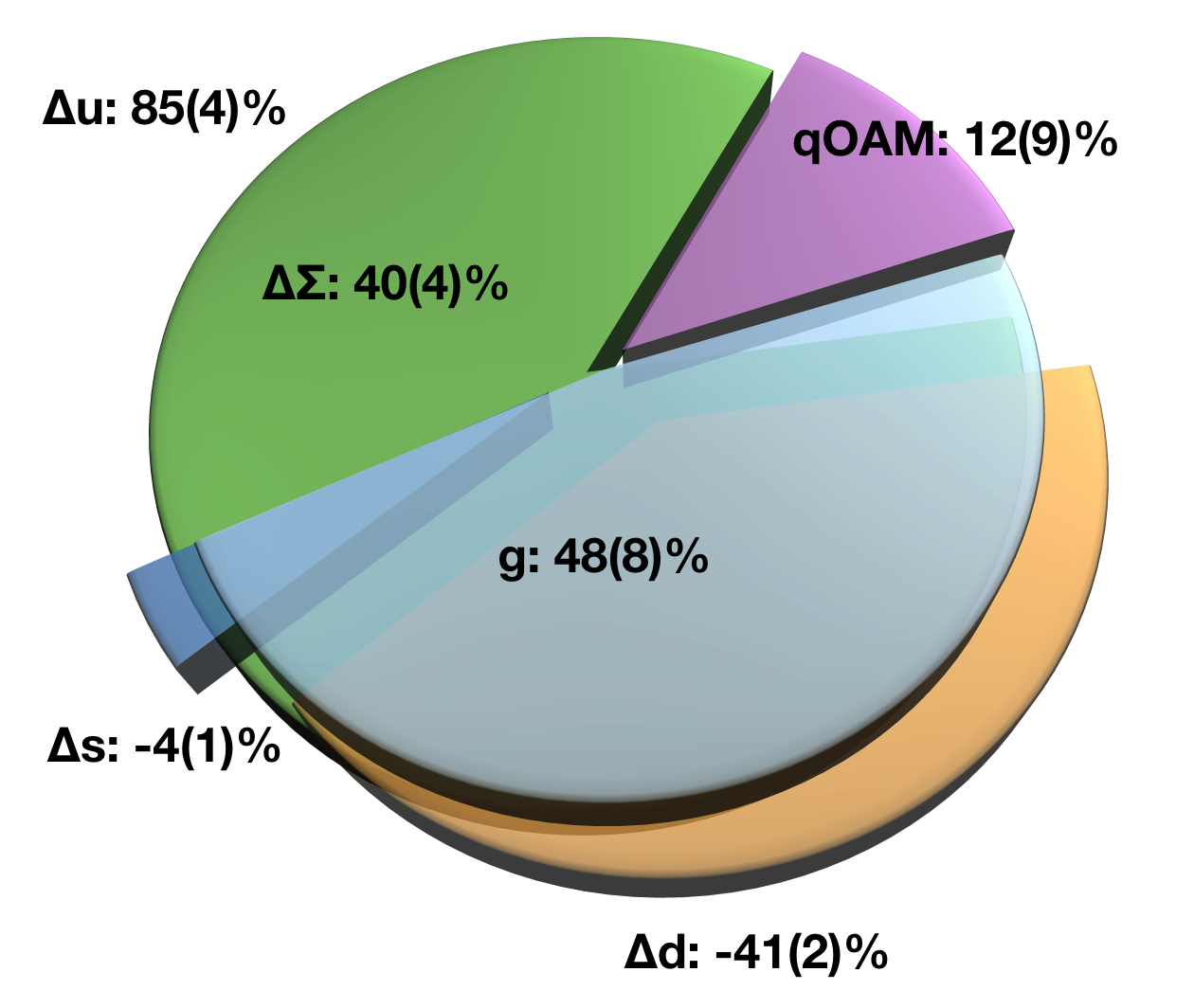}
  \label{2+1_AM}}
\caption{Left panel: The momentum fractions for the $u,d,s$ quarks and the gluon at $\mu = 2$ GeV from a lattice calculation of $n_f = 2+1$~\cite{Yang:2018nqn}. Right panel: The quark spins $\Delta u, \Delta d, \Delta s$, the total quark spin $\Delta \Sigma$, the gluon angular momentum $J_g$, and quark orbital angular momentum $L_q^{Ji}$ from the
same $n_f = 2+1$ configurations. Plot taken from Ref.~\cite{Yang:2019dha}. \label{fig.nf2+1}}
  \end{figure}

The momentum fractions $\langle x\rangle$ for the quarks and gluons are plotted in the left panel of Fig.~\ref{fig.nf2+1}. The quark spins for the $u,d,s$ flavors, the total quark spin $\Delta \Sigma$, the gluon
angular momentum $J_g$, and the quark orbital angular momentum $L_q^{Ji}$ are plotted in the right panel. They are
from lattice calculations with $2+1$ flavors~\cite{Yang:2018nqn,Yang:2019dha}

\subsection{Structure Functions and PDFs}
\label{sec:lattice:xdep}

There are several approaches to calculating the structure
functions and PDFs on the lattice. Calculating the structure functions from the hadronic tensor~\cite{Liu:1993cv,Liu:1998um,Liu:1999ak,Aglietti:1998mz,Detmold:2005gg,Liu:2016djw} or from the Compton amplitude~~\cite{Chambers:2017dov,Can:2020sxc} require confronting an inverse problem, either through a Laplace transform or from reconstruction of the moments.
Over the past few years, several new approaches have been proposed to directly extract the $x$-dependence of PDFs on the lattice. In this section, these approaches are  reviewed  since methods to access TMDs and TMD related quantities are built upon them. When the PDFs
are successfully calculated with these approaches, their moments can be cross-checked against those from the three-point functions as discussed in Sec.~\ref{sec:lattice:structure}.

\subsubsection{Hadronic tensor}  \label{sec:hadronic:tensor}

Since deep inelastic scattering measures the absorptive part of the 
Compton scattering amplitude, i.e. the hadronic tensor $W_{\mu\nu}$, it is the imaginary part of the forward amplitude and
can be expressed in terms of the current-current commutator in the nucleon,
\begin{equation}
 W_{\mu\nu} = \frac{1}{4\pi} \int d^4 x e^{i q\cdot x}\langle p(P,S)|[J_{\mu} (x), J_{\nu}(0)] | p(P,S)\rangle 
\end{equation}

Being related to an inclusive reaction, the hadronic tensor includes all intermediate states 
\begin{equation}   \label{eq:w}
W_{\mu\nu}(q^2, \nu) = \frac{1}{4\pi}  \sum_n \int \prod_{i =1}^n \left[\frac{d^3 p_i}{(2\pi)^3 2E_{pi}}\right]  \langle p(P,S)|J_{\mu}(0)|n\rangle
\langle n|J_{\nu}(0) | p(P,S)\rangle(2\pi)^3 \delta^{(4)} (p_n - P - q) . 
\end{equation}
It has been shown~\cite{Liu:1993cv,Liu:1998um,Liu:1999ak,Aglietti:1998mz,Detmold:2005gg,Liu:2016djw,Chambers:2015bka,Chambers:2017dov} that the 
hadronic tensor $W_{\mu\nu}(q^2, \nu)$ can be obtained from the Euclidean path-integral formalism through an inverse problem. 
In this case, one first defines an Euclidean hadronic tensor ${W}_{\mu\nu}^E(\vec{q},\tau)$ which involves a 3-momentum transfer and is defined from a ratio of a 4-point function to a 2-point function  

\begin{eqnarray}  \label{wmunu_tilde}
{W}_{\mu\nu}^E(\vec{q},\tau) &=&
 \frac{E_p }{m_N} \frac{{\rm Tr} (\Gamma^+ G_{pWp})}{{\rm Tr} (\Gamma^+ G_{pp})}
 \begin{array}{|l} \\  \\  t _f -t_2,  t_1 - t_0 \gg 1/\Delta E_p \end{array} \nonumber \\  
 &=& \frac{1}{2} \sum_S\left\langle p(P,S)\left|\sum_{\vec{x}} \frac{e^{-i \vec{q}\cdot \vec{x}}}{4\pi}
J_{\mu}(\vec{x},\tau) J_{\nu}(0,0)\right|p(P,S)\right\rangle,
\end{eqnarray}
where $G_{pWp}$ is the 4-pt function for the current-current correlator in the nucleon and $G_{pp}$ is the 2-pt function for the nucleon correlator. $\Gamma^+$ is the projector for the positive parity nucelon state as before, and  $t_0$ and $t_f$ are the source and sink times of the nucleon interpolation field, $t_1$ and the $t_2$ are the
 current insertion time slices, and $\tau = t_2 - t_1$. $\Delta E_p$ is the energy gap between the nucleon energy $E_p$ and that of the first excited state with the same quantum numbers  (i.e., the threshold of a nucleon and a pion in the $p$-wave). 
 Inserting intermediate states, 
${W}_{\mu\nu}^E(\vec{q},\tau)$ becomes
\begin{equation}   \label{eq:wtilde}
{W}_{\mu\nu}^E(\vec{q}^{\,2},\tau)
= \frac{1}{2} \sum_S \frac{1}{4 \pi}\sum_n \left(\frac{ m_N}{ E_n}\right)  \langle p(P,S)|J_{\mu}(0)|n\rangle
\langle n|J_{\nu}(0) | p(P,S)\rangle \delta (\vec{p}_n - \vec{P} - \vec{q})e^{- (E_n - E_p) \tau}.
\end{equation}
This approach does not require the initial nucleon momentum $\vec{P}$ to be large, and no renormalization is needed for the hadronic tensor constructed from conserved currents.

Formally, to recover the delta function $\delta(E_n - E_p - \nu)$ in Eq.~\eqref{eq:angle_integral} in  Minkowski space, one can consider the inverse Laplace transform with $\tau$ being treated as a dimensionful continuous variable
\begin{equation}  \label{eq:wmunu} 
W_{\mu\nu}(q^2,\nu) = \frac{1}{2m_Ni} \int_{c-i \infty}^{c+i \infty} d\tau\,
e^{\nu\tau} {W}_{\mu\nu}^E(\vec{q}, \tau),
\end{equation} 
with $c > 0$.  However, as there is no lattice data on the integration contour parallel to the imaginary $\tau$ axis, this can not be done. Instead, one can address it through an inverse problem with Laplace transform.  The task is to ``solve'' the inverse problem in order to find  the spectral density $W_{\mu\nu}(q^2,\nu)$ from its spectral representation in the Laplace integral
\begin{equation}   \label{eq:inverse}
  {W}_{\mu\nu}^E(\vec{q},\tau) = \int e^{- \nu\, \tau} \,W_{\mu\nu}(q^2,\nu)\, d \nu.
\end{equation}
Some  approaches to this inverse problem include the Maximum Entropy Method (MEM)~\cite{Bryan1990,Jarrell:1996rrw},
 the Bayesian Reconstruction (BR)~\cite{Burnier:2013nla} and 
 the Backus-Gilbert Method (BG)~\cite{Backus1968}.  All three approaches have been investigated in Ref.~\cite{Liang:2019frk}.

In order to study parton physics, another challenge of this approach is to access energy transfers such that the calculation can access the DIS region. To determine how large a $\nu$ is needed for DIS, one can look at $W$, the total invariant mass of the hadronic state for the nucleon target at rest
\begin{equation}
W^2 = (q+p)^2 = m_p^2 - Q^2 + 2 m_p \nu
\end{equation}
The global analyses of the high energy lepton-nucleon and Drell-Yan experiments to extract the parton distribution functions (PDFs) usually make a cut with $W^2 > 10\, {\rm GeV^2}$ to avoid the
 elastic and inelastic regions. Thus, to be in the DIS region, the energy transfer $\nu > 7$ GeV is needed  
 for $Q^2 = 4\, {\rm GeV^2}$ and a small lattice spacing (e.g. $a \le 0.04$ fm) is needed to reach such high energy 
 excitations~\cite{Liang:2019frk}. 
 Another aspect of the hadronic tensor is that it is valid in all  energy ranges from elastic scattering to inelastic scattering and on to deep inelastic scattering.
 It can be employed to study the neutrino-nucleon scattering
 cross section at low energies, such as at neutrino energies relevant to the DUNE experiments. 

\begin{figure}[t!]
  \centering
  \includegraphics[width=.5\textwidth,page=5]{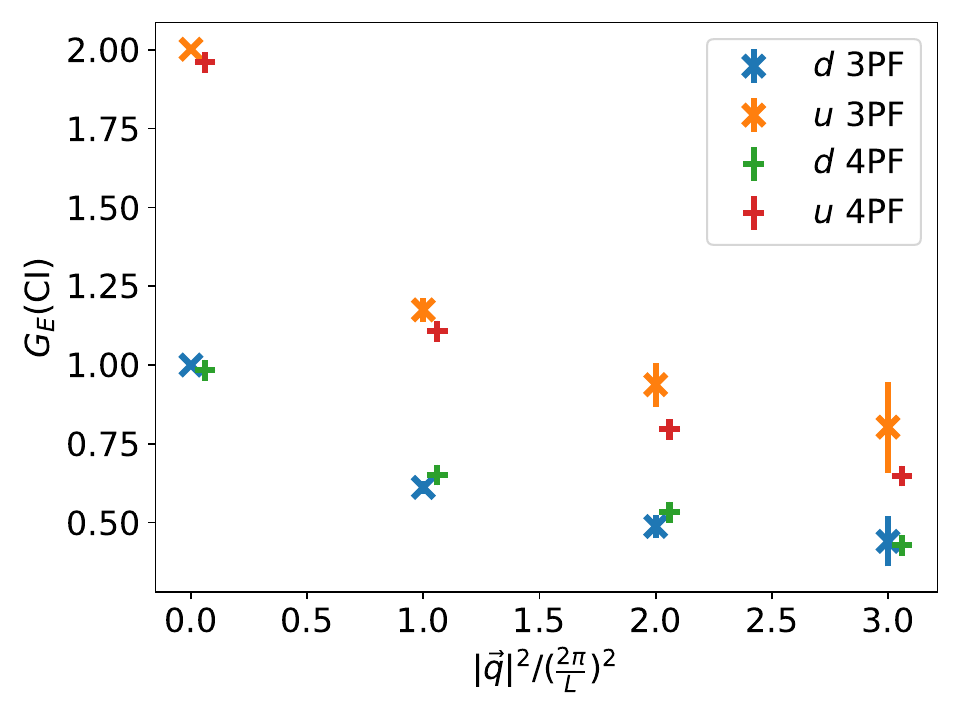}
  \caption{Comparison of the electric form factors (connected insertions only) calculated by using three-point functions (3PF) and four-point functions (4PF) for the lowest four momentum transfers (including zero) and for both $u$ and $d$ quarks. Plot taken from Ref.~\cite{Liang:2019frk}.
  \label{fig:check}}
\end{figure}

 For elastic scattering, the structure function from the hadronic tensor, a 4-point function, is the sum of the products of elastic nucleon form factors for the currents involved~\cite{Liang:2019frk}.
The elastic form factors can be calculated in  nucleon
3-pt functions. As a check of the lattice calculation, the
structure function for the $J_{4} J_{4}$ correlator is calculated for the elastic scattering case which can be obtained from the ground state of $W_{44}^E$ in Eqs.~(\ref{eq:wtilde}) and (\ref{eq:inverse}) with $E_n - E_p= \sqrt{m_N^2 + |\vec{q}|^2}$ for $\vec{P} =0$.
This should correspond to the product to the electric form factor $G_E(q^2)$. Fig.~{\ref{fig:check}} shows the lattice
calculation of the electric form factors (connected insertions only) calculated by means of the  3-point functions for both $u$ and $d$ quarks and compared that those from the corresponding structure functions of the elastic scattering from the hadronic tensor. As can be seen, they agree for the $u$ and $d$ quark within errors.

\vspace{0.75cm}
\noindent
\textit{\textbf{{Auxiliary heavy quark}}}
\vspace{0.25cm}

Another method to access PDFs is via fermion  bilinear currents which couple light quarks with a purely fictitious valence heavy quark~\cite{Detmold:2005gg,Detmold:2020lev,Detmold:2021uru}, and have the form:
\begin{equation}
  \label{eq:DLcurrent}
  J_{\Psi,\psi}^{\mu}(x)=\overline{\Psi}(x)\Gamma^\mu \psi(x) +  \overline{\psi}(x)\Gamma^\mu \Psi(x)\,, 
\end{equation}
with $\psi(x)$ ($\Psi(x)$) the light (fictitious heavy) quark field, and a general Dirac structure $\Gamma^\mu$. This approach has the advantage that in the continuum limit it removes power divergent mixing with lower-dimensionality operators, which is unavoidable with standard techniques. Also, the presence of the heavy fictitious quark results in suppression of the long-range correlations between the currents and  higher-twist contamination. One of the technical constraints of this method is the requirement of small lattice spacings (considerably smaller than 0.1 fm), so that heavy quark discretization effects are controllable.

In this approach, the Euclidean Compton  tensor with heavy quark currents is written as
\begin{eqnarray}
\label{eq:DLtensor}
  T^{\mu\nu}_{\Psi,\psi}(P,q)&\equiv& \sum_{S} \langle p(P^\prime, S)|\hspace*{0.03cm}
  t^{\mu\nu}_{\Psi,\psi}(q)|p(P, S)\rangle \\\nonumber
&=& \sum_{S} \int d^4x\ {\mathrm e}^{i q\cdot x} \langle p(P^\prime, S)|
  T \left [ J_{\Psi,\psi}^{\mu}(x) J_{\Psi,\psi}^{\nu}(0)\right ] |p(P, S) \rangle\,,
\end{eqnarray}
with certain constraints on the momenta so that the momentum transfer $q$ is $\simeq {\cal O}(m_\Psi)$, and $(p_M+q_M)^2<(m_\Psi+{\Lambda_{\rm QCD}})^2$, the latter in Minkowski space as indicated by the subscript ${}_M$. With such constraints, analytic continuation of the hadronic tensor to Euclidean spacetime is achieved with $q_4 \to i q_0$. LQCD calculations of the tensor $ T^{\mu\nu}_{\Psi,\psi}$ can be extrapolated to the continuum and then related to moments of the PDFs via the OPE. The Wilson coefficients entering this  OPE depend on the heavy quark mass and are presented at one loop in Ref.~\cite{Detmold:2021uru}.

This approach has been recently studied in Ref.~\cite{Detmold:2018kwu,Detmold:2021qln} for the pion distribution amplitude (DA), using three quenched ensembles at different lattice spacings so that the continuum limit can be taken. The Euclidean hadronic tensor defined by 
\begin{equation}
{U}^{[\mu\nu]}_A(q,p){=}\int_{\tau_{{\mathrm{min}}}}^{\tau_{{\mathrm{max}}}} \mbox{ } d\tau ~ e^{iq_4\tau}~R_3^{[\mu\nu]}(\tau,\vec{q},\vec{p})\,,
\end{equation}
where the quantity $R^{\mu\nu}_3(\vec{p},\vec{q},\tau)$ is accessed by spatial Fourier transform of three-point functions of two heavy-light currents separated in spacetime with temporal separation of $\tau$. One can then obtain the moments of the DA by variation of $q_4$.

\vspace{0.75cm}
\noindent
\textit{\textbf{{OPE without OPE}}}
\vspace{0.25cm}

An alternative method to access hadronic structure functions is proposed in Refs.~\cite{Chambers:2017dov,Can:2020sxc} and uses elements of earlier ideas~\cite{Martinelli:1998hz,Aglietti:1998ur}. In such an approach one calculates the time-ordered product of two currents, which have a small enough spacetime separation for perturbation theory to be valid, but at the same time large enough to suppress discretization effects.

In this method one utilizes the forward Compton amplitude, $T_{\mu\nu}$, which can be decomposed into the structure functions $F_1$ and $F_2$. For example, the $\mu=\nu=3$ component
\begin{equation}
\label{eq:OPE1}
T_{33}(p,q) = \sum_{n=2,4,\cdots}^\infty 4\omega^n \int_0^1 dx\, x^{n-1} F_1(x,q^2) = 4\omega \int_0^1 dx\, \frac{\omega x}{1-(\omega x)^2} F_1(x,q^2) \,, \quad 
\end{equation}
where $ \omega = 2p\cdot q/q^2$.
This can be used to extract the moments of $F_1(x,q^2)$. By construction, the computation of $T_{33}$ requires four-point functions, which makes the calculation computationally very demanding. To avoid such difficulties, the Feynman-Hellmann \index{Feynman-Hellmann method} method~\cite{Horsley:2012pz} is utilized in Ref.~\cite{Chambers:2017dov}, which adds a term $\lambda \mathcal{J}_{3f}(x;q)$ in the QCD Lagrangian, where $\lambda$ is a parameter which is treated perturbatively, and $\mathcal{J}_{3f}(x;q)=Z_V\cos(\vec{q}\cdot\vec{x})\; e_f \,\bar{\psi}_f(x)\gamma_3 \psi_f(x)$.
Taking the derivative of the nucleon energy with respect to $\lambda$ gives $T_{33}$ as shown in Ref.~\cite{Chambers:2017dov}:
\begin{equation}
\label{eq:FH}
T_{33}(p,q) = -2 E_\lambda(p,q)\, \frac{\partial^2}{\partial\lambda^2}  E_\lambda(p,q)\,\big|_{\lambda=0} \,.
\end{equation}

\begin{figure}[t!]
\begin{center}
\includegraphics[width=0.45\textwidth]{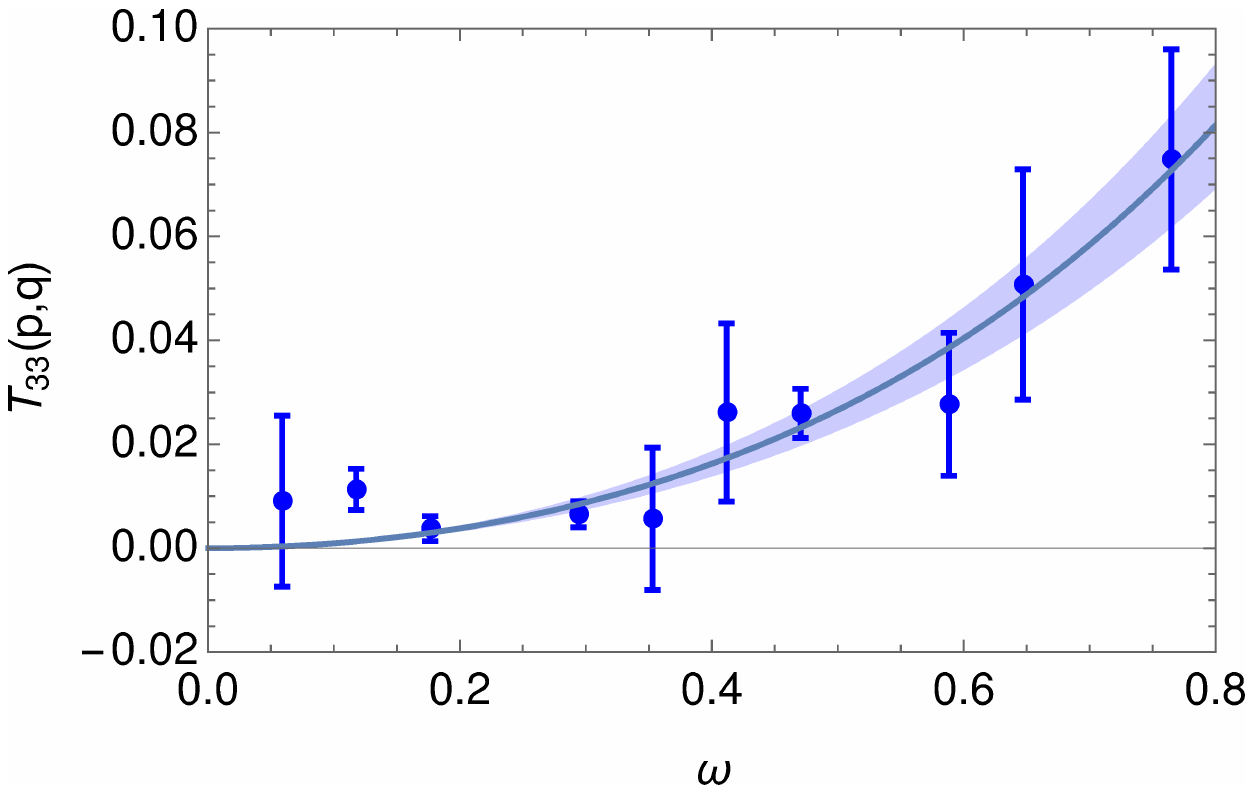}
\vspace*{-0.5cm}
\end{center}  
\caption{Compton amplitude obtained from the lattice computation of Ref.~\cite{Chambers:2017dov} (blue points), fitted to a sixth order polynomial (solid line). Plot taken from Ref.~\cite{Chambers:2017dov}.
}
\label{fig:OPEwOPE}
\end{figure}

Note that the addition of the extra term in the Lagrangian requires dedicated simulations for each value of $\lambda$, with multiple values needed in order to take the $\lambda\to0$ limit. 
This approach has been investigated numerically in Refs.~\cite{Chambers:2015bka,Chambers:2017dov,Can:2020sxc}; Fig.~\ref{fig:OPEwOPE} shows an example of the hadronic tensor and fits to it to extract moments of the structure function $F_1(x)$.

\subsubsection{Quasi-PDFs\index{quasi-PDF} in large-momentum effective theory approach}
\label{sec:lamet}

As mentioned in \sec{latt_def_qtmd}, the large-momentum effective theory (LaMET)\index{large-momentum effective theory (LaMET)} has been proposed to calculate the TMD and its soft function from lattice QCD.
In fact, LaMET is a generic approach to calculate parton physics on the light-cone from Euclidean LQCD  correlation functions (see recent reviews in Refs.~\cite{Cichy:2018mum,Ji:2020ect,Constantinou:2020pek}). The idea is to approximate a light-cone observable by boosting a static or time-independent quasi observable (a Euclidean space  matrix element of a operator) to a large-momentum frame, and then perform a systematic expansion of the latter in inverse powers of the momentum to extract the observable of interest.

One of the first applications of LaMET has been to the lattice calculation of gluon helicity contribution to the proton spin~\cite{Ji:2013fga,Hatta:2013gta,Yang:2016plb}, as was discussed in Sec.~\ref{decom_spin}. A method to calculate the canonical parton orbital angular momentum has also been proposed in Refs.~\cite{Ji:2014lra,Zhao:2015kca} based on the LaMET approach. However, the best-studied development so far is in the lattice calculation of the $x$-dependence of collinear PDFs.

We focus first on the unpolarized quark PDF as an illustrative example. One can calculate the so called quasi-PDF \index{quasi-PDF} which is defined from an equal-time spatial correlator~\cite{Ji:2013dva},
\begin{align}\label{eq:qpdf}
\hat{f}(y,P^z,\Lambda) &\equiv \!\int\!\! \frac{dz}{4\pi} \, e^{iyP^z z} 
\! \big\langle p(P,S) \big| \tilde{O}_{\Gamma}(z,0)\Big|_{\Lambda}\big|p(P,S)\big\rangle \,,\\
\tilde{O}_\Gamma(z,0)&=\bar{\psi} (z) \Gamma W_{\hat{z}}(0;0,z)\psi(0)\,,\label{eq:quasio}
\end{align}
where $\Gamma=\gamma^t$ or $\gamma^z$, $\Lambda$ is the ultraviolet (UV) momentum cutoff, and the spacelike Wilson line is
\begin{align}\label{eq:qwl}
W_{\hat{z}}(0;0,z) &= P \exp\bigg(ig_0 \int_0^{z}dz' A^z(z'\hat{z}) \bigg) \,.
\end{align}
Under a Lorentz boost in the longitudinal direction, the spatially separated operator $\tilde{O}_\Gamma(z,0)$ will approach the light-cone direction, and the quasi-PDF depends dynamically on the hadron momentum $P^z$ accordingly. Unlike the light-cone PDF where  Bjorken-$x$ is restricted to the interval $x\in[-1,1]$, the quasi-PDF has a support for $y\in (-\infty,\infty)$. 

In LQCD, the UV divergences are regulated in momentum space by the  cutoff $\Lambda\sim a^{-1}$, and one can only calculate hadron matrix elements at $P^z\ll \Lambda$. The light-cone PDF, however, corresponds to the limit $P^z\gg \Lambda$ which does not commute with the lattice regularization. Nevertheless, as long as $P^z\gg \Lambda_{\rm QCD}$, the relative magnitudes of $P^z$ and $\Lambda\sim 1/a$  do not affect the contributions from infrared (IR) degrees of freedom, thus the difference between them are in the UV region which is under perturbative control. This separation gives rise to a factorization formula relating the PDF and the quasi-PDF, which has been studied extensively in the literature~\cite{Xiong:2013bka,Ma:2014jla,Ma:2017pxb,Izubuchi:2018srq}.
In the $\MSbar$ scheme, the factorization formula for the non-singlet quasi-PDF has been proven~\cite{Ma:2014jla,Ji:2020ect} and rigorously derived as~\cite{Izubuchi:2018srq}
\begin{align} \label{eq:msfac}
\hat{f} \left( y, P^z,\mu\right)
= \int_{-1}^1 \frac{dx}{|x|}\: C \Bigl(
\frac{y}{x}, \frac{\mu}{|x|P^z} \Bigr)\: f \left(x,\mu\right)+ \ldots
\,,
\end{align}
where $\mu$ is the $\overline{\rm MS}$ scale, and for $y<0$, $q(x)=-\bar{q}(-x)$ with $\bar{q}$ being the antiquark. $C$ is the perturbative matching coefficient which depends on the logarithm of the parton momentum. The relative power corrections indicated by the ellipsis include target mass $M$ corrections which are known to all orders of $M^2/(P^z)^2$~\cite{Chen:2016utp}, as well as higher-twist contributions of ${\cal O}(\Lambda_{\rm QCD}^2/(y^2P_z^2),\Lambda_{\rm QCD}^2/((1-y)^2P_z^2))$, whose enhancement at $y=0$ and $y=1$ has been argued in Refs.~\cite{Braun:2018brg,Ji:2020brr}. Similar factorization formulas have also been rigorously derived for the gluon and singlet quark quasi-PDFs~\cite{Wang:2019tgg}, as well as for the quasi-GPDs~\cite{Liu:2019urm}.
\index{lattice QCD calculations!matching}

The above factorization formula can be inverted order by order in perturbation theory~\cite{Ji:2020ect},
\begin{align} \label{eq:msfac2}
f \left( x, \mu\right)
= \int_{-\infty}^\infty \frac{dy}{|y|}\: C^{-1} \Bigl(
\frac{x}{y}, \frac{\mu}{|y|P^z} \Bigr)\: \hat{f} \left(y,P^z,\mu\right)+ \ldots
\,,
\end{align}
where $C^{-1}$ is the inverse of the matching coefficient $C$, and the power corrections take similar forms to Eq.~\eqref{eq:msfac} except that $y$ is replaced by $x$. Therefore, Eq.~\eqref{eq:msfac2} provides a point-by-point determination of the PDF, which has controlled power corrections within a range of $x$, i.e., $x\in [x_{\rm min}, x_{\rm max}]$.
Based on Eq.~\eqref{eq:msfac2}, the systematic procedure to calculate the PDFs can be laid out as the following: 1) calculate the lattice matrix elements for the bare quasi-PDF; 2) renormalize the quasi-PDF and extrapolate to the physical quark mass, continuum and infinite volume limits; 3) perturbatively match to the light-cone PDF in the $\overline{\rm MS}$ scheme; 4) estimate the power corrections. For the target mass correction, step 3) can be done before step 4), while for the genuine higher-twist correction, one can extrapolate to the $P^z\to\infty$ limit for each $x$ after matching. The renormalization and matching procedures are closely related to each other, which will be elaborated in the following subsection.

\vspace{0.75cm}
\noindent
\textit{\textbf{{{Renormalization and matching}}}}
\vspace{0.25cm}
\index{lattice QCD calculations!matching}

The self-energies of spacelike Wilson lines are subject to linear power divergences with a lattice regulator and must be renormalized before one can take the continuum limit of matrix elements of the operators in Eq.~\eqref{eq:quasio}. The renormalization of Wilson lines has been well studied in th literature~\cite{Dotsenko:1979wb,Craigie:1980qs,Dorn:1986dt}. For the nonlocal quark bilinear operator $\tilde O_\Gamma(z,0)$ in Eq.~\eqref{eq:quasio}, it has been rigorously proven~\cite{Ji:2017oey,Green:2017xeu,Ishikawa:2017faj} that it can be multiplicatively renormalized in coordinate space as
\beq \label{eq:ren}
\tilde{O}^B_\Gamma(z,0) = \left[\bar{\psi}_i (z) \Gamma W_{\hat{z}}(0;0,z)\psi_i(0)\right]^B = Z_{\tilde O}\ e^{-\delta m|z|} \left[\bar{\psi_i} (z) \Gamma W_{\hat{z}}(0;0,z)\psi_i(0)\right]^R\,.
\eeq
where $i$ is an unsummed quark flavor index,
$\delta m$ has mass dimension one and absorbs the linear power divergence from the Wilson-line self-energy, and $Z_{\tilde O}$ includes additional logarithmic divergences associated with the end points as well as the wavefunction renormalizations. The superscripts $B$ and $R$ indicate bare and renormalized quantities.
Moreover, the renormalization is independent of the quark flavor and Dirac matrix $\Gamma$, and there is no mixing between quark and gluon sectors.
The proof has also been generalized to the gluon quasi-PDF, as the nonlocal gluon bilinear operator is shown to be multiplicatively renormalizable~\cite{Zhang:2018diq,Li:2018tpe} up to a contact term~\cite{Zhang:2018diq}.

\index{lattice QCD calculations!nonperturbative renormalization}
Based on Eq.~\eqref{eq:ren}, one can renormalize the quasi-PDF in lattice perturbation theory or perform a nonperturbative renormalization. In the former case, there have been a number of one-loop studies of the quasi-PDF with lattice regularization~\cite{Ishikawa:2016znu,Xiong:2017jtn,Constantinou:2017sej}, and then one can match the lattice regularized quasi-PDF to the $\MSbar$ scheme in continuum theory. However, lattice perturbation theory is known to converge slowly, while renormalization of $\tilde O_\Gamma$ or the quasi-PDF is now available in the $\MSbar$ scheme up to three loops~\cite{Chen:2020arf,Chen:2020iqi,Li:2020xml,Chen:2020ody,Braun:2020ymy}.

In order to implement a nonperturbative renormalization, one approach is to determine $\delta m$ in Eq.~\eqref{eq:ren} independently from the static quark-antiquark potential~\cite{Musch:2010ka,Ishikawa:2016znu,Zhang:2017bzy,Green:2017xeu,Green:2020xco,Alexandrou:2020qtt}. After the nonperturbative subtraction of linear power divergences from the quasi-PDF, one can then renormalize the logarithmic divergences using either lattice perturbation theory~\cite{Ishikawa:2016znu} or other nonperturbative schemes for local composite operators~\cite{Green:2017xeu}.

As has been mentioned in Sec.~\ref{sec:lqcd}, one can perform a nonperturbative renormalization in the regularization-independent momentum subtraction (RI/MOM) scheme~\cite{Martinelli:1994ty}, which has been widely used for local composite operators. Since the nonlocal quark bilinear $\tilde{O}_\Gamma(z,0)$ is of the lowest mass dimension, it does not have power divergent mixings resulting from the reduced symmetry of the hypercubic lattice. Therefore, it remains multiplicatively renormalizable. A RI/MOM scheme can be implemented as first proposed in Refs.~\cite{Constantinou:2017sej,Alexandrou:2017huk}, and an independent formalism to renormalize the quasi-PDF in the RI/MOM scheme and match it to the PDF was developed and carried out in Refs.~\cite{Stewart:2017tvs,Chen:2017mzz}. However, because of the breaking of chiral symmetry in certain fermion actions, there is additional operator mixing between  $\tilde{O}_\Gamma(z,0)$~\cite{Constantinou:2017sej,Green:2017xeu,Chen:2017mie} for different $\Gamma$ structures. For $\Gamma=\gamma^z$, the operator mixes with the scalar case with $\Gamma= I$ at ${\cal O}(a^0)$, while for $\Gamma=\gamma^t$, there is no mixing at ${\cal O}(a^0)$. To avoid such mixing, $\Gamma=\gamma^t$should be chosen instead of $\Gamma=\gamma^z$.

In the RI/MOM scheme, one defines the renormalization factor $Z_{\rm OM}(z,p^R_z,\mu_R,a)$ by imposing a momentum subtraction condition on the matrix element of $\tilde{O}_{\gamma^t}(z,0)$ at some kinematic point for each value of $z$,
\beq \label{eq:Z}
\left.Z^{-1}_{\rm OM}(z,p^R_z,\mu_R,a)\langle p|\tilde{O}^B_{\gamma^t}(z,0)|p\rangle \right|_{\tiny\begin{matrix}p^2=-\mu_R^2 \\ \!\!\!\!p_z=p^R_z\end{matrix}}= \langle p|\tilde{O}_{\gamma^t}(z,0)|p\rangle_{\rm tree}\,,
\eeq
where the condition is defined at an off-shell quark momentum $p^R_\mu=(p^R_t,p^R_x,p^R_y,p^R_z)$, and the renormalization scale $\mu_R^2=-p_R^2\gg \Lambda_{\rm QCD}^2$.
In the LQCD calculation, the momentum $p_\mu^R=(p^R_4,p^R_x,p^R_y,p^R_z)$ is Euclidean, and $\mu_R^2 = (p^R_4)^2+(p^R_x)^2+(p^R_y)^2+(p^R_z)^2\ge (p_z^R)^2$.
The bare matrix element $\langle p|\tilde{O}^B_{\gamma^t}(z,0)|p\rangle$ is defined from the amputated Green's function $\Lambda_{\gamma^t}(p,z)$ of $\tilde{O}^B_{\gamma^t}(z,0)$ with a  projection operator $\cal P$,
\begin{align}
\Lambda_{\Gamma}(z,p,a) & \equiv \left[S^{-1}_0(p,a)\right]^\dagger \sum_{x,y}e^{ip\cdot (x-y)} \langle 0| T\left[\psi_0(x) \tilde O^B_{\Gamma}(z,0) \bar{\psi}_0(y)\right]|0\rangle S^{-1}_0(p,a)\,,
\end{align}
\begin{align}
\sum_s\langle p,s|\tilde{O}^B_{\gamma^t}(z,0)|p,s\rangle &= \mbox{Tr}\left[ \Lambda_{\gamma^t}(z,p){\cal P}\right]\,.
\end{align}
Then, the bare hadron matrix element of $\tilde{O}^B_{\gamma^t}(z,0)$
\begin{align}
\tilde{h}_B(z,P_z,a^{-1}) 
  = \frac{1}{2P^t} 
  \langle P |\tilde{O}_{\gamma^t}^B(z,0)| P \rangle
\end{align}
is renormalized in coordinate space as
\beq \label{eq:rimomh}
\tilde{h}_R(z,P_z, p^R_z,\mu_R)=\lim_{a\to0}Z^{-1}_{\rm OM}(z,p^R_z,a^{-1},\mu_{\tiny R})\tilde{h}_B(z,P_z,a^{-1}) \,,
\eeq
where $\tilde{h}_R(z,P_z, p^R_z,\mu_R)$ is the renormalized matrix element. At finite lattice spacing, $\tilde{h}_R$  still has discretization errors, so calculations at different spacings are required extrapolate to the continuum limit, as indicated in Eq.~\eqref{eq:rimomh}.

Given the renormalized matrix elements in position space, the quasi-PDF can be constructed through the Fourier transform of Eq.~\eqref{eq:rimomh}.
The next step is to match the renormalized quasi-PDF to the $\overline{\rm MS}$ PDF. According to UV regularization independence, the RI/MOM matrix elements should be the same in dimensional regularization with $D=4-2\epsilon$,
\beq
\lim_{a\to0}Z^{-1}_{\rm OM}(z,p^R_z,\mu_R,a^{-1})\tilde{h}_B(z,P_z,a^{-1}) = \lim_{\epsilon\to0}Z^{-1}_{\rm OM}(z,p^R_z,\mu_R,\mu,\epsilon)\tilde{h}_B(z,P_z,\mu,\epsilon)\,,
\eeq
where $\tilde{h}_B(z,P_z,\mu,\epsilon)$ and $Z_{\rm OM}(z,p^R_z,\mu_R,\mu,\epsilon)$ are the bare matrix element and RI/MOM renormalization factor in the continuum theory, and $\mu$ is the UV scale introduced in dimensional regularization.
In this way, the matching coefficients can be computed in continuum perturbation theory, which is much easier than that in lattice regularization.

There are two strategies developed to carry out the matching for the RI/MOM quasi-PDF~\cite{Constantinou:2017sej,Stewart:2017tvs}. One is to convert $\tilde{h}_R(z,P_z, p^R_z,\mu_R)$ from RI/MOM to the $\overline{\rm MS}$ scheme first,
\begin{align}	\label{eq:conversion}
	\tilde{h}_{\overline{\rm MS}}(z,P_z, \mu) = \tilde{h}_R(z,P_z, p^R_z,\mu_R)\frac{Z_{\rm OM}(z,p^R_z,\mu_R,\mu,\epsilon)}{ Z_{\overline{\rm MS}}(\epsilon)}\,,
\end{align}
where $Z_{\overline{\rm MS}}$ is the $\MSbar$ renormalization factor, and $Z_{\rm OM}$ has been calculated at one-loop order~\cite{Constantinou:2017sej} in $z$-space and at two-loop order in the Fourier space of $z$~\cite{Chen:2020ody}. Then, one transforms the $\overline{\rm MS}$ matrix element $\tilde{h}_{\overline{\rm MS}}(z,P_z, \mu)$ into momentum space to obtain the quasi-PDF and match the latter to the $\overline{\rm MS}$ PDF using Eq.~\eqref{eq:msfac}, where the matching coefficient has been calculated at two-loop order~\cite{Chen:2020arf,Chen:2020iqi,Li:2020xml,Chen:2020ody}. Since the conversion factor in Eq.~\eqref{eq:conversion} is logarithmically divergent as $|z|\to0$, and the $\overline{\rm MS}$ quasi-PDF does not satisfy vector current conservation~\cite{Izubuchi:2018srq}, it was also proposed that one can modify the $\overline{\rm MS}$ scheme renormalization constant by a perturbative factor that cancels the singular terms in the $|z|\to0$ limit and restores the conservation law. Such schemes include the ratio scheme in Refs.~\cite{Izubuchi:2018srq,Zhao:2018fyu} and the modified $\overline{\rm MS}$ (MMS) scheme in Ref.~\cite{Alexandrou:2019lfo}. Since the $\overline{\rm MS}$ matrix element $\tilde{h}_{\overline{\rm MS}}(z,P_z, \mu)$ should be independent of the RI/MOM scales $\mu_R$ and $p_z^R$, its remnant dependence on them can in principle be fitted as polynomial lattice discretization effects~\cite{Alexandrou:2017huk}.
This two-step matching procedure has been implemented in the lattice calculations of iso-vector quark PDFs in Refs.~\cite{Alexandrou:2017huk,Alexandrou:2018pbm,Alexandrou:2018eet,Alexandrou:2019lfo,Bhattacharya:2020cen,Alexandrou:2020uyt}.

The other strategy for matching the quasi-PDF in the RI/MOM scheme is more straightforward~\cite{Stewart:2017tvs}. First, one Fourier transforms the RI/MOM matrix element $\tilde{h}_R(z,P_z, p^R_z,\mu_R)$ to momentum space to obtain the quasi-PDF $\tilde{f}(x,P_z, p^R_z,\mu_R)$, and then directly matches it onto the $\overline{\rm MS}$ PDF~\cite{Stewart:2017tvs,Liu:2018uuj,Wang:2019tgg,Liu:2018tox} through the formula below,
\begin{align}
	f \left(x,\mu\right)
	= \int_{-\infty}^\infty \frac{dy}{|y|}\: C_{\rm OM}\left(\frac{x}{y},\frac{\mu_R}{p_z^R},\frac{yP_z}{\mu},\frac{yP_z}{p_z^R}\right)\: \hat{f} \left(y,P_z, p^R_z,\mu_R\right)+\ldots
	\,,
\end{align}
where $C_{\rm OM}$ is the matching coefficient and the ellipsis denotes the power corrections.

This strategy was implemented in the lattice calculations of the isovector quark PDFs in the proton and pion in Refs.~\cite{Chen:2017mzz,Lin:2017ani,Chen:2018fwa,Liu:2018uuj}, as well as in recent calculations in Refs.~\cite{Zhang:2020gaj,Fan:2020nzz,Zhang:2020dkn,Lin:2020ssv,Lin:2020fsj} 

Apart from the RI/MOM scheme, it has also been proposed to renormalize the operator $\tilde O_\Gamma(z,0)$ by forming ratios of bare matrix elements in different external states, for example, the $P^z=0$ hadron state~\cite{Orginos:2017kos} or the vacuum~\cite{Braun:2018brg,Li:2020xml} in the denominator, and the corresponding matching coefficients to light-cone PDF has been derived up to two-loop order~\cite{Li:2020xml}. Since all these matrix elements become nonperturbative at large $z$, such ratio schemes only work at small distances when $z\ll \Lambda_{\rm QCD}^{-1}$. Therefore, they are only applicable in coordinate space based on an equivalent short-distance factorization or OPE of the equal-time correlation in Eq.~\eqref{eq:qpdf}, such as the pseudo-distribution approach to be discussed below.

Since the factorization formula for the quasi-PDF is proven in the $\overline{\rm MS}$ scheme in momentum space, while the lattice renormalization is performed in the coordinate space, any scheme that can not be perturbatively matched to the $\MSbar$ scheme will affect the validity of factorization. 
In fact, both the ratio and RI/MOM schemes suffer from this issue, for their conversion factors to $\overline{\rm MS}$ include logarithms of $z^2$ that become IR at large $z$~\cite{Constantinou:2017sej,Izubuchi:2018srq}.
Additionally, the ratio and RI/MOM schemes could introduce nonperturbative effects at large $z$ which can not be controlled systematically.
In contrast, the Wilson-line mass-subtraction scheme avoids this issue. Nevertheless, lattice discretization effects at $z\sim a$ will obscure the continuum limit of the renormalized matrix element to reproduce the divergent $\ln z^2$ behavior at small $z$ in the $\MSbar$ scheme. Such discretization effects, however, are cancelled in the RI/MOM and ratio schemes, which leads to a finite $z\to0$ limit of the renormalized matrix elements. 

To reconcile the advantages and disadvantages of the above schemes, the hybrid scheme was proposed in Ref.~\cite{Ji:2020brr} to renormalize the bare matrix elements at small and large $z$. At short distance $z\le z_{\rm S}$ where $z_{\rm S}\sim 0.2-0.3\ {\rm fm} \ll \Lambda_{\rm QCD}^{-1}$ is smaller than the distance at which the uncertainty in perturbation theory becomes too large, one uses either the RI/MOM or ratio scheme where the lattice discretization effects cancel; for $z>z_{\rm S}$, one uses the Wilson-line-mass-subtraction scheme, with the logarithmic renormalization factor determined by a continuity condition at $z=z_{\rm S}$. After the subtraction, one needs to match the lattice hybrid scheme to the continuum theory, which can be done using the method developed in Ref.~\cite{LatticePartonCollaborationLPC:2021xdx}. The $z_{\rm S}$-dependence will be cancelled by the perturbative matching in the final result. The perturbative matching for the hybrid renormalized quasi-PDF with ratio scheme at short distance has been derived at one-loop order~\cite{Ji:2020brr}.

\begin{figure}[t!]
\centering
\includegraphics[width=0.9\textwidth]{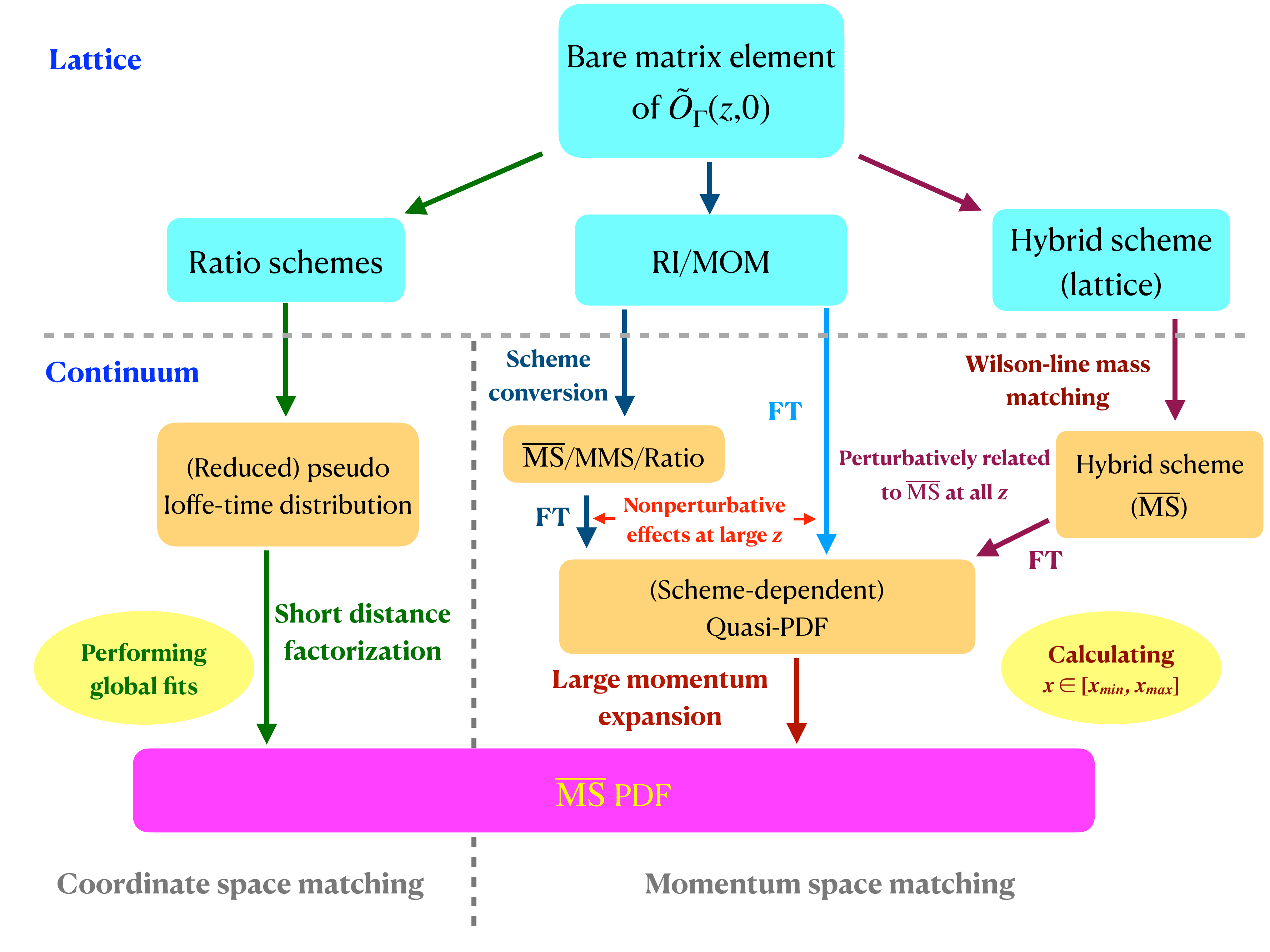}
\caption{Comparison of the ratio, RI/MOM and hybrid renormalization schemes in the lattice calculation of PDFs.
}
\label{fig:comparison_ren}
\end{figure}

Due to finite lattice size and decreasing signal-to-noise ratios at large $z$, the lattice results are only well determined for distances less than a truncation scale, $z_{\rm L}$, which causes difficulties in the Fourier transform required to obtain the quasi-PDF. Since the spacelike correlations in the $\overline{\rm MS}$ quantity must decay exponentially at large distance due to confinement, a physically motivated extrapolation model beyond $z_{\rm L}$ can be used to remove the unphysical oscillations in a truncated Fourier transform. In return this will introduce systematic uncertainties in the small-$x$ region, but it generally does not overlap with the region $x\in [x_{\rm min}, x_{\rm max}]$ where the LaMET expansion in Eq.~\eqref{eq:msfac2} has systematic control.
A comparison of ratio, RI/MOM and hybrid scheme analyses is shown in \fig{comparison_ren}.

There is a further way to renormalize the quasi-PDF on lattice, which is based on a redefinition of the quasi-PDF using the gradient flow method~\cite{Monahan:2016bvm}. The redefined quasi-PDF remains finite in the continuum limit, which is free from the power divergences on the lattice and can be perturbatively matched onto the $\overline{\rm MS}$ PDF~\cite{Monahan:2017hpu}.

\vspace{0.75cm}
\noindent
\textit{\textbf{{{Lattice Calculations}}}}
\vspace{0.25cm}

The LaMET methodology has been studied intensively on the lattice soon after its proposal~\cite{Lin:2014zya,Alexandrou:2014pna,Alexandrou:2015rja}. A lot of improvementshave been made regarding the renormalization, the matching, and the parameters of the ensembles employed. To date, lattice calculations are well beyond the exploratory phase, with investigations of twist-2 and twist-3 PDFs, as well as GPDs and TMDs. Systematic uncertainties such as excited-states contamination, volume effects, and cutoff effects are being addressed carefully. Also, the $x$-dependence of PDFs have been calculated directly at the physical quark masses~\cite{Alexandrou:2018pbm,Alexandrou:2018eet,Lin:2018pvv}.

\begin{figure}[t!]
\begin{center}
\includegraphics[width=0.45\textwidth]{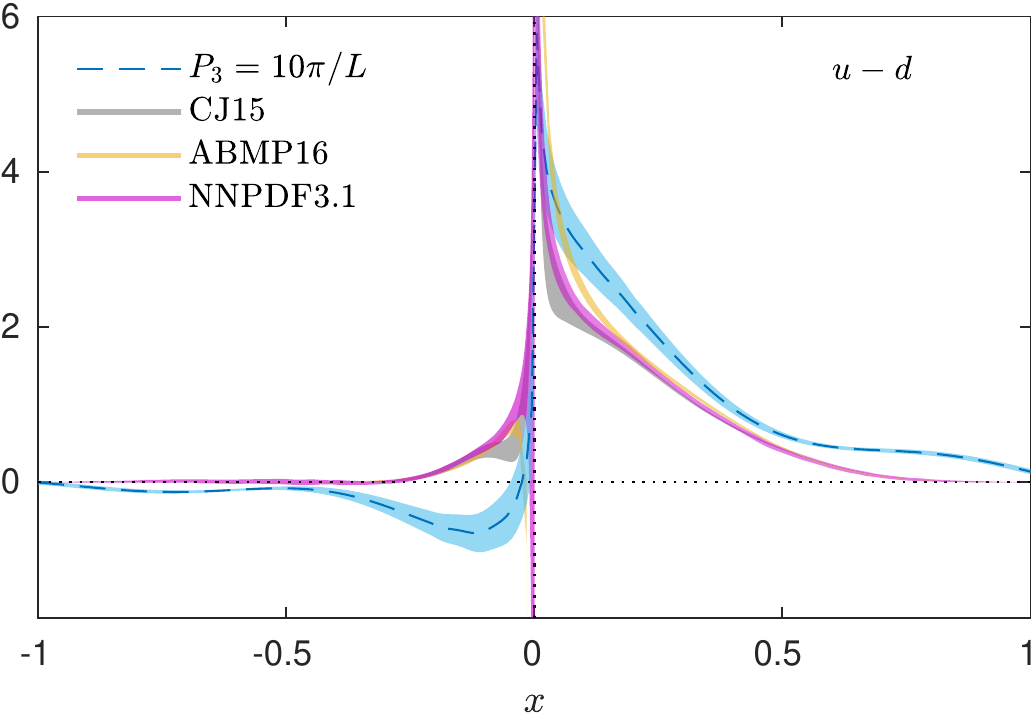}
\includegraphics[width=0.45\textwidth]{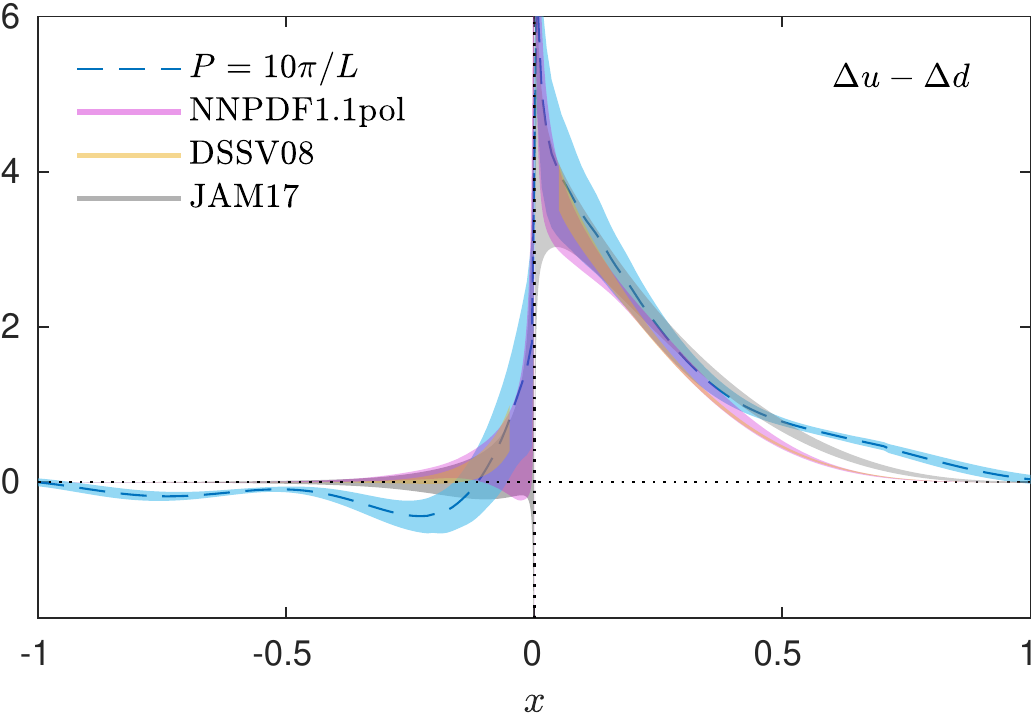}
\includegraphics[width=0.45\textwidth]{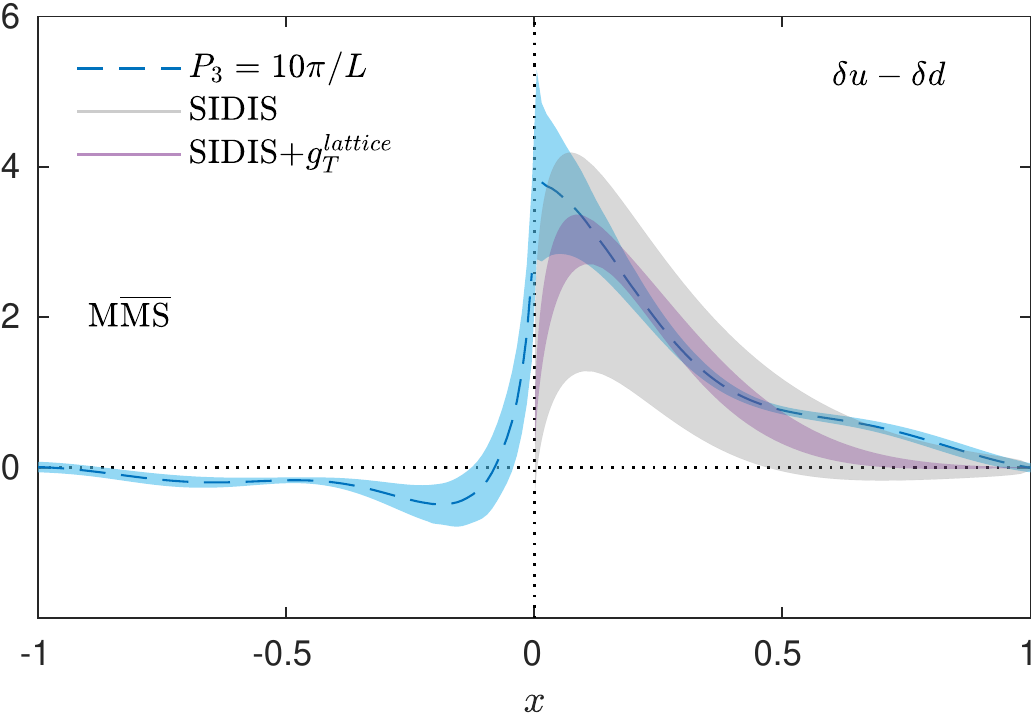}
\caption{The proton unpolarized (top left), helicity (top right) and transversity (bottom) PDFs at the physical point and $P_3=1.38$ GeV from Ref.~\cite{Alexandrou:2019lfo}. A comparison with with global fits~\cite{Alekhin:2017kpj,Ball:2017nwa,Accardi:2016qay,deFlorian:2009vb,Nocera:2014gqa,Ethier:2017zbq,Lin:2017stx} is also shown. Plot taken from Ref.~\cite{Constantinou:2020hdm}.}
\label{fig:PDFpheno}
\end{center}
\end{figure}

\begin{figure}[t!]
\centering
\includegraphics[scale=.975]{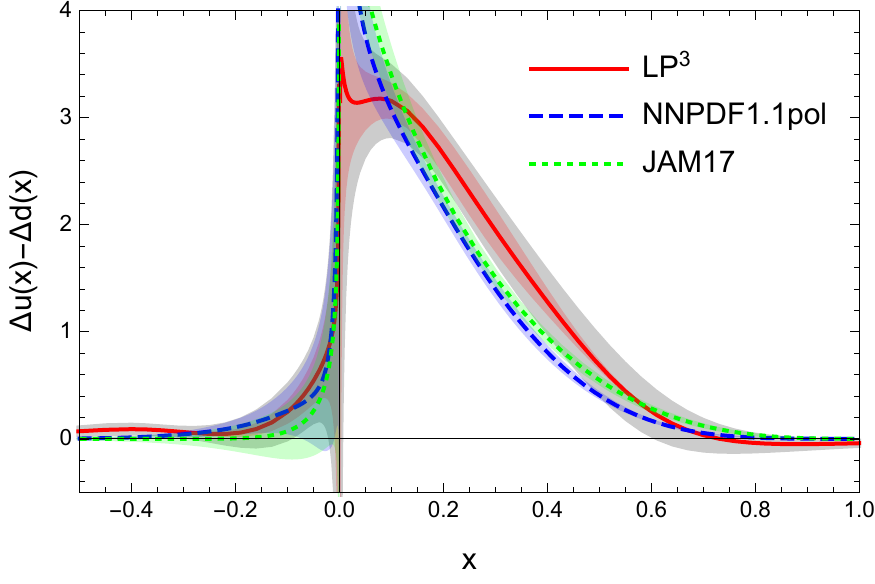}
\caption{The helicity PDF calculated in Ref.~\cite{Lin:2018pvv} using $P_3=3$~GeV (red curve), compared to global fits~\cite{Nocera:2014gqa,deFlorian:2014yva,Ethier:2017zbq}. Plot taken from Ref.~\cite{Lin:2018pvv}.}
\label{fig:finalPDF}
\end{figure}

The first complete calculations at the physical point for the unpolarized, helicity and transversity isovector PDFs appear in Refs.~\cite{Alexandrou:2018pbm,Alexandrou:2018eet}, followed by an analysis of selected sources of systematic uncertainties~\cite{Alexandrou:2019lfo}. These calculations use an $N_f=2$ ensemble of the twisted-mass~\cite{Frezzotti:2000nk,Shindler_2008} lattice discretization with physical light-quark mass and spatial extent of 4.5 fm. The results for all types of collinear PDFs using quasi-PDFs at $P_3=1.38$ GeV are shown in Fig.~\ref{fig:PDFpheno}. 
The helicity PDFs are extracted in Ref.~\cite{Lin:2018pvv} using a mixed action setup of clover fermions on a $N_f=2+1+1$ HISQ ensemble with spatial lattice extent $L\approx 5.8$~fm and a pion mass ${\approx}135$~MeV are shown in Fig.~\ref{fig:finalPDF}. 

In the aforementioned calculations at the physical quark masses, various nonperturbative renormalization schemes were applied as discussed above, followed by a matching kernel appropriate for the choice of renormalization. However, the calculations differ in the reconstruction of the $x$-dependence. This is an important aspect of the calculation that may introduce systematic uncertainties due to the limited number of lattice data entering the Fourier transform (FT). Refs.~\cite{Alexandrou:2018pbm,Alexandrou:2018eet,Alexandrou:2019lfo} use a standard discretized FT, while Ref.~\cite{Lin:2018pvv} applies the ``derivative method'' which relies on integration by parts and neglecting the surface term~\cite{Lin:2017ani}. While none of the methods overcomes the ill-defined inverse problem, the derivative method has been shown to lead to uncontrolled uncertainties in the small-$x$ region~\cite{Karpie:2019eiq,Alexandrou:2019lfo}.

\begin{figure}[t!]
    \centering
\includegraphics[scale=0.65]{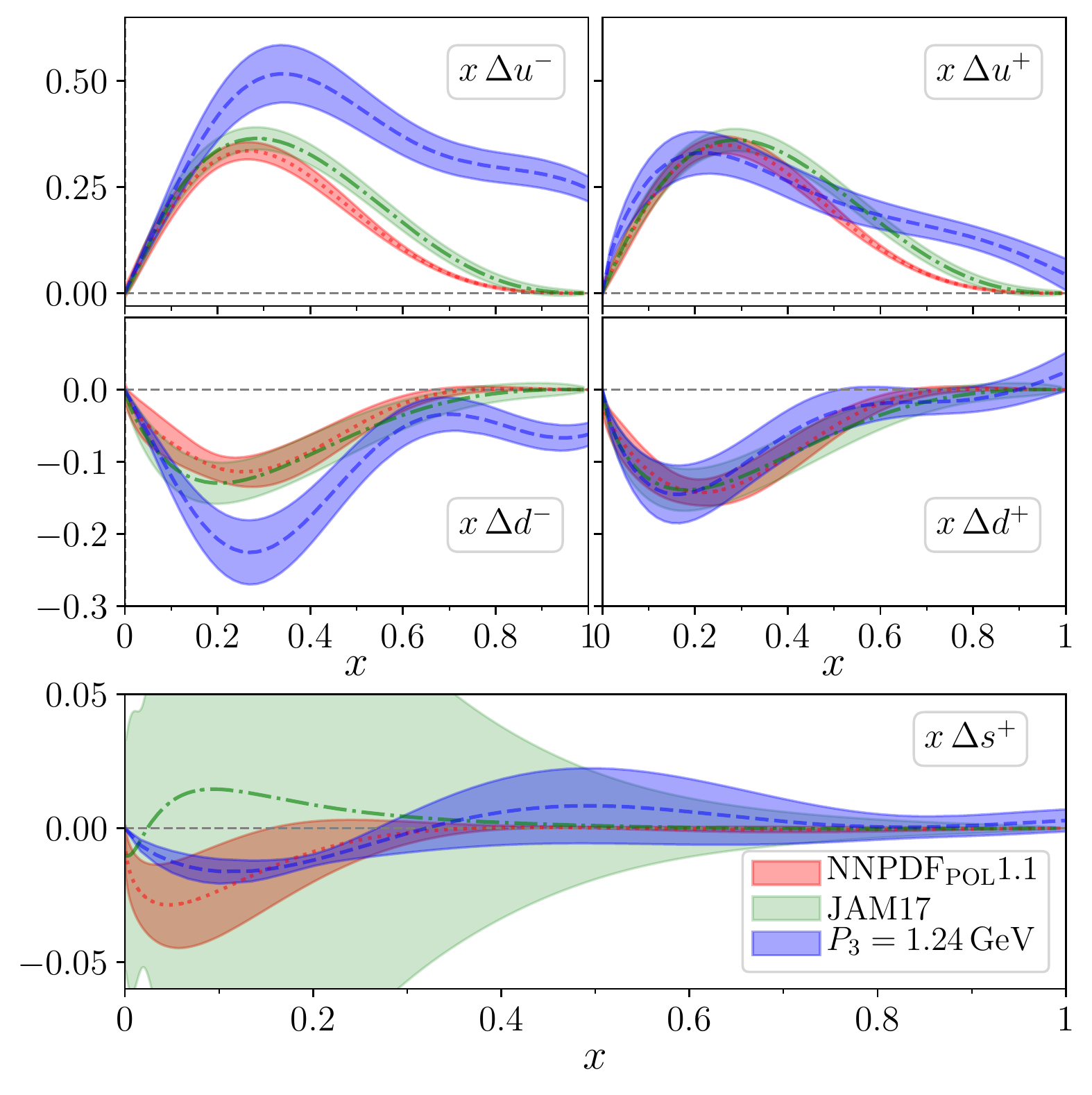}
    \caption{Lattice data on the $|x| \Delta u$ (top), $|x| \Delta d$ (center), and $|x| \Delta s^+$ (bottom) quark helicity PDFs (blue) with momentum boost $P_3$ = 1.24 GeV renormalized in the $\MSbar$ scheme at a scale of $2\,{\rm GeV}$. The global fits JAM17~\cite{Ethier:2017zbq} (green) and NNPDF$_{\rm POL}$1.1~\cite{Buckley:2014ana,Nocera:2014gqa} (red) are shown for comparison.  Plot taken from Ref.~\cite{Alexandrou:2020uyt}.}
    \label{fig:flavor_decomposition}
\end{figure}

The quasi-distributions formulation has been used to calculate the PDFs of other particles, such as the pion~\cite{Izubuchi:2019lyk,Gao:2020ito,Lin:2020ssv,Zhang:2020gaj}, kaon and $\Delta^+$~\cite{Chai:2020nxw}. Sources of systematic uncertainties using ensembles with quark masses larger than their physical values have been studied in Refs.~\cite{Lin:2019ocg,Alexandrou:2020qtt}. Another direction is the inclusion of the disconnected diagram contributions for the strange and charm unpolarized PDFs~\cite{Zhang:2020dkn} and the up, down and strange unpolarized, helicity and transversity PDFs~\cite{Alexandrou:2020uyt,Alexandrou:2021oih}. The flavor decomposition of the up and down quark PDFs, as well as the strange quark PDFs, are presented in Fig.~\ref{fig:flavor_decomposition} obtained in Ref.~\cite{Alexandrou:2020uyt} using an ensemble corresponding to a pion mass of 260 MeV. Results for $|x| \Delta q^+(x)\equiv |x| \left(\Delta q + \Delta\bar{q}\right)$ and $|x| \Delta q^-(x)\equiv |x| \left(\Delta q - \Delta\bar{q}\right)$ are shown for $q=u,\,d,\,s$, and compared with the JAM17~\cite{Ethier:2017zbq} and NNPDF$_{\rm POL}$1.1~\cite{Nocera:2014gqa,Buckley:2014ana} global fits. As can be seen, there is a tension for the case of $\Delta u^-$, and a mild disagreement for $\Delta d^-$. $\Delta s^+$ is compatible with the global fits and is more precise, suggesting a nonzero value for small values of $x$ which would be valuable input to global fits. 

\begin{figure}[t!]
\centering
\includegraphics[scale=0.56]{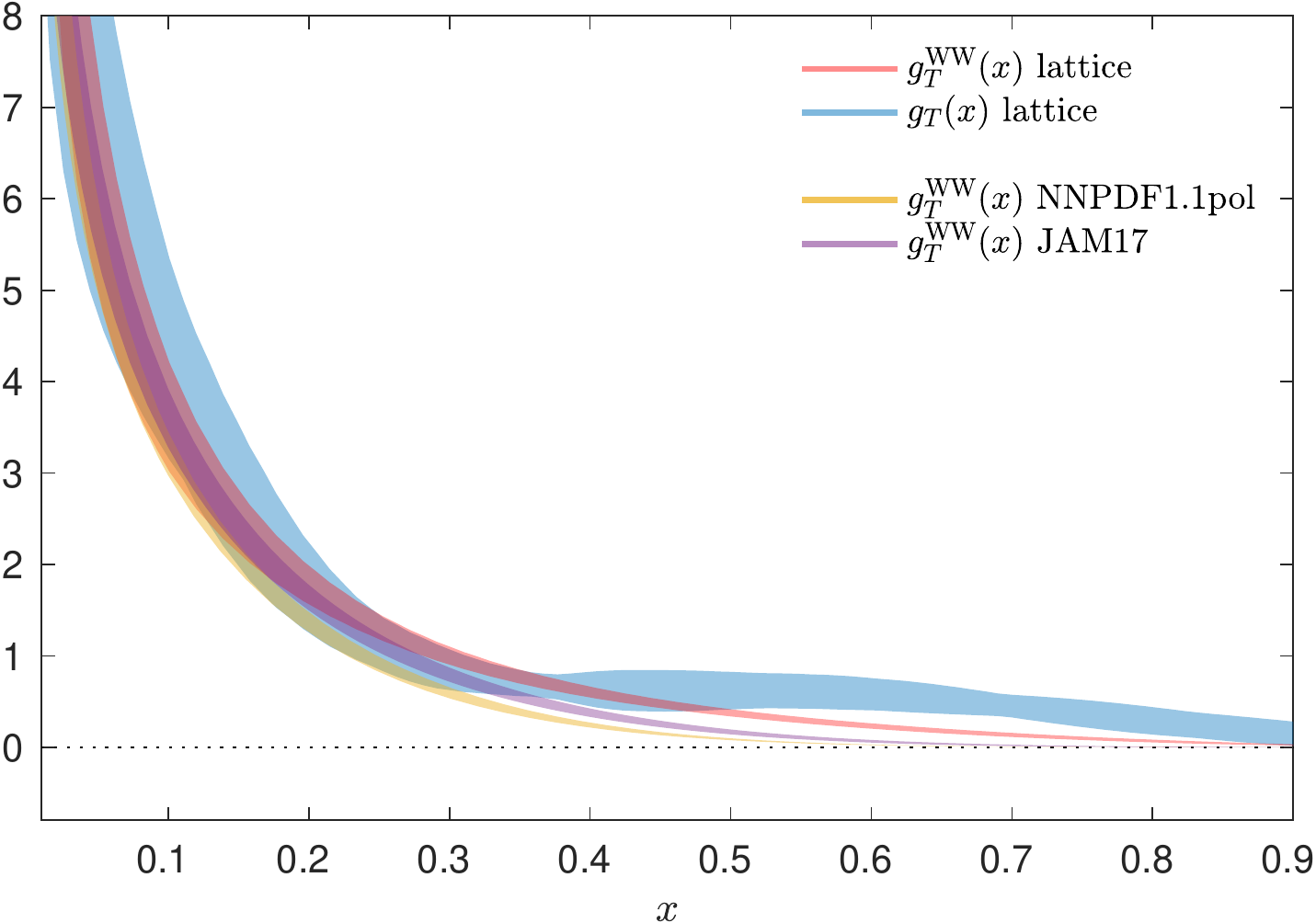}
\includegraphics[scale=0.56]{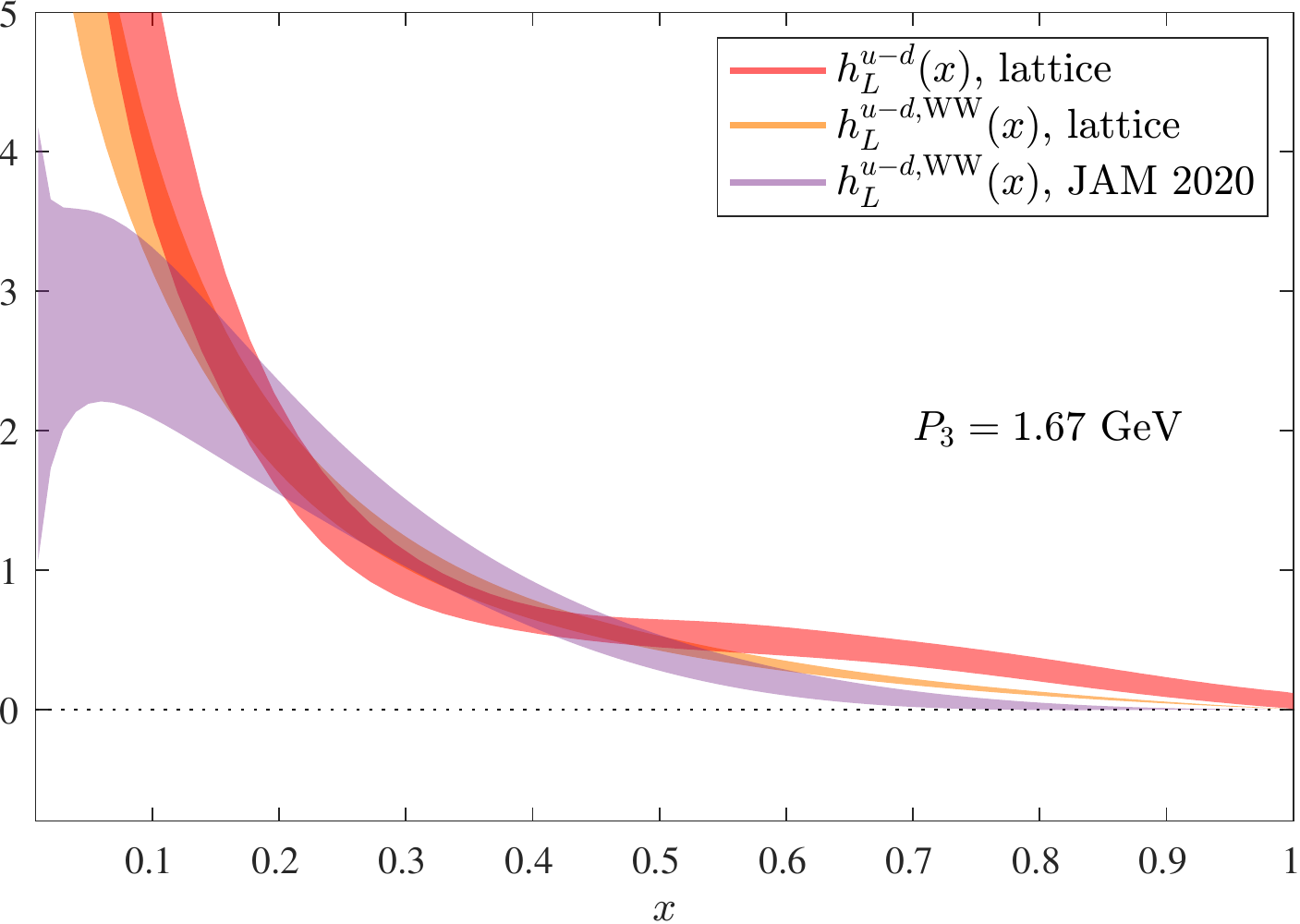}
	\caption{Left: Comparison of lattice results on $g_T(x)$ (blue band) with its WW estimates: lattice-extracted $g_T^{\rm WW}$ (red band) and global fits-extracted (NNPDF1.1pol~\cite{Nocera:2014gqa} orange band, JAM17~\cite{Ethier:2017zbq} purple band). Plot taken from Ref.~\cite{Bhattacharya:2020cen}. Right: The WW approximation for $h_L(x)$, for boosts $P_3=1.67$ GeV. The lattice estimate of $h_L(x)$ (red band) is compared with its WW-approximation (orange band) extracted on the same gauge ensemble and the one obtained from global fits (violet band) from the JAM collaboration~\cite{Cammarota:2020qcw}.  Plot taken from Ref.~\cite{Bhattacharya:2021moj}.}
 	\label{fig:WW_approx}
 \end{figure}

The quasi-PDFs approach has also been extended to the twist-3 PDFs, in particular for $g_T(x)$~\cite{Bhattacharya:2020cen,Bhattacharya:2020xlt}, $h_L(x)$~\cite{Bhattacharya:2021moj} and $e(x)$~\cite{Bhattacharya:2020jfj}. One of the important uses of the lattice results is as a test of the Wandzura-Wilczek (WW) approximation~\cite{Wandzura:1977qf}, according to which the twist-3 $g_T(x)$ and its corresponding twist-2 $g_1(x)$ are connected through
\begin{equation}
\label{eq:gT_WW}
g_T^{\rm WW}(x)=\int_x^1 \frac{dy}{y} g_1(y)\,.
\end{equation}
In the WW approximation, $g_T(x)$ is fully determined by the twist-2 $g_1(x)$. 
\index{Wandzura-Wilczek (type) approximation}
An analogous relation exists for $h_L(x)$~\cite{Jaffe:1991ra,Jaffe:1991kp}. The WW approximation has been implemented for both $g_T(x)$ and $h_L(x)$, which may provide qualitative understanding on the significance of the contribution due to quark-gluon correlations. The results are shown in Fig.~\ref{fig:WW_approx} for the quark region. The  lattice data for $g_T(x)$ are consistent with $g_T^{\rm WW}(x)$ for a considerable $x$-range, even though the uncertainties permit violations up to 40\% for $x\lesssim0.4$. Also, the slopes of $g_T$ and $g_T^{\rm WW}$ are the same up to $x\approx0.4$. The lattice results on $g_T^{\rm WW}$ are also compared to the estimate obtained using $g_1$ from global fits by the NNPDF \cite{Nocera:2014gqa} and JAM17 \cite{Ethier:2017zbq} collaborations, and a good agreement is found up to $x\approx 0.3$. For the $h_L(x)$ case, there is an agreement between  $h_L(x)$ and $h^{\rm WW}_L(x)$ for $x\lesssim 0.55$. Furthermore, the lattice results on $h^{\rm WW}_L(x)$ in the region $0.15 \lesssim x\lesssim 0.55$ are in agreement with $h^{\rm WW}_L(x)$ obtained from the JAM17 global fit~\cite{Cammarota:2020qcw}. It should be mentioned that the lattice calculations of twist-3 PDFs do not consider the mixing with quark-gluon-quark correlators, which requires considerable theoretical development, as well as computational resources. Exploration of twist-3 GPDs is a natural development given the progress in twist-3 PDFs, as well as twist-2 GPDs~\cite{Alexandrou:2020zbe,Alexandrou:2021bbo}. Preliminary results can be found in Ref.~\cite{Dodson_Lattice:2021}.

\subsubsection{Pseudo-distributions}
\label{sec:pseudo}
\index{pseudo-PDF|(}
An approach closely related to the quasi-distributions discussed above, is that of the pseudo-distributions introduced  in a series of publications~\cite{Radyushkin:2016hsy,Radyushkin:2017ffo,Radyushkin:2017cyf}. In this approach,  one calculates the same matrix elements as for  quasi-distributions, but now views them as functions of two Lorentz invariants, the ``Ioffe time''~\cite{Ioffe:1969kf}, $\nu{\equiv} p\cdot z$, and $z^2 $. The matrix element is written as ${\cal M}(\nu,z^2)=\langle P\vert \, \overline{\psi}(0,z)\,\gamma_0 W(z,0)\,\psi(0,0)\,\vert P\rangle$
and the ratio 
\begin{equation}
\label{eq:reduced}
\mathfrak{M}(\nu,z^2) = \frac{\mathcal{M}(\nu,z^2)\,/\,\mathcal{M}(\nu,0)}
{\mathcal{M}(0,z^2)\,/\,\mathcal{M}(0,0)},\,
\end{equation}
is called reduced Ioffe time pseudo-distribution (pseudo-ITD), and defines a gauge-invariant renormalization scheme. $\mathfrak{M}(\nu,z^2)$ is matched to the light-cone ITDs, $Q(\nu,\mu^2)$ via
\begin{eqnarray}
\mathfrak{M}(\nu,z^2) &=& Q(\nu,\mu^2) + \frac{\alpha_s C_F}{2\pi} \int_0^1 du \\
&\times& \left[ \ln \left(z^2\mu^2 \frac{e^{2\gamma_E +1}}4\right) B(u)
+ L(u) \right] Q(u\nu,\mu^2)\,,\nonumber
\end{eqnarray}
in which the kernel $B(u)$ evolves the pseudo-ITDs to a common scale $\mu$ and  $L(u)$ converts to the $\MSbar$ scheme. For more details see Refs.~\cite{Radyushkin:2018cvn,Zhang:2018ggy,Izubuchi:2018srq,Radyushkin:2018nbf}. Note that unlike in the quasi-distributions approach, here one relies on short-distance factorization. The light-cone PDFs may be extracted via a Fourier transform in Ioffe time.
\begin{equation}
\label{eq:PDF2ITD}
 q(x,\mu^2)  =\int d\nu \, e^{-i\nu x}Q(\nu,\mu^2),
\end{equation}

\begin{figure}[t!]
\begin{center}
\includegraphics[width=0.45\textwidth]{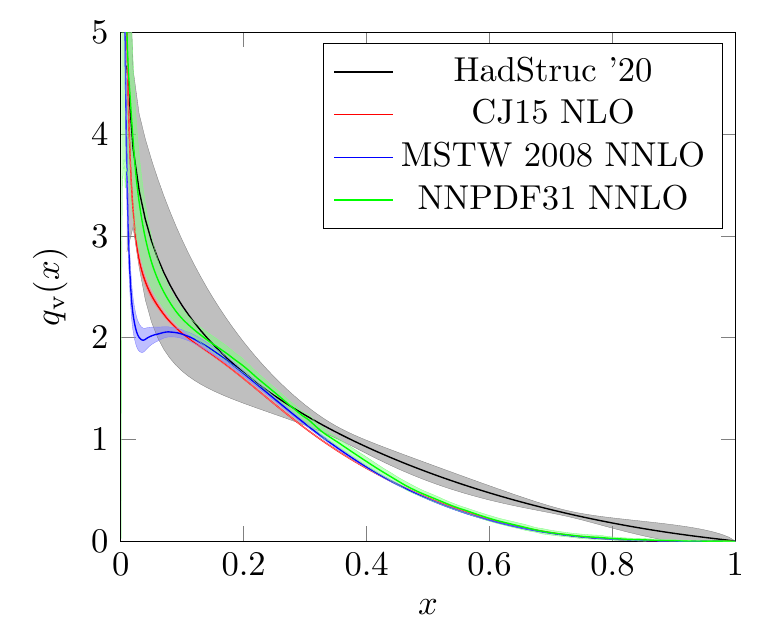}
\includegraphics[width=0.45\textwidth]{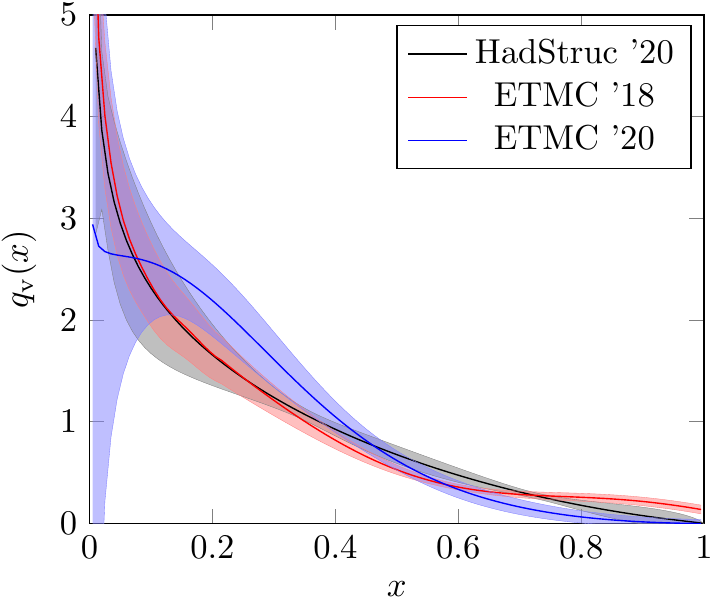}
\caption{Left: The nucleon isovector valence PDF (gray band) and the phenomenological determinations from CJ15~\cite{Accardi:2016qay} (green), MSTW2008~\cite{Martin:2009iq} (red) and NNPDF31~\cite{Ball:2017nwa} (blue). Plot taken from Ref.~\cite{Joo:2020spy}. Left: Lattice results on the unpolarized PDF using the quasi-PDFs method~\cite{Alexandrou:2019lfo} (red band) and pseudo-ITDs from Ref.~\cite{Joo:2020spy} (gray band) and Ref.~\cite{Bhat:2020ktg} (blue band).}
\label{fig:pdf_extrap}
\end{center}
\end{figure}

The pseudo-distribution approach has been studied in several publications with promising results~\cite{Orginos:2017kos,Radyushkin:2017sfi,Karpie:2017bzm,Radyushkin:2018cvn,Karpie:2018zaz,Karpie:2019eiq,Joo:2019bzr,Joo:2019jct,Joo:2020spy}.  Results on the nucleon pseudo-PDFs are presented in Ref.~\cite{Joo:2020spy} for the valence unpolarized PDF. Three ensembles have been used with the lightest quark mass corresponding to a pion mass of  170 MeV. Fig.~\ref{fig:pdf_extrap} shows the  results extrapolated to the physical quark masses compared to the phenomenological fits~\cite{Accardi:2016qay,Martin:2009iq,Ball:2017nwa}. Agreement is seen for $x \sim 0.25$, with the lattice results being significantly larger than the global fits at intermediate and large $x$ values. It is interesting to compare data at the physical point from different lattice formulations and/or  methodologies. In Fig.~\ref{fig:pdf_extrap} we show the unpolarized isovector valence PDF for the proton as obtained from the pseudo-PDFs method: HadStruc '20~\cite{Joo:2020spy},  ETMC '20~\cite{Bhat:2020ktg}, and the quasi-PDFs method: ETMC '18~\cite{Alexandrou:2018pbm}. The results exhibit agreement for a wide range of values for $x$. However, systematic effects are not fully quantified that potentially causes some tension in the large $x$ region.

\begin{figure}[t!]
\begin{center}
\includegraphics[scale=0.6]{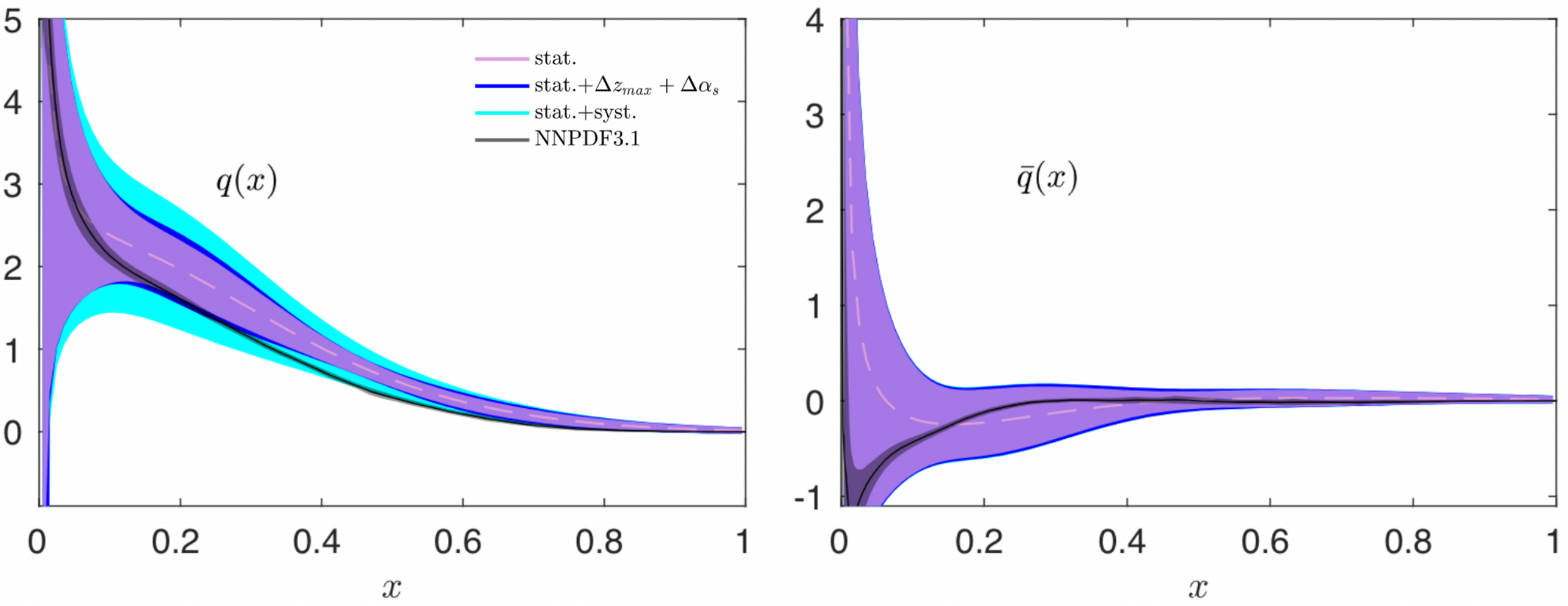}
\caption{Lattice estimates for the unpolarized PDFs for the full (left) and sea (right) contributions~\cite{Bhat:2020ktg}. The global fits of NNPDF~\cite{Ball:2017nwa} are shown with a dark grey band. The bands in the lattice data represent: the statistical error (purple), the combination of statistical and systematic due to the choice of $\nu_{\rm max}$ and $\alpha_s$ (blue), and the total error including also an estimate for the uncertainties related to cutoff effects, finite-volume effects, excited states contamination, truncation and higher-twist effects (cyan). Plot taken from Ref.~\cite{Bhat:2020ktg}.}
\label{fig:final}
\end{center}
\end{figure}

The full and sea-quark  PDFs have been obtained in Ref.~\cite{Bhat:2020ktg} using the pseudo-ITD method using one ensemble at the physical quark masses. Three reconstruction methods were implemented,  the standard FT, the Backus-Gilbert method, and fitting reconstruction~\cite{Bhat:2020ktg}. The latter performs better than the other approaches, and the increase in the uncertainties at small $x$ reflects the challenges of the inverse problems. The final results are shown in Fig.~\ref{fig:final}, and agreement is found with the phenomenological PDFs for both distributions.
\index{pseudo-PDF|)}

\subsubsection{Current-current correlator}
\label{sec:auxlq}

A different method to extract parton structure  was proposed in Ref.~\cite{Braun:2007wv} using an auxiliary light quark field, and applied for the pion distribution amplitude~\cite{Bali:2017gfr,Bali:2018spj}. The calculation relies on current-current correlators, with the currents positioned at points $z$ and $-z$ and the pion boosted with momentum $p$. An OPE may be used for small values of $z$ and in such a case the correlator is related to the Fourier transform of the pion DA. The momentum $p$ can be in any spatial direction, and ideally with a large component in the direction of the current separation ($z$), so that the Ioffe time $p\cdot z$, can take large values, which is an important condition to access the full pion DA. 

In the work of Ref.~\cite{Bali:2017gfr}, a first study is performed using $N_f{=}2$ clover fermions and a pion mass of 295 MeV, with a pion momentum $\sim2$ GeV, with the pion DA being extracted from the scalar-pseudoscalar channel. From this study, it is seen that there is a need to boost the pion to higher momentum to reach higher values $p\cdot z$. Having larger values for $p$ will allow smaller $z$ while still keeping the Ioffe time large. This is crucial, as the approach relies on small values of $z$, so that the perturbative expansions are meaningful. However, $p$ can not be increased arbitrarily due to the lattice cutoff, and thus, smaller values of the lattice spacing are needed.
In Ref.~\cite{Bali:2018spj} the pion DA was studied using different channels, that is, vector-vector, axial-axial, vector-axial, axial-vector, scalar-pseudoscalar and pseudoscalar-scalar, and their linear combinations. Given the findings of Refs.~\cite{Bali:2017gfr,Bali:2018spj}, the constraint  $|\vec{z}|>3a$ is imposed to suppress lattice artifacts, and the values used for $|\vec{z}|$ are relatively small. 
Results from these exploratory studies show that further investigation is needed to eliminate systematic uncertainties related to the unphysical quark masses, momentum boost, finite lattice spacing, and truncation of the perturbative expansion.

\subsubsection{Good lattice cross-sections}
\label{sec:goodlattxsec}
\index{good lattice cross-sections}

Light-cone distribution functions from LQCD can be related to matrix elements calculable on the lattice using the ``lattice cross sections'' (LCSs) approach~\cite{Ma:2014jla,Ma:2014jga,Ma:2017pxb}. The main idea of this approach is to calculate a large class of factorizable matrix elements within LQCD, which can be used in a global fit to extract PDFs, as done with experimental data and phenomenological fits. The matrix elements must be calculable in LQCD, renormalizable, and share the same factorizable logarithmic collinear divergences as the light-cone distribution functions. Quasi-PDFs, pseudo-PDFs and the Compton amplitude $T_{\mu\nu}$ are examples of good LCSs~\cite{Ma:2017pxb}. In general, good LCSs are related to hadronic matrix elements of operators  ${\cal O}_n$, where the hadron $h$ has momentum $P$ :
\begin{equation}
\label{eq:LCS}
{\sigma}_{n}(\omega,\xi^2,P^2,S,\mu)=\langle h(P,S)| {T}\{{\cal O}_n({\xi},\mu)\}|h(P,S)\rangle\,,\quad \omega{\equiv} P\cdot\xi\,.
\end{equation}
One possibility explored for the operator is a current-current correlators separated by distance $\xi$ ($\xi^2\neq0$), that is
\begin{equation}
\label{eq:JJ}
{\cal O}_{J_1 J_2}(\xi) \equiv \xi^{D_{J_1}+D_{J_2}-2}\, J_1^R(\xi)J_2^R(0)\, ,
\end{equation}
with $D_{J_i}$ the dimension of the renormalized current $J_i^R{=}Z_{J_i}J_i$, with $Z_{J_i}$ the renormalization function of $J_i$. This particular case is similar to the hadronic tensor approaches discussed above, but more general scenarios can also be considered.

The method was employed in Ref.~\cite{Sufian:2019bol} for an ensemble of $N_f=2+1$ clover fermions with pion mass of 430 MeV, to calculate current-current correlators for the vector and axial currents, and momentum boost up to $\sim$1.5 GeV. The work focuses on the antisymmetric combination of vector and axial-vector operators, which is directly linked to the pion quark distribution. More recently, the calculation improved with four ensembles with three pion masses (278, 358, 413 MeV) and two volumes~\cite{Sufian:2020vzb}. A chiral, continuum, volume, and higher-twist extrapolation has been applied, followed by the factorization and a parameterization of the lattice data on the PDF. The fits are in agreement within errors and are shown in the left panel of Fig.~\ref{fig:evolPDF}, and are compared with using E615 data~\cite{Conway:1989fs,Aicher:2010cb}.

\subsubsection{Comparison of methods}

It is interesting to compare results for the quark distribution in the pion using the various different approaches. Note that some the calculations use different fermion actions and analysis approach, such as the quasi-PDFs~\cite{Chen:2018fwa,Izubuchi:2019lyk}, pseudo-ITDs~\cite{Joo:2019bzr} and current-current correlators~\cite{Sufian:2019bol}. 

\begin{figure}[t!]
\begin{center}
\includegraphics[scale=0.62]{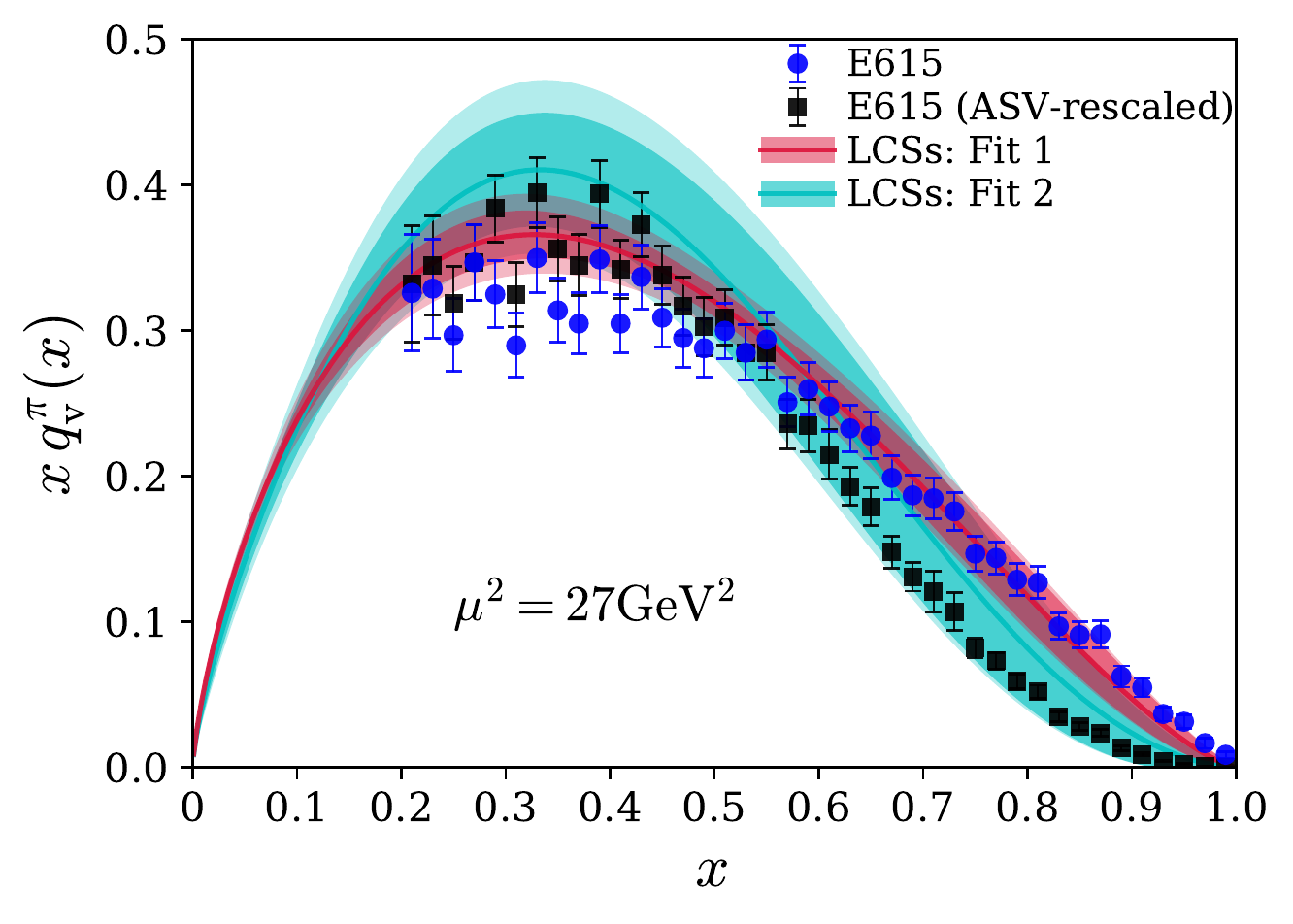}
\includegraphics[scale=0.4]{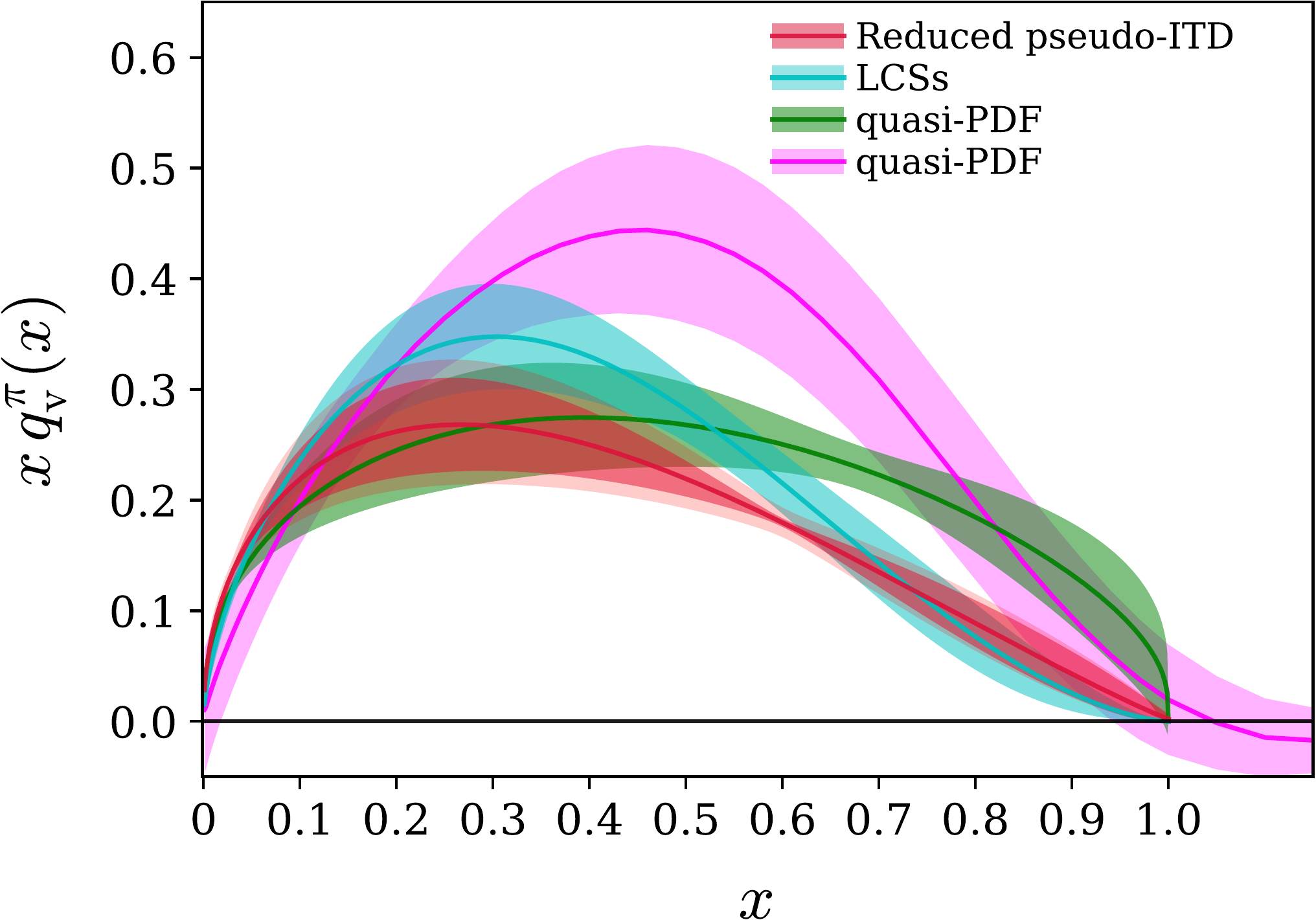}
\end{center}
\caption{
Left: Lattice data of pion $xq^\pi_{\rm v}(x)$-distribution using two parameterizations (cyan and red bands), and the E615 data from Ref.~\cite{Conway:1989fs} (blue) and Ref.~\cite{Chang:2014lva} (black). Plot taken from Ref.~\cite{Sufian:2020vzb}. Right: Lattice data for the pion PDF from Ref.~\cite{Chen:2018fwa} using quasi-PDF (pink band), Ref.~\cite{Izubuchi:2019lyk} also with quasi-PDF (green band), Ref.~\cite{Joo:2019bzr} using pseudo-ITDs (red band), and Ref.~\cite{Sufian:2019bol} using current-current correlators (cyan band). Plot taken from Ref.~\cite{Joo:2019bzr}. }
\label{fig:evolPDF}
\end{figure}

Ref.~\cite{Chen:2018fwa} uses an $N_f=2+1+1$ mixed-action ensemble of clover on HISQ fermions with a pion mass of 310 MeV and volume $24^3\times 64$. The analysis follows the quasi-PDF approach, and the $x$-dependence reconstruction is performed using the derivative method~\cite{Lin:2017ani}. The derivative method is based on integration by parts of the Fourier transform and neglects the surface term, which introduces uncontrolled uncertainties~\cite{Karpie:2018zaz}. Ref.~\cite{Izubuchi:2019lyk} makes use of the quasi-PDFs method on a mixed action of clover fermions in the valence sector and $N_f=2+1$ HISQ fermions with pion mass 300 MeV. The volume is $48^3\times 64$, corresponding to a spatial extent of $2.9$ fm ($a=0.06$ fm). Instead of a standard Fourier transform, two types of fits are applied to the lattice data in the coordinate space, similar to the methods of Refs.~\cite{Sufian:2019bol,Joo:2019bzr,Sufian:2020vzb}. Ref.~\cite{Joo:2019bzr} combines two $N_f=2+1$ ensembles of clover fermions with a pion mass of 415 MeV and different volumes (3 fm and 4 fm). Ref.~\cite{Sufian:2019bol} uses the same large-volume ensemble (4 fm). The comparison of the two calculation is shown in the left panel of Fig.~\ref{fig:evolPDF}, where an excellent agreement is observed.

The results from  the four different calculations are shown in the right panel of Fig.~\ref{fig:evolPDF}. The pseudo and current-current correlators data of Refs.~\cite{Sufian:2019bol,Joo:2019bzr} are in reasonable agreement with the calculation of Ref.~\cite{Izubuchi:2019lyk}. A tension is observed with the results of Ref.~\cite{Chen:2018fwa}, which uses the quasi-PDFs approach. This tension could possibly originate from the use of the derivative method to reconstruct the $x$-dependence, which neglects the surface term. Note that the calculations are evolved to different scales. However, the scale dependence is expected to be small for the values used. 

\subsection{Lattice QCD Calculations of TMD Observables}
\label{sec:lattice_tmd_calcs}

Having presented an overview of progress on understanding the longitudinal momentum dependence  of PDFs, a basis has been laid for discussing LQCD approaches to transverse momentum-dependent hadron structure. A number of different aspects have been investigated. They include a longer-established calculational program employing the Lorentz-invariant approach introduced in Sec.~\ref{sec:latt_def_lorentz}, focusing on 
TMD ratios, as well as calculations of TMD soft functions, and calculations of the Collins-Soper kernel.

\subsubsection{Lorentz-invariant approach}
\label{sec:TMDratios}
\index{lattice QCD calculations!TMDs}

\vspace{0.35cm}
\noindent
\textit{\textbf{{{Calculational scheme}}}}
\vspace{0.25cm}

As already indicated in Sec.~\ref{sec:latt_def_connection}, lattice calculations of TMD (and GTMD) observables are based on the evaluation of the fundamental hadronic matrix elements, cf. Eq.~\eqref{eq:latt_corr_def},
\begin{equation}
\widetilde{\Phi}^{[\Gamma ]}_{i} (b,P^{\prime },P,S,v,\eta,a) = \frac{1}{2}
\Bigl\langle p(P^{\prime },S) \Big| \bar\psi^{0}_{i} (b^{\mu }/2) \Gamma W_{\sqsupset \eta}^{v} (b^{\mu }/2,-b^{\mu }/2) \psi^{0}_{i} (-b^{\mu }/2) \Big| p(P,S) \Bigr\rangle
\label{eq:latt_corr_def_2}
\end{equation}
in states characterized by their momentum and spin; TMDs are derived from diagonal matrix elements, $P^{\prime } =P$, whereas GTMDs, to be discussed further in Chap.~\ref{sec:gtmd}, additionally depend on the momentum transfer $\Delta = P^{\prime } -P$. $\Gamma $ stands for an arbitrary Dirac matrix structure and $i$ labels the quark flavor.
As discussed in detail in Chap.~\ref{sec:TMDdefn}, the presence of the gauge connection $W_{\sqsupset\eta}^v$ introduces divergences additional to the wave function renormalizations of the quark operators; these can be absorbed into a multiplicative soft factor. In the calculational scheme described in the following, the explicit evaluation of soft factors is avoided by considering appropriate ratios in which they cancel. A method to evaluate soft factors in LQCD, which would allow one to extend lattice calculations beyond ratio observables, is discussed in Sec.~\ref{sec:latt_soft_func}.

As laid out in Chap.~\ref{sec:TMDdefn}, standard TMDs describing, e.g., the SIDIS and Drell-Yan processes are obtained using a staple-shaped gauge connection path,\footnote{More complex paths can also become relevant when one extends considerations beyond the simplest processes~\cite{Boer:2015kxa}.} as exhibited in Fig.~\ref{fig:stapleLattice}.
The path is characterized not only by the separation of the quark operators $b$, but also the direction of the staple $v$, and the length of the staple $\eta $. In a LQCD calculation, $\eta $ is finite, and one must extrapolate the data to the $\eta \rightarrow \infty$ limit.
In addition, $v$ is chosen to be space-like, in order to be able to connect the definition in Eq.~\eqref{eq:latt_corr_def_2} to a Lorentz frame in which $v$ is purely spatial, and in which therefore the lattice calculation can be performed.
As already discussed in Sec.~\ref{sec:latt_def_connection}, a useful parameter characterizing the rapidity of the staple direction $v$ relative to the average hadron momentum $\bar{P} =(P^{\prime } +P)/2$ is the Collins-Soper type evolution parameter $\hat{\zeta } = v\cdot \bar{P} / (\sqrt{|v^2 |} \, \sqrt{\bar{P}^{2} }) $. The connection with the modern Collins definition of TMDs is established in the limit $\hat{\zeta } \rightarrow \infty $.

\begin{figure}[t!]
\begin{center}
\includegraphics[width=8.4cm]{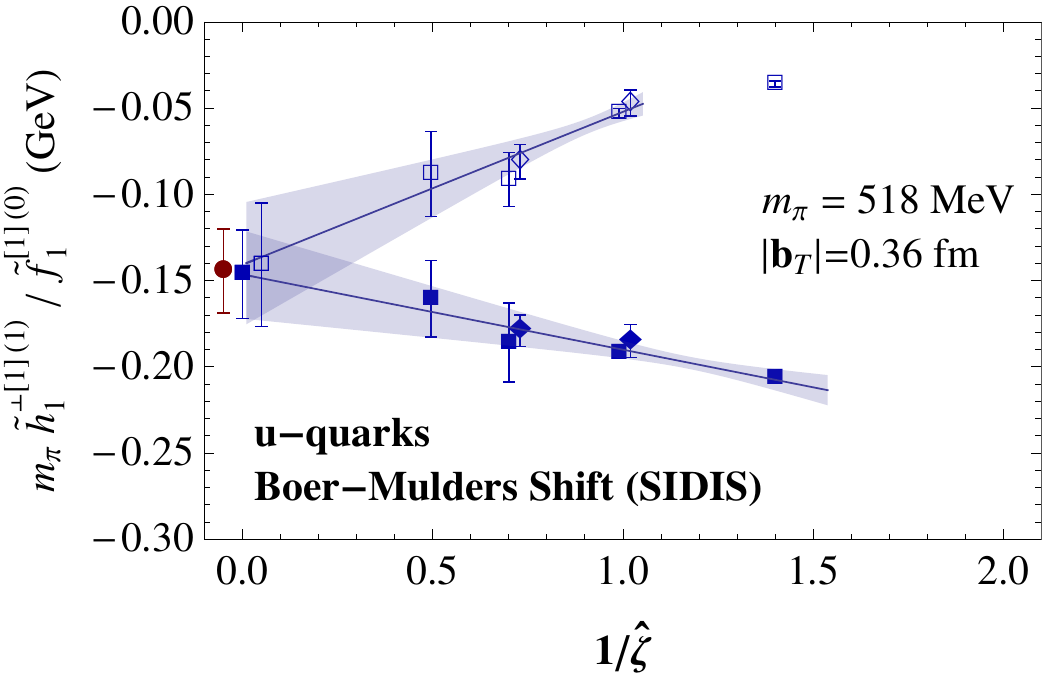}
\end{center}
\vspace{-0.4cm}
\caption{Pion $u$-quark SIDIS generalized Boer-Mulders shift as a
function of $1/\hat{\zeta } $ at a fixed $b_T $. Full symbols show full
shift, open symbols a partial contribution that dominates the shift at large
$\hat{\zeta } $; extrapolations of the two data sets (blue data points at $1/\hat{\zeta } =0$) coincide, indicating
that a stable description of the large $\hat{\zeta } $ evolution has been
achieved. Red data point at $1/\hat{\zeta } =0$ results from a combined fit to both data sets. Only connected contributions to the Boer-Mulders shift were included. Plot taken from  Ref.~\cite{Engelhardt:2015xja}.}
\label{bmfig}
\end{figure}

In practice, reaching values of $\hat{\zeta } $ in the range 1--2 in
lattice calculations appears to be sufficient to enter a regime in which
the data fit a power law behavior that can be extrapolated to the
$\hat{\zeta } \rightarrow \infty $ limit;
an illustration is provided by Fig.~\ref{bmfig}.
For a light particle such as the pion, this regime has been reached,
whereas for the nucleon, current calculations as of this writing are still concentrated at lower values and only beginning to enter the aforementioned regime. The extrapolation $\hat{\zeta } \rightarrow \infty $ therefore
appears feasible with continually improving calculations, but does figure among the chief systematic uncertainties of lattice TMD calculations. It persists as a challenge for future LQCD TMD investigations.

\begin{figure}[t!]
\includegraphics[width=8.4cm]{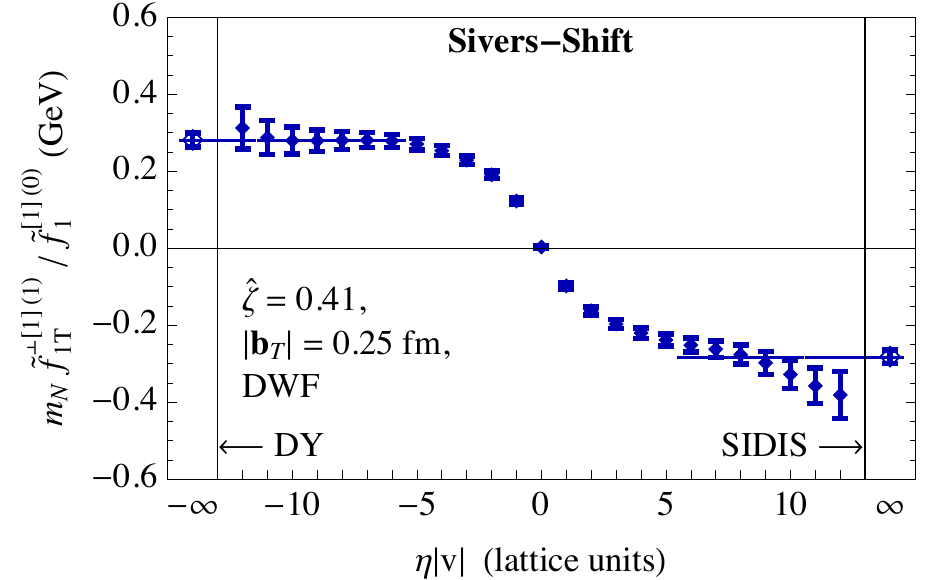}
\hspace{0.4cm}
\includegraphics[width=8.4cm]{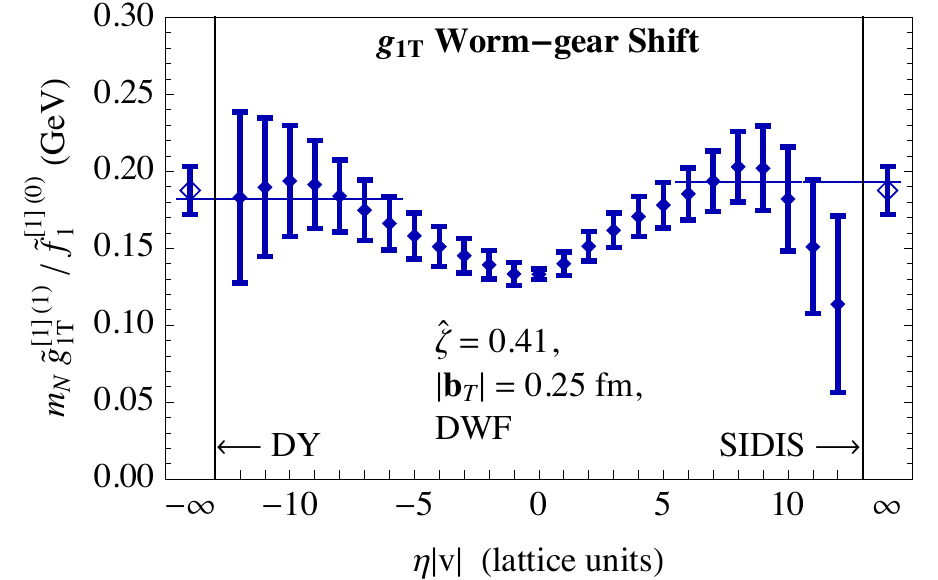}
\caption{Dependence of TMD observables on the staple length.  
Left: T-odd isovector ($u-d$ quark) generalized Sivers shift at fixed $b_T $
and $\hat{\zeta } $.
Right: T-even isovector generalized $g_{1T} $ worm-gear shift at fixed $b_T $ 
and $\hat{\zeta } $.
Data were obtained on a domain wall fermion (DWF) ensemble at pion mass
$m_{\pi } \approx 300\, \mbox{MeV} $ and lattice spacing
$a=0.084\, \mbox{fm} $.
Horizontal lines indicate averages of the data points in the ranges $\eta |v| \ge 6a$ and $\eta |v| \le -6a$, respectively, where plateau behavior is expected. Extrapolations at $\eta |v| = \pm \infty $ are obtained as mean values of the aforementioned averages (with a relative minus sign in the case of the Sivers shift). Plot taken from Ref.~\cite{Yoon:2017qzo}.}
\label{etafig}
\end{figure}

\begin{figure}[t!]
\begin{center}
\includegraphics[width=8.4cm]{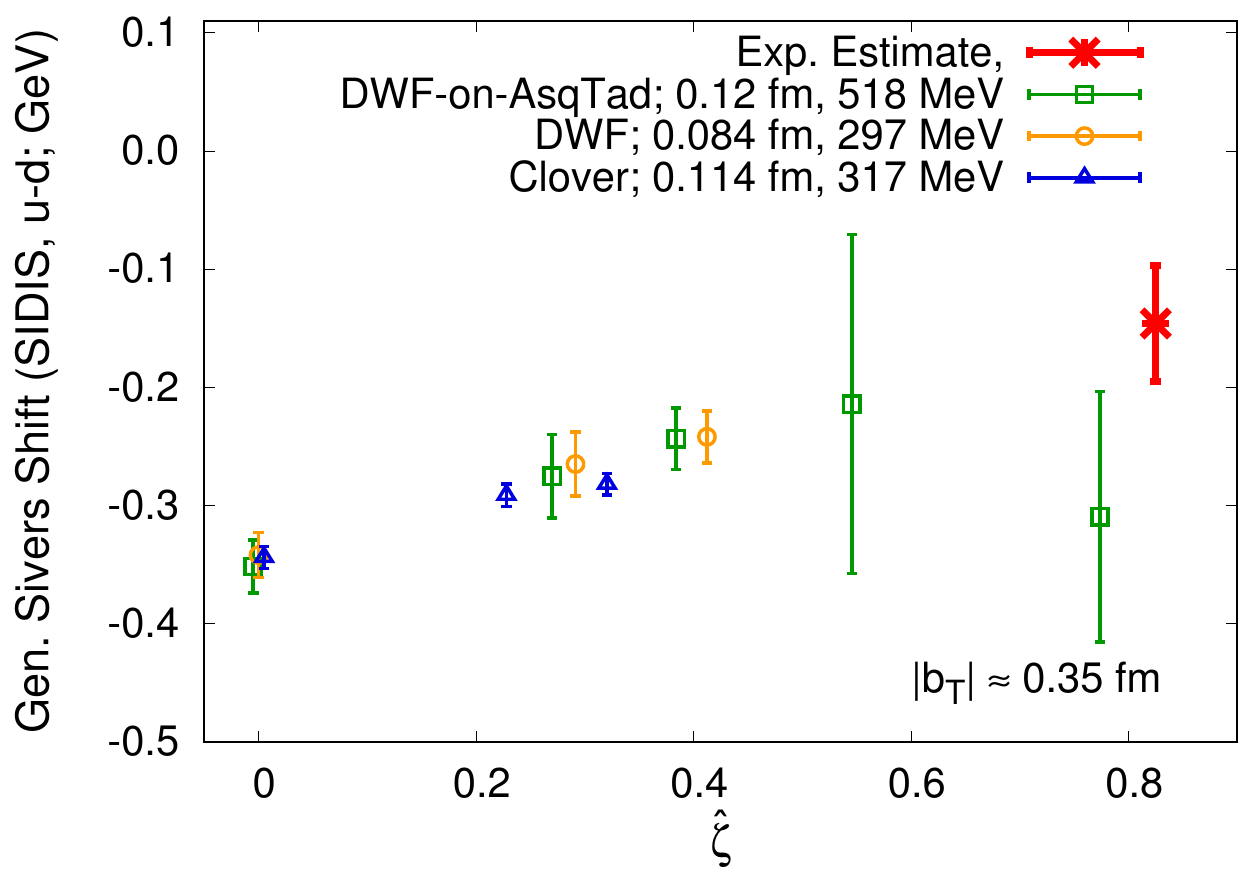}
\end{center}
\vspace{-0.4cm}
\caption{Compilation of LQCD results for the Sivers shift, compared to a phenomenological estimate obtained by constructing the Sivers shift from the results of the phenomenological analysis~\cite{Echevarria:2014xaa}, as described in Ref.~\cite{Yoon:2017qzo}. Lattice results from several studies combine to a consistent picture, with no significant dependence on the pion mass apparent in the range covered. The trend of the lattice data as a function of the Collins-Soper-type parameter $\hat{\zeta} $ suggests that agreement between lattice and phenomenological estimates is within reach as lattice studies progress towards larger $\hat{\zeta } $. Plot taken from Ref.~\cite{Yoon:2017qzo}.}
\label{sivcomp}
\end{figure}

To facilitate the transformation of the results obtained in the Lorentz frame in which the lattice calculation is performed back to the original frame in which TMDs are defined, it is useful to employ a decomposition of Eq.~\eqref{eq:latt_corr_def_2} into Lorentz invariants. Once determined in the lattice frame from the lattice data, these invariants are immediately valid also in the original frame. The full decomposition is discussed in Ref.~\cite{Musch:2011er}; it is analogous to the decomposition defining TMDs in momentum space. For a nucleon, at leading twist, one has the forms\footnote{Note that the convention for the operator separation $b$ used here has the opposite sign relative to the convention used in Ref.~\cite{Musch:2011er}.}
{\allowdisplaybreaks
\begin{eqnarray}
\frac{1}{2P^{+} }
\widetilde{\Phi }^{[\gamma^{+} ]}
&=& \widetilde{A}_{2B} -im_N \epsilon_{ij} b_i S_j
\widetilde{A}_{12B} , \label{adecomp1} \\
\frac{1}{2P^{+} }
\widetilde{\Phi }^{[\gamma^{+} \gamma^{5} ]}
&=& -S_L \widetilde{A}_{6B} -i((b\cdot P) S_L -m_N (b_T \cdot S_T ))
\widetilde{A}_{7B} ,  \\
\frac{1}{2P^{+} }
\widetilde{\Phi }^{[i\sigma^{i+} \gamma^{5} ]}
&=& -im_N \epsilon_{ij} b_j \widetilde{A}_{4B} -S_i
\widetilde{A}_{9B}
+im_N S_L b_i \widetilde{A}_{10B} \nonumber \\*
& & +m_N ((b\cdot P) S_L
-m_N (b_T \cdot S_T )) b_i \widetilde{A}_{11B} .
\end{eqnarray}
}The Lorentz invariant amplitude combinations $\widetilde{A}_{iB} $ are already suitable linear combinations of the amplitudes one finds in the most general decomposition~\cite{Musch:2011er}. They essentially correspond to Fourier-transformed TMDs, cf.~also Eq.~\eqref{eq:tmd_decomposition_2}. For the following, it is useful to introduce a notation for Mellin moments of Fourier-transformed TMDs, where $f(x,k_T^2 ,\ldots )$ stands for a generic TMD,
\begin{equation}
\tilde{f}^{[m](n)} ( b_T^2 ,\ldots ) \equiv
n!\left( -\frac{2}{m_N^2 }\partial_{b_T^2 } \right)^n \
\int_{-1}^{1} dx\, x^{m-1} \int d^2 k_T \,
e^{-ib_T \cdot k_T } f(x,k_T^2 ,\ldots ) \ .
\end{equation}
Through the invariant amplitudes $\widetilde{A}_{iB} $, one can then finally define observables constructed as ratios; for example, for transverse polarization, the following quantities have been studied~\cite{Musch:2011er}:
\begin{figure}[t]
\begin{center}
\includegraphics[width=8.4cm]{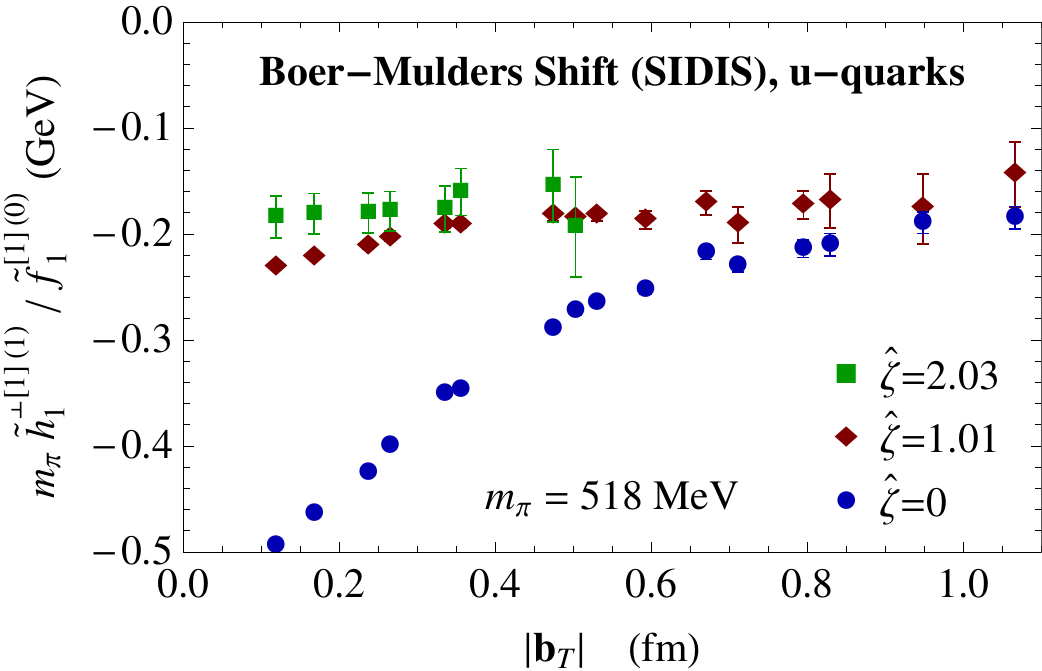}
\end{center}
\caption{Pion $u$-quark SIDIS generalized Boer-Mulders shift as a function of $b_T $, for several values of $\hat{\zeta } $. Only connected contributions to the Boer-Mulders shift were included. \index{Boer-Mulders effect}
Plot taken from Ref.~\cite{Engelhardt:2015xja}.}
\label{bdepfig}
\end{figure}
\begin{itemize}
\item
The generalized Sivers shift
\index{Sivers effect}
\begin{equation}
\langle k_{\perp } \rangle_{TU} (b_T^2 , \ldots )
= m_N \frac{\tilde{f}_{1T}^{\perp[1](1)} }{\tilde{f}_1^{[1](0)} }
= -m_N \frac{\widetilde{A}_{12B} (-b_T^2 ,b\cdot P=0,\hat{\zeta } ,
\eta v\cdot P)}{\widetilde{A}_{2B} (-b_T^2 ,b\cdot P=0,\hat{\zeta } ,
\eta v\cdot P)}  , 
\label{gsshift}
\end{equation}
which formally, in the $b_T \rightarrow 0$ limit, represents the average transverse momentum $k_{\perp } $ of unpolarized (``$U$'') quarks orthogonal to the transverse (``$T$'') spin of the proton, normalized to the corresponding number of valence quarks. It is ``generalized'' in the sense of being defined for arbitrary $b_T^2 $, not only $b_T \rightarrow 0$. This regulates ultraviolet divergences associated with the latter limit; also, the dependence on $b_T^2 $ of course encodes information about the $k_T $-dependence of the TMDs appearing in the ratio. Note that, in the numerator, the contributions from quarks and antiquarks are summed over~\cite{Mulders:1995dh}, whereas the denominator corresponds to the difference of quark and antiquark contributions (thus, the number of valence quarks in the $b_T \rightarrow 0$ limit). The generalized Sivers shift is T-odd, i.e., differs in sign between the SIDIS and Drell-Yan limits, cf.~Fig.~\ref{etafig} (left). A compilation of existing LQCD results for the generalized Sivers shift, compared to a phenomenological extraction, is presented in Fig.~\ref{sivcomp}~\cite{Yoon:2017qzo}.

\item
The generalized Boer-Mulders shift \index{Boer-Mulders effect}
\begin{equation}
\langle k_{\perp } \rangle_{UT} (b_T^2 , \ldots )
= m_N \frac{\tilde{h}_{1}^{\perp[1](1)} }{\tilde{f}_1^{[1](0)} }
= m_N \frac{\widetilde{A}_{4B} (-b_T^2 ,b\cdot P=0,\hat{\zeta } ,
\eta v\cdot P)}{\widetilde{A}_{2B} (-b_T^2 ,b\cdot P=0,\hat{\zeta } ,
\eta v\cdot P)}
\label{gbmshift}
\end{equation}
akin to the Sivers shift, is T-odd and formally, in the $b_T \rightarrow 0$ limit, represents the average transverse momentum $k_{\perp } $ of transversely polarized (``$T$'') quarks in the direction orthogonal to their spin, in an unpolarized (``$U$'') hadron. It can therefore also be defined for spinless hadrons such as the pion. Fig.~\ref{bdepfig} shows results for the pion generalized Boer-Mulders shift in the SIDIS limit~\cite{Engelhardt:2015xja}. The generalized Boer-Mulders shift is again normalized to the
corresponding number of valence quarks.

\item
The generalized $g_{1T} $ worm-gear shift
\index{worm-gear functions!lattice}
\begin{equation}
\langle k_{\perp } \rangle_{TL} (b_T^2 , \ldots )
= m_N \frac{\tilde{g}_{1T}^{[1](1)} }{\tilde{f}_1^{[1](0)} }
= -m_N \frac{\widetilde{A}_{7B} (-b_T^2 ,b\cdot P=0,\hat{\zeta } ,
\eta v\cdot P)}{\widetilde{A}_{2B} (-b_T^2 ,b\cdot P=0,\hat{\zeta } ,
\eta v\cdot P)}
\label{gwgshift}
\end{equation}
formally, in the $b_T \rightarrow 0$ limit, represents the average transverse momentum $k_{\perp } $ of longitudinally polarized (``$L$'') quarks in the direction of the transverse (``$T$'') spin of the proton, again normalized to the corresponding number of valence quarks. Unlike the Sivers and the Boer-Mulders shifts, this is a T-even quantity, i.e., the SIDIS and Drell-Yan limits coincide, cf.~Fig.~\ref{etafig} (right) from Ref.~\cite{Yoon:2017qzo}. In the displayed case, the bulk of the $g_{1T} $ worm-gear shift is induced already in the presence of a straight gauge link, and there is only a moderate modification through the final state interactions encoded in the SIDIS/Drell-Yan staple link structures.

\item
The generalized tensor charge
\index{tensor charge}
\begin{equation}
\frac{\tilde{h}_{1}^{[1](0)} }{\tilde{f}_1^{[1](0)} }
= -\frac{\widetilde{A}_{9B} (-b_T^2 ,b\cdot P=0,\hat{\zeta } ,
\eta v\cdot P) -\frac{1}{2} m_N^2 b^2 \widetilde{A}_{11B}
(-b_T^2 ,b\cdot P=0,\hat{\zeta } ,\eta v\cdot P)}{\widetilde{A}_{2B}
(-b_T^2 ,b\cdot P=0,\hat{\zeta } ,
\eta v\cdot P)}
\label{gtcharge}
\end{equation}
is also a T-even quantity. In contradistinction to the previous observables, it does not involve any weighting with $k_T $ and is directly related to the well-known transversity and unpolarized distribution functions. It is interpreted as a generalized tensor charge because, in the formal $b_T \rightarrow 0$ limit, the numerator corresponds to the  integral of the transversity distribution, i.e., the standard tensor charge. It is again normalized to the corresponding number of valence quarks. It should however be emphasized that the additional divergences that arise in the $b_T \rightarrow 0$ limit require further renormalization, as a consequence of which the ratio of tensor to vector renormalization constants, $Z_T /Z_V $, has to be accounted for when connecting the generalized tensor charge to the standard tensor charge.
\end{itemize}

\begin{figure}[t!]
\includegraphics[width=0.95\textwidth]{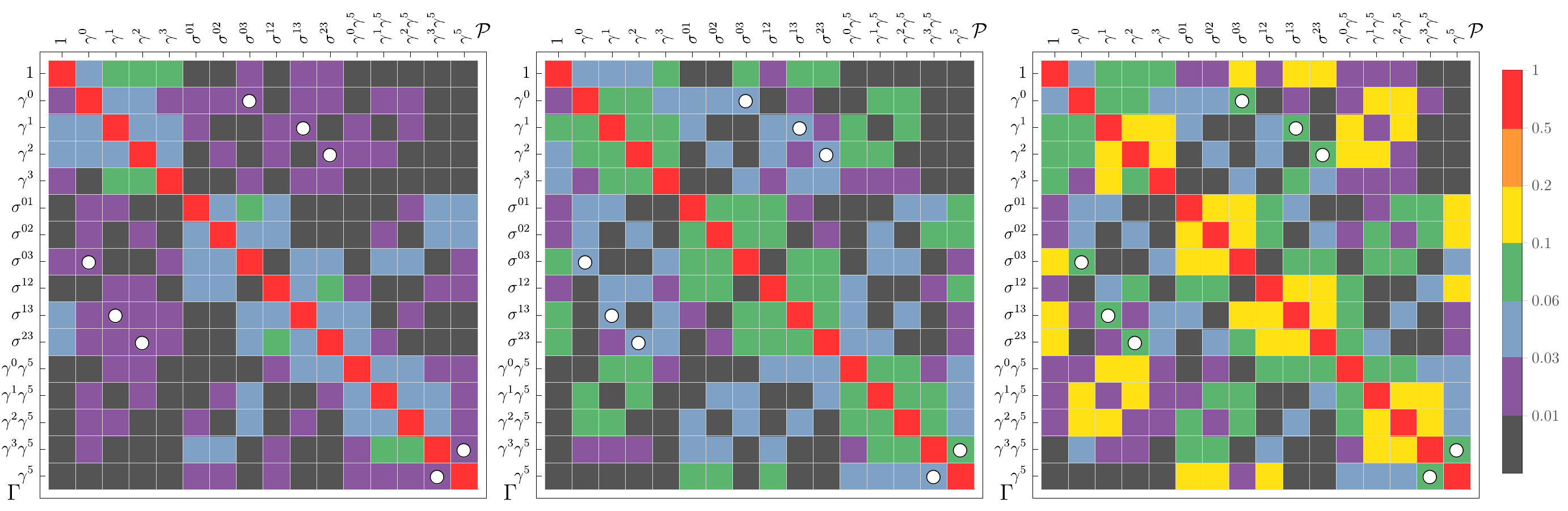}
\caption{Mixing pattern in the RI'/MOM scheme for quark bilinear operators with staple-shaped gauge links constructed using improved Wilson fermions. The quark operator separation $b$ is purely transverse, with $b_T/a=3,7,11$ from left to right, where $a=0.06$~fm denotes the lattice spacing. The staple length is given by $\eta /a =14$. Colors indicate mixing strengths. White circles indicate mixings already obtained in one-loop lattice perturbation theory~\cite{Constantinou:2019vyb}. Plot taken from Ref.~\cite{Shanahan:2019zcq}.}
\label{fig:mixing_ps}
\end{figure}

Note, in particular, that the ratios considered in Eqs.~\eqref{gsshift}-\eqref{gtcharge} cancel any multiplicative renormalization and soft factors associated with the $\widetilde{A}_{iB} $ amplitudes at finite $b_T $. It should be emphasized, however, that the multiplicative nature of the renormalization and soft factors obtained in the continuum theory is not immediately guaranteed to transfer to the lattice formulation; the renormalization pattern of the lattice quantities requires separate consideration depending on the concrete discretization employed, as is discussed below in connection with Fig.~\ref{fig:mixing_ps}.

\vspace{0.75cm}
\noindent
\textit{\textbf{{{Systematic behavior of lattice TMD observables -- numerical studies}}}}
\vspace{0.25cm}

As already indicated in the above discussion, a number of challenges have to
be addressed in order to arrive at controlled predictions for TMD observables
that can be connected to phenomenology. For one, whereas the extrapolation
to infinite staple length $\eta $ is fairly straightforward, accessing the
relevant $\hat{\zeta } $ regime is more difficult, since it requires data
at sufficiently high hadron momenta. Secondly, the purported cancellation
of renormalization and soft factors in ratios such as in Eqs.~\eqref{gsshift}-\eqref{gtcharge} requires
reexamination in the context of LQCD. Thirdly, progress towards the
physical quark masses must be made in lattice TMD calculations; initial studies
were performed at artificially large quark masses for reasons of computational
cost. In addition, early explorations of TMD observables focused on the
point $b\cdot P=0$, see Eqs.~\eqref{gsshift}-\eqref{gtcharge}; since the
longitudinal component of $b$ is Fourier conjugate to the longitudinal
momentum fraction $x$, setting $b\cdot P=0$ corresponds to evaluating only
the $x$-integral of TMDs. To access the $x$-dependence of TMD observables,
the numerical studies must be extended to include scans of the $b\cdot P$
direction. Furthermore, it is necessary to buttress these lattice TMD
investigations by performing quantitative studies of the scaling with the
lattice spacing $a$, in order to gain nonperturbative understanding of
TMD evolution (lattice calculations of the CS kernel governing rapidity evolution are discussed in Section \ref{sec:latticeCS}). In addition, the finite lattice size effects influencing the behavior
of nonlocal operators such as the one in Eq.~\eqref{eq:latt_corr_def_2} remain to be
understood, cf.~related considerations in Ref.~\cite{Briceno:2018lfj}.

\begin{figure}[t!]
\begin{center}
\includegraphics[width=8.4cm]{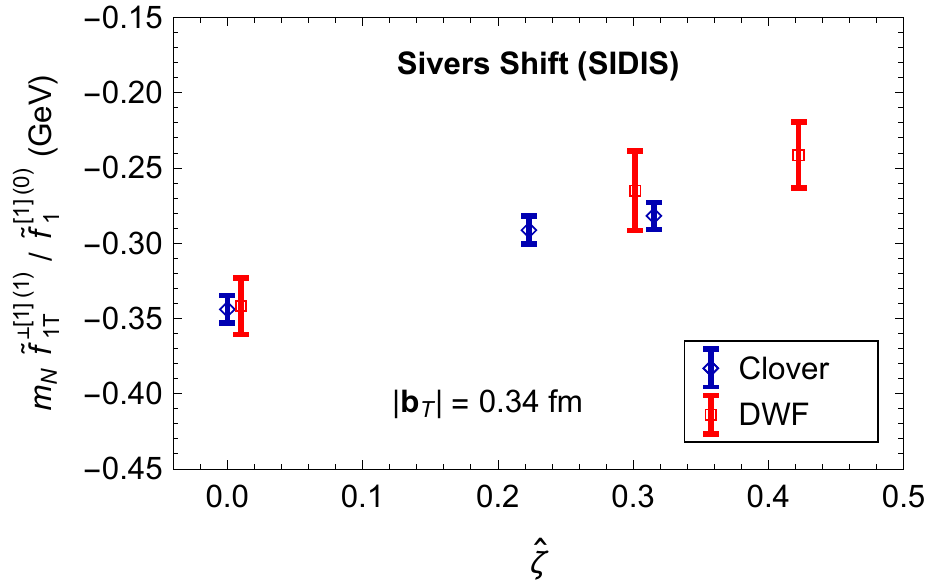}
\end{center}
\vspace{-0.4cm}
\caption{Nucleon isovector ($u-d$ quark) SIDIS generalized Sivers
shift as a function of $\hat{\zeta } $ at a fixed $b_T $. Shown are results
for an $a=0.114\, \mbox{fm} $ clover ensemble and an $a=0.084\, \mbox{fm} $
domain wall fermion ensemble at pion masses near 300 MeV; the results
are compatible with one another, indicating that the effects of
renormalization and soft factors are successfully canceled in the
Sivers shift ratio in Eq.~\eqref{gsshift}. Plot taken from Ref.~\cite{Yoon:2017qzo}.}
\label{siv2ens}
\end{figure}

Significant progress has been made in addressing these challenges. Fig.~\ref{bmfig} displays a result of a dedicated study~\cite{Engelhardt:2015xja} of the large $\hat{\zeta } $ regime using the example of the Boer-Mulders shift in the pion. \index{Boer-Mulders effect} The Boer-Mulders shift measures the average transverse momentum of quarks polarized in the transverse direction orthogonal to the given momentum, in an unpolarized hadron. The pion, by virtue of its lower mass compared with that of the nucleon, allows one to access higher $\hat{\zeta } $ (note that the hadron mass enters the denominator of $\hat{\zeta } $). This case demonstrates a stable extrapolation to the large $\hat{\zeta } $ limit, with the signal surviving in the limit. To obtain data of similar quality for the nucleon, it is necessary to employ
the momentum smearing method~\cite{Bali:2016lva}. Lattice TMD studies underway at the time of this writing incorporate this technique.

\begin{figure}[t!]
\includegraphics[width=8.4cm]{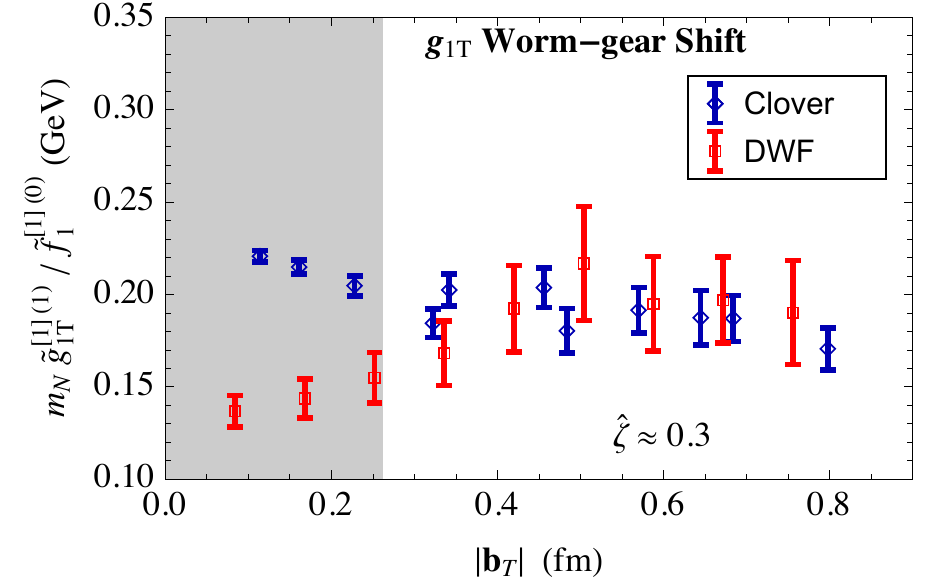}
\hspace{0.4cm}
\includegraphics[width=8.4cm]{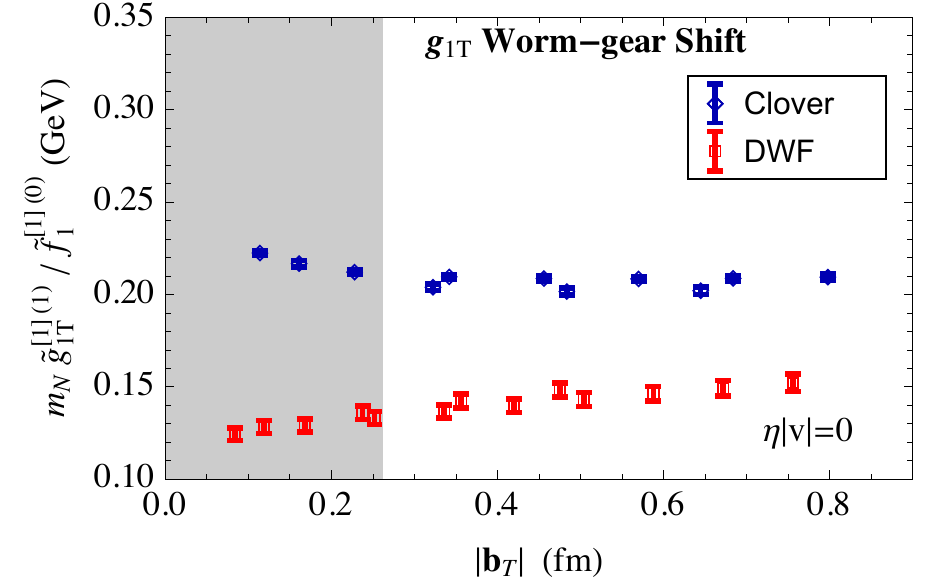}
\caption{Left: Isovector ($u-d$ quark) SIDIS generalized $g_{1T} $
worm-gear shift as a function of $b_T $ at a fixed $\hat{\zeta } $,
comparing results obtained using clover and domain wall fermions.
Right: Isovector straight-link generalized $g_{1T} $ worm-gear shift
as a function of $b_T $, comparing results
obtained using clover and domain wall fermions; the two panels were obtained using the same nucleon momenta in the lattice calculation. The shaded areas indicate the regions which may be subject to significant lattice artefacts even in the absence of operator mixing. Plot taken from Ref.~\cite{Yoon:2017qzo}.}
\label{mixfig}
\end{figure}

On the other hand, the question to what extent the multiplicative nature of renormalization and soft factors carries over from the continuum theory to the lattice formulation was investigated empirically in Ref.~\cite{Yoon:2017qzo} by varying the lattice discretization scheme. If lattice calculations are beset by deviations from purely multiplicative behavior of the renormalization factors, then the latter would cease to cancel in TMD ratios such as Eqs.~(\ref{gsshift})-(\ref{gtcharge}). Being a discretization effect, this would be expected to depend significantly on the type of discretization employed, and therefore manifest itself in a dependence of TMD ratios on the discretization scheme. In Ref.~\cite{Yoon:2017qzo}, calculations were performed on two ensembles at pion masses close to 300 MeV which differ substantially in discretization: A domain wall fermion ensemble with lattice spacing $a=0.084\, \mbox{fm} $, and a clover fermion ensemble with $a=0.114\, \mbox{fm} $. Fig.~\ref{siv2ens} displays a result obtained for the Sivers shift, exhibiting consistent results, corroborating the cancellation of renormalization factors in the ratio in Eq.~\eqref{gsshift} expected from continuum QCD. On the other hand, cf.~Fig.~\ref{mixfig}, in the case of the $g_{1T} $ worm-gear shift, a significant discrepancy is observed at small separations $b_T $, which is exacerbated, extending to all $b_T $, if one instead uses an operator with a straight gauge connection, as employed, e.g., in the PDF studies discussed in Sec.~\ref{sec:lattice:xdep}.

Significant progress has been made in understanding the operator mixing effects underlying these observations in more detail. Triggered by the breaking of chiral symmetry in fermion discretization schemes such as the clover discretization, operator mixing invalidates the simple cancellation of renormalization factors in TMD ratios such as in Eqs.~\eqref{gsshift}-\eqref{gtcharge}. The mixing pattern for clover fermions in lattice perturbation theory was derived both for the straight gauge link~\cite{Constantinou:2017sej} and the staple-link~\cite{Constantinou:2019vyb} cases; the fact that a discrepancy between clover fermion and domain wall fermion results is seen specifically in the $g_{1T} $ worm gear shift, as discussed above and displayed in Fig.~\ref{mixfig}, is consistent with this mixing pattern obtained in lattice perturbation theory.  The pattern of mixing can be further understood using an auxiliary field approach to recast bilocal quark operators in terms of local operators, as laid out for straight gauge links in Ref.~\cite{Green:2017xeu}, and extended to staple links in Ref.~\cite{Green:2020xco}. The nonperturbative mixing pattern for quark bilinear operators with staple-shaped gauge connections in the RI'/MOM scheme was explored in Ref.~\cite{Shanahan:2019zcq} where mixing patterns were found that extend beyond those found in one-loop perturbative calculations; a sample result for purely transverse quark operator separation $b$ is shown in Fig.~\ref{fig:mixing_ps}.
Lattice TMD calculations must take into account these more complex renormalization patterns. One avenue is the use of chirally symmetric formulations such as domain wall fermions in order to avoid certain operator mixings, another is to use a scheme along the lines put forward in Refs.~\cite{Green:2017xeu,Green:2020xco} to correctly incorporate mixing effects in clover fermion calculations.

\begin{figure}[t]
\includegraphics[width=8.4cm]{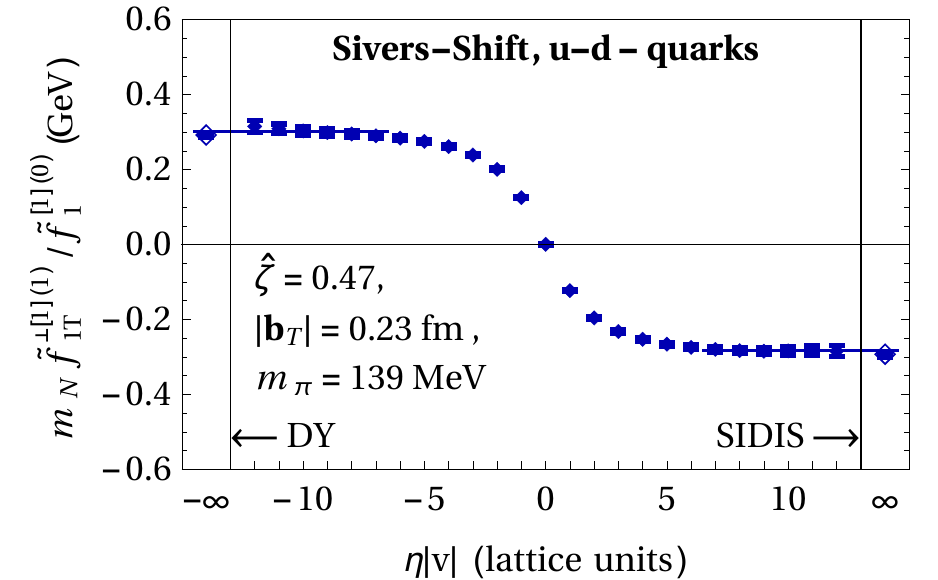}
\hspace{0.3cm}
\includegraphics[width=8.4cm]{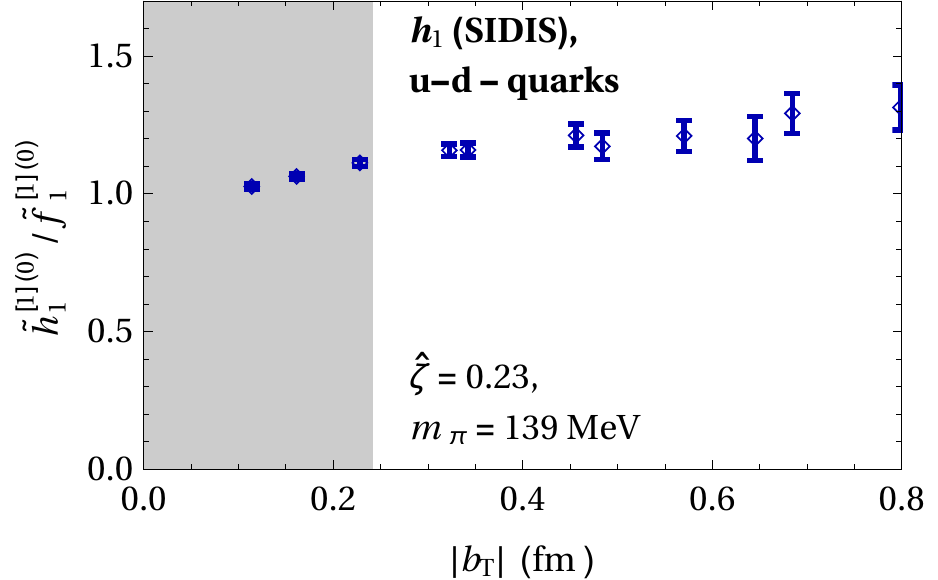}
\caption{Preliminary analysis of nucleon lattice TMD data at the physical quark masses. Left: Isovector generalized Sivers shift \index{Sivers effect} as a function of staple length $\eta $ at fixed $b_T $ and $\hat{\zeta } $. Right: Isovector generalized tensor \index{tensor charge} in the SIDIS limit as a function of $b_T $ for fixed $\hat{\zeta } $. Shaded area indicates
region which may be subject to significant lattice artefacts. Data were obtained using domain wall fermions at lattice spacing $a=0.114\, \mbox{fm} $. Plot taken from Ref.~\cite{Engelhardt:tobepublished.1}.}
\label{fig:mpi139}
\end{figure}

Progress has also been achieved in terms of extending lattice TMD calculations to the physical quark masses. Recent calculations have yielded the first results for TMD observables at the physical values of the quark masses, employing a RBC/UKQCD domain wall fermion ensemble with a lattice spacing $a=0.114\, \mbox{fm} $ \cite{Engelhardt:tobepublished.1}. Results from a preliminary analysis are exhibited in Fig.~\ref{fig:mpi139}.
Comparing the left panel with the left panel of Fig.~\ref{etafig}, displayed at similar values of $b_T $ and $\hat{\zeta } $, there appears to be no marked dependence of the isovector generalized Sivers shift on the quark masses in the explored range, extending all the way to the physical quark masses. The right panel of Fig.~\ref{fig:mpi139} shows results for the generalized tensor charge, cf. Eq.~\eqref{gtcharge}, in the SIDIS limit as a function of $b_T $.

\begin{figure}[t]
\centering
\includegraphics[width=9.4cm]{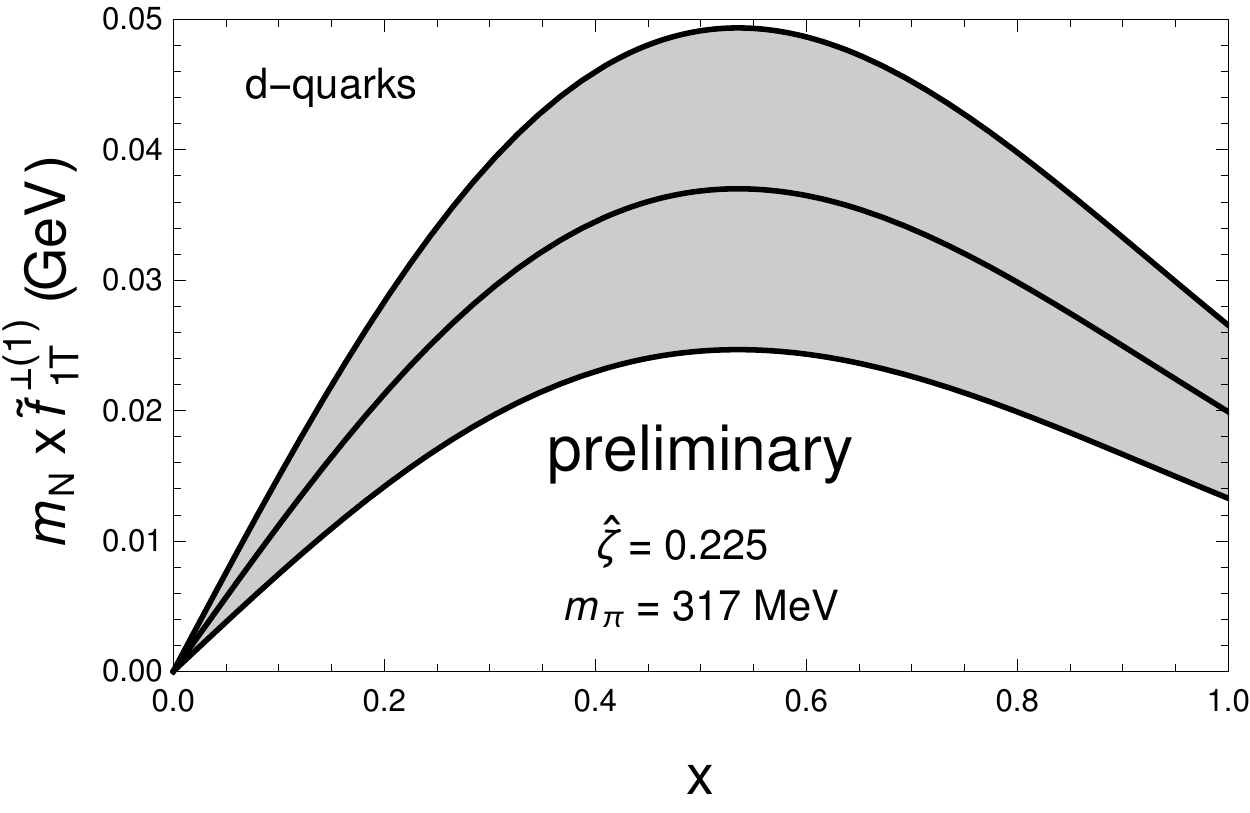}
\caption{Nucleon SIDIS $d$-quark generalized Sivers shift as a function of
momentum fraction $x$, multiplied by $x$, evaluated at $b_T =0.34\, \mbox{fm} $ at fixed
$\hat{\zeta } =0.225$. Data were obtained using a clover fermion
ensemble at $m_{\pi } =317\, \mbox{MeV} $.
This preliminary analysis, performed at rather low $\hat{\zeta } $, still significantly violates constraints such as the limit of support to $x\leq 1$; comprehensive studies in progress as of this writing are anticipated to properly account for these properties.}
\label{figsiversx}
\end{figure}

Lattice TMD calculations have also been extended to include the dependence
on the longitudinal momentum fraction $x$, by performing scans of the matrix element in Eq.~\eqref{eq:latt_corr_def_2} in the $b\cdot P$ direction;
$b\cdot P$ is Fourier conjugate to $x$. The geometries employed in performing this scan must obey the relation~\cite{Musch:2011er}
\begin{equation}
\frac{v\cdot b}{v\cdot P} = \frac{b\cdot P}{m_N^2 } \left( 1-\sqrt{1+1/\hat{\zeta }^{2} } \right)    , 
\end{equation}
which constitutes a Lorentz-invariant expression of the standard TMD kinematics. This forces one to use general off-axis directions on the lattice, which significantly complicates the analysis. On the other hand, an important simplification that arises is that the soft factors depend only on the transverse separation $b_T $. This is due to the staple-link structure of the gauge connection for TMDs. Consequently, the soft factors can be factored outside the longitudinal Fourier transformation to $x$-space. As a result, the cancellation of soft factors in ratios in $b$-space extends to ratios of longitudinal Fourier transforms, i.e., one can obtain renormalized $x$- and $b_T $-dependent TMD ratios, without having to construct the soft factors explicitly. The results of a preliminary exploration displayed in Fig.~\ref{figsiversx} indicate that it is feasible to obtain the $x$-dependence of TMD ratios in this fashion. This has motivated a new calculation underway as of this writing~\cite{Engelhardt:tobepublished.2}.

\begin{figure}[t]
\centering
\includegraphics[width=8.4cm]{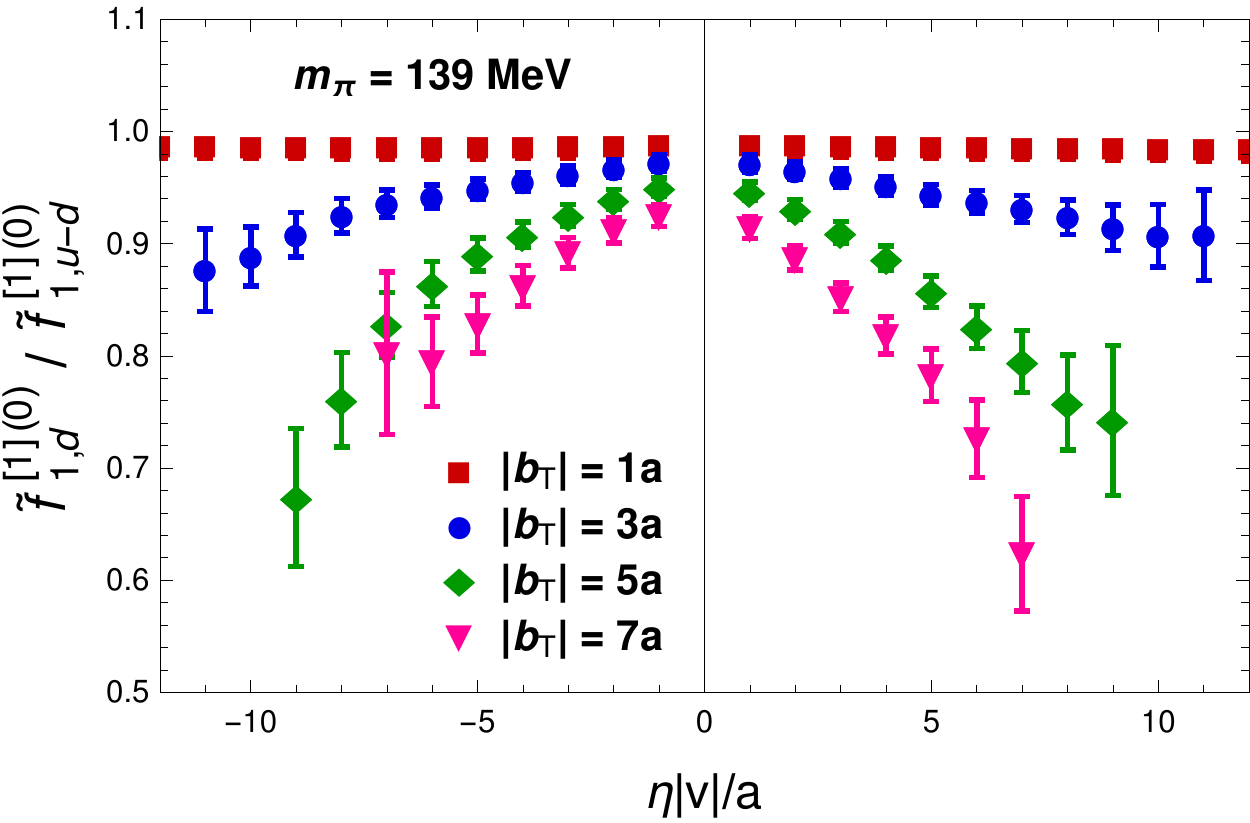}
\hspace{0.2cm}
\includegraphics[width=8.4cm]{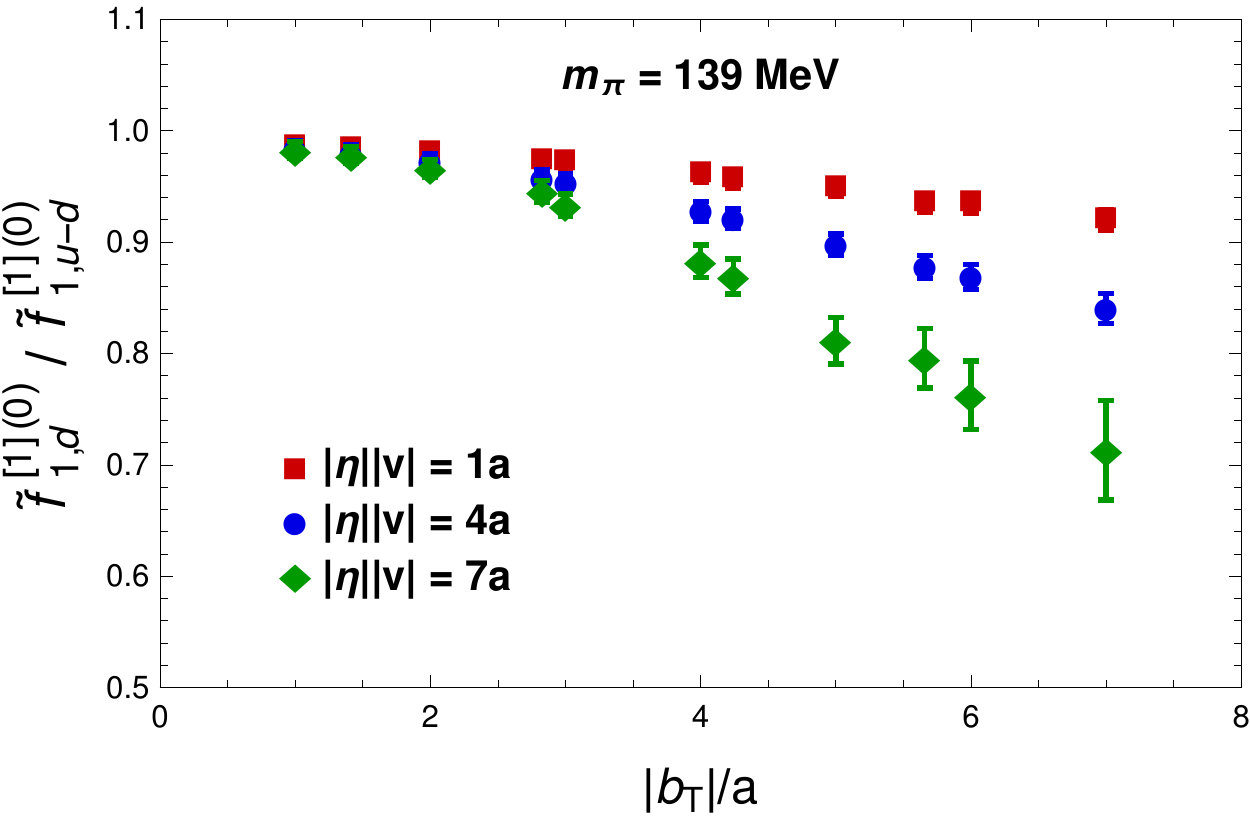}
\caption{Ratio of unpolarized TMD moments of different flavor content $f_{1,d}^{[1](0)} / f_{1,u-d}^{[1](0)} $ obtained at the physical quark masses for $\hat{\zeta } =0.23$. Left: as a function of staple length $\eta $, for selected $|b_T |$; right: as a function of $|b_T |$ for selected $|\eta |$, where for each choice of $|\eta |$, the data for $\eta =\pm |\eta |$ have been averaged. Disconnected contributions to the $d$-quark distribution are omitted, cf.~main text. Data were obtained using domain wall fermions at lattice spacing $a=0.114\, \mbox{fm}$ \cite{Engelhardt:tobepublished.1}.}
\label{fig:flav_latt}
\end{figure}

The above discussion has focused on TMD spin structure, evaluating ratios of TMD moments of different spin content, chiefly for the isovector, $u-d$, flavor combination. The reason for the isovector combination being favored in the presentation of results lies in the fact that the computationally expensive contributions from disconnected diagrams, which have hitherto not been evaluated in the lattice TMD program, exactly cancel in the isovector case. Flavor-separated quantities are subject to an additional systematic uncertainty unless these contributions are evaluated.
Nonetheless, input on the relative $u$-quark vs.~$d$-quark behavior of TMDs is highly desirable for phenomenological studies~\cite{Bacchetta:2018lna}, and can be presented in the form of ratios of TMD moments of different flavor content rather than spin content. Taking recourse to the recently obtained lattice TMD data at the physical quark masses~\cite{Engelhardt:tobepublished.1} already highlighted above, obtained using domain wall fermions at lattice spacing $a=0.114\, \mbox{fm} $, Fig.~\ref{fig:flav_latt} displays preliminary results for the ratio of unpolarized TMD moments $f_{1,d}^{[1](0)} / f_{1,u-d}^{[1](0)} $. Note that the ratio $f_{1,u}^{[1](0)} / f_{1,u-d}^{[1](0)}$ contains no additional independent information, since
\begin{equation}
\frac{f_{1,u}^{[1](0)} }{f_{1,u-d}^{[1](0)} } = \frac{f_{1,d}^{[1](0)} }{f_{1,u-d}^{[1](0)} } +1
\label{eq:latt_flav_rat}
\end{equation}
Note furthermore that the correction due to the omitted disconnected contributions would be identical for both flavors, such that Eq.~\eqref{eq:latt_flav_rat} would continue to hold. Fig.~\ref{fig:flav_latt} implies that the $d$-quark distribution is suppressed compared to the $u$-quark distribution as either $|b_T |$ or $|\eta |$ increases. For $\eta $ close to zero, i.e., in the absence of final state interactions, the dependence on $b_T $ appears fairly weak, whereas it strengthens as final state interactions are included. Conversely, in terms of transverse momentum $k_T $, one therefore expects the $d$-quark distribution to decay more slowly with rising $k_T $ than the $u$-quark distribution. The displayed data were obtained for $\hat{\zeta } = 0.23$, but analogous results for $\hat{\zeta } = 0.47$ and $\hat{\zeta } = 0.70$ do not differ significantly from those shown. In contrast to ratios with different spin content, the large-$|\eta |$ asymptotic limit is not readily reached in these ratios with different flavor content. This behavior remains to be understood and calls for further investigation.

TMD observables, as discussed in this section, are derived from the matrix element in Eq.~\eqref{eq:latt_corr_def_2} in the forward limit, $P^{\prime } =P$. On the other hand, lattice studies of the type presented here can also be generalized to nonzero momentum transfer in the transverse direction utilizing largely the same techniques, thus yielding GTMD observables. Such studies have been carried out, e.g., with a view to extracting information about quark orbital angular momentum in the proton. They are discussed in further detail in Sec.~\ref{sec:GTMD_OAM_lattice}.

\subsubsection{Calculation of soft function and TMDs}
\label{sec:latt_soft_func}

Over the past few years, much progress has been made towards the theoretical development of direct calculations of TMDs using the \index{large-momentum effective theory (LaMET)}LaMET approach~\cite{Ji:2014hxa,Ji:2018hvs,Ebert:2018gzl,Ebert:2019okf,Ebert:2019tvc,Ji:2019sxk,Ji:2019ewn,Vladimirov:2020ofp,Ebert:2020gxr,Ji:2020jeb}.
To calculate TMDs in this approach, one starts by constructing a quasi TMD\index{quasi TMD}~\cite{Ji:2014hxa,Ji:2018hvs,Ebert:2019okf}. For example, for quark of flavor $i$, the (naive) $\MSbar$ quasi TMD is defined in Eq.~\eqref{eq:qtmd0} as 
\begin{align}
 \hat{f}_i^{\rm n}(x, \bt,\mu,P^z) = \int \frac{\df b^z}{2\pi} \, e^{\img b^z (x P^z)}\,
 &\hat{Z}'_i(b^z,\mu,\tilde \mu) \hat{Z}_{\rm uv}^i(b^z,\tilde \mu, a)
 \nn\\&\times
 \hat{B}_i(b^z, \bt, a, P^z, \eta) \Big/\sqrt{{\hat S}^i(b_T, a, \eta)}
\, ,\nn\end{align}
where $b^\mu = (0, \bt, b^z)$. Here, $\hat{B}_i $ and ${\hat S}^i$ are the quasi beam and quasi soft functions, which are the analogs of the unsubtracted beam and soft functions.
The lattice renormalization factor is $\hat{Z}_{\rm uv}^i$, and $\hat{Z}^\prime_i$ converts from the lattice renormalization scheme to the $\MSbar$ scheme. 
Here ${\hat S}^i$ is naively defined from a Eulcidean Wilson loop~\cite{Ji:2018hvs,Ebert:2019okf}, we label $\hat{f}_i^{\rm n}$ with the superscript ``n'' indicating the choice of $\hat{S}^i$.
The lattice renormalization scale $\tilde \mu$ is distinguished from the $\MSbar$ scale $\mu$.
In lattice calculations, the Wilson lines that enter $\hat{B}_i $ and ${\hat S}^i$ necessarily have a finite extension $\eta$ chosen to be in the $\hat{z}$ direction, which is associated with rapidity regularization and its dependence cancels between $\hat{B}_i $ and ${\hat S}^i$.
Finally, $\hat{f}_i$ also depends on the proton momentum $P^z$ which acts as the analog of the Collins-Soper scale $\zeta$, and it was suggested that one can access information of the Collins-Soper evolution through the $P^z$-dependence~\cite{Ji:2014hxa}.

The quasi beam function is defined in Eq.~\eqref{eq:qbeam_def1} as
\begin{align} 
	\hat B_i(b^z,\bt,a,P^z,\eta) &= \frac{1}{2} \Bigl\langle p(P,S) \Big| 
    \bar\psi^{0}_i (b^{\mu }/2) \Gamma W_{{\sqsupset}\eta}^{\hat{z}} 
    (b^{\mu }/2,-b^{\mu }/2) \psi^{0}_i (-b^{\mu }/2) 
     \Big| p(P,S) \Bigr\rangle \,,\nn
\end{align}
where $\Gamma$ can be chosen as either $\Gamma = \gamma^0$ or $\Gamma = \gamma^z$.
As for the quasi soft function, its definition is not unique, and a naive choice is the vacuum matrix element of a rectangle-shaped Wilson loop along the $z$ direction,
\begin{align} \label{eq:qsoft}
 \hat{S}^q(b_T, a, \eta) &= \frac{1}{N_c} \bigl< 0 \big|
  {\rm Tr} \bigl\{ W_{-\hat z}(\bt;0,-\eta) W_{-\hat z}(\bt;\eta,0)
   W_{\hat{b}_T}(\eta \hat z;0,b_T)
   \nn\\&\hspace{2cm}\times
 W_{\hat z}(0;0,\eta) W_{\hat z}(0;-\eta,0)
 W_{\hat{b}_T}\bigl(-\eta \hat z;b_T,0\bigr) \bigr\}
 \bigl|0 \bigr>
\,,\end{align}
where the soft Wilson lines $W_{\pm\hat z}$ and $W_{\hat{b}_T}$ are along the $\pm z$ and transverse directions respectively.

\index{lattice QCD calculations!matching}
According to the boost argument in LaMET, the quasi beam function approaches the unsubtracted beam function in the infinite momentum limit, so a perturbative matching is possible between the two. However, the naive quasi soft
function fails this argument as it can only be boosted along  a single light-cone direction, which is not related to any soft function in TMD factorization. This is demonstrated by an explicit one-loop check~\cite{Ji:2018hvs,Ebert:2019okf}, where one finds that the quasi and physical TMD differs by an IR logarithm of $b_T$, which becomes nonperturbative when $b_T\sim \Lambda_{\rm QCD}^{-1}$. Although by bending the soft Wilson lines by ninety degrees removes this IR logarithm at one loop~\cite{Ji:2014hxa,Ebert:2019okf}, it was argued that it still exists at two loops due to the mismatch of cusp anomalous dimensions~\cite{Ji:2019sxk}.

Nevertheless, with the constraints from RG and Collins-Soper evolutions, as well as from one-loop results~\cite{Ebert:2019okf}, it was argued that the non-singlet naive quasi TMD is related to the physical TMD through
\begin{align} \label{eq:matching_result}
 \hat{f}_{\rm ns}^{\rm n}(x, \bt, \mu, P^z)
 = &~C^\TMD_{\rm ns}\bigl(\mu, x P^z\bigr)
   g^S_{q}(b_T, \mu)
   \exp\biggl[ \frac12  \gamma_\zeta^q(\mu, b_T) \ln\frac{(2 x P^z)^2}{\zeta} \biggr]
 \nn\\&\times
 \tilde f_{\rm ns}(x, \bt, \mu, \zeta)+ \cO\biggl(\frac{b_T}{\eta}, \frac{1}{b_T P^z}, \frac{1}{P^z \eta}\biggr)
\,,\end{align}
where $\gamma_\zeta^q(\mu, b_T)$ is the Collins-Soper evolution kernel. The perturbative matching coefficient $C^\TMD_{\rm ns}$ is diagonal in $x$-space and is also independent of the spin structure~\cite{Vladimirov:2020ofp,Ebert:2020gxr,Ji:2020jeb,Ebert:2022fmh}, and has been derived at one-loop order~\cite{Ji:2014hxa,Ebert:2018gzl,Ebert:2019okf}. The nonperturbative factor $g^S_{q}(b_T, \mu)$ reflects the failure of the naive quasi soft function, but is independent of the external hadron state or quark flavor. The power corrections follow from the hierarchy of scales $b^z\sim 1/P^z \ll b_T \ll \eta$.

With the above relation, it was proposed that one can calculate the Collins-Soper kernel~\cite{Ebert:2018gzl} and ratios of spin-dependent TMDs by forming ratios of the quasi TMDs in different hadron momentum states and with different spin structures~\cite{Ebert:2019okf,Ebert:2020gxr}, as both $g^S_{q}(b_T, \mu)$ and the quasi soft function cancel out.

Later on, it was found out that by replacing ${\hat S}^i$ with the Collins soft function the revised quasi TMD is equivalent to the LR scheme (see \sec{latt_schemes}) under a large proton rapidity expansion~\cite{Ebert:2022fmh}. Since the LR and Collins schemes differ by the $\eps\to0$ an $y_B\to -\infty$ limits~\cite{Ebert:2022fmh}, they can be perturbatively matched to each other according to the LaMET formalism, which is also the general EFT principle. This allows one to derive and generalize the factorization formula for the quasi TMD $\hat f_i$ as~\cite{Ebert:2022fmh},
\begin{align} \label{eq:qTMD_fact}
\hat{f}_i(x, \bt, \mu, P^z)\equiv \frac{ \hat{f}_i^{\rm n}(x, \bt, \mu, P^z)}{g^S_{\kappa_i}(b_T, \mu)}
 = &~C^\TMD_{\kappa_i}\bigl(\mu, x P^z\bigr)
   \exp\biggl[ \frac12  \gamma_\zeta^{\kappa_i}(\mu, b_T) \ln\frac{(2 x P^z)^2}{\zeta} \biggr]
 \nn\\&\times
 \tilde f_i(x, \bt, \mu, \zeta)+ \cO\biggl(\frac{b_T}{\eta}, \frac{1}{b_T P^z}, \frac{1}{P^z \eta}\biggr)
\,,\end{align}
where $i$ refers to either a gluon ($i=g$) or a specific quark flavor ($i=u,d,s,\ldots$), and $\kappa_i=q$ is universal for all quarks, but differs from $\kappa_i=g$ for gluons. Motivated by this derivation, the matching coefficient $C^\TMD_g\bigl(\mu, x P^z\bigr)$ for gluon quasi TMDs has also been recently derived at one-loop order~\cite{Schindler:2022eva}, which is different from $C^\TMD_q$ by replacing the SU(3) Casimir $C_F$ with $C_A$. Using the momentum RG equation, one can resum the matching coefficient $C^\TMD_{\kappa_i}\bigl(\mu, x P^z\bigr)$ at NLL accuracy~\cite{Ji:2019ewn,Ebert:2022fmh}. Notably, there is no mixing between the gluon and singlet quark channels, which will greatly simplify the lattice calculation of gluon TMDs.

Moreover, the function $g^S_{\kappa_i}(b_T, \mu)$ was found to be~\cite{Schindler:2022eva}
\begin{align}\label{eq:gs}
    g^S_{\kappa_i}(b_T, \mu) = \lim_{y_B\to-\infty} e^{-\gamma_\zeta^{\kappa_i}(\mu, b_T)(y_n-y_B)}\sqrt{\frac{\tilde S^i_{n_A(y_n)n_B(y_B)}(b_T,\mu,2y_n-2y_B)}{{\hat S}^i(b_T,\mu)}}
\,,\end{align}
where $\tilde S^i$ is the Collins soft function in \eq{ColinsSoft}, and ${\hat S}^i(b_T,\mu)$ is the $\MSbar$ renormalized naive quasi soft function in the continuum with $\eta=\infty$. This result exactly agrees with the reduced soft function $S^q_r(b_T,\mu)$ proposed in Ref.~\cite{Ji:2019sxk} as the rapidity-independent part of the CSS soft function.

More recently, an important step forward has been made with the proposal~\cite{Ji:2019sxk} to calculate the reduced soft function $S^q_r(b_T,\mu)$ through a time-like soft factor in heavy-quark effective theory on the lattice or through the pion form factor of a current-current correlator.
The latter method features a form factor defined as
\begin{align}
    F(b_T, P^z)&=\langle \pi(-P)|j_1(b_T)j_2(0)|\pi(P)\rangle\,,
\end{align}
where $j_1$ and $j_2$ are light-quark currents separated in the transverse plane, and the initial- and final-state pions travel with opposite momenta. In the large momentum limit, it is proposed that the above form factor can be factorized as~\cite{Ji:2019sxk}
\begin{align}
    F(b_T, P^z)&= S^q_r(b_T,\mu)\int dx dx'\ H(x,x',\mu)\ \Phi^\dagger(x,b_T,P^z,\mu)\ \Phi(x',b_T,P^z,\mu) + \ldots\,,
\end{align}
where $H(x,x',\mu)$ is a matching coefficient, $\Phi$ is a quasi TMD wave function defined with the same operator for the quasi beam function, and $\ldots$ are power corrections. The square root of the reduced soft function $S^q_r(b_T,\mu)$ can be identified as $g^S_{q}(b_T, \mu)$.

\begin{figure}[!t]
    \centering
    \includegraphics[width=0.6\textwidth]{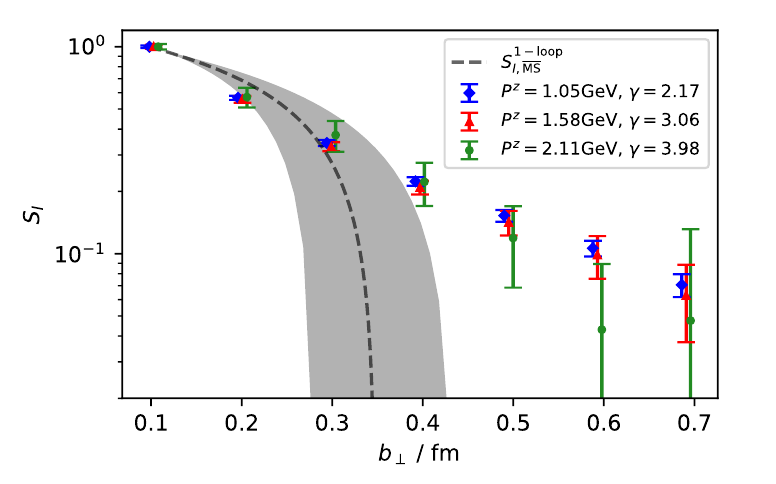}
    \caption{The first lattice results of the TMD soft function extracted from the pion form factor at different momentum $P^z$ with tree-level matching~\cite{Zhang:2020dbb}. The dashed line is the one-loop perturbative prediction, which becomes unreliable at $b_T\sim 0.3$ fm due to reaching the Landau pole. Plot taken from Ref.~\cite{Zhang:2020dbb}.}
    \label{fig:sr}
\end{figure}

This method has been applied for the first lattice calculation of the reduced soft function in Ref.~\cite{Zhang:2020dbb}, with the result shown in \fig{sr}. The result was obtained with tree-level matching at different pion momentum. The agreement with perturbative prediction at small $b_T$ and insensitivity to the pion momentum $P^z$ is a  promising sign of the effectiveness of this method. 
A new calculation with reduced operator mixing was carried out in Ref.~\cite{Li:2021wvl}, and similar agreement with perturbation theory has also been observed.
Finally, with the calculation of the quasi TMD, the physical TMD as well as the Drell-Yan cross section can be completely determined from LQCD~\cite{Ji:2019ewn}. This has also facilitated the method to calculate phenomenologically interesting TMDs such as the Sivers function~\cite{Ji:2020jeb}, as well as light-cone wavefunctions~\cite{Ji:2021znw}.

\subsubsection{Lattice QCD input to TMD evolution}
\label{sec:latticeCS}

\index{Collins-Soper evolution kernel!lattice calculations|(}
In addition to calculations of key TMD observables, as described in the preceding subsections, LQCD can provide important constraints on the Collins-Soper evolution kernel, also known as the rapidity anomalous dimension, which governs TMD evolution as discussed in \chap{evolution}. This kernel is nonperturbative for small parton transverse momentum $q_T\sim\Lambda_\text{QCD}$, and first-principles calculations of this quantity would provide insight into the  discrepancies in phenomenological determinations of the kernel in this region~\cite{Vladimirov:2020umg}.
 
\begin{figure}[t!]
    \centering
    \includegraphics[width=0.45\textwidth]{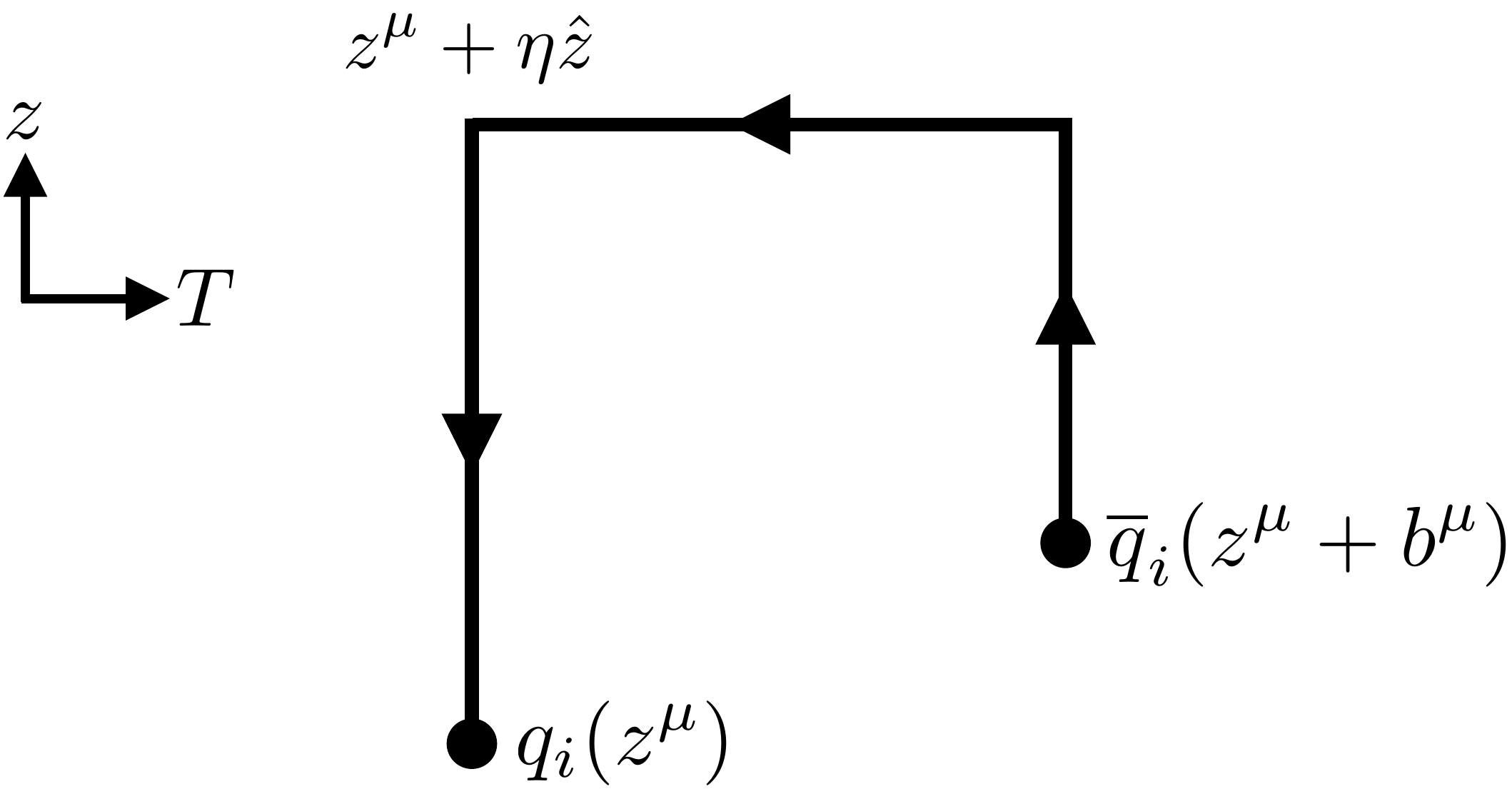}
    \caption{Illustration of the staple-shaped Wilson line structure of the nonlocal quark bilinear operators defining quasi beam functions $\hat B^\Gamma_{i}(b^\mu,a,\eta,P^z)$.}
    \label{fig:staple}
\end{figure}

Methods to determine the Collins-Soper kernel from LQCD have been developed in Refs.~\cite{Ebert:2018gzl,Ebert:2019okf,Ebert:2019tvc,Vladimirov:2020ofp}, based on the identification of the kernel with the ratio of quasi TMDs $\hat f_{\text{ns}}$:
\begin{align}\label{eq:gamma_zeta1}
	\gamma^q_\zeta(\mu, b_T) = \frac{1}{\ln(P^z_1/P^z_2)} \ln \frac{C^\TMD_q (\mu,x P_2^z)\, \hat f_{\text{ns}}(x, \bt, \mu, P_1^z)}	{C^\TMD_q (\mu,x P_1^z)\, \hat f_{\text{ns}}(x, \bt, \mu, P_2^z)}+\mathcal{O}\left({1}/({b_T P^z_i})\right)\,.
\end{align}
In this expression, $P^z_i \gg \Lambda_\text{QCD}$ are the $z$-components of the hadron momenta and $C^\TMD_\text{ns}$ is a perturbative matching coefficient that has been obtained at one-loop order~\cite{Ebert:2018gzl,Ebert:2019okf}. 
The quasi TMD $\hat f_i$ is defined as
\begin{align}\label{eq:quasiTMD}
&\hat{f}_{i}\big(x, \bt, \mu, P^{z}\big)\equiv  \lim_{\substack{a\to 0 \\ \eta\to \infty}}
\int \frac{\mathrm{d} b^{z}}{2 \pi} e^{-\mathrm{i}b^{z}\left(x P^{z}\right)} \mathcal{Z}^{\MSbar}_{\gamma^4\Gamma}(\mu,b^z\!,a){P^z\over E_{\vec{P}}}\hat{B}^{\Gamma}_{i}\big(b^{z},\bt,a,\eta,P^{z}\big) \hat{\Delta}_{S}\left(b_{T},a,\eta\right)\,.
\end{align}
Here $a$ denotes the lattice spacing, the subscript $i$ is the flavor index, and summation over Dirac structures $\Gamma$, is implied. 
The quasi beam function $\hat B^\Gamma_{i}(b^\mu,a,\eta,P^z)$ is defined as the matrix element of a quark bilinear operator with a staple-shaped Wilson line and Dirac structure as in Eq.~\eqref{eq:latt_corr_def_2}, where in this context the spacetime coordinates have been shifted and the closure of the staple is effected in an asymmetric fashion, as illustrated in Fig.~\ref{fig:staple}. 
The quasi soft factor $\hat{\Delta}_S$~\cite{Ji:2014hxa,Ji:2018hvs,Ebert:2018gzl,Ebert:2019okf} is also calculable in LQCD, but cancels in the ratio of Eq.~\eqref{eq:gamma_zeta1}.
The factor $\mathcal{Z}^{\MSbar}_{\gamma^4\Gamma}(\mu,b^z,a)$ renormalizes the quasi TMD and matches it to the $\overline{\text{MS}}$-scheme quasi TMD with Dirac structure $\gamma^4$ (where `4' indexes the temporal direction) at scale $\mu$~\cite{Constantinou:2019vyb,Ebert:2019tvc,Shanahan:2019zcq} (the Dirac structure $\gamma^3$ can also be used to match to the spin-independent TMD in the infinite-momentum limit).

\begin{figure*}[t!]
        \centering
\includegraphics[width=0.6\textwidth]{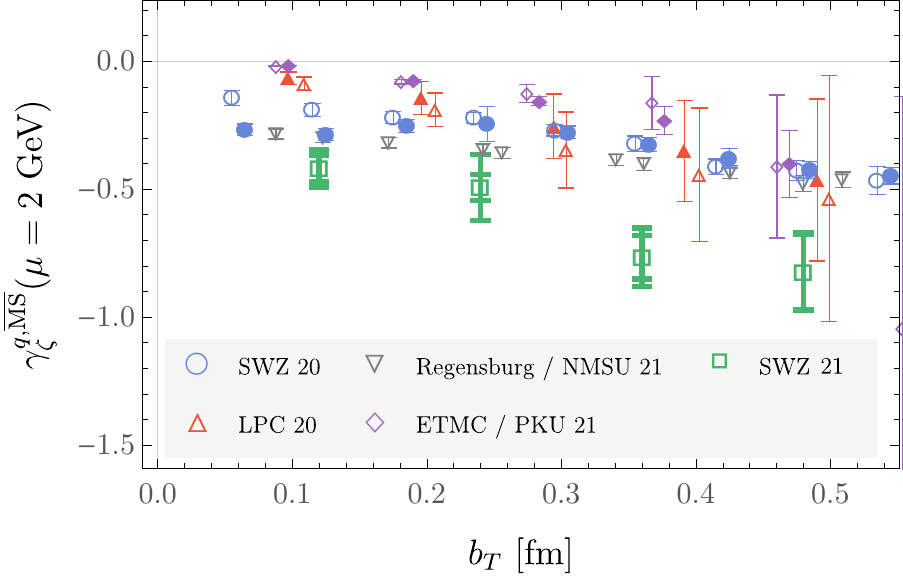}
        \label{CSvsbT}
        \caption{\label{fig:CS} Comparison between the Collins-Soper evolution kernel obtained from LQCD calculations in Ref.~\cite{Shanahan:2020zxr} (SWZ 20), Ref.~\cite{Zhang:2020dbb} (LPC 20), Ref.~\cite{Schlemmer:2021aij} (Regensburg/NMSU 21), and Ref.~\cite{Li:2021wvl} (ETMC/PKU 21), and Ref.~\cite{Shanahan:2021tst} (SWZ 21). Different sets of points with the same color show different sets of results from the same collaboration. Figure adapted from Ref.~\cite{Shanahan:2021tst}. }
\end{figure*}

In Refs.~\cite{Shanahan:2020zxr,Shanahan:2019zcq}, an exploratory calculation of the nonperturbative Collins-Soper kernel was undertaken in quenched LQCD, based on the method developed in Refs.~\cite{Ebert:2018gzl,Ebert:2019okf,Ebert:2019tvc}. In that calculation, the kernel was extracted over a range of scales $b_T\in (0.1,0.8)~\text{fm}$. The final results relied on modeling the $b^z$-space quasi beam functions to control truncation effects in the Fourier transform; nevertheless, the determination of the Collins-Soper kernel was found to be robust under the variation of models considered. More recently, the calculation of Ref.~\cite{Shanahan:2021tst} refined that exploratory study with an updated investigation following the same approach, but using dynamical fermions and more general functional forms as models in $b^z$-space.

Complementing the approach of determining the Collins-Soper kernel directly from Eq.~\eqref{eq:gamma_zeta1}, an alternative strategy using the Mellin moments of the expressions was proposed in Ref.~\cite{Vladimirov:2020ofp} and implemented in a fully-dynamical calculation in Ref.~\cite{Schlemmer:2021aij} in a study in which the Collins-Soper kernel was determined from three different TMDs $(f_1, g_{1T} , h_1)$ for the first time. In that approach, one only needs to calculate the quasi beam function or its derivatives at $b^z=0$. In comparison to the more direct approach of Refs.~\cite{Ebert:2018gzl,Ebert:2019okf,Ebert:2019tvc}, which requires the numerically-challenging integral over $b^z$ in the Fourier transform of Eq.~\eqref{eq:quasiTMD}, this reduces the computational cost and has the advantage that renormalization factors cancel in the ratio. However, this approach also requires a non-trivial integration over the TMD that is extracted from experiments, or theory, over a limited kinematic range. Similar methods~\cite{Ji:2021znw} have been pursued in Ref.~\cite{Zhang:2020dbb}, which presented the first dynamical calculation of the Collins-Soper kernel, and in Ref.~\cite{Li:2021wvl}, each of which also obtained the kernel via (different) ratios of bare quasi TMD wave functions at $b^z=0$ (i.e., under the assumption that mixing between quasi beam functions with different Dirac structures is negligible, and using leading-order matching). Comparison of results of the various methods will be valuable as calculations advance to phenomenologically relevant precision; the analysis of Ref.~\cite{Shanahan:2021tst} includes an analysis of several approaches, revealing significant systematic differences between the methods even when applied to the same LQCD beam function dataset. Very recently, an improved calculation of the Collins-Soper kernel with the quasi TMD wave function approach and with NLO matching was reported in Ref.~\cite{LPC:2022ibr}.

\begin{figure*}[t!]
        \centering
        \includegraphics[width=0.9\textwidth]{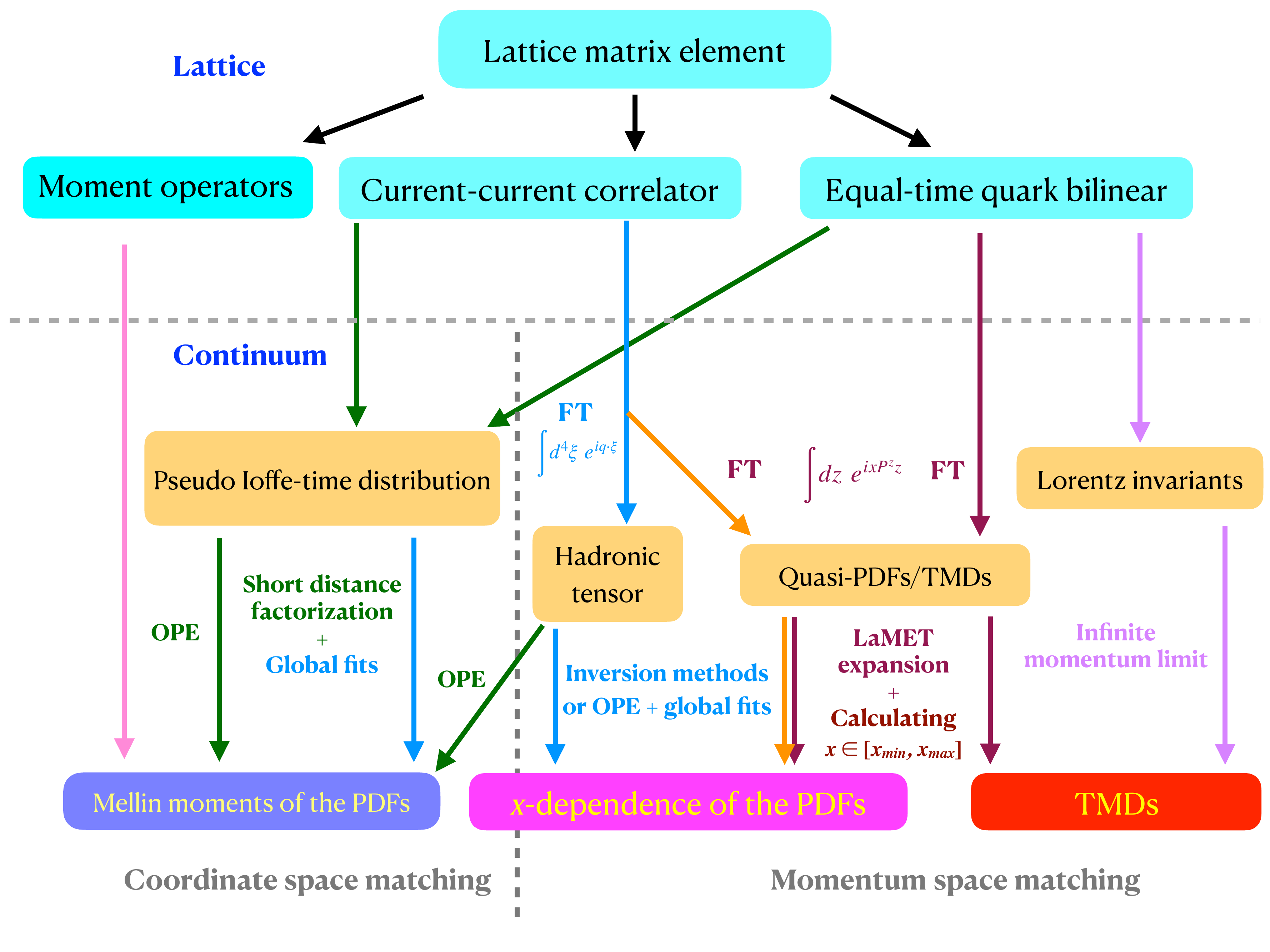}
\vspace{-0.4cm}
        \caption{\label{fig:comp_methods} Comparison of different approaches to calculate the PDFs and TMDs from lattice QCD.}
\end{figure*}

Fig.~\ref{fig:CS} summarizes the existing state-of-the-art LQCD calculations of the Collins-Soper kernel. Although the systematic uncertainties remain to be fully controlled, a qualitative conclusion can also already be drawn from the existing results; all LQCD calculations exhibit mild $b_T$ dependence in the Collins-Soper kernel at large values approaching $b_T\sim 1$fm, and it is clear that controlled first-principles calculations of the Collins-Soper kernel at nonperturbative scales as large as $b_T\sim 1$~\text{fm} are tractable with current methods.

\index{Collins-Soper evolution kernel!lattice calculations|)}

\subsubsection{Summary}
\label{sec:lattice:tmd:summary}
To summarize LQCD approaches to TMDs and the relation of those used to access PDFs, the different methods are compared in Fig.~\ref{fig:comp_methods}. With further development, it is expected that  lattice QCD will provide systematically controlled predictions for TMD physics.

%% file: sec-models/sec-models.tex
\newpage
\section{Models}
\label{sec:models}

\subsection{Why Models?}
\label{Sec:models-why}

In order to describe the structure of hadrons, it is
necessary to solve QCD in the nonperturbative regime.
The state-of-the-art
first-principle tool for that is LQCD where impressive 
progress has been made, see Sec.~\ref{sec:lattice}.
Models are not in competition with lattice studies but provide 
important complementary tools,
and are used in two conceptually different ways. 

(i) The first is expository. If there is a point that needs to 
be made that is independent of the details of the theory, then using a 
"toy model" can be effective to circumvent technical details of the full
theory, and elucidate the underlying physics. Here one often is happy
with a "proof-of-principle demonstration" and is not concerned how
realistic the used model is, as long as the model shares with QCD 
the essential features for the considered aspect.

(ii) The second is descriptive. Here the goal is to "approximate QCD" 
and determine, e.g., the nonperturbative properties of TMD functions 
as reliably as possible, e.g, in order to produce estimates for cross
sections. For that it is important to understand the range of 
applicability and the limitations of the used models. 

Regarding (i), it is worth recalling that model calculations 
have made a number of important contributions to the understanding of 
TMD physics. To name a few examples, let us mention the one-loop model 
calculation in a spectator model with an abelian gauge field 
\cite{Brodsky:2002cx} which paved the way towards the understanding 
of T-odd TMDs in QCD \cite{Collins:2002kn} and is reviewed in 
Sec.~\ref{Subsec-models:review-Brodsky-Hwang-Schmidt}. 
Similarly, model studies of the fragmentation process
\cite{Metz:2002iz} provided a basis for the understanding of the
universality of TMD fragmentation functions \cite{Collins:2004nx}
which is reviewed in Sec.~\ref{Sec:model-universality-FFs}.
As a last example, let us mention that calculations in quark-target 
models \cite{Kundu:2001pk} helped to establish that in QCD no 
relations exist between different TMD functions
\cite{Goeke:2003az,Meissner:2009ww} which will be 
reviewed in Sec.~\ref{Sec:models-qLIRs}.

Regarding (ii), 
let us highlight the many important
practical applications of models which range from  
predictions of new observables, to projections for future 
experiments, to guiding educated Ans\"atze for TMD fits, 
to building Monte Carlo event generators \cite{Avakian:2015vha}.
When phenomenological extractions of TMD functions are available, 
models allow us to train our physical intuition and interpret the 
results. If it is possible to explain a certain observation in a
model, this can shed valuable light on the underlying physics
because in models one can focus on specific aspects of hadronic 
physics and determine in simplified theoretical frameworks the 
roles these aspects play for a given process or partonic property.

Progress in TMD physics arises from combined efforts in experiment,
perturbative QCD, lattice QCD and phenomenology, and the work in
models contributes its share to this.

\subsection{The Brodsky-Hwang-Schmidt Calculation of a Transverse SSA}
\label{Subsec-models:review-Brodsky-Hwang-Schmidt}
\index{model!spectator model}
\index{Sivers function $f_{1T}^{\perp}$!model calculations}

\begin{figure}[t!]
\begin{center}
\includegraphics[width=12.0cm]{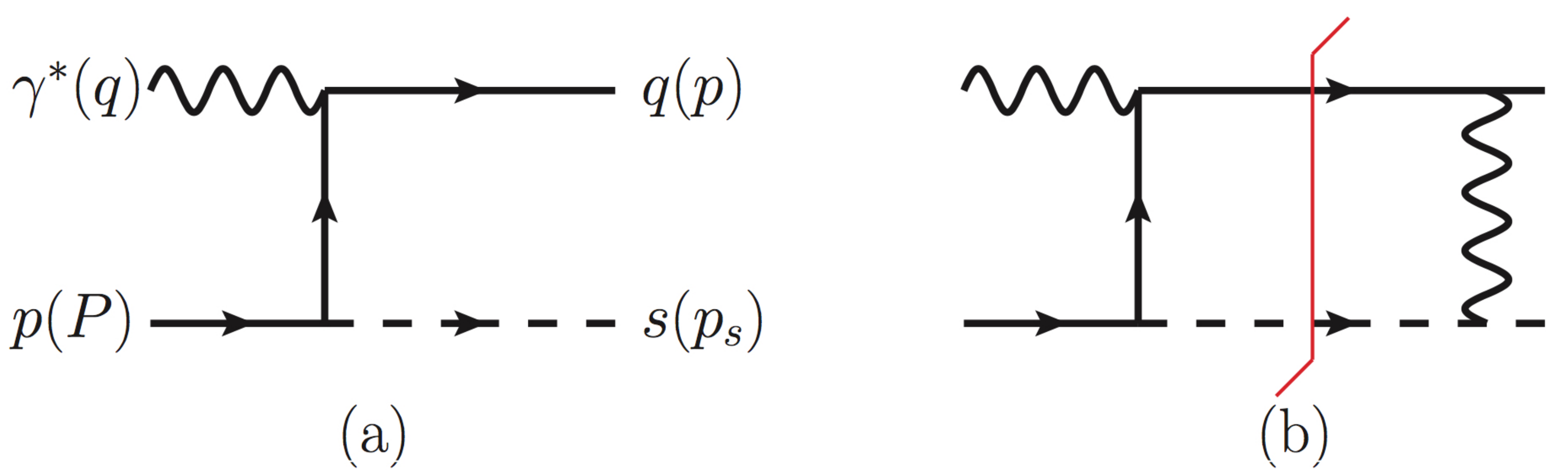} 
\end{center}
\vspace{-0.4cm}
\caption{Tree-level (left panel) and specific one-loop contribution (right panel) to the process in~(\ref{e:BHS_process}). 
The spectator is indicated by a dashed line. 
The red line in diagram (b) is the possible on-shell cut, which is essential for generating a transverse SSA. 
The interaction between the struck quark and the spectator is modeled through the exchange of a single (Abelian) gauge boson.
In the model of Ref.~\cite{Brodsky:2002cx} the proton carries no electric charge, while the quark charge is $e_1$ and the spectator charge $-e_1$.
} 
\label{f:BHS}
\end{figure}

In this section, we present a brief discussion of the model calculation by Brodsky, Hwang and Schmidt (BHS)~\cite{Brodsky:2002cx} of a transverse SSA, which played an
essential role in our understanding of T-odd TMD PDFs and as such had a significant influence on the TMD field. 
(See also the related discussions in Sec.~\ref{sec:beyondparton} and Sec.~\ref{sec:universality}.)
Specifically, the process
\begin{equation} 
\gamma^{\ast}(q) + p(P,S_T) \to q(p) + s(p_s)
\label{e:BHS_process}
\end{equation}
was considered, that is, a virtual photon hits a proton producing a quark ($q$) and a spectator ($s$).
This reaction was studied in a simple scalar diquark spectator model which, in particular, is characterized by a point-like proton-quark-diquark interaction.
The lowest-order diagram of the process is displayed in Fig.~\ref{f:BHS}(a).
The process in~(\ref{e:BHS_process}) can be viewed as a subprocess of SIDIS.
Of course, in the real world, the final-state quark will hadronize, where one may consider either semi-inclusive hadron or jet production.
However, extending the model to include the hadronization of the quark would not affect the main conclusion of the calculation in Ref.~\cite{Brodsky:2002cx}, namely, that for the reaction in~(\ref{e:BHS_process}) there is a nonzero SSA for a transversely polarized proton.
While it had been known since the 1960s that for processes like the one in~(\ref{e:BHS_process}) one can have nonzero transverse SSAs~\cite{Barut:1960zz}, the significance of this result for semi-inclusive reactions had not been realized before the BHS paper and a follow-up work by Collins~\cite{Collins:2002kn}.

For definiteness, we consider the Breit frame of the virtual photon, with the photon moving along the negative $z$-direction.
The proton has the large plus momentum $Q / \sqrt{2} x$, where $x = x_{\rm Bj} + {\cal O}(1/Q^2)$.
The quark carries the large minus momentum $p^- \approx q^-$ and a small transverse momentum $\pt$.
These requirements specify the kinematics according to 
\begin{eqnarray} 
& & q = \bigg( - \frac{Q}{\sqrt{2}}, \, \frac{Q}{\sqrt{2}}, \, \mathbf{0}_T \bigg) , \qquad
    P = \bigg( \frac{Q}{\sqrt{2} \, x}, \, \frac{xM^2}{\sqrt{2} \, Q}, \, \mathbf{0}_T \bigg) ,
\label{e:BHS_kin} \\
& & p = \bigg( \frac{p_T^2 + m_q^2}{\sqrt{2} \, Q}, \, \frac{Q}{\sqrt{2}}, \ \pt \bigg) , \qquad
    p_s = \bigg( \frac{Q (1-x)}{\sqrt{2} \, x}, \, 
                 \frac{x (p_T^2 + m_s^2)}{\sqrt{2} \, Q (1-x)}, \, - \pt \bigg) .
\nonumber
\end{eqnarray}
The expressions for $q$ and $P$ are exact, while for $p$ and $p_s$ just the leading terms are listed. 
The lowest-order diagram in Fig.~\ref{f:BHS}(a) alone leads to a vanishing transverse SSA, since this diagram does not have an imaginary part which is a necessary condition for such a spin observable.
However, a nonzero transverse SSA can be obtained through the interference of the diagrams in Fig.~\ref{f:BHS}(a) and Fig.~\ref{f:BHS}(b), with the latter providing the required imaginary part.
Averaging over the transverse photon polarizations, summing over the polarizations of the final-state quark and taking a transverse target SSA (with polarization of the proton in the $y$-direction) one finds~\cite{Brodsky:2002cx} 
\begin{equation} 
A_{UT,y} = \frac{(e_1)^2}{8 \pi} \, 
            \frac{2 (Mx + m_q) \, p_T^{x}}
            {(Mx + m_q)^2 + p_T^2} \,
            \frac{p_T^2 + \tilde{M}^2}{p_T^2} \,
            \ln \frac{p_T^2 + \tilde{M}^2}{\tilde{M}^2} \,,
\label{e:BHS_SSA}
\end{equation}
with $\tilde{M}^2 = x (1 - x) \big( - M^2 + m_q^2 / x + m_s^2 / (1 - x) \big).$\footnote{The overall sign of the transverse SSA reported in Ref.~\cite{Brodsky:2002cx} was incorrect as pointed out in Ref.~\cite{Burkardt:2003je}.}
The asymmetry vanishes if the transverse momentum of the quark vanishes.
We emphasize that $A_{UT}$ would be zero if there was no interaction between the struck quark and the spectator particle.
Initially, $A_{UT}$ in~(\ref{e:BHS_SSA}) was considered a new leading-twist effect which shows up in the TMD regime of semi-inclusive DIS and which may not even be factorizable~\cite{Brodsky:2002cx}.
Soon afterwards, however, the non-vanishing asymmetry was shown to be neither a new effect nor in contradiction with QCD factorization~\cite{Collins:2002kn}.
It rather can be understood as a model calculation for the T-odd Sivers function~\cite{Sivers:1989cc, Sivers:1990fh}, if the gauge link is included in its definition~\cite{Collins:2002kn}.
Therefore, in Ref.~\cite{Brodsky:2002cx} it was actually demonstrated for the first time explicitly that T-odd parton distributions can be nonzero.
A calculation of the Drell-Yan counterpart of 
the transverse SSA in Eq.~(\ref{e:BHS_SSA}) 
showed that this quantity reverses its sign~\cite{Brodsky:2013oya},\footnote{The same conclusion 
    was reached in an earlier work~\cite{Brodsky:2002rv},
    but the way the result was obtained could not be
    justified~\cite{Brodsky:2013oya}.}
in full agreement with the interpretation of the SSA as 
a model for the Sivers function and the model-independent
prediction of the relative sign of the Sivers function
between SIDIS and Drell-Yan~\cite{Collins:2002kn}.
For more discussion of the sign reversal of T-odd TMD PDFs we refer to Secs.~\ref{sec:beyondparton}, \ref{sec:universality} and \ref{sec:leadingTMDPDF}.

\subsection{Limits in QCD}
\label{Sec:models-limits-in-QCD}

This section discusses limits in QCD which can be
understood as specific models providing guidelines for 
the understanding of the nonperturbative properties of 
TMD PDFs. 

\index{parton model}
\subsubsection{The parton model}
\label{Sec:parton-model-general}

Based on Feynman's intuitive ideas \cite{Feynman:1969ej}, 
the parton model played an important role in establishing 
QCD as the theory of strong interactions
\cite{Gross:1973id,Politzer:1973fx}.
The formal connection of the parton model to QCD was
elucidated in \cite{Ellis:1978ty}. In many situations,
the parton model can be considered the "zeroth order
approximation" to QCD although this can not be understood as a rigorous limit. Nevertheless, 
owing to the asymptotic freedom of QCD, it is not 
surprising to obtain in this way useful (zeroth order) 
descriptions of cross sections
of many high-energy processes. The calculations of TMD
processes in such parton model frameworks are often a 
good first starting point for phenomenology, see the 
historical remarks in Secs.~\ref{sec:naivedef} and
\ref{sec:phenoTMDs_intro}. 

One such approach is the generalized parton model of
Refs.~\cite{Anselmino:2005sh,Anselmino:2011ch}, where, 
assuming factorization, various processes were studied 
at tree-level taking into account the 
transverse motions of partons in the initial hadrons
and/or of hadrons originating from a fragmenting parton. 
Making use of the helicity formalism, the cross sections
for the partonic subprocesses were computed at LO with 
exact (non-collinear) kinematics. This introduces phases 
in the expressions of the helicity amplitudes describing
a process which may lead to cancellations not present
when the kinematics in the partonic subprocess is 
strictly collinear. The purpose of the generalized parton model was not to compute or predict TMD PDFs
which were determined by fits to the data.
Rather the approach was of value for the phenomenological
exploration of TMD processes at early stages when, e.g., 
the exact TMD PDF definitions were not yet known. 
A systematic development of parton model concepts is 
the covariant parton model, Sec.~\ref{Sec:parton-model}.

\FloatBarrier
\subsubsection[  
The large-\texorpdfstring{$N_c$}{Nc} limit]{\boldmath The large-$N_c$ limit}
\FloatBarrier
\label{subsubsec:largeNc}
\index{large-$N_c$ limit}

\begin{table}[b!]
\renewcommand{\arraystretch}{1.4}
\begin{center}
\begin{tabular}{c|cccccccc}
TMD PDF & $f_1^q$ & $g_1^q$ & $h_1^q$ & $f_{1T}^{q\perp}$ & $g_{1T}^{\perp q}$ & 
          $h_{1L}^{\perp q}$ & $h_{1}^{\perp q}$ & $h_{1T}^{\perp q}$ \cr 
          \hline
$u+d$ & $N_c^2$ & $N_c$ & $N_c$ & $N_c^2$ & $N_c^2$ & $N_c^2$ & $N_c^3$ & $N_c^3$
        \cr
$u-d$ & $N_c$ & $N_c^2$ & $N_c^2$ & $N_c^3$ & $N_c^3$ & $N_c^3$ & $N_c^2$ & $N_c^4$
\end{tabular}
\vspace{-4mm}
\end{center}
\caption{\label{Tab:large-Nc} The large-$N_c$ behavior of the $u\pm d$ 
  flavor combinations of the nucleon TMD PDFs \cite{Pobylitsa:2003ty}.}
\end{table}

The limit $N_c\to\infty$ is a powerful theoretical tool \cite{tHooft:1973alw}.
In this limit, baryons are described as classical solitons of 
mesonic fields \cite{Witten:1979kh,Witten:1983tx} and their masses grow 
as $M\sim N_c$. While its exact solution in QCD is unknown in 3D, 
the symmetries of this large-$N_c$ soliton field are known.  
This information is sufficient \cite{Pobylitsa:2000tt} to derive 
relations for the flavor dependence of TMDs \cite{Pobylitsa:2003ty}.  
In the situation that $xN_c$ and $k_T$ are kept fixed as $N_c$ grows,
the results are summarized in Table~\ref{Tab:large-Nc}.
Analogous relations hold for antiquarks. 

In unpolarized (polarized) TMD PDFs, the $u+d$ ($u-d$) flavor combinations 
are leading in the large-$N_c$ expansion. Notice that TMD PDFs with a
$\perp$-label appear with 1 or 2 powers of $k_T/M$ in the quark 
correlator, see Eq.~(\ref{eq:tmd_decomposition}).
Due to $M\sim N_c$ this enhances the large-$N_c$ counting 
of the corresponding TMD PDFs in Table~\ref{Tab:large-Nc}. 
Observables defined as ratios, 
like spin or azimuthal asymmetries, are generically of order $N_c^0$. 
This is in particular the case for all proton or neutron asymmetries. 
In the case of the isoscalar deuteron target, however, all spin asymmetries 
are of order $1/N_c$. Even though in nature the number of colors is $N_c=3$,
this suppression is seen in experiment, where deuteron spin asymmetries are
observed to be systematically smaller than proton (or neutron)
spin asymmetries, see Chapter~\ref{sec:phenoTMDs}.

Let us discuss the prediction  
$|(f_{1T}^{\perp u}-f_{1T}^{\perp d})(x,k_T)| \sim N_c^3 \gg
|(f_{1T}^{\perp u}+f_{1T}^{\perp d})(x,k_T)| \sim N_c^2$
from Table~\ref{Tab:large-Nc} as an example. Remarkably,
in the first extraction of the Sivers function from SIDIS data, where
the Sivers effect was clearly seen but the error bars still sizable,
this prediction was implemented as a theoretical constraint 
$f_{1T}^{\perp u}(x,k_T) = -f_{1T}^{\perp d}(x,k_T)$
\index{Sivers function $f_{1T}^{\perp}$!model calculations}
neglecting
$1/N_c$-corrections and gave a very good description of 
the data \cite{Efremov:2004tp}. 
The latest extractions of the Sivers function based on the more 
precise data support this prediction from \cite{Pobylitsa:2003ty},
see Sec.~\ref{sec:Sivers_SIDIS}. 

The large-$N_c$ scaling of gluon distribution functions was also
discussed in Ref.~\cite{Efremov:2000ar}. For instance, it was shown that 
$f_{1T}^{\perp g}(x,k_T)\sim N_c^2$ is suppressed with respect 
to quark Sivers functions \cite{Efremov:2004tp}, as 
independently concluded in \cite{Brodsky:2006ha} and  
supported by phenomenology \cite{Anselmino:2006yq}.

\subsubsection{Non-relativistic limit}
\index{model!non-relativistic quark model}

Heisenberg's uncertainty principle implies that the constituents 
of a quantum system move. In systems of the size 
of ${\cal O}(1\,\mbox{\AA})$ the motion is non-relativistic
to a good approximation.
It is instructive to compute the "velocity" and "radius" of a 
"classical circular orbit" of an electron in Bohr's semi-classical 
model of hydrogen atom: 
the "radius" and "velocity" in the $n^{\rm th}$ orbit are 
$r_n = \frac1\alpha\,\lambda_e\,n^2$ and $v_n = \alpha\,c\,\frac1n$
where $\lambda_e = \hbar/(m_e c)$ is the Compton wavelength of the
electron ($m_e$ is strictly speaking the reduced mass). 
While not valid in a quantum treatment, such "semi-classical"
considerations correctly explain why atoms are relatively large 
and why they can be treated in non-relativistic quantum mechanics: 
namely because the electromagnetic interaction 
is relatively weak with $\alpha \simeq \frac1{137}$. 
In QCD, for hadrons made of light quarks, like nucleons, 
one deals with $\alpha_s(1\,{\rm GeV})={\cal O}(1)$
and a non-relativistic treatment is unjustified.

\index{transversity!models}
One may nevertheless ask the question: how would parton
distributions look like in a nucleon if the system could be
treated in a non-relativistic way? Investigating such questions
can give us valuable intuition. For instance, one
non-relativistic prediction which gained a lot of popularity,
is that transversity and helicity PDFs become equal in the
non-relativistic limit, 
\be\label{Eq:g1-h1-non-rel-limit}
    \lim\limits_{\rm non\!-\!rel} h_1^q(x)=
    \lim\limits_{\rm non\!-\!rel} g_1^q(x)\,.
\ee
This conclusion was derived in Ref.~\cite{Jaffe:1991ra}
within the bag model (to be discussed below) and has
been used to predict observables involving $h_1^q(x)$
until the first data on this PDF became available, see, e.g., 
the review article \cite{Barone:2001sp} and references therein.

One can introduce a non-relativistic limit for TMD PDFs by 
working in the constituent quark model limit where quark momenta
$|\mathbf{k}|\ll m_q$ become small, the nucleon size grows, 
and the constituent quark mass determines the nucleon 
mass as $M\to N_c m_q$ in the limit.
Introducing the SU(4) spin-flavor symmetry factors
$N_u=\frac12(N_c+1)$, $N_d=\frac12(N_c-1)$, 
$P_u=\frac16(N_c+5)$, $P_d=\frac16(1-N_c)$ for general 
$N_c$ \cite{Karl:1984cz}, the non-relativistic limit for
the T-even proton TMD PDFs is given by \cite{Efremov:2009ze}
\begin{align}
    \lim\limits_{\rm non\!-\!rel} f_1^q(x,k_T)&=
    N_q\,\delta\biggl(x-\frac{1}{N_c}\biggr)\,\delta{ }^{(2)}(\kt), 
    &\quad
    \lim\limits_{\rm non\!-\!rel}g_{1T}^{\perp q}(x,k_T)&=
    P_q\,N_c\delta\biggl(x-\frac{1}{N_c}\biggr)\,\delta{ }^{(2)}(\kt),
    \nonumber\\
    \lim\limits_{\rm non\!-\!rel} g_1^q(x,k_T)&=
    P_q\,\delta\biggl(x-\frac{1}{N_c}\biggr)\,\delta{ }^{(2)}(\kt), 
    &\quad
    \lim\limits_{\rm non\!-\!rel}h_{1L}^{\perp q}(x,k_T)&=
    -\,P_q\,N_c\delta\biggl(x-\frac{1}{N_c}\biggr)\,\delta{ }^{(2)}(\kt),
    \nonumber\\
    \lim\limits_{\rm non\!-\!rel} h_1^q(x,k_T)&=
    P_q\,\delta\biggl(x-\frac{1}{N_c}\biggr)\,\delta{ }^{(2)}(\kt), 
    &\quad
    \lim\limits_{\rm non\!-\!rel}h_{1T}^{\perp q}(x,k_T)&=
    -\,P_q\,\frac{N_c^2}{2}\delta\biggl(x-\frac{1}{N_c}\biggr)\,
    \delta{ }^{(2)}(\kt)
    \label{Eq:non-rel-lim}
\end{align}
If the system is not strictly non-relativistic, the motion 
of the quarks "smears out" the $\delta$-functions. 
Imagining the transverse motion of quarks to be due to random 
motion in the transverse plane, one might be tempted to 
"smear out" the $\delta{ }^{(2)}(\kt)$ in terms of Gaussians. 
While this does not prove anything, it makes the success 
of the Gaussian Ansatz to some extent plausible.
In practical calculations in non-relativistic models  
the "smearing" of the $\delta$-functions in $x$ and $\kt$ 
is considerable, and we will comment on this below in 
Sec.~\ref{Sec:models-T-odd-1-gluon-exchange}.

\subsection{Modelling of T-even TMD PDFs}
\label{Sec:models-T-even-quark-TMDs}

T-even TMD PDFs do not require explicit gauge field degrees
of freedom in order to be modelled. In this section we will 
review several such models. 

\subsubsection{Covariant parton model}
\label{Sec:parton-model}
\index{model!covariant parton model}

A consequent exploration of the parton model approach 
discussed in Sec.~\ref{Sec:parton-model-general} leads to the
covariant parton model. In this model, one assumes that
the QCD coupling constant $g(\mu)=0$ at any scale $\mu$. 
As a consequence, the partons are non-interacting and 
on-shell making the parton picture, within this model, 
valid not only in the infinite-momentum frame but in 
any frame. 
This is the essence of the covariant 
parton model~\cite{Zavada:1996kp,Efremov:2009ze,Zavada:2009ska,Efremov:2010mt,Efremov:2010cy,Bastami:2020rxn,Aslan:2022wqc,Aslan:2022kmd}. 

In this model, due to the absence of explicit gauge degrees of freedom, the Wilson-lines are replaced by unit matrices in color space, and T-odd TMD PDFs vanish. 
The quark correlator entering the definition of TMDs is largely simplified and given, in momentum space, by \cite{Bastami:2020rxn}
\be\label{eq:CPM2}
        \Phi^q(k,P,S) =
        M \, \Theta(k^0)\,\delta(k^2 - m^2)\;
        (\slashed{k}+m)\,
        \bigl(\mathcal{G}^q(P\cdot k)
        +\mathcal{H}^q(P\cdot k)\,\gamma^5\slashed{\omega}\bigr) \,. \;
\ee
Here the notation for the unintegrated quark correlator follows
Refs.~\cite{Tangerman:1994eh,Boer:1997nt} and is such that from
$\frac12\,\iint dk^+ dk^- \delta(k^+-xP^+)\,
{\rm tr}\,\Gamma\Phi^q(k,P,S)$ with
$\Gamma= \gamma^+,\,\gamma^+\gamma_5,\,i\sigma^{\alpha+}\gamma_5$, 
one recovers the expressions on the right-hand sides of
Eq.~(\ref{eq:tmd_decomposition}). 
In Eq.~(\ref{eq:CPM2}), $P$ and $S$ denote the nucleon
momentum and polarization vector,  
$k$ is the quark momentum with the onshellness of the quarks 
implemented by $\Theta(k^0) \delta(k^2-m^2)$, and 
$\slashed{\omega}=\gamma^\mu\omega_\mu$
with the quark polarization vector
$\omega^\mu$ satisfying $\omega\cdot k = 0$, $\omega^2=-1$ and given by
\be\label{eq:omega}
	\omega^\mu = S^\mu- \frac{M}{m} \, \frac{k \cdot S}{k \cdot P + m M} 
        \, k^\mu - \frac{k \cdot S}{k \cdot P + m M} \, P^\mu \,.
\ee
The nucleon structure is described in terms of two covariant 
functions of $P\cdot k$:~$\mathcal{G}^q(P\cdot k)$ describes the momentum distribution 
of unpolarized quarks of flavor $q=u,\,d,\,\dots$ inside the nucleon, and
$\mathcal{H}^q(P\cdot k)$ describes the distribution of polarized
quarks.
As a result all TMD PDFs are determined in terms of these two functions
\cite{Efremov:2009ze,Efremov:2010mt,Bastami:2020rxn,Aslan:2022wqc,Aslan:2022kmd}.

In the nucleon rest frame, $P\cdot k = M\,(\mathbf{k}^2+m^2)^{1/2}$ and the 
3D spherical symmetry becomes apparent which connects longitudinal and 
transverse quark momenta. As a consequence, in this model it is possible
to unambiguously predict TMD PDFs from collinear PDFs \cite{Efremov:2010mt}
which gives predictive power to the approach. The model automatically satisfies 
the Callan-Gross relation between the unpolarized DIS structure functions,
and the Wandzura-Wilczek approximation for the twist-3 collinear PDF
$g_T^a(x)$ becomes exact, namely $g_T^a(x) = \int_x^1 \frac{dy}{y}\,g_1^a(y)$ 
(more on subleading twist in Ch.~\ref{sec:twist3}).
The Wandzura-Wilczek approximation for $g_T^a(x)$ is supported by 
data with a good accuracy, see e.g.\ Ref.~\cite{Accardi:2009au} 
for a brief review.
This provides phenomenological support for the covariant parton 
model. The model can also describe qualitatively the Cahn effect
\cite{Zavada:2009ska}, although one of its limitations is that the
restriction to onshellness implies unrealistically small transverse
parton momenta \cite{Zavada:2009ska}. 

\index{worm-gear functions!models|(}
The covariant parton model relates the transverse moments of the 
Kotzinian-Mulders worm-gear functions to the helicity and 
transversity PDFs as follows \cite{Efremov:2009ze}
\begin{subequations}\ba
   	g_{1T}^{\perp(1)a}(x) &=& 
        \phantom{-} x\,\int_x^1\frac{d y}{y\;}\,g_1^a(y) \;,
	\label{Eq:WW-approx-g1T}\\
    	h_{1L}^{\perp(1)a}(x)&=& -
	x^2\!\int_x^1\frac{d y}{y^2\;}\,h_1^a(y)\;,
	\label{Eq:WW-approx-h1L}
\ea\end{subequations}
where current quark mass terms are neglected. In QCD these relations 
are spoiled by the appearance of matrix elements of quark-gluon operators.
\index{Wandzura-Wilczek (type) approximation}
Assuming these contributions to be small constitutes the WW-type 
approximation for TMD PDFs. (It is called WW-type approximation to be
distinguished from the WW approximation for collinear PDFs because 
different quark-gluon operators are neglected in both cases.) 
Based on the positive experience with WW approximation for $g_T^a(x)$, 
one may hope that the approximations 
(\ref{Eq:WW-approx-g1T},~\ref{Eq:WW-approx-h1L}) are useful
for the Kotzinian-Mulders worm-gear functions, 
though this remains to be tested by data.
Presently, little is known about these functions and the WW-type
approximations (\ref{Eq:WW-approx-g1T},~\ref{Eq:WW-approx-h1L}) 
have been explored for phenomenological applications
\cite{Kotzinian:2006dw,Avakian:2007mv,Bastami:2018xqd},
cf.\ Sec.~\ref{sec:phenomelology-other}. 
We stress that in the covariant parton model, the WW-type 
approximations are exact.\index{worm-gear functions!models|)}

For other studies of transverse parton momentum effects 
in similar parton model frameworks we refer to Refs.~\cite{Jackson:1989ph,Roberts:1996ub,
Bourrely:2005tp,Bourrely:2010ng,DAlesio:2009cps,Mirjalili:2022cal}.
Parton model applications addressing target mass
corrections or gluon polarization effects were reported in 
Refs.~\cite{Blumlein:1996tp,Blumlein:1996vs,Soffer:1997zy,Blumlein:1998nv}.
The free-quark ensemble model of Ref.~\cite{Tangerman:1994eh}
is another implementation of the parton model concept.

\subsubsection{Bag model}

\index{model!bag model}

This model was introduced in the early 1970s and continues to be useful. 
In its simplest version, 
non-interacting quarks are confined inside a spherical cavity with radius 
$R$ due to boundary conditions which ``simulate'' confinement. 
The nucleon is modelled by placing $N_c$ quarks in the ground state wave 
function which has positive parity and is given for massless quarks in 
momentum space by
\begin{equation}
    \Phi_{m}(\mathbf{k})=i\sqrt{4\pi}N R^3
    \left(\begin{array}{r} t_0(k)\chi_m\\
    { }\bm{\sigma}\cdot\hat{\mathbf{k}} \;t_1(k)\chi_m
    \end{array} \right ) \ , 
    \quad N=\frac{\omega^{3/2}}{(2R^3(\omega-1)\sin^2\omega)^{1/2}},
    \quad \hat{\mathbf{k}} = \mathbf{k}/k, \quad
    \quad k=|\mathbf{k}|.
    \label{Eq:wf-bag-model}
    \end{equation}
The $t_i(k)$ are defined as 
$t_i(k)=\int_0^1 u^2 du j_i(ukR)j_i(u\omega)$
in terms of spherical Bessel functions, 
$\bm{\sigma}$ denotes the Pauli matrices, $\chi_m$ the Pauli spinor,
$\omega\approx 2.04$ is the lowest solution of the transcendental
bag equation $\omega_i = (1-\omega_i) \tan\omega_i$.
The bag model wave function in Eq.~(\ref{Eq:wf-bag-model}) contains an 
$S$-wave (upper) component with orbital angular momentum $L=0$ 
accompanied by $t_0$, and a 
$P$-wave (lower) component with orbital angular momentum $L=1$, 
accompanied by $t_1$.
The results for the T-even leading TMD PDFs are given by
\cite{Avakian:2008dz,Avakian:2010br}
\begin{align}
   f_1^q(x,k_T) &= N_q  A[t_0^2+2\widehat{k}_z\,t_0t_1+t_1^2],             
               &\quad
   g_{1T}^{\perp q}(x,k_T) &= P_q\,A[\phantom{-}2\widehat{M}
    (t_0t_1+\widehat{k}_z\,t_1^2)] \nonumber\\
   g_1^q(x,k_T) &= P_q\,A[t_0^2+2\widehat{k}_z\,t_0t_1
                    +(2\widehat{k}_z^2-1)\,t_1^2],
               &\quad
      h_{1L}^{\perp q}(x,k_T) &= P_q\,A[-2\widehat{M}_N
    (t_0t_1+\widehat{k}_z\,t_1^2)] \nonumber\\
   h_1^q(x,k_T) &= P_q\,A[t_0^2+2\widehat{k}_z\,t_0t_1+\widehat{k}_z^2\,t_1^2],
               &\quad
   h_{1T}^{\perp q}(x,k_T) &= P_q\,A[-2\widehat{M}_N^{\,2} \,t_1^2]
    \label{Eq:h1Tperp}
\end{align}
with the SU(4) spin-flavor symmetry factors as defined below 
Eq.~(\ref{Eq:non-rel-lim}) and
\ba
    A=\frac{16\omega^4}{\pi^2(\omega -1)j_0^2(\omega)\,M^2}\,,\;\;\;
    k=\sqrt{k_z^2+k_T^2}\;,\;\;\;
    k_z=xM-\omega/R\;,\;\;\;\widehat{k}_z=\frac{k_z}{k}\;,\;\;\;
    \widehat{M}=\frac{M}{k}\;, \nonumber
\ea
where $M$ is the proton mass, and the bag radius is fixed such that 
$R M=4\omega$. All leading and subleading T-even TMD PDFs were studied 
in this model, and a complete set of linear and non-linear relations
among 
TMD PDFs was derived \cite{Avakian:2008dz,Avakian:2010br}. The bag model
supports the phenomenologically observed Gaussian $k_T$-dependence 
of TMD PDFs \cite{Schweitzer:2010tt}. 

One drawback of this model is that the bag boundary condition
violates chiral symmetry, a feature that can be improved using the 
so-called cloudy bag model \cite{Thomas:1982kv}. Another drawback 
is that it generates unphysical antiquark distributions and the 
TMD PDFs receive very small but nonzero support for $x>1$. The 
latter problem can be fixed by employing Peierls-Yoccoz projection
techniques, see Ref.~\cite{Signal:2021aum} for a recent study.

\subsubsection{Lightfront constituent quark models}

\index{model!lightfront constituent models}

In these models, the nucleon structure is modelled in terms 
of 3-quark light-cone wave functions (LCWFs) which contain
information on the bound state properties of the nucleon 
in terms of process- and frame-independent amplitudes 
which are eigenstates of the total quark orbital-angular momentum $L_z^q$ \cite{Brodsky:2000ii,Ji:2002xn}.
The TMD PDFs exhibit multipole patterns \cite{Pasquini:2008ax,Pasquini:2009bv}: 
for instance,
$f_1^q,$ $g_1^q$, $h_1^q$ are "monopole structures" associated with 
$\Delta L_z^q=0$ (i.e.\ diagonal in the quark angular momentum
components).
In contrast, $g_{1T}^{\perp q}$ and $h_{1L}^{\perp q}$ 
correspond to "dipole structures" arising from the interference of
S- and P-waves with $\Delta L_z^q=1$, and $h_{1T}^{\perp q}$ is 
a "quadrupole structure" associated with $\Delta L_z^q=2$ due
to the interference of two P-waves or one S-wave and one D-wave
\cite{Miller:2007ae,Burkardt:2007rv,Pasquini:2008ax,Pasquini:2009bv}.
Often used approaches for LCWFs include the lightfront constituent quark 
\cite{Pasquini:2008ax} and chiral quark-soliton \cite{Lorce:2011dv}
models.
The lightfront constituent quark model was applied to 
nucleon and pion TMD PDFs with phenomenological success 
\cite{Boffi:2009sh,Pasquini:2011tk,Pasquini:2014ppa}.
Many observables in SIDIS or Drell-Yan were described
typically within an accuracy of 10--40$\,\%$ in the 
region of $x \gtrsim 0.1$ where quark models can be
expected to work. The approach was extended to 
subleading functions~\cite{Lorce:2014hxa,Lorce:2016ugb}.

A light cone quark model was applied to the structure of pions and kaons in Ref.~\cite{Kaur:2020vkq}. Another model formulated in lightfront quantization is the basis light-front quantization approach \cite{Hu:2022ctr}.

\subsubsection{Spectator models}

\index{model!spectator model}

The first quark model applied to 
TMD PDFs 
was the quark-diquark spectator model~\cite{Jakob:1997wg}. 
Here the correlator defining TMDs is evaluated
by replacing the sum over all intermediate states with a single 
on-shell spectator thought to be an effective degree of freedom 
with the nonperturbative effects due to sea quarks and gluons 
effectively resummed. In models of the nucleon, the spectator 
can be a spin-0 isoscalar or spin-1 isovector diquark 
(when modelling pion TMD PDFs, the spectator is another quark or antiquark).
The effective nucleon-quark-diquark vertex may be modeled 
in terms of form factors. Various vertex functions and different 
choices for diquark masses and axial-vector polarization states 
have been used in literature 
\cite{Gamberg:2003ey,Bacchetta:2008af,She:2009jq,Lu:2010dt,Zhu:2011zza,Muller:2014tqa,Liu:2021ype}. 
The results can be interpreted in terms of the overlap of light-cone 
wave functions (LCWFs) for the diquark~\cite{Brodsky:2000ii}. 
Several versions of light-cone quark-diquark models were
discussed \cite{She:2009jq,Lu:2010dt,Zhu:2011zza,Muller:2014tqa}.
In these models it is in general not possible to satisfy simultaneously 
the quark-number and momentum sum rules, a limitation which can be 
remedied by resolving the internal diquark structure in a dynamical
framework \cite{Cloet:2007em}. 
For a review of the diquark concept in hadronic physics, we refer to \cite{Barabanov:2020jvn}.

\subsubsection{Nambu--Jona-Lasinio framework}

\index{model!Nambu--Jona-Lasinio model}

The Nambu--Jona-Lasinio model is based on an effective, 
non-renormalizable 4-quark interaction. The model incorporates 
one important low-energy aspect of QCD, namely chiral  symmetry 
and its  dynamical  breaking. Hadronic correlators are evaluated
by solving the Faddeev equation in a quark-diquark approximation, 
including both dynamical scalar and axial vector diquarks. The Nambu--Jona-Lasinio
framework can be used to model diquark correlations more realistically. 
The framework was used to study the transversity parton distribution
function \cite{Cloet:2007em}, and TMD PDFs of $\rho$-mesons 
\cite{Ninomiya:2017ggn} and pions \cite{Shi:2018zqd}.\index{transversity!models}

\subsubsection{AdS/QCD inspired models}

\index{model!AdS/QCD model}

The correspondence between 10-dimensional string theories 
in AdS$_5\times\,$S$^5$ space and conformal $N=4$ supersymmetric Yang-Mills
theories in $3+1$ spacetime has opened new ways to model 
QCD in the strong-coupling regime \cite{Brodsky:2006uqa}. 
A light-cone scalar diquark model exploring predictions from 
a soft-wall AdS/QCD model for the LCWF of the valence quark 
and the diquark was used to describe TMDs of the pion and 
nucleon in Refs.~\cite{Maji:2015vsa,Maji:2017bcz,Bacchetta:2017vzh}.
The results from this approach met phenomenology with 
success in Refs.~\cite{Maji:2017zbx,Maji:2017wwd}. A soft-wall AdS/QCD-motivated light-front quark-diquark model was explored in Ref.~\cite{Gurjar:2021dyv}.

\subsubsection{Chiral quark soliton model}

\index{model!chiral quark soliton model}

This model is based on a low-energy chiral theory describing 
the interaction of effective quark and antiquark degrees of 
freedom with Goldstone bosons of the 
spontaneous chiral symmetry breaking. The Lagrangian is given by
${\cal L}=\overline{\Psi} [i\slashed{\partial} + M\,\exp(i\gamma_5\tau^a\pi^a/f_\pi)]\Psi$
in the SU(2) version of the model where $\pi^a$ denote the 
pion fields, 
$M$ is the dynamically generated quark constituent mass, and 
$f_\pi=93\,{\rm MeV}$ the pion decay constant. The UV cutoff
of the theory $\mu_0 \sim \rho^{-1} \sim 600\,{\rm MeV}$ 
is associated with the nonperturbative short-distance scale $\rho$
at which chiral symmetry breaking occurs. Two distinct 
nonperturbative scales play important roles for the description
of the nucleon structure, namely the scale $\rho\sim 0.3\,{\rm fm}$
associated with chiral symmetry breaking and the scale 
$R_{\rm had}\sim 1\,{\rm fm}$ associated with the nucleon size. 
The interplay of these two distinct nonperturbative scales and their
hierarchy, $\rho\sim 0.3\,{\rm fm} \ll R_{\rm had}\sim1\,{\rm fm}$,
have profound consequences on TMDs: at a low scale $\mu_0$ and
small $k_T\lesssim R_{\rm had}^{-1}$, valence quarks dominate
the $k_T$-behavior of TMD PDFs.
But in the region $R_{\rm had}^{-1} < k_T < \mu_0$ the $k_T$-behavior
of the TMD PDFs $f_1^a(x,k_T)$, $g_1^a(x,k_T)$ is dominated by sea 
quarks, which exhibit slow power-like decays and overwhelm the 
contribution of valence quarks which decay exponentially in this
region~\cite{Schweitzer:2012hh,Wakamatsu:2009fn}.
In contrast, the transversity TMD PDF exhibits valence-quark type 
$k_T$-behavior in the entire $k_T$-region \cite{Schweitzer:2012hh}.\index{transversity!models}

\subsubsection{Predictions from quark models for T-even TMD PDFs}

In this section, we discuss results for T-even TMD PDFs from
several representative quark models. Let us begin with the
$k_T$-dependence of the unpolarized TMD PDF. In Chapter~\ref{sec:phenoTMDs}
we have seen that a lot of phenomenology related to $k_T$-effects 
has been successfully done assuming the Gaussian Ansatz. 
While there is general consensus that it is merely an approximation, 
it is a good question to ask why this Gaussian approximation 
works so well. While we do not know the answer to this question, 
it is interesting to see what models can teach us. 

\begin{figure}[tp]
\begin{center}
\includegraphics[width=0.32\textwidth]{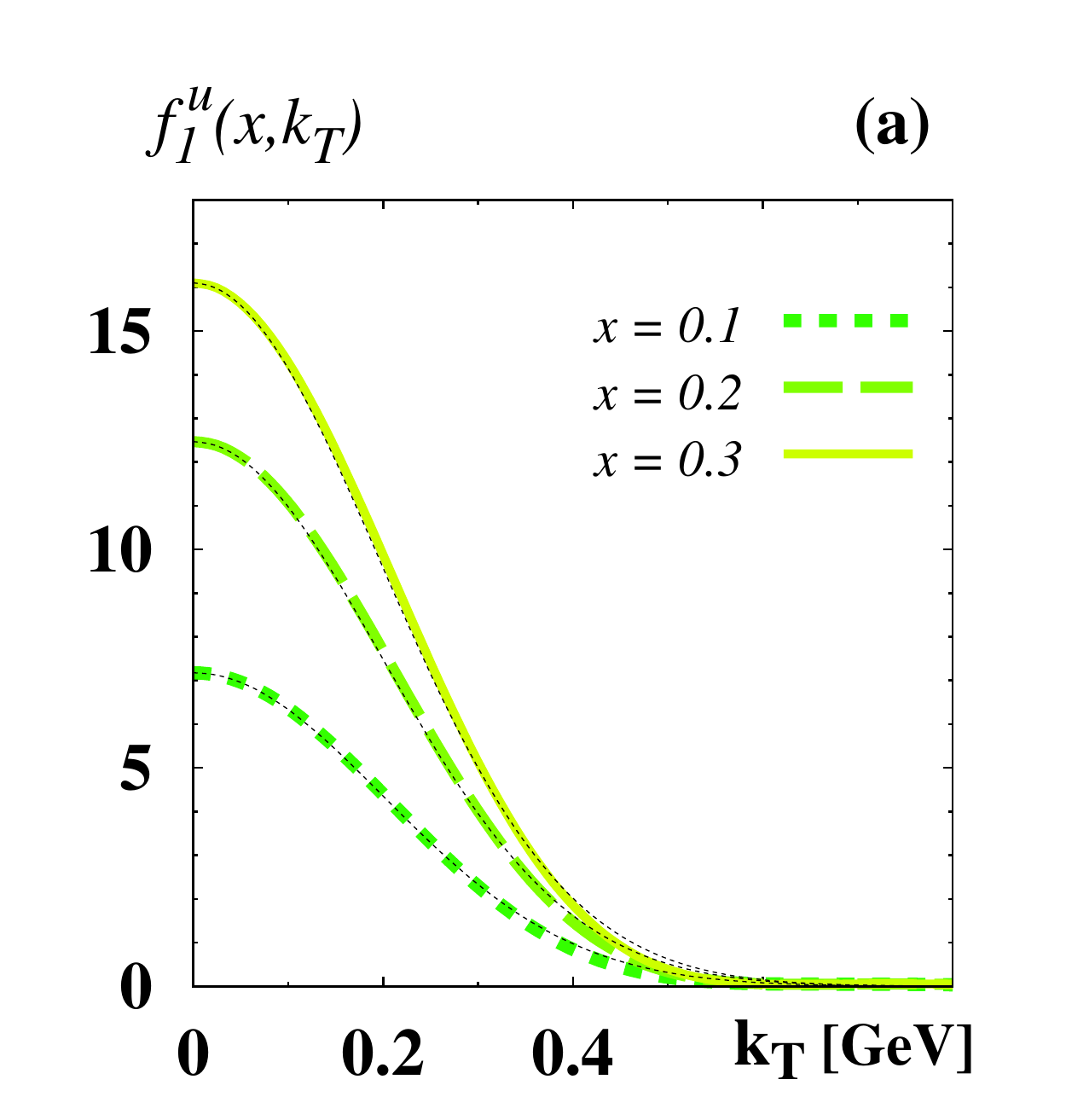}
\includegraphics[width=0.32\textwidth]{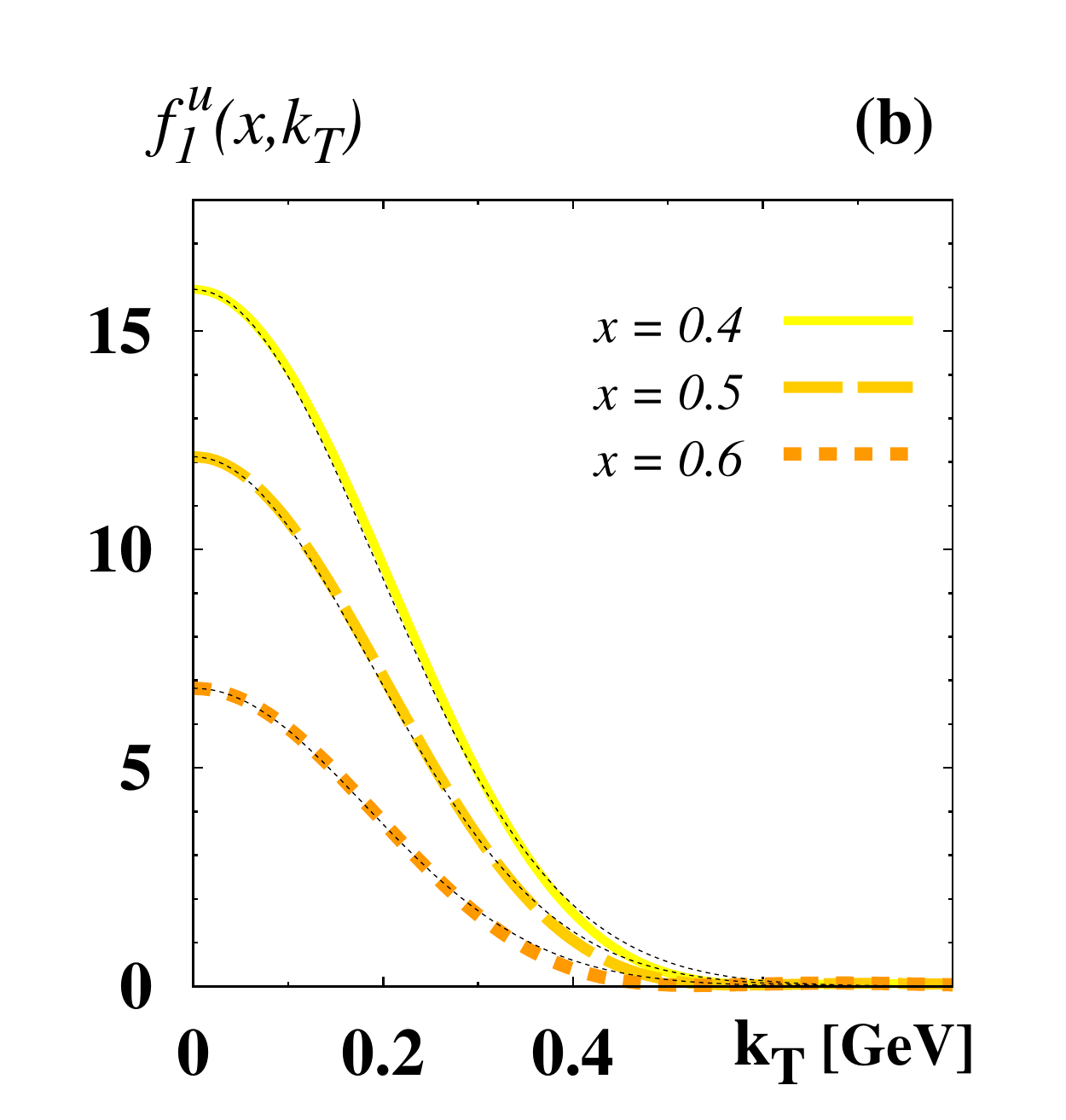}
\includegraphics[width=0.32\textwidth]{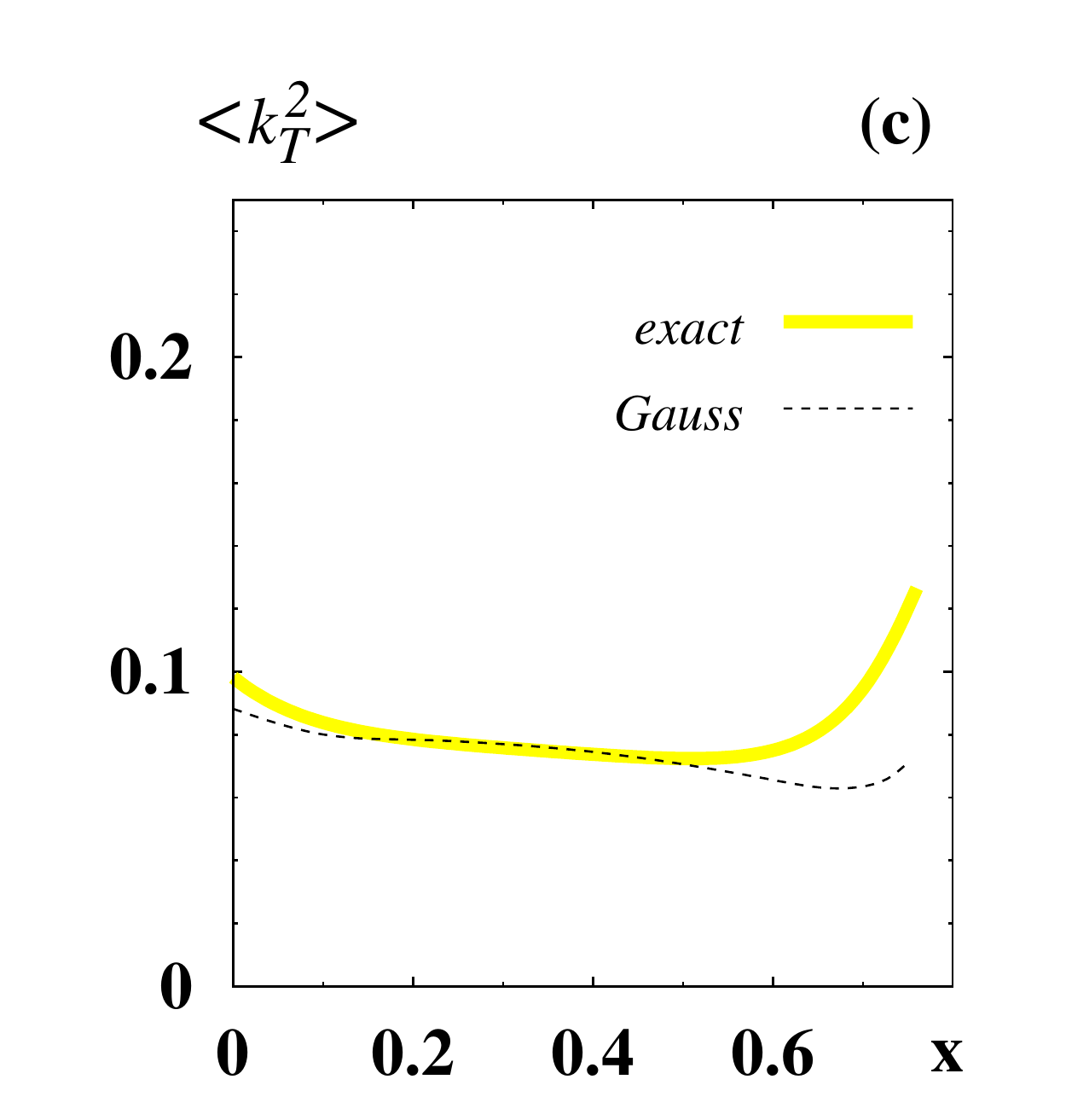}
\end{center}
\vspace{-0.4cm}
\caption{The $k_T$-dependence of $f_1^u(x,k_T)$ at 
(a) $x=0.1$, $0.2$, $0.3$, and 
(b) $x=0.4$, $0.5$, $0.6$ in the bag model at a low scale 
from Ref.~\cite{Avakian:2010br}.
The colored lines are the exact model results. The thin 
black-dotted lines are the respective Gaussian approximations. 
(c) The $x$-dependence of the exact model results 
for $\la k_T^2\ra$ (colored line) vs the Gaussian
widths (thin dotted lines) used in parts (a) 
and (b) of the figure.
\label{fig:models-TMD-PDFs-Gauss}}
\end{figure}

Let us first stress that no model studied so far exhibits 
exact Gaussian $k_T$-dependence. However, in several models 
the Gaussian Ansatz appears to be a useful 
approximation for the exact $k_T$-dependence. The 
Figs.~\ref{fig:models-TMD-PDFs-Gauss}a and
\ref{fig:models-TMD-PDFs-Gauss}b
show the results from the bag model \cite{Avakian:2010br}
for $f_1^u(x,k_T)$ as function of $k_T$ at selected values 
of $0.1 \le x \le 0.6$. The colored lines show exact model results
super-imposed on the respective Gaussian approximations. At the
low scale of the bag model, the exact numerical results are very well
approximated by the Gaussian Ansatz up to $k_T\lesssim 0.4\,{\rm GeV}$
($k_T$ larger than that can not be reliably studied in a low energy 
model with an initial scale of the order of 
$\mu_0\lesssim 0.4 \,{\rm GeV}$).

The Gaussian widths used in 
Figs.~\ref{fig:models-TMD-PDFs-Gauss}a and
\ref{fig:models-TMD-PDFs-Gauss}b to depict the Gauss model
approximations exhibit a moderate dependence on $x$ 
shown in Fig.~\ref{fig:models-TMD-PDFs-Gauss}c. These widths
are defined such that the Gaussian approximations of the true
model results is exact in the vicinity of $k_T=0$.
One could define the Gaussian widths also as 
$\la k_T^2(x)\ra = \int d^2 k_T\,k_T^2\,f_1^q(x,k_T)/\int d^2 k_T\,f_1^q(x,k_T)$.
If the $k_T$-dependence in the model was exactly Gaussian,
the two definitions would give numerically the same results. 
The Fig.~\ref{fig:models-TMD-PDFs-Gauss}c shows that the two
definitions of $\la k_T^2(x)\ra$ do not give the same results,
but the approximation is very good in the region of
$0.1\lesssim x\lesssim 0.6$.
 
The Gaussian widths of $f_1^u(x,k_T)$ 
in Fig.~\ref{fig:models-TMD-PDFs-Gauss}c are
$\la k_T^2\ra \sim 0.1\,{\rm GeV}^2$, i.e.\ about factor 
2--3 smaller than what is needed in phenomenology of typical 
SIDIS experiments \cite{Schweitzer:2010tt}. This is to be
expected since the model results refer to a low initial scale
and evolution broadens the $k_T$-dependence.
Attempts to implement (approximate or exact) $k_T$-evolution
starting at low initial scales can be found, e.g., in 
Refs.~\cite{Boffi:2009sh,Pasquini:2011tk,Pasquini:2014ppa,
Ceccopieri:2018nop,Bastami:2020rxn}.

\begin{figure}[t!]
\begin{center}
\includegraphics[width=0.32\textwidth]{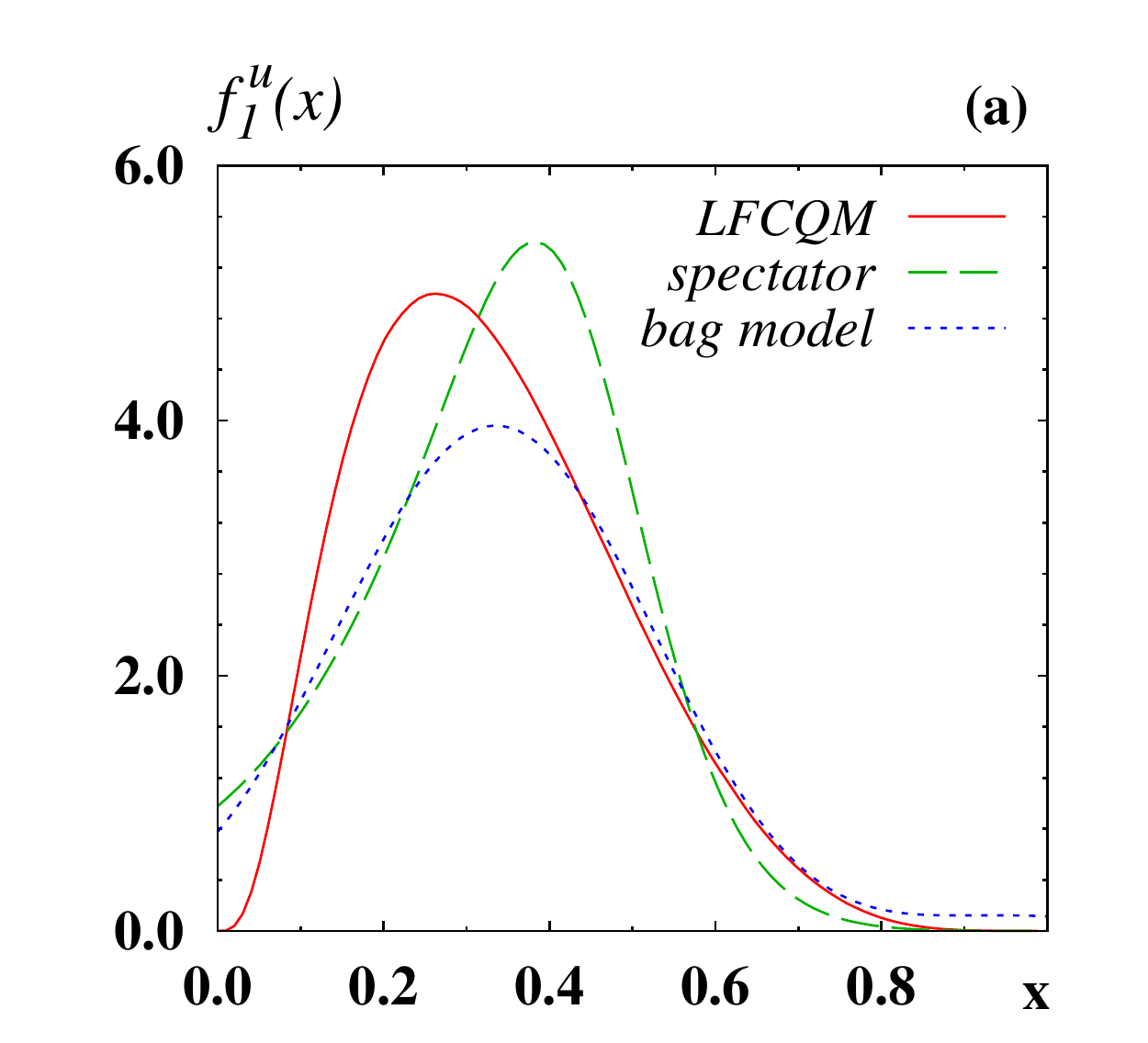}
\includegraphics[width=0.32\textwidth]{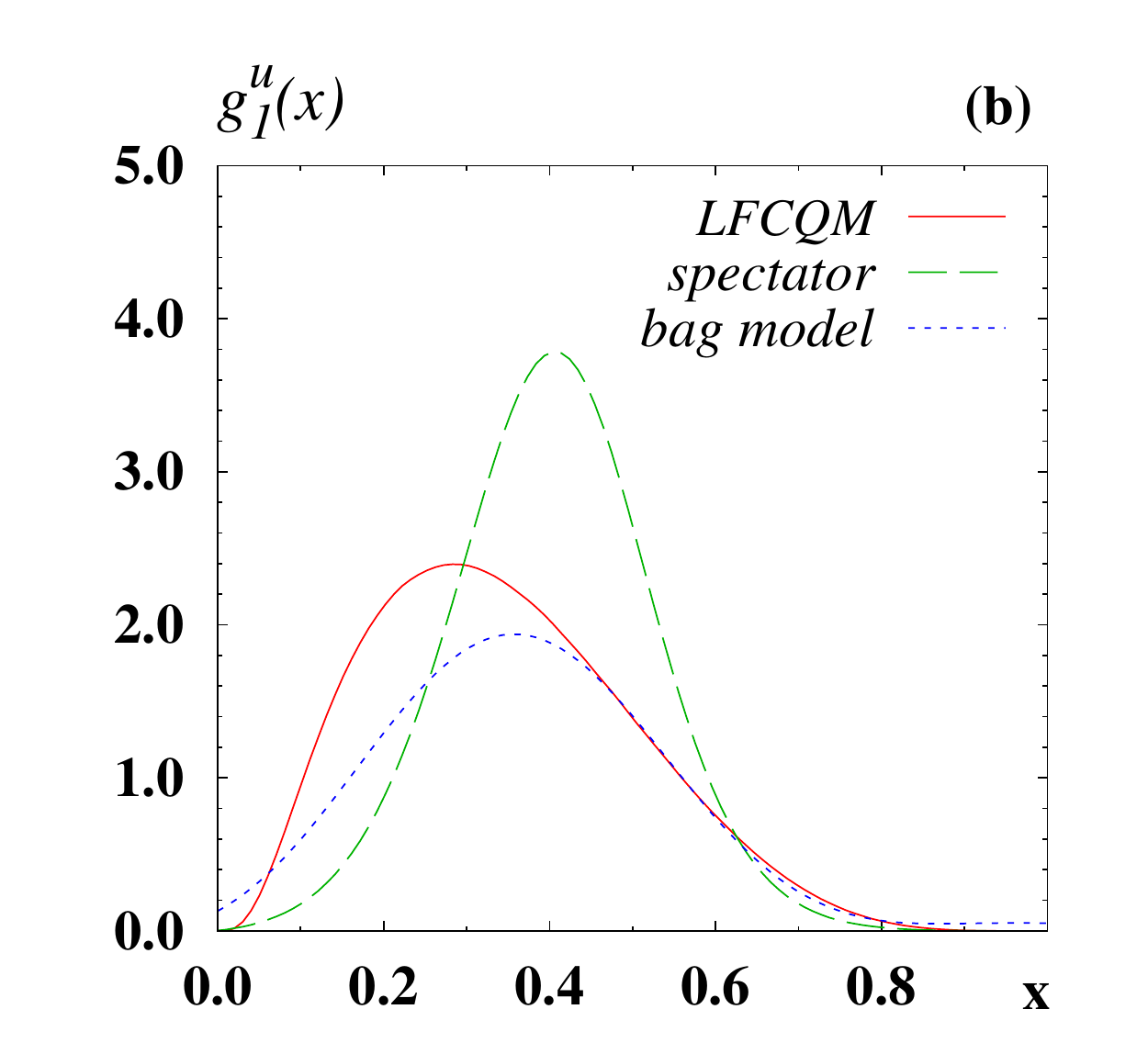}
\includegraphics[width=0.32\textwidth]{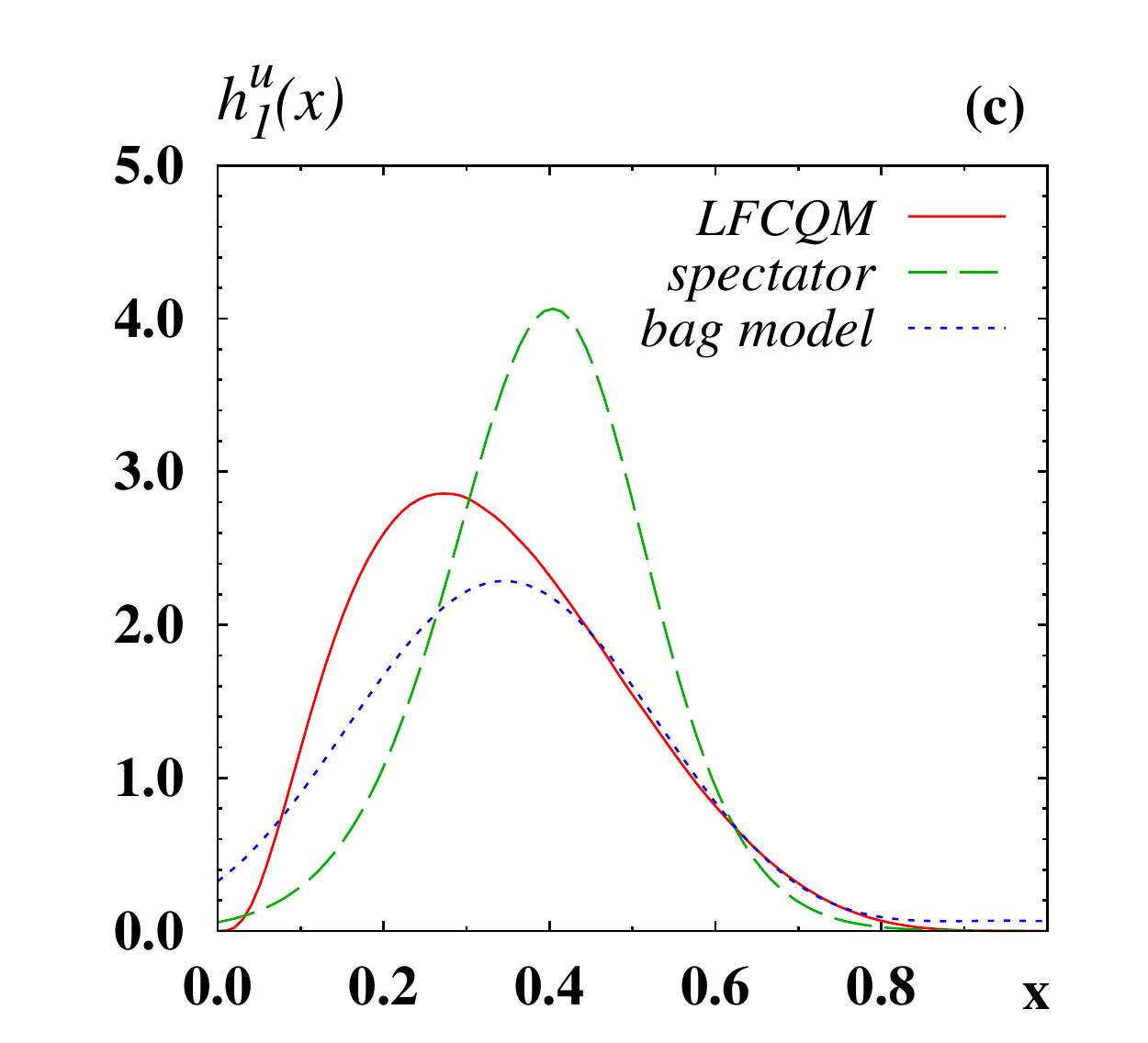} \\
\includegraphics[width=0.32\textwidth]{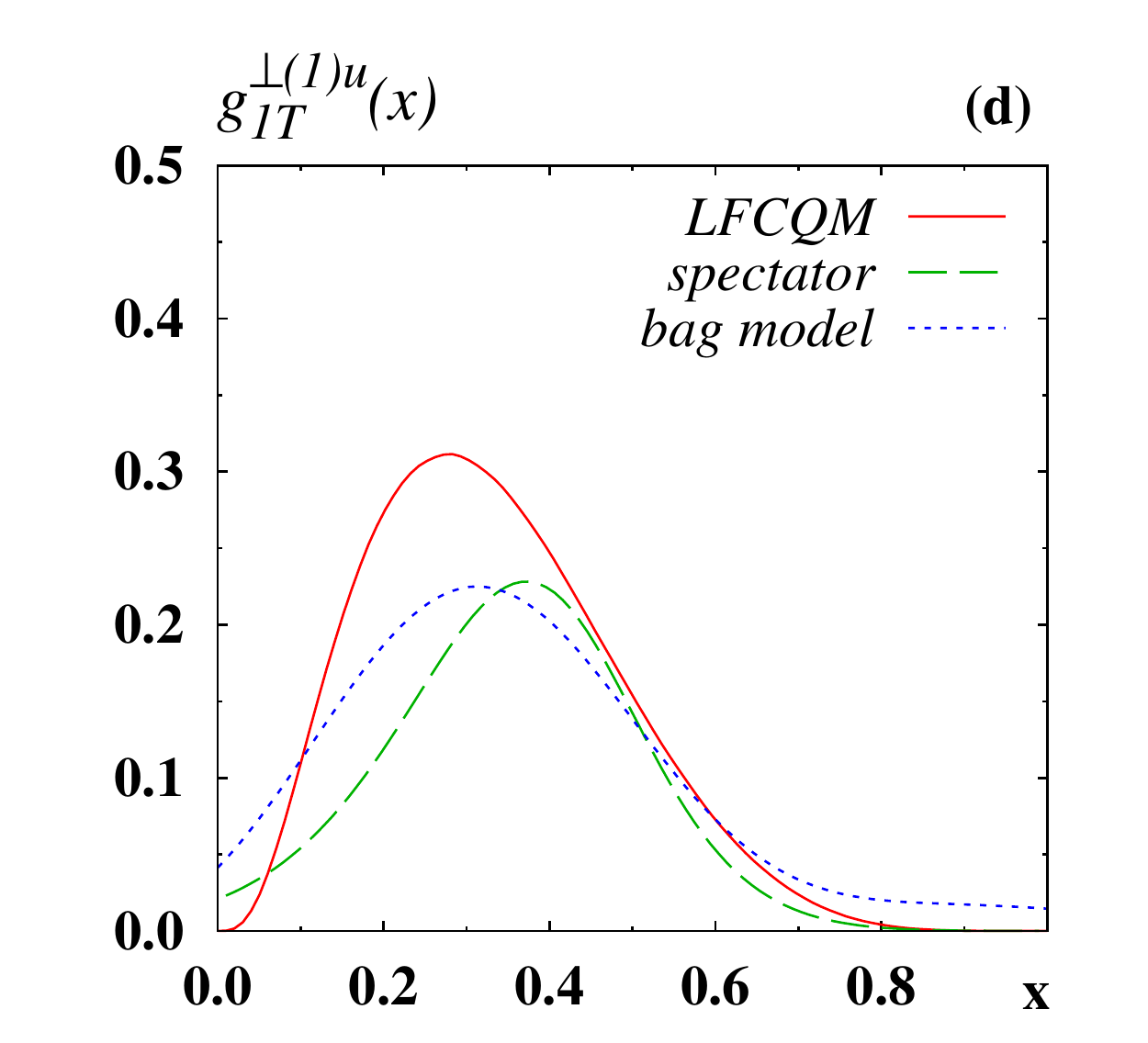} 
\includegraphics[width=0.32\textwidth]{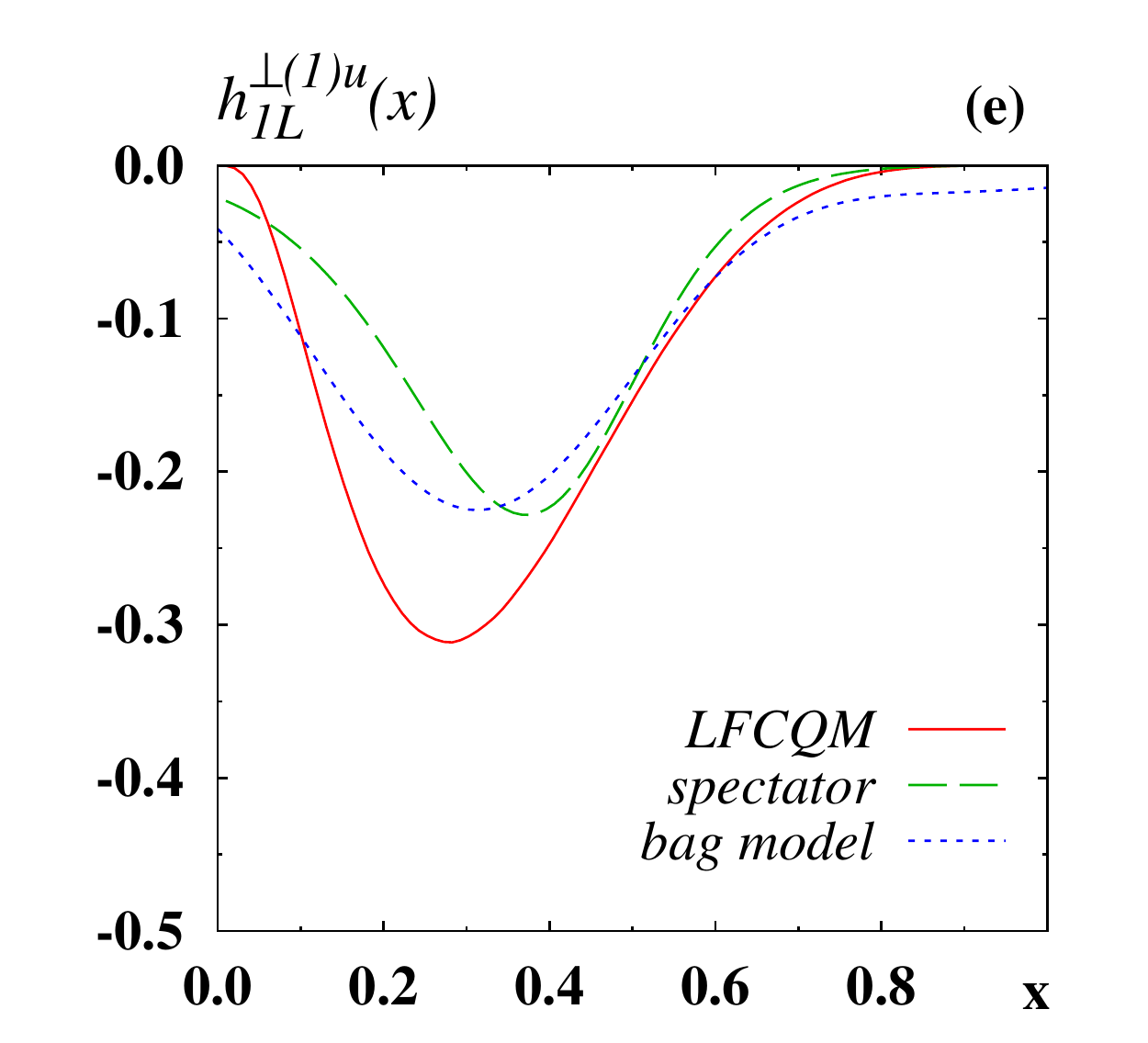}
\includegraphics[width=0.32\textwidth]{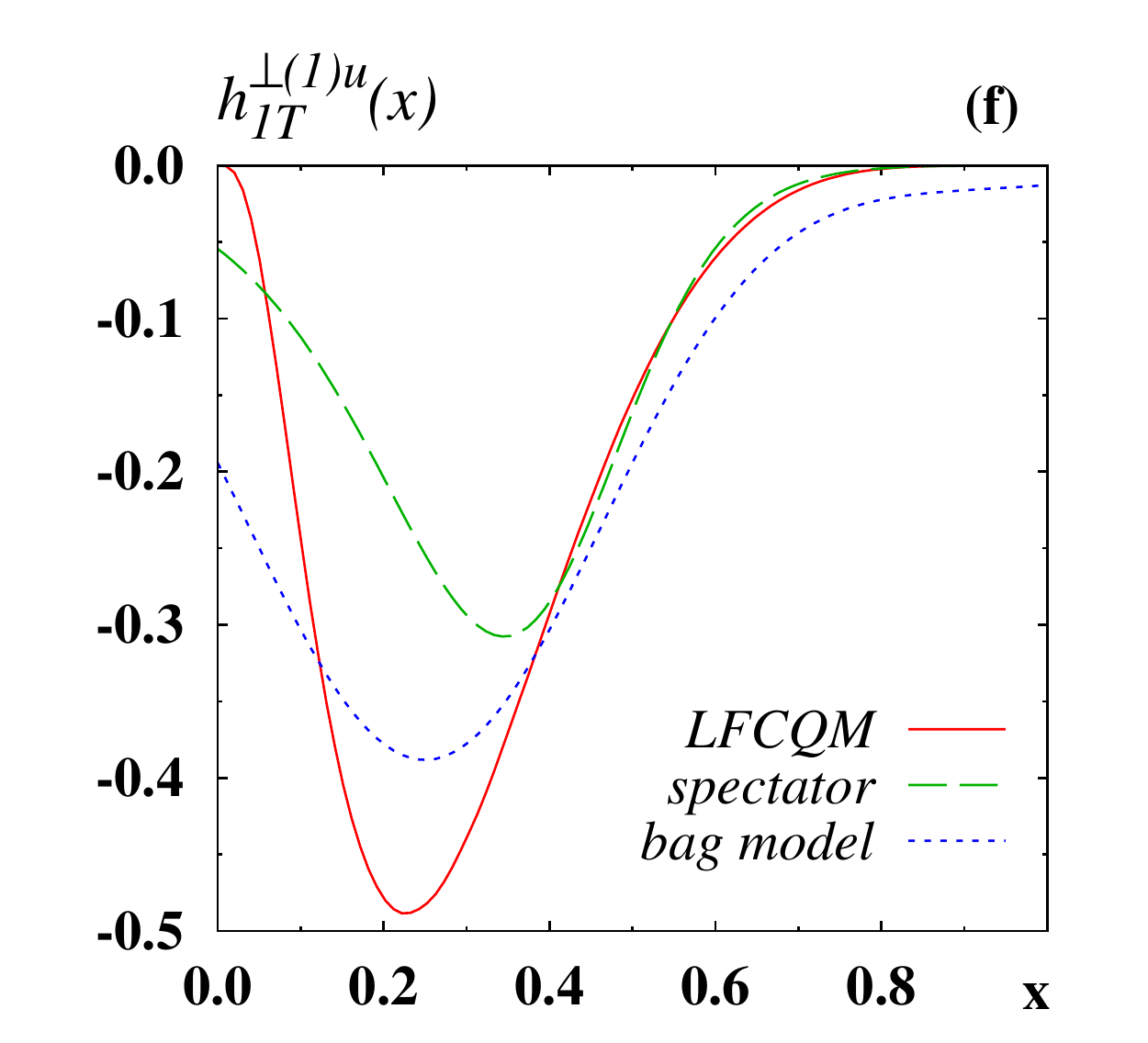}
\end{center}
\vspace{-0.4cm}
\caption{T-even TMD PDFs or their (1)-moments for 
$u$-quarks in proton from representative quark models:
light-front constituent quark model (LFCQM) \cite{Pasquini:2008ax},
spectator model \cite{Jakob:1997wg}, and
bag model \cite{Avakian:2010br}. 
The results refer to the low initial scales of the models.
\label{fig:models-TMD-PDFs-T-even-u}}
\end{figure}

In the bag model study of \cite{Avakian:2010br}, due to SU(4) 
spin-flavor symmetry, $f_1^d(x,k_T)$ is exactly one half of 
$f_1^u(x,k_T)$, and exhibits the same $k_T$-dependence. 
The other T-even TMDs 
$g_1^q(x,k_T)$, $h_1^q(x,k_T)$, 
$g_{1T}^{\perp q}(x,k_T)$, $h_{1L}^{\perp q}(x,k_T)$, 
$h_{1T}^{\perp q}(x,k_T)$ similarly exhibit approximate Gaussian 
$k_T$-behaviors \cite{Avakian:2010br}. In conclusion, the relativistic 
description of the nucleon as a 3-quark bound state in the bag model 
naturally supports the Gaussian approximation. In other models, the 
Gaussian $k_T$-dependence is also supported to a good approximation,
see e.g.\ \cite{Lorce:2016ugb,Maji:2017bcz}.

\begin{figure}[tp]
\begin{center}
\includegraphics[width=0.32\textwidth]{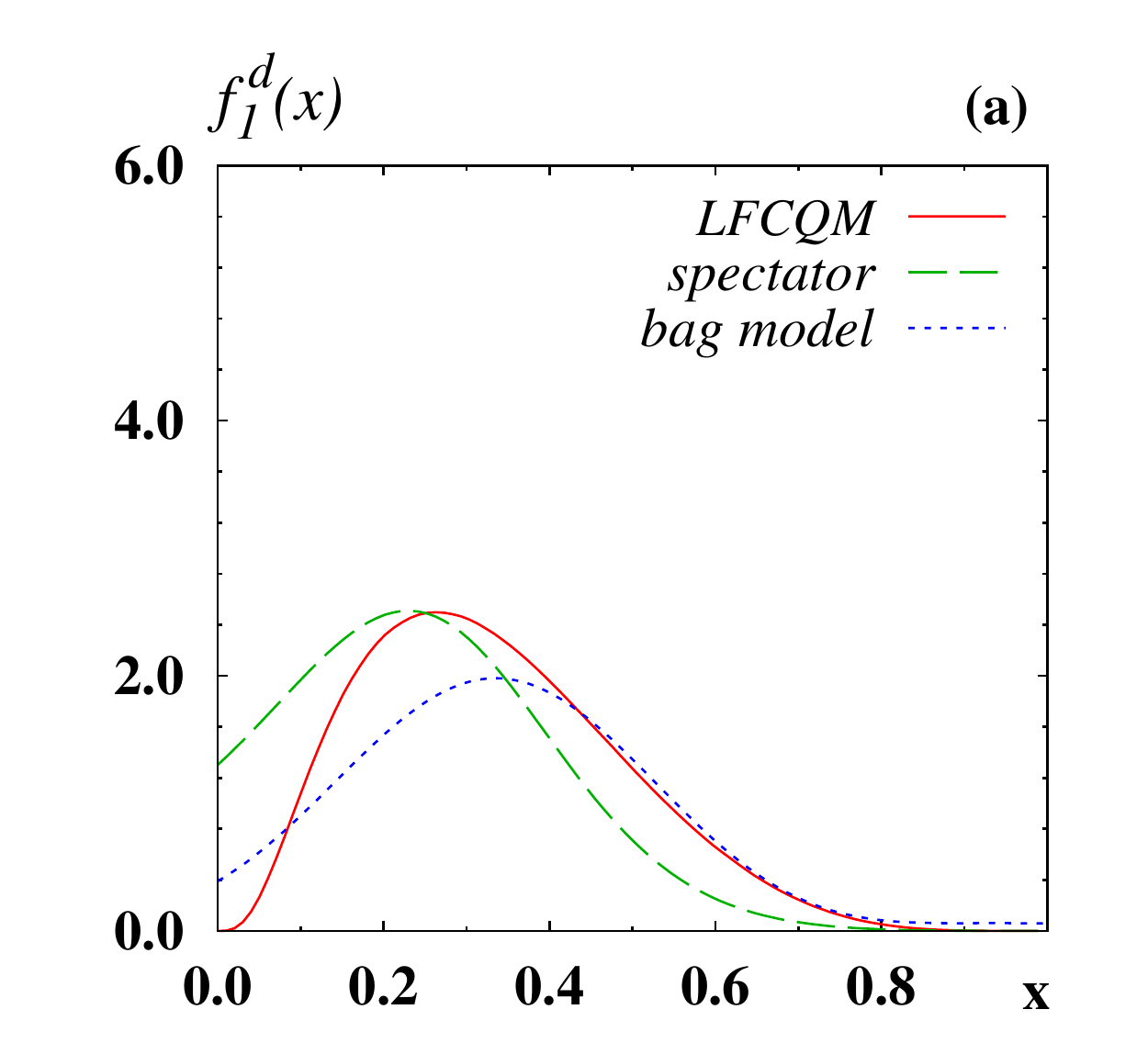}
\includegraphics[width=0.32\textwidth]{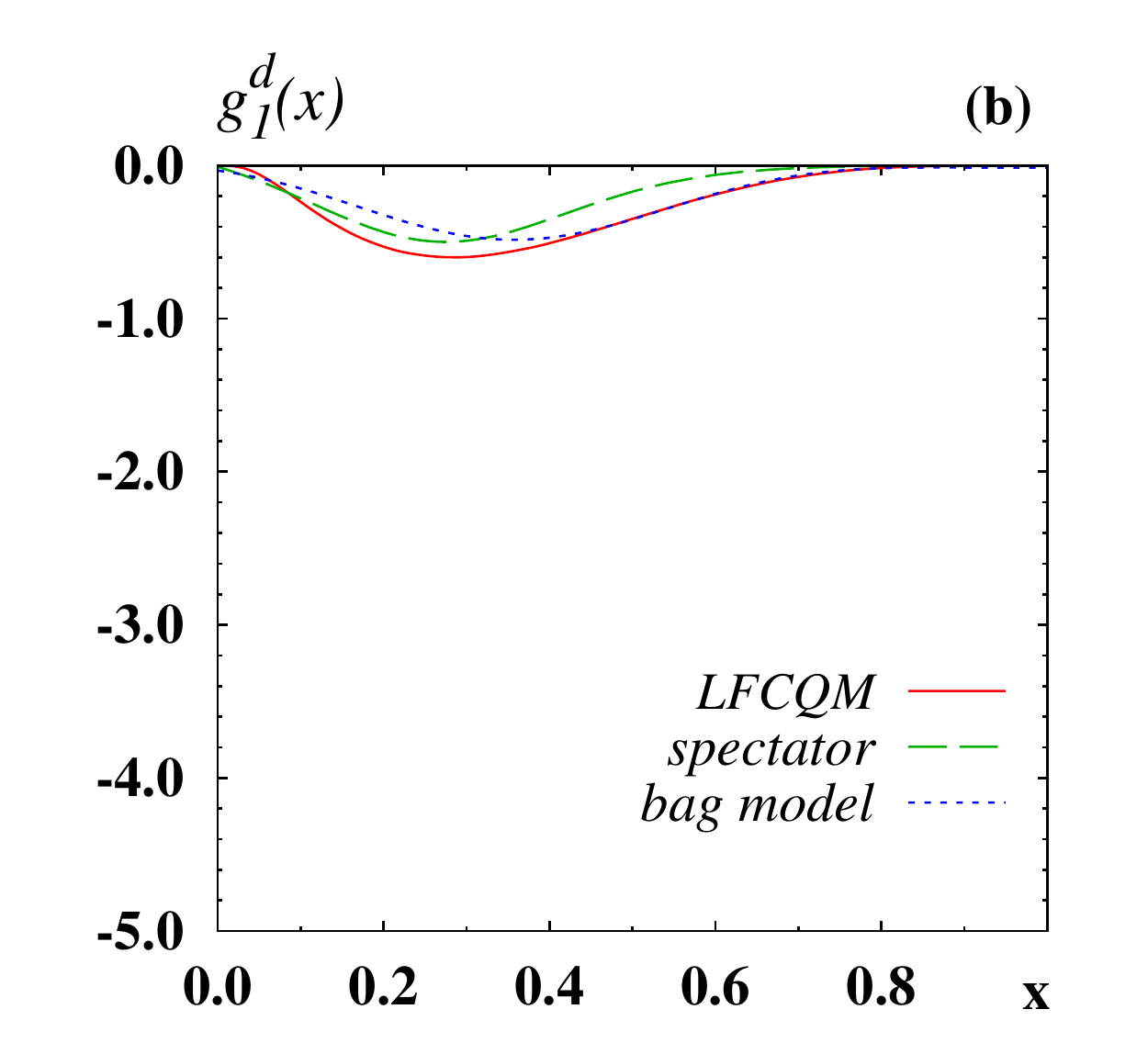}
\includegraphics[width=0.32\textwidth]{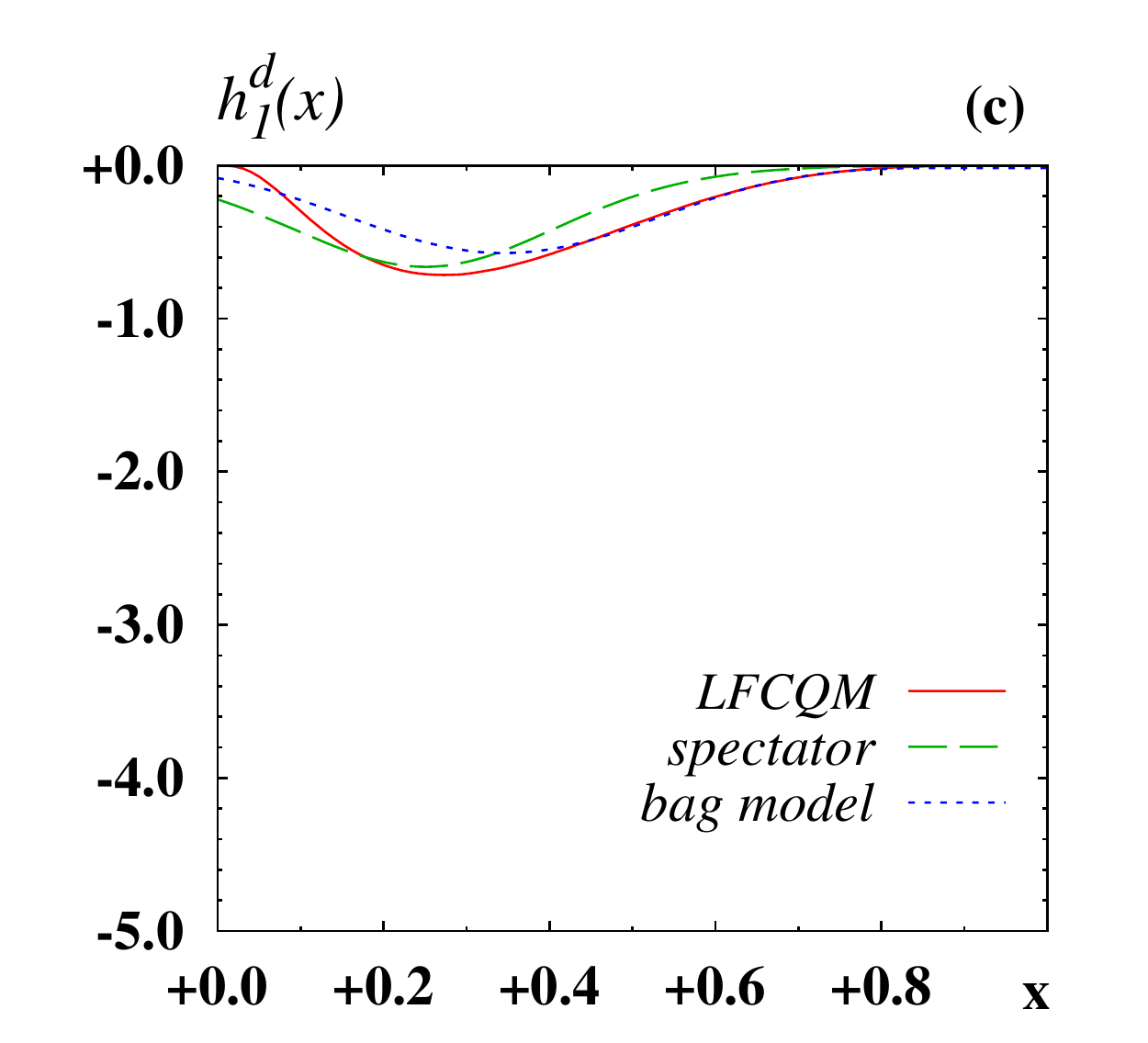} \\
\includegraphics[width=0.32\textwidth]{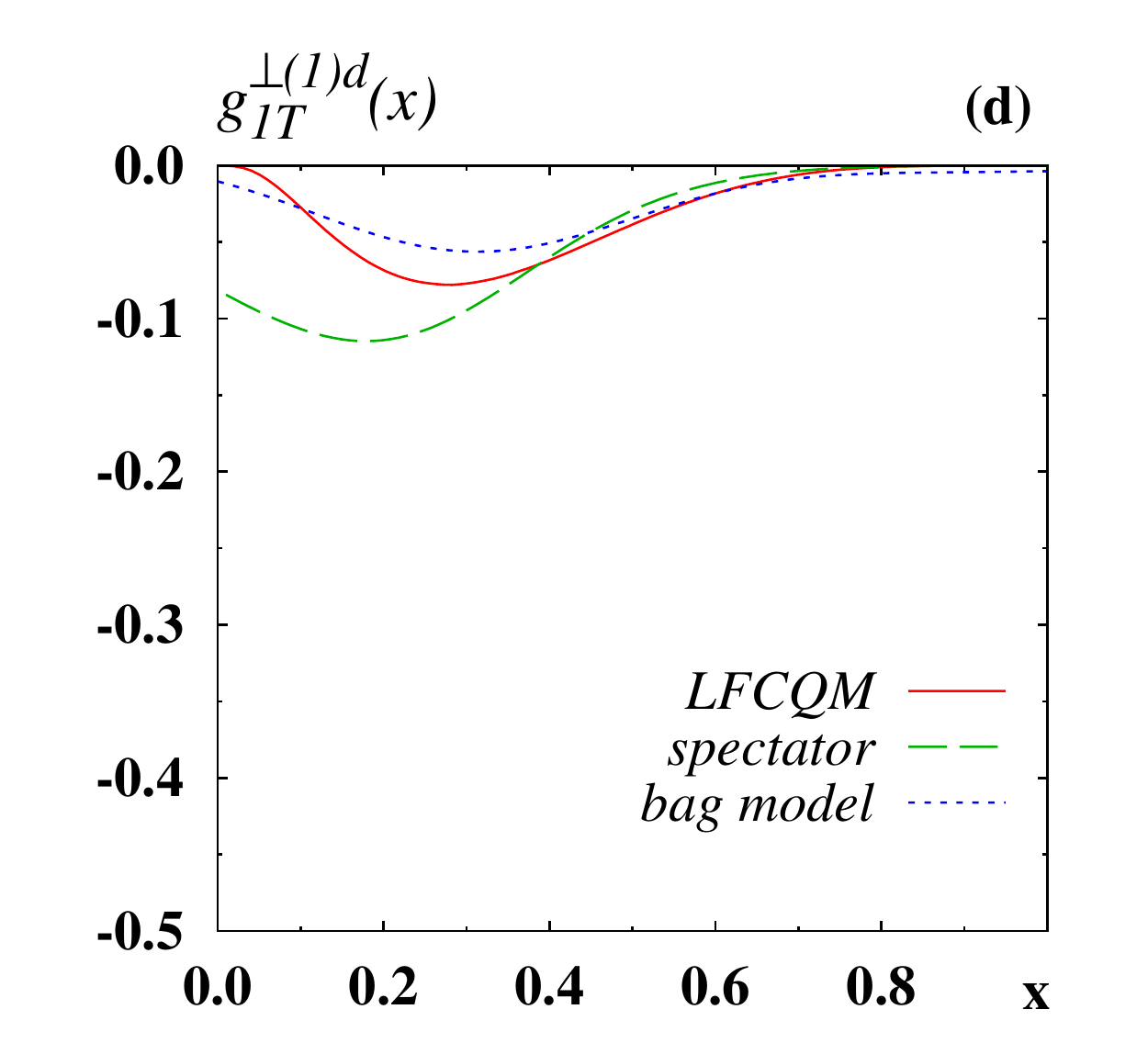} 
\includegraphics[width=0.32\textwidth]{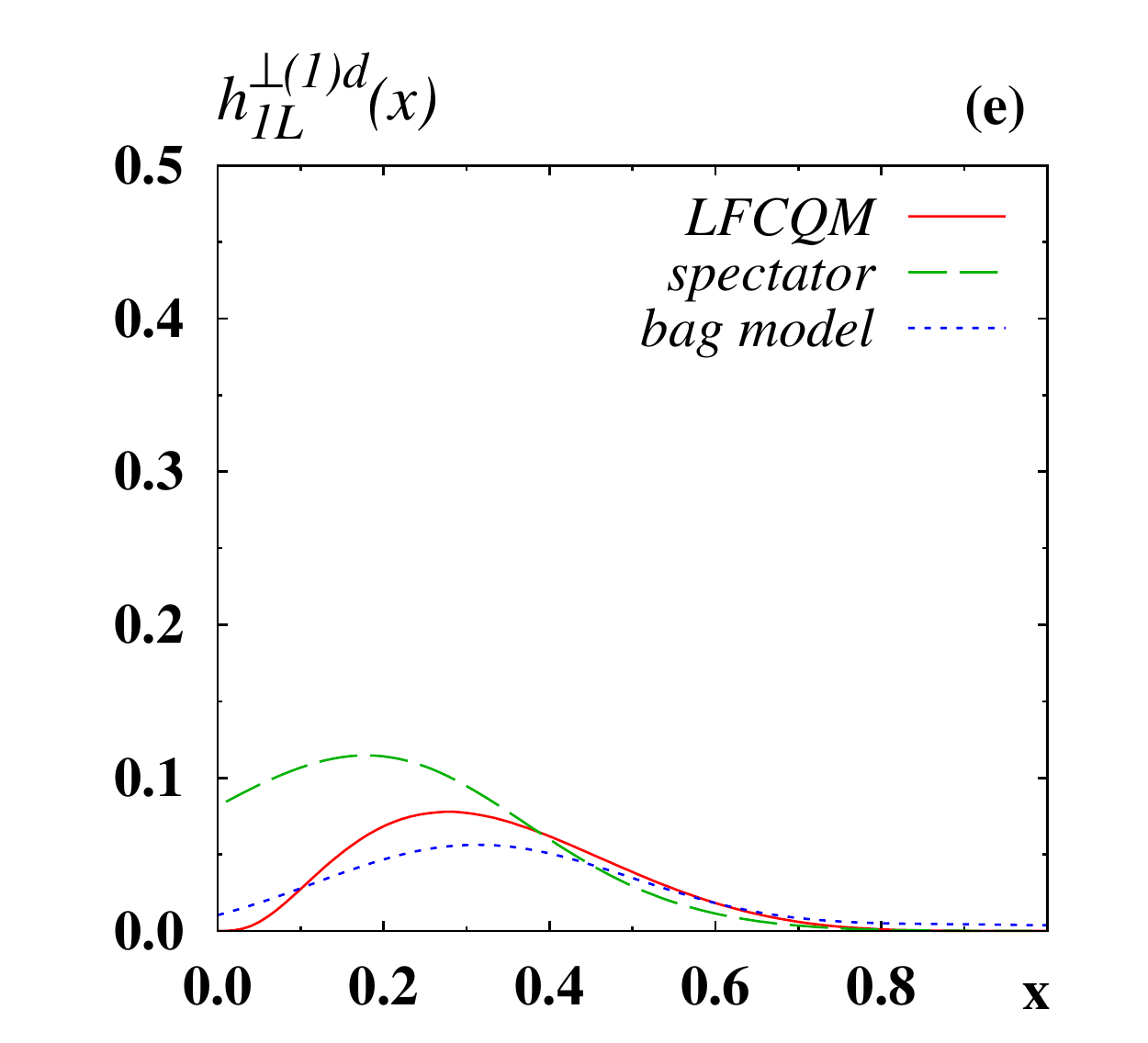}
\includegraphics[width=0.32\textwidth]{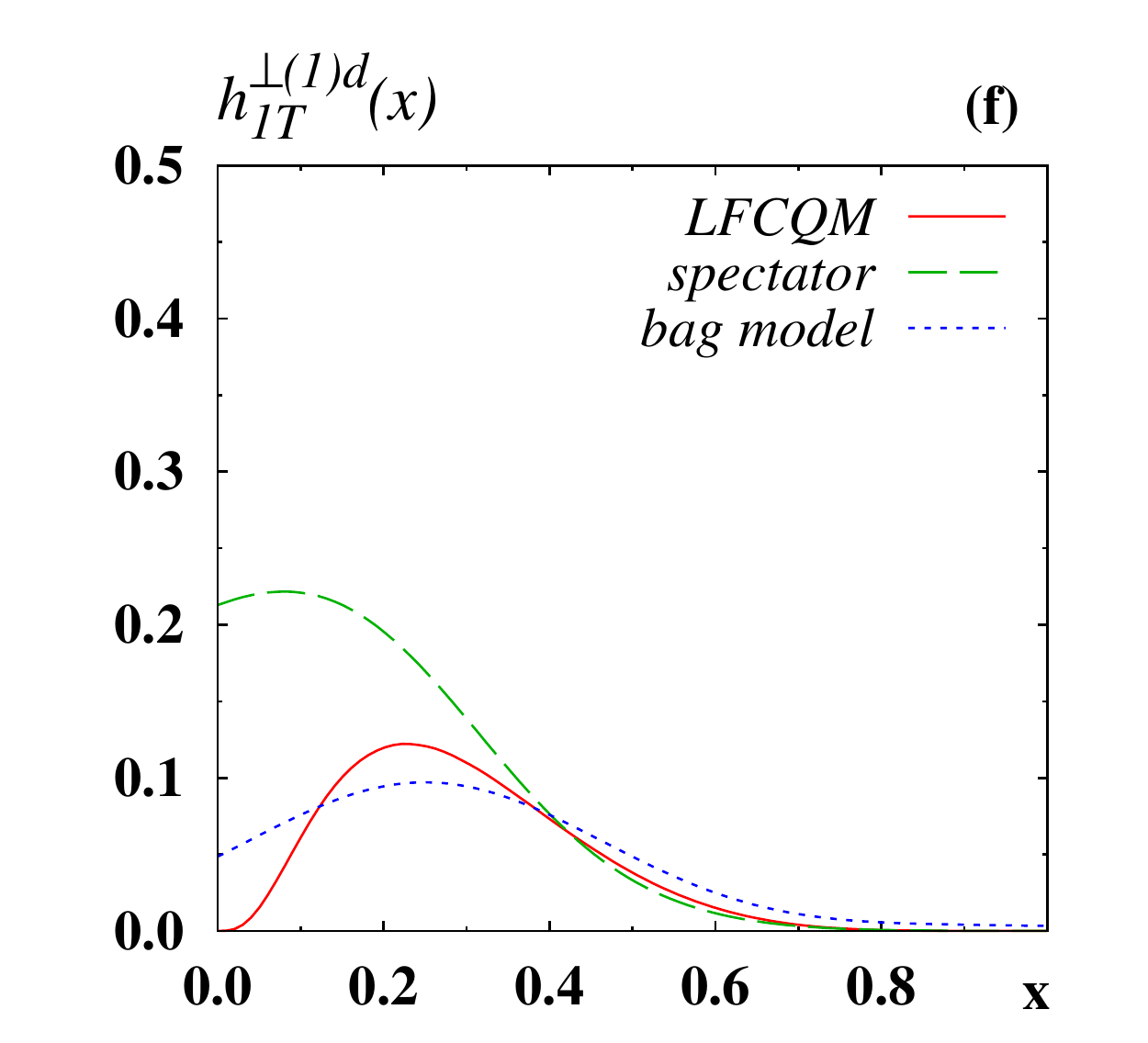}
\end{center}
\vspace{-4mm}
\caption{T-even TMD PDFs or their (1)-moments for 
$d$-quarks in proton from the same representative quark models
as in Fig.~\ref{fig:models-TMD-PDFs-T-even-u}. To facilitate the 
comparison, the same scales are chosen in both figures.
\label{fig:models-TMD-PDFs-T-even-d}}
\vspace{-4mm}
\end{figure}

Let us discuss next the $x$-dependence of TMD functions. 
In Figs.~\ref{fig:models-TMD-PDFs-T-even-u} and 
\ref{fig:models-TMD-PDFs-T-even-d} we show results from 
3 different models of the nucleon for 
$f_1^q(x)$, $g_1^q(x)$, $h_1^q(x)$, 
$g_{1T}^{\perp(1)q}(x)$, $h_{1L}^{\perp(1)q}(x)$, 
$h_{1T}^{\perp(1)q}(x)$. 
In models, the integrals over $k_T$ are convergent or 
can be simply regularized and, e.g., it is literally 
$f_1^q(x)=\int d^2k_T\,f_1^q(x,k_T)$. In QCD, there is no 
such simple connection between TMD PDFs and collinear PDFs,
see Sec.~\ref{Sec:formal-constraints}.
Notice that $f_1^q(x)$, $g_1^q(x)$, $h_1^q(x)$ are collinear 
PDFs and, especially in the case of $f_1^q(x)$ and $g_1^q(x)$,
well known from parametrizations. It is nevertheless interesting
to include them in the comparison. As explained in 
Chapter~\ref{sec:phenoTMDs}, the usage of transverse moments is
convenient in phenomenology (for pretzelosity actually the
(2)-transverse moment is more convenient, see
Sec.~\ref{sec:phenomelology-other},
but here we prefer to show also the (1)-moment because of
its relation to orbital angular momentum in models to be
discussed below in 
Sec.~\ref{subsec:pretzel-and-orbital-angular-momentum}).

The Fig.~\ref{fig:models-TMD-PDFs-T-even-u} shows the 
corresponding distributions of $u$-quarks in the proton,
while in Fig.~\ref{fig:models-TMD-PDFs-T-even-d} we show the 
results for $d$-quark distributions. 
The 3 models are the light-cone constituent quark model 
of Ref.~\cite{Pasquini:2008ax}, the bag model study of
Ref.~\cite{Avakian:2010br}, and the spectator model of 
Ref.~\cite{Jakob:1997wg}. All results refer to the low 
initial scales of the models estimated to be around 
$\mu_0\sim (0.3$--$0.5)\,{\rm GeV}$. In order to 
facilitate the comparison of the flavor dependence,
we have chosen the same scales for respectively the same 
TMD PDFs in Figs.~\ref{fig:models-TMD-PDFs-T-even-u} and 
\ref{fig:models-TMD-PDFs-T-even-d}. 

It should be stressed that these models have limitations.
For instance, they can be expected to be applicable in the 
valence-$x$ region $0.1\lesssim x \lesssim 0.6$ but, e.g., 
not at small $x$ where different principles govern the 
modelling of $x$- and $k_T$-effects, see Chapter~\ref{sec:smallx}.

Clearly, the different models give different results for the
TMD distributions in Figs.~\ref{fig:models-TMD-PDFs-T-even-u} 
and \ref{fig:models-TMD-PDFs-T-even-d}. This is to be expected. 
The models are not precision tools, and each of them has its 
own limitations.  
Considering how different these models are, it is remarkable 
that they agree on many features. E.g., the models agree on 
the magnitudes and signs of the TMD functions. 
(We stress that these models are representative.
Many other models give results similar to those 
shown in Figs.~\ref{fig:models-TMD-PDFs-T-even-u}
and \ref{fig:models-TMD-PDFs-T-even-d}.)

The common features include that transversity $h_1^q(x)$ 
is as large as the helicity PDF $g_1^q(x)$. This is compatible 
with information from phenomenology, see Chapter~\ref{sec:phenoTMDs}, 
and lattice QCD, see Chapter~\ref{sec:lattice}.\index{transversity!models}\index{worm-gear functions!models}
The Kotzinian-Mulders worm-gear functions $g_{1T}^{\perp q}$ 
and $h_{1L}^{\perp q}$ have the same magnitudes but opposite 
signs, an interesting observation to which we will come back 
in more detail in Sec.~\ref{Sec:relations-in-models}.
Also, in all 3 models, the pretzelosity function $h_{1T}^{\perp q}$ is  
larger than the Kotzinian-Mulders worm-gear function and has 
opposite sign with respect to transversity. 
It will be interesting to test these predictions in 
phenomenology.

The models discussed so far describe quark TMD PDFs. 
In fact, in some of these models, it is more appropriate 
to speak about distributions of valence quarks. 
The chiral quark soliton model is one of the few models 
where TMD PDFs of quarks and antiquarks can be defined 
and consistently evaluated \cite{Schweitzer:2012hh,Wakamatsu:2009fn}.
The remarkable prediction of the chiral quark soliton model 
are the distinctly different valence- and sea-quark
$k_T$-dependencies. 

\begin{figure}[tp]
\begin{center}
\includegraphics[width=0.32\textwidth]{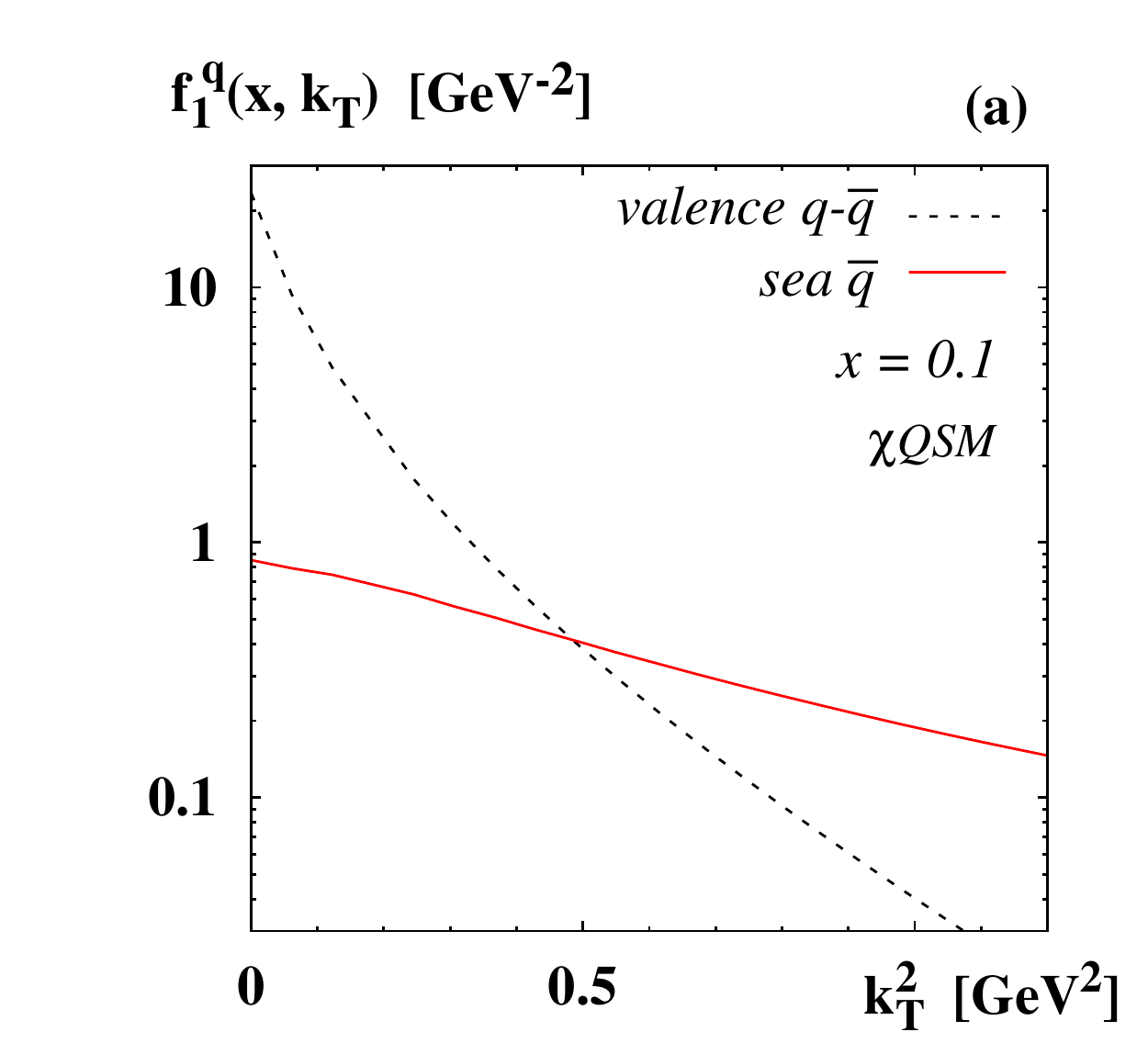} \ 
\includegraphics[width=0.32\textwidth]{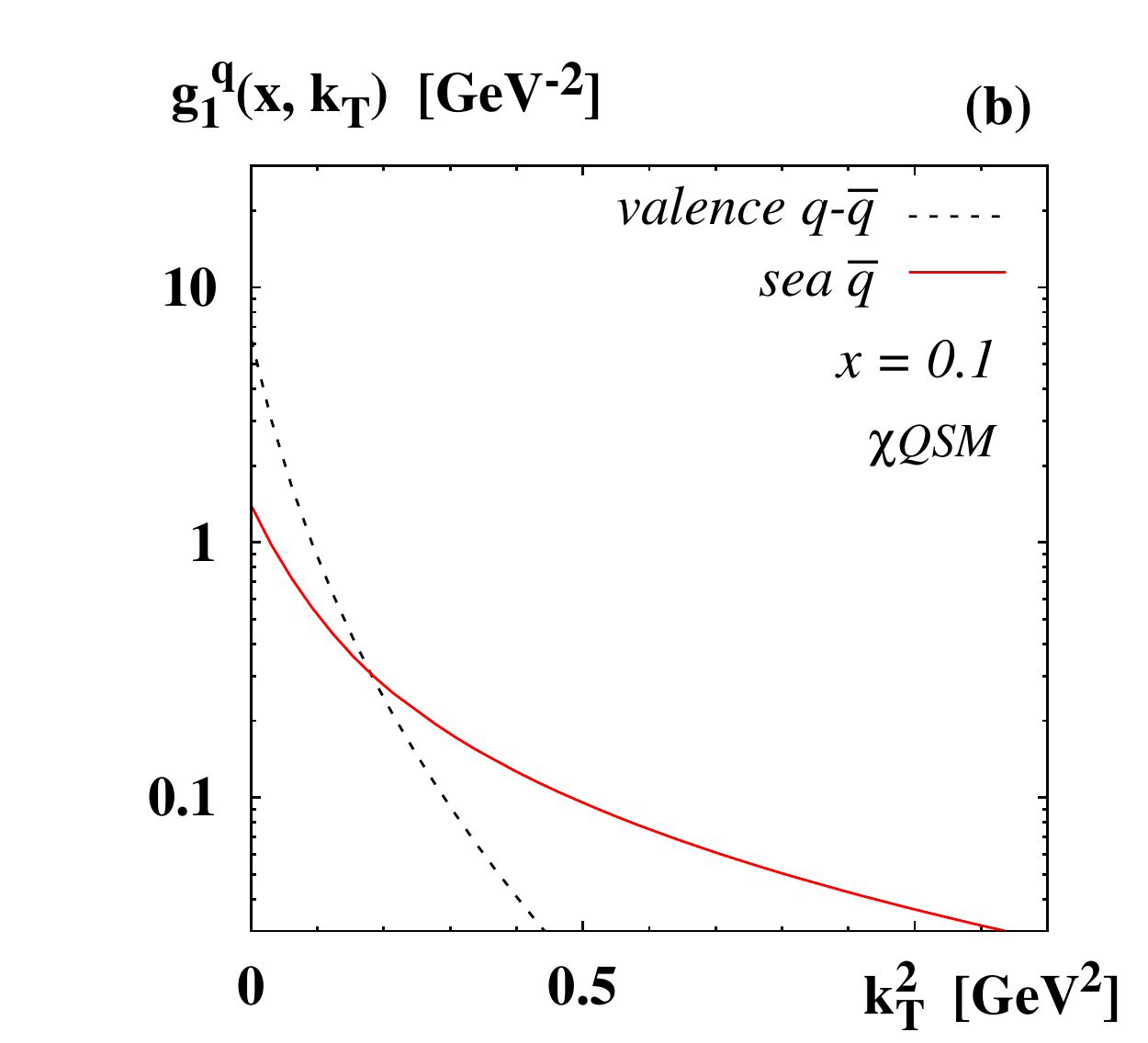}
\end{center}
\vspace{-0.4cm}
\caption{$f_1^q(x,k_T)$ and $g_1^q(x,k_T)$ at $x=0.1$
as functions of $k_T$ from the chiral quark
soliton model ($\chi$QSM) at low scale 
$\mu_0=0.6\,{\rm GeV}$ in leading order of 
large $N_c$.
Dashed lines: valence quarks.
Solid lines: sea quarks. 
The extended tails of sea quark distributions are due
short-range correlations between $q\bar q$-pairs caused
by chiral symmetry breaking. From Ref.~\cite{Schweitzer:2012hh}.
\label{fig:models-TMD-f1-val-sea-CQSM}}
\end{figure}

The Fig.~\ref{fig:models-TMD-f1-val-sea-CQSM} shows results 
on $f_1^q(x,k_T)$ and $g_1^q(x,k_T)$ at $x=0.1$ from the 
leading order of the large-$N_c$ expansion at the low 
scale $\mu_0\sim\rho_{\rm av}^{-1}\sim 0.6\,{\rm GeV}$
set by the average instanton size $\rho_{\rm av}$
\cite{Schweitzer:2012hh}.
In Fig.~\ref{fig:models-TMD-f1-val-sea-CQSM}, the $q$ and $\bar q$
stand for the respective leading large-$N_c$ flavor combinations.
In the case of $f_1^q(x,k_T)$ in Fig.~\ref{fig:models-TMD-f1-val-sea-CQSM}a
it is $q=u+d$ and $\bar q=\bar u+\bar d$.
In the case of $g_1^q(x,k_T)$ in Fig.~\ref{fig:models-TMD-f1-val-sea-CQSM}b
it is $q=u-d$ and $\bar q=\bar u-\bar d$.
The large-$N_c$ behavior of TMDS PDFs in the chiral
quark soliton model is in agreement with general 
prediction in QCD, see Sec.~\ref{subsubsec:largeNc}.

Remarkably, the valence quark distribution, $q-\bar q$,
falls of with $k_T$ steeply, while the sea quark distribution,
$\bar q$, has an extended power-like tail which in this model
is approximately described as
$f_1^{\bar q}(x,k_T) \sim f_1^{\bar q}(x)\,/\,k_T^2$
in the region $M^2\ll k_T^2 < \mu_0^2$ where $M=350\,{\rm MeV}$
is the constituent quark mass. The picture for $g_1^{\bar q}(x,k_T)$
is analogous. This is not accidental but, given the relation of
the unpolarized and helicity distributions to vector (V)
and axial-vector (A) currents, a consequence of the spontaneous
breaking of chiral symmetry. For instance, in the case of  transversity --- which is not related to the V- or A-currents
--- the sea quarks exhibit the same $k_T$-behavior as valence
quarks \cite{Schweitzer:2012hh}.

The distinct behavior of valence- vs sea-quark distributions
at a low scale $\mu_0$ is an interesting signature of chiral 
symmetry breaking for the nucleon structure. 
The $k_T$-dependence of valence-quark distributions is governed
by the hadronic scale $R_{\rm had}\sim M^{-1}$ which sets the size 
of light hadrons $R_{\rm had}\sim 1\,{\rm fm}$. The valence-quark
distributions in the chiral quark soliton model are qualitatively
similar to those in the other quark models discussed above.
In contrast to this, the $k_T$-dependence of sea quarks in the 
unpolarized and helicity distributions is governed by the much 
shorter length scale $\rho_{\rm av}\sim 0.3\,{\rm fm}$ which set 
the scale at which chiral symmetry is spontaneously  broken. The 
sea quarks experience short-range correlations at the length 
scale  $\rho_{\rm av}$ which cause this characteristic behavior
\cite{Schweitzer:2012hh}. There is a certain analogy to the 
short-range nucleon-nucleon correlations observed in nuclear 
physics \cite{Arrington:2011xs}. 
It is interesting to observe in Fig.~\ref{fig:models-TMD-f1-val-sea-CQSM}
that the $k_T$ of sea quarks is of more relative importance in
the case of polarized 
as compared to unpolarized sea quarks. It will be interesting to 
explore the phenomenological consequences from these predictions.

\subsection{Modelling of T-odd TMDs PDFs}
\label{Sec:models-T-odd-quark-TMDs}

This section is devoted to the modelling of T-odd TMD PDFs 
where the challenge lies in a consistent inclusion of the
final/initial state interactions encoded in the Wilson line.

\subsubsection{A no-go theorem}

In QCD the T-odd character of the Sivers and Boer-Mulders function 
is ultimately rooted in the process-dependent Wilson line encoding 
the initial or final state interactions. This means that in order 
to obtain nonzero results for T-odd TMD PDFs, a model must contain 
gauge field degrees of freedom. Let us consider a generic quark model,
which we define as a model with no explicit gauge field degrees of 
freedom. Any such realistic model will respect the basic symmetries 
of QCD, and most notably C, P, and T. Unless one deals with an exotic 
situation where time-reversal invariance is spontaneously broken
(which would be an unrealistic model as this is not the case in QCD),
the T-odd TMDs PDFs vanish. A corresponding no-go theorem was explicitly 
proven in the chiral sigma model in Ref.~\cite{Pobylitsa:2002fr}, and
it is straightforward to extend this proof to other models with no 
gauge field degrees of freedom. 

\subsubsection{Including gauge field degrees of freedom}
\label{Sec:models-T-odd-1-gluon-exchange}

In order to model T-odd TMD PDFs, it is necessary to include gauge field
degrees of freedom which can be done in various ways. One way of
modelling 
T-odd TMD PDFs is to consider the one gluon-exchange
mechanism which can be an ``abelian gluon'' or a ``QCD-type gluon.''
In such model calculations, one takes into account only the 
contribution from expanding the Wilson line to leading order in the
strong coupling and neglects higher orders. As a result, the obtained
T-odd TMD PDFs are proportional to the strong coupling constant 
$\alpha_s(\mu)$ where $\mu$ is the low initial scale of the 
considered model. 
After the first pioneering calculation of this type within a 
simple scalar-diquark model framework \cite{Brodsky:2002cx}, 
see Sec.~\ref{Subsec-models:review-Brodsky-Hwang-Schmidt} 
for a review, the Sivers and Boer-Mulders functions of the 
nucleon have been studied in more elaborate versions of spectator
\cite{Goldstein:2002vv, Gamberg:2003ey, Lu:2006kt, Gamberg:2007wm, Meissner:2007rx, Ellis:2008in, Bacchetta:2008af},
bag models \cite{Yuan:2003wk,Courtoy:2008dn,Courtoy:2009pc}, 
and lightfront constituent quark models
\cite{Pasquini:2010af,Pasquini:2011tk}.
The Boer-Mulders function of the pion was studied in
\cite{Lu:2004hu, Meissner:2008ay, Gamberg:2009uk}.
T-odd TMDs were also studied in non-relativistic 
models \cite{Courtoy:2008vi} (notice that in practical 
non-relativistic quark model calculations \cite{Courtoy:2008vi} 
the $\delta$-functions of the strict non-relativistic limit in Eq.~(\ref{Eq:non-rel-lim}) are considerably smeared out).

In these studies, different models are employed for the nucleon
structure, but conceptually the same 1-gluon exchange mechanism
is invoked to take into account the initial/final state
interactions.
Common to all these approaches is that the Sivers and 
Boer-Mulders functions require orbital angular momentum in 
the nucleon wave function: the matrix elements of these TMD 
PDFs involve a transitions between quark wave functions with 
orbital angular momentum components which differ by 
$\Delta L_z=\pm 1$.

\subsubsection{Quark-target model}

\index{model!quark-target model}

The Lagrangian of the quark target model 
is basically the QCD Lagrangian except that it is considered for 
one single flavor, and the model is solved in perturbative QCD. 
It is not intended to be a realistic model for a hadron, as the 
S-matrix contains colored states and the current quark mass
terms, which are negligible in QCD, have $100\,\%$ strength 
since the proton state is replaced by a quark. Nevertheless, 
it is an interesting testing ground, e.g., for relations
among TMD PDFs obtained from quark models with no gluonic
degrees of freedom. Such relations can be investigated in this
model under more realistic conditions thanks to the QCD-like 
color structure which includes quark-gluon-quark and 
three-gluon vertices \cite{Meissner:2007rx}.

\subsubsection{Lensing function}
\label{subsec:models-lensing-function}

\index{lensing function}

An attractive modelling approach is based on the observation 
that in some models, the same LCWF components enter in the
description of T-odd TMD PDFs and certain generalized parton
distribution functions
\cite{Burkardt:2002ks,Burkardt:2003uw,Burkardt:2003je,Meissner:2007rx}.
The difference is, of course, that T-odd TMD PDFs contain the effects 
of initial/final state interactions which, in simple models, can be
expressed in terms of so-called "chromodynamic lensing functions."
More precisely, this allows one to effectively express the leading 
T-odd TMD PDFs in terms of convolutions of the lensing functions with
generalized parton distribution functions in the impact parameter
representation which are introduced in Chapter \ref{sec:gtmd}.
This connection is rooted in the relations of TMD PDFs and 
generalized parton distribution functions to the overarching 
Wigner functions and generalized TMD functions
\cite{Diehl:2005jf,Burkardt:2005hp}, see Chapter \ref{sec:gtmd} 
for more discussion on these topics.

The generalized parton distribution function needed in this
picture to model the Sivers function is not known, but it is 
related to the known quark contributions to the anomalous magnetic
moment of proton and neutron. In this way one can conclude the overall
signs of the Sivers function for $u$- and $d$-flavors. This is possible
because the final state interactions, while in detail complicated to
describe, can be expected to be attractive on average. The signs of 
the Sivers function for $u$-quarks and $d$-quarks concluded in 
this way are in agreement with phenomenology \cite{Burkardt:2015qoa}.
In the case of the Boer-Mulders function certain chiral-odd
generalized parton distributions enter which are also not known.
But on the basis of information from models and lattice QCD, and
under the assumption that the final state interactions are on average
attractive, one finds also in this case the signs of the $u$- and
$d$-quark Boer-Mulders functions in agreement with phenomenologyical
information \cite{Burkardt:2015qoa}.

The lensing mechanism is realized to lowest order in spectator or
quark-target models \cite{Burkardt:2003je,Meissner:2007rx} but it
is not model-independent
\cite{Meissner:2008ay, Meissner:2009ww, Pasquini:2019evu}.
Two general conditions needed for the lensing function concept 
to be applicable are (i) that the coupling between the gauge 
boson and the spectator system conserves helicity, and (ii) 
that the model describes the considered hadron as a 2-body
system \cite{Pasquini:2019evu}.
For instance, the lensing function concept works in conceptually
similar models for the Boer-Mulders function of the pion
described as a 2-body system in terms of the minimal $\bar{q}q$ 
Fock state component, but it does not work for the nucleon 
whose minimal $qqq$ Fock-state component constitutes a
3-body system \cite{Pasquini:2019evu}.
The relation of TMD PDFs to GTMDs (the latter are discussed in detail in \chap{gtmd}) was studied in Ref.~\cite{Gurjar:2021dyv} in a soft-wall AdS/QCD-motivated light-front quark-diquark model. This model 
supports the lensing function approach \cite{Gurjar:2021dyv}.

\subsubsection{Augmented LCWFs, eikonal methods, instantons and other approaches}

An interesting approach consists in introducing augmented LFWFs 
which incorporate the initial/final state interaction effects in 
imaginary process-dependent phases. This approach was worked out in 
light-front time-ordered perturbation theory in Ref.~\cite{Brodsky:2010vs}.

The so far discussed approaches were largely based on the 1-gluon
rescattering mechanism for initial/final state interactions, i.e.,\
the expansion of the Wilson link to the lowest non-trivial order.
Attempts to go beyond that were undertaken in
Refs.~\cite{Gamberg:2007wm,Gamberg:2009ma,Gamberg:2009uk} using
nonperturbative eikonal methods to evaluate the Wilson line to all
orders in U(1), SU(2), SU(3) gauge groups. The results were used
to study T-odd TMDs in the lensing function approach. 

Further nonperturbative approaches to T-odd single-spin effects
based on the nonperturbative interactions induced by instantons
\index{model!instanton model} were explored in
Ref.~\cite{Ostrovsky:2004pd,Cherednikov:2006zn,Qian:2011ya}.
A potentially related soft rescattering mechanism was considered 
in \cite{Hoyer:2005ev} due to the possibility of an anomalous 
chromomagnetic moment of quarks. Such a Pauli coupling at the
vertex between the struck quark and exchanged gluon can be 
potentially generated by instanton effects \cite{Hoyer:2005ev}.

For completeness we remark that a way to circumvent the no-go
theorem concerning modelling of T-odd functions in chiral quark
models \cite{Pobylitsa:2002fr} was discussed in
\cite{Drago:2005gz} where the role of gluons was played by 
a ``hidden vector-meson gauge symmetry.'' An attempt
to implement this in a practical calculation was
presented in \cite{He:2019fzn}.

\subsubsection{Predictions from quark models for T-odd TMD PDFs}

\begin{figure}[t!]
\begin{center}
\includegraphics[width=0.32\textwidth]{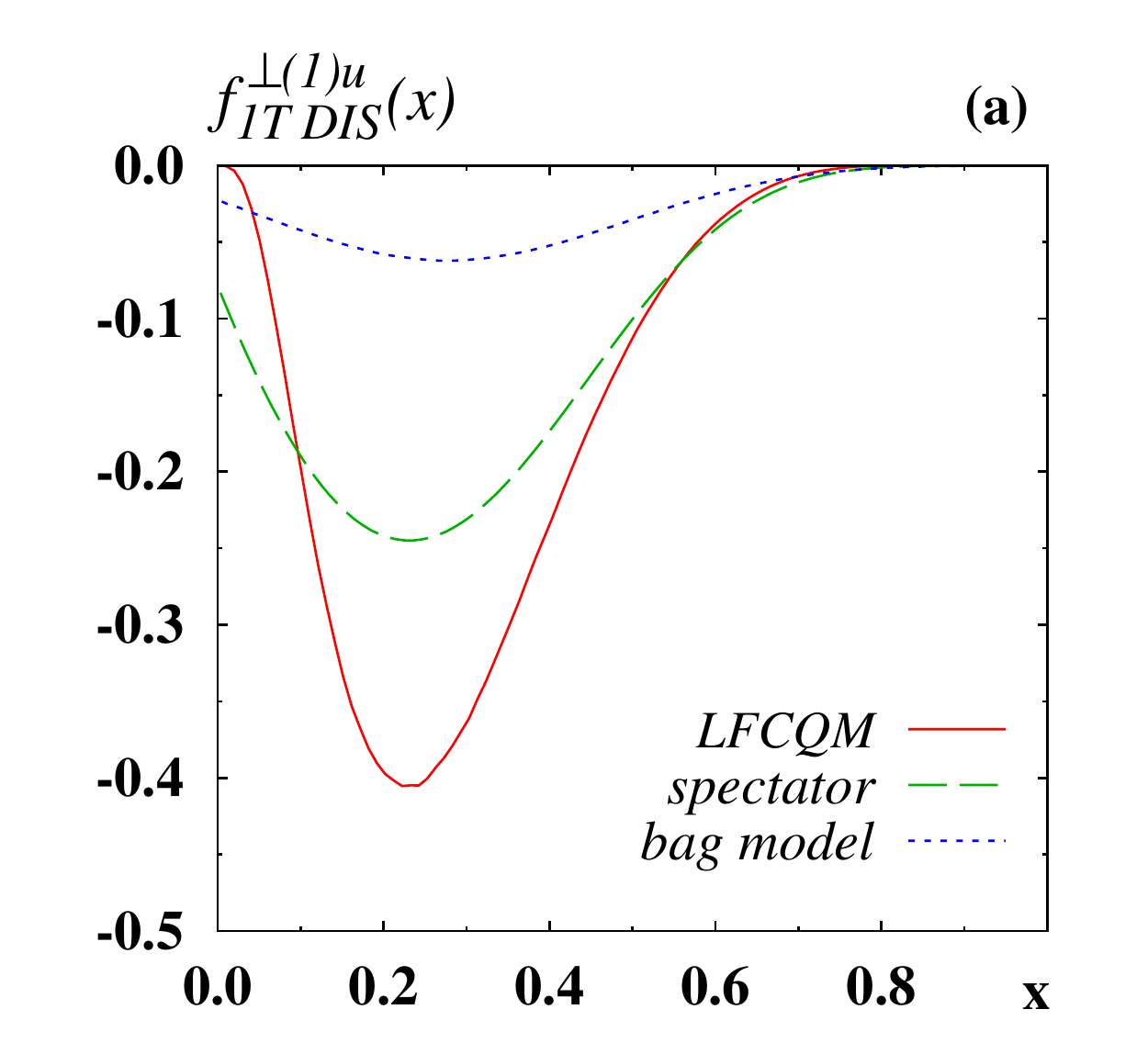}
\includegraphics[width=0.32\textwidth]{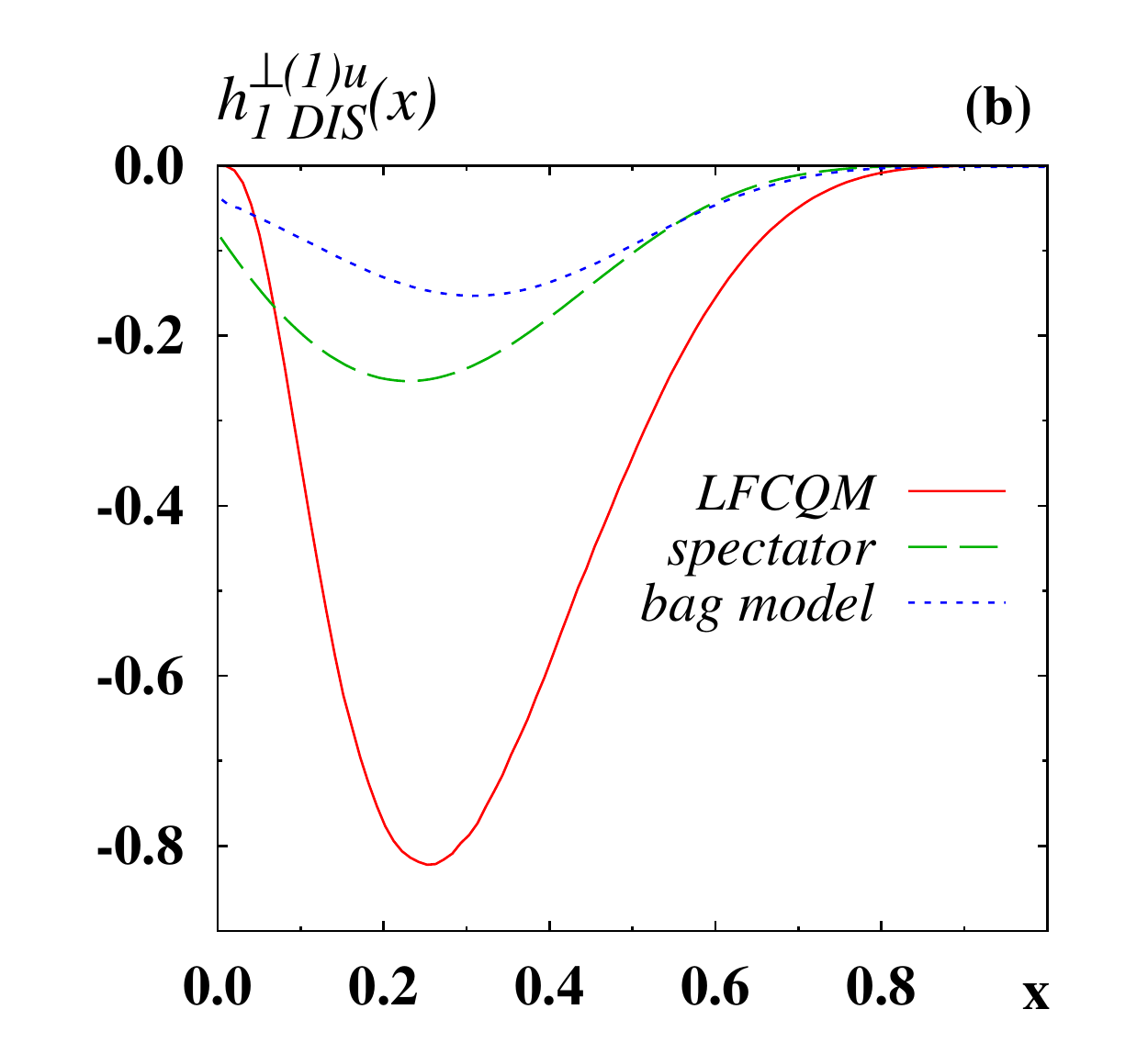}
\end{center}
\vspace{-0.4cm}
\caption{Results for (1)-moments of the T-odd  
Sivers and Boer-Mulders functions for $u$-quarks 
in a proton from representative quark models:
light-front constituent quark model (LFCQM) \cite{Pasquini:2010af},
spectator model \cite{Gamberg:2007wm}, and
bag model \cite{Courtoy:2008dn,Courtoy:2009pc}. 
The results refer to the low initial scales of the models.
\label{fig:models-TMD-PDFs-T-odd-u}}

\begin{center}
\includegraphics[width=0.32\textwidth]{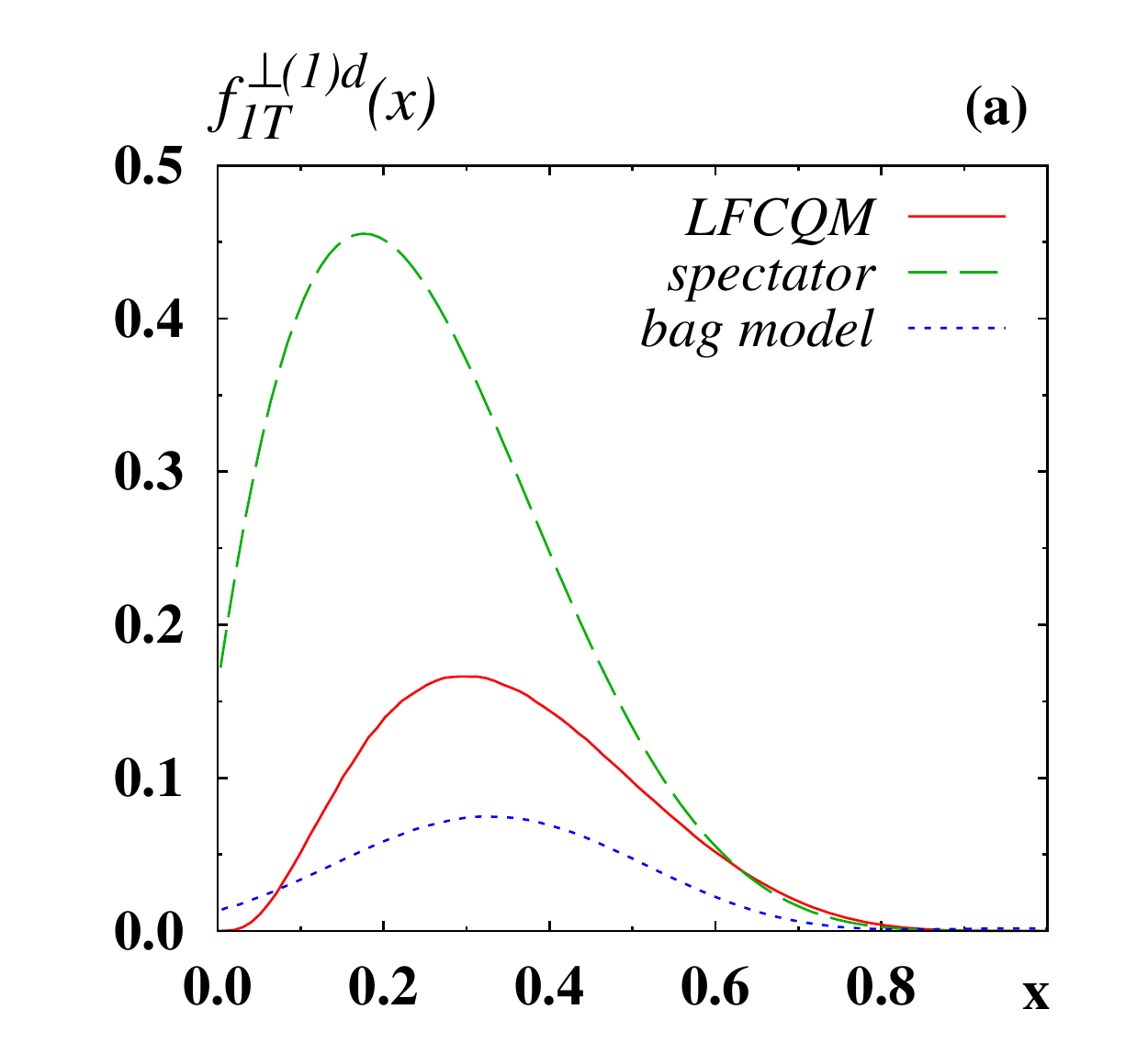}
\includegraphics[width=0.32\textwidth]{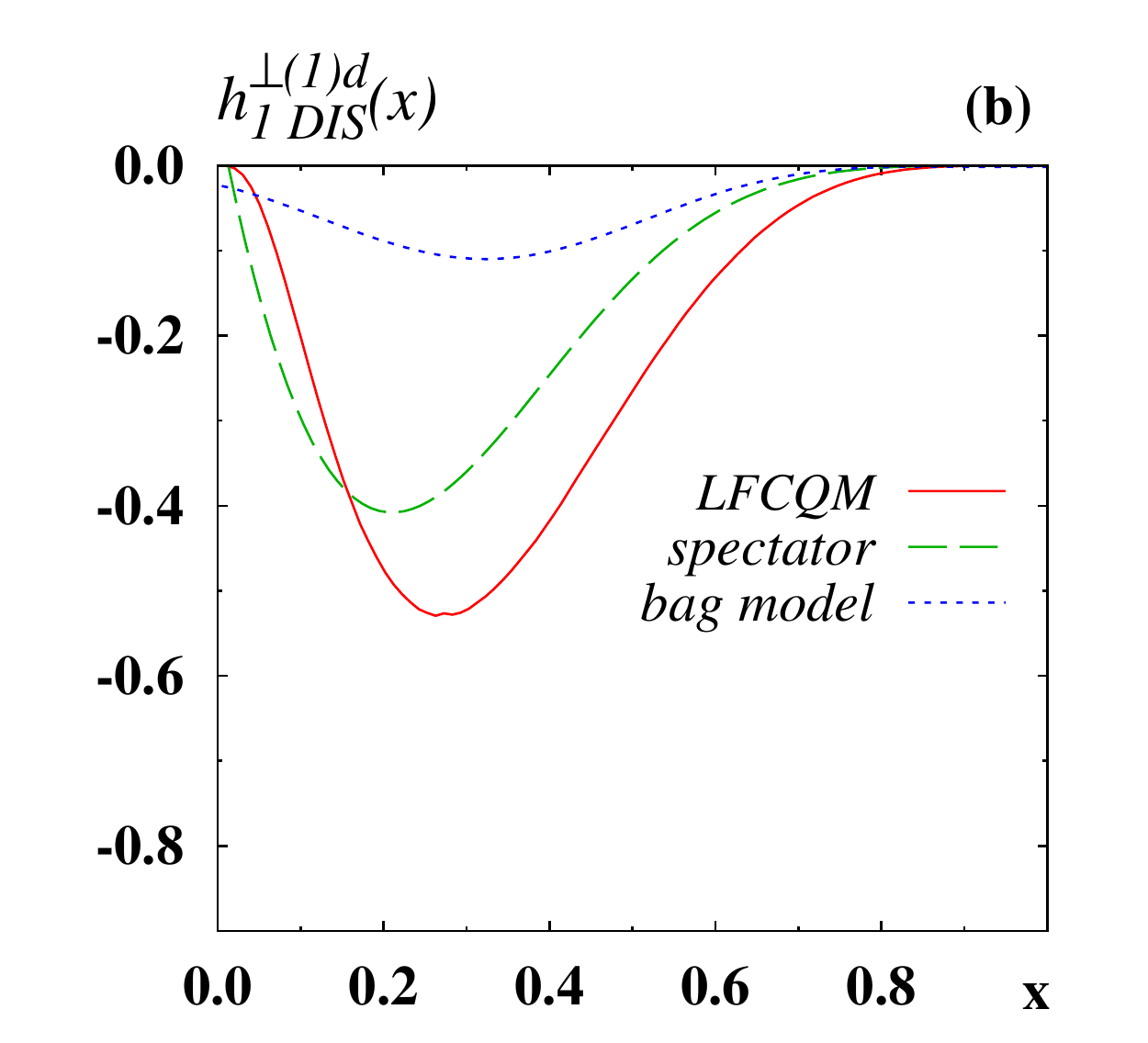}
\end{center}
\vspace{-0.4cm}
\caption{Results for (1)-moments of the T-odd  
Sivers and Boer-Mulders functions for $d$-quarks 
in the same models as in Fig.~\ref{fig:models-TMD-PDFs-T-odd-u}. 
To facilitate the  comparison, the same scales are chosen in both
figures.
\label{fig:models-TMD-PDFs-T-odd-d}}

\end{figure}

The aim of this section is to present model results for the leading 
T-odd TMD PDFs of the nucleon from several representative models.
In Fig.~\ref{fig:models-TMD-PDFs-T-odd-u} and 
\ref{fig:models-TMD-PDFs-T-odd-d} we show results from the
light-front constituent quark model (LFCQM) \cite{Pasquini:2010af},
spectator model \cite{Gamberg:2007wm}, and bag model
\cite{Courtoy:2008dn,Courtoy:2009pc}. The signs of the T-odd 
functions refer to DIS. All model results refer to  the low 
initial scales of the models which are below $0.5\,{\rm GeV}$
in all cases.

The results from the different models show a significant spread. 
At the same time, the models also agree on several features. 
The models predict the same signs and similar flavor dependencies.
For instance, the Sivers functions for $u$-quarks and $d$-quarks 
have opposite signs and comparable magnitudes. 
The Boer-Mulders functions for $u$-quarks and $d$-quarks have
the same signs and comparable magnitudes. The sign patterns 
are compatible with predictions from the large-$N_c$ limit
\cite{Pobylitsa:2003ty} discussed in Sec.~\ref{subsubsec:largeNc}.

There is also agreement that, within a given model, 
the Boer-Mulders function is comparable to the Sivers function or 
even larger. 
\index{Sivers function $f_{1T}^{\perp}$!model calculations}
However, the results on the magnitudes of the functions
computed in different models span a wide range: 
for $f_{1T}^{\perp u}$, $f_{1T}^{\perp d}$, $h_{1}^{\perp d}$,  
the light-front constituent quark model gives the largest results
\cite{Pasquini:2010af}, 
bag model the smallest results \cite{Courtoy:2008dn,Courtoy:2009pc},
while the spectator model results lie in between --- being sometimes 
closer to one or the other of the above-mentioned models 
\cite{Gamberg:2007wm}.
For $h_{1}^{\perp u}$, 
the spectator model gives the largest results, and 
the light-front constituent quark model gives results in-between
\cite{Pasquini:2010af}, while the bag model gives consistently
the smallest results also in this case \cite{Courtoy:2009pc}.

The results from the models have been compared with
phenomenology showing good qualitative agreement. For the 
Sivers function, the tendency of the light-front constituent 
quark model results is to be at the upper edge of the uncertainty 
band of the extractions 
\cite{Pasquini:2010af}, while the bag model results tend
to underestimate the lower bounds of the parametrizations 
\cite{Courtoy:2009pc} (the extractions are discussed in
Secs.~\ref{sec:Sivers_SIDIS} and 
\ref{sec:phenomelology-Boer-Mulders-and-beyond}). 

The comparison to phenomenological extractions for TMD PDFs is 
not straightforward, because it is necessary to evolve the
model results from the low initial scales to the scales of 
the parametrizations. While there is significant experience 
with applying DGLAP evolution down to initial scales as low 
as $\mu_0\sim 0.4\,{\rm GeV}^2$ (the effective expansion
parameter $\frac{\alpha_s(\mu_0)}{4\pi} \sim 0.1$ may be 
considered small enough to warrant LO or NLO evolution)
\cite{Gluck:1994uf,Gluck:1998xa,Gluck:1995yr,Gluck:2000dy,Gluck:2007ck},
much less is known about the application of TMD evolution
down to such low scales. 
The comparisons to extractions and phenomenological applications of
model results relied on estimates of TMD evolution effects
\cite{Boffi:2009sh,Pasquini:2011tk,Pasquini:2014ppa,Courtoy:2009pc}
until recently \cite{Ceccopieri:2018nop,Bastami:2020rxn}.

\subsection{Gluon TMDs}
\label{Sec:models-gluon-TMDs}

Gluon TMDs can in principle be studied through a variety of processes and at different facilities, see Sec.~\ref{sec:gluonTMD_obs}.
Nevertheless, compared to quark TMDs, very little is presently known about gluon TMDs for moderate parton momentum fractions.
The situation is different in the small-$x$ region for which a number of (theoretical) studies exist.
Here we will not discuss this regime of gluon saturation but refer to Ch.~\ref{sec:smallx} for more details.

At the present stage, it is important to gather as much information as possible about gluon TMDs through models and also LQCD which could help to guide the experiments.
Here we give a very brief account of available model calculations in this field, all of which are concerned with leading-power gluon TMD PDFs. 
In Ref.~\cite{Goeke:2006ef} a one-loop calculation of the gluon Sivers function in the 
quark-target model was presented.
The result was used to check the Burkardt sum rule~\cite{Burkardt:2004ur} for the Sivers function in that model. (The Burkardt sum rule is discussed
in Sec.~\ref{Sec:Burkardt-sum-rule}.) 
Furthermore, in Ref.~\cite{Meissner:2007rx} all gluon TMDs were then computed in the 
quark-target model.
Six out of the eight TMDs are nonzero, while two T-odd gluon TMDs vanish in the one-loop approximation.
This result does of course not imply that, in general, those two gluon TMDs vanish or are small.
Interestingly, the Burkardt sum rule, in combination with results from a large-$N_c$ analysis, according to which the up-quark and down-quark Sivers functions are almost equal in magnitude but opposite in sign, leads to the prediction that the gluon Sivers function is suppressed relative to the quark Sivers function~\cite{Efremov:2004tp}.
Current phenomenology is compatible with this prediction, see also the discussion in Sec.~\ref{subsubsec:largeNc}.

A spectator model calculation of the gluon Sivers function of the proton was performed in Ref.~\cite{Lu:2016vqu}, 
\index{Sivers function $f_{1T}^{\perp}$!model calculations}
leading to results in reasonable agreement with information from experiment.
Another spectator model calculation of gluon TMDs can be found in Ref.~\cite{Rodrigues:2001PhD} and, in particular, in the recent work in Ref.~\cite{Bacchetta:2020vty}.
The latter study addresses the four T-even gluon TMDs, where the model parameters are fixed, to the extent possible, by information on the integrated unpolarized and helicity gluon PDFs.
Very recently, the same spectator-model approach was extended to the T-odd sector, and results for the gluon Sivers function were reported~\cite{Bacchetta:2021lvw}.
The T-even gluon TMDs were also explored in an AdS/QCD approach~\cite{Lyubovitskij:2020xqj}, where the results satisfy the positivity bounds on the gluon TMDs derived previously~\cite{Mulders:2000sh}.
In a follow-up study, the same authors addressed the T-odd gluon TMDs, for which they derived a parameterization in terms of light-front wave functions~\cite{Lyubovitskij:2021qza}.
It is important and encouraging to see those works on models for gluon TMD PDFs in the region of moderate $x$.
Further developments in this area can be expected for the future.

\subsection{Quark TMD Fragmentation Functions}
\label{Sec:models-fragmentation}

Generally, it is difficult to compute (collinear and transverse-momentum-dependent) FFs for light hadrons in nonperturbative approaches. 
So far it has not been possible to address FFs in lattice QCD because of the state $| P_h, S_h; X \rangle$ that shows up in their definition.  
Also, obtaining realistic estimates of FFs in models is more challenging than for PDFs.  
Nevertheless, a number of model calculations for TMDFFs exist where mostly two classes of models have been explored --- spectator models in which a parton fragments into a hadron and a spectator in a single step, and models for multiple-hadron emission.  
In some recent works the two approaches have been combined.  
A brief account of those models is given here, with some emphasis on the frequently-used spectator models.  
At the end of this section we also discuss the universality of (T-odd) TMDFFs in a spectator model.
\begin{figure}[t]
\begin{center}
\includegraphics[width=7.0cm]{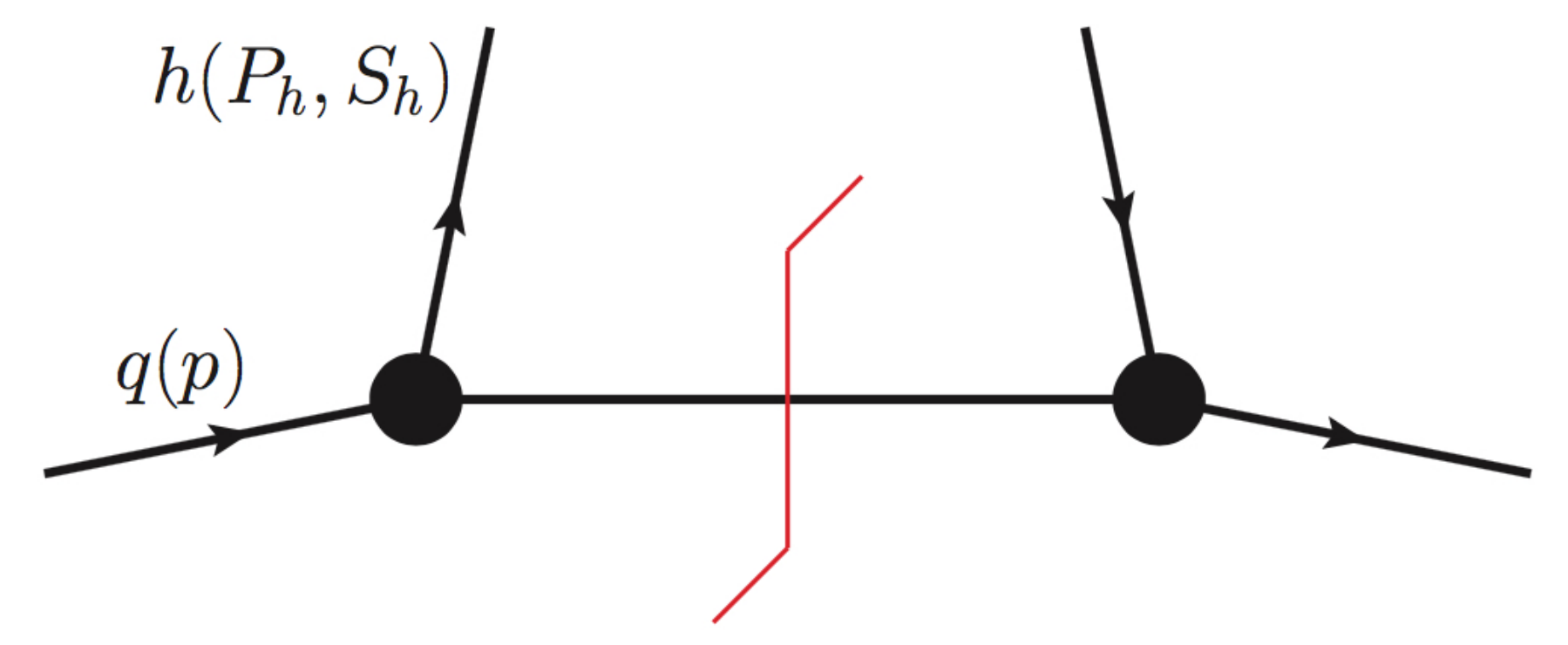} 
\end{center}
\vspace{-0.4cm}
\caption{Representation of quark fragmentation into a hadron (which is characterized by the four-momentum $P_h$ and, potentially, the covariant spin vector $S_h$) in a spectator model to leading order in the quark-hadron-spectator coupling.
The inclusive system $X$ is given by a single spectator particle. 
The red line in the diagram is the on-shell cut. 
} 
\label{f:spectator}
\end{figure}

\subsubsection{Spectator models for TMDFFs} 

\index{model!spectator model}

An early application of a spectator model for quark fragmentation has been discussed in Ref.~\cite{Jakob:1997wg}. 
The main idea of spectator models for FFs is illustrated in Fig.$\,$\ref{f:spectator}: 
a (time-like off-shell) quark fragments into a hadron and a single (on-shell) spectator particle. 
One finds that the squared four-momentum $p^2$ of the fragmenting quark is given by the transverse quark momentum $p_T$ (or, equivalently, the transverse hadron momentum) and $z$ according to
\begin{equation}
p^2 = p_T^{\,2}\frac{z}{1-z} + \frac{m_s^2}{1-z} + \frac{M_h^2}{z} \,,
\label{e:parton_virtuality}
\end{equation}
with $m_s$ denoting the spectator mass.
The fragmentation process determines the spectator type.
For instance, the spectator is a down quark in the case of fragmentation of an up quark into a $\pi^+$.
Or the spectator is a $\bar{u}\bar{d}$ anti-diquark if an up quark fragments into a proton.
Let us, as an example, consider the fragmentation of light quarks into pions.
For a pseudoscalar interaction between quarks and pions, one finds for the favored $u \to \pi^+$ fragmentation~\cite{Collins:1992kk,Bacchetta:2001di}
\begin{equation} 
D_1^{\pi^+/u}(z,z p_T) = \frac{1}{z} \, \frac{g_{\pi q}^2}{8 \pi^3} \, \frac{p_T^2 + m_q^2}{\Big(p_T^2 + m_q^2 + \frac{1-z}{z^2}m_{\pi}^2 \Big)^2} \,,
\label{e:ps_D1_kT}
\end{equation}
where $g_{\pi q}$ is the quark-pion coupling constant and $m_q$ the quark mass.
In this model, the fragmenting quark and the spectator have the same mass.
Nonzero disfavored FFs can be obtained by considering higher-order diagrams.
(For completeness, we remark that the unpolarized TMDFF for a quark into a sigma meson was calculated in the linear sigma model in Ref.~\cite{Collins:1994ax}.)

Integrating the result in Eq.~(\ref{e:ps_D1_kT}) upon $k_T$ provides $D_1^{\pi^+/u}(z)$, a function that is rather well known from fits to data.
This integral leads to a logarithmic divergence which, according to Eq.~\eqref{e:parton_virtuality}, may be avoided by putting a cut-off on the virtuality of the fragmenting quark.
The numerical result shows the right qualitative feature in that it decreases with increasing $z$, but there is no quantitative agreement with FF parametrizations based on data~\cite{Bacchetta:2002tk}.
In an attempt to improve the phenomenology also the pseudo-vector quark-pion coupling, as used in the chiral-invariant Georgi-Manohar model~\cite{Manohar:1983md}, was explored in Ref.~\cite{Bacchetta:2002tk}.
While a good description of the $z$ behavior of the FF is obtained, the magnitude is just about half of typical fit results~\cite{Bacchetta:2002tk}. 
Also the Nambu-Jona-Lasinio model~\cite{Ito:2009zc, Matevosyan:2010hh, Matevosyan:2011ey} and a lonlocal chiral quark model~\cite{Nam:2011hg, Nam:2012af, Yang:2013cza} were used to compute FFs in a spectator approach. \index{model!Nambu--Jona-Lasinio model}
Overall one again finds just qualitative agreement with existing fits, which implies that also results for the TMDFFs in such models are qualitative only, even though the results for the transverse momentum dependence may be reasonable. 
A better phenomenology of spectator models for FFs can be obtained by introducing more free parameters, where often two types of modifications are considered; see, for instance, Refs.~\cite{Jakob:1997wg, Bacchetta:2007wc}.
First, a form factor is used for the quark-hadron-spectator vertex.
Second, the spectator mass is allowed to vary.
Of course such approaches are no longer related to an underlying Lagrange density.
In Ref.~\cite{Bacchetta:2007wc} good results were obtained for integrated pion and kaon FFs in a spectator model with five parameters.
Once the parameters are fixed, other (TMD) FFs can be computed.
We note in passing that the momentum sum rule for the unpolarized FF has been discussed in Refs.~\cite{Ito:2009zc,Meissner:2010cc,Collins:2011zzd} in spectator models.

\begin{figure}[t]
\begin{center}
\includegraphics[width=12.0cm]{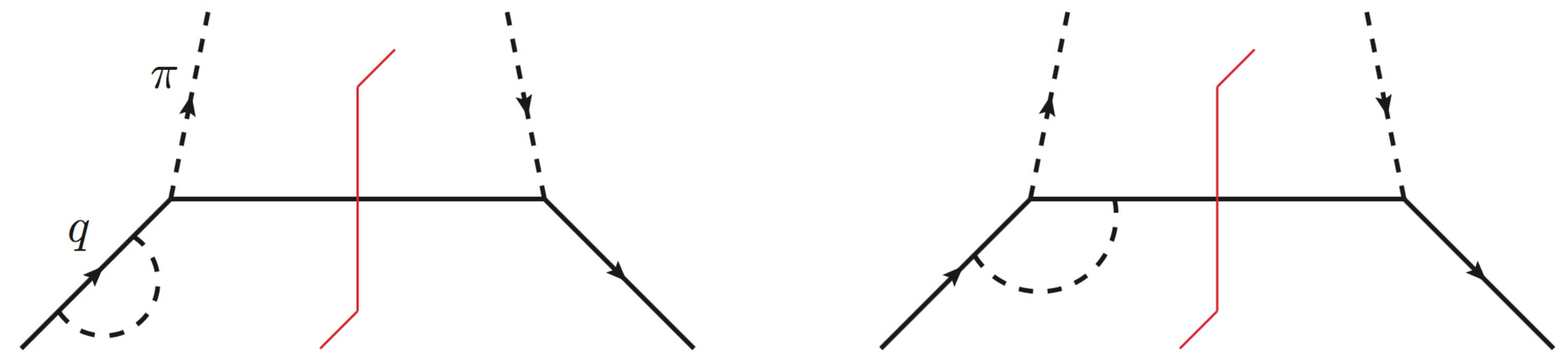} 
\end{center}
\vspace{-0.4cm}
\caption{One-loop corrections to the fragmentation of a quark into a pion for a pseudoscalar quark-pion coupling.
Shown are all the graphs contributing to the Collins function: quark self-energy (left) and vertex correction (right).
(Hermitean conjugate diagrams are not shown.)}
\label{f:ps_Collins}
\end{figure}

Spectator models were also used to compute the Collins function, with a first calculation already presented in the original paper by Collins~\cite{Collins:1992kk}.
In that study, a nonzero $H_1^\perp$ was found by taking into consideration the imaginary part of the propagator of the fragmenting quark.
Despite this result, it was speculated whether the Collins function might actually vanish due to a cancellation between different contributions to the fragmentation process, see, for instance, Ref.~\cite{Jaffe:1997hf}.
Therefore, in Ref.~\cite{Bacchetta:2001di} the aforementioned pseudoscalar quark-pion coupling was used for a complete field-theoretic model calculation.
The lowest-order diagrams in Fig.~\ref{f:ps_Collins} provide~\cite{Bacchetta:2001di}
\begin{equation} \label{e:ps_Collins}
H_1^{\perp \, \pi^+/q}(z, z p_T) = - \, \frac{g_{\pi q}^2}{4\pi^3} \frac{m_{\pi}}{1-z} 
\bigg( \frac{m_q \, {\rm Im} \, \tilde{\Sigma}(p^2)}{(p^2 - m_q^2)^2} 
+ \frac{{\rm Im} \, \tilde{\Gamma}(p^2)}{p^2 - m_q^2}\bigg) \bigg|_{p^2 = p_T^2 \frac{z}{1-z} + \frac{m_q^2}{1-z} + \frac{m_{\pi}^2}{z}} \,,
\end{equation}
with ${\rm Im} \, \tilde{\Sigma}$ and ${\rm Im} \, \tilde{\Gamma}$ indicating the imaginary part of the quark self-energy and the vertex correction, respectively.
The final result for the Collins function in Eq.~(\ref{e:ps_Collins}) is nonzero, which gave support to its existence from the theoretical point of view~\cite{Bacchetta:2001di}.
In the meantime, there exists compelling experimental evidence for a non-vanishing Collins function, see Sec.~\ref{subsubsec:Collins_pheno}.
In Ref.~\cite{Bacchetta:2002tk} the Georgi-Manohar model~\cite{Manohar:1983md} with pion loops was used to calculate the Collins function, while
later also gluon loops were considered in different spectator models~\cite{Bacchetta:2003xn, Amrath:2005gv, Bacchetta:2007wc, Matevosyan:2012ga}.
In such approaches, even predicting the sign of the Collins function is difficult because the individual diagrams can contribute with different signs~\cite{Amrath:2005gv}.
In the latest phenomenological papers just gluon loops have been used~\cite{Bacchetta:2007wc, Matevosyan:2012ga}.
In Ref.~\cite{Bacchetta:2007wc}, for instance, the Collins function for pions and kaons was computed, with the model parameters fixed by means of
the integrated unpolarized FFs.
In the case of pions, reasonable agreement with information from experimental data was obtained.
A spectator model was even used to compute the Collins function for $\Lambda$ hyperons~\cite {Wang:2018wqo}.

\subsubsection{Models for multi-step fragmentation process}

\index{multi-step fragmentation}

\begin{figure}[t]
\begin{center}
\includegraphics[width=9.0cm]{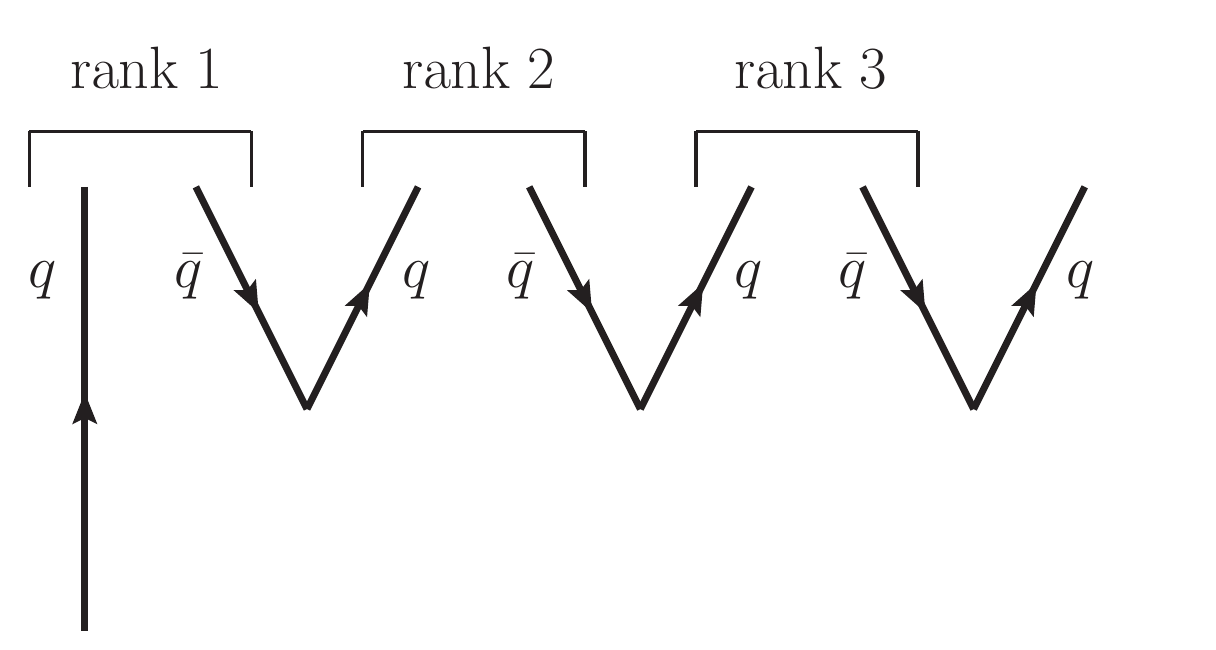} 
\end{center}
\vspace{-0.4cm}
\caption{Hierarchy of mesons that emerges when an initial quark combines with an antiquark from a produced $q\bar{q}$ pair to form the meson of rank 1.  The leftover quark then combines with an antiquark from another $q\bar{q}$ pair to form the meson of rank 2 and so on.}
\label{f:Feynman_Field}
\end{figure}

The second general class of models for FFs considers hadron production as a multi-step process, with the Feynman-Field model being an important representative~\cite{Field:1977fa}.
The underlying principle of that approach is shown in Fig.$\,$\ref{f:Feynman_Field}, namely, a high-energetic quark combines with an antiquark of a $q\bar{q}$ pair from the vacuum, where the combination process repeats until the remaining energy falls below some cut-off.
In the Feynman-Field model, multiple-hadron emission originating from a single parton is given by just one function, $f(\eta)$, characterizing a single emission, where $f(\eta)$ describes the probability that the first hierarchy (rank 1) meson leaves fractional momentum $\eta$ to the remaining cascade.
This model was quite successful in describing data from early $e^+ e^-$ annihilation experiments with a very limited set of parameters~\cite{Hoyer:1979ta,Ali:1979em}.
Nowadays the Feynman-Field model can still provide guidance when trying to parametrize FFs at an initial scale.

A similar approach is the string fragmentation model~\cite{Artru:1974hr,Bowler:1981sb,Andersson:1983jt,Andersson:1983ia}, where hadrons are also produced in a hierarchy as indicated in Fig.$\,$\ref{f:Feynman_Field}, but the treatment of the kinematics differs for the two models.
Moreover, in the string model, applied to $e^+ e^-$ annihilation for instance, one considers hadronization of the $q\bar{q}$ pair as opposed to independent fragmentation of the quark and antiquark employed in the Feynman-Field model.
The quark and antiquark lose energy to the color field between them, which is treated as a string-like configuration.
Once that energy exceeds a certain threshold, the string breaks up into hadrons.

A string fragmentation model was also used in an attempt to capture the main features of the Collins function~\cite{Artru:1995bh}.
Here we briefly repeat the main idea of that work.
Let's consider fragmentation of a transversely polarized quark into a spin-0 particle like a pion.
In the string model, it is assumed that a $q\bar{q}$ pair originating from string breaking has the quantum numbers of the vacuum, that is, $J^{P} = 0^+$~\cite{Artru:1995bh}. 
This situation is possible if the pair has spin $S = 1$ and orbital angular momentum $L = 1$, where the spin and orbital angular momentum point in opposite directions. 
Therefore, a correlation exists between the spin of the antiquark of the $q \bar{q}$ pair and its orbital angular momentum.
Because the antiquark and the original quark form a spin-0 meson, there is also a correlation between the orbital angular momentum of the antiquark, which gets transferred to the meson, and the transverse polarization of the fragmenting quark.
This leads to a nonzero Collins effect, namely, that the meson has a preferred direction relative to the plane which is given by the momentum and the spin of the fragmenting quark.
Originally, the model was confronted with data for the transverse SSA in processes like $p^{\uparrow} p \to h X$ where it provides the correct sign for the asymmetry~\cite{Artru:1995bh}.
The model also agrees in sign with the Collins function extracted from data in SIDIS and $e^+e^-$ annihilation.
Disfavored fragmentation requires rank 2 (and higher-rank) mesons which, in particular, leads to opposite signs for the favored and disfavored Collins functions~\cite{Artru:1995bh}, in accordance with phenomenology.
Further developments of the string fragmentation model for the Collins effect have been discussed in Refs.~\cite{Czyzewski:1996ih, Artru:2010st, Kerbizi:2018qpp, Kerbizi:2019ubp}.

\subsubsection{Combining models for FFs}

Another line of research in this area combines spectator models with the main idea underlying the Feynman-Field model.
Specifically, single-hadron emission is computed in a spectator model defined through a Lagrange density, and then iterated according to the Feynman-Field approach.
Such calculations for $D_1 (z)$ were carried out in the Nambu-–Jona-Lasinio model \cite{Ito:2009zc,Matevosyan:2010hh,Matevosyan:2011ey} and in a non-local chiral-quark model \cite{Nam:2011hg,Nam:2012af,Yang:2013cza}, while results for TMDFFs were reported in \cite{Matevosyan:2012ga,Matevosyan:2011vj,Bentz:2016rav} along with discussion about how model-independent constraints such as the Sch\"afer-Teryaev sum rule \cite{Schafer:1999kn,Meissner:2010cc} can be satisfied in such a model. 
A major goal of those works is to obtain quantitative results for FFs with as few parameters as possible. 
Presently, it is not fully clear if this goal can be met or if more flexible parametrizations for the single-hadron production are needed.

\subsubsection{Universality of TMD fragmentation functions}
\label{Sec:model-universality-FFs}

\index{universality of TMDs}

\begin{figure}[t]
\begin{center}
\includegraphics[width=6.0cm]{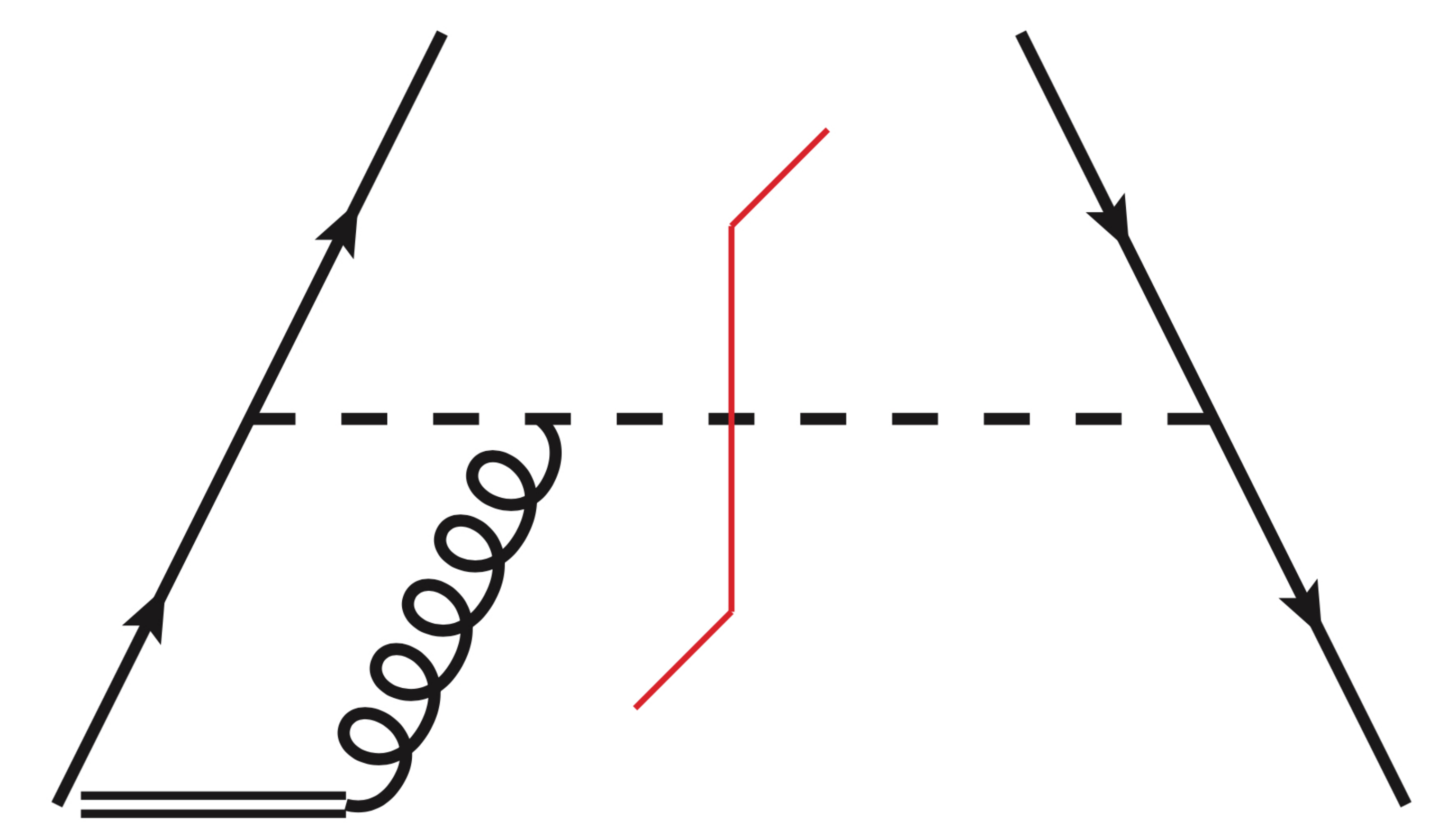} 
\end{center}
\vspace{-0.4cm}
\caption{Specific contribution to fragmentation correlator for the fragmentation of a quark into a hadron and a spectator particle.}
 \label{f:FF_universality}
\end{figure}

Here we would like to add some discussion about the universality of TMDFFs.
We remind the reader that TMD PDFs defined with future-pointing and past-pointing Wilson lines can be related through the parity and time-reversal operations.
This led to the crucial finding that T-even TMD PDFs do not depend on the direction of the Wilson line, while T-odd TMD PDFs have reversed signs in the two cases~\cite{Collins:2002kn}.
The question about universality also arises for TMDFFs as for those objects in the SIDIS process, {\it a priori}, one obtains past-pointing Wilson lines when eikonalizing the relevant propagators, while future-pointing Wilson lines emerge in $e^+e^-$ annihilation.
Since those two definitions can not be related via parity and time reversal, it remained at first unclear whether TMDFFs are universal or not.

To explore this situation, in Ref.~\cite{Metz:2002iz} a transverse SSA for fragmentation in both SIDIS and $e^+e^-$ annihilation was computed in a spectator model, including one-loop gluon exchange which is associated with the Wilson line of TMDFFs.
The factorized description of the SSA is proportional to the TMDFF $D_{1T}^\perp$.
The calculation of Ref.~\cite{Metz:2002iz} can be considered the counterpart of the Brodsky-Hwang-Schmidt calculation of the Sivers SSA for the fragmentation process~\cite{Brodsky:2002cx, Brodsky:2002rv, Brodsky:2013oya}.
It was found that the SSAs in the two processes are identical, suggesting universality of $D_{1T}^\perp$.

A more general investigation of the universality of TMDFFs was carried out in Ref.~\cite{Collins:2004nx}.
There it was argued that, when deriving TMD factorization, in the case of the fragmentation process one is actually not sensitive to the direction of the Wilson line.
This led to the conclusion that TMDFFs in both aforementioned processes can be defined through, for instance, future-pointing Wilson lines~\cite{Collins:2004nx} and, hence, they are universal.

To make the argument explicit in a simple example, we consider again a spectator model calculation and evaluate the diagram Fig.~\ref{f:FF_universality}~\cite{Collins_Metz_unpublished}.
Specifically, we just focus on the calculation for the difference between a future-pointing and a past-pointing Wilson line, where the essential part of the computation is given by 
\begin{eqnarray}
I_{\rm future} - I_{\rm past} & \sim & \int d^4l \, \frac{N(l^+,\lt)}{[(p-l)^2 - m_q^2 + i0][(p - P_h - l)^2 - m_s^2 + i0][l^2 + i0]} 
\nonumber \\
&& \times \; \bigg( \frac{1}{l^+ + i0} - \frac{1}{l^+ - i0}\bigg) 
\nonumber \\
& = & 0 \,.
\end{eqnarray}
To arrive at this result, we first carried out the $l^+$ integral by using $1/(l^+ + i0) - 1/(l^+ -i0) = -2 \pi i \delta(l^+)$.
Then the $l^-$ integral can be evaluated via contour integration.
However, since all the $l^-$ poles are on the same side of the real axis, the integral vanishes.
The calculation shows that, for this example, the direction of the Wilson line is irrelevant.
Note that the numerator of the integrand, $N(l^+,\lt)$, depends on the TMDFF under consideration.
However, for all leading-power TMDFFs this numerator does not depend on $l^-$, which is all what matters for the argument to hold.
Generally, the fact that TMDFFs are not sensitive to the direction of the Wilson line can be traced back to the specific kinematics of the fragmentation process.

The universality of TMDFFs has also been studied by considering transverse-momentum moments of the TMD correlator.
It was found that a potential non-universality of moments of TMDFFs can be related to certain higher-twist collinear (soft-gluon-pole) matrix elements.
But, through model-dependent and model-independent analyses, it was concluded that such matrix elements vanish~\cite{Gamberg:2008yt, Meissner:2008yf, Gamberg:2010uw, Kanazawa:2015ajw}; see also Ref.~\cite{Belitsky:1997ay}.
While those findings further corroborated the universality of TMDFFs, it must be pointed out that the moments were not taken for properly renormalized TMDFFs.
As already explained in Sec.~\ref{sec:integratedTMDs}, renormalization and taking transverse-momentum moments generally do not commute.
The potential implications of this feature on the results discussed in this paragraph remain to be studied.

\subsection{Formal Constraints on TMD Functions}
\label{Sec:formal-constraints}

In this section we discuss several general results which
hold formally for bare TMD functions. While they are  
generally not considered to be model-dependent, 
presently it is not known how to prove these relations
in terms of renormalized TMD functions. These general
constraints are, however, expected to be valid in models
and can be used to test model results. In addition,
some of these general constraints have been used 
in phenomenology.

\subsubsection{Positivity constraints}
\label{Sec:positivity-constraints}

\index{positivity of TMDs}

Cross sections of physical processes must be positive ---
no matter which spin configurations or azimuthal asymmetries
one considers. In order to guarantee this, one can introduce
a spin density matrix for the nucleon which is formally related 
to TMD PDFs and which obeys certain conditions. 
For a spin-$\frac12$ particle like the nucleon, this yields 
the following inequalities for the leading quark and antiquark
TMD PDFs \cite{Bacchetta:1999kz}
\begin{subequations}\ba
    f_1^a(x,k_T)  \ge 0 \;,\;\;\;
    |g_1^a(x,k_T)|\le f_1^a(x,k_T)  \;,\;\;\; 
    |h_1^a(x,k_T)|\le f_1^a(x,k_T)  \;,\;\;\;\label{Ineq:f1-g1-h1}&&
    \phantom{\biggl|}\\
    |h_1^a(x,k_T)|\le\frac12\biggl(f_1^a(x,k_T)+g_1^a(x,k_T)\biggr)\;,
    \;\;\; \label{Ineq:Soffer}&&\\
    |h_{1T}^{\perp(1) a}(x,k_T)|
    \le\frac12\biggl(f_1^a(x,k_T)-g_1^a(x,k_T)\biggr)\;,
    \;\;\; \label{Ineq:pretzel}&&\\
    g_{1T}^{\perp(1)a}(x,k_T)^2+f_{1T}^{\perp(1)a}(x,k_T)^2 \le
    \frac{k_T^2}{4M_N^2}\,\biggl(f_1^a(x,k_T)^2-g_1^a(x,k_T)^2\biggr)
    \;,\;\; \label{Ineq:g1Tperp}&&\\
    h_{1L}^{\perp(1)a}(x,k_T)^2+h_1^{\perp(1)a}(x,k_T)^2 \le
    \frac{k_T^2}{4M_N^2}\,\biggl(f_1^a(x,k_T)^2-g_1^a(x,k_T)^2\biggr)
    \;.\;\; \label{Ineq:h1Lperp}&&
\ea\end{subequations}
The meaning of the inequalities in (\ref{Ineq:f1-g1-h1}) is
obvious from the partonic interpretation, e.g., 
$f_1^a(x,k_T)\ge 0$ expresses the expectation that 
a probability (to find a parton carrying fraction $x$ of 
nucleon's $P^+$-momentum component and $k_T$) can not be
negative.
Similarly, the difference of probability densities of 
partons with positive helicities and negative helicities 
can not exceed their sum which implies 
$|g_1^a(x,k_T)|\le f_1^a(x,k_T)$, and analogously
for transversely polarized quarks.
The analog of Eq.~(\ref{Ineq:Soffer}) for PDFs is known as  the Soffer
inequality \cite{Soffer:1994ww} 
\index{Soffer bound}
and was generalized to TMD PDFs
in \cite{Bacchetta:1999kz} where also the inequalities 
(\ref{Ineq:pretzel}-\ref{Ineq:h1Lperp}) were derived.
For fragmentation functions analogous inequalities can be 
derived \cite{Bacchetta:1999kz}, as well as those for 
gluon TMD PDFs \cite{Mulders:2000sh}.

Despite being intuitive, the status of such inequalities 
remains unclear for renormalized TMD functions. 
It is natural to expect that some positivity constraints 
should hold also for renormalized TMD functions in order to 
guarantee the positivity of cross sections. But it is not
straightforward to prove this rigorously, see for instance
the recent attempt to prove the positivity of $f_1^a(x,k_T)$ 
in $\overline{\rm MS}$ scheme in \cite{Candido:2020yat}, 
and the critical review of this attempt in \cite{Collins:2021vke}.

In most model studies one typically does not face many of the
subtleties occurring in QCD in the definition of TMD PDFs,
and the inequalities (\ref{Ineq:f1-g1-h1}-\ref{Ineq:h1Lperp})
are routinely used to double check the internal consistency of 
the model calculations. The positivity inequalities have been 
also implemented in many phenomenological studies, see 
Chapter~\ref{sec:phenoTMDs}. For a discussion of the importance 
of the Soffer bound for the extraction of transversity, we
refer to Ref.~\cite{DAlesio:2020vtw}.

\subsubsection{Burkardt sum rule}
\label{Sec:Burkardt-sum-rule}

\index{Burkardt sum rule}

A non-trivial constraint in modelling (or fitting) 
of the Sivers function is given by the Burkardt sum
rule~\cite{Burkardt:2004ur}. This sum rule states that
the average transverse momentum induced in the
Sivers effect vanishes after summing over all partons
(cf.~Eq.~(\ref{eq:define-(1)-mom-in-kT-space}) for the notation for $k_T$ moments of TMDs),
\be
    \sum_a
    \int dx \int d^2k_T\,f_{1T}^{\perp(1)a}(x,k_T)
    = 0\,.
\ee
Due to its relation to the conservation of
(transverse) momentum, the sum rule can also
be proven formally through studies of the
energy-momentum tensor \cite{Lorce:2015lna}.

Also in the case of this sum rule, it is not obvious
how to formulate a rigorous proof in terms of renormalized
TMDs. Despite its formal character in QCD, the sum rule
was shown to be valid, e.g.,\ in one-loop calculations
in the quark-target model and a scalar diquark 
model of the nucleon \cite{Goeke:2006ef} as well as in 
light-cone constituent quark models~\cite{Pasquini:2010af}.
In the bag model the sum rule was found violated by a few
percent~\cite{Yuan:2003wk,Courtoy:2008dn,Courtoy:2009pc}
which has been attributed to the fact that the bag states 
are not good momentum eigenstates.
Also in non-relativistic calculations in constituent 
quark models a small violation of the sum rule was observed
due to similar reasons \cite{Courtoy:2008dn,Courtoy:2009pc}.

In the leading order of the large-$N_c$ limit, the 
Burkardt sum rule is saturated by the Sivers functions
of $u$- and $d$-flavors having the same magnitude but
opposite signs with the contributions from gluons and
other quark flavors appearing only at subleading order of 
the large $N_c$ expansion, see Sec.~\ref{subsubsec:largeNc}.

For completeness let us mention here also the formal connection of
the transverse moment of the Sivers function to the
collinear twist-3 Qiu-Sterman function \cite{Qiu:1991pp},
see Sec.~\ref{sec:largeqT}, 
which could allow one to study (formally) the scale 
dependence of the Burkardt sum rule 
\cite{Zhou:2015lxa}.

\subsubsection{Sch\"afer-Teryaev sum rule}
\label{Sec:Schafer-Teryaev-sum-rule}

\index{Sch\"afer-Teryaev sum rule}

The Sch\"afer-Teryaev sum rule is based on the conservation 
of the transverse momentum acquired by the hadrons during the
fragmentation process of a transversely polarized quark,
\be
    \sum_h (2S_h+1)
    \int dz \,z \int d^2K_T\,H_{1}^{\perp(1)q/h}(z,K_T)
    = 0\,.
\ee
The sum rule was proven in \cite{Schafer:1999kn} on the
basis of intuitive momentum conservation arguments. A 
more rigorous formal proof was given in \cite{Meissner:2010cc},
see also the review article \cite{Metz:2016swz} and 
Ref.~\cite{Accardi:2019luo}. 

It is difficult to test this sum rule in model calculations.
Strictly speaking, it requires the consideration of "all"
possible hadrons a quark can fragment into. Nevertheless,
the Sch\"afer-Teryaev sum rule was shown to hold in a
Manohar-Georgi type-model study \cite{Meissner:2010cc}, and 
the quark-jet model of Ref.~\cite{Bentz:2016rav}.

\subsection{Relations in Models}
\label{Sec:relations-in-models}

After reviewing model-independent formal constraints, 
here we discuss relations among TMD PDFs in models. 
While model-dependent, these relations are nevertheless 
of interest because they are observed in a wide class of 
models based on much different dynamics.

\subsubsection{Independence of TMD PDFs in QCD}
\label{Sec:no-TMD-relations-QCD}

In QCD no relations exist among different TMD functions
which are independent functions, each of which describing
different characteristics of the nucleon structure. 
This can be established by considering the fully unintegrated
quark correlator $\Phi^q$ for which we will use the definition
of Refs.~\cite{Tangerman:1994eh,Boer:1997nt} (see the text 
below Eq.~(\ref{eq:CPM2}) for an explanation of the notation). 

For a Lorentz-decomposition of the quark correlator, one 
can make use of four linearly independent 4-vectors:
quark momentum $k^\mu$, nucleon momentum $P^\mu$, nucleon
polarization vector $S^\mu$, and the gauge-link vector
$\vmod^\mu$, i.e.\ $\Phi^q=\Phi^q(k,P,S,v)$  
where we do not indicate that the renormalized correlator 
would depend also on renormalization and other scales.
These four linearly independent vectors allow one to carry out a
Lorentz-decomposition of the quark correlator in terms of 
32 Lorentz-scalar valued amplitudes: 12 $A^q_i$-amplitudes 
and 20 $B^q_i$-amplitudes. This naming scheme has historical
reasons, and will be clarified shortly.
The amplitudes depend on the scalars $P\cdot k$ and $k^2$
which will not be indicated for brevity in the following.
The fully unintegrated quark-quark correlator can be
decomposed according to
\cite{Goeke:2005hb,Goeke:2003az,Meissner:2009ww} 
\begin{eqnarray}
\label{eq:models-correlator-decompose}
&& \hspace{-8mm}
\Phi^q(k,P,S,\vmod) \; = \; 
  MA^q_1 
+ \slashed{P} A^q_2 
+ \slashed{k}A^q_3 
+ \frac{i}{2M} \; [\slashed{P},\slashed{k}] \; A^q_4 
+ i  (k \cdot S) \gamma_5 \; A^q_5 
+ M\slashed{S} \gamma_5 \; A^q_6 
+ \frac{k \cdot S}{M} \slashed{P} \gamma_5 \; A^q_7
\nonumber\\
&& \quad
+ \frac{k \cdot S}{M} \slashed{k} \gamma_5 \; A^q_8 
+ \frac{[\slashed{P},\slashed{S}]}{2} \gamma_5 \; A^q_9 
+ \frac{[\slashed{k},\slashed{S}]}{2} \gamma_5 \; A^q_{10} + \frac{k \cdot S}{2M^2} [\slashed{P},\slashed{k}] \gamma_5 \; A^q_{11} 
+ \frac{1}{M} \varepsilon^{\mu\nu\rho\sigma} 
  \gamma_{\mu} P_{\nu} k_{\rho} S_{\sigma} \; A^q_{12} 
  \nonumber \\
&& \quad
+ \frac{M^2}{P\cdot\vmod} \slashed{\vmod} \; B^q_1 
+ \frac{ i  M}{2 P \cdot   \vmod} [\slashed{P},\slashed{\vmod}] \; B^q_2 
+ \frac{ i  M}{2 P \cdot   \vmod} [\slashed{k},\slashed{\vmod}] \; B^q_3 
+ \frac{1}{P \cdot   \vmod} \varepsilon^{\mu\nu\rho\sigma} \gamma_{\mu} \gamma_5 P_{\nu} p_{\rho} \vmod_{\sigma} \; B^q_4 \nonumber \\
&& \quad
+ \frac{1}{P \cdot   \vmod} \varepsilon^{\mu\nu\rho\sigma} P_{\mu} k_{\nu} \vmod_{\rho} S_{\sigma} \; B^q_5 
+ \frac{i M^2}{P \cdot   \vmod} (  \vmod \cdot S) \gamma_5 \; B^q_6 
+ \frac{M}{P \cdot   \vmod} \varepsilon^{\mu\nu\rho\sigma} \gamma_{\mu} P_{\nu} \vmod_{\rho} S_{\sigma} \; B^q_7 
\nonumber \\
&& \quad
+ \frac{M}{P \cdot   \vmod} \varepsilon^{\mu\nu\rho\sigma} \gamma_{\mu} k_{\nu} \vmod_{\rho} S_{\sigma} \; B^q_8 
+ \frac{p \cdot S}{M(P \cdot   \vmod)} \varepsilon^{\mu\nu\rho\sigma} \gamma_{\mu} P_{\nu} k_{\rho} \vmod_{\sigma} \; B^q_9 
+ \frac{M(  \vmod \cdot S)}{(P \cdot   \vmod)^2} \varepsilon^{\mu\nu\rho\sigma} \gamma_{\mu} P_{\nu} k_{\rho} \vmod_{\sigma} \; B^q_{10} \nonumber \\
&& \quad+ \frac{M}{P \cdot   \vmod} (  \vmod \cdot S) \slashed{P} \gamma_5 \; B^q_{11} + \frac{M}{P \cdot   \vmod} (  \vmod \cdot S) \slashed{k} \gamma_5 \; B^q_{12} 
+ \frac{M}{P \cdot   \vmod} (p \cdot S) \slashed{\vmod} \gamma_5 \; B^q_{13} 
+ \frac{M^3}{(P \cdot   \vmod)^2} (  \vmod \cdot S) \slashed{\vmod} \gamma_5 \; B^q_{14} \nonumber \\
&& \quad
+ \frac{M^2}{2P \cdot   \vmod} [\slashed{\vmod},\slashed{S}] \gamma_5 \; B^q_{15} 
+ \frac{p \cdot S}{2P \cdot   \vmod} [\slashed{P},\slashed{\vmod}] \gamma_5 \; B^q_{16} 
+ \frac{p \cdot S}{2P \cdot   \vmod} [\slashed{k},\slashed{\vmod}] \gamma_5 \; B^q_{17} 
+ \frac{  \vmod \cdot S}{2P \cdot   \vmod} [\slashed{P},\slashed{k}] \gamma_5 B^q_{18} 
\nonumber \\
&& \quad
+ \frac{M^2(  \vmod \cdot S)}{2(P \cdot   \vmod)^2} [\slashed{P},\slashed{\vmod}] \gamma_5 B^q_{19} 
+ \frac{M^2(  \vmod \cdot S)}{2(P \cdot   \vmod)^2} [\slashed{k},\slashed{\vmod}] \gamma_5 B^q_{20} \,, 
\end{eqnarray}
where $\varepsilon^{0123} = 1$ is used. 
The naming scheme for the amplitudes is such that the 
$A^q_i$-amplitudes in (\ref{eq:models-correlator-decompose})
are accompanied by Dirac-structures contracted with the
4-vectors $k^\mu$, $P^\mu$, $S^\mu$, while the $B^q_i$-amplitudes
are associated in addition to that with the gauge-link 
vector $\vmod^\mu$.

The 32 amplitudes are independent structures, as 
there is no model-independent way to relate 
them to each other. At the same time,
there are 32 quark TMD PDFs: 8 at leading, 
16 at subleading, and 8 at subsubleading order.
The subleading functions will be discussed in 
Sec.~\ref{sec:twist3}. The subsubleading functions 
(associated with the Dirac structures $\gamma^-$,
$\gamma^-\gamma_5$, $i\sigma^{\alpha -}\gamma_5$)
are of rather academical character \cite{Goeke:2005hb}.
The crucial point is: there are as many 
independent amplitudes as there are overall (leading, 
subleading, subsubleading) TMD PDFs which implies
that no relations among TMD functions exist in
QCD.

\subsubsection{Quark-model Lorentz-invariance relations}
\label{Sec:models-qLIRs}

\index{Quark-model Lorentz-invariance relations (qLIRs)}

In the early works, the role of the gauge-link
vector $\vmod^\mu$ was not recognized, and the correlator
(\ref{eq:models-correlator-decompose}) was decomposed with the
$B^q_i$ amplitudes missing
\cite{Tangerman:1994eh,Tangerman:1994bb,Mulders:1995dh,Boer:1997nt}.
As mentioned in Sec.~\ref{Sec:models-why}, calculations in a
quark-target model \cite{Kundu:2001pk} helped to realize and
fix this oversight in \cite{Goeke:2003az,Meissner:2009ww}.
(As reviewed in Sec.~\ref{Subsec-models:review-Brodsky-Hwang-Schmidt}, 
the importance of the gauge link for the understanding of T-odd TMD
PDFs was also recognized thanks to a model calculation
\cite{Brodsky:2002cx}.)

What is an oversight in QCD, however, becomes strictly correct 
in quark models with no explicit gauge field degrees of freedom.
In these models, no gauge link is present in the model expressions
and the Lorentz decomposition rightly contains no $B^q_i$ amplitudes.
In quark models without explicit gluons, also T-odd structures are
absent, i.e.,\ in addition to $B^q_i$ amplitudes also the T-odd
$A_i^q$ amplitudes (namely $A^q_4$, $A_5^q$, $A^q_{12}$) are 
absent. One therefore ends up with more TMD functions than
amplitudes, and this implies the so-called "quark model 
Lorentz Invariance Relations" (qLIRs). The qLIRs in general
connect leading and subleading TMD functions. 
The subleading TMD functions will be discussed in detail 
in Chapter~\ref{sec:twist3}, but it is convenient to
include these model relations here for completeness.

More precisely, one has overall 6 leading and 8 subleading 
TMD PDFs, i.e.,\ 14 functions. At the same time, one has
9 T-even $A^q_i$ amplitudes in quark models without 
explicit gauge field degrees of freedom. 
This implies 5 qLIRs which are given by
\sub{\label{eq:qLIRs}
\ba \label{eq:LIR1}
g_T^q(x) \; &\stackrel{\text{qLIR}}{=}& \; 
  g_1^q(x) + \frac{d}{d x} g^{\perp(1)q}_{1T}(x) \, , \\
\label{eq:LIR2}
h_L^q(x) \; &\stackrel{\text{qLIR}}{=}& \; 
  h_1(x) - \frac{d}{d x} h^{\perp(1)q}_{1L}(x) \, , \\
\label{eq:LIR3}
h_T^q(x) \; &\stackrel{\text{qLIR}}{=}& \; 
  - \frac{d}{d x} h^{\perp(1)q}_{1T}(x) \, , \\
\label{eq:LIR4}
g_L^{\perp q}(x) + \frac{d}{d x} g^{\perp(1)q}_{T}(x)
\; &\stackrel{\text{qLIR}}{=}& 0 \, , \\
\label{eq:LIR5}
h_T^q(x,k_T) - 
h_T^{\perp q}(x,k_T)
\; &\stackrel{\text{qLIR}}{=}& 
h^{\perp q}_{1L}(x,k_T).
\ea}
The subleading functions 
$g_T^q$, $g_L^{\perp q}$, $h_L^q$, $h_T^q$, $h_T^{\perp q}$ will 
be defined and discussed in Sec.~\ref{sec:twist3}.

Being based on Lorentz invariance, the relations (\ref{eq:qLIRs}) 
must be obeyed in relativistic models without gluons which 
provides a powerful cross check for the numerics. Care may 
be in order in models where UV-divergences could spoil the
qLIRs. It may \cite{Avakian:2010br} or may not 
\cite{Aslan:2022zkz} be possible to find regularization
schemes in a given model which preserve qLIRs.
It will be interesting to see whether the qLIRs are
supported approximately in phenomenology, and learn 
about the size of the $B^q_i$-amplitudes.

\subsubsection{Relations among TMDs in quark models}

\index{transversity!models}In some quark models with wave functions obeying
SU(4) spin-flavor symmetry, the following relation can hold
between the unpolarized, helicity, and transversity TMD PDFs
\be
 \frac{f_1^q(x,k_T)}{N_q} + 
 \frac{g_1^q(x,k_T)}{P_q} = 2 \:
 \frac{h_1^q(x,k_T)}{P_q}\,, \label{Eq:rel-I} 
 \ee
where the SU(4) spin-flavor factors are given by 
$N_u=2$, $N_d=1$, $P_u=\frac43$, $P_d=-\frac13$
for $N_c=3$
(below Eq.~(\ref{Eq:non-rel-lim}) these spin-flavor
factors are given for general $N_c$). The model relation 
(\ref{Eq:rel-I}) holds in bag and light-cone constituent models
\cite{Pasquini:2008ax,Avakian:2010br}. Its $k_T$-integrated 
analog was discussed in the bag model even earlier in
\cite{Jaffe:1991ra,Signal:1996ct,Barone:2001sp}.

\index{worm-gear functions!models}
The following model relation connects the two Kotzinian-Mulders 
worm-gear functions,
\be
 g_{1T}^{\perp q}(x,k_T) = -\, h_{1L}^{\perp q}(x,k_T)\;.
 \label{Eq:rel-IV}
 \ee
First observed in the spectator model \cite{Jakob:1997wg}, 
it holds also in light-cone constituent quark model
\cite{Pasquini:2008ax}, covariant parton model \cite{Efremov:2009ze}, 
bag model \cite{Avakian:2010br}, light-cone quark-diquark model
\cite{Zhu:2011zza}, and the light-cone version of the chiral 
quark-soliton model \cite{Lorce:2011dv}. The partonic interpretation of 
(\ref{Eq:rel-IV}) is that the distributions of longitudinally polarized
quarks in a transversely polarized nucleon ($g_{1T}^{\perp a}$) are 
exactly opposite to the distributions of transversely polarized quarks 
in a longitudinally polarized nucleon ($h_{1L}^{\perp q}$).

\index{transversity!models}
Next, we discuss the interesting model relation which connects
helicity, transversity and pretzelosity, namely
\be
g_1^q(x,k_T) - h_1^q(x,k_T) = h_{1T}^{\perp(1)q}(x,k_T)\,.
\label{Eq:measure-of-relativity}
\ee
The relation (\ref{Eq:measure-of-relativity}) was first observed 
in the bag model \cite{Avakian:2008dz}. It is valid also in 
spectator model \cite{Jakob:1997wg}, different light-cone models 
\cite{Pasquini:2008ax,She:2009jq,Lorce:2011dv} and 
covariant parton model \cite{Efremov:2009ze}.
Recalling the non-relativistic model prediction
(\ref{Eq:g1-h1-non-rel-limit}), the relation 
(\ref{Eq:measure-of-relativity}) implies that in these 
quark models the transverse moment of pretzelosity can be
considered as a "measure" of relativistic effects.

Finally, let us mention the following non-linear relation 
which connects all three T-even chiral-odd TMD PDFs and is given by
\ba\label{Eq:non-lin-2}
      h_1^q(x,k_T)\,h_{1T}^{\perp q}(x,k_T) &=& -
      \frac{1}{2}\,\biggl[h_{1L}^{\perp q}(x,k_T)\biggr]^2.
\ea
This non-linear relation was derived in the covariant parton 
model \cite{Efremov:2009ze} and holds also in bag model 
\cite{Avakian:2008dz}, light-cone constituent quark model
\cite{Pasquini:2008ax} or the light-cone version of the chiral 
quark-soliton model \cite{Lorce:2011dv}.
Interestingly, it holds in the light-cone quark-diquark model
\cite{Zhu:2011zza} for $d$- but not for $u$-quarks.
More linear and non-linear model relations are known 
when subleading functions are included 
\cite{Avakian:2008dz,Bastami:2020rxn,Lorce:2014hxa,Lorce:2016ugb}.

The deeper reasons underlying the emergence of these model
relations in such a variety of conceptually very different 
models have been elucidated in  Ref.~\cite{Lorce:2011zta}. 
The common features of these models are that the quarks can
basically be considered bound in a mean field, and the
nucleon wave functions exhibit spherical symmetry. 
When these conditions are fulfilled, the relations 
(\ref{Eq:rel-IV}-\ref{Eq:non-lin-2}) hold. If one imposes
in addition to that SU(4) spin-flavor symmetry of the nucleon
wave function, then also the relation (\ref{Eq:rel-I}) holds.

Notice that SU(4) spin-flavor symmetry is necessary but not 
sufficient for Eq.~(\ref{Eq:rel-I}) to hold, which is valid 
only in models with nucleon wave-functions constructed from
'flavor-blind' quark wave-functions multiplied by the spin-flavor 
factors $N_q$ or $P_q$. 
The SU(4) spin-flavor symmetry can, however, be realized in 
more sophisticated ways, e.g., in the spectator model of Ref.~\cite{Jakob:1997wg} the SU(4) symmetry is implemented, but
(\ref{Eq:rel-I}) is spoiled by the different masses of the 
(scalar, axial-vector) diquarks. In this model, it is possible 
to restore (\ref{Eq:rel-I}) in the large-$N_c$ limit (where the 
scalar- and axial-vector diquark masses become equal; notice that 
in general large $N_c$ does not imply SU(4) spin-flavor symmetry).
In nature, SU(4) spin-flavor symmetry is supported only roughly
and one should not expect more from (\ref{Eq:rel-I}).

Relations connecting only polarized TMDs, like
(\ref{Eq:rel-IV}--\ref{Eq:non-lin-2}), do not require SU(4) 
spin-flavor symmetry, are supported by a larger class of models,
and may be more reliable.
The quark-target model \cite{Meissner:2007rx}, though, 
does not support (\ref{Eq:rel-IV}-\ref{Eq:non-lin-2})
which is not surprising: including gluonic fields brings us 
a step closer to QCD where one can not expect such relations
to hold. (Notice that even if such relations were valid in QCD 
at some scale, they would not be valid at other scales because 
the different functions obey different evolution equations.)

Two relatively robust conclusions concern the signs of the TMD PDFs:~the Kotzinian-Mulders worm-gear functions can be expected to have
opposite signs based on Eq.~(\ref{Eq:rel-IV}). One can conclude 
the same about pretzelosity and transversity from (\ref{Eq:non-lin-2}).

It remains to be seen, whether the model relations
(\ref{Eq:rel-IV}--\ref{Eq:non-lin-2}) will turn out to hold
at least approximately within some reasonable accuracy in some 
region of valence-like $x$. Future data, phenomenological work 
and lattice QCD studies will give insights in that respect.

\subsubsection{Connection of pretzelosity to orbital angular
momentum}
\label{subsec:pretzel-and-orbital-angular-momentum}

\index{orbital angular momentum (OAM)}

Since EMC measurements of polarized structure functions
triggered the "proton spin crisis", see Sec.~\ref{decom_spin} and 
Ref.~\cite{Leader:2013jra}, one important motivation to go beyond 
the collinear approximation and study TMD physics was to learn 
about the quark orbital motion and the role of orbital angular 
momentum in the spin structure of the nucleon.
But how are TMD PDFs related to quark orbital angular momentum? 

In QCD, there is no connection between orbital angular momentum 
and TMD PDFs (though there is one involving generalized transverse
momentum dependent parton distribution functions and Wigner functions, see Sec.~\ref{sec:gtmd}). In quark models the situation is different. 
In a wide class of
quark models incorporating different model dynamics, the 
following relation of the pretzelosity function $h_{1T}^{\perp q}$ 
to quark orbital angular momentum was found
\begin{align}
L_{z}^q = -\int d x \,  d^2 k_{T} \, \frac{k_T^2}{2M_N^2} h_{1T}^{\perp q}(x, k_T 
^2) = -\int d x  \; h_{1T}^{\perp (1) q}(x)
\qquad \mbox{(in quark models)}.
\label{oam}
\end{align}
This relation is supported in the spectator model,
bag model, light-front constituent quark model, and
the light-front version of the chiral quark soliton
model restricted to the 3-quark Fock-state sector,
or the covariant parton model 
 \cite{She:2009jq,Avakian:2008dz,Avakian:2010br,Efremov:2009ze,Efremov:2010cy,Lorce:2011kd}.
Despite being supported by many models, the 
connection of pretzelosity to orbital angular momentum
is, for instance, spoiled by the contributions of axial
diquarks in the model of \cite{Liu:2014zla} and is not 
valid in the AdS/QCD-based light-front quark-diquark model 
of Ref.~\cite{Gurjar:2021dyv}. 

In the models where it is supported, Eq.~\eqref{oam} holds for 
the expectation values of operators and not on operator level. 
None of the supporting models has explicit gluonic degrees of 
freedom. In these models, the spin contribution of a quark of 
flavor $q$ to the nucleon spin is given by
$2S_z^q = \int dx\,g_1^q(x)$ 
and the nucleon spin budget 
takes the simple form
\be
      \sum_q S_z^q  + \sum_q L_z^q = \frac12\,
      \quad \mbox{(in quark models)}.
\ee
In Ref.~\cite{Lorce:2011kn} the origins of the relation \eqref{oam} 
were elucidated. For the connection between quark orbital 
angular momentum and the pretzelosity to be valid in a model,
a key ingredient is a certain spherical symmetry of the quark
wave functions in the nucleon rest frame. It was furthermore 
shown that the quark orbital angular momentum defined by 
Eq.~\eqref{oam} contains contributions from the transverse 
center of momentum which cancel out in the total quark 
orbital angular momentum, i.e.,\ after summing over all 
quark flavours present in a model.

In QCD, orbital angular momentum can be described 
in terms of Wigner functions, see \sec{gtmd}. 
It is an open question whether the model relation of orbital angular 
momentum and pretzelosity, Eq.~(\ref{oam}), has
a connection to the expression for angular momentum defined in 
terms of Wigner functions in Eq.~(\ref{eq:F14OAM}) in \sec{gtmd}.
In QCD, no such relation can be expected. But it will be
interesting to address this question in models.

While model-dependent, the relation Eq.~\eqref{oam} remains the
closest connection of TMD PDFs to quark orbital angular momentum
uncovered thus far. 

\subsection{Summary and Outlook}

Model studies are needed and well motivated, as argued in
Sec.~\ref{Sec:models-why}, and had important impact on 
the progress in TMD physics. For instance, as reviewed in 
Sec.~\ref{Subsec-models:review-Brodsky-Hwang-Schmidt},
a model calculation paved the way towards clarifying the 
QCD foundations of T-odd TMD PDFs.
After discussing limits in QCD 
in Sec.~\ref{Sec:models-limits-in-QCD} 
(parton model, large-$N_c$, non-relativistic limit),
we reviewed models of 
T-even (in Sec.~\ref{Sec:models-T-even-quark-TMDs}) and
T-odd (in Sec.~\ref{Sec:models-T-odd-quark-TMDs}) TMD PDFs
of quarks. The Sec.~\ref{Sec:models-gluon-TMDs} was devoted 
to gluon TMD PDFs. Model studies of TMD fragmentation 
functions were discussed in Sec.~\ref{Sec:models-fragmentation}.
The Sec.~\ref{Sec:formal-constraints} addressed formal
inequalities and sum rules among TMD functions which, 
while not model-dependent, have not yet been proven 
rigorously in terms of renormalized TMD functions.
Finally, in Sec.~\ref{Sec:relations-in-models} we have 
reviewed relations among TMD PDFs valid and supported 
in a wide class of quark models without explicit gauge 
field degrees of freedom. 

Many model predictions remain to be tested in
experiment or lattice QCD. While there
are no reasons to believe that, e.g.,\ the model
relations among TMD functions are exact, it may well
turn out that some of them hold approximately
within a good accuracy. In such cases, it will be 
interesting to understand the exact reasons for that in QCD. 
At such instances, the understanding of TMDs and nucleon
structure is likely to make significant progress. 
Models are likely to yield  future surprises 
and new, unexpected and unanticipated insights and 
will continue to contribute their share to the 
progress in the field.

Before concluding, let us stress that this chapter was not intended to present a detailed and complete review of all model studies in the literature which would require far more space. Our goal was to highlight the important lessons learned from model studies and their applications. As stressed at the beginning of this chapter, progress in TMD physics arises from combined efforts in experiment, perturbative QCD, lattice QCD, phenomenology,  {\it and} models.

%% file: sec-smallx/sec-smallx.tex
\section{Small-x TMDs}
\label{sec:smallx}

\subsection{Gluon Saturation and TMDs at Small x}
\label{sec:saturation}

\index{small-x region}We will begin this section with a discussion of gluon saturation in the Regge asymptotics of QCD and an effective field theory (EFT) description of this regime. This discussion is important because the EFT description, called the Color Glass Condensate (CGC) \index{Color Glass Condensate (CGC)},  strongly constrains the structure of small-$x$ TMDs. For instance, as we shall discuss, the BFKL equation that describes the evolution of unintegrated gluon distributions at small $x$ can be recovered straightforwardly within the CGC EFT. Conversely, small-$x$ TMDs computed by extrapolating the TMD framework to small $x$ must have a regime of overlap with the CGC EFT which can help test and refine the dynamical assumptions within this framework. 
  
There are compelling theoretical arguments and strong experimental hints that suggest that  gluon distributions saturate at small Bjorken-$x$~\cite{Gribov:1984tu,Mueller:1985wy,McLerran:1993ni,McLerran:1993ka,Iancu:2003xm,Kovchegov:2012mbw}. Gluon saturation occurs when the nonlinear terms in the field strength tensor are of the same magnitude as the kinetic terms which is the case when the gauge fields are $O(1/g)$, or when the occupancy of field modes is $O(1/\alpha_S)$. In QCD's Regge limit, a probe with arbitrarily fine resolution $Q^2\gg \Lambda_{\rm QCD}^2$ will encounter such large field strengths at sufficiently small $x_{\rm Bj}$; the corresponding scale $Q\rightarrow Q_S(x)$ is appropriately called the saturation scale.  
This classicalization scale is also the scale which unitarizes the interaction of the probe with the target; the $S$-matrix for a probe of inverse size $\geq 1/Q$ goes rapidly to zero  
and its unitarization is accompanied by a significant slowing down in the growth rate of the cross section. Since $Q_S(x) \gg \Lambda_{\rm QCD}$ in Regge asymptotics, one can have $\alpha_S(Q_S)\ll 1$, which self-consistently satisfies the condition of high occupancy. 

As the very large $O(1/g)$ field strengths suggest, gluon saturation at small $x$ is an emergent nonperturbative phenomenon. Its dynamical origin is due to many-body screening and recombination higher twist effects that become large with increasing energy at fixed resolution and compete with the bremsstrahlung of soft gluons that is the dominant effect for weak field strengths. Because the coupling governing the emergent nonperturbative dynamics is  weak, one can systematically study how strong fields dynamically modify the landscape of many-body parton distributions inside a nucleon or nucleus. In particular, in this region of high parton densities, the effective degrees of freedom and their dynamics are qualitatively  different from those in the dilute ``Bjorken limit'' of QCD. In the latter, leading twist  DGLAP~~\cite{Gribov:1972ri,Lipatov:1974qm,Altarelli:1977zs,Dokshitzer:1977sg} evolution can be employed to understand the QCD evolution of parton distributions. 

The necessity for the emergence of gluon saturation can already be deduced from perturbative QCD. The dynamics of QCD evolution in this framework is governed by phase-space logarithms in $Q^2$ and $x$, that arise at each rung of the evolution ladder, and have the generic structure $\alpha_S \ln(Q^2) \ln(1/x)$. 
In the small-$x$ Regge limit of QCD, large logarithms $\alpha_S\ln(1/x)\sim O(1)$  dominate over the DGLAP logs $\alpha_S\ln(Q^2)$, which suggests that the dominant contributions to QCD evolution at small $x$ are obtained by  organizing the perturbative series accordingly to resum such ``leading logs in $x$" (LLx) contributions. 
The renormalization group equation describing this LLx evolution  is the BFKL \index{BFKL evolution equation} equation~\cite{Kuraev:1977fs,Balitsky:1978ic}, the solution of which, as anticipated,  demonstrates a very rapid growth of the gluon distribution, far more so than obtained by solving the DGLAP equation which does not fully account for the large $\alpha_S\ln(1/x)$ contributions. 

The $\alpha_S^{n+1}\ln^n(1/x)$ resummation of next-to-leading logs in $x$ (NLLx), to each $n$th order in perturbation theory generates the NLLx BFKL equation~\cite{Fadin:1998py,Ciafaloni:1998gs}; careful treatment of collinear poles that appear in the kernel of the NLLx BFKL equation leads to robust results that give a significantly slower growth in gluon distributions at small $x$ relative to the LLx BFKL equation~\cite{Ciafaloni:2003rd}. It is nevertheless significantly faster than DGLAP evolution at small $x$, growing at rate that will violate unitarity asympototically if unchecked. Since this growth leads to growing occupancy of field modes,  gluon saturation provides a dynamical self-regulating nonperturbative unitarization mechanism in QCD at weak coupling.  

Since first principles perturbative computations quickly run into the problem of dealing with
all-order twist contributions~\cite{Mueller:1985wy}, an alternate approach is to construct an effective field theory  that captures the many-body dynamics of the saturation regime and can be matched to perturbative computations at large-$x$ and momentum resolutions in their overlapping regime of validity. As noted earlier, the EFT describing the gluon saturation regime is the CGC~\cite{Iancu:2003xm,Gelis:2010nm,Kovchegov:2012mbw,Blaizot:2016qgz}, whose construction~\cite{McLerran:1993ka} relies on the following ingredients \index{Color Glass Condensate (CGC)}:
\begin{enumerate}
    \item A Born-Oppenheimer separation between large-$x$ and and small-$x$ modes; the former can be treated as heavy static modes on the light front while the latter are dynamical modes~\cite{Susskind:1967rg,Bardakci:1968zqb}. 
    \item Due to large coherence lengths at small-$x$, the correspondingly large number of colored static modes constitute higher dimensional (classical) representations of color charge. An explicit construction demonstrates that summations over the color charges of large-$x$ modes can be replaced by a path integral over classical color sources whose mean color charge density is zero, but its variance scales as (for a nucleus with atomic number $A\gg 1$) $\sim A^{1/3}$ for $x\ll A^{-1/3}$~\cite{McLerran:1993ka,McLerran:1993ni,Jeon:2004rk}.
    \item The large-$x$ static color sources (represented by a source density $\rho$) has the most general gauge invariant coupling~\cite{Jalilian-Marian:2000pwi} to the small-$x$ degreees of freedom, represented by the Yang-Mills action. 
\end{enumerate}

This CGC EFT implicitly contains a scale $x_0$ that separates the large-$x$ static degrees of freedom that are distributed according to a nonperturbative gauge invariant weight functional $W_{x_0}[\rho]$. As we will discuss shortly, the requirement that physics be independent of this scale generates the small-$x$ renormalization group equations describing QCD evolution in the saturation regime. 

But before we discuss these, we note that the CGC effective action has a classical saddle point 
$A_{\rm cl.}^\mu \equiv A_{\rm cl.}^\mu[\rho]$ which is an explicit functional of $\rho$ and is of $O(1/g)$. In the McLerran-Venugopalan (MV) \index{model!McLerran-Venugopalan (MV) model} model~\cite{McLerran:1993ka,McLerran:1993ni,McLerran:1994vd} for a large nucleus, where $W_{x_0}[\rho]$ is a Gaussian in $\rho$, the nonperturbative classical dynamics of $n$-point correlators of saturated gauge fields can be computed explicitly. For example, the number distribution in light-cone gauge is seen straightforwardly to give the Weizs\"{a}cker-Williams distribution \index{Weizs\"{a}cker-Williams gluon distribution} at large $k_\perp > Q_S$ but for $k_\perp \leq Q_S$ demonstrates a softer logarithmic dependence on $k_\perp$. Thus even at the classical level in light-cone gauge one sees a clear manifestation, in this non-Abelian  Weizs\"{a}cker-Williams distribution, of the role of non--linearities in taming the growth of gluon distributions. Since the Weizs\"{a}cker-Williams distribution provides the bremsstrahlung kernel for QCD evolution in the ``linear" regime, its tree-level modification due to gluon saturation provides a preview of a qualitatively different corresponding QCD evolution in this regime. 

In light-cone gauge, the non-trivial classical gauge fields $A_{\rm cl}^i$ are so-called ``pure gauge" fields carrying zero field strength that are 
separated by a discontinuity at $x^-=0$~\cite{McLerran:1993ka,McLerran:1993ni}, corresponding to highly singular field strengths (transverse electric and magnetic fields) that only have support at $x^-=0$. In contrast, in Lorenz gauge, the only nonzero component of the gauge field is $A_{\rm cl}^+$, which itself is singular at $x^-=0$, and are therefore often called ``shock wave" background field configurations~\cite{Balitsky:1995ub}.  By solving the Dirac equation in such a background, and likewise the Yang-Mills small fluctuation equations, one can construct respectively quark and gluon propagators in this shock wave background~\cite{McLerran:1994vd,Balitsky:1995ub}. Remarkably, in Lorenz gauge, these quark and gluon propagators have a very simple structure; in momentum space, they can be expressed as the convolution of free propagators with the insertion respectively of non-local momentum-dependent quark and gluon effective vertices~\cite{McLerran:1998nk,Ayala:1995hx,Balitsky:2001mr}. These effective vertices are proportional to Fourier transforms of the respective fundamental and adjoint Wilson lines of 
the shock wave field $A_{\rm cl}^{+,a} = -\frac{\rho^a}{\nabla_\perp^2}\delta(x^-)$. Note that this implies a dependence to all orders in powers of  the large-$x$ color charge densities $\rho$. 

The structure of these effective propagators allows one to establish an 
exact correspondence~\cite{Caron-Huot:2013fea,Bondarenko:2017vfc,Ayala:2017rmh,Hentschinski:2018rrf,Bondarenko:2018kqs,Bondarenko:2018eid} of the CGC EFT to Lipatov's Reggeon field theory~\cite{Lipatov:1996ts}. The color charge densities $\rho$ can be related to the Reggeon degrees of freedom~\cite{Caron-Huot:2013fea} and color singlet combinations of these 
to Pomeron\index{pomeron} and Odderon\index{odderon} degrees of freedom~\cite{Hatta:2005as,Jeon:2005cf}. Historically, 
such nonperturbative effective degrees of freedom were found to provide successful descriptions of high energy scattering but a first principles understanding from QCD has remained elusive. The correspondence noted here may therefore provide a useful link between the CGC EFT and intrinsically nonperturbative~\cite{Brower:2006ea} modern approaches to Reggeon/Pomeron physics\index{pomeron}.  

We now turn to the computation of physical observables in the CGC EFT and the renormalization group (RG) evolution equations that emerge from their proper treatment. A simple example is provided by the inclusive structure functions $F_2$ and $F_L$, which in general can be expressed in terms of bilinears of the quark propagators, ${\rm Tr}\left(S_A(x,y)\gamma^\mu S_A(y,x)\gamma^\nu\right)$ in arbitrary background fields. In the CGC EFT, the leading order result for $F_{2,L}$ is obtained by replacing $A\rightarrow A_{\rm cl}$, and $S_A$ by the shock wave propagators we mentioned earlier. One immediately recovers the Glauber-Mueller dipole model~\cite{Mueller:1989st}, with $F_{2,L} \propto \left(1-\langle{\cal S}\rangle\right)$, where ${\cal S}=\frac{1}{N_c}{\rm Tr}\left(V(x_\perp) V^\dagger(y_\perp)\right)$ is the dipole S-matrix, $V(x_\perp)= P\exp\left(i\frac{\rho}{\nabla_\perp^2}(x_\perp)\right)$ and $\langle {\cal S}\rangle = 
\int[D\rho] \,W_{x_0}[\rho]\, {\cal S}$. 

For the MV model with Gaussian distributed color sources $W_{x_0}[\rho]$ with weight $\mu_A^2$ (corresponding to the color charge squared per unit area),
one obtains 
\begin{equation}
    \langle {\cal S}\rangle (r_\perp) = \exp\left(- \frac{r_\perp^2 Q_S^2}{4}\ln\left(\frac{1}{r_\perp^2 \Lambda_{\rm QCD}^2}\right) \right)\,,
\end{equation}
where $Q_S^2 = \alpha_S C_F \mu_A^2$ with $C_F=(N_c^2-1)/2N_c$. For $r_\perp^2 Q_S^2\ll 1$, one recovers the color transparency limit of QCD for the dipole cross section; for $r_\perp^2 Q_S^2\gg 1$, $\langle{\cal S}\rangle\rightarrow 0$, corresponding to the ``color opacity" or black disc limit of QCD~\cite{Frankfurt:2011cs}.   

At next-to-leading order, including the contribution of real and virtual slow gluons in the shock wave background, leads to the relation~\cite{Balitsky:1995ub,Kovchegov:1998bi}:
\begin{equation}
    \frac{d\langle{\cal S}\rangle}{dY} = -\frac{\alpha_s N_c}{2 \pi^2} \int d^2 z_\perp\, \frac{(x_\perp -y_\perp)^2}{(x_\perp-z_\perp)^2 (z_\perp-y_\perp)^2}\, \langle {\cal S}(x_\perp,y_\perp)
    - {\cal S}(x_\perp,z_\perp)\,{\cal S}(z_\perp,y_\perp)\rangle \,,
    \label{eq:JIMWLK-dipole}
\end{equation}
where $Y= \ln(x_0/x)$ denotes the rapidity. In the large $N_c$ limit, and for $A\gg 1$, $\langle {\cal S}\,{\cal S} \rangle\approx \langle {\cal S}\rangle \langle {\cal S}\rangle$, leading to a closed form non-linear equation, the Balitsky-Kovchegov (BK) equation~\cite{Balitsky:1995ub,Kovchegov:1998bi} \index{Balitsky-Kovchegov (BK) equation}, for the dipole S-matrix. When $\langle {\cal S}\rangle \sim 1$, one can write $\langle {\cal S}\rangle\sim 1-\langle {\cal N}\rangle$, where the dipole amplitude $\langle{\cal N}\rangle$ ($\ll 1$) satisfies the BFKL equation. In the opposite limit, $\langle S\rangle\approx 0$, the BK equation unitarizes the cross section, as noted previously. In between these two regimes, lies a ``geometrical scaling" regime, where the dipole amplitude obeys leading twist shadowing; in other words, it satisfies leading twist BFKL evolution but is still sensitive to the presence of a saturation scale~\cite{Iancu:2002tr}. Interestingly, small-$x$ data from HERA data exhibit this geometric scaling phenomenon~\cite{Stasto:2000er}.

In general, Eq.~(\ref{eq:JIMWLK-dipole}) represents the RG equation for the evolution of the two-point correlator of Wilson lines. One can similarly write down the corresponding evolution equation for an arbitrary number of Wilson line correlators, which generates the 
Balitsky-JIMWLK hierarchy~\cite{Balitsky:1995ub,JalilianMarian:1997gr,JalilianMarian:1997dw,Iancu:2000hn,Ferreiro:2001qy}. The entire content of this hierarchy, and of the CGC EFT to LLx,  can alternatively be written as an evolution equation for the weight functional $W_Y[\rho]$:
\begin{equation}
    \frac{d W_Y[\rho]}{dY} = {\cal H}_{\rm JIMWLK}\otimes W_{Y}[\rho]\,,
\end{equation}
where ${\cal H}_{\rm JIMWLK}$ represents the JIMWLK Hamiltonian\index{JIMWLK evolution equation}. This equation has the structure of a Fokker-Planck equation in the space of functions $\rho$ and can therefore be equivalently written as a Langevin equation for the Wilson lines $V$~\cite{Blaizot:2002np}.
This Langevin equation can be solved numerically~\cite{Rummukainen:2003ns,Dumitru:2011vk} which allows us to determine the solution to Eq.~(\ref{eq:JIMWLK-dipole}) for finite $N_c$ as well as the evolution equations for higher point Wilson line correlators to LLx accuracy. 

\begin{figure}[htb]
 \begin{center}
 \includegraphics[width=100mm]{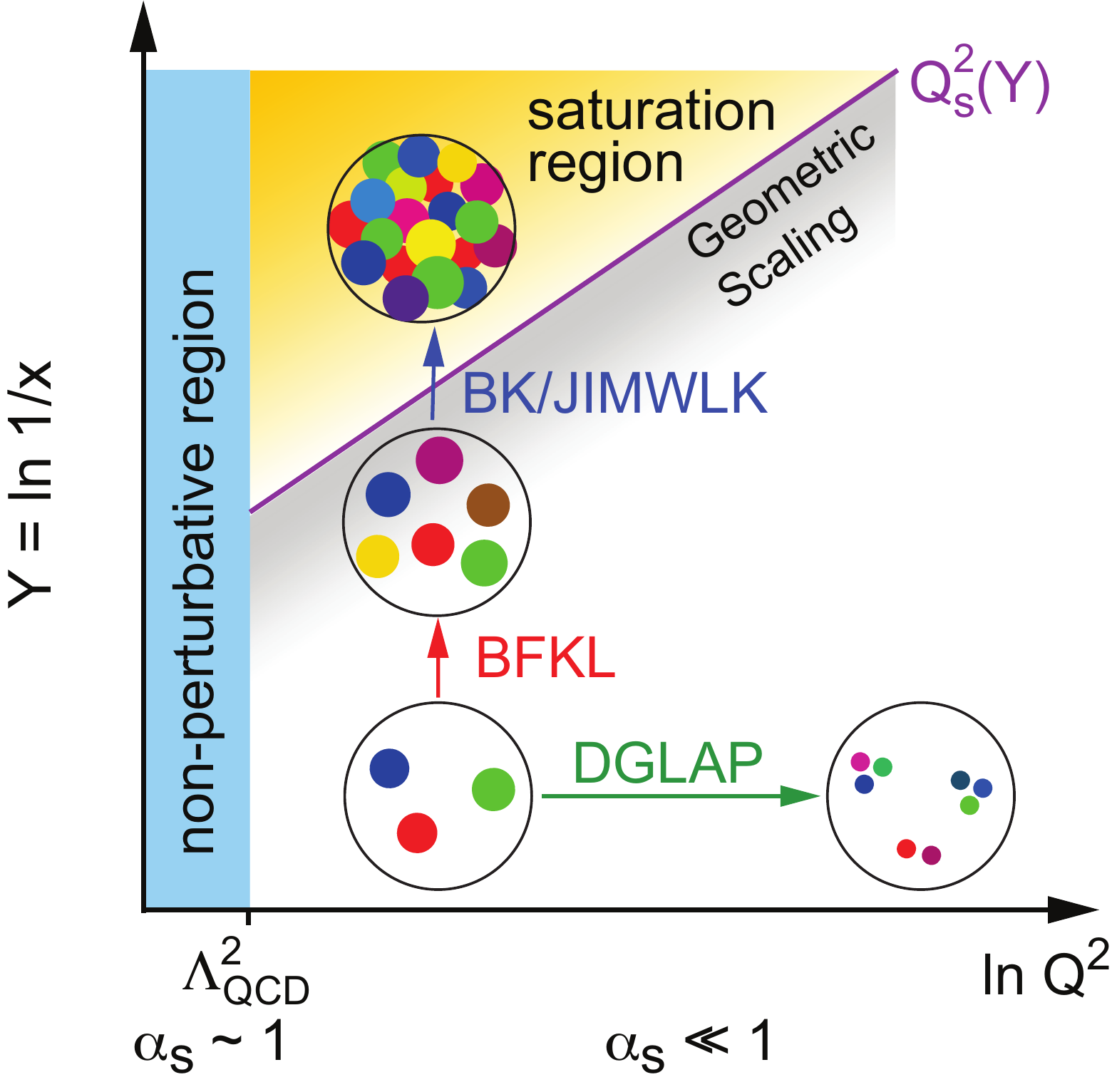}
 \end{center}
 \caption{\label{Smallx_Fig0} The map of high energy QCD in the ($Q^2$, $Y = \ln (1/x)$) plane. (This plot is adopted from Ref.~\cite{Accardi:2012qut}).}
\end{figure}
 
Our discussion up to this point can be summarized in Fig.~\ref{Smallx_Fig0} which shows the different regimes of QCD evolution at small-$x$. The line corresponding to the saturation scale $Q_S^2(Y)$, as noted corresponds to the boundary where classicalization and unitarization occurs. In nuclei, it is estimated that $Q_S^2(Y) \sim (A/x)^\delta$, where $\delta\approx 0.3$~\cite{Kowalski:2007rw}. At small-$x$, the CGC EFT predicts that distributions are universal after appropriate scaling of $Q_S$ with $A$. This suggests that 
saturation effects can be observed precociously in DIS off nuclei at lower energies than in DIS off protons.

This LLx framework in the CGC EFT has now been extended to next-to-leading-log (NLLx) accuracy 
for both BK~\cite{Balitsky:2008zza,Balitsky:2013fea,Kovchegov:2006vj}  and JIMWLK evolution~\cite{Lublinsky:2016meo,Kovner:2013ona,Kovner:2014lca,Grabovsky:2013mba,Caron-Huot:2013fea}. In particular, the NLL evolution kernels for the Wilson lines~\cite{Balitsky:2013fea}, for dipoles~\cite{Balitsky:2008zza}, for 3-point operators~\cite{Balitsky:2014mca} and for quadrupoles~\cite{Grabovsky:2015cza} have been used to derive the full NLL JIMWLK Hamiltonian~\cite{Kovner:2014lca}, which was then confirmed by a more explicit computation~\cite{Lublinsky:2016meo}. An unanticipated synergy between non-global logarithms and small-$x$ evolution has also enabled one to use computations to NNLO in the former to extract parts of the NNL BK kernel~\cite{Caron-Huot:2015bja}. Complete analytical expressions have been obtained for a varied range of physical observables. The fully inclusive DIS cross section was computed in~\cite{Balitsky:2010ze} and in~\cite{Beuf:2017bpd}, then confirmed in~\cite{Hanninen:2017ddy}. Semi-inclusive processes have been studied both for $pA$ collisions in the dilute-dense hybrid factorization ansatz~\cite{Chirilli:2011km,Altinoluk:2014eka}, and for $eA$ collisions~\cite{Caucal:2021ent,Roy:2019hwr,Roy:2019cux}, where the  NLO impact factor has been computed for di-jet and photon+di-jet final states. Finally, exclusive $eA$ collisions have been studied for dijet production~\cite{Boussarie:2016ogo} and for the production of a light vector meson~\cite{Boussarie:2016bkq}.

This general picture of interactions at small-$x$, in terms of the RG evolution of the shockwave classical fields, does not depend on the type of scattering reaction. It should therefore be valid for transverse momentum dependent (TMD) processes. The first step of TMD studies at small-$x$ would be the implementation of the available theoretical tools most effectively represented by the CGC formalism to a variety of TMD observables. In particular, there have been important developments that have brought to the fore the connections between the TMD formalism and the small-$x$ CGC formalism in various contexts~\cite{Marquet:2009ca,Dominguez:2010xd,Dominguez:2011wm,Metz:2011wb,Balitsky:2016dgz,Altinoluk:2019fui, Altinoluk:2019wyu}. It was realized for instance that TMD-like hard processes which involve a hard scale $Q$ in addition to the transverse momenta of the observables offer unique possibilities to probe the saturation regime. 

The most important point of convergence is the fact that unintegrated gluon distributions are important ingredients in both the TMD and CGC frameworks. In the latter, there is a classification of 
scatterings into dilute-dilute, dilute-dense and dense-dense depending on the field strengths of the color sources in the projectile and target and the transverse momenta of interest~\cite{Gelis:2007kn,Gelis:2010nm}. Unintegrated distributions appear at small-$x$ in dilute-dilute ($p+p$ scattering for instance) and dilute-dense ($p+A$ scattering being a natural example). In the latter case, the unintegrated distributions in the target are sensitive to coherent multi-parton interactions. Several processes have been proposed in the literature to study the unintegrated gluon distributions 
including semi-inclusive DIS~\cite{Marquet:2009ca}, low $p_T$ Drell-Yan~\cite{Baier:2004tj}, and back-to-back di-hadron correlations in forward $pA$ processes~\cite{Marquet:2007vb}. Recently, considerable progress has  been made in computing Sudakov double logarithms that can be resummed consistently in the small-$x$ formalism~\cite{Mueller:2012uf,Mueller:2013wwa,Balitsky:2015qba,Xiao:2017yya,Balitsky:2020jzt}. These computations provide a solid theoretical foundation for further rigorous investigations that probe the dynamics of the saturation regime with hard processes.

\subsection{Weizs\"acker-Williams and Dipole Gluon Distributions}
\label{sec:WW-DGD}

It is important to emphasize 
that the properties of the QCD dynamics typical for small $x$ lead to a very different picture of TMD scattering compared to the standard TMD framework at large $x$.  
Indeed, it was in the context of the CGC framework that the existence of {\it two} different unintegrated gluon distributions (UGDs) was first proven~\cite{Kharzeev:2003wz}. This observation is related to the question of \textit{non-universality} for TMD distributions due to the process-dependence of their gauge link structures, since then taken into account in the standard TMD approach~\cite{Dominguez:2010xd,Dominguez:2011wm}. The topic of non-universality at small $x$ is discussed in great detail in~\cite{Petreska:2018cbf}. 

Two types of gluon TMD distributions are the most common. The first such distribution, the Weizs\"{a}cker-Williams (WW) \index{Weizs\"{a}cker-Williams gluon distribution} gluon distribution, is calculated from the correlator of two classical gluon fields of relativistic hadrons (the non-Abelian Weizs\"{a}cker-Williams fields) ~\cite{McLerran:1993ni,McLerran:1993ka,JalilianMarian:1996xn,Kovchegov:1996ty}. The WW gluon distribution can be defined following the conventional gluon distribution~\cite{Collins:1981uw,{Ji:2005nu}}. Following Eq.~(\ref{eq:fg}), we can define the WW gluon distribution as,
\begin{eqnarray}
xG_{WW}(x,k_{\perp })&=&\int \frac{d\xi ^{-}d^2\xi _{\perp }}{(2\pi
)^{3}P^{+}}e^{-ixP^{+}\xi^{-}+ik_{\perp }\cdot \xi _{\perp }}   \langle P|G^{+i}(\xi ^{-},\xi _{\perp }){\cW}_\sqsubset(\xi^\mu,0) G^{+i}(0)|P\rangle \ ,  \label{g1}
\end{eqnarray}
where $G^{\mu \nu }$ is the gauge field strength tensor and ${\cW}_\sqsubset(\xi^\mu,0)$ represents the gauge link in the adjoint representation and points to the past ($-\infty$). 
This gluon distribution can also be defined in the fundamental representation~\cite{Bomhof:2006dp},
\begin{eqnarray}
xG_{WW}(x, k_\perp) &=& \frac{2}{P^+} \int \frac{d\xi^- d^2\xi_\perp}{(2\pi)^3}~ e^{-ixP^+\xi^- + ik_\perp \xi_\perp} 
\nonumber\\
&&\times\, {\rm Tr}~ \langle P|  G^{+i}(\xi^-,\xi_\perp){ W}_\sqsubset(\xi^\mu,0) G^{+i}(0^-, 0_\perp){ W}_\sqsubset^\dagger(\xi^\mu,0) | P\rangle\, .
 \label{GWW}
\end{eqnarray}
Here, ${ W}_\sqsubset(\xi^\mu,0)$ represents the gauge link in the fundamental representation. We note that for the WW gluon distribution, the two gauge links in the above definition point to the same direction (to $-\infty$). This gluon distribution 
corresponds to the gluon distributions associated with Higgs Boson production in hadronic collisions as described in \sec{TMDHiggs}. 
The above $G_{WW}(x,k_\perp)$ is referred to as $f_1^g$ there. 
If we want to study the gluon distribution associated with semi-inclusive deep inelastic scattering, the gauge links will point to the future ($+\infty$). The universality of the gluon distributions in the different processes will follow the discussions in previous sections, see, e.g., Sec.~\ref{sec:universality} and \ref{sec:procdep}. 
In the following, for brevity we do not include explicitly the transverse gauge links which connects the Wilson lines in Eqs. (\ref{GWW}, \ref{g1}) at $\xi^- \to +\infty$. Also note that all transverse gauge links are subdominant in the small-$x$ regime, 
which is why we will neglect them hereafter~\footnote{In a singular gauge, such as the light-cone gauge, we need to consider the transverse gauge link contributions at the spatial infinity~\cite{Belitsky:2002sm,Hatta:2006ci}.}.

The structure of the WW distribution used in the CGC approach coincides with the gluon TMD distribution  at large $x$ and has a clear physical interpretation as the number density of gluons inside the hadron in light-cone gauge. This makes it the primary candidate to study the transition region between dilute and dense regimes. Since the WW distribution is constructed from semi-infinite future-pointing Wilson lines (and past-pointing lines for the Drell-Yan process) it takes into account only final state interactions (initial state for DY) 
which occur after (before) 
the initial interaction of the hard probe with the target.

This makes it qualitatively different from the second gluon distribution, defined as the Fourier transformation of the color dipole cross section \index{dipole gluon distribution}:
\begin{eqnarray}
xG_{dip.}(x, k_\perp) &=& \frac{2}{P^+} \int \frac{d\xi^- d^2\xi_\perp}{(2\pi)^3}~ e^{-ixP^+\xi^- + ik_\perp \xi_\perp} 
\nonumber\\
&&\times {\rm Tr}~ \langle P| G^{+i}(\xi^-,\xi_\perp){ W}_\sqsubset(\xi^\mu,0) G^{+i}(0^-, 0_\perp){ W}_\sqsupset^\dagger(\xi^\mu,0)  | P\rangle\, , \label{GD}
\end{eqnarray}
where the two gauge links point to the opposite directions and they form a loop. These gauge links 
stretch between minus and plus infinity and take into account both final and initial interactions, which reflects in the shockwave nature of scattering at small-$x$ and separation of scales between projectile and target. 
Unlike the WW distribution, the dipole gluon distribution does not have a clear parton interpretation.
  
Within the CGC framework, the WW gluon distribution can be written in terms of the correlator of four Wilson lines as, 
\begin{equation}
xG_{WW}(x,k_\perp)=-\frac{2}{\alpha_S}\int\frac{d^2v_\perp}{(2\pi)^2}\frac{d^2v'_\perp}{(2\pi)^2}\;e^{-ik_\perp\cdot(v_\perp-v'_\perp)}\left\langle\text{Tr}\left[\partial_iU(v_\perp)\right]U^\dagger(v'_\perp)\left[\partial_iU(v_\perp')\right]U^\dagger(v_\perp)\right\rangle_{x},\label{GWW1}
\end{equation}   
where the Wilson line $U(v_{\perp})$ is defined as a gauge link from $(-\infty)$ to $(+\infty)$. By using the notation of Eq.~(\ref{eq:Wilson_lines}), we have $U(v_{\perp})=W_n(v_\perp,-\infty,+\infty)$. In the above equation, subscript $x$ represents the momentum fraction carried by the gluon when we evaluate the matrix element. The precise $x$ value is determined by the kinematics of the process. In addition, the normalizations of the states are different in the CGC computations. For example, Eq.~(\ref{GWW}) is normalized covariantly and the hadronic state $|P\rangle$ is relativistically normalized to $\langle P'|P\rangle=(2\pi)^32P^+\delta(P^+-P^{\prime+})\delta^{(2)}(P_\perp-P'_\perp)$, while the average in Eq. (\ref{GWW1}) and Eq.~(\ref{GD1}) below is taken over the CGC wave function and is normalized that $\langle1\rangle_{x}=1$, so that $\langle \hat {\cal O}\rangle_x=\frac{\langle P|\hat {\cal O}|P\rangle}{\langle P|P\rangle}$.

Similarly, the dipole gluon distribution can be directly evaluated in the CGC framework, 
\begin{equation}
xG_{dip.}(x,k_\perp)= k_\perp^2\frac{2}{\alpha_S}\int\frac{d^2v_\perp}{(2\pi)^2}\frac{d^2v'_\perp}{(2\pi)^2}\;e^{-ik_\perp\cdot(v_\perp-v'_\perp)}\left\langle\text{Tr}U(v_\perp)U^\dagger(v_\perp')\right\rangle_{x} .\label{GD1}
\end{equation}
To make the connections between the CGC results of Eqs.~(\ref{GWW1},\ref{GD1}) and those in Eqs.~(\ref{GWW},\ref{GD}) more clearly, one can apply the derivative on the Wilson line,
\begin{equation}
\partial_iU(v_\perp)=ig_S\int_{-\infty}^\infty \text{d}v^-\, W_n(v_\perp,-\infty,v^-)\,\left(\partial_iA^+(v^-,v_\perp)\right)\,W_n(v_\perp,v^-,\infty) \ . \label{eq:derivativelink}
\end{equation}
Notice that $\left(\partial_iA^+(v^-,v_\perp)\right)$ is the leading part of the
gauge invariant field strength tensor $G^{i+}(v_\perp)$ at small-$x$. 
Therefore, the above correlator can be written in terms of
a gauge invariant matrix element~\cite{Dominguez:2010xd,Dominguez:2011wm},
\begin{eqnarray}
&&-\langle\text{Tr}\left[\partial_iU(v_\perp)\right]U^\dagger(v'_\perp)\left[\partial_jU(v'_\perp)\right]U^\dagger(v_\perp)
\rangle_{x}\nonumber\\
&&~~~~=g_S^2\int_{-\infty}^\infty\text{d}v^-\text{d}v^{\prime-}
\left\langle\text{Tr}\left[G^{i+}(\vec{v})
{ W}_\sqsubset(v,v') G^{j+}(\vec{v}'){ W}_\sqsubset^\dagger(v,v') 
\right]\right\rangle_{x} \ .
\end{eqnarray}
To recover the gluon distribution function as written in Eq.~(\ref{GWW}), it is necessary to account for the different normalizations used to calculate the average of Wilson lines, see the discussions after Eq.~(\ref{GWW1}).

The above two gluon distributions form the fundamental building blocks of all unpolarized TMD gluon distributions at small $x$ in the planar limit~\cite{Dominguez:2011wm,Dominguez:2012ad}. It was realized that the WW gluon distribution could be directly accessed in the dijet production process in DIS while the photon-jet correlations measurement in $pA$ collisions can access the dipole gluon distribution directly~\cite{Dominguez:2010xd}. More complicated dijet production processes in $pA$ collisions will involve both of these gluon distributions through a convolution in transverse momentum space~\cite{Dominguez:2011wm}. Related phenomena have also been intensively investigated in the TMD factorization framework~\cite{Bomhof:2006dp,Collins:2007nk,Vogelsang:2007jk,Rogers:2010dm}, where the associated parton distributions are found to be non-universal. Detailed analyses~\cite{Dominguez:2011wm} have shown that these  results in the TMD formalism can be related to the small-$x$ calculations for dijet production~\cite{Marquet:2007vb}. Phenomenological applications of this formalism to the RHIC data on forward di-hadron productions in $dA$/$pA$ collisions have been carried out in Refs.~\cite{Albacete:2010pg,Stasto:2011ru,Stasto:2018rci,Albacete:2018ruq}.
More importantly, precision studies of dijet/di-hadron process in DIS at the future EIC will provide a unique perspective to probe the gluon saturation at small $x$ in large nuclei~\cite{Accardi:2012qut,Zheng:2014vka}. 

In addition, the azimuthal correlated (linearly polarized) TMD gluon distribution has played an important role in describing cross sections in hard processes at small $x$~\cite{Metz:2011wb}. For example, the linearly polarized WW gluon distribution  
$xh_{WW}^\perp(x,k_\perp)$ is identical to $xG_{WW}(x,k_\perp)$ at large transverse momentum and  
agrees with the perturbative QCD results. For the case $\Lambda ^{2}\ll k_{\perp}^{2}\ll Q_{s}^{2}$ one finds  
$xh_{WW}^\perp(x,k_\perp)$ is suppressed as compared to $xG_{WW}(x,k_\perp)$.
On the other hand, for the dipole gluon distribution, we have the following simple result,
\begin{equation}
    xh_{dip.}^{\perp }(x,k_{\perp }) = xG_{dip.}(x,k_\perp) \ ,
\end{equation}
for all $k_\perp$ region, 
which means that it has as many linearly polarized gluon pairs as unpolarized gluon pairs. The phenomenological implication of the above discussed linearly polarized gluon distributions have been investigated in a number of papers, 
in particular, that we may study them in great detail at the future EIC~\cite{Boer:2010zf,Boer:2016fqd,Boer:2017xpy,Dumitru:2015gaa,Dumitru:2018kuw,Mantysaari:2019csc,Mantysaari:2019hkq}.

\subsection{TMD Evolution and Resummation}
\label{sec:TMDsmallxevol}

The QCD evolution effects play an important role in describing the scale dependence of these gluon distributions. 
This includes the small-$x$ evolution, i.e., the BFKL/BK
evolution~\cite{Balitsky:1995ub,Kovchegov:1999yj}, and the so-called TMD evolution, i.e., the
Collins-Soper evolution~\cite{Collins:1981uk,Collins:1981uw}. With the small-$x$ approximations
applied in Eq.~(\ref{GWW1}, \ref{GD1}), the small-$x$ evolution effects are taken into account with
the associated evolution equations. However, from those equations, the Collins-Soper evolution
effects are not explicit. Recent developments have paved the way to perform resummation of large logarithms in the TMD gluon distributions at small  $x$~\cite{Zhou:2016tfe,Balitsky:2015qba,Marzani:2015oyb,Mueller:2012uf,Mueller:2013wwa,Xiao:2017yya}. It has been shown the above two resummations (Sudakov and small $x$) can be
performed consistently at the cross section level. 

To study the scale dependence of TMDs at small $x$, we go back to the full QCD definitions of the TMDs, in which the scale dependence naturally shows up in the associated TMD factorization for hard scattering processes. In the gauge invariant definitions of the gluon distributions, as shown in Eqs.~(\ref{GWW}, \ref{GD}), there are un-cancelled light-cone singularities from high order gluon radiations. The regularization introduces the scheme dependence for the un-subtracted gluon TMDs. However, the final result for the subtracted gluon TMDs will be independent of the rapidity regulator and the scheme, 
see more discussions in Sec.~\ref{sec:TMDdefn} and \ref{sec:evolution}.  

Similar to the case of the hard scattering processes studied in Refs.~\cite{Mueller:2012uf,Mueller:2013wwa}, the most important high order gluon radiation come from two regions: (1) soft gluon and (2) collinear gluon. The soft gluon radiation leads to the Collins-Soper evolution, whereas the collinear gluon contributes to the DGLAP resummation formulated in terms of the integrated parton distributions in the CSS resummation formalism. In the current case, these collinear gluon radiation contributions actually become the small-$x$ evolution contributions, which are described by the associated BK/JIMWLK equations~\cite{Balitsky:1995ub,Kovchegov:1999yj,JalilianMarian:1997jx, Iancu:2003xm}. The above two contributions are well separated in phase space. That is the reason that we can achieve resummations of large logarithms from these two sources consistently. The final results for the TMDs can be written as~\cite{Xiao:2017yya},
\begin{eqnarray}
xG_{WW}(x,k_\perp,\zeta_c=\mu_F=Q)&=&-\frac{2}{\alpha_S}
\int \frac{d^2v_\perp d^2v_\perp'}{(2\pi)^4}e^{ik_\perp\cdot r_\perp}
{\cal H}^{WW}(\alpha_s(Q))e^{-{\cal S}_{sud}(Q^2,r_\perp^2)} \nonumber\\
&&\times\, {\cal F}^{WW}_{Y=\ln 1/x}(v_\perp,v'_\perp)\ ,\label{resum}
\end{eqnarray}
where $r_\perp=v_\perp-v'_\perp$, $\zeta_c$ is the regulator for the end-point singularity in the TMD distributions in the Collins 2011 scheme~\cite{Collins:2011zzd}, and $\mu_F$ is the associated factorization scale. In the final factorization formula, these two scales are usually taken as the same as the hard momentum scale $Q$ in hard scattering processes.  Meanwhile, ${\cal F}^{WW}_Y$ is the Fourier transform of the WW gluon distribution, as in Eq.~(\ref{GWW1}),
\begin{equation}
{\cal F}^{WW}_Y(v_\perp,v'_\perp)=
\left\langle{\rm Tr}\left[
\partial_\perp^\beta U(v_\perp) U^\dagger(v'_\perp)\partial_\perp^\beta
U(v'_\perp)U^\dagger(v_\perp)\right]\right\rangle_x \ ,\label{fww}
\end{equation}
and $Y$ represents the rapidity of the gluon from the nucleus, $Y\sim \ln (1/x)$. The Sudakov form factor contains an all order resummation
\begin{eqnarray}
{\cal S}_{sud}=\int_{c_0^2/r_\perp^2}^{Q^2}\frac{d\mu^2}{\mu^2}\left[A\ln\frac{Q^2}{\mu^2}+B\right] \ ,  \label{sud1}
\end{eqnarray}
where $c_0=2e^{-\gamma_E}$ with $\gamma_E$ the Euler constant. The hard coefficients $A$ and $B$ can be calculated perturbatively~\cite{Collins:1984kg}: $A=\sum\limits_{i=1}^\infty A^{(i)}\left(\frac{\alpha_s}{\pi}\right)^i$ and $B=\sum\limits_{i=1}^\infty B^{(i)}\left(\frac{\alpha_s}{\pi}\right)^i$. One-loop results for these coefficients can be found in Ref.~\cite{Xiao:2017yya}. It is interesting to note that $B^{(1)}=0$ which is different from the TMD gluon distribution in the collinear framework. Reconciling these two frameworks has been a theoretical challenge in small-$x$ physics, see discussions in Ref.~\cite{Mueller:2012uf,Mueller:2013wwa,Xiao:2017yya, Balitsky:2015qba,Balitsky:2016dgz}.  
Similarly, we can write down the result for the dipole-gluon TMD~\cite{Xiao:2017yya},
\begin{eqnarray}
xG_{dip.}(x,k_\perp,\zeta_c=\mu_F=Q)&=&-\frac{2}{\alpha_S}
\int \frac{d^2v_\perp d^2v'_\perp}{(2\pi)^4}e^{ik_\perp\cdot r_\perp}{\cal H}^{DP}(\alpha_s(Q))e^{-{\cal S}_{sud}(Q^2,r_\perp^2)}\nonumber\\
&&\times\, {\vec{\nabla}}^2_{r_\perp}{\cal F}^{DP}_{Y=\ln 1/x}(v_\perp,v'_\perp)\
,\label{resum2}
\end{eqnarray}
where  ${\cal F}^{DP}_{Y}(v_\perp,v'_\perp)$ 
is defined as,
\begin{equation}
{\cal F}^{DP}_Y(v_\perp,v'_\perp)=\left\langle{\rm Tr}\left[ U(v_\perp)
U^\dagger(v'_\perp)\right]\right\rangle_x \ .\label{fdp}
\end{equation}
In the above equations, both ${\cal F}^{WW}_Y$ and ${\cal F}_Y^{DP}$ are the renormalized quadrupole and dipole amplitudes, respectively, which obey the associated small-$x$ evolution equations. The TMD evolution effects are included in the Sudakov factor. The remaining factors, ${\cal H}^{WW}(\alpha_s(Q))$ and ${\cal H}^{DP}(\alpha_s(Q))$, which are of order $1$, are the perturbatively calculable finite hard parts.

\subsection{Spin-dependent TMDs}
\label{sec:smallxSDtmds}
The problem of identifying the basic TMD distributions at small $x$ is of course not limited to the case of unpolarized scattering. In fact, it is essential to the study of the spin structure of the hadron. And, in particular, resolution to the so-called ``spin crisis" which states that while we know the value of the proton's spin it is not known how different components of the proton's dynamics contribute to it, see detailed discussions in Sec.~\ref{decom_spin}. 

A major source of uncertainty is the spin content of the proton at small $x$.  However, this study is highly non-trivial because, to leading eikonal order, longitudinal spin effects at small $x$ are highly suppressed. For example, the MV model~\cite{McLerran:1993ni,McLerran:1993ka} provides the following solution for the classical field formed by the small-$x$ gluons:
\begin{eqnarray}
A_{\rm cl}^+(x) = -\frac{1}{\partial^2_\perp} \rho(x_\perp)\delta(x^-);\ \ \ A_{\rm cl}^- = A_{\rm cl}^\perp = 0\,.
\label{MV}
\end{eqnarray}
 As noted, this solution has the form of a shockwave, with the small-$x$ gluons of the hadron shrunk to a single point $x^-=0$. To study longitudinal spin at small $x$, one has to, in general, extend the leading order eikonal solution of Eq.~(\ref{MV}) to include sub-eikonal effects.
 
\begin{figure}[t!]
\begin{center}
\includegraphics[width=120mm]{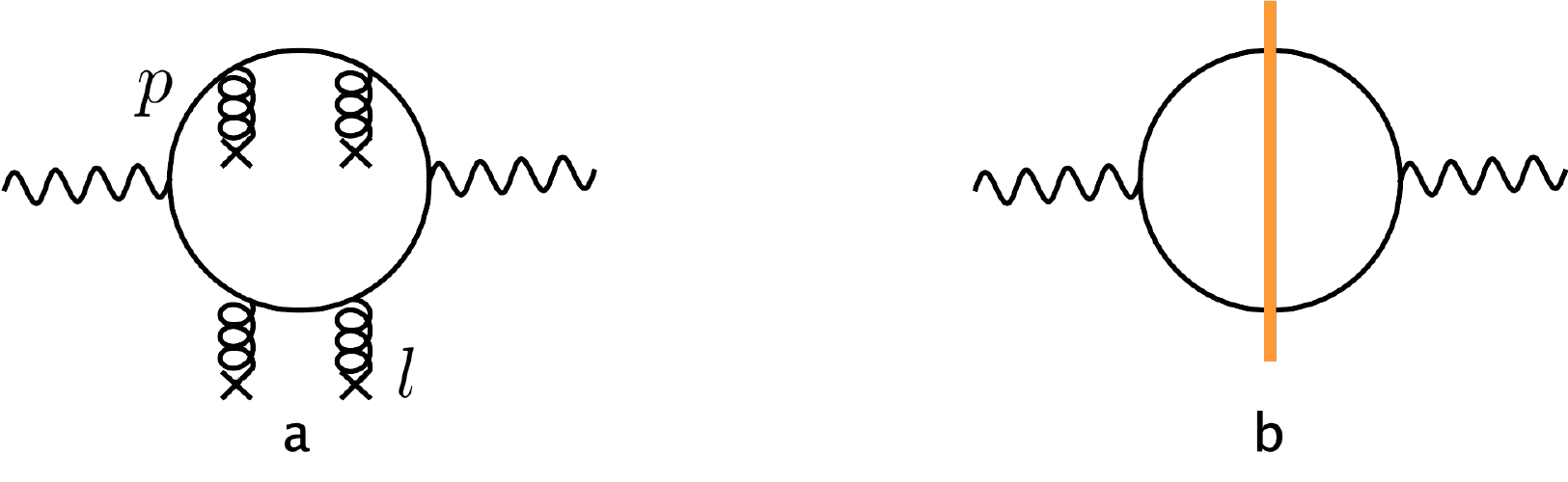}
\end{center}
\vspace{-0.4cm}
\caption{\label{Smallx_Fig1} a) Current-current correlator in an arbitrary background field; b) The same in the CGC shockwave background, where the spatial separation in $x^-$ shrinks to a point (represented by the orange vertical bar). }
\end{figure}

To illustrate this property, let us look at the leading diagram contributing to DIS at small~$x$ presented in Fig. \ref{Smallx_Fig1}a. At leading order, the interaction of the virtual photon with the target is realized through the photon splitting into a $q\bar{q}$ pair which subsequently interacts with a background field formed by partons of the target. Those partons are schematically depicted as vertical lines in Fig. \ref{Smallx_Fig1}a.
The structure of the background is of course defined by the small-$x$ QCD dynamics of the target and can be characterized by a typical momentum $l$. The momentum scale $p$ of the $q\bar{q}$ is different and in general is defined by the virtuality of the incoming photon $Q^2$ (hard scale).  As a result, the condition of the strict ordering of longitudinal components of momenta at small $x$ can be formulated by the condition
\begin{eqnarray}
l^+ \gg p^+,\ \ \ \ l^- \ll p^-\,.
\label{prop1}
\end{eqnarray}
This ordering of momenta can only be realized if the virtuality of the small-$x$ partons is dominated by the transverse momentum component:
\begin{eqnarray}
|l^+ l^-| \ll l^2_\perp\sim Q_s^2.
\label{prop2}
\end{eqnarray}

There are three types of sub-eikonal corrections to the shockwave approximation, as can be seen in Eq.~(\ref{MV}). First, there could be a non-trivial dependence on $x^-$ in lieu of the $\delta$ function. There could also be nonzero components to $A_{\rm cl}$ besides $A^+_{\rm cl}$, and finally there could be a dependence on $x^+$ as well, which we will not discuss here since it was found to be subsubleading~\cite{Chirilli:2018kkw}. The first type of correction is taken into account in computations with finite longitudinal size for the background field and is the most straightforward extension of the shockwave picture in Fig.~\ref{Smallx_Fig1}b. Going beyond the shockwave approximation in this case means that the kinematic conditions in Eq.~(\ref{prop1}) are loosened and there is some transfer of longitudinal momenta from the small-$x$ background to the hard part of the scattering. Loosening these conditions allows for TMD distributions at small $x$ to be defined with nonzero longitudinal phases, contrary to expressions such as Eqs.~(\ref{GWW1},\ref{GD1}), and thus overcome the cancellation of spin and angular momentum terms observed at small $x$~\cite{Hatta:2016aoc}. More frequently, these types of non-eikonal corrections are studied in describing the  transition between dilute and dense regions, specifically, the connection between the standard large-$x$ TMD framework and its small-$x$ counterpart we discussed previously \cite{Balitsky:2015qba,Balitsky:2016dgz}.

Spin effects are naturally described by the second type of non-eikonal corrections which correspond to taking nonzero transverse components of the background field $A_\perp$ into account. Strictly speaking, this type of non-eikonal correction can only arise in a background of the finite width as well. This makes computations with nonzero $A_\perp$ extremely difficult because of the competing non-eikonal effects of both first and second types, though there is considerable recent progress in this direction~\cite{Altinoluk:2014oxa,Altinoluk:2015gia,Chirilli:2018kkw,Agostini:2019avp,Cougoulic:2022gbk}.

In small-$x$ spin physics, and in its TMD applications, both types of corrections are related to the missing longitudinal phase in
Eqs.~(\ref{GWW1})-(\ref{GD1}), but for formal computations the second type of correction allows one to cast distributions in a form where the nonzero longitudinal width for the background field is not required. Then one can take the formal limit of the zero-width background field while  keeping nonzero $A_\perp$, thereby taking into account spin effects from the transfer of the polarization from the target in a more obvious way.

These strategies have been applied recently to spin-dependent TMDs and there are two realizations available. The first method \cite{Kirschner:1983di,Kirschner:1994rq,Kirschner:1994vc,Blumlein:1995jp,Bartels:1995iu,Bartels:1996wc,Blumlein:1996hb,Boussarie:2019icw} involves imposing the kinematic constraint 
at the level of Feynman diagrams.  In the second approach \cite{Kovchegov:2015pbl,Kovchegov:2016weo,Kovchegov:2018znm,Cougoulic:2022gbk}, the shockwave approximation is used. One can in the latter approach identify the structure of the operator describing the transfer of polarization from target to projectile at small $x$.

The spin dependent interaction is a sub-eikonal effect which is suppressed at high energies. For this reason the dominant contribution at small x has to be described by an operator which contains the smallest number of spin dependent interactions. In particular, such {\it helicity-dependent} operators have been constructed in \cite{Kovchegov:2015pbl,Kovchegov:2016weo} by calculating the quark production cross section in SIDIS on a longitudinally polarized proton or a nucleus. Similar to the unpolarized case the latter can be defined as a forward scattering amplitude for a color singlet longitudinally polarized quark-antiquark pair propagating in a background of a polarized target, see Fig. \ref{Smallx_Fig1}b. However, now the quark propagation contains exactly one sub-eikonal interaction carrying spin information from the target. As a result a concept of a polarized dipole amplitude can be introduced. This amplitude can be related to the quark helicity TMD, see discussion in \cite{Kovchegov:2015pbl}.

In the operator language this amplitude is defined by a correlator of a trace of a Wilson line which describes eikonal interactions with the target and a so-called polarized Wilson line operator. The latter is an extension of a regular light-cone Wilson line which contains a sub-eikonal helicity-dependent local operator insertion between two semi-infinite eikonal Wilson lines \cite{Kovchegov:2018znm}. In the case of gluon exchanges the sub-eikonal insertion is the $F_{12}$ component of the gluon field strength tensor. This form, which is counterintuitive in terms of TMD distributions since it resembles a very subleading twist distribution while actually appearing at leading twist, can be interpreted as arising from the scalar $\vec{\mu}\cdot \vec{B}=-\mu_z F^{12}$ for a quark with chromomagnetic moment $\vec{\mu}$ travelling through the chromomagnetic field $\vec{B}$. This operator structure was also obtained in \cite{Altinoluk:2014oxa,Jalilian-Marian:2017ttv,Chirilli:2018kkw,Tarasov:2019rfp}.

A significant difference between polarized and unpolarized scattering at small $x$ is that in the former case contribution of quark exchanges should be included already at the leading order. The helicity-dependent quark exchanges are of the same order as the sub-eikonal spin dependent gluon exchanges. In the case of quark exchanges the corresponding sub-eikonal insertion is provided by the non-local quark ''axial current" $\bar{\psi}(x_2)\gamma^+\gamma_5\psi(x_1)$ operator. The resulting operator which takes into account both effects \cite{Kovchegov:2016weo,Kovchegov:2018znm} is the polarized Wilson line operator\index{polarized Wilson line operator}
\begin{eqnarray}\label{VqG_decomp}
U^{\textrm{pol} [1]}(v_\perp) = U^{\textrm{G} [1]}(v_\perp) + U^{\textrm{q} [1]}(v_\perp),
\end{eqnarray}
where
\begin{eqnarray}
& U^{\textrm{G} [1]}(v_\perp)  = \frac{i \, g \, P^+}{s} \int\limits_{-\infty}^{\infty} d{v}^-\, W_n (v_\perp, +\infty, v^-) \, G^{12} (v^-, v_\perp) \, \, W_n(v_\perp, v^-, -\infty) \label{VG1}
\end{eqnarray}
and
\begin{eqnarray}
&U^{\textrm{q} [1]}(v_\perp)  &= \frac{g^2 P^+}{2 \, s} \int\limits_{-\infty}^{\infty} \!\! d{v}_1^- \! \int\limits_{v_1^-}^\infty d v_2^- W_n(v_\perp, \infty, v_2^-) \, t^b \, \psi_{\beta} (v_2^-,v_\perp) \, {\cW}_n^{ba} (v_\perp, v_2^-, v_1^-)
\nonumber\\
&&\times\left[ \gamma^+ \gamma^5 \right]_{\alpha \beta} \, \bar{\psi}_\alpha (v_1^-,v_\perp) \, t^a \, W_n(v_\perp, v_1^-, -\infty) . \label{Vq1}
\end{eqnarray}
This operator gives rise to the flavor singlet polarized dipole amplitude
\begin{eqnarray}\label{Q_def2}
&Q(v_\perp, w_\perp, x) &= \frac{s}{8 N_c P^+ V^-} \, \textrm{Re} \, \sum_{S_L} S_L \, \langle P, S_L| {\rm T} \, {\rm Tr} \left[ U(w_\perp) \, U^{\textrm{pol} [1] \, \dagger}(v_\perp) \right] 
\nonumber\\
&&+ {\rm T} \, {\rm Tr} \left[ U^{\textrm{pol} [1]}(v_\perp) \, U^\dagger(w_\perp) \right] | P, S_L\rangle_x\, ,
\end{eqnarray}
with the volume factor $V^- = \int d x^- d^2 x$.

Evolving this amplitude in the shockwave framework one can obtain the corresponding small-$x$ helicity evolution equations \cite{Kovchegov:2015pbl,Kovchegov:2016weo,Kovchegov:2018znm}. JIMWLK-type treatment of this evolution was constructed in~\cite{Cougoulic:2019aja}. The equations were obtained in both flavor singlet and non-singlet channels.\footnote{For the definition of the flavor non-singlet polarized dipole amplitude see \cite{Kovchegov:2016zex}.} The flavour singlet helicity evolution equations were solved numerically and analytically in large $N_c$ limit \cite{Kovchegov:2016weo,Kovchegov:2017jxc}. Flavor non-singlet equations at large $N_c$ were solved analytically in \cite{Kovchegov:2016zex}. A numerical solution of the large $N_c$ \& $N_f$ equations was obtained in~\cite{Kovchegov:2020hgb}.

However, operators (\ref{VG1}) and (\ref{Vq1}) are not the only sub-eikonal operators which contribute to the helicity evolution. As was shown in \cite{Cougoulic:2022gbk}, there is another operator
\begin{eqnarray}\label{Vi}
U^{i \, \textrm{G} [2]}(v_\perp) = \frac{P^+}{2 s} \, \int\limits_{-\infty}^{\infty} d {v}^- \, W_n(v_\perp, \infty, v^-) \, \left[ {D}^i (v^-, v_\perp) - \overleftarrow{D}^i (v^-, v_\perp) \right] \, W_n(v_\perp, v^-, -\infty)
\end{eqnarray}
which mixes with operators (\ref{VG1}), (\ref{Vq1}) in the helicity evolution. With this operator one can construct the polarized dipole amplitude
\begin{eqnarray}\label{Gj2}
&G^i (v_\perp, w_\perp, x) &= \frac{s}{8 N_c P^+ V^-} \, \sum_{S_L} S_L \, \langle P, S_L|  {\rm T} \, {\rm Tr} \left[  U^\dagger(w_\perp) \, U^{i \, \textrm{G} [2]}(v_\perp)\right] 
\nonumber\\
&&+  {\rm T} \, {\rm Tr} \left [U^{i \, \textrm{G} [2]\dagger}(v_\perp)  U(w_\perp) \right] | P, S_L\rangle_x.
\end{eqnarray}
which is a counterpart of the amplitude (\ref{Q_def2}).

Note that operator (\ref{Vi}) is {\it helicity-independent}. Indeed this operator arises naturally even in a scalar particle propagator in a shock-wave background of finite width when canonical momentum squared term $P^2_\perp$ is taken into account \cite{Hatta:2016aoc,Balitsky:2015qba,Chirilli:2018kkw,Altinoluk:2020oyd,Kovchegov:2021iyc}.

Operator (\ref{Vi}) is related to the Jaffe-Manohar gluon helicity PDF and can be obtained from the latter in the small-x limit, see \cite{Cougoulic:2022gbk} for details. As a result, evolution of the corresponding dipole amplitude by itself satisfies the pure-glue small-x polarized DGLAP evolution.

The small-x helicity evolution equations for amplitudes (\ref{Q_def2},\ref{Gj2}) which contain mixing between all three operators (\ref{VG1}, \ref{Vq1}, \ref{Vi}) were derived in \cite{Cougoulic:2022gbk}. In general, the evolution equations are not closed. They contain not only mixing between polarized quark and gluon exchanges but also include non-linear (saturation) terms with higher-order operators in the evolution kernel. Fortunately, the equations become closed in the large $N_c$ and large $N_c$ \& $N_f$ limits. 

The helicity evolution equations contain leading logarithms in the longitudinal integral in their kernels and the exact transverse integrations. In the double-logarithmic approximation (DLA) the large logarithm which is resumed by the helicity evolution equations is $\alpha_s \ln^2\big(\frac{1}{x}\big)$, i.e., two logarithms of energy for each power of the coupling constant. This is very different from the unpolarized small-x evolution where at the leading order the powers of $\alpha_s \ln\big(\frac{1}{x}\big)$ are resummed. \footnote{Recently a single-logarithmic correction $\alpha_s \ln\big(\frac{1}{x}\big)$  to the double-logarithmic kernel of the helicity evolution equations has been calculated as well \cite{Kovchegov:2021lvz}.}

The numerical solution of the large $N_c$ helicity evolution equations in the DLA approximation \cite{Cougoulic:2022gbk} gives the following asymptotic of the structure function
\begin{eqnarray}
&&g_1 (x, Q^2) \sim \left( \frac{1}{x} \right)^{3.66 \, \sqrt{\frac{\alpha_s \, N_c}{2 \pi}}} ,
\label{eq:g1-KPS}
\end{eqnarray}
 which is in complete agreement with the asymptotic obtained in the infrared evolution equations (IREE) approach of \cite{Bartels:1995iu,Bartels:1996wc}.
 
\index{transversity!small $x$}
Meanwhile, the shockwave approximation approach of Refs.~\cite{Kovchegov:2015pbl,Kovchegov:2016weo,Kovchegov:2018znm} has been extended to other spin-dependent distributions at small $x$, including the quark/gluon orbital angular momentum distribution~\cite{Kovchegov:2019rrz} and the quark transversity distribution~\cite{Kovchegov:2018zeq}. 
There has already been some progress on the phenomenology front, where a recent analysis of polarized inclusive DIS data incorporated the small-$x$ helicity evolution of the shockwave approach~\cite{Adamiak:2021ppq}.  More developments in this direction will be important to resolve the spin crisis with data from the EIC.

On the other hand, for a transversely polarized nucleon, the spin effects are not sub-eikonal and one finds that the naive-time-reversal-odd dipole gluon distributions can be described by a universal function~\cite{Boer:2015pni},
\begin{eqnarray}
xf_{1T}^{\perp g}=xh_{1T}^{ g}=x h_{1T}^{\perp g}=
\frac{-k_\perp^2 N_c}{4 \pi^2 \alpha_s } O_{1T}^\perp(x,k_\perp^2) \,,
 \end{eqnarray}
which is related to the so-called spin-dependent odderon $O_{1T}^\perp(x,k_\perp^2)$\index{odderon!spin-dependent}. The latter is defined through the dipole odderon operator of ${\rm Tr} \left [ U^{[\Box]}(0_T, y_T)
- U^{[\Box] \dag}(0_T,y_T) \right ]$~\cite{Hatta:2005as}. The spin-dependent odderon has been
considered in this way in~\cite{Zhou:2013gsa} and in many studies of
elastic scattering~\cite{Ryskin:1987ya,Buttimore:1998rj,Leader:1999ua}. Based on these developments, Ref.~\cite{Boussarie:2019vmk} has proposed to measure the small-$x$ gluon Sivers function through exclusive pion production in unpolarized electron-proton scattering in the forward region due to its connection to the QCD odderon\index{odderon}.

An important caveat to the above discussion is the possible role of topological effects due to the chiral anomaly, which has provoked considerable debate in the literature~\cite{Altarelli:1988nr,Carlitz:1988ab,Jaffe:1989jz}. The role of the chiral anomaly can be deduced from the first moment of $g_1$, which is equal to the quark helicity $\Delta \Sigma (Q^2)$ (plus a nearly constant term arising from a linear combination of iso-triplet and iso-octet axial charges). This quantity is given by the matrix element of the iso-singlet axial vector current $J_\mu^5$, which is not conserved, and in fact satisfies the famous anomaly equation 
\begin{equation}
    \partial^\mu J_\mu^5 = \frac{n_f \alpha_s}{2\pi} {\rm Tr}\left( G_{\mu\nu} {\tilde G}^{\mu\nu}\right)\,,
\end{equation}
with $n_f$ the number of light flavors, $G_{\mu\nu}$ the field strength tensor, and ${\tilde G_{\mu\nu}} = \frac{1}{2}\epsilon^{\mu\nu\rho\sigma} G_{\rho\sigma}$ is its dual. This is a statement of the explicit breaking of the $U_A(1)$ axial symmetry of QCD by quantum/topological effects. It was first argued by Veneziano that the problem of understanding the quark helicity of the proton is deeply tied to the $U_A(1)$ problem~\cite{Veneziano:1989ei}. Specifically, using anomalous chiral Ward identities, Shore and Veneziano~\cite{Shore:1990zu,Shore:1991dv} showed that 
\begin{equation}
    \Delta\Sigma(Q^2) \propto \sqrt{\chi^\prime(0)}\,,
    \label{eq:Shore-Veneziano}
\end{equation}
where $\chi^\prime (0)$ is the slope of the QCD topological susceptibility in the forward limit. Phenomenological estimates in this approach using QCD sum rules give estimates for $\Delta \Sigma$ that are in good agreement with HERMES and COMPASS data~\cite{Narison:1998aq,Narison:2021kny}. The computation of $\chi^\prime$ on the lattice has been discussed previously~\cite{Giusti:2001xh}; for a recent summary of computations of the topological susceptibility on the lattice, see \cite{Bali:2021qem}.

This ``topological screening" picture (specifically Eq.~(\ref{eq:Shore-Veneziano})) was recovered recently in a QFT worldline formalism~\cite{Tarasov:2020cwl,Tarasov:2021yll}. A remarkable result is that the chiral anomaly dominates $g_1$ not only in the Bjorken limit (which is consistent with an OPE analysis) but also in the Regge limit of $x_B\rightarrow 0$. In this framework, since the anomaly also dominates at small $x_B$, it is argued that its coupling to zero modes of the Dirac operator causes a breakdown of the eikonal expansion; instead, the cross-talk between the axial vector and pseudoscalar sectors in the form of a Goldberger-Treiman relation~\cite{Veneziano:1989ei,Shore:1990zu}, leads to spin diffusion through emergent axion-like dynamics. Here the axion is a primordial ${\bar \eta}$ meson which, through its coupling to the topological charge density, acquires mass and becomes the physical $\eta^\prime$ meson. Its dynamics concretely illustrates the connection between the $U_A(1)$ problem and the spin puzzle. 

Saturation at small $x_B$ introduces a novel twist to this picture. In 't Hooft's~\cite{tHooft:1976snw} formulation of the $U_A(1)$ problem, instanton-anti-instanton configurations saturate the topological charge density and thereby generate the mass of the $\eta^\prime$. However at small $x_B$, the topological susceptibility couples to the large density $\rho$ of color charges. These can cause ``over-the-barrier" sphaleron 
transitions~\cite{Klinkhamer:1984di}, previously suggested as a mechanism for electroweak baryogenesis~\cite{Kuzmin:1985mm}. 
While sphaleron-like transitions do not affect the $\eta^\prime$ mass, they introduce a drag effect~\cite{McLerran:1990de} that suppresses spin diffusion mediated by the ${\bar \eta}$. The sphaleron transition rate is governed by $Q_S$~\cite{Mace:2016svc}, and the corresponding drag on spin diffusion leads to a strong suppression of the isosinglet contribution to $g_1$ at small $x_B$ with an exponential dependence on the saturation scale. 

This suppression is qualitatively different from Eq.~(\ref{eq:g1-KPS}), which does not presume the existence of topological effects due to the chiral anomaly. Thus in principle, it 
should be possible to distinguish the two mechanisms at the EIC. 
If such a suppression is observed at the 
EIC, and confirmed by other non-inclusive measurements sensitive to the anomaly, it could provide first evidence for the existence of sphaleron transitions in QCD~\cite{Tarasov:2021yll}.

\subsection{Saturation and Multiple Scattering Effects for TMDs}
\label{sec:satMS}

Recent investigations~\cite{Altinoluk:2019fui, Altinoluk:2019wyu} have further extended the correspondence between small-$x$ observables and TMD physics by showing that any dilute-dense low-$x$ observable involving at most two colored particles in the final state can be rewritten entirely in terms of gluon TMD distributions. For example, the inclusive production of a forward dijet in $pA$ collisions is given as the convolution in transverse momentum transfer $k_\perp$ of hard scattering amplitudes with 2, 3 or 4 physical gluons with the corresponding 2, 3 and 4 gluon TMD distributions with the appropriate gauge link structures. 
This correspondence was extended using more fundamental gauge invariance arguments~\cite{Boussarie:2020vzf} in a way that can be systematically generalized. It relies on rewriting Wilson line operators into transverse strings built from so-called twisted field strength tensors $WG^{\mu\nu}(x)W^{\dagger}$,
where $W$ is a Wilson line of which $x$ is an end. In the Regge
limit, transverse gluon fields are pure gauges and they are the integrals
of twisted $WGW^{\dagger}$ tensors:
\begin{equation}
	A_{\perp}^{\mu}(z_{\perp})\equiv\frac{i}{g}U(z_{\perp})\partial_{\perp}^{\mu}U^{\dagger}(z_{\perp})=\int{\rm d}z^{-}[z^-,\infty]_{z_\perp} G^{\mu+}(z^{-},z_{\perp})[z^-,-\infty]_{z_\perp}.\label{eq:Aperp}
\end{equation}
As the reader can infer from Eq.~\ref{eq:derivativelink},  
these quantities are the fundamental building blocks which construct TMD
distributions at small longitudinal momentum transfer: for example,
the WW gluon distribution is none other than the Fourier transform
of $A_{\perp}^{\mu}(x_{\perp})A_{\perp}^{\nu}(y_{\perp})$. Pairs
of Wilson line operators can take the form of transverse strings built
from these pure gauge gluons. In the simplest case of a fundamental
dipole operator, one has:
\begin{equation}
	U(x_{\perp})U^{\dagger}(y_{\perp})={\cal P}\exp\left[ig\int_{y_{\perp}}^{x_{\perp}}{\rm d}z_{\perp}\cdot A_{\perp}(z_{\perp})\right],\label{eq:dip-cor}
\end{equation}
which defines the transverse string $\left[x_{\perp},y_{\perp}\right]$
built from pure gauge gluons. This relation is the small-$x$ limit
of a particular case of the non-Abelian Stokes formula,
\begin{equation}
	{\cal P}\exp\left[\oint_{{\cal C}}{\rm d}x_{\mu}A^{\mu}(x)\right]={\cal P}\exp\left[\int_{{\cal S}}{\rm d}\sigma_{\mu\nu}WF^{\mu\nu}W^{\dagger}\right],\label{eq:Stokes}
\end{equation}
where ${\cal C}$ is the square loop which links the points $(\infty^{+},x_{\perp})$, $(-\infty^{+},x_{\perp})$, $(\infty^{+},y_{\perp})$ and $(-\infty^{+},y_{\perp})$, and ${\cal S}$ is an appropriately chosen surface enclosed in ${\cal C}$\footnote{See~\cite{Boussarie:2020vzf} for details.}. Once we have the transverse
strings, we can use simple formulae such as
\begin{equation}
	\left[x_{\perp},y_{\perp}\right]=1+ig\int_{y_{\perp}}^{x_{\perp}}dz_{\perp}\cdot A_{\perp}(z_{\perp})[z_{\perp},y_{\perp}],\label{eq:left-inser}
\end{equation}
in order to perform an expansion in the powers of $gA_{\perp}$. A
remarkable recursive formula for the dipole operator can be deduced
from the results of~\cite{Boussarie:2020vzf}:
\begin{align}
	U(b_{\perp}+r_{\perp})U^{\dagger}(b_{\perp}) & =1+\int d^{2}v_{1\perp}\int\frac{d^{2}k_{1\perp}}{(2\pi)^{2}}e^{ik_{1\perp}\cdot(v_{1\perp}-b_{\perp})}ig(r_{\perp}\cdot A_{\perp})(v_{1\perp}){\cal H}_{1}(k_{1\perp},r_{\perp})\nonumber \\
	& +\int d^{2}v_{1\perp}d^{2}v_{2\perp}\int\frac{d^{2}k_{1\perp}}{(2\pi)^{2}}\frac{d^{2}k_{2\perp}}{(2\pi)^{2}}{\rm e}^{ik_{1\perp}\cdot(v_{1\perp}-b_{\perp})+ik_{2\perp}\cdot(v_{2\perp}-b_{\perp})}\label{eq:recursive-expansion}\\
	& \times ig(r_{\perp}\cdot A_{\perp})(v_{1\perp})U(v_{1\perp})U^{\dagger}(v_{2\perp})ig(r_{\perp}\cdot A_{\perp})(v_{2\perp}){\cal H}_{2}(k_{1\perp},k_{2\perp},r_{\perp}),\nonumber 
\end{align}
with
\begin{align}
 \label{eq:H1-int-form}
	{\cal H}_{1}(k_{1\perp},r_{\perp}) &=\int_{0}^{1}d\alpha e^{-i\alpha(k_{1\perp}\cdot r_{\perp})}
  , \\
\label{eq:H2-int-form}
	{\cal H}_{2}(k_{1\perp},k_{2\perp},r_{\perp}) & =\int_{0}^{1}d\alpha e^{-i\alpha(k_{1\perp}\cdot r_{\perp})}\int_{0}^{\alpha}d\beta e^{-i\beta(k_{2\perp}\cdot r_{\perp})}
.
\end{align}
This recursive relation allows for a straightforward expansion in
powers of $gA_{\perp}$. Once squared, it yields the complete rewriting
of the quadrupole operator into TMD distributions with infinite power
accuracy and takes into account all powers of the transverse momenta
and all twist corrections that are not suppressed by a power of the
center-of-mass energy. 

When discussing power expansions, it is customary to distinguish kinematic
effects from genuine higher twist effects. Eq.~(\ref{eq:recursive-expansion})
readily distinguishes both kinds of effects: genuine higher twists
come from $gA_{\perp}$ corrections, whereas kinematic power corrections
come from the expansion in the transverse momenta of the gluons $k_{1\perp},k_{2\perp}$...
which are intrinsic transverse momenta in the target hadron. In dense
targets, the CGC requires to resum all powers of $gA_{\perp}\sim1$,
while when dealing with dilute targets it is usually assumed to be
safe to neglect such powers. In that sense, all dilute frameworks
neglect the genuine higher twist corrections which are resummed by
the CGC EFT. There are, however, some interesting subtleties when comparing
dilute frameworks. Once higher powers of $gA_{\perp}$ in the CGC
formulas have been neglected, one recovers the so-called small-$x$
improved TMD framework which was constructed in \cite{Kotko:2015ura}.
The more standard dilute framework known as BFKL can be obtained from the improved TMD (iTMD) 
limit\index{improved TMD (iTMD) at small-$x$} by switching off all multiple scatterings from the gauge links that define the TMD distributions. 

It is actually expected that at large $k_\perp$, the gauge link structure of the distributions can be neglected as observed numerically in~\cite{Marquet:2016cgx, Marquet:2017xwy} and proven in~\cite{Altinoluk:2019wyu}, hence cancelling the first kind of multiple scatterings. As can be observed from the definition of the distributions~(\ref{GWW}) and~(\ref{GD}), the large $k_\perp$ regime corresponds to the regime of small transverse separation $r_\perp\sim 1/k_\perp$ between the physical gluons, while the small (semi-hard) $k_\perp$ corresponds to the regime of large transverse separation $r_\perp$, where the transverse distance will be filled by multiple gluons in the form of a gauge link. 

In this sense, the saturation scale $Q_s$  can be understood as the scale at which the separation $|r_\perp| \sim 1/|k_\perp|$ starts to be large enough for gauge links to matter in the distributions. Furthermore, the proof in~\cite{Altinoluk:2019wyu} leads to a subtle addition to the notion of the dilute limit articulated in \cite{Gelis:2010nm}: low-$x$ observables are only dilute at large $k_\perp$ when one applies the Wandzura-Wilczek approximation~\cite{Wandzura:1977qf}. This approximation amounts to neglecting higher genuine twist corrections and thus assuming $g_s A^\mu \ll 1$ in the projectile/target; this is the essence of the hybrid formalism developed in the context of the phenomenology of $p+A$ collisions~\cite{Dumitru:2005gt,Chirilli:2011km,Chirilli:2012jd}. Within the Wandzura-Wilczek approximation, and at small $k_\perp$, one should still expect multiple scatterings from the TMD in a dilute target, although the emergence of the gauge link structure would be postponed to lower values of $k_\perp$. In the CGC approach, this can be understood as the transition between dilute-dense and dilute-dilute regimes.

The dense-dense regime in the CGC does not have a $k_\perp$ factorized form~\cite{Gelis:2008rw}; however, results for single inclusive gluon distributions can be obtained numerically from solutions of the Yang-Mills equations with appropriate boundary conditions~\cite{Krasnitz:1998ns}. A similar transition in the context of quark pair production from dilute-dilute~\cite{Gelis:2003vh} to dilute-dense~\cite{Blaizot:2004wv,Fujii:2006ab} to dense-dense~\cite{Gelis:2005mn,Gelis:2015eua} have been worked out explicitly. 

The discussion above gives us an example of how powerful small-$x$ twist resummation tools are, and how much insight it gives for TMD physics. 
The non-universality of distributions can be fully understood as an effect due to multiple scatterings at small momentum transfer, which are very naturally taken into account in the small-$x$ effective theories. The decomposition of small-$x$ physics into different types of twists, which is uniquely written in a QCD gauge invariant way, leads to very interesting reinterpretations of well understood saturation effects, now in terms of TMD distributions. In the CGC, these distributions can correspond to dipoles but also quadrupoles, sextupoles and higher point Wilson line correlators that appear in semi-inclusive processes~\cite{JalilianMarian:2004da,Blaizot:2004wv,Dominguez:2011wm}; to leading logs in $x$, their evolution in $x$ can be computed by reformulating the JIMWLK equation as a Langevin equation~\cite{Blaizot:2002np,Dumitru:2011vk}.

\subsection{Outlook}
\label{sec:smallxOutlook}

In summary, there has been great progress in the last few years on TMDs at small-$x$, mainly on the connection between the TMD factorization and the small-$x$ CGC formalism. The ultimate goal is to extend the theoretical and phenomenological investigations of the two frameworks with the aim of obtaining a unified picture of parton distributions in the high parton density regime. A number of challenging issues need further investigations:
\begin{enumerate}

\item {\it Proton spin at small-$x$.} 
Recent progress has generated strong interest in the community to understand the proton spin structure at small-$x$ from the associated small-$x$ evolution equations. More theoretical efforts are needed to resolve the issues raised in these derivations which do not take into account topological effects~\cite{Kovchegov:2015pbl,Kovchegov:2016weo,Kovchegov:2018znm,Boussarie:2019icw,Cougoulic:2019aja} or instead take these into account~\cite{Tarasov:2020cwl,Tarasov:2021yll}. The final answer to these questions will provide important guidance for novel physics at the future EIC, where proton spin rum rule is a major focus.  

\item {\it Small-$x$ evolution of the TMD gluon distributions~\cite{Dumitru:2011vk}.} The theoretical framework exists to solve the small-$x$ evolution equations for the dipole and WW gluon distributions. One needs to develop an efficient program to numerically solve these equations and gain insight into the TMD gluon distributions at different $x$. The combination of theory developments and phenomenological applications to the experimental data will help clarify the role of parton dynamics relative to those of ``dipoles" and ``quadrupole" effective degrees of freedom in the high parton density regime. 

\item {\it Systematic study of gluon distributions at small $x$ to reach a quantitative level.} There has been tremendous progress in small-$x$ phenomenology in the last decade. It is important to continue these studies, but focus on the relevant TMD gluon distributions. In particular, one needs to investigate the role played by the polarization (of the gluon or the target nucleon) in the small-$x$ gluon TMDs. It has been shown that WW distribution of linearly polarized gluons is suppressed at small $k_\perp$ as compared to the dipole gluon distribution~\cite{Metz:2011wb}. It was also shown in~\cite{Boer:2017xpy, Altinoluk:2019wyu} that at large $k_\perp$ linearly polarized gluons are extremely important since the unpolarized and Boer-Mulders TMD become equal in that limit regardless of their gauge link structure\footnote{Note that Ref.~\cite{Boer:2017xpy}, however, shows a suppression from Sudakov resummations in the TMD evolution.}. Similarly, the target polarization may also affect the gluon distribution, such as the gluon Sivers function at small $x$~\cite{Kovchegov:2013cva,Kovchegov:2020kxg}.
\index{Sivers function $f_{1T}^{\perp}$!gluon}
There is much to explore along these directions, in particular, in light of future experiments at the EIC. 

\item {\it Further exploration of  probes for the TMD quark/gluon distributions in the small-$x$ region.} With the EIC on the horizon, one needs to address critical questions concerning  direct probes for the TMD gluon distributions at small $x$.  In particular, one of the key issues is the universality of distributions in the CGC formalism, as well as the matching of computations in the small-$x$ formalism to those in the TMD formalism at large transverse momenta. Next-to-leading order computations are now available for diffractive dijet~\cite{Boussarie:2016bkq} as well as inclusive photon+dijet production~\cite{Roy:2019hwr,Roy:2019cux} in $e+A$ collisions:~these results will be useful in extending the matching of the two formalisms to higher orders in perturbation theory. A specific example where such matching studies has led to significant phenomenological progress is in quarkonium production at collider energies~\cite{Kang:2013hta,Ma:2014mri,Ma:2015sia,Qiu:2013qka}. Such studies can be extended to DIS where quantitative predictions and comparisons with data will also provide crucial tests of the universality of QCD dynamics in the saturation regime of the theory. 
\end{enumerate}

%% file: sec-jets/sec-jets.tex
\section{Jet Fragmentation}
\label{sec:JetFrag}

\index{hadronic jets} Hadronic jets~\cite{Sterman:1977wj}, collimated showers of energetic 
final-state particles, have long been regarded as an essential tool to understand hard scattering  ($Q^2 \gg \lqcd^2$) processes in $e^+e^-$ collisions,  semi-inclusive deep inelastic scattering, and hadron-hadron collisions  from first principles in QCD. As they are copiously produced~\cite{Feynman:1978dt}, jets are easily accessible by experiment, and their discovery has stimulated some of the most important developments in the perturbation theory of strong interactions. At present,  cross sections for processes involving jets are  routinely calculated at  next-to-leading order, and next-to-next-to-leading order results are also becoming available~\cite{Campbell:2006wx,Ellis:2007ib,Boughezal:2015ded,Boughezal:2015dra}. This remarkable theoretical accuracy combined with careful uncertainty analysis~\cite{Olness:2009qd} has  made precision  jet studies  a promising  method  to search for new physics  beyond the  Standard Model at very high energies. 
In more complex environments, jet observables can differentiate between models and theories of parton shower formation~\cite{Vitev:2008rz,Vitev:2009rd}. This is exemplified by the recent CMS measurements of the radius dependence of the suppression of inclusive jets~\cite{CMS}.

In addition to inclusive and tagged jet cross sections, studies of \index{jet substructure} jet substructure provide precision tests of perturbative QCD in high energy processes. They originate from the studies of event shapes in $e^+e^-$ collisions, which helped test and confirm the gauge  structure of QCD~\cite{Farhi:1977sg, Georgi:1977sf, PhysRevLett.41.1581, PhysRevLett.41.1585, Heister:2003aj, Abdallah:2003xz, Achard:2004sv, Abbiendi:2004qz}. Accurate event shape calculations have allowed for   some of the most precise extractions of the strong coupling constant~\cite{Becher:2008cf, Chien:2010kc, Davison:2008vx,Abbate:2010xh,Gehrmann:2012sc,Hoang:2015hka,Kardos:2018kqj}. At hadron colliders, due to the presence of beam remnants, underlying event and pileup, the studies  of jet observables  become much more complicated.  Considerable effort  and progress have been made in the direction of  more  efficient jet reconstruction and the development of grooming techniques to  achieve this goal~\cite{Altheimer:2013yza}. 
In the past decades jets have increasingly been used to constrain essential perturbative and  nonperturbative aspects of QCD. 
For example,  jets are now routinely used  to constrain the PDFs in hadronic collisions~\cite{Ball:2010de,Dulat:2015mca}.  They are particularly useful in constraining the large-$x$ gluon distributions~\cite{Stump:2003yu}.  

Jets are not  fundamental objects in nature in the way that  hadrons are, but are reconstructed by grouping final-state particles via an algorithm. Different \index{jet algorithms} jet algorithms, so long as they maintain  infrared and collinear safety, provide different opportunities to probe QCD dynamics. It is desirable that such algorithms exhibit reduced sensitivity to the physics of hadronization, are  applicable at the detector level, and can be identically implemented for partons and final-state particles.
For one of the earliest examples of an analytic calculation with a fixed cone radius see~\cite{Sterman:1977wj}.  It is possible to classify most modern jet algorithms into one of two broad classes:  cone algorithms and  sequential clustering algorithms~\cite{Cacciari:2011ma}. Examples of the former are the Midpoint Cone, Iterative Cone, and Seedless Cone\cite{Salam:2007xv,Cacciari:2008gp}. Only the Seedless Cone is infrared and collinear safe. The sequential recombination algorithms  include the $k_T$, Cambridge/Aachen, and the anti-$k_T$~\cite{Catani:1993hr}. All of these satisfy the above criteria in addition to being clean and simple.

The $k_T$-class of algorithms in $pp$ collisions are based on a pair of distance measures, $d_{ij}$ measuring an inter-particle distance and $d_{iB}$ measuring a particle-beam distance. 
\begin{align}
\label{eq:ktmeaspp}
    d_{ij} &= \min(p_{Ti}^{2p},p_{Tj}^{2p})\frac{\Delta R_{ij}^2}{R^2}\,,\quad\text{where } \Delta R_{ij}^2 = (y_i-y_j)^2+(\phi_i-\phi_j)^2\,, \\
    d_{iB} &= p_{Ti}^{2p} \,, \nn
\end{align}
where $p_{T}$ is the transverse momentum, $y$ is the rapidity, and $\phi$ the azimuthal angle, all with respect to the hadronic beam axis. These measures are designed to be boost-invariant along the beam axis. Here $p$ is a parameter, which we  will choose to be $p=-1$, which yields the \emph{anti-$k_t$} algorithm \cite{Cacciari:2008gp}.
The anti-$k_t$ algorithm favors grouping energetic collinear particles with one another first, before collecting soft particles into the jets. By finding the minimum distance measure the particles can recombined or identified as a jet if this minimum is given by $d_{iB}$.

What all jets have in common is  a finite radius parameter $R$,  setting a transverse scale $\omega_J R$,  where $\omega_J$ is the light-cone energy of the jet.
Thus,  evaluation of jet production and jet substructure always requires control over the transverse momentum QCD dynamics.  One important 
problem related to the use of jets as probes is to develop improved methods to distinguish quark-initiated from gluon-initiated jets~\cite{Gallicchio:2011xq,Gallicchio:2012ez}. The jet charge is one observable~\cite{Berge:1980dx,Albanese:1984nv}
 that is sensitive to the flavor origin and has recently been measured at the LHC~\cite{Aad:2015cua,Sirunyan:2017tyr}.  Individual 
flavor jet charges remain distinct even in collisions with heavy nuclei~\cite{Chen:2019gqo,Li:2019dre} and first steps toward their measurement have been taken in such collisions~\cite{CMScharge}.  Jet substructure and jet fragmentation functions in particular can be used to 
probe the nonperturbative physics of hadronization in ways not possible with more inclusive 
measurements~\cite{Bain:2017wvk,Anderle:2017cgl}.

In the presence of nuclear matter, jet production is  sensitive to its transport properties.  In general, jet substructure observables are primarily dependent on the details of the final state  \index{jet-medium interactions} jet-medium interactions. They allow us to disentangle the initial state  cold nuclear matter effects and are cleaner 
probes of the medium properties when compared to inclusive cross sections~\cite{Chien:2015hda}. 
At the same time, different jet substructure observables are sensitive to radiation at different energy scales. By measuring jet cross sections, jet shapes and jet fragmentation functions, 
jet masses and particle multiplicities, the in-medium jet formation mechanism across a wide range of energy scales can be examined~\cite{Adamczyk:2017yhe,Sirunyan:2018gct,Sirunyan:2018qec,Aaboud:2019oac,Sirunyan:2018ncy,Aaboud:2018hpb}. Jet substructure observables and their medium modifications are also highly dependent on the partonic origin of jets.

\subsection{Jets as Probes of TMD PDFs}
\label{sec:jet-TMDPDFs}

\begin{figure}[htb]
    \centering
    \includegraphics[width=3.0in]{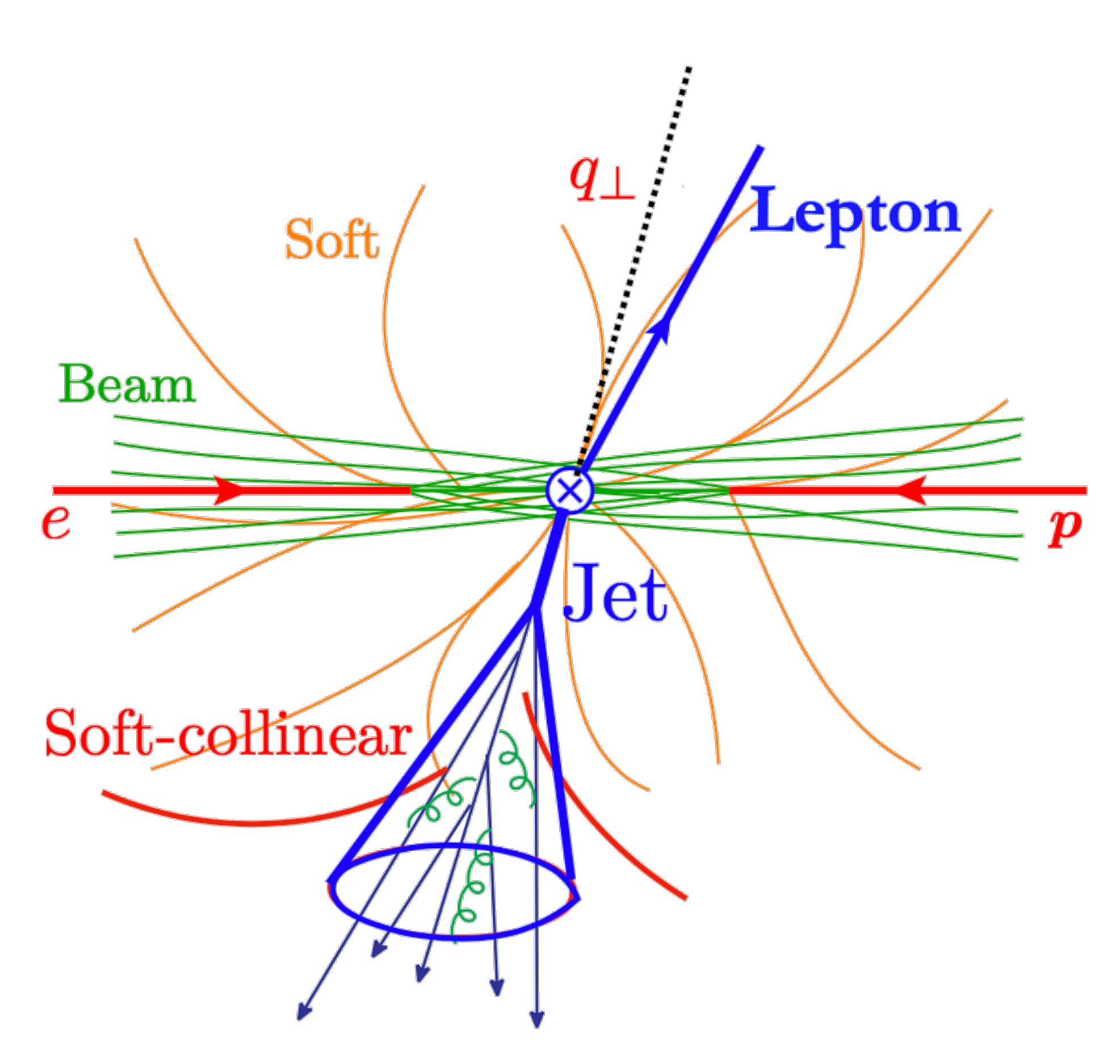}
    \caption{Illustration of back-to-back dijet production in transversely polarized $ep$ collisions: $p(P, \bm{S}_T) + e(\ell) \to J(y_J, \bm{P}_{JT})+e(\ell') +X$. The lepton-jet transverse momentum imbalance is defined as $\bm{q}_T = \bm{P}_{JT}+\bm{\ell}'_{T}$.}
    \label{fig:epjet}
\end{figure}

Jet production in unpolarized and polarized $ep$ collisions can be sensitive to the transverse motion of the partons inside the nucleon. For example, recently, production of back-to-back electron+jet in $ep$ collisions has been proposed as a probe of both unpolarized and polarized TMD PDFs~\cite{Liu:2018trl,Arratia:2020nxw,Liu:2020dct,Kang:2021ffh}, such as quark Sivers functions. In such a process, $p(P, \bm{S}_T) + e(\ell)+ \to J(y_J, \bm{P}_{JT})+e(\ell')+X$, one defines the transverse momentum imbalance, $\bm{q}_T = \bm{P}_{JT}+\bm{\ell}'_{T}$ and the average transverse momentum of the electron-jet, $\bm{P}_T = \left(\bm{P}_{JT}-\bm{\ell}'_{T}\right)/2$, as shown in Fig.~\ref{fig:epjet}.
In the back-to-back region where $q_T\ll P_T$, one can derive a TMD factorization as follows~\cite{Kang:2021ffh},
\begin{align}
    \frac{d\sigma}{dy_J dP_T^2 d^2\bm{q}_T} =  \hat{\sigma}_0 H(Q, \mu)\,\sum_q e_q^2 \, J_q(P_{T} R, \mu) & \int\frac{d^2\bm{b}_T}{(2\pi)^2}\, e^{i\bm{b}_T\cdot \bm{q}_T}\, x\, \tilde{B}_{q/p}(x, \bm{b}_T, \mu, \zeta/\nu^2) 
    \nonumber \\
    &\times \tilde{S}_q^{\rm global}(\bm{b}_T, \mu, \nu)\, \tilde{S}_q^{\rm cs}(\bm{b}_T, R, \mu)\,,
    \label{eq:jet-bare}
\end{align}
for electron-jet production in $ep$ collisions. Here, $\hat{\sigma}_0$ is the Born cross section for the unpolarized electron and quark scattering process, while $H(Q, \mu)$ is the hard function taking into account virtual corrections at the scale $Q$, with $Q^2 = -(\ell - \ell')^2$ denoting the virtuality of the exchanged photon. On the other hand, $J_q(P_{T}R, \mu)$ is the quark jet function~\cite{Ellis:2010rwa} which describes the production of the outgoing jet from a hard interaction. $\tilde{B}_{q/p}(x,b_T, \mu, \zeta/\nu^2)$ is the quark beam function given in Eq.~\eqref{eq:sigma_new_b}, $\tilde{S}_q^{\rm global}(\bm{b}_T, \mu, \nu)$ is a global soft function describing soft gluons of momentum $\sim q_T$ at arbitrary angles while $\tilde{S}_q^{\rm cs}(\bm{b}_T, R, \mu)$ is the collinear-soft function that describes soft gluon radiation close to the jet direction and able to probe the boundary of radius $R$. Note that the global soft function has rapidity divergence as indicated by the $\nu$-dependence, while the collinear-soft function does not. 

In general, the above factorization formula is more complex in its structure in comparison with the standard TMD processes such as SIDIS, Drell-Yan and $e^+e^-$ collisions. In particular, additional soft functions are involved in the formalism where jets are produced, while only a single soft function is required for the standard TMD processes. This provides additional complications in establishing rigorously the relationship between the TMD PDFs probed in the jet process and those standard TMD PDFs, in particular the role of these additional soft functions in the nonperturbative (or small transverse momentum) region. On the other hand it is precisely because of the richer structure in the soft functions that jet production might provide novel insights into TMD dynamics and the TMD PDFs in the nonperturbative region, which otherwise can not be extracted from the standard TMD processes. 

\begin{figure}[htb]
    \centering
    \includegraphics[width=4.2in]{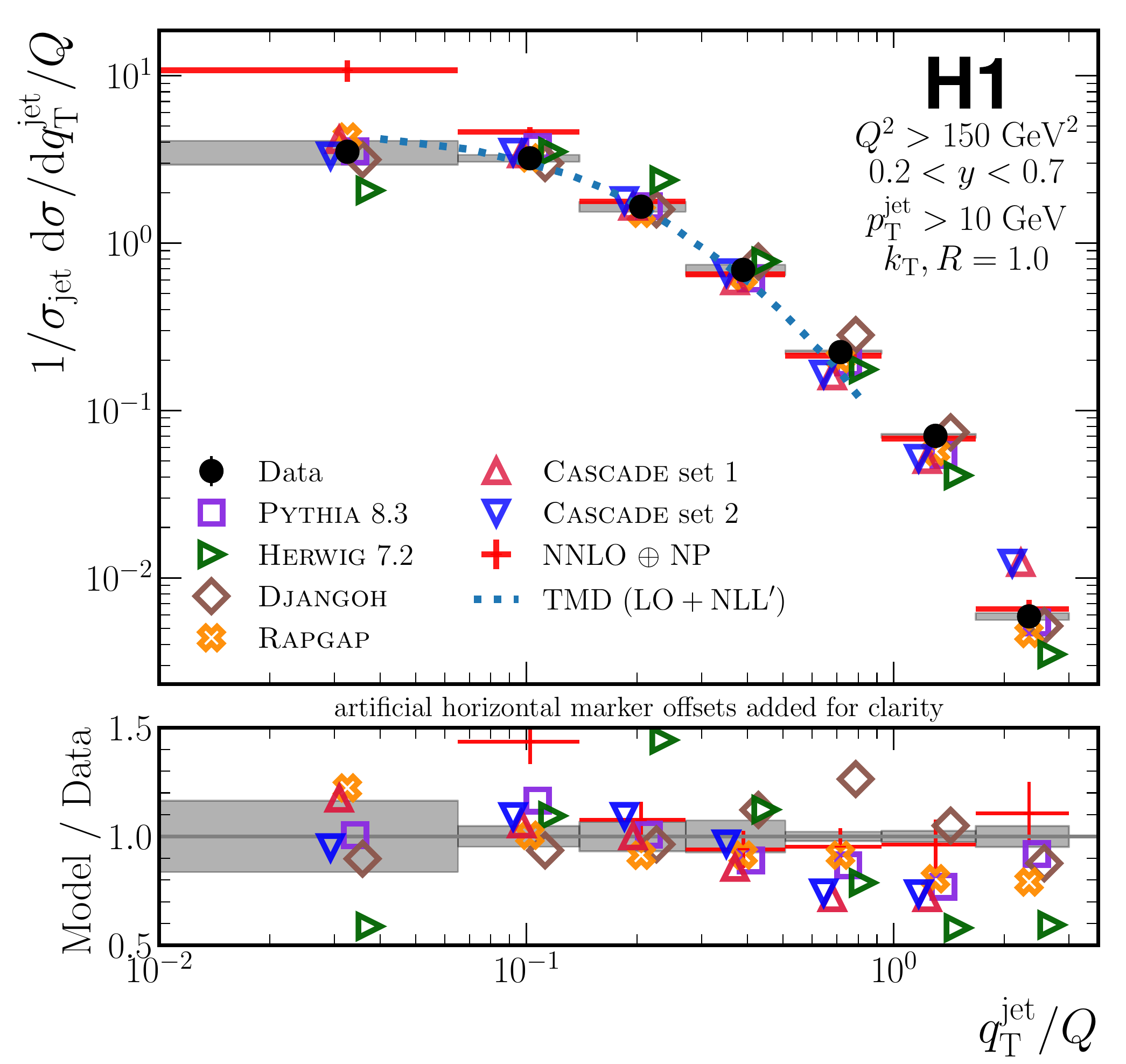}
    \caption{Theoretical comparison with the experimental data from HERA for lepton-jet back-to-back production. The prediction of TMD factorization is shown by the dashed blue line.  Figure from Ref.~\cite{H1:2021wkz}.}
    \label{fig:HERA_ejet}
\end{figure}

In the perturbative region ($1/b_T \gg \Lambda_{\rm QCD}$), one can show that at the next-to-leading order, 
\begin{align}
    \tilde{B}_{q/p}(x, \bm{b}_T, \mu, \zeta/\nu^2) \tilde{S}_q^{\rm global}(\bm{b}_T, \mu, \nu) = \tilde{f}_{q/p}(x, \bm{b}_T, \mu, \zeta) \tilde{S}_q^{\rm global}(\bm{b}_T, \mu)\,,
\end{align}
where we have used Eq.~\eqref{eq:fisBS} and redefined a global soft function $\tilde{S}_q^{\rm global}(\bm{b}_T, \mu)$ that is free of rapidity divergence, 
\begin{align}
   \tilde{S}_q^{\rm global}(\bm{b}_T, \mu, \nu) =  \tilde{S}_q^{\rm global}(\bm{b}_T, \mu) \,
    \sqrt{\tilde S_{n_a n_b}(b_T,\mu,\nu)}\,,
\end{align}
with the standard soft function $\tilde S_{n_a n_b}(b_T,\mu,\nu)$ given in Eq.~\eqref{eq:fisBS} and the NLO expression for $\tilde{S}_q^{\rm global}(\bm{b}_T, \mu)$ given in~\cite{Kang:2021ffh}. With such a procedure, we can then rewrite the factorization formula in Eq.~\eqref{eq:jet-bare} in terms of a standard TMD PDF $\tilde{f}_{q/p}(x, \bm{b}_T, \mu, \zeta)$ as follows 
\begin{align}
    \frac{d\sigma}{dy_J dP_T^2 d^2\bm{q}_T} =  \hat{\sigma}_0 H(Q, \mu)\,\sum_q e_q^2 \, J_q(P_{T} R, \mu) & \int\frac{d^2\bm{b}_T}{(2\pi)^2}\, e^{i\bm{b}_T\cdot \bm{q}_T}\, x\, \tilde{f}_{q/p}(x, \bm{b}_T, \mu, \zeta) 
    \nonumber \\
    &\times \tilde{S}_q^{\rm global}(\bm{b}_T, \mu)\, \tilde{S}_q^{\rm cs}(\bm{b}_T, R, \mu)\,.
    \label{eq:jet-standard-TMD}
\end{align}
Most recently, the H1 collaboration at HERA has performed the first measurement of lepton-jet momentum imbalance in lepton-proton scattering~\cite{H1:2021wkz}. As shown in the dashed curve in Fig.~\ref{fig:HERA_ejet}, the above TMD factorization formula gives a decent description of the experimental data at low momentum imbalance $q_T$. 

\begin{figure}[t!]
    \centering
    \includegraphics[width=4.2in]{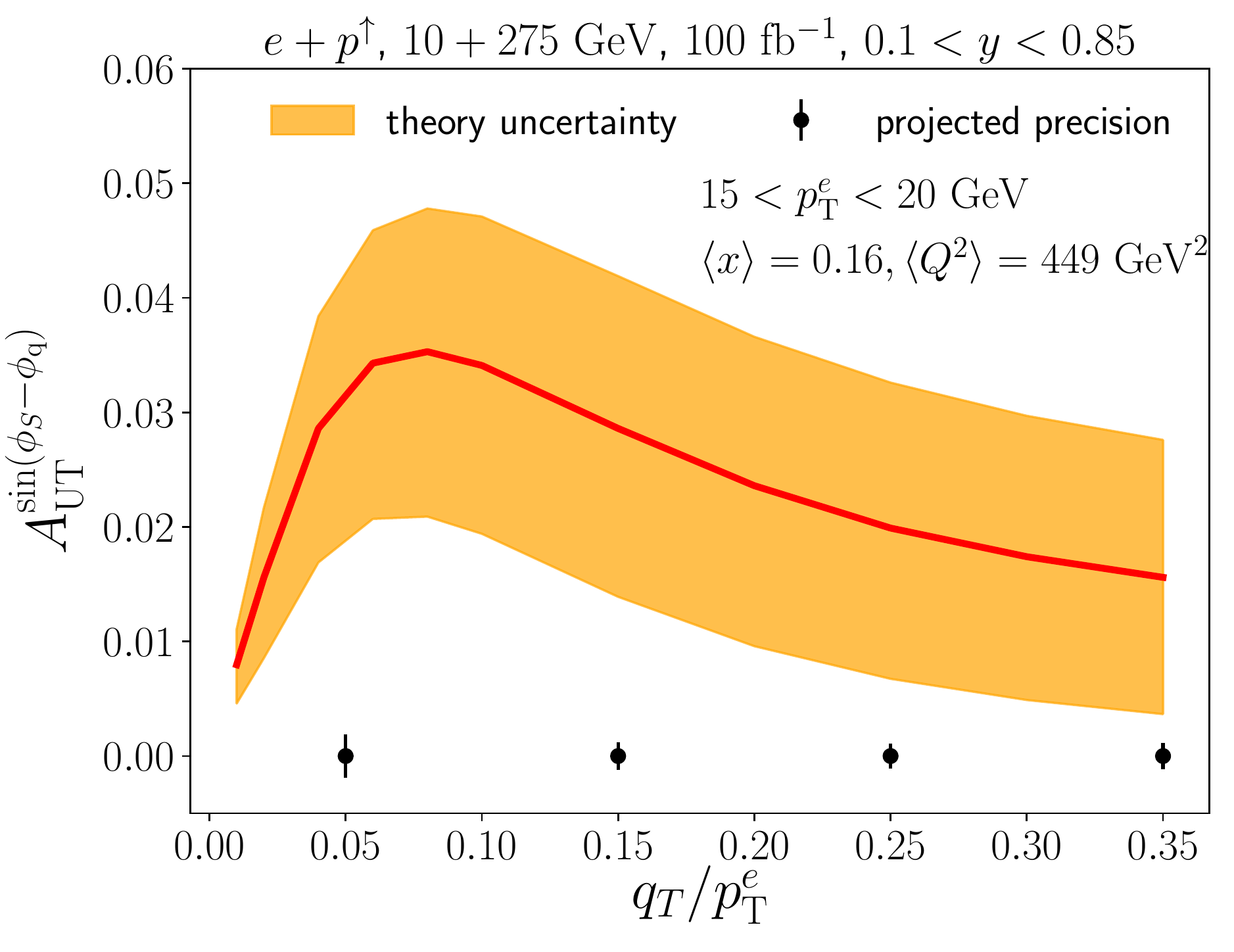}
    \caption{Theoretical result for the electron-jet asymmetry sensitive to the Sivers distribution (red). The uncertainty band (orange) displays the current uncertainty of the Sivers function. In addition, we show projections of statistical uncertainties for an EIC measurement (black error bars). Figure from Ref.~\cite{Arratia:2020nxw}.}
    \label{fig:Sivers_ep}
\end{figure}

In addition, it has been demonstrated in~\cite{Liu:2018trl,Arratia:2020nxw,Liu:2020dct} that lepton-jet production in $ep$ collisions also shows great potential in constraining the quark Sivers functions. 
\index{Sivers function $f_{1T}^{\perp}$!phenomenology}
See a recent study as shown in Fig.~\ref{fig:Sivers_ep}. Here the theoretical uncertainty in the Sivers asymmetry $A_{UT}^{\sin(\phi_S-\phi_q)}$ is computed using current knowledge of the quark Sivers functions and is shown as the orange band. On the other hand, the projections of statistical uncertainties for the EIC measurements are also shown as black error bars. Obviously there is much room for improvement in the accuracy of the theoretical predictions in order to meet the challenge of the anticipated experimental precision.  In order to perform flavor separation for different quark Sivers functions at the EIC, jet charge observables have been proposed in~\cite{Kang:2020fka}, which shows that jet charge measurements can substantially enhance the sensitivity of spin asymmetries to different partonic flavors in the nucleon. 

Besides lepton-jet production in $ep$ collisions where TMD factorization holds, one can also study back-to-back dijet and vector-boson-jet (such as $Z+{\rm jet}$ or $\gamma+{\rm jet}$) production in $pp$ collisions. These processes are usually even more involved, and TMD factorization typically breaks~\cite{Collins:2007nk,Rogers:2010dm}. Nevertheless, theoretical studies have been performed~\cite{Buffing:2018ggv,Chien:2019gyf,Kang:2020xez,Liu:2020jjv} and experimental measurements have also been undertaken~\cite{Aaij:2019ctd,bnltalk} for such processes. This would allow for probing potential TMD factorization breaking, or for constraining TMD PDFs if the breaking is relatively small. 

\subsection{Jet Substructure and Jet Fragmentation}
\label{sec:jetsubjetfrag}

The momentum distribution of hadrons inside a fully reconstructed jet, commonly referred to as the \index{jet fragmentation function (JFF)} jet fragmentation function (JFF)~\cite{Procura:2009vm,Jain:2011xz,Jain:2011iu,Chien:2015ctp,Kang:2019ahe,Arleo:2013tya,Kaufmann:2015hma,Kang:2016ehg,Dai:2016hzf,Kang:2017yde,Bain:2017wvk,Wang:2020kar}, has received increasing attention in recent years. The JFF probes the parton-to-hadron fragmentation function at a differential level and can thus provide new insights for the hadronization process. Jet fragmentation functions can be measured for either {\it inclusive} jet production or {\it exclusive} jet processes. Single inclusive jet production correspond to the process $AB\to \mathrm{jet}+X$, where incoming particles $A$ (or $B$) can be either a lepton or a proton, and one sums over all particles in the final state $X$ besides the observed jet. The factorization formalism for single inclusive jet production has a similar form as that for single inclusive hadron production, where one replaces the usual collinear fragmentation function $D_{h/i}(z, \mu)$ by a \index{semi-inclusive jet function} semi-inclusive jet function $J_i(z,p_T R, \mu)$. For example, the differential cross section for single inclusive jet production in $pp$ collisions can be written as~\cite{Kang:2016mcy,Kaufmann:2015hma}
\begin{align}
\frac{d\sigma^{pp\to {\rm jet}X}}{dp_Td\eta}  = &\sum_{a,b,c}
\int_{\xi_a^{\rm min}}^1\frac{d\xi_a}{\xi_a} f_a(\xi_a,\mu)
\int_{\xi_b^{\rm min}}^1\frac{d\xi_b}{\xi_b} f_b(\xi_b,\mu) 
\nonumber\\
&\times \int^1_{z_c^{\rm min}} \frac{dz_c}{z_c^2} H_{ab\to c}(\hat s,\hat p_T,\hat \eta,\mu) \,J_c(z_c,p_T R,\mu)\,,
\label{eq:inclusive-jet}
\end{align}
where $p_T$ and $\eta$ are the transverse momentum and the rapidity for the jet. The hard function $H_{ab\to c}$ depends on the partonic CM energy $\hat{s} = \xi_a \xi_b s$, the partonic transverse momentum $\hat{p}_T =p_T/z_c$ and the partonic rapidity $\hat{\eta} = \eta - \ln(\xi_a/\xi_b)/2$. The semi-inclusive jet function $J_c(z_c,p_T R,\mu)$ describes the transition from a parton $c$ with transverse momentum $\hat{p}_T$ to the jet with transverse momentum $p_T$ and jet radius $R$. Note that since the only measured hard momentum scale is the jet $p_T$, the process is sensitive to the collinear PDFs $f_a(\xi_a, \mu)$ and $f_b(\xi_b, \mu)$, just like the case for single inclusive hadron production~\cite{Owens:1986mp}. 

On the other hand, for exclusive jet processes $AB\to n\,\mathrm{jets}$, one measures a fixed number of signal jets but vetoes additional jets. For example, when measuring dijet production, by selecting the kinematics to be in the back-to-back configuration, we restrict the events to be those with exactly two jets in the selected kinematic region. Just as shown in \sec{jet-TMDPDFs}, the factorization formalism for such exclusive jet production processes are different from that of single inclusive jet production. For example, we see clearly that the back-to-back dijet production in $ep$ collisions is sensitive to the TMD PDFs. One also notices that the semi-inclusive jet function $J_i(z, p_T R, \mu)$ is replaced with the exclusive jet function $J_i(p_T R, \mu)$ in Eq.~\eqref{eq:jet-standard-TMD}. 

In both single inclusive jet and exclusive jet production cases, one can further measure the distribution of hadrons inside the jet. One usually characterizes such a hadron distribution by the longitudinal momentum fraction $z_h$ of the jet carried by the hadron and the transverse momentum $j_\perp$ with respect to the jet direction. For example, for single inclusive jet production in $pp$ collisions, $pp\to \left({\rm jet}(\eta, p_T, R)\, h(z_h, {\bm j}_\perp)\right)+X$, one measures the hadron distribution inside the jet
\begin{align}
F(z_h, {\bm j}_\perp; \eta, p_T, R) = \left.\frac{d\sigma^{pp\to (\text{jet}\,h)X}}{dp_Td\eta dz_h d^2 {\bm j}_\perp}
\right/\frac{d\sigma^{pp\to\text{jet}X}}{dp_Td\eta }\,, 
\end{align}
where $F(z_h, {\bm j}_\perp; \eta, p_T, R)$ is commonly referred to as the JFF, and the numerator and denominator are the differential jet cross sections with and without the reconstruction of the hadron $h$ inside the jet.  
The large light-cone momentum fraction of the jet carried by the hadron $h$ is denoted by $z_h$ and ${\bm j}_\perp$ is the transverse momentum of the hadron with respect to the standard jet axis. The factorization formula for the hadron distribution inside the single inclusive jet production can be written as
\begin{align}
\frac{d\sigma^{pp\to (\text{jet}\,h)X}}{dp_Td\eta dz_h d^2 {\bm j}_\perp}  = &\sum_{a,b,c}
\int_{\xi_a^{\rm min}}^1\frac{d\xi_a}{\xi_a} f_a(\xi_a,\mu)
\int_{\xi_b^{\rm min}}^1\frac{d\xi_b}{\xi_b} f_b(\xi_b,\mu) 
\nonumber\\
&\times \int^1_{z_c^{\rm min}} \frac{dz_c}{z_c^2} H_{ab\to c}(\hat s,\hat p_T,\hat \eta,\mu)\, {\cal G}_c^h(z_c,p_T R, z_h, {\bm j}_\perp,\mu, \zeta_J)\,.
\label{eq:inclusive-jet-h}
\end{align}
In other words, the factorizations for the numerator and the denominator are very similar to each other. For jet production with hadron distribution inside the jet, one simply replaces the semi-inclusive jet function $J_c(z_c, p_T R, \mu)$ in Eq.~\eqref{eq:inclusive-jet} by the semi-inclusive \index{TMD fragmenting jet function} TMD fragmenting jet function (TMD FJF) ${\cal G}_c^h(z_c, p_T R, z_h, {\bm j}_\perp,\mu, \zeta_J)$ in Eq.~\eqref{eq:inclusive-jet-h} to be defined below. As expected, since this is a TMD observable, we have a Collins-Soper scale $\zeta_J$. 

Jet fragmentation functions have been measured for single inclusive jets produced in unpolarized proton-proton collisions at the Large Hadron Collider (LHC) for light hadrons~\cite{Aad:2011sc,Aaboud:2017tke}, for open heavy flavor mesons~\cite{Aad:2011td,CMS:2018ovh,Acharya:2019zup}, and for heavy quarkonium~\cite{Aaij:2017fak,CMS:2018mjn}. Such measurements have already started to constrain the fragmentation functions for open heavy flavor mesons~\cite{Chien:2015ctp,Anderle:2017cgl}, and to pin down non-relativistic QCD (NRQCD) long-distance matrix elements, which characterize the hadronization process for heavy quarkonium production~\cite{Kang:2017yde,Bain:2017wvk}, see \sec{jetquarkonia}. At the same time, there are also important exclusive-type jet measurements at the LHC, e.g., exclusive jet production associated with vector bosons. See~\cite{Aaboud:2019oac} for a recent JFF measurement for photon-tagged jets. More recently the LHCb collaboration has measured both longitudinal and transverse momentum distributions of charged hadrons produced inside $Z$-tagged jets in the forward rapidity region in proton-proton collisions~\cite{Aaij:2019ctd}, $p+p\to Z+{\rm jet}+X$. At the same time, there have been recent studies for hadron distributions inside the jet in the back-to-back lepton-jet production in $ep$ collisions~\cite{Arratia:2020nxw,Kang:2021ffh}, a process that is very promising at the future EIC.

\FloatBarrier
\subsection[Hadron longitudinal distribution inside jets: \texorpdfstring{$z_h$}{zh} dependence]{\boldmath Hadron longitudinal distribution inside jets: $z_h$ dependence}
\FloatBarrier

If one measures only the longitudinal $z_h$ distribution of hadrons inside a fully reconstructed jet, with $z_h=\omega_h/\omega_J$, where $\omega_h$ and $\omega_J$ are the light-cone energy of the identified hadron and jet, respectively, then the JFF is sensitive to the standard collinear fragmentation functions. See an illustration in Fig.~\ref{fig:jff}. 
\bef
\includegraphics[width=0.5\textwidth]{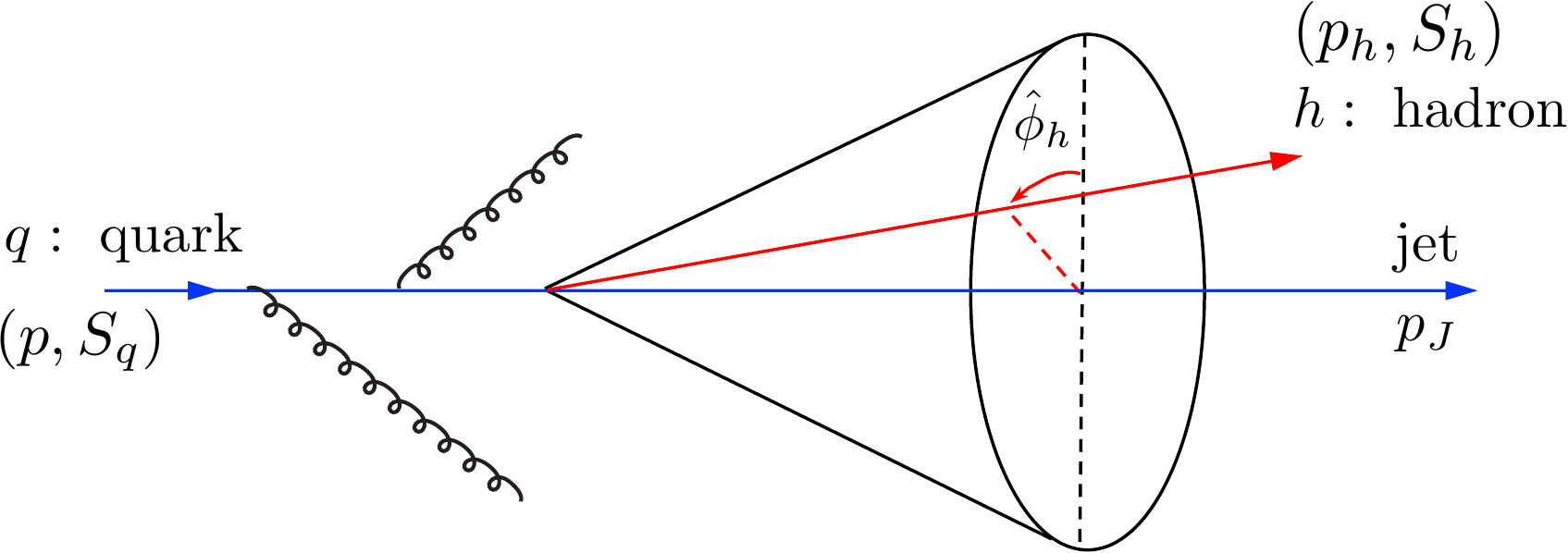} 
\caption{Illustration for the distribution of hadrons inside a fully reconstructed jet, that is initiated by a quark.}
\label{fig:jff}
\eef
For inclusive jet production, one further introduces a momentum fraction $z$ of the initiating parton carried by the jet, $z=\omega_J/\omega$, with $\omega$ representing the light-cone energy of the parton which initiates the jet. In this case, one has the so-called \index{semi-inclusive fragmenting jet function} semi-inclusive fragmenting jet function (FJF), ${\cal G}_i^h(z, z_h, \omega_J R, \mu)$,\index{fragmenting jet function (FJF)} whose operator definition is given in~\cite{Kang:2016ehg}. Note that in the $pp$ collisions where usually the jet transverse momentum $p_T$ is measured and particle transverse momenta are used to construct the jets, we have semi-inclusive FJF written as ${\cal G}_i^h(z, z_h, p_T R, \mu)$, just like in Eqs.~\eqref{eq:inclusive-jet} and \eqref{eq:inclusive-jet-h}. We will use both conventions below interchangeably. It can be shown~\cite{Kang:2016ehg} that such a semi-inclusive FJF follows a time-like DGLAP evolution equation, just like the usual collinear fragmentation functions
\begin{align}
\mu \frac{d}{d\mu} {\cal G}_i^h(z,z_h, p_TR, \mu) = \frac{\alpha_s(\mu)}{\pi} \sum_j \int_z^1  \frac{dz'}{z'} P_{ji}\left(\frac{z}{z'} \right) {\cal G}_j^h(z',z_h, p_T R,\mu)\,,
\end{align}
where $P_{ji}$ are the usual parton splitting functions. At the same time, for the scale $\mu\gg\Lambda_{\mathrm{QCD}}$, we can match the semi-inclusive FJF ${\cal G}_i^j(z,z_h,\omega_J,\mu)$ onto the collinear fragmentation functions $D_i^h(z,\mu)$ as follows:
\begin{align}
\label{eq:matching}
{\cal G}_i^h(z,z_h,p_T R,\mu) = \sum_j \int_{z_h}^1 \frac{dz_h'}{z_h'} {\mathcal J}_{ij}^{\rm incl}\left(z,z_h',p_TR,\mu\right) D_j^h\left(\frac{z_h}{z_h'},\mu\right)\left[1+{\mathcal O}\left(\frac{\Lambda^2_{\rm QCD}}{p_T^2R^2} \right)\right] \, ,
\end{align}
where the superscript ``incl'' in the matching coefficients ${\mathcal J}_{ij}^{\rm incl}$ emphasizes that they are for the semi-inclusive FJF, to be distinguished from the matching coefficients to be defined below for exclusive FJF. The expressions for ${\mathcal J}_{ij}^{\rm incl}$ are different for different jet algorithms and are given in~\cite{Kang:2016ehg}.

In exclusive jet production, one has a similar exclusive fragmenting jet function, ${\cal G}_i^h(z_h, p_TR, \mu)$. In such a set-up, one identifies only a certain number of signal jets and vetoes any additional jets. The only difference between the semi-inclusive FJF and the exclusive FJF lies in the fact that any out-of-jet radiation is power suppressed in the calculations of the exclusive FJF and can be neglected. Of course, the contribution of such out-of-jet radiation is characterized by the soft functions, see e.g. Eq.~\eqref{eq:jet-standard-TMD} for exclusive jet production. As a consequence, for the exclusive jet production, the energy of the initiating parton is fully contained inside the final jet, and thus the momentum fraction of the parton carried by the jet, $z$, is equal to one. Hence, $z$ is dropped in the definition and only $z_h$ is maintained. It can be shown that such exclusive FJF satisfies the following renormalization group equation
\begin{align}
\mu\frac{d}{d\mu} {\cal G}_i^h(z_h, p_TR,\mu) = \gamma_J^i(\mu) \, {\cal G}_i^h(z_h, p_TR,\mu),
\end{align}
The anomalous dimensions $\gamma_J^i$ are given by
\begin{align}
\gamma^i_J(\mu) = \Gamma^i_{\rm cusp}\left[\alpha_s(\mu)\right] \ln\left(\frac{\mu^2}{p_T^2R^2}\right)+ \gamma^i \left[\alpha_s(\mu)\right],
\label{eq:gamma_J}
\end{align}
where $\Gamma^i_{\rm cusp}$ and $\gamma^i$ are the cusp and non-cusp anomalous dimensions~\cite{Jain:2011xz,Becher:2006mr,Becher:2009th,Echevarria:2012pw,Moch:2004pa} with their expansions defined in Eq.~\eqref{eq:cusp-non} and they are the same as those for the exclusive jet functions in Eq.~\eqref{eq:jet-standard-TMD}, also referred to as the unmeasured jet function in~\cite{Ellis:2010rwa,Chien:2015cka}. The exclusive FJF can also be matched onto the standard collinear fragmentation functions, 
\begin{align}
\label{eq:matching_exc}
{\cal G}_i^h(z_h,p_T R,\mu) = \sum_j \int_{z_h}^1 \frac{dz_h'}{z_h'} {\mathcal J}_{ij}\left(z_h',p_TR,\mu\right) D_j^h\left(\frac{z_h}{z_h'},\mu\right)\left[1+{\mathcal O}\left(\frac{\Lambda^2_{\rm QCD}}{p_T^2R^2}\right)\right] \, ,
\end{align}
where the matching coefficients ${\mathcal J}_{ij}$ can be perturbatively computed~\cite{Procura:2011aq,Waalewijn:2012sv,Chien:2015vja} and are different from ${\mathcal J}_{ij}^{\rm incl}$ in semi-inclusive FJF case in Eq.~\eqref{eq:matching}.

\FloatBarrier
\subsection[Hadron transverse momentum distribution inside jets: \texorpdfstring{$j_\perp$}{jp}-dependence]{\boldmath Hadron transverse momentum distribution inside jets: $j_\perp$-dependence}
\FloatBarrier
 
If one measures both the longitudinal $z_h$ and transverse momentum $j_\perp$ distribution of hadrons inside the jet, such a measurement will be sensitive to the TMD FFs introduced in \sec{TMDFFs}. We again distinguish between inclusive jet production and exclusive jet processes. For single inclusive jet production, one introduces the so-called semi-inclusive TMD fragmenting jet functions (TMD FJFs), ${\cal G}_{i}^h(z, p_T R, z_h, {\bm j}_\perp, \mu, \zeta_J)$. In the TMD region where $j_\perp\ll p_T R$, we have the following factorized form for ${\cal G}_{i}^h$~\cite{Kang:2017glf}, 
\begin{align}
{\cal G}_{i}^h(z, p_T R, z_h, {\bm j}_\perp, \mu, \zeta_J) =  & {\cal H}_{c\to i}(z, p_T R, \mu) 
\int d^2 {\bm k}_\perp d^2{\bm \lambda}_\perp \delta^2\left(z_h {\bm \lambda}_\perp + {\bm k}_\perp - {\bm j}_\perp\right)
\nnu
&\times D_{h/i}(z_h, {\bm k}_\perp, \mu, \zeta/\nu^2) S_i({\bm \lambda}_\perp, \mu, \nu R)\; ,
\end{align}
where $S_i({\bm \lambda}_\perp, \mu, \nu R)$ is a collinear-soft function. One can show at the NLO that the collinear-soft function is related to the standard soft function $S_i({\bm \lambda}_\perp, \mu, \nu)$ in Eq.~\eqref{eq:soft_nlo_3} as follows:
\begin{align}
S_i({\bm \lambda}_\perp, \mu, \nu R) = \left.\sqrt{S_i({\bm \lambda}_\perp, \mu, \nu)}\right|_{\nu\to \nu R/2}\,.
\end{align}
Taking advantage of this relation, one can eventually show~\cite{Kang:2017glf,Kang:2021ffh}
\begin{align}
\label{eq:final-pert-momentum}
{\cal G}_{i}^h(z, p_T R, z_h, {\bm j}_\perp, \mu, \mu^2)  = {\mathcal C}_{i\to j}(z, p_T R, \mu)\, D_{h/j}(z_h, {\bm j}_\perp, \mu_J, \mu_J^2)\,, 
\end{align}
where $D_{h/j}(z_h, {\bm j}_\perp, \mu_J, \mu_J^2)$ on the right-hand side is the standard TMD FF at the scales $\mu_J = \sqrt{\zeta_J} = p_T R$ as probed in the usual semi-inclusive deep inelastic scattering (SIDIS) or $e^+e^-$ collisions in Sec.~\ref{sec:LQTMDFF}. On the other hand, ${\mathcal C}_{i\to j}$ are the coefficient functions that can be computed perturbatively as long as $\mu\gg \Lambda_{\rm QCD}$. It is important to realize that the semi-inclusive TMD FJFs satisfy the DGLAP evolution equations,
\begin{align}
\mu \frac{d}{d\mu} {\cal G}_i^h(z, p_T R, z_h, {\bm j}_\perp, \mu, \zeta_J) = \frac{\alpha_s(\mu)}{\pi} \sum_j \int_z^1  \frac{dz'}{z'} P_{ji}\left(\frac{z}{z'} \right) {\cal G}_j^h(z', p_T R, z_h, {\bm j}_\perp, \mu, \zeta_J)\,,
\end{align}
and thus when one evolves the above equations from the natural scale $\mu \sim \sqrt{\zeta_J}\sim p_T R$ to the hard scales $\mu \sim \sqrt{\zeta_J} \sim p_T$, one resums the series of logs $\ln R$ for small radius $R\ll 1$ jets. 

On the other hand, for exclusive jet processes, e.g., the hadron transverse momentum distribution inside $Z$-tagged jets, where the $Z$-boson and the jet are produced back-to-back, one introduces the exclusive TMD fragmenting jet function, ${\cal G}_{i}^h(p_T R, z_h, {\bm j}_\perp, \mu, \zeta_J)$~\cite{Kang:2019ahe}. They follow the same renormalization group equation like $ {\cal G}_i^h(z_h, p_TR,\mu)$ above, 
\begin{align}
\mu\frac{d}{d\mu} {\cal G}_i^h(p_T R, z_h, {\bm j}_\perp,R,\mu, \zeta_J) = \gamma_J^i(\mu) \, {\cal G}_i^h(p_T R, z_h, {\bm j}_\perp, \mu, \zeta_J)\,.
\label{eq:fjf-evo-excl}
\end{align}
At the same time, it can be related to the TMD fragmentation functions as follows
\begin{align}
{\cal G}_{i}^{h}(p_T R, z_h, {\bm j}_\perp, \mu, \mu^2) = D_{h/i}(z_h, {\bm j}_\perp, \mu_J, \mu_J^2) \exp\left[\int_{\mu_J}^{\mu} \frac{d\mu'}{\mu'} \gamma_J^i(\mu')\right]\,,
\end{align}
where the exponential factor is simply reflecting the fact that it follows the renormalization group equation as given in Eq.~\eqref{eq:fjf-evo-excl}. 

In general, jet substructure can receive contamination from both underlying event and non-global color correlations. Both types of contamination would lead to complications in establishing the relations between TMD FFs probed via jet substructure and those via standard TMD processes. Modern grooming techniques can be applied to remove these sources of contamination~\cite{Larkoski:2014wba}. For example, Refs.~\cite{Makris:2017arq,Gutierrez-Reyes:2019msa} have investigated how soft-drop grooming can be used to reduce the non-global logarithms. In addition, it has been shown there that the TMD hadron distribution with respect to the groomed jet axis is particularly sensitive to
nonperturbative physics of the TMD evolution at low values
of $j_\perp$, which can be probed in the variation of the cut-off parameter, $z_{\rm cut}$, of the groomer.

\subsubsection{Polarized jet fragmentation functions}

Our discussion above mainly deals with {\it unpolarized} hadron distributions inside the jet, which allows us to probe unpolarized collinear FFs or TMD FFs via jets. One can naturally ask questions if jets can also be used to study {\it polarized} TMD FFs. Ref.~\cite{Kang:2020xyq} provides a general theoretical framework for studying the distribution of hadrons inside a jet by taking full advantage of the polarization effects. The key development, referred to as polarized jet fragmentation functions, describes the situation where the parton that initiates the jet and the hadron that is inside the jet can both be polarized, as illustrated in Fig.~\ref{fig:jff}. For example, with polarized jet fragmentation functions, one could study $\Lambda$ hyperon polarization inside a jet produced in unpolarized proton-proton collisions, where one would be able to probe the so-called TMD polarized fragmentation functions (TMD PFFs). Such TMD PFFs have been recently measured by the Belle collaboration~\cite{Guan:2018ckx,Callos:2020qtu,DAlesio:2020wjq}. 

\begin{figure}[hbt]
\centering
\includegraphics[width=0.86\textwidth]{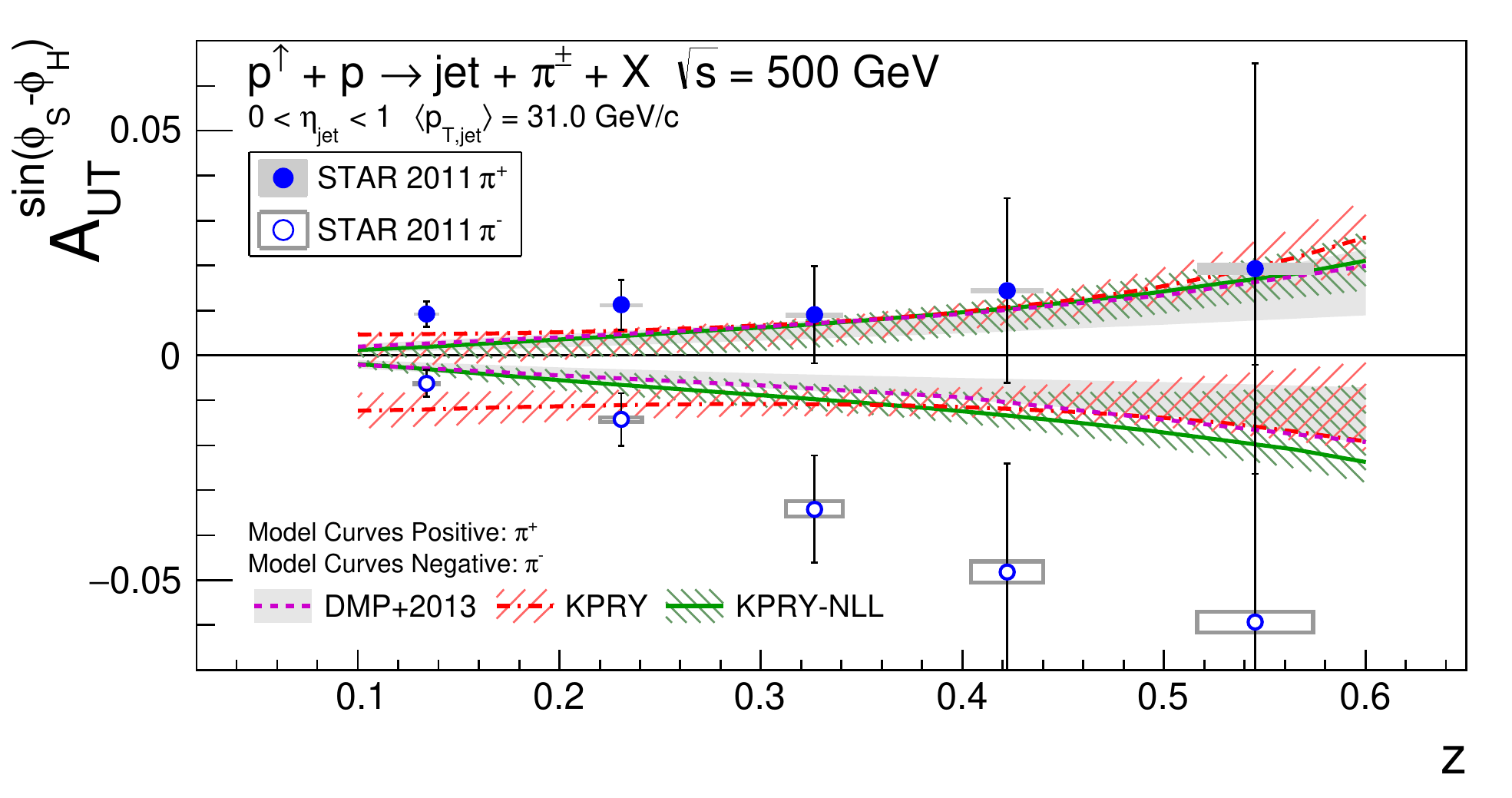} 
\caption{Collins asymmetries as a function of pion $z_h$ for jets reconstructed with $22.7 < p_T < 55.0$ GeV and rapidity $0 < \eta < 1$. The asymmetries are shown in comparison with model calculations from Refs.~\cite{DAlesio:2017bvu,Kang:2017btw}. Plot from~\cite{Adamczyk:2017wld}.}
\label{fig:star_collins}
\end{figure}

Another well-known example is the so-called Collins hadron asymmetry inside a jet. Here, if one studies the distribution of an unpolarized hadron inside the jet which is initiated by a transversely-polarized quark, then the transverse spin of the quark and the transverse momentum $j_\perp$ of the hadron with respect to the jet would be correlated, resulting in a non-trivial azimuthal modulation for the hadron distribution. This was first proposed in~\cite{Yuan:2007nd} to explore the Collins fragmentation functions, with further developments in~\cite{DAlesio:2010sag,Kang:2017btw}. The STAR collaboration at RHIC has since studied such a Collins asymmetry for $\pi^\pm$ production inside jets in transversely-polarized proton-proton collisions. Theoretical predictions from~\cite{DAlesio:2017bvu,Kang:2017btw} with the Collins fragmentation functions taken from a fit of SIDIS and $e^+e^-$ data has shown to give a good description of the experimental data. This indicates the universality of the Collins function among SIDIS,
\index{Collins function $H_1^{\perp}$!phenomenology}
\index{transversity!phenomenology|}
$e^+e^-$, and $pp$ collisions, although the experimental uncertainty is still quite large. Future measurements with improved statistics have been planned~\cite{Aschenauer:2016our}.

\subsection{Jets with Heavy Quarkonium}
\label{sec:jetquarkonia}

Heavy quarkonia \index{heavy quarkonia} are an interesting laboratory in which to apply the formalism for FJFs, as first pointed out in Ref.~\cite{Baumgart:2014upa}. As shown in \eq{matching} and \eq{matching_exc} the FJF can be written as a convolution of a perturbatively calculable matching coefficient and a fragmentation function. For light hadrons and singly heavy hadrons, the fragmentation functions are nonperturbative and must be extracted from data. 
An old idea from the 90's is that the Non-Relativistic QCD \index{non-relativistic QCD} factorization formalism (NRQCD)~\cite{Bodwin:1994jh} can be used to calculate heavy quarkonium fragmentation functions because the heavy quark mass provides a large scale justifying the use of perturbation theory~\cite{Braaten:1993rw,Braaten:1993mp,Braaten:1993jn,Braaten:1994kd}. 
In NRQCD the conjectured factorization for the fragmentation functions take the form (here we use $J/\psi$ as an example)
\begin{align}
\label{NRQCD}
D_i^{J/\psi}(z,m_c,\mu) = \sum_n C^n_i (z, \alpha_S, m_c, \mu) \langle {\cal O}_n^{J/\psi}\rangle
\end{align} 
where $n$ denotes the color and angular quantum numbers of the heavy charm-anticharm pair produced in the short distance process, $i\to c\bar{c}(n) + X$, and 
$\langle {\cal O}_n^{J/\psi}\rangle$ is a
long-distance matrix element (LDME)
\index{non-relativistic QCD! long-distance matrix element (LDME)} describing the nonperturbative transition of the $c\bar{c}$ in a state of definite color and angular momentum to the final state including the $J/\psi$. 
$C^n_i (z, \alpha_S,m_c,\mu)$ is a perturbatively calculable function  of $z$, $\alpha_S$, and $m_c$. 

If we identify a quarkonium inside a jet, we can combine the FJF formalism with NRQCD calculations of fragmentation functions to predict the distribution in $z$, where $z$ is the fraction of the energy carried by the quarkonium in the jet. For example, if we wish to calculate the cross section for $e^+e^-$ to two jets with a $J/\psi$ carrying a fraction $z$ of its jet energy, the cross section
is
\begin{eqnarray}
\frac{1}{\sigma_0}\frac{d\sigma}{dz} = \sum_{i,j} H_{ij}(\mu) \times J_i(\mu)
\times S^{\rm unmeas}(\mu) \times 
\int^1_0 \frac{dz'}{z'} S^{\rm meas}\left(\frac{z}{z'},\mu\right) \, {\cal G}_j^{J/\psi}(z',
 E R,\mu)
 \,.
\end{eqnarray}
Here $H_{ij}(\mu)$ is the hard cross section
for producing the partons $i$ and $j$ that initiate the jets, $J_i(\mu)$ is the jet function for the jet not containing the quarkonium, $S^{\rm unmeas}(\mu)$
is the soft function describing soft radiation outside the jets, $S^{\rm meas}(\mu)$ describes soft radiation in the jet with the quarkonium, and ${\cal G}_j^{J/\psi} (z, E R,\mu)$ is the   FJF for a jet of energy $E$ with a $J/\psi$ with energy fraction $z$. (Note that in $e^+e^-$ collisions the jet energy rather than the jet $p_T$ is typically measured.) Then Eq.~(\ref{NRQCD}) is used in \eq{matching_exc} to calculate the quarkonium FJF in terms of the LDME. Ref.~\cite{Baumgart:2014upa} showed that the FJF is well approximated by evaluating the NRQCD fragmentation at the scale $2 m_c$ then evolving that fragmentation function up to the jet energy scale. At that scale, perturbative corrections in the matching coefficients in \eq{matching_exc} are small.

Various extractions of the LDME exist in the literature, for reviews of the status of quarkonium production theory, see \cite{Chapon:2020heu}.
Global fits to the world's data on $J/\psi$ production provide a reasonable fit, but predict transverse polarization of $J/\psi$ at large $p_T$ at hadron colliders,
which is not seen in experiments~\cite{Butenschoen:2011yh,Butenschoen:2012qr}. Alternative fits which focus extensively on high $p_T$ data can do a better job of describing $J/\psi$ production in these
experiments \cite{Chao:2012iv,Bodwin:2014gia} , but at the expense of statistical accuracy as well as ignoring much of the world's data on $J/\psi$ production. Different NRQCD production mechanisms yield
different $z$ 
dependence for the $C^n_i (z, \alpha_S,m_c,\mu)$ so the $z$ distribution of $J/\psi$ within a jet is sensitive to the underlying production mechanism.
Ref.~\cite{Baumgart:2014upa} proposed the study of quarkonium production within jets as an alternative way to test NRQCD at high $p_T$ and extract LDMEs. 

Ref.~\cite{Bain:2016clc} performed analytical studies of heavy mesons and quarkonia produced in jets in an $e^+e^-$ collider using the FJF formalism. These were compared to the results of Monte Carlo simulations using Herwig and \textsc{Pythia}. The dependence of the cross section on the jet angularities~\cite{Berger:2003iw} and the fraction of the energy carried by the heavy meson, $z$, were studied. Ref.~\cite{Bain:2016clc} found agreement between Monte Carlo and the FJF formalism for heavy mesons. However, the $z$ dependence of the cross sections for quarkonia in jets is not well reproduced by Monte Carlo. Monte Carlo predicts a much harder distribution than the FJF formalism. This was attributed to incorrect modelling of radiation from color-octet pairs in default \textsc{Pythia}. These results were later confirmed by experiment when the LHCb experiment \cite{Aaij:2017fak} measured the distribution of $J/\psi$ within a jet for the first time. 

\begin{figure}[t!]
\centering
\includegraphics[width=1.0\textwidth]{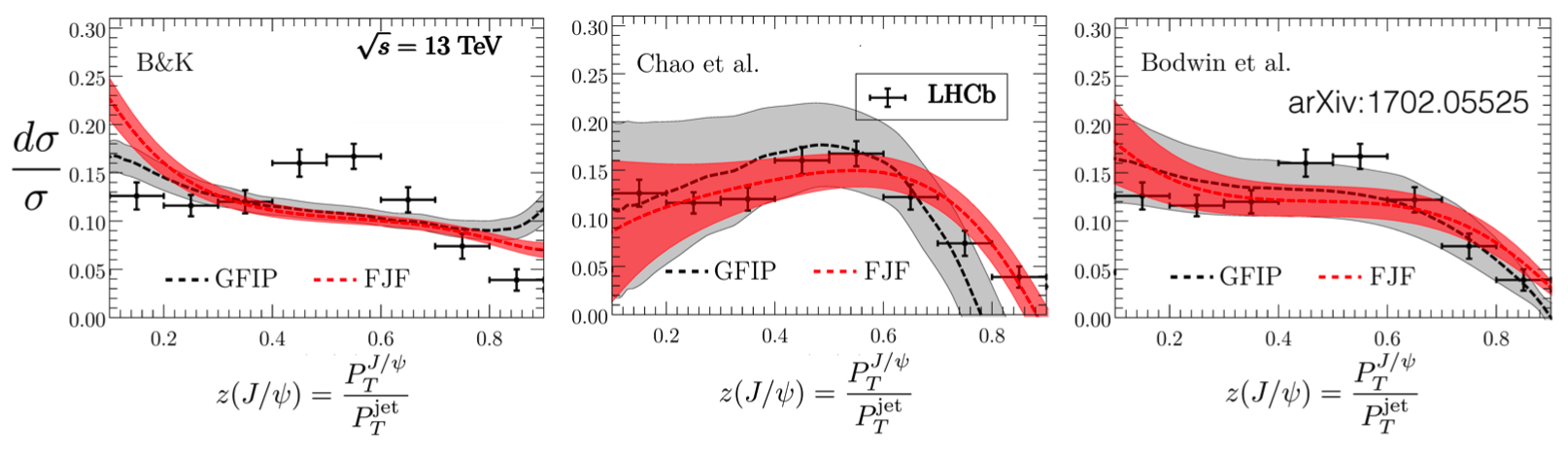} 
\caption{Comparisons of $z(J/\psi)$ measured by LHCb with FJF (red) and GFIP (gray) for three different choice of LDME. Figure from Ref.~\cite{Chapon:2020heu}, the plots originally appeared in Ref.~\cite{Bain:2017wvk}. }
\label{fig:zJpsi}
\end{figure}

LHCb~\cite{Aaij:2017fak} measured the distribution in $z(J/\psi)$, where $z(J/\psi) =p_T^{J/\psi}/p_T^{\rm jet}$, shown in \fig{zJpsi}.
The $z(J/\psi)$ distributions predicted by default \textsc{Pythia} (not shown) were much harder than observed, peaking at $z(J/\psi)> 0.8$. A  description of the LHCb data  obtained in Ref.~\cite{Bain:2017wvk} is also shown in \fig{zJpsi}. FJF is the calculation of the $z(J/\psi)$ distribution using the quarkonium FJF, calculated in the approximation mentioned above of evolving  the NRQCD fragmentation function from the scale $2m_c$ to the jet energy scale. Gluon Fragmentation Improved \textsc{Pythia} (GFIP) is a modified implementation of \textsc{Pythia}  described in Refs.~\cite{Bain:2016clc,Bain:2017wvk}.
NRQCD LDMEs extracted from high $p_T$ data did a better job of describing the $z(J/\psi)$ distributions of $J/\psi$ in jets than LDMEs from global fits.

Ref.~\cite{Bain:2016rrv} was the first to consider the transverse
momentum of the quarkonium within a jet. The TMD FJFs that appeared in Ref.~\cite{Bain:2016rrv} are very similar to those discussed in the previous subsections. The transverse momentum is defined relative to the jet axis. RGE and rapidity RGE (RRGE) were used to resum logarithms. nonperturbative effects were not taken into account in Ref.~\cite{Bain:2016rrv} and this  remains to be done before comparison with experiment can be made. Nonetheless, Ref.~\cite{Bain:2016rrv} showed that different NRQCD production mechanisms give rise to different $j_T$ distributions for the quarkonia within the jet, so the $j_T$ spectrum would allow for novel tests of NRQCD and a framework for extracting LDMEs. At present, the $j_T$ spectrum of quarkonia within jets has not been measured, but it would be interesting to study in the future. 

Finally important recent theoretical developments that merit  attention are the 
TMD quarkonium shape function
\index{TMD quarkonium shape function}, introduced in Refs.~\cite{Echevarria:2019ynx,Fleming:2019pzj},
and the TMD fragmentation function for quarkonia
\cite{Echevarria:2020qjk}.
These   objects appear in factorized cross sections that are relevant when the observable is sensitive to soft ($p\sim m_Q v$, where $m_Q$ is the heavy quark mass and $v$ its velocity)  radiation interacting with the heavy quark-antiquark.
Ref.~\cite{Echevarria:2019ynx} studied the process $pp\to \eta_c$ with only color-singlet mechanisms. 
Ref.~\cite{Fleming:2019pzj} studied $\Upsilon$ to two jets with identified heavy hadrons whose transverse momentum is measured. The TMD fragmentation function
was discovered in a study of the transverse momentum distribution of  $J/\psi$ coming from the fragmentation of light partons \cite{Echevarria:2020qjk} in SIDIS. This paper also determined what regions in phase space this process will dominate at the EIC and discusses the role of NRQCD TMD fragmentation. TMD  observables featuring quarkonia will figure prominently at the EIC, so the quarkonium  shape functions and TMD fragmentation functions will be relevant for future studies. For example, Ref.~\cite{Boer:2020bbd} studies asymmetries in $J/\psi$ plus jet production for extracting the gluon TMDs.  For other recent work on quarkonium production which utilizes the TMD formalism, see Refs.~\cite{DAlesio:2020eqo,Boer:2020bbd}.

\subsection{Transverse Energy-Energy Correlations}
\label{sec:TEEC}

Transverse-energy-energy correlations (TEEC)
\index{transverse-energy-energy correlations (TEEC)} are event shape observables that provide new ways to probe TMD dynamics. TEEC at hadronic colliders~\cite{Ali:1984yp} is an extension of the 
energy-energy correlation (EEC)
\index{energy-energy correlation (EEC)}~\cite{Basham:1978bw} variable introduced decades ago in $e^+e^-$ collisions to describe the global event shape. It is defined as  
\begin{align} 
    \text{TEEC} = \sum_{a,b} \int d\sigma_{pp\to a+b+X} \frac{2 E_{T,a} E_{T,b}}{|\sum_{i}E_{T,i}|^2} \delta(\cos\phi_{ab}-\cos\phi)\, ,
    \label{teec}
\end{align}
where $E_{T,i}$ is the transverse  energy of hadron $i$ relative to the collision axis and $\phi_{ab}$ is the azimuthal angle between hadrons $a$ and $b$. The  NLO QCD corrections for the TEEC observable were calculated in Ref.~\cite{Ali:2012rn}. In the back-to-back dijet limit TEEC exhibits remarkable perturbative simplicity~\cite{Gao:2019ojf}.  This observable can be generalized to DIS   by considering the transverse-energy and transverse-energy correlation between the lepton and hadrons in the final state~\cite{Li:2020bub}
\begin{align}~\label{eq:teec_dis}
    \text{TEEC} =&  \sum_{a} \int d\sigma_{lp\to l+a+X} \frac{ E_{T,l}  E_{T,a}}{E_{T,l} \sum_{i} E_{T,i}}  \delta(\cos\phi_{la}-\cos\phi)\, ,  
\end{align}
where the sum runs over all the hadrons in the final state and $\phi_{la}$ is the  azimuthal angle between the final-state lepton $l$ and hadron $a$.

Taking DIS as an example, the underlying partonic Born process is 
$ e(k_1) + q(k_2) \to e(k_3) + q(k_4) $ and the first order non-trivial contribution to TEEC begins from one order higher. Similarly to TEEC in hadronic collisions, the cross section in the back-to-back limit is factorized into the convolution of a hard function, beam function, soft function, and jet function.
Specifically, up to leading power in SCET in terms of the variable $\tau = [1+\cos(\phi)]/2$ the cross section can be written as  
\begin{align}\label{eq:sing}
    \frac{d\sigma^{(0)}}{d\tau} =& \sum_{f}  \int\frac{d\xi  dQ^2 }{\xi Q^2}  Q_{f}^2 \sigma_0 \frac{p_T}{\sqrt{\tau}}\int \frac{db}{2\pi} e^{-2ib\sqrt{\tau} p_T}  B_{f/N}(b,E_2, \xi, \mu, \nu ) H(Q, \mu )
    \nonumber \\ & \times 
    S\left(b,\frac{n_2\cdot n_4}{2},\mu,\nu\right)J_{f}(b,E_4, \mu, \nu) \,,
\end{align}
where $\sigma_0= \frac{2 \pi \alpha^2}{Q^2}[1+(1-y)^2] $,  $b$ is the conjugate variable to $k_y$,  $Q^2$ is the invariant mass of the virtual photon, and $y=Q^2/(\xi s)$.  Four-vectors  $n_2$ and $n_4$ represent the momentum directions of the momenta $k_2$ and $k_4$, respectively. $E_2$ and $E_4$ are the energies of $k_2$ and $k_4$. $\nu$ is the rapidity scale associated with the rapidity regulator for which we adopt the exponential regulator  introduced in Ref.~\cite{Li:2016axz} and reviewed in \sec{tmd_defs}.

\begin{figure}
    \centering
    \includegraphics[width=0.49 \textwidth]{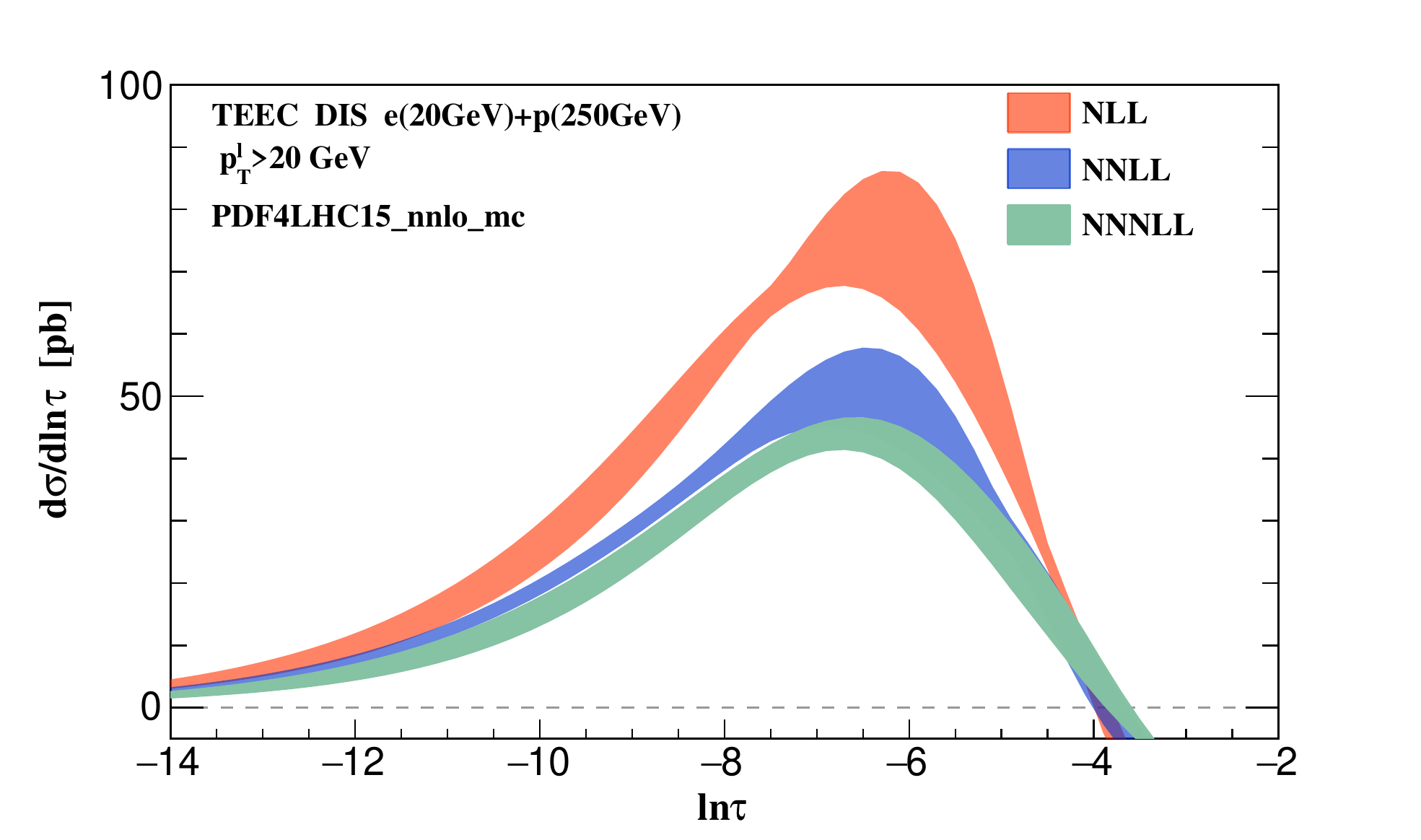}
    \includegraphics[width=0.49 \textwidth]{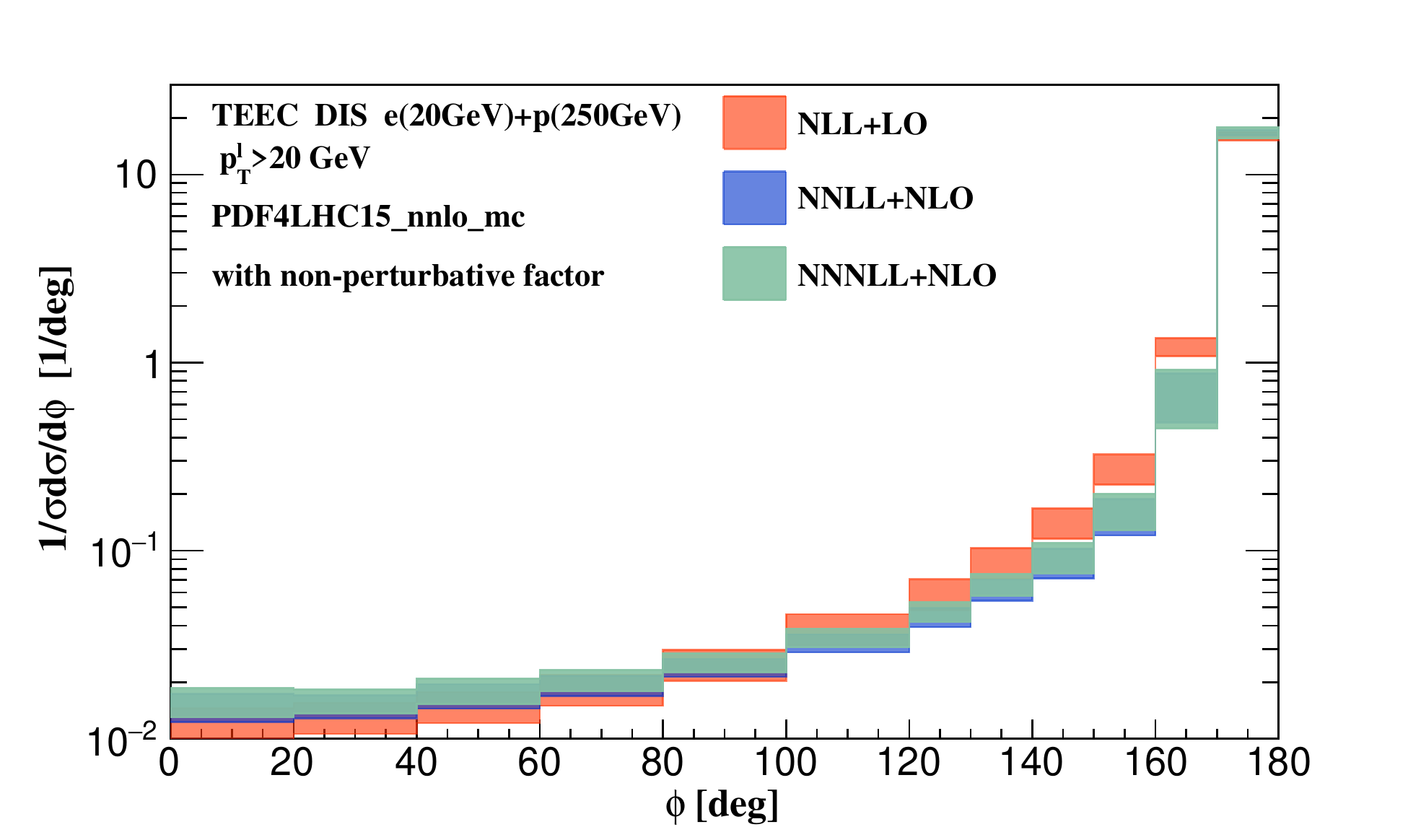}
    \caption{Left: resummed distributions in the back-to-back limit up to N$^3$LL accuracy. Note that results are not normalized by $\sigma$ in the $\tau$ interval shown.
 Right: TEEC $\phi$ distribution matched with a nonperturbative model. The orange, blue and green bands are the final predictions with scale uncertainties up to N$^3$LL+NLO. Plots are from Ref.~\cite{Li:2020bub}. }
    \label{fig:TEEC}
\end{figure}

The TMD beam functions have been calculated up to three loops for quark beam functions and two loops for gluon beam functions~\cite{Luo:2019hmp,Luo:2019bmw,Lubbert:2016rku,Luo:2019szz}  
The jet function $J_f$ is defined as the second Mellin  moment of the matching coefficients of the TMD fragmentation function. The soft function $S$ is the same as the TMD soft function. In addition to the close connection to TMD physics, TEEC in DIS has the advantage that it can be computed to high accuracy. The left panel of \fig{TEEC} presents the resummed predictions at NLL, NNLL,  and N$^3$LL accuracy in the back-to-back limit with scale uncertainties~\cite{Li:2020bub}. Ref.~\cite{Li:2020bub} finds good perturbative convergence 
There is about 30\% suppression in the peak region from NLL to NNLL, while it is about 5-6\% from NNLL to N$^3$LL. The reason is that these are absolute cross sections rather than ones normalized over a finite $\tau$ interval. The NLL uncertainty might also be underestimated. In general  the nonperturbative  (NP) corrections can be important in the infrared region and can be studied with the help of TEEC in DIS. The results  for the normalized TEEC $\phi$ distributions are shown in the right panel of Fig.~\ref{fig:TEEC}, where the nonperturbative Sudakov factor is also implemented~\cite{Li:2020bub}. The matching region is  chosen to be $160^{o}<\phi<175^{o}$ and for $\phi<160^{o}$ the distributions are generated by fixed-order calculations. The fixed-order predictions are calculated with $\mu_r=\mu_f=\kappa Q$ with $\kappa=(0.5,1,2)$. In the back-to-back limit, the predictions are significantly improved.

\begin{figure}
\includegraphics[width = 1.\textwidth]{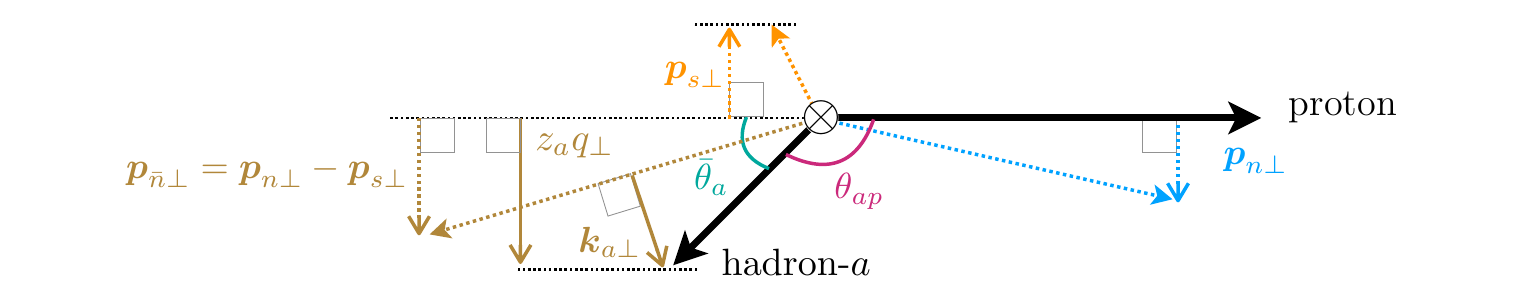}
 \caption{Illustration of the measurement of the transverse momentum $\boldmath{q}_{\perp}$ of the hadron-$a$ w.r.t. the proton axis in the Breit frame. }
\label{fig:measurment} 
\end{figure}

Measurements of QCD observables in DIS are often done in the Breit frame. 
Recently, a new definition of EEC in the
Breit frame
\index{Breit frame}, which is a natural frame for the study of TMD physics~\cite{Collins:2011zzd}, was presented~\cite{Li:2021txc}.
In this frame, the target hadron moves along $\hat{z}$ and the  virtual photon  moves in the opposite direction. The Born-level process is described by the lepton-parton scattering $e+q_i\to e+q_f$, 
where the outgoing quark $q_f$ back-scatters in the direction opposite to the proton.  Hadronization of the struck quark will form a collimated spray of radiation close to the $- \hat{z}$ direction. On the other hand, initial state radiation and beam remnants are moving in the opposite direction close to the proton's direction of motion.  It is this feature of the Breit frame, which leads to the clean separation of target and current fragmentation that we utilize to construct the novel EEC observable in DIS.
The kinematics,  together with the contributions from the collinear and soft momenta to the transverse momentum of the hadron $q_\perp$ is illustrated in Figure~\ref{fig:measurment}.    

\begin{figure}[t!]
    \centering
    \includegraphics[width=0.45\textwidth]{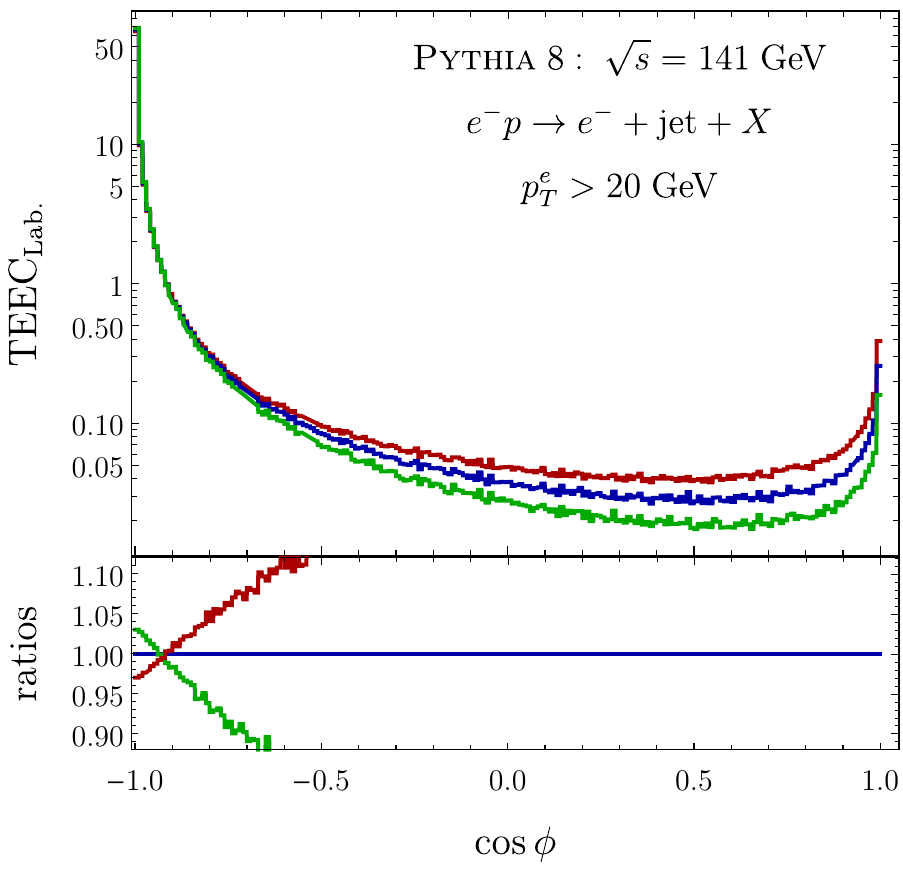}
\;\;\hspace{0.2cm}
    \includegraphics[width=0.45\textwidth]{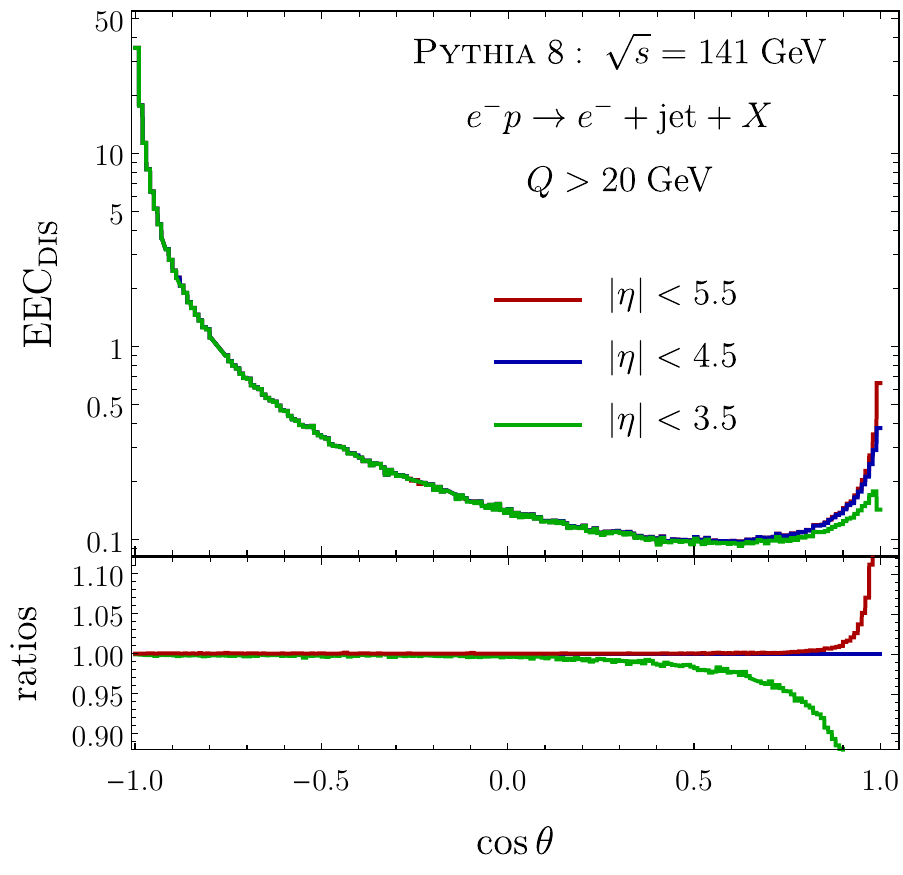}
    \vspace{-0.2cm}
    \caption{TEEC (left) and EEC (right) distributions from \textsc{Pythia} 8 with different rapidity cuts in the lab frame.  The ratio relative to the  
    $|\eta| < 4.5$ case is also shown. Plots originally appeared in Refs.~\cite{Li:2021txc}.}
    \label{fig:pythia}
\end{figure}

We denote the new event shape variable EEC$_{\text{DIS}}$ to avoid confusion with the conventional observable. Our definition reads,
\begin{equation} \label{eq:definition-DIS}
    \text{EEC}_{\text{DIS}}= \sum_a \int\, \frac{d\sigma_{e p \to e+ a+X}}{\sigma}\, z_a\, \delta(\cos\theta_{ap} - \cos\theta)\;,
\end{equation}
where
\begin{equation}\label{eq:weigth-DIS}
    z_a \equiv \frac{P\cdot p_a}{P\cdot (\sum_i p_i)}\;,
\end{equation}
and $p_a^{\mu}$ and $P^{\mu}$ are the momenta of the hadron $a$ and the incoming proton respectively. The sum over $i$ includes all final state hadrons,  including $a$. The angle $\theta_{ap}$ is the polar angle of hadron $a$, which is measured with respect to the incoming proton. Note that the asymmetric weight function, $z_a$, is Lorentz invariant and is suppressed for soft radiation and radiation close to the beam direction. Furthermore, this definition of EEC in the Breit frame naturally separates the contribution to the $\cos\theta$ spectrum from: i) wide angle soft radiation, ii) initial state radiation and beam remnants, and iii)  radiation from the hadronization of the struck quark. This unique feature makes the new observable in the back-to-back limit ($\theta \to \pi$) insensitive to experimental cuts on the particle pseudorapidity (in the Laboratory frame) due to detector acceptance limitations in the backward and forward regions, making the comparison of theory and experiment in this region even more accurate. This definition of EEC is spherically invariant, however, definitions that are fully Lorentz invariant and can be measured directly in any frame are also possible.

To illustrate the reduced sensitivity of the new observable to kinematics, we present the TEEC$_{\rm Lab}$~\cite{Li:2020bub} and EEC$_{\rm DIS}$ distributions predicted by \textsc{Pythia} 8~\cite{Sjostrand:2007gs,Sjostrand:2014zea} in Fig.~\ref{fig:pythia}. The  red,  blue, and  green lines represent the results with pseudorapidity cuts $|\eta|<5.5$, $|\eta|<4.5$, and $|\eta|<3.5$ in the lab frame, respectively, which imitates detector limitations in the backward and forward regions. In order to compare the results with different pseudorapidity cuts,  all the distributions in Fig.~\ref{fig:pythia} are normalized by the event number with $|\eta|<5.5$.   Because TEEC measures the correlation between hadrons and the final state lepton in the lab frame, pseudorapidity cuts have an impact on the full $\cos\phi$ range, as shown in left panel of Fig.~\ref{fig:pythia}. EEC is defined as the correlation between the final state hadrons and incoming proton in the Breit frame, and the pseudorapidity cuts only remove particles in the forward region where the weighted cross section is small. In the backward region the distribution is independent of the pseudorapidity cuts.

\subsection{Medium Modification of Jets}
\label{sec:mediumjets}

\begin{figure}[t!]
\vspace{-1em}
\begin{picture}(0,155)(0,185)
\hspace*{.8in}\includegraphics[width=5.5in]{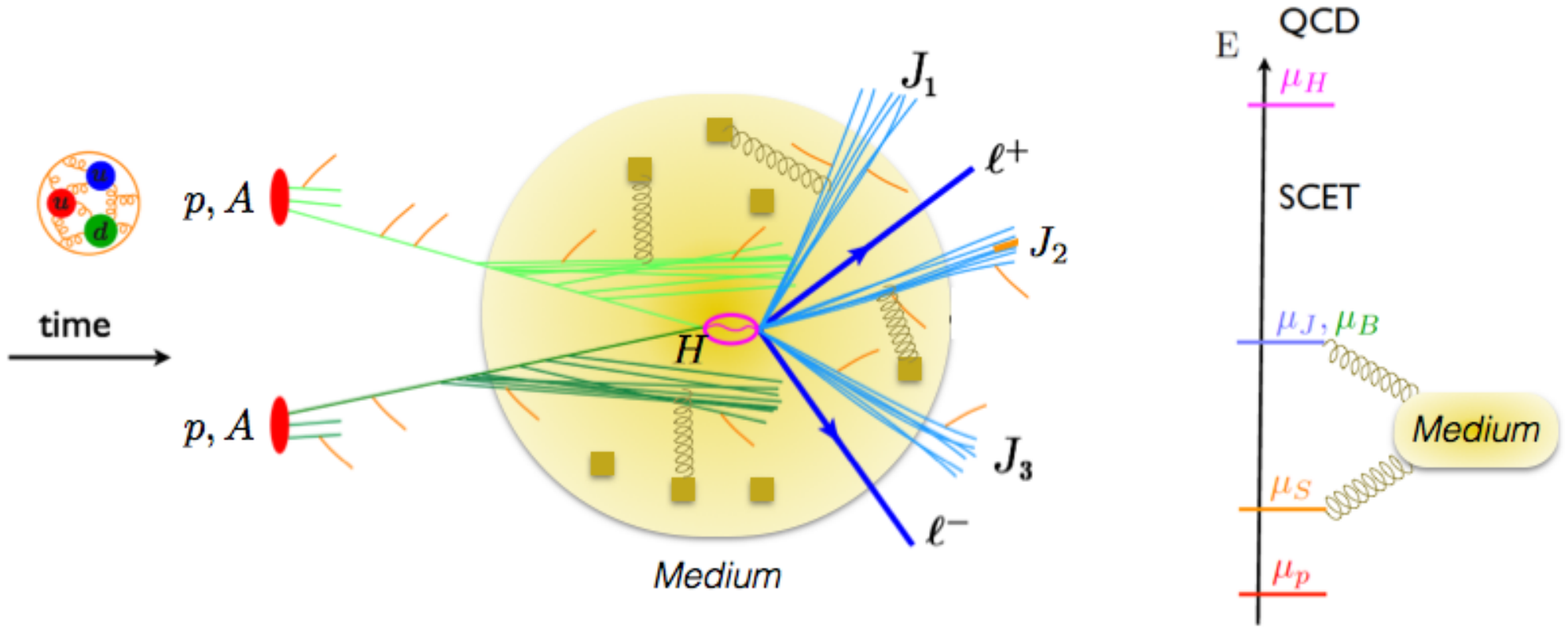}
\end{picture}
\vspace*{0cm}
\caption{\small Degrees of freedom in jet cross sections in p+p, p+A or A+A collisions.  At short distance there is the partonic hard collision described by the hard function $H(\mu_H)$, obtained from matching full QCD onto an EFT of collinear and soft modes. The collinear splitting and emissions of partons in jets are described by jet and beam functions $J(\mu_J)$, $B(\mu_B)$. Low energy soft particles connect beams and jets and mediate color exchange, described by the soft function $S(\mu_S)$. Nonperturbative dynamics of binding in the initial or final state are described by PDFs or nonperturbative matrix elements at the scale $\mu_p$. When a dense medium is created in heavy-ion collisions, interactions between the collinear or soft modes and the quarks/gluons in the medium occur through exchange of Glauber modes, which must be included in the EFT.}
\vspace{-0.1cm}
\label{fig:modes}
\end{figure}

The key theoretical tool to disentangle the different physics effects on jets and predict each of their contributions to high accuracy is factorization \cite{Collins:1985ue}.  The cross sections with a jet final state  in vacuum can be written in the form~\cite{Bauer:2008dt,Jouttenus:2011wh}
\begin{equation}
\label{eq:factorization}
\sigma = \mathrm{Tr}(HS) \otimes \prod_{i=1}^{n_B} B_i \otimes \prod_{j=1}^N J_j + \text{power corrections}\,,
\end{equation}
for a process with $n_B$ incoming hadronic beams and $N$ outgoing hadronic jets. The hard function $H$ is perturbative and contains information on the partonic hard scattering of $n_B$ incoming and $N$ outgoing partons, and the soft function $S$ contains the soft radiation between these hard partons. They are in general color matrices, and the trace is over color indices. The beam functions $B_i$ contain the PDFs for the colliding hadrons and also the effects of perturbative collinear radiation from them, while the jet functions $J_j$ contain the collinear splittings of the outgoing hard partons. These functions and the hierarchy of their scales in a typical $p+p$ or heavy-ion collisions are illustrated in Figure~\ref{fig:modes}. The same extension of the perturbative theory can be achieved from $e+p$ to $e+A$ collisions. 

\begin{figure}[t]
\begin{center}
\includegraphics[width= 0.8\textwidth]{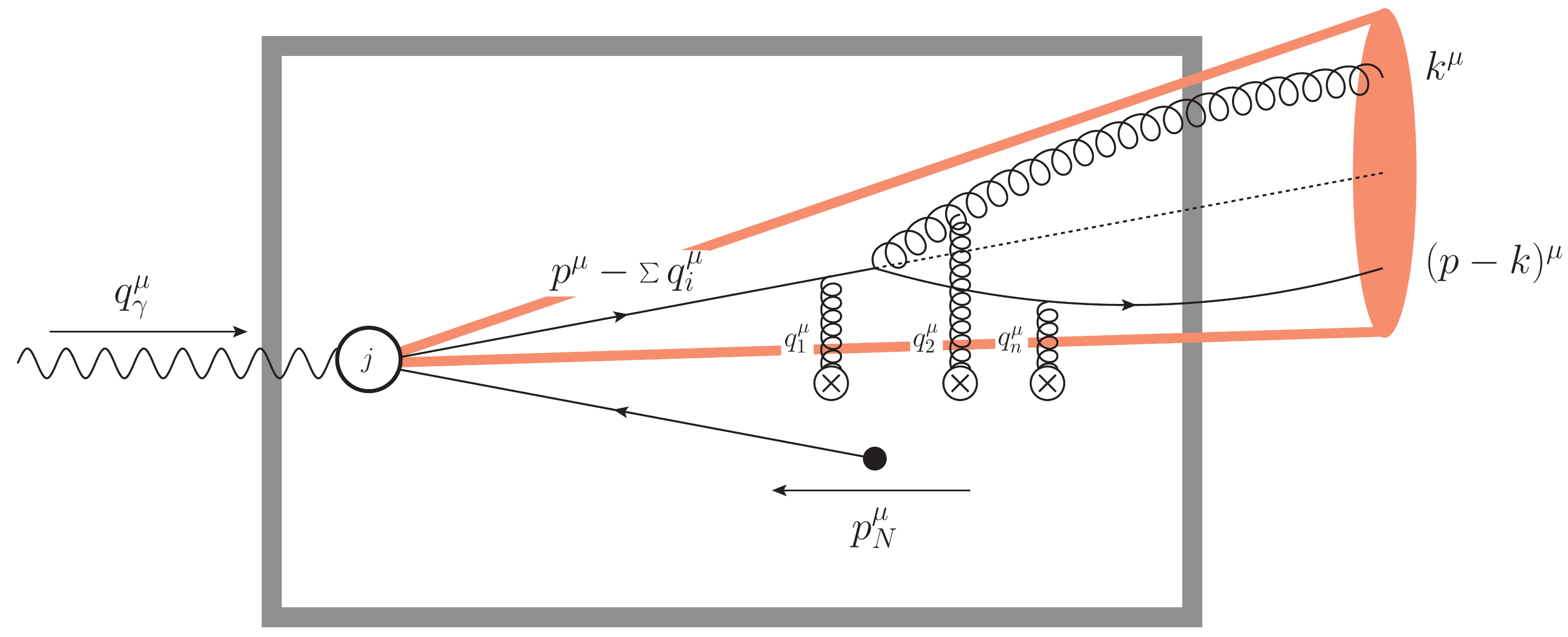} 
\caption{Illustration of parton splitting processes in matter  for SIDIS case in the Breit frame.  The dark box represents the medium and the red cone represents the jet.}
\label{f:Jet_Kinematics}
\end{center}
\end{figure}

Historically, first studies in the field focused on the production cross section of energetic particles and jets in high energy reactions with nuclei. This is one of the primary signatures of inelastic parton scattering in dense nuclear matter~\cite{Gyulassy:1990ye,Gyulassy:2003mc}. The rapid development of heavy ion programs at fixed target and collider experiments fueled tremendous interest in 
medium-induced bremsstrahlung
\index{medium-induced bremsstrahlung} processes and radiative 
parton energy loss
\index{parton energy loss} in QCD~\cite{Gyulassy:1993hr}, often discussed in analogy with the 
\index{Landau-Pomeranchuk-Migdal (LPM) effect \phantom{Landau} } Landau-Pomeranchuk-Migdal (LPM) effect 
for photon emission in QED~\cite{Landau:1953um,Migdal:1956tc}.  Radiative energy loss in QCD is synonymous with soft gluon bremsstrahlung, a process in which hard quarks and gluons shed energy in small quanta during propagation in a nuclear medium. As a result, the leading parton always remains the most energetic. This does not preclude the possibility  that it may  dissipate a sizable fraction of its  energy, but this is achieved through multiple gluon emission. All radiative parton energy loss approaches rely on perturbative techniques and treat the interactions of the jet with the quasi-particles of the medium primarily through $t$-channel  gluon exchanges~\cite{Gyulassy:1993hr}. Theoretical calculations differ in their assumptions about the kinematic regimes in which the parton system is produced and the size of the nuclear medium~\cite{Zakharov:1996fv,Zakharov:1997uu,Baier:1996sk,Baier:1998kq,Arnold:2002ja,Gyulassy:1999zd,Gyulassy:2000fs,Gyulassy:2000er,Wang:2001ifa,Guo:2006kz}.

In the past decade important progress was made in understanding the full longitudinal and transverse structure of in-medium parton showers. 
The Altarelli-Parisi splitting functions~\cite{Altarelli:1977zs} are the key ingredients in all modern high-precision calculations in QCD and in Monte-Carlo event generators.  For jet physics, quark and gluon branching processes play an essential role in understanding the radius dependence of inclusive  and tagged jet cross sections and of jet substructure. In Eq.~(\ref{eq:factorization}) the splitting kernels enter into the calculation of  beam and jet functions.  
The collision of ions introduces additional variables, such as the centrality and the nuclear species, in addition to the transverse momentum and rapidity 
of the jets. More importantly,  the vacuum splitting kernels receive medium-induced contributions~\cite{Wang:2009qb,Ovanesyan:2011xy,Ovanesyan:2011kn,Blaizot:2012fh,Fickinger:2013xwa,Apolinario:2014csa,Ovanesyan:2015dop,Kang:2016ofv,Sievert:2018imd,Sievert:2019cwq,Sadofyev:2021ohn} which depend on the centrality and the colliding system, see Figure~\ref{f:Jet_Kinematics}.  To the lowest non-trivial order for the double differential branching distributions  we have 
\begin{eqnarray}
&&\frac{dN^{\rm vac}(x,{\bf k}_\perp)}{ dxd^2\vc{k}_{\perp}     }
\rightarrow 
\frac{dN^{\rm vac}(x,{\bf k}_\perp)}{ dxd^2\vc{k}_{\perp} }
+  \frac{dN^{\rm med}(x,{\bf k}_\perp)}{ dxd^2\vc{k}_{\perp} } \, , 
\label{eq:medsplit}
\end{eqnarray} 
where $x$ is the longitudinal momentum fraction and ${\bf k}_\perp$ is the transverse momentum of the splitting relative to the
parent parton direction. 

One way of calculating in-medium branching processes is in terms of the correlations between multiple scattering centers, known as the 
opacity expansion \index{opacity expansion} or the \index{Gyulassy-Levai-Vitev (GLV) approach} Gyulassy-Levai-Vitev (GLV) approach. 
To first order in opacity defined as $L/\lambda$, where $L$ is the typical medium size and $\lambda$ is the scattering length,  the 
in-medium splitting kernels 
\index{in-medium splitting kernels} were explicitly
calculated  in~\cite{Ovanesyan:2011xy,Ovanesyan:2011kn} and shown to be gauge invariant:
\begin{eqnarray}
 &&  \left( \frac{dN^{\rm med}}{ dxd^2\vc{k}_{\perp} }\right)_{q\rightarrow qg}  =  \frac{\alpha_s}{2\pi^2}
C_F  \frac{1+(1-x)^2}{x}  
\int \frac{d\Delta z}{\lambda_g(z)}  
\int d^2{\bf q}_\perp  \frac{1}{\sigma_{el}} \frac{d\sigma_{el}^{\; {\rm medium}}}{d^2 {\bf q}_\perp} \; 
 \Bigg[  \frac{\vc{B}_{\perp}}{\vc{B}_{\perp}^2} \mcdot \left( \frac{\vc{B}_{\perp}}{\vc{B}_{\perp}^2}  -  \frac{\vc{C}_{\perp}}{\vc{C}_{\perp}^2}   \right)
   \nonumber \\
&&  \qquad \qquad  
  \times \big( 1-\cos[(\Omega_1 -\Omega_2)\Delta z] \big)  + \frac{\vc{C}_{\perp}}{\vc{C}_{\perp}^2} \mcdot \left( 2 \frac{\vc{C}_{\perp}}{\vc{C}_{\perp}^2}   
-    \frac{\vc{A}_{\perp}}{\vc{A}_{\perp}^2} - \frac{\vc{B}_{\perp}}{\vc{B}_{\perp}^2}  \right) \big(1- \cos[(\Omega_1 -\Omega_3)\Delta z] \big) \nonumber \\  
&&
   \qquad \qquad    + \frac{\vc{B}_{\perp}}{\vc{B}_{\perp}^2} \mcdot \frac{\vc{C}_{\perp}}{\vc{C}_{\perp}^2} 
\big( 1 -  \cos[(\Omega_2 -\Omega_3)\Delta z] \big)  
+ \frac{\vc{A}_{\perp}}{\vc{A}_{\perp}^2} \mcdot \left( \frac{\vc{D}_{\perp}}{\vc{D}_{\perp}^2} - \frac{\vc{A}_{\perp}}{\vc{A}_{\perp}^2} \right) 
\big(1-\cos[\Omega_4\Delta z]\big)  \nonumber \\
&& \qquad \qquad  -\frac{\vc{A}_{\perp}}{\vc{A}_{\perp}^2} \mcdot \frac{\vc{D}_{\perp}}{\vc{D}_{\perp}^2}\big(1-\cos[\Omega_5\Delta z]\big)   
+  \frac{1}{N_c^2}  \frac{\vc{B}_{\perp}}{\vc{B}_{\perp}^2} \mcdot  \left( \frac{\vc{A}_{\perp}}{\vc{A}_{\perp}^2}  -   
\frac{\vc{B}_{\perp}}{\vc{B}_{\perp}^2}      \right)
\big( 1-\cos[(\Omega_1 -\Omega_2)\Delta z] \big)   \Bigg] \, .  \qquad   \quad    \label{CohRadSX1} 
\end{eqnarray} 
Here,  $x$ is the large light-cone momentum fraction taken by the daughter parton. This choice corresponds to having the soft gluon emission limit when   $x\ll 1$.
In Eq.~(\ref{CohRadSX1})  $\lambda_g(z)$ is the scattering length of a gluon in the medium and $\left(1/\sigma_{el}\right) \,{d\sigma_{el}^{\; {\rm medium}}}/{d^2 {\bf q}_\perp}$ stands for normalized elastic scattering cross section of a parton in nuclear matter. The kinematics of the LO branching processes and interactions with the medium mediated by Glauber gluons 
enter through 
\begin{eqnarray}
&&\vc{A}_{\perp}=\vc{k}_{\perp},\,\, \vc{B}_{\perp}=\vc{k}_{\perp} + x \vc{q}_{\perp} , \,\,
\vc{C}_{\perp}=\vc{k}_{\perp} -  (1-x)\vc{q}_{\perp},  \,\,
\vc{D}_{\perp}=\vc{k}_{\perp}-\vc{q}_{\perp},\,\, \\
&&\Omega_1-\Omega_2=\frac{\vc{B}_{\perp}^2}{p_0^+ x(1-x)}, \, \, \Omega_1-\Omega_3=\frac{\vc{C}_{\perp}^2}{p_0^+x(1-x)},
\Omega_2-\Omega_3=\frac{\vc{C}_{\perp}^2-\vc{B}_{\perp}^2}{p_0^+x(1-x)}, \nonumber\\
&&  \,\,
\Omega_4=\frac{\vc{A}_{\perp}^2}{p_0^+x(1-x)}, \, \,
 \Omega_5=\frac{\vc{A}_{\perp}^2-\vc{D}_{\perp}^2}{p_0^+x(1-x)}.
\end{eqnarray}
The medium-induced 
splitting for $q \rightarrow gq$ can be obtained from Eq.~(\ref{CohRadSX1}) with the substitution 
$x\rightarrow 1-x$. 
The gluon splitting kernels are:   
{\allowdisplaybreaks
\begin{eqnarray}
&&   \left( \frac{dN^{\rm med}}{ dxd^2\vc{k}_{\perp} }\right)_{ \left\{ \begin{array}{c}   g \rightarrow gg\\     g\rightarrow q\bar{q}  \end{array} \right\} } 
= 
 \left\{ \begin{array}{c}     \frac{\alpha_s}{2\pi^2} \, 2 C_A \left(\frac{x}{1-x}+\frac{1-x}{x}+x(1-x) \right)  \\[1ex]
                           \frac{\alpha_s}{2\pi^2}  T_R \left( x^2+(1-x)^2 \right)  \end{array} \right\}
\int {d\Delta z}   \left\{ \begin{array}{c}    \frac{1}{\lambda_g(z)} \\[1ex]  \frac{1}{\lambda_q(z)}    \end{array} \right\}
\nonumber \\
&& \qquad \qquad
\times \int d^2{\bf q}_\perp  \frac{1}{\sigma_{el}} \frac{d\sigma_{el}^{\; {\rm medium}}}{d^2 {\bf q}_\perp} \;  
\Bigg[  2\, \frac{\vc{B}_{\perp}}{\vc{B}_{\perp}^2} \mcdot \left(\frac{\vc{B}_{\perp}}{\vc{B}_{\perp}^2}-\frac{\vc{A}_{\perp}}{\vc{A}_{\perp}^2}\right) \big( 1-\cos[(\Omega_1 -\Omega_2)\Delta z]  \big)  \nonumber \\ 
&& \qquad \qquad +2\, \frac{\vc{C}_{\perp}}{\vc{C}_{\perp}^2} \mcdot \left(\frac{\vc{C}_{\perp}}{\vc{C}_{\perp}^2}-\frac{\vc{A}_{\perp}}{\vc{A}_{\perp}^2}\right) \big( 1-\cos[(\Omega_1 -\Omega_3)\Delta z]  \big)
\nonumber \\
 &&\qquad \qquad +  \left\{ \begin{array}{c}   - \frac{1}{2}   \\[1ex] \frac{1}{N_c^2-1} \end{array} \right\}
\Bigg(2 \frac{\vc{B}_{\perp}}{\vc{B}_{\perp}^2}\mcdot\left(\frac{\vc{C}_{\perp}}{\vc{C}_{\perp}^2}-\frac{\vc{A}_{\perp}}{\vc{A}_{\perp}^2}\right)\big(1-\cos[(\Omega_1-\Omega_2)\Delta z ]\big)
\nonumber\\
 &&\qquad \qquad  +2\,\frac{\vc{C}_{\perp}}{\vc{C}_{\perp}^2}\mcdot\left(\frac{\vc{B}_{\perp}}{\vc{B}_{\perp}^2}-\frac{\vc{A}_{\perp}}{\vc{A}_{\perp}^2}\right)\big(1-\cos[(\Omega_1-\Omega_3)\Delta z]\big)-2\,\frac{\vc{C}_{\perp}}{\vc{C}_{\perp}^2}\mcdot \frac{\vc{B}_{\perp}}{\vc{B}_{\perp}^2}\big(1-\cos[(\Omega_2-\Omega_3)\Delta z]\big)
\nonumber\\
 &&\qquad \qquad +2\,\frac{\vc{A}_{\perp}}{\vc{A}_{\perp}^2}\mcdot\left(\frac{\vc{A}_{\perp}}{\vc{A}_{\perp}^2}-\frac{\vc{D}_{\perp}}{\vc{D}_{\perp}^2}\right)\big(1-\cos[\Omega_4\Delta z]\big)+2\,\frac{\vc{A}_{\perp}}{\vc{A}_{\perp}^2}\mcdot \frac{\vc{D}_{\perp}}{\vc{D}_{\perp}^2}\big(1-\cos[\Omega_5\Delta z]\big)\Bigg) \Bigg] \, ,      
\label{CohRadSX2} 
\end{eqnarray} 
}Note that the longitudinal and transverse momentum dependencies of the branching processes in matter do not factorize 
and for phenomenological applications $dN^{\rm med}(x,{\bf k}_\perp)/dxd^2\vc{k}_{\perp} $ are usually obtained as numerical  grids. Analytic studies of the $x\rightarrow 0, 1$  endpoint divergences have  recently appeared~\cite{Ke:2023ixa}, allowing to write down renormalization group equations for the evolution of medium-induced parton showers and to identify the large QCD matter-specific logarithms being resummed.  

\begin{figure}[t]
\begin{center}
\includegraphics[width= 0.8\textwidth]{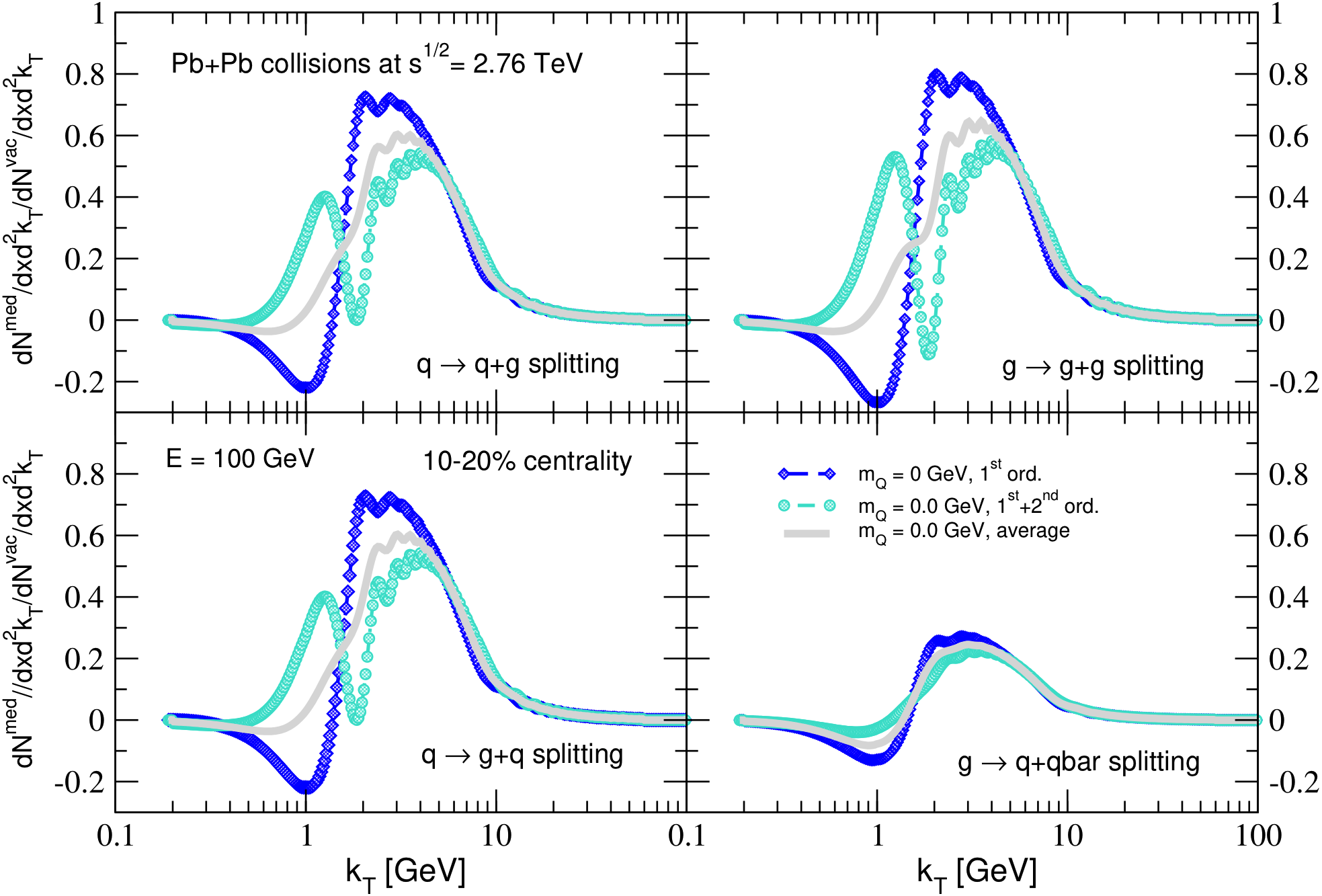} 
\caption{The transverse momentum distribution of the medium-induced radiation, as a ratio to the vacuum radiation spectrum.  Here the distribution is shown for a $100\, \mathrm{GeV}$ jet and $x = 0.3$.  We have chosen a $10-20\%$ centrality cut of $\sqrt{s_{NN}} = 2.76 \, \mathrm{TeV}$ PbPb collisions \cite{Sievert:2019cwq}. }
\label{f:KTspectra}
\end{center}
\end{figure}

One example that illustrates how medium induced showers can affect the TMD structure of jets is shown in Fig.~\ref{f:KTspectra}. It presents the ratio of the transverse momentum $k_T$ dependence of the medium-induced splitting kernel to the vacuum Altarelli-Parisi one - $ \frac{dN^{\rm med}}{dx d^2k_T} \Big/ \frac{dN^{\rm vac}}{dx d^2k_T}$. The blue and cyan symbols represent calculations to different orders in the interaction of the jet with the medium (opacity) and the grey line is the average. In all parton branching channels $i\rightarrow jk$  there is distinct broadening in the transverse momentum and, correspondingly, angular distributions of parton showers and jet constituents. These will manifest themselves in jet cross section and jet substructure modification in reactions with nuclei.

\subsubsection{Jet cross sections }
A recently developed framework to calculate jet cross sections is based on semi-inclusive jet functions  $J_{i}\left(z, p_T R, \mu \right)$, which describe the fragmentation of parton $i$  into a jet of radius $R$~\cite{Kang:2016mcy}, see Sec.~\ref{sec:jetsubjetfrag}.
Since medium induced parton showers emerge from branching processes that have longitudinal and transverse momentum structure different from the one in the vacuum, 
the essential many-body QCD physics is captured in the ratio of observables measured in nucleus collisions relative to the simpler proton ones. 
\begin{align}
R_{eA}^{\langle {\cal O} \rangle} (PS) &=   \langle {\cal O}_{eA} \rangle (PS) / \langle {\cal O}_{ep} \rangle (PS)  \quad : & &  \text{$e+A$ relative to $e+p$} \;,  \\
R_{AB}^{\langle {\cal O} \rangle} (PS) &=   \langle {\cal O}_{AB} \rangle (PS) / \langle {\cal O}_{pp} \rangle (PS) \quad :  & & \text{$A+B$ relative to $p+p$} \;, 
\label{eq:observables}
\end{align}
where $\langle {\cal O} \rangle$ is the observable, $e$, $p$, $A$ and  $B$ are the incoming leptons, hadrons or nuclei, and $PS$ is the phase space variable, such as the transverse momentum, 
rapidity,  subject radius or  fragmentation fraction.
The in-medium splitting functions in Eq.~(\ref{eq:medsplit}) have been used to obtain the  suppression of inclusive 
light and heavy meson production in  heavy ion collision using fixed-order~\cite{Kang:2016ofv}  and resummed calculations~\cite{Chang:2014fba,Kang:2014xsa,Chien:2015vja}.   For the
case of jets, the in-medium effects have been included at fixed order in the semi-inclusive jet functions~\cite{Kang:2017frl,Li:2018xuv,Li:2020rqj,Li:2021gjw}.  

If we denote for brevity
$   f_{i \rightarrow j k }^{\mathrm{med}} \left(z, \mathbf{k}_{\perp}\right) = {dN_{i \rightarrow j k }^{\mathrm{med}} }/{d^2\mathbf{k}_{\perp}  dz}$,  
at one loop the medium correction to the semi-inclusive jet functions read

{\allowdisplaybreaks
\begin{align}\label{eq:sp}
   J_{q}^{\rm{med}}\left(z, p_T R, \mu \right)&=\left[\int_{z(1-z) p_T R}^{\mu} d^2\mathbf{k}_{\perp} f_{q \rightarrow q g}^{\mathrm{med}} \left(z, \mathbf{k}_{\perp}\right)\right]_{+}
 +\int_{z(1-z) p_T R}^{\mu} d^2\mathbf{k}_{\perp} f_{q \rightarrow gq}^{\mathrm{med}} \left(z, \mathbf{k}_{\perp}\right)  \;, 
     \\
    J_{g}^{\rm{med}}\left(z, p_T R, \mu \right)&=
    \left[\int_{z(1-z) p_T R}^{\mu} d^2\mathbf{k}_{\perp} \left(h_{gg} \left(z, \mathbf{k}_{\perp}\right)\left(\frac{z}{1-z}+z(1-z)\right)\right)\right]_{+}
    \nonumber \\ & 
  \ \   + n_f \left[ \int_{z(1-z) p_T R}^{\mu} d^2\mathbf{k}_{\perp} f_{g\to q\bar{q}} \left(z, \mathbf{k}_{\perp}\right) \right]_+ 
    \nonumber \\& 
  \ \   + \int_{z(1-z) p_T R}^{\mu}d^2\mathbf{k}_{\perp} \Bigg( h_{gg}(z, \mathbf{k}_{\perp})\left(\frac{1-z}{z}+\frac{z(1-z)}{2} \right)
    + n_f  f_{g\to q\bar{q}}(z,\mathbf{k}_{\perp}) \Bigg) \;. \label{eq:sp1} 
\end{align}
}Recall that the plus prescription definition is given in Eq.~(\ref{eq:plusdef}).
Here we give the gluon semi-inclusive jet function explicitly and  note that
\begin{align}
    h_{gg} \left(z, \mathbf{k}_{\perp}\right)  =& \frac{ f_{g \rightarrow gg }^{\mathrm{med}}
   \left(z, \mathbf{k}_{\perp}\right)}{ \frac{z}{1-z} + \frac{1-z}{z}+z(1-z)} \;.  
\end{align}
The result for $J_q^{\mathrm{med}}$ $J_g^{\mathrm{med}}$ written in this form is finite for $z\to 1$ and we only need the upper UV cut-off $\mu$, which is suitable for numerical 
implementations and integrations. The contribution of the in-medium shower to jet cross sections depends on its transverse momentum structure.

\begin{figure}[t]
    \centering
       \includegraphics[width=0.5\textwidth]{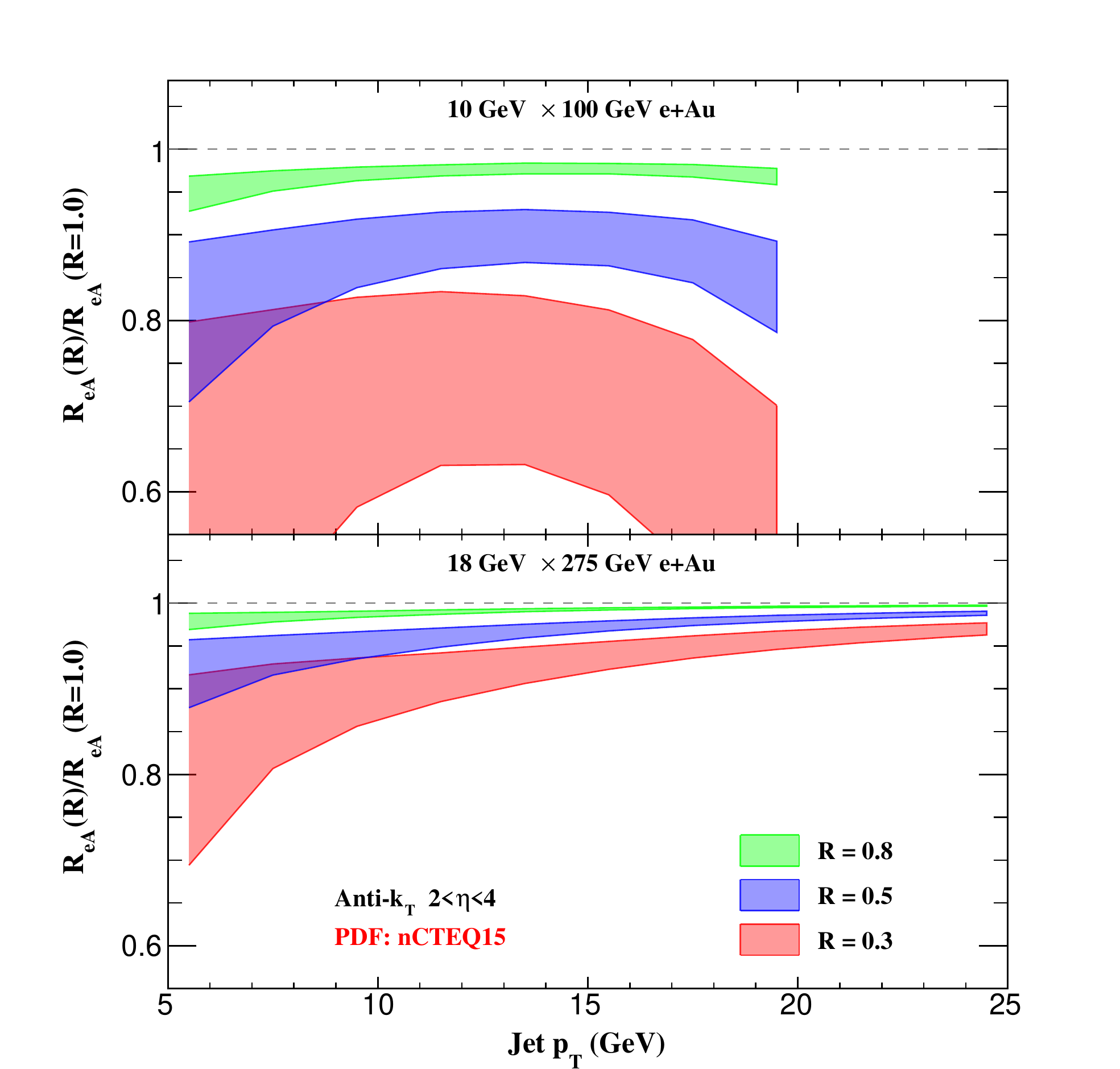}
       \includegraphics[width=0.47\textwidth]{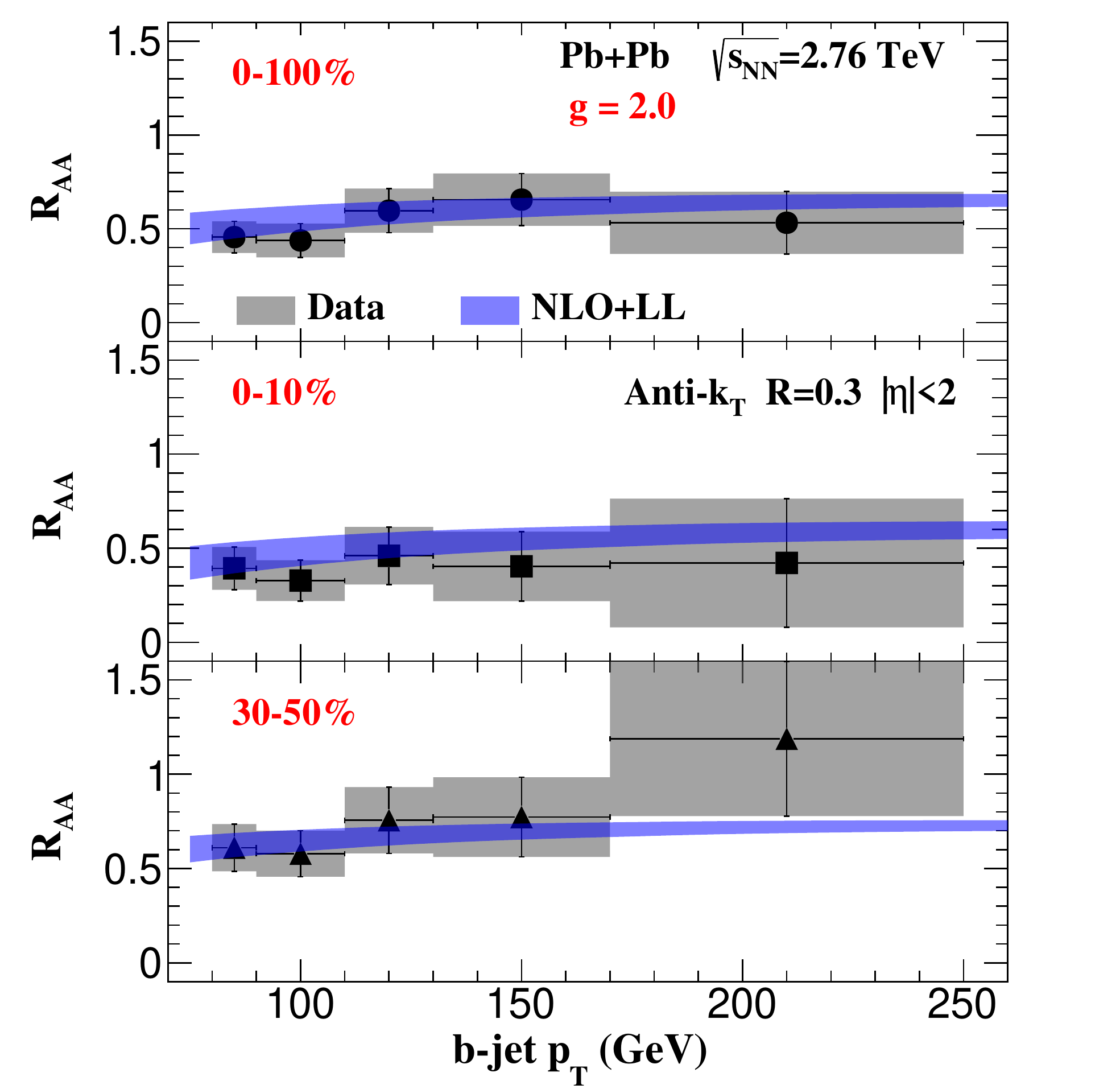}
    \vspace{-0.1cm}
    \caption{Left: ratio of jet  cross section modifications for different  radii $R_{eA}(R)/R_{eA}(R=1.0)$ in 10 $\times$ 100 GeV (upper) and  18 $\times$ 275 GeV (lower) $e+Au$ collisions, where the smaller jet radius is $R=0.3, 0.5$, and $0.8$,  and the jet rapidity  interval is $2<\eta<4$.  Right: the nuclear modification factor  $R_{AA}$  of  $b$-jets , defined as the ratio of the inclusive cross section in heavy ion reactions normalized by the number of binary collisions
    to the cross section in  proton collisions,   for different centrality classes (0-100\%, 0-10\% and 30-50\% ), as indicated  in the legend.  Data is from CMS measurements~\cite{Chatrchyan:2013exa}. Figures originally appeared in Refs.~\cite{Li:2020rqj,Li:2018xuv}.    }
    \label{fig:RpA}
\end{figure}

The  formalism of semi-inclusive jet functions in nuclear matter has been applied to light jet cross sections~\cite{Kang:2017frl,Li:2020rqj}. Very recently, the semi-inclusive jet 
functions for partons fragmenting into heavy flavor jets were computed for proton collisions~\cite{Dai:2018ywt}. This approach has also been extended to 
$c$-jet and $b$-jet production in heavy ion collisions~\cite{Li:2018xuv}.  
Examples of  jet cross section modification in different types of collisions including nuclei is shown in Fig.~\ref{fig:RpA}. These are defined as
\begin{equation}
R_{AA}= \frac{1}{\langle N_{\rm bin} \rangle}\frac{d\sigma_{AA}/dy dp_T}{d\sigma_{pp}/dy dp_T}\,, \qquad   R_{eA}= \frac{1}{A}\frac{d\sigma_{eA}/dy dp_T}{d\sigma_{ep}/dy dp_T}\,, 
\end{equation}
for heavy ion and electron-nucleus reactions. To study cold 
nuclear matter transport properties
\index{nuclear matter transport properties} with jets at the Electron-Ion Collider, it is essential reduce the role of nuclear PDFs and enhance the effects due to final-state interactions.  An efficient strategy  is to measure the ratio of the modifications with different jet radii, $R_{ eA}(R)/R_{eA}(R=1)$,  which is also an observable  very sensitive to the details of  in-medium branching processes~\cite{Vitev:2008rz} and greatly discriminating with respect to theoretical models~\cite{CMS:2019btm}.  Furthermore, it is  very beneficial  to explore smaller center-of-mass energies.  Predictions for the ratio of jet cross section suppressions for different radii at the EIC  is presented in Fig.~\ref{fig:RpA} (left),  where the upper and lower panels correspond to  results for 10 GeV ($e$) $\times$ 100 GeV ($A$) and 18 GeV ($e$) $\times$ 275 GeV ($A$) collisions, respectively.  The plot in the upper panel is truncated around $p_T \sim 20$~GeV because of phase space  constraints  in the lower energy  collisions. The red, blue, and green bands denote ratios with $R=0.3\,,0.5\,,0.8$, respectively. 
Since medium-induced parton showers are broader than the ones in the vacuum, for smaller jet radii the suppression from final-state interactions is more significant.   
Even though the scale uncertainties also grow,  the nuclear effect is clear and its magnitude is further enhanced  by the steeper $p_T$ spectra at lower $\sqrt{s}$. Centrality-dependent measurements can provide further insights into the path length dependence of final-state interactions in nuclear matter~\cite{Li:2023dhb} and  centrality class determination has been shown to be feasible via neutron tagging~\cite{Chang:2022hkt}  at the EIC.

A different type of nuclear modification is  shown in Fig.~\ref{fig:RpA} (right) --- $R_{AA}$ in lead-lead collisions at the LHC at $\sqrt{s} = 2.76$~TeV. 
Numerical calculations of $b$-jet suppression are compared to data~\cite{Chatrchyan:2013exa}  from the CMS collaboration. The properly normalized cross section in $A+A$ relative to $p+p$ collisions denoted $R_{AA}$ decreases, indicating larger suppression, with increasing collision centrality. The attenuation factor is less dependent on the  centrality when compared to the  light jet modification. Theoretical predictions agree very well with the data for both  the inclusive cross sections and the nuclear modification factors. Importantly, this framework can also be applied to heavy flavor in DIS and provide further insight to the transport properties of large nuclei and the physics of hadronization~\cite{Li:2020rqj,Li:2020zbk}.

\subsubsection{Jet substructure }

\begin{figure}[!t]
    \centering
        \includegraphics[width=0.45\textwidth]{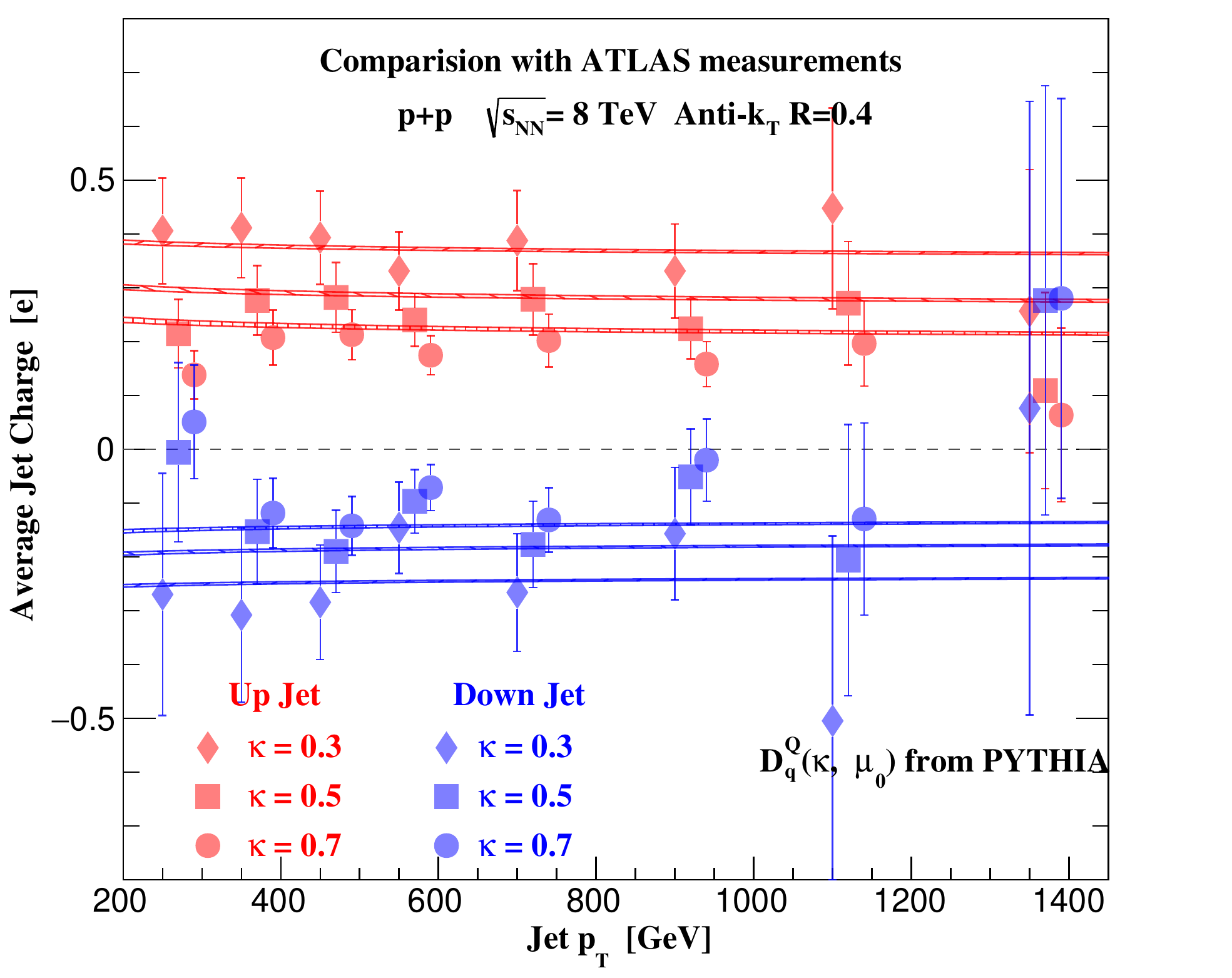}
    \includegraphics[width=0.45\textwidth]{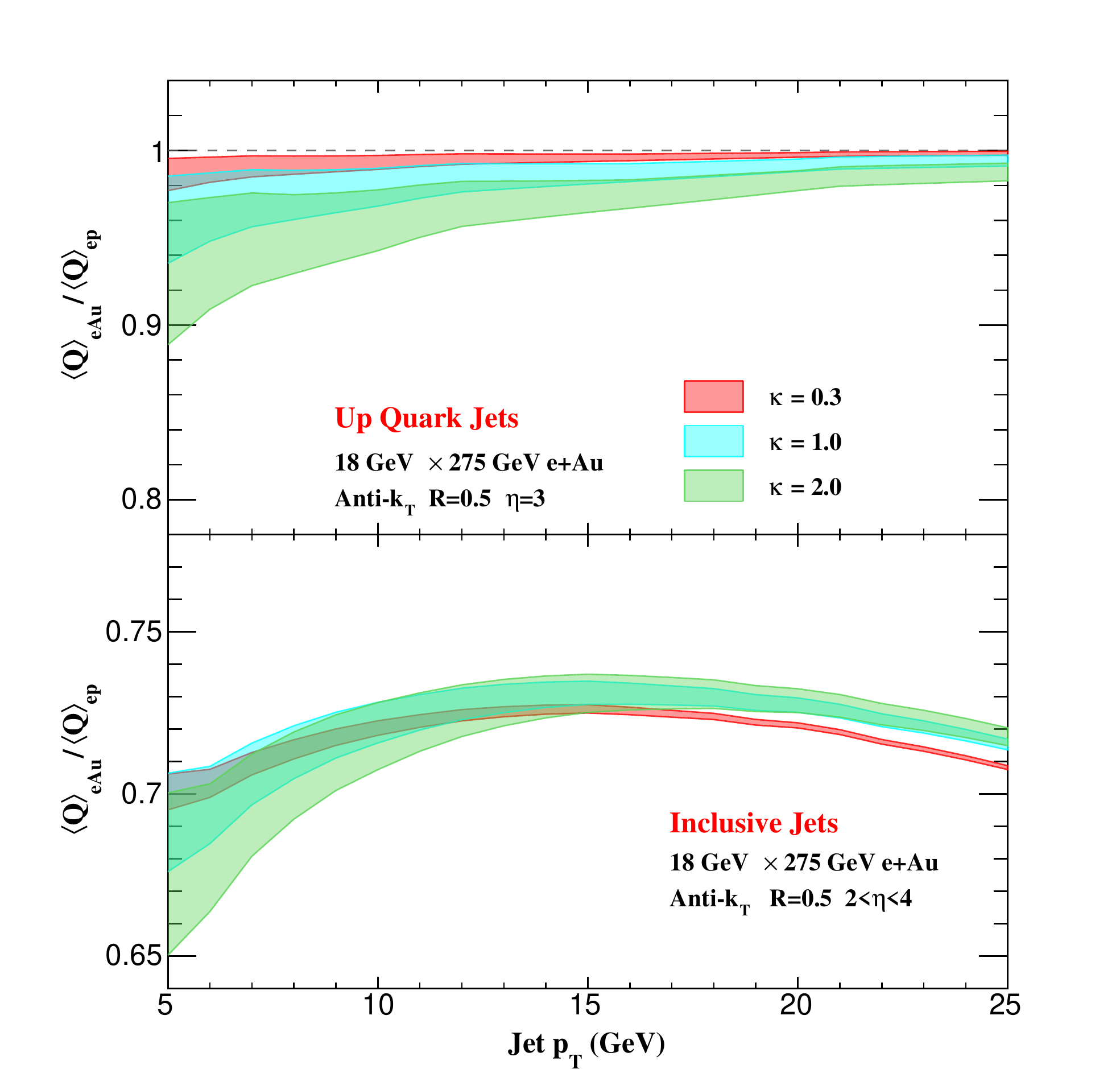}
    \caption{Left: theoretical calculations of up and down jet charges in p+p collisions for various values of $\kappa$ are compared to ATLAS measurements~\cite{Aad:2015cua}.  Right: modifications of the jet charge in $e+Au$ collisions. The upper panel is the modification for up-quark jet with $\eta=3$ 
    in the lab frame. The lower panel is the results for inclusive jet with $2<\eta<4$ in 18 $\times$ 275 GeV $e+Au$ collisions. These figures originally appeared in Refs.~\cite{Li:2019dre,Li:2020rqj}. }
    \label{fig:charge}
\end{figure}

The transverse and longitudinal structure of parton showers can further be studied with jet substructure. One such observable is the average jet charge, defined as the transverse momentum $p_T^{i}$  weighted sum of the charges $Q_i $ of the jet constituents
\begin{equation} \label{eq:charge}
    Q_{\kappa, {\rm jet}}  =  \left( p_T^{\rm jet} \right)^{-\kappa } \sum_{\rm i\in jet} Q_i \left(p_T^{i} \right)^{\kappa } \; ,  \quad \kappa > 0 \; .
\end{equation}
Studies in proton and heavy-ion collisions~\cite{Krohn:2012fg,Li:2019dre,Sirunyan:2020qvi} have found that the jet charge is strongly correlated with the electric charge of the  parent parton and can be used to separate  quark jets from anti-quark jets and to pinpoint their flavor origin. In the framework of soft-collinear effective theory the average jet charge can be expressed as follows \cite{Krohn:2012fg,Li:2019dre}:
\begin{align} \label{eq:Q}
     \langle Q_{\kappa, q} \rangle =& \frac{\tilde{\mathcal{J}}_{qq}(E,R,\kappa,\mu)}{J_{q}(E,R,\mu)} \tilde{D}_q^{Q}(\kappa) 
\  \exp\left[\int_{1{\rm GeV}}^{\mu}\frac{d{\mu'}}{{\mu'}}\frac{\alpha_s({\mu'})}{\pi} \tilde{P}_{qq}(\kappa) \right] \;, 
\end{align}
where $J_{q}(E,R,\mu)$ is a jet function and $\tilde{\mathcal{J}}_{qq}(E,R,\kappa,\mu)$ is the $(\kappa +1)$-th  Mellin moment of the Wilson coefficient for  matching  the quark  fragmenting  jet  function  onto a quark fragmentation  function. Note that up to NLO gluons do not contribute to the average jet charge. The $p_T$ dependence of $\langle Q_{\kappa, q} \rangle$ arises from scaling violations in QCD and comparison of theory~\cite{Aad:2015cua} to ATLAS experimental measurements~\cite{Aad:2015cua} for $\kappa = 0.3, 0.5,0.7$ is shown in the left panel of Fig.~\ref{fig:charge}.

Nuclear matter effects on the jet charge were studied in Refs.~\cite{Chen:2019gqo,Li:2019dre} for the case of heavy-ion collisions. Following  the derivations in Ref.~\cite{Li:2019dre} the average jet charge at the EIC can also be calculated and written as
\begin{align} \label{eq:AAQ}
    \langle Q_{q, \kappa}^{\rm eA} \rangle = &\langle Q_{q, \kappa}^{\rm ep} \rangle  \exp\left[ \int_{\mu_0}^{\mu} \frac{d{\mu'}}{{\mu'}}  
      \frac{\alpha_s({\mu'})}{2\pi^2} (2 \pi {\mu'}^2) \tilde{f}^{\rm med}_{qq}(\kappa,{\mu'}) \right] 
    \left(1+\tilde{\mathcal{J}}^{\rm med}_{qq}-J_{q}^{\rm med}\right) 
   + \mathcal{O}(\alpha_s^2)\,.
\end{align}
Here,  the exponential term comes from the medium-modified DGLAP evolution from $\mu_0\approx \Lambda_{\rm QCD}$ to the jet scale 
and $\tilde{f}^{\rm med}_{qq}(\kappa,\mu)=\int_0^1\,dx\, (x^\kappa-1) \, f^{\rm med}_{qq}(x,\mu)$. 
Finally, for the medium-induced jet functions contributions   in Eq.~(\ref{eq:AAQ}) we have explicitly
\begin{equation}
    \tilde{\mathcal{J}}^{\rm med}_{qq}-J_{q}^{\rm med} = 
    \int_{0}^{2 E x(1-x)\tan R/2} d^2 \mathbf{k}_{\perp} 
    f_{q \rightarrow  qg }^{\rm{med}}\left(x, \mathbf{k}_{\perp} \right)\,.
\end{equation}
Fig.~\ref{fig:charge} (right) presents the jet charge results at the EIC   in 18 GeV $\times$ 275 GeV $e+Au$ collision and for radius parameter $R=0.5$.  
The red, blue and green bands correspond to the jet charge parameter $\kappa=0.3\,,1.0\,,2.0$, see Eq.~(\ref{eq:charge}),   respectively.  
The upper panel shows the modification for the average charge of up-quark initialed jets, where the rapidity is fixed to be  $\eta=3$. It  is defined as  $\langle Q_{q, \kappa}^{\rm eA} \rangle/\langle Q_{q, \kappa}^{\rm ep} \rangle$ and predicted by Eq.~(\ref{eq:AAQ}),  which is independent of the jet flavor and 
originates purely  from final-state interactions.  Flavor separation for jets has  been accomplished at the LHC~\cite{Aad:2015cua} and can be pursued at the EIC.
For a larger $\kappa$, the $(\kappa+1)$-th Mellin moment of the splitting function is more sensitive to  soft-gluon emission, this is the  $x\sim 1$ region in the splitting function where medium enhancement for soft-gluon radiation is the largest.  As shown in the upper panel of Fig.~\ref{fig:charge}, the modification is more significant for larger $\kappa$. The overall corrections are of order 10\% or smaller and  decrease with increasing $p_T$. Measurements of jet charge modification in reactions with nuclei open the possibility for direct observation of medium-induced scaling violations in QCD. 
The modification of the average charge for inclusive jets behaves very differently because there is a  cancellation between contributions from jets initiated by  different flavor partons, in particular from up quarks and down quarks. The lower panel of Fig.~\ref{fig:charge} shows the ratio of average charges for inclusive jets with $R=0.5$ and $2<\eta<4$ for $e+A$ and $e+p$ collisions.  The modification is about 30\% and the $\kappa$ dependence is small due to the large difference between up/down quark density between proton and gold PDFs. Precision measurement of the charge for inclusive jets will be an excellent way to constrain isospin effects and  the up/down quark PDFs in the nucleus.

\begin{figure*}[t]
    \centering
    \includegraphics[scale=0.50]{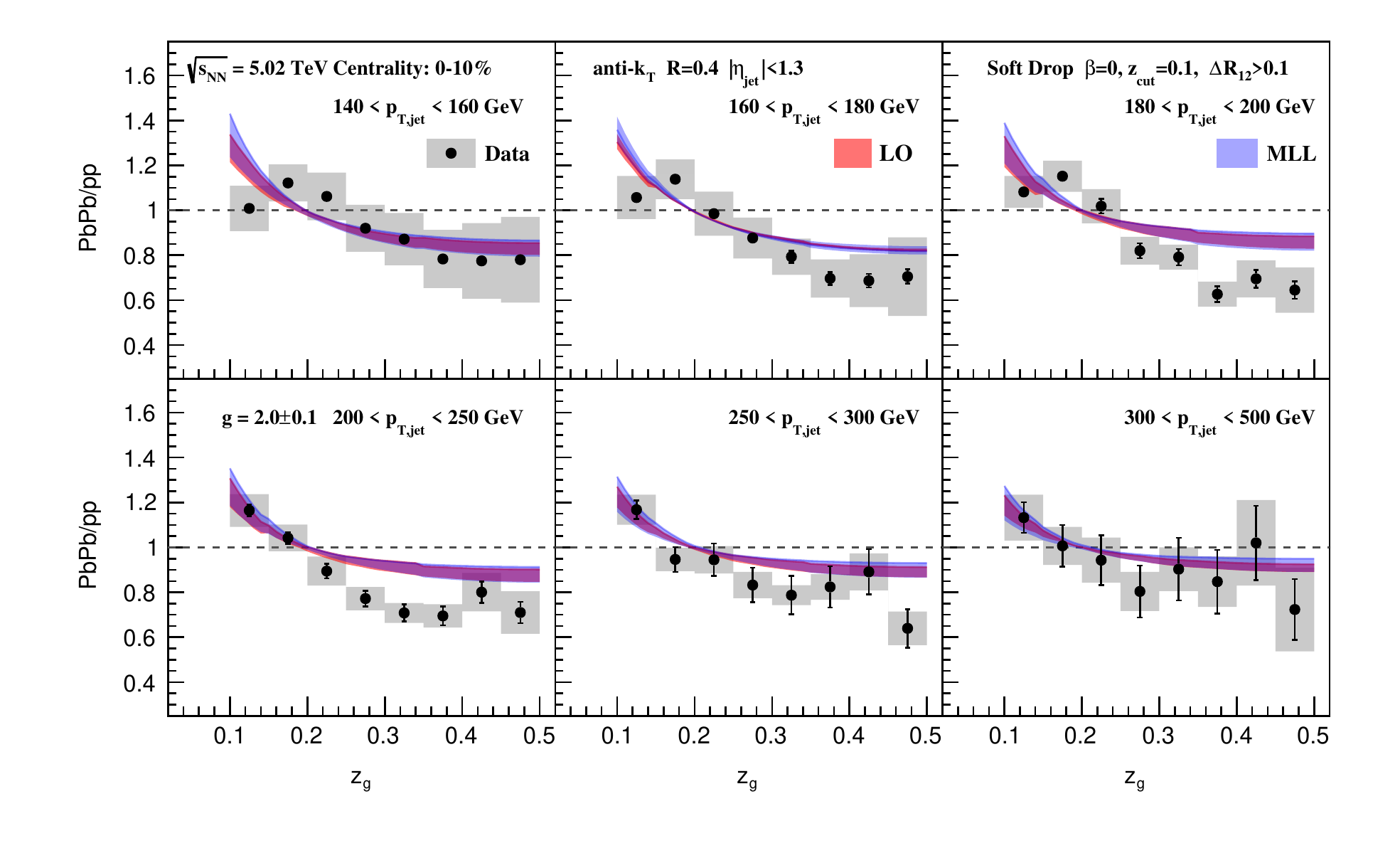} \; 
        \includegraphics[scale=0.5]{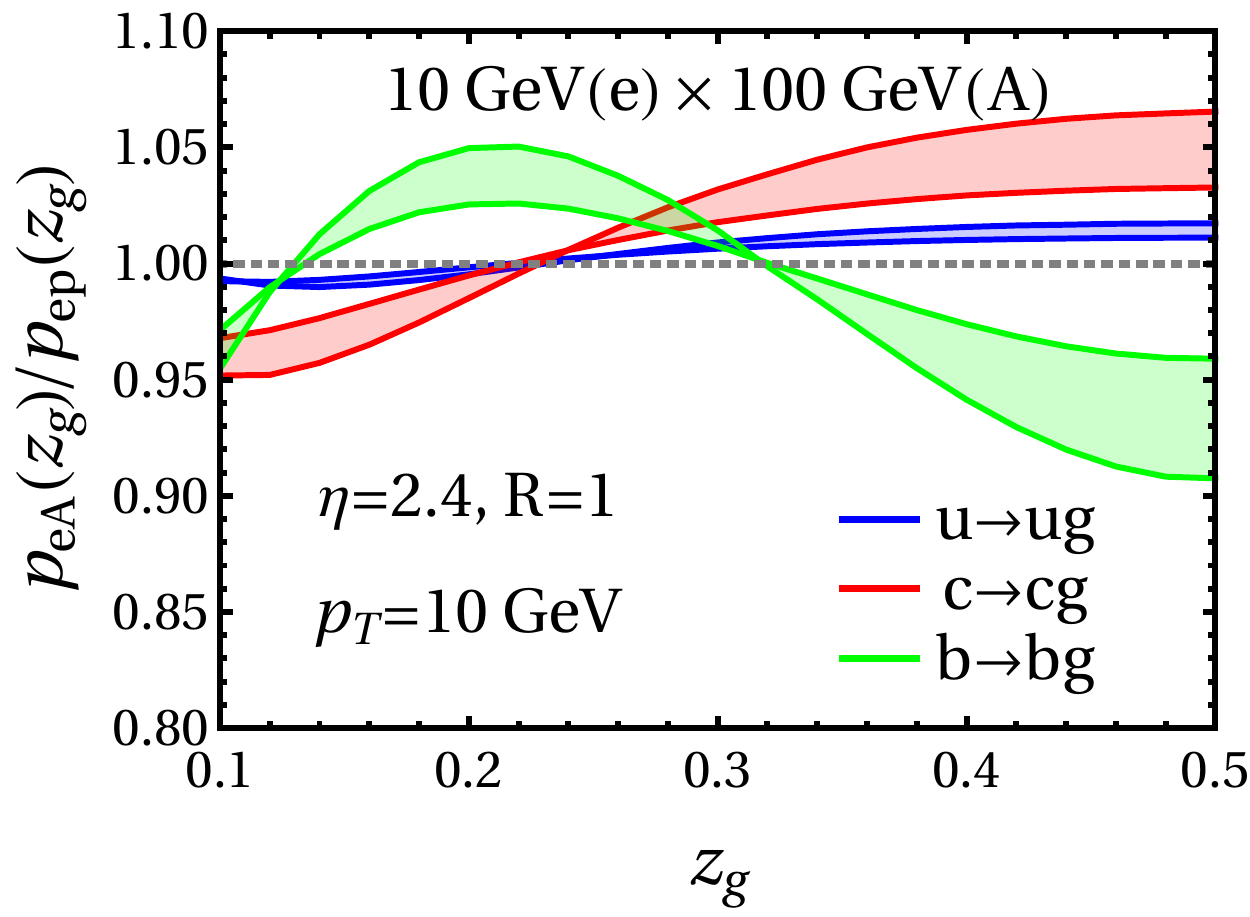}
    \vspace{-0.1cm}
    \caption{Left: theoretical predictions for the in-medium $z_g$ distribution modification  in $Pb+Pb$ collisions with different jet $p_T$ intervals at $\sqrt{s_{\rm NN}}=5.02$ TeV
    are compared to  CMS measurements~\cite{CMS:2017qlm}. Right: calculated light and heavy flavor jet $z_g$ modification in $e+A$ relative to e+p reactions at the EIC at forward rapidity $\eta =2.4$. Note the difference in shape relative to heavy ion collisions and the dependence on the heavy quark mass. Plots are taken from Refs.~\cite{Li:2017wwc,Li:2021gjw}.   }
    \label{fig:cms_all}
\end{figure*}

Another illuminating observable is the
groomed soft-dropped momentum sharing distribution
\index{groomed jet observable} $z_g$ of the two leading subjets inside a reconstructed jet~\cite{Larkoski:2017bvj},
as it can give first-hand information about  the QCD splitting   functions. Given a jet reconstructed using the anti-$k_T$ algorithm with radius $R$, one reclusters the jet using the Cambridge/Aachen algorithm and goes through the branching history, grooming away the soft branch at each step until the following condition is satisfied,
\begin{equation}
    z_\text{cut} < \frac{\min(p_{T_1},p_{T_2})}{p_{T_1}+p_{T_2}} \equiv z_g\;,
    \label{SD}
\end{equation}
i.e., the soft branch must carry more than a $z_\text{cut}$ fraction of the sum of the transverse momenta to not  be dropped. Note that by definition $z_\text{cut}<z_g<\frac{1}{2}$ and the groomed momentum sharing is not sensitive to soft radiation by design. Due to detector granularity one also demands that the angular separation between the two branches $\Delta R_{12} \equiv r_g$, which is also called the groomed jet radius, be greater than the angular detector resolution.
More generally, one can also study the subjet distribution as a function of  the angular separation $r_g$ as proposed in~\cite{Chien:2016led}.  This generalization provides access to the transverse momentum dependent physics of the branching processes. If one can distinguish the splitting process involving heavy flavor, for example by tagging jets and subjets 
with leading charm and beauty mesons ($D$, $B$),  such studies can be extended to heavy quark splitting processes~\cite{Ilten:2017rbd,Li:2017wwc}. It is convenient to rewrite the groomed jet radius $r_g=\theta_g R$ and the double differential 
distribution of subjets inside a reconstructed jet of radius $R$ can be calculated as follows
\begin{equation} \label{eq:mll}
    \frac{dN_j^{\rm vac,MLL}}{ dz_g d\theta_g} = \sum_{i} \left(\frac{dP^{\rm vac}}{dz_g d\theta_g}\right)_{j\to i \bar{i}}
     \underbrace{\exp \left[-\int_{\theta_g}^1 d\theta \int_{z_{\rm cut}}^{1/2} dz  \sum_{i} \left(\frac{dN^{\rm vac}}{dz d\theta}\right)_{j\to i \bar{i}}  \right]}_{\rm Sudakov~Factor}~.
\end{equation}
By integrating over the angular variable one can recover the subjet momentum sharing observable Eq.~(\ref{SD}). In the presence of QCD matter the full splitting functions include both a vacuum and medium-induced components.  Fig.~\ref{fig:cms_all} (left)  presents the modifications for jets of different transverse momenta $p_T$, defined as the ratio of the $z_g$ distributions in the medium and the vacuum. The groomed light jet momentum sharing distributions  are compared to CMS measurements over different kinematic ranges in 0-10\% central $Pb+Pb$ collisions  at $\sqrt{s_{\rm NN}}=5.02$~TeV~\cite{CMS:2017qlm}.  Jets are reconstructed using anti-$k_T$ algorithm with $R=0.4$ and $|\eta|<1.3$ in both $p+p$ and $Pb+Pb $ collisions.  Besides the jet $p_{T}$ and rapidity cut, an additional cut on the distance between the two subjets $\Delta R_{12}>0.1$ is applied due  to the detector resolution effect on the measurements. Thus, the data can be described  by both fixed order~\cite{Chien:2016led} and resummed predictions.  
For heavy flavor jets of relatively small energy, this observable provides a unique opportunity to understand the effects of the heavy quark mass on in-medium parton showers~\cite{Li:2017wwc}. This is illustrated in the left panel of  Fig.~\ref{fig:cms_all} on the example of  a calculation for the EIC~\cite{Li:2021gjw}, where both the shape and magnitude   of the  $z_g$ distribution modification are sensitive to the quark mass.  

To gain further insight into the transverse and longitudinal physics of parton showers and fragmentation,  calculations of jet substructure in matter can be extended to other observables such as jet shapes or jet fragmentation functions~\cite{Chien:2015hda,Sirunyan:2018ncy,Acharya:2018edi,Aaboud:2018hpb,Aaboud:2019oac,Li:2019dre}. We finally point out that an effective field theory of quarkonia in matter, 
NRQCD with Glauber gluons,
\index{non-relativistic QCD!with Glauber gluons}  has been developed~\cite{Makris:2019ttx,Makris:2019kap}. Applications to $e+A$ collisions at the EIC are being investigated. 
Furthermore, an intriguing description of the jet and the medium as an open quantum system has been proposed~\cite{Vaidya:2020cyi} - a direction that can be further pursued.     

\subsection{Outlook}
In this Chapter we discussed jet production and correlated observables. We further showed examples of their longitudinal and transverse substructure in more elementary hadronic collisions as well as nuclear collisions. 

In the introduction to this chapter we gave an introduction to jets and a  description of the algorithms used to define jets. We then described a cross section for jets recoiling against a lepton in SIDIS which gives an alternative way to extract the TMD PDFs. We emphasized that the lepton jet asymmetry is sensitive to the Sivers distribution. Identifying a hadron with a jet is one way to explore jet substructure. Depending on whether the jet is measured exclusively or inclusively this gives rise to distributions such as the Jet Fragmentation Function (JFF) and the Fragmenting Jet Function (FJF). These can be studied in a collinear approximation or the hadron's momentum transverse to the jet axis can also be measured. In the latter case the factorization is similar to that of the TMDs and similar evolution equations can be derived. Measuring these distributions gives novel ways of extracting the fragmentation functions of hadrons. If the identified hadron is heavy quarkonium, these functions can be calculated in the NRQCD factorization formalism, enabling novel tests of this theory of quarkonium production. Finally, we discussed TEEC correlators with jets as well as the modification of jet properties within a nuclear medium.

Jet studies are also of great interest in reactions with a nucleus in the initial state. We showed explicitly analytic and numerical results for the in-medium parton branching processes in the GLV approach,  and discussed selected applications to phenomenology.
In the future, it will be important to study further the modification of jet substructure in collisions involving nuclei relative to simpler reactions (for example $e+A$ relative to $e+p$). Such observables may include jet shapes and transverse to the jet axis distribution of fragmentation functions. The opportunities that the EIC offers in this respect are particularly interesting. In contrast to heavy ion collisions, the energy of the jet's parent parton, which determines the characteristics and relative contribution of the medium-induced parton shower, and the jet transverse momentum, which determines together with the jet radius the available phase space for jet substructure development, are very different. This, in turn may lead to a very different modification, as was shown in the right panel of  Fig.~\ref{fig:cms_all} for heavy flavor jet splitting functions at the EIC~\cite{Li:2021gjw}. Even though jet substructure modification is noticeably smaller that the modification of jet cross sections, we hope that the high-luminosity nature of the future EIC will enable these important measurements. Last but not least the first renormalization group analysis of in-medium parton shower evolution has become available~\cite{Ke:2023ixa}, bringing analytic insight  into the resummation of medium-induced radiation.

%% file: sec-twist3/sec-twist3.tex
\section{Subleading TMDs}
\label{sec:twist3}

\subsection{Introduction} 
\label{sec:subTMDintro}
\index{subleading power TMDs}

In this chapter we consider the subleading-power TMDs (which for brevity we will also refer to as subleading TMDs). 
Unless stated otherwise, by these quantities we mean functions which appear in semi-inclusive reactions suppressed by one inverse power of the hard scale $Q$ of the process. 
Generally, we indicate that subleading contributions are suppressed by powers of $\Lambda/Q$, where $\Lambda$ is a typical hadronic scale which could be the target mass $M$, the mass of a produced hadron $M_h$, a transverse momentum $P_T$, or $\Lambda_{\textrm QCD}$. (Note that quark mass effects can always be eliminated through the QCD equation of motion.) 
In the literature these subleading TMDs are often referred to as twist-3 TMDs, but we will not do so here in order to avoid confusion with the expansion of TMDs for perturbative $p_T$ in terms of longitudinal distributions that are categorized by their twist.

Subleading TMDs are important for a number of reasons.
First, their understanding is required for a complete description of SIDIS and similar semi-inclusive reactions.
Second, they may be relevant for a proper extraction of the leading-power effects from data. 
Third, subleading TMDs can be as sizeable as leading-power TMDs in some situations, particularly when $Q$ is not that large.   
Fourth, those functions are of interest in their own right as they, for instance, offer a mechanism to probe the physics of quark-gluon-quark correlations, which provide novel information about the partonic structure of hadrons, and are largely unexplored.
Such correlations may be considered quantum interference effects, and they could be related to average transverse forces acting on partons inside (polarized) hadrons~\cite{Burkardt:2008ps} as well as other phenomena.
As we will review, experimental information from SIDIS on effects related to subleading TMDs is available already. In the future, the EIC with its large kinematical coverage will be ideal for making further groundbreaking progress in this area. 

From a historical perspective it is very interesting that the subleading-power $\cos \phi_h$ azimuthal modulation of the unpolarized SIDIS cross section was important for the development of the TMD field, since one of the earliest discussions of transverse parton momenta in DIS is related to this observable~\cite{Ravndal:1973kt,Cahn:1978se,Cahn:1989yf}; see also Sec.~\ref{sec:phenoTMDs_intro} for more details.
Generally, although suppressed by $\Lambda/Q$ with respect to leading-power observables, subleading TMD observables are typically not small, especially in the kinematics of fixed-target experiments.  
In fact, the first-ever observed SSA in SIDIS was a sizeable power-suppressed longitudinal target SSA for pion production from the HERMES Collaboration~\cite{HERMES:1999ryv}. 
Those measurements, which triggered many theoretical studies and preceded the first measurements of the (leading-power) Sivers and Collins SSAs, were critical for the growth of TMD-related research.

The theory for subleading-power TMD observables is challenging and still in the early stage of development in comparison to the current state-of-the-art of leading power observables.
Treatments in the literature are mostly limited to a tree-level formalism~\cite{Tangerman:1994bb,Tangerman:1994eh,Mulders:1995dh,Boer:1997mf,Bacchetta:2006tn,Lu:2011pt}; however, early studies beyond tree level can be found in Refs.~\cite{Bacchetta:2008xw,Chen:2016hgw,Bacchetta:2019qkv}.
More recently
results beyond tree level based on the background field method~\cite{Vladimirov:2021hdn,Rodini:2022wki}, SCET~\cite{Ebert:2021jhy},  and the CSS factorization formalisms~\cite{Gamberg:2022lju} have appeared.
In \sec{subTMDobs} we discuss observables in SIDIS which are directly sensitive to subleading TMDs, defining them in terms of general QCD structure functions. 
In \sec{subTMDdistns} we provide definitions for subleading power TMD distributions, including those arising from quark-gluon-quark correlators (referred to as $qgq$ correlators), subleading quark distributions, and corrections associated to simple kinematic expansions. 
In \sec{subTMDfact} we present the current status for factorization formulas that relate the structure functions to leading and subleading TMDs, and then in \sec{subTMDexpt} we give a review of experimental measurements of subleading power TMD observables. 
Lattice QCD and model based determinations of subleading TMDs are taken up in 
\sec{subTMDcalc}. Finally, \sec{subTMDoutlook} gives a summary and outlook.

\subsection{Observables for Subleading TMDs}
\label{sec:subTMDobs}
\index{SIDIS}
\index{subleading power structure functions}

Since the earliest treatments of transverse motion of partons in the nucleon emerged from studies of power-suppressed contributions in SIDIS~\cite{Ravndal:1973kt,Cahn:1978se,Cahn:1989yf}, we will focus our discussion on the general structure of the subleading-power SIDIS cross section. 
In so doing, we consider both unpolarized and polarized targets. 
When the transverse hadron momentum $P_{hT}$
of the final-state hadron is much smaller than $Q$, 
a treatment in a TMD framework is appropriate.\footnote{In a frame in which both the target particle and the final-state hadron have no transverse momentum, one requires $q_T \ll Q$ for TMD factorization to work, where $\qt$ is the transverse momentum of the virtual photon. Since $q_T = P_{hT}/z$, from the point of view of power counting the conditions $q_T \ll Q$ and $P_{hT} \ll Q$ are equivalent.  However, depending on the numerical value for $z$, data which satisfy $P_{hT} \ll Q$ may not satisfy $q_T \ll Q$ and therefore be difficult to describe in a TMD approach.}

The fully differential SIDIS cross section --- assuming a one-photon exchange between the lepton and the nucleon, and unpolarized produced hadrons in the final state --- can be decomposed into 18 structure functions~\cite{Diehl:2005pc, Bacchetta:2006tn}. 
For low transverse momenta of the final-state hadron, eight of those structure functions are leading in a $\Lambda/Q$ expansion; see Eq.~\eqref{eq:SIDIS-leading}. 
Another eight are suppressed by a factor $\Lambda/Q$, while the remaining two are suppressed by a factor $\Lambda^2/Q^2$. 
Focusing on the ten subleading contributions we have, in the notation of Refs.~\cite{Bacchetta:2006tn, Bastami:2018xqd},
\begin{align} \label{e:SIDIS_subleading}
\frac{\df^6\sigma_{\rm subleading}}{\df\xbj \, \df y \, \df \zh \, \df \phi_S \, \df\phi_h \, \df\Phperp^2}
   & =
   	\frac{\alpha_{em}^2}{x\,y\,Q^2}\biggl(1-y+\frac12y^2\biggr)
        \biggl\{p_1 F_{UU,L}+
          \cos(\phi_h)\,p_3 \, F_{UU}^{\cos(\phi_h)}
	\nonumber\\
   & + \lambda\sin(\phi_h) \, p_4 \, F_{LU}^{\sin(\phi_h)}
	+ S_L\sin(\phi_h) \, p_3 \, F_{UL}^{\sin(\phi_h)}
	+ \lambda\, S_L \cos(\phi_h) \, p_4 \, F_{LL}^{\cos(\phi_h)}\phantom{\frac11}
	\nonumber\\
   & + S_T\sin(2\phi_h-\phi_S)\,p_3\,F_{UT}^{\sin(2\phi_h-\phi_S)}
        + S_T\sin(\phi_S)\,p_3\,F_{UT}^{\sin(\phi_S)} 
        \nonumber\\
   & + S_T \sin( \phi_h-\phi_S) \,p_1\, F_{UT,L}^{\sin( \phi_h-\phi_S)} \phantom{\frac11}\nonumber\\
   &  	+ \lambda\,S_T\cos(\phi_S)\,p_4\,F_{LT}^{\cos(\phi_S)}
        + \lambda\,S_T\cos(2\phi_h-\phi_S)\,p_4\,F_{LT}^{\cos(2\phi_h-\phi_S)}
	\biggr\} \,,
\end{align}
where the kinematic prefactors $p_i$ in Eq.~\eqref{e:SIDIS_subleading} are given in Eq.~\eqref{eq:y-prefactors}.  
We refer the reader to Sec.~\ref{sec:TMDSIDIS} for more details about the notation. The structure functions $F_{UU,L}$ and $F_{UT,L}^{\sin( \phi_h-\phi_S)}$ are of ${\cal O}(\Lambda^2/Q^2)$ for small transverse momenta of the final-state hadron. In this chapter we will focus on the remaining eight which are ${\cal O}(\Lambda/Q)$.

Although we use some structure functions from Eq.~\eqref{e:SIDIS_subleading} as benchmark observables for subleading-power TMDs, we would like to mention that there are several other observables of this kind.
For example, Ref.~\cite{Yang:2016qsf} addresses the production of polarized hadrons, e.g., lambda baryons, in SIDIS within the TMD formalism through $\mathcal{O}(\Lambda/Q)$. 
Ref.~\cite{Wei:2016far} even discusses $\mathcal{O}(\Lambda^2/Q^2)$ effects in SIDIS within the TMD formalism.
Observables sensitive to subleading TMDs may also be found in other processes such as the Drell-Yan dilepton production~\cite{Lu:2011th} and electron-positron annihilation into two almost back-to-back hadrons~\cite{Boer:1997mf, Boer:2008fr, Wei:2014pma, Chen:2016moq}.

In the parton-model approximation, the structure functions in Eq.~\eqref{e:SIDIS_subleading} can be expressed through subleading quark TMDs.
These subleading TMDs are defined in \sec{subTMDdistns}, while the corresponding factorization-based cross section formulas can be found in \sec{subTMDfact}.
The very fact that we have a considerable amount of data 
for the structure functions in Eq.~\eqref{e:SIDIS_subleading} alone gives a strong justification to study subleading TMDs in detail.

\subsection{Subleading TMD Distribution Functions} 
\label{sec:subTMDdistns}

Various sources for power suppressed terms have been identified and discussed in the literature. This includes corrections associated to kinematic prefactors involving contractions between the leptonic and hadronic tensors, which are sometimes referred to as kinematic power corrections.  Another type of contribution involve subleading terms in quark-quark correlators involving Dirac structures that differ from the leading power ones in \eq{Gamma_LP}, which are sometimes called intrinsic power corrections~\cite{Kanazawa:2015ajw}.
Finally there are contributions from hadronic matrix elements of (interaction dependent) quark-gluon-quark operators~\cite{Mulders:1995dh}, referred to as  quark-gluon-quark correlators, or $qgq$ correlators for short. These  are sometimes also referred to as dynamic power corrections.
Below we will explain that only the $qgq$ correlators actually introduce new independent subleading power TMDs, while all other $\Lambda/Q$ suppressed power corrections can be expressed in terms of leading power TMDs~\cite{Bacchetta:2006tn,Ebert:2021jhy,Gamberg:2022lju}. For this reason we start our discussion with the $qgq$ correlators.

\subsubsection{Quark-gluon-quark correlators}
\label{sec:qgqcorrelators}

\index{quark-gluon-quark correlators}
\index{dynamic power corrections}

Beyond leading power we begin to probe the structure of partons inside hadrons in greater depth. For observables that involve quark TMDs at leading power, the most important new operators have a gluon field strength in addition to the two quark fields present at leading power.  Matrix elements of these operators give rise to subleading power TMDs, called $qgq$ correlators, which will be defined in this section.  Since to-date the most complete discussion of factorization in subleading power SIDIS has been carried out using SCET in Ref.~\cite{Ebert:2021jhy}, we will introduce a bit of SCET formalism in our 
presentation.\footnote{We continue to follow our conventions, such as for the  normalization of the lightlike basis vectors, so some of the expressions here will differ slightly from Ref.~\cite{Ebert:2021jhy}.}
Where appropriate we also provide a translation to the notation for the $qgq$ correlators used in earlier literature~\cite{Mulders:1995dh,Bacchetta:2004zf,Goeke:2005hb,Bacchetta:2006tn}.
The general structure of these generalized $qgq$ correlators has also been studied in Refs.~\cite{Vladimirov:2021hdn,Rodini:2022wki,Gamberg:2022lju}.

The most general TMD $qgq$ correlators for PDFs and FFs are defined by the following matrix elements~\cite{Ebert:2021jhy}
\begin{align} \label{eq:def_Bb_hat}
 \hat B_{\cB\,i/p_S}^{\rho \beta \beta'}(x, \xi, \bt,\omega_a) &
 = \theta(\w_a)
   \MAe{p(P,S)}{ \big[ \bar\chi_{\na}^{\beta'i}(b_\perp) T_{n_a}(b_\perp,0)\, g{\cB}_{\na\perp,\,-\xi\w_a}^{\rho}(0)
            \,\chi_{\na,\,(1-\xi)\w_a}^{\beta i}(0)\big]_\tau }{p(P,S)}
\,,\nn\\
 \hat \cG_{\cB \, h/i}^{\rho\,\alpha'\alpha}(z,\xi,\bt,\omega_b) &
 = \frac{1}{2z N_c} \theta(\w_b) 
   \SumInt_{X_\bn} {\rm tr} \MAe{0}{ 
   \big[ Z_{\nb}^\dagger(b_\perp)\chi_{\nb}^{\alpha'i}(b_\perp) \big]_\tau }
   {h,X_\bn}
   \nn\\&\qquad\times
   \MAe{h,X_\bn}{ \big[ \bar\chi_{\nb,\,-(1-\xi)\w_b}^{\alpha i}(0)\,g{\cB}_{\nb\perp,\,\xi\w_b}^{\rho}(0)Z_{\nb}(0_\perp)\big]_\tau }{0}
\,,\end{align}
where $x=\omega_a/(\nb\cdot P)$ and $z=(\na\cdot P_h)/\omega_b$, and we recall that the superscripts $i$ are flavor indices. 
Here we make use of the SCET building block field for quarks, $\chi_n$, which involves the good components of the quark field attached to a Wilson line that extends off to infinity. Likewise, for gluons we have the building block field $\cB_{n\perp}^\rho$, which involves a gluon field strength attached to an adjoint Wilson line, where the index $\rho$ is transverse. They are defined by 
\begin{align} \label{eq:chiandB}
  \chi_{\na}^i(x) &= 
   W_{\nb}(\infty,x)\, \frac{\slashed{n}_a\slashed{n}_b}{2} \psi^i(x)  
  \,,\qquad\quad
  \chi_{\na,\w}^i(x) 
    = \bigl[\delta(\w - i\nb\cdot\partial)\,\chi_{\na}^i(x)\bigr]
\,,\\
	\cB_{\na\perp}^\rho(x)
	&
     = \frac{\img}{g} \,\frac{1}{i\nb\cdot\partial} {\nb}_\nu G^{B\nu\rho_\perp}(x) \cW_{\nb}^{BA}(\infty,x) T^A
  \,,\qquad
  \cB_{\na\perp,\w}^\rho(x)
    = \bigl[ \delta(\w + i\nb\cdot\partial)\cB_{\na\perp}^\rho(x) \bigr] 
	\,.\nn
\end{align}
All fields here should be considered to be bare even though we have not indicated this explicitly with an extra superscript $(0)$. Expanded in the gluon field, ${\cal B}_{n_a\perp}^\rho = A_\perp^\rho - (i\partial_\perp^\rho/i\nb\cdot\partial) \nb\cdot A + \ldots$.
The presence of the extra subscripts $\w$ in \eq{chiandB} indicates that the total $\nb\cdot p$ momentum component of the product of fields is fixed to $\w$, as shown. In \eq{def_Bb_hat} the momentum $\w_a$ gives the overall momentum of the fields at position $0$ (and at $b_\perp$), while $\xi$ determines how this momentum is shared between the quark and gluon fields that are at the same transverse position. The presence of $\xi$ corresponds in position space to allowing the quark and gluon fields that are at the same transverse position to be at different positions along the light-cone. 
The results in \eq{def_Bb_hat} are referred to as ``quark'' correlators since the lowest order term in the field without a momentum subscript would create or annihilate a quark.  Analogous formulas also exist for the ``anti-quark'' case where $\omega_a<0$ or $\omega_b<0$. 
 In \eq{def_Bb_hat} the $\alpha$, $\alpha'$, $\beta$, $\beta'$ are spinor indices, $i$ is a flavor index, and all color indices are traced over. 
Just like at leading power, the $[\cdots]_\tau$ notation indicates the presence of additional rapidity regulators. 
Finally, we have transverse Wilson line gauge links $T_{\na}(b_\perp,0)=W[\infty n_a+ 0_\perp \to \infty n_a + b_\perp]$, $Z_{\nb}(0_\perp)=W[\infty n_b + \infty a_\perp\to \infty n_b + 0_\perp]$ and $Z_{\nb}^\dagger(b_\perp)=W[\infty n_b + 0_\perp\to \infty n_b + \infty a_\perp]$.

The configuration space geometry of the $qgq$ correlators is actually quite similar to that of the TMDs at leading power. For the PDF, comparing to the staple shaped Wilson line path with two quarks on each end shown in \fig{wilsonlines}, the additional ingredient for the $qgq$ correlators is essentially that we add an extra gluon field strength at a new position on one of the light-cone paths. Similarly for the FF, a field strength is also added on one of the light-cone paths.

Contributions to the factorized hadronic tensor that involve the $qgq$ correlators also contain the same soft function as at leading power~\cite{Ebert:2021jhy}. This occurs because the soft gluons probe the $qg$ pair at the same transverse position, and only see the corresponding product of operators in its combined color triplet state. 
Therefore the bare soft function can be absorbed into these correlators just like at leading power, motivating the bare redefinitions
\begin{align}\label{eq:def_BB}
 \tilde B_{\cB \, i/p_S}^{\rho\,\alpha'\alpha}(x,\xi,\bt,\ldots) &
 = 
 \hat{B}_{\cB \, i/p_S}^{\rho\,\alpha'\alpha}(x,\xi,\bt,\ldots)
  \frac{\sqrt{\tilde S_{\na\nb}^0(b_T,\ldots)}}
   {\tilde S^{0\subt}_{\na\nb}(b_T,\ldots)}
\,,\nn\\ 
 \tilde \cG_{\cB \, h/i}^{\rho\,\beta'\beta}(z,\xi,\bt,\ldots) &
 = 
 \hat{\cG}_{\cB \, h/i}^{\rho\,\beta'\beta}(z,\xi,\bt,\ldots) 
  \frac{\sqrt{\tilde S_{\na\nb}^0(b_T,\ldots)}}
  {\tilde S^{0\subt}_{\na\nb}(b_T,\ldots)}
\,,
\end{align}
where $\tilde S^0_{\na\nb}$ is the leading power soft function given in \eq{softfunc} and $\tilde S^{0\subt}_{\na\nb}$ are the soft subtractions. The ellipses in various arguments indicate dependence on the UV regulator $\epsilon$, rapidity regulator $\tau$, etc.  We will not go into detail here on how to define the renormalized $qgq$ functions, which is significantly more complicated than at leading power.  Renormalized versions of these functions will depend on additional arguments like $\mu$ and $\zeta$.

\begin{figure}[t]
\centering
\includegraphics[width=0.45\textwidth]{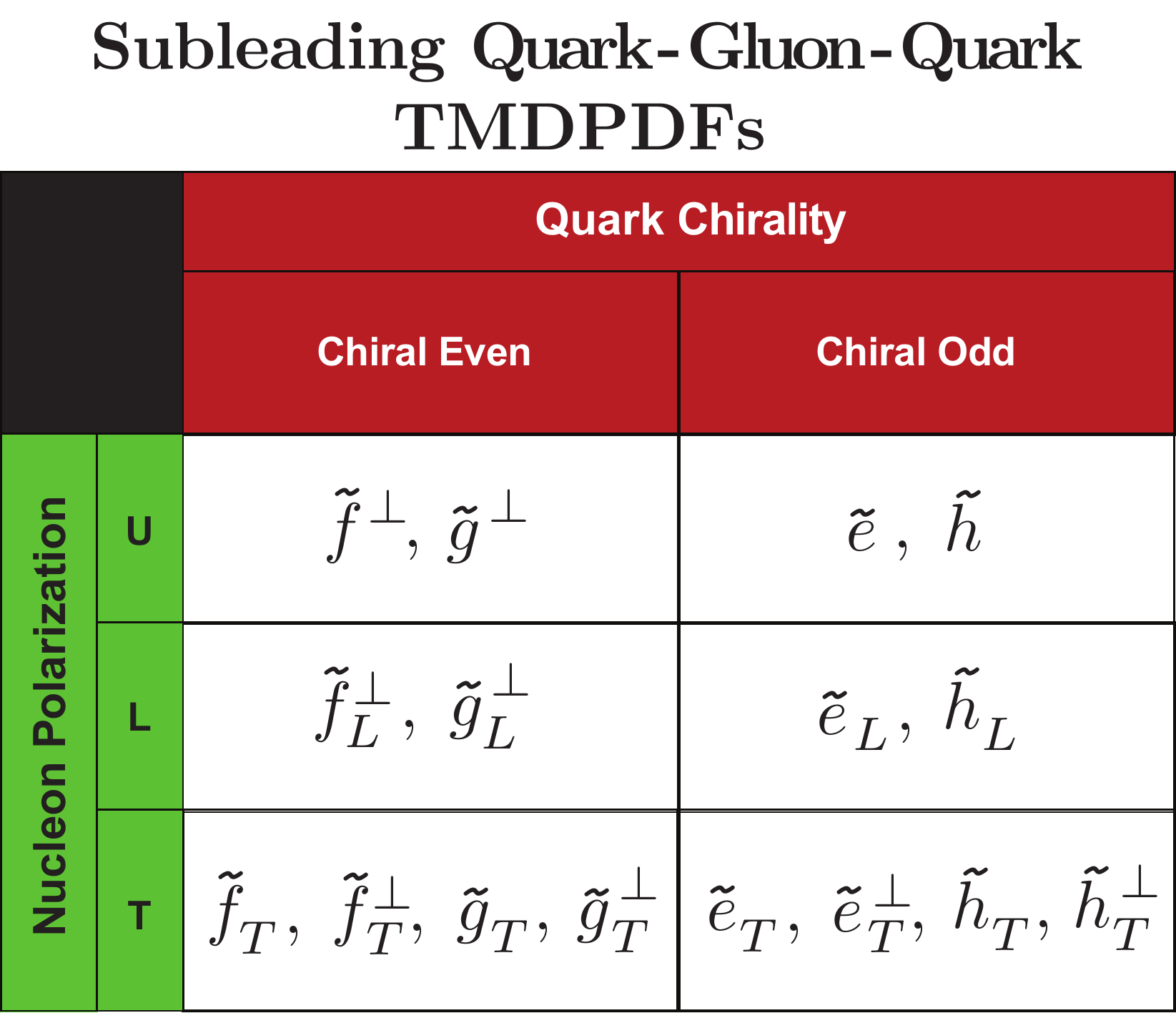}
\hspace{0.5cm}
\includegraphics[width=0.45\textwidth]{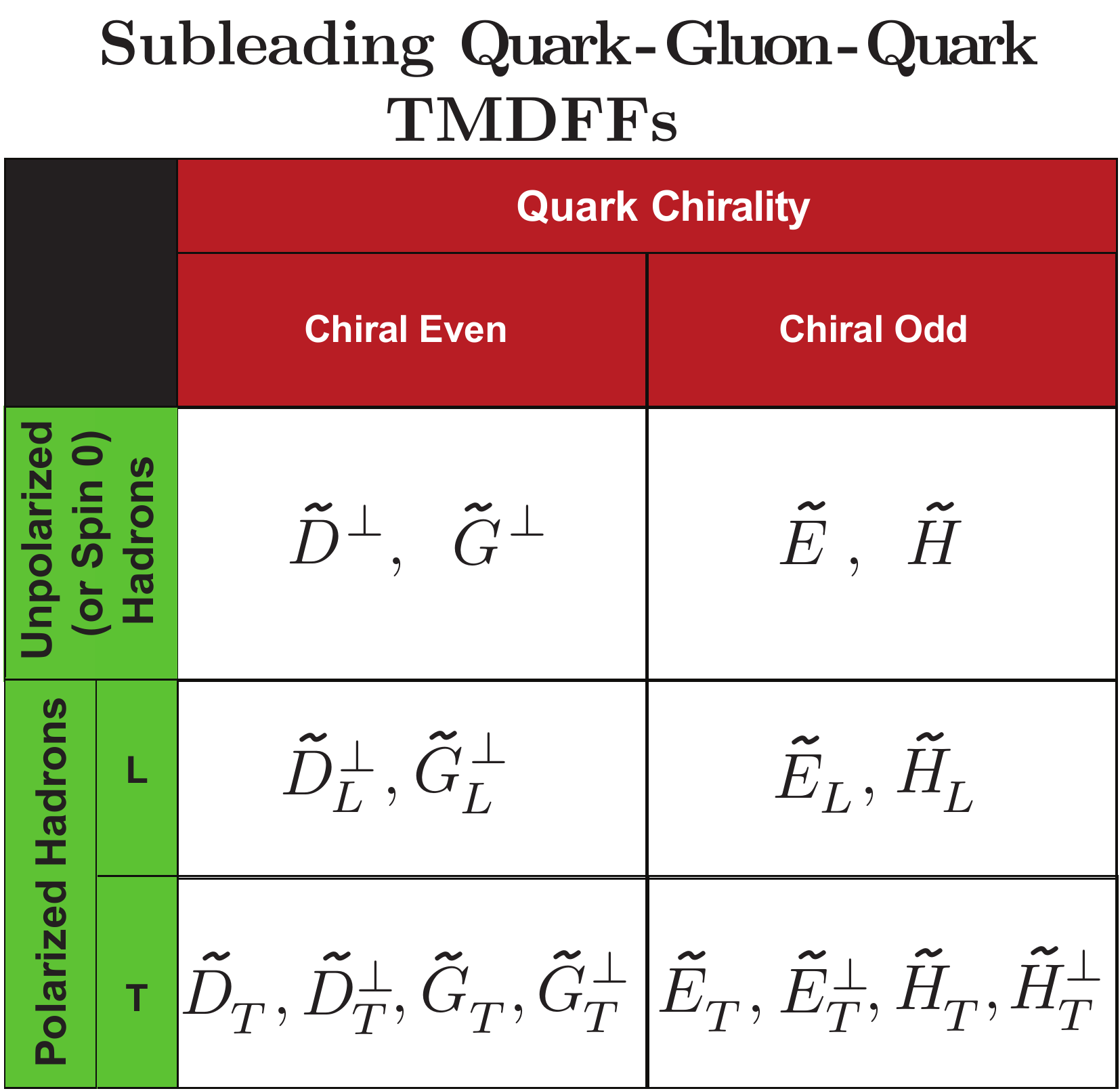}
\caption{
Table of the subleading quark-gluon-quark ($qgq$) TMDPDFs for the nucleon, which are suppressed in observables by the factor $\Lambda/Q$. 
The columns indicate the quark chirality, and rows the nucleon polarization~\cite{Mulders:1995dh,Bacchetta:2004zf,Goeke:2005hb,Bacchetta:2006tn}. 
\label{fig:TMDPDFs_qgq}}
\end{figure}

The general Lorentz decomposition of the $qgq$ TMD PDF with a polarized spin-$1/2$ hadron $H$ was first studied in~\cite{Mulders:1995dh}, the T-odd case was investigated in~\cite{Boer:2003cm}, and the complete decomposition was given in Ref.~\cite{Bacchetta:2006tn}.
In these papers the analysis was carried out for the correlator integrated over $\xi$, but the same Lorentz decomposition holds equally well for the generalized case discussed here. The result is
{\allowdisplaybreaks
\begin{align}\label{eq:qgqdecomp}
 \tilde B_{\cB\,i/H}^\rho(x,\xi,\bt) &
 = \frac{\Ma}{4 \Pa^+} \biggl\{
    \biggl[ -\img\Ma \bigl(\gtilde{f}{}^{\perp(1)}
        +\img\kappa \gtilde{g}{}^{\perp(1)}\bigr) b_{\perp \sigma}
            + \bigl(\kappa\gtilde{f}_T - \img \gtilde{g}_T \bigr) \eps_{\perp \sigma\delta} S_\perp^{\delta}
            \nn\\&\hspace{2cm}
            -\img\Ma S_L\bigl(\kappa\gtilde{f}{}_L^{\perp(1)}-\img\,\gtilde{g}{}_L^{\perp(1)}\bigr) \eps_{\perp \sigma \delta} b_{\perp}^{\delta}
            \nn\\&\hspace{2cm}
            - \frac12 \Ma^2\bigl(\kappa\gtilde{f}{}_T^{\perp(2)}-\img\,\gtilde{g}{}_T^{\perp(2)}\bigr) \eps_{\perp\sigma\delta}
              \Bigl(\frac12 b_\perp^2 S_\perp^\delta -b_\perp\cdot S_\perp b_\perp^\delta \Bigr)
    \biggr] \bigl(g_\perp^{\rho \sigma} - \img \eps_\perp^{\rho\sigma} \gamma_5\bigr)
   \nn\\&\hspace{1.8cm}
   -
\left[S_L\bigl(\gtilde{h}_L - \img\kappa\,\gtilde{e}_L\bigr)
  +\img\Ma \,b_\perp\cdot S_\perp \bigl(\gtilde{h}{}_T^{(1)} 
  - \img\kappa\,\gtilde{e}{}_T^{(1)}\bigr) \right]
           \gamma_\perp^{\rho}\,\gamma_5
   \nn\\&\hspace{1.8cm}
   +
\Bigl[\bigl(-\kappa \gtilde{h} + \img\,\gtilde{e}\,\bigr)
     -\img\Ma\bigl( \gtilde{h}{}_T^{\perp(1)}
           + \img\kappa\,\gtilde{e}{}_T^{\perp(1)}\bigr)\,
      \eps_\perp^{\sigma \delta} b_{\perp\sigma}\, S_{\perp\delta}
    \,\Bigr]
     \img \gamma_\perp^{\rho}
   \nn\\&\hspace{1.8cm}
   + \ldots \bigl(g_\perp^{\rho \sigma}
                  + \img \eps_\perp^{\rho\sigma} \gamma_5\bigr)
   \biggr\} \frac{\slashed \na}{2}
  \,.
\end{align}
}
For brevity, we suppress the arguments on the right-hand side. 
Our notation here uses two tildes to indicate the scalar $qgq$ correlators in $b_T$ space, while we reserve the notation with a single wide tilde for the $qgq$ correlators in $k_T$ space, such as $\widetilde f^\perp \equiv \widetilde f^\perp_{i/H}(x, \xi, k_T)$.
Only the displayed terms in \eq{qgqdecomp} contribute in the subleading power factorization formula for SIDIS. These sixteen $qgq$ TMD PDFs can be organized by which hadron polarization channel they contribute to, and by the quark chirality of the spinor indices, as shown in \fig{TMDPDFs_qgq}.  In \eq{qgqdecomp} the $\kappa=\mp 1$ according to \eq{kappaDYSIDIS}, and indicate the terms that are odd under time-reversal. These terms flip sign when considering contributions to the SIDIS versus Drell-Yan processes at subleading power. For SIDIS we can simply set $\kappa=-1$.

For the $qgq$ TMD FF with an unpolarized hadron $h$ we have 
\begin{align} \label{eq:qgqdecompFF}
 \tilde \cG_{\cB\, h/i}^{\rho}(z, \xi, \bt) &
 = \frac{M_h}{4 \Pb^-} \Bigl\{
   \img\Mb\bigl(\gtilde{D}^{\perp(1)} + \img\, \gtilde{G}^{\perp(1)} \bigr)
   b_{\perp \sigma} \bigl(g_\perp^{\rho \sigma} - \img \eps_\perp^{\rho\sigma} \gamma_5\bigr)
  + \bigl(\gtilde{H} - \img\,\gtilde{E}\bigr)\, \img \gamma_\perp^{\rho}
 \nn\\&\hspace{1.5cm}
  + \ldots
    \bigl(g_\perp^{\rho \sigma} + \img \eps_\perp^{\alpha\sigma} \gamma_5\bigr)
  \Bigr\} \frac{\slashed \nb}{2}
\,.\end{align}
Again, for brevity we suppress the arguments on the right-hand side,
so for example
$\gtilde D^\perp\equiv \gtilde D^\perp_{h/i}(z,\xi,b_T,\mu,\zeta_b)$,
and likewise for the all other TMDs. Here only the TMD FFs for a spin-$0$ or unpolarized final state hadron are shown~\cite{Mulders:1995dh}.  A more extensive enumeration of subleading power $qgq$ TMD FFs is shown in \fig{TMDPDFs_qgq}, and the decomposition that includes the hadron spin-dependent terms can be found in Ref.~\cite{Wei:2014pma,Chen:2016moq}.

It is useful to relate the above definitions of $qgq$ correlators to those in earlier literature, for which the most complete discussion was given in 
Ref.~\cite{Bacchetta:2006tn}.  The starting point in this construction are matrix elements with a covariant derivative $D^\mu=\partial^\mu+igA^\mu$, 
which are used to define a gauge-invariant "D-type" TMD correlator of the nucleon $\Phi_D^\rho (x,\bm k_T)$ as well as a fragmentation correlator $\Delta_D^\rho(z,\bm p_T)$ as follows (with spinor indices $\beta$, $\beta'$),
\begin{align}
\label{eq:PhiD}
    (\Phi_D^\rho)^{\beta\beta'}(x,\bm k_T) &= \int \frac{db^-\,d^2 \bm b_T}{(2\pi)^3}\,\mathrm{e}^{ib\cdot k}\langle p(P,S) |\,\bar{\psi}^{\beta'}(0)\,W_{\sqsupset}\,iD^\rho (b)\,\psi^{\beta}(b)\,|p(P,S)\rangle \Big|_{b^+=0}
   \,,\\
\label{eq:DeltaD}
    (\Delta_D^\rho)^{\beta\beta'}(z,\bm p_T)&=\frac{1}{2N_c z}\sum_X \int \frac{d b^+\,d^2 \bm b_T}{(2\pi)^3}\,\mathrm{e}^{ip\cdot b}\langle  0|\,W_{\halfstapleu}\,iD^\rho(b)\,\psi^{\beta}(b)\,|h(P);X\rangle\,\nonumber\\
    &\hspace{5cm}\times \; \langle h(P);X |\,\bar{\psi}^{\beta'}(0)\,W_{\halfstaple}\,|0\rangle\Big|_{b^-=0}
  \,.
\end{align}
The form of the Wilson lines here are the same as at leading power, in particular $W_{\sqsupset}$ is given in Eq.~(\ref{eq:stapleinout}) while $W_{\halfstaple}$, $W_{\halfstapleu}$ are defined in Eq.~(\ref{eq:half_staple_Wilson_line}). 
The definitions are similar to the ordinary TMD objects $f_{i/p_s}^{[\Gamma]}(x,\kt)$ (spin-dependent TMDPDF in Eqs.~(\ref{eq:unsubTMDPDFspin}), (\ref{eq:tmd_decomposition})) and  $\Delta_{h/i}^{[\Gamma]}(z, \pt=- z \pt')$  (TMDFF in Eqs.~(\ref{eq:unsubTMDFFspin}), (\ref{eq:unsubTMDFFspin_pT})), except for an additional insertion of the covariant derivative. 
Following Ref.~\cite{Bacchetta:2006tn} 
one can obtain $qgq$ correlators by subtracting the $\partial^\mu$ part of the $D^\mu$ correlators in a gauge invariant manner, defining~\cite{Mulders:1995dh}
\begin{align} \label{eq:PhiADeltaA}
\Phi_A^\rho(x,\bm k_T) 
&\equiv \Phi_D^\rho(x,\bm k_T)-\bm k_T^\rho\,f_{i/p_s}(x,\bm k_T) 
  \,, \\
 \Delta_A^\rho(z,\bm p_T)
 &\equiv \Delta_D^\rho(z,\bm p_T)-\bm p_T^\rho\,\Delta_{h/i}(z,\bm p_T)
  \,.\nn
\end{align}
The Lorentz decomposition for $\Phi_A^\rho$ and $\Delta_A^\rho$ in terms of scalar functions is identical to that already shown in \eq{qgqdecomp},
for which we use the same notation, so in position space 
\begin{align} \label{eq:QGQTMDPDFParam}
\tilde \Phi_A^\rho(x,\bm b_T) 
&= \frac{x\Ma}{4}\int\!\! \df\xi\, \Big[-i\Ma (\gtilde{f}^{\perp(1)}+i\kappa\gtilde{g}^{\perp(1)})
 b_{T\alpha} (g_T^{\rho\alpha}-i \epsilon_T^{\rho \alpha}\gamma_5)+...\Big] {\slashed \na} \,,
  \nn\\
\tilde \Delta_A^\rho(z,\bm b_T) 
& = \frac{m_h}{4z} \int\!\! \df \xi\, 
 \Big[i\Mb (\gtilde{D}^\perp-i\gtilde{G}^\perp) b_{T\alpha}  (g_T^{\rho\alpha}+i\epsilon_T^{\rho \alpha}\gamma_5)+(\gtilde{H}+i\gtilde{E})i\gamma_T^\rho\!+\!...\Big]
 {\slashed \nb} 
  \,.
\end{align}
where the ellipses denote the remaining terms in \eqs{qgqdecomp}{qgqdecompFF}. 
The key difference here is that in \eq{QGQTMDPDFParam} the $qgq$ correlators are integrated over $\xi$.  For example, for the PDF we have the relation
\begin{align} \label{eq:BbhattoPhiA}
 \int\!\df\xi\, \hat B_{\cB}^{\rho \beta \beta'}(x, \xi, \bt)
 = \frac{(\slashed{n}_a \slashed{n}_b)^{\beta\alpha}}{2}
   \frac{(\tilde\Phi_A^\rho)^{\alpha\alpha'}(x, \bt)}{2 x \Pa^+}
   \frac{(\slashed{n}_b \slashed{n}_a)^{\alpha'\beta'}}{2}
\,.\end{align}
Here the matrices on the far left and far right project onto the good quark components, accounting for the fact that the unprojected fermion field is used in \eqs{PhiD}{DeltaD}. They do not have a practical effect once the Lorentz decomposition has been done, since only the displayed terms in \eqs{qgqdecomp}{qgqdecompFF} that are nonzero with good components are kept. Finally, the $x \Pa^+$ is a simple normalization factor.
As we will discuss below, the $qgq$ correlators integrated over $\xi$ suffice at leading order in the perturbative hard coefficient, but not once $\alpha_s(Q)$ corrections are included.

\subsubsection{Subleading quark-quark correlators and equations of motion}
\label{sec:sublquarkquark}

\index{intrinsic power corrections}
 
When enumerating the hadronic functions that appear in power suppressed factorization formulas, an important concept is that of using a minimal operator basis, thus avoiding redundant operators and functions. An important tool for this reduction is the use of the field equations of motion, which goes back as far as the early analysis of inclusive DIS at leading power~\cite{Georgi:1976ve,Politzer:1980me}, and also play an important role in the enumeration of subleading power contributions in DIS, such as those at twist-4~\cite{Jaffe:1982pm,Ellis:1982cd}. In processes that are more complicated than inclusive DIS, such as those involving TMDs, the reduction of the operator basis may involve more sophisticated relations and be more subtle. 

In SCET the general construction of minimal operator bases at leading and subleading power has a long history, going back to early papers such as~\cite{Beneke:2002ph,Manohar:2002fd,Pirjol:2002km,Beneke:2002ni,Bauer:2003mga}.  
A dedicated effort to find the complete set of operator relations was carried out in~\cite{Marcantonini:2008qn}, resulting in the demonstration that at any power operators can be constructed from three collinear field building blocks $\{\chi_{\na}^i,\cB_{\na\perp}^\rho, \cP_\perp^\rho \}$, where $\cP_\perp$ is a transverse momentum operator, plus additional building blocks needed for a basis of soft field contributions. This reduction of operators is carried out at the bare level, after which renormalization is considered for the minimal basis.  Key ingredients in such constructions include the use of Wilson line unitarity, $W_{\nb}^\dagger W_{\nb}=1$, and the operator relation
\begin{align} \label{eq:Drelation}
 [W_{\nb}(\infty,x)]^\dagger i D_\perp^\rho W_{\nb}(\infty,x) = \cP_\perp^\rho + g \cB_{\na\perp}^\rho 
  \,,
\end{align}
which converts transverse covariant derivatives into $\cP_\perp$ and $\cB_{\na\perp}$ operators. Another important ingredient is the use of only the good fermion field components $\chi_{\na}^i$ for the construction of the subleading power operator basis. In particular, a fermion equation of motion can be used to eliminate the so-called bad fermion components, $\varphi_{\na}$, which satisfy the opposite projection relation to the good components, $\frac12 \gamma^+\gamma^-\varphi^i = \varphi^i$.  This results in the replacement
\begin{align}\label{eq:goodbad}
  \psi^i \to \Big( 1 + \frac{1}{\nb\cdot P}  [W_{\nb}(\infty,x)]^\dagger i\slashed{D}_\perp W_{\nb}(\infty,x) \frac{\slashed n_b}{2} \Big) \chi_{n_a}^i 
  \,.
\end{align}
Together with \eq{Drelation} this reduces operators involving bad fermion components to the $\{\chi_{\na}^i,\cB_{\na\perp}^\rho, \cP_\perp^\rho \}$\ basis. In TMD matrix elements, this same elimination of terms involving bad fermion components has been known since the earliest analyses of subleading power TMDs~\cite{Tangerman:1994bb,Bacchetta:2004zf,Bacchetta:2006tn}, and we discuss these relations in more detail below.
Further advances in operator basis construction in SCET were made in Refs.~\cite{Moult:2015aoa,Kolodrubetz:2016uim,Feige:2017zci,Moult:2017rpl,Chang:2017atu} by showing that modern amplitude helicity techniques can be used to carry out the operator enumeration entirely in terms of scalar building blocks, without the need to enumerate independent contractions of Lorentz indices.

For SIDIS at subleading power, a complete enumeration and analysis of the required operators has been carried out using SCET in Ref.~\cite{Ebert:2021jhy}, building on the above advances. To all orders in $\alpha_s$, the nonzero contributions include contributions involving the $qgq$ correlators discussed in \sec{qgqcorrelators}, 
kinematic power corrections from subleading terms in projecting the hadronic tensor $W_{\mu\nu}$ onto scalar structure functions $W_i = P_i^{\mu\nu} W_{\mu\nu}$,  
and contributions from $\bar\chi_{\na}^i \cP_\perp \chi_{\nb}^i$ type operators. Both of the latter two types of corrections can be entirely expressed in terms of the leading power TMDs discussed in \chap{TMDdefn}. For the kinematic power corrections this is immediately obvious, since they involve the same leading power hadronic matrix elements. For the operators with a $\cP_\perp$ this follows from   relations like~\cite{Ebert:2021jhy}
\begin{align} \label{eq:def_BP_f_a}
 & \theta(\w_a)
  \biggl\{
         \MAe{H}{\bar\chi_{\na}^{\beta' i}(b_\perp) \left[\slashed\cP_\perp{\slashed n}_b\, \chi_{\na,\w_a}^i(0)\right]^{\beta}}{H}
         + \MAe{H}{\left[\bar\chi_{\na}^i(b_\perp)\,
       {\slashed n}_b\,\slashed\cP_\perp^\dagger\right]^{\beta'} \,
  \chi_{\na,\w_a}^{\beta i}(0)}{H}\biggr\}
 \nn \\
 &\quad  
 = -\img \frac{\partial}{\partial b_\perp^\rho}
   \left[ \gamma_\perp^\rho\, {\slashed n}_b\,,\,\hat B_{i/H_S}(x,\bt) \right]^{\beta\beta'}
\,,
\end{align}
where $(P,S)$ labels in the $|H\rangle$ states have been suppressed and we have defined a leading power TMD quark correlator with open spinor indices $\beta, \beta'$
\begin{align}
 \hat B_{i/H_S}^{\beta\beta'}(x = \w_a/P_N^-, \bt) &
 = \theta(\w_a) \, \langle H(P,S) | \bar\chi_{n}^{\beta' i}(b_\perp) \chi_{n,\w_a}^{\beta i}(0) | H(P,S) \rangle
 \,.
\end{align}
The $\cP_\perp$ operators also come multiplied by the same soft Wilson line structures as the operators at leading power. 
Furthermore, for subleading power factorization formula, the relation in \eq{def_BP_f_a} continues to hold when the resulting soft functions in  \eq{softfunc} are absorbed into the TMD PDF correlators on both the right- and left-hand sides, since the difference is of higher order in the power expansion~\cite{Ebert:2021jhy}. 
Together with similar formulas for anti-quark contributions, and for the matrix elements yielding fragmentation functions, this enables all contributions from operators involving a $\cP_\perp$ to be written in terms of leading power TMDs.
In part, this explains why the subleading power factorization formula for SIDIS can be entirely written in terms of leading power TMDs and $qgq$ correlators, as discussed further in \sec{subTMDfact}.

At face value, the relation between the operators resulting in $qgq$ correlators and subleading power quark-quark correlators involving bad fermion components, leaves open the option of using the relation in either direction, for example to eliminate the $qgq$ correlators as opposed to the matrix elements with bad fermion components. Indeed, this is the case when working at tree level in the hard interactions, and the choice of expressing results in terms of the ``intrinsic TMDs'' defined from the subleading power quark-quark correlators has often been used in the literature~\cite{Mulders:1995dh,Bacchetta:2004zf,Goeke:2005hb,Bacchetta:2006tn,Gamberg:2022lju,Rodini:2022wki}. 
For this reason we will give the general basis decomposition for these subleading quark-quark correlator TMDs, and discuss their relation to the $qgq$ correlators in more detail. As we will see below, these relations only involve the $qgq$ correlators integrated over the variable $\xi$ as in \eq{BbhattoPhiA}.  
Thus, when working beyond tree level in the hard interactions, where the $\xi$ dependence becomes relevant, the freedom to choose which TMDs to work with is broken in favor of using the more general $qgq$ correlators~\cite{Ebert:2021jhy}. Nevertheless, results from phenomenology, lattice, or models for the subleading quark-quark correlators can always be expressed as constraints on the $\xi$ integrated $qgq$ correlators, and hence provide valuable information on these objects that is useful even beyond tree level.

\begin{figure}[t!]
\centering
\includegraphics[width=0.45\textwidth]{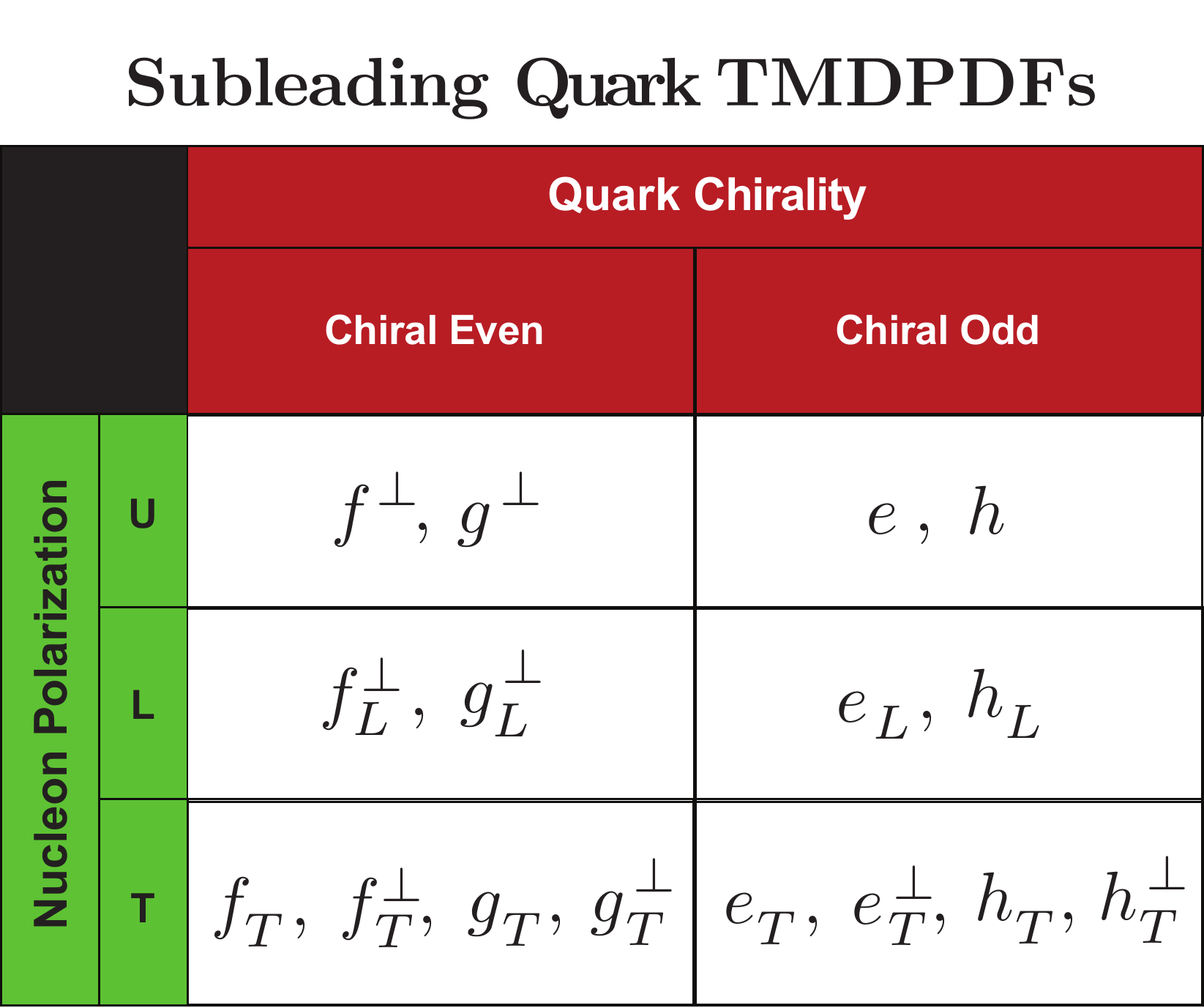} 
\hspace{0.45cm}
\includegraphics[width=0.45\textwidth]{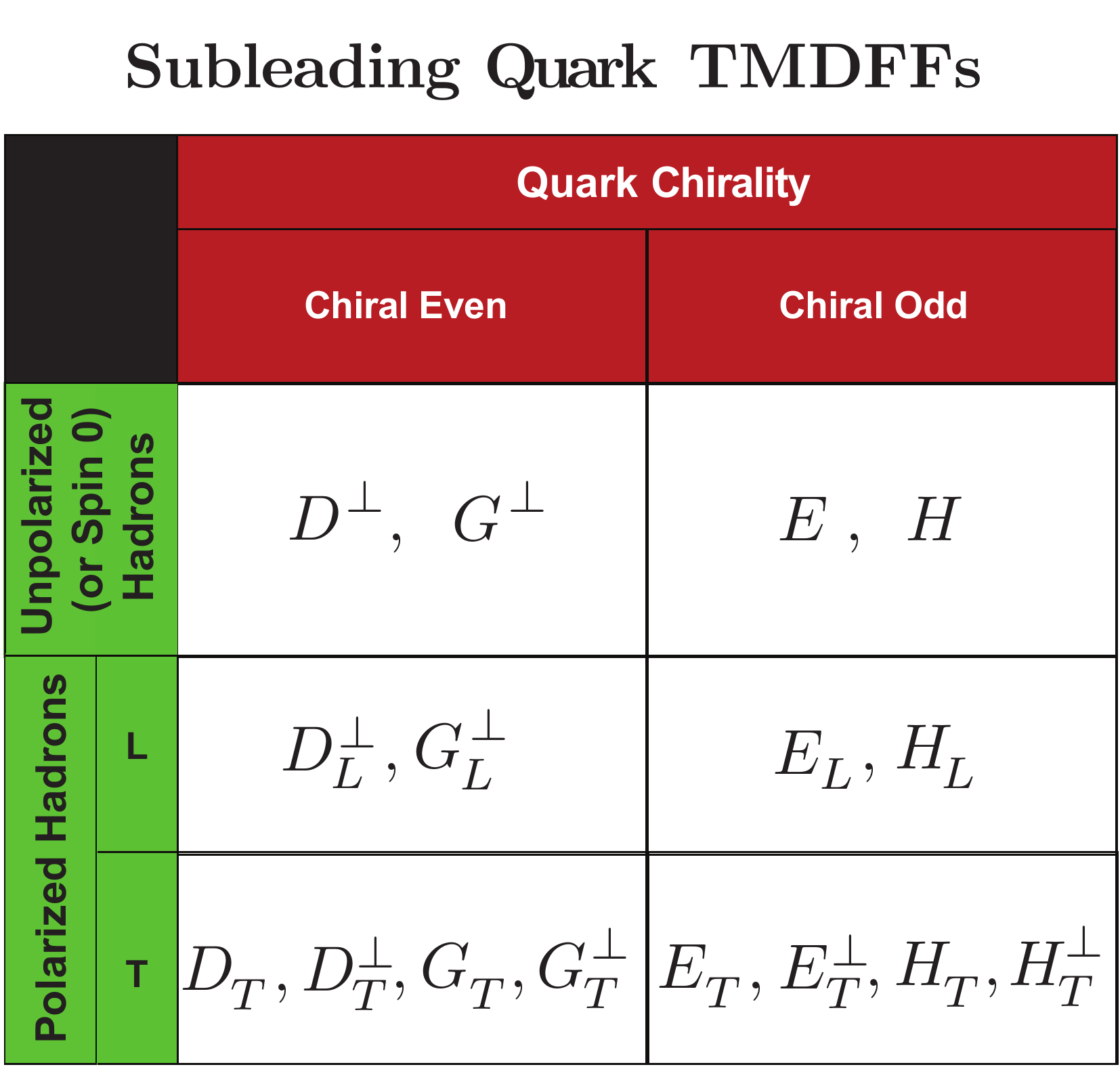}
\caption{
Table of the subleading quark TMDPDFs and TMDFFs, which are suppressed in observables by the factor $\Lambda/Q$. 
The columns indicate the quark chirality, and rows the hadron polarization~\cite{Mulders:1995dh,Bacchetta:2004zf,Goeke:2005hb,Bacchetta:2006tn}.}
\label{fig:TMDPDFs_tw3}
\end{figure}

The subleading power quark-quark correlators, which involve bad fermion components, are obtained by projecting with the Dirac structures
$\Gamma=\left\{ 1, i\gamma_5, \gamma^{\alpha},\gamma^{\alpha}\gamma_5, i\sigma^{\alpha\beta}\gamma_5,i\sigma^{+-}\gamma_5\right\}$.
This yields a total of 16 quark intrinsic TMDPDFs~\cite{Mulders:1995dh,Bacchetta:2004zf,Goeke:2005hb,Bacchetta:2006tn}, which are shown in Fig.~\ref{fig:TMDPDFs_tw3}. Working now in $k_T$ space, and using the generic quark-quark correlator notation in \eq{unsubTMDPDFspin} they are defined by
{\allowdisplaybreaks
\begin{align} \label{eq:sub-TMD-pdfs}
		 f_{i/p_S}^{[1]}(x,\kt)        &=
    	\frac{M}{P^+}\biggl[
	 e(x,k_T)
	- \frac{\eps_T^{\rho\sigma} k_{T\rho} S_{T\sigma}}{M}
	\, \kappa \, e_T^\perp(x,k_T)
    	\biggr], 
  \nn \\
	f_{i/p_S}^{[i\gamma_5]}(x,\kt)       &=
        \frac{M}{P^+}\biggl[
    	S_L \, \kappa \, e_L(x,k_T) -\frac{\kperp \cdot S_T}{M}\, \kappa \, e_T(x,k_T)
    	\biggr], 
 \nn \\
 f_{i/p_S}^{[\gamma^\alpha]} (x,\kt)      &=
        \frac{M}{P^+}\biggl[
    	\frac{\kperp^\alpha}{M} f^\perp(x,k_T) - \epsilon_T^{\alpha\rho}S_{T\rho} \, \kappa \,
    	f_T(x,k_T)
  \nonumber\\
	&\hskip 1cm \!-\!S_L\frac{\epsilon_T^{\alpha\rho}k_{T\rho}}{M} \, \kappa \, f_L^\perp(x,k_T)
	- \frac{\kt^2}{M^2} \biggl(\frac12 g_T^{\alpha\rho} + \frac{k_T^\alpha k_T^\rho}{\kt^2}\biggr) \epsilon_{T \rho \sigma} S_T^\sigma \, \kappa \, f_T^\perp(x,k_T)\!
	\biggr], 
  \nn \\
 f_{i/p_S}^{[\gamma^\alpha\gamma_5]}(x,\kt)  &=
    	\frac{M}{P^+}\biggl[
    	S_T^\alpha \, g_T(x,k_T)
	+ S_L\,\frac{k_T^\alpha}{M} g_L^\perp(x,k_T)
   \nonumber \\ 
		&\hskip 1cm \! - \frac{\kt^2}{M^2} \biggl(\frac12 g_T^{\alpha\rho} + \frac{k_T^\alpha k_T^\rho}{\kt^2}\biggr) S_{T\rho}
    	\, g_T^\perp(x,k_T)
	-\frac{\epsilon_T^{\alpha\rho}k_{T\rho}}{M}\, \kappa \, g^\perp(x,k_T)
	\biggr], 
  \nn\\
	f_{i/p_S}^{[i\,\sigma^{\alpha\beta}\gamma_5]}(x,\kt) &=
    	\frac{M}{P^+}\biggl[
    	\frac{S_T^\alpha \kperp^\beta-S_T^\beta \kperp^\alpha}{M}\,h_T^\perp(x,k_T)
    	-\epsilon_T^{\alpha\beta}\, \kappa \, h(x,k_T)
	\biggr], 
   \nn \\
	f_{i/p_S}^{[i\,\sigma^{+-}\,\gamma_5]}(x,\kt)
	&= \frac{M}{P^+}\biggl[
    	S_L\,h_L(x,k_T) - \frac{k_T\cdot S_T}{M}\,h_T(x,k_T)
    	\biggr]
   \,. 
\end{align}
}
For further details about the notation we refer to Sec.~\ref{sec:qgspinTMDFF}.  
Among them, 8 are chiral-even and 8 are chiral-odd.
Likewise, 8 subleading quark TMDPDFs are T-even and 8 are T-odd.
Unlike in the case of leading TMDs, it is not possible to assign a parton polarization to subleading TMDs as they have no density interpretation. 
Also, for a spin-$1/2$ hadron the same number of subleading TMDFFs exist, see Fig.~\ref{fig:TMDPDFs_tw3}. 
Given that the structure of the equations which define TMDPDFs and TMDFFs is very similar --- compare Eq.~\eqref{eq:tmd_decomposition} and Eq.~\eqref{eq:tmdff_decomposition} for the leading-power functions --- we don't give a set of equations for the subleading TMDFFs but just mention that their definition can be obtained from
\eq{sub-TMD-pdfs}
by replacing on the l.h.s.~$f_{i/p_S}^{[\Gamma]}(x,\kt)$ by $\Delta_{h/i}^{[\Gamma]}(z, -z\pt')$, and on the r.h.s.~the target mass $M$ by the mass of the produced hadron $M_h$, as well as $k_T$ by $p_T'$.  
Furthermore, the lower case letters for the TMDPDFs become upper case letters for the TMDFFs, with the exception of the projector $\Gamma = \gamma^\alpha$, where the symbol $D$ is used for the FFs (instead of $F$).
Recall also that the TMDFFs are functions of $(z, -zp_T')$, and that $\kappa$, the indicator of a non-trivial universality behavior, is absent for TMDFFs.

It was also found that 16 subleading gluon-gluon TMDPDFs and TMDFFs can be identified involving choices for the Lorentz indices on the field strengths $G^{\mu\nu}$ in \eq{fg} that lead to subleading power terms~\cite{Mulders:2000sh, Lorce:2013pza}.
Presently, for these objects not much is known beyond their classification, 
but based on the relations for this type of operator present in SCET~\cite{Marcantonini:2008qn}, we anticipate that they can also be related to a combination of leading power gluon-gluon TMDs, and subleading power TMD correlators involving 3-gluons (operators with 3 $\cB_\perp$s).
Finally, for information on subleading quark-quark TMDs in the case of spin-1 particles we refer to Refs.~\cite{Chen:2015ora, Kumano:2020ijt}.

We remind the reader that
relations exist between the correlators  at subleading power 
which leaves open the option of
expressing the structure functions in terms of various combinations of  kinematic, intrinsic, and dynamic contributions.
As stated above, when  working at tree level in the hard interactions,
where the $qgq$ correlators are trivially integrated over the variable $\xi$ as in \eq{BbhattoPhiA}
the choice of expressing results in terms of the ``intrinsic TMDs'' defined from the subleading power quark-quark correlators
has often been used in the literature~\cite{Tangerman:1994bb,Mulders:1995dh,Bacchetta:2004zf,Goeke:2005hb,Bacchetta:2006tn}.
In this approximation, 
a series of so called ``EOM relations''~\cite{Tangerman:1994bb,Mulders:1995dh,Bacchetta:2006tn} result from
expressing the fermion fields in the correlation function in terms of 
in terms of good and bad light-cone components~\cite{Jaffe:1996zw} as in~\eq{goodbad}.
It is important to emphasize these relations only involve the $qgq$ correlators integrated over the variable $\xi$ as in \eq{BbhattoPhiA}.  
Prominent relations include those that enter the structure function $F_{UU}^{\cos \phi_h}$ (see Sec.~\ref{sec:subSIDISsf})
in the chiral even sector~\cite{Tangerman:1994bb, Mulders:1995dh, Bacchetta:2006tn} 
\begin{align} \label{eq:EOMrltn}
  x f^\perp(x,k_T) &= x \tilde{f}^\perp(x,k_T) + f_1(x,k_T) \,, \quad  x h(x,k_T) = x \tilde{h}(x,k_T) + \frac{k_T^2}{M^2}h_1^\perp(x,k_T),\\ \nn
  \frac{D^\perp(z,p_T)}{z} &=  \frac{\tilde D^\perp(z,p_T)}{z} + D_1(z,p_T)
  \,, \quad \frac{H(z,p_T)}{z}=
  \frac{\tilde H(z,p_T)}{z} + \frac{p_T^2}{z^2M_h^2}H_1^\perp(z,p_T)
  \,.
\end{align} 
It is important to emphasize that these constraints are bare operator relations, and beyond leading order, are subject to renormalization.

\subsection{Factorization for SIDIS with Subleading Power TMDs} 
\label{sec:subTMDfact}

\subsubsection{Status of SIDIS factorization at next-to-leading power}
\label{sec:subpowerfact}

Beyond leading order in the TMD power expansion $\Lambda/Q\ll 1$, the derivation of factorization theorems for SIDIS cross sections becomes much more complex.  
The first predictions for the form of the factorization at next-to-leading power were derived at tree-level in perturbative QCD with a TMD parton model by Mulders and Tangerman in a classic paper~\cite{Mulders:1995dh}.  As is the case for all power-suppressed observables, the theoretical analysis, even at lowest order, is more involved compared to leading-power observables. 
A particular complication within a pQCD description is that different subleading effects, for both PDFs and FFs, contribute to the same power-suppressed observable, see, for instance, Ref.~\cite{Kanazawa:2015ajw}.
The work in Refs.~\cite{Boer:2003cm,Metz:2004je, Bacchetta:2004zf,Bacchetta:2006tn,Goeke:2005hb,Gamberg:2006ru}  provided further insight the nature of the subleading TMDs appearing in these results. 
A discussion about subleading-power TMD factorization in Drell-Yan has been presented in Refs.~\cite{Chen:2016hgw,Balitsky:2017gis,Balitsky:2017flc}.
Early on, model calculations for the SIDIS beam spin asymmetry $F_{LU}^{\sin\phi_h}$~\cite{Metz:2004je} and a corresponding T-odd twist-3 TMDPDF $g^\perp$ \cite{Gamberg:2006ru} indicated a problem with a light-cone divergence that did not cancel in tree-level formulas. 
It has recently been observed that the removal of light-cone divergences in subleading power TMDs involves both additive and multiplicative terms~\cite{Rodini:2022wki,Gamberg:2022lju,AnjieREF}.
An important cross-check on the form of subleading power TMD factorization can be obtained by considering the intermediate transverse momentum region $\Lambda_{\mathrm{QCD}}\ll |{\bm P}_{h\perp}|\ll Q$, where the TMDs can be matched onto collinear PDFs and FFs, see \sec{largeqT}, and cross section results must agree with those obtained directly from collinear factorization.
In Ref.~\cite{Bacchetta:2008xw} it was reported that simplest tree level subleading TMDs failed to satisfy this cross-check for the $\cos \phi_h$ modulation of the unpolarized SIDIS cross section.  Later on, Ref.~\cite{Bacchetta:2019qkv} argued that this cross-check could be satisfied for subleading power TMD factorization by the inclusion of the same soft subtractions as for TMDs at leading power.
A more detailed study of this matching has been carried out recently in Refs.~\cite{Rodini:2022wki,Gamberg:2022lju}, with a careful analysis of the issues with earlier literature, resulting in a demonstration that this cross-check is satisfied. 
Currently the most in-depth studies of factorization in subleading power SIDIS can be found in Refs.~\cite{Vladimirov:2021hdn,Ebert:2021jhy,Rodini:2022wki,Gamberg:2022lju}.

Here we provide a brief description of the all-orders in $\alpha_s$ analysis of next-to-leading power SIDIS factorization carried out with SCET in~\cite{Ebert:2021jhy}, which is the most advanced treatment on the issues associated to subleading power factorization to date. An introduction to factorization in SCET can be found in \sec{factSCET}. For the analysis that has been carried out so far it has been {\it assumed} that interactions involving the Glauber Lagrangian ${\cal L}_G^{(0)}$ do not spoil factorization at subleading power.  A proof of factorization would require a direct analysis demonstrating that this is indeed the case. Nevertheless, after making this assumption it is still possible to provide an all orders factorization of hard, collinear, and soft dynamics in subleading power SIDIS.

In SCET the contributions to the structure functions $F_{XY}^Z$ in Eq.~(\ref{e:SIDIS_subleading}) that start at subleading power can be divided into three main categories:
\begin{itemize}
 \item kinematic power corrections associated to the fact that the structure functions are defined by using an angular decomposition in the Trento frame which differs from the frame where the hadrons travel in the same direction where factorization is simplest,
 \item hard scattering power corrections which correct the short distance operator associated to the interactions localized by the virtual photon,
 \item subleading power Lagrangian  corrections associated to interactions between soft and collinear particles beyond leading power.
\end{itemize}
The kinematic power corrections are trivial, arising from subleading terms in the contraction of the leptonic and hadronic tensors, and involve only leading power TMDs. In Ref.~\cite{Ebert:2021jhy} it was shown that the subleading power Lagrangian corrections vanish for SIDIS at this order. They all involve leading power hard scattering currents which are transverse and hence give vanishing contribution to structure functions that start at subleading power. In a generic process subleading power corrections corrections can occur from either soft or collinear regions, and correspondingly involve subleading soft functions or subleading unsubtracted collinear functions (TMD PDFs or beam functions).   At ${\cal O}(\Lambda/Q)$ next-to-leading power subleading soft function can be built from a soft analog of ${\cal B}_{n_a\perp}^\rho$ in \eq{chiandB} dressed with Wilson lines similar to those in the leading power soft function. However, it has been shown in~\cite{Ebert:2021jhy} that all direct contributions from subleading soft operators give vanishing contributions to  SIDIS structure functions at next-to-leading power. However, at one higher order in the power expansion, ${\cal O}(\Lambda^2/Q^2)$, both subleading soft and subleading collinear contributions are known to contribute~\cite{Ebert:2018gsn}.  

The remaining nonzero contributions for next-to-leading power SIDIS are the kinematic power corrections, hard scattering operators involving a ${\cal P}_\perp$ which are related to those at leading power by \eq{def_BP_f_a}, and hard scattering operators involving a ${\cal B}_{n_a\perp}^\rho$ or ${\cal B}_{n_b\perp}^\rho$ which give $qgq$ correlators as in \eq{def_Bb_hat}. Of these the first two contributions involve the same short distance hard scattering coefficients as for leading power TMDs, while the $qgq$ correlators involve a single hard scattering coefficient~\cite{Vladimirov:2021hdn,Ebert:2021jhy}, which enters through a new hard function $H_{ii}^{(1)}(Q^2,\xi)$. For simplicity of the presentation we will assume that $H_{ii}^{(1)}(Q^2,\xi)$ is real here, an approximation that holds for leading-logarithmic resummation and for almost all contributions at next-to-leading order.  Results for these contributions are presented below, and it is trivial to add additional terms that result from imaginary parts in $H_{ii}^{(1)}(Q^2,\xi)$.

\begin{figure}[t]
\centering
\includegraphics[width=0.35\textwidth]{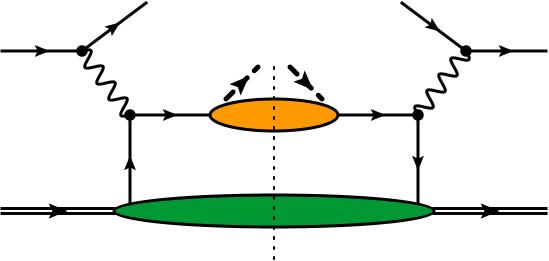}
\\[0.2cm]
\includegraphics[width=0.35\textwidth]{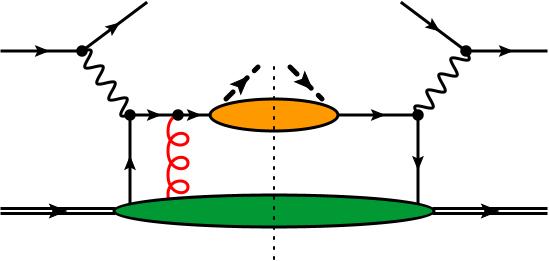}
\,\,\,\,\,
\includegraphics[width=0.35\textwidth]{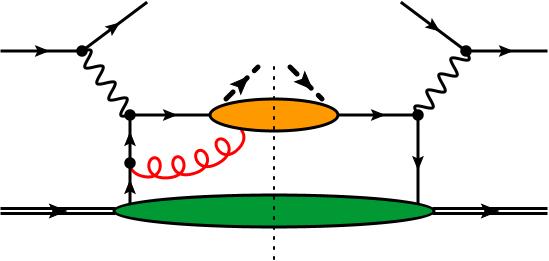}
\caption{Examples of SIDIS tree-level diagrams relevant for subleading-power observables. The upper diagram represent kinematical and ${\cal P}_\perp$ (or intrinsic) contributions, while the lower diagrams represent dynamical $qgq$ correlators contributions (mirror diagrams should also be included). }
\label{fig:Tw3DiagramsSIDIS}
\end{figure}

\subsubsection{SIDIS structure functions in terms of next-to-leading power TMDs}
\label{sec:subSIDISsf}

Under the assumptions outlined in \sec{subpowerfact}, the form of the factorization formula including $\alpha_s$ corrections, are as follows~\cite{Ebert:2021jhy}.
\noindent
For the unpolarized structure functions we have
\begin{align} \label{eq:FNLPunpolbT}
 F_{UU}^{\cos\phi_h} &
 = \cF\biggl\{ \frac{q_{T}}{Q}\, H_{ii}(Q^2) \left[- \tilde f_1 \tilde D_1
  + \tilde h_1^{\perp(1)} \tilde H_1^{\perp(1)}\right]
  \\
  &\qquad
  + H_{ii}(Q^2)\left[-\frac{\Ma}{Q} \tilde f_1^{(1)} \tilde D_1
   -\frac{\Mb}{Q} \tilde f_1 \tilde D_1^{(1)}
   +\frac{\Ma}{Q} \tilde h_1^{\perp (0')} \tilde H_1^{\perp (1)} +  \frac{\Mb}{Q} \tilde h_1^{\perp (1)} \tilde H_{1}^{\perp(0')}\right]
  \nn\\
  &\qquad 
  - H^{(1)}_{ii}(Q^2,\xi)\left[\frac{2x\Ma}{Q}\left(\gtilde{f}{}^{\perp(1)} D_1+\gtilde{h}\,\tilde H_1^{\perp(1)}\right)
  +\frac{2\Mb}{zQ}\left(\tilde f_1\gtilde{D}{}^{\perp(1)}
  + \tilde h_1^{\perp(1)} \gtilde{H}\right)\right]\biggl\}
\,,\nn\\
 F_{LU}^{\sin\phi_h} &
 = \cF\biggl\{
    H^{(1)}_{ii}(Q^2,\xi)
    \left[\frac{2x\Ma}{Q}\left(\gtilde{g}{}^{\perp(1)} \tilde D_1
    - \gtilde{e}\,\tilde H_1^{\perp(1)}\right)
    - \frac{2\Mb}{zQ}\left( \tilde f_1\gtilde{G}{}^{\perp(1)}
    - \tilde h_1^{\perp(1)} \gtilde{E}\right)\right]\biggr\}
\,.\nn
\end{align}
Here the terms with $qgq$ correlators, the functions with two tildes, are generated by diagrams like those shown on the last line of \fig{Tw3DiagramsSIDIS}, while the remaining terms are generated by the $\cP_\perp$ operators or kinematic power corrections, and can be represented by the diagram on the top line.
In a similar fashion the factorization formula for the longitudinally polarized structure functions are
\begin{align}
 F_{UL}^{\sin\phi_h} &
 = \cF\biggl\{ \frac{q_{T}}{Q}\,H_{ii}(Q^2)\,  \tilde h_{1L}^{\perp(1)} \tilde H_1^{\perp(1)}
    +H_{ii}(Q^2) \left(\frac{\Ma}{Q} \tilde h_{1L}^{\perp (0')} \tilde H_1^{\perp (1)}
   + \frac{\Mb}{Q} \tilde h_{1L}^{\perp (1)} \tilde H_{1}^{\perp(0')}\right)
 \\&\qquad
   +H_{ii}^{(1)}(Q^2,\xi)
    \left[\frac{2x\Ma}{Q}\left(\gtilde{f}{}^{\perp(1)}_{L} \tilde D_1
   - \gtilde{h}_{L}\tilde H_1^{\perp(1)}  \right)
   -\frac{2\Mb}{zQ}\left( \tilde g_{1L}\gtilde{G}{}^{\perp(1)}
   + \tilde h_{1L}^{\perp(1)} \gtilde{H}\right) \right] \biggr\}
\,,\nn\\
 F_{LL}^{\cos\phi_h}
 &= \cF\biggl\{-\frac{q_T}{Q} H_{ii}(Q^2)\, \tilde g_{1L} \tilde D_1
  -H_{ii}(Q^2)\left(\frac{\Ma}{Q}\, \tilde g_{1L}^{(1)} \tilde D_1 
  + \frac{\Mb}{Q}\, \tilde g_{1L} \tilde D_1^{(1)}\right)
  \nn\\ &\qquad 
   +  H_{ii}^{(1)}(Q^2,\xi)
    \left[-\frac{2x\Ma}{Q}\left(\gtilde{g}^{\perp(1)}_{L} \tilde D_1
   -\gtilde{e}_L \tilde H_1^{\perp(1)}\right)
   -\frac{2\Mb}{zQ}\left(\tilde g_{1L} \gtilde{D}^{\perp(1)}
   + \tilde h_{1L}^{\perp(1)} \gtilde{E}\right)\right] \biggr\}
\nn\,,
\end{align}
while for the transversely polarized structure functions
\begin{align} \label{eq:FNLPtranvbT}
 F_{UT}^{\sin\phi_S}
 &= \cF\biggl\{  -\frac{q_T}{2Q}\, H_{ii}(Q^2)\left( 
   \tilde f_{1T}^{\perp(1)} \tilde D_1 
   - 2 \tilde h_1 \tilde H_1^{\perp(1)}\right)\biggr\}
  \\&
  +\cF' \biggl\{
   H_{ii}(Q^2)\left(-\frac{\Ma}{2Q} \tilde f_{1T}^{\perp(0')}\tilde D_1
   -\frac{\Mb}{2Q} \tilde f_{1T}^{\perp(1)} \tilde D_1^{(1)}  
   +\frac{\Mb}{Q} \tilde h_{1} \tilde H_1^{\perp(0')}
  +\frac{\Ma}{Q} \,\tilde h_{1}^{(1)} \tilde H_1^{\perp(1)}\right)
  \nn\\&\qquad
   +  H_{ii}^{(1)}(Q^2,\xi)\biggl[\frac{2x\Ma}{Q}\gtilde{f}_T \tilde D_1 
    -\frac{2\Mb}{zQ} \tilde h_1 \gtilde{H}
    - \frac{x\Ma}{Q} \Bigl(\gtilde{h}_T^{(1)} -\gtilde{h}_T^{\perp(1)}\Bigr) \tilde H_1^{\perp(1)} 
  \nn\\&\hspace{2cm}
   - \frac{\Mb}{zQ} \left(\tilde g_{1T}^{(1)} \gtilde{G}^{\perp(1)} + \tilde f_{1T}^{\perp(1)} \gtilde{D}^{\perp(1)}\right) \biggr]  \biggr\}
 \,, \nn \\
 \ldots & 
  \,.\nn
\end{align}
The ellipses in \eq{FNLPtranvbT} denote analogous results for the transversely polarized structure functions 
$F_{UT}^{\sin\left(2\phi_h -\phi_S\right)}$, $F_{LT}^{\cos\phi_S}$, and $F_{LT}^{\cos\left(2\phi_h -\phi_S\right)}$, 
which also can be found in \cite{Ebert:2021jhy}.
Here all TMDs are in $b_T$ space as indicated by the presence of a tilde, while functions with a double tilde are the $qgq$ correlators. The definition of these $b_T$ space correlators, and their relation to momentum space correlators, are given in \app{Fourier_transform}.

Due to the presence of an additional convolution in $\xi$ we have a slightly modified definition of the convolution integral for the results at NLP in Eqs.~(\ref{eq:FNLPunpolbT}--\ref{eq:FNLPtranvbT}), which varies depending on whether or not it is acting on a function arising from a $qgq$ correlator:
\begin{align} \label{eq:cF_NLP}
 \cF[\cH\, \tilde g^{(n)} \tilde D^{(m)}] &
 = x \sum_{i} H_{ii}(q^+ q^-)
   \int_0^\infty \frac{\df b_T \, b_T}{2\pi} (\Ma b_T)^n (-\Mb b_T)^m J_{n+m}(b_T q_T) \,
   \nn\\&\hspace{3.5cm}\times
   \, g_i^{(n)}(x,b_T) \, D_i^{(m)}(z,b_T) + (i\to\bar i)
\,,\nn\\
 \cF[\cH\, \gtilde g^{(n)} \tilde D^{(m)}] &
 = x \sum_{i} \int\!\df \xi \, H_{ii}^{(1)}(q^+ q^-, \xi)
   \int_0^\infty \frac{\df b_T \, b_T}{2\pi} (\Ma b_T)^n (-\Mb b_T)^m J_{n+m}(b_T q_T) \,
   \nn\\&\hspace{3.5cm}\times
   \, \tilde g_i^{(n)}(x, \xi, b_T) \, D_i^{(m)}(z,b_T) + (i\to\bar i)
\,,\nn\\
 \cF[\cH\, \tilde g^{(n)} \gtilde D^{(m)}] &
 = x \sum_{i} \int\!\df \xi \, H_{ii}^{(1)}(q^+ q^-, \xi)
   \int_0^\infty \frac{\df b_T \, b_T}{2\pi} (\Ma b_T)^n (-\Mb b_T)^m J_{n+m}(b_T q_T) \,
   \nn\\&\hspace{3.5cm}\times
   \, g_{i}^{(n)}(x, b_T) \, \tilde D_{i}^{(m)}(z, \xi, b_T) + (i\to\bar i)
\,.\end{align}
At subleading power the powers of $b_T$ do not always come along with a $\cos\varphi =  -{\bf b_T}\cdot{\bf q_T}/(b_Tq_T)$ dependence on the azimuthal angle, so we also need to make use of the operator $\cF'[\cdots]$ that has the same definition as those in \eq{cF_NLP}, except for having the Bessel function $J_{n+m}(b_T q_T)$ replaced by $J_0(b_T q_T)$.

While the form of the all orders factorization results is theoretically simplest in $b_T$ space, experimentally we measure momentum space TMD structure functions, so it is natural to also consider the form that these factorization formulas take when written in terms of transverse momentum space TMDs. To keep this discussion contained we will focus on $F_{UU}^{\cos\phi_h}$, for which the equivalent result to that in \eq{FNLPunpolbT} is
\begin{align} \label{eq:W_UU_cosphi_msp}
 F_{UU}^{\cos\phi_h} &
 = \tilde \cF\biggl\{
   \cH^{(0)} \biggl[ - \frac{2 \bfhp\cdot\bfkperp}{Q} f_1 D_1
                     + \frac{2 k_T^2 \bfhp\cdot\bfpperp' }{\Ma \Mb Q} h_1^{\perp} H_1^{\perp} \biggr]
  \\ \nn
  &\qquad {
  - \cH^{(1)} \left[ \frac{2x\Ma}{Q} \left(\frac{\bfhp\cdot\bfkperp}{\Ma} \tilde{f}^{\perp} D_1 + \frac{\bfhp\cdot\bfpperp'}{\Mb} \tilde{h}\,H_1^{\perp}\right)
                    +\frac{2\Mb}{zQ} \left(\frac{\bfhp\cdot\bfpperp'}{\Mb} f_1\tilde{D}^{\perp} + \frac{\bfhp\cdot\bfkperp}{\Ma} h_1^{\perp} \tilde{H}\right)\right]}\biggl\}
\,.\end{align}
Here all TMDs are in transverse momentum space and the vector $\bfhp\equiv{\bm P}_{h T}/|{\bm P}_{h T}|$ is a unit vector in the direction of the transverse momentum of the produced hadron in the Trento frame. 
Analogous momentum space results for the other seven structure functions whose contributions start at subleading power can be found in Ref.~\cite{Ebert:2021jhy}.
The operation $\tilde\cF$
includes both the transverse momentum integral between PDFs and FFs, and the integral over $\xi$ between the hard function and $qgq$ TMDs, 
{\allowdisplaybreaks
\begin{align}\label{eq:cF_tilde_NLP}
 \tilde\cF[\w\, \cH\, g\, D] &
 = x \sum_i H_{ii}(q^+q^-) \int\!\df^2\kt\, \df^2\pt' \, \delta^{(2)}\bigl(\qt+\kt-\pt'\bigr)
  \nn\\ &\qquad\quad
  \times\w(\kt, \pt')\, g_i(x, k_T) \, D_i(z,z p_T')+ (i\to\bar i)
\,,\nn\\
 \tilde\cF[\w\, \cH\, \tilde g\, D] &
 = x \sum_f \int\!\df\xi \, H_{ii}^{(1)}(q^+q^-, \xi) \int\!\df^2\kt\, \df^2\pt' \, \delta^{(2)}\bigl(\qt+\kt-\pt'\bigr)
  \nn\\ &\qquad\quad
  \times\w(\kt, \pt')\, \tilde g_i(x, \xi, k_T) \, D_i(z,z p_T')+ (i\to\bar i)
\,,\nn\\
 \tilde\cF[\w\, \cH\, g \, \tilde D] &
 = x \sum_i \int\!\df\xi \, H_{ii}^{(1)}(q^+q^-, \xi) \int\!\df^2\kt\, \df^2\pt' \, \delta^{(2)}\bigl(\qt+\kt-\pt'\bigr)
  \nn\\ &\qquad\quad
  \times\w(\kt, \pt')\, g_i(x, k_T) \, \tilde D_i(z, \xi, z p_T')+ (i\to\bar i)
\,,\end{align}
}where the $\tilde g$ and $\tilde D$ are placeholders for the momentum space PDF and FF $qgq$ TMDs, respectively.

Formulating the cross section in the factorization frame carried out in Ref.~\cite{Ebert:2021jhy} results in structure functions expressed in terms of the kinematic and dynamical distributions which form a complete basis as in~\eqref{eq:W_UU_cosphi_msp}, and the intrinsic subleading distributions only enter by employing the equations of motion. On the other hand, in  Ref.~\cite{Gamberg:2022lju}
the cross section was formulated using the intrinsic and dynamical subleading basis.
For example, calculating the structure function  $F_{UU}^{\cos\phi_{h}}$ 
by contracting  the leptonic and hadronic tensors in the Breit frame in SIDIS,
using the intrinsic and dynamic subleading basis yields, 
\begin{align}\label{eq:FUUcosphi_tree}
	F_{UU}^{\cos\phi_{h}}
	&=
	{\cal C}\biggl[
   	 - \left(\frac{\bfhp\cdot \bfPhperp}{zQ}f_1D_1 +
          \frac{\bfhp\cdot \bfPhperp}{zQ}\frac{2\bfkperp\cdot \bfpperp'}{MM_h}
          h_{1}^{\perp }\,H_{1}^{\perp}\right)\\ \nn
          & -  \frac{\bfhp\cdot\bfpperp'}{Q}\left(f_1\frac{D^\perp}{z}
          +\frac{M}{M_h}xh H_1^\perp\right)
          -\frac{\bfhp\cdot\bfkperp}{Q}\left(xf^\perp D_1
        +\frac{M_h}{M} h_1^\perp \frac{H}{z}\right)\\ \nn
        & -  \frac{\bfhp\cdot\bfpperp'}{Q}\left(f_1\frac{\tilde D^\perp}{z} 
          +\frac{M}{M_h}x\tilde h H_1^\perp\right)
          -\frac{\bfhp\cdot\bfkperp}{Q}\left(x\tilde f^\perp D_1
        +\frac{M_h}{M} h_1^\perp \frac{\tilde H}{z}\right)\biggl]
  \,.
\end{align}
Here for simplicity we have given the expression at tree level and
the convolution integral $\cal C$ is given by~\eqref{eq:def-convolution-integral}.
This result has been extended beyond
leading order in Ref.~\cite{Gamberg:2022lju}.

It is useful to consider the simplifications that occur at tree level for the full set of subleading power structure functions that will be relevant for our phenomenological discussion. 
As stated above, at tree level the common hard function $H^{(1)}(Q^2,\xi)$ is independent of $\xi$, so we can freely integrate over this variable in the $qgq$ correlators. This enables us to make use of the integrated correlators $\tilde \Phi_Q^\rho(x,{\bf b_T})$ and $\tilde\Delta_A^\rho(z,{\bf b_T})$ in \eq{PhiADeltaA}, by using \eq{BbhattoPhiA} and the analog for the $qgq$ FF.
As stated above in Sec.~\ref{sec:sublquarkquark} the use of the equations of motion relations~\eqref{eq:EOMrltn}
  is also simpler at tree level. Then the results can be expressed so that only intrinsic subleading TMDPDFs and dynamical subleading TMDFFs appear.
  This leads to results that are fully consistent with the pioneering expressions derived at tree level in the parton model in  Refs.~\cite{Mulders:1995dh,Bacchetta:2006tn},

{\allowdisplaybreaks
\begin{align}
	F_{UU}^{\cos\phi_h}
	&=
	\frac{2M}{Q}\,{\cal C}\biggl[
   	-\frac{\bfhp\cdot\bfpperp^{\prime}}{ M_h}
	\biggl( x h\,H_{1}^{\perp }
   	+ \frac{M_h}{M}\,\,f_1 \frac{\tilde{D}^{\perp }}{z}\biggr)
	- \frac{\bfhp\cdot\bfkperp^{ }}{M} \biggl( x  f^{\perp } D_1
   	+ \frac{M_h}{M}\,\,h_{1}^{\perp } \frac{\tilde{H}}{z}\biggr)\biggr] ,
	\label{eq:FUUcosphi}\\
 	F_{UL}^{\sin\phi_h}
 	&=
	\frac{2M}{Q}\,{\cal C}\biggl[ -
   	\frac{\bfhp\cdot\bfpperp^{\prime}}{ M_h}
    	\biggl( x   h_L  H_1^{\perp } \!
   	+ \frac{M_h}{M}\,\,g_{1}\frac{\tilde{G}^{\perp } }{z}\biggr)
   	+\frac{\bfhp\cdot\bfkperp^{ }}{M}
    	\biggl( x  f_{L}^{\perp }  D_1 \!
   	- \frac{M_h}{M}\,\, h_{1L}^{\perp }  \frac{\tilde{H}}{z}\biggr)\biggr] ,
	\label{eq:FULsinphi}\\
 F_{LU}^{\sin\phi_h}
	&=
	\frac{2M}{Q}\,{\cal C}\biggl[ -
   	\frac{\bfhp\cdot\bfpperp^{\prime}}{ M_h}
   	\biggl( x \, e \, H_1^{\perp }
   	+ \frac{M_h}{M}\,\,f_1\frac{\tilde{G}^{\perp }}{z}\,\biggr)
   	+\frac{\bfhp\cdot\bfkperp^{ }}{M}
   	\biggl( x   g^{\perp }  D_1
   	+ \frac{M_h}{M}\,\, h_1^{\perp } \,\frac{\tilde{E}}{z} \biggr)\biggr] ,
	\label{eq:FLUsinphi}\;\;\;\;\;\\
	F_{UT}^{\sin \phi_S }
	&=
	\frac{2M}{Q}\,{\cal C}\biggl[ x   f_T   D_1 \;
   	- \frac{M_h}{M}\, \; h_{1} \, \frac{\tilde{H}}{z} \nonumber\\
   	&\hspace{1.2cm}
   	-\frac{\bfpperp^{\prime}\cdot\bfkperp^{ }}{2 M M_h}
	\biggl( x   h_{T}  H_{1}^{\perp }
   	+ \frac{M_h}{M}\, g_{1T}^\perp \,\frac{\tilde{G}^{\perp }}{z}
   	- x   h_{T}^{\perp }  H_{1}^{\perp }
	+ \frac{M_h}{M}\, f_{1T}^{\perp } \,\frac{\tilde{D}^{\perp }}{z}
   	\biggr) \biggr]\, , \label{eq:FUTsinphiS}
\end{align}
}where the simpler convolution of transverse momenta ${\cal C}$ is defined in 
\eq{def-convolution-integral} and used with the replacement $H_{ii}(Q^2,\mu) = e_i^2$.
At tree level similar results are also found for the other four structure functions that have not been shown. 
Here we have focused on four prominent subleading effects in SIDIS that have been measured at, for instance, HERMES, COMPASS, and Jefferson Lab: 
the structure functions $F_{UU}^{\cos \phi_h}$ (Cahn effect), 
$F_{UL}^{\sin \phi_h}$ (longitudinal target-spin asymmetry),
$F_{LU}^{\sin \phi_h}$ (longitudinal beam-spin asymmetry),  and $F_{UT}^{\sin \phi_s}$ (transverse target asymmetry). 
The phenomenological applications of subleading power factorization formulas for SIDIS have so far focused on results at tree-level, so we will make use of these expressions in our discussion of the analysis of experimental data in \sec{subTMDexpt}.

\begin{figure}[t]
\centering
\includegraphics[width=0.70\textwidth]{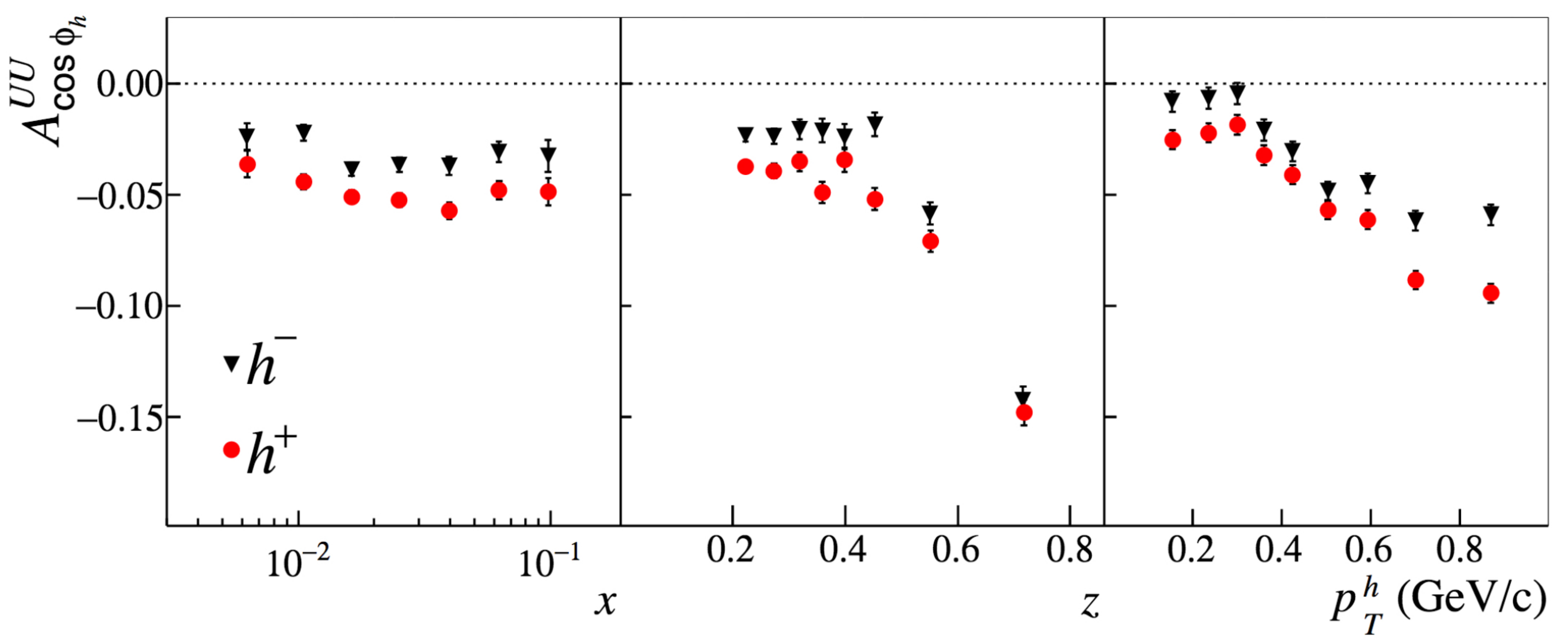}
\caption{COMPASS data, for a $^6$LiD target, of the Cahn asymmetry $A^{UU}_{\cos\phi_h} |_{\rm fig.} \propto F_{UU}^{\cos \phi_h}/F_{UU}$ for positively and negatively charged hadrons as a function of $x$, $z$ and 
$P_{hT} = p_T^h |_{\rm fig.} $~\cite{COMPASS:2014kcy}.}
\label{fig:Cahn_COMPASS}
\end{figure}

\subsection{Experimental Results for Subleading-Power TMD Observables}
\label{sec:subTMDexpt}

Let us now discuss in a bit more detail the experimental significance of power-suppressed TMD observables in SIDIS, where we refer to a recent review ~\cite{Avakian:2019drf} as well as~\cite{Bastami:2018xqd} for more details.
Here we concentrate on the four structure functions in Eqs.~(\ref{eq:FUUcosphi})--(\ref{eq:FUTsinphiS}) by beginning with the (Cahn effect) structure function $F_{UU}^{\cos\phi_h}$.
This was first measured by the EMC Collaboration in muon-proton scattering for the production of charged hadrons~\cite{EuropeanMuon:1983tsy, EuropeanMuon:1986ulc}, and a nonzero effect of up to $~10\%$ was reported.
Those results were followed by measurements from the E665 Collaboration at Fermilab~\cite{E665:1993pov} and from the ZEUS Collaboration at HERA~\cite{ZEUS:2000esx}.
The latter are the only collider data on this fundamental SIDIS observable.
Data also exist from the Hall-C Collaboration at Jefferson Lab for charged-pion production with both a proton and deuteron target~\cite{Mkrtchyan:2007sr}, and from the CLAS Collaboration for a proton target~\cite{CLAS:2008nzy}. 
The HERMES Collaboration measured the Cahn effect for charged pions, kaons and unidentified hadrons~\cite{HERMES:2012kpt}, while the COMPASS Collaboration reported data for the production of charged hadrons off a deuteron ($^6$LiD) target~\cite{COMPASS:2014kcy}; see Fig.~\ref{fig:Cahn_COMPASS}.
All the experimental results for the Cahn effect show that this observable can be as large as the leading-power effects in SIDIS such as the Sivers and Collins asymmetries.

\begin{figure}[t]
\centering
\includegraphics[width=0.50\textwidth]{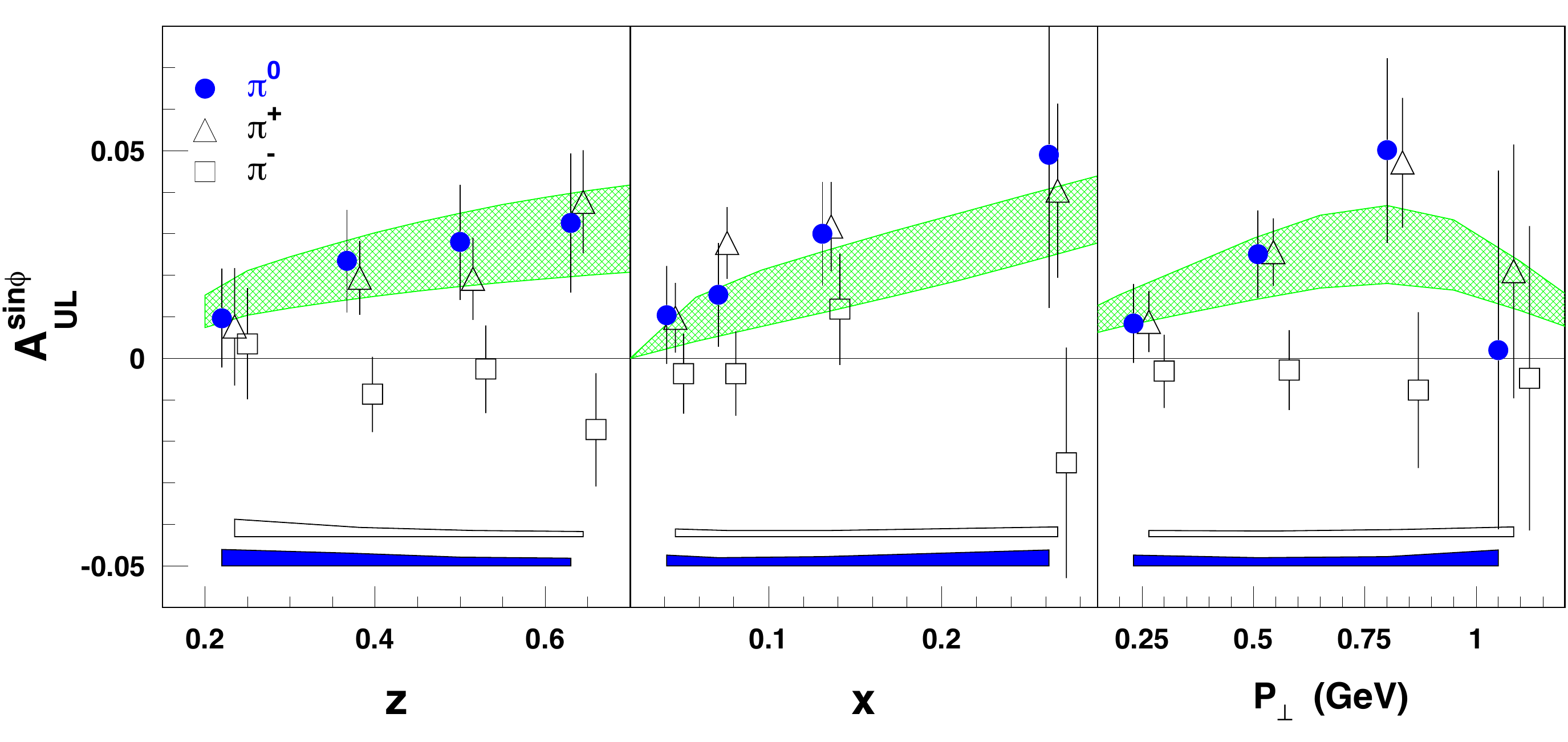}
 \hspace{1.0cm}
\includegraphics[width=0.36\textwidth]{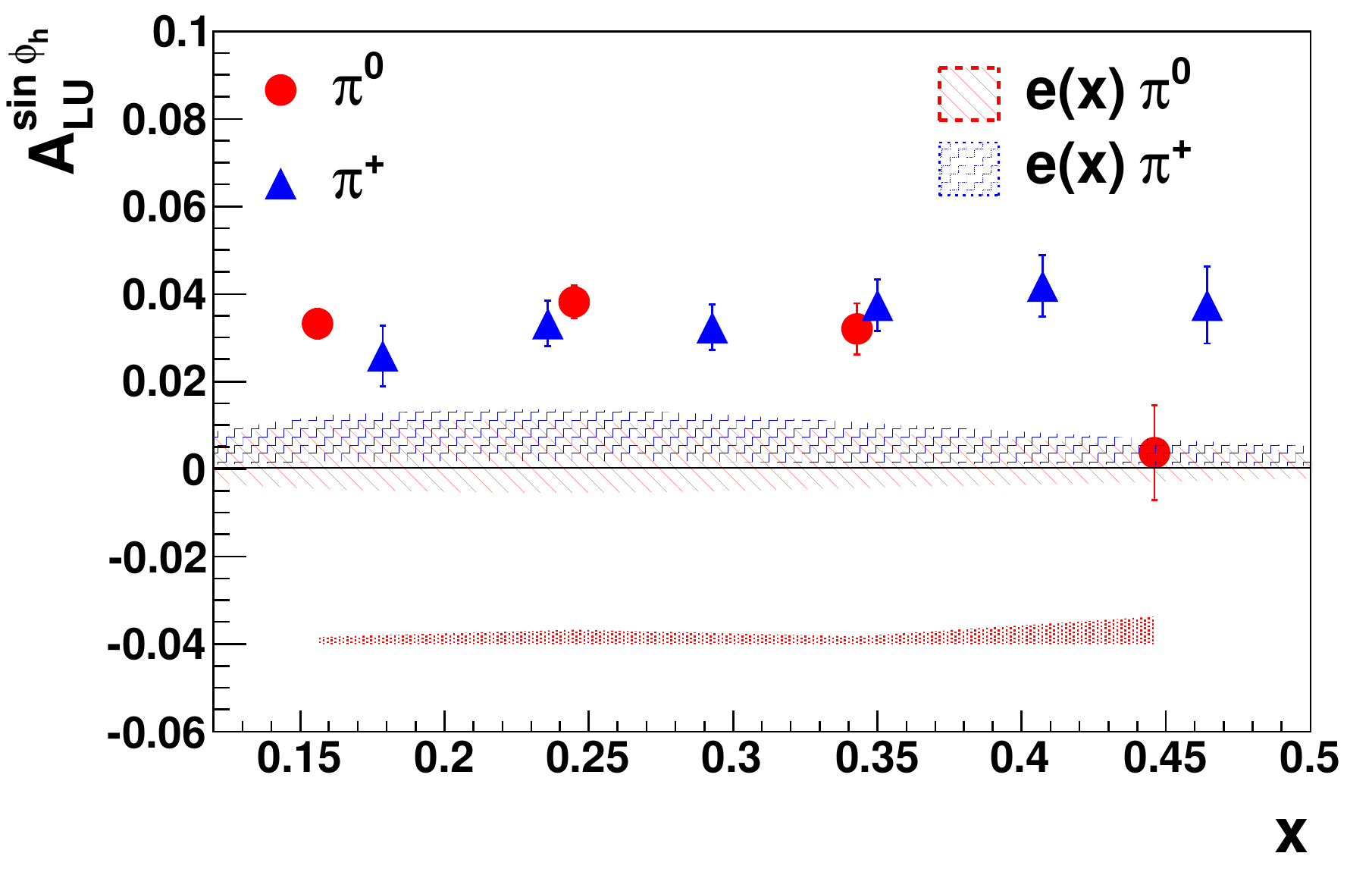}
\caption{\textbf{Left:} HERMES data 
for the longitudinal target-spin asymmetry $A_{UL}^{\sin\phi_h}\propto F_{UL}^{\sin\phi_h}/F_{UU}$ for pion production~\cite{HERMES:2001hbj}.
Error bars include the statistical uncertainties only. The filled (blue) and open (white) bands at the bottom of the panels represent the systematic uncertainties for neutral and charged pions, respectively. The shaded (green) areas show a range of predictions of a model calculation~\cite{Oganessian:1998ma, DeSanctis:2000fh} applied to the case of $\pi^0$ production.
\textbf{Right:} CLAS data for the beam-spin asymmetry $A_{LU}^{\sin\phi_h}\propto F_{LU}^{\sin\phi_h}/F_{UU}$ for $\pi^0$ and $\pi^+$ as a function of $x$ at an average $P_{hT}=0.38\,\mathrm{GeV}$ and for $0.4<z<0.7$~\cite{Aghasyan:2011ha}. 
The error bars correspond to statistical uncertainties, and the red error band at the bottom of the plot corresponds to systematic uncertainties. 
The red and blue hatched bands show model calculations involving only the term proportional to $e \otimes H_1^\perp$ in Eq.~\eqref{eq:FLUsinphi} for $\pi^0$ and $\pi^+$, respectively.}
\label{fig:AULLU}
\end{figure}

We now proceed to the longitudinal target-spin asymmetry $A_{UL}^{\sin\phi_h}$, the numerator of which at tree level is given by Eq.~\eqref{eq:FULsinphi}. 
We repeat that this asymmetry was the first-ever measured SSA in SIDIS.
Specifically, the HERMES Collaboration studied this observable for charged-pion production off a proton target, and reported effects of up to 5\% for the $\pi^+$ final state~\cite{HERMES:1999ryv}.
Afterwards, HERMES measured this asymmetry also for neutral pions~\cite{HERMES:2001hbj}, where the results are shown on the left panel of Fig.~\ref{fig:AULLU}, along with the data for charged pions from Ref.~\cite{HERMES:1999ryv}.
Additionally, HERMES published more precise data for a proton target~\cite{HERMES:2005mov} and results for a deuteron target~\cite{HERMES:2002buj}, including kaon final states.
Also the CLAS Collaboration reported data for $A_{UL}^{\sin\phi_h}$ with a proton target and for all three charge states of the pion~\cite{CLAS:2010fns, CLAS:2017yrm}.
Preliminary results obtained by the COMPASS experiment in single-hadron muon-production off protons are available as well~\cite{Parsamyan:2018ovx}.

The structure function in Eq.~\eqref{eq:FLUsinphi} is the numerator of the longitudinal beam-spin asymmetry $A_{LU}^{\sin\phi_h}$ in SIDIS, which also has been studied extensively in experiment. 
The HERMES Collaboration carried out the pioneering measurement of this observable for charged-pion production using a proton target, finding results compatible with zero within errors~\cite{HERMES:1999ryv}.
The first nonzero results (up to 5\%) for the beam-spin asymmetry were observed by the CLAS Collaboration for $\pi^+$ production~\cite{CLAS:2003qum}.
For a proton target, nonzero effects for both $\pi^+$ and $\pi^0$ final states were (also) found in later measurements by HERMES~\cite{ HERMES:2006pof, HERMES:2019zll} and by the CLAS experiment~\cite{Aghasyan:2011ha, CLAS:2014dmz, CLAS:2021opg}, while the COMPASS Collaboration published data for charged-hadron production off a deuteron target~\cite{COMPASS:2014kcy}. 
CLAS data for $A_{LU}^{\sin\phi_h}$ are shown on the right panel in Fig.~\ref{fig:AULLU}.
One motivation for studying this asymmetry has been to obtain information on the subleading TMD $e(x,k_T)$ which, according to the tree-level formula~\eqref{eq:FLUsinphi} couples to the Collins function.
However, the results in Fig.~\ref{fig:AULLU} show that other contributions to this asymmetry can be large and even dominating; see also the corresponding discussion in~\cite{CLAS:2021opg}.

\begin{figure}[t]
\centering
\includegraphics[width=0.55\textwidth]{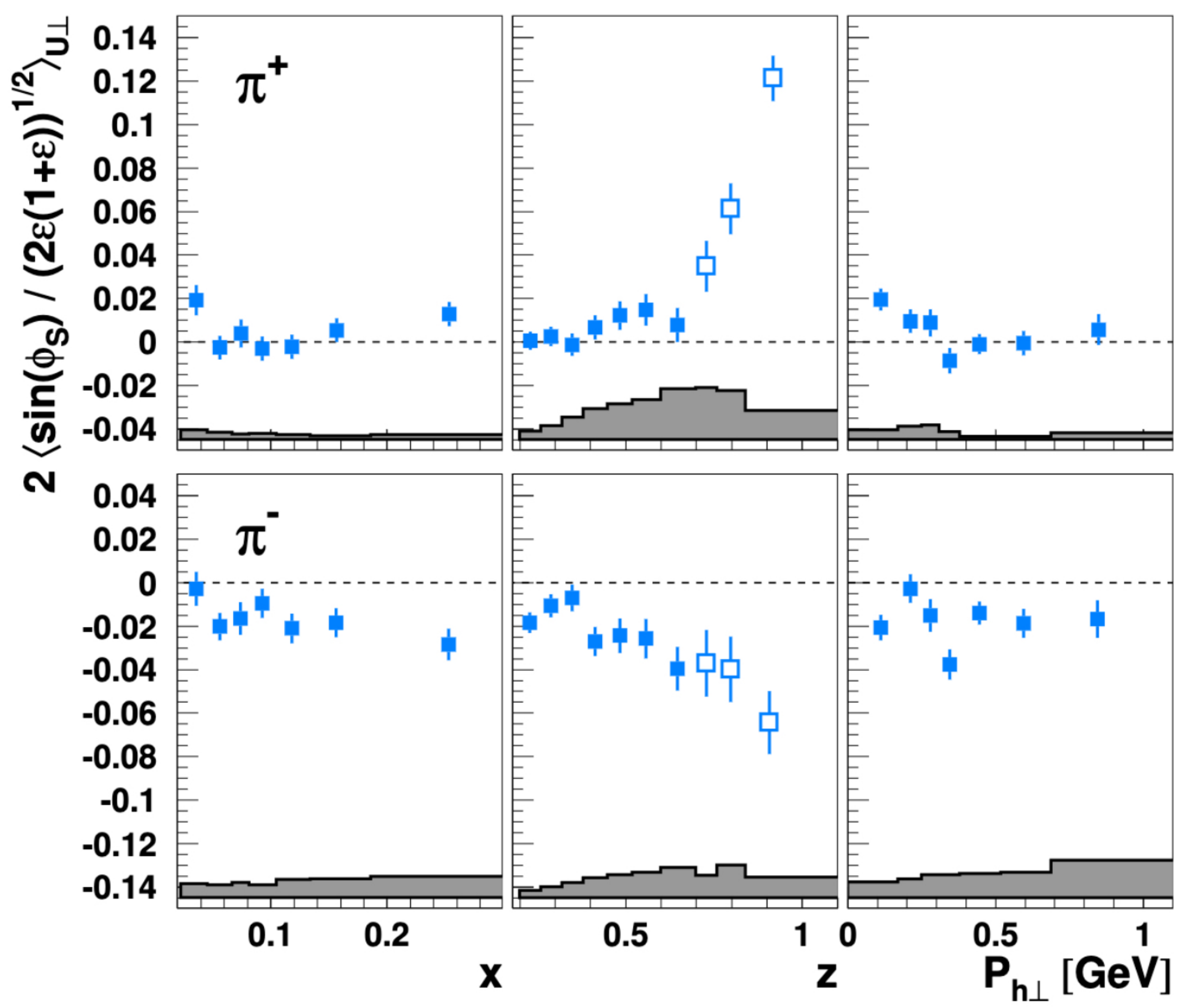}
\caption{HERMES data for charged pions of the $\sin \phi_S$ modulation of the SIDIS cross section for a proton target~\cite{HERMES:2020ifk}.
The shown observable is directly proportional to the structure function $F_{UT}^{\sin \phi_S}$ in Eq.~\eqref{eq:FUTsinphiS}.
Systematic uncertainties are given as bands.
Data at large values of $z$, marked by open points in the $z$ projection, are not included in the other projections; see~\cite{HERMES:2020ifk} for more details.} 
\label{fig:AUTsinphiS}
\end{figure}

Finally, we briefly discuss the transverse target SSA $A_{UT}^{\sin \phi_S}$ for which COMPASS, using a proton target, reported nonzero results for negatively charged hadrons, while their results for positively charged hadrons are compatible with zero.
Also the HERMES Collaboration measured this effect~\cite{HERMES:2020ifk}, with the results shown in Fig.~\ref{fig:AUTsinphiS}.
In qualitative agreement with the COMPASS results, HERMES finds somewhat larger effects for $\pi^-$ production.

We finish this section by emphasizing that the experimental results for the subleading effects discussed here are too large and too precise to simply ignore them. 
On the contrary, their detailed understanding requires dedicated theoretical efforts and analyses.
This is the main motivation for studying $\Lambda/Q$-suppressed observables within the TMD formalism, as this is the only rigorous QCD-based approach that can be applied to such observables.
The significant recent progress in this area has been summarized above.

\subsection{Estimating Subleading TMDs and Related Observables}
\label{sec:subTMDcalc}

We now turn our attention to calculations of subleading TMD effects.
As is the case for integrated PDFs, we repeat that subleading TMDs are not necessarily smaller than leading TMDs.  
While most calculations of TMDs address the leading sector (see Ch.~\ref{sec:lattice} and Ch.~\ref{sec:models} for an overview), quite a few estimates exist for subleading TMDs as well.
The vast majority of those studies are based on model calculations. 
Details about the main features of the relevant models can be found in Ch.~\ref{sec:models} and references therein.

\subsubsection{Generalized scalar charge from Lattice QCD}
\label{sec:latt_tmd_e}
\index{lattice QCD calculations!twist-3 TMDs}

Preliminary information is available for the subleading intrinsic TMD $e(x,k_T )$ from LQCD.
The LQCD methodology to evaluate selected TMD observables, described in detail in Sec.~\ref{sec:TMDratios}, can also be employed to access quantities involving subleading TMDs, in analogy to the leading TMD observables discussed there. 
In the case of the scalar Dirac structure $\Gamma = 1$, the fundamental hadronic matrix element in Eq.~(\ref{eq:latt_corr_def_2}) can be decomposed
into Lorentz-invariant amplitudes $\widetilde{A}_{i} $, $\widetilde{B}_{i}$ as follows (the complete decomposition is given in Ref.~\cite{Musch:2011er}),
\begin{equation}
\frac{1}{2 M}
\widetilde{\Phi }^{[1]}
= \widetilde{A}_{1}
+\frac{i M}{v\cdot P} \epsilon^{\mu \nu \rho \sigma} P_{\mu } b_{\nu }
v_{\rho } S_{\sigma }
\widetilde{B}_{5}\,.
\end{equation}
The Lorentz-invariant amplitudes are closely related to Fourier-transformed TMDs \cite{Musch:2011er}, and can thus be utilized to define the chiral-odd generalized scalar charge
\begin{equation}
\frac{\tilde{e}^{[1](0)} }{\tilde{f}_{1}^{[1](0)} } =
\frac{\widetilde{A}_{1} (-b_T^2 ,b\cdot P=0,\hat{\zeta } ,
\eta v\cdot P)}{\widetilde{A}_{2B} (-b_T^2 ,b\cdot P=0,\hat{\zeta } , \eta v\cdot P)}\,,
\label{gscalcharge}
\end{equation}
in analogy to the generalized tensor charge in Eq.~(\ref{gtcharge}); the arguments $v$, $\hat{\zeta} $ and $\eta $ describing the geometry of the staple-shaped gauge link are defined in Sec.~\ref{sec:latt_def_lorentz}.
The unpolarized amplitude $\widetilde{A}_{2B} $ in the denominator was introduced in Eq.~(\ref{adecomp1}). 
The ratio in Eq.~(\ref{gscalcharge}) is interpreted as a generalized scalar charge because, in the formal $b_T \rightarrow 0$ limit, i.e., upon complete integration over quark momentum components, the numerator corresponds to the standard scalar charge. It is normalized to the corresponding number of valence quarks by the denominator. It should, however, be emphasized that additional divergences arise in the $b_T \rightarrow 0$ limit (which corresponds to unrestricted integration over transverse momentum $k_T $) that require further renormalization. As a consequence, the ratio of scalar to vector renormalization constants, $Z_S /Z_V $, has to be accounted for when connecting the generalized scalar charge to the standard scalar charge.

\begin{figure}[t]
\includegraphics[width=8.4cm]{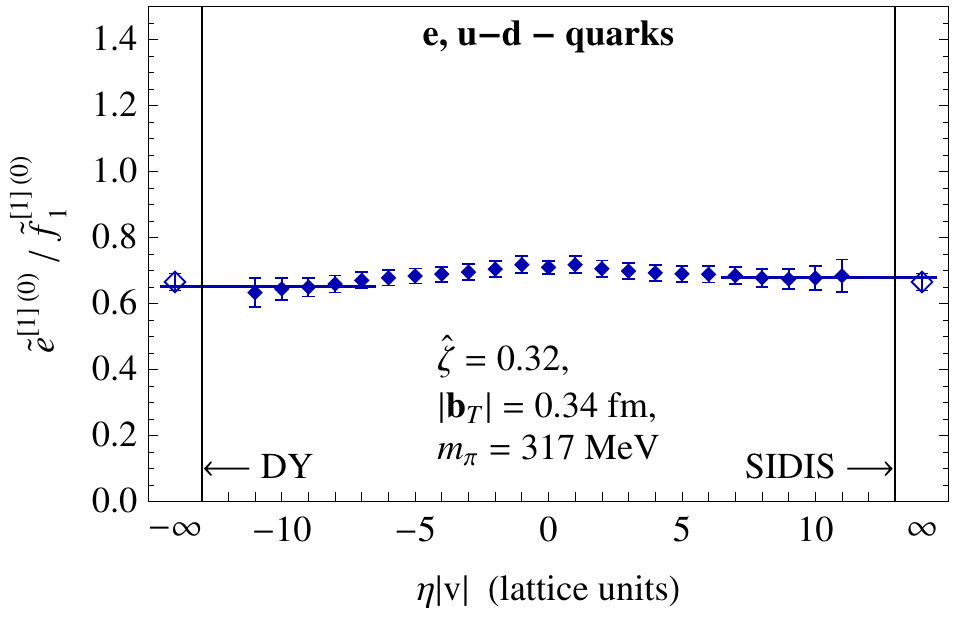}
\hspace{0.3cm}
\includegraphics[width=8.4cm]{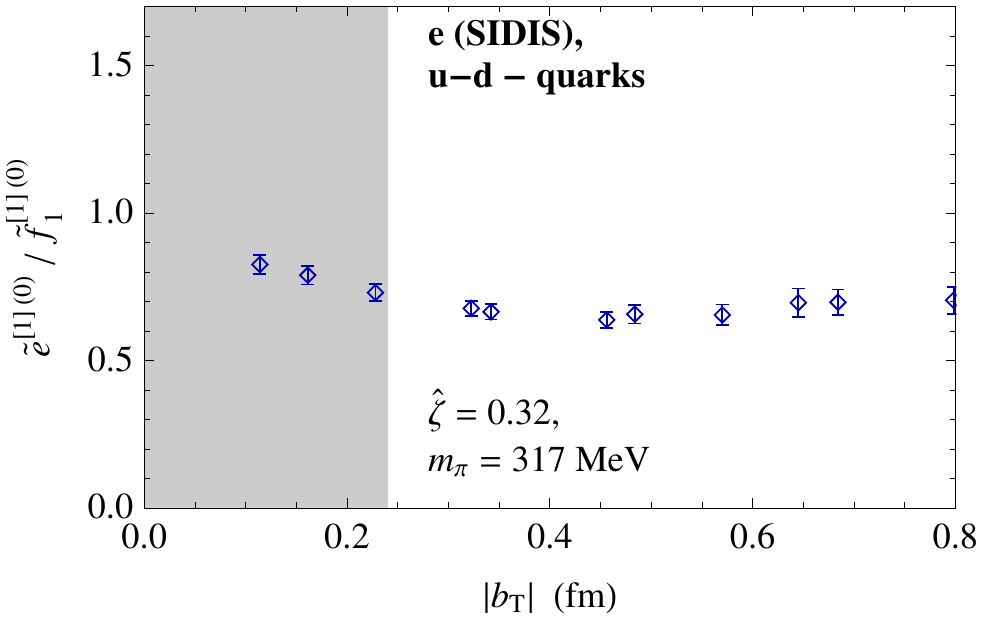}
\caption{Isovector generalized scalar charge in the nucleon, obtained using a clover fermion ensemble at the pion mass $m_{\pi } = 317\, \mbox{MeV} $.
Left: As a function of the staple length $\eta $ at fixed $b_T $ and $\hat{\zeta } $. Right: SIDIS/DY limit as a function of $b_T $ for fixed $\hat{\zeta } $, also presented previously in Ref.~\cite{Avakian:2019drf}. The shaded area indicates the region which may be subject to significant lattice artifacts.}
\label{fig:latt_tmd_e}
\end{figure}

In Fig.~\ref{fig:latt_tmd_e} we show results for the generalized scalar charge of the nucleon obtained using a clover fermion ensemble at the pion mass $m_{\pi } =317\, \mbox{MeV} $. No appreciable variation is seen as a function of the staple length $\eta $, indicating that the final-state interaction effects in the scalar and $\gamma^{+}$ nucleon matrix elements closely track one another. Also as a function of the quark operator separation $b_T$ in the SIDIS/DY limit, the variation of the generalized scalar charge appears to be weak. 
This stands in contrast to the significant variations seen for the Sivers, Boer-Mulders and $g_{1T}^\perp $ worm-gear shifts exhibited in Sec.~\ref{sec:TMDratios}.

\subsubsection{Model calculations of subleading TMDs}
Most model calculations of subleading TMDs have been performed in diquark spectator models, where many studies have included both scalar and vector diquarks in order to obtain results for up quarks and down quarks. \index{model!spectator model}
Analytical results in such a model for all 8 T-even subleading  intrinsic TMDPDFs can be found in Ref.~\cite{Jakob:1997wg}. 
In Ref.~\cite{Gamberg:2006ru}, the subleading T-odd PDF $g^\perp(x,k_T)$ was computed in the scalar diquark model, with an emphasis on scrutinizing the mere definition of subleading intrinsic TMDs.
It was found that, to lowest non-trivial order in perturbation theory, one encounters a light-cone singularity, a feature which does not show up for the corresponding calculation of leading T-odd PDFs.
Moreover, it was argued that actually all T-odd subleading intrinsic TMDs exhibit a light-cone singularity in the same model and in the quark-target model in QCD~\cite{Gamberg:2006ru}, with implications on QCD factorization involving renormalized subleading TMDs; see also Sec.~\ref{sec:subTMDfact}. \index{model!quark-target model} 

The subleading beam SSA $A_{LU}^{\sin \phi_h}$ in SIDIS, which is related to $g^\perp(x,k_T)$ in a tree-level analysis (see Eq.~\eqref{eq:FLUsinphi}), was also computed in the scalar diquark model~\cite{Afanasev:2006gw}.
By assuming factorization to be valid, from the finite result for the asymmetry, a finite expression for $g^\perp(x,k_T)$ was extracted.
However, that study did not address the direct calculation of $g^\perp(x,k_T)$ based on its operator definition, which explains the qualitatively different finding compared to Ref.~\cite{Gamberg:2006ru}. For related work see also Refs.~\cite{Rodini:2022wki,Gamberg:2022lju,AnjieREF}.
Another computation of $g^\perp(x,k_T)$ in a diquark spectator model for both up quarks and down quarks is discussed in Ref.~\cite{Mao:2012dk}, while further spectator model results for (T-even and T-odd) subleading TMDPDFs can be found in Refs.~\cite{Lu:2012gu, Mao:2014aoa, Mao:2014fma, Lu:2014fva, Mao:2016hdi, Yang:2018aue}. 
In some papers, subleading FFs were also studied in spectator models.
Specifically, calculations of the chiral-odd integrated FF $E(z)$ have been presented in Refs.~\cite{Ji:1993qx,Gamberg:2003pz}.
Furthermore, (T-odd) subleading TMDFFs, some of which are relevant for the QCD description of transverse SSAs in processes like $p^{\uparrow} p \to h X$, have been addressed in Refs.~\cite{Lu:2015wja, Yang:2016mxl}. 

All T-even subleading intrinsic TMDPDFs were also computed in the bag model~\cite{Avakian:2010br}.
\index{model!bag model}
Like in the case of leading TMDs, the results agree quite well with a Gaussian $k_T$-dependence.
Another interesting approach for estimating (subleading) TMDs is the light-front constituent model (LFCM). \index{model!lightfront constituent models}
In that framework, the TMDs are first represented through the overlap of light-front wave functions in a model-independent manner.
In a second step, one can exploit different models for the wave functions to obtain numerical results for the TMDs. 
The LFCM was applied to the T-even subleading intrinsic TMDPDFs for both the nucleon~\cite{Lorce:2014hxa} and the pion~\cite{Lorce:2016ugb}, with the treatment limited to the 3-quark ($3q$) sector.
In this approach, the analysis becomes quite cumbersome when including higher Fock states.
On the other hand, going beyond the $3q$ Fock state is expected to be very important in order to find realistic results for subleading TMDs.
For $e(x, k_T)$, in Ref.~\cite{Pasquini:2018oyz} the $3q + g$ Fock state has actually been included in the analysis in the LFCM.
We note that Ref.~\cite{Lorce:2014hxa} also contains a discussion of the T-even TMD $f^\perp(x,k_T)$ in the chiral quark soliton model, which presently is the only available result for a subleading TMD in this model.
\index{model!chiral quark soliton model}
(Studies of the collinear twist-3 PDFs $g_T(x)$, $h_L(x)$ and $e(x)$ in the chiral quark soliton model can be found in Refs.~\cite{Wakamatsu:2000ex, Schweitzer:2003uy, Wakamatsu:2003uu, Ohnishi:2003mf, Cebulla:2007ej}.) 
According to Eq.~\eqref{eq:FUUcosphi}, this TMD plays a critical role for the understanding of the $\cos \phi_h$ modulation of the unpolarized SIDIS cross section.
Furthermore, results for all T-even subleading TMDPDFs in the covariant parton model have been reported recently in Ref.~\cite{Bastami:2020rxn}.
\index{model!covariant parton model}

Calculations of TMDs (and related observables) in pQCD using the quark-target model are often used to study factorization and TMD evolution.
They can also shed light on the status of relations/constraints for TMDs that appear in other models.
A calculation of the subleading TMD $f^\perp(x,k_T)$ in the quark-target model was presented in Ref.~\cite{Mukherjee:2010iw}, whereas in Ref.~\cite{Chen:2016hgw} factorization for subleading TMDs for the Drell-Yan process was considered by focusing on the contribution related to $f^\perp(x,k_T)$, along with the evolution of that TMD.
Related work dealing with the twist-3 functions $g_T(x)$, $e(x)$ and $h_L(x)$ in the quark-target model can be found in, for instance, Refs.~\cite{Kundu:2001pk, Burkardt:2001iy, Aslan:2018tff, Bhattacharya:2020xlt, Bhattacharya:2020jfj, Bhattacharya:2021boh}.
We also want to briefly mention an interesting general feature for subleading parton distributions: 
They can exibit singular zero-mode contributions, that is, terms which are proportional to $\delta(x)$.
Such terms have been identified in model calculations but also in model-independent analyses, see~\cite{Burkardt:1995ts, Burkardt:2001iy, Efremov:2002qh, Schweitzer:2003uy, Wakamatsu:2003uu, Ohnishi:2003mf, Cebulla:2007ej, Pasquini:2018oyz, Aslan:2018tff, Bhattacharya:2021boh} and references therein for more details.

Subleading TMDs obtained in models without gluonic degrees of freedom may satisfy a number of so-called quark-model Lorentz invariance relations (qLIRs)~\cite{Mulders:1995dh, Boer:1997nt, Metz:2008ib, Teckentrup:2009tk, Avakian:2009jt}.
The qLIRs can provide a reasonable approximation for those TMDs and the corresponding subleading-power observables.
From a practical point of view, they allow for important cross checks of the analytical and numerical model results.
The qLIRs are discussed in more detail in Sec.~\ref{Sec:models-qLIRs} to which we refer the reader.
They must be distinguished from the LIRs which hold in full QCD and typically involve {\it qgq} correlations~\cite{Bukhvostov:1984rns, Bukhvostov:1984zhx, Belitsky:1997ay, Belitsky:1997zw, Kanazawa:2015ajw}.

A frequently-used approach for estimating subleading integrated PDFs is the Wandzura-Wilczek (WW) approximation, which was originally derived for $g_T(x)$~\cite{Wandzura:1977qf} but can also be applied to $h_L(x)$~\cite{Jaffe:1991ra}.
\index{Wandzura-Wilczek (type) approximation}
Here one makes use of the fact that, for instance, $g_T(x)$ can be decomposed into a term which is fixed by the twist-2 helicity distribution $g_1(x)$, plus a term that is given by a {\it qgq} correlator, where the WW approximation consists of neglecting the latter contribution.
At present, we are lacking very robust information about the quality of the WW approximation.
(More details about this point can be found in~\cite{Bastami:2018xqd} and references therein.)
However, for the lowest non-trivial $x$-moment of $g_T$ and $h_L$ instanton-vacuum model calculations~\cite{Balla:1997hf, Dressler:1999hc}, 
\index{model!instanton model} as well as a study in LQCD~\cite{Gockeler:2000ja}, suggest that the WW approximation works very well.
It has also been argued that, based on experimental data, a violation of the WW approximation for $g_T(x)$ at the level of 15-40\% is possible~\cite{Accardi:2009au}.
We also point out that the very first calculations of $g_T(x)$ and $h_L(x)$ in LQCD are compatible with this finding~\cite{Bhattacharya:2020cen, Bhattacharya:2021moj}; see Sec.~\ref{sec:lamet} for more details. 

\begin{figure}[t]
\centering
\includegraphics[width=0.30\textwidth]{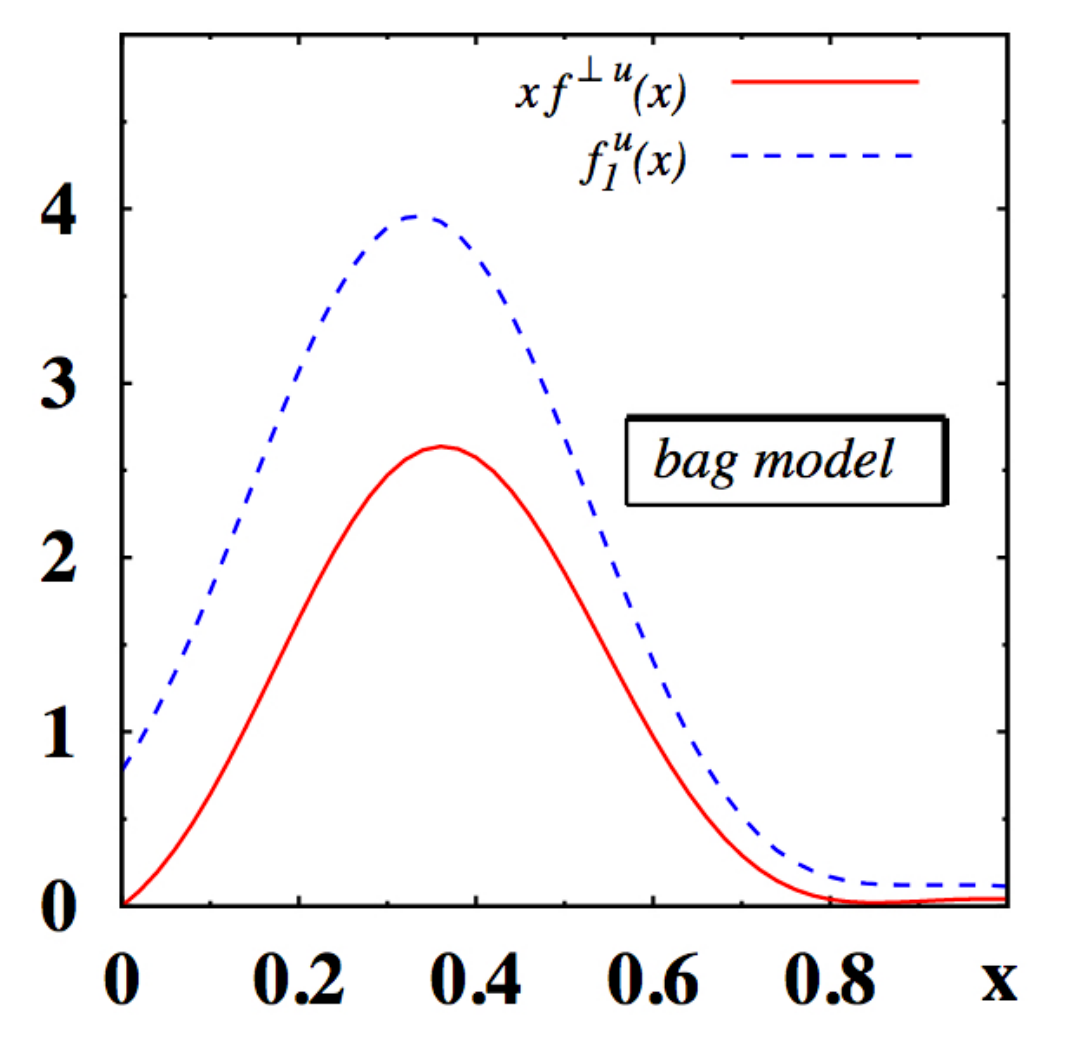}
\caption{Bag model results for the functions $x f^\perp(x)$ and $f_1(x)$ for up quarks in the proton~\cite{Avakian:2009jt, Avakian:2010br}.
These functions would be equal in the WW-type approximation in Eq.~(\ref{e:f_perp_WWapp}).}
\label{fig:f_perp_WWapp}
\end{figure}

A very similar approximation, which is typically called WW-type approximation, can be made for subleading intrinsic TMDs~\cite{Mulders:1995dh, Bacchetta:2006tn}.
We refer to~\cite{Bastami:2018xqd} for a comprehensive review of the WW-type approximation, where all parton model results for the subleading SIDIS structure functions at low transverse hadron momenta have been expressed in the WW-type approximation.
To discuss just one example of this approximation we use the first EOM relation~\cite{Mulders:1995dh, Bacchetta:2006tn} in~\eqref{eq:EOMrltn},
where  $\tilde{f}^\perp$ is  the  twist-3 term that is defined through the {\it qgq} correlator. 
The integrated WW-type approximation for the TMD $f^\perp$ is then given by
\begin{equation}
x f^\perp(x) \big|_{\rm WW-type} =  f_1(x) \,.
\label{e:f_perp_WWapp}
\end{equation}  
The curves in Fig.~\ref{fig:f_perp_WWapp} show bag model results for both the twist-3 function $x f^\perp(x)$ and the twist-2 function $f_1(x)$.
Obviously, in this model the quality of the WW-type approximation (strongly) depends on $x$, with the approximation working best for intermediate to large values of $x$ which is the region where quark model results are expected to be more reliable.
It also depends on the TMD under consideration, where we refer to~\cite{Avakian:2010br} for more numerical results.
The same general features apply to all models~\cite{Bastami:2018xqd}. 
We note in passing that, interestingly, the relation~\eqref{e:f_perp_WWapp} is exact in the chiral quark soliton model~\cite{Lorce:2014hxa}.

\subsubsection{Model calculations of subleading-power observables}
The calculation of the $\cos \phi_h$ dependence of the unpolarized SIDIS cross section by Cahn in the framework of a generalized parton model has similarities with applying the WW-type approximation to the tree level expression for $F_{UU}^{\cos \phi_h}$ in Eq.~\eqref{eq:FUUcosphi}~\cite{Cahn:1978se, Cahn:1989yf}.
A very similar approach for this structure function was employed in Ref.~\cite{Anselmino:2005nn}.
More specifically, the authors of that paper used a Gaussian ansatz for the TMDPDF $f_1(x, k_T)$ and the TMDFF $D_1(z,z p_T')$, and extracted (approximate) values for the respective average transverse momenta from data of the EMC Collaboration~\cite{Aubert:1983cz, Arneodo:1986cf} and the E665 Collaboration~\cite{Adams:1993hs}.
(For a related discussion we refer to~\cite{Schweitzer:2010tt}.)
They found the values $\langle k_T^2 \rangle = 0.25 \; \textrm{GeV}^2$ and $\langle p_T^2 \rangle = 0.20 \, \textrm{GeV}^2$~\cite{Anselmino:2005nn}, which compare reasonably well with the widths extracted from leading-power observables; see Sec.~\ref{sec:SIDISmult}.

We repeat that further interest in subleading SIDIS structure functions arose with the first observation of a nonzero longitudinal target SSA by the HERMES Collaboration~\cite{HERMES:1999ryv}.
As a consequence of that measurement, this SSA, as well as other subleading effects in SIDIS, were explored in a number of studies which made use of the WW-type approximation in one form or another~\cite{Oganessian:1998ma, Kotsinian:2000td, Boglione:2000jk, DeSanctis:2000fh, Ma:2000ip, Oganessian:2000um, Efremov:2001cz, Ma:2001ie, Efremov:2001ia, Oganessyan:2002pc, Efremov:2002ut, Oganessyan:2002er, Ma:2002ns, Efremov:2003tf, Efremov:2003eq, Yuan:2003gu, Schweitzer:2003yr}.
The goals of those works included describing the experimental data, extracting information on PDFs and FFs from the data in the WW-type approximation, and making predictions for different kinematics and experiments and/or other structure functions.
Recently, in Ref.~\cite{Bastami:2018xqd} a comprehensive numerical analysis of the SIDIS structure functions in the WW-type approximation was presented.

Another series of papers made use of spectator models in order to estimate subleading effects in SIDIS~\cite{Gamberg:2003pz, Mao:2012dk, Mao:2014aoa, Mao:2014fma, Lu:2014fva, Mao:2016hdi, Yang:2018aue}. 
Furthermore, in Ref.~\cite{Song:2010pf} the unpolarized SIDIS cross section was studied through $\Lambda^2/Q^2$ accuracy, with a particular focus on the $\cos 2\phi_h$ modulation.
Using a generalized parton model in the spirit of the work by Cahn~\cite{Cahn:1978se, Cahn:1989yf} provides an (important) nonzero $\Lambda^2/Q^2$ contribution to this structure function which is related to the leading unpolarized TMDPDF $f_1(x,k_T)$ and TMDFF $D_1(z,zp_T')$.
However, such a treatment does not lead to a full tree-level result in QCD which was aimed at in Ref.~\cite{Song:2010pf}.
Another interesting aspect of that work is a comparison between scattering off a proton versus a nuclear target.
Related studies, dealing with subleading-power observables for semi-inclusive reactions and nuclear targets, can be found in Refs.~\cite{Gao:2011mf, Song:2013sja, Chen:2013zpy}.

Calculations of power-suppressed observables in SIDIS were also instrumental for obtaining a complete list of subleading TMDs.
Specifically, Ref.~\cite{Afanasev:2003ze} addressed the structure function $F_{LU}^{\sin \phi_h}$ in Eq.~\eqref{eq:FLUsinphi} in a scalar diquark model.
This work was revisited and also extended to the structure function $F_{UL}^{\sin \phi_h}$ in Eq.~\eqref{eq:FULsinphi} in Ref.~\cite{Metz:2004je}.
Based on the results it was argued that the list of subleading intrinsic TMDs known at that time was incomplete~\cite{Metz:2004je}.
This development indeed led to the discovery of a new T-odd subleading intrinsic TMD for an unpolarized target, namely $g^\perp(x,k_T)$, in Ref.~\cite{Bacchetta:2004zf}. 
Later on, further studies uncovered two additional subleading TMDs for a spin-$\frac{1}{2}$ hadron~\cite{Goeke:2005hb}, completing the list of the 16 subleading intrinsic TMDPDFs in the table in Fig.~\ref{fig:TMDPDFs_tw3} above.

\subsection{Summary and Outlook}
\label{sec:subTMDoutlook}

So far, the main focus of the TMD community has been on the leading-power TMDs.
However, as we have emphasized in this chapter, subleading TMDs are important for the theoretical description of a variety of structure functions which only start at this order, and give access to novel TMD probes of PDFs and FFs.
These subleading TMD probes are related to quark-gluon-quark correlations which allow for studies of the hadron structure that are complementary to the investigation of parton densities described by leading TMDs.
While this alone provides a strong motivation for the field, it is important to try further reveal the physics encoded in the subleading TMDs.
For a long time, the unclear status of factorization has been a serious impediment in the field of subleading TMDs, but as discussed above, 
considerable progress in this area has recently been reported~\cite{Vladimirov:2021hdn,Ebert:2021jhy,Rodini:2022wki,Gamberg:2022lju}.
At the time of writing, important areas of active research include demonstrating the cancellation of potential factorization violating contributions from the Glauber region at subleading power, and constructing definitions for renormalized subleading power TMDs that can be shown to be  valid beyond the one-loop order that has been considered so far.

These recent developments hold promise to the put the studies of subleading TMDs on very safe ground.
They will generate renewed interest in this field and can initiate additional calculations of those functions and related observables in various approaches, including LQCD.
It is certainly worthwhile to take a fresh look at what information on subleading TMDs can be extracted from existing data and how future experiments, in particular the EIC, can move this field forward.

%% file: sec-gtmds/sec-gtmds.tex
\section{Generalized TMDs and Wigner Phase Space Distributions}
\label{sec:gtmd}

TMDs provide, on the one hand, the most complete description of hadronic structure as far as its dependence on quark and gluon longitudinal and transverse momentum components is concerned. On the other hand, however, a full representation of hadron dynamics is only attained by addressing, in addition to the quark and gluon 3D momentum structure, the correlation between their 
momenta and spatial coordinates. Through this correlation we can study rotational motion and, in particular, angular momentum as well as other mechanical properties of the proton. 

\begin{figure}[t!]
\begin{center}
\includegraphics[width=15cm]{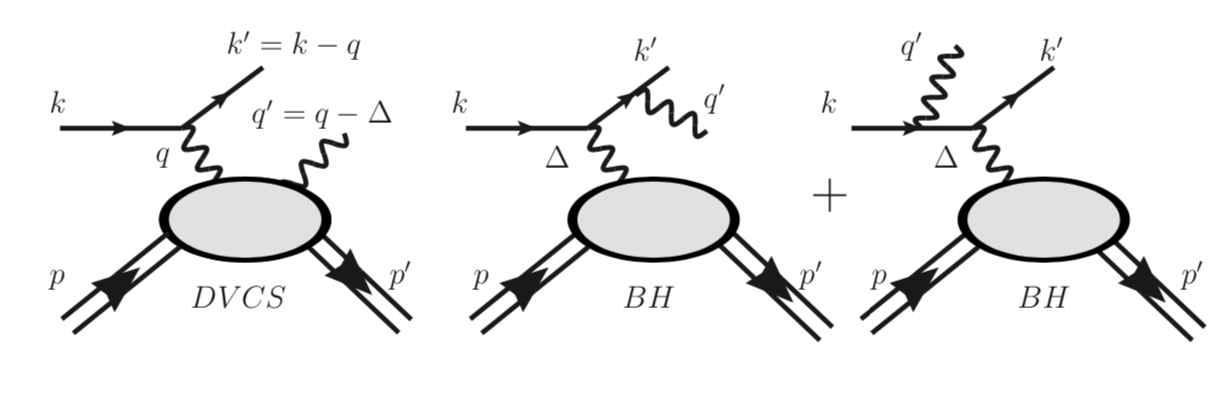}
\vspace{-0.3cm}
\caption{(adapted from Ref.~\cite{Kriesten:2019jep}) Exclusive electroproduction of a photon through the DVCS and Bethe-Heitler processes.}
\label{fig:BHDVCS}
\end{center}
\end{figure}

The idea of a phase-space distribution for a quantum mechanical system was first introduced by Wigner \cite{Wigner:1932eb}. Wigner distribution based approaches were subsequently applied to a large variety of systems; in the context of nuclear physics, this includes the description of parton showers  \cite{Geiger:1998yk} and  heavy ion collisions \cite{Heinz:2009xj}. To study the structure of the proton an approach was developed in \cite{Belitsky:2003nz} where it was shown that Wigner distributions can reduce to positive definite probability density distributions in particular limits. As we explain in detail in what follows, by taking the integral of the Wigner distribution over transverse momentum, one obtains a so-called impact parameter distribution (IPD)
\index{impact parameter distribution}
\cite{Soper:1976jc,Burkardt:2000za,Ralston:2001xs, Diehl:2002he, Burkardt:2002hr} describing the longitudinal momentum fraction $x$ distribution of partons located at a given transverse distance from the hadron center of momentum.\footnote{The ``impact parameter'' entering the Wigner distribution must be distinguished from the coordinate $\bt $ introduced for TMDs in this handbook. In the present treatment, we use the notation $\rt $ for the impact parameter to avoid confusion with the coordinate space quantity $\bt $ conjugate to $\kt $ used throughout.}
Furthermore, performing a Fourier transform with respect to the transverse coordinate variable, one obtains a generalized parton distribution (GPD).
\index{GPD}
GPDs bring the study of momentum-coordinate-space correlations inside the proton within experimental grasp since, as observed in Refs.~\cite{Muller:1994ses,Ji:1996ek,Ji:1996nm,Radyushkin:1996nd,Radyushkin:1996ru,Radyushkin:1997ki},
they are key observables parametrizing the matrix elements of deeply virtual exclusive scattering experiments. The prototype of a deeply virtual exclusive scattering experiment is deeply virtual Compton scattering (DVCS),
\index{deeply virtual Compton scattering (DVCS)}
where a photon is produced in the hard scattering while the initial proton recoils intact, cf.~Fig.~\ref{fig:BHDVCS}. In particular, Ji provided a connection between the DVCS scattering amplitude and angular momentum as described by the matrix elements of the QCD energy momentum tensor \cite{Ji:1996ek}.
Other deeply virtual exclusive experiments include meson electroproduction and crossed channel experiments such as timelike Compton scattering.

\begin{figure}[t!]
\centering 
\includegraphics[width=0.75\textwidth]{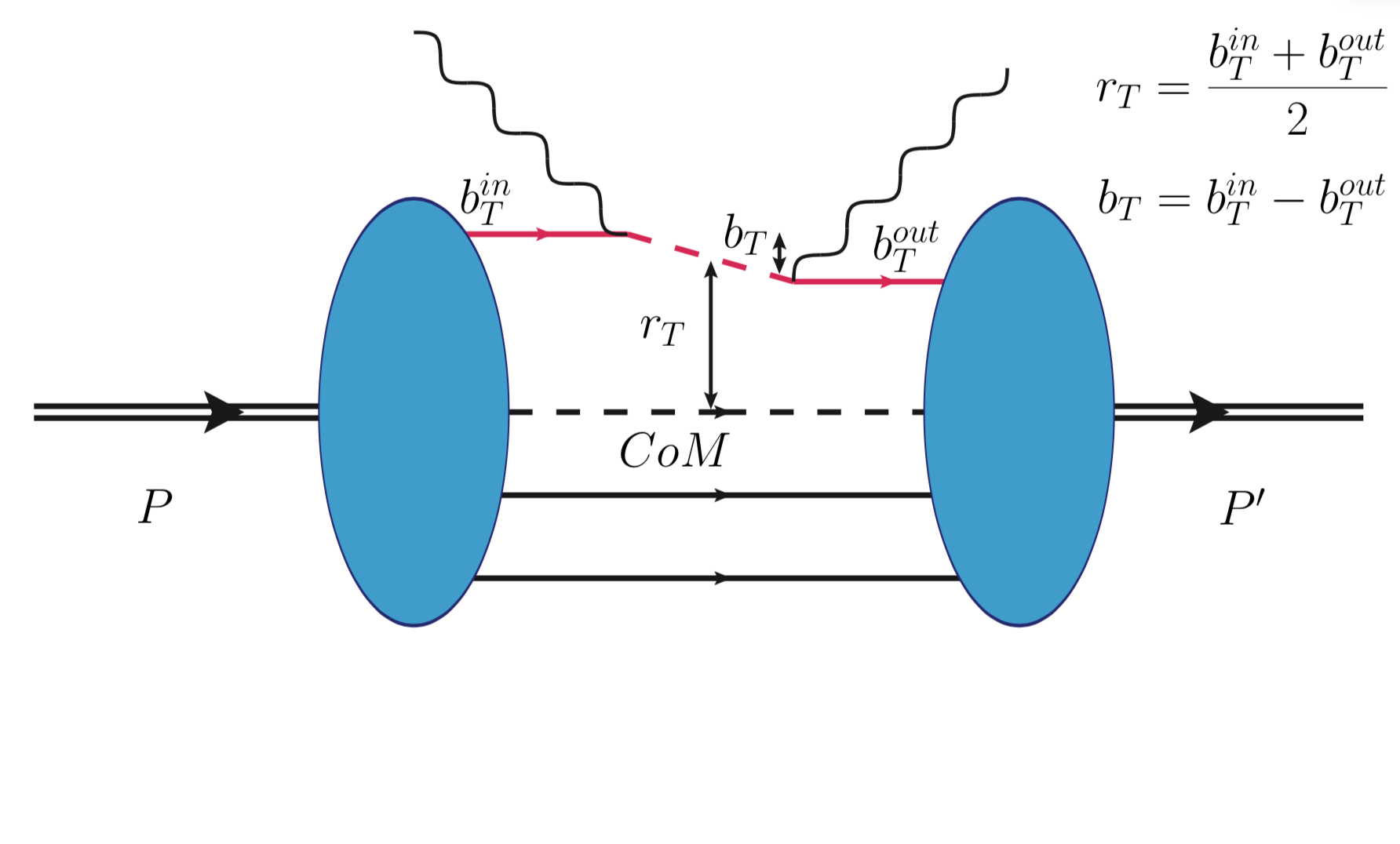} 
\phantom{x}\vspace{-1cm}
\caption{Transverse spatial coordinates entering the definition of GPDs through the correlation function in Eq.~\eqref{eq:GPDcorr_def}, cf.~also Ref.~\cite{Diehl:2002he}.}
\label{fig:GPDcorrelator}
\end{figure}
GPDs can be viewed as hybrid objects that, on the one hand, similarly  to the collinear PDFs, describe quark and gluon distributions in the longitudinal momentum fraction, $x$, at a given scale, $Q^2$. On the other hand, similarly to the nucleon elastic form factors, they give insight into the internal spatial distribution of the quark and gluon constituents through two additional kinematic variables: $\xi$, known as the skewness parameter, and the Mandelstam invariant $t=\Delta^{2} $. These, respectively, describe the longitudinal component and the square of the proton momentum transfer variable, $\Delta=P'-P$. GPDs parametrize the following collinear correlation function,
\index{GPD}
\begin{equation} 
\label{eq:GPDcorr_def}
{\cal F}^{q [\Gamma ]}_{S^{\prime } , S} 
  = \left. \frac{1}{2} \int  \frac{db^-}{2 \pi} \, e^{ix(P^+ + P'^+)b^-/2} \,  \Bigl\langle p(P^{\prime} ,S') \Big| 
    \bar\psi_{q} (b^{out}) \,  \Gamma \, W(b^{out} , b^{in } ) \,
     \psi_{q} (b^{in}) 
     \Big| p(P,S) \Bigr\rangle \right|_{b_T=0,b^+=0} \, ,
\end{equation}
where the Wilson line $W$ takes a straight path between $b^{in} $ and $b^{out} $ on the light cone. For example, in the particular case where the quark helicity is conserved, the structures $\Gamma=\gamma^+,\gamma^+\gamma_5$ are relevant; various combinations of operators and proton polarizations then lead to the correlation function being parametrized by four independent twist-two GPDs, $H^q $ and $E^q $ for $\gamma^+$, and $\widetilde{H}^{q} $ and $\widetilde{E}^{q} $ for $\gamma^+\gamma_5$.
For detailed reviews on GPDs and their experimental access, we refer the reader to Refs.~\cite{Ji:1998pc,Diehl:2003ny,Belitsky:2005qn,Boffi:2007yc,Kumericki:2016ehc}.

TMDs represent another limit of the Wigner distribution obtained by integrating over the transverse coordinate, ${\bf r}_T$. 
TMDs and GPDs can be seen, therefore, as different ``slices" of Wigner distributions, giving  complementary information on the distributions of partonic transverse momentum on the one hand, and transverse spatial coordinates on the other. 

Two sets of coordinate space quantities are needed to describe a phase-space distribution in QCD: ${\bf r}_T= ({\bf b}_T^{in}+{\bf b}_T^{out})/2$, which is Fourier conjugate to ${\bf \Delta}_T$, and ${\bf b}_T= {\bf b}_T^{in}-{\bf b}_T^{out}$, which is Fourier conjugate to the transverse momentum, ${\bf k}_T$.  All quantities are measured with respect to the proton center of momentum (CoM). By considering the collinear $\kt $ integrated quantity,  setting ${\bf b}_T=0$ as in Eq.~(\ref{eq:GPDcorr_def}), one has that $\rt$ can be interpreted as the average position of the parton inside the proton with respect to the CoM.  
Quark and gluon spatial probability distributions in the transverse coordinate,
$\rt $, $f(x,\rt )$,
are obtained by Fourier transformation with respect to the transverse component ${\bf \Delta}_T $, where ${\bf \Delta}_T^{2} = -t$ by setting the skewness parameter, $\xi=0$. The transverse coordinate space variables for the GPD correlator are shown in Figure \ref{fig:GPDcorrelator}, cf.~also Ref.~\cite{Diehl:2002he}. 

\begin{figure}[t!]
\centering
\includegraphics[width=0.80\textwidth]{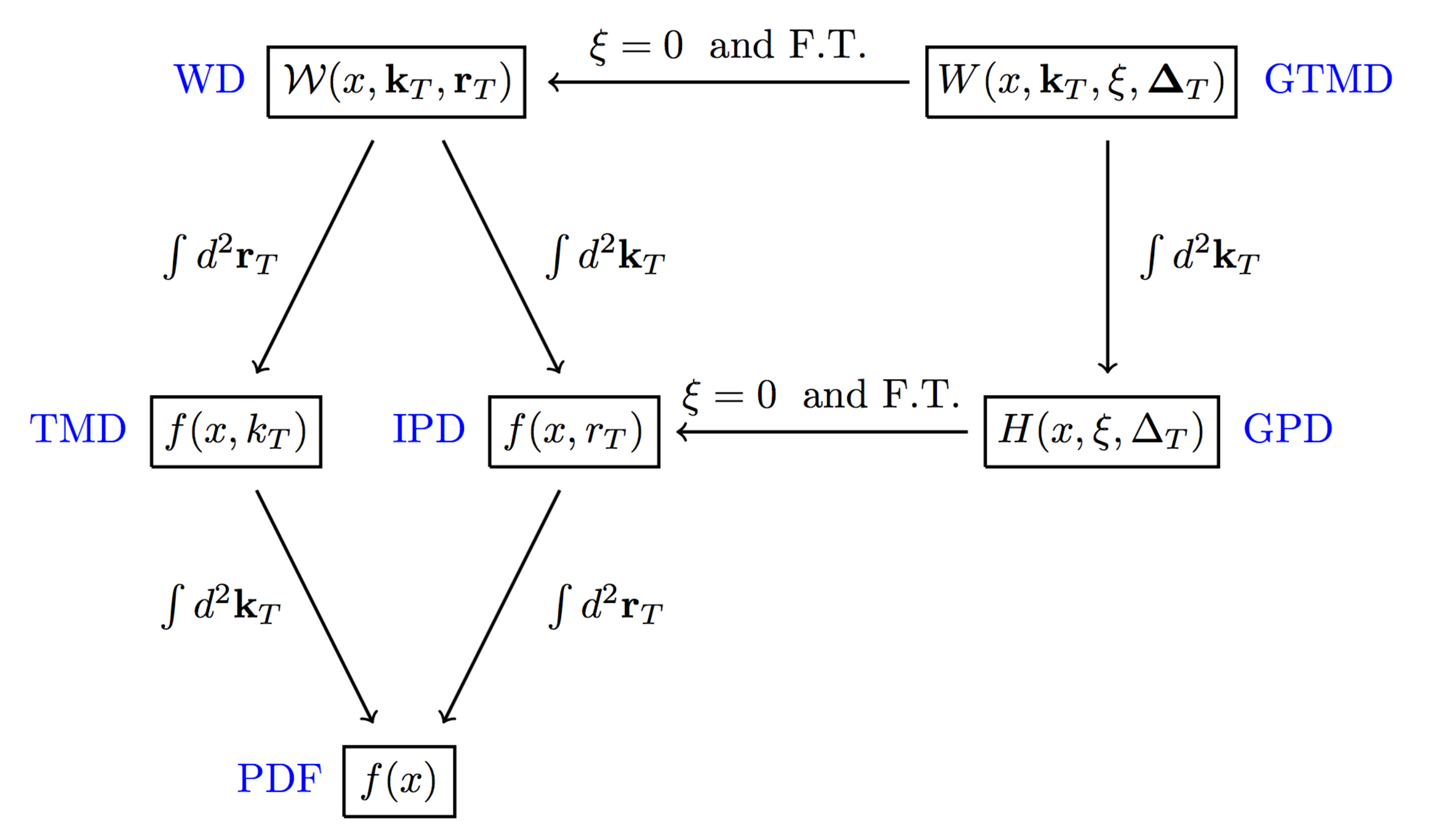} 
\caption{Quantities characterizing the multi-dimensional parton structure of hadrons and the relations between them.~\protect\footnotemark
In order to arrive through a Fourier transform~(F.T.) from GPDs and GTMDs, which (in principle) can be measured, at impact parameter distributions (IPDs) and Wigner distributions (WDs), respectively, an extrapolation to the kinematical point $\xi = 0$ is needed.}
\label{f:multi_dim_imaging}
\end{figure}

Wigner distributions encompass both types of distributions, TMDs and GPDs. In what follows, we elaborate on the complementary role of TMDs and GPDs using the concept of Wigner distributions as illustrated in the scheme in Figure \ref{f:multi_dim_imaging}.

\subsection{Wigner Distributions}
\label{sec:Wigner}
\index{Wigner distribution}

\footnotetext{These relationships all hold for the bare versions of these functions. For the renormalized versions the correspondence can be more complicated; see \sec{integratedTMDs} for a discussion of the $\int d^2{\bf k}_T$ integral relation between the TMD and PDF.}

Wigner distributions were first introduced in non-relativistic quantum mechanics \cite{Wigner:1932eb}.
For quantum particles, they are also known as quasi-distributions\footnote{Note this is a different concept than the quasi-distributions introduced in Chapter \ref{sec:lattice}.}
since they are affected by the uncertainty principle; in general, they are not positive definite, and therefore they do not have a straightforward probability interpretation.
For a 1D system, the relation between a Wigner distribution and the wave function in position space or momentum space reads \cite{Hillery:1983ms},
\begin{equation}
{\cal W}(x,k) = \int \frac{dx'}{2\pi} \, e^{i \, k \, x'} \psi^{\ast} \Big( x + \frac{x'}{2} \Big) \, \psi \Big( x - \frac{x'}{2} \Big)
= \int \frac{dk'}{2\pi} \, e^{- \, i \, k' x} \, \tilde{\psi}^{\ast} \Big( k + \frac{k'}{2} \Big) \, \tilde{\psi} \Big( k - \frac{k'}{2} \Big) \,,
\label{e:wigner_qm}
\end{equation}
which readily implies that integrating the Wigner distribution ${\cal W}(x,k)$ upon $k$ gives the position space density $|\psi(x)|^2 $, while integrating upon $x$ gives the momentum space density $|\tilde{\psi}(k)|^2$.
The calculation of the expectation value of an
observable $O$ is very appealing in the Wigner distribution framework. In particular,
\begin{equation}
\big\langle O \big\rangle = \int dx \, dk \, O(x,k) \, {\cal W}(x,k) \,,
\label{e:wigner_qm_obs}
\end{equation}
which is identical to the calculation of an expectation value using a classical phase space distribution.
In other areas of physics, such as quantum optics, Wigner distributions have been frequently used.
They can give deeper insights into the relation between quantum mechanics and classical mechanics.
The generalization from 1D to 3D is straightforward in non-relativistic quantum mechanics and leads to 6D Wigner distributions.

Partonic Wigner distributions can be defined in quantum field theory in terms of correlation functions in analogy to Eq.~(\ref{e:wigner_qm}).
To this end, we consider the Wigner operator~\cite{Ji:2003ak, Belitsky:2003nz, Lorce:2011kd}
\begin{equation}
\widehat{\cal W}^{q \, [{\Gamma}] \, W}(x, \kt, \rt)  = 
\int \frac{db^- \, d^2 \bt}{ 2 \, (2\pi)^3} \, e^{-i k \cdot b} \, 
\bar{\psi}_q \Big( \rt + \frac{b}{2} \Big) \, \Gamma \, W_{\sqsupset \eta}^{v}
\Big( \rt + \frac{b}{2} , \rt - \frac{b}{2} \Big) \, \psi_q \Big( \rt - \frac{b}{2} \Big) \Big|_{b^+ = 0} \,,
\label{e:wigner_op}
\end{equation}
where the analogy to Eq.~\eqref{e:wigner_qm} is obvious.
This operator depends on the longitudinal and transverse parton momenta, the transverse parton position $\rt$ (impact parameter), the Dirac structure $\Gamma$, and on the path of the Wilson line, denoted by the superscript $W$ on the left-hand side of Eq.~(\ref{e:wigner_op}). Note that, in the description of the Wilson line, the flexible notation of Eq.~(\ref{eq:stapleLattice}) is adopted, which includes the option of choosing a straight path between the quark operators, by setting $\eta =0$. The reason is that, in the context of Wigner distributions and GTMDs, both staple-shaped and straight Wilson line paths constitute physically interesting cases, as will be discussed further in Sec.~\ref{sec:GTMD_OAM} in the context of quark Orbital Angular Momentum (OAM).

We limit our description to 2D spatial Wigner distributions 
which can be unambiguously extracted from experiment through Fourier transformation in the transverse momentum transfer $\deltat $. 
It is well known
that a full 3D spatial description is hampered by relativistic proton recoil effects~\cite{Licht:1970pe,Licht:1970de}, while these effects are mitigated in a heavy nucleus. Whether it is possible to define meaningful 6D Wigner distributions for partons has generated some debate with various approaches addressing this problem \cite{Belitsky:2003nz,Lorce:2011kd}. 

The Wigner operator in Eq.~\eqref{e:wigner_op} can now be used to define the correlator for 5D Wigner distributions of quarks \cite{Ji:2003ak, Belitsky:2003nz, Lorce:2011kd},
\begin{align}
{\cal W}^{q \, [\Gamma] \, W}_{S^{\prime } ,S} (x, \kt, \rt )  & =
 \int\frac{d^2 \deltat}{(2\pi)^2} \; \bigg\langle
 p(P^{\prime } ,S^{\prime } ) \,\bigg| \, \widehat{\cal W}^{q \, [{\Gamma}] \, W}(x, \kt, \rt) \, \bigg| \, p(P,S) \bigg\rangle \,,
\label{e:wigner_corr1}
\end{align}
where $S^{\prime } $, $S$ denote the spins of the external states,
the average proton momentum $\bar{P} = (P^{\prime } +P)/2$ defines the longitudinal direction, and the momentum transfer is purely transverse, $P^{\prime } = \bar{P} + \deltat /2$, $P=\bar{P} -\deltat /2$.

The full 5D information contained in the Wigner distribution can be reduced in several ways by integrating out some of the variables. In particular, one can extract
\begin{align}
{\cal F}^{q \, [\Gamma]}_{S^{\prime } ,S} (x, \rt ) & = \int d^2 \kt \, {\cal W}^{q \, [\Gamma] \, W}_{S^{\prime } ,S} (x, \kt, \rt ) \,,
\label{e:Wigner_position} \\
\Phi^{q \, [\Gamma] \, W}_{S^{\prime } ,S} (x, \kt ) & = \int d^2 \rt \, {\cal W}^{q \, [\Gamma] \, W}_{S^{\prime } ,S} (x, \kt, \rt ) \,,
\label{e:Wigner_momentum} \\
\big \langle O \big \rangle^{[\Gamma ] W}_{S^{\prime } ,S}
& = \int dx \, d^2 \kt \, d^2 \rt \, O(x, \kt, \rt) \, {\cal W}^{q [\Gamma] \, W}_{S^{\prime } ,S} (x, \kt, \rt ) \,,
\label{e:Wigner_expectation_value}
\end{align}
where ${\cal F}^{q \, [\Gamma]}_{S^{\prime } ,S} (x,\rt )$ is the density of quarks in longitudinal momentum and transverse position space, while $\Phi^{q \, [\Gamma] W}_{S^{\prime } ,S} (x,\kt )$ is the density in momentum space. The quantity ${\cal F}^{q \, [\Gamma]}_{S^{\prime } ,S} (x,\rt )$ is what defines the so-called 
\index{impact parameter distribution}
impact parameter distributions $f_{S^{\prime } ,S} (x,\rt )$, cf.~\cite{Soper:1976jc}, which are related to GPDs, taken at $\xi =0$, through a Fourier transform~\cite{Burkardt:2000za}. 
Note that, upon taking the $\kt $ integral in Eq.~\eqref{e:Wigner_position}, the dependence on the path of the Wilson line disappears.  
We also point out that, strictly speaking, in this relation the same complications can arise that one has when integrating TMDs in order to get to PDFs. In particular, these relations are true at a bare level and must be reconsidered after renormalization has been carried out, cf.~the detailed discussion of the relation between TMDs and PDFs given in Sec.~\ref{sec:integratedTMDs}.
Note the very close analogy of Eqs.~\eqref{e:Wigner_position}-\eqref{e:Wigner_expectation_value} with the situation in non-relativistic quantum mechanics.
In Sec.~\ref{sec:GTMD_OAM}, we return to Eq.~(\ref{e:Wigner_expectation_value}) in the context of partonic orbital angular momentum and spin-orbit correlations.

The range of physical information contained in the full set of leading-twist nucleon Wigner distributions, including, in particular, the dependence on the nucleon spins $S^{\prime } $,
$S$ and the quark polarization encoded in the $\Gamma $ structure, has been further elucidated  in Ref.~\cite{Lorce:2015sqe}.
By decomposing these Wigner distributions into multipoles in the transverse phase space, correlations between target polarization, quark polarization and quark orbital angular momentum can be isolated and exhibited, and the structure of the different components visualized.

\subsection{Momentum Space Definition -- Generalized TMDs (GTMDs)}
\label{sec:GTMD}
\index{GTMD}

Similar to the relationship between impact parameter distributions and GPDs already mentioned above, the full
Wigner distributions defined in Eq.~\eqref{e:wigner_corr1}
can be connected to distributions that depend on transverse momentum transfer, ${\bf \Delta}_T$,
through Fourier transformation,
 \begin{align}
\mathbb{W}^{q \, [\Gamma] \, W}_{S^{\prime } ,S} (\bar{P} ,\Delta ,x, \kt ) \Big|_{\xi = 0} & = \int\frac{d^2 \deltat}{(2\pi)^2} \,
 e^{i \, \deltat \cdot \rt} \, {\cal W}^{q \, [\Gamma] \, W}_{S^{\prime } ,S} (x, \kt, \rt )  \,.
\label{e:wigner_corr2}
\end{align}
Note that in Eq.~\eqref{e:wigner_corr2} the Wigner operator is evaluated between states which have the same plus-momentum, that is, for $\xi = 0$.
On the other hand, the correlator on the left-hand side, $\mathbb{W}^{q \, [\Gamma] \, W}_{S^{\prime } ,S}$, can be defined for general momentum transfer $\Delta $,
including a longitudinal component, cf.~Eq.~(\ref{e:gtmd_corr}) below; only its $\xi =0 $ limit enters the relation (\ref{e:wigner_corr2}). Also the dependence on the average hadron momentum $\bar{P} $ has been made explicit on the left-hand side.

The correlator $\mathbb{W}^{q \, [\Gamma] \, W}_{S^{\prime } ,S} $ serves to define generalized TMDs (GTMDs) \cite{Meissner:2008ay, Meissner:2009ww,Lorce:2013pza},
which can be considered a natural extension of the concept of TMDs.
As will be discussed below, 
GTMDs are important for the definition of Orbital Angular Momentum (OAM) carried by partons.
As discussed in Sec.~\ref{sec:GTMD_observables}, the presently available information on GTMDs from experimental data is still extremely sparse, despite considerable progress towards identifying scattering processes from which GTMDs could, in principle, be extracted.

For a spin-$\frac{1}{2}$-target, the GTMD correlator in the quark sector in a helicity basis, i.e., in terms of longitudinal spin components $S^{\prime }_{L} $, $S_L $ can be written as~\cite{Meissner:2009ww}, cf. Eqs.~(\ref{e:wigner_op}) and (\ref{e:wigner_corr1}),
\begin{equation} 
\mathbb{W}_{S^{\prime }_{L} , S_L }^{q \, [\Gamma] \, W} (\bar{P}, \Delta, x, \kt) = 
\int \frac{db^- \, d^2 \bt}{2 (2\pi)^3} \, e^{-i k \cdot b} \, 
\langle p(P^{\prime } , S^{\prime }_{L} ) | \, \bar{\psi}_q(\tfrac{b}{2}) \, \Gamma \, W_{\sqsupset \eta}^{v} (\tfrac{b}{2},-\tfrac{b}{2} ) \, \psi_q(-\tfrac{b}{2}) \, | p(P, S_L ) \rangle \Big|_{b^+ = 0}
\label{e:gtmd_corr}
\end{equation}
with $q$ indicating the quark flavor, $\Gamma$ a generic gamma matrix, and the superscript $W$ the dependence on the choice of the Wilson line $W_{\sqsupset \eta}^{v} (b/2,-b/2)$.
Similar to the definition of GPDs, the matrix element is taken between states with, in general, different four-momenta and spins.
The correlator in Eq.~\eqref{e:gtmd_corr} can be parametrized in terms of Dirac bilinears multiplied by GTMDs, where at leading power $(\Gamma = \gamma^+, \; \gamma^+ \gamma_5, \; i\sigma^{+i} \gamma_{5} , i=1,2)$ a total of 16 quark GTMDs exist for a spin-$\frac{1}{2}$ hadron~\cite{Meissner:2009ww, Lorce:2013pza}.
As examples, we list the expressions for the vector and the axial-vector operators~\cite{Meissner:2009ww},\footnote{For ease of notation, in Eqs.~(\ref{e:gammap}),~(\ref{e:gammap5})~we suppress the arguments of the GTMD correlators and the GTMDs.}
\begin{eqnarray} 
\label{e:gammap}
\mathbb{W}_{S^{\prime }_{L} , S_L }^{q \, [\gamma^+] W} & = & \frac{1}{2M} \, \bar{u}(P^{\prime } ,S^{\prime }_{L} ) \bigg[ 
F_{1,1}^{qW} + \frac{i  \sigma^{i+}  k_T^i}{P^+} \, F_{1,2}^{qW} + \frac{i \, \sigma^{i+} \Delta_T^i}{P^+} \, F_{1,3}^{qW} 
+ \frac{i \sigma^{ij} k_T^i \Delta_T^j}{M^2} \, F_{1,4}^{qW}  \bigg] u(P,S_L ) \,, \nonumber \\
 \\
\mathbb{W}_{S^{\prime }_{L} ,S_L }^{q \, [\gamma^+ \gamma_5] W} & = & \frac{1}{2M} \, \bar{u}(P^{\prime } ,S^{\prime }_{L} ) \bigg[ 
- \frac{i \varepsilon^{ij} k_T^i \Delta_T^j}{M^2} \, G_{1,1}^{qW}
+ \frac{i  \sigma^{i+}  \gamma_5 k_T^i}{P^+} \, G_{1,2}^{qW} + \frac{i  \sigma^{i+}  \gamma_5 \Delta_T^i}{P^+} \, G_{1,3}^{qW}
\nonumber \\
&& \hspace{2.7cm} + \, i \sigma^{+-} \gamma_5 \, G_{1,4}^{qW}  \bigg] u(P,S_L ) \,,
\label{e:gammap5}
\end{eqnarray}
where the indices $i, j=1,2$ represent transverse components;  $F_{1,1}^{qW}, F_{1,2}^{qW}, F_{1,3}^{qW}, F_{1,4}^{qW}$ and $G_{1,1}^{qW}$, $G_{1,2}^{qW}$, $G_{1,3}^{qW}$, $G_{1,4}^{qW}$ are the quark GTMDs.
A generic GTMD $X(x,\kt, \xi, \deltat )$ depends on the (average) longitudinal $(x)$ and transverse $(\kt)$ parton momentum, as well as the longitudinal $(\xi)$ and transverse $(\deltat)$ momentum transfer to the target.  
We point out that, in general, GTMDs are complex functions, where the real part is invariant ($T$-even) under reversal of the staple direction,
$\eta \rightarrow -\eta $, whereas the imaginary part is $T$-odd, i.e., changes sign under reversal of the staple direction, cf.~also the corresponding discussion of $T$-even vs.~$T$-odd TMDs in Sec.~\ref{sec:qgspinTMDFF}. In Ref.~\cite{Meissner:2009ww}, 
the real and imaginary parts of the GTMD $X$ are correspondingly denoted as $X= X^e + i X^o$.
For a straight gauge link, $\eta =0$, the imaginary parts of the GTMDs vanish.
For gluons, 16 leading-power GTMDs also exist~\cite{Lorce:2013pza}.
Furthermore, the subleading quark and gluon GTMDs have been classified as well~\cite{Meissner:2009ww, Lorce:2013pza}.

As in the case of TMDs, the subtraction of a soft factor is required in Eq.~\eqref{e:gtmd_corr} for a proper definition of GTMDs, in extension of the detailed discussion in Ch.~\ref{sec:TMDdefn}.
In Ref.~\cite{Echevarria:2016mrc}, it was shown that the soft factor
\index{soft factor}
used for TMDs is also appropriate for GTMDs for $\xi = 0$, while the case of nonzero $\xi$ still needs to be explored.
Note also that, for brevity, we omitted two auxiliary scales in Eq.~\eqref{e:gtmd_corr} that are needed in QCD.
Studies of the scale dependence of GTMDs can be found in Refs.~\cite{Echevarria:2016mrc,Balitsky:2019ayf,Bertone:2022awq,Echevarria:2022ztg}.

The reductions of Wigner distributions exhibited in Eqs.~\eqref{e:Wigner_position} and~\eqref{e:Wigner_momentum}
have counterparts in momentum space, as a result of which
all GPDs and TMDs are projections of certain GTMDs.
Therefore, GTMDs, and Wigner distributions, can be considered partonic ``mother functions'', where it should be emphasized that only the GTMDs include a dependence on the skewness $\xi $ and therefore generate the full GPDs upon integrating out transverse momenta; by contrast, the Wigner distributions, because of their restriction to the 2D transverse spatial plane, only generate the GPDs evaluated specifically at $\xi =0$.

On the other hand, it should also be noted that various GTMDs (or Wigner distributions) disappear due to symmetry constraints when taking the GPD limit or the TMD limit of the respective correlator~\cite{Meissner:2008ay, Meissner:2009ww, Lorce:2013pza}. 
One important example is the GTMD $F_{1,4}^{qW}$, which is closely related to the orbital angular momentum of partons, as will be discussed further in Sec.~\ref{sec:GTMD_OAM}. 
Therefore, GTMDs (or Wigner distributions) contain considerably more information than their GPD and TMD projections alone.
They provide 6D (or 5D) images of hadrons, even though such images have to be interpreted with some care; see Sec.~\ref{sec:GTMD_models} below.
The relationship between the various quantities characterizing the (multi-dimensional) parton structure of hadrons is displayed in Fig.~\ref{f:multi_dim_imaging}.

The fact that all GPDs and TMDs are kinematical projections of GTMDs was used in Refs.~\cite{Meissner:2008ay, Meissner:2009ww} to explore possible non-trivial relations
between TMDs and GPDs that can be seen to hold in semi-classical approaches~\cite{Burkardt:2002ks} and in certain spectator model calculations~\cite{Burkardt:2003je, Lu:2006kt, Meissner:2007rx}. The relation between the quark Sivers function $f_{1T}^{\perp q}$ and the GPD $E^q$ is the best known example of such a connection~\cite{Burkardt:2002ks},
cf.~the discussion in Sec.~\ref{subsec:models-lensing-function}.
Several additional non-trivial relations can be identified as well~\cite{Meissner:2007rx}.
However, since the involved TMDs and GPDs appear as projections of different GTMDs, none of those
relations is model-independent~\cite{Meissner:2008ay, Meissner:2009ww}. Indeed, these
relations typically break down in more sophisticated model calculations~\cite{Meissner:2007rx, Pasquini:2019evu}.
On the other hand, such relations reveal some general qualitative features of certain TMDs and GPDs, and of observables in which those functions appear~\cite{Burkardt:2002ks}.

\subsection{Observables for GTMDs}
\label{sec:GTMD_observables}

After the first discussion of partonic Wigner distributions appeared in Refs.~\cite{Ji:2003ak, Belitsky:2003nz}, it took more than a decade until a scattering process was identified in which GTMDs could be measured directly~\cite{Hatta:2016dxp}; see also Ref.~\cite{Altinoluk:2015dpi}.
Specifically, in Ref.~\cite{Hatta:2016dxp} it was shown that gluon GTMDs at small $x$
\index{small-x region}
can be accessed through hard exclusive diffractive di-jet production in DIS,
\index{diffractive di-jet production}
a reaction which in the future should be measurable at the EIC.
\index{electron-ion collider (EIC)}
That work on observables for GTMDs was followed by several related studies~\cite{Zhou:2016rnt, Ji:2016jgn, Hatta:2016aoc, Hagiwara:2017ofm, Iancu:2017fzn, Hagiwara:2017fye, Boer:2018vdi, Mantysaari:2019csc, Salazar:2019ncp, Boussarie:2019vmk, Hagiwara:2021xkf}, all of which deal with gluon GTMDs, and all but one~\cite{Ji:2016jgn} focus on the small-$x$ region.
In the following we provide some details of the analysis presented in Ref.~\cite{Hatta:2016dxp} and briefly summarize what is presently known about observables for quark GTMDs.

To begin with, we note that in analogy to the quark Wigner distributions defined above, the gluon Wigner distributions are defined through the matrix element 
\begin{eqnarray}
x {\cal W}^{g} (x,\kt ;\rt )&=&\int\frac{db^-d^2\bt }{(2\pi)^3P^+}\int \frac{d^2\deltat }{(2\pi)^2} e^{-ixP^+ b^--i\kt \cdot \bt } 
\bigg \langle p(P^{\prime}) \bigg| G^{+i} \bigg( \rt +\frac{b}{2} \bigg)
\label{e:gluon_Wigner} \\
&&\times \;
W_{\sqsubset} \bigg( \rt +\frac{b}{2}, \rt -\frac{b}{2} \bigg) 
G^{+i} \bigg( \rt -\frac{b}{2} \bigg) W_{\sqsupset}^\dagger \bigg(\rt +\frac{b}{2}, \rt -\frac{b}{2} \bigg) \bigg| p(P) \bigg \rangle \bigg|_{b^{+} =0} \,,
\nonumber
\end{eqnarray}
where, as previously, $P^{\prime } = \bar{P} + \deltat /2$, $P=\bar{P} - \deltat /2$.
Here we consider unpolarized gluons and no target polarization.
$G^{\mu\nu}$ represents the gluon field strength tensor, $x$ and $\kt $ the (average) longitudinal momentum fraction and the transverse momentum for the gluon, respectively, and $\rt $ the coordinate space variable (gluon impact parameter).
As in the case of quarks discussed above, the Fourier transforms of gluon Wigner distributions w.r.t.~$\rt $ are GTMDs for gluons~\cite{Meissner:2009ww, Lorce:2013pza},
where in Eq.~(\ref{e:gluon_Wigner}) the gauge links associated with the gluon fields are such that one obtains the dipole gluon GTMD (see Ch.~\ref{sec:smallx} for more discussion) that is needed for the present purposes.

The dipole gluon GTMD correlator takes the form
\begin{eqnarray}
xG_{dip.}(x,\kt , \deltat )&=&2\int \frac{db^{-}
d^2 \bt}{\left(
2\pi \right) ^{3}P^{+}}
e^{-ixP^{+} b^{-} - i \kt \cdot \bt}
\label{gtd}\\
&&\times \left\langle  p(P^{\prime } )
\left|\text{Tr}\left[ G^{+i}\left(b/2\right)
W_{\sqsubset } (b/2,-b/2)
G^{+i}\left( -b/2\right) W_{\sqsupset }^\dagger(b/2,-b/2)
\right] \right| p(P)
\right\rangle\,, 
\nonumber
\end{eqnarray}
which, following the derivations in Ch.~\ref{sec:smallx}, in the small-$x$ region reduces to
\begin{eqnarray}
xG_{dip.}(x,\kt , \deltat )&=&\frac{2N_{c}}{\alpha_s}\int
\frac{d^{2}{\bf R}_T d^{2}{\bf R}_{T}^{\prime }}{(2\pi)^4}e^{i\kt \cdot \left(
{\bf R}_T -{\bf R}_{T}^{\prime }\right) +i \frac{\deltat }{2}\cdot({\bf R}_T +{\bf R}_{T}^{\prime } )} \nonumber\\
&&\times\left(\bm{\nabla}_{{\bf R}_T }\cdot
\bm{\nabla}_{{\bf R}_{T}^{\prime }}\right)\frac{1}{N_{c}}\left\langle\text{Tr}\left[
U\left( {\bf R}_{T}\right) U^{\dagger }\left( {\bf R}_{T}^{\prime }\right)
\right]\right\rangle_x \,. 
\label{op}
\end{eqnarray}
In Eq.~\eqref{op} we have used the Wilson line $U({\bf R}_T) = W_n({\bf R}_T, - \infty, + \infty)$, while the subscript $x$ indicates the momentum fraction of the gluon at which the matrix element is evaluated.
(For more information about the averaging procedure in the small-$x$ CGC formalism, indicated by $\langle \ldots \rangle_x$, we refer to the paragraph after Eq.~\eqref{GWW1}.)
The last factor in Eq.~\eqref{op} is the well-known impact-parameter-dependent dipole amplitude. 
Defining its double Fourier transform through
\begin{equation}
\frac{1}{N_{c}}\text{Tr}\left[
U\left( \rt +\frac{\bt }{2}\right) U^{\dagger }\left( \rt -\frac{\bt }{2} \right)\right] \equiv \int d^2 \kt  d^2 \deltat e^{-i\kt \cdot \bt -i \deltat \cdot \rt } \mathcal{F}_x(\kt , \deltat )\,,\label{g2}
\end{equation}
allows us to write
\begin{equation}
xG_{dip.}(x,\kt , \deltat )=(\kt^{2} -\deltat^{2}/4)\frac{2N_c}{\alpha_s}  \mathcal{F}_x(\kt , \deltat ) \,.
\end{equation}

\begin{figure}[t!]
\centering
\includegraphics[width=0.44\textwidth]{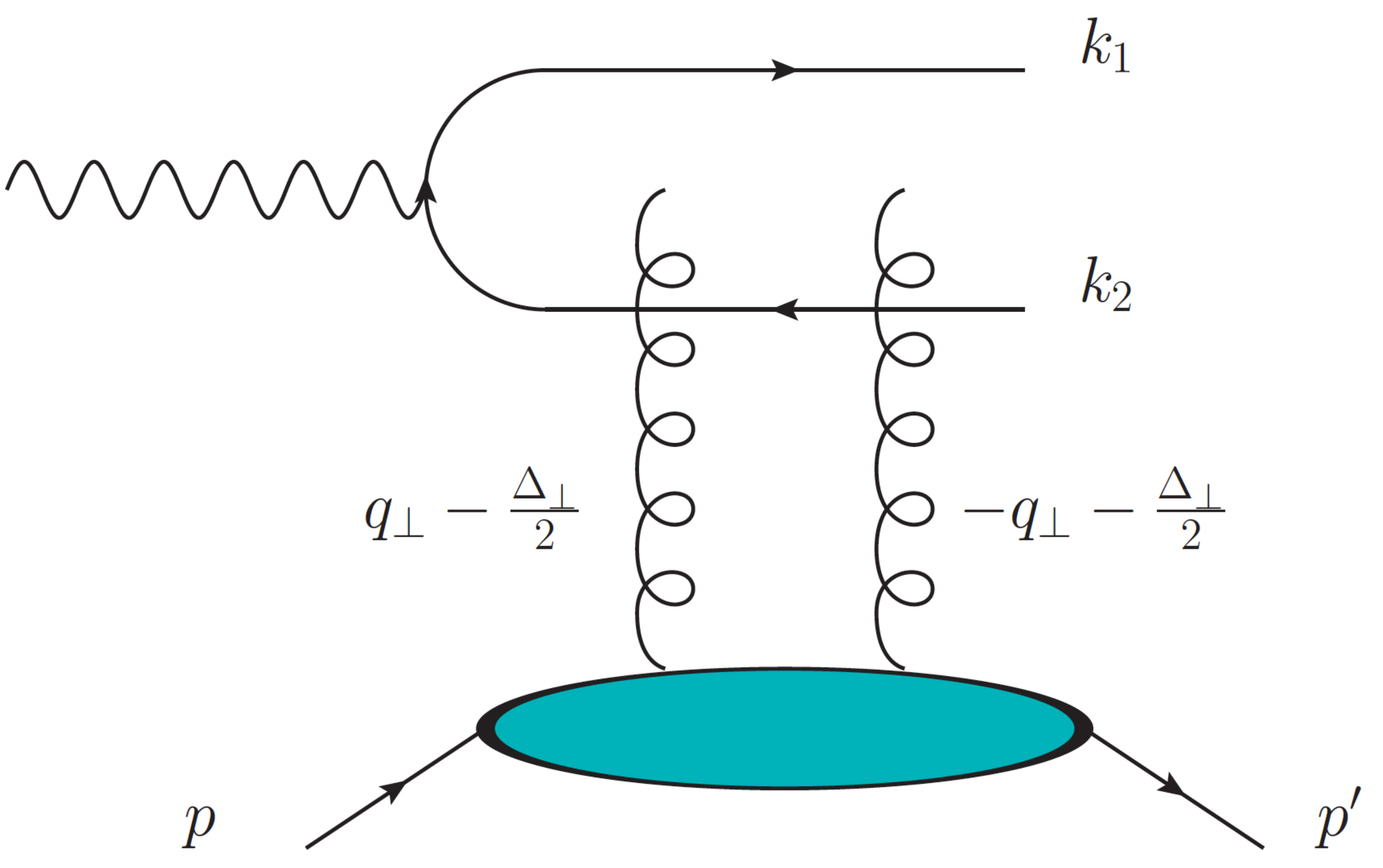} 
\hspace{1.0cm}
\includegraphics[width=0.44\textwidth]{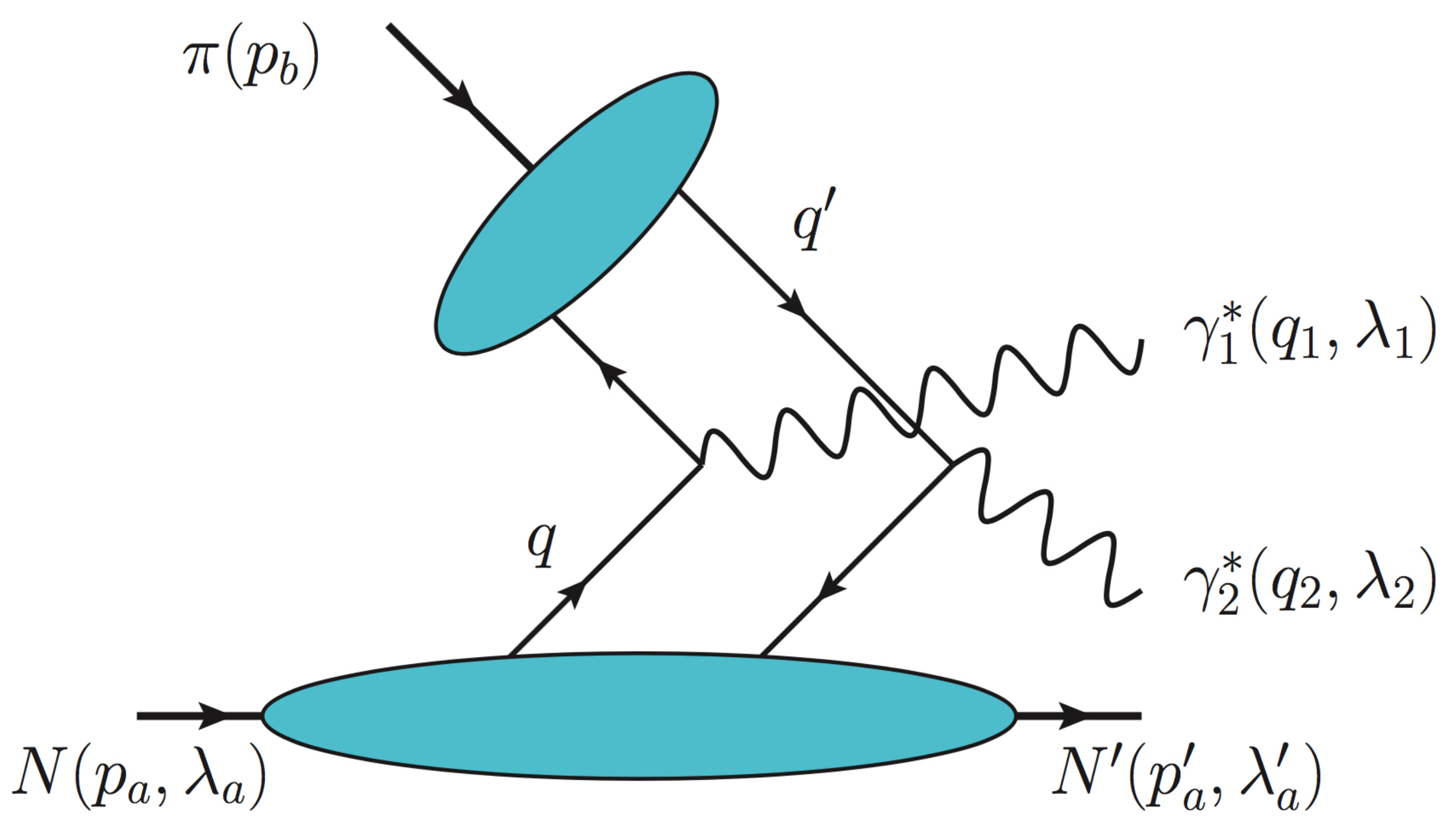}
\caption{Processes that can provide information on GTMDs. 
Left: Sketch of exclusive diffractive di-jet production which is sensitive to gluon GTMDs; figure from Ref.~\cite{Hatta:2016dxp}.
(The conventions for the momenta in the figure differ from the text. In particular, the 4-momenta of the produced quark and antiquark are $q_1 = k_1|_{\rm fig.}$ and $q_2 = k_2|_{\rm fig.}$, respectively. Moreover, the average transverse gluon momentum is $\kt = {\bf q}_\perp|_{\rm fig.}$.)
Right: One of two lowest-order diagrams for the exclusive pion-nucleon double Drell-Yan process which is sensitive to quark GTMDs; figure from Ref.~\cite{Bhattacharya:2017bvs}.
}
\label{f:wigner_process}
\end{figure}

It is $\mathcal{F}_x(\kt , \deltat )$ which shows up in the cross section for diffractive di-jet production in electron-ion collisions. Such processes probe the dipole gluon GTMD in the small-$x$ limit where the quark contribution is negligible~\cite{Hatta:2016dxp}. 
For the calculation of the cross section, one requires that the final-state quark-antiquark pair (see Fig.~\ref{f:wigner_process}~(left)) forms a color singlet state, leading to 
\begin{eqnarray}
\frac{d\sigma ^{\gamma_{T}^{\ast }A\rightarrow q\bar{q}X}}{dy_1d^2{\bf q}_{1T}dy_2d^2{\bf q}_{2T}}
&=&2N_{c}\alpha _{em}e_{q}^{2}\delta(x_{\gamma^*}-1) z(1-z) [z^2+(1-z)^2] 
\notag \\
&& \times \, \int d^2 \kt d^2 \kt^{\prime }  \mathcal{F}_x(\kt , \deltat )\mathcal{F}_x(\kt^{\prime } , \deltat ) 
\notag \\
&&\times \, \bigg[  \frac{{\bf P}_T}{{\bf P}_T^2 +\epsilon_f^2 } - \frac{{\bf P}_T-\kt }{({\bf P}_T -\kt )^2 +\epsilon_f^2} \bigg] \cdot  \bigg[  \frac{{\bf P}_T}{{\bf P}_T^2 +\epsilon_f^2 }  - \frac{{\bf P}_T -\kt^{\prime }}{({\bf P}_T -\kt^{\prime } )^2 +\epsilon_f^2 } \bigg] \,,   
\label{dct}
\phantom{aaaaa}
\end{eqnarray}
for transversely polarized photons, where $A$ indicates the target which can be a proton or any nucleus.
In Eq.~\eqref{dct}, $x_{\gamma^*} = z_q + z_{\bar{q}}$ with $z_q = z$ and $z_{\bar{q}} = 1 - z$ the momentum fractions of the virtual photon carried by the quark and antiquark, respectively.
Furthermore, $y_{1,2}$ and ${\bf q}_{1,2T}$ are the rapidities and transverse momenta of the quark and antiquark jets, respectively,  defined in the center of mass frame of the incoming photon and target, whereas ${\bf P}_T \equiv \frac{1}{2} ({\bf q}_{2T}-{\bf q}_{1T})$ represents the typical di-jet transverse momentum, and $\epsilon_f^2\equiv z(1-z)Q^2$. 
We are interested in the back-to-back kinematic region for the two jets where $|{\bf P}_T | \gg |{\bf q}_{1T}+{\bf q}_{2T}|$. 
Suppose that $\epsilon_f^2$ is not too large as compared to ${\bf P}_T^2$, then we expect that the $\kt $ integrals in Eq.~\eqref{dct} are dominated by the region $\kt \sim {\bf P}_T $ and the cross section is roughly  proportional to $\mathcal{F}_x^2({\bf P}_T , \deltat )$.
Thus, the diffractive di-jet production will be sensitive to the correlations between ${\bf P}_T$ and $\deltat $. 
Of particular interest is the angular correlation of the form $\cos2(\phi_{P_T }-\phi_{\Delta_{T} })$, which originates from the $\cos2\phi$ correlation in the GTMD and the Wigner distribution. 
With the detector capability at the EIC~\cite{Accardi:2012qut, AbdulKhalek:2021gbh}, we will be able to identify both ${\bf P}_T$ and $\deltat$ and measure the angular correlation between them.

Later on, the same process was considered in the small-$x$ region for a (longitudinally) polarized nucleon~\cite{Hatta:2016aoc} --- see also the recent update discussed in Ref.~\cite{Bhattacharya:2022vvo}.
The specific interest of that work was the gluon GTMD $F_{1,4}^g$, which is directly related to the orbital angular momentum of gluons,
\index{orbital angular momentum (OAM)}
as discussed in detail in the following section.
Interestingly, in Ref.~\cite{Ji:2016jgn} it was argued that di-jet production could even be used to address $F_{1,4}^g$ and, therefore, the gluon orbital angular momentum at moderate $x$.

In Refs.~\cite{Mantysaari:2019csc,Salazar:2019ncp}, the small-$x$ JIMWLK evolution effects for the dipole scattering amplitude and the associated diffractive di-jet production have been investigated. It was found that the elliptic angular correlation is sensitive to the small-$x$ evolution effects. This leads to an interesting probe for the small-$x$ physics at the future EIC.

In a very interesting recent work, a small-$x$ model for gluon GTMDs was fitted to HERA data on diffractive di-jet production in electron-proton collisions~\cite{Boer:2021upt}. 
The data were described well with a small number of fit parameters, and predictions were made for both photo-production and electro-production at the EIC.

GTMDs for gluons can also play an important role in exclusive $\pi^0$ production at high energies~\cite{Boussarie:2019vmk}.
It was shown that this process is related to a particular gluon GTMD, which in the forward limit reduces to the gluon Sivers function $f_{1T}^{\perp g}$.
In turn, at small $x$, the latter is intimately related to the QCD odderon~\cite{Zhou:2013gsa}.

Let us finally discuss potential observables for quark GTMDs. 
Presently, the only known process that is sensitive to quark GTMDs is the exclusive pion-nucleon double Drell-Yan reaction, $\pi N \to (\ell_1^- \ell_1^+) (\ell_2^- \ell_2^+) N'$~\cite{Bhattacharya:2017bvs}; see Fig.~\ref{f:wigner_process} (right). Using SCET, it has been argued recently that this process factorizes when going beyond the parton model approximation~\cite{Echevarria:2022ztg}.
(Note that for the exclusive nucleon-nucleon double Drell-Yan process, spectator-spectator interactions enter which pose a challenge for factorization.)
At leading order, the process would allow one to probe the Efremov-Radyushkin-Brodsky-Lepage region~\cite{Efremov:1979qk, Lepage:1979zb} of GTMDs which is characterized by $-\xi < x < \xi$, while the DGLAP region $(x \le - \xi \; {\rm or} \; x \ge \xi)$ is not accessible~\cite{Bhattacharya:2017bvs}. 
Through the double Drell-Yan process, in principle, all leading-power quark GTMDs could be addressed by making use of suitable polarization observables~\cite{Bhattacharya:2017bvs}.
However, the count rate for this reaction is very small since the cross section is proportional to $\alpha_{\rm em}^4$.
Furthermore, higher-order corrections for this process need to be computed for a thorough test of factorization. 
Closely related work~\cite{Bhattacharya:2018lgm, Boussarie:2018zwg} deals with addressing gluon GTMDs through double production of charge-parity even quarkonia such as the $\eta_c$ and $\eta_b$.
While those reactions have sufficiently large count rates and may allow  study of gluon GTMDs at moderate $x$~\cite{Bhattacharya:2018lgm}, detecting charge-parity even quarkonia is very challenging.

\subsection{Connection with Orbital Angular Momentum of Partons} 
\label{sec:GTMD_OAM}
\index{orbital angular momentum (OAM)}

A central topic in the investigation of hadron structure is the decomposition of the spin of the proton into the spins and OAM of its quark and gluon constituents. As will be seen in the course of the following discussion, an important application of GTMDs is the study of the OAM component.

The starting point for the discussion of angular momentum is its definition through the QCD energy momentum tensor (EMT). As already introduced in Sec.~\ref{decom_spin}, both the proton's total momentum as well as its angular momentum are encoded in the matrix elements of the EMT between proton helicity states. The latter are parametrized in terms of form factors which are functions of the four-momentum transfer squared, $t$, between the initial and the final proton \cite{Ji:1996ek}.
In 1996, Ji \cite{Ji:1996ek} made the key observation that the form factors of the EMT can be accessed experimentally, since they coincide, through the operator product expansion (OPE), with the expressions for the second Mellin moments of certain GPDs.
This led to the definition of the Ji sum rule that provides an initial decomposition of the proton spin into the quark and gluon total angular momenta, cf.~also the discussion in Sec.~\ref{decom_spin}, while furthermore expressing these contributions in terms of GPDs,
\begin{eqnarray}
J^{q (Ji)} +J^{g (Ji)} &=& \frac{1}{2} \int dx\, x \left(H^q (x,0,0) + E^q (x,0,0) \right) + \frac{1}{2} \int dx\, \left( H^g (x,0,0) + E^g (x,0,0) \right) \nonumber \\
&=& \frac{1}{2} \ .
\label{eq:ji_decomp_h_e}
\end{eqnarray}
Here, $H^{q,g}$ and $E^{q,g}$ are GPDs corresponding to different quark/gluon-proton helicity configurations, as already introduced in the discussion after Eq.~(\ref{eq:GPDcorr_def}).
Remarkably, the first Mellin moment of $H^q +E^q $ is given by the magnetic form factor, $G_M^q =F_1^q +F_2^q $, thus uncovering an interesting connection between partonic angular momentum and the magnetization density of the nucleon. Moreover, $F_1^q $ and $F_2^q $ are comparatively well-determined from experiment, which can be useful in constraining phenomenological models for $H^q $ and $E^q $. A similar relation is found for partonic angular momentum in a spin one target, {\it e.g.}, the deuteron \cite{Taneja:2011sy}.

Whereas Eq.~(\ref{eq:ji_decomp_h_e}) provides an initial decomposition of the proton spin into $J^q $ and $J^g $, a full analysis of proton spin
requires, on the one hand, to further identify the respective operators for parton spin and orbital angular momentum (OAM), and on the other, to give a physical interpretation of the components of the sum rule while simultaneously preserving the gauge invariance of the theory.
We recall, in what follows, the two main frameworks which have been adopted for the  decomposition of the total quark and gluon angular momenta, $J^{q}$ and $J^g$, into 
their respective spin and orbital components (a discussion related to the evaluation of these terms in LQCD was already given in Chapter \ref{sec:lattice}).

Before proceeding, note that the focus of most studies and measurements to date has been on the longitudinal/helicity components. In that case, as will be discussed in more detail below, the quark longitudinal OAM can be obtained directly from an appropriate quark-quark correlation function \cite{Lorce:2011kd}, leading to the evaluation of a moment in both $x$ and transverse momentum, $\kt $, of the GTMD $F_{1,4}^{qW} $ \cite{Meissner:2009ww}. By contrast, transverse angular momentum is more subtle, since transverse boosts are dynamical and the definition of OAM depends specifically on the point about which it is evaluated. Furthermore, transverse spin is represented by a twist three structure function, $g_T$, with a non-trivial $qgq$ structure (cf.~also the discussion in Chap.~\ref{sec:models}). For these reasons, transverse angular momentum is still an intensely debated subject as of this writing. For ongoing studies and literature on the subject we refer the reader to \cite{Ji:2020ena,Ji:2021znw,Guo:2021aik,Rajan3,Liu:2015xha} and references therein. 
 
Restricting our discussion to longitudinal angular momentum (taken along the $z$ axis), there is, on the one hand, the Ji decomposition \cite{Ji:1996ek},
\begin{equation}
\frac{1}{2}  \Delta \Sigma + L_z^{q (Ji)} +  J^{g (Ji)}_z = \frac{1}{2},
\label{eq:Jidecomp}
\end{equation}
while, on the other, the Jaffe and Manohar (JM) decomposition \cite{Jaffe:1989jz} reads,
\begin{equation}
\frac{1}{2} \Delta \Sigma + L^{q (JM)}_z + \Delta G + L^{g(JM)}_z = \frac{1}{2}
\label{eq:JMdecomp}
\end{equation}
(see also Eqs.\eqref{eq:JM} and \eqref{eq:Ji} and discussion in Chap.~\ref{sec:lattice}).
Various other pictures have been given in the literature that can be seen as variations of the two main frameworks represented by Eqs.~(\ref{eq:Jidecomp}) and (\ref{eq:JMdecomp}). In Ref.~\cite{Wakamatsu:2000fd}, for instance, $L^{q (Ji)}$ includes a potential term attributed to gluon angular momentum.
On the other side, in Ref.~\cite{Bashinsky:1998if}, a gauge invariant extension of Eq.~(\ref{eq:JMdecomp}) was proposed that led to several further developments. For reviews of the various decompositions we refer the reader to Refs.~\cite{Leader:2013jra,Ji:2012gc,Wakamatsu:2011mb}.

It should be stressed that, in order to validate the angular momentum decomposition in the quark sector, one needs three separate evaluations or experimental observations, for $J_z^{q (Ji)}$, $L_z^{q (Ji)}$ and $\Delta \Sigma$/2. While the LQCD calculations described in Chap.~\ref{sec:lattice} accessed $J_z^{q,g (Ji)}$ and spin, the orbital component, $L_z^{q (Ji)} $, was only obtained indirectly by subtraction. A direct quantification of the OAM contributions has been more elusive. This is the point where Wigner functions and GTMDs, as well as twist-3 GPDs connected to them, can provide new insights; they provide direct definitions of the OAM contributions and tie them to partonic distribution functions.

The key observation that has led to a more thorough understanding of the quark longitudinal OAM contribution, $L_z^q $, is that the corresponding correlation between coordinate space and momentum components can be written in terms of Wigner distributions, as initially alluded to in \cite{Belitsky:2003nz} and discussed explicitly in detail in Ref.~\cite{Lorce:2011kd}. There, it was furthermore observed that the specific GTMDs obtained by Fourier transforming the appropriate Wigner distributions could be identified with the functions given in the general parametrization of correlation functions introduced in Ref.~\cite{Meissner:2009ww}.
Specifically, longitudinal OAM in the quark sector is identified with a Wigner distribution weighted by the cross product of position and momentum in the transverse plane, $\rt \times \kt $ 
\cite{Belitsky:2003nz,Lorce:2011kd,Lorce:2011ni},
\begin{eqnarray}
\label{eq:WignerOAM}
L_z^{q,W} &=& \int dx \int d^2 \kt \int d^2 \rt \ (\rt \times \kt )_z \ \frac{1}{2} \left( {\cal W}^{q [\gamma^{+} ] \, W}_{++} - {\cal W}^{q [\gamma^{+} ] \, W}_{--} \right) \ .
\end{eqnarray}
The fact that both $\rt $ and $\kt $ lie in the transverse plane with respect to the proton momentum renders the longitudinal component of OAM more straightforward than the transverse one. As shown in \cite{Soper:1976jc}, the transverse plane is invariant under longitudinal boosts, i.e., transformations in the transverse plane can be effectively described by the Galilean group in 2D.

By a 2D Fourier transform in $\rt $, Eq.~(\ref{eq:WignerOAM}) is related to the corresponding GTMD description \cite{Lorce:2011kd,Lorce:2011ni,Hatta:2011ku},
\begin{eqnarray}
L_z^{q,W}  &=& \left. \int dx \int d^2 \kt \left({\bf k}_T \times i \frac{\partial}{ \partial {\bf \Delta}_T}\right)_z \ \frac{1}{2} \left( \mathbb{W}_{++}^{q [\gamma^{+}]\, W} - \mathbb{W}_{--}^{q [\gamma^{+}]\, W } \right) \right|_{\Delta =0} \\ &=& \left. -\int dx \int d^2 \kt  \frac{\kt^{2} }{M^2} F_{1,4}^{qW} \right|_{\Delta =0} ,
\label{eq:F14OAM}
\end{eqnarray}
where $\mathbb{W}_{\lambda\lambda'}^{[\Gamma]}$ was defined in Eq.~\eqref{e:gammap}. Note that the distributions are evaluated in the forward limit; below, this specification will be omitted for conciseness of notation. Eq.~(\ref{eq:WignerOAM}) on the one hand provides a very intuitive definition of OAM while, on the other, it corresponds to a specific parton helicity configuration which can be evaluated on the lattice (Secs.~\ref{sec:latt_def_lorentz}, \ref{sec:GTMD_OAM_lattice}), and, in principle, measured in experiments (Sec.~\ref{sec:GTMD_observables}).
Ref.~\cite{Burkardt:2012sd} discusses how, in analogy with the Sivers function, the $\kt^2$ moment of $F_{1,4}^{qW} $ depends on the gauge link structure $W$ entering its evaluation: in particular, for a straight link one obtains the OAM term entering Ji's definition, as already observed in Ref.~\cite{Ji:2012sj}, while a staple link yields the JM definition, cf.~\cite{Hatta:2011ku}. In other words, Ji's picture gives the intrinsic quark angular momentum, independent from spectator interactions, while the JM picture includes interactions with the spectators.    
The representation of OAM through \eq{F14OAM} can also be used to obtain an intuitive semi-classical interpretation of the difference $L_z^{q(JM)} - L_z^{q(Ji)}$ in terms of the torque acting on the active quark due to its interaction with the spectator partons of the target~\cite{Burkardt:2012sd}. 
Formally, the additional term is equivalent to a Qiu-Sterman type term for a longitudinally polarized proton \cite{Hatta:2011ku,Raja:2017xlo}.

The expression of OAM in terms of a Wigner distribution, Eq.~(\ref{eq:WignerOAM}), furthermore suggests defining the $x$-integrand in Eq.~(\ref{eq:WignerOAM}) to represent the corresponding partonic OAM density $L_z^{q,W} (x)$. This point of view will be taken in the further discussion to follow, with the caveat that, in general, $L_z^{q,W} (x)$ cannot be interpreted too literally as an average over the partonic OAM, $\rt \times \kt $, of bare quark partons, since the definition of the Wigner distribution includes the gauge link $W$. Therefore, it encodes a distribution of composite quark-gluon fields rather than bare quark partons. The light-cone gauge form of the Jaffe-Manohar definition of OAM, $L_z^{q (JM)} $, suggests that, in that case, a density interpretation can be justified, although the sensitivity of fields in the light-cone gauge to boundary conditions at infinity needs to be kept in mind.

Further insight into the decomposition of angular momentum was obtained in Refs.~\cite{Raja:2017xlo,Rajan:2016tlg}, by connecting the GTMDs representing OAM to twist-3 GPDs via
generalized Lorentz invariance relations (LIRs), similar to the ones introduced in Chap.~\ref{sec:twist3}, but involving off-forward proton states with $P\neq P'$.  
The following relation was derived for $F_{1,4}^{qW} $, 
\begin{equation}
\label{eq:LIR3_alt}
L_z^{q,W} (x) \equiv - \int d^2 \kt  \, \frac{ \kt^2}{M^2} \, F_{1,4}^{qW} =
 \int_x^1 dy \, \left( \widetilde{E}_{2T}^q + H^q + E^q + {\cal A}_{F_{14}} \right)
\end{equation}
where, following the notation of Ref.~\cite{Meissner:2009ww}, the {\it rhs} involves the twist-2 GPD combination $H^q+E^q$,  and  introduces the twist-3 GPD $\widetilde{E}_{2T}^{q} $. Furthermore,  ${\cal A}_{F_{14}}$ is an explicit $qgq$ term containing the dependence of the equation on the gauge link structure, the form of which is given in detail further below. All expressions are given in the forward limit $(\xi,t) \rightarrow 0$, similar to the integrands in the sum rule in Eq.~(\ref{eq:ji_decomp_h_e}), although these relations are valid point by point in the kinematic variables $x$ and $t$ and can be easily extended to the $\xi \neq 0$ case \cite{Rajan3}.
Note that the combination $\widetilde{E}_{2T}^q + H^q + E^q$ is expressed
in a format reminiscent of the one in the original Wandzura-Wilczek (WW) relation, where the twist-3 PDF, $g_T$, was decomposed into a twist-2 PDF, $g_1$, and a twist-3 PDF, $g_2$, as: $g_T = g_1 + g_2$ \cite{Wandzura:1977qf,Mulders:1996dh}. 
\index{Wandzura-Wilczek (type) approximation}
In the off-forward case considered here, we have the decomposition into the twist-2 combination, $(H^q + E^q) \leftrightarrow g_1$, and the twist-3 GPD, $\widetilde{E}_{2T}^q \leftrightarrow g_T$.

Eq.~(\ref{eq:LIR3_alt}) establishes a relation between the $\kt^2$ moment of $F_{1,4}^{qW} $, representing quark OAM, and the twist-3 GPD $\widetilde{E}_{2T}^{q} $.
The latter had been previously connected to Ji OAM in Refs.~\cite{Kiptily:2002nx,Hatta:2012cs}, within a derivation using OPE \cite{Balitsky:1987bk,Kivel:2000rb}, showing that (minus) the second Mellin moment of the twist-3 GPD $G_2 $ defined in Ref.~\cite{Kiptily:2002nx} yields $J^{q (Ji)}_z - (1/2) \Sigma \equiv L^{q (Ji)}_z $; the second Mellin moments of $G_2 $ and $\widetilde{E}_{2T}^{q} $ are related as
$\int dx \, x G_2 = -\int dx \, x (\widetilde{E}_{2T}^q + H^q + E^q)$.

Using the QCD equations of motion, it was further shown in Refs.~\cite{Rajan:2016tlg,Raja:2017xlo} that the Ji decomposition of longitudinal angular momentum for a proton target in the quark sector can be written both in terms of the GTMD $F_{1,4}^{qW} $ (with, specifically, a straight gauge link $W$) and the twist-3 GPD $\widetilde{E}_{2T}^q$ as,
\begin{eqnarray}
J_z^{q (Ji)} \hspace{1cm} &=& \hspace{2cm} L_z^{q (Ji)} \hspace{1.95cm}
+ \hspace{1cm} S_z^q \nonumber \\
\frac{1}{2} \int dx\, x \, (H^q+E^q) &=& \int dx\, x (\widetilde{E}_{2T}^q +H^q+E^q)
\ \ \ \ \ \ + \ \ \ \ \ \ \frac{1}{2} \int dx\, \widetilde{H}^q
\label{eq:gpddecomp} \\
&=& \hspace{0.cm} -\int dx\, \int d^2 \kt \, \frac{\kt^2}{M^2} F_{1,4}^{qW} \ \ \ \  + \ \ \ \
\frac{1}{2} \int dx\, \widetilde{H}^q.
\label{eq:gtmddecomp}
\end{eqnarray}
The work in Refs.~\cite{Raja:2017xlo,Rajan:2016tlg} describes the connection between the two descriptions while introducing the following generalized WW relation for $\widetilde{E}_{2T}^q$,  obtained by extending to the off-forward case the set of QCD relations involving transverse momentum first introduced in  Refs.~\cite{Mulders:1995dh,Tangerman:1994bb}, and using the equations of motion,
\begin{eqnarray}
\label{F14_WW1}
\widetilde{E}_{2T}^q(x) = \hspace{-0.5cm}
&& - \int_x^1 \frac{dy}{y}(H^q(y) + E^q(y)) - \left[ \frac{\widetilde{H}^q(x)}{x} -\int_x^1 \frac{dy}{y^2} \widetilde{H}^q(y)\right]  - \left[ \frac{1}{x}\mathcal{M}_{F_{14} }(x) - \int_x^1 \frac{dy}{y^2} \mathcal{M}_{F_{14} }(y)  \right] \nonumber \\
&& - \int_x^1 \frac{dy}{y} {\cal A}_{F_{14} }(y) 
\end{eqnarray}
In Eq.\eqref{F14_WW1}, $\mathcal{M}_{F_{14}}$ is the 
$qgq$ interaction term stemming from the gauge field $A$ in the covariant derivative in the equations of motion, whereas ${\cal A}_{F_{14}}$ parametrizes the dependence of the Lorentz invariance relation on the gauge link. For a straight link one has ${\cal A}_{F_{14}} = 0$. 
For a staple link, it can be written in terms of unintegrated in $k^-$ invariant amplitudes \cite{Meissner:2009ww} as \cite{Raja:2017xlo},  
\begin{eqnarray}
{\cal A}_{F_{14} } &=& v^{-} \frac{(2P^{+} )^2 }{M^2 } \int d^2 \kt \int dk^{-} \nonumber \\
&& \times \left[\frac{\kt \cdot \deltat }{\deltat^{2} } (A_{11}^F +xA_{12}^F )
+A_{14}^F + \frac{\kt^2 \deltat^{2} - (\kt \cdot \deltat )^{2} }{\deltat^{2} } \left(
\frac{\partial A_8^F }{\partial (k\cdot v)} +
x\frac{\partial A_9^F }{\partial (k\cdot v)} \right) \right] \nonumber \\
\label{eq:lir_staple}
\end{eqnarray}
where the 4-vector
$v=(0,v^{-} ,0,0)$ describes the direction of the staple, which here is taken to extend along the light cone. The completely unintegrated invariant amplitudes named $A^F_i$ depend on all possible scalar products of the vectors $P, k, \Delta$, and $v$, and were introduced already in initial studies of TMDs \cite{Mulders:1996dh}. These amplitudes are essential for formulating the LIR that identifies partonic OAM with a twist-3 GPD, since they define common structures underlying both twist-2 and twist-3 distributions. A detailed description for both the staple and straight link cases is given in Refs.~\cite{Meissner:2009ww,Raja:2017xlo}.
The relation between Ji and JM OAM using a straight and a staple link, respectively, can be evaluated through Eq.~\eqref{eq:lir_staple}. An explicitly calculable form of the difference between the two definitions was obtained in Ref.~\cite{Raja:2017xlo} in terms of ${\cal M}_{F_{14}}$ or ${\cal A}_{F_{14}}$ as,  
\begin{equation}
\label{eq:keypoint}
 L_z^{q (JM)}(x) - L_z^{q (Ji)}(x) = \mathcal{M}_{F_{14}}(x) - \mathcal{M}_{F_{14}}(x)\Big|_{v=0} =
    - \int_x^1 dy \, \mathcal{A}_{F_{14}}(y). 
\end{equation}
It is instructive to point out the similarity with the composition of the intrinsic $qgq$ contribution to the twist-3 structure function $g_T$ \cite{Accardi:2009au}. 
Two different contributions were singled out in Ref.~\cite{Accardi:2009au}, named $\hat{g}_T$ and $\tilde{g}_T$.
These play similar roles to  ${\cal M}_{F_{14}}$ and ${\cal A}_{F_{14}}$ for OAM. The immediate consequence of there being two separate $qgq$ terms in Eq.~(\ref{F14_WW1}) is that 
 an experimental determination of the difference between JM and Ji OAM, described by ${\cal A}_{F_{14} } $, is possible only by measuring separately the GPDs $\widetilde{E}_{2T}^{q} $, $H^q $ and $E^q $, and the GTMD $F_{1,4}^{qW} $ for a staple-shaped gauge link $W$. 
 
GTMDs and Wigner distributions are also fundamental for investigating spin-orbit correlations. 
\index{spin-orbit correlation}
The correlation between the quark longitudinal spin and OAM, denoted by $C_z^q$ in Ref.~\cite{Lorce:2011kd}, can be computed according to Refs.~\cite{Lorce:2011kd, Lorce:2014mxa} as
\begin{align}
\label{e:q_SO_corr}
\big \langle C_z^{q,W} \big \rangle & = \int dx \int d^2 \kt \int d^2 \rt \, \big( \rt \times \kt \big)_z \, {\cal W}_{++}^{q \, [\gamma^+ \gamma_5] \, W}(x, \kt, \rt) 
\nonumber \\
& = \int dx \int d^2 \kt  \, \frac{\kt^2}{M^2} \, G_{1,1}^{qW}(x, \kt, \xi, \deltat) \, \big|_{\Delta = 0} \,.
\end{align}
In Ref.~\cite{Raja:2017xlo}, cf.~also Ref.~\cite{Lorce:2014mxa}, a relation analogous to the one for OAM in Eq.~(\ref{eq:LIR3_alt}) was obtained for $C_z^q$, namely,
\begin{eqnarray}
\big \langle C_z^{q,W} \big \rangle (x) \equiv \int d^2\kt \, \frac{\kt^2}{M^2} \, G_{1,1}^{qW}  & = & \int_x^1 dy \, \left(2\widetilde{H}_{2T}^{\prime \, q} + E_{2T}^{\prime \, q}
  + \widetilde{H}^{q} - {\cal A}_{G_{11} } \right) ,
  \label{eq:pres_so} 
  \end{eqnarray}
where ${\cal A}_{G_{11} }$ is a gauge link term that can be evaluated similarly to the OAM one in Eq.~(\ref{eq:lir_staple}).

Finally, also the OAM component in the proton transverse spin decomposition \cite{Burkardt:2005hp,Guo:2021aik,Ji:2012vj,Rajan3} can be identified by deriving analogous relations involving transverse polarization. 
Additional spin-orbit correlations can be identified when considering transverse polarization effects~\cite{Bhoonah:2017olu}.

In conclusion, while $J^{q,g}$ and OAM measurements through collinear GPDs are feasible, 
GTMDs, providing in principle the density distributions for OAM, remain experimentally more difficult to extract \cite{Bhattacharya:2017bvs,Hagiwara:2016kam,Liuti:2017uxp}.  
On the other hand, twist-2 and twist-3 GPDs and GTMDs can be evaluated in ab initio calculations \cite{Engelhardt:2020qtg}, as illustrated in the following section.

\subsection{GTMD observables from LQCD: Quark orbital angular momentum in the proton}
\label{sec:GTMD_OAM_lattice}
\begin{figure}[t!]
\includegraphics[width=8.4cm]{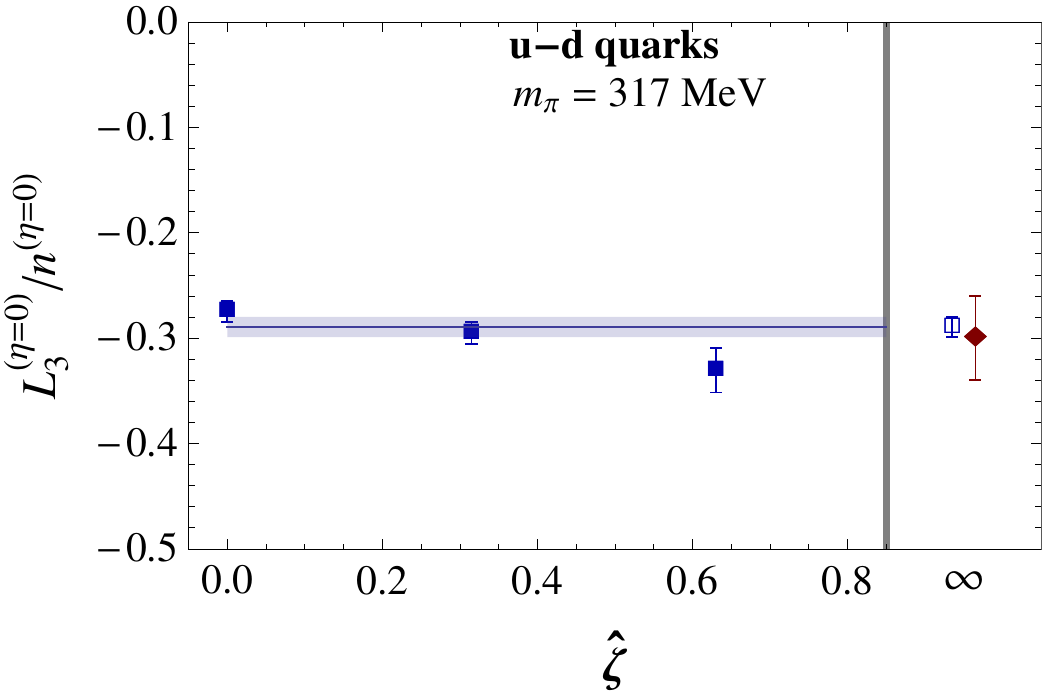}
\hspace{0.3cm}
\includegraphics[width=8.4cm]{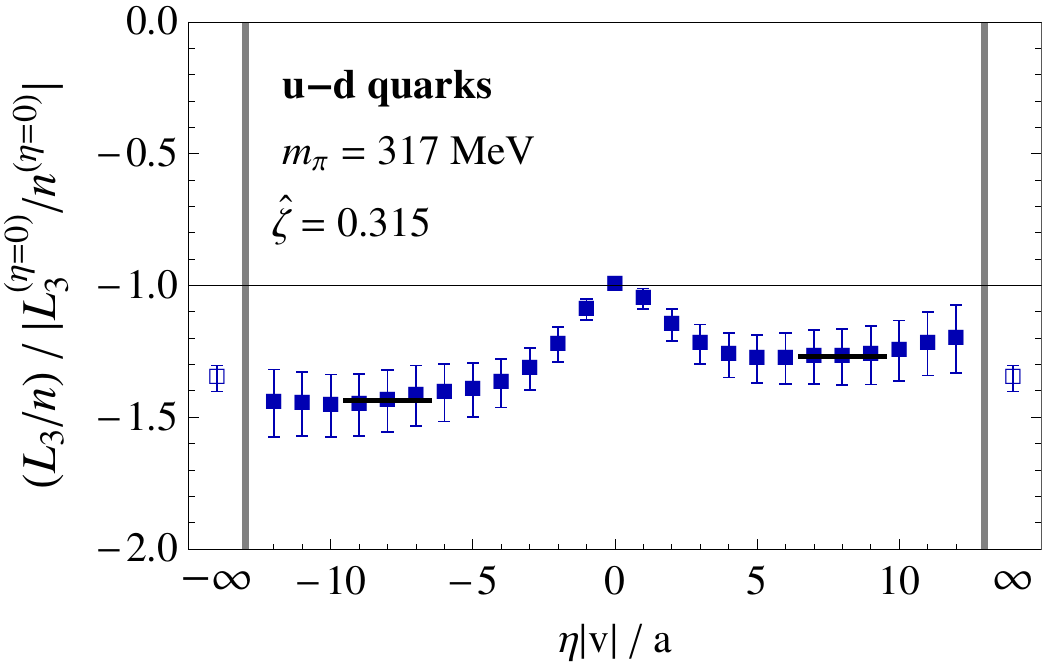}
\caption{Isovector ($u-d$ quark) longitudinal orbital angular momentum in the proton in units of the number of valence quarks $n$, cf.~the definition Eq.~(\ref{lratio}), obtained using a clover fermion ensemble at $m_{\pi } =317\, \mbox{MeV} $; from Ref.~\cite{Engelhardt:2020qtg}.
Left: Ji OAM obtained using a straight gauge link; in this case,
$\hat{\zeta} $ essentially quantifies the proton momentum in the
3-direction (the direction in which the staple link also extends once
the staple length is varied), but the obtained results can not in fact depend
on $\hat{\zeta} $, since there is no physical staple direction $v$ in the
straight-link case. Fitting, therefore, a constant value to the data yields
the extrapolated blue data point. The red data point is the value obtained
from Ji's sum rule at the same pion mass (since this value was not available on the same ensemble, an interpolation of data from Ref.~\cite{Bratt:2010jn} was used instead). Right: Varying the staple length $\eta $ allows for a
continuous, gauge-invariant interpolation between Ji OAM ($\eta =0$) and
Jaffe-Manohar OAM ($|\eta |\rightarrow \infty $). Data are shown in units
of the magnitude of Ji OAM. The sign of this ratio reflects the fact that isovector quark OAM is negative.}
\label{oamplot}
\end{figure}

\index{orbital angular momentum (OAM)}
\index{lattice QCD calculations!GTMDs}
Lattice QCD calculations of TMD observables were discussed in Sec.~\ref{sec:lattice_tmd_calcs}. They are based on evaluating the fundamental matrix element in Eq.~(\ref{eq:latt_corr_def_2})
in the forward limit, $P^{\prime } =P$. By generalizing such calculations to include a momentum
transfer $\deltat =P^{\prime } -P$ in the transverse direction, one can furthermore access GTMD observables; since $\deltat $ is
Fourier conjugate to the impact parameter $\rt $ of the struck quark in a deep inelastic scattering process, one thus supplements the transverse momentum information with transverse position information. In effect, one can access information about Wigner distributions
${\cal W}^{q[\Gamma]W} (x,\kt ,\rt )$ simultaneously characterizing quark position
and momentum. As discussed in Sec.~\ref{sec:GTMD_OAM}, a prime application immediately offering itself is the direct
evaluation of quark orbital angular momentum (OAM) in the proton,
associated with the GTMD $F_{1,4}^{qW} $,
cf.~Eq.~(\ref{eq:F14OAM}).
Casting this, via Eq.~(\ref{e:gammap}), in terms of the matrix element in Eq.~(\ref{eq:latt_corr_def_2}) \cite{Engelhardt:2017miy}, one can evaluate the longitudinal component $L_z^{q,W} $ of quark OAM
in a longitudinally polarized proton, normalized to the number of valence
quarks $n$, in the form \cite{Engelhardt:2017miy,Engelhardt:2020qtg}
\begin{equation}
\frac{L^{q,W}_z }{n} \! = \!
\frac{-\epsilon_{jk} \frac{\partial }{\partial b_{T,j} }
\frac{\partial }{\partial \Delta_{T,k} }
\left. \langle p(P^{\prime} , S_L ) \! \mid \! \bar{\psi }_i^0 (\tfrac{b}{2})
\gamma^+ W_{\sqsupset \eta }^{v} (\tfrac{b}{2},-\tfrac{b}{2}) \psi_i^0 (-\tfrac{b}{2})  \! \mid \! p(P,S_L ) \rangle
\right|_{b^+ = b^- =0\, , \
\deltat =0\, , \ \bt \rightarrow 0} }{\hspace{2.15cm} \left.
\langle p(P^{\prime}, S_L ) \! \mid \! \bar{\psi }_i^0 (\tfrac{b}{2}) \gamma^+
W_{\sqsupset \eta }^{v} (\tfrac{b}{2},-\tfrac{b}{2}) \psi_i^0 (-\tfrac{b}{2}) \! \mid \! p(P,S_L ) \rangle
\right|_{b^+ = b^- =0\, , \ \deltat =0\, , \ \bt \rightarrow 0} }
\label{lratio}
\end{equation}
where the index $i$ specifies the quark flavor under consideration.
The ratio Eq.~(\ref{lratio}) serves to cancel soft factors associated with the gauge
links, in analogy to the TMD studies described in Sec.~\ref{sec:TMDratios}. The limit $\bt \rightarrow 0$ has to be taken with care, since it engenders additional divergences; this is analogous to the subtlety involved in relating TMDs to PDFs discussed in detail in Sec.~\ref{sec:integratedTMDs}. Through
its dependence on the gauge link\footnote{In the GTMD case, the Collins-Soper type evolution parameter $\hat{\zeta } $ characterizing the staple direction $v$ of the gauge link is defined using the average hadron momentum $\bar{P} = (P^{\prime } +P)/2$ as $\hat{\zeta} = v\cdot \bar{P} /(\sqrt{|v^2|} \sqrt{\bar{P}^{2} } )$.} $W_{\sqsupset \eta }^{v} $, the ratio Eq.~(\ref{lratio}) allows one to access
both the quark OAM of the Ji decomposition of proton spin
\index{spin decomposition!Ji}
(by choosing zero staple length $\eta $ in Fig.~\ref{fig:stapleLattice}, i.e., a straight gauge link between the quark
operators) as well as the quark OAM of the Jaffe-Manohar decomposition
of proton spin
\index{spin decomposition!Jaffe-Manohar}
(by choosing infinite staple length $\eta $)
\cite{Lorce:2011kd,Ji:2012sj,Hatta:2011ku,Burkardt:2012sd}. As far as LQCD calculations are
concerned, this formulation thus offers the opportunity to go beyond
previous work, which has been restricted to Ji quark OAM, evaluated
as $L_z^q =J_z^q -S_z^q $ via Ji's sum rule \cite{Ji:1996ek} (LQCD calculations employing Ji's sum rule are discussed in Sec.~\ref{decom_spin}).
Fig.~\ref{oamplot} shows
results for Ji OAM using the GTMD approach, compared with the Ji sum rule value, as well as a continuous, gauge-invariant interpolation between Ji OAM and
Jaffe-Manohar OAM, achieved by varying the staple length $\eta $.

An important technical aspect encountered in carrying out such calculations is the need to construct an unbiased estimate of the derivative with respect to transverse momentum transfer $\deltat $ in Eq.~(\ref{lratio}). In the initial exploration \cite{Engelhardt:2017miy}, the derivative
was evaluated as a finite difference using a rather large interval in $\deltat $, which led to a significant systematic
bias, and consequently a discrepancy between the value obtained for
Ji OAM from Eq.~(\ref{lratio}) and the one obtained from the Ji sum rule.
The data shown in Fig.~\ref{oamplot}, taken from Ref.~\cite{Engelhardt:2020qtg}, were instead obtained using a direct derivative method \cite{deDivitiis:2012vs} which eliminates this bias; essentially, one samples directly the $\deltat $-derivative of the proton matrix element instead of evaluating the proton matrix element itself and numerically extracting its derivative a posteriori (details are given in Ref.~\cite{Engelhardt:2020qtg}). With this methodological improvement, agreement between the results obtained using Eq.~(\ref{lratio}) and using the Ji sum rule
is indeed achieved, as shown in Fig.~\ref{oamplot} (left).

Since, as already noted in Sec.~\ref{sec:GTMD_OAM}, a staple-shaped gauge link incorporates the final state interactions
\index{final state interactions}
experienced by the struck quark in a deep-inelastic
scattering process, the data shown in Fig.~\ref{oamplot} (right)
elucidate the consequent torque \cite{Burkardt:2012sd} experienced by a struck quark leaving
the proton remnant, beginning with Ji OAM and approaching Jaffe-Manohar
OAM at asymptotic distances. The difference between Jaffe-Manohar and
Ji OAM, i.e., the accumulated torque, can be clearly resolved and is
sizeable, amounting to roughly 1/3 of the originally present Ji OAM
at the pion mass $m_{\pi } \approx 317\, \mbox{MeV} $. The torque is
directed such as to enhance OAM as the quark leaves the proton.

Besides the above Wigner function approach and Ji's sum rule, a third
avenue of accessing Ji quark OAM in the proton is via the twist-3 GPD $\widetilde{E}_{2T}^{q} $ \cite{Kiptily:2002nx,Rajan:2016tlg,Raja:2017xlo}, cf.~the discussion in Sec.~\ref{sec:GTMD_OAM},
or its twist-3 GTMD ``mother distributions''
$F_{2,7}^{qW} $ and $F_{2,8}^{qW} $, cf.~\cite{Meissner:2009ww}.
In this case, one
is led to evaluate a correlator of the type in Eq.~(\ref{eq:latt_corr_def_2}) specifically
for $\Gamma =\gamma^{i} $, where $i$ denotes one of the transverse
directions. A preliminary analysis of corresponding lattice data as of this writing indicates that this avenue is feasible and yields results compatible with the other methods, albeit with larger numerical uncertainties at comparable numerical effort.

The study of GTMD observables using LQCD can moreover be extended
to encompass further characteristics of the nucleon, such as the quark spin-orbit correlations
\index{spin-orbit correlation} quantified by the GTMD $G_{1,1}^{qW} $, cf.~Eq.~(\ref{e:q_SO_corr}) and the associated discussion. This case again requires employing
the direct derivative method mentioned above, in order to evaluate a derivative with respect to momentum transfer. A first calculation in this direction, reported in \cite{Engelhardt:LATTICE2021},
employs the domain wall fermion (DWF) discretization, which mitigates possible operator mixing effects by preserving chiral symmetry, as discussed in Sec.~\ref{sec:TMDratios}.

\subsection{Model results and their interpretation}
\label{sec:GTMD_models}

\begin{figure}[t!]
\centering
\includegraphics[width=0.90\textwidth]{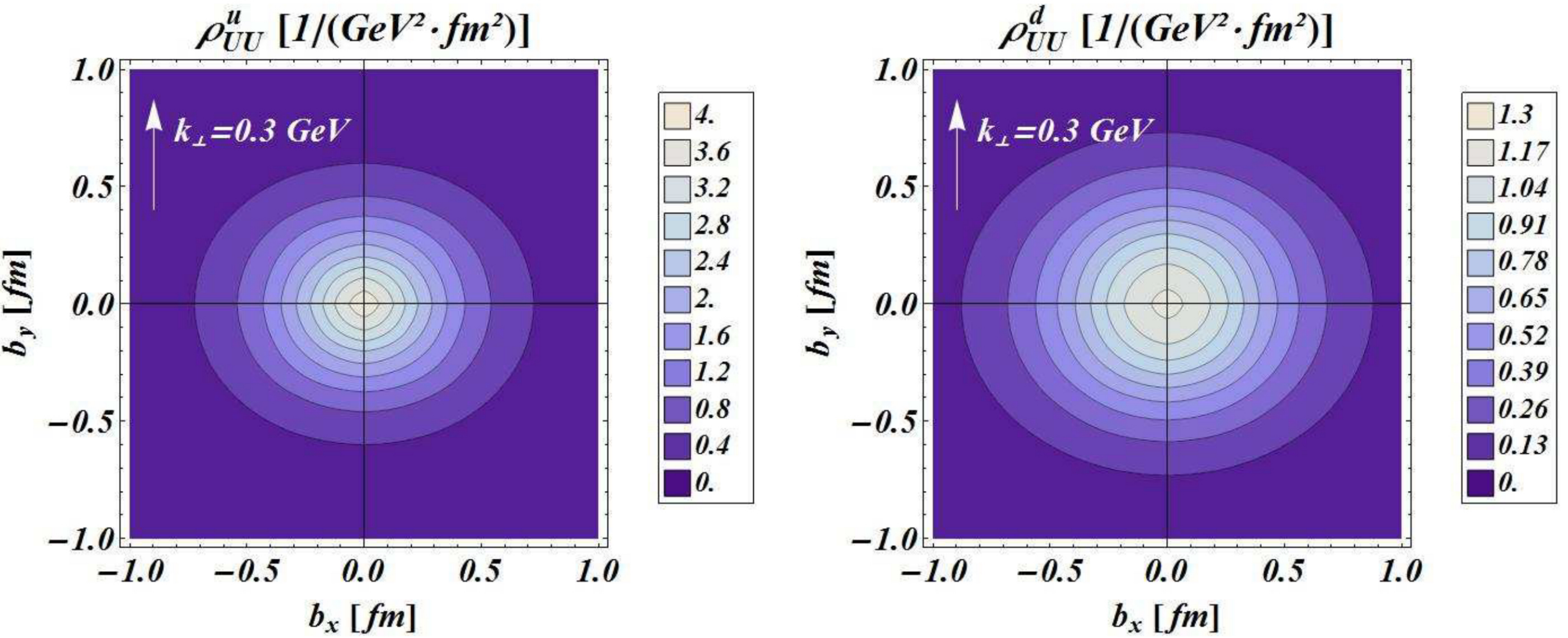} 
\caption{Wigner distributions for unpolarized up quarks (left) and down quarks (right), as defined in Eq.~\eqref{e:Wigner_unpol}, in a light-front constituent quark model; figure from Ref.~\cite{Lorce:2011kd}.
The results are shown for a fixed $\kt$ with $k_T = 0.3$ GeV and pointing in the positive $y$-direction, as a function of $r_x = b_x|_{\rm fig.}$ and $r_y = b_y|_{\rm fig.}$.
Comparing up quarks and down quarks, the same color corresponds to different values of the Wigner distributions.} 
\label{f:wigner_model}
\end{figure}

GTMDs and Wigner distributions of the nucleon have been computed in different models, including diquark spectator models~\cite{Meissner:2009ww, Kanazawa:2014nha, Liu:2014vwa, Liu:2015eqa, Kaur:2018dns, Maji:2022tog}, light-front quark models~\cite{Lorce:2011dv, Lorce:2011kd, Gutsche:2016gcd}, the light-cone version of the chiral quark-soliton model~\cite{Lorce:2011dv, Lorce:2011kd}, the quark-target model~\cite{Kanazawa:2014nha, Mukherjee:2014nya, Miller:2014vla, Hagiwara:2014iya, Mukherjee:2015aja, More:2017zqq, Kumar:2017xcm, More:2017zqp}, the bag model~\cite{Courtoy:2016des}, and models which make use of the AdS/QCD correspondence~\cite{Chakrabarti:2016yuw, Chakrabarti:2017teq, Chakrabarti:2019wjx}.  
Model calculations of those quantities for the pion are available as well~\cite{Meissner:2008ay, Ma:2018ysi, Kaur:2019kpi, Luo:2020yqj, Tan:2021osk}.
We refer to Ch.~\ref{sec:models} for the salient features of the pertinent models and for further related references.
Moreover, there exist various papers on (model) calculations of gluon GTMDs and Wigner distributions in the small-$x$ region~\cite{Hatta:2016dxp, Hagiwara:2016kam, Zhou:2016rnt, Hatta:2016aoc, Kovner:2017vro, Hagiwara:2017fye, Kovner:2017ssr, Boer:2018vdi, Boussarie:2018zwg, ReinkePelicer:2018gyh, Mantysaari:2019csc, Boussarie:2019vmk, Hagiwara:2020mqb}.
Several of those studies are closely related to treatments of gluon TMDs at small $x$ which are discussed in Ch.~\ref{sec:smallx}.

As one example, Fig.~\ref{f:wigner_model} displays results for the Wigner distribution
\index{Wigner distribution}
of unpolarized quarks in an unpolarized proton, obtained in a light-front constituent quark model~\cite{Lorce:2011kd}.
To be precise, the figure shows the quantity
\begin{equation}
\rho_{UU}^{q} (\kt,\rt) = \int dx \, {\cal F}_{1,1}^{q} (x,\kt,\rt) \,,
\label{e:Wigner_unpol}
\end{equation}    
where ${\cal F}_{1,1}^{q}$ is the Wigner distribution which is the Fourier transform of the GTMD $F_{1,1}^{q}$.
(Here we have dropped the dependence of the Wigner distribution and the GTMD on the Wilson line since the model employed in Ref.~\cite{Lorce:2011kd} does not contain gluons.) 
The distributions for up quarks and down quarks, with a fixed transverse momentum in the $y$-direction, are plotted as a function of $(r_x,r_y)$.
The most important qualitative result is that the distributions are not axially symmetric.
Interpreting the results as densities, one concludes that a configuration with large $\rt \perp \kt$ is more likely than a configuration with large $\rt \parallel \kt$, 
which can be understood based on the finite extension of the proton~\cite{Lorce:2011kd}.
The left-right symmetry of the densities is a model-independent result while the top-bottom symmetry could be traced back to the lack of gluons in the model~\cite{Lorce:2011kd}.
Note also that the spread of the distributions is smaller for up quarks than for down quarks, reflecting the fact that up quarks are more concentrated at the center of the proton than down quarks.

While the results of the previous paragraph and other similar findings suggest that Wigner distributions can be used for 5D imaging of hadrons, one must exercise some care in this context. 
(For discussions concerning 6D imaging of hadrons we refer to~\cite{Ji:2003ak, Belitsky:2003nz, Maji:2022tog, Han:2022tlh} and references therein.)
It is already known from non-relativistic quantum mechanics that Wigner distributions are quasi-probability distributions only, and as such they can become negative.
Considering the quark-target model to lowest non-trivial order in pQCD, it has been made explicit that  partonic Wigner distributions can also become negative~\cite{Hagiwara:2014iya} which implies that interpretations of results for Wigner distributions in the sense of multi-dimensional densities are not always straightforward.
In order to address this shortcoming, the authors of Ref.~\cite{Hagiwara:2014iya} suggested to use the so-called Husimi distribution~\cite{Husimi:1940} instead of the Wigner distribution.
\index{Husimi distribution}
Like the Wigner distribution, the Husimi phase space distribution is used in non-relativistic quantum mechanics.
The main underlying idea is a Gaussian smearing for both position and momentum in such a manner that positivity of the distribution is maintained.
(It is expected that also partonic Husimi distributions are positive semi-definite, but a rigorous proof of this property is still lacking~\cite{Hagiwara:2014iya}.)
The Gaussian smearing, however, implies that for Husimi distributions the connections to the densities in position and momentum space are lost, in contrast to Wigner distributions where these connections are expressed through Eqs.~\eqref{e:Wigner_position} and \eqref{e:Wigner_momentum}.
Further research is required in order to better understand the opportunities as well as the limitations of a 5D imaging of hadrons.

%% file: sec-outlook/sec-outlook.tex
\section{Summary and Outlook}

Deep inelastic scattering (DIS) experiments in the early 1970s first revealed the internal structure of the nucleon through the phenomenon of Bjorken scaling. The parton model gave an intuitive explanation of this phenomenon as the consequence of the nucleon being constituted of quasi-free partons, now known to be quarks and gluons. The structure functions measured in DIS were calculated in terms of parton distribution function (PDFs) which described the distribution in the fraction, $x$, of the nucleon's momentum carried by each parton. Initially only the momentum fraction of the parton along the lightlike direction of the large component of the nucleon's momentum  was considered. As QCD developed it became apparent the PDFs were scale dependent and obeyed evolution (DGLAP) equations which allow one to calculate how  the PDFs change as the scale at which they are probed is varied. The ability to measure the PDFs in
various reactions in experiments across a wide range scales
and successfully describe the evolution of the PDFs using DGLAP is a major achievement of QCD. Collinear factorization - convolving collinear PDFs with perturbatively calculable hard cross sections  - has been the main tool for making predictions for high energy physics experiments for decades. Today most cross sections for searches for new physics at the LHC, for example, are calculated in the collinear factorization approximation. 

However, neglecting the transverse motion of the partons within the nucleon misses much of the rich internal structure of the proton. It is like studying the Solar System and knowing only the average distance of each planet from the sun and not the shapes or periods of their orbits. In the last couple of decades a huge amount of both experimental and theoretical work has gone into studying the transverse structure of the nucleons and nuclei. The relevant PDFs which depend on the lightlike momentum fraction and the transverse momentum of the partons, $k_T$, are called transverse momentum dependent PDFs (TMD PDFs). The TMD PDFs along with the transverse momentum dependent fragmentation functions (TMD FFs), collectively known as TMDs, are the main subject of this handbook.

Processes which are sensitive to the $k_T$ distribution of partons inside the hadron are necessarily less inclusive than DIS. They also involve at least two scales, a hard scale justifying the application of perturbation theory along with the transverse momentum which can typically be much closer to $\Lambda_{\rm QCD}$. The three main processes for which TMD factorization is relevant are: semi-inclusive DIS (SIDIS), in which a specific hadron in the final state is measured, the Drell-Yan process in which the transverse momentum of the lepton pair is measured, and di-hadron production in $e^+e^-$ collisions. The factorization theorems for these processes each involve different TMD PDFs and TMD FFs. SIDIS 
involves a TMD PDF for the initial state proton and a TMD FF for the final state hadron, Drell-Yan involves two TMD PDFs for the initial state hadrons and di-hadron production involves two TMD FFs for each final state hadron. These processes were studied in detail in the first five chapters of the Handbook. 

The TMD PDFs also probe the spin structure of the nucleon in a more complex way than is possible in collinear factorization. In collinear factorization, there are three leading twist PDFs. We can study the distribution of unpolarized quarks in unpolarized nucleons, longitudinally polarized quarks in longitudinally polarized nucleons and transversely polarized quarks in transversely polarized nucleons, for example. 
Introducing dependence on the transverse momentum allows one to study correlations in spin that are not possible in collinear factorization. 
For quarks, there are eight different TMD PDFs at leading power and one can study the distribution of unpolarized quarks in transversely polarized nucleons (Sivers function), the distribution of transversely polarized quarks in unpolarized nucleons (Boer-Mulders function),  longitudinally  polarized quarks in  transversely polarized nucleons, and vice versa, (both known as Worm-gear functions) and transversely polarized quarks whose polarization is orthogonal to  the transverse polarization of the nucleon (Pretzelosity function). The situation is summarized in Figure \ref{fig:qTMDPDFsLP} in \chap{Intro}. Various angular modulations of the cross sections turn out to be proportional to convolutions of TMDs, see, e.g., 
Eq.~\ref{e:sidis-modulation}. Measuring these angular modulations allows us to extract the various TMDs.  \chap{Intro} gives an overview of the field, discusses the aforementioned quark TMD PDFs, and also gives some cross section formulae for SIDIS, Drell-Yan, and di-hadron production, while the full versions of these formulae can be found in \chap{TMDdefn}.

 It is important to extend this discussion of the TMD PDFs to include gluons as well as quarks, include TMD FFs for both,   and  give well-defined field theoretical definitions of these functions. 
The definition of TMDs is a rather involved topic as unlike collinear PDFs the ingredients involved in the construction of TMD PDFs must be setup to handle both ultraviolet and rapidity divergences when computed in perturbation theory. The Wilson line structure in the definitions of the TMD PDFs
is also considerably more complicated than the Wilson line structures appearing in collinear PDFs. Instead of a single lightlike Wilson line connecting the two partonic field operators in the correlator, there are two lightlike Wilson lines attached to each parton field. These are separated in transverse position space and connected by a transverse Wilson line at infinity to create a gauge invariant object. 
The definitions of the eight leading TMD PDFs and TMD FFs for both quarks and gluons are given in \chap{TMDdefn}. There are number of approaches to defining the ingredients necessary to construct the TMDs, leading to the same final TMDs, and these constructions are also reviewed. 
A one-loop calculation exhibiting the ultraviolet and rapidity divergences is performed in this chapter. Subtraction of soft Wilson lines is needed to properly define the TMD PDFs in order to remove all divergences.  Finally, full cross section formulae for SIDIS, Drell-Yan, and di-hadron production are also given in this chapter. 

The proof of the factorization theorems provide the QCD basis for a description of these reactions in terms of TMD PDFs and/or TMD FFs, and can be found in \chap{Factorization}. Factorization theorems not only guarantee that the UV- and rapidity divergences can be tamed to provide well-defined definitions of the TMD functions. They also ensure that the same universal TMD functions enter the descriptions of different processes and give predictive power to the approach. Another important subject for TMDs are their renormalization group equations (RGE) and rapidity renormalization group equations (RRGE), which are used to sum large logarithms, as discussed in \chap{evolution}. A thorough review of the phenomenology and extraction of TMDs from data can be found in \chap{phenoTMDs}. A striking observation in this chapter is that early fits with simplistic models of TMDs, e.g., the collinear PDF times a  Gaussian in $k_T$, have given way to more sophisticated parametrizations that are consistent  with the evolution discussed in \chap{evolution}. It is clear that we have made substantial progress in extracting some of the quark TMDs from SIDIS, DY, $W^\pm/Z$ and $pp$ scattering data from a variety of experiments including HERMES (DESY), COMPASS (CERN), Belle, BaBar, RHIC (BNL), Tevatron (Fermilab), LHC (CERN) and JLab. Substantial progress has been achieved in understanding the unpolarised TMD PDFs and FFs as well as many polarised TMDs such as the Sivers functions, transversity, and Boer-Mulders functions.   However, some of the TMDs, e.g., worm gear and pretzelosity, are not yet as well constrained. Another important issue going forward is the extraction of gluon TMDs.  

A very important development in recent years is the application of the methods of lattice field theory to both collinear and TMD PDFs. For a long time it was thought the only meaningful quantities one could compute using a lattice approach were matrix elements of local operators that correspond to Mellin moments of the collinear PDFs. It has been recently realized that by computing Euclidean matrix elements in highly boosted states one could access the collinear PDFs via a matching calculation. For a while it was thought that  it would be difficult to extract the soft matrix elements with more than one lightlike Wilson line needed for computing TMD PDFs but this problem has recently been solved. There are a variety of different schemes for performing lattice TMD calculations. Progress in theory and comparison with experimentally determined collinear PDFs and TMD PDFs are discussed in \chap{lattice}.    

 An interesting aspect of TMD physics are the so called T-odd distributions. Originally, thought to be vanishing because of the time reversal invariance of QCD, it was later realized because the time reversal operation reverses the orientation of the Wilson line, that these functions are non zero and take the opposite sign in Drell-Yan and SIDIS. The Sivers asymmetry is an example of this, and recently this prediction became amenable to experimental checks. That QCD could generate such an asymmetry was first realized  in a model calculation that is described in  \chap{models}. Other models for TMDs are described in this chapter as well. These are useful  for estimating the size and sign of asymmetries and testing conjectured relations between TMDs. Predictions from models can be tested with first-principle lattice QCD results and phenomenological extractions of TMD functions.

The small-$x$ limit for both collinear and TMD PDFs is the subject of extensive theoretical studies. Here the rapid growth of the gluon distribution at small-$x$ predicted by both DGLAP and BFKL evolution must eventually saturate due to unitarity. The physics of saturation is described by an effective theory called the Color Glass Condensate (CGC). Here two types of gluon distributions appear. The Weizs\"acker-Williams distributions are the small-$x$ limit of the TMDs studied earlier in the handbook. At small-$x$, DIS can be physically understood as scattering of color dipole fields off classical shock wave background fields. This scattering is described by a gluon distribution with a novel Wilson line structure. Reconciling the TMD description of small-$x$ physics with the approach of the CGC, along with evolution and resummation, spin-dependent physics,  as well as the outlook for this field are discussed in \chap{smallx}.  

Novel tests of TMD physics arise if one considers final states with jets, which are collimated beams of energetic hadrons in the final state. More details on how jets are defined and the algorithms used to reconstruct them are given in the beginning of \chap{JetFrag} and the rest of the chapter describes a variety of studies involving jets. In SIDIS, final states with a jet instead of an identified hadron can be used to extract TMDs including the Sivers function. Identifying a hadron in a jet probes jet substructure and can studied either in a collinear approximation or including the transverse momentum of the hadron relative to the jet axis. In the latter case, factorization theorems and evolution equations are analogous to those for the TMDs. These observables are calculable in terms of the collinear fragmentation functions so they provide a new mechanism for extraction of fragmentation functions. If the hadron is a heavy quarkonium the fragmentation functions are calculable in Non-Relativistic QCD (NRQCD) and new tests
of this approach to quarkonium production have been obtained. Another interesting test of TMD dynamics comes from studying Transverse Energy-Energy Correlations (TEEC)  in global event shapes in $e^+e^-$. Finally, the modification of jet properties is also sensitive to transverse momentum dynamics and provides a new probe of nuclear media such as cold nuclear matter and the quark gluon plasma. 

While the bulk of this document deals with the leading TMDs, subleading TMDs are an important subject that cannot be ignored. By subleading we mean that these TMDs are suppressed by $\Lambda/Q$ where $Q$ is the underlying hard scale and $\Lambda$ is a hadronic scale. Despite this suppression, these are important because new effects arise at subleading order which are not present at leading order. The Cahn effect, a $\cos \phi_h$ modulation of the SIDIS cross section, where $\phi_h$ is the angle between the lepton and hadron planes, is a subleading TMD effect and also one of the earliest important results in the field of TMD physics. 
The first observations of single-spin asymmetries  in SIDIS were subleading power, and were made by Hermes and JLab. At subleading order one encounters quark-gluon-quark correlations that are not present in the leading TMDs. \chap{twist3} of this handbook is devoted to subleading TMDs.
The subleading TMD contribution to the SIDIS cross section is given and the  16 subleading TMDPDFs and TMDFFs are classified and defined. Factorization for subleading TMDs, more subtle than for leading TMDs, is discussed. Experimental data on the Cahn effect and other asymmetries due to subleading TMDs is presented. The chapter closes with calculations of subleading TMDs from the lattice and models.

Finally, it is also possible to study generalized distributions that contain information about the spatial distribution of partons within the nucleon in addition to the momentum distributions.  This can be achieved by considering matrix elements in which the protons
in the in-state and the out-state have different four-momenta. In this way, one arrives at the generalized TMDs (GTMDs), which in addition to depending on the collinear momentum fraction and transverse momentum, as TMDs do, also depend on the longitudinal and transverse momentum {\em transfer}, as Generalized Parton Distributions (GPDs) do. Since the transverse momentum transfer to a parton is Fourier conjugate to its impact parameter, both the transverse motion and the transverse position of partons are thus encoded in GTMDs. Setting the longitudinal momentum transfer to zero and Fourier transforming GTMDs with respect to the transverse momentum transfer, one obtains the Wigner distributions which are well known in many areas of physics. If one integrates out the information about the transverse position of partons from the Wigner distributions, one obtains the 
TMD PDFs which are the main subject of this handbook. Alternatively, starting with GTMDs and instead integrating out the information on the transverse motion of partons, one obtains the GPDs. 
All of these functions are the subject of \chap{gtmd} and the relationships between these functions are summarized in Fig.~\ref{f:multi_dim_imaging}. GPDs can be accesssed as the matrix elements appearing in Deeply Virtual Compton scattering and the GTMDs are accesssible through exclusive double diffractive dijet production and the exclusive double Drell-Yan processes. Because the Wigner distributions contain both the spatial and the momentum information of the partons, they can be used to directly calculate the orbital angular momentum. This has been 
exploited to perform a lattice calculation of the isovector quark orbital angular momentum in the proton. Models have also been used to calculate Wigner functions. 

TMD physics will continue to focus on precise extraction of quark TMDs from the classic TMD processes: SIDIS, Drell-Yan, and di-hadron production in $e^+e^-$ collisions. We should also emphasize the importance of TMD factorization for the calculation of the $p_T$ spectrum of heavy particles such as the Higgs boson and particles  from beyond the standard model physics  at the LHC. In the future, final states with jets and heavy flavor will be of great interest as well. These will be especially important for extracting gluon TMDs which are presently not tightly constrained. Rigorous proofs of TMD factorization for processes involving heavy flavor and jets and other novel TMD processes will be needed.  The exciting developments in lattice QCD which allow direct calculation of TMD PDFs will become even more impressive as algorithms improve and computing power increases. Model calculations will continue to provide insight and make predictions for the TMDs which will be tested. An important open question is whether one can develop nonperturbative methods for TMD FFs as well. Ultimately we hope to gain a complete 3D picture of how partons are distributed within the nucleon and nucleus. We would also like to obtain a precise understanding of the decomposition of the nucleon's spin and mass, spin-orbit correlations, and   the orbital motion of partons within the nucleon.
Finally, we expect to go beyond TMDs to extract higher dimensional functions like the Wigner distributions which contain information about the partons' distribution in both position and momentum space.  Accomplishing all of this will require advances in theory, numerical simulations, and more data from existing experiments as well as future experiments like the Electron-Ion Collider.

%% file: sec-acknowledgement/sec-acknowledgement.tex
\section{Acknowledgement}

We thank Ted Rogers for collaboration on certain aspects of this Handbook, and in particular Chapter 3.  
We would like to warmly thank the students of the 2022 TMD Winter School for reading a draft version of this Handbook and offering many comments and edits to help improve it.
We would like to thank Tatiana Donskova for designing the logo of the TMD Collaboration.
%
%
This work has been supported within the framework of the TMD Topical Collaboration of the U.S. Department of Energy, Office of Nuclear Physics.
This work has also been supported in part by the National Science Foundation under Grants No.~PHY-1945471 (Kang), No.~PHY-2110472 (Metz), No.~PHY-2011763 (Pitonyak), No.~PHY-2012002 (Prokudin), No.~PHY-1812423 and No.~PHY-2111490 (Schweitzer). In addition, this work has been supported in part by the U.S. Department of Energy, under contracts No.~DE-FG02-96ER40965 (Burkardt, Engelhardt), No.~DE-SC0020405  (Constantinou), No.~DE-SC0011090 (Detmold, Ebert, Negele, Shanahan, and Stewart), No.~DE-FG02-04ER41338 (Fleming), No.~DE-FG02-07ER41460 (Gamberg), No.~DE-SC0020682 (Ji), No.~DE-SC0013065 (Liu), No.~DE-SC0016286 (Liuti), No.~DE-FG02-05ER41367 (Mehen), No.~DE-SC0020081 (Tarasov), No.~DE-SC0012704 (Venugopalan), No.~DE-AC02-05CH11231 (Yuan), No.~DE-AC02-06CH11357 (Zhao), and No.~DE-AC05-06OR23177 (Prokudin, Qiu) under which Jefferson Science Associates, LLC, manages and operates Jefferson Lab; through the DOE Office of Nuclear Physics and the LDRD Program of Los Alamos National Laboratory, which is operated by Triad National Security, LLC, for the National Nuclear Security Administration under contract No.~89233218CNA000001 (Lee, Vitev).  Stewart was also supported in part by the Simons Foundation through the Investigator grant 327942, and Ebert was supported in part by the Alexander von Humboldt Foundation through a Feodor Lynen Research Fellowship.

%% file: sec-appendix/app-glossary.tex

\section{Conventions}
\label{app:glossary}

This appendix discusses notational conventions we have adopted. For the conventions for light-cone coordinates we refer to the discussion in \sec{TMDdefn}.  Our convention for the sign of the coupling in covariant derivatives is $iD^\mu = i\partial^\mu - g A^\mu$. We use
\begin{enumerate}
    \item a metric convention with $g_{00}=+1$ and $g_{ii}=-1$.
    \item ${\bf q}_T$ for Euclidean transverse momentum, ${\bf b}_T$ for Euclidean transverse coordinate, non-bold face for magnitudes such as $q_T=|{\bf q}_T|$ and $b_T=|{\bf b}_T|$, and also non-bold face for Minkowski four vectors, $q_T^\mu$ etc.
    \item ${\bf k}_T$ for transverse momentum in TMDPDF, and ${\bf p}_T$ for that in TMDFF, saving ${\bf q}_T$ for leptonic transverse momentum. 
    \item $Y$ for lepton rapidity, and small $y$ with subscripts for other rapidities.
    \item either $i/H$ or $i/p$ for a parton of type $i$ inside a hadron $H$ or proton $p$, and $h/i$ for a hadron $h$ produced by a primary parton of type $i$.
    \item $f_{i/p}(x,{\bf b}_T,\mu,\zeta)$ for renormalized TMDPDF\\
          $f_{i/p}^{0\unsub} (=B_{i/p}^{0{\rm naive}})$ for bare unsubtracted TMDPDF  (equal to bare naive beam function)\\
          $B_{i/p}^0$ for bare beam function, where  $B_{i/p}^0=f_{i/p}^{0(\rm u)}/S_i^{0\rm subt}$\\
          $B_{i/p}(x,{\bf b}_T,\mu,\zeta/\nu^2)$ for renormalized beam function\\
          $S_i^0$ for bare soft function\\
          $S_i^{0\rm subt}$ for overlap factor = soft subtraction factor.
    \item  $W_n(x; a,b)$ for a straight Wilson line where $n$ is the direction and the path goes from $x^\mu + a n^\mu$ to $x^\mu + b n^\mu$.  For a generic path $\gamma$ we use the notation $W[\gamma]$.
    \item the notation $|H(P,S)\rangle$ for a hadron $H$ state with momentum $P^\mu$ and spin $S$.
    \item the constant $b_0 = 2 e^{-\gamma_E}$ which commonly occurs when taking Fourier transforms in the $\MSbar$ scheme. 
\end{enumerate}

%% file: sec-appendix/app-Feynman-rules.tex

\section{Feynman rules}
\label{app:feynman_rules}

In order to evaluate \eq{beam_nlo_1} perturbatively, we need to know the Feynman rules of the Wilson line $W_{\sqsubset}(b^\mu,0)$.
Since it is composed of several straight Wilson line segments, it suffices to consider the straight Wilson line defined in \eq{Wilson_lines}.
We can perturbatively expand it as
\begin{align}
 W_n(x; a,b) &
 = P \exp\biggl[ -\img g_0 \int_a^b \df s \, n \cdot A^{a\,0}(x^\mu + s n^\mu) t^a\biggr]
 \nn\\&
 = 1 - \img g_0 n^\mu t^a \int_a^b \df s \, A^{a\,0}_\mu(x^\mu + s n^\mu)
   + \cO(g_0^2)
\,.\end{align}
At this order, the path ordering $P$ has no effect. The corresponding Feynman rule can be obtained using standard techniques
for the Feynman rules of the gluon field $A^a_\mu$ itself. We obtain
\begin{align}
 \raisebox{-2ex}{
\includegraphics[height=1cm]{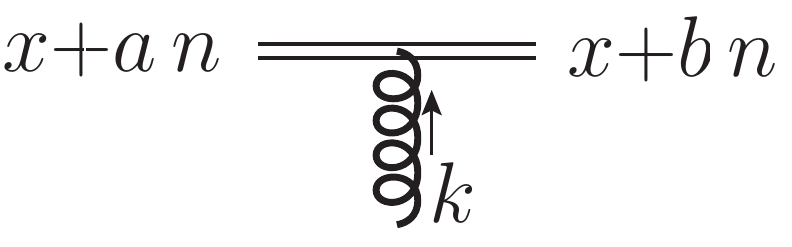}
}~
 &= - \img g_0 n^\mu t^a \int_a^b \df s \, e^{-\img k \cdot (x + s n)}
 \nn\\&
 = g_0 n^\mu t^a e^{-\img k \cdot x} \frac{e^{-\img b (k \cdot n)} - e^{-\img a (k \cdot n)}}{k \cdot n}
\,.\end{align}
Care has to be taken when taking the limit of either $a \to \infty$ or $b \to \infty$,
in which case one has to give momentum $k$ a small imaginary part to make the pure phase vanish.
For concreteness, the one-loop Wilson rules for the linear segments in \eq{Wilson_lines} are given by
\begin{align}
 W_{n_b}(b^\mu;-\infty,0) \,\text{:}\qquad&
 -g_0 n_b^\mu t^a \frac{1}{k^+ + \img0} e^{-\img k \cdot b}
\,,\nn\\
 W_{n_b}^{\dagger}(0;-\infty,0) \,\text{:}\qquad&
 +g_0 n_b^\mu t^a \frac{1}{k^+ - \img0}
\,,\end{align}
and the transverse Wilson vanishes at light-cone infinity.
The relative sign between the two results reflects the inverse direction of $W_{n_b}$ and $W_{n_b}^\dagger$.

%% file: sec-appendix/app-Fourier.tex
\section{Fourier transforms}
\label{app:Fourier_transform}

Here, we collect some useful definitions and identities for Fourier transforms in transverse space.
As discussed in \sec{TMDdefn}, the sign convention of the Fourier transform differs
between TMD PDFs and TMD FFs, and thus we will discuss both cases seperately.

\subsection{Conventions for the TMD PDF}
\label{app:FT_TMDPDF}

In the case of the TMD PDF, our convention for the Fourier transform and its inverse is
\begin{align} \label{eq:app:Fourier_definition}
 \tilde f(\bt) = \int\df^2\pt \, e^{-\img \bt \cdot \pt} f(\pt)
\,,\qquad
 f(\pt) = \int\frac{\df^2\bt}{(2\pi)^2} e^{+\img \bt \cdot \pt} \tilde f(\bt)
\,,\end{align}
where $\tilde f(\bt)$ is the function in Fourier or position space,
and $f(\pt)$ is the function in momentum space.
If $f(\pt)$ is independent of the azimuthal angle, i.e.~$f(\pt) \equiv f(|\pt|)$,
then one can use the identity
\begin{align} \label{eq:app:Fourier_J0}
 \tilde f(b_T) &
 = \int_0^\infty \df p_T \, p_T \int_0^{2\pi}\df\phi \, e^{-\img b_T p_T \cos\phi} f(p_T)
 = 2\pi \int_0^\infty \df p_T \, p_T J_0(b_T p_T) f(p_T)
\,,\end{align}
where $J_0(x)$ is the $0$-th order Bessel function of the first kind.
In this case, $\tilde f(\bt) \equiv \tilde f(b_T)$ is independent of the azimuthal angle as well,
which yields the corresponding identity for the inverse transform
\begin{align} \label{eq:app:Fourier_J0_inv}
 f(p_T) &
 = \frac{1}{(2\pi)^2} \int_0^\infty \df b_T \, b_T \int_0^{2\pi}\df\phi \, e^{\img b_T p_T \cos\phi} \tilde f(b_T)
 = \frac{1}{2\pi} \int_0^\infty \df b_T \, b_T J_0(b_T p_T) \tilde f(b_T)
\,.\end{align}
From \eqs{app:Fourier_J0}{app:Fourier_J0_inv}, it is clear that the Fourier transform $\tilde f(b_T)$ of a real function $f(p_T)$ is real,
and likewise for the inverse Fourier transform.

A key feature of the Fourier transform is that it turns convolutions in momentum space into simple products,
\begin{align}
 \int\df^2{\bf k}_1 \df^2 {\bf k}_2 \, \delta^{(2)}(\pt - {\bf k}_1 - {\bf k}_2) f({\bf k}_1) g({\bf k}_2)
 = \int\frac{\df^2\bt}{(2\pi)^2} e^{\img \bt \cdot \pt} \tilde f(\bt) \tilde g(\bt)
\,,\end{align}
which can be easily seen by inserting \eq{app:Fourier_definition} together with the distributional identity
\begin{align}
 \delta^{(2)}(\pt - {\bf k}_1 - {\bf k}_2)  = \int\frac{\df^2\bt}{(2\pi)^2} e^{\img \bt \cdot (\pt - {\bf k}_1 - {\bf k}_2)}
\,.\end{align}
In \sec{qgspinTMDFF}, we also need Fourier transforms of functions of the form $p_T^\mu f(p_T)$,
which can be obtained as
\begin{align} \label{eq:app:FT_1}
 \int\df^2\pt e^{-\img \pt \cdot \bt} \, (p_T^\mu \cdots p_T^\nu) f(p_T) &
 = \Bigl(-\img \frac{\partial}{\partial b_{T \mu}}\Bigr) \cdots \Bigl(-\img \frac{\partial}{\partial b_{T \nu}}\Bigr)
   \int\df^2\pt e^{-\img \pt \cdot \bt} f(p_T)
 \nn\\&
 = (-\img \partial^\mu) \cdots (-\img \partial^\nu) \tilde f(b_T)
 \nn\\&
 = (-\img \partial^\mu) \cdots (-\img \partial^\nu) \, 2 \pi \int_0^\infty \df p_T \, p_T J_0(b_T p_T) f(p_T)
\,.\end{align}
By acting with the partial derivative
\begin{align}
 \partial^\mu
 \equiv \frac{\partial}{\partial b_{T\mu}}
 = - \frac{b_T^\mu}{b_T} \frac{\partial}{\partial b_T}
\end{align}
on the exponential phase, one induces the desired tensor structure $p_T^\mu \dots p_T^\nu$ in the Fourier integral.
(Recall that $\pt \cdot \bt = - p_T^\mu b_{T\mu}$, which fixes the sign of the derivative factors.)
Thus, we can conveniently express this Fourier transform as derivatives acting on the Fourier transform $\tilde f(b_T)$,
which in the last line was expressed using \eq{app:Fourier_J0}.
Using \eq{app:FT_1} together with the Bessel function identity
\begin{align} \label{eq:Jn_derivative_relation}
 \frac{\df}{\df z} z^{-m} J_m(z) & = - z^{-m} J_{m+1}(z)
\,,\end{align}
we easily obtain the explicit results
\begin{align} \label{eq:app:FT_2}
 \int\df^2\pt \, e^{-\img \pt \cdot \bt} \, \frac{p_T^\mu}{p_T} f(p_T) &
 = (-\img) \frac{b_T^\mu}{b_T} \times 2\pi\int_0^\infty \df p_T \, p_T \, J_1(b_T p_T) f(p_T)
\,,\\\nn
  \int\df^2\pt \, e^{-\img \pt \cdot \bt} \, \left(\frac{g_T^{\mu\nu}}{2} + \frac{p_T^\mu p_T^\nu}{\pt^2}\right) \, f(p_T) &
  = (-\img)^2 \left( \frac{g_T^{\mu\nu}}{2}  +  \frac{b_T^\mu b_T^\nu}{\bt^2}  \right)
      \times 2\pi \int_0^\infty \df p_T \, p_T \, J_2(b_T p_T) \, f(p_T)
\,.\end{align}
The integrals over $p_T$ have the same structure as in \eq{app:Fourier_J0}, up to exchanging $J_0(x)$ by $J_1(x)$ and $J_2(x)$, respectively.
From \eq{app:FT_2}, we easily obtain the relations
\begin{alignat}{2} \label{eq:app:FT_3}
 \int\df^2\pt \, e^{-\img \pt \cdot \bt} \, \frac{p_T^\mu}{M} f(p_T) &
 = (-\img) b_T^\mu M \tilde f^{(1)}(b_T)
\,,\\\nn
  \int\df^2\pt \, e^{-\img \pt \cdot \bt} \, \frac{\pt^2}{M^2} \left(\frac{g_T^{\mu\nu}}{2} + \frac{p_T^\mu p_T^\nu}{\pt^2}\right) \, f(p_T) &
  = \frac{(-\img)^2}{2} b_T^2 M^2 \left( \frac{g_T^{\mu\nu}}{2}  +  \frac{b_T^\mu b_T^\nu}{\bt^2}  \right) \tilde f^{(2)}(b_T)
\,,\end{alignat}
where the $\tilde f^{(n)}$ denote derivatives with respect to $b_T$ as defined in \eq{TMD_bt_derivative},
\begin{align} \label{eq:app:TMD_bt_derivative}
 \tilde f^{(n)}(b_T) &
 \equiv n! \left(\frac{-1}{M^2 b_T} \partial_{b_T} \right)^n \tilde f(b_T)
 = \frac{2\pi\, n!}{(b_T M)^n} \int_0^\infty \df p_T \, p_T \left(\frac{p_T}{M}\right)^n J_n(b_T p_T) \, f(p_T)
\,.\end{align}
The equality in the second step follows directly from \eq{Jn_derivative_relation}.
The factor of $n!$ arises from following the convention of \cite{Boer:2011xd}.
Also note that the \eq{app:TMD_bt_derivative} is manifestly real if $f(p_T)$ is real,
and hence the explicit factors of $\img$ have been extracted in \eq{app:FT_3}.

For the gluon TMD, we also need the Fourier transform
\begin{align}
 \int\df^2\pt e^{-\img \pt \cdot \bt} \, \frac{p_T^\mu p_T^\nu p_T^\sigma}{p_T^3} f(p_T) &
 = \frac{1}{b_T}(b_T^\mu g_\perp^{\nu\sigma} + b_T^\nu g_\perp^{\sigma\mu} + b_T^\sigma g_\perp^{\mu\nu})
   2\pi (-\img)^3  \int_0^\infty \df p_T \, p_T  \frac{J_2(b_T p_T)}{p_T b_T} f(p_T)
 \nn\\&\quad
 +  \frac{b_T^\nu b_T^\sigma b_T^\mu}{b_T^3}  2\pi (-\img)^3 \int_0^\infty \df p_T \, p_T J_3(b_T p_T) f(p_T)
\,.\end{align}
More precisely, we will only need the second term, while the first one that is completely
symmetry under exchange of the indices $\mu,\nu,\sigma$ will drop out. Then, using
\eq{app:TMD_bt_derivative}, we have
\begin{align}
 \int\df^2\pt e^{-\img \pt \cdot \bt} \, \frac{p_T^\mu p_T^\nu p_T^\sigma}{M^3} f(p_T) &
 = \frac{(-\img)^3}{6} M^3 \, b_T^\nu b_T^\sigma b_T^\mu \, \tilde f^{(3)}(b_T)
 + (\mathrm{symmetric~in~} \mu, \nu, \sigma)
\,.\end{align}

\eq{app:TMD_bt_derivative} can be inverted using the orthogonality relation of Bessel functions,
\begin{align}
 \int_0^\infty \df b_T \, b_T \, J_n(p_T b_T) \, J_n(p'_T b_T) = \frac{1}{p_T} \delta(p_T-p'_T)
\,,\end{align}
from which one easily finds that
\begin{align} \label{eq:app:TMD_bt_derivative_inv}
 f(p_T) &
 = \frac{M^{2n}}{ 2 \pi\, n!} \int_0^\infty \df b_T \, b_T  \left(\frac{b_T}{p_T}\right)^n J_n(b_T p_T) \, \tilde f^{(n)}(b_T)
\,.\end{align}

For TMDs appearing at subleading power, we also encounter Fourier transformations of the form~\cite{Ebert:2021jhy}
\begin{align} \label{eq:app:FT_4}
 f^{(0')}(b_T) &\equiv
 \int\df^2\kt \,  e^{-\img \bt \cdot \kt} \frac{k_T^2}{\Ma^2} f(k_T)
 = 2 \pi \int_0^\infty \df k_T \, k_T J_0(b_T k_T) \frac{k_T^2}{\Ma^2} f(k_T)
 \nn\\&
 = \frac{(-\img)^2}{2}\Ma^2\, b_T^2 \tilde f^{(2)}(b_T)
     - 2 (-\img)^2  \tilde f^{(1)}(b_T)
\,,\nn\\
 -\img \Ma b_\perp^\mu f^{(1')}(b_T) &\equiv
 \int\df^2\kt \,  e^{-\img \bt \cdot \kt} \frac{k_\perp^\mu k_T^2}{\Ma^3} f(k_T)
 = -\img \frac{b_\perp^\mu}{b_T}  2\pi \int_0^\infty \df k_T \, k_T J_1(b_T k_T) \frac{k_T^3}{\Ma^3} f(k_T)
 \nn\\&
 = -\img \Ma b_\perp^\mu \Bigl[ \frac{(-\img)^2}{6}\Ma^2\, b_T^2 \, \tilde f^{(3)}(b_T)
  - 2 (-\img)^2 \tilde f^{(2)}(b_T) \Bigr]
\,.
\end{align}

\subsection{Conventions for the TMD FF}
\label{app:FT_TMDFF}

In the case of the TMD FF, our convention for the Fourier transform and its inverse is
\begin{align} \label{eq:app:Fourier_definition_FF}
 \tilde D(\bt) = \int\df^2\kt \, e^{+\img \bt \cdot \kt} D(\kt)
\,,\qquad
 D(\kt) = \int\frac{\df^2\bt}{(2\pi)^2} e^{-\img \bt \cdot \kt} \tilde D(\bt)
\,,\end{align}
where $\tilde D(\bt)$ is the function in Fourier or position space.
Compared to the convention for the TMD PDF in \eq{app:Fourier_definition},
this differs by the sign of the Fourier phase. Thus, all identities
derived in \app{FT_TMDPDF} can be applied to the TMD FF by simply letting $\bt \to -\bt$.
Also note that the Fourier transform  of the TMD FF is defined with respect to the
hadron frame, i.e.~$\kt$ is the transverse momentum of the parton fragmenting
into a hadron relative to the hadron momenta, see \sec{TMDFFs} for more details.

%% file: sec-appendix/app-definitions.tex
\section{Explicit definitions of TMD PDFs}
\label{app:TMDdefn}

In the following, we give more details on all rapidity regulators employed in the literature that give rise to the result in \eq{tmdpdf_nlo}, i.e.\ those that correspond to the $\MSbar$ scheme.
We do not give explicit results for the regulated results of the bare unsubtracted TMD PDF and soft function,
but these can be found in \cite{Ebert:2019okf} for all considered regulators.

\tocless\subsection{Wilson lines off the light-cone}

In the modern definition by Collins~\cite{Collins:2011zzd}, the lightlike direction $n_a$ and $n_b$ defined in \eq{nab} are replaced by spacelike reference vectors,
\begin{align} \label{eq:Collins_rap_app}
 n_a^\mu = \frac{1}{\sqrt2}(1,0,0,+1) &\quad\to\quad n_A^\mu(y_A) \equiv n_a^\mu - e^{-2 {y_A}} n_b^\mu = (1,-e^{-2 y_A},{\bf 0}_T)
\,,\nn\\
 n_b^\mu = \frac{1}{\sqrt2}(1,0,0,-1) &\quad\to\quad n_B^\mu(y_B) \equiv n_b^\mu - e^{+2 {y_B}} n_a^\mu = (-e^{+2 y_B},1,{\bf 0}_T)
\,.\end{align}
The bare unsubtracted TMD PDF in \eq{beamfunc} for a proton close to the $n_a$ direction is then defined by replacing $n_b \to n_B(y_B)$,
\begin{align} \label{eq:beamfunc_Collins}
 f_{i/P}^{0\,\unsub}(x,\bt,\epsilon,y_B,x P^-) &= \int\frac{\df b^+}{2\pi} e^{-\img b^+ (x P^-)}
 \Bigl< p(P) \Bigr| \bar q(b^\mu) W_{n_B(y_B)}(b^\mu; -\infty, 0) \frac{\gamma^-}{2} W_{n_\perp}(-\infty n_B(y_B), b_T,0)
 \nn\\&\hspace{5cm}\times
 W_{n_B(y_B)}^\dagger(0; -\infty, 0) q(0) \Bigl| p(P) \Bigr>
\,.\end{align}
Similarly, the soft function in \eq{softfunc} is modified as
\begin{align} \label{eq:softfunc_Collins}
 S^{0}_{n_a n_b}(b_T,\epsilon,y_A-y_B) &= \frac{1}{N_c} \bigl< 0 \bigr| {\rm Tr} \bigl[
   W_{n_A(y_A)}^\dagger(\bt;-\infty,0) W_{n_B(y_B)}(\bt;-\infty,0) W_{n_\perp}(-\infty n_B(y_B); b_T,0)
   \nn\\&\hspace{1.7cm}\times
   W_{n_B(y_B)}^\dagger(\bt;-\infty,0) W_{n_A(y_A)}(\bt;-\infty,0) W_{n_\perp}^\dagger(-\infty n_A(y_A); b_T,0)
 \bigr]_\tau \bigl|0 \bigr>
\,.\end{align}
By Lorentz invariance, the regulated bare soft function only depends on the difference $y_A-y_B$~\cite{Buffing:2017mqm}.
The renormalized TMD PDF is finally constructed as~\cite{Collins:2011zzd}
\begin{align} \label{eq:tmdpdf_collins}
 f_{i/P}(x, \bt, \mu, \zeta) &
 = \lim_{\substack{y_A\to+\infty\\y_B\to-\infty}} Z_{\rm uv}
   f_{i/P}^{0\,\unsub}(x,\bt,\epsilon,y_B,x P^-)
   \sqrt{\frac{S^{0}_{n_a n_b}(b_T,\epsilon,y_A-y_n)}{S^{0}_{n_a n_b}(b_T,\epsilon,y_A-y_B) S^{0}_{n_a n_b}(b_T,\epsilon,y_n-y_B)}}
\nn\\&
 = \lim_{y_B\to-\infty} Z_{\rm uv}
   \frac{f_{i/P}^{0\,\unsub}(x,\bt,\epsilon,y_B,x P^-)}{\sqrt{S^{0}_{n_a n_b}(b_T,\epsilon, 2y_n-2y_B)}}
\,,\end{align}
where the result in the last line was derived in \cite{Buffing:2017mqm}.
The UV renormalization factor $Z_{\rm uv}$ is often further split into a field strength renormalization $Z_2$ and the operator renormalization $Z_F$, \emph{i.e.}~$Z_{\rm uv} = Z_2 Z_F$.
In \eq{tmdpdf_collins}, $y_{A,B}$ are the Wilson line rapidities as defined in \eq{Collins_rap_app},
and $y_n$ is an additional rapidity parameter that controls the split of soft radiation into the two TMD PDFs.
The $\zeta$ scale is defined as
\begin{align} \label{eq:zeta_Collins}
 \zeta &= (x P^- e^{- y_n})^2  = (x m_P e^{y_P - y_n})^2
\,,\end{align}
where $y_P$ is the rapidity of the proton.

\tocless\subsection{$\delta$ regulator}

The $\delta$ regulator was introduced by Echevariia, Idilbi and Scimemi (EIS) for TMD PDFs in \cite{GarciaEchevarria:2011rb}
and used to defined TMD PDFs in the notation of \sec{tmdpdfs_new} in \cite{Echevarria:2012js}.
Here, we briefly present the $\delta$ regulator as modified in \cite{Echevarria:2015usa,Echevarria:2015byo,Echevarria:2016scs}, which is necessary to be applicable beyond next-to-leading order.
For more details on the regulator, we refer to \cite{Echevarria:2015byo}.

The $\delta$ regularization scheme consists of modifying the lightlike Wilson lines appearing in the collinear and soft matrix elements, while the transverse gauge links are not modified.
The Wilson lines $W_n$ appearing in the unsubtracted TMD PDF, see \eq{beamfunc}, are modified as
\begin{align} \label{eq:coll_Wilson_lines_delta}
 W_n(x^\mu; -\infty, 0) &= P \exp\biggl[ -\img g_s \int_{-\infty}^0 \df s \, n \cdot A^{a\,0}(x^\mu + s n^\mu) t^a e^{\delta^- s x} \biggr]
\,.\end{align}
Here, $\delta^-$ is the regulator, which plays the role of $\tau$ in the unsubtracted TMD PDF $f_{i/P}^{0\,\unsub}(x,\bt,\epsilon,\tau,x P^-)$,
and the $x$ in $e^{\delta^- s x}$ is the Bjorken momentum fraction of the struck parton.
In the soft function defined in \eq{softfunc}, the lightlike Wilson lines originally defined in \eq{Wilson_lines} are changed as
\begin{align} \label{eq:soft_wilson_line_delta}
 W_n(x^\mu; -\infty,0)
 =&~ P \exp\biggl[ -\img g_s \int_{-\infty}^0 \df s \, n \cdot A^{a\,0}(x^\mu + s n^\mu) t^a\biggr]
 \nn\\
 \to&~ P \exp\biggl[ -\img g_s \int_{-\infty}^0 \df s \, n \cdot A^{a\,0}(x^\mu + s n^\mu) t^a \, e^{\delta^- s}\biggr]
\,,\end{align}
and likewise for the other lightlike Wilson line $W_\bn$, up to replacing $\delta^- \to \delta^+$.

Note that the $\delta$ regulator violates gauge invariance, but gauge violation is power suppressed in $\delta^\pm$ and thus gauge invariance holds as long as $\delta^\pm$ is kept infinitesimal.
In perturbation theory, this regularization procedures amounts to shifting Wilson line vertices as
\begin{align}
 \frac{1}{(k_1^- +\img0) \, (k_2^- + \img0) \cdots}
 \quad\to\quad
 \frac{1}{(k_1^- +\img\delta^-) \, (k_2^- + 2\img\delta^-) \cdots}
\,,\end{align}
where the $k_i$ are the momenta of the gluons emitted from a Wilson line $W_n$, ordered such that $k_1$ is closest to $-\infty$.
The shift in these propagators fully regulates rapidity divergences, such as those in the example integrals in \eqs{beam_nlo_5}{soft_nlo_3} are regulated.
Note that the exponential form of introducing $\delta^-$ in \eq{soft_wilson_line_delta} is crucial for important properties such as non-Abelian exponentiation, see \cite{Echevarria:2015byo} for more details.

With the $\delta$ regulator, the soft function can be symmetrically split into $n$-collinear and $\bn$-collinear component as
\begin{align} \label{eq:S_delta_split}
 S^q_{\rm EIS}\bigl(b_T, \eps, \sqrt{\delta^+ \delta^-}\bigr) &
 = \sqrt{ S^q_{\rm EIS}(b_T,\eps, \delta^- e^{-y_n})} \sqrt{ S^q_{\rm EIS}(b_T,\eps, \delta^+ e^{+y_n})}
\,.\end{align}
Here, $y_n$ is an arbitrary parameter that governs the split of the soft function into the two beam directions.
With this regulator, the zero-bin subtraction is equal to the soft function itself, $S^{0 subt} = S^0$, and hence one can define the TMD PDF by
\begin{align}
  f_{i/p}(x, \bt, \mu, \zeta) &
 =  \lim_{\substack{\epsilon\to 0 \\ \delta^-\to 0}} Z_{\rm uv}^i(\mu,\zeta,\epsilon) \,
    \frac{f_{i/P}^{0\,\unsub}\bigl(x, \bt, \epsilon, \delta^-/(x P^-) \bigr)}
         {\sqrt{ S^q_{\rm EIS}(b_T,\eps, \delta^- e^{-y_n})}}
\,,\end{align}
and likewise for the other proton. The Collins-Soper scales are defined as
\begin{align}
 \zeta_a = \bigl(x_a P_a^- e^{-y_n} \bigr)^2
\,,\qquad
 \zeta_b = \bigl(x_b P_b^+ e^{+y_n} \bigr)^2
\,,\qquad
 \zeta_a \zeta_b = Q^4
\,,\end{align}
where $x_{a,b}$ and $P_{a,b}$ are the momentum fractions and proton momenta entering the two two TMD PDFs, see \eq{sigma_new}.
To relate these results to the generic notation used in \sec{tmdpdfs_new}, one can identify $1/\tau = \ln(\delta^- e^{-y_n})$.

\tocless\subsection{$\eta$ regulator}

The $\eta$ regulator was introduced by Chiu, Jain, Neill and Rothstein (CJNR) in \cite{Chiu:2011qc,Chiu:2012ir}.
It is defined to modify Wilson lines in momentum space, i.e.\ the Fourier transforms of \eq{Wilson_lines}.
The lightlike Wilson lines entering in the unsubtracted beam function and soft functions, \eqs{beamfunc}{softfunc}, are modified as
\begin{align} \label{eq:eta_regulator}
 f^{0\unsub}_{i/P}: \qquad &
 W_n \to \sum_\text{perms} \exp\biggl[ - g_s w^2 \frac{|\bn \cdot {\cal P}_g|^{-\eta}}{\nu^{-\eta}} \frac{\bn \cdot A_n}{\bn \cdot {\cal P}} \biggr]
 \,,\qquad
 R(k,\eta) = w^2 \biggl|\frac{k^-}{\nu}\biggr|^{-\eta}
\,,\nn\\
 S^0_{n_a n_b}: \qquad &
 W_n \to \sum_\text{perms} \exp\biggl[ - g_s w \frac{|2 {\cal P}_{g3}|^{-\eta/2}}{\nu^{-\eta/2}} \frac{n \cdot A_s}{n \cdot {\cal P}} \biggr]
 \,,\qquad R(k,\eta) = w \biggl|\frac{k^z}{\nu}\biggr|^{-\eta/2}
\,.\end{align}
Here, the function $R(k,\eta)$ shows the resulting regulating factor as entering the examples in \eqs{beam_nlo_5}{soft_nlo_3}.
In \eq{eta_regulator}, the momentum operator ${\cal P}$ picks up the momentum of the gluon fields $A$, and $\eta$ is the rapidity regulator with an associated rapidity scale $\nu$.
The different powers of $\eta$ arise because the soft function involves double the number of Wilson lines than the beam function.

A key feature of this regulator is that rapidity divergences manifest themselves as poles in $1/\eta$ as $\eta\to0$, which can be removed with a counterterm at the cost of leaving a dependence on the ``rapidity scale'' $\nu$.
This is analogous the ultraviolet renormalization, where poles in $1/\eps$ are removed, giving rise to the $\mu$ dependence.
The bookkeeping parameter $w$ in \eq{eta_regulator} plays the role of running coupling, and will be set to $w\to1$ after renormalization.
In this approach, the Collins-Soper evolution is identical to the evolution in $\nu$, and the CS kernel is obtained as the anomalous dimension associated with removing poles in $\eta$.

In the $\eta$ regulator, the soft zero-bin subtraction is absent, as $S^{0\,\subt}_{n_a n_b} = 1$.
In terms of the notation of \sec{TMDforDY}, we have
\begin{align}
\eta = \tau \,,\qquad \zeta = (x P^-)^2 \,, \quad y_n = 0
\,.\end{align}
The choice of fixing $y_n=0$ arises because of the symmetric treatment of the two beam functions, but can be relaxed as in the other definitions if so desired.

Finally, we remark that while the $\eta$ regulator can be used to combine unsubtracted beam and soft functions into the TMD PDF, as in \eq{tmdpdf_1}, it is usually applied such that these functions are renormalized separately, see \eqs{B_renorm}{S_renorm}.
\tocless\subsection{Exponential regulator}

In contrast to the previous regulators, the exponential regulator introduced in \cite{Li:2016axz} does not directly modify the lightlike Wilson lines appearing in \eqs{beamfunc}{softfunc}, but modifies the phase space of each real emissions in the perturbative calculation by a factor
\begin{equation} \label{eq:R_exp_reg}
 R(k,\tau) = \exp\Bigl[ - k^0 \tau e^{-\gamma_E} \Bigr]
\,.\end{equation}
One then takes the $\tau\to0$ limit, keeping only divergent terms.
The individual beam and soft functions obtained in this manner are not $\tau$ independent.
Instead, the $\tau$ evolution is identical to the rapidity RGE of the $\eta$ regulator.
Hence, similar to a Wilsonian approach, the cutoff $\tau$ plays both the role of regulating divergences and being the evolution variable.
Of course, as usual the $\tau$ dependence cancels after combining beam and soft functions into the TMD PDF, exposing the standard CS evolution.

The exponential regulator can also be viewed as extending the unsubtracted TMD PDF and soft functions in position space, which only depend on $(b^+,\bt)$ and $\bt$, respectively, to depend on the full four-vector $b^\mu$.
To be precise, for the soft function one can write
\begin{align} \label{eq:soft_exp_reg}
 S^0_{n_a n_b}(b_T,\eps,\tau) = \lim_{\tau\to0} S^0_{n_a n_b}\bigl(b^+ = i \tau e^{-\gamma_E}, b^- = i \tau e^{-\gamma_E}, b_T, \eps, \tau \Bigr)
\,,\end{align}
where the soft function of the right-hand sides depends on $b^\mu = (b^+,b^-,\bt)$, and one takes the light-cone momenta to zero. Its definition is analogous to that in \eq{softfunc}, but with all Wilson lines ending at $\bt$ now being shifted to $b^\mu$.
A similar equation holds for the unsubtracted TMD PDF, where the matrix element is extended to the $b^+ = i \tau e^{-\gamma_E}$ direction.
In this approach, it is clear that the $\tau$ regulator is equivalent to modifying Wilson lines, and thus by construction is gauge-invariant even before taking the limit $\tau\to0$.

Another advantage of \eq{soft_exp_reg} is that it connects the TMD soft function to the fully-differential soft function, which allowed for the calculation of the soft function using the exponential regulator to three loops~\cite{Li:2016ctv}.
Recently, also the quark beam functions has been calculated at N$^3$LO in this regulator~\cite{Luo:2019szz}, completing the three-loop calculation of the TMD PDF.

\tocless\subsection{Analytic and pure rapidity regulator}

The analytic regulator was first introduced in Becher and Neubert (BN) in \cite{Becher:2010tm} for TMDs and later modified in \cite{Becher:2011dz}.
In the latter formulation, it is implemented by modifying the phase space for all real emissions by
\begin{align} \label{eq:R_anal_reg}
 R(k,\alpha) = \Bigl(\frac{\nu}{k^+}\Bigr)^\alpha
\,,\end{align}
and then letting $\alpha\to0$, which exposes poles in $1/\alpha$.
%
In this approach, the soft function is absent, $S^0_{n_a n_b} \equiv 1$ to all orders in perturbation theory.
Thus, in order to obtain a well-defined TMD PDF, one has to calculate both the $n_a$-collinear and $n_b$-collinear unsubtracted TMD PDFs, which can be combined to obtain the physical TMD PDFs,
\begin{align} \label{eq:BBSBN}
 \lim_{\substack{\eps\to0\\\alpha\to0}} & \Bigl[ f_{q/n_a}^{0 \unsub, \rm BN}(x_1,\bt,\eps,\alpha) f_{q/n_b}^{0 \unsub, \rm BN}(x_2,\bt,\eps,\alpha) \Bigr]
\nn\\ &
 = \biggl(\frac{b_T^2 Q^2}{b_0^2}\biggr)^{-\gamma_\zeta^q(\mu,b_T)} f_{q/n_a}^{\rm BN}(x_1, \bt, \mu) f_{q/n_b}^{\rm BN}(x_2, \bt, \mu)
\,.\end{align}
Note that in this formulation, the TMD PDFs are explicitly independent of $\zeta$, as the combined $\zeta$ dependence is pulled out in the form of the prefactor depending on $Q^2 = \sqrt{\zeta_a \zeta_b}$.
In the language of \cite{Becher:2010tm}, the origin of this factor is attributed to the collinear anomaly, which is equivalent to the occurrence of rapidity divergences.

Since \eq{R_anal_reg} only depends on $k^+$, not on $k^-$, the resulting TMD PDFs $f_{q/n_a}^{0 \unsub, \rm BN}$ and $f_{q/n_b}^{0 \unsub, \rm BN}$ are not symmetric.
A symmetric formulation of \eq{R_anal_reg} was given in \cite{Ebert:2018gsn} as
\begin{align} \label{eq:R_vita_reg}
 R(k,\eta) = w^2 \upsilon^2 \biggl|\frac{k^+}{k^-}\biggr|^{-\eta/2}
\,.\end{align}
This regulator was named ``pure rapidity regulator'', as the combination $y_k = \frac12 \ln(k^+/k^-)$ precisely corresponds to the rapidity of a real emission with momentum $k$.
Employing \eq{R_vita_reg}, one obtains symmetric results for the TMD PDFs in the $n_a$ and $n_b$ direction, and poles in $1/\eta$ as $\eta\to0$ can be renormalized, which yields TMD PDFs to obtain TMD PDFs identical to those of \eq{eta_regulator}.

%
%

%% file: sec-appendix/app-RGkernels.tex
\section{Expansions for evolution kernels}
\label{app:kernelexpansions}




In this appendix we collect formulas for perturbative expansions of the evolution kernels that enter the solution of RGEs and rapidity RGEs in \sec{evolution}. The key anomalous dimensions controlling TMD evolution are the RG anomalous dimension $\gamma_q=\gamma_\mu$ in \eq{RG.TMD.pdf}, rapidity anomalous dimension $\tilde K = \gamma_\zeta^q$ in \eq{CSS.evol}, and the Collins-Soper kernel or cusp anomalous dimension $\gamma_K = 2\GammaC$ in \eq{RG.K}. The RG anomalous dimension takes the form \eq{mu_anom_dim}, containing both a cusp and a ``non-cusp'' part, which have perturbative expansions in $\as$:
\begin{equation}\label{eq:cusp-non}
\GammaC[\as(\mu)] = \sum_{n=0}^\infty \Bigl(\frac{\as(\mu)}{4\pi}\Bigr)^{n+1} \Gamma_n\,,\quad 
    \gamma_\mu[\as(\mu)] = \sum_{n=0}^\infty \Bigl(\frac{\as(\mu)}{4\pi}\Bigr)^{n+1} \gamma_n\,,
\end{equation}
where the coefficients $\Gamma_n,\gamma_n$ are constants. Equivalently,
\begin{equation}
\label{eq:gamma.CSS.expansions}
\gamma_K =\sum_{n=1}^\infty \gamma_K^{(n)} \left(\frac{\alpha_s}{\pi}\right)^n\,,\quad
\gamma_q =\sum_{n=1}^\infty \gamma_q^{(n)} \left(\frac{\alpha_s}{\pi}\right)^n \,.
\end{equation}
Note the difference in indexing by 1 in \eqs{cusp-non}{gamma.CSS.expansions}, an accident of history in the CSS and SCET literatures.
The solutions to the RG and RRGE evolution equations in \eq{TMD.evol} are expressed in terms of integrals over these anomalous dimensions in $\mu$ or $\zeta$, for which we will give explicit expressions below. The accuracy of resummation that is achieved, then, is determined by the accuracy to which these anomalous dimensions \eq{cusp-non} or \eq{gamma.CSS.expansions} are known and the integrals over them are evaluated. We summarized these orders of accuracy in Table~\ref{tbl:resum_orders}; for instance, to next-to-leading-log (NLL) accuracy the required anomalous dimension coefficients are $\Gamma_{0,1}$ and $\gamma_0$, or equivalently, $\gamma_K^{(1,2)}$ and $\gamma_i^{(1)}$.
They are spin-independent~\cite{Collins:1984kg,Qiu:2000ga,Landry:2002ix,Moch:2005id,Kang:2011mr,Aybat:2011zv,Echevarria:2012pw,Grozin:2014hna,Collins:2014jpa},  and are given for quark TMD PDFs by
\bea
 \gamma_K^{(1)} = \frac{\Gamma_0^q}{2} = 2 C_F,
\quad
\gamma_K^{(2)} = \frac{\Gamma_1^q}{8} ={C_F}\left[C_A\left(\frac{67}{18}-\frac{\pi^2}{6}\right)-\frac{10}{9}T_F \, n_f\right],
\quad
\gamma_q^{(1)} = \frac{\gamma_0^{q}}{4} = \frac{3}{2} C_F
\, ,
\eea
where $C_F=4/3$, $C_A=3$, $T_F=1/2$ and $n_f$ is the number of active flavors. Higher-order coefficients and results for gluons TMD PDFs are given below. 

The solutions to the TMD evolution equations \eq{TMD.evol}, developed in \sec{CoordEvol} and \sec{evolution_SCET}, are all expressed in terms of integrals of anomalous dimensions, e.g.,
\begin{align}
\label{eq:Keta}
K_\Gamma(\mu_L,\mu) &= \int_{\mu_L}^\mu \frac{d\mu'}{\mu'} \GammaC[\as(\mu')]\ln\frac{\mu'}{\mu_L} \\
\eta_\Gamma(\mu_L,\mu) &= \int_{\mu_L}^\mu \frac{d\mu'}{\mu'} \GammaC[\as(\mu')] \,,\quad K_\gamma(\mu_L,\mu) = \int_{\mu_L}^\mu \frac{d\mu'}{\mu'} \gamma[\as(\mu')]\,. \nn
\end{align}
Direct integration of these expressions is complicated by the $\mu$ evolution of $\as(\mu)$, which must be taken into account to all orders when there are large logs.

The integrals over $\mu$ in \eq{Keta} can be evaluated nicely in closed form at each order of resummed accuracy by changing integration variables\footnote{See \cite{Bell:2018gce,Billis:2019evv} for commentary on how using \eq{dmu} at a truncated perturbative order may affect explicit RG invariance (i.e. $\mu$ independence) \cite{Bell:2018gce} or numerical accuracy \cite{Billis:2019evv} of a resummed cross section using resulting expansions of \eq{Keta}. See also \cite{Ebert:2021aoo} for a beautiful method to evaluate the integrals over $\alpha_s$ fully analytically with no such truncation errors.
} in \eq{Keta} \cite{Ligeti:2008ac,Abbate:2010xh}:
\begin{equation}
\label{eq:dmu}
\frac{d\mu}{\mu} = \frac{d\as}{\beta[\as]}\,,\qquad \ln\frac{\mu}{\mu_L} = \int_{\as(\mu_L)}^{\as(\mu)}\frac{d\as}{\beta[\as]} \,.
\end{equation}
The $\beta$ function has the expansion
\begin{equation}
\label{eq:beta_expansion}
\beta[\as(\mu)] = -2\as(\mu)\sum_{n=0}^\infty \Bigl(\frac{\as(\mu)}{4\pi}\Bigr)^{n+1} \beta_n\,.
\end{equation}
Using these expansions, we can evaluate \eq{Keta} order by order (see, e.g., \cite{Abbate:2010xh,Almeida:2014uva,Bell:2018gce}). For example, to evaluate $\eta_\Gamma$ in \eq{Keta}, we evaluate:
\begin{align}
\label{eq:eta_integral}
\eta_\Gamma(\mu_L,\mu) &= \int_{\as(\mu_L)}^{\alpha_s(\mu)} \frac{d\alpha_s}{-2\alpha_s}\frac{\Gamma_0 (\as/4\pi) + \Gamma_1(\as/4\pi)^2 + \Gamma_2(\as/4\pi)^3 + \cdots }{\beta_0 (\as/4\pi)+ \beta_1(\as/4\pi)^2+ \beta_2(\as/4\pi)^3 + \cdots} \\
&= -\frac{\Gamma_0}{2\beta_0}\int_{\as(\mu_L)}^{\alpha_s(\mu)} d\alpha_s \biggl[ \frac{1}{\alpha_s} +\Bigl(\frac{\Gamma_1}{\Gamma_0} - \frac{\beta_1}{\beta_0}\Bigr) + \Bigl(\frac{\Gamma_2}{\Gamma_0} + B_2 - \frac{\Gamma_1\beta_1}{\Gamma_0\beta_0} \Bigr)\alpha_s + \cdots\biggr]\,,
\end{align}
up to higher-order terms. This integrand, resulting from Taylor expanding the denominator in \eq{eta_integral}, is truncated at the desired order (but see the footnote above, in particular \cite{Ebert:2021aoo} for a method to avoid such truncation). Performing the integral then yields:
\begin{align}
\label{eq:etaexp}
\eta_\Gamma(\mu_L,\mu) &= -\frac{\Gamma_0}{2\beta_0}\biggl[\ln r + \frac{\as(\mu_L)}{4\pi}\biggl(\frac{\Gamma_1}{\Gamma_0}-\frac{\beta_1}{\beta_0}\biggr)(r-1) + \biggl(\frac{\as(\mu_L)}{4\pi}\biggr)^2 \biggl(B_2 + \frac{\Gamma_2}{\Gamma_0}-\frac{\Gamma_1\beta_1}{\Gamma_0\beta_0}\biggr)\frac{r^2-1}{2} \\
&\qquad\qquad +\biggl(\frac{\as(\mu_L)}{4\pi}\biggr)^3 \biggl(B_3 + \frac{\Gamma_1}{\Gamma_0}B_2  - \frac{\Gamma_2\beta_1}{\Gamma_0\beta_0} + \frac{\Gamma_3}{\Gamma_0}\biggr)\frac{r^3-1}{3} + \cdots\biggr]\,, \nn
\end{align}
the coefficients $B_{2,3}$ are given by
\begin{equation}
B_2 \equiv \frac{\beta_1^2}{\beta_0^2}-\frac{\beta_2}{\beta_0}\,,\qquad B_3 \equiv -\frac{\beta_1^3}{\beta_0^3}+\frac{2\beta_1\beta_2}{\beta_0^2}-\frac{\beta_3}{\beta_0}\,.
\end{equation}
Performing the similar steps for $K_\Gamma$ in \eq{Keta}, which involves one more integral using \eq{dmu} twice, yields
\begin{align}
\label{eq:Kexp}
K_\Gamma(\mu_L,\mu) &=\frac{\Gamma_0}{4\beta_0^2}  \biggl\{ \frac{4\pi}{\as(\mu_L)}\Bigl(\ln r + \frac{1}{r} -1\Bigr) + \biggl(\frac{\Gamma_1}{\Gamma_0} - \frac{\beta_1}{\beta_0}\biggl) (r - 1 -\ln r) - \frac{\beta_1}{2\beta_0}\ln^2 r \\
&+ \frac{\as(\mu_L)}{4\pi} \biggl[ B_2 \Bigl( \frac{r^2 - 1}{2}-\ln r\Bigr) + \biggl(\frac{\beta_1\Gamma_1}{\beta_0\Gamma_0} - \frac{\beta_1^2}{\beta_0^2}\biggr) (r-1-r\ln r) + \biggl(\frac{\Gamma_2}{\Gamma_0}-\frac{\beta_1\Gamma_1}{\beta_0\Gamma_0}\biggr)\frac{(1- r)^2}{2}\biggr] \nn\\
&+ \biggl(\frac{\as(\mu_L)}{4\pi}\biggr)^2 \biggl[ \biggl( \frac{\Gamma_3}{\Gamma_0}-\frac{\Gamma_2\beta_1}{\Gamma_0\beta_0}+\frac{\Gamma_1}{\Gamma_0}B_2 + B_3\biggr) \frac{r^3-1}{3} - \frac{B_3}{2}\ln r - B_2\biggl(\frac{\Gamma_1}{\Gamma_0}-\frac{\beta_1}{\beta_0}\biggr)(r-1)\nn \\
&\quad - \frac{\beta_1}{2\beta_0}\biggl(\frac{\Gamma_2}{\Gamma_0} - \frac{\Gamma_1\beta_1}{\Gamma_0\beta_0} + B_2\biggr)r^2\ln r + \biggl(-\frac{2\Gamma_3}{\Gamma_0} + \frac{3\Gamma_2\beta_1}{\Gamma_0\beta_0} - \frac{\Gamma_1\beta_1^2}{\Gamma_0\beta_0^2} + \frac{\beta_3}{\beta_0}-\frac{\beta_1\beta_2}{\beta_0^2}\biggr)\frac{r^2\minus 1}{4}\biggr] \plus \cdots\biggr\}.\nn
\end{align}
organized in groups of terms of order $1/\as$ (LL), 1 (NLL), $\as$ (NNLL), and $\as^2$ (N$^3$LL) in log counting, with the $\cdots$ denoting terms of higher order. Large logs of $\mu/\mu_L$ are essentially captured in the ratio $r$:
\begin{equation}
    r \equiv r(\mu_L,\mu) = \frac{\as(\mu)}{\as(\mu_L)}\,.
\end{equation}

Note that the terms in $\eta_\Gamma$ in \eq{etaexp} are one power of $\as$ smaller than the corresponding terms in $K_\Gamma$ in \eq{Kexp}, but should be kept to the same corresponding order, i.e. in $\eta_\Gamma$, we keep the $\cO(1)$ term at LL, the $\cO(\as)$ terms at NLL, $\cO(\as^2)$ at NNLL, and $\cO(\as^3)$ at N$^3$LL. This is because of the way the combination of $K_\Gamma$ and $\eta_\Gamma$ appears in the evolution kernels, e.g. \eq{UBS} or \eq{UH}, with $\eta_\Gamma$ always multiplied by another log. See also \cite{Almeida:2014uva}.

The non-cusp kernel $K_\gamma$ in \eq{Keta} has the same expansion as \eq{etaexp} with $\Gamma_n\to \gamma_n$. For $K_\gamma$, the expansion can be truncated according to the standard counting, $\cO(1/\as)$ at LL (in this case, zero), $\cO(1)$ at NLL, $\cO(\as)$ at NNLL, etc.

Finally, we collect expressions for the cusp and non-cusp anomalous dimension coefficients above needed to N$^3$LL accuracy, and beta function coefficients. For the cusp~\cite{Korchemsky:1987wg, Moch:2004pa},
\begin{subequations}
\begin{align}
\Gamma_0 &= 4C_i \\
\Gamma_1 &= 4C_i\Bigl[ \Bigl(\frac{67}{9} - \frac{\pi^2}{3}\Bigr) C_A - \frac{20}{9}T_Fn_f\Bigr] \\
\Gamma_2 &= 4C_i \Bigl[ \Bigl( \frac{245}{6} - \frac{134}{27}\pi^2 + \frac{11}{45}\pi^4 + \frac{22}{3}\zeta_3\Bigr) C_A^2 + \Bigl(-\frac{418}{27} + \frac{40}{27}\pi^2 - \frac{56}{3}\zeta_3\Bigr) C_A T_F n_f \\
& \quad + \Bigl(-\frac{55}{3}+16\zeta_3\Bigr)C_F T_F n_f - \frac{16}{27}T_F^2 n_f^2\Bigr]\,, \nonumber
\end{align}
\end{subequations}
where $C_i = C_{F,A}$ for quark or gluon TMD PDFs. The exact analytic four-loop coefficient $\Gamma_3$ has also been recently obtained in \cite{Henn:2019swt}, while the 5-loop has been obtained approximately in \cite{Herzog:2018kwj}. The non-cusp part of the $\mu$-anomalous dimension of the quark TMD PDF in \eq{SCET.mu.anomdim} has the coefficients
\begin{subequations}
\begin{align}
\gamma_{0}^q &= 6C_F \\
\gamma_1^q &= C_F \Bigl[ \Bigl( \frac{146}{9} - 80\zeta_3\Bigr)C_A + (3-4\pi^2+48\zeta_3)C_F + \Bigl(\frac{121}{9} + \frac{2}{3}\pi^2\Bigr)\beta_0\Bigr]\,,
\end{align}
\end{subequations}
up to two loops \cite{Becher:2006mr}.  Also up to two loops, the coefficients of the non-cusp part of the rapidity anomalous dimension in \eq{gammanuBS} are given by:
\begin{subequations}
\begin{align}
\gamma_{\nu,0} &= 0 \\
\gamma_{\nu,1} &= C_i \Bigl[ \Bigl( \frac{64}{9} - 28\zeta_3\Bigr)C_A +\frac{56}{9}\beta_0\Bigr]\,,
\end{align}
\end{subequations}
where $C_i=C_{F,A}$ for quarks or gluons. The three-loop expression has been found in \cite{Li:2016ctv}, and even the four-loop in \cite{Duhr:2022yyp}.

For completeness we also collect the coefficients of the beta function in \eq{beta_expansion}, in the $\MSbar$ scheme,
\begin{subequations}
\begin{align}
\beta_0 &= \frac{11}{3}C_A - \frac{4}{3}T_F n_f \\
\beta_1 &= \frac{34}{3} C_A^2 - \Bigl(\frac{20}{3}C_A + 4C_F\Bigr) T_F n_f \\
\beta_2 &= \frac{2857}{54} C_A^3 + \Bigl( C_F^2 - \frac{205}{18} C_F C_A - \frac{1415}{54}C_A^2\Bigr) + \Bigl(\frac{11}{9} C_F + \frac{79}{54}C_A\Bigr)4T_F^2 n_f^2 \,,
\end{align}
\end{subequations}
up to three loops~\cite{Tarasov:1980au,Larin:1993tp}. The four-loop expression has been found and confirmed in \cite{vanRitbergen:1997va,Czakon:2004bu}, and even the five-loop in \cite{Baikov:2016tgj}.

%% file: sec-appendix/acronyms.tex
\newpage

\section*{List of acronyms}
\addcontentsline{toc}{section}{List of acronyms}
    \begin{longtable}{ll} 
   
      AdS & Anti-DeSitter \\
      BFKL & Balitsky-Fadin-Kuraev-Lipatov  \\
      BH & Bethe-Heitler \\
      BN & Becher-Neubert \\
      CGC & Color glass condensate \\
      CM & Center of momentum \\
      CJNR & Chiu-Jain-Neill-Rothstein \\
      CSS & Collins-Soper-Sterman \\
      CS & Collins-Soper \\
      DGLAP & Dokshitzer-Gribov-Lipatov-Altarelli-Parisi \\
      DA & Distribution amplitude \\ 
      dof & Degrees of freedom \\
      DVCS & Deeply virtual Compton scattering \\
      DWF & Domain wall fermion \\
      DY & Drell-Yan \\
      EEC & Energy-energy-correlations \\
      EIC & Electron-Ion collider \\
      EIS & Echevarria-Idilbi-Scimemi \\
      EMSTVZ & Ebert-Moult-Stewart-Tackmann-Vita-Zhu \\
      EMT & Energy-momentum tensor \\
      FF & Fragmentation function\\
      FJF & Fragmenting Jet Function \\
      GFIP & Gluon Fragmentation Improved \textsc{Pythia} \\
      GLV & Gyulassy-Levai-Vitev \\
      GPD & Generalized parton distribution function\\
      GTMD & Generalized transverse momentum dependent parton distribution function\\
      ITD & Ioffe time distribution \\
      iTMD & improved TMD \\
      IPD & Impact parameter distribution \\
      JFF & Jet Fragmentation Function \\
      JIMWLK & Jalilian-Marian-Iancu-McLerran-Weigert-Leonidov-Kovner\\
      JM &  Jaffe-Manohar \\
      JMY & Ji-Ma-Yuan \\
      LaMET & large-momentum effective theory \\
      LCWF & Light-cone wave function\\
      LCS & Lattice cross sections \\
      LDME & Long distance matrix elements \\
      LFCM & Light front constituent model \\
      LIR & Lorentz-invariance relation\\
      LL & Leading log\\
      LLx & Leading Log in x \\
      LNZ & Li-Neill-Zhu \\
      LO & Leading order\\
      LPM & Landau-Pomeranchuk-Migdal \\
      LQCD & Lattice Quantum Chromodynamics \\
      MHENS & Musch-H\"agler-Engelhardt-Negele-Sch\"afer\\
      MS ($\rm \overline{MS}$) & Minimal Subtraction (MS-bar) \\
      MV & McLerran-Venugopalan \\
      NLL & Next-to-leading log\\
      NLLx & Next-to-leading log in \\
      NNLL & Next-to-next-to-leading log\\
      N$^3$LL & Next-to-next-to-next-to-leading log\\
      NLO & Next-to-leading order \\
      NNLO & Next-to-next-to-leading order \\
      N3LO & Next-to-next-to-next-to-leading order \\
      NRQCD & Non-relativistic QCD \\
      nTMDs & nuclear TMDs \\
      OAM & Orbital angular momentum \\
      OPE & Operator product expansion\\
      PDF & Parton distribution function\\
      pQCD & perturbative quantum chromodynamics \\
      QCD & quantum chromodynamics \\
      qLIR & Quark model Lorentz invariance relations \\
      QS & Qiu-Sterman (function)\\
      QGP & Quark-gluon plasma \\
      RI/MOM & regularization independent momentum subtraction \\
      RGE & Renormalization group equation \\
      RRGE & Rapidity RGE \\
      SCET & Soft Collinear Effective Theory \\
      SIDIS & Semi-inclusive deep-inelastic scattering\\
      SISCone & SeedlessCone \\
      SSA & Single-spin asymmetry\\
      TEEC & Transverse-energy-energy-correlations \\
      TMD & Tranverse momentum dependent\\
      TMD FJFs & TMD Fragmenting jet functions \\
      TMD PFFs & TMD Polarized fragmentation functions \\
      TMDs & Transverse momentum dependent distributions\\
      UGD & Unintegrated gluon distributions \\

      WW & Wandzura-Wilczek 
      (or 
      Weizs\"acker-Williams, in Ch. 8 only)
    \end{longtable}

%% file: sec-appendix/chaptercontacts.tex
\newpage

\section*{Errata and Chapter Contacts}
\addcontentsline{toc}{section}{Erratum and Chapter Contact Emails}

Please report any typos or errors that you spot in the Handbook in this 
\href{https://docs.google.com/document/d/13tT0f8Ib-pzMRQ4Tdzgtbz6ivFTPc2iKo7EJRI15-X4/edit?usp=sharing}{Google Document}.\\ 

\noindent
For questions and comments on specific chapters in the handbook, which are not just typos, please use the following links:

\begin{itemize}[itemindent=3em] 

\item[Chapter 1:] 
{\href{mailto:qjianw@gmail.com,spf@arizona.edu?subject=TMD Handbook Chapter 1}{Send email to Sean Fleming and Jian-Wei Qiu}}

\item[Chapter 2:]
\href{mailto:iains@mit.edu,ebert@mit.edu,avp5627@psu.edu?subject=TMD Handbook Chapter 2}{Send email to Markus Ebert, Alexei Prokudin, and Iain Stewart} 

\item[Chapter 3:]
\href{mailto:iains@mit.edu,qjianw@gmail.com?subject=TMD Handbook Chapter 3}{Send email to Iain Stewart}

\item[Chapter 4:]
\href{mailto:lpg10@psu.edu,clee@lanl.gov,mehen@phy.duke.edu?subject=TMD Handbook Chapter 4}{Send email to Leonard Gamberg, Christopher Lee, and Thomas Mehen}

\item[Chapter 5:] \raisebox{-0.26cm}{\parbox{5.2in}{
\href{mailto:avp5627@psu.edu,pitonyak@lvc.edu,marc.schlegel@uni-tuebingen.de,zkang@g.ucla.edu?subject=TMD Handbook Chapter 5}{Send email to Zhong-Bo Kang, Daniel Pitonyak, Alexei Prokudin, and Marc Schlegel}
}}

\item[Chapter 6:] \raisebox{-0.26cm}{\parbox{5.5in}{
\href{mailto:wdetmold@mit.edu,marthac@temple.edu,yong.zhao@anl.gov,engel@physics.nmsu.edu,liu@g.uky.edu,pshana@mit.edu?subject=TMD Handbook Chapter 6}{Send email to Martha Constantinou, William Detmold, Michael Engelhardt, Keh-Fei Liu, Phiala Shanahan, and Yong Zhao}
}}

\item[Chapter 7:]
\href{mailto:peter.schweitzer@uconn.edu,andreas.metz@temple.edu,mburkardt@physics.nmsu.edu?subject=TMD Handbook Chapter 7}{Send email to Matthias Burkardt, Andreas Metz, and Peter Schweitzer}

\item[Chapter 8:] \raisebox{-0.26cm}{\parbox{5.5in}{
\href{mailto:renaud.boussarie@gmail.com,fyuan@lbl.gov,raju.venugopalan@gmail.com,tarasov.3@osu.edu?subject=TMD Handbook Chapter 8}{Send email to Renaud Boussarie, Andrey Tarasov, Raju Venugopalan, and Feng Yuan}
}}

\item[Chapter 9:]
\href{mailto:mehen@phy.duke.edu,ivitev@lanl.gov,zkang@g.ucla.edu?subject=TMD Handbook General Comment}{Send email to Zhong-Bo Kang, Thomas Mehen, and Ivan Vitev}

\item[Chapter 10:]\raisebox{-0.26cm}{\parbox{5.2in}{
\href{mailto:andreas.metz@temple.edu,lpg10@psu.edu,marc.schlegel@uni-tuebingen.de,iains@mit.edu?subject=TMD Handbook Chapter 10}{Send email to Leonard Gamberg, Andreas Metz, Marc Schlegel, and Iain Stewart}
}}

\item[Chapter 11:] \raisebox{-0.26cm}{\parbox{5.2in}{
\href{mailto:sl4y@virginia.edu,andreas.metz@temple.edu,engel@physics.nmsu.edu,fyuan@lbl.gov?subject=TMD Handbook Chapter 11}{Send email to Michael Engelhardt, Simonetta Liuti, Andreas Metz, and Feng Yuan}
}}

\item[General:] 
\href{mailto:mehen@phy.duke.edu,iains@mit.edu?subject=TMD Handbook General Comment}{Send email to Thomas Mehen and Iain Stewart}

\end{itemize}

%% file: tmdbook.bbl
\providecommand{\href}[2]{#2}\begingroup\raggedright\begin{thebibliography}{1000}

\bibitem{Brambilla:2014jmp}
N.~Brambilla et~al., \emph{{QCD and Strongly Coupled Gauge Theories: Challenges
  and Perspectives}}, {\emph{Eur. Phys. J.} {\bfseries C74} (2014) 2981}
  [\href{https://arxiv.org/abs/1404.3723}{{\ttfamily 1404.3723}}].

\bibitem{Gross:1973id}
D.~J. Gross and F.~Wilczek, \emph{{Ultraviolet Behavior of Nonabelian Gauge
  Theories}}, {\emph{Phys. Rev. Lett.} {\bfseries 30} (1973) 1343}.

\bibitem{Politzer:1973fx}
H.~D. Politzer, \emph{{Reliable Perturbative Results for Strong
  Interactions?}}, {\emph{Phys. Rev. Lett.} {\bfseries 30} (1973) 1346}.

\bibitem{Collins:1989gx}
J.~C. Collins, D.~E. Soper and G.~F. Sterman, \emph{{Factorization of Hard
  Processes in QCD}}, {\emph{Adv. Ser. Direct. High Energy Phys.} {\bfseries 5}
  (1989) 1} [\href{https://arxiv.org/abs/hep-ph/0409313}{{\ttfamily
  hep-ph/0409313}}].

\bibitem{Accardi:2012qut}
A.~Accardi et~al., \emph{{Electron Ion Collider: The Next QCD Frontier}},
  {\emph{Eur. Phys. J.} {\bfseries A52} (2016) 268}
  [\href{https://arxiv.org/abs/1212.1701}{{\ttfamily 1212.1701}}].

\bibitem{muon-Times}
M.~Bartusiak, \emph{{Who ordered the Muon?}},
  \href{https://arxiv.org/abs/{https://www.nytimes.com/1987/09/27/books/science-technology-who-ordered-the-muon.html}}{{\ttfamily
  {https://www.nytimes.com/1987/09/27/books/science-technology-who-ordered-the-muon.html}}}.

\bibitem{Bloom:1969kc}
E.~D. Bloom et~al., \emph{{High-Energy Inelastic e p Scattering at 6-Degrees
  and 10-Degrees}}, {\emph{Phys. Rev. Lett.} {\bfseries 23} (1969) 930}.

\bibitem{Feynman:1969ej}
R.~P. Feynman, \emph{{Very high-energy collisions of hadrons}}, {\emph{Phys.
  Rev. Lett.} {\bfseries 23} (1969) 1415}.

\bibitem{Drell:1970wh}
S.~Drell and T.-M. Yan, \emph{{Massive Lepton Pair Production in Hadron-Hadron
  Collisions at High-Energies}}, {\emph{Phys. Rev. Lett.} {\bfseries 25} (1970)
  316}.

\bibitem{Berger:1982vs}
E.~L. Berger, \emph{{Massive Lepton Pair Production--What has QCD Done to the
  Classical Drell-Yan Model?}}, {\emph{AIP Conf. Proc.} {\bfseries 98} (1983)
  312}.

\bibitem{Collins:2011zzd}
J.~Collins, \emph{{Foundations of perturbative QCD}}, Cambridge monographs on
  particle physics, nuclear physics, and cosmology. Cambridge Univ. Press, New
  York, NY, 2011.

\bibitem{Constantinou:2020hdm}
M.~Constantinou et~al., \emph{{Parton distributions and lattice-QCD
  calculations: Toward 3D structure}}, {\emph{Prog. Part. Nucl. Phys.}
  {\bfseries 121} (2021) 103908}
  [\href{https://arxiv.org/abs/2006.08636}{{\ttfamily 2006.08636}}].

\bibitem{She:2009jq}
J.~She, J.~Zhu and B.-Q. Ma, \emph{{Pretzelosity $h_{1T}^\perp$ and quark
  orbital angular momentum}}, {\emph{Phys. Rev.} {\bfseries D79} (2009) 054008}
  [\href{https://arxiv.org/abs/0902.3718}{{\ttfamily 0902.3718}}].

\bibitem{Avakian:2010br}
H.~Avakian, A.~V. Efremov, P.~Schweitzer and F.~Yuan, \emph{{The transverse
  momentum dependent distribution functions in the bag model}}, {\emph{Phys.
  Rev.} {\bfseries D81} (2010) 074035}
  [\href{https://arxiv.org/abs/1001.5467}{{\ttfamily 1001.5467}}].

\bibitem{Efremov:2010cy}
A.~Efremov, P.~Schweitzer, O.~Teryaev and P.~Zavada, \emph{{Images of Quark
  Intrinsic Motion in Covariant Parton Model}}, {\emph{PoS} {\bfseries DIS2010}
  (2010) 253} [\href{https://arxiv.org/abs/1008.3827}{{\ttfamily 1008.3827}}].

\bibitem{Ji:2004wu}
X.-d. Ji, J.-p. Ma and F.~Yuan, \emph{{QCD factorization for semi-inclusive
  deep-inelastic scattering at low transverse momentum}}, {\emph{Phys.Rev.}
  {\bfseries D71} (2005) 034005}
  [\href{https://arxiv.org/abs/hep-ph/0404183}{{\ttfamily hep-ph/0404183}}].

\bibitem{Collins:1981va}
J.~C. Collins and D.~E. Soper, \emph{{Back-To-Back Jets: Fourier Transform from
  $b$ to $k_T$}}, {\emph{Nucl.Phys.} {\bfseries B197} (1982) 446}.

\bibitem{Cammarota:2020qcw}
{\scshape Jefferson Lab Angular Momentum} collaboration, J.~Cammarota,
  L.~Gamberg, Z.-B. Kang, J.~A. Miller, D.~Pitonyak, A.~Prokudin et~al.,
  \emph{{Origin of single transverse-spin asymmetries in high-energy
  collisions}}, {\emph{Phys. Rev. D} {\bfseries 102} (2020) 054002}
  [\href{https://arxiv.org/abs/2002.08384}{{\ttfamily 2002.08384}}].

\bibitem{Bacchetta:2004jz}
A.~Bacchetta, U.~D'Alesio, M.~Diehl and C.~Miller, \emph{{Single-spin
  asymmetries: The Trento conventions}}, {\emph{Phys. Rev. D} {\bfseries 70}
  (2004) 117504} [\href{https://arxiv.org/abs/hep-ph/0410050}{{\ttfamily
  hep-ph/0410050}}].

\bibitem{Liu:2020rvc}
T.~Liu, W.~Melnitchouk, J.-W. Qiu and N.~Sato, \emph{{Factorized approach to
  radiative corrections for inelastic lepton-hadron collisions}}, {\emph{Phys.
  Rev. D} {\bfseries 104} (2021) 094033}
  [\href{https://arxiv.org/abs/2008.02895}{{\ttfamily 2008.02895}}].

\bibitem{Liu:2021jfp}
T.~Liu, W.~Melnitchouk, J.-W. Qiu and N.~Sato, \emph{{A new approach to
  semi-inclusive deep-inelastic scattering with QED and QCD factorization}},
  {\emph{JHEP} {\bfseries 11} (2021) 157}
  [\href{https://arxiv.org/abs/2108.13371}{{\ttfamily 2108.13371}}].

\bibitem{Aoki:2021kgd}
Y.~Aoki et~al., \emph{{FLAG Review 2021}},
  \href{https://arxiv.org/abs/2111.09849}{{\ttfamily 2111.09849}}.

\bibitem{Liu:1993cv}
K.-F. Liu and S.-J. Dong, \emph{{Origin of difference between anti-d and anti-u
  partons in the nucleon}},
  \href{https://doi.org/10.1103/PhysRevLett.72.1790}{\emph{Phys. Rev. Lett.}
  {\bfseries 72} (1994) 1790}
  [\href{https://arxiv.org/abs/hep-ph/9306299}{{\ttfamily hep-ph/9306299}}].

\bibitem{Liu:1999ak}
K.-F. Liu, \emph{{Parton degrees of freedom from the path integral formalism}},
  {\emph{Phys.Rev.} {\bfseries D62} (2000) 074501}
  [\href{https://arxiv.org/abs/hep-ph/9910306}{{\ttfamily hep-ph/9910306}}].

\bibitem{Liu:2016djw}
K.-F. Liu, \emph{{Parton Distribution Function from the Hadronic Tensor on the
  Lattice}}, \href{https://doi.org/10.22323/1.251.0115}{\emph{PoS} {\bfseries
  LATTICE2015} (2016) 115} [\href{https://arxiv.org/abs/1603.07352}{{\ttfamily
  1603.07352}}].

\bibitem{Aglietti:1998ur}
U.~Aglietti, M.~Ciuchini, G.~Corbo, E.~Franco, G.~Martinelli and
  L.~Silvestrini, \emph{{Model independent determination of the light cone wave
  functions for exclusive processes}}, {\emph{Phys. Lett.} {\bfseries B441}
  (1998) 371} [\href{https://arxiv.org/abs/hep-ph/9806277}{{\ttfamily
  hep-ph/9806277}}].

\bibitem{Abada:2001if}
A.~Abada, P.~Boucaud, G.~Herdoiza, J.~P. Leroy, J.~Micheli, O.~Pene et~al.,
  \emph{{Preliminaries on a lattice analysis of the pion light cone wave
  function: A Partonic signal?}}, {\emph{Phys. Rev. D} {\bfseries 64} (2001)
  074511} [\href{https://arxiv.org/abs/hep-ph/0105221}{{\ttfamily
  hep-ph/0105221}}].

\bibitem{Detmold:2005gg}
W.~Detmold and C.~D. Lin, \emph{{Deep-inelastic scattering and the operator
  product expansion in lattice QCD}}, {\emph{Phys.Rev.} {\bfseries D73} (2006)
  014501} [\href{https://arxiv.org/abs/hep-lat/0507007}{{\ttfamily
  hep-lat/0507007}}].

\bibitem{Braun:2007wv}
V.~Braun and D.~M{\"u}ller, \emph{{Exclusive processes in position space and
  the pion distribution amplitude}}, {\emph{Eur. Phys. J.} {\bfseries C55}
  (2008) 349} [\href{https://arxiv.org/abs/0709.1348}{{\ttfamily 0709.1348}}].

\bibitem{Musch:2010ka}
B.~U. Musch, P.~H\"agler, J.~W. Negele and A.~Sch\"afer, \emph{{Exploring quark
  transverse momentum distributions with lattice QCD}}, {\emph{Phys.Rev.}
  {\bfseries D83} (2011) 094507}
  [\href{https://arxiv.org/abs/1011.1213}{{\ttfamily 1011.1213}}].

\bibitem{Ji:2013dva}
X.~Ji, \emph{{Parton Physics on a Euclidean Lattice}}, {\emph{Phys. Rev. Lett.}
  {\bfseries 110} (2013) 262002}
  [\href{https://arxiv.org/abs/1305.1539}{{\ttfamily 1305.1539}}].

\bibitem{Ji:2014gla}
X.~Ji, \emph{{Parton Physics from Large-Momentum Effective Field Theory}},
  {\emph{Sci.China Phys.Mech.Astron.} {\bfseries 57} (2014) 1407}
  [\href{https://arxiv.org/abs/1404.6680}{{\ttfamily 1404.6680}}].

\bibitem{Ji:2020ect}
X.~Ji, Y.-S. Liu, Y.~Liu, J.-H. Zhang and Y.~Zhao, \emph{{Large-momentum
  effective theory}},
  \href{https://doi.org/10.1103/RevModPhys.93.035005}{\emph{Rev. Mod. Phys.}
  {\bfseries 93} (2021) 035005}
  [\href{https://arxiv.org/abs/2004.03543}{{\ttfamily 2004.03543}}].

\bibitem{Radyushkin:2017cyf}
A.~V. Radyushkin, \emph{{Quasi-parton distribution functions, momentum
  distributions, and pseudo-parton distribution functions}}, {\emph{Phys. Rev.}
  {\bfseries D96} (2017) 034025}
  [\href{https://arxiv.org/abs/1705.01488}{{\ttfamily 1705.01488}}].

\bibitem{Chambers:2017dov}
A.~J. Chambers, R.~Horsley, Y.~Nakamura, H.~Perlt, P.~E.~L. Rakow,
  G.~Schierholz et~al., \emph{{Nucleon Structure Functions from Operator
  Product Expansion on the Lattice}}, {\emph{Phys. Rev. Lett.} {\bfseries 118}
  (2017) 242001} [\href{https://arxiv.org/abs/1703.01153}{{\ttfamily
  1703.01153}}].

\bibitem{Ma:2017pxb}
Y.-Q. Ma and J.-W. Qiu, \emph{{Exploring Partonic Structure of Hadrons Using ab
  initio Lattice QCD Calculations}}, {\emph{Phys. Rev. Lett.} {\bfseries 120}
  (2018) 022003} [\href{https://arxiv.org/abs/1709.03018}{{\ttfamily
  1709.03018}}].

\bibitem{Ma:2014jla}
Y.-Q. Ma and J.-W. Qiu, \emph{{Extracting Parton Distribution Functions from
  Lattice QCD Calculations}}, {\emph{Phys. Rev.} {\bfseries D98} (2018) 074021}
  [\href{https://arxiv.org/abs/1404.6860}{{\ttfamily 1404.6860}}].

\bibitem{Brodsky:2002cx}
S.~J. Brodsky, D.~S. Hwang and I.~Schmidt, \emph{Final-state interactions and
  single-spin asymmetries in semi-inclusive deep inelastic scattering},
  {\emph{Phys. Lett.} {\bfseries B530} (2002) 99}
  [\href{https://arxiv.org/abs/hep-ph/0201296}{{\ttfamily hep-ph/0201296}}].

\bibitem{Gardiner:1970wy}
C.~W. Gardiner and D.~P. Majumdar, \emph{{Effect of a transverse momentum
  distribution in the parton model}}, {\emph{Phys. Rev. D} {\bfseries 2} (1970)
  2040}.

\bibitem{Gribov:1972ri}
V.~N. Gribov and L.~N. Lipatov, \emph{{Deep inelastic e p scattering in
  perturbation theory}}, {\emph{Sov. J. Nucl. Phys.} {\bfseries 15} (1972)
  438}.

\bibitem{Lipatov:1974qm}
L.~N. Lipatov, \emph{{The parton model and perturbation theory}}, {\emph{Sov.
  J. Nucl. Phys.} {\bfseries 20} (1975) 94}.

\bibitem{Altarelli:1977zs}
G.~Altarelli and G.~Parisi, \emph{{Asymptotic Freedom in Parton Language}},
  {\emph{Nucl. Phys.} {\bfseries B126} (1977) 298}.

\bibitem{Dokshitzer:1977sg}
Y.~L. Dokshitzer, \emph{{Calculation of the Structure Functions for Deep
  Inelastic Scattering and $e^+e^-$ Annihilation by Perturbation Theory in
  Quantum Chromodynamics.}}, {\emph{Sov. Phys. JETP} {\bfseries 46} (1977)
  641}.

\bibitem{Collins:2003fm}
J.~C. Collins, \emph{What exactly is a parton density?}, {\emph{Acta Phys.
  Polon.} {\bfseries B34} (2003) 3103}
  [\href{https://arxiv.org/abs/hep-ph/0304122}{{\ttfamily hep-ph/0304122}}].

\bibitem{Soper:1979fq}
D.~E. Soper, \emph{{Partons and Their Transverse Momenta in {QCD}}},
  {\emph{Phys. Rev. Lett.} {\bfseries 43} (1979) 1847}.

\bibitem{Ralston:1980pp}
J.~P. Ralston and D.~E. Soper, \emph{{D}rell-{Y}an model at measured {$Q_T$}:
  Asymptotic smallness of one loop corrections}, {\emph{Nucl. Phys.} {\bfseries
  B172} (1980) 445}.

\bibitem{Collins:1980ih}
J.~C. Collins, \emph{Algorithm to compute corrections to the {S}udakov
  form-factor}, {\emph{Phys. Rev.} {\bfseries D22} (1980) 1478}.

\bibitem{Collins:1981tt}
J.~C. Collins, D.~E. Soper and G.~Sterman, \emph{Does the {D}rell-{Y}an
  cross-section factorize?}, {\emph{Phys. Lett.} {\bfseries B109} (1982) 388}.

\bibitem{Collins:1982wa}
J.~C. Collins, D.~E. Soper and G.~F. Sterman, \emph{{Factorization for One Loop
  Corrections in the {Drell-Yan} Process}}, {\emph{Nucl. Phys.} {\bfseries
  B223} (1983) 381}.

\bibitem{Belitsky:2002sm}
A.~V. Belitsky, X.~Ji and F.~Yuan, \emph{{Final state interactions and gauge
  invariant parton distributions}}, {\emph{Nucl. Phys.} {\bfseries B656} (2003)
  165} [\href{https://arxiv.org/abs/hep-ph/0208038}{{\ttfamily
  hep-ph/0208038}}].

\bibitem{Sivers:1990cc}
D.~W. Sivers, \emph{Single spin production asymmetries from the hard scattering
  of point-like constituents}, {\emph{Phys. Rev.} {\bfseries D41} (1990) 83}.

\bibitem{Klem:1976ui}
R.~Klem, J.~Bowers, H.~Courant, H.~Kagan, M.~Marshak, E.~Peterson et~al.,
  \emph{{Measurement of Asymmetries of Inclusive Pion Production in Proton
  Proton Interactions at 6 GeV/c and 11.8 GeV/c}}, {\emph{Phys. Rev. Lett.}
  {\bfseries 36} (1976) 929}.

\bibitem{Dragoset:1978gg}
W.~Dragoset, J.~Roberts, J.~Bowers, H.~Courant, H.~Kagan et~al.,
  \emph{{Asymmetries in Inclusive Proton-Nucleon Scattering at 11.75 GeV/c}},
  \href{https://doi.org/10.1103/PhysRevD.18.3939}{\emph{Phys.Rev.} {\bfseries
  D18} (1978) 3939}.

\bibitem{Antille:1980th}
J.~Antille, L.~Dick, L.~Madansky, D.~Perret-Gallix, M.~Werlen et~al.,
  \emph{{spin dependence of the inclusive reaction p p (polarized) $\to \pi^0
  X$ at 24 GeV/c for high-$p_T$ $\pi^0$ produced in the central region}},
  \href{https://doi.org/10.1016/0370-2693(80)90933-8}{\emph{Phys.Lett.}
  {\bfseries B94} (1980) 523}.

\bibitem{Apokin:1990ik}
V.~Apokin, Y.~Arestov, O.~Astafev, N.~Belikov, B.~Chuiko et~al.,
  \emph{{Observation of significant spin effects in hard collisions at 40
  GeV/c}},
  \href{https://doi.org/10.1016/0370-2693(90)91414-7}{\emph{Phys.Lett.}
  {\bfseries B243} (1990) 461}.

\bibitem{Saroff:1989gn}
S.~Saroff, B.~Baller, G.~Blazey, H.~Courant, K.~J. Heller et~al., \emph{{Single
  spin asymmetry in inclusive reactions: $p_\uparrow p\to \pi^+ + X$, $\pi^- +
  X$, and $p + X$ at 13.3 GeV/c and 18.5 GeV/c}},
  \href{https://doi.org/10.1103/PhysRevLett.64.995}{\emph{Phys.Rev.Lett.}
  {\bfseries 64} (1990) 995}.

\bibitem{Adams:1991rw}
{\scshape E581 Collaboration, E704 Collaboration} collaboration, D.~Adams
  et~al., \emph{{Comparison of spin asymmetries and cross-sections in pi0
  production by 200 GeV polarized anti-protons and protons}},
  \href{https://doi.org/10.1016/0370-2693(91)91351-U}{\emph{Phys.Lett.}
  {\bfseries B261} (1991) 201}.

\bibitem{Adams:1991cs}
{\scshape FNAL-E704} collaboration, D.~L. Adams et~al., \emph{{Analyzing power
  in inclusive pi+ and pi- production at high $x_F$ with a 200 GeV polarized
  proton beam}},
  \href{https://doi.org/10.1016/0370-2693(91)90378-4}{\emph{Phys. Lett.}
  {\bfseries B264} (1991) 462}.

\bibitem{Aidala:2012mv}
C.~A. Aidala, S.~D. Bass, D.~Hasch and G.~K. Mallot, \emph{{The Spin Structure
  of the Nucleon}}, {\emph{Rev. Mod. Phys.} {\bfseries 85} (2013) 655}
  [\href{https://arxiv.org/abs/1209.2803}{{\ttfamily 1209.2803}}].

\bibitem{Chen:2012taa}
J.-P. Chen, \emph{{QCD evolution and TMD/Spin experiments}},
  {\emph{Int.J.Mod.Phys.Conf.Ser.} {\bfseries 20} (2012) 45}.

\bibitem{Collins:1992kk}
J.~C. Collins, \emph{{Fragmentation of transversely polarized quarks probed in
  transverse momentum distributions}}, {\emph{Nucl.Phys.} {\bfseries B396}
  (1993) 161} [\href{https://arxiv.org/abs/hep-ph/9208213}{{\ttfamily
  hep-ph/9208213}}].

\bibitem{Collins:2002kn}
J.~C. Collins, \emph{{Leading twist single transverse-spin asymmetries:
  Drell-Yan and deep inelastic scattering}}, {\emph{Phys.Lett.} {\bfseries
  B536} (2002) 43} [\href{https://arxiv.org/abs/hep-ph/0204004}{{\ttfamily
  hep-ph/0204004}}].

\bibitem{Boer:1997nt}
D.~Boer and P.~J. Mulders, \emph{{Time reversal odd distribution functions in
  leptoproduction}}, {\emph{Phys. Rev.} {\bfseries D57} (1998) 5780}
  [\href{https://arxiv.org/abs/hep-ph/9711485}{{\ttfamily hep-ph/9711485}}].

\bibitem{Tangerman:1994eh}
R.~D. Tangerman and P.~J. Mulders, \emph{{Intrinsic transverse momentum and the
  polarized Drell-Yan process}}, {\emph{Phys. Rev.} {\bfseries D51} (1995)
  3357} [\href{https://arxiv.org/abs/hep-ph/9403227}{{\ttfamily
  hep-ph/9403227}}].

\bibitem{Mulders:1996dh}
P.~J. Mulders and R.~D. Tangerman, \emph{The complete tree-level result up to
  order {$1/Q$} for polarized deep-inelastic leptoproduction}, {\emph{Nucl.
  Phys.} {\bfseries B461} (1996) 197}
  [\href{https://arxiv.org/abs/hep-ph/9510301}{{\ttfamily hep-ph/9510301}}].

\bibitem{Collins:1984kg}
J.~C. Collins, D.~E. Soper and G.~Sterman, \emph{{Transverse Momentum
  Distribution in Drell-Yan Pair and W and Z Boson Production}}, {\emph{Nucl.
  Phys.} {\bfseries B250} (1985) 199}.

\bibitem{deFlorian:2000pr}
D.~de~Florian and M.~Grazzini, \emph{{Next-to-next-to-leading logarithmic
  corrections at small transverse momentum in hadronic collisions}},
  {\emph{Phys. Rev. Lett.} {\bfseries 85} (2000) 4678}
  [\href{https://arxiv.org/abs/hep-ph/0008152}{{\ttfamily hep-ph/0008152}}].

\bibitem{deFlorian:2001zd}
D.~de~Florian and M.~Grazzini, \emph{{The Structure of large logarithmic
  corrections at small transverse momentum in hadronic collisions}},
  {\emph{Nucl. Phys.} {\bfseries B616} (2001) 247}
  [\href{https://arxiv.org/abs/hep-ph/0108273}{{\ttfamily hep-ph/0108273}}].

\bibitem{Catani:2000vq}
S.~Catani, D.~de~Florian and M.~Grazzini, \emph{{Universality of nonleading
  logarithmic contributions in transverse momentum distributions}},
  {\emph{Nucl.Phys.} {\bfseries B596} (2001) 299}
  [\href{https://arxiv.org/abs/hep-ph/0008184}{{\ttfamily hep-ph/0008184}}].

\bibitem{Catani:2009sm}
S.~Catani, L.~Cieri, G.~Ferrera, D.~de~Florian and M.~Grazzini, \emph{{Vector
  boson production at hadron colliders: a fully exclusive QCD calculation at
  NNLO}}, {\emph{Phys. Rev. Lett.} {\bfseries 103} (2009) 082001}
  [\href{https://arxiv.org/abs/0903.2120}{{\ttfamily 0903.2120}}].

\bibitem{Catani:2012qa}
S.~Catani, L.~Cieri, D.~de~Florian, G.~Ferrera and M.~Grazzini, \emph{{Vector
  boson production at hadron colliders: hard-collinear coefficients at the
  NNLO}}, {\emph{Eur. Phys. J.} {\bfseries C72} (2012) 2195}
  [\href{https://arxiv.org/abs/1209.0158}{{\ttfamily 1209.0158}}].

\bibitem{Bozzi:2010xn}
G.~Bozzi, S.~Catani, G.~Ferrera, D.~de~Florian and M.~Grazzini,
  \emph{{Production of Drell-Yan lepton pairs in hadron collisions:
  Transverse-momentum resummation at next-to-next-to-leading logarithmic
  accuracy}}, {\emph{Phys. Lett.} {\bfseries B696} (2011) 207}
  [\href{https://arxiv.org/abs/1007.2351}{{\ttfamily 1007.2351}}].

\bibitem{Catani:2013tia}
S.~Catani, L.~Cieri, D.~de~Florian, G.~Ferrera and M.~Grazzini,
  \emph{{Universality of transverse-momentum resummation and hard factors at
  the NNLO}}, {\emph{Nucl. Phys.} {\bfseries B881} (2014) 414}
  [\href{https://arxiv.org/abs/1311.1654}{{\ttfamily 1311.1654}}].

\bibitem{Collins:1999dz}
J.~C. Collins and F.~Hautmann, \emph{Infrared divergences and non-lightlike
  eikonal lines in {S}udakov processes}, {\emph{Phys. Lett.} {\bfseries B472}
  (2000) 129} [\href{https://arxiv.org/abs/hep-ph/9908467}{{\ttfamily
  hep-ph/9908467}}].

\bibitem{Boer:2003cm}
D.~Boer, P.~J. Mulders and F.~Pijlman, \emph{Universality of t-odd effects in
  single spin and azimuthal asymmetries}, {\emph{Nucl. Phys.} {\bfseries B667}
  (2003) 201} [\href{https://arxiv.org/abs/hep-ph/0303034}{{\ttfamily
  hep-ph/0303034}}].

\bibitem{Bomhof:2004aw}
C.~Bomhof, P.~Mulders and F.~Pijlman, \emph{{Gauge link structure in
  quark-quark correlators in hard processes}}, {\emph{Phys.Lett.} {\bfseries
  B596} (2004) 277} [\href{https://arxiv.org/abs/hep-ph/0406099}{{\ttfamily
  hep-ph/0406099}}].

\bibitem{Ji:2004xq}
X.-d. Ji, J.-P. Ma and F.~Yuan, \emph{{QCD factorization for spin-dependent
  cross sections in DIS and Drell-Yan processes at low transverse momentum}},
  {\emph{Phys. Lett. B} {\bfseries 597} (2004) 299}
  [\href{https://arxiv.org/abs/hep-ph/0405085}{{\ttfamily hep-ph/0405085}}].

\bibitem{Bomhof:2006dp}
C.~Bomhof, P.~Mulders and F.~Pijlman, \emph{{The Construction of gauge-links in
  arbitrary hard processes}}, {\emph{Eur.Phys.J.} {\bfseries C47} (2006) 147}
  [\href{https://arxiv.org/abs/hep-ph/0601171}{{\ttfamily hep-ph/0601171}}].

\bibitem{Cherednikov:2008uk}
I.~Cherednikov and N.~Stefanis, \emph{{New results on gauge-invariant TMD PDFs
  in QCD}},  \href{https://arxiv.org/abs/0809.1315}{{\ttfamily 0809.1315}}.

\bibitem{Cherednikov:2007tw}
I.~Cherednikov and N.~Stefanis, \emph{Renormalization, {W}ilson lines, and
  transverse-momentum dependent parton distribution functions}, {\emph{Phys.
  Rev.} {\bfseries D77} (2008) 094001}
  [\href{https://arxiv.org/abs/0710.1955}{{\ttfamily 0710.1955}}].

\bibitem{Cherednikov:2008ua}
I.~Cherednikov and N.~Stefanis, \emph{Wilson lines and transverse-momentum
  dependent parton distribution functions: {A} renormalization-group analysis},
  {\emph{Nucl. Phys.} {\bfseries B802} (2008) 146}
  [\href{https://arxiv.org/abs/0802.2821}{{\ttfamily 0802.2821}}].

\bibitem{Hautmann:2007uw}
F.~Hautmann, \emph{{Endpoint singularities in unintegrated parton
  distributions}},
  \href{https://doi.org/10.1016/j.physletb.2007.08.081}{\emph{Phys.Lett.}
  {\bfseries B655} (2007) 26}
  [\href{https://arxiv.org/abs/hep-ph/0702196}{{\ttfamily hep-ph/0702196}}].

\bibitem{Collins:2008ht}
J.~Collins, \emph{{Rapidity divergences and valid definitions of parton
  densities}}, {\emph{PoS} {\bfseries LC2008} (2008) 028}
  [\href{https://arxiv.org/abs/0808.2665}{{\ttfamily 0808.2665}}].

\bibitem{Hautmann:2009zzb}
F.~Hautmann, \emph{{Unintegrated parton distributions and applications to jet
  physics}}, {\emph{Acta Phys.Polon.} {\bfseries B40} (2009) 2139}.

\bibitem{Aybat:2011zv}
S.~Aybat and T.~C. Rogers, \emph{{TMD Parton Distribution and Fragmentation
  Functions with QCD Evolution}}, {\emph{Phys.Rev.} {\bfseries D83} (2011)
  114042} [\href{https://arxiv.org/abs/1101.5057}{{\ttfamily 1101.5057}}].

\bibitem{Ebert:2019zkb}
M.~A. Ebert and F.~J. Tackmann, \emph{{Impact of isolation and fiducial cuts on
  q$_{T}$ and N-jettiness subtractions}}, {\emph{JHEP} {\bfseries 03} (2020)
  158} [\href{https://arxiv.org/abs/1911.08486}{{\ttfamily 1911.08486}}].

\bibitem{Ebert:2020dfc}
M.~A. Ebert, J.~K.~L. Michel, I.~W. Stewart and F.~J. Tackmann,
  \emph{{Drell-Yan $q_{T}$ resummation of fiducial power corrections at
  N$^{3}$LL}}, \href{https://doi.org/10.1007/JHEP04(2021)102}{\emph{JHEP}
  {\bfseries 04} (2021) 102}
  [\href{https://arxiv.org/abs/2006.11382}{{\ttfamily 2006.11382}}].

\bibitem{Collins:1981uk}
J.~C. Collins and D.~E. Soper, \emph{{Back-To-Back Jets in QCD}}, {\emph{Nucl.
  Phys.} {\bfseries B193} (1981) 381}.

\bibitem{Collins:1985ue}
J.~C. Collins, D.~E. Soper and G.~F. Sterman, \emph{{Factorization for Short
  Distance Hadron - Hadron Scattering}}, {\emph{Nucl.Phys.} {\bfseries B261}
  (1985) 104}.

\bibitem{Collins:1988ig}
J.~C. Collins, D.~E. Soper and G.~F. Sterman, \emph{{Soft Gluons and
  Factorization}}, {\emph{Nucl. Phys.} {\bfseries B308} (1988) 833}.

\bibitem{Diehl:2015bca}
M.~Diehl, J.~R. Gaunt, D.~Ostermeier, P.~Pl{\"o}{\ss}l and A.~Sch{\"a}fer,
  \emph{{Cancellation of Glauber gluon exchange in the double Drell-Yan
  process}}, {\emph{JHEP} {\bfseries 01} (2016) 076}
  [\href{https://arxiv.org/abs/1510.08696}{{\ttfamily 1510.08696}}].

\bibitem{Catani:2010pd}
S.~Catani and M.~Grazzini, \emph{{QCD transverse-momentum resummation in gluon
  fusion processes}}, {\emph{Nucl. Phys.} {\bfseries B845} (2011) 297}
  [\href{https://arxiv.org/abs/1011.3918}{{\ttfamily 1011.3918}}].

\bibitem{Bauer:2000ew}
C.~W. Bauer, S.~Fleming and M.~E. Luke, \emph{{Summing Sudakov logarithms in $B
  \to X_s \gamma$ in effective field theory}}, {\emph{Phys.Rev.} {\bfseries
  D63} (2000) 014006} [\href{https://arxiv.org/abs/hep-ph/0005275}{{\ttfamily
  hep-ph/0005275}}].

\bibitem{Bauer:2000yr}
C.~W. Bauer, S.~Fleming, D.~Pirjol and I.~W. Stewart, \emph{{An Effective field
  theory for collinear and soft gluons: Heavy to light decays}},
  {\emph{Phys.Rev.} {\bfseries D63} (2001) 114020}
  [\href{https://arxiv.org/abs/hep-ph/0011336}{{\ttfamily hep-ph/0011336}}].

\bibitem{Bauer:2001ct}
C.~W. Bauer and I.~W. Stewart, \emph{{Invariant operators in collinear
  effective theory}}, {\emph{Phys.Lett.} {\bfseries B516} (2001) 134}
  [\href{https://arxiv.org/abs/hep-ph/0107001}{{\ttfamily hep-ph/0107001}}].

\bibitem{Bauer:2001yt}
C.~W. Bauer, D.~Pirjol and I.~W. Stewart, \emph{{Soft collinear factorization
  in effective field theory}}, {\emph{Phys.Rev.} {\bfseries D65} (2002) 054022}
  [\href{https://arxiv.org/abs/hep-ph/0109045}{{\ttfamily hep-ph/0109045}}].

\bibitem{Becher:2010tm}
T.~Becher and M.~Neubert, \emph{{{Drell-Yan} Production at Small $q_T$,
  Transverse Parton Distributions and the Collinear Anomaly}},
  {\emph{Eur.Phys.J.} {\bfseries C71} (2011) 1665}
  [\href{https://arxiv.org/abs/1007.4005}{{\ttfamily 1007.4005}}].

\bibitem{Becher:2011xn}
T.~Becher, M.~Neubert and D.~Wilhelm, \emph{{Electroweak Gauge-Boson Production
  at Small $q_T$: Infrared Safety from the Collinear Anomaly}}, {\emph{JHEP}
  {\bfseries 02} (2012) 124} [\href{https://arxiv.org/abs/1109.6027}{{\ttfamily
  1109.6027}}].

\bibitem{Becher:2012yn}
T.~Becher, M.~Neubert and D.~Wilhelm, \emph{{Higgs-Boson Production at Small
  Transverse Momentum}}, {\emph{JHEP} {\bfseries 05} (2013) 110}
  [\href{https://arxiv.org/abs/1212.2621}{{\ttfamily 1212.2621}}].

\bibitem{GarciaEchevarria:2011rb}
M.~G. Echevarria, A.~Idilbi and I.~Scimemi, \emph{{Factorization Theorem For
  Drell-Yan At Low $q_T$ And Transverse Momentum Distributions
  On-The-Light-Cone}}, {\emph{JHEP} {\bfseries 1207} (2012) 002}
  [\href{https://arxiv.org/abs/1111.4996}{{\ttfamily 1111.4996}}].

\bibitem{Echevarria:2012js}
M.~G. Echevarria, A.~Idilbi and I.~Scimemi, \emph{{Soft and Collinear
  Factorization and Transverse Momentum Dependent Parton Distribution
  Functions}}, {\emph{Phys.Lett.} {\bfseries B726} (2013) 795}
  [\href{https://arxiv.org/abs/1211.1947}{{\ttfamily 1211.1947}}].

\bibitem{Echevarria:2014rua}
M.~G. Echevarria, A.~Idilbi and I.~Scimemi, \emph{{Unified treatment of the QCD
  evolution of all (un-)polarized transverse momentum dependent functions:
  Collins function as a study case}}, {\emph{Phys.Rev.} {\bfseries D90} (2014)
  014003} [\href{https://arxiv.org/abs/1402.0869}{{\ttfamily 1402.0869}}].

\bibitem{Chiu:2012ir}
J.-Y. Chiu, A.~Jain, D.~Neill and I.~Z. Rothstein, \emph{{A Formalism for the
  Systematic Treatment of Rapidity Logarithms in Quantum Field Theory}},
  {\emph{JHEP} {\bfseries 1205} (2012) 084}
  [\href{https://arxiv.org/abs/1202.0814}{{\ttfamily 1202.0814}}].

\bibitem{Collins:2017oxh}
J.~Collins and T.~C. Rogers, \emph{{Connecting Different TMD Factorization
  Formalisms in QCD}}, {\emph{Phys. Rev.} {\bfseries D96} (2017) 054011}
  [\href{https://arxiv.org/abs/1705.07167}{{\ttfamily 1705.07167}}].

\bibitem{Ebert:2019okf}
M.~A. Ebert, I.~W. Stewart and Y.~Zhao, \emph{{Towards Quasi-Transverse
  Momentum Dependent PDFs Computable on the Lattice}},
  \href{https://doi.org/10.1007/JHEP09(2019)037}{\emph{JHEP} {\bfseries 09}
  (2019) 037} [\href{https://arxiv.org/abs/1901.03685}{{\ttfamily
  1901.03685}}].

\bibitem{Li:2016axz}
Y.~Li, D.~Neill and H.~X. Zhu, \emph{{An exponential regulator for rapidity
  divergences}},
  \href{https://doi.org/10.1016/j.nuclphysb.2020.115193}{\emph{Nucl. Phys. B}
  {\bfseries 960} (2020) 115193}
  [\href{https://arxiv.org/abs/1604.00392}{{\ttfamily 1604.00392}}].

\bibitem{Stewart:2009yx}
I.~W. Stewart, F.~J. Tackmann and W.~J. Waalewijn, \emph{{Factorization at the
  LHC: From PDFs to Initial State Jets}}, {\emph{Phys. Rev. D} {\bfseries 81}
  (2010) 094035} [\href{https://arxiv.org/abs/0910.0467}{{\ttfamily
  0910.0467}}].

\bibitem{Manohar:2006nz}
A.~V. Manohar and I.~W. Stewart, \emph{{The Zero-Bin and Mode Factorization in
  Quantum Field Theory}}, {\emph{Phys. Rev.} {\bfseries D76} (2007) 074002}
  [\href{https://arxiv.org/abs/hep-ph/0605001}{{\ttfamily hep-ph/0605001}}].

\bibitem{Collins:1992tv}
J.~C. Collins and F.~V. Tkachov, \emph{{Breakdown of dimensional regularization
  in the Sudakov problem}}, {\emph{Phys. Lett.} {\bfseries B294} (1992) 403}
  [\href{https://arxiv.org/abs/hep-ph/9208209}{{\ttfamily hep-ph/9208209}}].

\bibitem{Chiu:2011qc}
J.-y. Chiu, A.~Jain, D.~Neill and I.~Z. Rothstein, \emph{{The Rapidity
  Renormalization Group}}, {\emph{Phys. Rev. Lett.} {\bfseries 108} (2012)
  151601} [\href{https://arxiv.org/abs/1104.0881}{{\ttfamily 1104.0881}}].

\bibitem{Ji:2002aa}
X.-d. Ji and F.~Yuan, \emph{{Parton distributions in light cone gauge: Where
  are the final state interactions?}}, {\emph{Phys. Lett.} {\bfseries B543}
  (2002) 66} [\href{https://arxiv.org/abs/hep-ph/0206057}{{\ttfamily
  hep-ph/0206057}}].

\bibitem{Idilbi:2010im}
A.~Idilbi and I.~Scimemi, \emph{{Singular and Regular Gauges in Soft Collinear
  Effective Theory: The Introduction of the New Wilson Line T}}, {\emph{Phys.
  Lett.} {\bfseries B695} (2011) 463}
  [\href{https://arxiv.org/abs/1009.2776}{{\ttfamily 1009.2776}}].

\bibitem{GarciaEchevarria:2011md}
M.~Garcia-Echevarria, A.~Idilbi and I.~Scimemi, \emph{{SCET, Light-Cone Gauge
  and the T-Wilson Lines}}, {\emph{Phys. Rev.} {\bfseries D84} (2011) 011502}
  [\href{https://arxiv.org/abs/1104.0686}{{\ttfamily 1104.0686}}].

\bibitem{Buffing:2017mqm}
M.~G.~A. Buffing, M.~Diehl and T.~Kasemets, \emph{{Transverse momentum in
  double parton scattering: factorisation, evolution and matching}},
  {\emph{JHEP} {\bfseries 01} (2018) 044}
  [\href{https://arxiv.org/abs/1708.03528}{{\ttfamily 1708.03528}}].

\bibitem{Echevarria:2015usa}
M.~G. Echevarria, I.~Scimemi and A.~Vladimirov, \emph{{Transverse momentum
  dependent fragmentation function at next-to--next-to--leading order}},
  {\emph{Phys. Rev.} {\bfseries D93} (2016) 011502}
  [\href{https://arxiv.org/abs/1509.06392}{{\ttfamily 1509.06392}}].

\bibitem{Echevarria:2015byo}
M.~G. Echevarria, I.~Scimemi and A.~Vladimirov, \emph{{Universal transverse
  momentum dependent soft function at NNLO}}, {\emph{Phys. Rev.} {\bfseries
  D93} (2016) 054004} [\href{https://arxiv.org/abs/1511.05590}{{\ttfamily
  1511.05590}}].

\bibitem{Echevarria:2016scs}
M.~G. Echevarria, I.~Scimemi and A.~Vladimirov, \emph{{Unpolarized Transverse
  Momentum Dependent Parton Distribution and Fragmentation Functions at
  next-to-next-to-leading order}}, {\emph{JHEP} {\bfseries 09} (2016) 004}
  [\href{https://arxiv.org/abs/1604.07869}{{\ttfamily 1604.07869}}].

\bibitem{Becher:2011dz}
T.~Becher and G.~Bell, \emph{{Analytic Regularization in Soft-Collinear
  Effective Theory}}, {\emph{Phys. Lett.} {\bfseries B713} (2012) 41}
  [\href{https://arxiv.org/abs/1112.3907}{{\ttfamily 1112.3907}}].

\bibitem{Ebert:2018gsn}
M.~A. Ebert, I.~Moult, I.~W. Stewart, F.~J. Tackmann, G.~Vita and H.~X. Zhu,
  \emph{{Subleading power rapidity divergences and power corrections for
  q$_{T}$}}, {\emph{JHEP} {\bfseries 04} (2019) 123}
  [\href{https://arxiv.org/abs/1812.08189}{{\ttfamily 1812.08189}}].

\bibitem{tHooft:1973wag}
G.~'t~Hooft and M.~J.~G. Veltman, \emph{{DIAGRAMMAR}}, {\emph{NATO Sci. Ser. B}
  {\bfseries 4} (1974) 177}.

\bibitem{Collins:1981uw}
J.~C. Collins and D.~E. Soper, \emph{{Parton Distribution and Decay
  Functions}}, {\emph{Nucl.Phys.} {\bfseries B194} (1982) 445}.

\bibitem{Boglione:2019nwk}
M.~Boglione, A.~Dotson, L.~Gamberg, S.~Gordon, J.~O. Gonzalez-Hernandez,
  A.~Prokudin et~al., \emph{{Mapping the Kinematical Regimes of Semi-Inclusive
  Deep Inelastic Scattering}}, {\emph{JHEP} {\bfseries 10} (2019) 122}
  [\href{https://arxiv.org/abs/1904.12882}{{\ttfamily 1904.12882}}].

\bibitem{Manohar:1992tz}
A.~V. Manohar, \emph{{An Introduction to spin dependent deep inelastic
  scattering}},  in \emph{{Lake Louise Winter Institute: Symmetry and Spin in
  the Standard Model}}, 3, 1992,
  \href{https://arxiv.org/abs/hep-ph/9204208}{{\ttfamily hep-ph/9204208}}.

\bibitem{Bacchetta:2006tn}
A.~Bacchetta, M.~Diehl, K.~Goeke, A.~Metz, P.~J. Mulders and M.~Schlegel,
  \emph{{Semi-inclusive deep inelastic scattering at small transverse
  momentum}}, {\emph{JHEP} {\bfseries 02} (2007) 093}
  [\href{https://arxiv.org/abs/hep-ph/0611265}{{\ttfamily hep-ph/0611265}}].

\bibitem{Collins:2004nx}
J.~C. Collins and A.~Metz, \emph{{Universality of soft and collinear factors in
  hard-scattering factorization}}, {\emph{Phys.Rev.Lett.} {\bfseries 93} (2004)
  252001} [\href{https://arxiv.org/abs/hep-ph/0408249}{{\ttfamily
  hep-ph/0408249}}].

\bibitem{Kogut:1969xa}
J.~B. Kogut and D.~E. Soper, \emph{{Quantum Electrodynamics in the Infinite
  Momentum Frame}}, {\emph{Phys. Rev. D} {\bfseries 1} (1970) 2901}.

\bibitem{Bauer:2002nz}
C.~W. Bauer, S.~Fleming, D.~Pirjol, I.~Z. Rothstein and I.~W. Stewart,
  \emph{{Hard scattering factorization from effective field theory}},
  {\emph{Phys. Rev. D} {\bfseries 66} (2002) 014017}
  [\href{https://arxiv.org/abs/hep-ph/0202088}{{\ttfamily hep-ph/0202088}}].

\bibitem{Anselmino:2005sh}
M.~Anselmino, M.~Boglione, U.~D'Alesio, E.~Leader, S.~Melis and F.~Murgia,
  \emph{{The general partonic structure for hadronic spin asymmetries}},
  {\emph{Phys. Rev. D} {\bfseries 73} (2006) 014020}
  [\href{https://arxiv.org/abs/hep-ph/0509035}{{\ttfamily hep-ph/0509035}}].

\bibitem{DAlesio:2007bjf}
U.~D'Alesio and F.~Murgia, \emph{{Azimuthal and Single Spin Asymmetries in Hard
  Scattering Processes}}, {\emph{Prog. Part. Nucl. Phys.} {\bfseries 61} (2008)
  394} [\href{https://arxiv.org/abs/0712.4328}{{\ttfamily 0712.4328}}].

\bibitem{Anselmino:2011ch}
M.~Anselmino, M.~Boglione, U.~D'Alesio, S.~Melis, F.~Murgia, E.~R. Nocera
  et~al., \emph{{General Helicity Formalism for Polarized Semi-Inclusive Deep
  Inelastic Scattering}}, {\emph{Phys. Rev.} {\bfseries D83} (2011) 114019}
  [\href{https://arxiv.org/abs/1101.1011}{{\ttfamily 1101.1011}}].

\bibitem{Ralston:1979ys}
J.~P. Ralston and D.~E. Soper, \emph{{Production of Dimuons from High-Energy
  Polarized Proton Proton Collisions}}, {\emph{Nucl. Phys.} {\bfseries B152}
  (1979) 109}.

\bibitem{Mulders:1995dh}
P.~J. Mulders and R.~D. Tangerman, \emph{{The Complete tree level result up to
  order 1/Q for polarized deep inelastic leptoproduction}}, {\emph{Nucl. Phys.}
  {\bfseries B461} (1996) 197}
  [\href{https://arxiv.org/abs/hep-ph/9510301}{{\ttfamily hep-ph/9510301}}].

\bibitem{Bacchetta:2004zf}
A.~Bacchetta, P.~J. Mulders and F.~Pijlman, \emph{{New observables in
  longitudinal single-spin asymmetries in semi-inclusive DIS}}, {\emph{Phys.
  Lett. B} {\bfseries 595} (2004) 309}
  [\href{https://arxiv.org/abs/hep-ph/0405154}{{\ttfamily hep-ph/0405154}}].

\bibitem{Goeke:2005hb}
K.~Goeke, A.~Metz and M.~Schlegel, \emph{{Parameterization of the quark-quark
  correlator of a spin-1/2 hadron}}, {\emph{Phys. Lett. B} {\bfseries 618}
  (2005) 90} [\href{https://arxiv.org/abs/hep-ph/0504130}{{\ttfamily
  hep-ph/0504130}}].

\bibitem{Sivers:1989cc}
D.~W. Sivers, \emph{Single spin production asymmetries from the hard scattering
  of point - like constituents}, {\emph{Phys. Rev.} {\bfseries D41} (1990) 83}.

\bibitem{Brodsky:2002rv}
S.~J. Brodsky, D.~S. Hwang and I.~Schmidt, \emph{{Initial state interactions
  and single spin asymmetries in Drell-Yan processes}}, {\emph{Nucl. Phys.}
  {\bfseries B642} (2002) 344}
  [\href{https://arxiv.org/abs/hep-ph/0206259}{{\ttfamily hep-ph/0206259}}].

\bibitem{Kotzinian:1997wt}
A.~Kotzinian and P.~Mulders, \emph{{Probing transverse quark polarization via
  azimuthal asymmetries in leptoproduction}}, {\emph{Phys. Lett. B} {\bfseries
  406} (1997) 373} [\href{https://arxiv.org/abs/hep-ph/9701330}{{\ttfamily
  hep-ph/9701330}}].

\bibitem{Kotzinian:1995cz}
A.~Kotzinian and P.~Mulders, \emph{{Longitudinal quark polarization in
  transversely polarized nucleons}}, {\emph{Phys. Rev. D} {\bfseries 54} (1996)
  1229} [\href{https://arxiv.org/abs/hep-ph/9511420}{{\ttfamily
  hep-ph/9511420}}].

\bibitem{Miller:2003sa}
G.~A. Miller, \emph{{Shapes of the proton}}, {\emph{Phys. Rev. C} {\bfseries
  68} (2003) 022201} [\href{https://arxiv.org/abs/nucl-th/0304076}{{\ttfamily
  nucl-th/0304076}}].

\bibitem{nytimes03}
K.~Chang, ``It's a ball. no, it's a pretzel. must be a proton..''
  \url{https://www.nytimes.com/2003/05/06/science/it-s-a-ball-no-it-s-a-pretzel-must-be-a-proton.html},
  May, 2003.

\bibitem{Kang:2009bp}
Z.-B. Kang and J.-W. Qiu, \emph{{Testing the Time-Reversal Modified
  Universality of the Sivers Function}}, {\emph{Phys. Rev. Lett.} {\bfseries
  103} (2009) 172001} [\href{https://arxiv.org/abs/0903.3629}{{\ttfamily
  0903.3629}}].

\bibitem{Boer:2011xd}
D.~Boer, L.~Gamberg, B.~Musch and A.~Prokudin, \emph{{Bessel-Weighted
  Asymmetries in Semi Inclusive Deep Inelastic Scattering}}, {\emph{JHEP}
  {\bfseries 1110} (2011) 021}
  [\href{https://arxiv.org/abs/1107.5294}{{\ttfamily 1107.5294}}].

\bibitem{Musch:2011er}
B.~Musch, P.~H\"agler, M.~Engelhardt, J.~Negele and A.~Sch\"afer, \emph{{Sivers
  and Boer-Mulders observables from lattice QCD}}, {\emph{Phys.Rev.} {\bfseries
  D85} (2012) 094510} [\href{https://arxiv.org/abs/1111.4249}{{\ttfamily
  1111.4249}}].

\bibitem{Engelhardt:2015xja}
M.~Engelhardt, P.~H{\"a}gler, B.~Musch, J.~Negele and A.~Sch{\"a}fer,
  \emph{{Lattice QCD study of the Boer-Mulders effect in a pion}}, {\emph{Phys.
  Rev.} {\bfseries D93} (2016) 054501}
  [\href{https://arxiv.org/abs/1506.07826}{{\ttfamily 1506.07826}}].

\bibitem{Yoon:2017qzo}
B.~Yoon, M.~Engelhardt, R.~Gupta, T.~Bhattacharya, J.~R. Green, B.~U. Musch
  et~al., \emph{{Nucleon Transverse Momentum-dependent Parton Distributions in
  Lattice QCD: Renormalization Patterns and Discretization Effects}},
  {\emph{Phys. Rev.} {\bfseries D96} (2017) 094508}
  [\href{https://arxiv.org/abs/1706.03406}{{\ttfamily 1706.03406}}].

\bibitem{Scimemi:2018mmi}
I.~Scimemi and A.~Vladimirov, \emph{{Matching of transverse momentum dependent
  distributions at twist-3}}, {\emph{Eur. Phys. J. C} {\bfseries 78} (2018)
  802} [\href{https://arxiv.org/abs/1804.08148}{{\ttfamily 1804.08148}}].

\bibitem{Gutierrez-Reyes:2017glx}
D.~Guti\'errez-Reyes, I.~Scimemi and A.~A. Vladimirov, \emph{{Twist-2 matching
  of transverse momentum dependent distributions}}, {\emph{Phys. Lett. B}
  {\bfseries 769} (2017) 84}
  [\href{https://arxiv.org/abs/1702.06558}{{\ttfamily 1702.06558}}].

\bibitem{Collins:2021vke}
J.~Collins, T.~C. Rogers and N.~Sato, \emph{{Positivity and renormalization of
  parton densities}},
  \href{https://doi.org/10.1103/PhysRevD.105.076010}{\emph{Phys. Rev. D}
  {\bfseries 105} (2022) 076010}
  [\href{https://arxiv.org/abs/2111.01170}{{\ttfamily 2111.01170}}].

\bibitem{Buffing:2013kca}
M.~G.~A. Buffing, A.~Mukherjee and P.~J. Mulders, \emph{{Generalized
  Universality of Definite Rank Gluon Transverse Momentum Dependent
  Correlators}}, {\emph{Phys. Rev. D} {\bfseries 88} (2013) 054027}
  [\href{https://arxiv.org/abs/1306.5897}{{\ttfamily 1306.5897}}].

\bibitem{Boer:2015vso}
D.~Boer, C.~Lorc\'e, C.~Pisano and J.~Zhou, \emph{{The gluon Sivers
  distribution: status and future prospects}}, {\emph{Adv. High Energy Phys.}
  {\bfseries 2015} (2015) 371396}
  [\href{https://arxiv.org/abs/1504.04332}{{\ttfamily 1504.04332}}].

\bibitem{Mulders:2000sh}
P.~Mulders and J.~Rodrigues, \emph{{Transverse momentum dependence in gluon
  distribution and fragmentation functions}}, {\emph{Phys.Rev.} {\bfseries D63}
  (2001) 094021} [\href{https://arxiv.org/abs/hep-ph/0009343}{{\ttfamily
  hep-ph/0009343}}].

\bibitem{Meissner:2007rx}
S.~Meissner, A.~Metz and K.~Goeke, \emph{{Relations between generalized and
  transverse momentum dependent parton distributions}}, {\emph{Phys.Rev.}
  {\bfseries D76} (2007) 034002}
  [\href{https://arxiv.org/abs/hep-ph/0703176}{{\ttfamily hep-ph/0703176}}].

\bibitem{Lyubovitskij:2021qza}
V.~E. Lyubovitskij and I.~Schmidt, \emph{{New findings in gluon TMD physics}},
  {\emph{Phys. Rev. D} {\bfseries 104} (2021) 014001}
  [\href{https://arxiv.org/abs/2105.07842}{{\ttfamily 2105.07842}}].

\bibitem{Echevarria:2015uaa}
M.~G. Echevarria, T.~Kasemets, P.~J. Mulders and C.~Pisano, \emph{{QCD
  evolution of (un)polarized gluon TMDPDFs and the Higgs $q_T$-distribution}},
  \href{https://doi.org/10.1007/JHEP07(2015)158}{\emph{JHEP} {\bfseries 07}
  (2015) 158} [\href{https://arxiv.org/abs/1502.05354}{{\ttfamily
  1502.05354}}].

\bibitem{Catani:2011kr}
S.~Catani and M.~Grazzini, \emph{{Higgs Boson Production at Hadron Colliders:
  Hard-Collinear Coefficients at the NNLO}}, {\emph{Eur. Phys. J.} {\bfseries
  C72} (2012) 2013} [\href{https://arxiv.org/abs/1106.4652}{{\ttfamily
  1106.4652}}].

\bibitem{Gehrmann:2014yya}
T.~Gehrmann, T.~Luebbert and L.~L. Yang, \emph{{Calculation of the transverse
  parton distribution functions at next-to-next-to-leading order}},
  {\emph{JHEP} {\bfseries 06} (2014) 155}
  [\href{https://arxiv.org/abs/1403.6451}{{\ttfamily 1403.6451}}].

\bibitem{Luebbert:2016itl}
T.~L{\"u}bbert, J.~Oredsson and M.~Stahlhofen, \emph{{Rapidity renormalized TMD
  soft and beam functions at two loops}}, {\emph{JHEP} {\bfseries 03} (2016)
  168} [\href{https://arxiv.org/abs/1602.01829}{{\ttfamily 1602.01829}}].

\bibitem{Li:2016ctv}
Y.~Li and H.~X. Zhu, \emph{{Bootstrapping Rapidity Anomalous Dimensions for
  Transverse-Momentum Resummation}}, {\emph{Phys. Rev. Lett.} {\bfseries 118}
  (2017) 022004} [\href{https://arxiv.org/abs/1604.01404}{{\ttfamily
  1604.01404}}].

\bibitem{Luo:2019szz}
M.-x. Luo, T.-Z. Yang, H.~X. Zhu and Y.~J. Zhu, \emph{{Quark Transverse Parton
  Distribution at the Next-to-Next-to-Next-to-Leading Order}},
  \href{https://doi.org/10.1103/PhysRevLett.124.092001}{\emph{Phys. Rev. Lett.}
  {\bfseries 124} (2020) 092001}
  [\href{https://arxiv.org/abs/1912.05778}{{\ttfamily 1912.05778}}].

\bibitem{Ebert:2020yqt}
M.~A. Ebert, B.~Mistlberger and G.~Vita, \emph{{Transverse momentum dependent
  PDFs at N$^3$LO}}, \href{https://doi.org/10.1007/JHEP09(2020)146}{\emph{JHEP}
  {\bfseries 09} (2020) 146}
  [\href{https://arxiv.org/abs/2006.05329}{{\ttfamily 2006.05329}}].

\bibitem{Luo:2020epw}
M.-x. Luo, T.-Z. Yang, H.~X. Zhu and Y.~J. Zhu, \emph{{Unpolarized quark and
  gluon TMD PDFs and FFs at N$^{3}$LO}},
  \href{https://doi.org/10.1007/JHEP06(2021)115}{\emph{JHEP} {\bfseries 06}
  (2021) 115} [\href{https://arxiv.org/abs/2012.03256}{{\ttfamily
  2012.03256}}].

\bibitem{Qiu:1991pp}
J.-w. Qiu and G.~F. Sterman, \emph{{Single transverse spin asymmetries}},
  {\emph{Phys. Rev. Lett.} {\bfseries 67} (1991) 2264}.

\bibitem{Braun:2009mi}
V.~Braun, A.~Manashov and B.~Pirnay, \emph{{Scale dependence of twist-three
  contributions to single spin asymmetries}}, {\emph{Phys. Rev. D} {\bfseries
  80} (2009) 114002} [\href{https://arxiv.org/abs/0909.3410}{{\ttfamily
  0909.3410}}].

\bibitem{Ebert:2020qef}
M.~A. Ebert, B.~Mistlberger and G.~Vita, \emph{{TMD fragmentation functions at
  N$^{3}$LO}}, \href{https://doi.org/10.1007/JHEP07(2021)121}{\emph{JHEP}
  {\bfseries 07} (2021) 121}
  [\href{https://arxiv.org/abs/2012.07853}{{\ttfamily 2012.07853}}].

\bibitem{Moos:2020wvd}
V.~Moos and A.~Vladimirov, \emph{{Calculation of transverse momentum dependent
  distributions beyond the leading power}}, {\emph{JHEP} {\bfseries 12} (2020)
  145} [\href{https://arxiv.org/abs/2008.01744}{{\ttfamily 2008.01744}}].

\bibitem{Balitsky:1990ck}
I.~I. Balitsky and V.~M. Braun, \emph{{The Nonlocal operator expansion for
  inclusive particle production in $e^+e^-$ annihilation}}, {\emph{Nucl. Phys.
  B} {\bfseries 361} (1991) 93}.

\bibitem{Luo:2019hmp}
M.-X. Luo, X.~Wang, X.~Xu, L.~L. Yang, T.-Z. Yang and H.~X. Zhu,
  \emph{{Transverse Parton Distribution and Fragmentation Functions at NNLO:
  the Quark Case}}, {\emph{JHEP} {\bfseries 10} (2019) 083}
  [\href{https://arxiv.org/abs/1908.03831}{{\ttfamily 1908.03831}}].

\bibitem{Chen:2020uvt}
H.~Chen, T.-Z. Yang, H.~X. Zhu and Y.~J. Zhu, \emph{{Analytic Continuation and
  Reciprocity Relation for Collinear Splitting in QCD}},
  \href{https://doi.org/10.1088/1674-1137/abde2d}{\emph{Chin. Phys. C}
  {\bfseries 45} (2021) 043101}
  [\href{https://arxiv.org/abs/2006.10534}{{\ttfamily 2006.10534}}].

\bibitem{Bacchetta:2013pqa}
A.~Bacchetta and A.~Prokudin, \emph{{Evolution of the helicity and transversity
  Transverse-Momentum-Dependent parton distributions}}, {\emph{Nucl. Phys. B}
  {\bfseries 875} (2013) 536}
  [\href{https://arxiv.org/abs/1303.2129}{{\ttfamily 1303.2129}}].

\bibitem{Kanazawa:2015ajw}
K.~Kanazawa, Y.~Koike, A.~Metz, D.~Pitonyak and M.~Schlegel, \emph{{Operator
  Constraints for Twist-3 Functions and Lorentz Invariance Properties of
  Twist-3 Observables}}, {\emph{Phys. Rev. D} {\bfseries 93} (2016) 054024}
  [\href{https://arxiv.org/abs/1512.07233}{{\ttfamily 1512.07233}}].

\bibitem{Ji:2006ub}
X.~Ji, J.-W. Qiu, W.~Vogelsang and F.~Yuan, \emph{{A Unified picture for single
  transverse-spin asymmetries in hard processes}}, {\emph{Phys.Rev.Lett.}
  {\bfseries 97} (2006) 082002}
  [\href{https://arxiv.org/abs/hep-ph/0602239}{{\ttfamily hep-ph/0602239}}].

\bibitem{Ji:2006vf}
X.~Ji, J.-w. Qiu, W.~Vogelsang and F.~Yuan, \emph{{Single Transverse-Spin
  Asymmetry in Drell-Yan Production at Large and Moderate Transverse
  Momentum}}, {\emph{Phys. Rev. D} {\bfseries 73} (2006) 094017}
  [\href{https://arxiv.org/abs/hep-ph/0604023}{{\ttfamily hep-ph/0604023}}].

\bibitem{Koike:2007dg}
Y.~Koike, W.~Vogelsang and F.~Yuan, \emph{{On the Relation Between Mechanisms
  for Single-Transverse-Spin Asymmetries}}, {\emph{Phys. Lett. B} {\bfseries
  659} (2008) 878} [\href{https://arxiv.org/abs/0711.0636}{{\ttfamily
  0711.0636}}].

\bibitem{Kang:2011mr}
Z.-B. Kang, B.-W. Xiao and F.~Yuan, \emph{{QCD Resummation for Single Spin
  Asymmetries}}, {\emph{Phys.Rev.Lett.} {\bfseries 107} (2011) 152002}
  [\href{https://arxiv.org/abs/1106.0266}{{\ttfamily 1106.0266}}].

\bibitem{Sun:2013hua}
P.~Sun and F.~Yuan, \emph{{Transverse momentum dependent evolution: Matching
  semi-inclusive deep inelastic scattering processes to Drell-Yan and W/Z boson
  production}}, {\emph{Phys.Rev.} {\bfseries D88} (2013) 114012}
  [\href{https://arxiv.org/abs/1308.5003}{{\ttfamily 1308.5003}}].

\bibitem{Dai:2014ala}
L.-Y. Dai, Z.-B. Kang, A.~Prokudin and I.~Vitev, \emph{{Next-to-leading order
  transverse momentum-weighted Sivers asymmetry in semi-inclusive deep
  inelastic scattering: the role of the three-gluon correlator}}, {\emph{Phys.
  Rev. D} {\bfseries 92} (2015) 114024}
  [\href{https://arxiv.org/abs/1409.5851}{{\ttfamily 1409.5851}}].

\bibitem{Scimemi:2019gge}
I.~Scimemi, A.~Tarasov and A.~Vladimirov, \emph{{Collinear matching for Sivers
  function at next-to-leading order}}, {\emph{JHEP} {\bfseries 05} (2019) 125}
  [\href{https://arxiv.org/abs/1901.04519}{{\ttfamily 1901.04519}}].

\bibitem{Gutierrez-Reyes:2018iod}
D.~Gutierrez-Reyes, I.~Scimemi and A.~Vladimirov, \emph{{Transverse momentum
  dependent transversely polarized distributions at
  next-to-next-to-leading-order}}, {\emph{JHEP} {\bfseries 07} (2018) 172}
  [\href{https://arxiv.org/abs/1805.07243}{{\ttfamily 1805.07243}}].

\bibitem{Luo:2019bmw}
M.-X. Luo, T.-Z. Yang, H.~X. Zhu and Y.~J. Zhu, \emph{{Transverse Parton
  Distribution and Fragmentation Functions at NNLO: the Gluon Case}},
  {\emph{JHEP} {\bfseries 01} (2020) 040}
  [\href{https://arxiv.org/abs/1909.13820}{{\ttfamily 1909.13820}}].

\bibitem{Gutierrez-Reyes:2019rug}
D.~Gutierrez-Reyes, S.~Leal-Gomez, I.~Scimemi and A.~Vladimirov,
  \emph{{Linearly polarized gluons at next-to-next-to leading order and the
  Higgs transverse momentum distribution}}, {\emph{JHEP} {\bfseries 11} (2019)
  121} [\href{https://arxiv.org/abs/1907.03780}{{\ttfamily 1907.03780}}].

\bibitem{Qiu:2000hf}
J.-w. Qiu and X.-f. Zhang, \emph{{Role of the nonperturbative input in QCD
  resummed Drell-Yan $Q_{T}$ distributions}}, {\emph{Phys. Rev.} {\bfseries
  D63} (2001) 114011} [\href{https://arxiv.org/abs/hep-ph/0012348}{{\ttfamily
  hep-ph/0012348}}].

\bibitem{Berger:2002ut}
E.~L. Berger and J.-w. Qiu, \emph{{Differential cross-section for Higgs boson
  production including all orders soft gluon resummation}}, {\emph{Phys. Rev.}
  {\bfseries D67} (2003) 034026}
  [\href{https://arxiv.org/abs/hep-ph/0210135}{{\ttfamily hep-ph/0210135}}].

\bibitem{Ebert:2022cku}
M.~A. Ebert, J.~K.~L. Michel, I.~W. Stewart and Z.~Sun, \emph{{Disentangling
  long and short distances in momentum-space TMDs}},
  \href{https://doi.org/10.1007/JHEP07(2022)129}{\emph{JHEP} {\bfseries 07}
  (2022) 129} [\href{https://arxiv.org/abs/2201.07237}{{\ttfamily
  2201.07237}}].

\bibitem{Hagler:2009mb}
P.~H\"agler, B.~Musch, J.~Negele and A.~Sch\"afer, \emph{{Intrinsic quark
  transverse momentum in the nucleon from lattice QCD}}, {\emph{Europhys.Lett.}
  {\bfseries 88} (2009) 61001}
  [\href{https://arxiv.org/abs/0908.1283}{{\ttfamily 0908.1283}}].

\bibitem{Yoon:2016dyh}
B.~Yoon, T.~Bhattacharya, M.~Engelhardt, J.~Green, R.~Gupta, P.~H{\"a}gler
  et~al., \emph{{Lattice QCD calculations of nucleon transverse
  momentum-dependent parton distributions using clover and domain wall
  fermions}},  in \emph{{Proceedings, 33rd International Symposium on Lattice
  Field Theory (Lattice 2015): Kobe, Japan, July 14-18, 2015}}, SISSA, SISSA,
  2015, \href{https://arxiv.org/abs/1601.05717}{{\ttfamily 1601.05717}}.

\bibitem{Ji:2014hxa}
X.~Ji, P.~Sun, X.~Xiong and F.~Yuan, \emph{{Soft factor subtraction and
  transverse momentum dependent parton distributions on the lattice}},
  {\emph{Phys.Rev.} {\bfseries D91} (2015) 074009}
  [\href{https://arxiv.org/abs/1405.7640}{{\ttfamily 1405.7640}}].

\bibitem{Ji:2018hvs}
X.~Ji, L.-C. Jin, F.~Yuan, J.-H. Zhang and Y.~Zhao, \emph{{Transverse Momentum
  Dependent Quasi-Parton-Distributions}}, {\emph{Phys. Rev.} {\bfseries D99}
  (2019) 114006} [\href{https://arxiv.org/abs/1801.05930}{{\ttfamily
  1801.05930}}].

\bibitem{Ebert:2018gzl}
M.~A. Ebert, I.~W. Stewart and Y.~Zhao, \emph{{Determining the Nonperturbative
  Collins-Soper Kernel From Lattice QCD}}, {\emph{Phys. Rev.} {\bfseries D99}
  (2019) 034505} [\href{https://arxiv.org/abs/1811.00026}{{\ttfamily
  1811.00026}}].

\bibitem{Ebert:2019tvc}
M.~A. Ebert, I.~W. Stewart and Y.~Zhao, \emph{{Renormalization and Matching for
  the Collins-Soper Kernel from Lattice QCD}},
  \href{https://doi.org/10.1007/JHEP03(2020)099}{\emph{JHEP} {\bfseries 03}
  (2020) 099} [\href{https://arxiv.org/abs/1910.08569}{{\ttfamily
  1910.08569}}].

\bibitem{Ji:2019sxk}
X.~Ji, Y.~Liu and Y.-S. Liu, \emph{{TMD soft function from large-momentum
  effective theory}}, {\emph{Nucl. Phys. B} {\bfseries 955} (2020) 115054}
  [\href{https://arxiv.org/abs/1910.11415}{{\ttfamily 1910.11415}}].

\bibitem{Ji:2019ewn}
X.~Ji, Y.~Liu and Y.-S. Liu, \emph{{Transverse-momentum-dependent parton
  distribution functions from large-momentum effective theory}}, {\emph{Phys.
  Lett. B} {\bfseries 811} (2020) 135946}
  [\href{https://arxiv.org/abs/1911.03840}{{\ttfamily 1911.03840}}].

\bibitem{Vladimirov:2020ofp}
A.~A. Vladimirov and A.~Sch\"afer, \emph{{Transverse momentum dependent
  factorization for lattice observables}}, {\emph{Phys. Rev. D} {\bfseries 101}
  (2020) 074517} [\href{https://arxiv.org/abs/2002.07527}{{\ttfamily
  2002.07527}}].

\bibitem{Ebert:2020gxr}
M.~A. Ebert, S.~T. Schindler, I.~W. Stewart and Y.~Zhao, \emph{{One-loop
  Matching for Spin-Dependent Quasi-TMDs}}, {\emph{JHEP} {\bfseries 09} (2020)
  099} [\href{https://arxiv.org/abs/2004.14831}{{\ttfamily 2004.14831}}].

\bibitem{Ji:2020jeb}
X.~Ji, Y.~Liu, A.~Sch\"afer and F.~Yuan, \emph{{Single Transverse-Spin
  Asymmetry and Sivers Function in Large Momentum Effective Theory}},
  \href{https://doi.org/10.1103/PhysRevD.103.074005}{\emph{Phys. Rev. D}
  {\bfseries 103} (2021) 074005}
  [\href{https://arxiv.org/abs/2011.13397}{{\ttfamily 2011.13397}}].

\bibitem{Shanahan:2019zcq}
P.~Shanahan, M.~L. Wagman and Y.~Zhao, \emph{{Nonperturbative renormalization
  of staple-shaped Wilson line operators in lattice QCD}},
  \href{https://doi.org/10.1103/PhysRevD.101.074505}{\emph{Phys. Rev. D}
  {\bfseries 101} (2020) 074505}
  [\href{https://arxiv.org/abs/1911.00800}{{\ttfamily 1911.00800}}].

\bibitem{Ebert:2022fmh}
M.~A. Ebert, S.~T. Schindler, I.~W. Stewart and Y.~Zhao, \emph{{Factorization
  connecting continuum \& lattice TMDs}},
  \href{https://doi.org/10.1007/JHEP04(2022)178}{\emph{JHEP} {\bfseries 04}
  (2022) 178} [\href{https://arxiv.org/abs/2201.08401}{{\ttfamily
  2201.08401}}].

\bibitem{Collins:1977iv}
J.~C. Collins and D.~E. Soper, \emph{{Angular Distribution of Dileptons in
  High-Energy Hadron Collisions}}, {\emph{Phys. Rev. D} {\bfseries 16} (1977)
  2219}.

\bibitem{Bastami:2020asv}
S.~Bastami, L.~Gamberg, B.~Parsamyan, B.~Pasquini, A.~Prokudin and
  P.~Schweitzer, \emph{{The Drell-Yan process with pions and polarized
  nucleons}}, \href{https://doi.org/10.1007/JHEP02(2021)166}{\emph{JHEP}
  {\bfseries 02} (2021) 166}
  [\href{https://arxiv.org/abs/2005.14322}{{\ttfamily 2005.14322}}].

\bibitem{Arnold:2008kf}
S.~Arnold, A.~Metz and M.~Schlegel, \emph{{Dilepton production from polarized
  hadron hadron collisions}}, {\emph{Phys. Rev.} {\bfseries D79} (2009) 034005}
  [\href{https://arxiv.org/abs/0809.2262}{{\ttfamily 0809.2262}}].

\bibitem{Aghasyan:2017jop}
{\scshape COMPASS} collaboration, M.~Aghasyan et~al., \emph{{First measurement
  of transverse-spin-dependent azimuthal asymmetries in the Drell-Yan
  process}}, {\emph{Phys. Rev. Lett.} {\bfseries 119} (2017) 112002}
  [\href{https://arxiv.org/abs/1704.00488}{{\ttfamily 1704.00488}}].

\bibitem{Aad:2014lwa}
{\scshape ATLAS} collaboration, G.~Aad et~al., \emph{{Measurements of fiducial
  and differential cross sections for Higgs boson production in the diphoton
  decay channel at $\sqrt{s} = 8$ TeV with ATLAS}}, {\emph{JHEP} {\bfseries 09}
  (2014) 112} [\href{https://arxiv.org/abs/1407.4222}{{\ttfamily 1407.4222}}].

\bibitem{Aad:2014tca}
{\scshape ATLAS} collaboration, G.~Aad et~al., \emph{{Fiducial and differential
  cross sections of Higgs boson production measured in the four-lepton decay
  channel in $pp$ collisions at $\sqrt{s}$ = 8 TeV with the ATLAS detector}},
  {\emph{Phys. Lett.} {\bfseries B738} (2014) 234}
  [\href{https://arxiv.org/abs/1408.3226}{{\ttfamily 1408.3226}}].

\bibitem{Aad:2016lvc}
{\scshape ATLAS} collaboration, G.~Aad et~al., \emph{{Measurement of fiducial
  differential cross sections of gluon-fusion production of Higgs bosons
  decaying to $WW^* \to e \nu \mu \nu$ with the ATLAS detector at $\sqrt{s} = 8
  $ TeV}}, {\emph{JHEP} {\bfseries 08} (2016) 104}
  [\href{https://arxiv.org/abs/1604.02997}{{\ttfamily 1604.02997}}].

\bibitem{Aaboud:2017oem}
{\scshape ATLAS} collaboration, M.~Aaboud et~al., \emph{{Measurement of
  inclusive and differential cross sections in the $H \rightarrow ZZ^*
  \rightarrow 4\ell$ decay channel in $pp$ collisions at $\sqrt{s}=13$ TeV with
  the ATLAS detector}},
  \href{https://doi.org/10.1007/JHEP10(2017)132}{\emph{JHEP} {\bfseries 10}
  (2017) 132} [\href{https://arxiv.org/abs/1708.02810}{{\ttfamily
  1708.02810}}].

\bibitem{Aaboud:2018xdt}
{\scshape ATLAS} collaboration, M.~Aaboud et~al., \emph{{Measurements of Higgs
  boson properties in the diphoton decay channel with 36 fb$^{-1}$ of $pp$
  collision data at $\sqrt{s} = 13$ TeV with the ATLAS detector}},
  \href{https://doi.org/10.1103/PhysRevD.98.052005}{\emph{Phys. Rev.}
  {\bfseries D98} (2018) 052005}
  [\href{https://arxiv.org/abs/1802.04146}{{\ttfamily 1802.04146}}].

\bibitem{ATLAS:2020wny}
{\scshape ATLAS} collaboration, G.~Aad et~al., \emph{{Measurements of the Higgs
  boson inclusive and differential fiducial cross sections in the 4$\ell$ decay
  channel at $\sqrt{s}$ = 13 TeV}}, {\emph{Submitted to: Eur. Phys. J.} (2020)
  } [\href{https://arxiv.org/abs/2004.03969}{{\ttfamily 2004.03969}}].

\bibitem{Khachatryan:2015rxa}
{\scshape CMS} collaboration, V.~Khachatryan et~al., \emph{{Measurement of
  differential cross sections for Higgs boson production in the diphoton decay
  channel in pp collisions at $\sqrt{s} = 8\,\text {TeV} $}}, {\emph{Eur. Phys.
  J.} {\bfseries C76} (2016) 13}
  [\href{https://arxiv.org/abs/1508.07819}{{\ttfamily 1508.07819}}].

\bibitem{Khachatryan:2015yvw}
{\scshape CMS} collaboration, V.~Khachatryan et~al., \emph{{Measurement of
  differential and integrated fiducial cross sections for Higgs boson
  production in the four-lepton decay channel in pp collisions at $ \sqrt{s}=7$
  and 8 TeV}}, {\emph{JHEP} {\bfseries 04} (2016) 005}
  [\href{https://arxiv.org/abs/1512.08377}{{\ttfamily 1512.08377}}].

\bibitem{Khachatryan:2016vnn}
{\scshape CMS} collaboration, V.~Khachatryan et~al., \emph{{Measurement of the
  transverse momentum spectrum of the Higgs boson produced in pp collisions at
  $ \sqrt{s}=8 $ TeV using $H \to WW$ decays}},
  \href{https://doi.org/10.1007/JHEP03(2017)032}{\emph{JHEP} {\bfseries 03}
  (2017) 032} [\href{https://arxiv.org/abs/1606.01522}{{\ttfamily
  1606.01522}}].

\bibitem{Sirunyan:2018kta}
{\scshape CMS} collaboration, A.~M. Sirunyan et~al., \emph{{Measurement of
  inclusive and differential Higgs boson production cross sections in the
  diphoton decay channel in proton-proton collisions at $\sqrt{s}=$ 13 TeV}},
  {\emph{JHEP} {\bfseries 01} (2019) 183}
  [\href{https://arxiv.org/abs/1807.03825}{{\ttfamily 1807.03825}}].

\bibitem{Sirunyan:2018sgc}
{\scshape CMS} collaboration, A.~M. Sirunyan et~al., \emph{{Measurement and
  interpretation of differential cross sections for Higgs boson production at
  $\sqrt{s} =$ 13 TeV}},
  \href{https://doi.org/10.1016/j.physletb.2019.03.059}{\emph{Phys. Lett. B}
  {\bfseries 792} (2019) 369}
  [\href{https://arxiv.org/abs/1812.06504}{{\ttfamily 1812.06504}}].

\bibitem{Mantry:2009qz}
S.~Mantry and F.~Petriello, \emph{{Factorization and Resummation of Higgs Boson
  Differential Distributions in Soft-Collinear Effective Theory}},
  {\emph{Phys.Rev.} {\bfseries D81} (2010) 093007}
  [\href{https://arxiv.org/abs/0911.4135}{{\ttfamily 0911.4135}}].

\bibitem{Kang:2015msa}
Z.-B. Kang, A.~Prokudin, P.~Sun and F.~Yuan, \emph{{Extraction of Quark
  Transversity Distribution and Collins Fragmentation Functions with QCD
  Evolution}}, {\emph{Phys. Rev.} {\bfseries D93} (2016) 014009}
  [\href{https://arxiv.org/abs/1505.05589}{{\ttfamily 1505.05589}}].

\bibitem{Bastami:2018xqd}
S.~Bastami et~al., \emph{{Semi-Inclusive Deep Inelastic Scattering in
  Wandzura-Wilczek-type approximation}}, {\emph{JHEP} {\bfseries 06} (2019)
  007} [\href{https://arxiv.org/abs/1807.10606}{{\ttfamily 1807.10606}}].

\bibitem{Boer:1997mf}
D.~Boer, R.~Jakob and P.~J. Mulders, \emph{{Asymmetries in polarized hadron
  production in $e^+e^-$ annihilation up to order 1/Q}}, {\emph{Nucl. Phys.}
  {\bfseries B504} (1997) 345}
  [\href{https://arxiv.org/abs/hep-ph/9702281}{{\ttfamily hep-ph/9702281}}].

\bibitem{Boer:2008fr}
D.~Boer, \emph{{Angular dependences in inclusive two-hadron production at
  BELLE}}, {\emph{Nucl. Phys.} {\bfseries B806} (2009) 23}
  [\href{https://arxiv.org/abs/0804.2408}{{\ttfamily 0804.2408}}].

\bibitem{Pitonyak:2013dsu}
D.~Pitonyak, M.~Schlegel and A.~Metz, \emph{{Polarized hadron pair production
  from electron-positron annihilation}}, {\emph{Phys. Rev.} {\bfseries D89}
  (2014) 054032} [\href{https://arxiv.org/abs/1310.6240}{{\ttfamily
  1310.6240}}].

\bibitem{Rothstein:2016bsq}
I.~Z. Rothstein and I.~W. Stewart, \emph{{An Effective Field Theory for Forward
  Scattering and Factorization Violation}}, {\emph{JHEP} {\bfseries 08} (2016)
  025} [\href{https://arxiv.org/abs/1601.04695}{{\ttfamily 1601.04695}}].

\bibitem{Libby:1978bx}
S.~B. Libby and G.~Sterman, \emph{Mass divergences in two-particle inelastic
  scattering}, {\emph{Phys. Rev.} {\bfseries D18} (1978) 4737}.

\bibitem{Glauber:1955qq}
R.~J. Glauber, \emph{{Cross-sections in deuterium at high-energies}},
  {\emph{Phys. Rev.} {\bfseries 100} (1955) 242}.

\bibitem{Collins:1998ps}
J.~C. Collins, D.~E. Soper and G.~F. Sterman, \emph{{Factorization is not
  violated}}, {\emph{Phys. Lett.} {\bfseries B438} (1998) 184}
  [\href{https://arxiv.org/abs/hep-ph/9806234}{{\ttfamily hep-ph/9806234}}].

\bibitem{Khoze:2000dj}
V.~A. Khoze, A.~D. Martin and M.~Ryskin, \emph{{The Compatibility of
  diffractive hard scattering in $p \bar{p}$ and $e p$ collisions}},
  \href{https://doi.org/10.1016/S0370-2693(01)00170-8}{\emph{Phys.Lett.}
  {\bfseries B502} (2001) 87}
  [\href{https://arxiv.org/abs/hep-ph/0007083}{{\ttfamily hep-ph/0007083}}].

\bibitem{Kaidalov:2001iz}
A.~Kaidalov, V.~A. Khoze, A.~D. Martin and M.~Ryskin, \emph{{Probabilities of
  rapidity gaps in high-energy interactions}},
  \href{https://doi.org/10.1007/s100520100751}{\emph{Eur.Phys.J.} {\bfseries
  C21} (2001) 521} [\href{https://arxiv.org/abs/hep-ph/0105145}{{\ttfamily
  hep-ph/0105145}}].

\bibitem{Klasen:2008ah}
M.~Klasen and G.~Kramer, \emph{{Review of factorization breaking in diffractive
  photoproduction of dijets}},
  \href{https://doi.org/10.1142/S0217732308027461}{\emph{Mod.Phys.Lett.}
  {\bfseries A23} (2008) 1885}
  [\href{https://arxiv.org/abs/0806.2269}{{\ttfamily 0806.2269}}].

\bibitem{Kaidalov:2009fp}
A.~Kaidalov, V.~Khoze, A.~Martin and M.~Ryskin, \emph{{Factorization breaking
  in diffractive dijet photoproduction at HERA}},
  \href{https://doi.org/10.1140/epjc/s10052-010-1260-3}{\emph{Eur.Phys.J.}
  {\bfseries C66} (2010) 373}
  [\href{https://arxiv.org/abs/0911.3716}{{\ttfamily 0911.3716}}].

\bibitem{Collins:2007nk}
J.~Collins and J.-W. Qiu, \emph{{$k_{T}$ factorization is violated in
  production of high- transverse-momentum particles in hadron-hadron
  collisions}}, {\emph{Phys. Rev.} {\bfseries D75} (2007) 114014}
  [\href{https://arxiv.org/abs/0705.2141}{{\ttfamily 0705.2141}}].

\bibitem{Rogers:2010dm}
T.~C. Rogers and P.~J. Mulders, \emph{{No Generalized TMD-Factorization in
  Hadro-Production of High Transverse Momentum Hadrons}}, {\emph{Phys. Rev. D}
  {\bfseries 81} (2010) 094006}
  [\href{https://arxiv.org/abs/1001.2977}{{\ttfamily 1001.2977}}].

\bibitem{Mulders:2011zt}
P.~Mulders and T.~Rogers, \emph{{Gauge Links, TMD-Factorization, and
  TMD-Factorization Breaking}},
  \href{https://arxiv.org/abs/1102.4569}{{\ttfamily 1102.4569}}.

\bibitem{arXiv:1001.2977}
T.~C. Rogers and P.~J. Mulders, \emph{{No Generalized TMD-Factorization in
  Hadro-Production of High Transverse Momentum Hadrons}}, {\emph{Phys.Rev.}
  {\bfseries D81} (2010) 094006}
  [\href{https://arxiv.org/abs/1001.2977}{{\ttfamily 1001.2977}}].

\bibitem{Rothstein:unpublished}
I.~Z. Rothstein and I.~W. Stewart, \emph{{A Proof of Factorization for
  Drell-Yan with SCET}}, {\emph{to appear. See
  https://indico.ph.ed.ac.uk/event/63
  /contributions/1040/attachments/799/980/Ira\_Rothstein.pdf} }.

\bibitem{Fleming:2014rea}
S.~Fleming, \emph{{The role of Glauber exchange in soft collinear effective
  theory and the
  Balitsky\textendash{}Fadin\textendash{}Kuraev\textendash{}Lipatov Equation}},
  {\emph{Phys. Lett. B} {\bfseries 735} (2014) 266}
  [\href{https://arxiv.org/abs/1404.5672}{{\ttfamily 1404.5672}}].

\bibitem{Ji:2005nu}
X.-d. Ji, J.-P. Ma and F.~Yuan, \emph{{Transverse-momentum-dependent gluon
  distributions and semi-inclusive processes at hadron colliders}},
  {\emph{JHEP} {\bfseries 07} (2005) 020}
  [\href{https://arxiv.org/abs/hep-ph/0503015}{{\ttfamily hep-ph/0503015}}].

\bibitem{Aybat:2011ta}
S.~M. Aybat, A.~Prokudin and T.~C. Rogers, \emph{{Calculation of TMD Evolution
  for Transverse Single Spin Asymmetry Measurements}}, {\emph{Phys.Rev.Lett.}
  {\bfseries 108} (2012) 242003}
  [\href{https://arxiv.org/abs/1112.4423}{{\ttfamily 1112.4423}}].

\bibitem{Mueller:1979ih}
A.~H. Mueller, \emph{{On the Asymptotic Behavior of the Sudakov Form-factor}},
  {\emph{Phys. Rev. D} {\bfseries 20} (1979) 2037}.

\bibitem{Sen:1981sd}
A.~Sen, \emph{{Asymptotic Behavior of the Sudakov Form-Factor in QCD}},
  {\emph{Phys. Rev. D} {\bfseries 24} (1981) 3281}.

\bibitem{Contopanagos:1996nh}
H.~Contopanagos, E.~Laenen and G.~F. Sterman, \emph{{Sudakov factorization and
  resummation}}, {\emph{Nucl. Phys. B} {\bfseries 484} (1997) 303}
  [\href{https://arxiv.org/abs/hep-ph/9604313}{{\ttfamily hep-ph/9604313}}].

\bibitem{Altarelli:1984pt}
G.~Altarelli, R.~K. Ellis, M.~Greco and G.~Martinelli, \emph{{Vector Boson
  Production at Colliders: A Theoretical Reappraisal}}, {\emph{Nucl. Phys. B}
  {\bfseries 246} (1984) 12}.

\bibitem{Davies:1984hs}
C.~Davies and W.~J. Stirling, \emph{{Nonleading Corrections to the Drell-Yan
  Cross-Section at Small Transverse Momentum}}, {\emph{Nucl.Phys.} {\bfseries
  B244} (1984) 337}.

\bibitem{Arnold:1990yk}
P.~B. Arnold and R.~P. Kauffman, \emph{{W and Z production at next-to-leading
  order: From large $q_T$ to small}}, {\emph{Nucl. Phys.} {\bfseries B349}
  (1991) 381}.

\bibitem{Abbate:2010xh}
R.~Abbate, M.~Fickinger, A.~H. Hoang, V.~Mateu and I.~W. Stewart, \emph{{Thrust
  at $N^3LL$ with Power Corrections and a Precision Global Fit for
  $\alpha_s(m_Z)$}}, {\emph{Phys. Rev.} {\bfseries D83} (2011) 074021}
  [\href{https://arxiv.org/abs/1006.3080}{{\ttfamily 1006.3080}}].

\bibitem{Almeida:2014uva}
L.~G. Almeida, S.~D. Ellis, C.~Lee, G.~Sterman, I.~Sung et~al.,
  \emph{{Comparing and counting logs in direct and effective methods of QCD
  resummation}}, {\emph{JHEP} {\bfseries 1404} (2014) 174}
  [\href{https://arxiv.org/abs/1401.4460}{{\ttfamily 1401.4460}}].

\bibitem{Scimemi:2018xaf}
I.~Scimemi and A.~Vladimirov, \emph{{Systematic analysis of double-scale
  evolution}}, {\emph{JHEP} {\bfseries 08} (2018) 003}
  [\href{https://arxiv.org/abs/1803.11089}{{\ttfamily 1803.11089}}].

\bibitem{Collins:2012ss}
J.~Collins, \emph{{CSS Equation, etc, Follow from Structure of TMD
  Factorization}},  \href{https://arxiv.org/abs/1212.5974}{{\ttfamily
  1212.5974}}.

\bibitem{Polyakov:1980ca}
A.~M. Polyakov, \emph{{Gauge Fields as Rings of Glue}}, {\emph{Nucl. Phys. B}
  {\bfseries 164} (1980) 171}.

\bibitem{Korchemsky:1985xj}
G.~P. Korchemsky and A.~V. Radyushkin, \emph{{Loop Space Formalism and
  Renormalization Group for the Infrared Asymptotics of {QCD}}}, {\emph{Phys.
  Lett. B} {\bfseries 171} (1986) 459}.

\bibitem{Korchemsky:1987wg}
G.~P. Korchemsky and A.~V. Radyushkin, \emph{{Renormalization of the Wilson
  Loops Beyond the Leading Order}}, {\emph{Nucl. Phys.} {\bfseries B283} (1987)
  342}.

\bibitem{Collins:1989bt}
J.~C. Collins, \emph{{Sudakov form-factors}}, {\emph{Adv. Ser. Direct. High
  Energy Phys.} {\bfseries 5} (1989) 573}
  [\href{https://arxiv.org/abs/hep-ph/0312336}{{\ttfamily hep-ph/0312336}}].

\bibitem{Moch:2004pa}
S.~Moch, J.~A.~M. Vermaseren and A.~Vogt, \emph{{The Three loop splitting
  functions in QCD: The Nonsinglet case}}, {\emph{Nucl. Phys.} {\bfseries B688}
  (2004) 101} [\href{https://arxiv.org/abs/hep-ph/0403192}{{\ttfamily
  hep-ph/0403192}}].

\bibitem{Henn:2019swt}
J.~M. Henn, G.~P. Korchemsky and B.~Mistlberger, \emph{{The full four-loop cusp
  anomalous dimension in $\mathcal{N} = 4$ super Yang-Mills and QCD}},
  {\emph{JHEP} {\bfseries 04} (2020) 018}
  [\href{https://arxiv.org/abs/1911.10174}{{\ttfamily 1911.10174}}].

\bibitem{Collins:2014jpa}
J.~Collins and T.~Rogers, \emph{{Understanding the large-distance behavior of
  transverse-momentum-dependent parton densities and the Collins-Soper
  evolution kernel}},
  \href{https://doi.org/10.1103/PhysRevD.91.074020}{\emph{Phys. Rev. D}
  {\bfseries 91} (2015) 074020}
  [\href{https://arxiv.org/abs/1412.3820}{{\ttfamily 1412.3820}}].

\bibitem{Bacchetta:2015ora}
A.~Bacchetta, M.~G. Echevarria, P.~J.~G. Mulders, M.~Radici and A.~Signori,
  \emph{{Effects of TMD evolution and partonic flavor on $e^+e^-$ annihilation
  into hadrons}}, {\emph{JHEP} {\bfseries 11} (2015) 076}
  [\href{https://arxiv.org/abs/1508.00402}{{\ttfamily 1508.00402}}].

\bibitem{Bacchetta:2017gcc}
A.~Bacchetta, F.~Delcarro, C.~Pisano, M.~Radici and A.~Signori,
  \emph{{Extraction of partonic transverse momentum distributions from
  semi-inclusive deep-inelastic scattering, Drell-Yan and Z-boson production}},
  {\emph{JHEP} {\bfseries 06} (2017) 081}
  [\href{https://arxiv.org/abs/1703.10157}{{\ttfamily 1703.10157}}].

\bibitem{Qiu:2000ga}
J.-W. Qiu and X.-f. Zhang, \emph{{QCD prediction for heavy boson transverse
  momentum distributions}}, {\emph{Phys.Rev.Lett.} {\bfseries 86} (2001) 2724}
  [\href{https://arxiv.org/abs/hep-ph/0012058}{{\ttfamily hep-ph/0012058}}].

\bibitem{Laenen:2000de}
E.~Laenen, G.~F. Sterman and W.~Vogelsang, \emph{{Higher order QCD corrections
  in prompt photon production}}, {\emph{Phys. Rev. Lett.} {\bfseries 84} (2000)
  4296} [\href{https://arxiv.org/abs/hep-ph/0002078}{{\ttfamily
  hep-ph/0002078}}].

\bibitem{Kulesza:2002rh}
A.~Kulesza, G.~F. Sterman and W.~Vogelsang, \emph{{Joint resummation in
  electroweak boson production}}, {\emph{Phys.Rev.} {\bfseries D66} (2002)
  014011} [\href{https://arxiv.org/abs/hep-ph/0202251}{{\ttfamily
  hep-ph/0202251}}].

\bibitem{Monni:2016ktx}
P.~F. Monni, E.~Re and P.~Torrielli, \emph{{Higgs Transverse-Momentum
  Resummation in Direct Space}}, {\emph{Phys. Rev. Lett.} {\bfseries 116}
  (2016) 242001} [\href{https://arxiv.org/abs/1604.02191}{{\ttfamily
  1604.02191}}].

\bibitem{Ebert:2016gcn}
M.~A. Ebert and F.~J. Tackmann, \emph{{Resummation of Transverse Momentum
  Distributions in Distribution Space}}, {\emph{JHEP} {\bfseries 02} (2017)
  110} [\href{https://arxiv.org/abs/1611.08610}{{\ttfamily 1611.08610}}].

\bibitem{Kang:2017cjk}
D.~Kang, C.~Lee and V.~Vaidya, \emph{{A fast and accurate method for
  perturbative resummation of transverse momentum-dependent observables}},
  \href{https://doi.org/10.1007/JHEP04(2018)149}{\emph{JHEP} {\bfseries 04}
  (2018) 149} [\href{https://arxiv.org/abs/1710.00078}{{\ttfamily
  1710.00078}}].

\bibitem{Idilbi:2004vb}
A.~Idilbi, X.-d. Ji, J.-P. Ma and F.~Yuan, \emph{Collins-{S}oper equation for
  the energy evolution of transverse-momentum and spin dependent parton
  distributions}, {\emph{Phys. Rev.} {\bfseries D70} (2004) 074021}
  [\href{https://arxiv.org/abs/hep-ph/0406302}{{\ttfamily hep-ph/0406302}}].

\bibitem{Bozzi:2005wk}
G.~Bozzi, S.~Catani, D.~de~Florian and M.~Grazzini, \emph{{Transverse-momentum
  resummation and the spectrum of the Higgs boson at the LHC}}, {\emph{Nucl.
  Phys.} {\bfseries B737} (2006) 73}
  [\href{https://arxiv.org/abs/hep-ph/0508068}{{\ttfamily hep-ph/0508068}}].

\bibitem{Nadolsky:1999kb}
P.~M. Nadolsky, D.~R. Stump and C.~P. Yuan, \emph{{Semiinclusive hadron
  production at HERA: The Effect of QCD gluon resummation}}, {\emph{Phys. Rev.}
  {\bfseries D61} (2000) 014003}
  [\href{https://arxiv.org/abs/hep-ph/9906280}{{\ttfamily hep-ph/9906280}}].

\bibitem{Koike:2006fn}
Y.~Koike, J.~Nagashima and W.~Vogelsang, \emph{{Resummation for polarized
  semi-inclusive deep-inelastic scattering at small transverse momentum}},
  {\emph{Nucl.Phys.} {\bfseries B744} (2006) 59}
  [\href{https://arxiv.org/abs/hep-ph/0602188}{{\ttfamily hep-ph/0602188}}].

\bibitem{Collins:2016hqq}
J.~Collins, L.~Gamberg, A.~Prokudin, T.~C. Rogers, N.~Sato and B.~Wang,
  \emph{{Relating Transverse Momentum Dependent and Collinear Factorization
  Theorems in a Generalized Formalism}}, {\emph{Phys. Rev.} {\bfseries D94}
  (2016) 034014} [\href{https://arxiv.org/abs/1605.00671}{{\ttfamily
  1605.00671}}].

\bibitem{Berger:2004cc}
E.~L. Berger, J.-W. Qiu and Y.-l. Wang, \emph{{Transverse momentum distribution
  of $\upsilon$ production in hadronic collisions}}, {\emph{Phys.Rev.}
  {\bfseries D71} (2005) 034007}
  [\href{https://arxiv.org/abs/hep-ph/0404158}{{\ttfamily hep-ph/0404158}}].

\bibitem{Bacchetta:2008xw}
A.~Bacchetta, D.~Boer, M.~Diehl and P.~J. Mulders, \emph{{Matches and
  mismatches in the descriptions of semi-inclusive processes at low and high
  transverse momentum}},
  \href{https://doi.org/10.1088/1126-6708/2008/08/023}{\emph{JHEP} {\bfseries
  08} (2008) 023} [\href{https://arxiv.org/abs/0803.0227}{{\ttfamily
  0803.0227}}].

\bibitem{Boglione:2014oea}
M.~Boglione, J.~O.~G. Hernandez, S.~Melis and A.~Prokudin, \emph{{A study on
  the interplay between perturbative QCD and CSS/TMD formalism in SIDIS
  processes}}, {\emph{JHEP} {\bfseries 1502} (2015) 095}
  [\href{https://arxiv.org/abs/1412.1383}{{\ttfamily 1412.1383}}].

\bibitem{Collins:2017ybb}
J.~Collins, L.~Gamberg, A.~Prokudin, T.~Rogers, N.~Sato and B.~Wang,
  \emph{{Combining TMD factorization and collinear factorization}},  in
  \emph{{22nd International Symposium on Spin Physics}}, 2, 2017,
  \href{https://arxiv.org/abs/1702.00387}{{\ttfamily 1702.00387}}.

\bibitem{Echevarria:2018qyi}
M.~G. Echevarria, T.~Kasemets, J.-P. Lansberg, C.~Pisano and A.~Signori,
  \emph{{Matching factorization theorems with an inverse-error weighting}},
  {\emph{Phys. Lett. B} {\bfseries 781} (2018) 161}
  [\href{https://arxiv.org/abs/1801.01480}{{\ttfamily 1801.01480}}].

\bibitem{Grewal:2020hoc}
M.~Grewal, Z.-B. Kang, J.-W. Qiu and A.~Signori, \emph{{Predictive power of
  transverse-momentum-dependent distributions}},
  \href{https://doi.org/10.1103/PhysRevD.101.114023}{\emph{Phys. Rev. D}
  {\bfseries 101} (2020) 114023}
  [\href{https://arxiv.org/abs/2003.07453}{{\ttfamily 2003.07453}}].

\bibitem{Ligeti:2008ac}
Z.~Ligeti, I.~W. Stewart and F.~J. Tackmann, \emph{{Treating the b quark
  distribution function with reliable uncertainties}}, {\emph{Phys. Rev. D}
  {\bfseries 78} (2008) 114014}
  [\href{https://arxiv.org/abs/0807.1926}{{\ttfamily 0807.1926}}].

\bibitem{Berger:2010xi}
C.~F. Berger, C.~Marcantonini, I.~W. Stewart, F.~J. Tackmann and W.~J.
  Waalewijn, \emph{{Higgs Production with a Central Jet Veto at NNLL+NNLO}},
  {\emph{JHEP} {\bfseries 04} (2011) 092}
  [\href{https://arxiv.org/abs/1012.4480}{{\ttfamily 1012.4480}}].

\bibitem{Neill:2015roa}
D.~Neill, I.~Z. Rothstein and V.~Vaidya, \emph{{The Higgs Transverse Momentum
  Distribution at NNLL and its Theoretical Errors}}, {\emph{JHEP} {\bfseries
  12} (2015) 097} [\href{https://arxiv.org/abs/1503.00005}{{\ttfamily
  1503.00005}}].

\bibitem{Lustermans:2019plv}
G.~Lustermans, J.~K.~L. Michel, F.~J. Tackmann and W.~J. Waalewijn,
  \emph{{Joint two-dimensional resummation in $q_{T}$ and $0$-jettiness at
  NNLL}}, \href{https://doi.org/10.1007/JHEP03(2019)124}{\emph{JHEP} {\bfseries
  03} (2019) 124} [\href{https://arxiv.org/abs/1901.03331}{{\ttfamily
  1901.03331}}].

\bibitem{Chay:1991jc}
J.-g. Chay, S.~D. Ellis and W.~J. Stirling, \emph{{Azimuthal asymmetry in
  lepton - photon scattering at high-energies}}, {\emph{Phys. Lett. B}
  {\bfseries 269} (1991) 175}.

\bibitem{Anselmino:2006rv}
M.~Anselmino, M.~Boglione, A.~Prokudin and C.~Turk, \emph{{Semi-Inclusive Deep
  Inelastic Scattering processes from small to large $P_T$}}, {\emph{Eur. Phys.
  J.} {\bfseries A31} (2007) 373}
  [\href{https://arxiv.org/abs/hep-ph/0606286}{{\ttfamily hep-ph/0606286}}].

\bibitem{Berger:2001wr}
E.~L. Berger, J.-w. Qiu and X.-f. Zhang, \emph{{QCD factorized Drell-Yan
  cross-section at large transverse momentum}},
  \href{https://doi.org/10.1103/PhysRevD.65.034006}{\emph{Phys. Rev. D}
  {\bfseries 65} (2002) 034006}
  [\href{https://arxiv.org/abs/hep-ph/0107309}{{\ttfamily hep-ph/0107309}}].

\bibitem{Bozzi:2003jy}
G.~Bozzi, S.~Catani, D.~de~Florian and M.~Grazzini, \emph{{The $q_T$ spectrum
  of the Higgs boson at the LHC in QCD perturbation theory}},
  {\emph{Phys.Lett.} {\bfseries B564} (2003) 65}
  [\href{https://arxiv.org/abs/hep-ph/0302104}{{\ttfamily hep-ph/0302104}}].

\bibitem{Gamberg:2017jha}
L.~Gamberg, A.~Metz, D.~Pitonyak and A.~Prokudin, \emph{{Connections between
  collinear and transverse-momentum-dependent polarized observables within the
  Collins\textendash{}Soper\textendash{}Sterman formalism}}, {\emph{Phys. Lett.
  B} {\bfseries 781} (2018) 443}
  [\href{https://arxiv.org/abs/1712.08116}{{\ttfamily 1712.08116}}].

\bibitem{Banfi:2004yd}
A.~Banfi, G.~P. Salam and G.~Zanderighi, \emph{{Principles of general
  final-state resummation and automated implementation}}, {\emph{JHEP}
  {\bfseries 03} (2005) 073}
  [\href{https://arxiv.org/abs/hep-ph/0407286}{{\ttfamily hep-ph/0407286}}].

\bibitem{Banfi:2014sua}
A.~Banfi, H.~McAslan, P.~F. Monni and G.~Zanderighi, \emph{{A general method
  for the resummation of event-shape distributions in $e^+ e^-$ annihilation}},
  {\emph{JHEP} {\bfseries 05} (2015) 102}
  [\href{https://arxiv.org/abs/1412.2126}{{\ttfamily 1412.2126}}].

\bibitem{Bizon:2017rah}
W.~Bizon, P.~F. Monni, E.~Re, L.~Rottoli and P.~Torrielli,
  \emph{{Momentum-space resummation for transverse observables and the Higgs
  p$_{\perp}$ at N$^{3}$LL+NNLO}}, {\emph{JHEP} {\bfseries 02} (2018) 108}
  [\href{https://arxiv.org/abs/1705.09127}{{\ttfamily 1705.09127}}].

\bibitem{Banfi:2012yh}
A.~Banfi, G.~P. Salam and G.~Zanderighi, \emph{{NLL+NNLO predictions for
  jet-veto efficiencies in Higgs-boson and Drell-Yan production}}, {\emph{JHEP}
  {\bfseries 06} (2012) 159} [\href{https://arxiv.org/abs/1203.5773}{{\ttfamily
  1203.5773}}].

\bibitem{Frixione:1998dw}
S.~Frixione, P.~Nason and G.~Ridolfi, \emph{{Problems in the resummation of
  soft gluon effects in the transverse momentum distributions of massive vector
  bosons in hadronic collisions}}, {\emph{Nucl. Phys.} {\bfseries B542} (1999)
  311} [\href{https://arxiv.org/abs/hep-ph/9809367}{{\ttfamily
  hep-ph/9809367}}].

\bibitem{Feynman:1977yr}
R.~P. Feynman, R.~D. Field and G.~C. Fox, \emph{{Correlations Among Particles
  and Jets Produced with Large Transverse Momenta}}, {\emph{Nucl. Phys.}
  {\bfseries B128} (1977) 1}.

\bibitem{Feynman:1978dt}
R.~P. Feynman, R.~D. Field and G.~C. Fox, \emph{{A Quantum Chromodynamic
  Approach for the Large Transverse Momentum Production of Particles and
  Jets}}, {\emph{Phys. Rev.} {\bfseries D18} (1978) 3320}.

\bibitem{Anselmino:1994tv}
M.~Anselmino, M.~Boglione and F.~Murgia, \emph{{Single spin asymmetry for
  $p^\uparrow p \to \pi X$ in perturbative QCD}}, {\emph{Phys. Lett.}
  {\bfseries B362} (1995) 164}
  [\href{https://arxiv.org/abs/hep-ph/9503290}{{\ttfamily hep-ph/9503290}}].

\bibitem{Anselmino:1999pw}
M.~Anselmino, M.~Boglione and F.~Murgia, \emph{{Phenomenology of single spin
  asymmetries in $p^\uparrow p \to \pi X$}}, {\emph{Phys. Rev.} {\bfseries D60}
  (1999) 054027} [\href{https://arxiv.org/abs/hep-ph/9901442}{{\ttfamily
  hep-ph/9901442}}].

\bibitem{Boglione:1999dq}
M.~Boglione and E.~Leader, \emph{{Reassessment of the Collins mechanism for
  single spin asymmetries and the behavior of $\Delta d(x)$ at large x}},
  {\emph{Phys. Rev.} {\bfseries D61} (2000) 114001}
  [\href{https://arxiv.org/abs/hep-ph/9911207}{{\ttfamily hep-ph/9911207}}].

\bibitem{Anselmino:2004nk}
M.~Anselmino, M.~Boglione, U.~D'Alesio, E.~Leader and F.~Murgia,
  \emph{{Accessing Sivers gluon distribution via transverse single spin
  asymmetries in $p^\uparrow p \to D\,X$ processes at RHIC}}, {\emph{Phys.
  Rev.} {\bfseries D70} (2004) 074025}
  [\href{https://arxiv.org/abs/hep-ph/0407100}{{\ttfamily hep-ph/0407100}}].

\bibitem{Georgi:1977tv}
H.~Georgi and H.~D. Politzer, \emph{{Clean Tests of QCD in mu p Scattering}},
  {\emph{Phys. Rev. Lett.} {\bfseries 40} (1978) 3}.

\bibitem{Mendez:1978zx}
A.~Mendez, \emph{{QCD Predictions for Semiinclusive and Inclusive
  Leptoproduction}}, {\emph{Nucl. Phys.} {\bfseries B145} (1978) 199}.

\bibitem{Kane:1978nd}
G.~L. Kane, J.~Pumplin and W.~Repko, \emph{{Transverse Quark Polarization in
  Large $p_T$ Reactions, $e^+e^-$ Jets, and Leptoproduction: A Test of QCD}},
  {\emph{Phys.Rev.Lett.} {\bfseries 41} (1978) 1689}.

\bibitem{Cahn:1978se}
R.~N. Cahn, \emph{{Azimuthal Dependence in Leptoproduction: A Simple Parton
  Model Calculation}}, {\emph{Phys. Lett. B} {\bfseries 78} (1978) 269}.

\bibitem{Cahn:1989yf}
R.~Cahn, \emph{{Critique of Parton Model Calculations of Azimuthal Dependence
  in Leptoproduction}}, {\emph{Phys. Rev. D} {\bfseries 40} (1989) 3107}.

\bibitem{Kotzinian:1994dv}
A.~Kotzinian, \emph{{New quark distributions and semiinclusive
  electroproduction on the polarized nucleons}}, {\emph{Nucl. Phys.} {\bfseries
  B441} (1995) 234} [\href{https://arxiv.org/abs/hep-ph/9412283}{{\ttfamily
  hep-ph/9412283}}].

\bibitem{Tangerman:1995hw}
R.~D. Tangerman and P.~J. Mulders, \emph{{Probing transverse quark polarization
  in deep inelastic leptoproduction}}, {\emph{Phys. Lett.} {\bfseries B352}
  (1995) 129} [\href{https://arxiv.org/abs/hep-ph/9501202}{{\ttfamily
  hep-ph/9501202}}].

\bibitem{Efremov:1981sh}
A.~Efremov and O.~Teryaev, \emph{{On Spin Effects in Quantum Chromodynamics}},
  {\emph{Sov.J.Nucl.Phys.} {\bfseries 36} (1982) 140}.

\bibitem{Efremov:1983eb}
A.~V. Efremov and O.~V. Teryaev, \emph{{THE TRANSVERSAL POLARIZATION IN QUANTUM
  CHROMODYNAMICS}}, {\emph{Sov. J. Nucl. Phys.} {\bfseries 39} (1984) 962}.

\bibitem{Qiu:1998ia}
J.-w. Qiu and G.~F. Sterman, \emph{{Single transverse spin asymmetries in
  hadronic pion production}}, {\emph{Phys. Rev.} {\bfseries D59} (1999) 014004}
  [\href{https://arxiv.org/abs/hep-ph/9806356}{{\ttfamily hep-ph/9806356}}].

\bibitem{Qiu:1991wg}
J.-W. Qiu and G.~Sterman, \emph{Single transverse spin asymmetries in direct
  photon production}, {\emph{Nucl. Phys.} {\bfseries B378} (1992) 52}.

\bibitem{Sivers:1990fh}
D.~W. Sivers, \emph{Hard scattering scaling laws for single spin production
  asymmetries}, {\emph{Phys. Rev.} {\bfseries D43} (1991) 261}.

\bibitem{Brodsky:2013oya}
S.~J. Brodsky, D.~S. Hwang, Y.~V. Kovchegov, I.~Schmidt and M.~D. Sievert,
  \emph{{Single-Spin Asymmetries in Semi-inclusive Deep Inelastic Scattering
  and Drell-Yan Processes}}, {\emph{Phys. Rev. D} {\bfseries 88} (2013) 014032}
  [\href{https://arxiv.org/abs/1304.5237}{{\ttfamily 1304.5237}}].

\bibitem{Nachtmann:1977ek}
O.~Nachtmann, \emph{{A New Tool for the Study of Fundamental Interactions:
  Parity Odd Correlations in Quark Fragmentation}}, {\emph{Nucl. Phys. B}
  {\bfseries 127} (1977) 314}.

\bibitem{Dalitz:1988ab}
R.~H. Dalitz, G.~R. Goldstein and R.~Marshall, \emph{{On the Helicity of Charm
  Jets}}, {\emph{Z. Phys. C} {\bfseries 42} (1989) 441}.

\bibitem{Dalitz:1988aq}
R.~H. Dalitz, G.~R. Goldstein and R.~Marshall, \emph{{Heavy Quark Spin
  Correlations in $e^+ e^-$ Annihilations}}, {\emph{Phys. Lett. B} {\bfseries
  215} (1988) 783}.

\bibitem{Efremov:1992pe}
A.~V. Efremov, L.~Mankiewicz and N.~A. Tornqvist, \emph{{Jet handedness as a
  measure of quark and gluon polarization}}, {\emph{Phys. Lett. B} {\bfseries
  284} (1992) 394}.

\bibitem{Collins:1993kq}
J.~C. Collins, S.~F. Heppelmann and G.~A. Ladinsky, \emph{{Measuring
  transversity densities in singly polarized hadron hadron and lepton - hadron
  collisions}}, {\emph{Nucl. Phys.} {\bfseries B420} (1994) 565}
  [\href{https://arxiv.org/abs/hep-ph/9305309}{{\ttfamily hep-ph/9305309}}].

\bibitem{Jaffe:1997hf}
R.~L. Jaffe, X.-m. Jin and J.~Tang, \emph{{Interference fragmentation functions
  and the nucleon's transversity}}, {\emph{Phys. Rev. Lett.} {\bfseries 80}
  (1998) 1166} [\href{https://arxiv.org/abs/hep-ph/9709322}{{\ttfamily
  hep-ph/9709322}}].

\bibitem{Bianconi:1999cd}
A.~Bianconi, S.~Boffi, R.~Jakob and M.~Radici, \emph{{Two hadron interference
  fragmentation functions. Part 1. General framework}}, {\emph{Phys. Rev. D}
  {\bfseries 62} (2000) 034008}
  [\href{https://arxiv.org/abs/hep-ph/9907475}{{\ttfamily hep-ph/9907475}}].

\bibitem{Radici:2001na}
M.~Radici, R.~Jakob and A.~Bianconi, \emph{{Accessing transversity with
  interference fragmentation functions}}, {\emph{Phys. Rev.} {\bfseries D65}
  (2002) 074031} [\href{https://arxiv.org/abs/hep-ph/0110252}{{\ttfamily
  hep-ph/0110252}}].

\bibitem{Bacchetta:2002ux}
A.~Bacchetta and M.~Radici, \emph{{Partial wave analysis of two hadron
  fragmentation functions}}, {\emph{Phys. Rev.} {\bfseries D67} (2003) 094002}
  [\href{https://arxiv.org/abs/hep-ph/0212300}{{\ttfamily hep-ph/0212300}}].

\bibitem{Metz:2016swz}
A.~Metz and A.~Vossen, \emph{{Parton Fragmentation Functions}}, {\emph{Prog.
  Part. Nucl. Phys.} {\bfseries 91} (2016) 136}
  [\href{https://arxiv.org/abs/1607.02521}{{\ttfamily 1607.02521}}].

\bibitem{EPJA}
M.~Anselmino, M.~Guidal and P.~Rossi, \emph{Topical issue on the 3-{D}
  structure of the nucleon},
  \href{https://doi.org/10.1140/epja/i2016-16164-4}{\emph{The European Physical
  Journal A} {\bfseries 52} (2016) 164}.

\bibitem{Bacchetta:2016ccz}
A.~Bacchetta, \emph{{Where do we stand with a 3-D picture of the proton?}},
  {\emph{Eur. Phys. J. A} {\bfseries 52} (2016) 163}
  [\href{https://arxiv.org/abs/2107.06772}{{\ttfamily 2107.06772}}].

\bibitem{Aschenauer:2015ndk}
E.~C. Aschenauer, U.~D'Alesio and F.~Murgia, \emph{{TMDs and SSAs in hadronic
  interactions}}, {\emph{Eur. Phys. J. A} {\bfseries 52} (2016) 156}
  [\href{https://arxiv.org/abs/1512.05379}{{\ttfamily 1512.05379}}].

\bibitem{Boglione:2015zyc}
M.~Boglione and A.~Prokudin, \emph{{Phenomenology of transverse spin: past,
  present and future}}, {\emph{Eur. Phys. J. A} {\bfseries 52} (2016) 154}
  [\href{https://arxiv.org/abs/1511.06924}{{\ttfamily 1511.06924}}].

\bibitem{Avakian:2016rst}
H.~Avakian, A.~Bressan and M.~Contalbrigo, \emph{{Experimental results on
  TMDs}}, {\emph{Eur. Phys. J. A} {\bfseries 52} (2016) 150}.

\bibitem{Airapetian:2012ki}
{\scshape HERMES} collaboration, A.~Airapetian et~al., \emph{{Multiplicities of
  charged pions and kaons from semi-inclusive deep-inelastic scattering by the
  proton and the deuteron}}, {\emph{Phys. Rev.} {\bfseries D87} (2013) 074029}
  [\href{https://arxiv.org/abs/1212.5407}{{\ttfamily 1212.5407}}].

\bibitem{Aghasyan:2017ctw}
{\scshape COMPASS} collaboration, M.~Aghasyan et~al.,
  \emph{{Transverse-momentum-dependent Multiplicities of Charged Hadrons in
  Muon-Deuteron Deep Inelastic Scattering}}, {\emph{Phys. Rev.} {\bfseries D97}
  (2018) 032006} [\href{https://arxiv.org/abs/1709.07374}{{\ttfamily
  1709.07374}}].

\bibitem{Anselmino:2005nn}
M.~Anselmino, M.~Boglione, U.~D'Alesio, A.~Kotzinian, F.~Murgia and
  A.~Prokudin, \emph{{The Role of Cahn and Sivers effects in deep inelastic
  scattering}}, {\emph{Phys. Rev. D} {\bfseries 71} (2005) 074006}
  [\href{https://arxiv.org/abs/hep-ph/0501196}{{\ttfamily hep-ph/0501196}}].

\bibitem{Schweitzer:2010tt}
P.~Schweitzer, T.~Teckentrup and A.~Metz, \emph{{Intrinsic transverse parton
  momenta in deeply inelastic reactions}}, {\emph{Phys. Rev.} {\bfseries D81}
  (2010) 094019} [\href{https://arxiv.org/abs/1003.2190}{{\ttfamily
  1003.2190}}].

\bibitem{Anselmino:2013lza}
M.~Anselmino, M.~Boglione, J.~O. Gonzalez~Hernandez, S.~Melis and A.~Prokudin,
  \emph{{Unpolarised Transverse Momentum Dependent Distribution and
  Fragmentation Functions from SIDIS Multiplicities}}, {\emph{JHEP} {\bfseries
  04} (2014) 005} [\href{https://arxiv.org/abs/1312.6261}{{\ttfamily
  1312.6261}}].

\bibitem{Signori:2013mda}
A.~Signori, A.~Bacchetta, M.~Radici and G.~Schnell, \emph{{Investigations into
  the flavor dependence of partonic transverse momentum}}, {\emph{JHEP}
  {\bfseries 11} (2013) 194} [\href{https://arxiv.org/abs/1309.3507}{{\ttfamily
  1309.3507}}].

\bibitem{Musch:2007ya}
{\scshape LHPC} collaboration, B.~U. Musch, P.~H\"agler, A.~Sch\"afer,
  M.~G\"ockeler, D.~B. Renner and J.~W. Negele, \emph{{Transverse momentum
  distributions of quarks from the lattice using extended gauge links}},
  {\emph{PoS} {\bfseries LATTICE2007} (2007) 155}
  [\href{https://arxiv.org/abs/0710.4423}{{\ttfamily 0710.4423}}].

\bibitem{Orginos:2017kos}
K.~Orginos, A.~Radyushkin, J.~Karpie and S.~Zafeiropoulos, \emph{{Lattice QCD
  exploration of parton pseudo-distribution functions}}, {\emph{Phys. Rev.}
  {\bfseries D96} (2017) 094503}
  [\href{https://arxiv.org/abs/1706.05373}{{\ttfamily 1706.05373}}].

\bibitem{Landry:2002ix}
F.~Landry, R.~Brock, P.~M. Nadolsky and C.~Yuan, \emph{{Tevatron Run-1 $Z$
  boson data and Collins-Soper-Sterman resummation formalism}},
  {\emph{Phys.Rev.} {\bfseries D67} (2003) 073016}
  [\href{https://arxiv.org/abs/hep-ph/0212159}{{\ttfamily hep-ph/0212159}}].

\bibitem{Su:2014wpa}
P.~Sun, J.~Isaacson, C.~P. Yuan and F.~Yuan, \emph{{Nonperturbative functions
  for SIDIS and Drell--Yan processes}}, {\emph{Int. J. Mod. Phys.} {\bfseries
  A33} (2018) 1841006} [\href{https://arxiv.org/abs/1406.3073}{{\ttfamily
  1406.3073}}].

\bibitem{Kulesza:2003wn}
A.~Kulesza, G.~F. Sterman and W.~Vogelsang, \emph{{Joint resummation for Higgs
  production}}, {\emph{Phys. Rev.} {\bfseries D69} (2004) 014012}
  [\href{https://arxiv.org/abs/hep-ph/0309264}{{\ttfamily hep-ph/0309264}}].

\bibitem{Catani:2003zt}
S.~Catani, D.~de~Florian, M.~Grazzini and P.~Nason, \emph{{Soft gluon
  resummation for Higgs boson production at hadron colliders}}, {\emph{JHEP}
  {\bfseries 0307} (2003) 028}
  [\href{https://arxiv.org/abs/hep-ph/0306211}{{\ttfamily hep-ph/0306211}}].

\bibitem{Bertone:2019nxa}
V.~Bertone, I.~Scimemi and A.~Vladimirov, \emph{{Extraction of unpolarized
  quark transverse momentum dependent parton distributions from
  Drell-Yan/Z-boson production}},
  \href{https://doi.org/10.1007/JHEP06(2019)028}{\emph{JHEP} {\bfseries 06}
  (2019) 028} [\href{https://arxiv.org/abs/1902.08474}{{\ttfamily
  1902.08474}}].

\bibitem{Scimemi:2019cmh}
I.~Scimemi and A.~Vladimirov, \emph{{Non-perturbative structure of
  semi-inclusive deep-inelastic and Drell-Yan scattering at small transverse
  momentum}}, {\emph{JHEP} {\bfseries 06} (2020) 137}
  [\href{https://arxiv.org/abs/1912.06532}{{\ttfamily 1912.06532}}].

\bibitem{Vladimirov:2020umg}
A.~A. Vladimirov, \emph{{Self-contained definition of the Collins-Soper
  kernel}}, {\emph{Phys. Rev. Lett.} {\bfseries 125} (2020) 192002}
  [\href{https://arxiv.org/abs/2003.02288}{{\ttfamily 2003.02288}}].

\bibitem{Konychev:2005iy}
A.~V. Konychev and P.~M. Nadolsky, \emph{{Universality of the
  Collins-Soper-Sterman nonperturbative function in gauge boson production}},
  {\emph{Phys.Lett.} {\bfseries B633} (2006) 710}
  [\href{https://arxiv.org/abs/hep-ph/0506225}{{\ttfamily hep-ph/0506225}}].

\bibitem{Aidala:2014hva}
C.~Aidala, B.~Field, L.~Gamberg and T.~Rogers, \emph{{Limits on TMD Evolution
  From Semi-Inclusive Deep Inelastic Scattering at Moderate $Q$}},
  {\emph{Phys.Rev.} {\bfseries D89} (2014) 094002}
  [\href{https://arxiv.org/abs/1401.2654}{{\ttfamily 1401.2654}}].

\bibitem{Vladimirov:2019bfa}
A.~Vladimirov, \emph{{Pion-induced Drell-Yan processes within TMD
  factorization}}, {\emph{JHEP} {\bfseries 10} (2019) 090}
  [\href{https://arxiv.org/abs/1907.10356}{{\ttfamily 1907.10356}}].

\bibitem{Vladimirov:2016dll}
A.~A. Vladimirov, \emph{{Correspondence between Soft and Rapidity Anomalous
  Dimensions}}, {\emph{Phys. Rev. Lett.} {\bfseries 118} (2017) 062001}
  [\href{https://arxiv.org/abs/1610.05791}{{\ttfamily 1610.05791}}].

\bibitem{Scimemi:2017etj}
I.~Scimemi and A.~Vladimirov, \emph{{Analysis of vector boson production within
  TMD factorization}}, {\emph{Eur. Phys. J.} {\bfseries C78} (2018) 89}
  [\href{https://arxiv.org/abs/1706.01473}{{\ttfamily 1706.01473}}].

\bibitem{Bacchetta:2019sam}
A.~Bacchetta, V.~Bertone, C.~Bissolotti, G.~Bozzi, F.~Delcarro, F.~Piacenza
  et~al., \emph{{Transverse-momentum-dependent parton distributions up to
  N$^{3}$LL from Drell-Yan data}}, {\emph{JHEP} {\bfseries 07} (2020) 117}
  [\href{https://arxiv.org/abs/1912.07550}{{\ttfamily 1912.07550}}].

\bibitem{Echevarria:2014xaa}
M.~G. Echevarria, A.~Idilbi, Z.-B. Kang and I.~Vitev, \emph{{QCD Evolution of
  the Sivers Asymmetry}}, {\emph{Phys. Rev.} {\bfseries D89} (2014) 074013}
  [\href{https://arxiv.org/abs/1401.5078}{{\ttfamily 1401.5078}}].

\bibitem{Moreno:1990sf}
G.~Moreno et~al., \emph{{Dimuon production in proton - copper collisions at
  $\sqrt{s}$ = 38.8 GeV}}, {\emph{Phys. Rev. D} {\bfseries 43} (1991) 2815}.

\bibitem{Ito:1980ev}
A.~Ito et~al., \emph{{Measurement of the Continuum of Dimuons Produced in
  High-Energy Proton - Nucleus Collisions}}, {\emph{Phys. Rev. D} {\bfseries
  23} (1981) 604}.

\bibitem{Affolder:1999jh}
{\scshape CDF} collaboration, T.~Affolder et~al., \emph{{The transverse
  momentum and total cross section of $e^+e^-$ pairs in the $Z$ boson region
  from $p\bar{p}$ collisions at $\sqrt{s} = 1.8$ TeV}}, {\emph{Phys. Rev.
  Lett.} {\bfseries 84} (2000) 845}
  [\href{https://arxiv.org/abs/hep-ex/0001021}{{\ttfamily hep-ex/0001021}}].

\bibitem{Aaltonen:2012fi}
{\scshape CDF} collaboration, T.~Aaltonen et~al., \emph{{Transverse momentum
  cross section of $e^+e^-$ pairs in the $Z$-boson region from $p\bar{p}$
  collisions at $\sqrt{s}=1.96$ TeV}}, {\emph{Phys. Rev. D} {\bfseries 86}
  (2012) 052010} [\href{https://arxiv.org/abs/1207.7138}{{\ttfamily
  1207.7138}}].

\bibitem{Abbott:1999wk}
{\scshape D0} collaboration, B.~Abbott et~al., \emph{{Measurement of the
  inclusive differential cross section for $Z$ bosons as a function of
  transverse momentum in $\bar{p}p$ collisions at $\sqrt{s} = 1.8$ TeV}},
  {\emph{Phys. Rev. D} {\bfseries 61} (2000) 032004}
  [\href{https://arxiv.org/abs/hep-ex/9907009}{{\ttfamily hep-ex/9907009}}].

\bibitem{Abazov:2007ac}
{\scshape D0} collaboration, V.~Abazov et~al., \emph{{Measurement of the shape
  of the boson transverse momentum distribution in $p \bar{p} \to Z /
  \gamma^{*} \to e^+ e^- + X$ events produced at $\sqrt{s}$ = 1.96 TeV}},
  {\emph{Phys. Rev. Lett.} {\bfseries 100} (2008) 102002}
  [\href{https://arxiv.org/abs/0712.0803}{{\ttfamily 0712.0803}}].

\bibitem{Abazov:2010kn}
{\scshape D0} collaboration, V.~M. Abazov et~al., \emph{{Measurement of the
  Normalized $Z/\gamma^* -> \mu^+\mu^-$ Transverse Momentum Distribution in
  $p\bar{p}$ Collisions at $\sqrt{s}=1.96$ TeV}}, {\emph{Phys. Lett. B}
  {\bfseries 693} (2010) 522}
  [\href{https://arxiv.org/abs/1006.0618}{{\ttfamily 1006.0618}}].

\bibitem{Aaij:2015gna}
{\scshape LHCb} collaboration, R.~Aaij et~al., \emph{{Measurement of the
  forward $Z$ boson production cross-section in $pp$ collisions at $\sqrt{s}=7$
  TeV}}, {\emph{JHEP} {\bfseries 08} (2015) 039}
  [\href{https://arxiv.org/abs/1505.07024}{{\ttfamily 1505.07024}}].

\bibitem{Aaij:2015zlq}
{\scshape LHCb} collaboration, R.~Aaij et~al., \emph{{Measurement of forward W
  and Z boson production in $pp$ collisions at $ \sqrt{s} = 8 $ TeV}},
  {\emph{JHEP} {\bfseries 01} (2016) 155}
  [\href{https://arxiv.org/abs/1511.08039}{{\ttfamily 1511.08039}}].

\bibitem{Aaij:2016mgv}
{\scshape LHCb} collaboration, R.~Aaij et~al., \emph{{Measurement of the
  forward Z boson production cross-section in pp collisions at $\sqrt{s} = 13$
  TeV}}, {\emph{JHEP} {\bfseries 09} (2016) 136}
  [\href{https://arxiv.org/abs/1607.06495}{{\ttfamily 1607.06495}}].

\bibitem{Chatrchyan:2011wt}
{\scshape CMS} collaboration, S.~Chatrchyan et~al., \emph{{Measurement of the
  Rapidity and Transverse Momentum Distributions of $Z$ Bosons in $pp$
  Collisions at $\sqrt{s}=7$ TeV}}, {\emph{Phys. Rev.} {\bfseries D85} (2012)
  032002} [\href{https://arxiv.org/abs/1110.4973}{{\ttfamily 1110.4973}}].

\bibitem{Khachatryan:2016nbe}
{\scshape CMS} collaboration, V.~Khachatryan et~al., \emph{{Measurement of the
  transverse momentum spectra of weak vector bosons produced in proton-proton
  collisions at $ \sqrt{s} = 8 $ TeV}}, {\emph{JHEP} {\bfseries 02} (2017) 096}
  [\href{https://arxiv.org/abs/1606.05864}{{\ttfamily 1606.05864}}].

\bibitem{Aad:2015auj}
{\scshape ATLAS} collaboration, G.~Aad et~al., \emph{{Measurement of the
  transverse momentum and $\phi ^*_{\eta }$ distributions of Drell--Yan lepton
  pairs in proton--proton collisions at $\sqrt{s} = 8$ TeV with the ATLAS
  detector}}, {\emph{Eur. Phys. J.} {\bfseries C76} (2016) 291}
  [\href{https://arxiv.org/abs/1512.02192}{{\ttfamily 1512.02192}}].

\bibitem{Boer:2011fh}
D.~Boer, M.~Diehl, R.~Milner, R.~Venugopalan, W.~Vogelsang et~al.,
  \emph{{Gluons and the quark sea at high energies: Distributions,
  polarization, tomography}},
  \href{https://arxiv.org/abs/1108.1713}{{\ttfamily 1108.1713}}.

\bibitem{Dudek:2012vr}
J.~Dudek et~al., \emph{{Physics Opportunities with the 12 GeV Upgrade at
  Jefferson Lab}}, {\emph{Eur. Phys. J.} {\bfseries A48} (2012) 187}
  [\href{https://arxiv.org/abs/1208.1244}{{\ttfamily 1208.1244}}].

\bibitem{Aschenauer:2015eha}
E.-C. Aschenauer, A.~Bazilevsky, M.~Diehl, J.~Drachenberg, K.~O. Eyser et~al.,
  \emph{{The RHIC SPIN Program: Achievements and Future Opportunities}},
  \href{https://arxiv.org/abs/1501.01220}{{\ttfamily 1501.01220}}.

\bibitem{Gautheron:2010wva}
{\scshape COMPASS} collaboration, F.~Gautheron et~al., \emph{{COMPASS-II
  Proposal}}, .

\bibitem{Bradamante:2018ick}
{\scshape COMPASS} collaboration, F.~Bradamante, \emph{{The future SIDIS
  measurement on transversely polarized deuterons by the COMPASS
  Collaboration}}, {\emph{PoS} {\bfseries SPIN2018} (2018) 045}
  [\href{https://arxiv.org/abs/1812.07281}{{\ttfamily 1812.07281}}].

\bibitem{Anselmino:2010bs}
M.~Anselmino, M.~Boglione, U.~D'Alesio, S.~Melis, F.~Murgia and A.~Prokudin,
  \emph{{New insight on the Sivers transverse momentum dependent distribution
  function}}, {\emph{J. Phys. Conf. Ser.} {\bfseries 295} (2011) 012062}
  [\href{https://arxiv.org/abs/1012.3565}{{\ttfamily 1012.3565}}].

\bibitem{Anselmino:2005ea}
M.~Anselmino, M.~Boglione, U.~D'Alesio, A.~Kotzinian, F.~Murgia and
  A.~Prokudin, \emph{{Extracting the Sivers function from polarized SIDIS data
  and making predictions}}, {\emph{Phys. Rev.} {\bfseries D72} (2005) 094007}
  [\href{https://arxiv.org/abs/hep-ph/0507181}{{\ttfamily hep-ph/0507181}}].

\bibitem{Anselmino:2005an}
M.~Anselmino et~al., \emph{{Comparing extractions of Sivers functions}},  in
  \emph{{Transversity. Proceedings, Workshop, Como, Italy, September 7-10,
  2005}}, pp.~236--243, 2005,
  \href{https://arxiv.org/abs/hep-ph/0511017}{{\ttfamily hep-ph/0511017}}.

\bibitem{Collins:2005ie}
J.~Collins, A.~Efremov, K.~Goeke, S.~Menzel, A.~Metz et~al., \emph{{Sivers
  effect in semi-inclusive deeply inelastic scattering}}, {\emph{Phys.Rev.}
  {\bfseries D73} (2006) 014021}
  [\href{https://arxiv.org/abs/hep-ph/0509076}{{\ttfamily hep-ph/0509076}}].

\bibitem{Vogelsang:2005cs}
W.~Vogelsang and F.~Yuan, \emph{{Single-transverse spin asymmetries: From DIS
  to hadronic collisions}}, {\emph{Phys. Rev.} {\bfseries D72} (2005) 054028}
  [\href{https://arxiv.org/abs/hep-ph/0507266}{{\ttfamily hep-ph/0507266}}].

\bibitem{Anselmino:2008sga}
M.~Anselmino, M.~Boglione, U.~D'Alesio, A.~Kotzinian, S.~Melis, F.~Murgia
  et~al., \emph{{Sivers Effect for Pion and Kaon Production in Semi-Inclusive
  Deep Inelastic Scattering}}, {\emph{Eur. Phys. J.} {\bfseries A39} (2009) 89}
  [\href{https://arxiv.org/abs/0805.2677}{{\ttfamily 0805.2677}}].

\bibitem{Bacchetta:2011gx}
A.~Bacchetta and M.~Radici, \emph{{Constraining quark angular momentum through
  semi-inclusive measurements}}, {\emph{Phys. Rev. Lett.} {\bfseries 107}
  (2011) 212001} [\href{https://arxiv.org/abs/1107.5755}{{\ttfamily
  1107.5755}}].

\bibitem{Bacchetta:2020gko}
A.~Bacchetta, F.~Delcarro, C.~Pisano and M.~Radici, \emph{{The 3-dimensional
  distribution of quarks in momentum space}},
  \href{https://doi.org/10.1016/j.physletb.2022.136961}{\emph{Phys. Lett. B}
  {\bfseries 827} (2022) 136961}
  [\href{https://arxiv.org/abs/2004.14278}{{\ttfamily 2004.14278}}].

\bibitem{Echevarria:2020hpy}
M.~G. Echevarria, Z.-B. Kang and J.~Terry, \emph{{Global analysis of the Sivers
  functions at NLO+NNLL in QCD}}, {\emph{JHEP} {\bfseries 01} (2021) 126}
  [\href{https://arxiv.org/abs/2009.10710}{{\ttfamily 2009.10710}}].

\bibitem{Bury:2020vhj}
M.~Bury, A.~Prokudin and A.~Vladimirov, \emph{{Extraction of the Sivers
  Function from SIDIS, Drell-Yan, and $W^{\pm}/Z$ Data at
  Next-to-Next-to-Next-to Leading Order}}, {\emph{Phys. Rev. Lett.} {\bfseries
  126} (2021) 112002} [\href{https://arxiv.org/abs/2012.05135}{{\ttfamily
  2012.05135}}].

\bibitem{Airapetian:2009ae}
{\scshape HERMES} collaboration, A.~Airapetian et~al., \emph{{Observation of
  the Naive-T-odd Sivers Effect in Deep-Inelastic Scattering}}, {\emph{Phys.
  Rev. Lett.} {\bfseries 103} (2009) 152002}
  [\href{https://arxiv.org/abs/0906.3918}{{\ttfamily 0906.3918}}].

\bibitem{Airapetian:2020zzo}
{\scshape HERMES} collaboration, A.~Airapetian et~al., \emph{{Azimuthal single-
  and double-spin asymmetries in semi-inclusive deep-inelastic lepton
  scattering by transversely polarized protons}},
  \href{https://arxiv.org/abs/2007.07755}{{\ttfamily 2007.07755}}.

\bibitem{Alekseev:2008aa}
{\scshape COMPASS} collaboration, M.~Alekseev et~al., \emph{{Collins and Sivers
  asymmetries for pions and kaons in muon-deuteron DIS}}, {\emph{Phys. Lett.}
  {\bfseries B673} (2009) 127}
  [\href{https://arxiv.org/abs/0802.2160}{{\ttfamily 0802.2160}}].

\bibitem{Adolph:2014zba}
{\scshape COMPASS} collaboration, C.~Adolph et~al., \emph{{Collins and Sivers
  asymmetries in muonproduction of pions and kaons off transversely polarised
  protons}}, {\emph{Phys. Lett.} {\bfseries B744} (2015) 250}
  [\href{https://arxiv.org/abs/1408.4405}{{\ttfamily 1408.4405}}].

\bibitem{Qian:2011py}
{\scshape Jefferson Lab Hall A} collaboration, X.~Qian et~al., \emph{{Single
  Spin Asymmetries in Charged Pion Production from Semi-Inclusive Deep
  Inelastic Scattering on a Transversely Polarized $^3$He Target}},
  {\emph{Phys. Rev. Lett.} {\bfseries 107} (2011) 072003}
  [\href{https://arxiv.org/abs/1106.0363}{{\ttfamily 1106.0363}}].

\bibitem{Adamczyk:2015gyk}
{\scshape STAR} collaboration, L.~Adamczyk et~al., \emph{{Measurement of the
  transverse single-spin asymmetry in $p^\uparrow+p \to W^{\pm}/Z^0$ at RHIC}},
  {\emph{Phys. Rev. Lett.} {\bfseries 116} (2016) 132301}
  [\href{https://arxiv.org/abs/1511.06003}{{\ttfamily 1511.06003}}].

\bibitem{Zhao:2014qvx}
{\scshape Jefferson Lab Hall A} collaboration, Y.~X. Zhao et~al., \emph{{Single
  spin asymmetries in charged kaon production from semi-inclusive deep
  inelastic scattering on a transversely polarized $^3He$ target}},
  {\emph{Phys. Rev.} {\bfseries C90} (2014) 055201}
  [\href{https://arxiv.org/abs/1404.7204}{{\ttfamily 1404.7204}}].

\bibitem{Adolph:2016dvl}
{\scshape COMPASS} collaboration, C.~Adolph et~al., \emph{{Sivers asymmetry
  extracted in SIDIS at the hard scales of the Drell--Yan process at COMPASS}},
  {\emph{Phys. Lett.} {\bfseries B770} (2017) 138}
  [\href{https://arxiv.org/abs/1609.07374}{{\ttfamily 1609.07374}}].

\bibitem{Anselmino:2016uie}
M.~Anselmino, M.~Boglione, U.~D'Alesio, F.~Murgia and A.~Prokudin, \emph{{Study
  of the sign change of the Sivers function from STAR Collaboration W/Z
  production data}}, {\emph{JHEP} {\bfseries 04} (2017) 046}
  [\href{https://arxiv.org/abs/1612.06413}{{\ttfamily 1612.06413}}].

\bibitem{Boglione:2018dqd}
M.~Boglione, U.~D'Alesio, C.~Flore and J.~Gonzalez-Hernandez, \emph{{Assessing
  signals of TMD physics in SIDIS azimuthal asymmetries and in the extraction
  of the Sivers function}}, {\emph{JHEP} {\bfseries 07} (2018) 148}
  [\href{https://arxiv.org/abs/1806.10645}{{\ttfamily 1806.10645}}].

\bibitem{Bury:2021sue}
M.~Bury, A.~Prokudin and A.~Vladimirov, \emph{{Extraction of the Sivers
  function from SIDIS, Drell-Yan, and $W^\pm/Z$ boson production data with TMD
  evolution}}, {\emph{JHEP} {\bfseries 05} (2021) 151}
  [\href{https://arxiv.org/abs/2103.03270}{{\ttfamily 2103.03270}}].

\bibitem{Qiu:2020oqr}
J.-W. Qiu, T.~C. Rogers and B.~Wang, \emph{{Intrinsic Transverse Momentum and
  Evolution in Weighted Spin Asymmetries}}, {\emph{Phys. Rev. D} {\bfseries
  101} (2020) 116017} [\href{https://arxiv.org/abs/2004.13193}{{\ttfamily
  2004.13193}}].

\bibitem{Jaffe:1991kp}
R.~L. Jaffe and X.-D. Ji, \emph{{Chiral odd parton distributions and polarized
  Drell-Yan}}, {\emph{Phys. Rev. Lett.} {\bfseries 67} (1991) 552}.

\bibitem{Goldstein:1974ym}
G.~R. Goldstein and M.~J. Moravcsik, \emph{{Optimally Simple Connection Between
  the Reaction Matrix and the Observables}}, {\emph{Annals Phys.} {\bfseries
  98} (1976) 128}.

\bibitem{Artru:1989zv}
X.~Artru and M.~Mekhfi, \emph{{Transversely Polarized Parton Densities, their
  Evolution and their Measurement}}, {\emph{Z. Phys.} {\bfseries C45} (1990)
  669}.

\bibitem{Soffer:1994ww}
J.~Soffer, \emph{{Positivity constraints for spin dependent parton
  distributions}}, {\emph{Phys. Rev. Lett.} {\bfseries 74} (1995) 1292}
  [\href{https://arxiv.org/abs/hep-ph/9409254}{{\ttfamily hep-ph/9409254}}].

\bibitem{Barone:1997fh}
V.~Barone, \emph{{On the QCD evolution of the transversity distribution}},
  {\emph{Phys. Lett.} {\bfseries B409} (1997) 499}
  [\href{https://arxiv.org/abs/hep-ph/9703343}{{\ttfamily hep-ph/9703343}}].

\bibitem{Vogelsang:1997ak}
W.~Vogelsang, \emph{{Next-to-leading order evolution of transversity
  distributions and Soffer's inequality}}, {\emph{Phys. Rev.} {\bfseries D57}
  (1998) 1886} [\href{https://arxiv.org/abs/hep-ph/9706511}{{\ttfamily
  hep-ph/9706511}}].

\bibitem{Airapetian:2010ds}
{\scshape HERMES} collaboration, A.~Airapetian et~al., \emph{{Effects of
  transversity in deep-inelastic scattering by polarized protons}},
  {\emph{Phys. Lett. B} {\bfseries 693} (2010) 11}
  [\href{https://arxiv.org/abs/1006.4221}{{\ttfamily 1006.4221}}].

\bibitem{Kang:2014zza}
Z.-B. Kang, A.~Prokudin, P.~Sun and F.~Yuan, \emph{{Nucleon tensor charge from
  Collins azimuthal asymmetry measurements}}, {\emph{Phys.Rev.} {\bfseries D91}
  (2015) 071501} [\href{https://arxiv.org/abs/1410.4877}{{\ttfamily
  1410.4877}}].

\bibitem{Anselmino:2013vqa}
M.~Anselmino, M.~Boglione, U.~D'Alesio, S.~Melis, F.~Murgia et~al.,
  \emph{{Simultaneous extraction of transversity and Collins functions from new
  SIDIS and $e^+e^-$ data}}, {\emph{Phys.Rev.} {\bfseries D87} (2013) 094019}
  [\href{https://arxiv.org/abs/1303.3822}{{\ttfamily 1303.3822}}].

\bibitem{Anselmino:2015sxa}
M.~Anselmino, M.~Boglione, U.~D'Alesio, J.~O. Gonzalez~Hernandez, S.~Melis,
  F.~Murgia et~al., \emph{{Collins functions for pions from SIDIS and new
  $e^+e^-$ data: a first glance at their transverse momentum dependence}},
  {\emph{Phys. Rev.} {\bfseries D92} (2015) 114023}
  [\href{https://arxiv.org/abs/1510.05389}{{\ttfamily 1510.05389}}].

\bibitem{Radici:2018iag}
M.~Radici and A.~Bacchetta, \emph{{First Extraction of Transversity from a
  Global Analysis of Electron-Proton and Proton-Proton Data}}, {\emph{Phys.
  Rev. Lett.} {\bfseries 120} (2018) 192001}
  [\href{https://arxiv.org/abs/1802.05212}{{\ttfamily 1802.05212}}].

\bibitem{Benel:2019mcq}
J.~Benel, A.~Courtoy and R.~Ferro-Hernandez, \emph{{A constrained fit of the
  valence transversity distributions from dihadron production}}, {\emph{Eur.
  Phys. J.} {\bfseries C80} (2020) 465}
  [\href{https://arxiv.org/abs/1912.03289}{{\ttfamily 1912.03289}}].

\bibitem{DAlesio:2020vtw}
U.~D'Alesio, C.~Flore and A.~Prokudin, \emph{{Role of the Soffer bound in
  determination of transversity and the tensor charge}}, {\emph{Phys. Lett.}
  {\bfseries B803} (2020) 135347}
  [\href{https://arxiv.org/abs/2001.01573}{{\ttfamily 2001.01573}}].

\bibitem{COMPASS:2005csq}
{\scshape COMPASS} collaboration, V.~Y. Alexakhin et~al., \emph{{First
  measurement of the transverse spin asymmetries of the deuteron in
  semi-inclusive deep inelastic scattering}}, {\emph{Phys. Rev. Lett.}
  {\bfseries 94} (2005) 202002}
  [\href{https://arxiv.org/abs/hep-ex/0503002}{{\ttfamily hep-ex/0503002}}].

\bibitem{Seidl:2008xc}
{\scshape Belle} collaboration, R.~Seidl et~al., \emph{{Measurement of
  Azimuthal Asymmetries in Inclusive Production of Hadron Pairs in $e^+e^-$
  Annihilation at $s^{1/2} = 10.58$ GeV}}, {\emph{Phys. Rev.} {\bfseries D78}
  (2008) 032011} [\href{https://arxiv.org/abs/0805.2975}{{\ttfamily
  0805.2975}}].

\bibitem{TheBABAR:2013yha}
{\scshape BaBar} collaboration, J.~P. Lees et~al., \emph{{Measurement of
  Collins asymmetries in inclusive production of charged pion pairs in $e^+e^-$
  annihilation at BABAR}}, {\emph{Phys. Rev.} {\bfseries D90} (2014) 052003}
  [\href{https://arxiv.org/abs/1309.5278}{{\ttfamily 1309.5278}}].

\bibitem{Wan:2020lps}
J.~Wan, C.~Tan and Z.~Lu, \emph{{The cos2$\phi$ azimuthal asymmetry in $e^+e^-
  \to \pi^+\pi^-$ X process}}, {\emph{Phys. Lett. B} {\bfseries 811} (2020)
  135884}.

\bibitem{Efremov:2006qm}
A.~V. Efremov, K.~Goeke and P.~Schweitzer, \emph{{Collins effect in
  semi-inclusive deeply inelastic scattering and in $e^+e^-$ annihilation}},
  {\emph{Phys. Rev. D} {\bfseries 73} (2006) 094025}
  [\href{https://arxiv.org/abs/hep-ph/0603054}{{\ttfamily hep-ph/0603054}}].

\bibitem{Goldstein:2014aja}
G.~R. Goldstein, J.~O. Gonzalez~Hernandez and S.~Liuti, \emph{{Flavor
  dependence of chiral odd generalized parton distributions and the tensor
  charge from the analysis of combined $\pi^0$ and $\eta$ exclusive
  electroproduction data}},  \href{https://arxiv.org/abs/1401.0438}{{\ttfamily
  1401.0438}}.

\bibitem{Radici:2015mwa}
M.~Radici, A.~Courtoy, A.~Bacchetta and M.~Guagnelli, \emph{{Improved
  extraction of valence transversity distributions from inclusive dihadron
  production}}, {\emph{JHEP} {\bfseries 05} (2015) 123}
  [\href{https://arxiv.org/abs/1503.03495}{{\ttfamily 1503.03495}}].

\bibitem{Gupta:2018qil}
R.~Gupta, Y.-C. Jang, B.~Yoon, H.-W. Lin, V.~Cirigliano and T.~Bhattacharya,
  \emph{{Isovector Charges of the Nucleon from 2+1+1-flavor Lattice QCD}},
  {\emph{Phys. Rev.} {\bfseries D98} (2018) 034503}
  [\href{https://arxiv.org/abs/1806.09006}{{\ttfamily 1806.09006}}].

\bibitem{Hasan:2019noy}
N.~Hasan, J.~Green, S.~Meinel, M.~Engelhardt, S.~Krieg, J.~Negele et~al.,
  \emph{{Nucleon axial, scalar, and tensor charges using lattice QCD at the
  physical pion mass}}, {\emph{Phys. Rev.} {\bfseries D99} (2019) 114505}
  [\href{https://arxiv.org/abs/1903.06487}{{\ttfamily 1903.06487}}].

\bibitem{Alexandrou:2019brg}
C.~Alexandrou, S.~Bacchio, M.~Constantinou, J.~Finkenrath, K.~Hadjiyiannakou,
  K.~Jansen et~al., \emph{{Nucleon axial, tensor, and scalar charges and
  $\sigma$-terms in lattice QCD}}, {\emph{Phys. Rev. D} {\bfseries 102} (2020)
  054517} [\href{https://arxiv.org/abs/1909.00485}{{\ttfamily 1909.00485}}].

\bibitem{Pitschmann:2014jxa}
M.~Pitschmann, C.-Y. Seng, C.~D. Roberts and S.~M. Schmidt, \emph{{Nucleon
  tensor charges and electric dipole moments}}, {\emph{Phys. Rev.} {\bfseries
  D91} (2015) 074004} [\href{https://arxiv.org/abs/1411.2052}{{\ttfamily
  1411.2052}}].

\bibitem{Jaffe:1991ra}
R.~L. Jaffe and X.-D. Ji, \emph{{Chiral odd parton distributions and Drell-Yan
  processes}}, {\emph{Nucl. Phys.} {\bfseries B375} (1992) 527}.

\bibitem{Cortes:1991ja}
J.~L. Cortes, B.~Pire and J.~P. Ralston, \emph{{Measuring the transverse
  polarization of quarks in the proton}}, {\emph{Z. Phys.} {\bfseries C55}
  (1992) 409}.

\bibitem{Gamberg:2001qc}
L.~P. Gamberg and G.~R. Goldstein, \emph{{Flavor spin symmetry estimate of the
  nucleon tensor charge}}, {\emph{Phys. Rev. Lett.} {\bfseries 87} (2001)
  242001} [\href{https://arxiv.org/abs/hep-ph/0107176}{{\ttfamily
  hep-ph/0107176}}].

\bibitem{Courtoy:2015haa}
A.~Courtoy, S.~Bae\ss{}ler, M.~Gonz\'alez-Alonso and S.~Liuti,
  \emph{{Beyond-Standard-Model Tensor Interaction and Hadron Phenomenology}},
  {\emph{Phys. Rev. Lett.} {\bfseries 115} (2015) 162001}
  [\href{https://arxiv.org/abs/1503.06814}{{\ttfamily 1503.06814}}].

\bibitem{Yamanaka:2017mef}
N.~Yamanaka, B.~K. Sahoo, N.~Yoshinaga, T.~Sato, K.~Asahi and B.~P. Das,
  \emph{{Probing exotic phenomena at the interface of nuclear and particle
  physics with the electric dipole moments of diamagnetic atoms: A unique
  window to hadronic and semi-leptonic CP violation}}, {\emph{Eur. Phys. J. A}
  {\bfseries 53} (2017) 54} [\href{https://arxiv.org/abs/1703.01570}{{\ttfamily
  1703.01570}}].

\bibitem{Gao:2017ade}
T.~Liu, Z.~Zhao and H.~Gao, \emph{{Experimental constraint on quark electric
  dipole moments}}, {\emph{Phys. Rev. D} {\bfseries 97} (2018) 074018}
  [\href{https://arxiv.org/abs/1704.00113}{{\ttfamily 1704.00113}}].

\bibitem{Gonzalez-Alonso:2018omy}
M.~Gonz\'alez-Alonso, O.~Naviliat-Cuncic and N.~Severijns, \emph{{New physics
  searches in nuclear and neutron $\beta$ decay}}, {\emph{Prog. Part. Nucl.
  Phys.} {\bfseries 104} (2019) 165}
  [\href{https://arxiv.org/abs/1803.08732}{{\ttfamily 1803.08732}}].

\bibitem{Yamanaka:2018uud}
{\scshape JLQCD} collaboration, N.~Yamanaka, S.~Hashimoto, T.~Kaneko and
  H.~Ohki, \emph{{Nucleon charges with dynamical overlap fermions}},
  {\emph{Phys. Rev. D} {\bfseries 98} (2018) 054516}
  [\href{https://arxiv.org/abs/1805.10507}{{\ttfamily 1805.10507}}].

\bibitem{Owens:1986mp}
J.~F. Owens, \emph{{Large Momentum Transfer Production of Direct Photons, Jets,
  and Particles}}, {\emph{Rev. Mod. Phys.} {\bfseries 59} (1987) 465}.

\bibitem{Kouvaris:2006zy}
C.~Kouvaris, J.-W. Qiu, W.~Vogelsang and F.~Yuan, \emph{{Single transverse-spin
  asymmetry in high transverse momentum pion production in p p collisions}},
  {\emph{Phys. Rev.} {\bfseries D74} (2006) 114013}
  [\href{https://arxiv.org/abs/hep-ph/0609238}{{\ttfamily hep-ph/0609238}}].

\bibitem{Kang:2013ufa}
Z.-B. Kang, I.~Vitev and H.~Xing, \emph{{Multiple scattering effects on
  inclusive particle production in the large-x regime}}, {\emph{Phys. Rev.}
  {\bfseries D88} (2013) 054010}
  [\href{https://arxiv.org/abs/1307.3557}{{\ttfamily 1307.3557}}].

\bibitem{Efremov:1984ip}
A.~Efremov and O.~Teryaev, \emph{{QCD Asymmetry and Polarized Hadron Structure
  Functions}}, {\emph{Phys.Lett.} {\bfseries B150} (1985) 383}.

\bibitem{Eguchi:2006qz}
H.~Eguchi, Y.~Koike and K.~Tanaka, \emph{{Single Transverse Spin Asymmetry for
  Large-$p_T$ Pion Production in Semi-Inclusive Deep Inelastic Scattering}},
  {\emph{Nucl.Phys.} {\bfseries B752} (2006) 1}
  [\href{https://arxiv.org/abs/hep-ph/0604003}{{\ttfamily hep-ph/0604003}}].

\bibitem{Eguchi:2006mc}
H.~Eguchi, Y.~Koike and K.~Tanaka, \emph{{Twist-3 Formalism for Single
  Transverse Spin Asymmetry Reexamined: Semi-Inclusive Deep Inelastic
  Scattering}}, {\emph{Nucl.Phys.} {\bfseries B763} (2007) 198}
  [\href{https://arxiv.org/abs/hep-ph/0610314}{{\ttfamily hep-ph/0610314}}].

\bibitem{Koike:2009ge}
Y.~Koike and T.~Tomita, \emph{{Soft-fermion-pole contribution to single-spin
  asymmetry for pion production in pp collisions}}, {\emph{Phys.Lett.}
  {\bfseries B675} (2009) 181}
  [\href{https://arxiv.org/abs/0903.1923}{{\ttfamily 0903.1923}}].

\bibitem{Metz:2012ct}
A.~Metz and D.~Pitonyak, \emph{{Fragmentation contribution to the transverse
  single-spin asymmetry in proton-proton collisions}}, {\emph{Phys.Lett.}
  {\bfseries B723} (2013) 365}
  [\href{https://arxiv.org/abs/1212.5037}{{\ttfamily 1212.5037}}].

\bibitem{Beppu:2013uda}
H.~Beppu, K.~Kanazawa, Y.~Koike and S.~Yoshida, \emph{{Three-gluon contribution
  to the single spin asymmetry for light hadron production in pp collision}},
  {\emph{Phys. Rev.} {\bfseries D89} (2014) 034029}
  [\href{https://arxiv.org/abs/1312.6862}{{\ttfamily 1312.6862}}].

\bibitem{Adam:2020edg}
{\scshape STAR} collaboration, J.~Adam et~al., \emph{{Measurement of transverse
  single-spin asymmetries of $\pi^0$ and electromagnetic jets at forward
  rapidity in 200 and 500 GeV transversely polarized proton-proton
  collisions}}, \href{https://doi.org/10.1103/PhysRevD.103.092009}{\emph{Phys.
  Rev. D} {\bfseries 103} (2021) 092009}
  [\href{https://arxiv.org/abs/2012.11428}{{\ttfamily 2012.11428}}].

\bibitem{Aybat:2011ge}
S.~M. Aybat, J.~C. Collins, J.-W. Qiu and T.~C. Rogers, \emph{{The QCD
  Evolution of the Sivers Function}}, {\emph{Phys.Rev.} {\bfseries D85} (2012)
  034043} [\href{https://arxiv.org/abs/1110.6428}{{\ttfamily 1110.6428}}].

\bibitem{Kanazawa:2000hz}
Y.~Kanazawa and Y.~Koike, \emph{{Chiral odd contribution to single transverse
  spin asymmetry in hadronic pion production}}, {\emph{Phys.Lett.} {\bfseries
  B478} (2000) 121} [\href{https://arxiv.org/abs/hep-ph/0001021}{{\ttfamily
  hep-ph/0001021}}].

\bibitem{Gamberg:2017gle}
L.~Gamberg, Z.-B. Kang, D.~Pitonyak and A.~Prokudin, \emph{{Phenomenological
  constraints on $A_N$ in $p^\uparrow p\to \pi\, X$ from Lorentz invariance
  relations}}, {\emph{Phys. Lett. B} {\bfseries 770} (2017) 242}
  [\href{https://arxiv.org/abs/1701.09170}{{\ttfamily 1701.09170}}].

\bibitem{Lee:2007zzh}
{\scshape BRAHMS} collaboration, J.~Lee and F.~Videbaek, \emph{{Single spin
  asymmetries of identified hadrons in polarized p + p at $s^{1/2} = $ 62.4 and
  200 GeV}}, {\emph{AIP Conf. Proc.} {\bfseries 915} (2007) 533}.

\bibitem{Adams:2003fx}
{\scshape STAR} collaboration, J.~Adams et~al., \emph{{Cross-sections and
  transverse single spin asymmetries in forward neutral pion production from
  proton collisions at $s^{1/2} = 200$ GeV}}, {\emph{Phys. Rev. Lett.}
  {\bfseries 92} (2004) 171801}
  [\href{https://arxiv.org/abs/hep-ex/0310058}{{\ttfamily hep-ex/0310058}}].

\bibitem{Abelev:2008af}
{\scshape STAR} collaboration, B.~Abelev et~al., \emph{{Forward Neutral Pion
  Transverse Single Spin Asymmetries in p+p Collisions at $s^{1/2} = 200$
  GeV}}, {\emph{Phys. Rev. Lett.} {\bfseries 101} (2008) 222001}
  [\href{https://arxiv.org/abs/0801.2990}{{\ttfamily 0801.2990}}].

\bibitem{Adamczyk:2012xd}
{\scshape STAR} collaboration, L.~Adamczyk et~al., \emph{{Transverse
  Single-Spin Asymmetry and Cross-Section for $\pi^0$ and $\eta$ Mesons at
  Large Feynman-$x$ in Polarized $p+p$ Collisions at $\sqrt{s}=200$ GeV}},
  {\emph{Phys.Rev.} {\bfseries D86} (2012) 051101}
  [\href{https://arxiv.org/abs/1205.6826}{{\ttfamily 1205.6826}}].

\bibitem{Bunce:1976yb}
G.~Bunce, R.~Handler, R.~March, P.~Martin, L.~Pondrom et~al., \emph{{Lambda0
  Hyperon Polarization in Inclusive Production by 300 GeV Protons on
  Beryllium.}}, {\emph{Phys.Rev.Lett.} {\bfseries 36} (1976) 1113}.

\bibitem{Krueger:1998hz}
K.~Krueger et~al., \emph{{Large analyzing power in inclusive $\pi^\pm$
  production at high $x_F$ with a 22 GeV/c polarized proton beam}},
  {\emph{Phys. Lett. B} {\bfseries 459} (1999) 412}.

\bibitem{Allgower:2002qi}
C.~Allgower et~al., \emph{{Measurement of analyzing powers of $\pi^+$ and
  $\pi^-$ produced on a hydrogen and a carbon target with a 22 GeV/c incident
  polarized proton beam}}, {\emph{Phys. Rev. D} {\bfseries 65} (2002) 092008}.

\bibitem{Adler:2005in}
{\scshape PHENIX} collaboration, S.~Adler et~al., \emph{{Measurement of
  transverse single-spin asymmetries for mid-rapidity production of neutral
  pions and charged hadrons in polarized p+p collisions at $s^{1/2} = 200$
  GeV}}, {\emph{Phys. Rev. Lett.} {\bfseries 95} (2005) 202001}
  [\href{https://arxiv.org/abs/hep-ex/0507073}{{\ttfamily hep-ex/0507073}}].

\bibitem{Arsene:2008aa}
{\scshape BRAHMS} collaboration, I.~Arsene et~al., \emph{{Single Transverse
  Spin Asymmetries of Identified Charged Hadrons in Polarized p+p Collisions at
  $s^{1/2} = 62.4$ GeV}}, {\emph{Phys. Rev. Lett.} {\bfseries 101} (2008)
  042001} [\href{https://arxiv.org/abs/0801.1078}{{\ttfamily 0801.1078}}].

\bibitem{Adamczyk:2012qj}
{\scshape STAR} collaboration, L.~Adamczyk et~al., \emph{{Longitudinal and
  transverse spin asymmetries for inclusive jet production at mid-rapidity in
  polarized $p+p$ collisions at $\sqrt{s}=200$ GeV}}, {\emph{Phys. Rev. D}
  {\bfseries 86} (2012) 032006}
  [\href{https://arxiv.org/abs/1205.2735}{{\ttfamily 1205.2735}}].

\bibitem{Bland:2013pkt}
{\scshape AnDY} collaboration, L.~Bland et~al., \emph{{Cross Sections and
  Transverse Single-Spin Asymmetries in Forward Jet Production from Proton
  Collisions at $\sqrt{s}=500$ GeV}}, {\emph{Phys. Lett. B} {\bfseries 750}
  (2015) 660} [\href{https://arxiv.org/abs/1304.1454}{{\ttfamily 1304.1454}}].

\bibitem{Adare:2013ekj}
{\scshape PHENIX} collaboration, A.~Adare et~al., \emph{{Measurement of
  transverse-single-spin asymmetries for midrapidity and forward-rapidity
  production of hadrons in polarized p+p collisions at $\sqrt{s}=$200 and 62.4
  GeV}}, {\emph{Phys.Rev.} {\bfseries D90} (2014) 012006}
  [\href{https://arxiv.org/abs/1312.1995}{{\ttfamily 1312.1995}}].

\bibitem{Adare:2014qzo}
{\scshape PHENIX} collaboration, A.~Adare et~al., \emph{{Cross section and
  transverse single-spin asymmetry of $\eta$ mesons in $p^{\uparrow}+p$
  collisions at $\sqrt{s}=200$ GeV at forward rapidity}}, {\emph{Phys. Rev. D}
  {\bfseries 90} (2014) 072008}
  [\href{https://arxiv.org/abs/1406.3541}{{\ttfamily 1406.3541}}].

\bibitem{Acharya:2020opv}
{\scshape PHENIX} collaboration, U.~A. Acharya et~al., \emph{{Transverse
  momentum dependent forward neutron single spin asymmetries in transversely
  polarized $p+p$ collisions at $\sqrt {s}$ = 200 GeV}},
  \href{https://doi.org/10.1103/PhysRevD.103.032007}{\emph{Phys. Rev. D}
  {\bfseries 103} (2021) 032007}
  [\href{https://arxiv.org/abs/2011.14187}{{\ttfamily 2011.14187}}].

\bibitem{Kanazawa:2014dca}
K.~Kanazawa, Y.~Koike, A.~Metz and D.~Pitonyak, \emph{{Towards an explanation
  of transverse single-spin asymmetries in proton-proton collisions: the role
  of fragmentation in collinear factorization}}, {\emph{Phys.Rev.} {\bfseries
  D89} (2014) 111501} [\href{https://arxiv.org/abs/1404.1033}{{\ttfamily
  1404.1033}}].

\bibitem{Boer:1999mm}
D.~Boer, \emph{{Investigating the origins of transverse spin asymmetries at
  RHIC}}, {\emph{Phys. Rev.} {\bfseries D60} (1999) 014012}
  [\href{https://arxiv.org/abs/hep-ph/9902255}{{\ttfamily hep-ph/9902255}}].

\bibitem{Boer:2006eq}
D.~Boer and W.~Vogelsang, \emph{{Drell-Yan lepton angular distribution at small
  transverse momentum}}, {\emph{Phys. Rev. D} {\bfseries 74} (2006) 014004}
  [\href{https://arxiv.org/abs/hep-ph/0604177}{{\ttfamily hep-ph/0604177}}].

\bibitem{Lam:1978pu}
C.~Lam and W.-K. Tung, \emph{{A Systematic Approach to Inclusive Lepton Pair
  Production in Hadronic Collisions}}, {\emph{Phys. Rev. D} {\bfseries 18}
  (1978) 2447}.

\bibitem{Lam:1980uc}
C.~Lam and W.-K. Tung, \emph{{A Parton Model Relation without QCD Modifications
  in Lepton Pair Productions}}, {\emph{Phys. Rev. D} {\bfseries 21} (1980)
  2712}.

\bibitem{Collins:1978yt}
J.~C. Collins, \emph{{Simple Prediction of QCD for Angular Distribution of
  Dileptons in Hadron Collisions}}, {\emph{Phys. Rev. Lett.} {\bfseries 42}
  (1979) 291}.

\bibitem{Mirkes:1994dp}
E.~Mirkes and J.~Ohnemus, \emph{{Angular distributions of Drell-Yan lepton
  pairs at the Tevatron: Order $\alpha_s^{2}$ corrections and Monte Carlo
  studies}}, {\emph{Phys. Rev. D} {\bfseries 51} (1995) 4891}
  [\href{https://arxiv.org/abs/hep-ph/9412289}{{\ttfamily hep-ph/9412289}}].

\bibitem{Gauld:2017tww}
R.~Gauld, A.~Gehrmann-De~Ridder, T.~Gehrmann, E.~W.~N. Glover and A.~Huss,
  \emph{{Precise predictions for the angular coefficients in Z-boson production
  at the LHC}}, {\emph{JHEP} {\bfseries 11} (2017) 003}
  [\href{https://arxiv.org/abs/1708.00008}{{\ttfamily 1708.00008}}].

\bibitem{Falciano:1986wk}
{\scshape NA10} collaboration, S.~Falciano et~al., \emph{{Angular Distributions
  of Muon Pairs Produced by 194 GeV/$c$ Negative Pions}}, {\emph{Z. Phys. C}
  {\bfseries 31} (1986) 513}.

\bibitem{Guanziroli:1987rp}
{\scshape NA10} collaboration, M.~Guanziroli et~al., \emph{{Angular
  Distributions of Muon Pairs Produced by Negative Pions on Deuterium and
  Tungsten}}, {\emph{Z. Phys. C} {\bfseries 37} (1988) 545}.

\bibitem{Conway:1989fs}
J.~Conway et~al., \emph{{Experimental Study of Muon Pairs Produced by 252 GeV
  Pions on Tungsten}}, {\emph{Phys. Rev. D} {\bfseries 39} (1989) 92}.

\bibitem{Zhu:2006gx}
{\scshape NuSea} collaboration, L.~Zhu et~al., \emph{{Measurement of Angular
  Distributions of Drell-Yan Dimuons in p + d Interaction at 800 GeV/c}},
  {\emph{Phys. Rev. Lett.} {\bfseries 99} (2007) 082301}
  [\href{https://arxiv.org/abs/hep-ex/0609005}{{\ttfamily hep-ex/0609005}}].

\bibitem{Zhu:2008sj}
{\scshape NuSea} collaboration, L.~Zhu et~al., \emph{{Measurement of Angular
  Distributions of Drell-Yan Dimuons in p + p Interactions at 800 GeV/c}},
  {\emph{Phys. Rev. Lett.} {\bfseries 102} (2009) 182001}
  [\href{https://arxiv.org/abs/0811.4589}{{\ttfamily 0811.4589}}].

\bibitem{Aaltonen:2011nr}
{\scshape CDF} collaboration, T.~Aaltonen et~al., \emph{{First Measurement of
  the Angular Coefficients of Drell-Yan $e^{+}e^{-}$ pairs in the Z Mass Region
  from $p\bar{p}$ Collisions at $\sqrt{s}$ = 1.96 TeV}}, {\emph{Phys. Rev.
  Lett.} {\bfseries 106} (2011) 241801}
  [\href{https://arxiv.org/abs/1103.5699}{{\ttfamily 1103.5699}}].

\bibitem{Khachatryan:2015paa}
{\scshape CMS} collaboration, V.~Khachatryan et~al., \emph{{Angular
  coefficients of Z bosons produced in pp collisions at $\sqrt{s}$ = 8 TeV and
  decaying to $\mu^+ \mu^-$ as a function of transverse momentum and
  rapidity}}, {\emph{Phys. Lett. B} {\bfseries 750} (2015) 154}
  [\href{https://arxiv.org/abs/1504.03512}{{\ttfamily 1504.03512}}].

\bibitem{Zhang:2008nu}
B.~Zhang, Z.~Lu, B.-Q. Ma and I.~Schmidt, \emph{{Extracting Boer-Mulders
  functions from p+D Drell-Yan processes}}, {\emph{Phys. Rev. D} {\bfseries 77}
  (2008) 054011} [\href{https://arxiv.org/abs/0803.1692}{{\ttfamily
  0803.1692}}].

\bibitem{Lu:2009ip}
Z.~Lu and I.~Schmidt, \emph{{Updating Boer-Mulders functions from unpolarized
  pd and pp Drell-Yan data}}, {\emph{Phys. Rev. D} {\bfseries 81} (2010)
  034023} [\href{https://arxiv.org/abs/0912.2031}{{\ttfamily 0912.2031}}].

\bibitem{Lambertsen:2016wgj}
M.~Lambertsen and W.~Vogelsang, \emph{{Drell-Yan lepton angular distributions
  in perturbative QCD}}, {\emph{Phys. Rev. D} {\bfseries 93} (2016) 114013}
  [\href{https://arxiv.org/abs/1605.02625}{{\ttfamily 1605.02625}}].

\bibitem{Peng:2015spa}
J.-C. Peng, W.-C. Chang, R.~E. McClellan and O.~Teryaev, \emph{{Interpretation
  of Angular Distributions of $Z$-boson Production at Colliders}}, {\emph{Phys.
  Lett. B} {\bfseries 758} (2016) 384}
  [\href{https://arxiv.org/abs/1511.08932}{{\ttfamily 1511.08932}}].

\bibitem{Chang:2017kuv}
W.-C. Chang, R.~E. McClellan, J.-C. Peng and O.~Teryaev, \emph{{Dependencies of
  lepton angular distribution coefficients on the transverse momentum and
  rapidity of $Z$ bosons produced in $pp$ collisions at the LHC}}, {\emph{Phys.
  Rev. D} {\bfseries 96} (2017) 054020}
  [\href{https://arxiv.org/abs/1708.05807}{{\ttfamily 1708.05807}}].

\bibitem{Peng:2018tty}
J.-C. Peng, D.~Boer, W.-C. Chang, R.~E. McClellan and O.~Teryaev, \emph{{On the
  rotational invariance and non-invariance of lepton angular distributions in
  Drell\textendash{}Yan and quarkonium production}}, {\emph{Phys. Lett. B}
  {\bfseries 789} (2019) 356}
  [\href{https://arxiv.org/abs/1808.04398}{{\ttfamily 1808.04398}}].

\bibitem{Chang:2018pvk}
W.-C. Chang, R.~E. McClellan, J.-C. Peng and O.~Teryaev, \emph{{Lepton Angular
  Distributions of Fixed-target Drell-Yan Experiments in Perturbative QCD and a
  Geometric Approach}}, {\emph{Phys. Rev. D} {\bfseries 99} (2019) 014032}
  [\href{https://arxiv.org/abs/1811.03256}{{\ttfamily 1811.03256}}].

\bibitem{Lyu:2020nul}
Y.~Lyu, W.-C. Chang, R.~E. Mcclellan, J.-C. Peng and O.~Teryaev, \emph{{Lepton
  angular distribution of $W$ boson productions}},
  \href{https://doi.org/10.1103/PhysRevD.103.034011}{\emph{Phys. Rev. D}
  {\bfseries 103} (2021) 034011}
  [\href{https://arxiv.org/abs/2010.01826}{{\ttfamily 2010.01826}}].

\bibitem{Osipenko:2008aa}
{\scshape CLAS} collaboration, M.~Osipenko et~al., \emph{{Measurement of
  unpolarized semi-inclusive pi+ electroproduction off the proton}},
  {\emph{Phys. Rev. D} {\bfseries 80} (2009) 032004}
  [\href{https://arxiv.org/abs/0809.1153}{{\ttfamily 0809.1153}}].

\bibitem{Yan:2016ods}
{\scshape Jefferson Lab Hall A} collaboration, X.~Yan et~al., \emph{{First
  measurement of unpolarized semi-inclusive deep-inelastic scattering cross
  sections from a $^3$He target}}, {\emph{Phys. Rev. C} {\bfseries 95} (2017)
  035209} [\href{https://arxiv.org/abs/1610.02350}{{\ttfamily 1610.02350}}].

\bibitem{Airapetian:2012yg}
{\scshape HERMES} collaboration, A.~Airapetian et~al., \emph{{Azimuthal
  distributions of charged hadrons, pions, and kaons produced in deep-inelastic
  scattering off unpolarized protons and deuterons}}, {\emph{Phys. Rev.}
  {\bfseries D87} (2013) 012010}
  [\href{https://arxiv.org/abs/1204.4161}{{\ttfamily 1204.4161}}].

\bibitem{Adolph:2014pwc}
{\scshape COMPASS} collaboration, C.~Adolph et~al., \emph{{Measurement of
  azimuthal hadron asymmetries in semi-inclusive deep inelastic scattering off
  unpolarised nucleons}}, {\emph{Nucl. Phys.} {\bfseries B886} (2014) 1046}
  [\href{https://arxiv.org/abs/1401.6284}{{\ttfamily 1401.6284}}].

\bibitem{Moretti:2020uee}
{\scshape COMPASS} collaboration, A.~Moretti, \emph{{Azimuthal asymmetries of
  hadrons produced in unpolarized SIDIS at COMPASS}}, {\emph{J. Phys. Conf.
  Ser.} {\bfseries 1435} (2020) 012043}.

\bibitem{Barone:2009hw}
V.~Barone, S.~Melis and A.~Prokudin, \emph{{The Boer-Mulders effect in
  unpolarized SIDIS: An Analysis of the COMPASS and HERMES data on the cos 2
  phi asymmetry}}, {\emph{Phys. Rev.} {\bfseries D81} (2010) 114026}
  [\href{https://arxiv.org/abs/0912.5194}{{\ttfamily 0912.5194}}].

\bibitem{Barone:2010gk}
V.~Barone, S.~Melis and A.~Prokudin, \emph{{Azimuthal asymmetries in
  unpolarized Drell-Yan processes and the Boer-Mulders distributions of
  antiquarks}}, {\emph{Phys. Rev.} {\bfseries D82} (2010) 114025}
  [\href{https://arxiv.org/abs/1009.3423}{{\ttfamily 1009.3423}}].

\bibitem{Barone:2015ksa}
V.~Barone, M.~Boglione, J.~O. Gonzalez~Hernandez and S.~Melis,
  \emph{{Phenomenological analysis of azimuthal asymmetries in unpolarized
  semi-inclusive deep inelastic scattering}}, {\emph{Phys. Rev.} {\bfseries
  D91} (2015) 074019} [\href{https://arxiv.org/abs/1502.04214}{{\ttfamily
  1502.04214}}].

\bibitem{Christova:2017zxa}
E.~Christova, E.~Leader and M.~Stoilov, \emph{{Consistency tests for the
  extraction of the Boer-Mulders and Sivers functions}}, {\emph{Phys. Rev. D}
  {\bfseries 97} (2018) 056018}
  [\href{https://arxiv.org/abs/1705.10613}{{\ttfamily 1705.10613}}].

\bibitem{Christova:2020ahe}
E.~Christova, D.~Kotlorz and E.~Leader, \emph{{New study of the Boer-Mulders
  function: Implications for the quark and hadron transverse momenta}},
  {\emph{Phys. Rev. D} {\bfseries 102} (2020) 014035}
  [\href{https://arxiv.org/abs/2004.02117}{{\ttfamily 2004.02117}}].

\bibitem{Bhattacharya:2021twu}
S.~Bhattacharya, Z.-B. Kang, A.~Metz, G.~Penn and D.~Pitonyak, \emph{{First
  global QCD analysis of the TMD $g_{1T}$ from semi-inclusive DIS data}},
  \href{https://doi.org/10.1103/PhysRevD.105.034007}{\emph{Phys. Rev. D}
  {\bfseries 105} (2022) 034007}
  [\href{https://arxiv.org/abs/2110.10253}{{\ttfamily 2110.10253}}].

\bibitem{COMPASS:2016led}
{\scshape COMPASS} collaboration, C.~Adolph et~al., \emph{{Sivers asymmetry
  extracted in SIDIS at the hard scales of the Drell\textendash{}Yan process at
  COMPASS}}, {\emph{Phys. Lett. B} {\bfseries 770} (2017) 138}
  [\href{https://arxiv.org/abs/1609.07374}{{\ttfamily 1609.07374}}].

\bibitem{Parsamyan:2018evv}
B.~Parsamyan, \emph{{Measurement of target-polarization dependent azimuthal
  asymmetries in SIDIS and Drell-Yan processes at COMPASS experiment}},
  {\emph{PoS} {\bfseries QCDEV2017} (2018) 042}.

\bibitem{HERMES:2020ifk}
{\scshape HERMES} collaboration, A.~Airapetian et~al., \emph{{Azimuthal single-
  and double-spin asymmetries in semi-inclusive deep-inelastic lepton
  scattering by transversely polarized protons}},
  \href{https://doi.org/10.1007/JHEP12(2020)010}{\emph{JHEP} {\bfseries 12}
  (2020) 010} [\href{https://arxiv.org/abs/2007.07755}{{\ttfamily
  2007.07755}}].

\bibitem{JeffersonLabHallA:2011vwy}
{\scshape Jefferson Lab Hall A} collaboration, J.~Huang et~al.,
  \emph{{Beam-Target Double Spin Asymmetry $A_{LT}$ in Charged Pion Production
  from Deep Inelastic Scattering on a Transversely Polarized $^{3}$He Target at
  1.4 \ensuremath{<} Q$^2$ \ensuremath{<} 2.7 GeV$^2$}}, {\emph{Phys. Rev.
  Lett.} {\bfseries 108} (2012) 052001}
  [\href{https://arxiv.org/abs/1108.0489}{{\ttfamily 1108.0489}}].

\bibitem{Tangerman:1994bb}
R.~Tangerman and P.~Mulders, \emph{{Polarized twist-3 distributions $g_T$ and
  $h_L$ and the role of intrinsic transverse momentum}},
  \href{https://arxiv.org/abs/hep-ph/9408305}{{\ttfamily hep-ph/9408305}}.

\bibitem{Kotzinian:2006dw}
A.~Kotzinian, B.~Parsamyan and A.~Prokudin, \emph{{Predictions for double spin
  asymmetry $A_{LT}$ in semi inclusive DIS}}, {\emph{Phys. Rev. D} {\bfseries
  73} (2006) 114017} [\href{https://arxiv.org/abs/hep-ph/0603194}{{\ttfamily
  hep-ph/0603194}}].

\bibitem{Avakian:2007mv}
H.~Avakian, A.~Efremov, K.~Goeke, A.~Metz, P.~Schweitzer and T.~Teckentrup,
  \emph{{Are there approximate relations among transverse momentum dependent
  distribution functions?}}, {\emph{Phys. Rev. D} {\bfseries 77} (2008) 014023}
  [\href{https://arxiv.org/abs/0709.3253}{{\ttfamily 0709.3253}}].

\bibitem{Metz:2008ib}
A.~Metz, P.~Schweitzer and T.~Teckentrup, \emph{{Lorentz invariance relations
  between parton distributions and the Wandzura-Wilczek approximation}},
  {\emph{Phys. Lett. B} {\bfseries 680} (2009) 141}
  [\href{https://arxiv.org/abs/0810.5212}{{\ttfamily 0810.5212}}].

\bibitem{Teckentrup:2009tk}
T.~Teckentrup, A.~Metz and P.~Schweitzer, \emph{{Lorentz invariance relations
  and Wandzura-Wilczek approximation}}, {\emph{Mod. Phys. Lett. A} {\bfseries
  24} (2009) 2950} [\href{https://arxiv.org/abs/0910.2567}{{\ttfamily
  0910.2567}}].

\bibitem{Pobylitsa:2003ty}
P.~V. Pobylitsa, \emph{{Transverse momentum dependent parton distributions in
  large $N_c$ QCD}},  \href{https://arxiv.org/abs/hep-ph/0301236}{{\ttfamily
  hep-ph/0301236}}.

\bibitem{HERMES:1999ryv}
{\scshape HERMES} collaboration, A.~Airapetian et~al., \emph{{Observation of a
  single spin azimuthal asymmetry in semiinclusive pion electro production}},
  {\emph{Phys. Rev. Lett.} {\bfseries 84} (2000) 4047}
  [\href{https://arxiv.org/abs/hep-ex/9910062}{{\ttfamily hep-ex/9910062}}].

\bibitem{HERMES:2002buj}
{\scshape HERMES} collaboration, A.~Airapetian et~al., \emph{{Measurement of
  single spin azimuthal asymmetries in semiinclusive electroproduction of pions
  and kaons on a longitudinally polarized deuterium target}}, {\emph{Phys.
  Lett. B} {\bfseries 562} (2003) 182}
  [\href{https://arxiv.org/abs/hep-ex/0212039}{{\ttfamily hep-ex/0212039}}].

\bibitem{Avakian:2010ae}
{\scshape CLAS} collaboration, H.~Avakian et~al., \emph{{Measurement of Single
  and Double Spin Asymmetries in Deep Inelastic Pion Electroproduction with a
  Longitudinally Polarized Target}}, {\emph{Phys. Rev. Lett.} {\bfseries 105}
  (2010) 262002} [\href{https://arxiv.org/abs/1003.4549}{{\ttfamily
  1003.4549}}].

\bibitem{Lu:2011pt}
Z.~Lu, B.-Q. Ma and J.~She, \emph{{The $\sin2\phi$ azimuthal asymmetry in
  single longitudinally polarized $\pi N$ Drell-Yan process}}, {\emph{Phys.
  Rev. D} {\bfseries 84} (2011) 034010}
  [\href{https://arxiv.org/abs/1104.5410}{{\ttfamily 1104.5410}}].

\bibitem{Zhu:2011zza}
J.~Zhu and B.-Q. Ma, \emph{{Proposal for measuring new transverse momentum
  dependent parton distributions $g_{1T}$ and $h_{1L}^\perp$ through
  semi-inclusive deep inelastic scattering}}, {\emph{Phys. Lett.} {\bfseries
  B696} (2011) 246} [\href{https://arxiv.org/abs/1104.4564}{{\ttfamily
  1104.4564}}].

\bibitem{Boffi:2009sh}
S.~Boffi, A.~V. Efremov, B.~Pasquini and P.~Schweitzer, \emph{{Azimuthal spin
  asymmetries in light-cone constituent quark models}}, {\emph{Phys. Rev.}
  {\bfseries D79} (2009) 094012}
  [\href{https://arxiv.org/abs/0903.1271}{{\ttfamily 0903.1271}}].

\bibitem{Ma:2000ip}
B.-Q. Ma, I.~Schmidt and J.-J. Yang, \emph{{Nucleon transversity distribution
  from azimuthal spin asymmetry in pion electroproduction}}, {\emph{Phys. Rev.
  D} {\bfseries 63} (2001) 037501}
  [\href{https://arxiv.org/abs/hep-ph/0009297}{{\ttfamily hep-ph/0009297}}].

\bibitem{Ma:2001ie}
B.-Q. Ma, I.~Schmidt and J.-J. Yang, \emph{{Azimuthal spin asymmetries of pion
  electroproduction}}, {\emph{Phys. Rev. D} {\bfseries 65} (2002) 034010}
  [\href{https://arxiv.org/abs/hep-ph/0110324}{{\ttfamily hep-ph/0110324}}].

\bibitem{Li:2021mmi}
H.~Li and Z.~Lu, \emph{{The $\sin2\phi_h$ azimuthal asymmetry of pion
  production in SIDIS within TMD factorization}},
  \href{https://arxiv.org/abs/2111.03840}{{\ttfamily 2111.03840}}.

\bibitem{Lefky:2014eia}
C.~Lefky and A.~Prokudin, \emph{{Extraction of the distribution function
  $h^{\perp}_{1T}$ from experimental data}}, {\emph{Phys. Rev.} {\bfseries D91}
  (2015) 034010} [\href{https://arxiv.org/abs/1411.0580}{{\ttfamily
  1411.0580}}].

\bibitem{xue:2021svd}
S.-C. Xue, X.~Wang, D.-M. Li and Z.~Lu, \emph{{Weighted $\sin(3\phi_h-\phi_S)$
  asymmetry from pretzelosity in SIDIS at electron ion colliders}},
  {\emph{Phys. Lett. B} {\bfseries 820} (2021) 136598}
  [\href{https://arxiv.org/abs/2104.05173}{{\ttfamily 2104.05173}}].

\bibitem{Angeles-Martinez:2015sea}
R.~Angeles-Martinez et~al., \emph{{Transverse Momentum Dependent (TMD) parton
  distribution functions: status and prospects}}, {\emph{Acta Phys. Polon.}
  {\bfseries B46} (2015) 2501}
  [\href{https://arxiv.org/abs/1507.05267}{{\ttfamily 1507.05267}}].

\bibitem{Echevarria:2019ynx}
M.~G. Echevarria, \emph{{Proper TMD factorization for quarkonia production:
  $pp\to\eta_{c,b}$ as a study case}}, {\emph{JHEP} {\bfseries 10} (2019) 144}
  [\href{https://arxiv.org/abs/1907.06494}{{\ttfamily 1907.06494}}].

\bibitem{Balazs:2006cc}
C.~Balazs, E.~L. Berger, P.~M. Nadolsky and C.~P. Yuan, \emph{{All-orders
  resummation for diphoton production at hadron colliders}}, {\emph{Phys.
  Lett.} {\bfseries B637} (2006) 235}
  [\href{https://arxiv.org/abs/hep-ph/0603037}{{\ttfamily hep-ph/0603037}}].

\bibitem{Nadolsky:2007ba}
P.~M. Nadolsky, C.~Balazs, E.~L. Berger and C.~P. Yuan, \emph{{Gluon-gluon
  contributions to the production of continuum diphoton pairs at hadron
  colliders}}, {\emph{Phys. Rev.} {\bfseries D76} (2007) 013008}
  [\href{https://arxiv.org/abs/hep-ph/0702003}{{\ttfamily hep-ph/0702003}}].

\bibitem{Qiu:2011ai}
J.-W. Qiu, M.~Schlegel and W.~Vogelsang, \emph{{Probing Gluonic Spin-Orbit
  Correlations in Photon Pair Production}}, {\emph{Phys.Rev.Lett.} {\bfseries
  107} (2011) 062001} [\href{https://arxiv.org/abs/1103.3861}{{\ttfamily
  1103.3861}}].

\bibitem{Boer:2011kf}
D.~Boer, W.~J. den Dunnen, C.~Pisano, M.~Schlegel and W.~Vogelsang,
  \emph{{Linearly Polarized Gluons and the Higgs Transverse Momentum
  Distribution}}, {\emph{Phys.Rev.Lett.} {\bfseries 108} (2012) 032002}
  [\href{https://arxiv.org/abs/1109.1444}{{\ttfamily 1109.1444}}].

\bibitem{Sun:2011iw}
P.~Sun, B.-W. Xiao and F.~Yuan, \emph{{Gluon Distribution Functions and Higgs
  Boson Production at Moderate Transverse Momentum}}, {\emph{Phys. Rev.}
  {\bfseries D84} (2011) 094005}
  [\href{https://arxiv.org/abs/1109.1354}{{\ttfamily 1109.1354}}].

\bibitem{Boer:2013fca}
D.~Boer, W.~J. den Dunnen, C.~Pisano and M.~Schlegel, \emph{{Determining the
  Higgs spin and parity in the diphoton decay channel}}, {\emph{Phys. Rev.
  Lett.} {\bfseries 111} (2013) 032002}
  [\href{https://arxiv.org/abs/1304.2654}{{\ttfamily 1304.2654}}].

\bibitem{Boer:2012bt}
D.~Boer and C.~Pisano, \emph{{Polarized gluon studies with charmonium and
  bottomonium at LHCb and AFTER}}, {\emph{Phys. Rev.} {\bfseries D86} (2012)
  094007} [\href{https://arxiv.org/abs/1208.3642}{{\ttfamily 1208.3642}}].

\bibitem{Ma:2012hh}
J.~P. Ma, J.~X. Wang and S.~Zhao, \emph{{Transverse momentum dependent
  factorization for quarkonium production at low transverse momentum}},
  {\emph{Phys. Rev.} {\bfseries D88} (2013) 014027}
  [\href{https://arxiv.org/abs/1211.7144}{{\ttfamily 1211.7144}}].

\bibitem{Ma:2014oha}
J.~P. Ma, J.~X. Wang and S.~Zhao, \emph{{Breakdown of QCD Factorization for
  P-Wave Quarkonium Production at Low Transverse Momentum}}, {\emph{Phys.
  Lett.} {\bfseries B737} (2014) 103}
  [\href{https://arxiv.org/abs/1405.3373}{{\ttfamily 1405.3373}}].

\bibitem{Dunnen:2014eta}
W.~J. den Dunnen, J.-P. Lansberg, C.~Pisano and M.~Schlegel, \emph{{Accessing
  the Transverse Dynamics and Polarization of Gluons inside the Proton at the
  LHC}}, {\emph{Phys.Rev.Lett.} {\bfseries 112} (2014) 212001}
  [\href{https://arxiv.org/abs/1401.7611}{{\ttfamily 1401.7611}}].

\bibitem{Lansberg:2017tlc}
J.-P. Lansberg, C.~Pisano and M.~Schlegel, \emph{{Associated production of a
  dilepton and a $\Upsilon(J/\psi)$ at the LHC as a probe of gluon transverse
  momentum dependent distributions}}, {\emph{Nucl. Phys.} {\bfseries B920}
  (2017) 192} [\href{https://arxiv.org/abs/1702.00305}{{\ttfamily
  1702.00305}}].

\bibitem{Lansberg:2017dzg}
J.-P. Lansberg, C.~Pisano, F.~Scarpa and M.~Schlegel, \emph{{Pinning down the
  linearly-polarised gluons inside unpolarised protons using quarkonium-pair
  production at the LHC}}, {\emph{Phys. Lett.} {\bfseries B784} (2018) 217}
  [\href{https://arxiv.org/abs/1710.01684}{{\ttfamily 1710.01684}}].

\bibitem{Scarpa:2019fol}
F.~Scarpa, D.~Boer, M.~G. Echevarria, J.-P. Lansberg, C.~Pisano and
  M.~Schlegel, \emph{{Studies of gluon TMDs and their evolution using
  quarkonium-pair production at the LHC}},
  \href{https://doi.org/10.1140/epjc/s10052-020-7619-1}{\emph{Eur. Phys. J. C}
  {\bfseries 80} (2020) 87} [\href{https://arxiv.org/abs/1909.05769}{{\ttfamily
  1909.05769}}].

\bibitem{Yuan:2008vn}
F.~Yuan, \emph{{Heavy Quarkonium Production in Single Transverse Polarized High
  Energy Scattering}}, {\emph{Phys. Rev.} {\bfseries D78} (2008) 014024}
  [\href{https://arxiv.org/abs/0801.4357}{{\ttfamily 0801.4357}}].

\bibitem{Boer:2014tka}
D.~Boer and W.~J. den Dunnen, \emph{{TMD evolution and the Higgs transverse
  momentum distribution}}, {\emph{Nucl. Phys. B} {\bfseries 886} (2014) 421}
  [\href{https://arxiv.org/abs/1404.6753}{{\ttfamily 1404.6753}}].

\bibitem{Aaij:2016bqq}
{\scshape LHCb} collaboration, R.~Aaij et~al., \emph{{Measurement of the
  J/$\psi$ pair production cross-section in pp collisions at $ \sqrt{s}=13 $
  TeV}}, {\emph{JHEP} {\bfseries 06} (2017) 047}
  [\href{https://arxiv.org/abs/1612.07451}{{\ttfamily 1612.07451}}].

\bibitem{Aaij:2011yc}
{\scshape LHCb} collaboration, R.~Aaij et~al., \emph{{Observation of $J/\psi$
  pair production in $pp$ collisions at $\sqrt{s}=7 TeV$}}, {\emph{Phys. Lett.
  B} {\bfseries 707} (2012) 52}
  [\href{https://arxiv.org/abs/1109.0963}{{\ttfamily 1109.0963}}].

\bibitem{Khachatryan:2014iia}
{\scshape CMS} collaboration, V.~Khachatryan et~al., \emph{{Measurement of
  Prompt $J/\psi$ Pair Production in pp Collisions at $ \sqrt{s} $ = 7 Tev}},
  {\emph{JHEP} {\bfseries 09} (2014) 094}
  [\href{https://arxiv.org/abs/1406.0484}{{\ttfamily 1406.0484}}].

\bibitem{Aaboud:2016fzt}
{\scshape ATLAS} collaboration, M.~Aaboud et~al., \emph{{Measurement of the
  prompt J/ $\psi $ pair production cross-section in pp collisions at $\sqrt{s}
  = 8$ TeV with the ATLAS detector}}, {\emph{Eur. Phys. J. C} {\bfseries 77}
  (2017) 76} [\href{https://arxiv.org/abs/1612.02950}{{\ttfamily 1612.02950}}].

\bibitem{Abazov:2014qba}
{\scshape D0} collaboration, V.~M. Abazov et~al., \emph{{Observation and
  Studies of Double $J/\psi$ Production at the Tevatron}}, {\emph{Phys. Rev. D}
  {\bfseries 90} (2014) 111101}
  [\href{https://arxiv.org/abs/1406.2380}{{\ttfamily 1406.2380}}].

\bibitem{Bacchetta:1999kz}
A.~Bacchetta, M.~Boglione, A.~Henneman and P.~J. Mulders, \emph{{Bounds on
  transverse momentum dependent distribution and fragmentation functions}},
  {\emph{Phys. Rev. Lett.} {\bfseries 85} (2000) 712}
  [\href{https://arxiv.org/abs/hep-ph/9912490}{{\ttfamily hep-ph/9912490}}].

\bibitem{Hadjidakis:2018ifr}
C.~Hadjidakis et~al., \emph{{A fixed-target programme at the LHC: Physics case
  and projected performances for heavy-ion, hadron, spin and astroparticle
  studies}}, \href{https://doi.org/10.1016/j.physrep.2021.01.002}{\emph{Phys.
  Rept.} {\bfseries 911} (2021) 1}
  [\href{https://arxiv.org/abs/1807.00603}{{\ttfamily 1807.00603}}].

\bibitem{Kikola:2017hnp}
D.~Kiko\l{}a, M.~G. Echevarria, C.~Hadjidakis, J.-P. Lansberg, C.~Lorc\'e,
  L.~Massacrier et~al., \emph{{Feasibility Studies for Single Transverse-Spin
  Asymmetry Measurements at a Fixed-Target Experiment Using the LHC Proton and
  Lead Beams (AFTER@LHC)}}, {\emph{Few Body Syst.} {\bfseries 58} (2017) 139}
  [\href{https://arxiv.org/abs/1702.01546}{{\ttfamily 1702.01546}}].

\bibitem{Aidala:2019pit}
C.~A. Aidala et~al., \emph{{The LHCSpin Project}},
  \href{https://arxiv.org/abs/1901.08002}{{\ttfamily 1901.08002}}.

\bibitem{Boer:2010zf}
D.~Boer, S.~J. Brodsky, P.~J. Mulders and C.~Pisano, \emph{{Direct Probes of
  Linearly Polarized Gluons inside Unpolarized Hadrons}},
  {\emph{Phys.Rev.Lett.} {\bfseries 106} (2011) 132001}
  [\href{https://arxiv.org/abs/1011.4225}{{\ttfamily 1011.4225}}].

\bibitem{Kang:2020xgk}
Z.-B. Kang, J.~Reiten, D.~Y. Shao and J.~Terry, \emph{{QCD evolution of the
  gluon Sivers function in heavy flavor dijet production at the Electron-Ion
  Collider}}, {\emph{JHEP} {\bfseries 05} (2021) 286}
  [\href{https://arxiv.org/abs/2012.01756}{{\ttfamily 2012.01756}}].

\bibitem{delCastillo:2020omr}
R.~F. del Castillo, M.~G. Echevarria, Y.~Makris and I.~Scimemi, \emph{{TMD
  factorization for dijet and heavy-meson pair in DIS}}, {\emph{JHEP}
  {\bfseries 01} (2021) 088}
  [\href{https://arxiv.org/abs/2008.07531}{{\ttfamily 2008.07531}}].

\bibitem{Zhang:2017uiz}
G.-P. Zhang, \emph{{Back-to-back heavy quark pair production in Semi-inclusive
  DIS}}, {\emph{JHEP} {\bfseries 11} (2017) 069}
  [\href{https://arxiv.org/abs/1709.08970}{{\ttfamily 1709.08970}}].

\bibitem{Zheng:2018ssm}
L.~Zheng, E.~C. Aschenauer, J.~H. Lee, B.-W. Xiao and Z.-B. Yin,
  \emph{{Accessing the gluon Sivers function at a future electron-ion
  collider}}, {\emph{Phys. Rev. D} {\bfseries 98} (2018) 034011}
  [\href{https://arxiv.org/abs/1805.05290}{{\ttfamily 1805.05290}}].

\bibitem{Mukherjee:2016qxa}
A.~Mukherjee and S.~Rajesh, \emph{{$J/\psi $ production in polarized and
  unpolarized ep collision and Sivers and $\cos 2\phi $ asymmetries}},
  {\emph{Eur. Phys. J. C} {\bfseries 77} (2017) 854}
  [\href{https://arxiv.org/abs/1609.05596}{{\ttfamily 1609.05596}}].

\bibitem{Kishore:2018ugo}
R.~Kishore and A.~Mukherjee, \emph{{Accessing linearly polarized gluon
  distribution in $J/\psi$ production at the electron-ion collider}},
  {\emph{Phys. Rev. D} {\bfseries 99} (2019) 054012}
  [\href{https://arxiv.org/abs/1811.07495}{{\ttfamily 1811.07495}}].

\bibitem{Kishore:2019fzb}
R.~Kishore, A.~Mukherjee and S.~Rajesh, \emph{{Sivers asymmetry in the
  photoproduction of a $J/\psi$ and a jet at the EIC}}, {\emph{Phys. Rev. D}
  {\bfseries 101} (2020) 054003}
  [\href{https://arxiv.org/abs/1908.03698}{{\ttfamily 1908.03698}}].

\bibitem{DAlesio:2019qpk}
U.~D'Alesio, F.~Murgia, C.~Pisano and P.~Taels, \emph{{Azimuthal asymmetries in
  semi-inclusive $J/\psi\,+\,\mathrm{jet}$ production at an EIC}}, {\emph{Phys.
  Rev. D} {\bfseries 100} (2019) 094016}
  [\href{https://arxiv.org/abs/1908.00446}{{\ttfamily 1908.00446}}].

\bibitem{Alrashed:2021csd}
M.~Alrashed, D.~Anderle, Z.-B. Kang, J.~Terry and H.~Xing,
  \emph{{Three-dimensional imaging in nuclei}},
  \href{https://doi.org/10.1103/PhysRevLett.129.242001}{\emph{Phys. Rev. Lett.}
  {\bfseries 129} (2022) 242001}
  [\href{https://arxiv.org/abs/2107.12401}{{\ttfamily 2107.12401}}].

\bibitem{EuropeanMuon:1983wih}
{\scshape European Muon} collaboration, J.~J. Aubert et~al., \emph{{The ratio
  of the nucleon structure functions $F_2^N$ for iron and deuterium}},
  {\emph{Phys. Lett. B} {\bfseries 123} (1983) 275}.

\bibitem{AbdulKhalek:2021gbh}
R.~Abdul~Khalek et~al., \emph{{Science Requirements and Detector Concepts for
  the Electron-Ion Collider: EIC Yellow Report}},
  \href{https://arxiv.org/abs/2103.05419}{{\ttfamily 2103.05419}}.

\bibitem{Stirling:1993gc}
W.~J. Stirling and M.~R. Whalley, \emph{{A Compilation of Drell-Yan
  cross-sections}}, {\emph{J. Phys. G} {\bfseries 19} (1993) D1}.

\bibitem{HERMES:2007plz}
{\scshape HERMES} collaboration, A.~Airapetian et~al., \emph{{Hadronization in
  semi-inclusive deep-inelastic scattering on nuclei}}, {\emph{Nucl. Phys. B}
  {\bfseries 780} (2007) 1} [\href{https://arxiv.org/abs/0704.3270}{{\ttfamily
  0704.3270}}].

\bibitem{Burkert:2008rj}
V.~D. Burkert, \emph{{CLAS12 and its initial Science Program at the Jefferson
  Lab Upgrade}},  in \emph{{CLAS 12 RICH Detector Workshop}}, 10, 2008,
  \href{https://arxiv.org/abs/0810.4718}{{\ttfamily 0810.4718}}.

\bibitem{CLAS:2021jhm}
{\scshape CLAS} collaboration, S.~Moran et~al., \emph{{Measurement of
  charged-pion production in deep-inelastic scattering off nuclei with the CLAS
  detector}}, \href{https://doi.org/10.1103/PhysRevC.105.015201}{\emph{Phys.
  Rev. C} {\bfseries 105} (2022) 015201}
  [\href{https://arxiv.org/abs/2109.09951}{{\ttfamily 2109.09951}}].

\bibitem{NuSea:1999egr}
{\scshape NuSea} collaboration, M.~A. Vasilev et~al., \emph{{Parton energy loss
  limits and shadowing in Drell-Yan dimuon production}}, {\emph{Phys. Rev.
  Lett.} {\bfseries 83} (1999) 2304}
  [\href{https://arxiv.org/abs/hep-ex/9906010}{{\ttfamily hep-ex/9906010}}].

\bibitem{Alde:1990im}
D.~M. Alde et~al., \emph{{Nuclear dependence of dimuon production at 800 GeV.
  FNAL-772 experiment}}, {\emph{Phys. Rev. Lett.} {\bfseries 64} (1990) 2479}.

\bibitem{Leung:2018tql}
{\scshape PHENIX} collaboration, Y.~H. Leung, \emph{{PHENIX measurements of
  charm, bottom, and Drell-Yan via dimuons in p+p and p+Au collisions at
  $\sqrt{s_{\rm NN}}$ = 200 GeV}}, {\emph{PoS} {\bfseries HardProbes2018}
  (2018) 160}.

\bibitem{CMS:2015zlj}
{\scshape CMS} collaboration, V.~Khachatryan et~al., \emph{{Study of Z boson
  production in pPb collisions at $\sqrt {s_{NN}} = 5.02$ TeV}}, {\emph{Phys.
  Lett. B} {\bfseries 759} (2016) 36}
  [\href{https://arxiv.org/abs/1512.06461}{{\ttfamily 1512.06461}}].

\bibitem{ATLAS:2015mwq}
{\scshape ATLAS} collaboration, G.~Aad et~al., \emph{{$Z$ boson production in
  $p+$Pb collisions at $\sqrt{s_{NN}}=5.02$ TeV measured with the ATLAS
  detector}}, {\emph{Phys. Rev. C} {\bfseries 92} (2015) 044915}
  [\href{https://arxiv.org/abs/1507.06232}{{\ttfamily 1507.06232}}].

\bibitem{Qiu:1990xy}
J.-w. Qiu and G.~F. Sterman, \emph{{Power corrections to hadronic scattering.
  2. Factorization}}, {\emph{Nucl. Phys. B} {\bfseries 353} (1991) 137}.

\bibitem{Rothe:1992nt}
H.~J. Rothe, \emph{{Lattice gauge theories: An Introduction}}, {\emph{World
  Sci. Lect. Notes Phys.} {\bfseries 43} (1992) 1}.

\bibitem{Gattringer:2010zz}
C.~Gattringer and C.~B. Lang, \emph{{Quantum chromodynamics on the lattice}},
  {\emph{Lect. Notes Phys.} {\bfseries 788} (2010) 1}.

\bibitem{Wilson:1974sk}
K.~G. Wilson, \emph{{Confinement of Quarks}}, {\emph{Phys. Rev.} {\bfseries
  D10} (1974) 2445}.

\bibitem{Celmaster:1982ht}
W.~Celmaster, \emph{{Gauge Theories on the Body - Centered Hypercubic
  Lattice}}, {\emph{Phys. Rev.} {\bfseries D26} (1982) 2955}.

\bibitem{Kogut:1974ag}
J.~B. Kogut and L.~Susskind, \emph{{Hamiltonian Formulation of Wilson's Lattice
  Gauge Theories}}, {\emph{Phys. Rev.} {\bfseries D11} (1975) 395}.

\bibitem{Frezzotti:2003ni}
R.~Frezzotti and G.~C. Rossi, \emph{{Chirally improving Wilson fermions. 1.
  O(a) improvement}}, {\emph{JHEP} {\bfseries 08} (2004) 007}
  [\href{https://arxiv.org/abs/hep-lat/0306014}{{\ttfamily hep-lat/0306014}}].

\bibitem{Kaplan:1992bt}
D.~B. Kaplan, \emph{{A Method for simulating chiral fermions on the lattice}},
  {\emph{Phys. Lett.} {\bfseries B288} (1992) 342}
  [\href{https://arxiv.org/abs/hep-lat/9206013}{{\ttfamily hep-lat/9206013}}].

\bibitem{Narayanan:1994gw}
R.~Narayanan and H.~Neuberger, \emph{{A Construction of lattice chiral gauge
  theories}}, {\emph{Nucl. Phys.} {\bfseries B443} (1995) 305}
  [\href{https://arxiv.org/abs/hep-th/9411108}{{\ttfamily hep-th/9411108}}].

\bibitem{Sheikholeslami:1985ij}
B.~Sheikholeslami and R.~Wohlert, \emph{{Improved Continuum Limit Lattice
  Action for QCD with Wilson Fermions}}, {\emph{Nucl. Phys.} {\bfseries B259}
  (1985) 572}.

\bibitem{Duane:1987de}
S.~Duane, A.~Kennedy, B.~Pendleton and D.~Roweth, \emph{{Hybrid Monte Carlo}},
  {\emph{Phys. Lett. B} {\bfseries 195} (1987) 216}.

\bibitem{Aoki:2019cca}
{\scshape Flavour Lattice Averaging Group} collaboration, S.~Aoki et~al.,
  \emph{{FLAG Review 2019: Flavour Lattice Averaging Group (FLAG)}},
  {\emph{Eur. Phys. J. C} {\bfseries 80} (2020) 113}
  [\href{https://arxiv.org/abs/1902.08191}{{\ttfamily 1902.08191}}].

\bibitem{Capitani:2002mp}
S.~Capitani, \emph{{Lattice perturbation theory}}, {\emph{Phys. Rept.}
  {\bfseries 382} (2003) 113}
  [\href{https://arxiv.org/abs/hep-lat/0211036}{{\ttfamily hep-lat/0211036}}].

\bibitem{Martinelli:1994ty}
G.~Martinelli, C.~Pittori, C.~T. Sachrajda, M.~Testa and A.~Vladikas, \emph{{A
  General method for nonperturbative renormalization of lattice operators}},
  {\emph{Nucl. Phys.} {\bfseries B445} (1995) 81}
  [\href{https://arxiv.org/abs/hep-lat/9411010}{{\ttfamily hep-lat/9411010}}].

\bibitem{Martinelli:1982cb}
G.~Martinelli, G.~Parisi, R.~Petronzio and F.~Rapuano, \emph{{The Proton and
  Neutron Magnetic Moments in Lattice {QCD}}}, {\emph{Phys. Lett. B} {\bfseries
  116} (1982) 434}.

\bibitem{Fucito:1982ff}
F.~Fucito, G.~Parisi and S.~Petrarca, \emph{{First Evaluation of $g_A/g_V$ in
  Lattice QCD in the Quenched Approximation}}, {\emph{Phys. Lett. B} {\bfseries
  115} (1982) 148}.

\bibitem{Martinelli:1987zd}
G.~Martinelli and C.~T. Sachrajda, \emph{{Pion Structure Functions From Lattice
  {QCD}}}, {\emph{Phys. Lett.} {\bfseries B196} (1987) 184}.

\bibitem{Martinelli:1988rr}
G.~Martinelli and C.~T. Sachrajda, \emph{{A Lattice Study of Nucleon
  Structure}}, {\emph{Nucl.Phys.} {\bfseries B316} (1989) 355}.

\bibitem{Martinelli:1988xs}
G.~Martinelli and C.~T. Sachrajda, \emph{{The Quark Distribution Amplitude of
  the Proton: A Lattice Computation of the Lowest Two Moments}}, {\emph{Phys.
  Lett.} {\bfseries B217} (1989) 319}.

\bibitem{Kronfeld:2019nfb}
{\scshape USQCD} collaboration, A.~S. Kronfeld, D.~G. Richards, W.~Detmold,
  R.~Gupta, H.-W. Lin, K.-F. Liu et~al., \emph{{Lattice QCD and
  Neutrino-Nucleus Scattering}}, {\emph{Eur. Phys. J. A} {\bfseries 55} (2019)
  196} [\href{https://arxiv.org/abs/1904.09931}{{\ttfamily 1904.09931}}].

\bibitem{Lin:2017snn}
H.-W. Lin et~al., \emph{{Parton distributions and lattice QCD calculations: a
  community white paper}}, {\emph{Prog. Part. Nucl. Phys.} {\bfseries 100}
  (2018) 107} [\href{https://arxiv.org/abs/1711.07916}{{\ttfamily
  1711.07916}}].

\bibitem{Ji:1996ek}
X.-D. Ji, \emph{{Gauge-Invariant Decomposition of Nucleon Spin}},
  {\emph{Phys.Rev.Lett.} {\bfseries 78} (1997) 610}
  [\href{https://arxiv.org/abs/hep-ph/9603249}{{\ttfamily hep-ph/9603249}}].

\bibitem{Ji:1996nm}
X.-D. Ji, \emph{{Deeply virtual Compton scattering}}, {\emph{Phys. Rev. D}
  {\bfseries 55} (1997) 7114}
  [\href{https://arxiv.org/abs/hep-ph/9609381}{{\ttfamily hep-ph/9609381}}].

\bibitem{Muller:1994ses}
D.~M\"uller, D.~Robaschik, B.~Geyer, F.~M. Dittes and J.~Ho\v{r}ej\v{s}i,
  \emph{{Wave functions, evolution equations and evolution kernels from light
  ray operators of QCD}}, {\emph{Fortsch. Phys.} {\bfseries 42} (1994) 101}
  [\href{https://arxiv.org/abs/hep-ph/9812448}{{\ttfamily hep-ph/9812448}}].

\bibitem{Radyushkin:1997ki}
A.~V. Radyushkin, \emph{{Nonforward parton distributions}}, {\emph{Phys. Rev.
  D} {\bfseries 56} (1997) 5524}
  [\href{https://arxiv.org/abs/hep-ph/9704207}{{\ttfamily hep-ph/9704207}}].

\bibitem{Ji:1998pc}
X.-D. Ji, \emph{{Off forward parton distributions}}, {\emph{J. Phys. G}
  {\bfseries 24} (1998) 1181}
  [\href{https://arxiv.org/abs/hep-ph/9807358}{{\ttfamily hep-ph/9807358}}].

\bibitem{Diehl:2003ny}
M.~Diehl, \emph{{Generalized parton distributions}}, {\emph{Phys. Rept.}
  {\bfseries 388} (2003) 41}
  [\href{https://arxiv.org/abs/hep-ph/0307382}{{\ttfamily hep-ph/0307382}}].

\bibitem{Belitsky:2005qn}
A.~V. Belitsky and A.~V. Radyushkin, \emph{{Unraveling hadron structure with
  generalized parton distributions}}, {\emph{Phys. Rept.} {\bfseries 418}
  (2005) 1} [\href{https://arxiv.org/abs/hep-ph/0504030}{{\ttfamily
  hep-ph/0504030}}].

\bibitem{Boffi:2007yc}
S.~Boffi and B.~Pasquini, \emph{{Generalized parton distributions and the
  structure of the nucleon}}, {\emph{Riv. Nuovo Cim.} {\bfseries 30} (2007)
  387} [\href{https://arxiv.org/abs/0711.2625}{{\ttfamily 0711.2625}}].

\bibitem{Kumericki:2016ehc}
K.~Kumericki, S.~Liuti and H.~Moutarde, \emph{{GPD phenomenology and DVCS
  fitting}: {Entering the high-precision era}},
  \href{https://doi.org/10.1140/epja/i2016-16157-3}{\emph{Eur. Phys. J. A}
  {\bfseries 52} (2016) 157}
  [\href{https://arxiv.org/abs/1602.02763}{{\ttfamily 1602.02763}}].

\bibitem{Martinelli:1998hz}
G.~Martinelli, \emph{{Hadronic weak interactions of light quarks}},
  {\emph{Nucl. Phys. Proc. Suppl.} {\bfseries 73} (1999) 58}
  [\href{https://arxiv.org/abs/hep-lat/9810013}{{\ttfamily hep-lat/9810013}}].

\bibitem{Gockeler:2000ja}
M.~G\"ockeler, R.~Horsley, W.~K\"urzinger, H.~Oelrich, D.~Pleiter, P.~E. Rakow
  et~al., \emph{{A Lattice calculation of the nucleon's spin dependent
  structure function $g_2$ revisited}}, {\emph{Phys. Rev. D} {\bfseries 63}
  (2001) 074506} [\href{https://arxiv.org/abs/hep-lat/0011091}{{\ttfamily
  hep-lat/0011091}}].

\bibitem{Martinelli:1996pk}
G.~Martinelli and C.~T. Sachrajda, \emph{{On the difficulty of computing higher
  twist corrections}},
  \href{https://doi.org/10.1016/0550-3213(96)00415-4}{\emph{Nucl. Phys.}
  {\bfseries B478} (1996) 660}
  [\href{https://arxiv.org/abs/hep-ph/9605336}{{\ttfamily hep-ph/9605336}}].

\bibitem{Davoudi:2012ya}
Z.~Davoudi and M.~J. Savage, \emph{{Restoration of Rotational Symmetry in the
  Continuum Limit of Lattice Field Theories}}, {\emph{Phys.Rev.} {\bfseries
  D86} (2012) 054505} [\href{https://arxiv.org/abs/1204.4146}{{\ttfamily
  1204.4146}}].

\bibitem{Teryaev:1999su}
O.~V. Teryaev, \emph{{Spin structure of nucleon and equivalence principle}},
  \href{https://arxiv.org/abs/hep-ph/9904376}{{\ttfamily hep-ph/9904376}}.

\bibitem{Jaffe:1989jz}
R.~Jaffe and A.~Manohar, \emph{{The $g_1$ Problem: Fact and Fantasy on the Spin
  of the Proton}}, {\emph{Nucl.Phys.} {\bfseries B337} (1990) 509}.

\bibitem{Lorce:2011kd}
C.~Lorc\'e and B.~Pasquini, \emph{{Quark Wigner Distributions and Orbital
  Angular Momentum}}, {\emph{Phys.Rev.} {\bfseries D84} (2011) 014015}
  [\href{https://arxiv.org/abs/1106.0139}{{\ttfamily 1106.0139}}].

\bibitem{Zhao:2015kca}
Y.~Zhao, K.-F. Liu and Y.~Yang, \emph{{Orbital Angular Momentum and Generalized
  Transverse Momentum Distribution}},
  \href{https://doi.org/10.1103/PhysRevD.93.054006}{\emph{Phys. Rev.}
  {\bfseries D93} (2016) 054006}
  [\href{https://arxiv.org/abs/1506.08832}{{\ttfamily 1506.08832}}].

\bibitem{Engelhardt:2017miy}
M.~Engelhardt, \emph{{Quark orbital dynamics in the proton from Lattice QCD --
  from Ji to Jaffe-Manohar orbital angular momentum}}, {\emph{Phys. Rev.}
  {\bfseries D95} (2017) 094505}
  [\href{https://arxiv.org/abs/1701.01536}{{\ttfamily 1701.01536}}].

\bibitem{Engelhardt:2020qtg}
M.~Engelhardt, J.~Green, N.~Hasan, S.~Krieg, S.~Meinel, J.~Negele et~al.,
  \emph{{From Ji to Jaffe-Manohar orbital angular momentum in Lattice QCD using
  a direct derivative method}}, {\emph{Phys. Rev. D} {\bfseries 102} (2020)
  074505} [\href{https://arxiv.org/abs/2008.03660}{{\ttfamily 2008.03660}}].

\bibitem{Leader:2013jra}
E.~Leader and C.~Lorc\'e, \emph{{The angular momentum controversy: What is it
  all about and does it matter?}}, {\emph{Phys.Rept.} {\bfseries 541} (2014)
  163} [\href{https://arxiv.org/abs/1309.4235}{{\ttfamily 1309.4235}}].

\bibitem{Liu:2015xha}
K.-F. Liu and C.~Lorc\'e, \emph{{The Parton Orbital Angular Momentum: Status
  and Prospects}}, \href{https://doi.org/10.1140/epja/i2016-16160-8}{\emph{Eur.
  Phys. J.} {\bfseries A52} (2016) 160}
  [\href{https://arxiv.org/abs/1508.00911}{{\ttfamily 1508.00911}}].

\bibitem{Ji:2020ena}
X.~Ji, F.~Yuan and Y.~Zhao, \emph{{What we know and what we
  don\textquoteright{}t know about the proton spin after 30 years}},
  \href{https://doi.org/10.1038/s42254-020-00248-4}{\emph{Nature Rev. Phys.}
  {\bfseries 3} (2021) 27} [\href{https://arxiv.org/abs/2009.01291}{{\ttfamily
  2009.01291}}].

\bibitem{Ashman:1987hv}
{\scshape European Muon} collaboration, J.~Ashman et~al., \emph{{A Measurement
  of the Spin Asymmetry and Determination of the Structure Function $g_1$ in
  Deep Inelastic Muon-Proton Scattering}},
  \href{https://doi.org/10.1016/0370-2693(88)91523-7}{\emph{Phys. Lett.}
  {\bfseries B206} (1988) 364}.

\bibitem{deFlorian:2009vb}
D.~de~Florian, R.~Sassot, M.~Stratmann and W.~Vogelsang, \emph{{Extraction of
  Spin-Dependent Parton Densities and Their Uncertainties}}, {\emph{Phys.Rev.}
  {\bfseries D80} (2009) 034030}
  [\href{https://arxiv.org/abs/0904.3821}{{\ttfamily 0904.3821}}].

\bibitem{Dong:1993pk}
S.-J. Dong and K.-F. Liu, \emph{{Stochastic estimation with $Z_2$ noise}},
  \href{https://doi.org/10.1016/0370-2693(94)90440-5}{\emph{Phys. Lett.}
  {\bfseries B328} (1994) 130}
  [\href{https://arxiv.org/abs/hep-lat/9308015}{{\ttfamily hep-lat/9308015}}].

\bibitem{deFlorian:2014yva}
D.~de~Florian, R.~Sassot, M.~Stratmann and W.~Vogelsang, \emph{{Evidence for
  polarization of gluons in the proton}}, {\emph{Phys.Rev.Lett.} {\bfseries
  113} (2014) 012001} [\href{https://arxiv.org/abs/1404.4293}{{\ttfamily
  1404.4293}}].

\bibitem{Nocera:2014gqa}
{\scshape NNPDF} collaboration, E.~R. Nocera, R.~D. Ball, S.~Forte, G.~Ridolfi
  and J.~Rojo, \emph{{A first unbiased global determination of polarized PDFs
  and their uncertainties}},
  \href{https://doi.org/10.1016/j.nuclphysb.2014.08.008}{\emph{Nucl. Phys.}
  {\bfseries B887} (2014) 276}
  [\href{https://arxiv.org/abs/1406.5539}{{\ttfamily 1406.5539}}].

\bibitem{Djawotho:2013pga}
{\scshape STAR} collaboration, P.~Djawotho, \emph{{Gluon polarization and jet
  production at STAR}},
  \href{https://doi.org/10.1393/ncc/i2013-11569-3}{\emph{Nuovo Cim.} {\bfseries
  C036} (2013) 35} [\href{https://arxiv.org/abs/1303.0543}{{\ttfamily
  1303.0543}}].

\bibitem{Adare:2014hsq}
{\scshape PHENIX} collaboration, A.~Adare et~al., \emph{{Inclusive
  double-helicity asymmetries in neutral-pion and eta-meson production in
  $\vec{p}+\vec{p}$ collisions at $\sqrt{s}=200$ GeV}},
  \href{https://doi.org/10.1103/PhysRevD.90.012007}{\emph{Phys. Rev.}
  {\bfseries D90} (2014) 012007}
  [\href{https://arxiv.org/abs/1402.6296}{{\ttfamily 1402.6296}}].

\bibitem{Ji:2013fga}
X.~Ji, J.-H. Zhang and Y.~Zhao, \emph{{Physics of the Gluon-Helicity
  Contribution to Proton Spin}}, {\emph{Phys.Rev.Lett.} {\bfseries 111} (2013)
  112002} [\href{https://arxiv.org/abs/1304.6708}{{\ttfamily 1304.6708}}].

\bibitem{Hatta:2013gta}
Y.~Hatta, X.~Ji and Y.~Zhao, \emph{{Gluon helicity $\Delta G$ from a
  universality class of operators on a lattice}}, {\emph{Phys.Rev.} {\bfseries
  D89} (2014) 085030} [\href{https://arxiv.org/abs/1310.4263}{{\ttfamily
  1310.4263}}].

\bibitem{Ji:2014lra}
X.~Ji, J.-H. Zhang and Y.~Zhao, \emph{{Justifying the Naive Partonic Sum Rule
  for Proton Spin}},
  \href{https://doi.org/10.1016/j.physletb.2015.02.054}{\emph{Phys. Lett.}
  {\bfseries B743} (2015) 180}
  [\href{https://arxiv.org/abs/1409.6329}{{\ttfamily 1409.6329}}].

\bibitem{Chen:2008ag}
X.-S. Chen, X.-F. Lu, W.-M. Sun, F.~Wang and T.~Goldman, \emph{{Spin and
  orbital angular momentum in gauge theories: Nucleon spin structure and
  multipole radiation revisited}}, {\emph{Phys.Rev.Lett.} {\bfseries 100}
  (2008) 232002} [\href{https://arxiv.org/abs/0806.3166}{{\ttfamily
  0806.3166}}].

\bibitem{Yang:2016plb}
Y.-B. Yang, R.~S. Sufian, A.~Alexandru, T.~Draper, M.~J. Glatzmaier, K.-F. Liu
  et~al., \emph{{Glue Spin and Helicity in the Proton from Lattice QCD}},
  \href{https://doi.org/10.1103/PhysRevLett.118.102001}{\emph{Phys. Rev. Lett.}
  {\bfseries 118} (2017) 102001}
  [\href{https://arxiv.org/abs/1609.05937}{{\ttfamily 1609.05937}}].

\bibitem{Mathur:1999uf}
N.~Mathur, S.~J. Dong, K.~F. Liu, L.~Mankiewicz and N.~C. Mukhopadhyay,
  \emph{{Quark orbital angular momentum from lattice QCD}},
  \href{https://doi.org/10.1103/PhysRevD.62.114504}{\emph{Phys. Rev.}
  {\bfseries D62} (2000) 114504}
  [\href{https://arxiv.org/abs/hep-ph/9912289}{{\ttfamily hep-ph/9912289}}].

\bibitem{Hagler:2003jd}
{\scshape LHPC, SESAM} collaboration, P.~H\"agler, J.~W. Negele, D.~B. Renner,
  W.~Schroers, T.~Lippert and K.~Schilling, \emph{{Moments of nucleon
  generalized parton distributions in lattice QCD}},
  \href{https://doi.org/10.1103/PhysRevD.68.034505}{\emph{Phys. Rev.}
  {\bfseries D68} (2003) 034505}
  [\href{https://arxiv.org/abs/hep-lat/0304018}{{\ttfamily hep-lat/0304018}}].

\bibitem{Bratt:2010jn}
{\scshape LHPC} collaboration, J.~D. Bratt et~al., \emph{{Nucleon structure
  from mixed action calculations using 2+1 flavors of asqtad sea and domain
  wall valence fermions}},
  \href{https://doi.org/10.1103/PhysRevD.82.094502}{\emph{Phys. Rev.}
  {\bfseries D82} (2010) 094502}
  [\href{https://arxiv.org/abs/1001.3620}{{\ttfamily 1001.3620}}].

\bibitem{Yang:2018nqn}
Y.-B. Yang, J.~Liang, Y.-J. Bi, Y.~Chen, T.~Draper, K.-F. Liu et~al.,
  \emph{{Proton Mass Decomposition from the QCD Energy Momentum Tensor}},
  \href{https://doi.org/10.1103/PhysRevLett.121.212001}{\emph{Phys. Rev. Lett.}
  {\bfseries 121} (2018) 212001}
  [\href{https://arxiv.org/abs/1808.08677}{{\ttfamily 1808.08677}}].

\bibitem{Alexandrou:2017oeh}
C.~Alexandrou, M.~Constantinou, K.~Hadjiyiannakou, K.~Jansen, C.~Kallidonis,
  G.~Koutsou et~al., \emph{{Nucleon Spin and Momentum Decomposition Using
  Lattice QCD Simulations}},
  \href{https://doi.org/10.1103/PhysRevLett.119.142002}{\emph{Phys. Rev. Lett.}
  {\bfseries 119} (2017) 142002}
  [\href{https://arxiv.org/abs/1706.02973}{{\ttfamily 1706.02973}}].

\bibitem{Alexandrou:2020sml}
C.~Alexandrou, S.~Bacchio, M.~Constantinou, J.~Finkenrath, K.~Hadjiyiannakou,
  K.~Jansen et~al., \emph{{Complete flavor decomposition of the spin and
  momentum fraction of the proton using lattice QCD simulations at physical
  pion mass}}, {\emph{Phys. Rev. D} {\bfseries 101} (2020) 094513}
  [\href{https://arxiv.org/abs/2003.08486}{{\ttfamily 2003.08486}}].

\bibitem{Yang:2019dha}
{\scshape Xqcd} collaboration, Y.-B. Yang, \emph{{A Lattice Story of Proton
  Spin}}, \href{https://doi.org/10.22323/1.334.0017}{\emph{PoS} {\bfseries
  LATTICE2018} (2019) 017} [\href{https://arxiv.org/abs/1904.04138}{{\ttfamily
  1904.04138}}].

\bibitem{Liu:1998um}
K.~F. Liu, S.~J. Dong, T.~Draper, D.~Leinweber, J.~H. Sloan, W.~Wilcox et~al.,
  \emph{{Valence QCD: Connecting QCD to the quark model}},
  \href{https://doi.org/10.1103/PhysRevD.59.112001}{\emph{Phys. Rev.}
  {\bfseries D59} (1999) 112001}
  [\href{https://arxiv.org/abs/hep-ph/9806491}{{\ttfamily hep-ph/9806491}}].

\bibitem{Aglietti:1998mz}
U.~Aglietti, M.~Ciuchini, G.~Corbo, E.~Franco, G.~Martinelli and
  L.~Silvestrini, \emph{{Model independent determination of the shape function
  for inclusive B decays and of the structure functions in DIS}},
  \href{https://doi.org/10.1016/S0370-2693(98)00677-7}{\emph{Phys. Lett.}
  {\bfseries B432} (1998) 411}
  [\href{https://arxiv.org/abs/hep-ph/9804416}{{\ttfamily hep-ph/9804416}}].

\bibitem{Can:2020sxc}
K.~Can et~al., \emph{{Lattice QCD evaluation of the Compton amplitude employing
  the Feynman-Hellmann theorem}}, {\emph{Phys. Rev. D} {\bfseries 102} (2020)
  114505} [\href{https://arxiv.org/abs/2007.01523}{{\ttfamily 2007.01523}}].

\bibitem{Chambers:2015bka}
A.~J. Chambers et~al., \emph{{Disconnected contributions to the spin of the
  nucleon}}, \href{https://doi.org/10.1103/PhysRevD.92.114517}{\emph{Phys.
  Rev.} {\bfseries D92} (2015) 114517}
  [\href{https://arxiv.org/abs/1508.06856}{{\ttfamily 1508.06856}}].

\bibitem{Bryan1990}
R.~Bryan, \emph{Maximum entropy analysis of oversampled data problems},
  {\emph{European Biophysics Journal} {\bfseries 18} (1990) 165}.

\bibitem{Jarrell:1996rrw}
M.~Jarrell and J.~E. Gubernatis, \emph{{Bayesian inference and the analytic
  continuation of imaginary-time quantum Monte Carlo data}},
  \href{https://doi.org/10.1016/0370-1573(95)00074-7}{\emph{Phys. Rept.}
  {\bfseries 269} (1996) 133}.

\bibitem{Burnier:2013nla}
Y.~Burnier and A.~Rothkopf, \emph{{Bayesian Approach to Spectral Function
  Reconstruction for Euclidean Quantum Field Theories}},
  \href{https://doi.org/10.1103/PhysRevLett.111.182003}{\emph{Phys. Rev. Lett.}
  {\bfseries 111} (2013) 182003}
  [\href{https://arxiv.org/abs/1307.6106}{{\ttfamily 1307.6106}}].

\bibitem{Backus1968}
G.~Backus and F.~Gilbert, \emph{The resolving power of gross earth data},
  \href{https://doi.org/10.1111/j.1365-246X.1968.tb00216.x}{\emph{Geophysical
  Journal of the Royal Astronomical Society} {\bfseries 16} (1968) 169}.

\bibitem{Liang:2019frk}
{\scshape XQCD} collaboration, J.~Liang, T.~Draper, K.-F. Liu, A.~Rothkopf and
  Y.-B. Yang, \emph{{Towards the nucleon hadronic tensor from lattice QCD}},
  \href{https://doi.org/10.1103/PhysRevD.101.114503}{\emph{Phys. Rev. D}
  {\bfseries 101} (2020) 114503}
  [\href{https://arxiv.org/abs/1906.05312}{{\ttfamily 1906.05312}}].

\bibitem{Detmold:2020lev}
W.~Detmold, A.~V. Grebe, I.~Kanamori, C.~J.~D. Lin, S.~Mondal, R.~J. Perry
  et~al., \emph{{A Preliminary Determination of the Second Mellin Moment of the
  Pion's Distribution Amplitude Using the Heavy Quark Operator Product
  Expansion}},  in \emph{{Asia-Pacific Symposium for Lattice Field Theory}}, 9,
  2020, \href{https://arxiv.org/abs/2009.09473}{{\ttfamily 2009.09473}}.

\bibitem{Detmold:2021uru}
{\scshape HOPE} collaboration, W.~Detmold, A.~V. Grebe, I.~Kanamori, C.~J.~D.
  Lin, R.~J. Perry and Y.~Zhao, \emph{{Parton physics from a heavy-quark
  operator product expansion: Formalism and Wilson coefficients}}, {\emph{Phys.
  Rev. D} {\bfseries 104} (2021) 074511}
  [\href{https://arxiv.org/abs/2103.09529}{{\ttfamily 2103.09529}}].

\bibitem{Detmold:2018kwu}
W.~Detmold, I.~Kanamori, C.~J.~D. Lin, S.~Mondal and Y.~Zhao, \emph{{Moments of
  pion distribution amplitude using operator product expansion on the
  lattice}}, {\emph{PoS} {\bfseries LATTICE2018} (2018) 106}
  [\href{https://arxiv.org/abs/1810.12194}{{\ttfamily 1810.12194}}].

\bibitem{Detmold:2021qln}
{\scshape HOPE} collaboration, W.~Detmold, A.~V. Grebe, I.~Kanamori, C.~J.~D.
  Lin, S.~Mondal, R.~J. Perry et~al., \emph{{Parton physics from a heavy-quark
  operator product expansion: Lattice QCD calculation of the second moment of
  the pion distribution amplitude}},
  \href{https://doi.org/10.1103/PhysRevD.105.034506}{\emph{Phys. Rev. D}
  {\bfseries 105} (2022) 034506}
  [\href{https://arxiv.org/abs/2109.15241}{{\ttfamily 2109.15241}}].

\bibitem{Horsley:2012pz}
{\scshape QCDSF, UKQCD} collaboration, R.~Horsley, R.~Millo, Y.~Nakamura,
  H.~Perlt, D.~Pleiter, P.~E.~L. Rakow et~al., \emph{{A Lattice Study of the
  Glue in the Nucleon}}, {\emph{Phys. Lett.} {\bfseries B714} (2012) 312}
  [\href{https://arxiv.org/abs/1205.6410}{{\ttfamily 1205.6410}}].

\bibitem{Cichy:2018mum}
K.~Cichy and M.~Constantinou, \emph{{A guide to light-cone PDFs from Lattice
  QCD: an overview of approaches, techniques and results}}, {\emph{Adv. High
  Energy Phys.} {\bfseries 2019} (2019) 3036904}
  [\href{https://arxiv.org/abs/1811.07248}{{\ttfamily 1811.07248}}].

\bibitem{Constantinou:2020pek}
M.~Constantinou, \emph{{The x-dependence of hadronic parton distributions: A
  review on the progress of lattice QCD}}, {\emph{Eur. Phys. J. A} {\bfseries
  57} (2021) 77} [\href{https://arxiv.org/abs/2010.02445}{{\ttfamily
  2010.02445}}].

\bibitem{Xiong:2013bka}
X.~Xiong, X.~Ji, J.-H. Zhang and Y.~Zhao, \emph{{One-loop matching for parton
  distributions: Nonsinglet case}}, {\emph{Phys. Rev.} {\bfseries D90} (2014)
  014051} [\href{https://arxiv.org/abs/1310.7471}{{\ttfamily 1310.7471}}].

\bibitem{Izubuchi:2018srq}
T.~Izubuchi, X.~Ji, L.~Jin, I.~W. Stewart and Y.~Zhao, \emph{{Factorization
  Theorem Relating Euclidean and Light-Cone Parton Distributions}},
  {\emph{Phys. Rev.} {\bfseries D98} (2018) 056004}
  [\href{https://arxiv.org/abs/1801.03917}{{\ttfamily 1801.03917}}].

\bibitem{Chen:2016utp}
J.-W. Chen, S.~D. Cohen, X.~Ji, H.-W. Lin and J.-H. Zhang, \emph{{Nucleon
  Helicity and Transversity Parton Distributions from Lattice QCD}},
  {\emph{Nucl. Phys.} {\bfseries B911} (2016) 246}
  [\href{https://arxiv.org/abs/1603.06664}{{\ttfamily 1603.06664}}].

\bibitem{Braun:2018brg}
V.~M. Braun, A.~Vladimirov and J.-H. Zhang, \emph{{Power corrections and
  renormalons in parton quasidistributions}}, {\emph{Phys. Rev.} {\bfseries
  D99} (2019) 014013} [\href{https://arxiv.org/abs/1810.00048}{{\ttfamily
  1810.00048}}].

\bibitem{Ji:2020brr}
X.~Ji, Y.~Liu, A.~Sch\"afer, W.~Wang, Y.-B. Yang, J.-H. Zhang et~al., \emph{{A
  Hybrid Renormalization Scheme for Quasi Light-Front Correlations in
  Large-Momentum Effective Theory}},
  \href{https://doi.org/10.1016/j.nuclphysb.2021.115311}{\emph{Nucl. Phys. B}
  {\bfseries 964} (2021) 115311}
  [\href{https://arxiv.org/abs/2008.03886}{{\ttfamily 2008.03886}}].

\bibitem{Wang:2019tgg}
W.~Wang, J.-H. Zhang, S.~Zhao and R.~Zhu, \emph{{Complete matching for
  quasidistribution functions in large momentum effective theory}},
  \href{https://doi.org/10.1103/PhysRevD.100.074509}{\emph{Phys. Rev. D}
  {\bfseries 100} (2019) 074509}
  [\href{https://arxiv.org/abs/1904.00978}{{\ttfamily 1904.00978}}].

\bibitem{Liu:2019urm}
Y.-S. Liu, W.~Wang, J.~Xu, Q.-A. Zhang, J.-H. Zhang, S.~Zhao et~al.,
  \emph{{Matching generalized parton quasidistributions in the RI/MOM scheme}},
  {\emph{Phys. Rev.} {\bfseries D100} (2019) 034006}
  [\href{https://arxiv.org/abs/1902.00307}{{\ttfamily 1902.00307}}].

\bibitem{Dotsenko:1979wb}
V.~Dotsenko and S.~Vergeles, \emph{{Renormalizability of Phase Factors in the
  Nonabelian Gauge Theory}}, {\emph{Nucl. Phys. B} {\bfseries 169} (1980) 527}.

\bibitem{Craigie:1980qs}
N.~S. Craigie and H.~Dorn, \emph{{On the Renormalization and Short Distance
  Properties of Hadronic Operators in {QCD}}}, {\emph{Nucl. Phys.} {\bfseries
  B185} (1981) 204}.

\bibitem{Dorn:1986dt}
H.~Dorn, \emph{{Renormalization of Path Ordered Phase Factors and Related
  Hadron Operators in Gauge Field Theories}}, {\emph{Fortsch. Phys.} {\bfseries
  34} (1986) 11}.

\bibitem{Ji:2017oey}
X.~Ji, J.-H. Zhang and Y.~Zhao, \emph{{Renormalization in Large Momentum
  Effective Theory of Parton Physics}}, {\emph{Phys. Rev. Lett.} {\bfseries
  120} (2018) 112001} [\href{https://arxiv.org/abs/1706.08962}{{\ttfamily
  1706.08962}}].

\bibitem{Green:2017xeu}
J.~Green, K.~Jansen and F.~Steffens, \emph{{Nonperturbative Renormalization of
  Nonlocal Quark Bilinears for Parton Quasidistribution Functions on the
  Lattice Using an Auxiliary Field}}, {\emph{Phys. Rev. Lett.} {\bfseries 121}
  (2018) 022004} [\href{https://arxiv.org/abs/1707.07152}{{\ttfamily
  1707.07152}}].

\bibitem{Ishikawa:2017faj}
T.~Ishikawa, Y.-Q. Ma, J.-W. Qiu and S.~Yoshida, \emph{{Renormalizability of
  quasiparton distribution functions}}, {\emph{Phys. Rev.} {\bfseries D96}
  (2017) 094019} [\href{https://arxiv.org/abs/1707.03107}{{\ttfamily
  1707.03107}}].

\bibitem{Zhang:2018diq}
J.-H. Zhang, X.~Ji, A.~Sch{\"a}fer, W.~Wang and S.~Zhao, \emph{{Accessing Gluon
  Parton Distributions in Large Momentum Effective Theory}}, {\emph{Phys. Rev.
  Lett.} {\bfseries 122} (2019) 142001}
  [\href{https://arxiv.org/abs/1808.10824}{{\ttfamily 1808.10824}}].

\bibitem{Li:2018tpe}
Z.-Y. Li, Y.-Q. Ma and J.-W. Qiu, \emph{{Multiplicative Renormalizability of
  Operators defining Quasiparton Distributions}}, {\emph{Phys. Rev. Lett.}
  {\bfseries 122} (2019) 062002}
  [\href{https://arxiv.org/abs/1809.01836}{{\ttfamily 1809.01836}}].

\bibitem{Ishikawa:2016znu}
T.~Ishikawa, Y.-Q. Ma, J.-W. Qiu and S.~Yoshida, \emph{{Practical quasi parton
  distribution functions}},  \href{https://arxiv.org/abs/1609.02018}{{\ttfamily
  1609.02018}}.

\bibitem{Xiong:2017jtn}
X.~Xiong, T.~Luu and U.-G. Mei{\ss}ner, \emph{{Quasi-Parton Distribution
  Function in Lattice Perturbation Theory}},
  \href{https://arxiv.org/abs/1705.00246}{{\ttfamily 1705.00246}}.

\bibitem{Constantinou:2017sej}
M.~Constantinou and H.~Panagopoulos, \emph{{Perturbative renormalization of
  quasi-parton distribution functions}}, {\emph{Phys. Rev.} {\bfseries D96}
  (2017) 054506} [\href{https://arxiv.org/abs/1705.11193}{{\ttfamily
  1705.11193}}].

\bibitem{Chen:2020arf}
L.-B. Chen, W.~Wang and R.~Zhu, \emph{{Quasi parton distribution functions at
  NNLO: flavor non-diagonal quark contributions}},
  \href{https://doi.org/10.1103/PhysRevD.102.011503}{\emph{Phys. Rev. D}
  {\bfseries 102} (2020) 011503}
  [\href{https://arxiv.org/abs/2005.13757}{{\ttfamily 2005.13757}}].

\bibitem{Chen:2020iqi}
L.-B. Chen, W.~Wang and R.~Zhu, \emph{{Master integrals for two-loop QCD
  corrections to quark quasi PDFs}},
  \href{https://doi.org/10.1007/JHEP10(2020)079}{\emph{JHEP} {\bfseries 10}
  (2020) 079} [\href{https://arxiv.org/abs/2006.10917}{{\ttfamily
  2006.10917}}].

\bibitem{Li:2020xml}
Z.-Y. Li, Y.-Q. Ma and J.-W. Qiu, \emph{{Extraction of
  Next-to-Next-to-Leading-Order Parton Distribution Functions from Lattice QCD
  Calculations}},
  \href{https://doi.org/10.1103/PhysRevLett.126.072001}{\emph{Phys. Rev. Lett.}
  {\bfseries 126} (2021) 072001}
  [\href{https://arxiv.org/abs/2006.12370}{{\ttfamily 2006.12370}}].

\bibitem{Chen:2020ody}
L.-B. Chen, W.~Wang and R.~Zhu, \emph{{Next-to-Next-to-Leading Order
  Calculation of Quasiparton Distribution Functions}},
  \href{https://doi.org/10.1103/PhysRevLett.126.072002}{\emph{Phys. Rev. Lett.}
  {\bfseries 126} (2021) 072002}
  [\href{https://arxiv.org/abs/2006.14825}{{\ttfamily 2006.14825}}].

\bibitem{Braun:2020ymy}
V.~M. Braun, K.~G. Chetyrkin and B.~A. Kniehl, \emph{{Renormalization of parton
  quasi-distributions beyond the leading order: spacelike vs. timelike}},
  {\emph{JHEP} {\bfseries 07} (2020) 161}
  [\href{https://arxiv.org/abs/2004.01043}{{\ttfamily 2004.01043}}].

\bibitem{Zhang:2017bzy}
J.-H. Zhang, J.-W. Chen, X.~Ji, L.~Jin and H.-W. Lin, \emph{{Pion Distribution
  Amplitude from Lattice QCD}}, {\emph{Phys. Rev.} {\bfseries D95} (2017)
  094514} [\href{https://arxiv.org/abs/1702.00008}{{\ttfamily 1702.00008}}].

\bibitem{Green:2020xco}
J.~R. Green, K.~Jansen and F.~Steffens, \emph{{Improvement, generalization, and
  scheme conversion of Wilson-line operators on the lattice in the auxiliary
  field approach}}, {\emph{Phys. Rev. D} {\bfseries 101} (2020) 074509}
  [\href{https://arxiv.org/abs/2002.09408}{{\ttfamily 2002.09408}}].

\bibitem{Alexandrou:2020qtt}
C.~Alexandrou, K.~Cichy, M.~Constantinou, J.~R. Green, K.~Hadjiyiannakou,
  K.~Jansen et~al., \emph{Lattice continuum-limit study of nucleon quasi-pdfs},
  \href{https://doi.org/10.1103/PhysRevD.103.094512}{\emph{Phys. Rev. D}
  {\bfseries 103} (2021) 094512}
  [\href{https://arxiv.org/abs/2011.00964}{{\ttfamily 2011.00964}}].

\bibitem{Alexandrou:2017huk}
C.~Alexandrou, K.~Cichy, M.~Constantinou, K.~Hadjiyiannakou, K.~Jansen,
  H.~Panagopoulos et~al., \emph{{A complete non-perturbative renormalization
  prescription for quasi-PDFs}}, {\emph{Nucl. Phys.} {\bfseries B923} (2017)
  394} [\href{https://arxiv.org/abs/1706.00265}{{\ttfamily 1706.00265}}].

\bibitem{Stewart:2017tvs}
I.~W. Stewart and Y.~Zhao, \emph{{Matching the quasiparton distribution in a
  momentum subtraction scheme}}, {\emph{Phys. Rev.} {\bfseries D97} (2018)
  054512} [\href{https://arxiv.org/abs/1709.04933}{{\ttfamily 1709.04933}}].

\bibitem{Chen:2017mzz}
J.-W. Chen, T.~Ishikawa, L.~Jin, H.-W. Lin, Y.-B. Yang, J.-H. Zhang et~al.,
  \emph{{Parton distribution function with nonperturbative renormalization from
  lattice QCD}}, {\emph{Phys. Rev.} {\bfseries D97} (2018) 014505}
  [\href{https://arxiv.org/abs/1706.01295}{{\ttfamily 1706.01295}}].

\bibitem{Chen:2017mie}
{\scshape LP3} collaboration, J.-W. Chen, T.~Ishikawa, L.~Jin, H.-W. Lin, J.-H.
  Zhang and Y.~Zhao, \emph{{Symmetry properties of nonlocal quark bilinear
  operators on a Lattice}},
  \href{https://doi.org/10.1088/1674-1137/43/10/103101}{\emph{Chin. Phys. C}
  {\bfseries 43} (2019) 103101}
  [\href{https://arxiv.org/abs/1710.01089}{{\ttfamily 1710.01089}}].

\bibitem{Zhao:2018fyu}
Y.~Zhao, \emph{{Unraveling high-energy hadron structures with lattice QCD}},
  {\emph{Int. J. Mod. Phys.} {\bfseries A33} (2019) 1830033}
  [\href{https://arxiv.org/abs/1812.07192}{{\ttfamily 1812.07192}}].

\bibitem{Alexandrou:2019lfo}
C.~Alexandrou, K.~Cichy, M.~Constantinou, K.~Hadjiyiannakou, K.~Jansen,
  A.~Scapellato et~al., \emph{{Systematic uncertainties in parton distribution
  functions from lattice QCD simulations at the physical point}}, {\emph{Phys.
  Rev.} {\bfseries D99} (2019) 114504}
  [\href{https://arxiv.org/abs/1902.00587}{{\ttfamily 1902.00587}}].

\bibitem{Alexandrou:2018pbm}
C.~Alexandrou, K.~Cichy, M.~Constantinou, K.~Jansen, A.~Scapellato and
  F.~Steffens, \emph{{Light-Cone Parton Distribution Functions from Lattice
  QCD}}, {\emph{Phys. Rev. Lett.} {\bfseries 121} (2018) 112001}
  [\href{https://arxiv.org/abs/1803.02685}{{\ttfamily 1803.02685}}].

\bibitem{Alexandrou:2018eet}
C.~Alexandrou, K.~Cichy, M.~Constantinou, K.~Jansen, A.~Scapellato and
  F.~Steffens, \emph{{Transversity parton distribution functions from lattice
  QCD}}, \href{https://doi.org/10.1103/PhysRevD.98.091503}{\emph{Phys. Rev.}
  {\bfseries D98} (2018) 091503}
  [\href{https://arxiv.org/abs/1807.00232}{{\ttfamily 1807.00232}}].

\bibitem{Bhattacharya:2020cen}
S.~Bhattacharya, K.~Cichy, M.~Constantinou, A.~Metz, A.~Scapellato and
  F.~Steffens, \emph{{Insights on proton structure from lattice QCD: The
  twist-3 parton distribution function $g_T(x)$}}, {\emph{Phys. Rev. D}
  {\bfseries 102} (2020) 111501}
  [\href{https://arxiv.org/abs/2004.04130}{{\ttfamily 2004.04130}}].

\bibitem{Alexandrou:2020uyt}
C.~Alexandrou, M.~Constantinou, K.~Hadjiyiannakou, K.~Jansen and F.~Manigrasso,
  \emph{{Flavor decomposition for the proton helicity parton distribution
  functions}},
  \href{https://doi.org/10.1103/PhysRevLett.126.102003}{\emph{Phys. Rev. Lett.}
  {\bfseries 126} (2021) 102003}
  [\href{https://arxiv.org/abs/2009.13061}{{\ttfamily 2009.13061}}].

\bibitem{Liu:2018uuj}
Y.-S. Liu, J.-W. Chen, L.~Jin, H.-W. Lin, Y.-B. Yang, J.-H. Zhang et~al.,
  \emph{{Unpolarized quark distribution from lattice QCD: A systematic analysis
  of renormalization and matching}},
  \href{https://doi.org/10.1103/PhysRevD.101.034020}{\emph{Phys. Rev. D}
  {\bfseries 101} (2020) 034020}
  [\href{https://arxiv.org/abs/1807.06566}{{\ttfamily 1807.06566}}].

\bibitem{Liu:2018tox}
Y.-S. Liu, W.~Wang, J.~Xu, Q.-A. Zhang, S.~Zhao and Y.~Zhao, \emph{{Matching
  the meson quasidistribution amplitude in the RI/MOM scheme}}, {\emph{Phys.
  Rev.} {\bfseries D99} (2019) 094036}
  [\href{https://arxiv.org/abs/1810.10879}{{\ttfamily 1810.10879}}].

\bibitem{Lin:2017ani}
{\scshape LP3} collaboration, H.-W. Lin, J.-W. Chen, T.~Ishikawa and J.-H.
  Zhang, \emph{{Improved parton distribution functions at the physical pion
  mass}}, {\emph{Phys. Rev.} {\bfseries D98} (2018) 054504}
  [\href{https://arxiv.org/abs/1708.05301}{{\ttfamily 1708.05301}}].

\bibitem{Chen:2018fwa}
J.-H. Zhang, J.-W. Chen, L.~Jin, H.-W. Lin, A.~Sch\"afer and Y.~Zhao,
  \emph{{First direct lattice-QCD calculation of the $x$-dependence of the pion
  parton distribution function}},
  \href{https://doi.org/10.1103/PhysRevD.100.034505}{\emph{Phys. Rev. D}
  {\bfseries 100} (2019) 034505}
  [\href{https://arxiv.org/abs/1804.01483}{{\ttfamily 1804.01483}}].

\bibitem{Zhang:2020gaj}
R.~Zhang, C.~Honkala, H.-W. Lin and J.-W. Chen, \emph{{Pion and kaon
  distribution amplitudes in the continuum limit}},
  \href{https://doi.org/10.1103/PhysRevD.102.094519}{\emph{Phys. Rev. D}
  {\bfseries 102} (2020) 094519}
  [\href{https://arxiv.org/abs/2005.13955}{{\ttfamily 2005.13955}}].

\bibitem{Fan:2020nzz}
Z.~Fan, X.~Gao, R.~Li, H.-W. Lin, N.~Karthik, S.~Mukherjee et~al.,
  \emph{{Isovector parton distribution functions of the proton on a superfine
  lattice}}, {\emph{Phys. Rev. D} {\bfseries 102} (2020) 074504}
  [\href{https://arxiv.org/abs/2005.12015}{{\ttfamily 2005.12015}}].

\bibitem{Zhang:2020dkn}
R.~Zhang, H.-W. Lin and B.~Yoon, \emph{{Probing nucleon strange and charm
  distributions with lattice QCD}},
  \href{https://doi.org/10.1103/PhysRevD.104.094511}{\emph{Phys. Rev. D}
  {\bfseries 104} (2021) 094511}
  [\href{https://arxiv.org/abs/2005.01124}{{\ttfamily 2005.01124}}].

\bibitem{Lin:2020ssv}
H.-W. Lin, J.-W. Chen, Z.~Fan, J.-H. Zhang and R.~Zhang, \emph{{Valence-Quark
  Distribution of the Kaon and Pion from Lattice QCD}},
  \href{https://doi.org/10.1103/PhysRevD.103.014516}{\emph{Phys. Rev. D}
  {\bfseries 103} (2021) 014516}
  [\href{https://arxiv.org/abs/2003.14128}{{\ttfamily 2003.14128}}].

\bibitem{Lin:2020fsj}
H.-W. Lin, J.-W. Chen and R.~Zhang, \emph{{Lattice Nucleon Isovector
  Unpolarized Parton Distribution in the Physical-Continuum Limit}},
  \href{https://arxiv.org/abs/2011.14971}{{\ttfamily 2011.14971}}.

\bibitem{LatticePartonCollaborationLPC:2021xdx}
{\scshape Lattice Parton Collaboration (LPC)} collaboration, Y.-K. Huo et~al.,
  \emph{{Self-renormalization of quasi-light-front correlators on the
  lattice}}, {\emph{Nucl. Phys. B} {\bfseries 969} (2021) 115443}
  [\href{https://arxiv.org/abs/2103.02965}{{\ttfamily 2103.02965}}].

\bibitem{Monahan:2016bvm}
C.~Monahan and K.~Orginos, \emph{{Quasi parton distributions and the gradient
  flow}}, \href{https://doi.org/10.1007/JHEP03(2017)116}{\emph{JHEP} {\bfseries
  03} (2017) 116} [\href{https://arxiv.org/abs/1612.01584}{{\ttfamily
  1612.01584}}].

\bibitem{Monahan:2017hpu}
C.~Monahan, \emph{{Smeared quasidistributions in perturbation theory}},
  {\emph{Phys. Rev.} {\bfseries D97} (2018) 054507}
  [\href{https://arxiv.org/abs/1710.04607}{{\ttfamily 1710.04607}}].

\bibitem{Lin:2014zya}
H.-W. Lin, J.-W. Chen, S.~D. Cohen and X.~Ji, \emph{{Flavor Structure of the
  Nucleon Sea from Lattice QCD}}, {\emph{Phys.Rev.} {\bfseries D91} (2015)
  054510} [\href{https://arxiv.org/abs/1402.1462}{{\ttfamily 1402.1462}}].

\bibitem{Alexandrou:2014pna}
C.~Alexandrou, K.~Cichy, V.~Drach, E.~Garcia-Ramos, K.~Hadjiyiannakou et~al.,
  \emph{{First results with twisted mass fermions towards the computation of
  parton distribution functions on the lattice}}, {\emph{PoS} {\bfseries
  LATTICE2014} (2014) 135} [\href{https://arxiv.org/abs/1411.0891}{{\ttfamily
  1411.0891}}].

\bibitem{Alexandrou:2015rja}
C.~Alexandrou, K.~Cichy, V.~Drach, E.~Garcia-Ramos, K.~Hadjiyiannakou,
  K.~Jansen et~al., \emph{{Lattice calculation of parton distributions}},
  {\emph{Phys. Rev.} {\bfseries D92} (2015) 014502}
  [\href{https://arxiv.org/abs/1504.07455}{{\ttfamily 1504.07455}}].

\bibitem{Lin:2018pvv}
H.-W. Lin, J.-W. Chen, X.~Ji, L.~Jin, R.~Li, Y.-S. Liu et~al., \emph{{Proton
  Isovector Helicity Distribution on the Lattice at Physical Pion Mass}},
  {\emph{Phys. Rev. Lett.} {\bfseries 121} (2018) 242003}
  [\href{https://arxiv.org/abs/1807.07431}{{\ttfamily 1807.07431}}].

\bibitem{Frezzotti:2000nk}
{\scshape ALPHA} collaboration, R.~Frezzotti, P.~A. Grassi, S.~Sint and
  P.~Weisz, \emph{Lattice {QCD} with a chirally twisted mass term},
  {\emph{JHEP} {\bfseries 08} (2001) 058}
  [\href{https://arxiv.org/abs/hep-lat/0101001}{{\ttfamily hep-lat/0101001}}].

\bibitem{Shindler_2008}
A.~Shindler, \emph{Twisted mass lattice {QCD}}, {\emph{Physics Reports}
  {\bfseries 461} (2008) 37}.

\bibitem{Alekhin:2017kpj}
S.~Alekhin, J.~Bl\"umlein, S.~Moch and R.~Placakyte, \emph{{Parton distribution
  functions, $\alpha_s$, and heavy-quark masses for LHC Run II}}, {\emph{Phys.
  Rev. D} {\bfseries 96} (2017) 014011}
  [\href{https://arxiv.org/abs/1701.05838}{{\ttfamily 1701.05838}}].

\bibitem{Ball:2017nwa}
{\scshape NNPDF} collaboration, R.~D. Ball et~al., \emph{{Parton distributions
  from high-precision collider data}}, {\emph{Eur. Phys. J.} {\bfseries C77}
  (2017) 663} [\href{https://arxiv.org/abs/1706.00428}{{\ttfamily
  1706.00428}}].

\bibitem{Accardi:2016qay}
A.~Accardi, L.~T. Brady, W.~Melnitchouk, J.~F. Owens and N.~Sato,
  \emph{{Constraints on large-$x$ parton distributions from new weak boson
  production and deep-inelastic scattering data}}, {\emph{Phys. Rev.}
  {\bfseries D93} (2016) 114017}
  [\href{https://arxiv.org/abs/1602.03154}{{\ttfamily 1602.03154}}].

\bibitem{Ethier:2017zbq}
J.~J. Ethier, N.~Sato and W.~Melnitchouk, \emph{{First simultaneous extraction
  of spin-dependent parton distributions and fragmentation functions from a
  global QCD analysis}}, {\emph{Phys. Rev. Lett.} {\bfseries 119} (2017)
  132001} [\href{https://arxiv.org/abs/1705.05889}{{\ttfamily 1705.05889}}].

\bibitem{Lin:2017stx}
H.-W. Lin, W.~Melnitchouk, A.~Prokudin, N.~Sato and H.~Shows, \emph{{First
  Monte Carlo Global Analysis of Nucleon Transversity with Lattice QCD
  Constraints}}, {\emph{Phys. Rev. Lett.} {\bfseries 120} (2018) 152502}
  [\href{https://arxiv.org/abs/1710.09858}{{\ttfamily 1710.09858}}].

\bibitem{Karpie:2019eiq}
J.~Karpie, K.~Orginos, A.~Rothkopf and S.~Zafeiropoulos, \emph{{Reconstructing
  parton distribution functions from Ioffe time data: from Bayesian methods to
  Neural Networks}}, {\emph{JHEP} {\bfseries 04} (2019) 057}
  [\href{https://arxiv.org/abs/1901.05408}{{\ttfamily 1901.05408}}].

\bibitem{Izubuchi:2019lyk}
T.~Izubuchi, L.~Jin, C.~Kallidonis, N.~Karthik, S.~Mukherjee, P.~Petreczky
  et~al., \emph{{Valence parton distribution function of pion from fine
  lattice}}, {\emph{Phys. Rev.} {\bfseries D100} (2019) 034516}
  [\href{https://arxiv.org/abs/1905.06349}{{\ttfamily 1905.06349}}].

\bibitem{Gao:2020ito}
X.~Gao, L.~Jin, C.~Kallidonis, N.~Karthik, S.~Mukherjee, P.~Petreczky et~al.,
  \emph{{Valence parton distribution of the pion from lattice QCD: Approaching
  the continuum limit}},
  \href{https://doi.org/10.1103/PhysRevD.102.094513}{\emph{Phys. Rev. D}
  {\bfseries 102} (2020) 094513}
  [\href{https://arxiv.org/abs/2007.06590}{{\ttfamily 2007.06590}}].

\bibitem{Chai:2020nxw}
Y.~Chai et~al., \emph{{Parton distribution functions of $\Delta^+$ on the
  lattice}}, {\emph{Phys. Rev. D} {\bfseries 102} (2020) 014508}
  [\href{https://arxiv.org/abs/2002.12044}{{\ttfamily 2002.12044}}].

\bibitem{Lin:2019ocg}
H.-W. Lin and R.~Zhang, \emph{{Lattice finite-volume dependence of the nucleon
  parton distributions}}, {\emph{Phys. Rev. D} {\bfseries 100} (2019) 074502}.

\bibitem{Alexandrou:2021oih}
C.~Alexandrou, M.~Constantinou, K.~Hadjiyiannakou, K.~Jansen and F.~Manigrasso,
  \emph{{Flavor decomposition of the nucleon unpolarized, helicity, and
  transversity parton distribution functions from lattice QCD simulations}},
  {\emph{Phys. Rev. D} {\bfseries 104} (2021) 054503}
  [\href{https://arxiv.org/abs/2106.16065}{{\ttfamily 2106.16065}}].

\bibitem{Buckley:2014ana}
A.~Buckley, J.~Ferrando, S.~Lloyd, K.~Nordstr{\"o}m, B.~Page, M.~R{\"u}fenacht
  et~al., \emph{{LHAPDF6: parton density access in the LHC precision era}},
  {\emph{Eur. Phys. J.} {\bfseries C75} (2015) 132}
  [\href{https://arxiv.org/abs/1412.7420}{{\ttfamily 1412.7420}}].

\bibitem{Bhattacharya:2020xlt}
S.~Bhattacharya, K.~Cichy, M.~Constantinou, A.~Metz, A.~Scapellato and
  F.~Steffens, \emph{{One-loop matching for the twist-3 parton distribution
  $g_T (x)$}}, {\emph{Phys. Rev.} {\bfseries D102} (2020) 034005}
  [\href{https://arxiv.org/abs/2005.10939}{{\ttfamily 2005.10939}}].

\bibitem{Bhattacharya:2021moj}
S.~Bhattacharya, K.~Cichy, M.~Constantinou, A.~Metz, A.~Scapellato and
  F.~Steffens, \emph{{Parton distribution functions beyond leading twist from
  lattice QCD: The $h_L(x)$ case}},
  \href{https://doi.org/10.1103/PhysRevD.104.114510}{\emph{Phys. Rev. D}
  {\bfseries 104} (2021) 114510}
  [\href{https://arxiv.org/abs/2107.02574}{{\ttfamily 2107.02574}}].

\bibitem{Bhattacharya:2020jfj}
S.~Bhattacharya, K.~Cichy, M.~Constantinou, A.~Metz, A.~Scapellato and
  F.~Steffens, \emph{{The role of zero-mode contributions in the matching for
  the twist-3 PDFs $e(x)$ and $h_{L}(x)$}}, {\emph{Phys. Rev. D} {\bfseries
  102} (2020) 114025} [\href{https://arxiv.org/abs/2006.12347}{{\ttfamily
  2006.12347}}].

\bibitem{Wandzura:1977qf}
S.~Wandzura and F.~Wilczek, \emph{{Sum Rules for Spin Dependent
  Electroproduction: Test of Relativistic Constituent Quarks}}, {\emph{Phys.
  Lett.} {\bfseries 72B} (1977) 195}.

\bibitem{Alexandrou:2020zbe}
C.~Alexandrou, K.~Cichy, M.~Constantinou, K.~Hadjiyiannakou, K.~Jansen,
  A.~Scapellato et~al., \emph{{Unpolarized and helicity generalized parton
  distributions of the proton within lattice QCD}}, {\emph{Phys. Rev. Lett.}
  {\bfseries 125} (2020) 262001}
  [\href{https://arxiv.org/abs/2008.10573}{{\ttfamily 2008.10573}}].

\bibitem{Alexandrou:2021bbo}
C.~Alexandrou, K.~Cichy, M.~Constantinou, K.~Hadjiyiannakou, K.~Jansen,
  A.~Scapellato et~al., \emph{{Transversity GPDs of the proton from lattice
  QCD}},  \href{https://arxiv.org/abs/2108.10789}{{\ttfamily 2108.10789}}.

\bibitem{Dodson_Lattice:2021}
S.~Bhattacharya, K.~Cichy, M.~Constantinou, J.~Dodson, A.~Metz, A.~Scapellato
  et~al., \emph{{First Lattice QCD Study of Proton Twist-3 GPDs}},  2021.

\bibitem{Radyushkin:2016hsy}
A.~Radyushkin, \emph{{Nonperturbative Evolution of Parton
  Quasi-Distributions}}, {\emph{Phys. Lett.} {\bfseries B767} (2017) 314}
  [\href{https://arxiv.org/abs/1612.05170}{{\ttfamily 1612.05170}}].

\bibitem{Radyushkin:2017ffo}
A.~Radyushkin, \emph{{Target Mass Effects in Parton Quasi-Distributions}},
  {\emph{Phys. Lett.} {\bfseries B770} (2017) 514}
  [\href{https://arxiv.org/abs/1702.01726}{{\ttfamily 1702.01726}}].

\bibitem{Ioffe:1969kf}
B.~L. Ioffe, \emph{{Space-time picture of photon and neutrino scattering and
  electroproduction cross-section asymptotics}}, {\emph{Phys. Lett.} {\bfseries
  30B} (1969) 123}.

\bibitem{Radyushkin:2018cvn}
A.~Radyushkin, \emph{{One-loop evolution of parton pseudo-distribution
  functions on the lattice}}, {\emph{Phys. Rev.} {\bfseries D98} (2018) 014019}
  [\href{https://arxiv.org/abs/1801.02427}{{\ttfamily 1801.02427}}].

\bibitem{Zhang:2018ggy}
J.-H. Zhang, J.-W. Chen and C.~Monahan, \emph{{Parton distribution functions
  from reduced Ioffe-time distributions}}, {\emph{Phys. Rev.} {\bfseries D97}
  (2018) 074508} [\href{https://arxiv.org/abs/1801.03023}{{\ttfamily
  1801.03023}}].

\bibitem{Radyushkin:2018nbf}
A.~V. Radyushkin, \emph{{Structure of parton quasi-distributions and their
  moments}}, {\emph{Phys. Lett.} {\bfseries B788} (2019) 380}
  [\href{https://arxiv.org/abs/1807.07509}{{\ttfamily 1807.07509}}].

\bibitem{Radyushkin:2017sfi}
A.~Radyushkin, \emph{{Quasi-PDFs and pseudo-PDFs}}, {\emph{PoS} {\bfseries
  QCDEV2017} (2017) 021} [\href{https://arxiv.org/abs/1711.06031}{{\ttfamily
  1711.06031}}].

\bibitem{Karpie:2017bzm}
J.~Karpie, K.~Orginos, A.~Radyushkin and S.~Zafeiropoulos, \emph{{Parton
  distribution functions on the lattice and in the continuum}}, {\emph{EPJ Web
  Conf.} {\bfseries 175} (2018) 06032}
  [\href{https://arxiv.org/abs/1710.08288}{{\ttfamily 1710.08288}}].

\bibitem{Karpie:2018zaz}
J.~Karpie, K.~Orginos and S.~Zafeiropoulos, \emph{{Moments of Ioffe time parton
  distribution functions from non-local matrix elements}}, {\emph{JHEP}
  {\bfseries 11} (2018) 178}
  [\href{https://arxiv.org/abs/1807.10933}{{\ttfamily 1807.10933}}].

\bibitem{Joo:2019bzr}
B.~Jo{\'o}, J.~Karpie, K.~Orginos, A.~V. Radyushkin, D.~G. Richards, R.~S.
  Sufian et~al., \emph{{Pion Valence Structure from Ioffe Time
  Pseudo-Distributions}}, {\emph{Phys. Rev.} {\bfseries D100} (2019) 114512}
  [\href{https://arxiv.org/abs/1909.08517}{{\ttfamily 1909.08517}}].

\bibitem{Joo:2019jct}
B.~Jo{\'o}, J.~Karpie, K.~Orginos, A.~Radyushkin, D.~Richards and
  S.~Zafeiropoulos, \emph{{Parton Distribution Functions from Ioffe time
  pseudo-distributions}}, {\emph{JHEP} {\bfseries 12} (2019) 081}
  [\href{https://arxiv.org/abs/1908.09771}{{\ttfamily 1908.09771}}].

\bibitem{Joo:2020spy}
B.~Jo\'o, J.~Karpie, K.~Orginos, A.~V. Radyushkin, D.~G. Richards and
  S.~Zafeiropoulos, \emph{{Parton Distribution Functions from Ioffe Time
  Pseudodistributions from Lattice Calculations: Approaching the Physical
  Point}}, \href{https://doi.org/10.1103/PhysRevLett.125.232003}{\emph{Phys.
  Rev. Lett.} {\bfseries 125} (2020) 232003}
  [\href{https://arxiv.org/abs/2004.01687}{{\ttfamily 2004.01687}}].

\bibitem{Martin:2009iq}
A.~Martin, W.~Stirling, R.~Thorne and G.~Watt, \emph{{Parton distributions for
  the LHC}}, {\emph{Eur. Phys. J. C} {\bfseries 63} (2009) 189}
  [\href{https://arxiv.org/abs/0901.0002}{{\ttfamily 0901.0002}}].

\bibitem{Bhat:2020ktg}
M.~Bhat, K.~Cichy, M.~Constantinou and A.~Scapellato, \emph{{Flavor nonsinglet
  parton distribution functions from lattice QCD at physical quark masses via
  the pseudodistribution approach}},
  \href{https://doi.org/10.1103/PhysRevD.103.034510}{\emph{Phys. Rev. D}
  {\bfseries 103} (2021) 034510}
  [\href{https://arxiv.org/abs/2005.02102}{{\ttfamily 2005.02102}}].

\bibitem{Bali:2017gfr}
G.~S. Bali et~al., \emph{{Pion distribution amplitude from Euclidean
  correlation functions}}, {\emph{Eur. Phys. J.} {\bfseries C78} (2018) 217}
  [\href{https://arxiv.org/abs/1709.04325}{{\ttfamily 1709.04325}}].

\bibitem{Bali:2018spj}
G.~S. Bali, V.~M. Braun, B.~Gl{\"a}{\ss}le, M.~G{\"o}ckeler, M.~Gruber,
  F.~Hutzler et~al., \emph{{Pion distribution amplitude from Euclidean
  correlation functions: Exploring universality and higher-twist effects}},
  {\emph{Phys. Rev.} {\bfseries D98} (2018) 094507}
  [\href{https://arxiv.org/abs/1807.06671}{{\ttfamily 1807.06671}}].

\bibitem{Ma:2014jga}
Y.-Q. Ma and J.-W. Qiu, \emph{{QCD Factorization and PDFs from Lattice QCD
  Calculation}}, {\emph{Int.J.Mod.Phys.Conf.Ser.} {\bfseries 37} (2015) 0041}
  [\href{https://arxiv.org/abs/1412.2688}{{\ttfamily 1412.2688}}].

\bibitem{Sufian:2019bol}
R.~S. Sufian, J.~Karpie, C.~Egerer, K.~Orginos, J.-W. Qiu and D.~G. Richards,
  \emph{{Pion Valence Quark Distribution from Matrix Element Calculated in
  Lattice QCD}}, {\emph{Phys. Rev.} {\bfseries D99} (2019) 074507}
  [\href{https://arxiv.org/abs/1901.03921}{{\ttfamily 1901.03921}}].

\bibitem{Sufian:2020vzb}
R.~S. Sufian, C.~Egerer, J.~Karpie, R.~G. Edwards, B.~Jo\'o, Y.-Q. Ma et~al.,
  \emph{{Pion Valence Quark Distribution from Current-Current Correlation in
  Lattice QCD}}, \href{https://doi.org/10.1103/PhysRevD.102.054508}{\emph{Phys.
  Rev. D} {\bfseries 102} (2020) 054508}
  [\href{https://arxiv.org/abs/2001.04960}{{\ttfamily 2001.04960}}].

\bibitem{Aicher:2010cb}
M.~Aicher, A.~Sch\"afer and W.~Vogelsang, \emph{{Soft-gluon resummation and the
  valence parton distribution function of the pion}}, {\emph{Phys. Rev. Lett.}
  {\bfseries 105} (2010) 252003}
  [\href{https://arxiv.org/abs/1009.2481}{{\ttfamily 1009.2481}}].

\bibitem{Chang:2014lva}
L.~Chang, C.~Mezrag, H.~Moutarde, C.~D. Roberts, J.~Rodr\'\i{}guez-Quintero and
  P.~C. Tandy, \emph{{Basic features of the pion valence-quark distribution
  function}}, {\emph{Phys. Lett. B} {\bfseries 737} (2014) 23}
  [\href{https://arxiv.org/abs/1406.5450}{{\ttfamily 1406.5450}}].

\bibitem{Boer:2015kxa}
D.~Boer, M.~Buffing and P.~Mulders, \emph{{Operator analysis of p$_{T}$ -widths
  of TMDs}}, {\emph{JHEP} {\bfseries 08} (2015) 053}
  [\href{https://arxiv.org/abs/1503.03760}{{\ttfamily 1503.03760}}].

\bibitem{Briceno:2018lfj}
R.~A. Brice\~no, J.~V. Guerrero, M.~T. Hansen and C.~J. Monahan,
  \emph{{Finite-volume effects due to spatially nonlocal operators}},
  {\emph{Phys. Rev. D} {\bfseries 98} (2018) 014511}
  [\href{https://arxiv.org/abs/1805.01034}{{\ttfamily 1805.01034}}].

\bibitem{Bali:2016lva}
G.~S. Bali, B.~Lang, B.~U. Musch and A.~Sch{\"a}fer, \emph{{Novel quark
  smearing for hadrons with high momenta in lattice QCD}}, {\emph{Phys. Rev.}
  {\bfseries D93} (2016) 094515}
  [\href{https://arxiv.org/abs/1602.05525}{{\ttfamily 1602.05525}}].

\bibitem{Constantinou:2019vyb}
M.~Constantinou, H.~Panagopoulos and G.~Spanoudes, \emph{{One-loop
  renormalization of staple-shaped operators in continuum and lattice
  regularizations}}, {\emph{Phys. Rev.} {\bfseries D99} (2019) 074508}
  [\href{https://arxiv.org/abs/1901.03862}{{\ttfamily 1901.03862}}].

\bibitem{Engelhardt:tobepublished.1}
M.~Engelhardt, J.~R. Green, N.~Hasan, T.~Izubuchi, C.~Kallidonis, S.~Krieg
  et~al.{\emph{, to be published} }.

\bibitem{Engelhardt:tobepublished.2}
M.~Engelhardt, J.~R. Green, S.~Krieg, S.~Meinel, J.~Negele, A.~Pochinsky
  et~al.{\emph{, to be published} }.

\bibitem{Bacchetta:2018lna}
A.~Bacchetta, G.~Bozzi, M.~Radici, M.~Ritzmann and A.~Signori, \emph{{Effect of
  Flavor-Dependent Partonic Transverse Momentum on the Determination of the $W$
  Boson Mass in Hadronic Collisions}}, {\emph{Phys. Lett. B} {\bfseries 788}
  (2019) 542} [\href{https://arxiv.org/abs/1807.02101}{{\ttfamily
  1807.02101}}].

\bibitem{Schindler:2022eva}
S.~T. Schindler, I.~W. Stewart and Y.~Zhao, \emph{{One-loop matching for gluon
  lattice TMDs}},  \href{https://arxiv.org/abs/2205.12369}{{\ttfamily
  2205.12369}}.

\bibitem{Zhang:2020dbb}
{\scshape Lattice Parton} collaboration, Q.-A. Zhang et~al., \emph{{Lattice QCD
  Calculations of Transverse-Momentum-Dependent Soft Function through
  Large-Momentum Effective Theory}}, {\emph{Phys. Rev. Lett.} {\bfseries 125}
  (2020) 192001} [\href{https://arxiv.org/abs/2005.14572}{{\ttfamily
  2005.14572}}].

\bibitem{Li:2021wvl}
Y.~Li et~al., \emph{{Lattice QCD Study of Transverse-Momentum Dependent Soft
  Function}}, \href{https://doi.org/10.1103/PhysRevLett.128.062002}{\emph{Phys.
  Rev. Lett.} {\bfseries 128} (2022) 062002}
  [\href{https://arxiv.org/abs/2106.13027}{{\ttfamily 2106.13027}}].

\bibitem{Ji:2021znw}
X.~Ji and Y.~Liu, \emph{{Computing light-front wave functions without
  light-front quantization: A large-momentum effective theory approach}},
  \href{https://doi.org/10.1103/PhysRevD.105.076014}{\emph{Phys. Rev. D}
  {\bfseries 105} (2022) 076014}
  [\href{https://arxiv.org/abs/2106.05310}{{\ttfamily 2106.05310}}].

\bibitem{Shanahan:2020zxr}
P.~Shanahan, M.~Wagman and Y.~Zhao, \emph{{Collins-Soper kernel for TMD
  evolution from lattice QCD}},
  \href{https://doi.org/10.1103/PhysRevD.102.014511}{\emph{Phys. Rev. D}
  {\bfseries 102} (2020) 014511}
  [\href{https://arxiv.org/abs/2003.06063}{{\ttfamily 2003.06063}}].

\bibitem{Schlemmer:2021aij}
M.~Schlemmer, A.~Vladimirov, C.~Zimmermann, M.~Engelhardt and A.~Sch\"afer,
  \emph{{Determination of the Collins-Soper Kernel from Lattice QCD}},
  \href{https://doi.org/10.1007/JHEP08(2021)004}{\emph{JHEP} {\bfseries 08}
  (2021) 004} [\href{https://arxiv.org/abs/2103.16991}{{\ttfamily
  2103.16991}}].

\bibitem{Shanahan:2021tst}
P.~Shanahan, M.~Wagman and Y.~Zhao, \emph{{Lattice QCD calculation of the
  Collins-Soper kernel from quasi-TMDPDFs}},
  \href{https://doi.org/10.1103/PhysRevD.104.114502}{\emph{Phys. Rev. D}
  {\bfseries 104} (2021) 114502}
  [\href{https://arxiv.org/abs/2107.11930}{{\ttfamily 2107.11930}}].

\bibitem{LPC:2022ibr}
{\scshape LPC} collaboration, M.-H. Chu et~al., \emph{{Nonperturbative
  Determination of Collins-Soper Kernel from Quasi Transverse-Momentum
  Dependent Wave Functions}},
  \href{https://arxiv.org/abs/2204.00200}{{\ttfamily 2204.00200}}.

\bibitem{Metz:2002iz}
A.~Metz, \emph{{Gluon-exchange in spin-dependent fragmentation}}, {\emph{Phys.
  Lett.} {\bfseries B549} (2002) 139}
  [\href{https://arxiv.org/abs/hep-ph/0209054}{{\ttfamily hep-ph/0209054}}].

\bibitem{Kundu:2001pk}
R.~Kundu and A.~Metz, \emph{{Higher twist and transverse momentum dependent
  parton distributions: A Light front Hamiltonian approach}}, {\emph{Phys. Rev.
  D} {\bfseries 65} (2002) 014009}
  [\href{https://arxiv.org/abs/hep-ph/0107073}{{\ttfamily hep-ph/0107073}}].

\bibitem{Goeke:2003az}
K.~Goeke, A.~Metz, P.~Pobylitsa and M.~Polyakov, \emph{{Lorentz invariance
  relations among parton distributions revisited}}, {\emph{Phys. Lett. B}
  {\bfseries 567} (2003) 27}
  [\href{https://arxiv.org/abs/hep-ph/0302028}{{\ttfamily hep-ph/0302028}}].

\bibitem{Meissner:2009ww}
S.~Meissner, A.~Metz and M.~Schlegel, \emph{{Generalized parton correlation
  functions for a spin-1/2 hadron}}, {\emph{JHEP} {\bfseries 0908} (2009) 056}
  [\href{https://arxiv.org/abs/0906.5323}{{\ttfamily 0906.5323}}].

\bibitem{Avakian:2015vha}
H.~Avakian, H.~Matevosyan, B.~Pasquini and P.~Schweitzer, \emph{{Studying the
  information content of TMDs using Monte Carlo generators}}, {\emph{J. Phys.
  G} {\bfseries 42} (2015) 034015}.

\bibitem{Barut:1960zz}
A.~O. Barut and C.~Fronsdal, \emph{{Spin-Orbit Correlations in mu-e and e--e-
  Scattering}}, {\emph{Phys. Rev.} {\bfseries 120} (1960) 1871}.

\bibitem{Burkardt:2003je}
M.~Burkardt and D.~S. Hwang, \emph{{Sivers asymmetry and generalized parton
  distributions in impact parameter space}}, {\emph{Phys. Rev.} {\bfseries D69}
  (2004) 074032} [\href{https://arxiv.org/abs/hep-ph/0309072}{{\ttfamily
  hep-ph/0309072}}].

\bibitem{Ellis:1978ty}
R.~K. Ellis, H.~Georgi, M.~Machacek, H.~D. Politzer and G.~G. Ross,
  \emph{{Perturbation Theory and the Parton Model in QCD}}, {\emph{Nucl. Phys.
  B} {\bfseries 152} (1979) 285}.

\bibitem{tHooft:1973alw}
G.~'t~Hooft, \emph{{A Planar Diagram Theory for Strong Interactions}},
  {\emph{Nucl. Phys. B} {\bfseries 72} (1974) 461}.

\bibitem{Witten:1979kh}
E.~Witten, \emph{{Baryons in the 1/N Expansion}}, {\emph{Nucl. Phys.}
  {\bfseries B160} (1979) 57}.

\bibitem{Witten:1983tx}
E.~Witten, \emph{{Current Algebra, Baryons, and Quark Confinement}},
  {\emph{Nucl. Phys. B} {\bfseries 223} (1983) 433}.

\bibitem{Pobylitsa:2000tt}
P.~V. Pobylitsa and M.~V. Polyakov, \emph{{New positivity bounds on parton
  distributions in multicolored QCD}}, {\emph{Phys. Rev. D} {\bfseries 62}
  (2000) 097502} [\href{https://arxiv.org/abs/hep-ph/0004094}{{\ttfamily
  hep-ph/0004094}}].

\bibitem{Efremov:2004tp}
A.~V. Efremov, K.~Goeke, S.~Menzel, A.~Metz and P.~Schweitzer, \emph{{Sivers
  effect in semi-inclusive DIS and in the Drell-Yan process}}, {\emph{Phys.
  Lett.} {\bfseries B612} (2005) 233}
  [\href{https://arxiv.org/abs/hep-ph/0412353}{{\ttfamily hep-ph/0412353}}].

\bibitem{Efremov:2000ar}
A.~V. Efremov, K.~Goeke and P.~V. Pobylitsa, \emph{{Gluon and quark
  distributions in large $N_c$ QCD: Theory versus phenomenology}}, {\emph{Phys.
  Lett.} {\bfseries B488} (2000) 182}
  [\href{https://arxiv.org/abs/hep-ph/0004196}{{\ttfamily hep-ph/0004196}}].

\bibitem{Brodsky:2006ha}
S.~J. Brodsky and S.~Gardner, \emph{{Evidence for the Absence of Gluon Orbital
  Angular Momentum in the Nucleon}}, {\emph{Phys. Lett.} {\bfseries B643}
  (2006) 22} [\href{https://arxiv.org/abs/hep-ph/0608219}{{\ttfamily
  hep-ph/0608219}}].

\bibitem{Anselmino:2006yq}
M.~Anselmino, U.~D'Alesio, S.~Melis and F.~Murgia, \emph{{Constraints on the
  gluon Sivers distribution via transverse single spin asymmetries at
  mid-rapidity in $p^\uparrow p \to \pi^0 X$ processes at RHIC}}, {\emph{Phys.
  Rev.} {\bfseries D74} (2006) 094011}
  [\href{https://arxiv.org/abs/hep-ph/0608211}{{\ttfamily hep-ph/0608211}}].

\bibitem{Barone:2001sp}
V.~Barone, A.~Drago and P.~G. Ratcliffe, \emph{{Transverse polarisation of
  quarks in hadrons}}, {\emph{Phys. Rept.} {\bfseries 359} (2002) 1}
  [\href{https://arxiv.org/abs/hep-ph/0104283}{{\ttfamily hep-ph/0104283}}].

\bibitem{Karl:1984cz}
G.~Karl and J.~E. Paton, \emph{{Naive Quark Model for an Arbitrary Number of
  Colors}}, {\emph{Phys. Rev.} {\bfseries D30} (1984) 238}.

\bibitem{Efremov:2009ze}
A.~V. Efremov, P.~Schweitzer, O.~V. Teryaev and P.~Zavada, \emph{{Transverse
  momentum dependent distribution functions in a covariant parton model
  approach with quark orbital motion}}, {\emph{Phys. Rev.} {\bfseries D80}
  (2009) 014021} [\href{https://arxiv.org/abs/0903.3490}{{\ttfamily
  0903.3490}}].

\bibitem{Zavada:1996kp}
P.~Zavada, \emph{{The Structure functions and parton momenta distribution in
  the hadron rest system}}, {\emph{Phys. Rev. D} {\bfseries 55} (1997) 4290}
  [\href{https://arxiv.org/abs/hep-ph/9609372}{{\ttfamily hep-ph/9609372}}].

\bibitem{Zavada:2009ska}
P.~Zavada, \emph{{Generalized Cahn effect and parton 3D motion in a covariant
  approach}}, {\emph{Phys. Rev. D} {\bfseries 83} (2011) 014022}
  [\href{https://arxiv.org/abs/0908.2316}{{\ttfamily 0908.2316}}].

\bibitem{Efremov:2010mt}
A.~V. Efremov, P.~Schweitzer, O.~V. Teryaev and P.~Zavada, \emph{{The relation
  between TMDs and PDFs in the covariant parton model approach}}, {\emph{Phys.
  Rev.} {\bfseries D83} (2011) 054025}
  [\href{https://arxiv.org/abs/1012.5296}{{\ttfamily 1012.5296}}].

\bibitem{Bastami:2020rxn}
S.~Bastami, A.~V. Efremov, P.~Schweitzer, O.~V. Teryaev and P.~Zavada,
  \emph{{Structure of the nucleon at leading and subleading twist in the
  covariant parton model}},
  \href{https://doi.org/10.1103/PhysRevD.103.014024}{\emph{Phys. Rev. D}
  {\bfseries 103} (2021) 014024}
  [\href{https://arxiv.org/abs/2011.06203}{{\ttfamily 2011.06203}}].

\bibitem{Aslan:2022wqc}
F.~Aslan, S.~Bastami and P.~Schweitzer, \emph{{Parton model description of
  quark and antiquark correlators and TMDs}},
  \href{https://doi.org/10.1016/j.nuclphysb.2022.115947}{\emph{Nucl. Phys. B}
  {\bfseries 984} (2022) 115947}
  [\href{https://arxiv.org/abs/2206.07273}{{\ttfamily 2206.07273}}].

\bibitem{Aslan:2022kmd}
F.~Aslan, S.~Bastami, A.~Mahabir, A.~Tandogan and P.~Schweitzer,
  \emph{{Quark-model relations among TMDs in the parton model}},
  \href{https://doi.org/10.1103/PhysRevD.106.096010}{\emph{Phys. Rev. D}
  {\bfseries 106} (2022) 096010}
  [\href{https://arxiv.org/abs/2209.02355}{{\ttfamily 2209.02355}}].

\bibitem{Accardi:2009au}
A.~Accardi, A.~Bacchetta, W.~Melnitchouk and M.~Schlegel, \emph{{What can break
  the Wandzura-Wilczek relation?}}, {\emph{JHEP} {\bfseries 11} (2009) 093}
  [\href{https://arxiv.org/abs/0907.2942}{{\ttfamily 0907.2942}}].

\bibitem{Jackson:1989ph}
J.~D. Jackson, G.~G. Ross and R.~G. Roberts, \emph{{Polarized Structure
  Functions in the Parton Model}}, {\emph{Phys. Lett. B} {\bfseries 226} (1989)
  159}.

\bibitem{Roberts:1996ub}
R.~G. Roberts and G.~G. Ross, \emph{{Quark model description of polarized deep
  inelastic scattering and the prediction of $g_2$}}, {\emph{Phys. Lett. B}
  {\bfseries 373} (1996) 235}
  [\href{https://arxiv.org/abs/hep-ph/9601235}{{\ttfamily hep-ph/9601235}}].

\bibitem{Bourrely:2005tp}
C.~Bourrely, J.~Soffer and F.~Buccella, \emph{{The Extension to the transverse
  momentum of the statistical parton distributions}}, {\emph{Mod. Phys. Lett.}
  {\bfseries A21} (2006) 143}
  [\href{https://arxiv.org/abs/hep-ph/0507328}{{\ttfamily hep-ph/0507328}}].

\bibitem{Bourrely:2010ng}
C.~Bourrely, F.~Buccella and J.~Soffer, \emph{{Semiinclusive DIS cross sections
  and spin asymmetries in the quantum statistical parton distributions
  approach}}, {\emph{Phys. Rev.} {\bfseries D83} (2011) 074008}
  [\href{https://arxiv.org/abs/1008.5322}{{\ttfamily 1008.5322}}].

\bibitem{DAlesio:2009cps}
U.~D'Alesio, E.~Leader and F.~Murgia, \emph{{On the importance of Lorentz
  structure in the parton model: Target mass corrections, transverse momentum
  dependence, positivity bounds}},
  \href{https://doi.org/10.1103/PhysRevD.81.036010}{\emph{Phys. Rev. D}
  {\bfseries 81} (2010) 036010}
  [\href{https://arxiv.org/abs/0909.5650}{{\ttfamily 0909.5650}}].

\bibitem{Mirjalili:2022cal}
A.~Mirjalili and S.~Tehrani~Atashbar, \emph{{Nucleon spin structure functions,
  considering target mass correction, and higher twist effects at the NNLO
  accuracy and their transverse momentum dependence}},
  \href{https://doi.org/10.1103/PhysRevD.105.074023}{\emph{Phys. Rev. D}
  {\bfseries 105} (2022) 074023}
  [\href{https://arxiv.org/abs/2203.13904}{{\ttfamily 2203.13904}}].

\bibitem{Blumlein:1996tp}
J.~Bl\"umlein and N.~Kochelev, \emph{{On the twist-2 contributions to polarized
  structure functions and new sum rules}}, {\emph{Phys. Lett. B} {\bfseries
  381} (1996) 296} [\href{https://arxiv.org/abs/hep-ph/9603397}{{\ttfamily
  hep-ph/9603397}}].

\bibitem{Blumlein:1996vs}
J.~Bl\"umlein and N.~Kochelev, \emph{{On the twist -2 and twist - three
  contributions to the spin dependent electroweak structure functions}},
  {\emph{Nucl. Phys. B} {\bfseries 498} (1997) 285}
  [\href{https://arxiv.org/abs/hep-ph/9612318}{{\ttfamily hep-ph/9612318}}].

\bibitem{Soffer:1997zy}
J.~Soffer and O.~V. Teryaev, \emph{{Gluon polarization in transversely
  polarized nucleons and jet spin asymmetries at RHIC}}, {\emph{Phys. Rev. D}
  {\bfseries 56} (1997) R1353}
  [\href{https://arxiv.org/abs/hep-ph/9702352}{{\ttfamily hep-ph/9702352}}].

\bibitem{Blumlein:1998nv}
J.~Bl\"umlein and A.~Tkabladze, \emph{{Target mass corrections for polarized
  structure functions and new sum rules}}, {\emph{Nucl. Phys. B} {\bfseries
  553} (1999) 427} [\href{https://arxiv.org/abs/hep-ph/9812478}{{\ttfamily
  hep-ph/9812478}}].

\bibitem{Avakian:2008dz}
H.~Avakian, A.~V. Efremov, P.~Schweitzer and F.~Yuan, \emph{{Transverse
  momentum dependent distribution function $h_{1T}^\perp$ and the single spin
  asymmetry $A_{UT}^{\sin(3\phi-\phi_S)}$}}, {\emph{Phys. Rev.} {\bfseries D78}
  (2008) 114024} [\href{https://arxiv.org/abs/0805.3355}{{\ttfamily
  0805.3355}}].

\bibitem{Thomas:1982kv}
A.~W. Thomas, \emph{{Chiral Symmetry and the Bag Model: A New Starting Point
  for Nuclear Physics}}, {\emph{Adv. Nucl. Phys.} {\bfseries 13} (1984) 1}.

\bibitem{Signal:2021aum}
A.~I. Signal and F.~G. Cao, \emph{{Transverse momentum and transverse momentum
  distributions in the MIT bag model}},
  \href{https://doi.org/10.1016/j.physletb.2022.136898}{\emph{Phys. Lett. B}
  {\bfseries 826} (2022) 136898}
  [\href{https://arxiv.org/abs/2108.12116}{{\ttfamily 2108.12116}}].

\bibitem{Brodsky:2000ii}
S.~J. Brodsky, D.~S. Hwang, B.-Q. Ma and I.~Schmidt, \emph{{Light cone
  representation of the spin and orbital angular momentum of relativistic
  composite systems}},
  \href{https://doi.org/10.1016/S0550-3213(00)00626-X}{\emph{Nucl. Phys.}
  {\bfseries B593} (2001) 311}
  [\href{https://arxiv.org/abs/hep-th/0003082}{{\ttfamily hep-th/0003082}}].

\bibitem{Ji:2002xn}
X.-d. Ji, J.-P. Ma and F.~Yuan, \emph{{Three quark light cone amplitudes of the
  proton and quark orbital motion dependent observables}}, {\emph{Nucl. Phys.}
  {\bfseries B652} (2003) 383}
  [\href{https://arxiv.org/abs/hep-ph/0210430}{{\ttfamily hep-ph/0210430}}].

\bibitem{Pasquini:2008ax}
B.~Pasquini, S.~Cazzaniga and S.~Boffi, \emph{{Transverse momentum dependent
  parton distributions in a light-cone quark model}}, {\emph{Phys. Rev.}
  {\bfseries D78} (2008) 034025}
  [\href{https://arxiv.org/abs/0806.2298}{{\ttfamily 0806.2298}}].

\bibitem{Pasquini:2009bv}
B.~Pasquini, S.~Boffi and P.~Schweitzer, \emph{{The Spin Structure of the
  Nucleon in Light-Cone Quark Models}}, {\emph{Mod. Phys. Lett.} {\bfseries
  A24} (2009) 2903} [\href{https://arxiv.org/abs/0910.1677}{{\ttfamily
  0910.1677}}].

\bibitem{Miller:2007ae}
G.~A. Miller, \emph{{Densities, Parton Distributions, and Measuring the
  Non-Spherical Shape of the Nucleon}}, {\emph{Phys. Rev.} {\bfseries C76}
  (2007) 065209} [\href{https://arxiv.org/abs/0708.2297}{{\ttfamily
  0708.2297}}].

\bibitem{Burkardt:2007rv}
M.~Burkardt, \emph{{Spin-orbit correlations and single-spin asymmetries}},  in
  \emph{{Workshop on Exclusive Reactions at High Momentum Transfer}}, 9, 2007,
  \href{https://arxiv.org/abs/0709.2966}{{\ttfamily 0709.2966}}.

\bibitem{Lorce:2011dv}
C.~Lorc\'e, B.~Pasquini and M.~Vanderhaeghen, \emph{{Unified framework for
  generalized and transverse-momentum dependent parton distributions within a
  3Q light-cone picture of the nucleon}}, {\emph{JHEP} {\bfseries 05} (2011)
  041} [\href{https://arxiv.org/abs/1102.4704}{{\ttfamily 1102.4704}}].

\bibitem{Pasquini:2011tk}
B.~Pasquini and P.~Schweitzer, \emph{{Naive time-reversal odd phenomena in
  semi-inclusive deep-inelastic scattering from light-cone constituent quark
  models}}, {\emph{Phys. Rev.} {\bfseries D83} (2011) 114044}
  [\href{https://arxiv.org/abs/1103.5977}{{\ttfamily 1103.5977}}].

\bibitem{Pasquini:2014ppa}
B.~Pasquini and P.~Schweitzer, \emph{{Pion transverse momentum dependent parton
  distributions in a light-front constituent approach, and the Boer-Mulders
  effect in the pion-induced Drell-Yan process}}, {\emph{Phys. Rev.} {\bfseries
  D90} (2014) 014050} [\href{https://arxiv.org/abs/1406.2056}{{\ttfamily
  1406.2056}}].

\bibitem{Lorce:2014hxa}
C.~Lorc\'{e}, B.~Pasquini and P.~Schweitzer, \emph{{Unpolarized transverse
  momentum dependent parton distribution functions beyond leading twist in
  quark models}}, {\emph{JHEP} {\bfseries 01} (2015) 103}
  [\href{https://arxiv.org/abs/1411.2550}{{\ttfamily 1411.2550}}].

\bibitem{Lorce:2016ugb}
C.~Lorc\'{e}, B.~Pasquini and P.~Schweitzer, \emph{{Transverse pion structure
  beyond leading twist in constituent models}}, {\emph{Eur. Phys. J.}
  {\bfseries C76} (2016) 415}
  [\href{https://arxiv.org/abs/1605.00815}{{\ttfamily 1605.00815}}].

\bibitem{Kaur:2020vkq}
S.~Kaur, N.~Kumar, J.~Lan, C.~Mondal and H.~Dahiya, \emph{{Tomography of light
  mesons in the light-cone quark model}},
  \href{https://doi.org/10.1103/PhysRevD.102.014021}{\emph{Phys. Rev. D}
  {\bfseries 102} (2020) 014021}
  [\href{https://arxiv.org/abs/2002.01199}{{\ttfamily 2002.01199}}].

\bibitem{Hu:2022ctr}
Z.~Hu, S.~Xu, C.~Mondal, X.~Zhao and J.~P. Vary, \emph{{Transverse momentum
  structure of proton within the basis light-front quantization framework}},
  \href{https://arxiv.org/abs/2205.04714}{{\ttfamily 2205.04714}}.

\bibitem{Jakob:1997wg}
R.~Jakob, P.~J. Mulders and J.~Rodrigues, \emph{{Modeling quark distribution
  and fragmentation functions}}, {\emph{Nucl. Phys.} {\bfseries A626} (1997)
  937} [\href{https://arxiv.org/abs/hep-ph/9704335}{{\ttfamily
  hep-ph/9704335}}].

\bibitem{Gamberg:2003ey}
L.~P. Gamberg, G.~R. Goldstein and K.~A. Oganessyan, \emph{{Novel transversity
  properties in semiinclusive deep inelastic scattering}}, {\emph{Phys. Rev.}
  {\bfseries D67} (2003) 071504}
  [\href{https://arxiv.org/abs/hep-ph/0301018}{{\ttfamily hep-ph/0301018}}].

\bibitem{Bacchetta:2008af}
A.~Bacchetta, F.~Conti and M.~Radici, \emph{{Transverse-momentum distributions
  in a diquark spectator model}}, {\emph{Phys. Rev.} {\bfseries D78} (2008)
  074010} [\href{https://arxiv.org/abs/0807.0323}{{\ttfamily 0807.0323}}].

\bibitem{Lu:2010dt}
Z.~Lu and I.~Schmidt, \emph{{Orbital structure of quarks inside the nucleon in
  the light-cone diquark model}}, {\emph{Phys. Rev.} {\bfseries D82} (2010)
  094005} [\href{https://arxiv.org/abs/1008.2684}{{\ttfamily 1008.2684}}].

\bibitem{Muller:2014tqa}
D.~M{\"u}ller and D.~S. Hwang, \emph{{The concept of phenomenological
  light-front wave functions -- Regge improved diquark model predictions}},
  \href{https://arxiv.org/abs/1407.1655}{{\ttfamily 1407.1655}}.

\bibitem{Liu:2021ype}
X.~Liu, W.~Mao, X.~Wang and B.-Q. Ma, \emph{{Leading and higher twist
  transverse momentum dependent parton distribution functions in the spectator
  model}}, {\emph{Phys. Rev. D} {\bfseries 104} (2021) 094043}
  [\href{https://arxiv.org/abs/2110.14070}{{\ttfamily 2110.14070}}].

\bibitem{Cloet:2007em}
I.~C. Cloet, W.~Bentz and A.~W. Thomas, \emph{{Transversity quark distributions
  in a covariant quark-diquark model}}, {\emph{Phys. Lett.} {\bfseries B659}
  (2008) 214} [\href{https://arxiv.org/abs/0708.3246}{{\ttfamily 0708.3246}}].

\bibitem{Barabanov:2020jvn}
M.~Y. Barabanov et~al., \emph{{Diquark correlations in hadron physics: Origin,
  impact and evidence}},
  \href{https://doi.org/10.1016/j.ppnp.2020.103835}{\emph{Prog. Part. Nucl.
  Phys.} {\bfseries 116} (2021) 103835}
  [\href{https://arxiv.org/abs/2008.07630}{{\ttfamily 2008.07630}}].

\bibitem{Ninomiya:2017ggn}
Y.~Ninomiya, W.~Bentz and I.~C. Clo{\"e}t, \emph{{Transverse-momentum-dependent
  quark distribution functions of spin-one targets: Formalism and covariant
  calculations}}, {\emph{Phys. Rev.} {\bfseries C96} (2017) 045206}
  [\href{https://arxiv.org/abs/1707.03787}{{\ttfamily 1707.03787}}].

\bibitem{Shi:2018zqd}
C.~Shi and I.~C. Clo{\"e}t, \emph{{Intrinsic Transverse Motion of the Pion's
  Valence Quarks}}, {\emph{Phys. Rev. Lett.} {\bfseries 122} (2019) 082301}
  [\href{https://arxiv.org/abs/1806.04799}{{\ttfamily 1806.04799}}].

\bibitem{Brodsky:2006uqa}
S.~J. Brodsky and G.~F. de~Teramond, \emph{{Hadronic spectra and light-front
  wavefunctions in holographic QCD}}, {\emph{Phys. Rev. Lett.} {\bfseries 96}
  (2006) 201601} [\href{https://arxiv.org/abs/hep-ph/0602252}{{\ttfamily
  hep-ph/0602252}}].

\bibitem{Maji:2015vsa}
T.~Maji, C.~Mondal, D.~Chakrabarti and O.~V. Teryaev, \emph{{Relating
  transverse structure of various parton distributions}}, {\emph{JHEP}
  {\bfseries 01} (2016) 165}
  [\href{https://arxiv.org/abs/1506.04560}{{\ttfamily 1506.04560}}].

\bibitem{Maji:2017bcz}
T.~Maji and D.~Chakrabarti, \emph{{Transverse structure of a proton in a
  light-front quark-diquark model}}, {\emph{Phys. Rev.} {\bfseries D95} (2017)
  074009} [\href{https://arxiv.org/abs/1702.04557}{{\ttfamily 1702.04557}}].

\bibitem{Bacchetta:2017vzh}
A.~Bacchetta, S.~Cotogno and B.~Pasquini, \emph{{The transverse structure of
  the pion in momentum space inspired by the AdS/QCD correspondence}},
  {\emph{Phys. Lett.} {\bfseries B771} (2017) 546}
  [\href{https://arxiv.org/abs/1703.07669}{{\ttfamily 1703.07669}}].

\bibitem{Maji:2017zbx}
T.~Maji, D.~Chakrabarti and O.~V. Teryaev, \emph{{Model predictions for
  azimuthal spin asymmetries for HERMES and COMPASS kinematics}}, {\emph{Phys.
  Rev.} {\bfseries D96} (2017) 114023}
  [\href{https://arxiv.org/abs/1711.01746}{{\ttfamily 1711.01746}}].

\bibitem{Maji:2017wwd}
T.~Maji, D.~Chakrabarti and A.~Mukherjee, \emph{{Sivers and cos2$\phi$
  asymmetries in semi-inclusive deep inelastic scattering in light-front
  holographic model}}, {\emph{Phys. Rev.} {\bfseries D97} (2018) 014016}
  [\href{https://arxiv.org/abs/1711.02930}{{\ttfamily 1711.02930}}].

\bibitem{Gurjar:2021dyv}
B.~Gurjar, D.~Chakrabarti, P.~Choudhary, A.~Mukherjee and P.~Talukdar,
  \emph{{Relations between generalized parton distributions and transverse
  momentum dependent parton distributions}}, {\emph{Phys. Rev. D} {\bfseries
  104} (2021) 076028} [\href{https://arxiv.org/abs/2107.02216}{{\ttfamily
  2107.02216}}].

\bibitem{Schweitzer:2012hh}
P.~Schweitzer, M.~Strikman and C.~Weiss, \emph{{Intrinsic transverse momentum
  and parton correlations from dynamical chiral symmetry breaking}},
  {\emph{JHEP} {\bfseries 1301} (2013) 163}
  [\href{https://arxiv.org/abs/1210.1267}{{\ttfamily 1210.1267}}].

\bibitem{Wakamatsu:2009fn}
M.~Wakamatsu, \emph{{Transverse momentum distributions of quarks in the nucleon
  from the Chiral Quark Soliton Model}}, {\emph{Phys. Rev.} {\bfseries D79}
  (2009) 094028} [\href{https://arxiv.org/abs/0903.1886}{{\ttfamily
  0903.1886}}].

\bibitem{Ceccopieri:2018nop}
F.~A. Ceccopieri, A.~Courtoy, S.~Noguera and S.~Scopetta, \emph{{Pion nucleus
  Drell\textendash{}Yan process and parton transverse momentum in the pion}},
  {\emph{Eur. Phys. J. C} {\bfseries 78} (2018) 644}
  [\href{https://arxiv.org/abs/1801.07682}{{\ttfamily 1801.07682}}].

\bibitem{Arrington:2011xs}
J.~Arrington, D.~W. Higinbotham, G.~Rosner and M.~Sargsian, \emph{{Hard probes
  of short-range nucleon-nucleon correlations}}, {\emph{Prog. Part. Nucl.
  Phys.} {\bfseries 67} (2012) 898}
  [\href{https://arxiv.org/abs/1104.1196}{{\ttfamily 1104.1196}}].

\bibitem{Pobylitsa:2002fr}
P.~V. Pobylitsa, \emph{{T odd quark distributions: QCD versus chiral models}},
  \href{https://arxiv.org/abs/hep-ph/0212027}{{\ttfamily hep-ph/0212027}}.

\bibitem{Goldstein:2002vv}
G.~R. Goldstein and L.~Gamberg, \emph{{Transversity and meson
  photoproduction}},  in \emph{{High energy physics. Proceedings, 31st
  International Conference, ICHEP 2002, Amsterdam, Netherlands, July 25-31,
  2002}}, pp.~452--454, 2002,
  \href{https://arxiv.org/abs/hep-ph/0209085}{{\ttfamily hep-ph/0209085}}.

\bibitem{Lu:2006kt}
Z.~Lu and I.~Schmidt, \emph{{Connection between the Sivers function and the
  anomalous magnetic moment}}, {\emph{Phys. Rev.} {\bfseries D75} (2007)
  073008} [\href{https://arxiv.org/abs/hep-ph/0611158}{{\ttfamily
  hep-ph/0611158}}].

\bibitem{Gamberg:2007wm}
L.~P. Gamberg, G.~R. Goldstein and M.~Schlegel, \emph{{Transverse Quark Spin
  Effects and the Flavor Dependence of the Boer-Mulders Function}},
  {\emph{Phys. Rev.} {\bfseries D77} (2008) 094016}
  [\href{https://arxiv.org/abs/0708.0324}{{\ttfamily 0708.0324}}].

\bibitem{Ellis:2008in}
J.~R. Ellis, D.~S. Hwang and A.~Kotzinian, \emph{{Sivers Asymmetries for
  Inclusive Pion and Kaon Production in Deep-Inelastic Scattering}},
  {\emph{Phys. Rev.} {\bfseries D80} (2009) 074033}
  [\href{https://arxiv.org/abs/0808.1567}{{\ttfamily 0808.1567}}].

\bibitem{Yuan:2003wk}
F.~Yuan, \emph{{Sivers function in the MIT bag model}}, {\emph{Phys. Lett.}
  {\bfseries B575} (2003) 45}
  [\href{https://arxiv.org/abs/hep-ph/0308157}{{\ttfamily hep-ph/0308157}}].

\bibitem{Courtoy:2008dn}
A.~Courtoy, S.~Scopetta and V.~Vento, \emph{{Model calculations of the Sivers
  function satisfying the Burkardt Sum Rule}}, {\emph{Phys. Rev.} {\bfseries
  D79} (2009) 074001} [\href{https://arxiv.org/abs/0811.1191}{{\ttfamily
  0811.1191}}].

\bibitem{Courtoy:2009pc}
A.~Courtoy, S.~Scopetta and V.~Vento, \emph{{Analyzing the Boer-Mulders
  function within different quark models}}, {\emph{Phys. Rev.} {\bfseries D80}
  (2009) 074032} [\href{https://arxiv.org/abs/0909.1404}{{\ttfamily
  0909.1404}}].

\bibitem{Pasquini:2010af}
B.~Pasquini and F.~Yuan, \emph{{Sivers and Boer-Mulders functions in Light-Cone
  Quark Models}}, {\emph{Phys. Rev.} {\bfseries D81} (2010) 114013}
  [\href{https://arxiv.org/abs/1001.5398}{{\ttfamily 1001.5398}}].

\bibitem{Lu:2004hu}
Z.~Lu and B.-Q. Ma, \emph{{Non-zero transversity distribution of the pion in a
  quark-spectator-antiquark model}}, {\emph{Phys. Rev. D} {\bfseries 70} (2004)
  094044} [\href{https://arxiv.org/abs/hep-ph/0411043}{{\ttfamily
  hep-ph/0411043}}].

\bibitem{Meissner:2008ay}
S.~Meissner, A.~Metz, M.~Schlegel and K.~Goeke, \emph{{Generalized parton
  correlation functions for a spin-0 hadron}}, {\emph{JHEP} {\bfseries 08}
  (2008) 038} [\href{https://arxiv.org/abs/0805.3165}{{\ttfamily 0805.3165}}].

\bibitem{Gamberg:2009uk}
L.~Gamberg and M.~Schlegel, \emph{{Final state interactions and the transverse
  structure of the pion using non-perturbative eikonal methods}}, {\emph{Phys.
  Lett.} {\bfseries B685} (2010) 95}
  [\href{https://arxiv.org/abs/0911.1964}{{\ttfamily 0911.1964}}].

\bibitem{Courtoy:2008vi}
A.~Courtoy, F.~Fratini, S.~Scopetta and V.~Vento, \emph{{A Quark model analysis
  of the Sivers function}}, {\emph{Phys. Rev.} {\bfseries D78} (2008) 034002}
  [\href{https://arxiv.org/abs/0801.4347}{{\ttfamily 0801.4347}}].

\bibitem{Burkardt:2002ks}
M.~Burkardt, \emph{{Impact parameter dependent parton distributions and
  transverse single spin asymmetries}}, {\emph{Phys. Rev.} {\bfseries D66}
  (2002) 114005} [\href{https://arxiv.org/abs/hep-ph/0209179}{{\ttfamily
  hep-ph/0209179}}].

\bibitem{Burkardt:2003uw}
M.~Burkardt, \emph{{Chromodynamic lensing and transverse single spin
  asymmetries}}, {\emph{Nucl. Phys.} {\bfseries A735} (2004) 185}
  [\href{https://arxiv.org/abs/hep-ph/0302144}{{\ttfamily hep-ph/0302144}}].

\bibitem{Diehl:2005jf}
M.~Diehl and P.~H\"agler, \emph{{Spin densities in the transverse plane and
  generalized transversity distributions}}, {\emph{Eur. Phys. J. C} {\bfseries
  44} (2005) 87} [\href{https://arxiv.org/abs/hep-ph/0504175}{{\ttfamily
  hep-ph/0504175}}].

\bibitem{Burkardt:2005hp}
M.~Burkardt, \emph{{Transverse deformation of parton distributions and
  transversity decomposition of angular momentum}}, {\emph{Phys. Rev. D}
  {\bfseries 72} (2005) 094020}
  [\href{https://arxiv.org/abs/hep-ph/0505189}{{\ttfamily hep-ph/0505189}}].

\bibitem{Burkardt:2015qoa}
M.~Burkardt and B.~Pasquini, \emph{{Modelling the nucleon structure}},
  {\emph{Eur. Phys. J. A} {\bfseries 52} (2016) 161}
  [\href{https://arxiv.org/abs/1510.02567}{{\ttfamily 1510.02567}}].

\bibitem{Pasquini:2019evu}
B.~Pasquini, S.~Rodini and A.~Bacchetta, \emph{{Revisiting model relations
  between T-odd transverse-momentum dependent parton distributions and
  generalized parton distributions}}, {\emph{Phys. Rev. D} {\bfseries 100}
  (2019) 054039} [\href{https://arxiv.org/abs/1907.06960}{{\ttfamily
  1907.06960}}].

\bibitem{Brodsky:2010vs}
S.~J. Brodsky, B.~Pasquini, B.-W. Xiao and F.~Yuan, \emph{{Phases of Augmented
  Hadronic Light-Front Wave Functions}}, {\emph{Phys. Lett.} {\bfseries B687}
  (2010) 327} [\href{https://arxiv.org/abs/1001.1163}{{\ttfamily 1001.1163}}].

\bibitem{Gamberg:2009ma}
L.~Gamberg and M.~Schlegel, \emph{{Final State Interactions and the Transverse
  Structure of the Pion}}, {\emph{Mod. Phys. Lett.} {\bfseries A24} (2009)
  2960} [\href{https://arxiv.org/abs/0912.5399}{{\ttfamily 0912.5399}}].

\bibitem{Ostrovsky:2004pd}
D.~Ostrovsky and E.~Shuryak, \emph{{Instanton-induced azimuthal spin asymmetry
  in deep inelastic scattering}}, {\emph{Phys. Rev. D} {\bfseries 71} (2005)
  014037} [\href{https://arxiv.org/abs/hep-ph/0409253}{{\ttfamily
  hep-ph/0409253}}].

\bibitem{Cherednikov:2006zn}
I.~O. Cherednikov, U.~D'Alesio, N.~I. Kochelev and F.~Murgia, \emph{{Instanton
  contribution to the Sivers function}}, {\emph{Phys. Lett.} {\bfseries B642}
  (2006) 39} [\href{https://arxiv.org/abs/hep-ph/0606238}{{\ttfamily
  hep-ph/0606238}}].

\bibitem{Qian:2011ya}
Y.~Qian and I.~Zahed, \emph{{Single Spin Asymmetry through QCD Instantons}},
  {\emph{Phys. Rev. D} {\bfseries 86} (2012) 014033}
  [\href{https://arxiv.org/abs/1112.4552}{{\ttfamily 1112.4552}}].

\bibitem{Hoyer:2005ev}
P.~Hoyer and M.~Jarvinen, \emph{{Soft rescattering in DIS: Effects of helicity
  flip}}, {\emph{JHEP} {\bfseries 10} (2005) 080}
  [\href{https://arxiv.org/abs/hep-ph/0509058}{{\ttfamily hep-ph/0509058}}].

\bibitem{Drago:2005gz}
A.~Drago, \emph{{Time-reversal odd distribution functions in chiral models with
  vector mesons}}, {\emph{Phys. Rev.} {\bfseries D71} (2005) 057501}
  [\href{https://arxiv.org/abs/hep-ph/0501282}{{\ttfamily hep-ph/0501282}}].

\bibitem{He:2019fzn}
F.~He and P.~Wang, \emph{{Sivers distribution functions of sea quark in proton
  with chiral Lagrangian}}, {\emph{Phys. Rev. D} {\bfseries 100} (2019) 074032}
  [\href{https://arxiv.org/abs/1904.06815}{{\ttfamily 1904.06815}}].

\bibitem{Gluck:1994uf}
M.~Gluck, E.~Reya and A.~Vogt, \emph{{Dynamical parton distributions of the
  proton and small x physics}}, {\emph{Z. Phys. C} {\bfseries 67} (1995) 433}.

\bibitem{Gluck:1998xa}
M.~Gl\"uck, E.~Reya and A.~Vogt, \emph{{Dynamical parton distributions
  revisited}}, {\emph{Eur. Phys. J. C} {\bfseries 5} (1998) 461}
  [\href{https://arxiv.org/abs/hep-ph/9806404}{{\ttfamily hep-ph/9806404}}].

\bibitem{Gluck:1995yr}
M.~Gluck, E.~Reya, M.~Stratmann and W.~Vogelsang, \emph{{Next-to-leading order
  radiative parton model analysis of polarized deep inelastic lepton - nucleon
  scattering}}, {\emph{Phys. Rev. D} {\bfseries 53} (1996) 4775}
  [\href{https://arxiv.org/abs/hep-ph/9508347}{{\ttfamily hep-ph/9508347}}].

\bibitem{Gluck:2000dy}
M.~Gluck, E.~Reya, M.~Stratmann and W.~Vogelsang, \emph{{Models for the
  polarized parton distributions of the nucleon}}, {\emph{Phys. Rev. D}
  {\bfseries 63} (2001) 094005}
  [\href{https://arxiv.org/abs/hep-ph/0011215}{{\ttfamily hep-ph/0011215}}].

\bibitem{Gluck:2007ck}
M.~Gluck, P.~Jimenez-Delgado and E.~Reya, \emph{{Dynamical parton distributions
  of the nucleon and very small-x physics}}, {\emph{Eur. Phys. J. C} {\bfseries
  53} (2008) 355} [\href{https://arxiv.org/abs/0709.0614}{{\ttfamily
  0709.0614}}].

\bibitem{Goeke:2006ef}
K.~Goeke, S.~Meissner, A.~Metz and M.~Schlegel, \emph{{Checking the Burkardt
  sum rule for the Sivers function by model calculations}}, {\emph{Phys. Lett.}
  {\bfseries B637} (2006) 241}
  [\href{https://arxiv.org/abs/hep-ph/0601133}{{\ttfamily hep-ph/0601133}}].

\bibitem{Burkardt:2004ur}
M.~Burkardt, \emph{{Sivers mechanism for gluons}}, {\emph{Phys. Rev.}
  {\bfseries D69} (2004) 091501}
  [\href{https://arxiv.org/abs/hep-ph/0402014}{{\ttfamily hep-ph/0402014}}].

\bibitem{Lu:2016vqu}
Z.~Lu and B.-Q. Ma, \emph{{Gluon Sivers function in a light-cone spectator
  model}}, {\emph{Phys. Rev. D} {\bfseries 94} (2016) 094022}
  [\href{https://arxiv.org/abs/1611.00125}{{\ttfamily 1611.00125}}].

\bibitem{Rodrigues:2001PhD}
J.~Rodrigues, \emph{{Modelling quark and gluon correlation functions}}, Ph.D.
  thesis, Vrije Univ. Amsterdam, Amsterdam, 2001.

\bibitem{Bacchetta:2020vty}
A.~Bacchetta, F.~G. Celiberto, M.~Radici and P.~Taels,
  \emph{{Transverse-momentum-dependent gluon distribution functions in a
  spectator model}}, {\emph{Eur. Phys. J. C} {\bfseries 80} (2020) 733}
  [\href{https://arxiv.org/abs/2005.02288}{{\ttfamily 2005.02288}}].

\bibitem{Bacchetta:2021lvw}
A.~Bacchetta, F.~G. Celiberto and M.~Radici, \emph{{Toward twist-2 $T$-odd
  transverse-momentum-dependent gluon distributions: the $f$-type Sivers
  function}},  in \emph{{European Physical Society Conference on High Energy
  Physics 2021}}, 11, 2021, \href{https://arxiv.org/abs/2111.01686}{{\ttfamily
  2111.01686}}.

\bibitem{Lyubovitskij:2020xqj}
V.~E. Lyubovitskij and I.~Schmidt, \emph{{Gluon parton densities in soft-wall
  AdS/QCD}}, \href{https://doi.org/10.1103/PhysRevD.103.094017}{\emph{Phys.
  Rev. D} {\bfseries 103} (2021) 094017}
  [\href{https://arxiv.org/abs/2012.01334}{{\ttfamily 2012.01334}}].

\bibitem{Bacchetta:2001di}
A.~Bacchetta, R.~Kundu, A.~Metz and P.~J. Mulders, \emph{{The Collins
  fragmentation function: A Simple model calculation}}, {\emph{Phys. Lett.}
  {\bfseries B506} (2001) 155}
  [\href{https://arxiv.org/abs/hep-ph/0102278}{{\ttfamily hep-ph/0102278}}].

\bibitem{Collins:1994ax}
J.~C. Collins and G.~A. Ladinsky, \emph{{On $\pi$-$\pi$ correlations in
  polarized quark fragmentation using the linear sigma model}},
  \href{https://arxiv.org/abs/hep-ph/9411444}{{\ttfamily hep-ph/9411444}}.

\bibitem{Bacchetta:2002tk}
A.~Bacchetta, R.~Kundu, A.~Metz and P.~J. Mulders, \emph{{Estimate of the
  Collins fragmentation function in a chiral invariant approach}}, {\emph{Phys.
  Rev.} {\bfseries D65} (2002) 094021}
  [\href{https://arxiv.org/abs/hep-ph/0201091}{{\ttfamily hep-ph/0201091}}].

\bibitem{Manohar:1983md}
A.~Manohar and H.~Georgi, \emph{{Chiral Quarks and the Nonrelativistic Quark
  Model}}, {\emph{Nucl. Phys.} {\bfseries B234} (1984) 189}.

\bibitem{Ito:2009zc}
T.~Ito, W.~Bentz, I.~C. Cloet, A.~W. Thomas and K.~Yazaki, \emph{{The NJL-jet
  model for quark fragmentation functions}}, {\emph{Phys. Rev.} {\bfseries D80}
  (2009) 074008} [\href{https://arxiv.org/abs/0906.5362}{{\ttfamily
  0906.5362}}].

\bibitem{Matevosyan:2010hh}
H.~H. Matevosyan, A.~W. Thomas and W.~Bentz, \emph{{Calculating Kaon
  Fragmentation Functions from NJL-Jet Model}}, {\emph{Phys. Rev.} {\bfseries
  D83} (2011) 074003} [\href{https://arxiv.org/abs/1011.1052}{{\ttfamily
  1011.1052}}].

\bibitem{Matevosyan:2011ey}
H.~H. Matevosyan, A.~W. Thomas and W.~Bentz, \emph{{Monte Carlo Simulations of
  Hadronic Fragmentation Functions using NJL-Jet Model}}, {\emph{Phys. Rev.}
  {\bfseries D83} (2011) 114010}
  [\href{https://arxiv.org/abs/1103.3085}{{\ttfamily 1103.3085}}].

\bibitem{Nam:2011hg}
S.-i. Nam and C.-W. Kao, \emph{{Fragmentation functions and parton distribution
  functions for the pion with the nonlocal interactions}}, {\emph{Phys. Rev.}
  {\bfseries D85} (2012) 034023}
  [\href{https://arxiv.org/abs/1111.4444}{{\ttfamily 1111.4444}}].

\bibitem{Nam:2012af}
S.-i. Nam and C.-W. Kao, \emph{{Fragmentation and quark distribution functions
  for the pion and kaon with explicit flavor-SU(3)-symmetry breaking}},
  {\emph{Phys. Rev.} {\bfseries D85} (2012) 094023}
  [\href{https://arxiv.org/abs/1202.3281}{{\ttfamily 1202.3281}}].

\bibitem{Yang:2013cza}
D.~J. Yang, F.~J. Jiang, C.~W. Kao and S.~i. Nam, \emph{{Quark-jet contribution
  to the fragmentation functions for the pion and kaon with the nonlocal
  interactions}}, {\emph{Phys. Rev.} {\bfseries D87} (2013) 094007}
  [\href{https://arxiv.org/abs/1304.0525}{{\ttfamily 1304.0525}}].

\bibitem{Bacchetta:2007wc}
A.~Bacchetta, L.~P. Gamberg, G.~R. Goldstein and A.~Mukherjee, \emph{{Collins
  fragmentation function for pions and kaons in a spectator model}},
  {\emph{Phys. Lett.} {\bfseries B659} (2008) 234}
  [\href{https://arxiv.org/abs/0707.3372}{{\ttfamily 0707.3372}}].

\bibitem{Meissner:2010cc}
S.~Meissner, A.~Metz and D.~Pitonyak, \emph{{Momentum sum rules for
  fragmentation functions}}, {\emph{Phys. Lett.} {\bfseries B690} (2010) 296}
  [\href{https://arxiv.org/abs/1002.4393}{{\ttfamily 1002.4393}}].

\bibitem{Bacchetta:2003xn}
A.~Bacchetta, A.~Metz and J.-J. Yang, \emph{{Collins fragmentation function
  from gluon rescattering}}, {\emph{Phys. Lett.} {\bfseries B574} (2003) 225}
  [\href{https://arxiv.org/abs/hep-ph/0307282}{{\ttfamily hep-ph/0307282}}].

\bibitem{Amrath:2005gv}
D.~Amrath, A.~Bacchetta and A.~Metz, \emph{{Reviewing model calculations of the
  Collins fragmentation function}}, {\emph{Phys. Rev.} {\bfseries D71} (2005)
  114018} [\href{https://arxiv.org/abs/hep-ph/0504124}{{\ttfamily
  hep-ph/0504124}}].

\bibitem{Matevosyan:2012ga}
H.~H. Matevosyan, A.~W. Thomas and W.~Bentz, \emph{{Collins Fragmentation
  Function within NJL-jet Model}}, {\emph{Phys. Rev.} {\bfseries D86} (2012)
  034025} [\href{https://arxiv.org/abs/1205.5813}{{\ttfamily 1205.5813}}].

\bibitem{Wang:2018wqo}
X.~Wang, Y.~Yang and Z.~Lu, \emph{{Double Collins effect in $e^+ e^-\to\Lambda
  \bar\Lambda X$ process in a diquark spectator model}}, {\emph{Phys. Rev. D}
  {\bfseries 97} (2018) 114015}
  [\href{https://arxiv.org/abs/1802.01843}{{\ttfamily 1802.01843}}].

\bibitem{Field:1977fa}
R.~D. Field and R.~P. Feynman, \emph{{A Parametrization of the Properties of
  Quark Jets}}, {\emph{Nucl. Phys.} {\bfseries B136} (1978) 1}.

\bibitem{Hoyer:1979ta}
P.~Hoyer, P.~Osland, H.~G. Sander, T.~F. Walsh and P.~M. Zerwas, \emph{{Quantum
  Chromodynamics and Jets in $e^+e^-$}}, {\emph{Nucl. Phys.} {\bfseries B161}
  (1979) 349}.

\bibitem{Ali:1979em}
A.~Ali, E.~Pietarinen, G.~Kramer and J.~Willrodt, \emph{{A QCD Analysis of the
  High-Energy $e^+e^-$ Data from PETRA}}, {\emph{Phys. Lett.} {\bfseries 93B}
  (1980) 155}.

\bibitem{Artru:1974hr}
X.~Artru and G.~Mennessier, \emph{{String model and multiproduction}},
  {\emph{Nucl. Phys.} {\bfseries B70} (1974) 93}.

\bibitem{Bowler:1981sb}
M.~G. Bowler, \emph{{$e^+e^-$ Production of Heavy Quarks in the String Model}},
  {\emph{Z. Phys.} {\bfseries C11} (1981) 169}.

\bibitem{Andersson:1983jt}
B.~Andersson, G.~Gustafson and B.~Soderberg, \emph{{A General Model for Jet
  Fragmentation}}, {\emph{Z. Phys.} {\bfseries C20} (1983) 317}.

\bibitem{Andersson:1983ia}
B.~Andersson, G.~Gustafson, G.~Ingelman and T.~Sjostrand, \emph{{Parton
  Fragmentation and String Dynamics}}, {\emph{Phys. Rept.} {\bfseries 97}
  (1983) 31}.

\bibitem{Artru:1995bh}
X.~Artru, J.~Czyzewski and H.~Yabuki, \emph{{Single spin asymmetry in inclusive
  pion production, Collins effect and the string model}}, {\emph{Z. Phys.}
  {\bfseries C73} (1997) 527}
  [\href{https://arxiv.org/abs/hep-ph/9508239}{{\ttfamily hep-ph/9508239}}].

\bibitem{Czyzewski:1996ih}
J.~Czyzewski, \emph{{Single spin asymmetry of vector meson production in the
  string model}}, {\emph{Acta Phys. Polon.} {\bfseries 27} (1996) 1759}
  [\href{https://arxiv.org/abs/hep-ph/9606390}{{\ttfamily hep-ph/9606390}}].

\bibitem{Artru:2010st}
X.~Artru, \emph{{Recursive fragmentation model with quark spin. Application to
  quark polarimetry}},  \href{https://arxiv.org/abs/1001.1061}{{\ttfamily
  1001.1061}}.

\bibitem{Kerbizi:2018qpp}
A.~Kerbizi, X.~Artru, Z.~Belghobsi, F.~Bradamante and A.~Martin,
  \emph{{Recursive model for the fragmentation of polarized quarks}},
  {\emph{Phys. Rev. D} {\bfseries 97} (2018) 074010}
  [\href{https://arxiv.org/abs/1802.00962}{{\ttfamily 1802.00962}}].

\bibitem{Kerbizi:2019ubp}
A.~Kerbizi, X.~Artru, Z.~Belghobsi and A.~Martin, \emph{{Simplified recursive
  $^3P_0$ model for the fragmentation of polarized quarks}}, {\emph{Phys. Rev.
  D} {\bfseries 100} (2019) 014003}
  [\href{https://arxiv.org/abs/1903.01736}{{\ttfamily 1903.01736}}].

\bibitem{Matevosyan:2011vj}
H.~H. Matevosyan, W.~Bentz, I.~C. Cloet and A.~W. Thomas, \emph{{Transverse
  Momentum Dependent Fragmentation and Quark Distribution Functions from the
  NJL-jet Model}}, {\emph{Phys. Rev.} {\bfseries D85} (2012) 014021}
  [\href{https://arxiv.org/abs/1111.1740}{{\ttfamily 1111.1740}}].

\bibitem{Bentz:2016rav}
W.~Bentz, A.~Kotzinian, H.~H. Matevosyan, Y.~Ninomiya, A.~W. Thomas and
  K.~Yazaki, \emph{{Quark-Jet model for transverse momentum dependent
  fragmentation functions}}, {\emph{Phys. Rev.} {\bfseries D94} (2016) 034004}
  [\href{https://arxiv.org/abs/1603.08333}{{\ttfamily 1603.08333}}].

\bibitem{Schafer:1999kn}
A.~Sch\"afer and O.~V. Teryaev, \emph{{Sum rules for the T - odd fragmentation
  functions}}, {\emph{Phys. Rev.} {\bfseries D61} (2000) 077903}
  [\href{https://arxiv.org/abs/hep-ph/9908412}{{\ttfamily hep-ph/9908412}}].

\bibitem{Collins_Metz_unpublished}
J.~Collins and A.~Metz, \emph{{unpublished}}, .

\bibitem{Gamberg:2008yt}
L.~P. Gamberg, A.~Mukherjee and P.~J. Mulders, \emph{{Spectral analysis of
  gluonic pole matrix elements for fragmentation}}, {\emph{Phys. Rev. D}
  {\bfseries 77} (2008) 114026}
  [\href{https://arxiv.org/abs/0803.2632}{{\ttfamily 0803.2632}}].

\bibitem{Meissner:2008yf}
S.~Meissner and A.~Metz, \emph{{Partonic pole matrix elements for
  fragmentation}}, {\emph{Phys. Rev. Lett.} {\bfseries 102} (2009) 172003}
  [\href{https://arxiv.org/abs/0812.3783}{{\ttfamily 0812.3783}}].

\bibitem{Gamberg:2010uw}
L.~P. Gamberg, A.~Mukherjee and P.~J. Mulders, \emph{{A model independent
  analysis of gluonic pole matrix elements and universality of TMD
  fragmentation functions}}, {\emph{Phys. Rev. D} {\bfseries 83} (2011) 071503}
  [\href{https://arxiv.org/abs/1010.4556}{{\ttfamily 1010.4556}}].

\bibitem{Belitsky:1997ay}
A.~V. Belitsky, \emph{{Leading order analysis of twist-3 space- and time-like
  cut vertices in QCD}}, {\emph{Int. J. Mod. Phys. A} {\bfseries 32} (2017)
  1730018} [\href{https://arxiv.org/abs/hep-ph/9703432}{{\ttfamily
  hep-ph/9703432}}].

\bibitem{Candido:2020yat}
A.~Candido, S.~Forte and F.~Hekhorn, \emph{{Can $ \overline{\mathrm{MS}} $
  parton distributions be negative?}}, {\emph{JHEP} {\bfseries 11} (2020) 129}
  [\href{https://arxiv.org/abs/2006.07377}{{\ttfamily 2006.07377}}].

\bibitem{Lorce:2015lna}
C.~Lorc\'e, \emph{{The light-front gauge-invariant energy-momentum tensor}},
  {\emph{JHEP} {\bfseries 08} (2015) 045}
  [\href{https://arxiv.org/abs/1502.06656}{{\ttfamily 1502.06656}}].

\bibitem{Zhou:2015lxa}
J.~Zhou, \emph{{Note on the scale dependence of the Burkardt sum rule}},
  {\emph{Phys. Rev. D} {\bfseries 92} (2015) 074016}
  [\href{https://arxiv.org/abs/1507.02819}{{\ttfamily 1507.02819}}].

\bibitem{Accardi:2019luo}
A.~Accardi and A.~Signori, \emph{{Quark fragmentation as a probe of dynamical
  mass generation}}, {\emph{Phys. Lett. B} {\bfseries 798} (2019) 134993}
  [\href{https://arxiv.org/abs/1903.04458}{{\ttfamily 1903.04458}}].

\bibitem{Aslan:2022zkz}
F.~Aslan, L.~Gamberg, J.~O. Gonzalez-Hernandez, T.~Rainaldi and T.~C. Rogers,
  \emph{{Basics of factorization in a scalar Yukawa field theory}},
  \href{https://arxiv.org/abs/2212.00757}{{\ttfamily 2212.00757}}.

\bibitem{Signal:1996ct}
A.~I. Signal, \emph{{Calculations of higher twist distribution functions in the
  MIT bag model}}, {\emph{Nucl. Phys. B} {\bfseries 497} (1997) 415}
  [\href{https://arxiv.org/abs/hep-ph/9610480}{{\ttfamily hep-ph/9610480}}].

\bibitem{Lorce:2011zta}
C.~Lorc\'e and B.~Pasquini, \emph{{On the Origin of Model Relations among
  Transverse-Momentum Dependent Parton Distributions}}, {\emph{Phys. Rev. D}
  {\bfseries 84} (2011) 034039}
  [\href{https://arxiv.org/abs/1104.5651}{{\ttfamily 1104.5651}}].

\bibitem{Liu:2014zla}
T.~Liu and B.-Q. Ma, \emph{{Quark angular momentum in a spectator model}},
  {\emph{Phys. Lett. B} {\bfseries 741} (2015) 256}
  [\href{https://arxiv.org/abs/1501.00062}{{\ttfamily 1501.00062}}].

\bibitem{Lorce:2011kn}
C.~Lorc\'e and B.~Pasquini, \emph{{Pretzelosity TMD and Quark Orbital Angular
  Momentum}}, {\emph{Phys. Lett. B} {\bfseries 710} (2012) 486}
  [\href{https://arxiv.org/abs/1111.6069}{{\ttfamily 1111.6069}}].

\bibitem{Gribov:1984tu}
L.~Gribov, E.~Levin and M.~Ryskin, \emph{{Semihard Processes in QCD}},
  {\emph{Phys.Rept.} {\bfseries 100} (1983) 1}.

\bibitem{Mueller:1985wy}
A.~H. Mueller and J.-W. Qiu, \emph{{Gluon Recombination and Shadowing at Small
  Values of x}}, {\emph{Nucl.Phys.} {\bfseries B268} (1986) 427}.

\bibitem{McLerran:1993ni}
L.~D. McLerran and R.~Venugopalan, \emph{{Computing quark and gluon
  distribution functions for very large nuclei}}, {\emph{Phys.Rev.} {\bfseries
  D49} (1994) 2233} [\href{https://arxiv.org/abs/hep-ph/9309289}{{\ttfamily
  hep-ph/9309289}}].

\bibitem{McLerran:1993ka}
L.~D. McLerran and R.~Venugopalan, \emph{{Gluon distribution functions for very
  large nuclei at small transverse momentum}}, {\emph{Phys. Rev.} {\bfseries
  D49} (1994) 3352} [\href{https://arxiv.org/abs/hep-ph/9311205}{{\ttfamily
  hep-ph/9311205}}].

\bibitem{Iancu:2003xm}
E.~Iancu and R.~Venugopalan, \emph{{The Color glass condensate and high-energy
  scattering in QCD}},  \href{https://arxiv.org/abs/hep-ph/0303204}{{\ttfamily
  hep-ph/0303204}}.

\bibitem{Kovchegov:2012mbw}
Y.~V. Kovchegov and E.~Levin, \emph{{Quantum chromodynamics at high energy}},
  {\emph{Camb. Monogr. Part. Phys. Nucl. Phys. Cosmol.} {\bfseries 33} (2012)
  1}.

\bibitem{Kuraev:1977fs}
E.~A. Kuraev, L.~N. Lipatov and V.~S. Fadin, \emph{{The Pomeranchuk Singularity
  in Nonabelian Gauge Theories}}, {\emph{Sov. Phys. JETP} {\bfseries 45} (1977)
  199}.

\bibitem{Balitsky:1978ic}
I.~I. Balitsky and L.~N. Lipatov, \emph{{The Pomeranchuk Singularity in Quantum
  Chromodynamics}}, {\emph{Sov. J. Nucl. Phys.} {\bfseries 28} (1978) 822}.

\bibitem{Fadin:1998py}
V.~S. Fadin and L.~N. Lipatov, \emph{{BFKL pomeron in the next-to-leading
  approximation}}, {\emph{Phys. Lett. B} {\bfseries 429} (1998) 127}
  [\href{https://arxiv.org/abs/hep-ph/9802290}{{\ttfamily hep-ph/9802290}}].

\bibitem{Ciafaloni:1998gs}
M.~Ciafaloni and G.~Camici, \emph{{Energy scale(s) and next-to-leading BFKL
  equation}}, {\emph{Phys. Lett. B} {\bfseries 430} (1998) 349}
  [\href{https://arxiv.org/abs/hep-ph/9803389}{{\ttfamily hep-ph/9803389}}].

\bibitem{Ciafaloni:2003rd}
M.~Ciafaloni, D.~Colferai, G.~P. Salam and A.~M. Stasto, \emph{{Renormalization
  group improved small x Green's function}}, {\emph{Phys. Rev. D} {\bfseries
  68} (2003) 114003} [\href{https://arxiv.org/abs/hep-ph/0307188}{{\ttfamily
  hep-ph/0307188}}].

\bibitem{Gelis:2010nm}
F.~Gelis, E.~Iancu, J.~Jalilian-Marian and R.~Venugopalan, \emph{{The Color
  Glass Condensate}}, {\emph{Ann. Rev. Nucl. Part. Sci.} {\bfseries 60} (2010)
  463} [\href{https://arxiv.org/abs/1002.0333}{{\ttfamily 1002.0333}}].

\bibitem{Blaizot:2016qgz}
J.-P. Blaizot, \emph{{High gluon densities in heavy ion collisions}},
  {\emph{Rept. Prog. Phys.} {\bfseries 80} (2017) 032301}
  [\href{https://arxiv.org/abs/1607.04448}{{\ttfamily 1607.04448}}].

\bibitem{Susskind:1967rg}
L.~Susskind, \emph{{Model of selfinduced strong interactions}}, {\emph{Phys.
  Rev.} {\bfseries 165} (1968) 1535}.

\bibitem{Bardakci:1968zqb}
K.~Bardakci and M.~B. Halpern, \emph{{Theories at infinite momentum}},
  {\emph{Phys. Rev.} {\bfseries 176} (1968) 1686}.

\bibitem{Jeon:2004rk}
S.~Jeon and R.~Venugopalan, \emph{{Random walks of partons in SU($N_c$) and
  classical representations of color charges in QCD at small x}}, {\emph{Phys.
  Rev. D} {\bfseries 70} (2004) 105012}
  [\href{https://arxiv.org/abs/hep-ph/0406169}{{\ttfamily hep-ph/0406169}}].

\bibitem{Jalilian-Marian:2000pwi}
J.~Jalilian-Marian, S.~Jeon and R.~Venugopalan, \emph{{Wong's equations and the
  small x effective action in QCD}}, {\emph{Phys. Rev. D} {\bfseries 63} (2001)
  036004} [\href{https://arxiv.org/abs/hep-ph/0003070}{{\ttfamily
  hep-ph/0003070}}].

\bibitem{McLerran:1994vd}
L.~D. McLerran and R.~Venugopalan, \emph{{Green's functions in the color field
  of a large nucleus}}, {\emph{Phys. Rev.} {\bfseries D50} (1994) 2225}
  [\href{https://arxiv.org/abs/hep-ph/9402335}{{\ttfamily hep-ph/9402335}}].

\bibitem{Balitsky:1995ub}
I.~Balitsky, \emph{{Operator expansion for high-energy scattering}},
  {\emph{Nucl. Phys.} {\bfseries B463} (1996) 99}
  [\href{https://arxiv.org/abs/hep-ph/9509348}{{\ttfamily hep-ph/9509348}}].

\bibitem{McLerran:1998nk}
L.~D. McLerran and R.~Venugopalan, \emph{{Fock space distributions, structure
  functions, higher twists and small x}}, {\emph{Phys. Rev.} {\bfseries D59}
  (1999) 094002} [\href{https://arxiv.org/abs/hep-ph/9809427}{{\ttfamily
  hep-ph/9809427}}].

\bibitem{Ayala:1995hx}
A.~Ayala, J.~Jalilian-Marian, L.~D. McLerran and R.~Venugopalan, \emph{{Quantum
  corrections to the Weizsacker-Williams gluon distribution function at small
  x}}, {\emph{Phys. Rev.} {\bfseries D53} (1996) 458}
  [\href{https://arxiv.org/abs/hep-ph/9508302}{{\ttfamily hep-ph/9508302}}].

\bibitem{Balitsky:2001mr}
I.~I. Balitsky and A.~V. Belitsky, \emph{{Nonlinear evolution in high density
  QCD}}, {\emph{Nucl. Phys.} {\bfseries B629} (2002) 290}
  [\href{https://arxiv.org/abs/hep-ph/0110158}{{\ttfamily hep-ph/0110158}}].

\bibitem{Caron-Huot:2013fea}
S.~Caron-Huot, \emph{{When does the gluon reggeize?}}, {\emph{JHEP} {\bfseries
  05} (2015) 093} [\href{https://arxiv.org/abs/1309.6521}{{\ttfamily
  1309.6521}}].

\bibitem{Bondarenko:2017vfc}
S.~Bondarenko, L.~Lipatov, S.~Pozdnyakov and A.~Prygarin, \emph{{One loop
  light-cone QCD, effective action for reggeized gluons and QCD RFT calculus}},
  {\emph{Eur. Phys. J.} {\bfseries C77} (2017) 630}
  [\href{https://arxiv.org/abs/1708.05183}{{\ttfamily 1708.05183}}].

\bibitem{Ayala:2017rmh}
A.~Ayala, M.~Hentschinski, J.~Jalilian-Marian and M.~E. Tejeda-Yeomans,
  \emph{{Spinor helicity methods in high-energy factorization: efficient
  momentum-space calculations in the Color Glass Condensate formalism}},
  {\emph{Nucl. Phys.} {\bfseries B920} (2017) 232}
  [\href{https://arxiv.org/abs/1701.07143}{{\ttfamily 1701.07143}}].

\bibitem{Hentschinski:2018rrf}
M.~Hentschinski, \emph{{Color glass condensate formalism, Balitsky-JIMWLK
  evolution, and Lipatov's high energy effective action}}, {\emph{Phys. Rev.}
  {\bfseries D97} (2018) 114027}
  [\href{https://arxiv.org/abs/1802.06755}{{\ttfamily 1802.06755}}].

\bibitem{Bondarenko:2018kqs}
S.~Bondarenko and S.~Pozdnyakov, \emph{{S-matrix and productions amplitudes in
  high energy QCD}}, {\emph{Phys. Lett.} {\bfseries B783} (2018) 207}
  [\href{https://arxiv.org/abs/1803.04131}{{\ttfamily 1803.04131}}].

\bibitem{Bondarenko:2018eid}
S.~Bondarenko and S.~Pozdnyakov, \emph{{On correlators of Reggeon fields and
  operators of Wilson lines in high energy QCD}}, {\emph{Int. J. Mod. Phys.}
  {\bfseries A33} (2018) 1850204}
  [\href{https://arxiv.org/abs/1806.02563}{{\ttfamily 1806.02563}}].

\bibitem{Lipatov:1996ts}
L.~N. Lipatov, \emph{{Small x physics in perturbative QCD}}, {\emph{Phys.
  Rept.} {\bfseries 286} (1997) 131}
  [\href{https://arxiv.org/abs/hep-ph/9610276}{{\ttfamily hep-ph/9610276}}].

\bibitem{Hatta:2005as}
Y.~Hatta, E.~Iancu, K.~Itakura and L.~McLerran, \emph{{Odderon in the color
  glass condensate}}, {\emph{Nucl. Phys.} {\bfseries A760} (2005) 172}
  [\href{https://arxiv.org/abs/hep-ph/0501171}{{\ttfamily hep-ph/0501171}}].

\bibitem{Jeon:2005cf}
S.~Jeon and R.~Venugopalan, \emph{{A Classical Odderon in QCD at high
  energies}}, {\emph{Phys. Rev. D} {\bfseries 71} (2005) 125003}
  [\href{https://arxiv.org/abs/hep-ph/0503219}{{\ttfamily hep-ph/0503219}}].

\bibitem{Brower:2006ea}
R.~C. Brower, J.~Polchinski, M.~J. Strassler and C.-I. Tan, \emph{{The Pomeron
  and gauge/string duality}}, {\emph{JHEP} {\bfseries 12} (2007) 005}
  [\href{https://arxiv.org/abs/hep-th/0603115}{{\ttfamily hep-th/0603115}}].

\bibitem{Mueller:1989st}
A.~H. Mueller, \emph{{Small x Behavior and Parton Saturation: A QCD Model}},
  {\emph{Nucl. Phys. B} {\bfseries 335} (1990) 115}.

\bibitem{Frankfurt:2011cs}
L.~Frankfurt, V.~Guzey and M.~Strikman, \emph{{Leading Twist Nuclear Shadowing
  Phenomena in Hard Processes with Nuclei}}, {\emph{Phys. Rept.} {\bfseries
  512} (2012) 255} [\href{https://arxiv.org/abs/1106.2091}{{\ttfamily
  1106.2091}}].

\bibitem{Kovchegov:1998bi}
Y.~V. Kovchegov and A.~H. Mueller, \emph{{Gluon production in current nucleus
  and nucleon - nucleus collisions in a quasiclassical approximation}},
  {\emph{Nucl.Phys.} {\bfseries B529} (1998) 451}
  [\href{https://arxiv.org/abs/hep-ph/9802440}{{\ttfamily hep-ph/9802440}}].

\bibitem{Iancu:2002tr}
E.~Iancu, K.~Itakura and L.~McLerran, \emph{{Geometric scaling above the
  saturation scale}}, {\emph{Nucl. Phys. A} {\bfseries 708} (2002) 327}
  [\href{https://arxiv.org/abs/hep-ph/0203137}{{\ttfamily hep-ph/0203137}}].

\bibitem{Stasto:2000er}
A.~M. Stasto, K.~J. Golec-Biernat and J.~Kwiecinski, \emph{{Geometric scaling
  for the total gamma* p cross-section in the low x region}}, {\emph{Phys. Rev.
  Lett.} {\bfseries 86} (2001) 596}
  [\href{https://arxiv.org/abs/hep-ph/0007192}{{\ttfamily hep-ph/0007192}}].

\bibitem{JalilianMarian:1997gr}
J.~Jalilian-Marian, A.~Kovner, A.~Leonidov and H.~Weigert, \emph{{The Wilson
  renormalization group for low x physics: Towards the high density regime}},
  {\emph{Phys. Rev.} {\bfseries D59} (1998) 014014}
  [\href{https://arxiv.org/abs/hep-ph/9706377}{{\ttfamily hep-ph/9706377}}].

\bibitem{JalilianMarian:1997dw}
J.~Jalilian-Marian, A.~Kovner and H.~Weigert, \emph{{The Wilson renormalization
  group for low x physics: Gluon evolution at finite parton density}},
  {\emph{Phys. Rev.} {\bfseries D59} (1998) 014015}
  [\href{https://arxiv.org/abs/hep-ph/9709432}{{\ttfamily hep-ph/9709432}}].

\bibitem{Iancu:2000hn}
E.~Iancu, A.~Leonidov and L.~D. McLerran, \emph{{Nonlinear gluon evolution in
  the color glass condensate. 1.}}, {\emph{Nucl. Phys.} {\bfseries A692} (2001)
  583} [\href{https://arxiv.org/abs/hep-ph/0011241}{{\ttfamily
  hep-ph/0011241}}].

\bibitem{Ferreiro:2001qy}
E.~Ferreiro, E.~Iancu, A.~Leonidov and L.~McLerran, \emph{{Nonlinear gluon
  evolution in the color glass condensate. 2.}}, {\emph{Nucl. Phys.} {\bfseries
  A703} (2002) 489} [\href{https://arxiv.org/abs/hep-ph/0109115}{{\ttfamily
  hep-ph/0109115}}].

\bibitem{Blaizot:2002np}
J.-P. Blaizot, E.~Iancu and H.~Weigert, \emph{{Nonlinear gluon evolution in
  path integral form}}, {\emph{Nucl. Phys.} {\bfseries A713} (2003) 441}
  [\href{https://arxiv.org/abs/hep-ph/0206279}{{\ttfamily hep-ph/0206279}}].

\bibitem{Rummukainen:2003ns}
K.~Rummukainen and H.~Weigert, \emph{{Universal features of JIMWLK and BK
  evolution at small x}}, {\emph{Nucl. Phys. A} {\bfseries 739} (2004) 183}
  [\href{https://arxiv.org/abs/hep-ph/0309306}{{\ttfamily hep-ph/0309306}}].

\bibitem{Dumitru:2011vk}
A.~Dumitru, J.~Jalilian-Marian, T.~Lappi, B.~Schenke and R.~Venugopalan,
  \emph{{Renormalization group evolution of multi-gluon correlators in high
  energy QCD}}, {\emph{Phys.Lett.} {\bfseries B706} (2011) 219}
  [\href{https://arxiv.org/abs/1108.4764}{{\ttfamily 1108.4764}}].

\bibitem{Kowalski:2007rw}
H.~Kowalski, T.~Lappi and R.~Venugopalan, \emph{{Nuclear enhancement of
  universal dynamics of high parton densities}}, {\emph{Phys. Rev. Lett.}
  {\bfseries 100} (2008) 022303}
  [\href{https://arxiv.org/abs/0705.3047}{{\ttfamily 0705.3047}}].

\bibitem{Balitsky:2008zza}
I.~Balitsky and G.~A. Chirilli, \emph{{Next-to-leading order evolution of color
  dipoles}}, {\emph{Phys. Rev.} {\bfseries D77} (2008) 014019}
  [\href{https://arxiv.org/abs/0710.4330}{{\ttfamily 0710.4330}}].

\bibitem{Balitsky:2013fea}
I.~Balitsky and G.~A. Chirilli, \emph{{Rapidity evolution of Wilson lines at
  the next-to-leading order}}, {\emph{Phys. Rev.} {\bfseries D88} (2013)
  111501} [\href{https://arxiv.org/abs/1309.7644}{{\ttfamily 1309.7644}}].

\bibitem{Kovchegov:2006vj}
Y.~V. Kovchegov and H.~Weigert, \emph{{Triumvirate of Running Couplings in
  Small-x Evolution}}, {\emph{Nucl. Phys.} {\bfseries A784} (2007) 188}
  [\href{https://arxiv.org/abs/hep-ph/0609090}{{\ttfamily hep-ph/0609090}}].

\bibitem{Lublinsky:2016meo}
M.~Lublinsky and Y.~Mulian, \emph{{High Energy QCD at NLO: from light-cone wave
  function to JIMWLK evolution}}, {\emph{JHEP} {\bfseries 05} (2017) 097}
  [\href{https://arxiv.org/abs/1610.03453}{{\ttfamily 1610.03453}}].

\bibitem{Kovner:2013ona}
A.~Kovner, M.~Lublinsky and Y.~Mulian, \emph{{Jalilian-Marian, Iancu, McLerran,
  Weigert, Leonidov, Kovner evolution at next to leading order}}, {\emph{Phys.
  Rev.} {\bfseries D89} (2014) 061704}
  [\href{https://arxiv.org/abs/1310.0378}{{\ttfamily 1310.0378}}].

\bibitem{Kovner:2014lca}
A.~Kovner, M.~Lublinsky and Y.~Mulian, \emph{{NLO JIMWLK evolution
  unabridged}}, {\emph{JHEP} {\bfseries 08} (2014) 114}
  [\href{https://arxiv.org/abs/1405.0418}{{\ttfamily 1405.0418}}].

\bibitem{Grabovsky:2013mba}
A.~V. Grabovsky, \emph{{Connected contribution to the kernel of the evolution
  equation for 3-quark Wilson loop operator}}, {\emph{JHEP} {\bfseries 09}
  (2013) 141} [\href{https://arxiv.org/abs/1307.5414}{{\ttfamily 1307.5414}}].

\bibitem{Balitsky:2014mca}
I.~Balitsky and A.~V. Grabovsky, \emph{{NLO evolution of 3-quark Wilson loop
  operator}}, {\emph{JHEP} {\bfseries 01} (2015) 009}
  [\href{https://arxiv.org/abs/1405.0443}{{\ttfamily 1405.0443}}].

\bibitem{Grabovsky:2015cza}
A.~V. Grabovsky, \emph{{On the low-x NLO evolution of 4 point colorless
  operators}}, {\emph{JHEP} {\bfseries 09} (2015) 194}
  [\href{https://arxiv.org/abs/1507.08622}{{\ttfamily 1507.08622}}].

\bibitem{Caron-Huot:2015bja}
S.~Caron-Huot, \emph{{Resummation of non-global logarithms and the BFKL
  equation}}, {\emph{JHEP} {\bfseries 03} (2018) 036}
  [\href{https://arxiv.org/abs/1501.03754}{{\ttfamily 1501.03754}}].

\bibitem{Balitsky:2010ze}
I.~Balitsky and G.~A. Chirilli, \emph{{Photon impact factor in the
  next-to-leading order}}, {\emph{Phys. Rev. D} {\bfseries 83} (2011) 031502}
  [\href{https://arxiv.org/abs/1009.4729}{{\ttfamily 1009.4729}}].

\bibitem{Beuf:2017bpd}
G.~Beuf, \emph{{Dipole factorization for DIS at NLO: Combining the $q\bar{q}$
  and $q\bar{q}g$ contributions}}, {\emph{Phys. Rev. D} {\bfseries 96} (2017)
  074033} [\href{https://arxiv.org/abs/1708.06557}{{\ttfamily 1708.06557}}].

\bibitem{Hanninen:2017ddy}
H.~H\"anninen, T.~Lappi and R.~Paatelainen, \emph{{One-loop corrections to
  light cone wave functions: the dipole picture DIS cross section}},
  {\emph{Annals Phys.} {\bfseries 393} (2018) 358}
  [\href{https://arxiv.org/abs/1711.08207}{{\ttfamily 1711.08207}}].

\bibitem{Chirilli:2011km}
G.~A. Chirilli, B.-W. Xiao and F.~Yuan, \emph{{One-loop Factorization for
  Inclusive Hadron Production in $pA$ Collisions in the Saturation Formalism}},
  {\emph{Phys. Rev. Lett.} {\bfseries 108} (2012) 122301}
  [\href{https://arxiv.org/abs/1112.1061}{{\ttfamily 1112.1061}}].

\bibitem{Altinoluk:2014eka}
T.~Altinoluk, N.~Armesto, G.~Beuf, A.~Kovner and M.~Lublinsky,
  \emph{{Single-inclusive particle production in proton-nucleus collisions at
  next-to-leading order in the hybrid formalism}}, {\emph{Phys. Rev. D}
  {\bfseries 91} (2015) 094016}
  [\href{https://arxiv.org/abs/1411.2869}{{\ttfamily 1411.2869}}].

\bibitem{Caucal:2021ent}
P.~Caucal, F.~Salazar and R.~Venugopalan, \emph{{Dijet impact factor in DIS at
  next-to-leading order in the Color Glass Condensate}},
  \href{https://doi.org/10.1007/JHEP11(2021)222}{\emph{JHEP} {\bfseries 11}
  (2021) 222} [\href{https://arxiv.org/abs/2108.06347}{{\ttfamily
  2108.06347}}].

\bibitem{Roy:2019hwr}
K.~Roy and R.~Venugopalan, \emph{{NLO impact factor for inclusive
  photon$+$dijet production in $e+A$ DIS at small $x$}}, {\emph{Phys. Rev. D}
  {\bfseries 101} (2020) 034028}
  [\href{https://arxiv.org/abs/1911.04530}{{\ttfamily 1911.04530}}].

\bibitem{Roy:2019cux}
K.~Roy and R.~Venugopalan, \emph{{Extracting many-body correlators of saturated
  gluons with precision from inclusive photon+dijet final states in deeply
  inelastic scattering}}, {\emph{Phys. Rev. D} {\bfseries 101} (2020) 071505}
  [\href{https://arxiv.org/abs/1911.04519}{{\ttfamily 1911.04519}}].

\bibitem{Boussarie:2016ogo}
R.~Boussarie, A.~V. Grabovsky, L.~Szymanowski and S.~Wallon, \emph{{On the one
  loop $ {\gamma}^{\left(\ast \right)}\to q\overline{q} $ impact factor and the
  exclusive diffractive cross sections for the production of two or three
  jets}}, {\emph{JHEP} {\bfseries 11} (2016) 149}
  [\href{https://arxiv.org/abs/1606.00419}{{\ttfamily 1606.00419}}].

\bibitem{Boussarie:2016bkq}
R.~Boussarie, A.~V. Grabovsky, D.~{\relax Yu}. Ivanov, L.~Szymanowski and
  S.~Wallon, \emph{{Next-to-Leading Order Computation of Exclusive Diffractive
  Light Vector Meson Production in a Saturation Framework}}, {\emph{Phys. Rev.
  Lett.} {\bfseries 119} (2017) 072002}
  [\href{https://arxiv.org/abs/1612.08026}{{\ttfamily 1612.08026}}].

\bibitem{Marquet:2009ca}
C.~Marquet, B.-W. Xiao and F.~Yuan, \emph{{Semi-inclusive Deep Inelastic
  Scattering at small x}}, {\emph{Phys.Lett.} {\bfseries B682} (2009) 207}
  [\href{https://arxiv.org/abs/0906.1454}{{\ttfamily 0906.1454}}].

\bibitem{Dominguez:2010xd}
F.~Dominguez, B.-W. Xiao and F.~Yuan, \emph{{$k_t$-factorization for Hard
  Processes in Nuclei}}, {\emph{Phys.Rev.Lett.} {\bfseries 106} (2011) 022301}
  [\href{https://arxiv.org/abs/1009.2141}{{\ttfamily 1009.2141}}].

\bibitem{Dominguez:2011wm}
F.~Dominguez, C.~Marquet, B.-W. Xiao and F.~Yuan, \emph{{Universality of
  Unintegrated Gluon Distributions at small x}}, {\emph{Phys. Rev.} {\bfseries
  D83} (2011) 105005} [\href{https://arxiv.org/abs/1101.0715}{{\ttfamily
  1101.0715}}].

\bibitem{Metz:2011wb}
A.~Metz and J.~Zhou, \emph{{Distribution of linearly polarized gluons inside a
  large nucleus}}, {\emph{Phys.Rev.} {\bfseries D84} (2011) 051503}
  [\href{https://arxiv.org/abs/1105.1991}{{\ttfamily 1105.1991}}].

\bibitem{Balitsky:2016dgz}
I.~Balitsky and A.~Tarasov, \emph{{Gluon TMD in particle production from low to
  moderate x}}, {\emph{JHEP} {\bfseries 06} (2016) 164}
  [\href{https://arxiv.org/abs/1603.06548}{{\ttfamily 1603.06548}}].

\bibitem{Altinoluk:2019fui}
T.~Altinoluk, R.~Boussarie and P.~Kotko, \emph{{Interplay of the CGC and TMD
  frameworks to all orders in kinematic twist}}, {\emph{JHEP} {\bfseries 05}
  (2019) 156} [\href{https://arxiv.org/abs/1901.01175}{{\ttfamily
  1901.01175}}].

\bibitem{Altinoluk:2019wyu}
T.~Altinoluk and R.~Boussarie, \emph{{Low $x$ physics as an infinite twist
  (G)TMD framework: unravelling the origins of saturation}},
  \href{https://doi.org/10.1007/JHEP10(2019)208}{\emph{JHEP} {\bfseries 10}
  (2019) 208} [\href{https://arxiv.org/abs/1902.07930}{{\ttfamily
  1902.07930}}].

\bibitem{Gelis:2007kn}
F.~Gelis, T.~Lappi and R.~Venugopalan, \emph{{High energy scattering in Quantum
  Chromodynamics}}, {\emph{Int. J. Mod. Phys.} {\bfseries E16} (2007) 2595}
  [\href{https://arxiv.org/abs/0708.0047}{{\ttfamily 0708.0047}}].

\bibitem{Baier:2004tj}
R.~Baier, A.~Mueller and D.~Schiff, \emph{{Saturation and shadowing in
  high-energy proton nucleus dilepton production}}, {\emph{Nucl.Phys.}
  {\bfseries A741} (2004) 358}
  [\href{https://arxiv.org/abs/hep-ph/0403201}{{\ttfamily hep-ph/0403201}}].

\bibitem{Marquet:2007vb}
C.~Marquet, \emph{{Forward inclusive dijet production and azimuthal
  correlations in p(A) collisions}}, {\emph{Nucl.Phys.} {\bfseries A796} (2007)
  41} [\href{https://arxiv.org/abs/0708.0231}{{\ttfamily 0708.0231}}].

\bibitem{Mueller:2012uf}
A.~Mueller, B.-W. Xiao and F.~Yuan, \emph{{Sudakov Resummation in Small-$x$
  Saturation Formalism}}, {\emph{Phys.Rev.Lett.} {\bfseries 110} (2013) 082301}
  [\href{https://arxiv.org/abs/1210.5792}{{\ttfamily 1210.5792}}].

\bibitem{Mueller:2013wwa}
A.~H. Mueller, B.-W. Xiao and F.~Yuan, \emph{{Sudakov double logarithms
  resummation in hard processes in the small-x saturation formalism}},
  {\emph{Phys. Rev.} {\bfseries D88} (2013) 114010}
  [\href{https://arxiv.org/abs/1308.2993}{{\ttfamily 1308.2993}}].

\bibitem{Balitsky:2015qba}
I.~Balitsky and A.~Tarasov, \emph{{Rapidity evolution of gluon TMD from low to
  moderate x}}, {\emph{JHEP} {\bfseries 10} (2015) 017}
  [\href{https://arxiv.org/abs/1505.02151}{{\ttfamily 1505.02151}}].

\bibitem{Xiao:2017yya}
B.-W. Xiao, F.~Yuan and J.~Zhou, \emph{{Transverse Momentum Dependent Parton
  Distributions at Small-x}}, {\emph{Nucl. Phys. B} {\bfseries 921} (2017) 104}
  [\href{https://arxiv.org/abs/1703.06163}{{\ttfamily 1703.06163}}].

\bibitem{Balitsky:2020jzt}
I.~Balitsky, \emph{{Gauge-invariant TMD factorization for Drell-Yan hadronic
  tensor at small x}},
  \href{https://doi.org/10.1007/JHEP05(2021)046}{\emph{JHEP} {\bfseries 05}
  (2021) 046} [\href{https://arxiv.org/abs/2012.01588}{{\ttfamily
  2012.01588}}].

\bibitem{Kharzeev:2003wz}
D.~Kharzeev, Y.~V. Kovchegov and K.~Tuchin, \emph{{Cronin effect and high $p_T$
  suppression in pA collisions}}, {\emph{Phys. Rev.} {\bfseries D68} (2003)
  094013} [\href{https://arxiv.org/abs/hep-ph/0307037}{{\ttfamily
  hep-ph/0307037}}].

\bibitem{Petreska:2018cbf}
E.~Petreska, \emph{{TMD gluon distributions at small x in the CGC theory}},
  {\emph{Int. J. Mod. Phys.} {\bfseries E27} (2018) 1830003}
  [\href{https://arxiv.org/abs/1804.04981}{{\ttfamily 1804.04981}}].

\bibitem{JalilianMarian:1996xn}
J.~Jalilian-Marian, A.~Kovner, L.~D. McLerran and H.~Weigert, \emph{{The
  Intrinsic glue distribution at very small x}}, {\emph{Phys. Rev.} {\bfseries
  D55} (1997) 5414} [\href{https://arxiv.org/abs/hep-ph/9606337}{{\ttfamily
  hep-ph/9606337}}].

\bibitem{Kovchegov:1996ty}
Y.~V. Kovchegov, \emph{{NonAbelian Weizsacker-Williams field and a
  two-dimensional effective color charge density for a very large nucleus}},
  {\emph{Phys. Rev.} {\bfseries D54} (1996) 5463}
  [\href{https://arxiv.org/abs/hep-ph/9605446}{{\ttfamily hep-ph/9605446}}].

\bibitem{Hatta:2006ci}
Y.~Hatta, \emph{{CGC formalism with two sources}}, {\emph{Nucl. Phys. A}
  {\bfseries 781} (2007) 104}
  [\href{https://arxiv.org/abs/hep-ph/0607126}{{\ttfamily hep-ph/0607126}}].

\bibitem{Dominguez:2012ad}
F.~Dominguez, C.~Marquet, A.~M. Stasto and B.-W. Xiao, \emph{{Universality of
  multiparticle production in QCD at high energies}}, {\emph{Phys. Rev. D}
  {\bfseries 87} (2013) 034007}
  [\href{https://arxiv.org/abs/1210.1141}{{\ttfamily 1210.1141}}].

\bibitem{Vogelsang:2007jk}
W.~Vogelsang and F.~Yuan, \emph{{Hadronic Dijet Imbalance and
  Transverse-Momentum Dependent Parton Distributions}}, {\emph{Phys.Rev.}
  {\bfseries D76} (2007) 094013}
  [\href{https://arxiv.org/abs/0708.4398}{{\ttfamily 0708.4398}}].

\bibitem{Albacete:2010pg}
J.~L. Albacete and C.~Marquet, \emph{{Azimuthal correlations of forward
  di-hadrons in d+Au collisions at RHIC in the Color Glass Condensate}},
  {\emph{Phys. Rev. Lett.} {\bfseries 105} (2010) 162301}
  [\href{https://arxiv.org/abs/1005.4065}{{\ttfamily 1005.4065}}].

\bibitem{Stasto:2011ru}
A.~Stasto, B.-W. Xiao and F.~Yuan, \emph{{Back-to-Back Correlations of
  Di-hadrons in dAu Collisions at RHIC}}, {\emph{Phys. Lett. B} {\bfseries 716}
  (2012) 430} [\href{https://arxiv.org/abs/1109.1817}{{\ttfamily 1109.1817}}].

\bibitem{Stasto:2018rci}
A.~Stasto, S.-Y. Wei, B.-W. Xiao and F.~Yuan, \emph{{On the Dihadron Angular
  Correlations in Forward $pA$ collisions}}, {\emph{Phys. Lett. B} {\bfseries
  784} (2018) 301} [\href{https://arxiv.org/abs/1805.05712}{{\ttfamily
  1805.05712}}].

\bibitem{Albacete:2018ruq}
J.~L. Albacete, G.~Giacalone, C.~Marquet and M.~Matas, \emph{{Forward dihadron
  back-to-back correlations in $pA$ collisions}}, {\emph{Phys. Rev. D}
  {\bfseries 99} (2019) 014002}
  [\href{https://arxiv.org/abs/1805.05711}{{\ttfamily 1805.05711}}].

\bibitem{Zheng:2014vka}
L.~Zheng, E.~Aschenauer, J.~Lee and B.-W. Xiao, \emph{{Probing Gluon Saturation
  through Dihadron Correlations at an Electron-Ion Collider}}, {\emph{Phys.
  Rev. D} {\bfseries 89} (2014) 074037}
  [\href{https://arxiv.org/abs/1403.2413}{{\ttfamily 1403.2413}}].

\bibitem{Boer:2016fqd}
D.~Boer, P.~J. Mulders, C.~Pisano and J.~Zhou, \emph{{Asymmetries in Heavy
  Quark Pair and Dijet Production at an EIC}}, {\emph{JHEP} {\bfseries 08}
  (2016) 001} [\href{https://arxiv.org/abs/1605.07934}{{\ttfamily
  1605.07934}}].

\bibitem{Boer:2017xpy}
D.~Boer, P.~J. Mulders, J.~Zhou and Y.-j. Zhou, \emph{{Suppression of maximal
  linear gluon polarization in angular asymmetries}}, {\emph{JHEP} {\bfseries
  10} (2017) 196} [\href{https://arxiv.org/abs/1702.08195}{{\ttfamily
  1702.08195}}].

\bibitem{Dumitru:2015gaa}
A.~Dumitru, T.~Lappi and V.~Skokov, \emph{{Distribution of Linearly Polarized
  Gluons and Elliptic Azimuthal Anisotropy in Deep Inelastic Scattering Dijet
  Production at High Energy}}, {\emph{Phys. Rev. Lett.} {\bfseries 115} (2015)
  252301} [\href{https://arxiv.org/abs/1508.04438}{{\ttfamily 1508.04438}}].

\bibitem{Dumitru:2018kuw}
A.~Dumitru, V.~Skokov and T.~Ullrich, \emph{{Measuring the
  Weizs\"acker-Williams distribution of linearly polarized gluons at an
  electron-ion collider through dijet azimuthal asymmetries}}, {\emph{Phys.
  Rev. C} {\bfseries 99} (2019) 015204}
  [\href{https://arxiv.org/abs/1809.02615}{{\ttfamily 1809.02615}}].

\bibitem{Mantysaari:2019csc}
H.~M{\"a}ntysaari, N.~Mueller and B.~Schenke, \emph{{Diffractive Dijet
  Production and Wigner Distributions from the Color Glass Condensate}},
  {\emph{Phys. Rev.} {\bfseries D99} (2019) 074004}
  [\href{https://arxiv.org/abs/1902.05087}{{\ttfamily 1902.05087}}].

\bibitem{Mantysaari:2019hkq}
H.~M\"antysaari, N.~Mueller, F.~Salazar and B.~Schenke, \emph{{Multigluon
  Correlations and Evidence of Saturation from Dijet Measurements at an
  Electron-Ion Collider}}, {\emph{Phys. Rev. Lett.} {\bfseries 124} (2020)
  112301} [\href{https://arxiv.org/abs/1912.05586}{{\ttfamily 1912.05586}}].

\bibitem{Kovchegov:1999yj}
Y.~V. Kovchegov, \emph{{Small x $F_2$ structure function of a nucleus including
  multiple pomeron exchanges}}, {\emph{Phys. Rev.} {\bfseries D60} (1999)
  034008} [\href{https://arxiv.org/abs/hep-ph/9901281}{{\ttfamily
  hep-ph/9901281}}].

\bibitem{Zhou:2016tfe}
J.~Zhou, \emph{{The evolution of the small x gluon TMD}}, {\emph{JHEP}
  {\bfseries 06} (2016) 151}
  [\href{https://arxiv.org/abs/1603.07426}{{\ttfamily 1603.07426}}].

\bibitem{Marzani:2015oyb}
S.~Marzani, \emph{{Combining $Q_T$ and small-$x$ resummations}}, {\emph{Phys.
  Rev.} {\bfseries D93} (2016) 054047}
  [\href{https://arxiv.org/abs/1511.06039}{{\ttfamily 1511.06039}}].

\bibitem{JalilianMarian:1997jx}
J.~Jalilian-Marian, A.~Kovner, A.~Leonidov and H.~Weigert, \emph{{The BFKL
  equation from the Wilson renormalization group}}, {\emph{Nucl. Phys.}
  {\bfseries B504} (1997) 415}
  [\href{https://arxiv.org/abs/hep-ph/9701284}{{\ttfamily hep-ph/9701284}}].

\bibitem{Chirilli:2018kkw}
G.~A. Chirilli, \emph{{Sub-eikonal corrections to scattering amplitudes at high
  energy}}, \href{https://doi.org/10.1007/JHEP01(2019)118}{\emph{JHEP}
  {\bfseries 01} (2019) 118}
  [\href{https://arxiv.org/abs/1807.11435}{{\ttfamily 1807.11435}}].

\bibitem{Hatta:2016aoc}
Y.~Hatta, Y.~Nakagawa, F.~Yuan, Y.~Zhao and B.~Xiao, \emph{{Gluon orbital
  angular momentum at small-$x$}}, {\emph{Phys. Rev.} {\bfseries D95} (2017)
  114032} [\href{https://arxiv.org/abs/1612.02445}{{\ttfamily 1612.02445}}].

\bibitem{Altinoluk:2014oxa}
T.~Altinoluk, N.~Armesto, G.~Beuf, M.~Mart{\~A}-nez and C.~A. Salgado,
  \emph{{Next-to-eikonal corrections in the CGC: gluon production and spin
  asymmetries in pA collisions}}, {\emph{JHEP} {\bfseries 07} (2014) 068}
  [\href{https://arxiv.org/abs/1404.2219}{{\ttfamily 1404.2219}}].

\bibitem{Altinoluk:2015gia}
T.~Altinoluk, N.~Armesto, G.~Beuf and A.~Moscoso,
  \emph{{Next-to-next-to-eikonal corrections in the CGC}}, {\emph{JHEP}
  {\bfseries 01} (2016) 114}
  [\href{https://arxiv.org/abs/1505.01400}{{\ttfamily 1505.01400}}].

\bibitem{Agostini:2019avp}
P.~Agostini, T.~Altinoluk and N.~Armesto, \emph{{Non-eikonal corrections to
  multi-particle production in the Color Glass Condensate}}, {\emph{Eur. Phys.
  J.} {\bfseries C79} (2019) 600}
  [\href{https://arxiv.org/abs/1902.04483}{{\ttfamily 1902.04483}}].

\bibitem{Cougoulic:2022gbk}
F.~Cougoulic, Y.~V. Kovchegov, A.~Tarasov and Y.~Tawabutr, \emph{{Quark and
  Gluon Helicity Evolution at Small $x$: Revised and Updated}},
  \href{https://arxiv.org/abs/2204.11898}{{\ttfamily 2204.11898}}.

\bibitem{Kirschner:1983di}
R.~Kirschner and L.~n. Lipatov, \emph{{Double Logarithmic Asymptotics and Regge
  Singularities of Quark Amplitudes with Flavor Exchange}}, {\emph{Nucl. Phys.
  B} {\bfseries 213} (1983) 122}.

\bibitem{Kirschner:1994rq}
R.~Kirschner, \emph{{Reggeon interactions in perturbative QCD}}, {\emph{Z.
  Phys. C} {\bfseries 65} (1995) 505}
  [\href{https://arxiv.org/abs/hep-th/9407085}{{\ttfamily hep-th/9407085}}].

\bibitem{Kirschner:1994vc}
R.~Kirschner, \emph{{Regge asymptotics of scattering with flavor exchange in
  QCD}}, {\emph{Z. Phys. C} {\bfseries 67} (1995) 459}
  [\href{https://arxiv.org/abs/hep-th/9404158}{{\ttfamily hep-th/9404158}}].

\bibitem{Blumlein:1995jp}
J.~Bl\"umlein and A.~Vogt, \emph{{On the behavior of nonsinglet structure
  functions at small x}}, {\emph{Phys. Lett. B} {\bfseries 370} (1996) 149}
  [\href{https://arxiv.org/abs/hep-ph/9510410}{{\ttfamily hep-ph/9510410}}].

\bibitem{Bartels:1995iu}
J.~Bartels, B.~I. Ermolaev and M.~G. Ryskin, \emph{{Nonsinglet contributions to
  the structure function $g_1$ at small x}}, {\emph{Z. Phys. C} {\bfseries 70}
  (1996) 273} [\href{https://arxiv.org/abs/hep-ph/9507271}{{\ttfamily
  hep-ph/9507271}}].

\bibitem{Bartels:1996wc}
J.~Bartels, B.~I. Ermolaev and M.~G. Ryskin, \emph{{Flavor singlet contribution
  to the structure function $g_1$ at small x}}, {\emph{Z. Phys.} {\bfseries
  C72} (1996) 627} [\href{https://arxiv.org/abs/hep-ph/9603204}{{\ttfamily
  hep-ph/9603204}}].

\bibitem{Blumlein:1996hb}
J.~Bl\"umlein and A.~Vogt, \emph{{The Singlet contribution to the structure
  function $g_1(x, Q^2)$ at small x}}, {\emph{Phys. Lett. B} {\bfseries 386}
  (1996) 350} [\href{https://arxiv.org/abs/hep-ph/9606254}{{\ttfamily
  hep-ph/9606254}}].

\bibitem{Boussarie:2019icw}
R.~Boussarie, Y.~Hatta and F.~Yuan, \emph{{Proton Spin Structure at
  Small-$x$}}, {\emph{Phys. Lett.} {\bfseries B797} (2019) 134817}
  [\href{https://arxiv.org/abs/1904.02693}{{\ttfamily 1904.02693}}].

\bibitem{Kovchegov:2015pbl}
Y.~V. Kovchegov, D.~Pitonyak and M.~D. Sievert, \emph{{Helicity Evolution at
  Small-x}}, {\emph{JHEP} {\bfseries 01} (2016) 072}
  [\href{https://arxiv.org/abs/1511.06737}{{\ttfamily 1511.06737}}].

\bibitem{Kovchegov:2016weo}
Y.~V. Kovchegov, D.~Pitonyak and M.~D. Sievert, \emph{{Small-$x$ asymptotics of
  the quark helicity distribution}}, {\emph{Phys. Rev. Lett.} {\bfseries 118}
  (2017) 052001} [\href{https://arxiv.org/abs/1610.06188}{{\ttfamily
  1610.06188}}].

\bibitem{Kovchegov:2018znm}
Y.~V. Kovchegov and M.~D. Sievert, \emph{{Small-$x$ Helicity Evolution: an
  Operator Treatment}}, {\emph{Phys. Rev.} {\bfseries D99} (2019) 054032}
  [\href{https://arxiv.org/abs/1808.09010}{{\ttfamily 1808.09010}}].

\bibitem{Jalilian-Marian:2017ttv}
J.~Jalilian-Marian, \emph{{Elastic scattering of a quark from a color field:
  longitudinal momentum exchange}}, {\emph{Phys. Rev. D} {\bfseries 96} (2017)
  074020} [\href{https://arxiv.org/abs/1708.07533}{{\ttfamily 1708.07533}}].

\bibitem{Tarasov:2019rfp}
A.~Tarasov and R.~Venugopalan, \emph{{Structure functions at small x from
  worldlines: Unpolarized distributions}}, {\emph{Phys. Rev. D} {\bfseries 100}
  (2019) 054007} [\href{https://arxiv.org/abs/1903.11624}{{\ttfamily
  1903.11624}}].

\bibitem{Cougoulic:2019aja}
F.~Cougoulic and Y.~V. Kovchegov, \emph{{Helicity-dependent generalization of
  the JIMWLK evolution}}, {\emph{Phys. Rev. D} {\bfseries 100} (2019) 114020}
  [\href{https://arxiv.org/abs/1910.04268}{{\ttfamily 1910.04268}}].

\bibitem{Kovchegov:2016zex}
Y.~V. Kovchegov, D.~Pitonyak and M.~D. Sievert, \emph{{Helicity Evolution at
  Small $x$: Flavor Singlet and Non-Singlet Observables}}, {\emph{Phys. Rev. D}
  {\bfseries 95} (2017) 014033}
  [\href{https://arxiv.org/abs/1610.06197}{{\ttfamily 1610.06197}}].

\bibitem{Kovchegov:2017jxc}
Y.~V. Kovchegov, D.~Pitonyak and M.~D. Sievert, \emph{{Small-$x$ Asymptotics of
  the Quark Helicity Distribution: Analytic Results}}, {\emph{Phys. Lett. B}
  {\bfseries 772} (2017) 136}
  [\href{https://arxiv.org/abs/1703.05809}{{\ttfamily 1703.05809}}].

\bibitem{Kovchegov:2020hgb}
Y.~V. Kovchegov and Y.~Tawabutr, \emph{{Helicity at Small $x$: Oscillations
  Generated by Bringing Back the Quarks}}, {\emph{JHEP} {\bfseries 08} (2020)
  014} [\href{https://arxiv.org/abs/2005.07285}{{\ttfamily 2005.07285}}].

\bibitem{Altinoluk:2020oyd}
T.~Altinoluk, G.~Beuf, A.~Czajka and A.~Tymowska, \emph{{Quarks at
  next-to-eikonal accuracy in the CGC: Forward quark-nucleus scattering}},
  \href{https://doi.org/10.1103/PhysRevD.104.014019}{\emph{Phys. Rev. D}
  {\bfseries 104} (2021) 014019}
  [\href{https://arxiv.org/abs/2012.03886}{{\ttfamily 2012.03886}}].

\bibitem{Kovchegov:2021iyc}
Y.~V. Kovchegov and M.~G. Santiago, \emph{{Quark Sivers function at small x:
  spin-dependent odderon and the sub-eikonal evolution}},
  \href{https://doi.org/10.1007/JHEP11(2021)200}{\emph{JHEP} {\bfseries 11}
  (2021) 200} [\href{https://arxiv.org/abs/2108.03667}{{\ttfamily
  2108.03667}}].

\bibitem{Kovchegov:2021lvz}
Y.~V. Kovchegov, A.~Tarasov and Y.~Tawabutr, \emph{{Helicity evolution at small
  x: the single-logarithmic contribution}},
  \href{https://doi.org/10.1007/JHEP03(2022)184}{\emph{JHEP} {\bfseries 03}
  (2022) 184} [\href{https://arxiv.org/abs/2104.11765}{{\ttfamily
  2104.11765}}].

\bibitem{Kovchegov:2019rrz}
Y.~V. Kovchegov, \emph{{Orbital Angular Momentum at Small $x$}}, {\emph{JHEP}
  {\bfseries 03} (2019) 174}
  [\href{https://arxiv.org/abs/1901.07453}{{\ttfamily 1901.07453}}].

\bibitem{Kovchegov:2018zeq}
Y.~V. Kovchegov and M.~D. Sievert, \emph{{Valence Quark Transversity at Small
  $x$}}, {\emph{Phys. Rev. D} {\bfseries 99} (2019) 054033}
  [\href{https://arxiv.org/abs/1808.10354}{{\ttfamily 1808.10354}}].

\bibitem{Adamiak:2021ppq}
{\scshape Jefferson Lab Angular Momentum} collaboration, D.~Adamiak, Y.~V.
  Kovchegov, W.~Melnitchouk, D.~Pitonyak, N.~Sato and M.~D. Sievert,
  \emph{{First analysis of world polarized DIS data with small-x helicity
  evolution}}, \href{https://doi.org/10.1103/PhysRevD.104.L031501}{\emph{Phys.
  Rev. D} {\bfseries 104} (2021) L031501}
  [\href{https://arxiv.org/abs/2102.06159}{{\ttfamily 2102.06159}}].

\bibitem{Boer:2015pni}
D.~Boer, M.~G. Echevarria, P.~Mulders and J.~Zhou, \emph{{Single spin
  asymmetries from a single Wilson loop}}, {\emph{Phys. Rev. Lett.} {\bfseries
  116} (2016) 122001} [\href{https://arxiv.org/abs/1511.03485}{{\ttfamily
  1511.03485}}].

\bibitem{Zhou:2013gsa}
J.~Zhou, \emph{{Transverse single spin asymmetries at small x and the anomalous
  magnetic moment}}, {\emph{Phys. Rev.} {\bfseries D89} (2014) 074050}
  [\href{https://arxiv.org/abs/1308.5912}{{\ttfamily 1308.5912}}].

\bibitem{Ryskin:1987ya}
M.~G. Ryskin, \emph{{Odderon and Polarization Phenomena in QCD}}, {\emph{Sov.
  J. Nucl. Phys.} {\bfseries 46} (1987) 337}.

\bibitem{Buttimore:1998rj}
N.~H. Buttimore, B.~Z. Kopeliovich, E.~Leader, J.~Soffer and T.~L. Trueman,
  \emph{{The spin dependence of high-energy proton scattering}}, {\emph{Phys.
  Rev.} {\bfseries D59} (1999) 114010}
  [\href{https://arxiv.org/abs/hep-ph/9901339}{{\ttfamily hep-ph/9901339}}].

\bibitem{Leader:1999ua}
E.~Leader and T.~L. Trueman, \emph{{The Odderon and spin dependence of
  high-energy proton proton scattering}}, {\emph{Phys. Rev.} {\bfseries D61}
  (2000) 077504} [\href{https://arxiv.org/abs/hep-ph/9908221}{{\ttfamily
  hep-ph/9908221}}].

\bibitem{Boussarie:2019vmk}
R.~Boussarie, Y.~Hatta, L.~Szymanowski and S.~Wallon, \emph{{Probing the Gluon
  Sivers Function with an Unpolarized Target: GTMD Distributions and the
  Odderons}}, {\emph{Phys. Rev. Lett.} {\bfseries 124} (2020) 172501}
  [\href{https://arxiv.org/abs/1912.08182}{{\ttfamily 1912.08182}}].

\bibitem{Altarelli:1988nr}
G.~Altarelli and G.~G. Ross, \emph{{The Anomalous Gluon Contribution to
  Polarized Leptoproduction}}, {\emph{Phys. Lett. B} {\bfseries 212} (1988)
  391}.

\bibitem{Carlitz:1988ab}
R.~D. Carlitz, J.~C. Collins and A.~H. Mueller, \emph{{The Role of the Axial
  Anomaly in Measuring Spin Dependent Parton Distributions}}, {\emph{Phys.
  Lett. B} {\bfseries 214} (1988) 229}.

\bibitem{Veneziano:1989ei}
G.~Veneziano, \emph{{Is There a QCD Spin Crisis?}}, {\emph{Mod. Phys. Lett. A}
  {\bfseries 4} (1989) 1605}.

\bibitem{Shore:1990zu}
G.~M. Shore and G.~Veneziano, \emph{{The U(1) {Goldberger-Treiman} Relation and
  the Two Components of the Proton 'Spin'}}, {\emph{Phys. Lett. B} {\bfseries
  244} (1990) 75}.

\bibitem{Shore:1991dv}
G.~M. Shore and G.~Veneziano, \emph{{The U(1) Goldberger-Treiman relation and
  the proton 'spin': A Renormalization group analysis}}, {\emph{Nucl. Phys. B}
  {\bfseries 381} (1992) 23}.

\bibitem{Narison:1998aq}
S.~Narison, G.~M. Shore and G.~Veneziano, \emph{{Topological charge screening
  and the 'proton spin' beyond the chiral limit}}, {\emph{Nucl. Phys. B}
  {\bfseries 546} (1999) 235}
  [\href{https://arxiv.org/abs/hep-ph/9812333}{{\ttfamily hep-ph/9812333}}].

\bibitem{Narison:2021kny}
S.~Narison, \emph{{Slope of the topological charge, proton spin and the
  $0^{-+}$ pseudoscalar di-gluonia spectra}},
  \href{https://doi.org/10.1016/j.nuclphysa.2022.122393}{\emph{Nucl. Phys. A}
  {\bfseries 1020} (2022) 122393}
  [\href{https://arxiv.org/abs/2111.02873}{{\ttfamily 2111.02873}}].

\bibitem{Giusti:2001xh}
L.~Giusti, G.~C. Rossi, M.~Testa and G.~Veneziano, \emph{{The $U_A(1)$ problem
  on the lattice with Ginsparg-Wilson fermions}}, {\emph{Nucl. Phys. B}
  {\bfseries 628} (2002) 234}
  [\href{https://arxiv.org/abs/hep-lat/0108009}{{\ttfamily hep-lat/0108009}}].

\bibitem{Bali:2021qem}
{\scshape RQCD} collaboration, G.~S. Bali, V.~Braun, S.~Collins, A.~Sch\"afer
  and J.~Simeth, \emph{{Masses and decay constants of the \ensuremath{\eta} and
  \ensuremath{\eta}' mesons from lattice QCD}}, {\emph{JHEP} {\bfseries 08}
  (2021) 137} [\href{https://arxiv.org/abs/2106.05398}{{\ttfamily
  2106.05398}}].

\bibitem{Tarasov:2020cwl}
A.~Tarasov and R.~Venugopalan, \emph{{The role of the chiral anomaly in
  polarized deeply inelastic scattering I: Finding the triangle graph inside
  the box diagram in Bjorken and Regge asymptotics}}, {\emph{Phys. Rev. D}
  {\bfseries 102} (2020) 114022}
  [\href{https://arxiv.org/abs/2008.08104}{{\ttfamily 2008.08104}}].

\bibitem{Tarasov:2021yll}
A.~Tarasov and R.~Venugopalan, \emph{{Role of the chiral anomaly in polarized
  deeply inelastic scattering. II. Topological screening and transitions from
  emergent axionlike dynamics}},
  \href{https://doi.org/10.1103/PhysRevD.105.014020}{\emph{Phys. Rev. D}
  {\bfseries 105} (2022) 014020}
  [\href{https://arxiv.org/abs/2109.10370}{{\ttfamily 2109.10370}}].

\bibitem{tHooft:1976snw}
G.~'t~Hooft, \emph{{Computation of the Quantum Effects Due to a
  Four-Dimensional Pseudoparticle}}, {\emph{Phys. Rev. D} {\bfseries 14} (1976)
  3432}.

\bibitem{Klinkhamer:1984di}
F.~R. Klinkhamer and N.~S. Manton, \emph{{A Saddle Point Solution in the
  Weinberg-Salam Theory}}, {\emph{Phys. Rev. D} {\bfseries 30} (1984) 2212}.

\bibitem{Kuzmin:1985mm}
V.~A. Kuzmin, V.~A. Rubakov and M.~E. Shaposhnikov, \emph{{On the Anomalous
  Electroweak Baryon Number Nonconservation in the Early Universe}},
  {\emph{Phys. Lett. B} {\bfseries 155} (1985) 36}.

\bibitem{McLerran:1990de}
L.~D. McLerran, E.~Mottola and M.~E. Shaposhnikov, \emph{{Sphalerons and Axion
  Dynamics in High Temperature {QCD}}}, {\emph{Phys. Rev. D} {\bfseries 43}
  (1991) 2027}.

\bibitem{Mace:2016svc}
M.~Mace, S.~Schlichting and R.~Venugopalan, \emph{{Off-equilibrium sphaleron
  transitions in the Glasma}}, {\emph{Phys. Rev. D} {\bfseries 93} (2016)
  074036} [\href{https://arxiv.org/abs/1601.07342}{{\ttfamily 1601.07342}}].

\bibitem{Boussarie:2020vzf}
R.~Boussarie and Y.~Mehtar-Tani, \emph{{On gauge invariance of transverse
  momentum dependent distributions at small x}}, {\emph{Phys. Rev. D}
  {\bfseries 103} (2021) 094012}
  [\href{https://arxiv.org/abs/2001.06449}{{\ttfamily 2001.06449}}].

\bibitem{Kotko:2015ura}
P.~Kotko, K.~Kutak, C.~Marquet, E.~Petreska, S.~Sapeta and A.~van Hameren,
  \emph{{Improved TMD factorization for forward dijet production in
  dilute-dense hadronic collisions}}, {\emph{JHEP} {\bfseries 09} (2015) 106}
  [\href{https://arxiv.org/abs/1503.03421}{{\ttfamily 1503.03421}}].

\bibitem{Marquet:2016cgx}
C.~Marquet, E.~Petreska and C.~Roiesnel, \emph{{Transverse-momentum-dependent
  gluon distributions from JIMWLK evolution}}, {\emph{JHEP} {\bfseries 10}
  (2016) 065} [\href{https://arxiv.org/abs/1608.02577}{{\ttfamily
  1608.02577}}].

\bibitem{Marquet:2017xwy}
C.~Marquet, C.~Roiesnel and P.~Taels, \emph{{Linearly polarized small-$x$
  gluons in forward heavy-quark pair production}}, {\emph{Phys. Rev.}
  {\bfseries D97} (2018) 014004}
  [\href{https://arxiv.org/abs/1710.05698}{{\ttfamily 1710.05698}}].

\bibitem{Dumitru:2005gt}
A.~Dumitru, A.~Hayashigaki and J.~Jalilian-Marian, \emph{{The Color glass
  condensate and hadron production in the forward region}}, {\emph{Nucl. Phys.}
  {\bfseries A765} (2006) 464}
  [\href{https://arxiv.org/abs/hep-ph/0506308}{{\ttfamily hep-ph/0506308}}].

\bibitem{Chirilli:2012jd}
G.~A. Chirilli, B.-W. Xiao and F.~Yuan, \emph{{Inclusive Hadron Productions in
  pA Collisions}}, {\emph{Phys. Rev. D} {\bfseries 86} (2012) 054005}
  [\href{https://arxiv.org/abs/1203.6139}{{\ttfamily 1203.6139}}].

\bibitem{Gelis:2008rw}
F.~Gelis, T.~Lappi and R.~Venugopalan, \emph{{High energy factorization in
  nucleus-nucleus collisions}}, {\emph{Phys. Rev. D} {\bfseries 78} (2008)
  054019} [\href{https://arxiv.org/abs/0804.2630}{{\ttfamily 0804.2630}}].

\bibitem{Krasnitz:1998ns}
A.~Krasnitz and R.~Venugopalan, \emph{{Nonperturbative computation of gluon
  minijet production in nuclear collisions at very high-energies}},
  {\emph{Nucl. Phys. B} {\bfseries 557} (1999) 237}
  [\href{https://arxiv.org/abs/hep-ph/9809433}{{\ttfamily hep-ph/9809433}}].

\bibitem{Gelis:2003vh}
F.~Gelis and R.~Venugopalan, \emph{{Large mass q anti-q production from the
  color glass condensate}}, {\emph{Phys. Rev.} {\bfseries D69} (2004) 014019}
  [\href{https://arxiv.org/abs/hep-ph/0310090}{{\ttfamily hep-ph/0310090}}].

\bibitem{Blaizot:2004wv}
J.~P. Blaizot, F.~Gelis and R.~Venugopalan, \emph{{High-energy pA collisions in
  the color glass condensate approach. 2. Quark production}}, {\emph{Nucl.
  Phys.} {\bfseries A743} (2004) 57}
  [\href{https://arxiv.org/abs/hep-ph/0402257}{{\ttfamily hep-ph/0402257}}].

\bibitem{Fujii:2006ab}
H.~Fujii, F.~Gelis and R.~Venugopalan, \emph{{Quark pair production in high
  energy pA collisions: General features}}, {\emph{Nucl. Phys.} {\bfseries
  A780} (2006) 146} [\href{https://arxiv.org/abs/hep-ph/0603099}{{\ttfamily
  hep-ph/0603099}}].

\bibitem{Gelis:2005mn}
F.~Gelis, K.~Kajantie and T.~Lappi, \emph{{Production of quark pairs from
  classical gluon fields}}, {\emph{Eur. Phys. J. A} {\bfseries 29} (2006) 89}
  [\href{https://arxiv.org/abs/hep-ph/0509363}{{\ttfamily hep-ph/0509363}}].

\bibitem{Gelis:2015eua}
F.~Gelis and N.~Tanji, \emph{{Quark production in heavy ion collisions:
  formalism and boost invariant fermionic light-cone mode functions}},
  {\emph{JHEP} {\bfseries 02} (2016) 126}
  [\href{https://arxiv.org/abs/1506.03327}{{\ttfamily 1506.03327}}].

\bibitem{JalilianMarian:2004da}
J.~Jalilian-Marian and Y.~V. Kovchegov, \emph{{Inclusive two-gluon and valence
  quark-gluon production in DIS and pA}}, {\emph{Phys. Rev.} {\bfseries D70}
  (2004) 114017} [\href{https://arxiv.org/abs/hep-ph/0405266}{{\ttfamily
  hep-ph/0405266}}].

\bibitem{Kovchegov:2013cva}
Y.~V. Kovchegov and M.~D. Sievert, \emph{{Sivers function in the quasiclassical
  approximation}}, {\emph{Phys.Rev.} {\bfseries D89} (2014) 054035}
  [\href{https://arxiv.org/abs/1310.5028}{{\ttfamily 1310.5028}}].

\bibitem{Kovchegov:2020kxg}
Y.~V. Kovchegov and M.~G. Santiago, \emph{{Lensing mechanism meets small- $x$
  physics: Single transverse spin asymmetry in $p^{\uparrow}+p$ and
  $p^{\uparrow}+A$ collisions}}, {\emph{Phys. Rev. D} {\bfseries 102} (2020)
  014022} [\href{https://arxiv.org/abs/2003.12650}{{\ttfamily 2003.12650}}].

\bibitem{Kang:2013hta}
Z.-B. Kang, Y.-Q. Ma and R.~Venugopalan, \emph{{Quarkonium production in high
  energy proton-nucleus collisions: CGC meets NRQCD}}, {\emph{JHEP} {\bfseries
  01} (2014) 056} [\href{https://arxiv.org/abs/1309.7337}{{\ttfamily
  1309.7337}}].

\bibitem{Ma:2014mri}
Y.-Q. Ma and R.~Venugopalan, \emph{{Comprehensive Description of $J/\psi$
  Production in Proton-Proton Collisions at Collider Energies}},
  {\emph{Phys.Rev.Lett.} {\bfseries 113} (2014) 192301}
  [\href{https://arxiv.org/abs/1408.4075}{{\ttfamily 1408.4075}}].

\bibitem{Ma:2015sia}
Y.-Q. Ma, R.~Venugopalan and H.-F. Zhang, \emph{{$J/\psi$ production and
  suppression in high energy proton-nucleus collisions}}, {\emph{Phys. Rev.}
  {\bfseries D92} (2015) 071901}
  [\href{https://arxiv.org/abs/1503.07772}{{\ttfamily 1503.07772}}].

\bibitem{Qiu:2013qka}
J.-W. Qiu, P.~Sun, B.-W. Xiao and F.~Yuan, \emph{{Universal Suppression of
  Heavy Quarkonium Production in pA Collisions at Low Transverse Momentum}},
  {\emph{Phys. Rev.} {\bfseries D89} (2014) 034007}
  [\href{https://arxiv.org/abs/1310.2230}{{\ttfamily 1310.2230}}].

\bibitem{Sterman:1977wj}
G.~Sterman and S.~Weinberg, \emph{{Jets from Quantum Chromodynamics}},
  {\emph{Phys. Rev. Lett.} {\bfseries 39} (1977) 1436}.

\bibitem{Campbell:2006wx}
J.~M. Campbell, J.~W. Huston and W.~J. Stirling, \emph{{Hard Interactions of
  Quarks and Gluons: A Primer for LHC Physics}}, {\emph{Rept. Prog. Phys.}
  {\bfseries 70} (2007) 89}
  [\href{https://arxiv.org/abs/hep-ph/0611148}{{\ttfamily hep-ph/0611148}}].

\bibitem{Ellis:2007ib}
S.~D. Ellis, J.~Huston, K.~Hatakeyama, P.~Loch and M.~Tonnesmann, \emph{{Jets
  in hadron-hadron collisions}}, {\emph{Prog. Part. Nucl. Phys.} {\bfseries 60}
  (2008) 484} [\href{https://arxiv.org/abs/0712.2447}{{\ttfamily 0712.2447}}].

\bibitem{Boughezal:2015ded}
R.~Boughezal, J.~M. Campbell, R.~K. Ellis, C.~Focke, W.~T. Giele, X.~Liu
  et~al., \emph{{Z-boson production in association with a jet at
  next-to-next-to-leading order in perturbative QCD}}, {\emph{Phys. Rev. Lett.}
  {\bfseries 116} (2016) 152001}
  [\href{https://arxiv.org/abs/1512.01291}{{\ttfamily 1512.01291}}].

\bibitem{Boughezal:2015dra}
R.~Boughezal, F.~Caola, K.~Melnikov, F.~Petriello and M.~Schulze, \emph{{Higgs
  boson production in association with a jet at next-to-next-to-leading
  order}}, {\emph{Phys. Rev. Lett.} {\bfseries 115} (2015) 082003}
  [\href{https://arxiv.org/abs/1504.07922}{{\ttfamily 1504.07922}}].

\bibitem{Olness:2009qd}
F.~I. Olness and D.~E. Soper, \emph{{Correlated theoretical uncertainties for
  the one-jet inclusive cross section}}, {\emph{Phys. Rev.} {\bfseries D81}
  (2010) 035018} [\href{https://arxiv.org/abs/0907.5052}{{\ttfamily
  0907.5052}}].

\bibitem{Vitev:2008rz}
I.~Vitev, S.~Wicks and B.-W. Zhang, \emph{{A Theory of jet shapes and cross
  sections: From hadrons to nuclei}}, {\emph{JHEP} {\bfseries 11} (2008) 093}
  [\href{https://arxiv.org/abs/0810.2807}{{\ttfamily 0810.2807}}].

\bibitem{Vitev:2009rd}
I.~Vitev and B.-W. Zhang, \emph{{Jet tomography of high-energy nucleus-nucleus
  collisions at next-to-leading order}}, {\emph{Phys. Rev. Lett.} {\bfseries
  104} (2010) 132001} [\href{https://arxiv.org/abs/0910.1090}{{\ttfamily
  0910.1090}}].

\bibitem{CMS}
M.~T. for~the CMS~Collaboration, \emph{{Mapping the redistribution of jet
  energy in PbPb collisions using jets with various radius parameters with
  CMS}}, {\emph{Quark Matter 2019 proceedings} (2019) }.

\bibitem{Farhi:1977sg}
E.~Farhi, \emph{{A QCD Test for Jets}}, {\emph{Phys.Rev.Lett.} {\bfseries 39}
  (1977) 1587}.

\bibitem{Georgi:1977sf}
H.~Georgi and M.~Machacek, \emph{{A Simple QCD Prediction of Jet Structure in
  $e^+e^-$ Annihilation}}, {\emph{Phys.Rev.Lett.} {\bfseries 39} (1977) 1237}.

\bibitem{PhysRevLett.41.1581}
G.~C. Fox and S.~Wolfram, \emph{Observables for the analysis of event shapes in
  $e^+ e^-$ annihilation and other processes}, {\emph{Phys. Rev. Lett.}
  {\bfseries 41} (1978) 1581}.

\bibitem{PhysRevLett.41.1585}
C.~L. Basham, L.~S. Brown, S.~D. Ellis and S.~T. Love, \emph{Energy
  correlations in electron-positron annihilation: Testing quantum
  chromodynamics}, {\emph{Phys. Rev. Lett.} {\bfseries 41} (1978) 1585}.

\bibitem{Heister:2003aj}
{\scshape ALEPH Collaboration} collaboration, A.~Heister et~al., \emph{{Studies
  of QCD at $e^+e^-$ centre-of-mass energies between 91 GeV and 209 GeV}},
  {\emph{Eur.Phys.J.} {\bfseries C35} (2004) 457}.

\bibitem{Abdallah:2003xz}
{\scshape DELPHI Collaboration} collaboration, J.~Abdallah et~al., \emph{{A
  Study of the energy evolution of event shape distributions and their means
  with the DELPHI detector at LEP}}, {\emph{Eur.Phys.J.} {\bfseries C29} (2003)
  285} [\href{https://arxiv.org/abs/hep-ex/0307048}{{\ttfamily
  hep-ex/0307048}}].

\bibitem{Achard:2004sv}
{\scshape L3 Collaboration} collaboration, P.~Achard et~al., \emph{{Studies of
  hadronic event structure in $e^{+} e^{-}$ annihilation from 30 GeV to 209 GeV
  with the L3 detector}}, {\emph{Phys.Rept.} {\bfseries 399} (2004) 71}
  [\href{https://arxiv.org/abs/hep-ex/0406049}{{\ttfamily hep-ex/0406049}}].

\bibitem{Abbiendi:2004qz}
{\scshape OPAL Collaboration} collaboration, G.~Abbiendi et~al.,
  \emph{{Measurement of event shape distributions and moments in $e^+e^-$ of
  hadrons at 91 GeV - 209 GeV and a determination of $\alpha_s$}},
  {\emph{Eur.Phys.J.} {\bfseries C40} (2005) 287}
  [\href{https://arxiv.org/abs/hep-ex/0503051}{{\ttfamily hep-ex/0503051}}].

\bibitem{Becher:2008cf}
T.~Becher and M.~D. Schwartz, \emph{{A precise determination of $\alpha_s$ from
  LEP thrust data using effective field theory}}, {\emph{JHEP} {\bfseries 0807}
  (2008) 034} [\href{https://arxiv.org/abs/0803.0342}{{\ttfamily 0803.0342}}].

\bibitem{Chien:2010kc}
Y.-T. Chien and M.~D. Schwartz, \emph{{Resummation of heavy jet mass and
  comparison to LEP data}}, {\emph{JHEP} {\bfseries 1008} (2010) 058}
  [\href{https://arxiv.org/abs/1005.1644}{{\ttfamily 1005.1644}}].

\bibitem{Davison:2008vx}
R.~Davison and B.~Webber, \emph{{Non-Perturbative Contribution to the Thrust
  Distribution in $e^+e^-$ Annihilation}}, {\emph{Eur.Phys.J.} {\bfseries C59}
  (2009) 13} [\href{https://arxiv.org/abs/0809.3326}{{\ttfamily 0809.3326}}].

\bibitem{Gehrmann:2012sc}
T.~Gehrmann, G.~Luisoni and P.~F. Monni, \emph{{Power corrections in the
  dispersive model for a determination of the strong coupling constant from the
  thrust distribution}}, {\emph{Eur. Phys. J. C} {\bfseries 73} (2013) 2265}
  [\href{https://arxiv.org/abs/1210.6945}{{\ttfamily 1210.6945}}].

\bibitem{Hoang:2015hka}
A.~H. Hoang, D.~W. Kolodrubetz, V.~Mateu and I.~W. Stewart, \emph{{Precise
  determination of $\alpha_s$ from the $C$-parameter distribution}},
  {\emph{Phys. Rev. D} {\bfseries 91} (2015) 094018}
  [\href{https://arxiv.org/abs/1501.04111}{{\ttfamily 1501.04111}}].

\bibitem{Kardos:2018kqj}
A.~Kardos, S.~Kluth, G.~Somogyi, Z.~Tulip\'ant and A.~Verbytskyi,
  \emph{{Precise determination of $\alpha _{S}(M_Z)$ from a global fit of
  energy\textendash{}energy correlation to NNLO+NNLL predictions}}, {\emph{Eur.
  Phys. J. C} {\bfseries 78} (2018) 498}
  [\href{https://arxiv.org/abs/1804.09146}{{\ttfamily 1804.09146}}].

\bibitem{Altheimer:2013yza}
A.~Altheimer et~al., \emph{{Boosted Objects and Jet Substructure at the LHC.
  Report of BOOST2012, held at IFIC Valencia, 23rd-27th of July 2012}},
  \href{https://doi.org/10.1140/epjc/s10052-014-2792-8}{\emph{Eur. Phys. J. C}
  {\bfseries 74} (2014) 2792}
  [\href{https://arxiv.org/abs/1311.2708}{{\ttfamily 1311.2708}}].

\bibitem{Ball:2010de}
R.~D. Ball, L.~Del~Debbio, S.~Forte, A.~Guffanti, J.~I. Latorre, J.~Rojo
  et~al., \emph{{A first unbiased global NLO determination of parton
  distributions and their uncertainties}}, {\emph{Nucl. Phys.} {\bfseries B838}
  (2010) 136} [\href{https://arxiv.org/abs/1002.4407}{{\ttfamily 1002.4407}}].

\bibitem{Dulat:2015mca}
S.~Dulat, T.-J. Hou, J.~Gao, M.~Guzzi, J.~Huston, P.~Nadolsky et~al.,
  \emph{{New parton distribution functions from a global analysis of quantum
  chromodynamics}}, {\emph{Phys. Rev.} {\bfseries D93} (2016) 033006}
  [\href{https://arxiv.org/abs/1506.07443}{{\ttfamily 1506.07443}}].

\bibitem{Stump:2003yu}
D.~Stump, J.~Huston, J.~Pumplin, W.-K. Tung, H.~L. Lai, S.~Kuhlmann et~al.,
  \emph{{Inclusive jet production, parton distributions, and the search for new
  physics}}, {\emph{JHEP} {\bfseries 10} (2003) 046}
  [\href{https://arxiv.org/abs/hep-ph/0303013}{{\ttfamily hep-ph/0303013}}].

\bibitem{Cacciari:2011ma}
M.~Cacciari, G.~P. Salam and G.~Soyez, \emph{{FastJet User Manual}},
  {\emph{Eur. Phys. J. C} {\bfseries 72} (2012) 1896}
  [\href{https://arxiv.org/abs/1111.6097}{{\ttfamily 1111.6097}}].

\bibitem{Salam:2007xv}
G.~P. Salam and G.~Soyez, \emph{{A Practical Seedless Infrared-Safe Cone Jet
  Algorithm}}, {\emph{JHEP} {\bfseries 05} (2007) 086}
  [\href{https://arxiv.org/abs/0704.0292}{{\ttfamily 0704.0292}}].

\bibitem{Cacciari:2008gp}
M.~Cacciari, G.~P. Salam and G.~Soyez, \emph{{The anti-$k_t$ jet clustering
  algorithm}}, {\emph{JHEP} {\bfseries 04} (2008) 063}
  [\href{https://arxiv.org/abs/0802.1189}{{\ttfamily 0802.1189}}].

\bibitem{Catani:1993hr}
S.~Catani, Y.~L. Dokshitzer, M.~H. Seymour and B.~R. Webber,
  \emph{{Longitudinally invariant $K_t$ clustering algorithms for hadron hadron
  collisions}}, {\emph{Nucl. Phys. B} {\bfseries 406} (1993) 187}.

\bibitem{Gallicchio:2011xq}
J.~Gallicchio and M.~D. Schwartz, \emph{{Quark and Gluon Tagging at the LHC}},
  {\emph{Phys.Rev.Lett.} {\bfseries 107} (2011) 172001}
  [\href{https://arxiv.org/abs/1106.3076}{{\ttfamily 1106.3076}}].

\bibitem{Gallicchio:2012ez}
J.~Gallicchio and M.~D. Schwartz, \emph{{Quark and Gluon Jet Substructure}},
  {\emph{JHEP} {\bfseries 1304} (2013) 090}
  [\href{https://arxiv.org/abs/1211.7038}{{\ttfamily 1211.7038}}].

\bibitem{Berge:1980dx}
J.~P. Berge et~al., \emph{{Quark Jets from anti-neutrino Interactions. 1. Net
  Charge and Factorization in the Quark Jets}}, {\emph{Nucl. Phys.} {\bfseries
  B184} (1981) 13}.

\bibitem{Albanese:1984nv}
{\scshape European Muon} collaboration, J.~P. Albanese et~al., \emph{{Quark
  Charge Retention in Final State Hadrons From Deep Inelastic Muon
  Scattering}}, {\emph{Phys. Lett.} {\bfseries 144B} (1984) 302}.

\bibitem{Aad:2015cua}
{\scshape ATLAS} collaboration, G.~Aad et~al., \emph{{Measurement of jet charge
  in dijet events from $\sqrt{s}$ = 8 TeV pp collisions with the ATLAS
  detector}}, {\emph{Phys. Rev.} {\bfseries D93} (2016) 052003}
  [\href{https://arxiv.org/abs/1509.05190}{{\ttfamily 1509.05190}}].

\bibitem{Sirunyan:2017tyr}
{\scshape CMS} collaboration, A.~M. Sirunyan et~al., \emph{{Measurements of jet
  charge with dijet events in pp collisions at $\sqrt{s} = 8$ TeV}},
  {\emph{JHEP} {\bfseries 10} (2017) 131}
  [\href{https://arxiv.org/abs/1706.05868}{{\ttfamily 1706.05868}}].

\bibitem{Chen:2019gqo}
S.-Y. Chen, B.-W. Zhang and E.-K. Wang, \emph{{Jet charge in high energy
  nuclear collisions}},
  \href{https://doi.org/10.1088/1674-1137/44/2/024103}{\emph{Chin. Phys. C}
  {\bfseries 44} (2020) 024103}
  [\href{https://arxiv.org/abs/1908.01518}{{\ttfamily 1908.01518}}].

\bibitem{Li:2019dre}
H.~T. Li and I.~Vitev, \emph{{Jet charge modification in dense QCD matter}},
  {\emph{Phys. Rev. D} {\bfseries 101} (2020) 076020}
  [\href{https://arxiv.org/abs/1908.06979}{{\ttfamily 1908.06979}}].

\bibitem{CMScharge}
D.~H. for~the CMS~Collaboration, \emph{{Studies of Quark and Gluon
  Contributions to Jets using Jet Charge Measurements in pp and PbPb
  Collisions}}, {\emph{Quark Matter 2019 proceedings} (2019) }.

\bibitem{Bain:2017wvk}
R.~Bain, L.~Dai, A.~Leibovich, Y.~Makris and T.~Mehen, \emph{{NRQCD Confronts
  LHCb Data on Quarkonium Production within Jets}}, {\emph{Phys. Rev. Lett.}
  {\bfseries 119} (2017) 032002}
  [\href{https://arxiv.org/abs/1702.05525}{{\ttfamily 1702.05525}}].

\bibitem{Anderle:2017cgl}
D.~P. Anderle, T.~Kaufmann, M.~Stratmann, F.~Ringer and I.~Vitev, \emph{{Using
  hadron-in-jet data in a global analysis of $D^{*}$ fragmentation functions}},
  {\emph{Phys. Rev.} {\bfseries D96} (2017) 034028}
  [\href{https://arxiv.org/abs/1706.09857}{{\ttfamily 1706.09857}}].

\bibitem{Chien:2015hda}
Y.-T. Chien and I.~Vitev, \emph{{Towards the understanding of jet shapes and
  cross sections in heavy ion collisions using soft-collinear effective
  theory}}, {\emph{JHEP} {\bfseries 05} (2016) 023}
  [\href{https://arxiv.org/abs/1509.07257}{{\ttfamily 1509.07257}}].

\bibitem{Adamczyk:2017yhe}
{\scshape STAR} collaboration, L.~Adamczyk et~al., \emph{{Measurements of jet
  quenching with semi-inclusive hadron+jet distributions in Au+Au collisions at
  $\sqrt{s_{NN}}$ = 200 GeV}}, {\emph{Phys. Rev.} {\bfseries C96} (2017)
  024905} [\href{https://arxiv.org/abs/1702.01108}{{\ttfamily 1702.01108}}].

\bibitem{Sirunyan:2018gct}
{\scshape CMS} collaboration, A.~M. Sirunyan et~al., \emph{{Measurement of the
  groomed jet mass in PbPb and pp collisions at $ \sqrt{s_{\mathrm{NN}}}=5.02 $
  TeV}}, {\emph{JHEP} {\bfseries 10} (2018) 161}
  [\href{https://arxiv.org/abs/1805.05145}{{\ttfamily 1805.05145}}].

\bibitem{Sirunyan:2018qec}
{\scshape CMS} collaboration, A.~M. Sirunyan et~al., \emph{{Observation of
  Medium-Induced Modifications of Jet Fragmentation in Pb-Pb Collisions at
  $\sqrt{s_{NN}}$ = 5.02~TeV Using Isolated Photon-Tagged Jets}}, {\emph{Phys.
  Rev. Lett.} {\bfseries 121} (2018) 242301}
  [\href{https://arxiv.org/abs/1801.04895}{{\ttfamily 1801.04895}}].

\bibitem{Aaboud:2019oac}
{\scshape ATLAS} collaboration, M.~Aaboud et~al., \emph{{Comparison of
  Fragmentation Functions for Jets Dominated by Light Quarks and Gluons from
  $pp$ and Pb+Pb Collisions in ATLAS}},
  \href{https://doi.org/10.1103/PhysRevLett.123.042001}{\emph{Phys. Rev. Lett.}
  {\bfseries 123} (2019) 042001}
  [\href{https://arxiv.org/abs/1902.10007}{{\ttfamily 1902.10007}}].

\bibitem{Sirunyan:2018ncy}
{\scshape CMS} collaboration, A.~M. Sirunyan et~al., \emph{{Jet Shapes of
  Isolated Photon-Tagged Jets in Pb-Pb and pp Collisions at
  $\sqrt{s_\mathrm{NN}} =$ 5.02 TeV}}, {\emph{Phys. Rev. Lett.} {\bfseries 122}
  (2019) 152001} [\href{https://arxiv.org/abs/1809.08602}{{\ttfamily
  1809.08602}}].

\bibitem{Aaboud:2018hpb}
{\scshape ATLAS} collaboration, M.~Aaboud et~al., \emph{{Measurement of jet
  fragmentation in Pb+Pb and $pp$ collisions at $\sqrt{s_{NN}} = 5.02$ TeV with
  the ATLAS detector}}, {\emph{Phys. Rev.} {\bfseries C98} (2018) 024908}
  [\href{https://arxiv.org/abs/1805.05424}{{\ttfamily 1805.05424}}].

\bibitem{Liu:2018trl}
X.~Liu, F.~Ringer, W.~Vogelsang and F.~Yuan, \emph{{Lepton-jet Correlations in
  Deep Inelastic Scattering at the Electron-Ion Collider}}, {\emph{Phys. Rev.
  Lett.} {\bfseries 122} (2019) 192003}
  [\href{https://arxiv.org/abs/1812.08077}{{\ttfamily 1812.08077}}].

\bibitem{Arratia:2020nxw}
M.~Arratia, Z.-B. Kang, A.~Prokudin and F.~Ringer, \emph{{Jet-based
  measurements of Sivers and Collins asymmetries at the future electron-ion
  collider}}, {\emph{Phys. Rev. D} {\bfseries 102} (2020) 074015}
  [\href{https://arxiv.org/abs/2007.07281}{{\ttfamily 2007.07281}}].

\bibitem{Liu:2020dct}
X.~Liu, F.~Ringer, W.~Vogelsang and F.~Yuan, \emph{{Lepton-jet Correlation in
  Deep Inelastic Scattering}}, {\emph{Phys. Rev. D} {\bfseries 102} (2020)
  094022} [\href{https://arxiv.org/abs/2007.12866}{{\ttfamily 2007.12866}}].

\bibitem{Kang:2021ffh}
Z.-B. Kang, K.~Lee, D.~Y. Shao and F.~Zhao, \emph{{Spin asymmetries in
  electron-jet production at the future electron ion collider}}, {\emph{JHEP}
  {\bfseries 11} (2021) 005}
  [\href{https://arxiv.org/abs/2106.15624}{{\ttfamily 2106.15624}}].

\bibitem{Ellis:2010rwa}
S.~D. Ellis, C.~K. Vermilion, J.~R. Walsh, A.~Hornig and C.~Lee, \emph{{Jet
  Shapes and Jet Algorithms in SCET}}, {\emph{JHEP} {\bfseries 11} (2010) 101}
  [\href{https://arxiv.org/abs/1001.0014}{{\ttfamily 1001.0014}}].

\bibitem{H1:2021wkz}
{\scshape H1} collaboration, V.~Andreev et~al., \emph{{Measurement of
  Lepton-Jet Correlation in Deep-Inelastic Scattering with the H1 Detector
  Using Machine Learning for Unfolding}},
  \href{https://doi.org/10.1103/PhysRevLett.128.132002}{\emph{Phys. Rev. Lett.}
  {\bfseries 128} (2022) 132002}
  [\href{https://arxiv.org/abs/2108.12376}{{\ttfamily 2108.12376}}].

\bibitem{Kang:2020fka}
Z.-B. Kang, X.~Liu, S.~Mantry and D.~Y. Shao, \emph{{Jet Charge: A Flavor Prism
  for Spin Asymmetries at the EIC}}, {\emph{Phys. Rev. Lett.} {\bfseries 125}
  (2020) 242003} [\href{https://arxiv.org/abs/2008.00655}{{\ttfamily
  2008.00655}}].

\bibitem{Buffing:2018ggv}
M.~G.~A. Buffing, Z.-B. Kang, K.~Lee and X.~Liu, \emph{{A transverse momentum
  dependent framework for back-to-back photon+jet production}},
  \href{https://arxiv.org/abs/1812.07549}{{\ttfamily 1812.07549}}.

\bibitem{Chien:2019gyf}
Y.-T. Chien, D.~Y. Shao and B.~Wu, \emph{{Resummation of Boson-Jet Correlation
  at Hadron Colliders}}, {\emph{JHEP} {\bfseries 11} (2019) 025}
  [\href{https://arxiv.org/abs/1905.01335}{{\ttfamily 1905.01335}}].

\bibitem{Kang:2020xez}
Z.-B. Kang, K.~Lee, D.~Y. Shao and J.~Terry, \emph{{The Sivers Asymmetry in
  Hadronic Dijet Production}}, {\emph{JHEP} {\bfseries 02} (2021) 066}
  [\href{https://arxiv.org/abs/2008.05470}{{\ttfamily 2008.05470}}].

\bibitem{Liu:2020jjv}
X.~Liu, F.~Ringer, W.~Vogelsang and F.~Yuan, \emph{{Factorization and its
  Breaking in Dijet Single Transverse Spin Asymmetries in $pp$ Collisions}},
  {\emph{Phys. Rev. D} {\bfseries 102} (2020) 114012}
  [\href{https://arxiv.org/abs/2008.03666}{{\ttfamily 2008.03666}}].

\bibitem{Aaij:2019ctd}
{\scshape LHCb} collaboration, R.~Aaij et~al., \emph{{Measurement of charged
  hadron production in $Z$-tagged jets in proton-proton collisions at $\sqrt{s}
  = 8$ TeV}}, {\emph{Phys. Rev. Lett.} {\bfseries 123} (2019) 232001}
  [\href{https://arxiv.org/abs/1904.08878}{{\ttfamily 1904.08878}}].

\bibitem{bnltalk}
H.~Liu, \emph{{Talk given at RIKEN BNL Workshop Jet Observables at the
  Electron-Ion Collider}}, 2020.

\bibitem{Procura:2009vm}
M.~Procura and I.~W. Stewart, \emph{{Quark Fragmentation within an Identified
  Jet}}, {\emph{Phys. Rev.} {\bfseries D81} (2010) 074009}
  [\href{https://arxiv.org/abs/0911.4980}{{\ttfamily 0911.4980}}].

\bibitem{Jain:2011xz}
A.~Jain, M.~Procura and W.~J. Waalewijn, \emph{{Parton Fragmentation within an
  Identified Jet at NNLL}}, {\emph{JHEP} {\bfseries 05} (2011) 035}
  [\href{https://arxiv.org/abs/1101.4953}{{\ttfamily 1101.4953}}].

\bibitem{Jain:2011iu}
A.~Jain, M.~Procura and W.~J. Waalewijn, \emph{{Fully-Unintegrated Parton
  Distribution and Fragmentation Functions at Perturbative $k_T$}},
  {\emph{JHEP} {\bfseries 04} (2012) 132}
  [\href{https://arxiv.org/abs/1110.0839}{{\ttfamily 1110.0839}}].

\bibitem{Chien:2015ctp}
Y.-T. Chien, Z.-B. Kang, F.~Ringer, I.~Vitev and H.~Xing, \emph{{Jet
  fragmentation functions in proton-proton collisions using soft-collinear
  effective theory}}, {\emph{JHEP} {\bfseries 05} (2016) 125}
  [\href{https://arxiv.org/abs/1512.06851}{{\ttfamily 1512.06851}}].

\bibitem{Kang:2019ahe}
Z.-B. Kang, K.~Lee, J.~Terry and H.~Xing, \emph{{Jet fragmentation functions
  for $Z$-tagged jets}},
  \href{https://doi.org/10.1016/j.physletb.2019.134978}{\emph{Phys. Lett. B}
  {\bfseries 798} (2019) 134978}
  [\href{https://arxiv.org/abs/1906.07187}{{\ttfamily 1906.07187}}].

\bibitem{Arleo:2013tya}
F.~Arleo, M.~Fontannaz, J.-P. Guillet and C.~L. Nguyen, \emph{{Probing
  fragmentation functions from same-side hadron-jet momentum correlations in
  p-p collisions}}, {\emph{JHEP} {\bfseries 04} (2014) 147}
  [\href{https://arxiv.org/abs/1311.7356}{{\ttfamily 1311.7356}}].

\bibitem{Kaufmann:2015hma}
T.~Kaufmann, A.~Mukherjee and W.~Vogelsang, \emph{{Hadron Fragmentation Inside
  Jets in Hadronic Collisions}}, {\emph{Phys. Rev. D} {\bfseries 92} (2015)
  054015} [\href{https://arxiv.org/abs/1506.01415}{{\ttfamily 1506.01415}}].

\bibitem{Kang:2016ehg}
Z.-B. Kang, F.~Ringer and I.~Vitev, \emph{{Jet substructure using
  semi-inclusive jet functions in SCET}}, {\emph{JHEP} {\bfseries 11} (2016)
  155} [\href{https://arxiv.org/abs/1606.07063}{{\ttfamily 1606.07063}}].

\bibitem{Dai:2016hzf}
L.~Dai, C.~Kim and A.~K. Leibovich, \emph{{Fragmentation of a Jet with Small
  Radius}}, {\emph{Phys. Rev. D} {\bfseries 94} (2016) 114023}
  [\href{https://arxiv.org/abs/1606.07411}{{\ttfamily 1606.07411}}].

\bibitem{Kang:2017yde}
Z.-B. Kang, J.-W. Qiu, F.~Ringer, H.~Xing and H.~Zhang, \emph{{$J/\psi$
  production and polarization within a jet}}, {\emph{Phys. Rev. Lett.}
  {\bfseries 119} (2017) 032001}
  [\href{https://arxiv.org/abs/1702.03287}{{\ttfamily 1702.03287}}].

\bibitem{Wang:2020kar}
L.~Wang, Z.-B. Kang, H.~Xing and B.-W. Zhang, \emph{{Semi-inclusive jet
  functions and jet substructure in $J_{E_T}^{(I)}$ and $J_{E_T}^{(II)}$
  algorithms}}, {\emph{Phys. Rev. D} {\bfseries 103} (2021) 054043}
  [\href{https://arxiv.org/abs/2003.03796}{{\ttfamily 2003.03796}}].

\bibitem{Kang:2016mcy}
Z.-B. Kang, F.~Ringer and I.~Vitev, \emph{{The semi-inclusive jet function in
  SCET and small radius resummation for inclusive jet production}},
  {\emph{JHEP} {\bfseries 10} (2016) 125}
  [\href{https://arxiv.org/abs/1606.06732}{{\ttfamily 1606.06732}}].

\bibitem{Aad:2011sc}
{\scshape ATLAS} collaboration, G.~Aad et~al., \emph{{Measurement of the jet
  fragmentation function and transverse profile in proton-proton collisions at
  a center-of-mass energy of 7 TeV with the ATLAS detector}}, {\emph{Eur. Phys.
  J.} {\bfseries C71} (2011) 1795}
  [\href{https://arxiv.org/abs/1109.5816}{{\ttfamily 1109.5816}}].

\bibitem{Aaboud:2017tke}
{\scshape ATLAS} collaboration, M.~Aaboud et~al., \emph{{Measurement of jet
  fragmentation in 5.02 TeV proton-lead and proton-proton collisions with the
  ATLAS detector}}, {\emph{Nucl. Phys.} {\bfseries A978} (2018) 65}
  [\href{https://arxiv.org/abs/1706.02859}{{\ttfamily 1706.02859}}].

\bibitem{Aad:2011td}
{\scshape ATLAS} collaboration, G.~Aad et~al., \emph{{Measurement of $D^{*+/-}$
  meson production in jets from pp collisions at $\sqrt(s)$ = 7 TeV with the
  ATLAS detector}}, {\emph{Phys. Rev.} {\bfseries D85} (2012) 052005}
  [\href{https://arxiv.org/abs/1112.4432}{{\ttfamily 1112.4432}}].

\bibitem{CMS:2018ovh}
{\scshape CMS} collaboration, C.~Collaboration, \emph{{Measurement of the
  radial profile of $D^{0}$ mesons in jets produced in pp and PbPb collisions
  at 5.02 TeV}}, .

\bibitem{Acharya:2019zup}
{\scshape ALICE} collaboration, S.~Acharya et~al., \emph{{Measurement of the
  production of charm jets tagged with D$^{0}$ mesons in pp collisions at $
  \sqrt{\mathrm{s}}=7 $ TeV}},
  \href{https://doi.org/10.1007/JHEP08(2019)133}{\emph{JHEP} {\bfseries 08}
  (2019) 133} [\href{https://arxiv.org/abs/1905.02510}{{\ttfamily
  1905.02510}}].

\bibitem{Aaij:2017fak}
{\scshape LHCb} collaboration, R.~Aaij et~al., \emph{{Study of J/$\psi$
  Production in Jets}}, {\emph{Phys. Rev. Lett.} {\bfseries 118} (2017) 192001}
  [\href{https://arxiv.org/abs/1701.05116}{{\ttfamily 1701.05116}}].

\bibitem{CMS:2018mjn}
{\scshape CMS} collaboration, \emph{{Production of prompt and nonprompt ${\rm
  J}\hspace{-.08em}/\hspace{-.14em}\psi$ mesons in jets in pp collisions at
  $\sqrt{s} = 5.02~\mathrm{TeV}$}}, .

\bibitem{Becher:2006mr}
T.~Becher, M.~Neubert and B.~D. Pecjak, \emph{{Factorization and Momentum-Space
  Resummation in Deep-Inelastic Scattering}}, {\emph{JHEP} {\bfseries 01}
  (2007) 076} [\href{https://arxiv.org/abs/hep-ph/0607228}{{\ttfamily
  hep-ph/0607228}}].

\bibitem{Becher:2009th}
T.~Becher and M.~D. Schwartz, \emph{{Direct photon production with effective
  field theory}}, {\emph{JHEP} {\bfseries 02} (2010) 040}
  [\href{https://arxiv.org/abs/0911.0681}{{\ttfamily 0911.0681}}].

\bibitem{Echevarria:2012pw}
M.~G. Echevarria, A.~Idilbi, A.~Sch{\"a}fer and I.~Scimemi,
  \emph{{Model-Independent Evolution of Transverse Momentum Dependent
  Distribution Functions (TMDs) at NNLL}}, {\emph{Eur.Phys.J.} {\bfseries C73}
  (2013) 2636} [\href{https://arxiv.org/abs/1208.1281}{{\ttfamily 1208.1281}}].

\bibitem{Chien:2015cka}
Y.-T. Chien, A.~Hornig and C.~Lee, \emph{{Soft-collinear mode for jet cross
  sections in soft collinear effective theory}}, {\emph{Phys. Rev. D}
  {\bfseries 93} (2016) 014033}
  [\href{https://arxiv.org/abs/1509.04287}{{\ttfamily 1509.04287}}].

\bibitem{Procura:2011aq}
M.~Procura and W.~J. Waalewijn, \emph{{Fragmentation in Jets: Cone and
  Threshold Effects}}, {\emph{Phys. Rev.} {\bfseries D85} (2012) 114041}
  [\href{https://arxiv.org/abs/1111.6605}{{\ttfamily 1111.6605}}].

\bibitem{Waalewijn:2012sv}
W.~J. Waalewijn, \emph{{Calculating the Charge of a Jet}}, {\emph{Phys. Rev. D}
  {\bfseries 86} (2012) 094030}
  [\href{https://arxiv.org/abs/1209.3019}{{\ttfamily 1209.3019}}].

\bibitem{Chien:2015vja}
Y.-T. Chien, A.~Emerman, Z.-B. Kang, G.~Ovanesyan and I.~Vitev, \emph{{Jet
  Quenching from QCD Evolution}}, {\emph{Phys. Rev. D} {\bfseries 93} (2016)
  074030} [\href{https://arxiv.org/abs/1509.02936}{{\ttfamily 1509.02936}}].

\bibitem{Kang:2017glf}
Z.-B. Kang, X.~Liu, F.~Ringer and H.~Xing, \emph{{The transverse momentum
  distribution of hadrons within jets}}, {\emph{JHEP} {\bfseries 11} (2017)
  068} [\href{https://arxiv.org/abs/1705.08443}{{\ttfamily 1705.08443}}].

\bibitem{Larkoski:2014wba}
A.~J. Larkoski, S.~Marzani, G.~Soyez and J.~Thaler, \emph{{Soft Drop}},
  {\emph{JHEP} {\bfseries 05} (2014) 146}
  [\href{https://arxiv.org/abs/1402.2657}{{\ttfamily 1402.2657}}].

\bibitem{Makris:2017arq}
Y.~Makris, D.~Neill and V.~Vaidya, \emph{{Probing Transverse-Momentum Dependent
  Evolution With Groomed Jets}}, {\emph{JHEP} {\bfseries 07} (2018) 167}
  [\href{https://arxiv.org/abs/1712.07653}{{\ttfamily 1712.07653}}].

\bibitem{Gutierrez-Reyes:2019msa}
D.~Gutierrez-Reyes, Y.~Makris, V.~Vaidya, I.~Scimemi and L.~Zoppi,
  \emph{{Probing Transverse-Momentum Distributions With Groomed Jets}},
  {\emph{JHEP} {\bfseries 08} (2019) 161}
  [\href{https://arxiv.org/abs/1907.05896}{{\ttfamily 1907.05896}}].

\bibitem{Kang:2020xyq}
Z.-B. Kang, K.~Lee and F.~Zhao, \emph{{Polarized jet fragmentation functions}},
  {\emph{Phys. Lett. B} {\bfseries 809} (2020) 135756}
  [\href{https://arxiv.org/abs/2005.02398}{{\ttfamily 2005.02398}}].

\bibitem{Guan:2018ckx}
{\scshape Belle} collaboration, Y.~Guan et~al., \emph{{Observation of
  Transverse $\Lambda/\bar{\Lambda}$ Hyperon Polarization in $e^+e^-$
  Annihilation at Belle}}, {\emph{Phys. Rev. Lett.} {\bfseries 122} (2019)
  042001} [\href{https://arxiv.org/abs/1808.05000}{{\ttfamily 1808.05000}}].

\bibitem{Callos:2020qtu}
D.~Callos, Z.-B. Kang and J.~Terry, \emph{{Extracting the transverse momentum
  dependent polarizing fragmentation functions}}, {\emph{Phys. Rev. D}
  {\bfseries 102} (2020) 096007}
  [\href{https://arxiv.org/abs/2003.04828}{{\ttfamily 2003.04828}}].

\bibitem{DAlesio:2020wjq}
U.~D'Alesio, F.~Murgia and M.~Zaccheddu, \emph{{First extraction of the
  $\Lambda$ polarizing fragmentation function from Belle $e^+e^-$ data}},
  {\emph{Phys. Rev. D} {\bfseries 102} (2020) 054001}
  [\href{https://arxiv.org/abs/2003.01128}{{\ttfamily 2003.01128}}].

\bibitem{DAlesio:2017bvu}
U.~D'Alesio, F.~Murgia and C.~Pisano, \emph{{Testing the universality of the
  Collins function in pion-jet production at RHIC}}, {\emph{Phys. Lett. B}
  {\bfseries 773} (2017) 300}
  [\href{https://arxiv.org/abs/1707.00914}{{\ttfamily 1707.00914}}].

\bibitem{Kang:2017btw}
Z.-B. Kang, A.~Prokudin, F.~Ringer and F.~Yuan, \emph{{Collins azimuthal
  asymmetries of hadron production inside jets}}, {\emph{Phys. Lett.}
  {\bfseries B774} (2017) 635}
  [\href{https://arxiv.org/abs/1707.00913}{{\ttfamily 1707.00913}}].

\bibitem{Adamczyk:2017wld}
{\scshape STAR} collaboration, L.~Adamczyk et~al., \emph{{Azimuthal transverse
  single-spin asymmetries of inclusive jets and charged pions within jets from
  polarized-proton collisions at $\sqrt{s} = 500$ GeV}}, {\emph{Phys. Rev. D}
  {\bfseries 97} (2018) 032004}
  [\href{https://arxiv.org/abs/1708.07080}{{\ttfamily 1708.07080}}].

\bibitem{Yuan:2007nd}
F.~Yuan, \emph{{Azimuthal asymmetric distribution of hadrons inside a jet at
  hadron collider}}, {\emph{Phys.Rev.Lett.} {\bfseries 100} (2008) 032003}
  [\href{https://arxiv.org/abs/0709.3272}{{\ttfamily 0709.3272}}].

\bibitem{DAlesio:2010sag}
U.~D'Alesio, F.~Murgia and C.~Pisano, \emph{{Azimuthal asymmetries for hadron
  distributions inside a jet in hadronic collisions}}, {\emph{Phys. Rev. D}
  {\bfseries 83} (2011) 034021}
  [\href{https://arxiv.org/abs/1011.2692}{{\ttfamily 1011.2692}}].

\bibitem{Aschenauer:2016our}
E.-C. Aschenauer et~al., \emph{{The RHIC Cold QCD Plan for 2017 to 2023: A
  Portal to the EIC}},  \href{https://arxiv.org/abs/1602.03922}{{\ttfamily
  1602.03922}}.

\bibitem{Baumgart:2014upa}
M.~Baumgart, A.~K. Leibovich, T.~Mehen and I.~Z. Rothstein, \emph{{Probing
  Quarkonium Production Mechanisms with Jet Substructure}}, {\emph{JHEP}
  {\bfseries 11} (2014) 003} [\href{https://arxiv.org/abs/1406.2295}{{\ttfamily
  1406.2295}}].

\bibitem{Bodwin:1994jh}
G.~T. Bodwin, E.~Braaten and G.~Lepage, \emph{{Rigorous QCD analysis of
  inclusive annihilation and production of heavy quarkonium}}, {\emph{Phys.
  Rev. D} {\bfseries 51} (1995) 1125}
  [\href{https://arxiv.org/abs/hep-ph/9407339}{{\ttfamily hep-ph/9407339}}].

\bibitem{Braaten:1993rw}
E.~Braaten and T.~C. Yuan, \emph{{Gluon fragmentation into heavy quarkonium}},
  {\emph{Phys. Rev. Lett.} {\bfseries 71} (1993) 1673}
  [\href{https://arxiv.org/abs/hep-ph/9303205}{{\ttfamily hep-ph/9303205}}].

\bibitem{Braaten:1993mp}
E.~Braaten, K.-m. Cheung and T.~C. Yuan, \emph{{Z0 decay into charmonium via
  charm quark fragmentation}}, {\emph{Phys. Rev. D} {\bfseries 48} (1993) 4230}
  [\href{https://arxiv.org/abs/hep-ph/9302307}{{\ttfamily hep-ph/9302307}}].

\bibitem{Braaten:1993jn}
E.~Braaten, K.-m. Cheung and T.~C. Yuan, \emph{{Perturbative QCD fragmentation
  functions for $B_c$ and $B_{c}$ * production}}, {\emph{Phys. Rev. D}
  {\bfseries 48} (1993) 5049}
  [\href{https://arxiv.org/abs/hep-ph/9305206}{{\ttfamily hep-ph/9305206}}].

\bibitem{Braaten:1994kd}
E.~Braaten and T.~C. Yuan, \emph{{Gluon fragmentation into P wave heavy
  quarkonium}}, {\emph{Phys. Rev. D} {\bfseries 50} (1994) 3176}
  [\href{https://arxiv.org/abs/hep-ph/9403401}{{\ttfamily hep-ph/9403401}}].

\bibitem{Chapon:2020heu}
E.~Chapon et~al., \emph{{Prospects for quarkonium studies at the
  high-luminosity LHC}}, {\emph{Prog. Part. Nucl. Phys.} {\bfseries 122} (2022)
  103906} [\href{https://arxiv.org/abs/2012.14161}{{\ttfamily 2012.14161}}].

\bibitem{Butenschoen:2011yh}
M.~Butenschoen and B.~A. Kniehl, \emph{{World data of J/psi production
  consolidate NRQCD factorization at NLO}}, {\emph{Phys. Rev. D} {\bfseries 84}
  (2011) 051501} [\href{https://arxiv.org/abs/1105.0820}{{\ttfamily
  1105.0820}}].

\bibitem{Butenschoen:2012qr}
M.~Butenschoen and B.~A. Kniehl, \emph{{Next-to-leading-order tests of NRQCD
  factorization with $J/\psi$ yield and polarization}}, {\emph{Mod. Phys.
  Lett.} {\bfseries A28} (2013) 1350027}
  [\href{https://arxiv.org/abs/1212.2037}{{\ttfamily 1212.2037}}].

\bibitem{Chao:2012iv}
K.-T. Chao, Y.-Q. Ma, H.-S. Shao, K.~Wang and Y.-J. Zhang, \emph{{$J/\psi$
  Polarization at Hadron Colliders in Nonrelativistic QCD}},
  {\emph{Phys.Rev.Lett.} {\bfseries 108} (2012) 242004}
  [\href{https://arxiv.org/abs/1201.2675}{{\ttfamily 1201.2675}}].

\bibitem{Bodwin:2014gia}
G.~T. Bodwin, H.~S. Chung, U.-R. Kim and J.~Lee, \emph{{Fragmentation
  contributions to $J/\psi$ production at the Tevatron and the LHC}},
  {\emph{Phys. Rev. Lett.} {\bfseries 113} (2014) 022001}
  [\href{https://arxiv.org/abs/1403.3612}{{\ttfamily 1403.3612}}].

\bibitem{Bain:2016clc}
R.~Bain, L.~Dai, A.~Hornig, A.~K. Leibovich, Y.~Makris and T.~Mehen,
  \emph{{Analytic and Monte Carlo Studies of Jets with Heavy Mesons and
  Quarkonia}}, {\emph{JHEP} {\bfseries 06} (2016) 121}
  [\href{https://arxiv.org/abs/1603.06981}{{\ttfamily 1603.06981}}].

\bibitem{Berger:2003iw}
C.~F. Berger, T.~Kucs and G.~F. Sterman, \emph{{Event shape / energy flow
  correlations}}, {\emph{Phys. Rev. D} {\bfseries 68} (2003) 014012}
  [\href{https://arxiv.org/abs/hep-ph/0303051}{{\ttfamily hep-ph/0303051}}].

\bibitem{Bain:2016rrv}
R.~Bain, Y.~Makris and T.~Mehen, \emph{{Transverse Momentum Dependent
  Fragmenting Jet Functions with Applications to Quarkonium Production}},
  {\emph{JHEP} {\bfseries 11} (2016) 144}
  [\href{https://arxiv.org/abs/1610.06508}{{\ttfamily 1610.06508}}].

\bibitem{Fleming:2019pzj}
S.~Fleming, Y.~Makris and T.~Mehen, \emph{{An effective field theory approach
  to quarkonium at small transverse momentum}}, {\emph{JHEP} {\bfseries 04}
  (2020) 122} [\href{https://arxiv.org/abs/1910.03586}{{\ttfamily
  1910.03586}}].

\bibitem{Echevarria:2020qjk}
M.~G. Echevarria, Y.~Makris and I.~Scimemi, \emph{{Quarkonium TMD fragmentation
  functions in NRQCD}}, {\emph{JHEP} {\bfseries 10} (2020) 164}
  [\href{https://arxiv.org/abs/2007.05547}{{\ttfamily 2007.05547}}].

\bibitem{Boer:2020bbd}
D.~Boer, U.~D'Alesio, F.~Murgia, C.~Pisano and P.~Taels,
  \emph{{J/\ensuremath{\psi} meson production in SIDIS: matching high and low
  transverse momentum}}, {\emph{JHEP} {\bfseries 09} (2020) 040}
  [\href{https://arxiv.org/abs/2004.06740}{{\ttfamily 2004.06740}}].

\bibitem{DAlesio:2020eqo}
U.~D'Alesio, L.~Maxia, F.~Murgia, C.~Pisano and S.~Rajesh, \emph{{Process
  dependence of the gluon Sivers function in $p^\uparrow p \to J/\psi + X$
  within a TMD scheme in NRQCD}}, {\emph{Phys. Rev. D} {\bfseries 102} (2020)
  094011} [\href{https://arxiv.org/abs/2007.03353}{{\ttfamily 2007.03353}}].

\bibitem{Ali:1984yp}
A.~Ali, E.~Pietarinen and W.~Stirling, \emph{{Transverse Energy-energy
  Correlations: A Test of Perturbative {QCD} for the Proton - Anti-proton
  Collider}}, {\emph{Phys.\ Lett.\ B} {\bfseries 141} (1984) 447}.

\bibitem{Basham:1978bw}
C.~Basham, L.~S. Brown, S.~D. Ellis and S.~T. Love, \emph{{Energy Correlations
  in electron - Positron Annihilation: Testing QCD}}, {\emph{Phys.\ Rev.\
  Lett.} {\bfseries 41} (1978) 1585}.

\bibitem{Ali:2012rn}
A.~Ali, F.~Barreiro, J.~Llorente and W.~Wang, \emph{{Transverse Energy-Energy
  Correlations in Next-to-Leading Order in $\alpha_s$ at the LHC}},
  {\emph{Phys. Rev.} {\bfseries D86} (2012) 114017}
  [\href{https://arxiv.org/abs/1205.1689}{{\ttfamily 1205.1689}}].

\bibitem{Gao:2019ojf}
A.~Gao, H.~T. Li, I.~Moult and H.~X. Zhu, \emph{{Precision QCD Event Shapes at
  Hadron Colliders: The Transverse Energy-Energy Correlator in the Back-to-Back
  Limit}}, {\emph{Phys. Rev. Lett.} {\bfseries 123} (2019) 062001}
  [\href{https://arxiv.org/abs/1901.04497}{{\ttfamily 1901.04497}}].

\bibitem{Li:2020bub}
H.~T. Li, I.~Vitev and Y.~J. Zhu, \emph{{Transverse-Energy-Energy Correlations
  in Deep Inelastic Scattering}}, {\emph{JHEP} {\bfseries 11} (2020) 051}
  [\href{https://arxiv.org/abs/2006.02437}{{\ttfamily 2006.02437}}].

\bibitem{Lubbert:2016rku}
T.~L\"ubbert, J.~Oredsson and M.~Stahlhofen, \emph{{Rapidity renormalized TMD
  soft and beam functions at two loops}}, {\emph{JHEP} {\bfseries 03} (2016)
  168} [\href{https://arxiv.org/abs/1602.01829}{{\ttfamily 1602.01829}}].

\bibitem{Li:2021txc}
H.~T. Li, Y.~Makris and I.~Vitev, \emph{{Energy-energy correlators in Deep
  Inelastic Scattering}}, {\emph{Phys. Rev. D} {\bfseries 103} (2021) 094005}
  [\href{https://arxiv.org/abs/2102.05669}{{\ttfamily 2102.05669}}].

\bibitem{Sjostrand:2007gs}
T.~Sjostrand, S.~Mrenna and P.~Z. Skands, \emph{{A Brief Introduction to PYTHIA
  8.1}}, {\emph{Comput. Phys. Commun.} {\bfseries 178} (2008) 852}
  [\href{https://arxiv.org/abs/0710.3820}{{\ttfamily 0710.3820}}].

\bibitem{Sjostrand:2014zea}
T.~Sj{\"o}strand, S.~Ask, J.~R. Christiansen, R.~Corke, N.~Desai, P.~Ilten
  et~al., \emph{{An Introduction to PYTHIA 8.2}}, {\emph{Comput. Phys. Commun.}
  {\bfseries 191} (2015) 159}
  [\href{https://arxiv.org/abs/1410.3012}{{\ttfamily 1410.3012}}].

\bibitem{Bauer:2008dt}
C.~W. Bauer, S.~Fleming, C.~Lee and G.~Sterman, \emph{Factorization of
  {$e^+e^-$} event shape distributions with hadronic final states in {Soft
  Collinear Effective Theory}}, {\emph{Phys. Rev.} {\bfseries D78} (2008)
  034027} [\href{https://arxiv.org/abs/0801.4569}{{\ttfamily 0801.4569}}].

\bibitem{Jouttenus:2011wh}
T.~T. Jouttenus, I.~W. Stewart, F.~J. Tackmann and W.~J. Waalewijn, \emph{{The
  Soft Function for Exclusive N-Jet Production at Hadron Colliders}},
  \href{https://doi.org/10.1103/PhysRevD.83.114030}{\emph{Phys.Rev.} {\bfseries
  D83} (2011) 114030} [\href{https://arxiv.org/abs/1102.4344}{{\ttfamily
  1102.4344}}].

\bibitem{Gyulassy:1990ye}
M.~Gyulassy and M.~Plumer, \emph{{Jet Quenching in Dense Matter}}, {\emph{Phys.
  Lett.} {\bfseries B243} (1990) 432}.

\bibitem{Gyulassy:2003mc}
M.~Gyulassy, I.~Vitev, X.-N. Wang and B.-W. Zhang, \emph{{Jet quenching and
  radiative energy loss in dense nuclear matter}},
  \href{https://arxiv.org/abs/nucl-th/0302077}{{\ttfamily nucl-th/0302077}}.

\bibitem{Gyulassy:1993hr}
M.~Gyulassy and X.-n. Wang, \emph{{Multiple collisions and induced gluon
  Bremsstrahlung in QCD}}, {\emph{Nucl. Phys.} {\bfseries B420} (1994) 583}
  [\href{https://arxiv.org/abs/nucl-th/9306003}{{\ttfamily nucl-th/9306003}}].

\bibitem{Landau:1953um}
L.~D. Landau and I.~Pomeranchuk, \emph{{Limits of applicability of the theory
  of bremsstrahlung electrons and pair production at high-energies}},
  {\emph{Dokl. Akad. Nauk Ser. Fiz.} {\bfseries 92} (1953) 535}.

\bibitem{Migdal:1956tc}
A.~B. Migdal, \emph{{Bremsstrahlung and pair production in condensed media at
  high-energies}}, {\emph{Phys. Rev.} {\bfseries 103} (1956) 1811}.

\bibitem{Zakharov:1996fv}
B.~Zakharov, \emph{{Fully quantum treatment of the Landau-Pomeranchuk-Migdal
  effect in QED and QCD}}, {\emph{JETP Lett.} {\bfseries 63} (1996) 952}
  [\href{https://arxiv.org/abs/hep-ph/9607440}{{\ttfamily hep-ph/9607440}}].

\bibitem{Zakharov:1997uu}
B.~G. Zakharov, \emph{{Radiative energy loss of high-energy quarks in finite
  size nuclear matter and quark - gluon plasma}}, {\emph{JETP Lett.} {\bfseries
  65} (1997) 615} [\href{https://arxiv.org/abs/hep-ph/9704255}{{\ttfamily
  hep-ph/9704255}}].

\bibitem{Baier:1996sk}
R.~Baier, Y.~L. Dokshitzer, A.~H. Mueller, S.~Peigne and D.~Schiff,
  \emph{{Radiative energy loss and $p_T$-broadening of high energy partons in
  nuclei}}, {\emph{Nucl. Phys.} {\bfseries B484} (1997) 265}
  [\href{https://arxiv.org/abs/hep-ph/9608322}{{\ttfamily hep-ph/9608322}}].

\bibitem{Baier:1998kq}
R.~Baier, Y.~L. Dokshitzer, A.~H. Mueller and D.~Schiff, \emph{{Medium induced
  radiative energy loss: Equivalence between the BDMPS and Zakharov
  formalisms}}, {\emph{Nucl. Phys.} {\bfseries B531} (1998) 403}
  [\href{https://arxiv.org/abs/hep-ph/9804212}{{\ttfamily hep-ph/9804212}}].

\bibitem{Arnold:2002ja}
P.~B. Arnold, G.~D. Moore and L.~G. Yaffe, \emph{{Photon and gluon emission in
  relativistic plasmas}}, {\emph{JHEP} {\bfseries 06} (2002) 030}
  [\href{https://arxiv.org/abs/hep-ph/0204343}{{\ttfamily hep-ph/0204343}}].

\bibitem{Gyulassy:1999zd}
M.~Gyulassy, P.~Levai and I.~Vitev, \emph{{Jet quenching in thin quark gluon
  plasmas. 1. Formalism}}, {\emph{Nucl. Phys.} {\bfseries B571} (2000) 197}
  [\href{https://arxiv.org/abs/hep-ph/9907461}{{\ttfamily hep-ph/9907461}}].

\bibitem{Gyulassy:2000fs}
M.~Gyulassy, P.~Levai and I.~Vitev, \emph{{NonAbelian energy loss at finite
  opacity}}, {\emph{Phys.Rev.Lett.} {\bfseries 85} (2000) 5535}
  [\href{https://arxiv.org/abs/nucl-th/0005032}{{\ttfamily nucl-th/0005032}}].

\bibitem{Gyulassy:2000er}
M.~Gyulassy, P.~Levai and I.~Vitev, \emph{{Reaction operator approach to
  non-Abelian energy loss}}, {\emph{Nucl.Phys.} {\bfseries B594} (2001) 371}
  [\href{https://arxiv.org/abs/nucl-th/0006010}{{\ttfamily nucl-th/0006010}}].

\bibitem{Wang:2001ifa}
X.-N. Wang and X.-f. Guo, \emph{{Multiple parton scattering in nuclei: Parton
  energy loss}}, {\emph{Nucl. Phys.} {\bfseries A696} (2001) 788}
  [\href{https://arxiv.org/abs/hep-ph/0102230}{{\ttfamily hep-ph/0102230}}].

\bibitem{Guo:2006kz}
Y.~Guo, B.-W. Zhang and E.~Wang, \emph{{Parton Energy Loss at Twist-Six in
  Deeply Inelastic e-A Scattering}}, {\emph{Phys. Lett.} {\bfseries B641}
  (2006) 38} [\href{https://arxiv.org/abs/hep-ph/0606312}{{\ttfamily
  hep-ph/0606312}}].

\bibitem{Wang:2009qb}
W.-t. Deng and X.-N. Wang, \emph{{Multiple Parton Scattering in Nuclei:
  Modified DGLAP Evolution for Fragmentation Functions}}, {\emph{Phys. Rev.}
  {\bfseries C81} (2010) 024902}
  [\href{https://arxiv.org/abs/0910.3403}{{\ttfamily 0910.3403}}].

\bibitem{Ovanesyan:2011xy}
G.~Ovanesyan and I.~Vitev, \emph{{An effective theory for jet propagation in
  dense QCD matter: jet broadening and medium-induced bremsstrahlung}},
  {\emph{JHEP} {\bfseries 06} (2011) 080}
  [\href{https://arxiv.org/abs/1103.1074}{{\ttfamily 1103.1074}}].

\bibitem{Ovanesyan:2011kn}
G.~Ovanesyan and I.~Vitev, \emph{{Medium-induced parton splitting kernels from
  Soft Collinear Effective Theory with Glauber gluons}}, {\emph{Phys.Lett.}
  {\bfseries B706} (2012) 371}
  [\href{https://arxiv.org/abs/1109.5619}{{\ttfamily 1109.5619}}].

\bibitem{Blaizot:2012fh}
J.-P. Blaizot, F.~Dominguez, E.~Iancu and Y.~Mehtar-Tani, \emph{{Medium-induced
  gluon branching}}, {\emph{JHEP} {\bfseries 01} (2013) 143}
  [\href{https://arxiv.org/abs/1209.4585}{{\ttfamily 1209.4585}}].

\bibitem{Fickinger:2013xwa}
M.~Fickinger, G.~Ovanesyan and I.~Vitev, \emph{{Angular distributions of higher
  order splitting functions in the vacuum and in dense QCD matter}},
  {\emph{JHEP} {\bfseries 07} (2013) 059}
  [\href{https://arxiv.org/abs/1304.3497}{{\ttfamily 1304.3497}}].

\bibitem{Apolinario:2014csa}
L.~Apolinerio, N.~Armesto, J.~G. Milhano and C.~A. Salgado,
  \emph{{Medium-induced gluon radiation and colour decoherence beyond the soft
  approximation}}, {\emph{JHEP} {\bfseries 02} (2015) 119}
  [\href{https://arxiv.org/abs/1407.0599}{{\ttfamily 1407.0599}}].

\bibitem{Ovanesyan:2015dop}
G.~Ovanesyan, F.~Ringer and I.~Vitev, \emph{{Initial-state splitting kernels in
  cold nuclear matter}}, {\emph{Phys. Lett.} {\bfseries B760} (2016) 706}
  [\href{https://arxiv.org/abs/1512.00006}{{\ttfamily 1512.00006}}].

\bibitem{Kang:2016ofv}
Z.-B. Kang, F.~Ringer and I.~Vitev, \emph{{Effective field theory approach to
  open heavy flavor production in heavy-ion collisions}}, {\emph{JHEP}
  {\bfseries 03} (2017) 146}
  [\href{https://arxiv.org/abs/1610.02043}{{\ttfamily 1610.02043}}].

\bibitem{Sievert:2018imd}
M.~D. Sievert and I.~Vitev, \emph{{Quark branching in QCD matter to any order
  in opacity beyond the soft gluon emission limit}}, {\emph{Phys. Rev.}
  {\bfseries D98} (2018) 094010}
  [\href{https://arxiv.org/abs/1807.03799}{{\ttfamily 1807.03799}}].

\bibitem{Sievert:2019cwq}
M.~D. Sievert, I.~Vitev and B.~Yoon, \emph{{A complete set of in-medium
  splitting functions to any order in opacity}}, {\emph{Phys. Lett.} {\bfseries
  B795} (2019) 502} [\href{https://arxiv.org/abs/1903.06170}{{\ttfamily
  1903.06170}}].

\bibitem{Sadofyev:2021ohn}
A.~V. Sadofyev, M.~D. Sievert and I.~Vitev, \emph{{Ab~initio coupling of jets
  to collective flow in the opacity expansion approach}},
  \href{https://doi.org/10.1103/PhysRevD.104.094044}{\emph{Phys. Rev. D}
  {\bfseries 104} (2021) 094044}
  [\href{https://arxiv.org/abs/2104.09513}{{\ttfamily 2104.09513}}].

\bibitem{Ke:2023ixa}
W.~Ke and I.~Vitev, \emph{{Understanding parton evolution in matter from
  renormalization group analysis}},
  \href{https://arxiv.org/abs/2301.11940}{{\ttfamily 2301.11940}}.

\bibitem{Chang:2014fba}
N.-B. Chang, W.-T. Deng and X.-N. Wang, \emph{{Initial conditions for the
  modified evolution of fragmentation functions in the nuclear medium}},
  {\emph{Phys. Rev.} {\bfseries C89} (2014) 034911}
  [\href{https://arxiv.org/abs/1401.5109}{{\ttfamily 1401.5109}}].

\bibitem{Kang:2014xsa}
Z.-B. Kang, R.~Lashof-Regas, G.~Ovanesyan, P.~Saad and I.~Vitev, \emph{{Jet
  quenching phenomenology from soft-collinear effective theory with Glauber
  gluons}}, {\emph{Phys. Rev. Lett.} {\bfseries 114} (2015) 092002}
  [\href{https://arxiv.org/abs/1405.2612}{{\ttfamily 1405.2612}}].

\bibitem{Kang:2017frl}
Z.-B. Kang, F.~Ringer and I.~Vitev, \emph{{Inclusive production of small radius
  jets in heavy-ion collisions}}, {\emph{Phys. Lett.} {\bfseries B769} (2017)
  242} [\href{https://arxiv.org/abs/1701.05839}{{\ttfamily 1701.05839}}].

\bibitem{Li:2018xuv}
H.~T. Li and I.~Vitev, \emph{{Inclusive heavy flavor jet production with
  semi-inclusive jet functions: from proton to heavy-ion collisions}},
  \href{https://doi.org/10.1007/JHEP07(2019)148}{\emph{JHEP} {\bfseries 07}
  (2019) 148} [\href{https://arxiv.org/abs/1811.07905}{{\ttfamily
  1811.07905}}].

\bibitem{Li:2020rqj}
H.~T. Li and I.~Vitev, \emph{{Nuclear Matter Effects on Jet Production at
  Electron-Ion Colliders}},
  \href{https://doi.org/10.1103/PhysRevLett.126.252001}{\emph{Phys. Rev. Lett.}
  {\bfseries 126} (2021) 252001}
  [\href{https://arxiv.org/abs/2010.05912}{{\ttfamily 2010.05912}}].

\bibitem{Li:2021gjw}
H.~T. Li, Z.~L. Liu and I.~Vitev, \emph{{Heavy flavor jet production and
  substructure in electron-nucleus collisions}},
  \href{https://doi.org/10.1016/j.physletb.2022.137007}{\emph{Phys. Lett. B}
  {\bfseries 827} (2022) 137007}
  [\href{https://arxiv.org/abs/2108.07809}{{\ttfamily 2108.07809}}].

\bibitem{Dai:2018ywt}
L.~Dai, C.~Kim and A.~K. Leibovich, \emph{{Heavy Quark Jet Fragmentation}},
  {\emph{JHEP} {\bfseries 09} (2018) 109}
  [\href{https://arxiv.org/abs/1805.06014}{{\ttfamily 1805.06014}}].

\bibitem{CMS:2019btm}
{\scshape CMS} collaboration, C.~Collaboration, \emph{{Measurement of Jet
  Nuclear Modification Factor in PbPb Collisions at $\sqrt{s_{NN}}$ = 5.02 TeV
  with CMS}}, .

\bibitem{Li:2023dhb}
H.~T. Li, Z.~L. Liu and I.~Vitev, \emph{{Centrality-dependent modification of
  hadron and jet production in electron-nucleus collisions}},
  \href{https://arxiv.org/abs/2303.14201}{{\ttfamily 2303.14201}}.

\bibitem{Chang:2022hkt}
W.~Chang, E.-C. Aschenauer, M.~D. Baker, A.~Jentsch, J.-H. Lee, Z.~Tu et~al.,
  \emph{{Benchmark eA generator for leptoproduction in high-energy
  lepton-nucleus collisions}}, {\emph{Phys. Rev. D} {\bfseries 106} (2022)
  012007} [\href{https://arxiv.org/abs/2204.11998}{{\ttfamily 2204.11998}}].

\bibitem{Chatrchyan:2013exa}
{\scshape CMS} collaboration, S.~Chatrchyan et~al., \emph{{Evidence of b-Jet
  Quenching in PbPb Collisions at $\sqrt{s_{NN}}=2.76$ TeV}}, {\emph{Phys. Rev.
  Lett.} {\bfseries 113} (2014) 132301}
  [\href{https://arxiv.org/abs/1312.4198}{{\ttfamily 1312.4198}}].

\bibitem{Li:2020zbk}
H.~T. Li, Z.~L. Liu and I.~Vitev, \emph{{Heavy meson tomography of cold nuclear
  matter at the electron-ion collider}},
  \href{https://doi.org/10.1016/j.physletb.2021.136261}{\emph{Phys. Lett. B}
  {\bfseries 816} (2021) 136261}
  [\href{https://arxiv.org/abs/2007.10994}{{\ttfamily 2007.10994}}].

\bibitem{Krohn:2012fg}
D.~Krohn, M.~D. Schwartz, T.~Lin and W.~J. Waalewijn, \emph{{Jet Charge at the
  LHC}}, {\emph{Phys. Rev. Lett.} {\bfseries 110} (2013) 212001}
  [\href{https://arxiv.org/abs/1209.2421}{{\ttfamily 1209.2421}}].

\bibitem{Sirunyan:2020qvi}
{\scshape CMS} collaboration, A.~M. Sirunyan et~al., \emph{{Measurement of
  quark- and gluon-like jet fractions using jet charge in PbPb and pp
  collisions at 5.02 TeV}}, {\emph{JHEP} {\bfseries 07} (2020) 115}
  [\href{https://arxiv.org/abs/2004.00602}{{\ttfamily 2004.00602}}].

\bibitem{Larkoski:2017bvj}
A.~Larkoski, S.~Marzani, J.~Thaler, A.~Tripathee and W.~Xue, \emph{{Exposing
  the QCD Splitting Function with CMS Open Data}},
  \href{https://doi.org/10.1103/PhysRevLett.119.132003}{\emph{Phys. Rev. Lett.}
  {\bfseries 119} (2017) 132003}
  [\href{https://arxiv.org/abs/1704.05066}{{\ttfamily 1704.05066}}].

\bibitem{Chien:2016led}
Y.-T. Chien and I.~Vitev, \emph{{Probing the Hardest Branching within Jets in
  Heavy-Ion Collisions}}, {\emph{Phys. Rev. Lett.} {\bfseries 119} (2017)
  112301} [\href{https://arxiv.org/abs/1608.07283}{{\ttfamily 1608.07283}}].

\bibitem{Ilten:2017rbd}
P.~Ilten, N.~L. Rodd, J.~Thaler and M.~Williams, \emph{{Disentangling Heavy
  Flavor at Colliders}}, {\emph{Phys. Rev.} {\bfseries D96} (2017) 054019}
  [\href{https://arxiv.org/abs/1702.02947}{{\ttfamily 1702.02947}}].

\bibitem{Li:2017wwc}
H.~T. Li and I.~Vitev, \emph{{Inverting the mass hierarchy of jet quenching
  effects with prompt $b$-jet substructure}}, {\emph{Phys. Lett.} {\bfseries
  B793} (2019) 259} [\href{https://arxiv.org/abs/1801.00008}{{\ttfamily
  1801.00008}}].

\bibitem{CMS:2017qlm}
{\scshape CMS} collaboration, A.~M. Sirunyan et~al., \emph{{Measurement of the
  Splitting Function in $pp$ and Pb-Pb Collisions at $\sqrt{s_{_{\mathrm{NN}}}}
  =$ 5.02 TeV}},
  \href{https://doi.org/10.1103/PhysRevLett.120.142302}{\emph{Phys. Rev. Lett.}
  {\bfseries 120} (2018) 142302}
  [\href{https://arxiv.org/abs/1708.09429}{{\ttfamily 1708.09429}}].

\bibitem{Acharya:2018edi}
{\scshape ALICE} collaboration, S.~Acharya et~al., \emph{{Jet fragmentation
  transverse momentum measurements from di-hadron correlations in $\sqrt{s}$ =
  7 TeV pp and $\sqrt{s_{\rm{NN}}}$ = 5.02 TeV p-Pb collisions}}, {\emph{JHEP}
  {\bfseries 03} (2019) 169}
  [\href{https://arxiv.org/abs/1811.09742}{{\ttfamily 1811.09742}}].

\bibitem{Makris:2019ttx}
Y.~Makris and I.~Vitev, \emph{{An Effective Theory of Quarkonia in QCD
  Matter}}, {\emph{JHEP} {\bfseries 10} (2019) 111}
  [\href{https://arxiv.org/abs/1906.04186}{{\ttfamily 1906.04186}}].

\bibitem{Makris:2019kap}
Y.~Makris and I.~Vitev, \emph{{An Effective Theory of Quarkonia in QCD
  Matter}}, {\emph{Nucl. Phys. A} {\bfseries 1005} (2021) 121848}
  [\href{https://arxiv.org/abs/1912.08008}{{\ttfamily 1912.08008}}].

\bibitem{Vaidya:2020cyi}
V.~Vaidya and X.~Yao, \emph{{Transverse momentum broadening of a jet in
  quark-gluon plasma: an open quantum system EFT}}, {\emph{JHEP} {\bfseries 10}
  (2020) 024} [\href{https://arxiv.org/abs/2004.11403}{{\ttfamily
  2004.11403}}].

\bibitem{Burkardt:2008ps}
M.~Burkardt, \emph{{Transverse force on quarks in deep-inelastic scattering}},
  \href{https://doi.org/10.1103/PhysRevD.88.114502}{\emph{Phys. Rev. D}
  {\bfseries 88} (2013) 114502}
  [\href{https://arxiv.org/abs/0810.3589}{{\ttfamily 0810.3589}}].

\bibitem{Ravndal:1973kt}
F.~Ravndal, \emph{{On the azimuthal dependence of semiinclusive, deep inelastic
  electroproduction cross-sections}}, {\emph{Phys. Lett. B} {\bfseries 43}
  (1973) 301}.

\bibitem{Chen:2016hgw}
A.~Chen and J.~Ma, \emph{{Light-Cone Singularities and
  Transverse-Momentum-Dependent Factorization at Twist-3}}, {\emph{Phys. Lett.
  B} {\bfseries 768} (2017) 380}
  [\href{https://arxiv.org/abs/1610.08634}{{\ttfamily 1610.08634}}].

\bibitem{Bacchetta:2019qkv}
A.~Bacchetta, G.~Bozzi, M.~G. Echevarria, C.~Pisano, A.~Prokudin and M.~Radici,
  \emph{{Azimuthal asymmetries in unpolarized SIDIS and Drell-Yan processes: a
  case study towards TMD factorization at subleading twist}},
  \href{https://doi.org/10.1016/j.physletb.2019.134850}{\emph{Phys. Lett. B}
  {\bfseries 797} (2019) 134850}
  [\href{https://arxiv.org/abs/1906.07037}{{\ttfamily 1906.07037}}].

\bibitem{Vladimirov:2021hdn}
A.~Vladimirov, V.~Moos and I.~Scimemi, \emph{{Transverse momentum dependent
  operator expansion at next-to-leading power}},
  \href{https://doi.org/10.1007/JHEP01(2022)110}{\emph{JHEP} {\bfseries 01}
  (2022) 110} [\href{https://arxiv.org/abs/2109.09771}{{\ttfamily
  2109.09771}}].

\bibitem{Rodini:2022wki}
S.~Rodini and A.~Vladimirov, \emph{{Definition and evolution of transverse
  momentum dependent distribution of twist-three}},
  \href{https://doi.org/10.1007/JHEP08(2022)031}{\emph{JHEP} {\bfseries 08}
  (2022) 031} [\href{https://arxiv.org/abs/2204.03856}{{\ttfamily
  2204.03856}}].

\bibitem{Ebert:2021jhy}
M.~A. Ebert, A.~Gao and I.~W. Stewart, \emph{{Factorization for azimuthal
  asymmetries in SIDIS at next-to-leading power}},
  \href{https://doi.org/10.1007/JHEP06(2022)007}{\emph{JHEP} {\bfseries 06}
  (2022) 007} [\href{https://arxiv.org/abs/2112.07680}{{\ttfamily
  2112.07680}}].

\bibitem{Gamberg:2022lju}
L.~Gamberg, Z.-B. Kang, D.~Y. Shao, J.~Terry and F.~Zhao,
  \emph{{Transverse-momentum-dependent factorization at next-to-leading
  power}},  \href{https://arxiv.org/abs/2211.13209}{{\ttfamily 2211.13209}}.

\bibitem{Diehl:2005pc}
M.~Diehl and S.~Sapeta, \emph{{On the analysis of lepton scattering on
  longitudinally or transversely polarized protons}}, {\emph{Eur. Phys. J.}
  {\bfseries C41} (2005) 515}
  [\href{https://arxiv.org/abs/hep-ph/0503023}{{\ttfamily hep-ph/0503023}}].

\bibitem{Yang:2016qsf}
Y.~Yang and Z.~Lu, \emph{{Polarized \ensuremath{\Lambda} hyperon production in
  semi-inclusive deep inelastic scattering off an unpolarized nucleon target}},
  {\emph{Phys. Rev. D} {\bfseries 95} (2017) 074026}
  [\href{https://arxiv.org/abs/1611.07755}{{\ttfamily 1611.07755}}].

\bibitem{Wei:2016far}
S.-y. Wei, Y.-k. Song, K.-b. Chen and Z.-t. Liang, \emph{{Twist-4 contributions
  to semi-inclusive deeply inelastic scatterings with polarized beam and
  target}}, {\emph{Phys. Rev. D} {\bfseries 95} (2017) 074017}
  [\href{https://arxiv.org/abs/1611.08688}{{\ttfamily 1611.08688}}].

\bibitem{Lu:2011th}
Z.~Lu and I.~Schmidt, \emph{{Transverse momentum dependent twist-three result
  for polarized Drell-Yan processes}}, {\emph{Phys. Rev. D} {\bfseries 84}
  (2011) 114004} [\href{https://arxiv.org/abs/1109.3232}{{\ttfamily
  1109.3232}}].

\bibitem{Wei:2014pma}
S.-Y. Wei, K.-b. Chen, Y.-k. Song and Z.-t. Liang, \emph{{Leading and higher
  twist contributions in semi-inclusive $e^{+}e^{-}$ annihilation at high
  energies}}, \href{https://doi.org/10.1103/PhysRevD.91.034015}{\emph{Phys.
  Rev. D} {\bfseries 91} (2015) 034015}
  [\href{https://arxiv.org/abs/1410.4314}{{\ttfamily 1410.4314}}].

\bibitem{Chen:2016moq}
K.-b. Chen, W.-h. Yang, S.-y. Wei and Z.-t. Liang, \emph{{Tensor polarization
  dependent fragmentation functions and $e^+e^-\to V\pi X$ at high energies}},
  {\emph{Phys. Rev. D} {\bfseries 94} (2016) 034003}
  [\href{https://arxiv.org/abs/1605.07790}{{\ttfamily 1605.07790}}].

\bibitem{Georgi:1976ve}
H.~Georgi and H.~D. Politzer, \emph{{Freedom at Moderate Energies: Masses in
  Color Dynamics}}, \href{https://doi.org/10.1103/PhysRevD.14.1829}{\emph{Phys.
  Rev. D} {\bfseries 14} (1976) 1829}.

\bibitem{Politzer:1980me}
H.~D. Politzer, \emph{{Power Corrections at Short Distances}},
  \href{https://doi.org/10.1016/0550-3213(80)90172-8}{\emph{Nucl. Phys. B}
  {\bfseries 172} (1980) 349}.

\bibitem{Jaffe:1982pm}
R.~L. Jaffe and M.~Soldate, \emph{{Twist Four in Electroproduction: Canonical
  Operators and Coefficient Functions}},
  \href{https://doi.org/10.1103/PhysRevD.26.49}{\emph{Phys. Rev. D} {\bfseries
  26} (1982) 49}.

\bibitem{Ellis:1982cd}
R.~K. Ellis, W.~Furmanski and R.~Petronzio, \emph{{Unraveling Higher Twists}},
  \href{https://doi.org/10.1016/0550-3213(83)90597-7}{\emph{Nucl. Phys. B}
  {\bfseries 212} (1983) 29}.

\bibitem{Beneke:2002ph}
M.~Beneke, A.~P. Chapovsky, M.~Diehl and T.~Feldmann, \emph{{Soft collinear
  effective theory and heavy to light currents beyond leading power}},
  {\emph{Nucl. Phys.} {\bfseries B643} (2002) 431}
  [\href{https://arxiv.org/abs/hep-ph/0206152}{{\ttfamily hep-ph/0206152}}].

\bibitem{Manohar:2002fd}
A.~V. Manohar, T.~Mehen, D.~Pirjol and I.~W. Stewart, \emph{{Reparameterization
  invariance for collinear operators}}, {\emph{Phys. Lett.} {\bfseries B539}
  (2002) 59} [\href{https://arxiv.org/abs/hep-ph/0204229}{{\ttfamily
  hep-ph/0204229}}].

\bibitem{Pirjol:2002km}
D.~Pirjol and I.~W. Stewart, \emph{{A Complete basis for power suppressed
  collinear ultrasoft operators}}, {\emph{Phys. Rev.} {\bfseries D67} (2003)
  094005} [\href{https://arxiv.org/abs/hep-ph/0211251}{{\ttfamily
  hep-ph/0211251}}].

\bibitem{Beneke:2002ni}
M.~Beneke and T.~Feldmann, \emph{{Multipole expanded soft collinear effective
  theory with non-abelian gauge symmetry}}, {\emph{Phys. Lett.} {\bfseries
  B553} (2003) 267} [\href{https://arxiv.org/abs/hep-ph/0211358}{{\ttfamily
  hep-ph/0211358}}].

\bibitem{Bauer:2003mga}
C.~W. Bauer, D.~Pirjol and I.~W. Stewart, \emph{{On Power suppressed operators
  and gauge invariance in SCET}}, {\emph{Phys. Rev.} {\bfseries D68} (2003)
  034021} [\href{https://arxiv.org/abs/hep-ph/0303156}{{\ttfamily
  hep-ph/0303156}}].

\bibitem{Marcantonini:2008qn}
C.~Marcantonini and I.~W. Stewart, \emph{{Reparameterization Invariant
  Collinear Operators}}, {\emph{Phys. Rev.} {\bfseries D79} (2009) 065028}
  [\href{https://arxiv.org/abs/0809.1093}{{\ttfamily 0809.1093}}].

\bibitem{Moult:2015aoa}
I.~Moult, I.~W. Stewart, F.~J. Tackmann and W.~J. Waalewijn, \emph{{Employing
  Helicity Amplitudes for Resummation}}, {\emph{Phys. Rev. D} {\bfseries 93}
  (2016) 094003} [\href{https://arxiv.org/abs/1508.02397}{{\ttfamily
  1508.02397}}].

\bibitem{Kolodrubetz:2016uim}
D.~W. Kolodrubetz, I.~Moult and I.~W. Stewart, \emph{{Building Blocks for
  Subleading Helicity Operators}},
  \href{https://doi.org/10.1007/JHEP05(2016)139}{\emph{JHEP} {\bfseries 05}
  (2016) 139} [\href{https://arxiv.org/abs/1601.02607}{{\ttfamily
  1601.02607}}].

\bibitem{Feige:2017zci}
I.~Feige, D.~W. Kolodrubetz, I.~Moult and I.~W. Stewart, \emph{{A Complete
  Basis of Helicity Operators for Subleading Factorization}},
  \href{https://doi.org/10.1007/JHEP11(2017)142}{\emph{JHEP} {\bfseries 11}
  (2017) 142} [\href{https://arxiv.org/abs/1703.03411}{{\ttfamily
  1703.03411}}].

\bibitem{Moult:2017rpl}
I.~Moult, I.~W. Stewart and G.~Vita, \emph{{A subleading operator basis and
  matching for gg \textrightarrow{} H}},
  \href{https://doi.org/10.1007/JHEP07(2017)067}{\emph{JHEP} {\bfseries 07}
  (2017) 067} [\href{https://arxiv.org/abs/1703.03408}{{\ttfamily
  1703.03408}}].

\bibitem{Chang:2017atu}
C.-H. Chang, I.~W. Stewart and G.~Vita, \emph{{A Subleading Power Operator
  Basis for the Scalar Quark Current}},
  \href{https://doi.org/10.1007/JHEP04(2018)041}{\emph{JHEP} {\bfseries 04}
  (2018) 041} [\href{https://arxiv.org/abs/1712.04343}{{\ttfamily
  1712.04343}}].

\bibitem{Lorce:2013pza}
C.~Lorc\'{e} and B.~Pasquini, \emph{{Structure analysis of the generalized
  correlator of quark and gluon for a spin-1/2 target}}, {\emph{JHEP}
  {\bfseries 09} (2013) 138} [\href{https://arxiv.org/abs/1307.4497}{{\ttfamily
  1307.4497}}].

\bibitem{Chen:2015ora}
K.-b. Chen, S.-y. Wei, W.-h. Yang and Z.-t. Liang, \emph{{Three dimensional
  fragmentation functions from the quark-quark correlator}},
  \href{https://arxiv.org/abs/1505.02856}{{\ttfamily 1505.02856}}.

\bibitem{Kumano:2020ijt}
S.~Kumano and Q.-T. Song, \emph{{Transverse-momentum-dependent parton
  distribution functions up to twist 4 for spin-1 hadrons}},
  \href{https://doi.org/10.1103/PhysRevD.103.014025}{\emph{Phys. Rev. D}
  {\bfseries 103} (2021) 014025}
  [\href{https://arxiv.org/abs/2011.08583}{{\ttfamily 2011.08583}}].

\bibitem{Jaffe:1996zw}
R.~L. Jaffe, \emph{{Spin, twist and hadron structure in deep inelastic
  processes}},  in \emph{{Ettore Majorana International School of Nucleon
  Structure: 1st Course: The Spin Structure of the Nucleon}}, pp.~42--129, 1,
  1996, \href{https://arxiv.org/abs/hep-ph/9602236}{{\ttfamily
  hep-ph/9602236}}.

\bibitem{Metz:2004je}
A.~Metz and M.~Schlegel, \emph{{Twist three single spin asymmetries in
  semiinclusive deep inelastic scattering}}, {\emph{Eur. Phys. J.} {\bfseries
  A22} (2004) 489} [\href{https://arxiv.org/abs/hep-ph/0403182}{{\ttfamily
  hep-ph/0403182}}].

\bibitem{Gamberg:2006ru}
L.~P. Gamberg, D.~S. Hwang, A.~Metz and M.~Schlegel, \emph{{Light-cone
  divergence in twist-3 correlation functions}}, {\emph{Phys. Lett.} {\bfseries
  B639} (2006) 508} [\href{https://arxiv.org/abs/hep-ph/0604022}{{\ttfamily
  hep-ph/0604022}}].

\bibitem{Balitsky:2017gis}
I.~Balitsky and A.~Tarasov, \emph{{Power corrections to TMD factorization for
  Z-boson production}},
  \href{https://doi.org/10.1007/JHEP05(2018)150}{\emph{JHEP} {\bfseries 05}
  (2018) 150} [\href{https://arxiv.org/abs/1712.09389}{{\ttfamily
  1712.09389}}].

\bibitem{Balitsky:2017flc}
I.~Balitsky and A.~Tarasov, \emph{{Higher-twist corrections to gluon TMD
  factorization}}, \href{https://doi.org/10.1007/JHEP07(2017)095}{\emph{JHEP}
  {\bfseries 07} (2017) 095}
  [\href{https://arxiv.org/abs/1706.01415}{{\ttfamily 1706.01415}}].

\bibitem{AnjieREF}
A.~Gao, J.~K.~L. Michel and I.~W. Stewart, \emph{Renormalization of tmd
  quark-gluon-quark correlators}, {\emph{talks given by A.G. at REF 2022 and
  SCET 2023.} }.

\bibitem{COMPASS:2014kcy}
{\scshape COMPASS} collaboration, C.~Adolph et~al., \emph{{Measurement of
  azimuthal hadron asymmetries in semi-inclusive deep inelastic scattering off
  unpolarised nucleons}},
  \href{https://doi.org/10.1016/j.nuclphysb.2014.07.019}{\emph{Nucl. Phys. B}
  {\bfseries 886} (2014) 1046}
  [\href{https://arxiv.org/abs/1401.6284}{{\ttfamily 1401.6284}}].

\bibitem{Avakian:2019drf}
H.~Avakian, B.~Parsamyan and A.~Prokudin, \emph{{Spin orbit correlations and
  the structure of the nucleon}}, {\emph{Riv. Nuovo Cim.} {\bfseries 42} (2019)
  1} [\href{https://arxiv.org/abs/1909.13664}{{\ttfamily 1909.13664}}].

\bibitem{EuropeanMuon:1983tsy}
{\scshape European Muon} collaboration, J.~J. Aubert et~al., \emph{{Measurement
  of Hadronic Azimuthal Distributions in Deep Inelastic Muon Proton
  Scattering}}, {\emph{Phys. Lett. B} {\bfseries 130} (1983) 118}.

\bibitem{EuropeanMuon:1986ulc}
{\scshape European Muon} collaboration, M.~Arneodo et~al., \emph{{Measurement
  of Hadron Azimuthal Distributions in Deep Inelastic Muon Proton Scattering}},
  {\emph{Z. Phys. C} {\bfseries 34} (1987) 277}.

\bibitem{E665:1993pov}
{\scshape E665} collaboration, M.~R. Adams et~al., \emph{{Perturbative QCD
  effects observed in 490 GeV deep inelastic muon scattering}},
  \href{https://doi.org/10.1103/PhysRevD.48.5057}{\emph{Phys. Rev. D}
  {\bfseries 48} (1993) 5057}.

\bibitem{ZEUS:2000esx}
{\scshape ZEUS} collaboration, J.~Breitweg et~al., \emph{{Measurement of
  azimuthal asymmetries in deep inelastic scattering}},
  \href{https://doi.org/10.1016/S0370-2693(00)00430-5}{\emph{Phys. Lett. B}
  {\bfseries 481} (2000) 199}
  [\href{https://arxiv.org/abs/hep-ex/0003017}{{\ttfamily hep-ex/0003017}}].

\bibitem{Mkrtchyan:2007sr}
H.~Mkrtchyan et~al., \emph{{Transverse momentum dependence of semi-inclusive
  pion production}}, {\emph{Phys. Lett.} {\bfseries B665} (2008) 20}
  [\href{https://arxiv.org/abs/0709.3020}{{\ttfamily 0709.3020}}].

\bibitem{CLAS:2008nzy}
{\scshape CLAS} collaboration, M.~Osipenko et~al., \emph{{Measurement of
  unpolarized semi-inclusive pi+ electroproduction off the proton}},
  \href{https://doi.org/10.1103/PhysRevD.80.032004}{\emph{Phys. Rev. D}
  {\bfseries 80} (2009) 032004}
  [\href{https://arxiv.org/abs/0809.1153}{{\ttfamily 0809.1153}}].

\bibitem{HERMES:2012kpt}
{\scshape HERMES} collaboration, A.~Airapetian et~al., \emph{{Azimuthal
  distributions of charged hadrons, pions, and kaons produced in deep-inelastic
  scattering off unpolarized protons and deuterons}},
  \href{https://doi.org/10.1103/PhysRevD.87.012010}{\emph{Phys. Rev. D}
  {\bfseries 87} (2013) 012010}
  [\href{https://arxiv.org/abs/1204.4161}{{\ttfamily 1204.4161}}].

\bibitem{HERMES:2001hbj}
{\scshape HERMES} collaboration, A.~Airapetian et~al., \emph{{Single spin
  azimuthal asymmetries in electroproduction of neutral pions in semiinclusive
  deep inelastic scattering}}, {\emph{Phys. Rev. D} {\bfseries 64} (2001)
  097101} [\href{https://arxiv.org/abs/hep-ex/0104005}{{\ttfamily
  hep-ex/0104005}}].

\bibitem{HERMES:2005mov}
{\scshape HERMES} collaboration, A.~Airapetian et~al., \emph{{Subleading-twist
  effects in single-spin asymmetries in semi-inclusive deep-inelastic
  scattering on a longitudinally polarized hydrogen target}},
  \href{https://doi.org/10.1016/j.physletb.2005.06.067}{\emph{Phys. Lett. B}
  {\bfseries 622} (2005) 14}
  [\href{https://arxiv.org/abs/hep-ex/0505042}{{\ttfamily hep-ex/0505042}}].

\bibitem{CLAS:2010fns}
{\scshape CLAS} collaboration, H.~Avakian et~al., \emph{{Measurement of Single
  and Double Spin Asymmetries in Deep Inelastic Pion Electroproduction with a
  Longitudinally Polarized Target}},
  \href{https://doi.org/10.1103/PhysRevLett.105.262002}{\emph{Phys. Rev. Lett.}
  {\bfseries 105} (2010) 262002}
  [\href{https://arxiv.org/abs/1003.4549}{{\ttfamily 1003.4549}}].

\bibitem{CLAS:2017yrm}
{\scshape CLAS} collaboration, S.~Jawalkar et~al., \emph{{Semi-Inclusive
  $\pi_0$ target and beam-target asymmetries from 6 GeV electron scattering
  with CLAS}},
  \href{https://doi.org/10.1016/j.physletb.2018.06.014}{\emph{Phys. Lett. B}
  {\bfseries 782} (2018) 662}
  [\href{https://arxiv.org/abs/1709.10054}{{\ttfamily 1709.10054}}].

\bibitem{Parsamyan:2018ovx}
B.~Parsamyan, \emph{{Measurement of longitudinal-target-polarization dependent
  azimuthal asymmetries in SIDIS at COMPASS experiment}},
  \href{https://doi.org/10.22323/1.297.0259}{\emph{PoS} {\bfseries DIS2017}
  (2018) 259} [\href{https://arxiv.org/abs/1801.01488}{{\ttfamily
  1801.01488}}].

\bibitem{Oganessian:1998ma}
K.~Oganessian, H.~Avakian, N.~Bianchi and A.~Kotzinian, \emph{{Sin($\phi$)
  azimuthal asymmetry in semiinclusive electroproduction on longitudinally
  polarized nucleon}},  in \emph{{8th International Conference on the Structure
  of Baryons}}, pp.~320--324, 8, 1998,
  \href{https://arxiv.org/abs/hep-ph/9808368}{{\ttfamily hep-ph/9808368}}.

\bibitem{DeSanctis:2000fh}
E.~De~Sanctis, W.~Nowak and K.~Oganesian, \emph{{Single spin azimuthal
  asymmetries in the `Reduced twist - three approximation'}}, {\emph{Phys.
  Lett. B} {\bfseries 483} (2000) 69}
  [\href{https://arxiv.org/abs/hep-ph/0002091}{{\ttfamily hep-ph/0002091}}].

\bibitem{Aghasyan:2011ha}
M.~Aghasyan et~al., \emph{{Precise measurements of beam spin asymmetries in
  semi-inclusive $\pi^0$ production}}, {\emph{Phys. Lett. B} {\bfseries 704}
  (2011) 397} [\href{https://arxiv.org/abs/1106.2293}{{\ttfamily 1106.2293}}].

\bibitem{CLAS:2003qum}
{\scshape CLAS} collaboration, H.~Avakian et~al., \emph{{Measurement of
  beam-spin asymmetries for $\pi^+$ electroproduction above the baryon
  resonance region}}, {\emph{Phys. Rev. D} {\bfseries 69} (2004) 112004}
  [\href{https://arxiv.org/abs/hep-ex/0301005}{{\ttfamily hep-ex/0301005}}].

\bibitem{HERMES:2006pof}
{\scshape HERMES} collaboration, A.~Airapetian et~al., \emph{{Beam-Spin
  Asymmetries in the Azimuthal Distribution of Pion Electroproduction}},
  \href{https://doi.org/10.1016/j.physletb.2007.03.015}{\emph{Phys. Lett. B}
  {\bfseries 648} (2007) 164}
  [\href{https://arxiv.org/abs/hep-ex/0612059}{{\ttfamily hep-ex/0612059}}].

\bibitem{HERMES:2019zll}
{\scshape HERMES} collaboration, A.~Airapetian et~al., \emph{{Beam-helicity
  asymmetries for single-hadron production in semi-inclusive deep-inelastic
  scattering from unpolarized hydrogen and deuterium targets}},
  \href{https://doi.org/10.1016/j.physletb.2019.134886}{\emph{Phys. Lett. B}
  {\bfseries 797} (2019) 134886}
  [\href{https://arxiv.org/abs/1903.08544}{{\ttfamily 1903.08544}}].

\bibitem{CLAS:2014dmz}
{\scshape CLAS} collaboration, W.~Gohn et~al., \emph{{Beam-spin asymmetries
  from semi-inclusive pion electroproduction}}, {\emph{Phys. Rev. D} {\bfseries
  89} (2014) 072011} [\href{https://arxiv.org/abs/1402.4097}{{\ttfamily
  1402.4097}}].

\bibitem{CLAS:2021opg}
{\scshape CLAS} collaboration, S.~Diehl et~al., \emph{{Multidimensional, High
  Precision Measurements of Beam Single Spin Asymmetries in Semi-inclusive
  $\pi^{+}$~Electroproduction off Protons in the Valence Region}},
  \href{https://doi.org/10.1103/PhysRevLett.128.062005}{\emph{Phys. Rev. Lett.}
  {\bfseries 128} (2022) 062005}
  [\href{https://arxiv.org/abs/2101.03544}{{\ttfamily 2101.03544}}].

\bibitem{Afanasev:2006gw}
A.~V. Afanasev and C.~E. Carlson, \emph{{Beam Single-Spin Asymmetry in
  Semi-Inclusive Deep Inelastic Scattering}}, {\emph{Phys. Rev. D} {\bfseries
  74} (2006) 114027} [\href{https://arxiv.org/abs/hep-ph/0603269}{{\ttfamily
  hep-ph/0603269}}].

\bibitem{Mao:2012dk}
W.~Mao and Z.~Lu, \emph{{Beam single spin asymmetry of neutral pion production
  in semi-inclusive deep inelastic scattering}}, {\emph{Phys. Rev. D}
  {\bfseries 87} (2013) 014012}
  [\href{https://arxiv.org/abs/1210.4790}{{\ttfamily 1210.4790}}].

\bibitem{Lu:2012gu}
Z.~Lu and I.~Schmidt, \emph{{T-odd quark-gluon-quark correlation function in
  the diquark model}}, {\emph{Phys. Lett. B} {\bfseries 712} (2012) 451}
  [\href{https://arxiv.org/abs/1202.0700}{{\ttfamily 1202.0700}}].

\bibitem{Mao:2014aoa}
W.~Mao, Z.~Lu and B.-Q. Ma, \emph{{Transverse single-spin asymmetries of pion
  production in semi-inclusive DIS at subleading twist}}, {\emph{Phys. Rev. D}
  {\bfseries 90} (2014) 014048}
  [\href{https://arxiv.org/abs/1405.3876}{{\ttfamily 1405.3876}}].

\bibitem{Mao:2014fma}
W.~Mao, Z.~Lu, B.-Q. Ma and I.~Schmidt, \emph{{Double spin asymmetries
  $A_{LT}^{\cos\phi_S}$ and $A_{LT}^{\cos(2\phi_h -\phi_S)}$ in semi-inclusive
  DIS}}, {\emph{Phys. Rev. D} {\bfseries 91} (2015) 034029}
  [\href{https://arxiv.org/abs/1412.7390}{{\ttfamily 1412.7390}}].

\bibitem{Lu:2014fva}
Z.~Lu, \emph{{Single-spin asymmetries in electroproduction of pions on the
  longitudinally polarized nucleon targets}}, {\emph{Phys. Rev. D} {\bfseries
  90} (2014) 014037} [\href{https://arxiv.org/abs/1404.4229}{{\ttfamily
  1404.4229}}].

\bibitem{Mao:2016hdi}
W.~Mao, X.~Wang, X.~Du, Z.~Lu and B.-Q. Ma, \emph{{On the cos$\phi_h$ asymmetry
  in electroproduction of pions in double longitudinally polarized process}},
  {\emph{Nucl. Phys. A} {\bfseries 945} (2016) 153}.

\bibitem{Yang:2018aue}
Y.~Yang, W.~Mao and Z.~Lu, \emph{{Single spin asymmetry in transverse polarized
  proton production and $\Lambda$ production in semi-inclusive DIS at
  twist-3}}, {\emph{Eur. Phys. J. Plus} {\bfseries 134} (2019) 259}
  [\href{https://arxiv.org/abs/1808.10565}{{\ttfamily 1808.10565}}].

\bibitem{Ji:1993qx}
X.-D. Ji and Z.-K. Zhu, \emph{{Quark fragmentation functions in low-energy
  chiral theory}},  \href{https://arxiv.org/abs/hep-ph/9402303}{{\ttfamily
  hep-ph/9402303}}.

\bibitem{Gamberg:2003pz}
L.~P. Gamberg, D.~S. Hwang and K.~A. Oganessyan, \emph{{Chiral odd
  fragmentation functions in single pion inclusive electroproduction}},
  {\emph{Phys. Lett.} {\bfseries B584} (2004) 276}
  [\href{https://arxiv.org/abs/hep-ph/0311221}{{\ttfamily hep-ph/0311221}}].

\bibitem{Lu:2015wja}
Z.~Lu and I.~Schmidt, \emph{{Twist-3 fragmentation functions in a spectator
  model with gluon rescattering}}, {\emph{Phys. Lett.} {\bfseries B747} (2015)
  357} [\href{https://arxiv.org/abs/1501.04379}{{\ttfamily 1501.04379}}].

\bibitem{Yang:2016mxl}
Y.~Yang, Z.~Lu and I.~Schmidt, \emph{{Twist-3 T-odd fragmentation functions
  $G^\perp$ and $\tilde{G}^\perp$ in a spectator model}}, {\emph{Phys. Lett. B}
  {\bfseries 761} (2016) 333}
  [\href{https://arxiv.org/abs/1607.01638}{{\ttfamily 1607.01638}}].

\bibitem{Pasquini:2018oyz}
B.~Pasquini and S.~Rodini, \emph{{The twist-three distribution $e^q(x,k_\perp)$
  in a light-front model}}, {\emph{Phys. Lett. B} {\bfseries 788} (2019) 414}
  [\href{https://arxiv.org/abs/1806.10932}{{\ttfamily 1806.10932}}].

\bibitem{Wakamatsu:2000ex}
M.~Wakamatsu, \emph{{Polarized structure functions $g_2(x)$ in the chiral quark
  soliton model}}, {\emph{Phys. Lett. B} {\bfseries 487} (2000) 118}
  [\href{https://arxiv.org/abs/hep-ph/0006212}{{\ttfamily hep-ph/0006212}}].

\bibitem{Schweitzer:2003uy}
P.~Schweitzer, \emph{{The Chirally odd twist three distribution function
  $e^a(x)$ in the chiral quark soliton model}}, {\emph{Phys. Rev. D} {\bfseries
  67} (2003) 114010} [\href{https://arxiv.org/abs/hep-ph/0303011}{{\ttfamily
  hep-ph/0303011}}].

\bibitem{Wakamatsu:2003uu}
M.~Wakamatsu and Y.~Ohnishi, \emph{{The Nonperturbative origin of
  delta-function singularity in the chirally odd twist three distribution
  function e(x)}}, {\emph{Phys. Rev. D} {\bfseries 67} (2003) 114011}
  [\href{https://arxiv.org/abs/hep-ph/0303007}{{\ttfamily hep-ph/0303007}}].

\bibitem{Ohnishi:2003mf}
Y.~Ohnishi and M.~Wakamatsu, \emph{{$\pi$-N sigma term and chiral odd twist
  three distribution function e(x) of the nucleon in the chiral quark soliton
  model}}, {\emph{Phys. Rev. D} {\bfseries 69} (2004) 114002}
  [\href{https://arxiv.org/abs/hep-ph/0312044}{{\ttfamily hep-ph/0312044}}].

\bibitem{Cebulla:2007ej}
C.~Cebulla, J.~Ossmann, P.~Schweitzer and D.~Urbano, \emph{{The Twist-3 parton
  distribution function $e^a(x)$ in large-$N_c$ chiral theory}}, {\emph{Acta
  Phys. Polon. B} {\bfseries 39} (2008) 609}
  [\href{https://arxiv.org/abs/0710.3103}{{\ttfamily 0710.3103}}].

\bibitem{Mukherjee:2010iw}
A.~Mukherjee and R.~Korrapati, \emph{{Twist Three Distribution
  $f^\perp(x,k^\perp)$ in Light-front Hamiltonian Approach}}, {\emph{Mod. Phys.
  Lett. A} {\bfseries 26} (2011) 2653}
  [\href{https://arxiv.org/abs/1005.2830}{{\ttfamily 1005.2830}}].

\bibitem{Burkardt:2001iy}
M.~Burkardt and Y.~Koike, \emph{{Violation of sum rules for twist three parton
  distributions in QCD}}, {\emph{Nucl. Phys. B} {\bfseries 632} (2002) 311}
  [\href{https://arxiv.org/abs/hep-ph/0111343}{{\ttfamily hep-ph/0111343}}].

\bibitem{Aslan:2018tff}
F.~Aslan and M.~Burkardt, \emph{{Singularities in Twist-3 Quark
  Distributions}}, {\emph{Phys. Rev. D} {\bfseries 101} (2020) 016010}
  [\href{https://arxiv.org/abs/1811.00938}{{\ttfamily 1811.00938}}].

\bibitem{Bhattacharya:2021boh}
S.~Bhattacharya and A.~Metz, \emph{{Burkhardt-Cottingham-type sum rules for
  light-cone and quasi-PDFs}},
  \href{https://doi.org/10.1103/PhysRevD.105.054027}{\emph{Phys. Rev. D}
  {\bfseries 105} (2022) 054027}
  [\href{https://arxiv.org/abs/2105.07282}{{\ttfamily 2105.07282}}].

\bibitem{Burkardt:1995ts}
M.~Burkardt, \emph{{On the possible violation of sum rules for higher twist
  parton distributions}}, {\emph{Phys. Rev. D} {\bfseries 52} (1995) 3841}
  [\href{https://arxiv.org/abs/hep-ph/9505226}{{\ttfamily hep-ph/9505226}}].

\bibitem{Efremov:2002qh}
A.~V. Efremov and P.~Schweitzer, \emph{{The Chirally odd twist 3 distribution
  e(a)(x)}}, {\emph{JHEP} {\bfseries 08} (2003) 006}
  [\href{https://arxiv.org/abs/hep-ph/0212044}{{\ttfamily hep-ph/0212044}}].

\bibitem{Avakian:2009jt}
H.~Avakian, A.~Efremov, P.~Schweitzer, O.~Teryaev, F.~Yuan and P.~Zavada,
  \emph{{Insights on non-perturbative aspects of TMDs from models}},
  {\emph{Mod. Phys. Lett. A} {\bfseries 24} (2009) 2995}
  [\href{https://arxiv.org/abs/0910.3181}{{\ttfamily 0910.3181}}].

\bibitem{Bukhvostov:1984rns}
A.~P. Bukhvostov, E.~A. Kuraev and L.~N. Lipatov, \emph{{Deep inelastic
  scattering by a polarized target in quantum chromodynamics}}, {\emph{Sov.
  Phys. JETP} {\bfseries 60} (1984) 22}.

\bibitem{Bukhvostov:1984zhx}
A.~P. Bukhvostov, E.~A. Kuraev and L.~N. Lipatov, \emph{{Deep inelastic
  scattering on a polarized target in Abelian gauge theory}}, {\emph{Sov. J.
  Nucl. Phys.} {\bfseries 39} (1984) 121}.

\bibitem{Belitsky:1997zw}
A.~V. Belitsky and D.~M\"uller, \emph{{Scale dependence of the chiral odd twist
  - three distributions $h_L(x)$ and $e(x)$}}, {\emph{Nucl. Phys. B} {\bfseries
  503} (1997) 279} [\href{https://arxiv.org/abs/hep-ph/9702354}{{\ttfamily
  hep-ph/9702354}}].

\bibitem{Balla:1997hf}
J.~Balla, M.~V. Polyakov and C.~Weiss, \emph{{Nucleon matrix elements of higher
  twist operators from the instanton vacuum}}, {\emph{Nucl. Phys. B} {\bfseries
  510} (1998) 327} [\href{https://arxiv.org/abs/hep-ph/9707515}{{\ttfamily
  hep-ph/9707515}}].

\bibitem{Dressler:1999hc}
B.~Dressler and M.~V. Polyakov, \emph{{On the twist - three contribution to
  h(L) in the instanton vacuum}}, {\emph{Phys. Rev. D} {\bfseries 61} (2000)
  097501} [\href{https://arxiv.org/abs/hep-ph/9912376}{{\ttfamily
  hep-ph/9912376}}].

\bibitem{Aubert:1983cz}
{\scshape European Muon} collaboration, J.~J. Aubert et~al., \emph{{Measurement
  of Hadronic Azimuthal Distributions in Deep Inelastic Muon Proton
  Scattering}}, {\emph{Phys. Lett.} {\bfseries 130B} (1983) 118}.

\bibitem{Arneodo:1986cf}
{\scshape European Muon} collaboration, M.~Arneodo et~al., \emph{{Measurement
  of Hadron Azimuthal Distributions in Deep Inelastic Muon Proton Scattering}},
  {\emph{Z. Phys. C} {\bfseries 34} (1987) 277}.

\bibitem{Adams:1993hs}
{\scshape E665} collaboration, M.~R. Adams et~al., \emph{{Perturbative QCD
  effects observed in 490 GeV deep inelastic muon scattering}}, {\emph{Phys.
  Rev.} {\bfseries D48} (1993) 5057}.

\bibitem{Kotsinian:2000td}
A.~Kotsinian, K.~Oganesian, H.~Avakian and E.~De~Sanctis, \emph{{Single target
  spin asymmetries in semiinclusive pion electroproduction on longitudinally
  polarized protons}}, {\emph{Nucl. Phys. B Proc. Suppl.} (1999) }
  [\href{https://arxiv.org/abs/hep-ph/9908466}{{\ttfamily hep-ph/9908466}}].

\bibitem{Boglione:2000jk}
M.~Boglione and P.~Mulders, \emph{{Azimuthal spin asymmetries in semiinclusive
  production from positron proton scattering}}, {\emph{Phys. Lett. B}
  {\bfseries 478} (2000) 114}
  [\href{https://arxiv.org/abs/hep-ph/0001196}{{\ttfamily hep-ph/0001196}}].

\bibitem{Oganessian:2000um}
K.~Oganessian, N.~Bianchi, E.~De~Sanctis and W.~Nowak, \emph{{Investigation of
  single spin asymmetries in pi+ electroproduction}}, {\emph{Nucl. Phys. A}
  {\bfseries 689} (2001) 784}
  [\href{https://arxiv.org/abs/hep-ph/0010261}{{\ttfamily hep-ph/0010261}}].

\bibitem{Efremov:2001cz}
A.~Efremov, K.~Goeke and P.~Schweitzer, \emph{{Azimuthal asymmetry in
  electroproduction of neutral pions in semiinclusive DIS}}, {\emph{Phys. Lett.
  B} {\bfseries 522} (2001) 37}
  [\href{https://arxiv.org/abs/hep-ph/0108213}{{\ttfamily hep-ph/0108213}}].

\bibitem{Efremov:2001ia}
A.~Efremov, K.~Goeke and P.~Schweitzer, \emph{{Predictions for azimuthal
  asymmetries in pion and kaon production in SIDIS off a longitudinally
  polarized deuterium target at HERMES}}, {\emph{Eur. Phys. J. C} {\bfseries
  24} (2002) 407} [\href{https://arxiv.org/abs/hep-ph/0112166}{{\ttfamily
  hep-ph/0112166}}].

\bibitem{Oganessyan:2002pc}
K.~Oganessyan, P.~Mulders and E.~De~Sanctis, \emph{{Double spin cos phi
  asymmetry in semiinclusive electroproduction}}, {\emph{Phys. Lett. B}
  {\bfseries 532} (2002) 87}
  [\href{https://arxiv.org/abs/hep-ph/0201061}{{\ttfamily hep-ph/0201061}}].

\bibitem{Efremov:2002ut}
A.~Efremov, K.~Goeke and P.~Schweitzer, \emph{{Azimuthal asymmetries at CLAS:
  extraction of $e^a(x)$ and prediction of A$_{UL}$}}, {\emph{Phys. Rev. D}
  {\bfseries 67} (2003) 114014}
  [\href{https://arxiv.org/abs/hep-ph/0208124}{{\ttfamily hep-ph/0208124}}].

\bibitem{Oganessyan:2002er}
K.~Oganessyan, L.~Asilyan, M.~Anselmino and E.~De~Sanctis, \emph{{Spin
  independent and double spin cos phi asymmetries in semiinclusive pion
  electroproduction}}, {\emph{Phys. Lett. B} {\bfseries 564} (2003) 60}
  [\href{https://arxiv.org/abs/hep-ph/0208208}{{\ttfamily hep-ph/0208208}}].

\bibitem{Ma:2002ns}
B.-Q. Ma, I.~Schmidt and J.-J. Yang, \emph{{Reanalysis of azimuthal spin
  asymmetries of meson electroproduction}}, {\emph{Phys. Rev. D} {\bfseries 66}
  (2002) 094001} [\href{https://arxiv.org/abs/hep-ph/0209114}{{\ttfamily
  hep-ph/0209114}}].

\bibitem{Efremov:2003tf}
A.~Efremov, K.~Goeke and P.~Schweitzer, \emph{{Sivers versus Collins effect in
  azimuthal single spin asymmetries in pion production in SIDIS}}, {\emph{Phys.
  Lett. B} {\bfseries 568} (2003) 63}
  [\href{https://arxiv.org/abs/hep-ph/0303062}{{\ttfamily hep-ph/0303062}}].

\bibitem{Efremov:2003eq}
A.~Efremov, K.~Goeke and P.~Schweitzer, \emph{{Collins effect and single spin
  azimuthal asymmetries in the HERMES and COMPASS experiments}}, {\emph{Eur.
  Phys. J. C} {\bfseries 32} (2003) 337}
  [\href{https://arxiv.org/abs/hep-ph/0309209}{{\ttfamily hep-ph/0309209}}].

\bibitem{Yuan:2003gu}
F.~Yuan, \emph{{The Beam single spin asymmetry in semiinclusive deep inelastic
  scattering}}, {\emph{Phys. Lett. B} {\bfseries 589} (2004) 28}
  [\href{https://arxiv.org/abs/hep-ph/0310279}{{\ttfamily hep-ph/0310279}}].

\bibitem{Schweitzer:2003yr}
P.~Schweitzer and A.~Bacchetta, \emph{{Azimuthal single spin asymmetries in
  SIDIS in the light of chiral symmetry breaking}}, {\emph{Nucl. Phys. A}
  {\bfseries 732} (2004) 106}
  [\href{https://arxiv.org/abs/hep-ph/0310318}{{\ttfamily hep-ph/0310318}}].

\bibitem{Song:2010pf}
Y.-k. Song, J.-h. Gao, Z.-t. Liang and X.-N. Wang, \emph{{Twist-4 contributions
  to the azimuthal asymmetry in SIDIS}}, {\emph{Phys. Rev. D} {\bfseries 83}
  (2011) 054010} [\href{https://arxiv.org/abs/1012.4179}{{\ttfamily
  1012.4179}}].

\bibitem{Gao:2011mf}
J.-H. Gao, A.~Sch\"afer and J.~Zhou, \emph{{Azimuthal asymmetry in SIDIS off
  nuclei as probe for $\hat q$}}, {\emph{Phys. Rev. D} {\bfseries 85} (2012)
  074027} [\href{https://arxiv.org/abs/1111.1633}{{\ttfamily 1111.1633}}].

\bibitem{Song:2013sja}
Y.-k. Song, J.-h. Gao, Z.-t. Liang and X.-N. Wang, \emph{{Azimuthal asymmetries
  in semi-inclusive DIS with polarized beam and/or target and their nuclear
  dependences}}, {\emph{Phys. Rev. D} {\bfseries 89} (2014) 014005}
  [\href{https://arxiv.org/abs/1308.1159}{{\ttfamily 1308.1159}}].

\bibitem{Chen:2013zpy}
L.~Chen, J.-h. Gao and Z.-t. Liang, \emph{{Nuclear dependencies of azimuthal
  asymmetries in the Drell-Yan process}}, {\emph{Phys. Rev. C} {\bfseries 89}
  (2014) 035204} [\href{https://arxiv.org/abs/1308.3746}{{\ttfamily
  1308.3746}}].

\bibitem{Afanasev:2003ze}
A.~Afanasev and C.~E. Carlson, \emph{{Single spin beam asymmetry in
  semiexclusive deep inelastic electroproduction}},  in \emph{{Intersections of
  particle and nuclear physics. Proceedings, 8th Conference, CIPANP 2003, New
  York, USA, May 19-24, 2003}}, 2003,
  \href{https://arxiv.org/abs/hep-ph/0308163}{{\ttfamily hep-ph/0308163}}.

\bibitem{Wigner:1932eb}
E.~P. Wigner, \emph{{On the quantum correction for thermodynamic equilibrium}},
  {\emph{Phys. Rev.} {\bfseries 40} (1932) 749}.

\bibitem{Geiger:1998yk}
K.~Geiger, J.~R. Ellis, U.~W. Heinz and U.~A. Wiedemann, \emph{{Bose-Einstein
  correlations in a space-time approach to $e^+e^-$ annihilation into
  hadrons}}, {\emph{Phys. Rev. D} {\bfseries 61} (2000) 054002}
  [\href{https://arxiv.org/abs/hep-ph/9811270}{{\ttfamily hep-ph/9811270}}].

\bibitem{Heinz:2009xj}
U.~W. Heinz, \emph{{Early collective expansion: Relativistic hydrodynamics and
  the transport properties of QCD matter}}, {\emph{Landolt-Bornstein}
  {\bfseries 23} (2010) 240} [\href{https://arxiv.org/abs/0901.4355}{{\ttfamily
  0901.4355}}].

\bibitem{Belitsky:2003nz}
A.~V. Belitsky, X.-d. Ji and F.~Yuan, \emph{{Quark imaging in the proton via
  quantum phase space distributions}}, {\emph{Phys. Rev.} {\bfseries D69}
  (2004) 074014} [\href{https://arxiv.org/abs/hep-ph/0307383}{{\ttfamily
  hep-ph/0307383}}].

\bibitem{Soper:1976jc}
D.~E. Soper, \emph{{The Parton Model and the Bethe-Salpeter Wave Function}},
  {\emph{Phys. Rev.} {\bfseries D15} (1977) 1141}.

\bibitem{Burkardt:2000za}
M.~Burkardt, \emph{{Impact parameter dependent parton distributions and off
  forward parton distributions for $\zeta \to 0$}}, {\emph{Phys. Rev. D}
  {\bfseries 62} (2000) 071503}
  [\href{https://arxiv.org/abs/hep-ph/0005108}{{\ttfamily hep-ph/0005108}}].

\bibitem{Ralston:2001xs}
J.~P. Ralston and B.~Pire, \emph{{Femtophotography of protons to nuclei with
  deeply virtual Compton scattering}},
  \href{https://doi.org/10.1103/PhysRevD.66.111501}{\emph{Phys. Rev. D}
  {\bfseries 66} (2002) 111501}
  [\href{https://arxiv.org/abs/hep-ph/0110075}{{\ttfamily hep-ph/0110075}}].

\bibitem{Diehl:2002he}
M.~Diehl, \emph{{Generalized parton distributions in impact parameter space}},
  \href{https://doi.org/10.1007/s10052-002-1016-9}{\emph{Eur. Phys. J. C}
  {\bfseries 25} (2002) 223}
  [\href{https://arxiv.org/abs/hep-ph/0205208}{{\ttfamily hep-ph/0205208}}].

\bibitem{Burkardt:2002hr}
M.~Burkardt, \emph{{Impact parameter space interpretation for generalized
  parton distributions}},
  \href{https://doi.org/10.1142/S0217751X03012370}{\emph{Int. J. Mod. Phys. A}
  {\bfseries 18} (2003) 173}
  [\href{https://arxiv.org/abs/hep-ph/0207047}{{\ttfamily hep-ph/0207047}}].

\bibitem{Radyushkin:1996nd}
A.~V. Radyushkin, \emph{{Scaling limit of deeply virtual Compton scattering}},
  \href{https://doi.org/10.1016/0370-2693(96)00528-X}{\emph{Phys. Lett. B}
  {\bfseries 380} (1996) 417}
  [\href{https://arxiv.org/abs/hep-ph/9604317}{{\ttfamily hep-ph/9604317}}].

\bibitem{Radyushkin:1996ru}
A.~V. Radyushkin, \emph{{Asymmetric gluon distributions and hard diffractive
  electroproduction}},
  \href{https://doi.org/10.1016/0370-2693(96)00844-1}{\emph{Phys. Lett. B}
  {\bfseries 385} (1996) 333}
  [\href{https://arxiv.org/abs/hep-ph/9605431}{{\ttfamily hep-ph/9605431}}].

\bibitem{Kriesten:2019jep}
B.~Kriesten, S.~Liuti, L.~Calero-Diaz, D.~Keller, A.~Meyer, G.~R. Goldstein
  et~al., \emph{{Extraction of generalized parton distribution observables from
  deeply virtual electron proton scattering experiments}}, {\emph{Phys. Rev. D}
  {\bfseries 101} (2020) 054021}
  [\href{https://arxiv.org/abs/1903.05742}{{\ttfamily 1903.05742}}].

\bibitem{Hillery:1983ms}
M.~Hillery, R.~F. O'Connell, M.~O. Scully and E.~P. Wigner, \emph{{Distribution
  functions in physics: Fundamentals}}, {\emph{Phys. Rept.} {\bfseries 106}
  (1984) 121}.

\bibitem{Ji:2003ak}
X.-d. Ji, \emph{{Viewing the proton through 'color' filters}}, {\emph{Phys.
  Rev. Lett.} {\bfseries 91} (2003) 062001}
  [\href{https://arxiv.org/abs/hep-ph/0304037}{{\ttfamily hep-ph/0304037}}].

\bibitem{Licht:1970pe}
A.~L. Licht and A.~Pagnamenta, \emph{{Wave functions and form-factors for
  relativistic composite particles. i}}, {\emph{Phys. Rev. D} {\bfseries 2}
  (1970) 1150}.

\bibitem{Licht:1970de}
A.~L. Licht and A.~Pagnamenta, \emph{{Wave functions and form-factors for
  relativistic composite particles. ii}}, {\emph{Phys. Rev. D} {\bfseries 2}
  (1970) 1156}.

\bibitem{Lorce:2015sqe}
C.~Lorc\'e and B.~Pasquini, \emph{{Multipole decomposition of the nucleon
  transverse phase space}}, {\emph{Phys. Rev. D} {\bfseries 93} (2016) 034040}
  [\href{https://arxiv.org/abs/1512.06744}{{\ttfamily 1512.06744}}].

\bibitem{Echevarria:2016mrc}
M.~G. Echevarria, A.~Idilbi, K.~Kanazawa, C.~Lorc\'{e}, A.~Metz, B.~Pasquini
  et~al., \emph{{Proper definition and evolution of generalized transverse
  momentum dependent distributions}}, {\emph{Phys. Lett.} {\bfseries B759}
  (2016) 336} [\href{https://arxiv.org/abs/1602.06953}{{\ttfamily
  1602.06953}}].

\bibitem{Balitsky:2019ayf}
I.~Balitsky and G.~A. Chirilli, \emph{{Conformal invariance of
  transverse-momentum dependent parton distributions rapidity evolution}},
  {\emph{Phys. Rev. D} {\bfseries 100} (2019) 051504}
  [\href{https://arxiv.org/abs/1905.09144}{{\ttfamily 1905.09144}}].

\bibitem{Bertone:2022awq}
V.~Bertone, \emph{{Matching generalised transverse-momentum-dependent
  distributions onto generalised parton distributions at one loop}},
  \href{https://arxiv.org/abs/2207.09526}{{\ttfamily 2207.09526}}.

\bibitem{Echevarria:2022ztg}
M.~G. Echevarria, P.~A.~G. Garcia and I.~Scimemi, \emph{{GTMDs and the
  factorization of exclusive double Drell-Yan}},
  \href{https://arxiv.org/abs/2208.00021}{{\ttfamily 2208.00021}}.

\bibitem{Hatta:2016dxp}
Y.~Hatta, B.-W. Xiao and F.~Yuan, \emph{{Probing the Small- x Gluon Tomography
  in Correlated Hard Diffractive Dijet Production in Deep Inelastic
  Scattering}}, {\emph{Phys. Rev. Lett.} {\bfseries 116} (2016) 202301}
  [\href{https://arxiv.org/abs/1601.01585}{{\ttfamily 1601.01585}}].

\bibitem{Altinoluk:2015dpi}
T.~Altinoluk, N.~Armesto, G.~Beuf and A.~H. Rezaeian, \emph{{Diffractive Dijet
  Production in Deep Inelastic Scattering and Photon-Hadron Collisions in the
  Color Glass Condensate}}, {\emph{Phys. Lett. B} {\bfseries 758} (2016) 373}
  [\href{https://arxiv.org/abs/1511.07452}{{\ttfamily 1511.07452}}].

\bibitem{Zhou:2016rnt}
J.~Zhou, \emph{{Elliptic gluon generalized transverse-momentum-dependent
  distribution inside a large nucleus}}, {\emph{Phys. Rev.} {\bfseries D94}
  (2016) 114017} [\href{https://arxiv.org/abs/1611.02397}{{\ttfamily
  1611.02397}}].

\bibitem{Ji:2016jgn}
X.~Ji, F.~Yuan and Y.~Zhao, \emph{{Hunting the Gluon Orbital Angular Momentum
  at the Electron-Ion Collider}}, {\emph{Phys. Rev. Lett.} {\bfseries 118}
  (2017) 192004} [\href{https://arxiv.org/abs/1612.02438}{{\ttfamily
  1612.02438}}].

\bibitem{Hagiwara:2017ofm}
Y.~Hagiwara, Y.~Hatta, B.-W. Xiao and F.~Yuan, \emph{{Elliptic Flow in Small
  Systems due to Elliptic Gluon Distributions?}}, {\emph{Phys. Lett.}
  {\bfseries B771} (2017) 374}
  [\href{https://arxiv.org/abs/1701.04254}{{\ttfamily 1701.04254}}].

\bibitem{Iancu:2017fzn}
E.~Iancu and A.~H. Rezaeian, \emph{{Elliptic flow from color-dipole orientation
  in pp and pA collisions}}, {\emph{Phys. Rev.} {\bfseries D95} (2017) 094003}
  [\href{https://arxiv.org/abs/1702.03943}{{\ttfamily 1702.03943}}].

\bibitem{Hagiwara:2017fye}
Y.~Hagiwara, Y.~Hatta, R.~Pasechnik, M.~Tasevsky and O.~Teryaev,
  \emph{{Accessing the gluon Wigner distribution in ultraperipheral $pA$
  collisions}}, {\emph{Phys. Rev.} {\bfseries D96} (2017) 034009}
  [\href{https://arxiv.org/abs/1706.01765}{{\ttfamily 1706.01765}}].

\bibitem{Boer:2018vdi}
D.~Boer, T.~Van~Daal, P.~J. Mulders and E.~Petreska, \emph{{Directed flow from
  C-odd gluon correlations at small $x$}}, {\emph{JHEP} {\bfseries 07} (2018)
  140} [\href{https://arxiv.org/abs/1805.05219}{{\ttfamily 1805.05219}}].

\bibitem{Salazar:2019ncp}
F.~Salazar and B.~Schenke, \emph{{Diffractive dijet production in impact
  parameter dependent saturation models}}, {\emph{Phys. Rev.} {\bfseries D100}
  (2019) 034007} [\href{https://arxiv.org/abs/1905.03763}{{\ttfamily
  1905.03763}}].

\bibitem{Hagiwara:2021xkf}
Y.~Hagiwara, C.~Zhang, J.~Zhou and Y.-j. Zhou, \emph{{Probing the gluon
  tomography in photoproduction of dipion}}, {\emph{Phys. Rev. D} {\bfseries
  104} (2021) 094021}.

\bibitem{Bhattacharya:2017bvs}
S.~Bhattacharya, A.~Metz and J.~Zhou, \emph{{Generalized TMDs and the exclusive
  double Drell--Yan process}}, {\emph{Phys. Lett.} {\bfseries B771} (2017) 396}
  [\href{https://arxiv.org/abs/1702.04387}{{\ttfamily 1702.04387}}].

\bibitem{Bhattacharya:2022vvo}
S.~Bhattacharya, R.~Boussarie and Y.~Hatta, \emph{{Signature of the Gluon
  Orbital Angular Momentum}},
  \href{https://doi.org/10.1103/PhysRevLett.128.182002}{\emph{Phys. Rev. Lett.}
  {\bfseries 128} (2022) 182002}
  [\href{https://arxiv.org/abs/2201.08709}{{\ttfamily 2201.08709}}].

\bibitem{Boer:2021upt}
D.~Boer and C.~Setyadi, \emph{{GTMD model predictions for diffractive dijet
  production at EIC}}, {\emph{Phys. Rev. D} {\bfseries 104} (2021) 074006}
  [\href{https://arxiv.org/abs/2106.15148}{{\ttfamily 2106.15148}}].

\bibitem{Efremov:1979qk}
A.~V. Efremov and A.~V. Radyushkin, \emph{{Factorization and Asymptotical
  Behavior of Pion Form-Factor in QCD}}, {\emph{Phys. Lett.} {\bfseries 94B}
  (1980) 245}.

\bibitem{Lepage:1979zb}
G.~P. Lepage and S.~J. Brodsky, \emph{{Exclusive Processes in Quantum
  Chromodynamics: Evolution Equations for Hadronic Wave Functions and the
  Form-Factors of Mesons}}, {\emph{Phys. Lett.} {\bfseries 87B} (1979) 359}.

\bibitem{Bhattacharya:2018lgm}
S.~Bhattacharya, A.~Metz, V.~K. Ojha, J.-Y. Tsai and J.~Zhou, \emph{{Exclusive
  double quarkonium production and generalized TMDs of gluons}},
  \href{https://doi.org/10.1016/j.physletb.2022.137383}{\emph{Phys. Lett. B}
  {\bfseries 833} (2022) 137383}
  [\href{https://arxiv.org/abs/1802.10550}{{\ttfamily 1802.10550}}].

\bibitem{Boussarie:2018zwg}
R.~Boussarie, Y.~Hatta, B.-W. Xiao and F.~Yuan, \emph{{Probing the
  Weizs{\"a}cker-Williams gluon Wigner distribution in $pp$ collisions}},
  {\emph{Phys. Rev.} {\bfseries D98} (2018) 074015}
  [\href{https://arxiv.org/abs/1807.08697}{{\ttfamily 1807.08697}}].

\bibitem{Taneja:2011sy}
S.~K. Taneja, K.~Kathuria, S.~Liuti and G.~R. Goldstein, \emph{{Angular
  momentum sum rule for spin one hadronic systems}}, {\emph{Phys. Rev. D}
  {\bfseries 86} (2012) 036008}
  [\href{https://arxiv.org/abs/1101.0581}{{\ttfamily 1101.0581}}].

\bibitem{Guo:2021aik}
Y.~Guo, X.~Ji and K.~Shiells, \emph{{Novel twist-three transverse-spin sum rule
  for the proton and related generalized parton distributions}}, {\emph{Nucl.
  Phys. B} {\bfseries 969} (2021) 115440}
  [\href{https://arxiv.org/abs/2101.05243}{{\ttfamily 2101.05243}}].

\bibitem{Rajan3}
O.~Alkasassbeh, A.~Rajan, M.~Engelhardt and S.~Liuti, \emph{{Transverse Spin
  Sum Rule, to be submitted}},  2022.

\bibitem{Wakamatsu:2000fd}
M.~Wakamatsu, \emph{{Chiral odd distribution functions in the chiral quark
  soliton model}}, {\emph{Phys. Lett.} {\bfseries B509} (2001) 59}
  [\href{https://arxiv.org/abs/hep-ph/0012331}{{\ttfamily hep-ph/0012331}}].

\bibitem{Bashinsky:1998if}
S.~Bashinsky and R.~L. Jaffe, \emph{{Quark and gluon orbital angular momentum
  and spin in hard processes}}, {\emph{Nucl. Phys. B} {\bfseries 536} (1998)
  303} [\href{https://arxiv.org/abs/hep-ph/9804397}{{\ttfamily
  hep-ph/9804397}}].

\bibitem{Ji:2012gc}
X.~Ji, Y.~Xu and Y.~Zhao, \emph{{Gluon Spin, Canonical Momentum, and Gauge
  Symmetry}}, {\emph{JHEP} {\bfseries 08} (2012) 082}
  [\href{https://arxiv.org/abs/1205.0156}{{\ttfamily 1205.0156}}].

\bibitem{Wakamatsu:2011mb}
M.~Wakamatsu, \emph{{Gauge-independence of gluon spin in the nucleon and its
  evolution}}, {\emph{Phys. Rev. D} {\bfseries 84} (2011) 037501}
  [\href{https://arxiv.org/abs/1104.1465}{{\ttfamily 1104.1465}}].

\bibitem{Lorce:2011ni}
C.~Lorc\'e, B.~Pasquini, X.~Xiong and F.~Yuan, \emph{{The quark orbital angular
  momentum from Wigner distributions and light-cone wave functions}},
  {\emph{Phys. Rev. D} {\bfseries 85} (2012) 114006}
  [\href{https://arxiv.org/abs/1111.4827}{{\ttfamily 1111.4827}}].

\bibitem{Hatta:2011ku}
Y.~Hatta, \emph{{Notes on the orbital angular momentum of quarks in the
  nucleon}}, {\emph{Phys.Lett.} {\bfseries B708} (2012) 186}
  [\href{https://arxiv.org/abs/1111.3547}{{\ttfamily 1111.3547}}].

\bibitem{Burkardt:2012sd}
M.~Burkardt, \emph{{Parton Orbital Angular Momentum and Final State
  Interactions}}, {\emph{Phys.Rev.} {\bfseries D88} (2013) 014014}
  [\href{https://arxiv.org/abs/1205.2916}{{\ttfamily 1205.2916}}].

\bibitem{Ji:2012sj}
X.~Ji, X.~Xiong and F.~Yuan, \emph{{Proton Spin Structure from Measurable
  Parton Distributions}}, {\emph{Phys. Rev. Lett.} {\bfseries 109} (2012)
  152005} [\href{https://arxiv.org/abs/1202.2843}{{\ttfamily 1202.2843}}].

\bibitem{Raja:2017xlo}
A.~Rajan, M.~Engelhardt and S.~Liuti, \emph{{Lorentz Invariance and QCD
  Equation of Motion Relations for Generalized Parton Distributions and the
  Dynamical Origin of Proton Orbital Angular Momentum}}, {\emph{Phys. Rev.}
  {\bfseries D98} (2018) 074022}
  [\href{https://arxiv.org/abs/1709.05770}{{\ttfamily 1709.05770}}].

\bibitem{Rajan:2016tlg}
A.~Rajan, A.~Courtoy, M.~Engelhardt and S.~Liuti, \emph{{Parton Transverse
  Momentum and Orbital Angular Momentum Distributions}}, {\emph{Phys. Rev.}
  {\bfseries D94} (2016) 034041}
  [\href{https://arxiv.org/abs/1601.06117}{{\ttfamily 1601.06117}}].

\bibitem{Kiptily:2002nx}
D.~Kiptily and M.~Polyakov, \emph{{Genuine twist three contributions to the
  generalized parton distributions from instantons}}, {\emph{Eur. Phys. J.}
  {\bfseries C37} (2004) 105}
  [\href{https://arxiv.org/abs/hep-ph/0212372}{{\ttfamily hep-ph/0212372}}].

\bibitem{Hatta:2012cs}
Y.~Hatta and S.~Yoshida, \emph{{Twist analysis of the nucleon spin in QCD}},
  {\emph{JHEP} {\bfseries 1210} (2012) 080}
  [\href{https://arxiv.org/abs/1207.5332}{{\ttfamily 1207.5332}}].

\bibitem{Balitsky:1987bk}
I.~I. Balitsky and V.~M. Braun, \emph{{Evolution Equations for QCD String
  Operators}}, {\emph{Nucl. Phys. B} {\bfseries 311} (1989) 541}.

\bibitem{Kivel:2000rb}
N.~Kivel, M.~V. Polyakov, A.~Sch\"afer and O.~V. Teryaev, \emph{{On the
  Wandzura-Wilczek approximation for the twist - three DVCS amplitude}},
  {\emph{Phys. Lett. B} {\bfseries 497} (2001) 73}
  [\href{https://arxiv.org/abs/hep-ph/0007315}{{\ttfamily hep-ph/0007315}}].

\bibitem{Lorce:2014mxa}
C.~Lorc\'{e}, \emph{{Spin--orbit correlations in the nucleon}}, {\emph{Phys.
  Lett.} {\bfseries B735} (2014) 344}
  [\href{https://arxiv.org/abs/1401.7784}{{\ttfamily 1401.7784}}].

\bibitem{Ji:2012vj}
X.~Ji, X.~Xiong and F.~Yuan, \emph{{Transverse Polarization of the Nucleon in
  Parton Picture}}, {\emph{Phys. Lett. B} {\bfseries 717} (2012) 214}
  [\href{https://arxiv.org/abs/1209.3246}{{\ttfamily 1209.3246}}].

\bibitem{Bhoonah:2017olu}
A.~Bhoonah and C.~Lorc\'e, \emph{{Quark transverse spin\textendash{}orbit
  correlations}}, {\emph{Phys. Lett. B} {\bfseries 774} (2017) 435}
  [\href{https://arxiv.org/abs/1703.08322}{{\ttfamily 1703.08322}}].

\bibitem{Hagiwara:2016kam}
Y.~Hagiwara, Y.~Hatta and T.~Ueda, \emph{{Wigner, Husimi, and generalized
  transverse momentum dependent distributions in the color glass condensate}},
  {\emph{Phys. Rev. D} {\bfseries 94} (2016) 094036}
  [\href{https://arxiv.org/abs/1609.05773}{{\ttfamily 1609.05773}}].

\bibitem{Liuti:2017uxp}
S.~Liuti, \emph{{DVCS and TCS in New Helicity Amplitudes Formalism}},  in
  \emph{{Mini-Proceedings of the Workshop on High-Intensity Photon Sources
  (HIPS2017), Catholic University of America, Washington, D.C., USA, February
  6-7, 2017}}, pp.~48--50, \href{https://arxiv.org/abs/1704.00816}{{\ttfamily
  1704.00816}}.

\bibitem{deDivitiis:2012vs}
G.~M. de~Divitiis, R.~Petronzio and N.~Tantalo, \emph{{On the extraction of
  zero momentum form factors on the lattice}}, {\emph{Phys. Lett.} {\bfseries
  B718} (2012) 589} [\href{https://arxiv.org/abs/1208.5914}{{\ttfamily
  1208.5914}}].

\bibitem{Engelhardt:LATTICE2021}
M.~Engelhardt, J.~Green, N.~Hasan, T.~Izubuchi, C.~Kallidonis, S.~Krieg et~al.,
  \emph{{Quark spin-orbit correlations in the proton}}, {\emph{PoS} {\bfseries
  LATTICE2021} (2021) 413}.

\bibitem{Kanazawa:2014nha}
K.~Kanazawa, C.~Lorc\'{e}, A.~Metz, B.~Pasquini and M.~Schlegel, \emph{{Twist-2
  generalized transverse-momentum dependent parton distributions and the
  spin/orbital structure of the nucleon}}, {\emph{Phys.Rev.} {\bfseries D90}
  (2014) 014028} [\href{https://arxiv.org/abs/1403.5226}{{\ttfamily
  1403.5226}}].

\bibitem{Liu:2014vwa}
T.~Liu, \emph{{Quark orbital motions from Wigner distributions}},
  \href{https://arxiv.org/abs/1406.7709}{{\ttfamily 1406.7709}}.

\bibitem{Liu:2015eqa}
T.~Liu and B.-Q. Ma, \emph{{Quark Wigner distributions in a light-cone
  spectator model}}, {\emph{Phys. Rev. D} {\bfseries 91} (2015) 034019}
  [\href{https://arxiv.org/abs/1501.07690}{{\ttfamily 1501.07690}}].

\bibitem{Kaur:2018dns}
S.~Kaur and H.~Dahiya, \emph{{Wigner distributions and GTMDs in a proton using
  light-front quark\textendash{}diquark model}}, {\emph{Nucl. Phys. B}
  {\bfseries 937} (2018) 272}
  [\href{https://arxiv.org/abs/1810.09099}{{\ttfamily 1810.09099}}].

\bibitem{Maji:2022tog}
T.~Maji, C.~Mondal and D.~Kang, \emph{{Leading twist GTMDs at nonzero skewness
  and Wigner distributions in boost-invariant longitudinal position space}},
  \href{https://doi.org/10.1103/PhysRevD.105.074024}{\emph{Phys. Rev. D}
  {\bfseries 105} (2022) 074024}
  [\href{https://arxiv.org/abs/2202.08635}{{\ttfamily 2202.08635}}].

\bibitem{Gutsche:2016gcd}
T.~Gutsche, V.~E. Lyubovitskij and I.~Schmidt, \emph{{Nucleon parton
  distributions in a light-front quark model}}, {\emph{Eur. Phys. J. C}
  {\bfseries 77} (2017) 86} [\href{https://arxiv.org/abs/1610.03526}{{\ttfamily
  1610.03526}}].

\bibitem{Mukherjee:2014nya}
A.~Mukherjee, S.~Nair and V.~K. Ojha, \emph{{Quark Wigner Distributions and
  Orbital Angular Momentum in Light-front Dressed Quark Model}}, {\emph{Phys.
  Rev. D} {\bfseries 90} (2014) 014024}
  [\href{https://arxiv.org/abs/1403.6233}{{\ttfamily 1403.6233}}].

\bibitem{Miller:2014vla}
G.~A. Miller, \emph{{Electron structure: Shape, size, and generalized parton
  distributions in QED}}, {\emph{Phys. Rev. D} {\bfseries 90} (2014) 113001}
  [\href{https://arxiv.org/abs/1409.7412}{{\ttfamily 1409.7412}}].

\bibitem{Hagiwara:2014iya}
Y.~Hagiwara and Y.~Hatta, \emph{{Use of the Husimi distribution for nucleon
  tomography}}, {\emph{Nucl. Phys.} {\bfseries A940} (2015) 158}
  [\href{https://arxiv.org/abs/1412.4591}{{\ttfamily 1412.4591}}].

\bibitem{Mukherjee:2015aja}
A.~Mukherjee, S.~Nair and V.~K. Ojha, \emph{{Wigner distributions for gluons in
  a light-front dressed quark model}}, {\emph{Phys. Rev. D} {\bfseries 91}
  (2015) 054018} [\href{https://arxiv.org/abs/1501.03728}{{\ttfamily
  1501.03728}}].

\bibitem{More:2017zqq}
J.~More, A.~Mukherjee and S.~Nair, \emph{{Quark Wigner Distributions Using
  Light-Front Wave Functions}}, {\emph{Phys. Rev. D} {\bfseries 95} (2017)
  074039} [\href{https://arxiv.org/abs/1701.00339}{{\ttfamily 1701.00339}}].

\bibitem{Kumar:2017xcm}
N.~Kumar and C.~Mondal, \emph{{Wigner distributions for an electron}},
  {\emph{Nucl. Phys. B} {\bfseries 931} (2018) 226}
  [\href{https://arxiv.org/abs/1705.03183}{{\ttfamily 1705.03183}}].

\bibitem{More:2017zqp}
J.~More, A.~Mukherjee and S.~Nair, \emph{{Wigner Distributions For Gluons}},
  {\emph{Eur. Phys. J. C} {\bfseries 78} (2018) 389}
  [\href{https://arxiv.org/abs/1709.00943}{{\ttfamily 1709.00943}}].

\bibitem{Courtoy:2016des}
A.~Courtoy and A.~S. Miramontes, \emph{{Quark Orbital Angular Momentum in the
  MIT Bag Model}}, {\emph{Phys. Rev. D} {\bfseries 95} (2017) 014027}
  [\href{https://arxiv.org/abs/1611.03375}{{\ttfamily 1611.03375}}].

\bibitem{Chakrabarti:2016yuw}
D.~Chakrabarti, T.~Maji, C.~Mondal and A.~Mukherjee, \emph{{Wigner
  distributions and orbital angular momentum of a proton}}, {\emph{Eur. Phys.
  J. C} {\bfseries 76} (2016) 409}
  [\href{https://arxiv.org/abs/1601.03217}{{\ttfamily 1601.03217}}].

\bibitem{Chakrabarti:2017teq}
D.~Chakrabarti, T.~Maji, C.~Mondal and A.~Mukherjee, \emph{{Quark Wigner
  distributions and spin-spin correlations}}, {\emph{Phys. Rev. D} {\bfseries
  95} (2017) 074028} [\href{https://arxiv.org/abs/1701.08551}{{\ttfamily
  1701.08551}}].

\bibitem{Chakrabarti:2019wjx}
D.~Chakrabarti, N.~Kumar, T.~Maji and A.~Mukherjee, \emph{{Sivers and
  Boer\textendash{}Mulders GTMDs in light-front holographic
  quark\textendash{}diquark model}}, {\emph{Eur. Phys. J. Plus} {\bfseries 135}
  (2020) 496} [\href{https://arxiv.org/abs/1902.07051}{{\ttfamily
  1902.07051}}].

\bibitem{Ma:2018ysi}
Z.-L. Ma and Z.~Lu, \emph{{Quark Wigner distribution of the pion meson in
  light-cone quark model}}, {\emph{Phys. Rev. D} {\bfseries 98} (2018) 054024}
  [\href{https://arxiv.org/abs/1808.00140}{{\ttfamily 1808.00140}}].

\bibitem{Kaur:2019kpi}
N.~Kaur and H.~Dahiya, \emph{{Quark Wigner Distributions and GTMDs of Pion in
  the Light-Front Holographic Model}}, {\emph{Eur. Phys. J. A} {\bfseries 56}
  (2020) 172} [\href{https://arxiv.org/abs/1909.10146}{{\ttfamily
  1909.10146}}].

\bibitem{Luo:2020yqj}
X.~Luo and H.~Sun, \emph{{T-odd generalized and quasi transverse momentum
  dependent parton distribution in a scalar spectator model}}, {\emph{Eur.
  Phys. J. C} {\bfseries 80} (2020) 828}
  [\href{https://arxiv.org/abs/2005.09832}{{\ttfamily 2005.09832}}].

\bibitem{Tan:2021osk}
C.~Tan and Z.~Lu, \emph{{Quark spin-orbit correlations in the pion in a
  light-cone quark model}},
  \href{https://doi.org/10.1103/PhysRevD.105.034004}{\emph{Phys. Rev. D}
  {\bfseries 105} (2022) 034004}
  [\href{https://arxiv.org/abs/2110.08502}{{\ttfamily 2110.08502}}].

\bibitem{Kovner:2017vro}
A.~Kovner and A.~H. Rezaeian, \emph{{Double parton scattering in the color
  glass condensate: Hanbury-Brown-Twiss correlations in double inclusive photon
  production}}, {\emph{Phys. Rev. D} {\bfseries 95} (2017) 114028}
  [\href{https://arxiv.org/abs/1701.00494}{{\ttfamily 1701.00494}}].

\bibitem{Kovner:2017ssr}
A.~Kovner and A.~H. Rezaeian, \emph{{Double parton scattering in the CGC:
  Double quark production and effects of quantum statistics}}, {\emph{Phys.
  Rev. D} {\bfseries 96} (2017) 074018}
  [\href{https://arxiv.org/abs/1707.06985}{{\ttfamily 1707.06985}}].

\bibitem{ReinkePelicer:2018gyh}
M.~Reinke~Pelicer, E.~Gr\"ave De~Oliveira and R.~Pasechnik, \emph{{Exclusive
  heavy quark-pair production in ultraperipheral collisions}}, {\emph{Phys.
  Rev. D} {\bfseries 99} (2019) 034016}
  [\href{https://arxiv.org/abs/1811.12888}{{\ttfamily 1811.12888}}].

\bibitem{Hagiwara:2020mqb}
Y.~Hagiwara, Y.~Hatta, R.~Pasechnik and J.~Zhou, \emph{{Spin-dependent Pomeron
  and Odderon in elastic proton-proton scattering}}, {\emph{Eur. Phys. J. C}
  {\bfseries 80} (2020) 427}
  [\href{https://arxiv.org/abs/2003.03680}{{\ttfamily 2003.03680}}].

\bibitem{Han:2022tlh}
Y.~Han, T.~Liu and B.-Q. Ma, \emph{{Six-dimensional light-front Wigner
  distribution of hadrons}},
  \href{https://doi.org/10.1016/j.physletb.2022.137127}{\emph{Phys. Lett. B}
  {\bfseries 830} (2022) 137127}
  [\href{https://arxiv.org/abs/2202.10359}{{\ttfamily 2202.10359}}].

\bibitem{Husimi:1940}
K.~Husimi, \emph{{Some formal aspects of the density matrix}}, {\emph{Proc.
  Phys. Math. Soc. Jpn.} {\bfseries 22} (1940) 264}.

\bibitem{Moch:2005id}
S.~Moch, J.~A.~M. Vermaseren and A.~Vogt, \emph{{The Quark form-factor at
  higher orders}}, {\emph{JHEP} {\bfseries 08} (2005) 049}
  [\href{https://arxiv.org/abs/hep-ph/0507039}{{\ttfamily hep-ph/0507039}}].

\bibitem{Grozin:2014hna}
A.~Grozin, J.~M. Henn, G.~P. Korchemsky and P.~Marquard, \emph{{Three Loop Cusp
  Anomalous Dimension in QCD}}, {\emph{Phys. Rev. Lett.} {\bfseries 114} (2015)
  062006} [\href{https://arxiv.org/abs/1409.0023}{{\ttfamily 1409.0023}}].

\bibitem{Bell:2018gce}
G.~Bell, A.~Hornig, C.~Lee and J.~Talbert, \emph{{$e^+ e^-$ angularity
  distributions at NNLL$^\prime$ accuracy}}, {\emph{JHEP} {\bfseries 01} (2019)
  147} [\href{https://arxiv.org/abs/1808.07867}{{\ttfamily 1808.07867}}].

\bibitem{Billis:2019evv}
G.~Billis, F.~J. Tackmann and J.~Talbert, \emph{{Higher-Order Sudakov
  Resummation in Coupled Gauge Theories}}, {\emph{JHEP} {\bfseries 03} (2020)
  182} [\href{https://arxiv.org/abs/1907.02971}{{\ttfamily 1907.02971}}].

\bibitem{Ebert:2021aoo}
M.~A. Ebert, \emph{{Analytic results for Sudakov form factors in QCD}},
  \href{https://doi.org/10.1007/JHEP02(2022)136}{\emph{JHEP} {\bfseries 02}
  (2022) 136} [\href{https://arxiv.org/abs/2110.11360}{{\ttfamily
  2110.11360}}].

\bibitem{Herzog:2018kwj}
F.~Herzog, S.~Moch, B.~Ruijl, T.~Ueda, J.~A.~M. Vermaseren and A.~Vogt,
  \emph{{Five-loop contributions to low-N non-singlet anomalous dimensions in
  QCD}}, \href{https://doi.org/10.1016/j.physletb.2019.01.060}{\emph{Phys.
  Lett. B} {\bfseries 790} (2019) 436}
  [\href{https://arxiv.org/abs/1812.11818}{{\ttfamily 1812.11818}}].

\bibitem{Duhr:2022yyp}
C.~Duhr, B.~Mistlberger and G.~Vita, \emph{{The Four-Loop Rapidity Anomalous
  Dimension and Event Shapes to Fourth Logarithmic Order}},
  \href{https://arxiv.org/abs/2205.02242}{{\ttfamily 2205.02242}}.

\bibitem{Tarasov:1980au}
O.~V. Tarasov, A.~A. Vladimirov and A.~{\relax Yu}. Zharkov, \emph{{The
  Gell-Mann-Low Function of QCD in the Three Loop Approximation}}, {\emph{Phys.
  Lett.} {\bfseries B93} (1980) 429}.

\bibitem{Larin:1993tp}
S.~A. Larin and J.~A.~M. Vermaseren, \emph{{The Three loop QCD Beta function
  and anomalous dimensions}}, {\emph{Phys. Lett.} {\bfseries B303} (1993) 334}
  [\href{https://arxiv.org/abs/hep-ph/9302208}{{\ttfamily hep-ph/9302208}}].

\bibitem{vanRitbergen:1997va}
T.~van Ritbergen, J.~A.~M. Vermaseren and S.~A. Larin, \emph{{The Four loop
  beta function in quantum chromodynamics}}, {\emph{Phys. Lett.} {\bfseries
  B400} (1997) 379} [\href{https://arxiv.org/abs/hep-ph/9701390}{{\ttfamily
  hep-ph/9701390}}].

\bibitem{Czakon:2004bu}
M.~Czakon, \emph{{The Four-loop QCD beta-function and anomalous dimensions}},
  {\emph{Nucl. Phys.} {\bfseries B710} (2005) 485}
  [\href{https://arxiv.org/abs/hep-ph/0411261}{{\ttfamily hep-ph/0411261}}].

\bibitem{Baikov:2016tgj}
P.~A. Baikov, K.~G. Chetyrkin and J.~H. K\"uhn, \emph{{Five-Loop Running of the
  QCD coupling constant}},
  \href{https://doi.org/10.1103/PhysRevLett.118.082002}{\emph{Phys. Rev. Lett.}
  {\bfseries 118} (2017) 082002}
  [\href{https://arxiv.org/abs/1606.08659}{{\ttfamily 1606.08659}}].

\end{thebibliography}\endgroup
